# Hole-doping induced ferromagnetism in 2D materials

R. Meng[1*], L.M.C. Pereira[1], J.P. Locquet[1], V.V. Afanas'ev[1], G. Pourtois[2], and M. Houssa[1,2*]

[1]Department of Physics and Astronomy, KU Leuven, Celestijnenlaan 200D, Leuven B-3001, Belgium

[2]imec, Kapeldreef 75, B-3001 Leuven, Belgium

[*]Email: ruishen.meng@kuleuven.be; michel.houssa@kuleuven.be

## Abstract

Two-dimensional (2D) ferromagnetic materials are considered as promising candidates for the future generations of spintronic devices. Yet, 2D materials with intrinsic ferromagnetism are scarce. High-throughput first-principles simulations are performed in order to screen 2D materials that present a non-magnetic to a ferromagnetic transition upon hole doping. A global evolutionary search is subsequently performed, in order to identify alternative possible atomic structures of the eligible candidates, and 122 materials exhibiting a hole-doping induced ferromagnetism are identified. Their energetic and dynamic stability, as well as their magnetic properties under hole doping are investigated systematically. Half of these 2D materials are metal halides, followed by chalcogenides, oxides and nitrides, some of them having predicted Curie temperatures above 300 K. The exchange interactions responsible for the ferromagnetic order in these 2D materials are also discussed. This work not only provides theoretical insights into hole-doped 2D ferromagnetic materials, but also enriches the family of 2D magnetic materials for possible spintronic applications.

## Introduction

Two-dimensional magnetic materials are serving as a powerful platform for the investigation and understanding of magnetism at the ultimate 2D limit.[1] Owing to the atomic thickness and controllable electron-spin degree of freedom, they are also considered to be promising candidates for the next-generation spintronic devices. Especially, 2D ferromagnetic materials with semiconducting or half-metallic properties (metallic on one spin channel and semiconducting on the other spin channel) draw particular interest.[2] By means of theoretical calculations, a variety of 2D ferromagnetic materials have been discovered.[3] Some of these 2D materials have been successfully synthesized and their fascinating magnetic properties have been explored subsequently.[4] For instance, 2D $CrI_3$ and $Cr_2Ge_2Te_6$ were predicted to be intrinsic ferromagnetic semiconductors, a prediction which was later confirmed experimentally.[5] Further investigation shows that 2D $CrI_3$ used as a spin-filter tunnel barrier sandwiched between graphene contacts can have ultra-high tunneling magnetoresistance.[6] However, the scarcity of 2D ferromagnetic semiconductors/half-metals and their rather low Curie temperature ($T_c$) may hamper practical applications as well as the further investigation of 2D magnetism. Continuous seeking for more 2D ferromagnetic semiconducting or half-metallic materials with relatively high Curie temperature is necessary.

In addition to these intrinsic 2D magnetic materials, there are also studies which focus on inducing magnetism into non-magnetic 2D materials. Quite a few theoretical reports have shown that 2D III-VI (gallium oxides/chalcogenides and indium oxides/chalcogenides)[7] and IV-VI (tin oxides/chalcogenides and lead oxides)[8] semiconductors, and some other 2D materials such as $InP_3$, etc.[9] would exhibit non-magnetic to ferromagnetic properties upon hole doping. Ferromagnetism in these hole-doped 2D materials arises from an exchange splitting of electronic states at the top of the valence band, where the density of states (DOS) exhibits a sharp Van Hove singularity, that would lead to an electronic instability. Controlling the magnetic state of these materials by e.g., a gate bias has potential interesting applications for novel spintronic devices.

Unlike most of intrinsic magnetic materials, whose magnetic moments are mainly originated from unsaturated *d* orbitals of transition metals, the magnetic moments of these hole-doped 2D ferromagnetic materials are mainly contributed by the delocalized *p* orbitals of anions, which could be advantageous for the long-range ferromagnetic order. However, a systematic investigation of the magnetic properties of these 2D materials (spin-polarization energies, magnetic exchange coupling strength, magnetic anisotropy, and Curie temperature) is lacking. Besides, the mechanism responsible for the exchange coupling and delocalization of the spin-polarized holes has not been discussed, to our knowledge. In addition, it is also of significance to find out whether there are other 2D materials which become ferromagnetic upon hole doping.

In this study, we screen thousands of 2D non-magnetic semiconductors/insulators from three databases,[3b, 3c, 10] followed by a high-throughput density functional theory calculation to identify potential 2D ferromagnetic materials induced by hole doping. We then verify the stability of the potential candidates with respect to competing atomic structures, obtained by global structure searching based on evolutionary algorithm[11], via exploring energy convex hulls and phonon spectrum. Afterwards, the magnetic behaviors of the stable candidates at different doping densities are investigated systematically. Eventually, 122 materials are recognized as stable 2D ferromagnetic materials upon hole doping, some of them having a computed Curie temperature near or above room temperature. The exchange interaction mechanisms responsible for the ferromagnetic coupling in these 2D materials are also discussed.

# RESULTS AND DISCUSSION

The general workflow of this research is illustrated by a funnel plot on the left panel of Figure 1, which is categorized into five processes; the number of materials used for screening and the screening criteria in each process is also given. As a starting point, to include as many 2D materials as possible, we gather about 8000 2D crystal structures from three databases, 2DMatPedia,[10] computational 2D materials database (C2BD)[3b] and Materials Cloud two-dimensional crystals database (MC2D)[3c] that contains exfoliable 2D materials. In the prescreening process, the magnetic as well as the metallic materials are excluded since we are only interested in the non-magnetic semiconductors. Besides, the repeated structures and the structures with low thermodynamic stability ($E_c$ < 0.2 eV/atom, $E_c$ is the cohesive energy defined by the total energy of the compound minus the total energies of its constituent atoms) are also screened out. As a result, about 3000 non-magnetic 2D semiconducting materials, with moderate stability, are picked out for the subsequent hole doping simulations.

The typical doping densities, ranging from $5\times10^{12}$ cm$^{-2}$ to $8\times10^{14}$ cm$^{-2}$ are used in the hole doping simulations, and structural relaxations are carried out for each doping density. The spin-polarization energy is obtained by $E_{spin} = E_{NM} - E_{FM}$, where $E_{FM}$ and $E_{NM}$ are the energy of the ferromagnetic and non-magnetic states, respectively at PBE level. In this case, a positive $E_{spin}$ suggests that the ferromagnetic state is more energetically stable. Typically, for the GaS, GaSe and InSe monolayer, the maximum $E_{spin}$ is about 10 meV/hole,[7a, 7b, 7c] whereas other materials, like GaO, SnO, etc.[7d, 8b] have relatively larger $E_{spin}$. Herein, we use this 10 meV/hole as a standard value, and materials with $E_{spin}$ smaller than 10 mev/hole are discarded. The induced magnetic moment is also compared with the number of holes used in the simulation. If the ratio of magnetic moment to hole number is ~1 (1 $\mu_B$/hole), it indicates that the injected holes can be fully spin polarized. Materials with induced magnetic moment much smaller than 1 $\mu_B$/hole are excluded. Finally, materials exhibiting ferromagnetism only at a hole density below $10^{13}$ cm$^{-2}$ or only after a high hole density of $6 \times 10^{14}$ cm$^{-2}$ are also discarded.

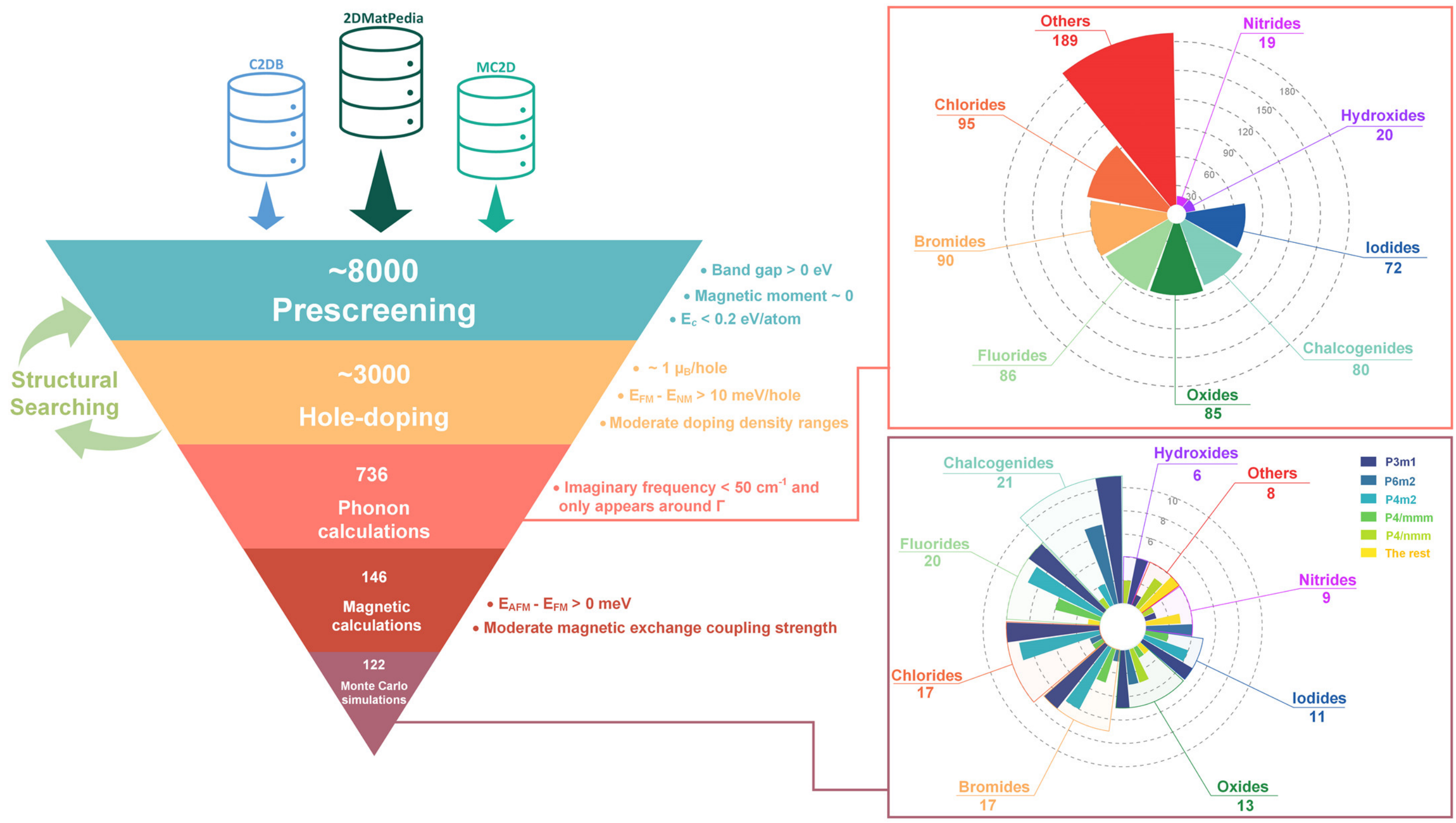

Figure 1. General prescreening and high-throughput screening strategies applied to 2D structures from 2DMatpedia Database, Computational 2D Materials Database (C2DB) and Materials Cloud two-dimensional crystals database (MC2D) to identify potential 2D ferromagnetic materials upon hole doping (left panel). The number of candidate materials and the screening

criteria used for each screening step are provided. Polar histogram showing the number of structures belonging to different chemical compositions after hole doping screening (top right panel); Polar histogram displaying the number of final stable 2D ferromagnetic materials in terms of chemical compositions and their crystal space groups (bottom right panel).

A globally evolutionary structure searching is subsequently carried out to explore atomic structures with different stoichiometries for a given material that shows promising ferromagnetic properties upon hole doping. In this step, the convex hull of a specific chemical composition is provided with the apex determined by the most stable materials in energy. If a new structure/phase on the convex hull is found, it is sent back to the hole doping simulations. Subsequently, the number of candidate materials is narrowed down to 736. According to their chemical formula, these candidate materials are classified into different prototypes, as shown by the polar histogram on the top right panel of Figure 1. The most prevalent chemical composition corresponds to metal halides (~47%), followed by other materials such as ternary compounds, phosphorides, carbides, etc. (~26%, grouped as "Others"). Oxides (~12%) and chalcogenides (~11%) also account for a part of the candidate materials. There is also a fraction of hydroxides (~3%) and nitrides (~3%). Phonons calculations on these candidates are next performed, to verify their dynamic stability. In this procedure, materials with imaginary frequency only appears around the Γ point in the first Brillouin zone and less than 50 $cm^{-1}$ are regarded as dynamically stable. Afterwards, 146 candidates are selected for further screening.

The next step consists in performing detailed calculations of the magnetic properties of these selected 2D materials, i.e., magnetic exchange coupling strength, magnetic anisotropy energy (MAE), and Curie temperature. Generally, the magnetic moments of 2DHDFM are well localized on the *p*-orbitals of the anion sites, which is also confirmed by the HSE06 calculations in Figure S1 that the spin density localized almost solely on the anion site. Therefore, their FM configuration as well as a few antiferromagnetic (AFM) configurations can be constructed and their energies obtained from first-principles calculations can be mapped to the spin Hamiltonian:

$$H = -\sum_{i \neq j} J_{ij} \mathrm{S}_i \cdot \mathrm{S}_j - A \sum_i (\mathrm{S}_i^z)^2$$

where $J_{ij}$ is the exchange interaction between $i$ and $j$ atomic sites, $\mathrm{S}_i$ is a unit vector denoting the local spin direction of atom *i*, and $S_j$ is the unit vector of the spin moment direction of atom *j*. For ferromagnetic materials where neighboring spins align in parallel, $J_{ij} > 0$, while for antiferromagnetic materials where the spins prefer to align anti-parallel, $J_{ij} < 0$. $A$ is the MAE per magnetic ion, where positive values correspond to a preferred alignment along the z-axis, while negative values correspond to a preferred alignment along the $x-y$ plane. The magnetic exchange interaction strength can then be obtained by comparing the total energies of different magnetic configurations.[12] Note that materials with negative magnetic exchange couplings, i.e., whose antiferromagnetic state is more energetically stable than the ferromagnetic one, or materials with weak FM coupling ($J_{ij} < 1$ meV) are not considered further.

Through the screening process described above, 122 candidates are identified as stable 2D hole-doped ferromagnetic materials upon hole doping, which are denoted as *2DHDFM* hereafter. These materials become half-metals after hole doping, with a metallic DOS at the Fermi level ($E_F$) for the spin-down channels and a simultaneous band gap for the spin-up channels. The magnetic moment of *2DHDFM* is primarily originating from the *p*-orbitals of the anions. In the following, we focus on the analysis of their structural and magnetic properties. More detailed information, including their formation energies in the convex hull, atomic structures and atomic coordinates, phonon dispersion spectra, band structures and electronic density of states, magnetic moment and spin polarization energy as a function of hole density, magnetic configurations and the corresponding spin Hamiltonian, the magnetic exchange coupling strength, MAE and temperature-dependent normalized magnetization curves at specific hole density are provided in the supplementary materials.

The classification of *2DHDFM* based on their chemical composition and space group is shown by the polar histogram on the bottom right panel of Figure 1. Clearly, 65 out of 122 candidates belong to metal halides, including 20 fluorides, 17 chlorides, 17 bromides and 11 iodides. The rest covers 21 chalcogenides, 13 oxides, 9 nitrides, 6 hydroxides and 8 other materials. Among those materials, P3m1 is the primary space group, followed by the P4m2 and P6m2. The representative atomic structures of these materials in different space groups are given in Figure 2.

The metal element in the 2D dihalides with P3m1 and P4m2 space groups are mainly from the group IIA Be, Mg, Ca, Sr and Ba, and group IIB Zn Cd and Hg. The atomic structure of the P3m1-dihalides is commonly known as the 1T structure, with one halogen atom bonded to three neighboring metal atoms. The ones with P4m2 space group have a tetragonal lattice with one halogen atom connecting to only two neighboring metal atoms. According to their convex hulls, P3m1-dihalides are usually more stable than the P4m2-ones, except for $ZnBr_2$. $PbCl_2$ and $PbBr_2$ are the only two halides with P6m2 space group, and their structure is commonly known as the 2H hexagonal structure. Nevertheless, they are less stable than the P3m1 counterparts. In addition, there are two different halide structures with tetragonal P4/mmm space group. They contain metal elements from group IA and group IVA, respectively. The former form metal monohalides, such as LiCl and NaBr, with both halogen and metal atoms being fourfold coordinated. The latter consist in tetrahalides, $XF_4$ (X=Si, Ge, Sn, Pb), with an atomic structure similar to the P4m2 dihalides, the two extra fluorine atoms forming bonds to one X atom along the out-of-plane direction in the unit-cell.

The majority of chalcogenides belongs to the P3m1 and P6m2 space groups, while a few of them belong to the P4m2 and P4/mmm space groups. The chalcogenides mainly consist of elements from the group IIIA Al, Ga, In, Tl, group IVA Si, Sn, Pb and group IIB Zn and Cd. The monochalcogenides, MX (M=Al, Ga, In, Tl; X=S, Se), which contain two metal atoms and two chalcogenide atoms in the unit cell, have both P3m1 and P6m2 space groups. Each X atom is bonded to three neighboring M atoms, and every M atom is fourfold coordinated, forming bonds with three adjacent X atoms and one M atom, with a characteristic X-M-M-X vertical stacking. The only difference between the two space groups is that the X atoms in the upper and bottom layer overlap with each other from the top view of the atomic plane for the P6m2 structures, while they have stagger position in the P3m1 structures. For a given material, the total energy difference between these two space groups is very small (typically less than 50 meV/unit cell). Note that these monochalcogenides gradually become non-magnetic when the hole doping density is typically larger than $2 \times 10^{14}$ cm$^{-2}$. There are two other P3m1 atomic structures for the chalcogenides. The first one corresponds to the dichalcogenides, including $SnS_2$, $PbS_2$ and $PdS_2$, having the 1T hexagonal structure. The second one corresponds to the monochalcogenides SiS, SnS, PbS, PbSe and ZnSe, holding the buckled hexagonal structure. The space group of CdS and ZnS, however, belongs to P6m2, because of the planar hexagonal atomic plane.

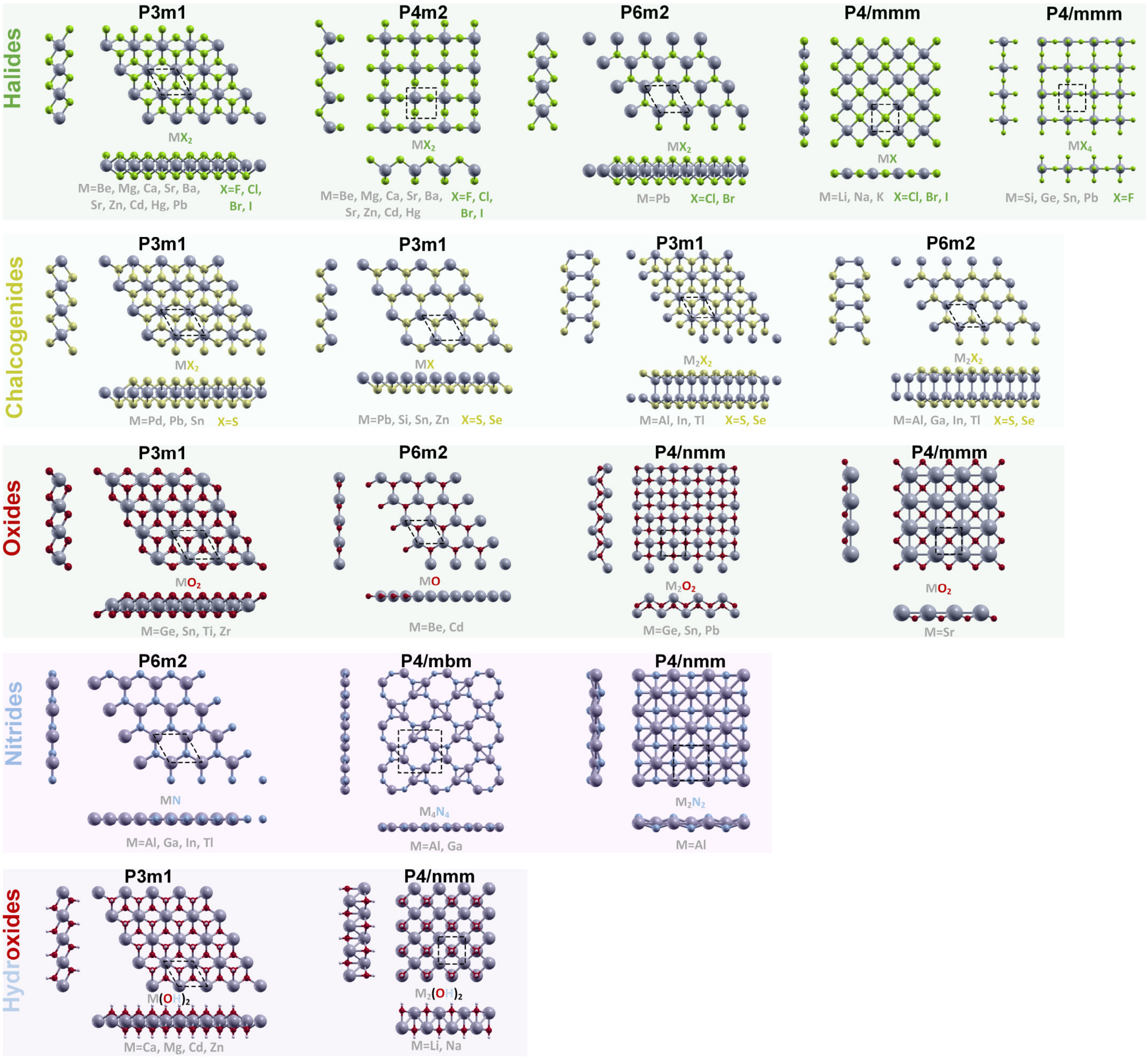

Figure 2. The representative atomic structures of the *2DHDFM*, sorted by chemical compositions and space groups. The chemical formula and the corresponding constituent elements of each atomic structures are provided. The unit cells are marked out by the black dotted lines.

Concerning the *2DHDFM* oxides, $GeO_2$, $SnO_2$, $TiO_2$ and $ZrO_2$ belong to the P3m1 space group with 1T structure, while the space group of BeO, CdO is P6m2 (similar to CdS and ZnS), with their atomic planes being also planar. Another space group of the oxides is P4/nmm, which includes GeO, SnO and PbO. In these three materials, each cation (anion) is bonded to four neighboring anions (cations) in the tetragonal structure. $SrO_2$ is the only oxide with P4/mmm space group, and each O atom bonds to four neighboring Sr atoms, while each Sr atom bonds to eight O atoms.

AlN, GaN, InN and TlN are the four *2DHDFM* nitrides with planar hexagonal P6m2 structure. The rest of the 2D nitrides include planar tetragonal $Al_4N_4$ and $Ga_4N_4$ with P4/mbm, two $Al_2N_2$ structures with either P3m1 or P4/nmm space group, and $C_3N_4$ with C2 space group. In addition, there are only six *2DHDFM* hydroxides, including $Ca(OH)_2$, $Mg(OH)_2$, $Cd(OH)_2$ and $Zn(OH)_2$ with P3m1 space group, and $Li_2(OH)_2$, $Na_2(OH)_2$ with P4/nmm space group. The P3m1 and P4/nmm-hydroxides have structures similar to the 1T structure, and the ones of GeO, SnO and PbO, respectively, with the two O atoms terminated by two H atoms in the unit cell. Note that these six hydroxides have all parent 3D bulk structures and can be easily exfoliated.[3c] The remaining *2DHDFM* mainly involve ternary compounds, such as $Pb_2Br_2F_2$, $Sc_2Br_2O_2$ and $Sr_2Br_2H_2$ with P4/nmm, $In_2Br_2O_2$ and $In_2Cl_2O_2$ with Pmmn, and $TiPbO_3$ with P4mm space group.

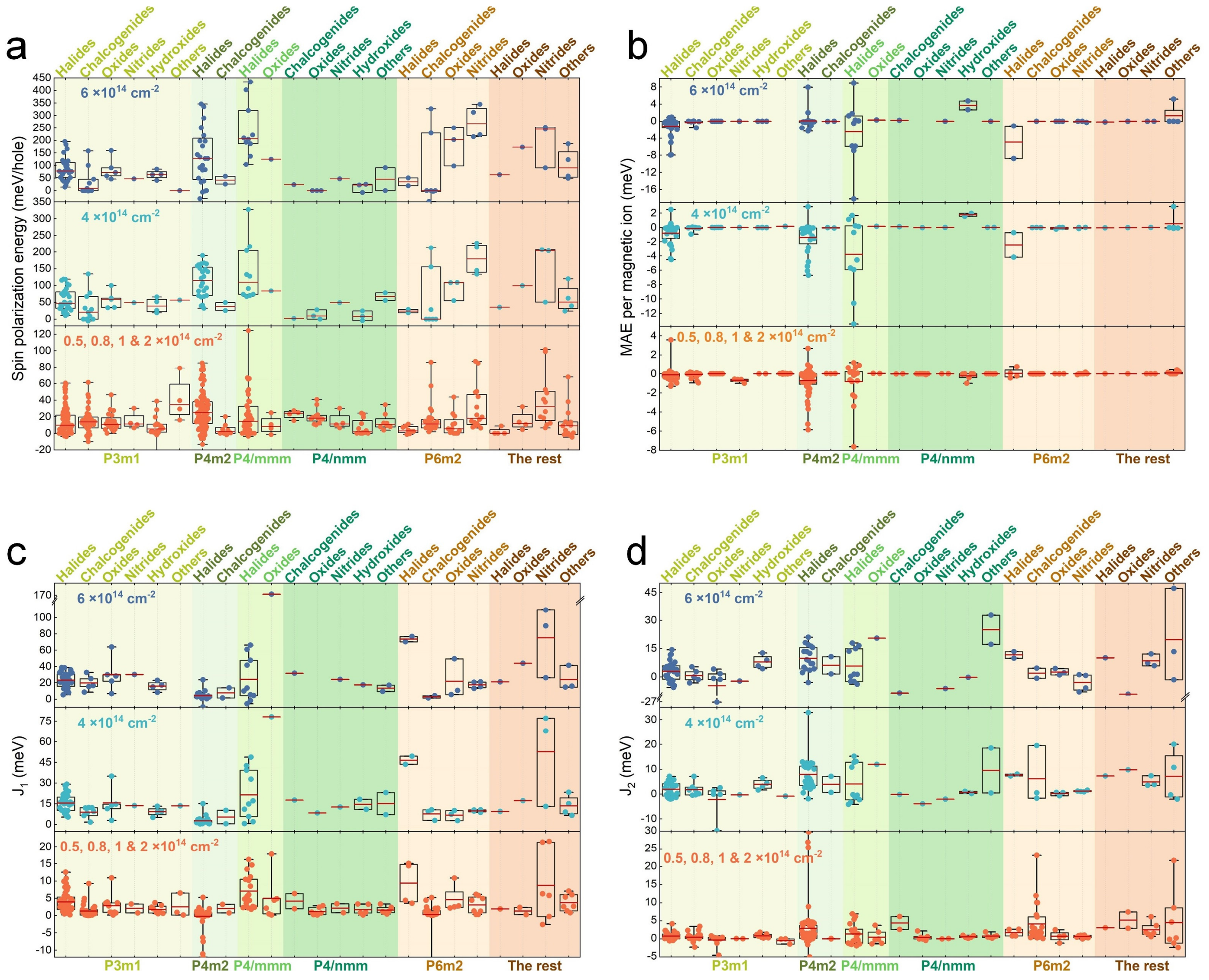

Figure 3. Box plots of the spin-polarization energy (a) magnetic anisotropic energy (MAE) per magnetic ion (b) nearest-neighbor exchange interaction parameter, $J_1$ (c) and next-nearest-neighbor exchange interaction parameter $J_2$ (d) at specific hole doping density of the 122 *2DHDFM*, classified by space group and chemical composition. The red lines inside the boxes represent the median values and the horizontal bars denote the low/high boundaries of the spreads.

The spin-polarization energies $E_{spin}$ of *2DHDFM* at different doping densities are presented in Figure 3 (a). Generally, $E_{spin}$ increases monotonically as the hole doping density increases. There is a clear distinction in the $E_{spin}$ distribution among different compounds with different space groups. For the P3m1, P4m2 and P4/mmm space groups, halides typically have high median as well as large $E_{spin}$ at a given hole doping density. Besides, the oxides and nitrides with P6m2 also have moderate to high $E_{spin}$, with medians of ~200 to ~250 meV/hole at $6 \times 10^{14}$ cm$^{-2}$. On the other hand, the *2DHDFM* with P4/nmm space group have relatively small $E_{spin}$, typically less than 100 meV/hole at $6 \times 10^{14}$ cm$^{-2}$.

$E_{spin}$ has the tendency of being relatively larger for the binary compounds with lighter anions in the same crystal structure. This is in agreement with values reported in the literature[7a, 7b, 7d, 7e] for the 2D gallium or indium oxides and chalcogenides, with $E_{spin}$ following the order of GaO > GaS > GaSe, and InO>InS>InSe. From our calculations, we find that for halides with the same space group, fluorides or chlorides usually have the largest $E_{spin}$, followed by bromides and iodides at the same hole density. As can be seen from Table 1, $MgF_2$ has larger $E_{spin}$ than $MgCl_2$, and for the calcium halides, their $E_{spin}$ follow the order $CaF_2$ > $CaCl_2$ > $CaBr_2$ > $CaI_2$. A similar trend is also clear for the strontium, barium, cadmium and zinc halides. Note, however, that the fluorides of barium, cadmium and mercury do not follow the same trend, their $E_{spin}$ being smaller than the ones of their chloride counterparts. In fact, for materials with a larger difference in electronegativity ($\Delta\chi$) between the anions and cations, one expects a larger coulomb interaction and thus a larger exchange splitting, and subsequently, a larger $E_{spin}$. This may explain why the metal halides have generally larger $E_{spin}$, since their $\Delta\chi$ is large. In particular, the P4/mmm-monohalides have the largest $E_{spin}$, because the group IA metals have small electronegativities, while the halides (especially fluoride) have large electronegativities. For instance, NaBr and KBr have quite high $E_{spin}$, over 400 meV/hole at $6 \times 10^{14}$ cm$^{-2}$. The trend mentioned above for Ga and In oxides, sulfides, and selenides is also consistent with the decrease of the electronegativity from O to Se.

The magnetic anisotropy energy (MAE) of *2DHDFM* varies from materials to materials, and generally grows as the hole doping density increases, as can be seen from Figure 3(b). The MAE of the majority of *2DHDFM* are below 1 meV per magnetic ion. Particularly, metal halides have relatively large absolute MAE, as compared to other compounds, especially the ones with P4/mmm and P4m2 space groups. Some of them even have absolute MAE larger than 10 meV per magnetic ion at 4 and $6 \times 10^{14}$ cm$^{-2}$. This is likely related to the large spin-orbit coupling strength of bromine and iodine. Other materials with lighter elements (and thus smaller spin-orbit coupling), like fluorides and oxides, have much smaller absolute MAE, usually less than 100 μeV/magnetic ion. Interestingly, the median of MAE of most *2DHDFM* are negative, which reveals that the spins prefer to align along the in-plane direction.

The magnetic exchange coupling parameters of the *2DHDFM* at different doping densities are summarized in Figure 3 (c) and (d), where $J_1$ and $J_2$ represents the nearest and next-nearest neighbor coupling, respectively. These parameters can be obtained from the total energy difference of various possible ferromagnetic and antiferromagnetic configurations.[12] Overall, the magnetic exchange coupling parameters also increase by increasing the hole doping density. $J_1$ is positive for most *2DHDFM*, which is characteristic of a ferromagnetic first-neighbor interaction. On the other hand, $J_2$ is typically smaller than $J_1$ (as expected), and $J_2$ is negative for a couple of *2DHDFM*, suggesting competing ferromagnetic and antiferromagnetic interactions in these compounds, as discussed further below. The computed Curie temperature $T_c$ of *2DHDFM* is summarized in Figure 4. Increasing the hole doping density also gives rise to a higher $T_c$ for most of *2DHDFM*. Upon a hole doping density of 4 × $10^{14}$ $cm^{-2}$, the medians $T_c$ of various *2DHDFM* are in the range of ~50 K to ~170 K, and further increase to ~70 K to ~270 K at 6 × $10^{14}$ $cm^{-2}$. As expected, the Curie temperature is correlated to the values of the magnetic exchange coupling parameters. Very interestingly, the sign of $J_2$ is found to play a very important role on $T_c$. Consider the typical case of $CaF_2$ ($T_c$=70 K) and $HgF_2$ ($T_c$=333 K), which have the same P3m1 crystal structure (see Table 1). While these two materials have similar $J_1$ (about 23 meV), $CaF_2$ has a negative $J_2$ (about -2 meV) and $HgF_2$ has a positive $J_2$ (about 12 meV). A ferromagnetic next-nearest neighbor interaction thus favors a large $T_c$ in these 2D materials, such as in $BaF_2$ (337 K), $CdF_2$ (320 K), $HgF_2$ (333 K), $PbCl_2$ (394 K) and $PbBr_2$ (301 K) at a hole doping density of 6 × $10^{14}$ $cm^{-2}$.

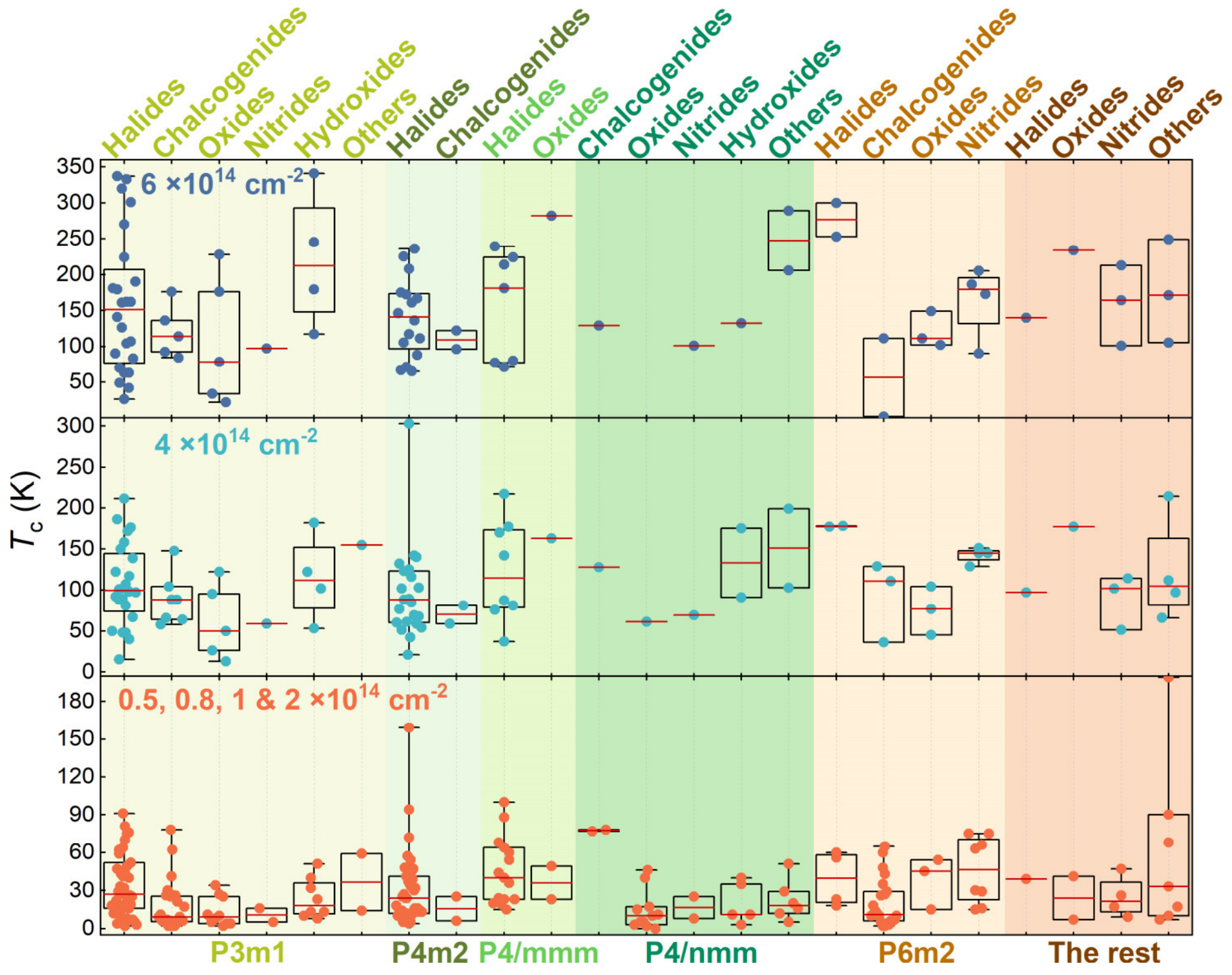

Figure 4. Box plots of the Curie temperatures ($T_c$) obtained from Monte Carlo simulations for *2DHDFM* at different hole doping densities. The red lines inside the boxes represent the median values and the horizontal bar denotes the low/high boundaries of the spreads.

In general, the magnetic moments of *2DHDFM* are mainly contributed by the anion *p*-orbitals, as shown in Figure 5 (a). Since a large portion of these 2D materials consists of P3m1-dihalides, the magnetic properties of these materials are first discussed below. In these compounds, $J_1$ corresponds to the out-of-plane exchange coupling between two halogens from the top and bottom atomic planes, as shown in Figure 6 (a); a local coordinate system is used for these P3m1-dihalides, where the metal ion is in the center of a distorted octahedra with six ligands at the vertices. Note that the shorter halogen-halogen distance is about 2.3 Å in fluorides, and gradually increases to about 5 Å from chlorides to iodides. Considering the delocalized nature of the *p* orbitals, the direct out-of-plane magnetic interaction mainly arises from the exchange coupling between the $p_z$ orbitals of the halogens, as illustrated in Figure 6 (b); this direct exchange coupling is ferromagnetic, leading to a positive value of $J_1$, which lies between about 5 and 39 meV (see Table 1). Note also that $J_1$ is smaller in iodides, due to the larger out-of-plane halogen-halogen distance in these compounds.

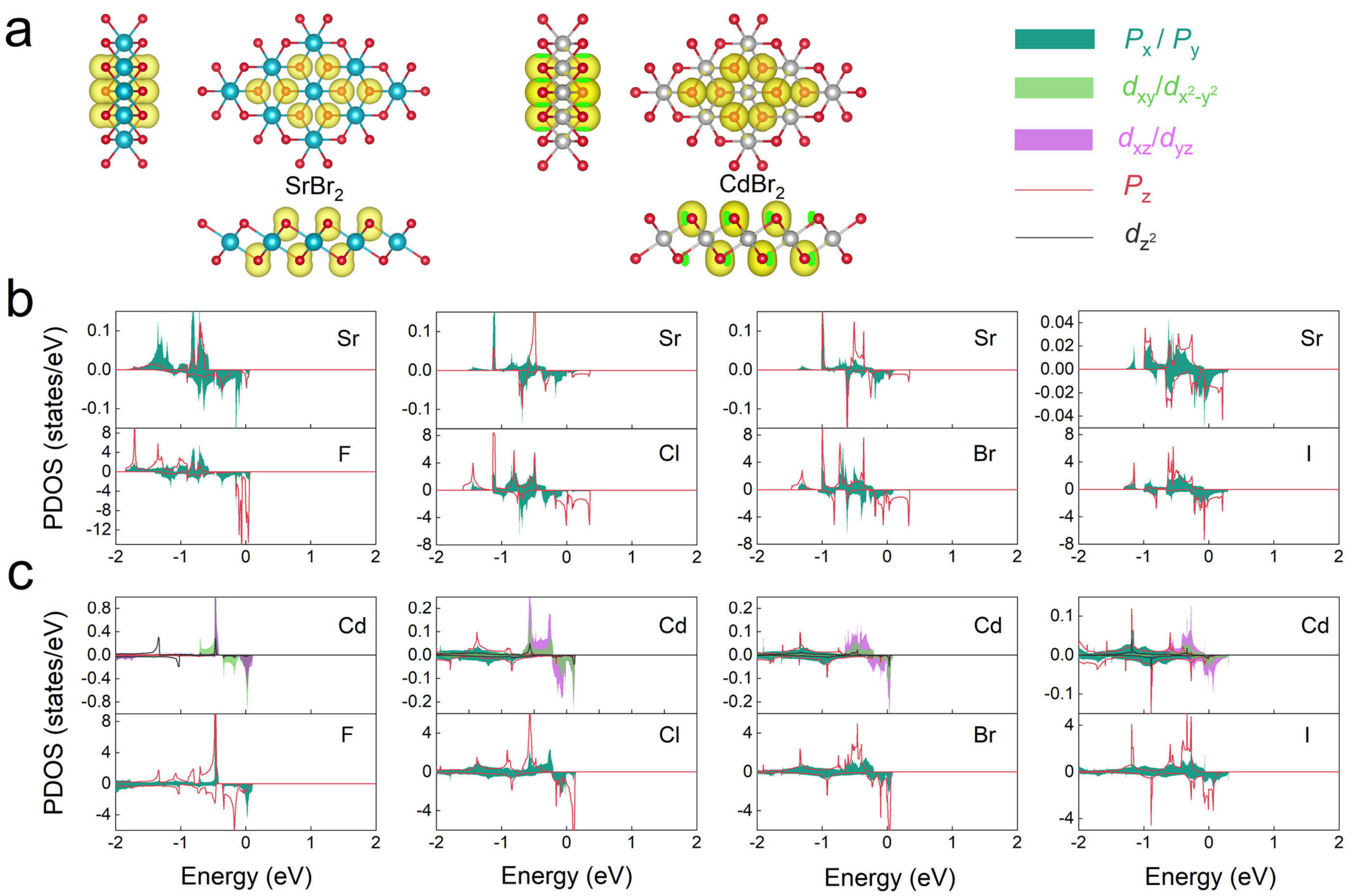

Figure 5. Spin density plots (a) and orbital projected density of states (b) and (c) for P3m1 strontium dihalides and cadmium dihalides monolayers at hole density of $6 \times 10^{14}$ cm$^{-2}$. Positive (negative) values refer to up (down) spins. The Fermi level is set at zero energy.

On the other hand, $J_2$ corresponds to the in-plane exchange coupling between two halogens from the same atomic plane, see Figure 6 (a). In this case, the direct exchange interaction between the anion *p*-orbitals is weaker since the in-plane halogen-halogen distance is larger. Consequently, the indirect exchange interaction, involving the coupling between the halogen-*p* orbitals with the metal orbitals, should also play an important role in the next-nearest neighbor interaction. Group IIA and IIB metals are transferring their two outmost valence electrons to the halogens *p*-orbitals, to form the dihalide compounds. In the indirect exchange mechanisms, it is assumed that some covalent mixing between the cation and anion orbitals is energetically favorable. From the projected electronic density of states (PDOS) of strontium and cadmium dihalides, shown in Figure 5 (b) and (c), respectively, one can indeed visualize the possible hybridization between the (spin-down) *p* orbitals of the halogen and the (spin-down) *p* or *d* orbitals of the metals near the Fermi level; note that for group IIB metal dihalides, the degeneracy between the metal *d* orbitals is partially lifted by the distorted octahedral crystal field, resulting in the formation of $e_1(d_{xz}$ and $d_{yz})$, $e_2(d_{x^2-y^2}$ and $d_{xy})$ and $a_1(d_{z^2})$ states. Specifically, the in-plane halogen-$p_x/p_y$ orbitals can couple to metal-$p_x/p_y$ orbitals and halogen-$p_z$ orbitals couple to metal-$p_z$ orbitals for strontium halides. For cadmium dihalides, the halogen-$p_x/p_y$ orbitals couple to metal-$d_{x^2-y^2}/d_{xy}$ and $d_{xz}/d_{yz}$ orbitals, and the halogen-$p_z$ orbitals also hybridize with the $d_{xz}/d_{yz}$ orbitals. From Table 1, $J_2$ of group IIA metal dihalides can be positive or negative, indicating the competition between different indirect exchange mechanisms, such as the super-exchange coupling between orbitals pointing in the same direction (antiferromagnetic, e.g., hybridization between halogen-$p_z$ orbitals with metal-$p_z$ orbitals, as illustrated in Figure 6 (c)), the super-exchange mechanism involving the coupling between orthogonal orbitals (ferromagnetic, e.g., hybridization between halogen-$p_x$/$p_y$ orbitals with metal-$p_x$/$p_y$ orbitals, as depicted in Figure 6 (c)) and the double exchange mechanism (ferromagnetic). This latter mechanism involves the interaction between anions orbitals with different charged (or valence) states, e.g., between partially filled halogen orbitals (where holes are localized) with filled halogen orbitals. Very interestingly, $J_2$ is found to be positive for all group IIB metal halides (see Table 1), the coupling between the halogen $p$ orbitals with the metal $e_1$ and $e_2$ states most likely favoring the ferromagnetic exchange mechanisms, as shown in Figure 6 (d)-(f); note that the contribution of the metal $d$-states to the spin density, observed on Figure 5 (a), also points towards the possible hybridization between the cation and anion orbitals near the Fermi level. Although GGA functional tends to over-delocalize electrons, the contribution of the metal $d$-states in $CdBr_2$ can still be confirmed in the spin density obtained from HSE06 functional.

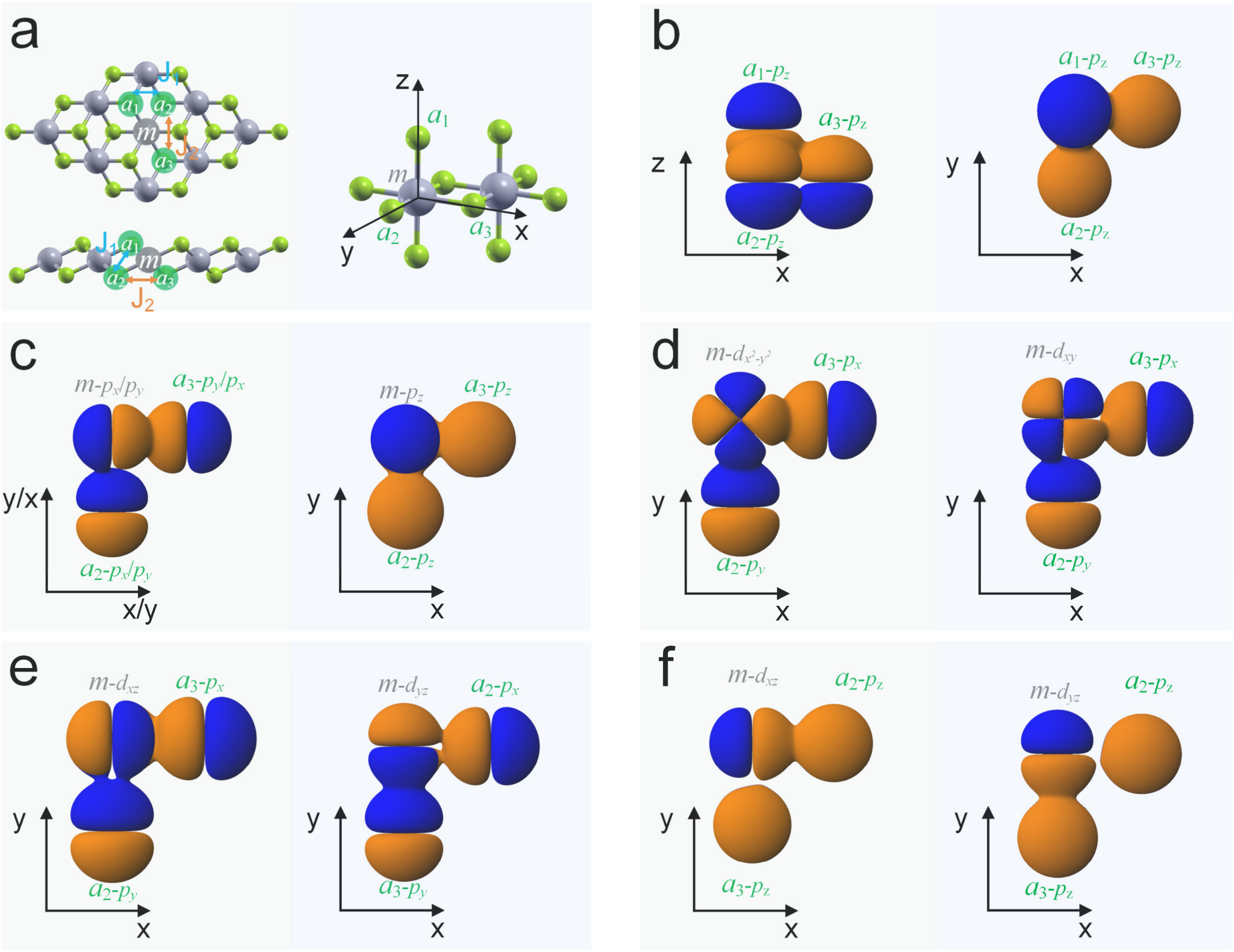

Figure 6. (a) Exchange interaction paths in P3m1 metal dihalides. $J_1$ is the exchange coupling parameter originated from the halide ions on the upper layer ($a_1$) and the bottom layer ($a_2$). $J_2$ is the exchange coupling parameter originated from the halide ions on the same atomic plane ($a_2$ and $a_3$), $m$ is the metal ions. Edge-sharing distorted $ma_6$ octahedra with the local xyz coordinates of the P3m1 structure are also given on the right. Schematic diagrams of (b) direct exchange between halogen-$p_z$ orbitals, (c) the super-exchange interactions between halogen-$p_x$/$p_y$ orbitals with metal-$p_x$/$p_y$ orbitals and halogen-$p_z$ orbitals

with metal-$p_z$ orbitals, (d) halogen-$p_x/p_y$ orbitals with metal-$d_{x^2-y^2}/d_{xy}$ orbitals, (e) halogen-$p_x/p_y$ orbitals with metal-$d_{xz}/d_{yz}$ orbitals, and (f) halogen-$p_z$ orbitals with metal-$d_{xz}/d_{yz}$ orbitals.

We now consider the possible magnetic exchange interactions in planar 2DHDFM, namely without out-of-plane anion interactions. Similar to the case of P3m1-dihalides, $J_1$ of the P4/mmm-lithium monohalides decreases from LiCl to LiI, as can be seen in Table S1. In this case, as shown in Figure S2, $J_1$ is mainly due to the direct exchange between the halogen $p_z$-orbitals. In addition, from the PDOS shown in Figure S3, one can see that there is a very limited possible hybridization between the Li-*s* orbitals with the halogen-*p* orbitals, pointing out to a relatively weak indirect exchange interaction in these materials. On the other hand, in the case of hole-doped 2D planar P6m2-sulfides (such as ZnS and CdS) and nitrides (such as AlN and GaN), the ferromagnetic order is essentially mediated by the indirect exchange interaction between the sulfur or nitrogen $p_z$-states and the metal-$d_{xz}/d_{yz}$ and $p_z$ states, as evidence by the PDOS in Figure S3.

Considering the possible use of these 2DHDFM in spintronic devices, the metal dihalides (like $BaF_2$ and $CdF_2$) appear as interesting candidates for high-temperature (above room temperature) applications. However, these materials might have some stability issues, possibly interacting/reacting with the ambient and being oxidized in air. On the other hand, potential more stable 2DHDFM have also been identified for low temperature applications. For example, AlN, GaN and ZnS crystallizes in the Wurtzite phase. In ultra-thin films (and at the 2D limit), such materials could be grown in a h-BN like phase[13]. In addition, these 2D materials should be much less reactive with the ambient. Other materials of interest are e.g., 2D $ZrO_2$ and $TiO_2$, which are predicted to be 2DHDFM with moderate Curie temperatures of 176 and 228 K, respectively.

**Table 1**. Spin polarization energy $E_{spin}$, electronegativity difference $\Delta\chi$, magnetic anisotropic energy MAE, nearest-neighbor exchange interaction parameter $J_1$, next-nearest-neighbor exchange interaction parameter $J_2$, and Curie temperature $T_c$ at hole doping density of 6 × $10^{14}$ $cm^{-2}$ for the P3m1-halides.

| | $E_{spin}$ (meV) | $\Delta\chi$ | MAE (μeV) | $J_1$ (meV) | $J_2$ (meV) | $T_c$ (K) |
|---|---|---|---|---|---|---|
| $BeF_2$ | 120 | 1.94 | -9 | 21.440 | -3.343 | 26 |
| $MgF_2$ | 152 | 2.67 | -15 | 28.624 | -3.931) | 63 |
| $MgCl_2$ | 101 | 1.85 | -163 | 34.845 | 3.081 | 224 |
| $CaF_2$ | 184 | 2.98 | -62 | 23.212 | -1.996 | 70 |
| $CaCl_2$ | 80 | 2.16 | -11 | 36.036 | -5.636 | 49 |
| $CaBr_2$ | 40 | 1.96 | -613 | 27.296 | -1.552 | 104 |
| $CaI_2$ | 14 | 1.66 | -620 | 5.415 | 0.663 | 42 |
| $SrF_2$ | 195 | 3.03 | -125 | 16.900 | -0.740 | 63 |
| $SrCl_2$ | 118 | 2.21 | -295 | 32.319 | -4.562 | 83 |
| $SrBr_2$ | 71 | 2.01 | 126 | 38.474 | -2.717 | 126 |
| $SrI_2$ | 35 | 1.71 | -1956 | 9.024 | 3.092 | 107 |
| $BaF_2$ | 96 | 3.09 | 946 | 38.685 | 7.868 | 337 |
| $BaCl_2$ | 167 | 2.27 | 195 | 20.222 | -0.710 | 90 |
| $BaBr_2$ | 110 | 2.07 | 320 | 32.384 | 1.431 | 181 |
| $BaI_2$ | 73 | 1.77 | -1776 | 11.161 | 5.215 | 162 |
| $ZnF_2$ | 76 | 2.33 | 314 | 14.890 | 4.560 | 155 |
| $ZnCl_2$ | 65 | 1.51 | -314 | 20.431 | 2.748 | 161 |
| $ZnBr_2$ | 42 | 1.31 | -3800 | 13.446 | 5.452 | 179 |
| $CdF_2$ | 107 | 2.29 | 816 | 29.730 | 9.200 | 320 |
| $CdCl_2$ | 114 | 1.47 | -171 | 31.241 | 0.785 | 162 |
| $CdBr_2$ | 78 | 1.27 | -5005 | 20.870 | 3.980 | 190 |
| $CdI_2$ | 37 | 0.97 | -7917 | 8.155 | 4.530 | 141 |
| $HgF_2$ | 71 | 1.98 | -1316 | 23.307 | 12.027 | 333 |
| $HgCl_2$ | 84 | 1.16 | 366 | 30.671 | 6.515 | 270 |
| $HgBr_2$ | 77 | 0.96 | -3391 | 27.471 | 6.280 | 264 |
| $HgI_2$ | 52 | 0.66 | -7939 | 12.72 | 5.672 | 185 |
| $PbCl_2$ | 53 | 0.83 | 0 | 27.041 | 14.43 | 394 |
| $PbBr_2$ | 23 | 0.63 | -1013 | 18.215 | 11.75 | 301 |

# Conclusions

Using high-throughput density functional theory calculations, we have identified 122 stable *2DHDFM* that exhibit a non-magnetic to a ferromagnetic phase transition upon hole doping. In these *2DHDFM*, metal halides take up the biggest proportion, followed by the sulfides, oxides and nitrides. The magnetic properties of these 2D materials, such as their spin polarization energy, magnetic anisotropic energy, magnetic exchange coupling parameters and Curie temperature typically increase with the hole doping density. Among these materials, metal dihalides with a P3m1 phase, like $BaF_2$, $CdF_2$, $PbCl_2$ and $PbBr_2$ are predicted to have Curie temperatures above 300 K at a typical hole density of $6 \times 10^{14}$ $cm^{-2}$, these materials being potentially interesting for spintronic devices operating above room temperature. In general, the ferromagnetic interaction in these materials is mediated by a direct exchange interaction between $p_z$-orbitals of anions in different atomic planes (out-of-plane coupling) as well as the indirect exchange interaction between the $p_z$-anion states and the metal *p* or *d* states (in-plane coupling). On the other hand, 2D planar materials with moderate Curie temperatures, typically ranging between 110 K and 230 K, have also been identified for possible spintronic applications at low temperature. Such materials are e.g., P3m1 sulfides and nitrides, like ZnS, AlN, and GaN. In these materials, the ferromagnetic coupling mainly arises from the indirect exchange interaction between the anion $p_z$-orbitals and the *p* or *d* metal states. These materials should be more stable than their halogen counterparts, especially considering their possible interaction with the ambient (oxidation), and could potentially be grown in a planar 2D form, like h-BN.

## Acknowledgments

Part of this work has been financially supported by the FLAG-ERA grant DIMAG, by the Research Foundation – Flanders (FWO) as well as the KU Leuven Research Fund, project C14/17/080 and C14/21/083. Part of the computational resources and services used in this work have been provided by the VSC (Flemish Supercomputer Center), funded by the FWO and the Flemish Government – department EWI.

**Supplementary Materials for**

# Hole-doping induced ferromagnetism in 2D materials

R. Meng[1*], L.M.C. Pereira[1], J.P. Locquet[1], V.V. Afanas'ev[1], G. Pourtois[2], and M. Houssa[1,2*]

[1]Department of Physics and Astronomy, KU Leuven, Celestijnenlaan 200D, Leuven B-3001, Belgium

[2]imec, Kapeldreef 75, B-3001 Leuven, Belgium

[*]Email: ruishen.meng@kuleuven.be; michel.houssa@kuleuven.be

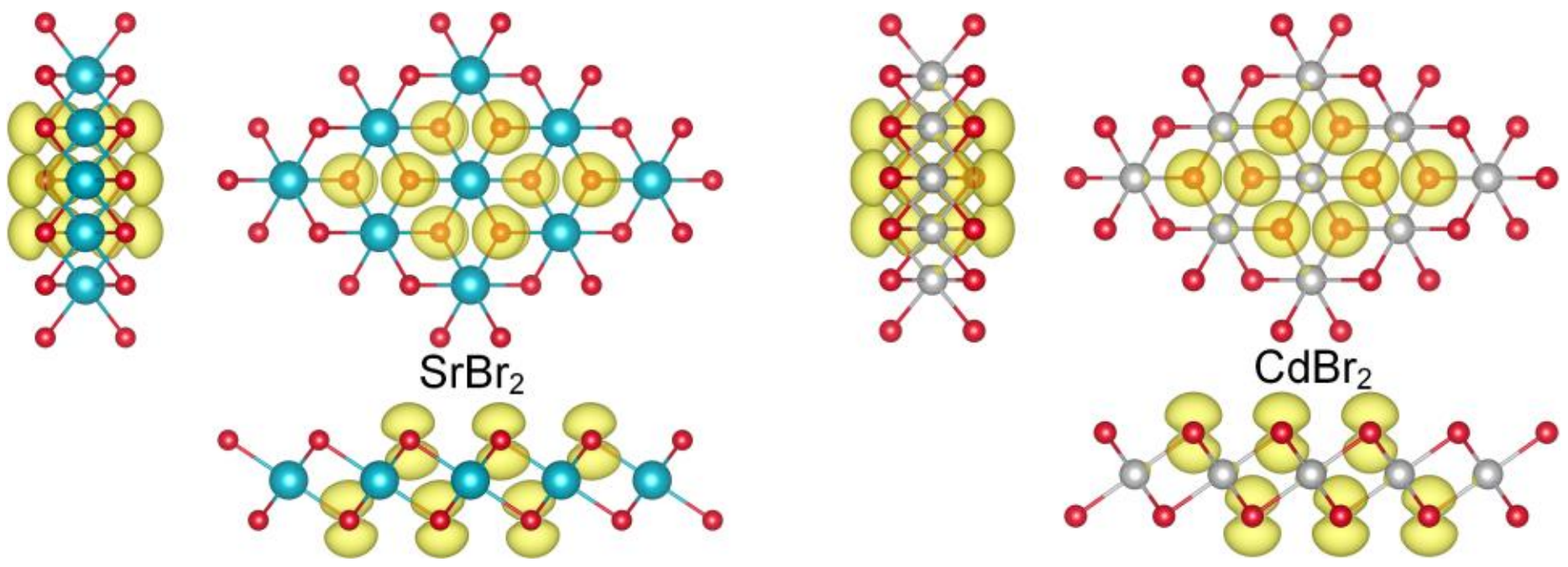

Figure S1. Spin density plots calculated by HSE06 functional for P3m1 strontium dihalides and cadmium dihalides monolayers at hole density of 6 × $10^{14}$ cm$^{-2}$.

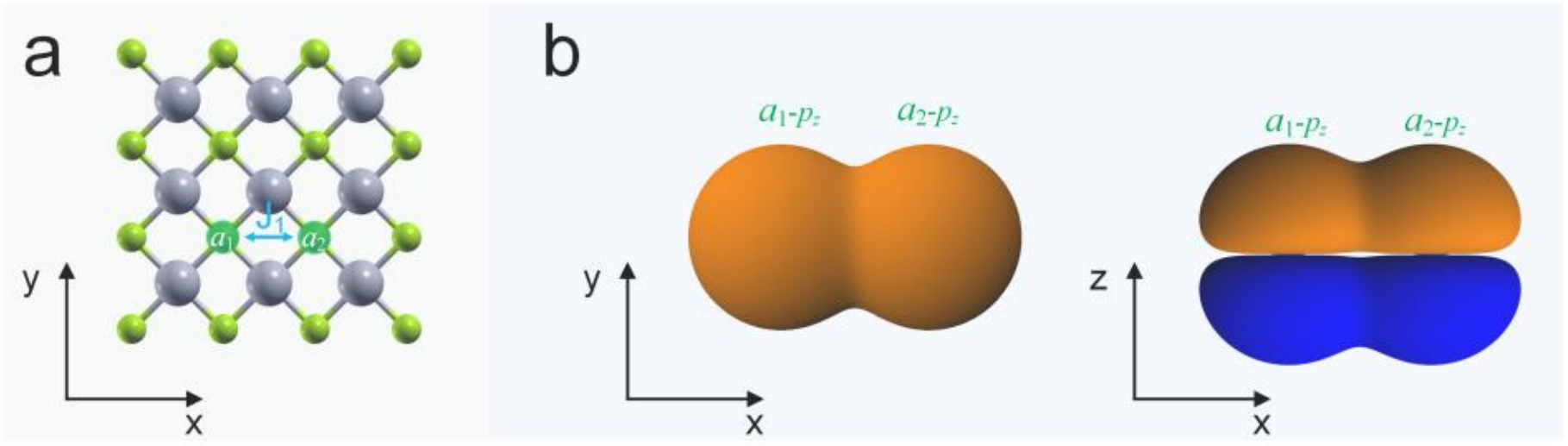

Figure S2. (a) Local xyz coordinates of the lithium monohalide with P4/mmm space group, $a_1$ and $a_2$ are the two nearest halide ions. (b) Schematic diagrams of the direct exchange coupling $J_1$ between $p_z$ orbitals of the nearest halide ions.

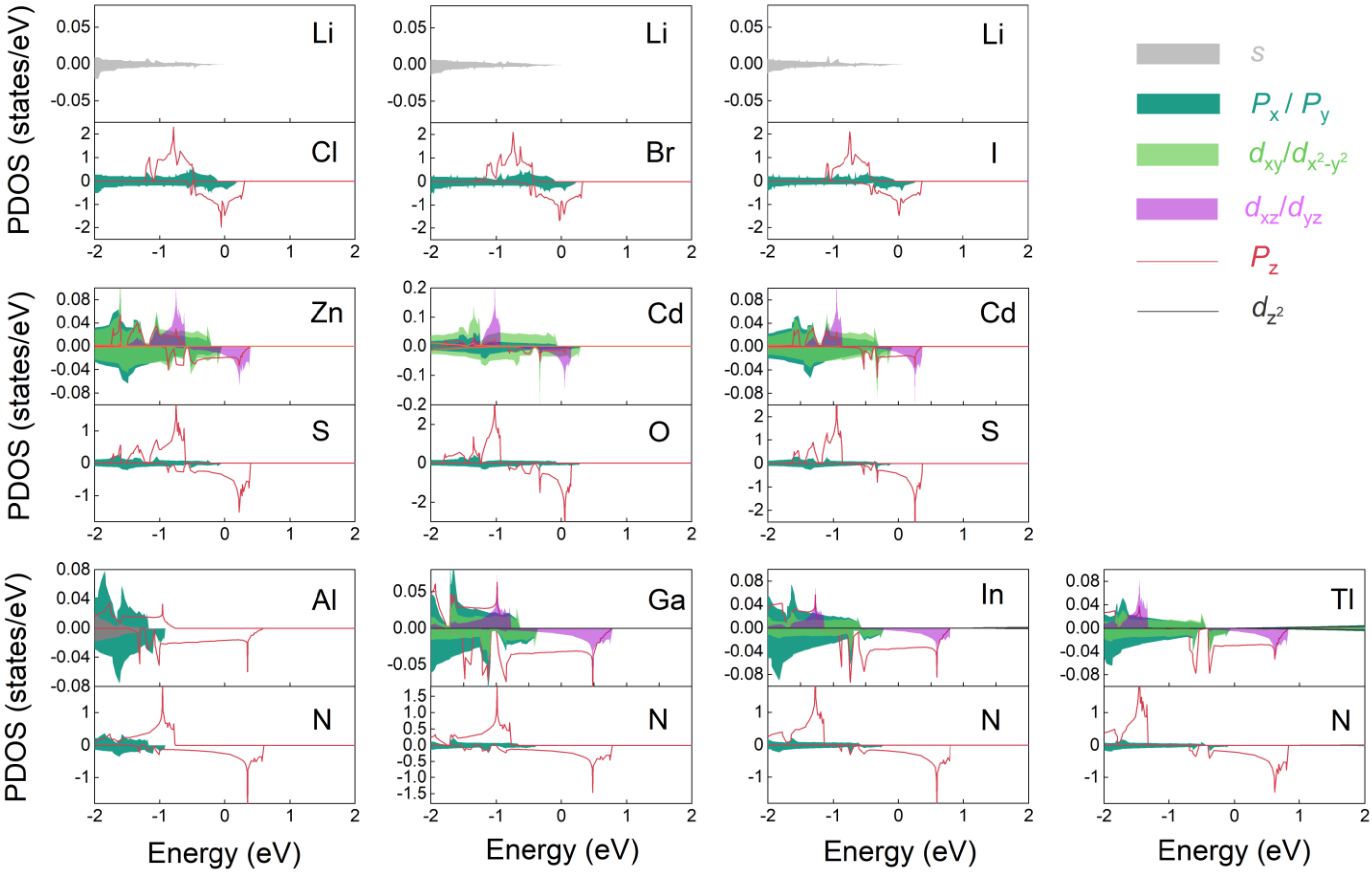

Figure S3. Orbital projected density of states for planar *2DHDFM* under hole doping. The doping densities are 4 × $10^{14}$ cm$^{-2}$ for lithium monohalides with P4/mmm space group, ZnS, CdO and CdS with P6m2 space group and 6 × $10^{14}$ cm$^{-2}$ for the nitrides with P6m2 space group. Positive (negative) values refer to up (down) spins. The Fermi level is set at zero energy.

**Table S1**. Spin polarization energy $E_{\mathrm{spin}}$, electronegativity difference $\Delta\chi$, magnetic anisotropic energy MAE, nearest-neighbor exchange interaction parameter $J_1$, next-nearest-neighbor exchange interaction parameter $J_2$, and Curie temperature $T_c$. unless otherwise stated, the hole doping density is $6 \times 10^{14}$ cm$^{-2}$.

| | | | $E_{\mathrm{spin}}$ (meV) | **MAE** (μeV) | $J_1$ (meV) | $J_2$ (meV) | $T_c$ (K) |
|---|---|---|---|---|---|---|---|
| Halides | P4/mmm | LiCl (@4× $10^{14}$) | 74 | -703 | 17.186 | -4.158 | 81 |
| | | LiBr (@4× $10^{14}$) | 70 | -5684 | 14.856 | -3.304 | 87 |
| | | LiI (@4× $10^{14}$) | 67 | -10596 | 11.105 | -2.303 | 76 |
| | | NaBr (@4× $10^{14}$) | 216 | -4569 | 5.733 | -1.239 | 37 |
| | | NaI (@4× $10^{14}$) | 204 | -13476 | 1.998 | -0.447 | 16 |
| | | KBr | 435 | 8932 | 4.438 | 1.330 | 80 |
| | | $SiF_4$ | 104 | 86 | 60.935 | 13.560 | 181 |
| | | $GeF_4$ | 136 | 262 | 66.101 | 17.979 | 239 |
| | | $SnF_4$ | 204 | 1254 | 41.674 | 16.566 | 224 |
| | | $PbF_4$ | 221 | 1777 | 47.632 | 14.948 | 214 |
| Chalcogenides | P6m2 | $Al_2S_2$ | 30 | 0 | 1.198 | 6.646 | 26 |
| | | $Al_2Se_2$ | 19 | -13 | 2.478 | 11.954 | 35 |
| | | $Ga_2S_2$ | 7 | -2 | 1.403 | 6.086 | 29 |
| | | $In_2S_2$ | 18 | -7 | 2.088 | 10.015 | 43 |
| | | $Tl_2S_2$ | 2 | -19 | 2.155 | 10.024 | 48 |
| | | CdS | 327 | -14 | 1.513 | -0.709 | 2 |
| | | ZnS | 230 | 2 | 4.198 | 4.633 | 111 |
| | P3m1 | $SnS_2$ | 45 | -160 | 15.621 | 2.803 | 136 |
| | | $PbS_2$ | 56 | -83 | 16.746 | 5.295 | 176 |
| | | $PdS_2$ (@4× $10^{14}$) | 5 | 163 | 1.965 | 0.940 | 28 |
| | | SiS | 100 | 11 | 25.120 | -0.052 | 114 |
| | | SnS | 30 | -81 | 32.824 | -3.198 | 84 |
| | | PbS (@4× $10^{14}$) | 9 | -921 | 12.006 | -0.356 | 148 |
| | | PbSe (@2× $10^{14}$) | 23 | -961 | 4.179 | -0.892 | 41 |
| | | ZnSe | 158 | -1593 | 8.749 | -1.448 | 92 |
| Chalcogenides | P3m1 | $GeO_2$ | 58 | 0 | 6.709 | -0.855 | 22 |
| | | $SnO_2$ | 160 | 6 | 22.170 | -1.288 | 78 |
| | | $TiO_2$ | 72 | -2 | 28.475 | 4.176 | 228 |
| | | $ZrO_2$ | 91 | -7 | 29.899 | 1.660 | 176 |
| | P4/nmm | $Ge_2O_2$ | 35 | -2 | 4.906 | -1.399 | 40 |
| | | $Sn_2O_2$ | 41 | -16 | 8.415 | -3.919 | 61 |
| | | $Pb_2O_2$ | 19 | -95 | 2.874 | 2.166 | 46 |
| | P6m2 | BeO | 203 | 0 | 10.600 | 1.074 | 149 |
| | | CdO | 250 | -4 | 5.751 | 2.175 | 102 |
| | P4/mmm | $SrO_2$ | 125 | 99 | 171.382 | 20.580 | 282 |
| Nitrides | P6m2 | AlN | 215 | 2 | 13.287 | 1.178 | 186 |
| | | GaN | 223 | -1 | 15.387 | 0.784 | 205 |
| | | InN | 313 | -18 | 21.218 | -5.844 | 173 |
| | | TlN | 345 | -269 | 20.763 | -7.956 | 90 |
| | P4/mbm | $Al_4N_4$ | 253 | -3 | 109.369 | 7.384 | 101 |
| | | $Ga_4N_4$ | 245 | -1 | 89.792 | 12.179 | 164 |
| | P4/nmm | $Al_2N_2$ | 46 | -2 | 24.138 | -6.191 | 101 |
| | P3m1 | $Al_2N_2$ | 97 | 0 | 30.102 | -2.211 | 97 |
| | C2 | $C_3N_4$ | 91 | -4 | 26.360 | 5.933 | 213 |
| Hydroxides | P3m1 | $Mg(OH)_2$ | 41 | -6 | 8.349 | 4.370 | 117 |
| | | $Ca(OH)_2$ | 65 | 2 | 15.051 | 5.868 | 179 |
| | | $Zn(OH)_2$ | 63 | -2 | 17.292 | 8.873 | 245 |
| | | $Cd(OH)_2$ | 85 | -12 | 23.053 | 12.691 | 341 |
| | P4/nmm | $Li_2(OH)_2$ | 91 | 7 | 17.512 | -0.214 | 132 |
| | | $Na_2(OH)_2$ | 78 | 8 | 18.275 | 1.077 | 175 |

## Computational Methods

In this work, all the density functional theory calculations were performed using the Vienna ab initio simulation package (VASP) package[1,2], with electron-ion interaction described by projector augmented wave (PAW) pseudopotentials. The generalized gradient approximation (GGA), parameterized by the Perdew-Burke-Ernzerhof (PBE)[3] approach was used as the exchange correlation functional. The energy cutoff of 550 eV and k-point meshes of $0.03\times2\pi$ and $0.02\times2\pi$ $Å^{-1}$ were used for structural optimizations and self-consistent calculations. Total energy convergence criterion of $10^{-6}$ eV/cell and force convergence criterion of 0.005 eV/Å were chosen for complete relaxations of lattice constants as well as atomic positions. The Heyd−Scuseria−Ernzerhof functional (HSE06),[4] which mixes 25% nonlocal exchange with the PBE functional, was also used for testing purpose. In the doping calculations, the hole density was tuned by removing electrons from the cell, with a jellium background with opposite charge added to maintain charge neutrality and the atomic positions are reoptimized at different hole densities.

Phonon dispersion curves were calculated by the PHONOPY package[5] on the basis of Density Functional Perturbation Theory (DFPT). Curie temperatures were estimated using Monte Carlo simulations, as implemented in the VAMPIRE package[6]. The simulated systems for all materials consist of a platelet with at least 10000 spins with a rectangular supercell. The spins were thermalized for 10000 equilibrium steps, followed by 20000 averaging steps for the calculation of the thermal equilibrium magnetization at every temperature.

The ab initio evolutionary algorithm was performed by USPEX[7,8], interfaced with VASP, was used to find the potential structures that are ferromagnetic under hole doping. The variable-composition searching was chosen with the total atom numbers of the 2D crystals set to be 2-8, and their thicknesses and vacuum spaces restricted to be 4 Å and 20 Å, respectively. 150 groups of symmetry were used to produce random symmetric structure generator for initial population. Then, the full structure relaxations were performed, and the most stable and metastable structures were screened and inherited into the next generation by comparing their formation enthalpy. The number of generation is set to 40.

## List of Structures

More details about the 122 *2DHDFM* are summarized in the tables and they are provided in term of their molecular formula, and classified by their space groups. The 2D materials are selected if their spin polarization energies are larger than 10 meV/hole. In the tables, the 2dmat-ID and MC2D-ID are the identification number of the 2D materials from the 2DMatPedia database[9] and the supplementary information of the article[10], respectively. Meanwhile, if the materials exist in the Computational 2D Materials Database (C2DB)[11], they will be marked with ✓. At the same time, if the material is solely found by the USPEX and not existed in the databases mentioned above, this material is also marked with ✓.

The convex hulls are defined as the enthalpy of formation versus the composition of different elements with all possible stoichiometries, which are searched by USPEX. The formation enthalpy in the convex hull figure is obtained by subtracting the total energy of the total energy of the compound by the total energies of the 2D structures of its constituent elements found by USPEX. The corresponding atomic structures with the unit-cell marked out by the black dashed line, as well as the lattice geometry and the atomic positions in fractional coordinates are also provided. The spin polarization energy is defined as the total energy of the non-magnetic state minus the one of the ferromagnetic state. Thus, a positive spin polarization energy indicates that the ferromagnetic state is more stable. The magnetic configurations, i.e., ferromagnetic (FM), and different antiferromagnetic (AFM) structures and their corresponding spin Hamiltonian are given to extract the nearest-neighbor and second nearest-neighbor and/or third nearest-neighbor exchange coupling parameters. The energy of the FM structure is calculated using the unit-cell while the supercell used to obtain the energies for the AFM structures are marked by the black dashed lines. Magnetic anisotropic energies (MAE) are determined by obtaining the energy difference by orienting the spins in the out-of-plane and the in-plane (from 0° to 360°) direction. Finally, the Curie temperatures ($T_c$) are evaluated using Monte Carlo calculations. The temperature dependent magnetization is fitted using the Curie-Bloch equation in the classical limit:

$$\mathrm{m}(T) = (1 - \frac{T}{T_c})^{\beta}$$

where $T$ is the temperature and β is a critical exponent.

## P3m1

| | | | | | | |
|---|---|---|---|---|---|---|
| 1. $BeF_2$ | 2. $MgF_2$ | 3. $MgCl_2$ | 4. $CaF_2$ | 5. $CaCl_2$ | 6. $CaBr_2$ | 7. $CaI_2$ |
| 8. $SrF_2$ | 9. $SrCl_2$ | 10. $SrBr_2$ | 11. $SrI_2$ | 12. $BaF_2$ | 13. $BaCl_2$ | 14. $BaBr_2$ |
| 15. $BaI_2$ | 16. $ZnF_2$ | 17. $ZnCl_2$ | 18. $ZnBr_2$ | 19. $CdF_2$ | 20. $CdCl_2$ | 21. $CdBr_2$ |
| 22. $CdI_2$ | 23. $HgF_2$ | 24. $HgCl_2$ | 25. $HgBr_2$ | 26. $HgI_2$ | 27. $PbCl_2$ | 28. $PbBr_2$ |
| 29. $Al_2S_2$ | 30. $Al_2Se_2$ | 31. $In_2S_2$ | 32. SiS | 33. SnS | 34. PbS | 35. PbSe |
| 36. ZnSe | 37. $PdS_2$ | 38. $SnS_2$ | 39. $PbS_2$ | 40. $Al_2O_3$ | 41. $GeO_2$ | 42. $SnO_2$ |
| 43. $TiO_2$ | 44. $ZrO_2$ | 45. $Al_2N_2$ | 46. $Mg(OH)_2$ | 47. $Ca(OH)_2$ | 48. $Zn(OH)_2$ | 49. $Cd(OH)_2$ |
| 50. TeC | | | | | | |

## P6m2

| | | | | | | |
|---|---|---|---|---|---|---|
| 51. $PbCl_2$ | 52. $PbBr_2$ | 53. $Al_2S_2$ | 54. $Al_2Se_2$ | 55. $Ga_2S_2$ | 56. $In_2S_2$ | 57. $Tl_2S_2$ |
| 58. CdS | 59. ZnS | 60. BeO | 61. CdO | 62. $TeO_3$ | 63. AlN | 64. GaN |
| 65. InN | 66. TlN | | | | | |

## P4m2

| | | | | | | |
|---|---|---|---|---|---|---|
| 67. $BeF_2$ | 68. $BeCl_2$ | 69. $MgF_2$ | 70. $MgCl_2$ | 71. $MgBr_2$ | 72. $CaF_2$ | 73. $CaCl_2$ |
| 74. $CaBr_2$ | 75. $CaI_2$ | 76. $SrF_2$ | 77. $SrCl_2$ | 78. $SrBr_2$ | 79. $SrI_2$ | 80. $BaF_2$ |
| 81. $BaCl_2$ | 82. $BaBr_2$ | 83. $BaI_2$ | 84. $ZnF_2$ | 85. $ZnCl_2$ | 86. $ZnBr_2$ | 87. $CdF_2$ |
| 88. $CdCl_2$ | 89. $CdBr_2$ | 90. $CdI_2$ | 91. $SnS_2$ | 92. $PbS_2$ | | |

## P4/mmm

| | | | | | | |
|---|---|---|---|---|---|---|
| 93. $SiF_4$ | 94. $GeF_4$ | 95. $SnF_4$ | 96. $PbF_4$ | 97. LiCl | 98. LiBr | 99. LiI |
| 100. NaBr | 101. NaI | 102. KBr | 103. $SrO_2$ | | | |

## P4/nmm

| | | | | | |
|---|---|---|---|---|---|
| 104. $Cd_2S_2$ | 105. $Ge_2O_2$ | 106. $Pb_2O_2$ | 107. $Sn_2O_2$ | 108. $Al_2N_2$ | 109. $Li_2(OH)_2$ |
| 110. $Na_2(OH)_2$ | 111. $Pb_2Br_2F_2$ | 112. $Sc_2Br_2O_2$ | 113. $Sr_2Br_2H_2$ | | |

## The rest

| | | | | | | |
|---|---|---|---|---|---|---|
| 114. $AlF_3$ | 115. $Al_4N_4$ | 116. $Ga_4N_4$ | 117. $B_2O_3$ | 118. $C_3N_4$ | 119. $In_2Cl_2O_2$ | 120. $In_2Br_2O_2$ |
| 121. $Mg_2Ge_2O_6$ | 122. $TiPbO_3$ | | | | | |

# 1. $BeF_2$

| MC2D-ID | C2DB | 2dmat-ID | USPEX | Space group | Band gap (eV) |
|---|---|---|---|---|---|
| - | - | 2dm-43 | - | P3m1 | 9.28 |

| Convex hull | Atomic structure | Atomic coordinates | Phonon dispersion curve |
|---|---|---|---|

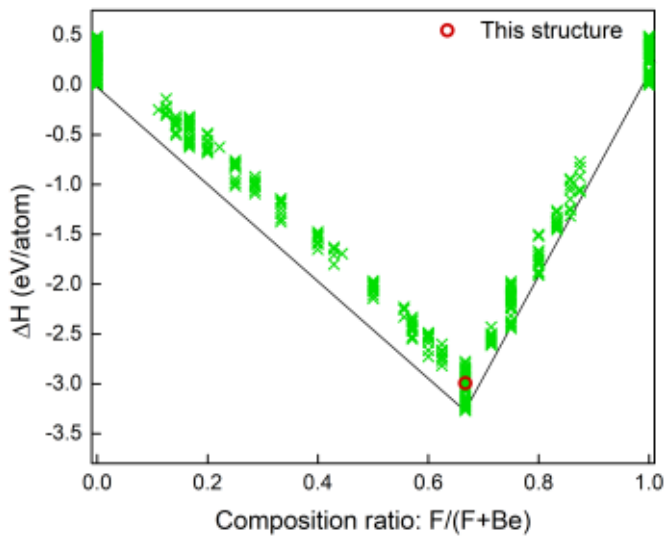

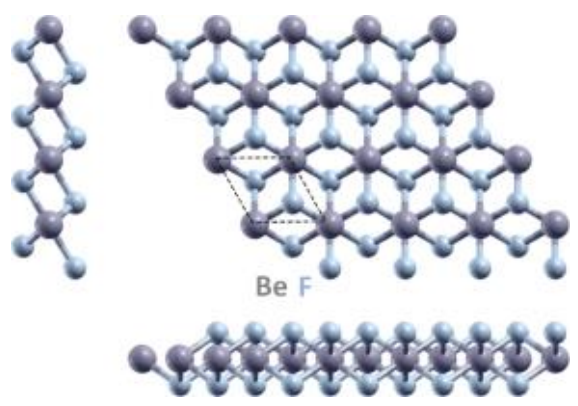

```
Be1F2
   1.00000000000000
   2.6145080605214228   0.0000000000000000   0.0000000000000000
  -1.3072541701761635   2.2642303665544037   0.0000000000000000
   0.0000000000000000   0.0000000000000000  23.5298450000000017
   Be   F
   1    2
Direct
  0.0000000084641769  0.9999999978600869  0.9999999098079044
  0.6666681049509435  0.3333319357115911  0.0384319141545930
  0.3333318865848796  0.6666680664283149  0.9615681760375026
```

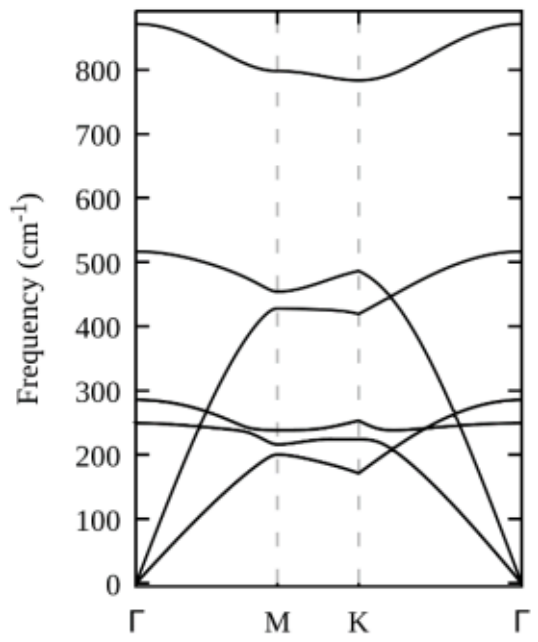

| Projected band structure and density of states | Magnetic moment and spin polarization energy as a function of hole doping concentration |
|---|---|

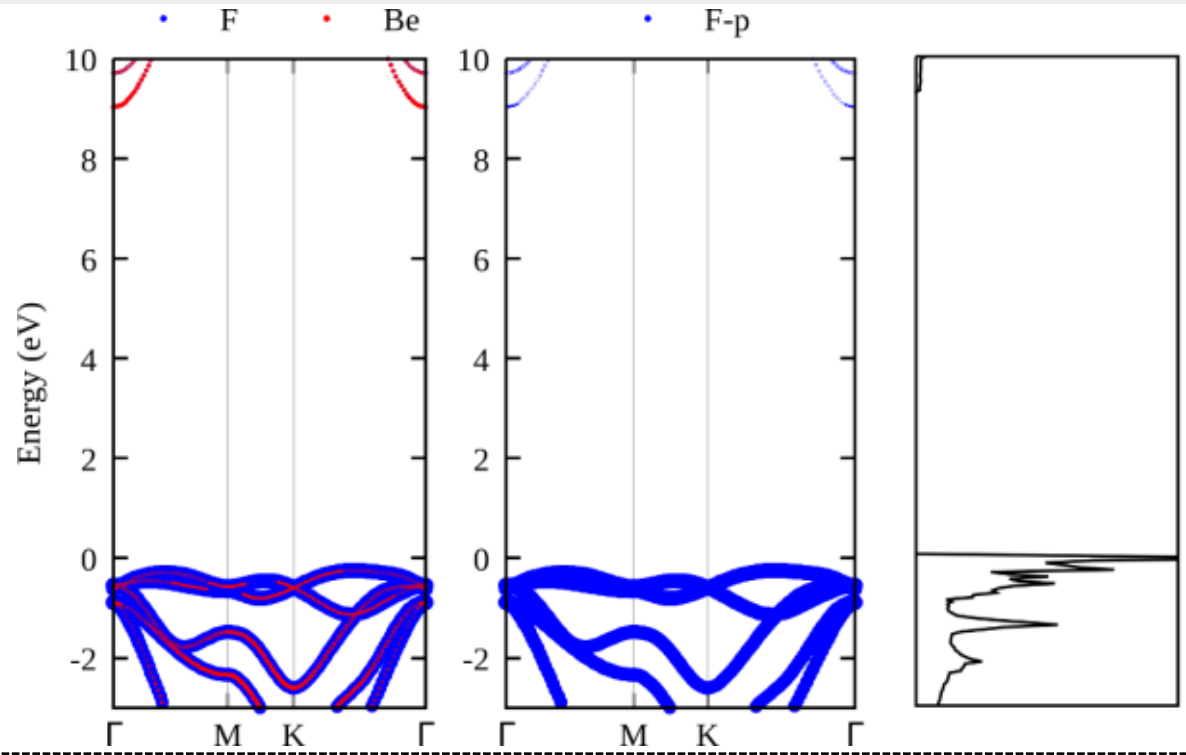

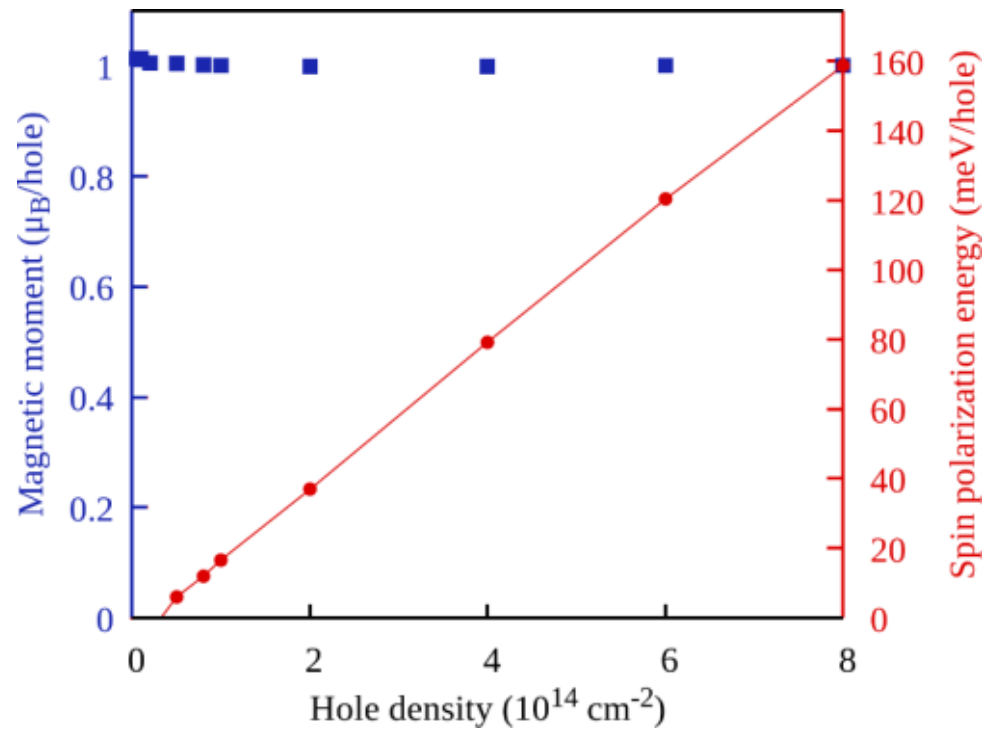

| Magnetic configurations and spin Hamiltonian | Magnetic exchange coupling parameters |
|---|---|

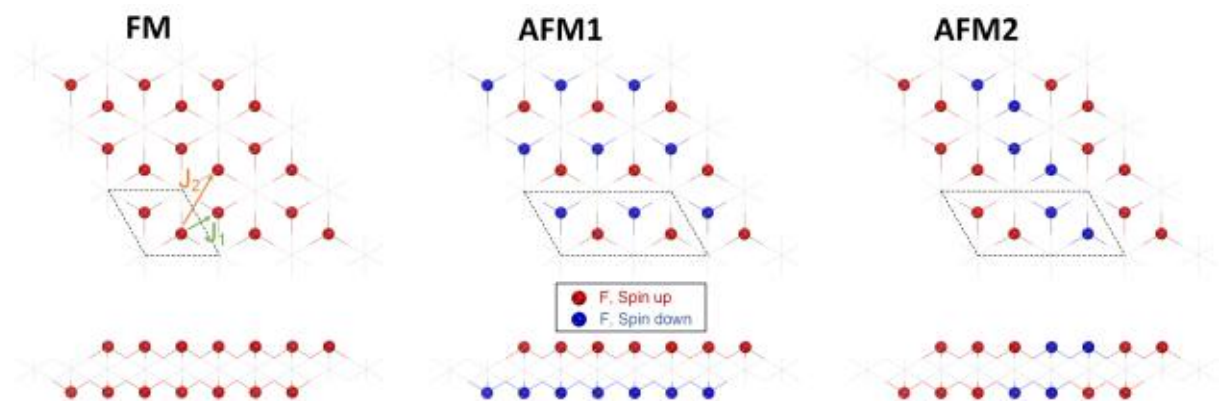

$$E_{FM} = E_0 - 3J_1S^2 - 6J_2S^2$$
$$E_{AFM1} = E_0 + 3J_1S^2 - 6J_2S^2$$
$$E_{AFM2} = E_0 - J_1S^2 + 2J_2S^2$$

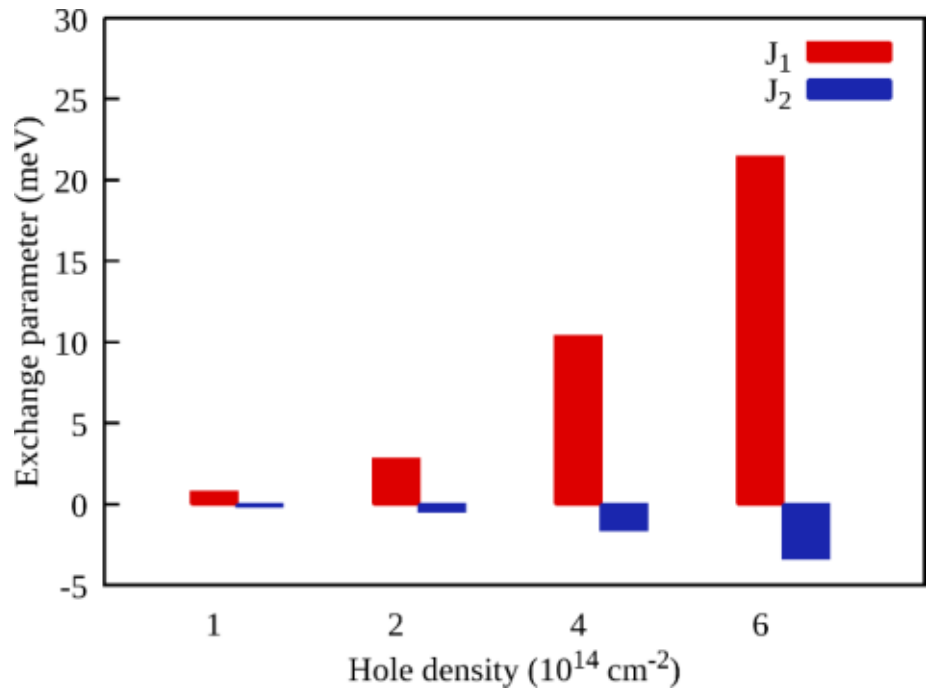

| Magnetic anisotropy energy (MAE, μeV) per magnetic atom | Monte Carlo simulations of the normalized magnetization of as a function of temperature |
|---|---|

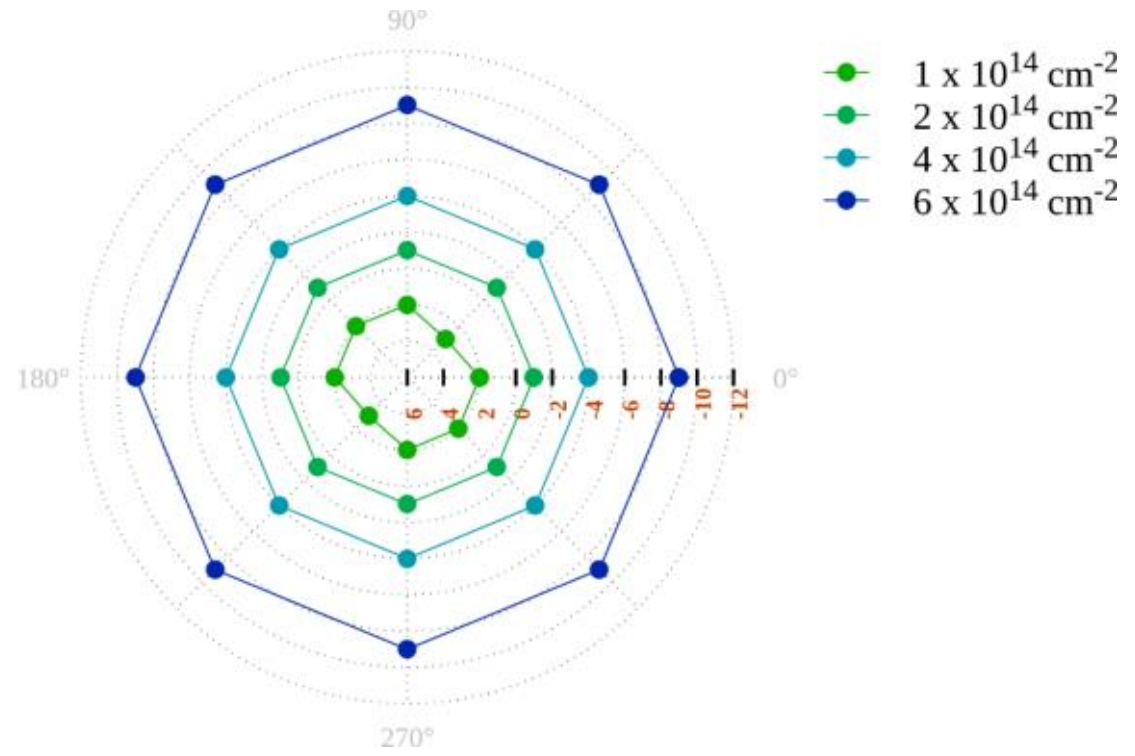

MAE = $E_{\parallel} - E_{\perp}$, a positive (negative) value of MAE indicates the off-plane (in-plane) easy axis.

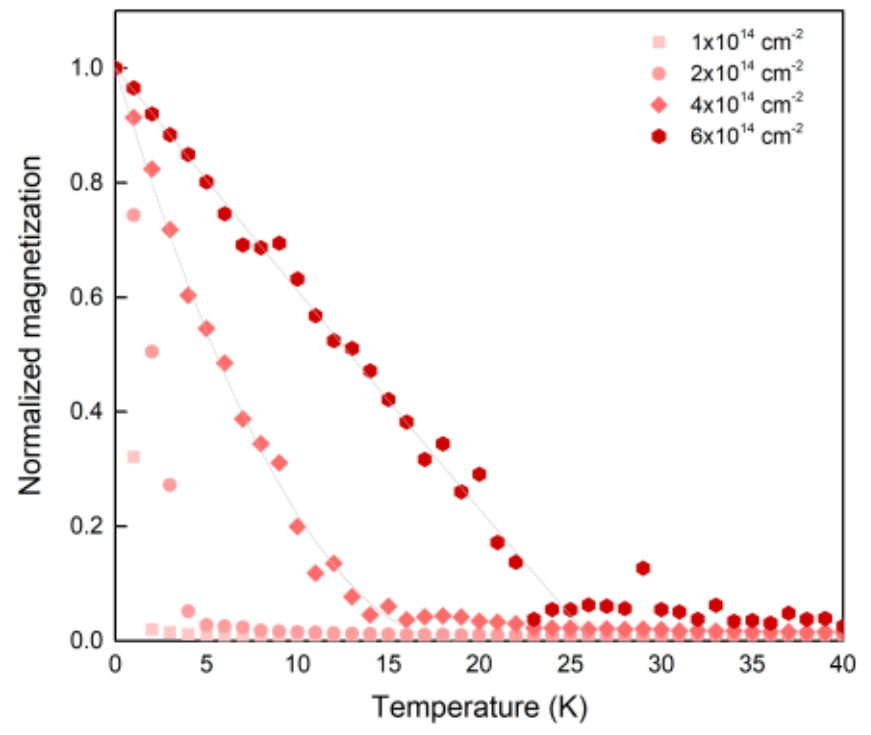

$T_c$: $1\times10^{14}$ cm$^{-2}$: 2 K  $2\times10^{14}$ cm$^{-2}$: 3 K  $4\times10^{14}$ cm$^{-2}$: 15 K  $6\times10^{14}$ cm$^{-2}$: 26 K

# 2. $MgF_2$

| MC2D-ID | C2DB | 2dmat-ID | USPEX | Space group | Band gap (eV) |
|---|---|---|---|---|---|
| - | - | 2dm-1016 | - | P3m1 | 7.43 |

**Convex hull** | **Atomic structure** | **Atomic coordinates** | **Phonon dispersion curve**

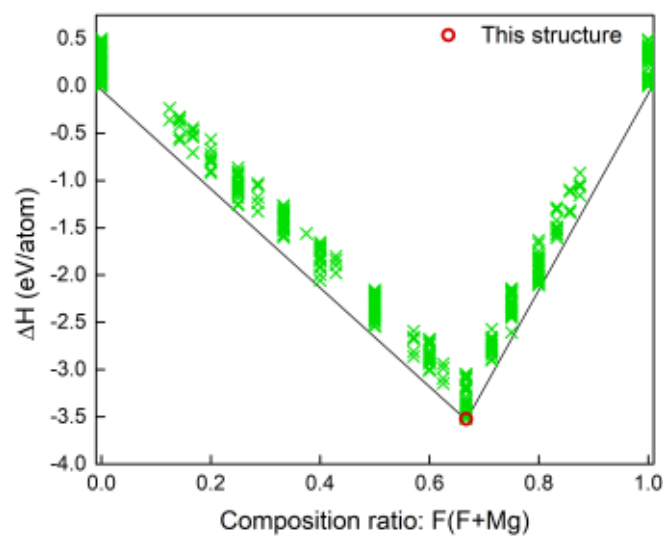

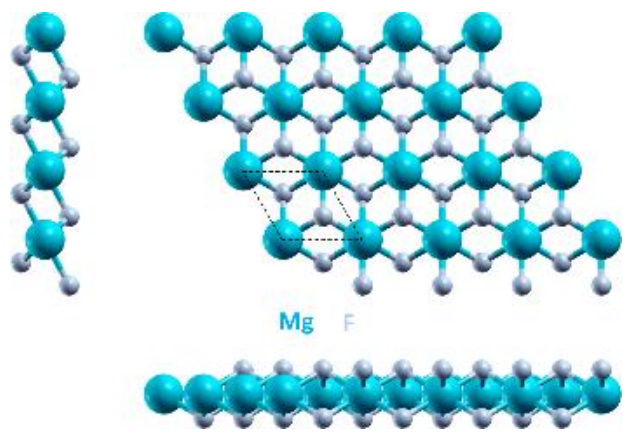

```
Mg1F2
   1.00000000000000
   3.1052391845157428    0.0000000000000000    0.0000000000000000
  -1.5526285592966793    2.6892109035038221    0.0000000000000000
   0.0000000000000000    0.0000000000000000   23.5298450000000017
   Mg    F
   1     2
Direct
  0.9999999999447198  0.9999999999375007  0.0000000006161187
  0.6666669873034792  0.3333329995349033  0.0399737101175219
  0.3333330127518010  0.6666670005275961  0.9600262892663594
```

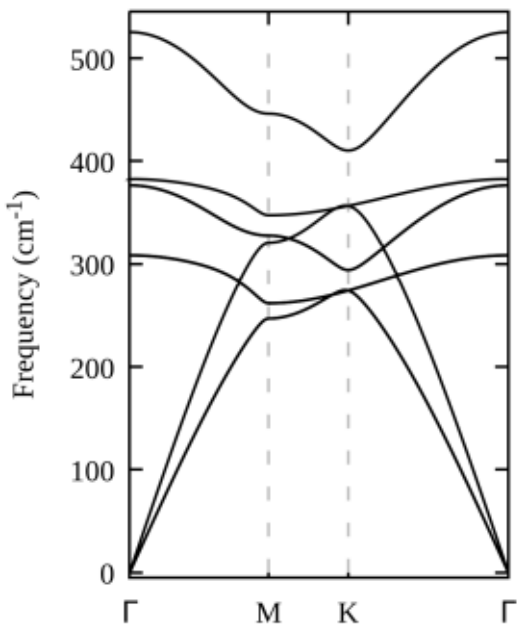

**Projected band structure and density of states** | **Magnetic moment and spin polarization energy as a function of hole doping concentration**

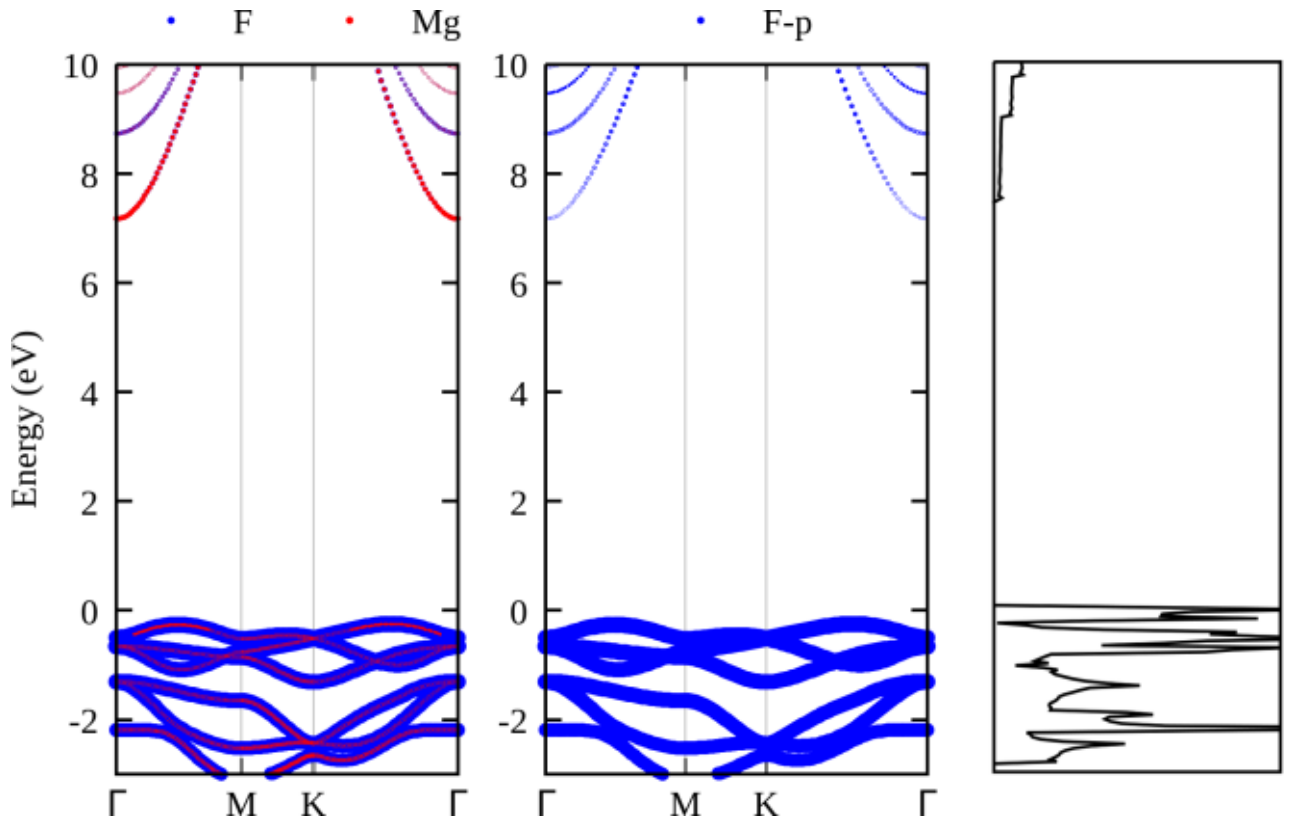

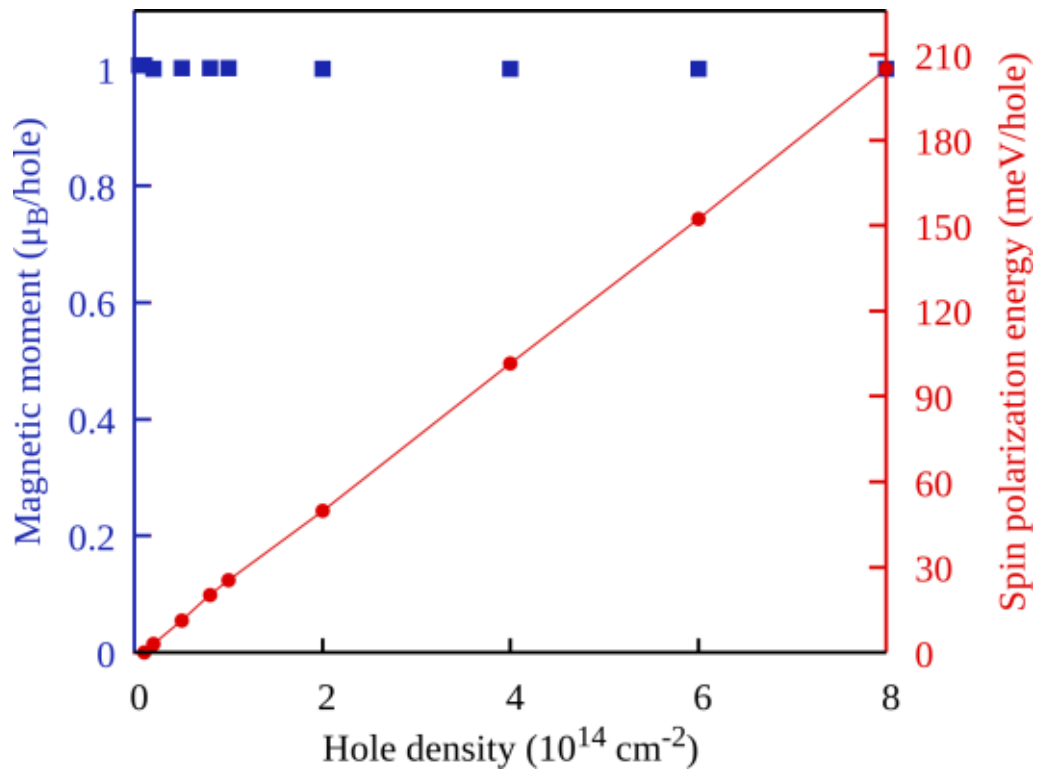

**Magnetic configurations and spin Hamiltonian** | **Magnetic exchange coupling parameters**

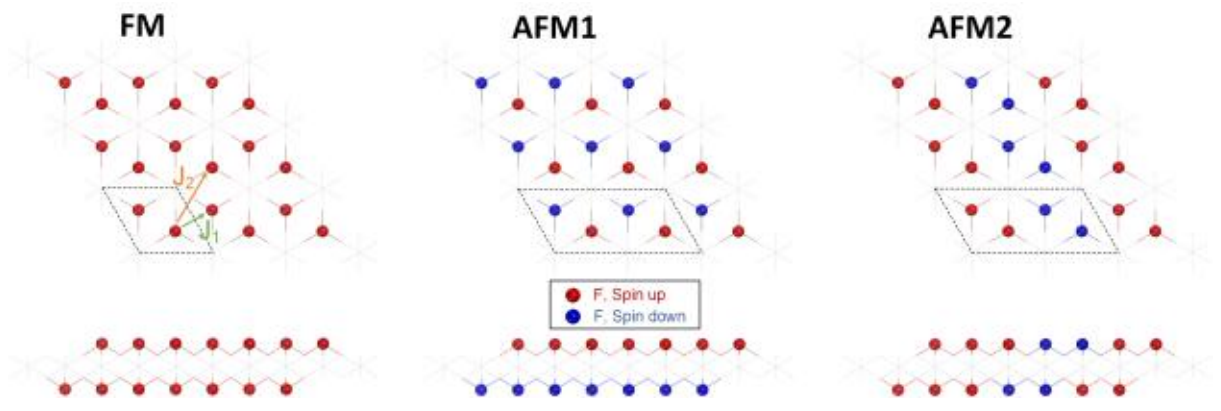

$$E_{FM} = E_0 - 3J_1S^2 - 6J_2S^2$$
$$E_{AFM1} = E_0 + 3J_1S^2 - 6J_2S^2$$
$$E_{AFM2} = E_0 - J_1S^2 + 2J_2S^2$$

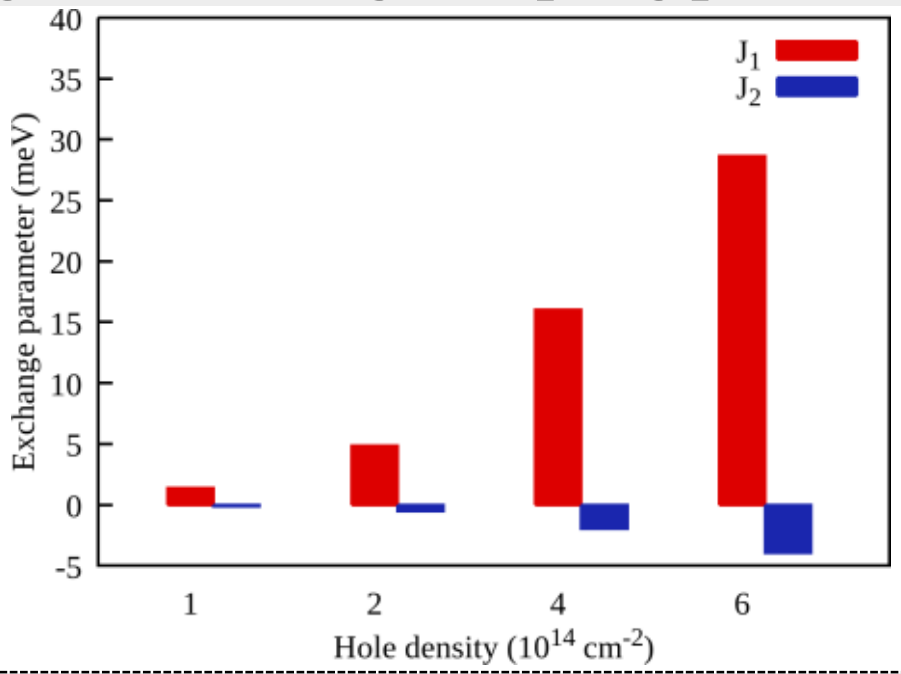

**Magnetic anisotropy energy (MAE, μeV) per magnetic atom** | **Monte Carlo simulations of the normalized magnetization of as a function of temperature**

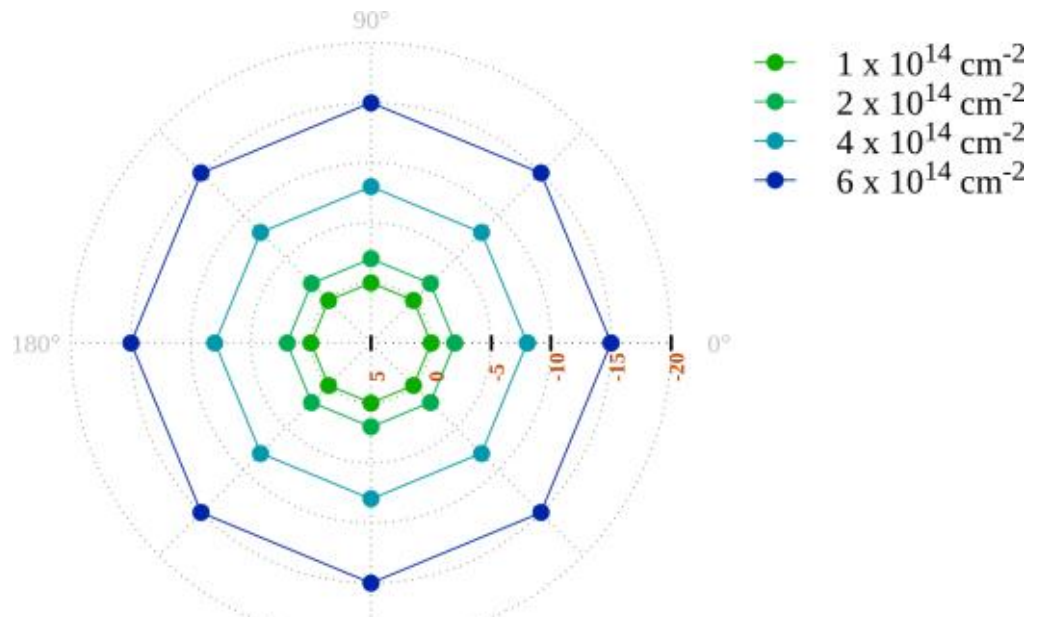

MAE = $E_{\parallel} - E_{\perp}$, a positive (negative) value of MAE indicates the off-plane (in-plane) easy axis.

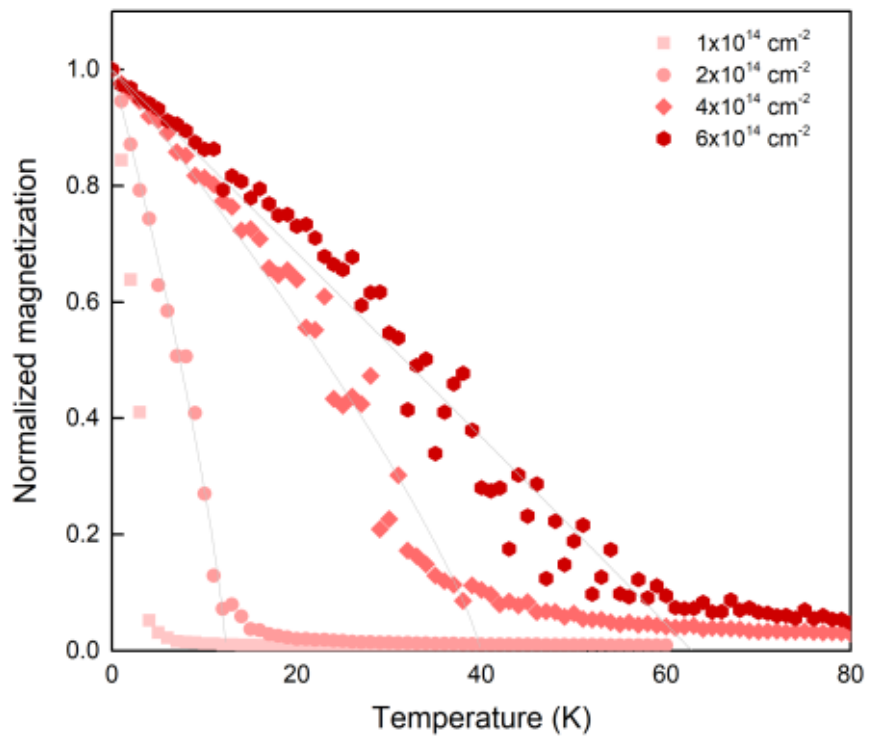

$T_c$: $1\times10^{14}$ cm$^{-2}$: 4 K  $2\times10^{14}$ cm$^{-2}$: 12 K  $4\times10^{14}$ cm$^{-2}$: 40 K  $6\times10^{14}$ cm$^{-2}$: 63 K

# 3. $MgCl_2$

| MC2D-ID | C2DB | 2dmat-ID | USPEX | Space group | Band gap (eV) |
|---|---|---|---|---|---|
| 118 | ✓ | 2dm-3734 | - | P3m1 | 6.00 |

**Convex hull**

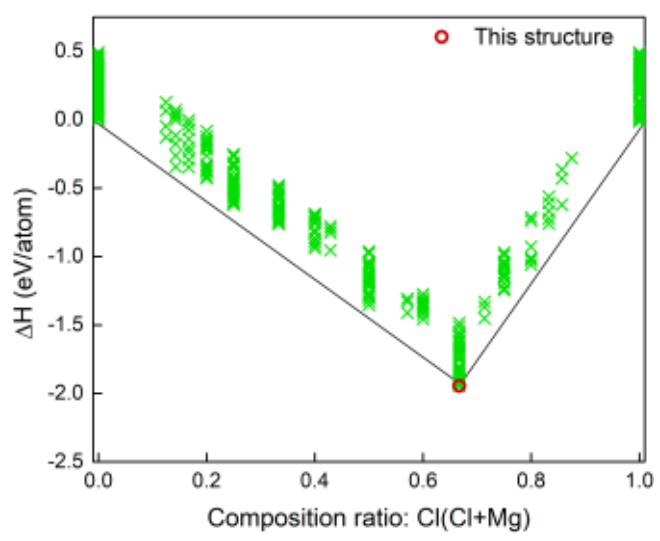

**Atomic structure**

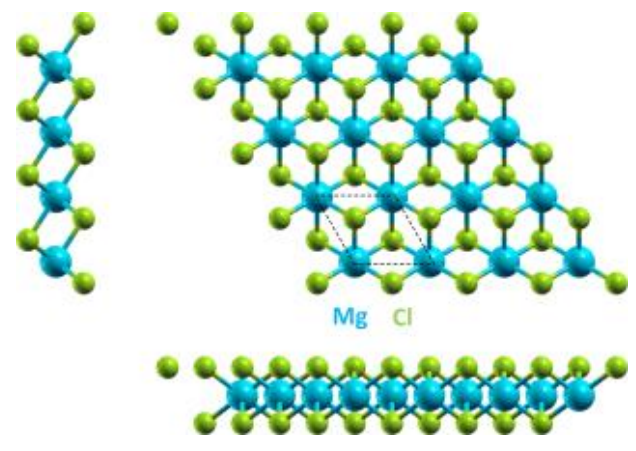

**Atomic coordinates**

Mg1Cl2
1.0000000000000
3.6665835364464066 0.0000000000000000 0.0000000000000000
-1.8332905255061329 3.1753551780468938 0.0000000000000000
0.0000000000000000 0.0000000000000000 22.7262264346999991
Mg Cl
1 2
Direct
0.6666666683536562 0.3333333367887548 0.4999999819558667
0.0000000213911591 0.0000000059911471 0.5603683136885778
0.3333333102710228 0.6666666572517741 0.4396317043577582

**Phonon dispersion curve**

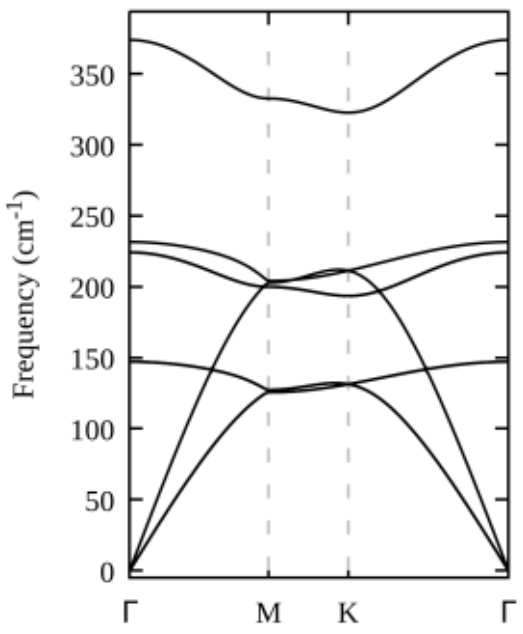

**Projected band structure and density of states**

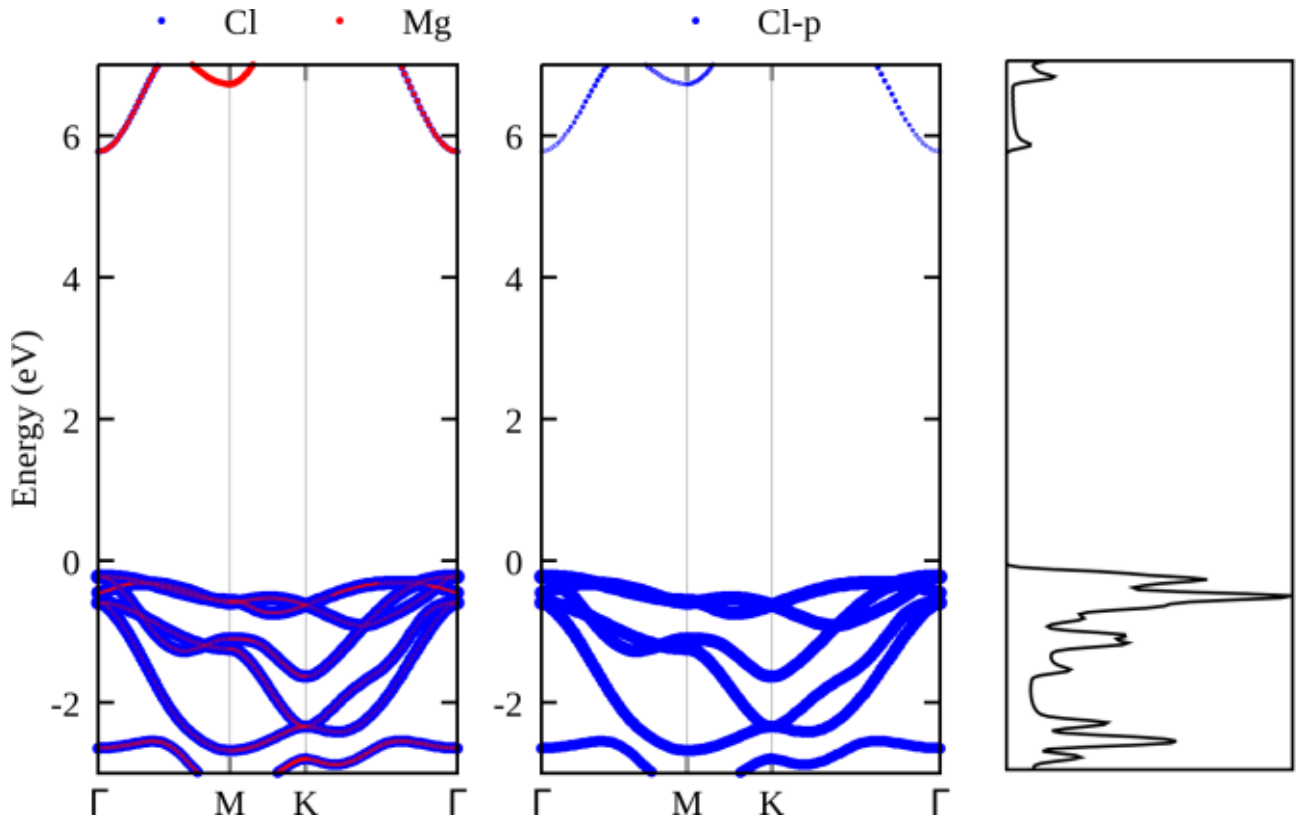

**Magnetic moment and spin polarization energy as a function of hole doping concentration**

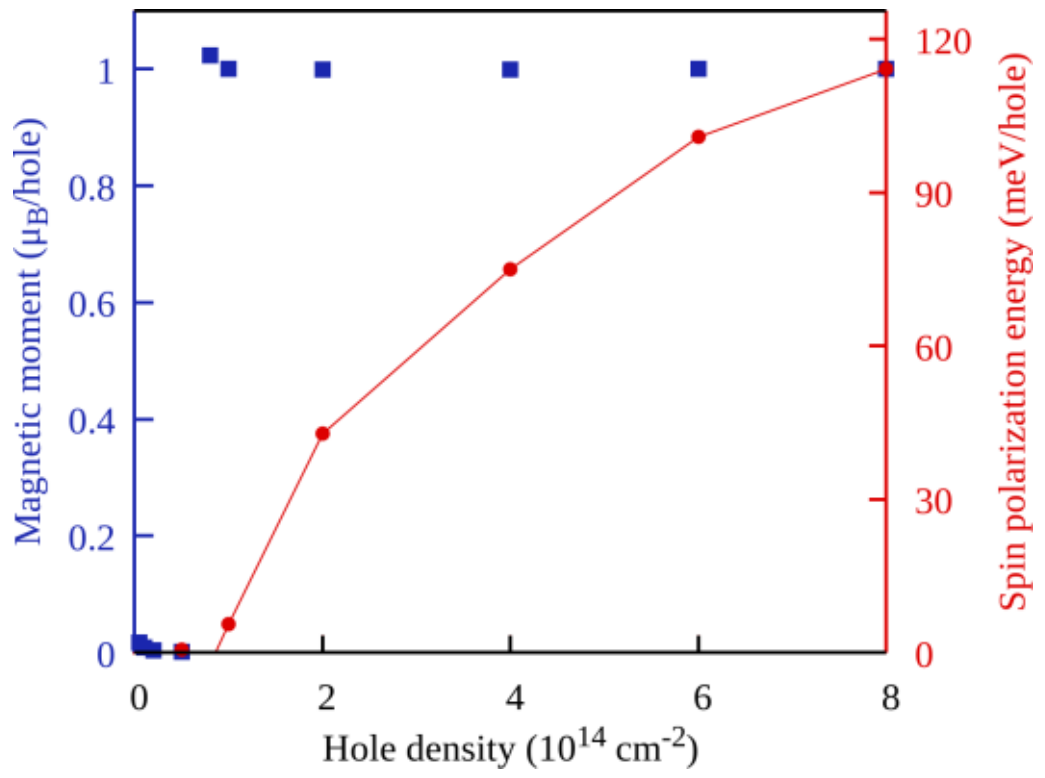

**Magnetic configurations and spin Hamiltonian**

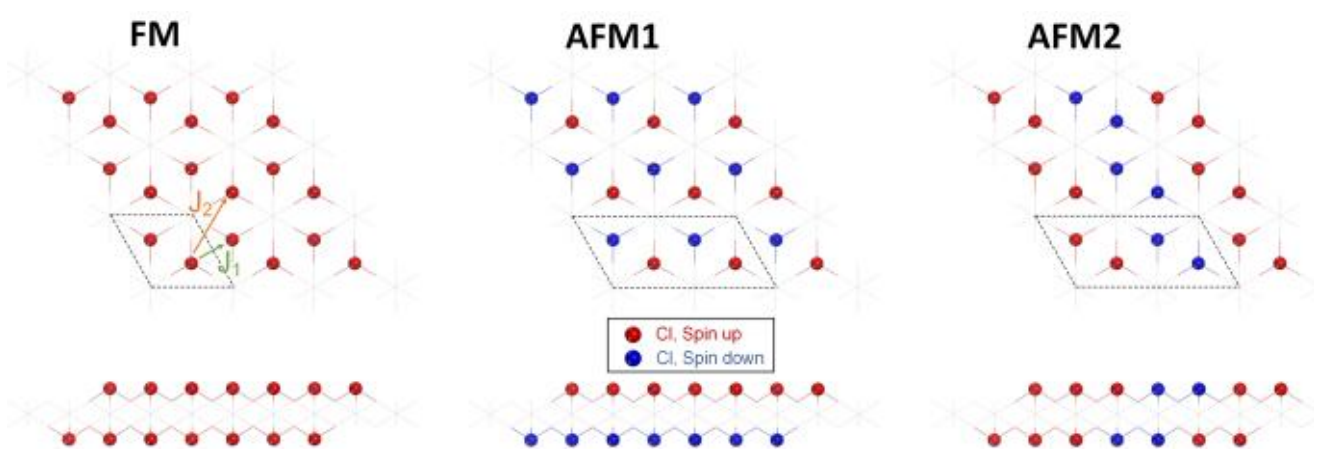

$$E_{FM} = E_0 - 3J_1S^2 - 6J_2S^2$$
$$E_{AFM1} = E_0 + 3J_1S^2 - 6J_2S^2$$
$$E_{AFM2} = E_0 - J_1S^2 + 2J_2S^2$$

**Magnetic exchange coupling parameters**

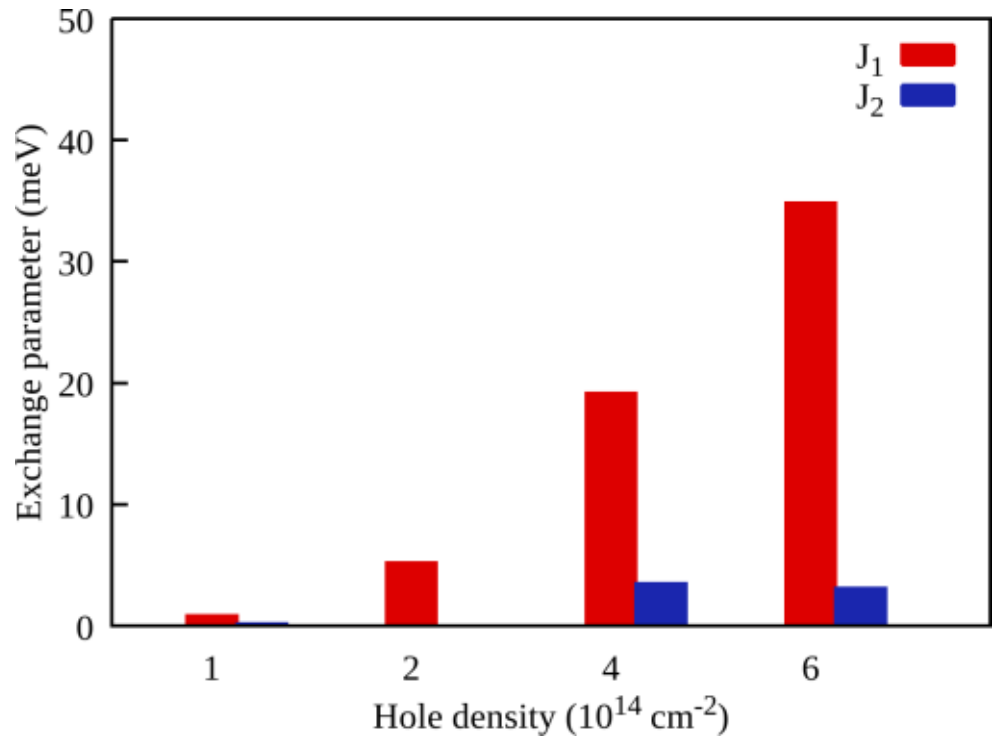

**Magnetic anisotropy energy (MAE, μeV) per magnetic atom**

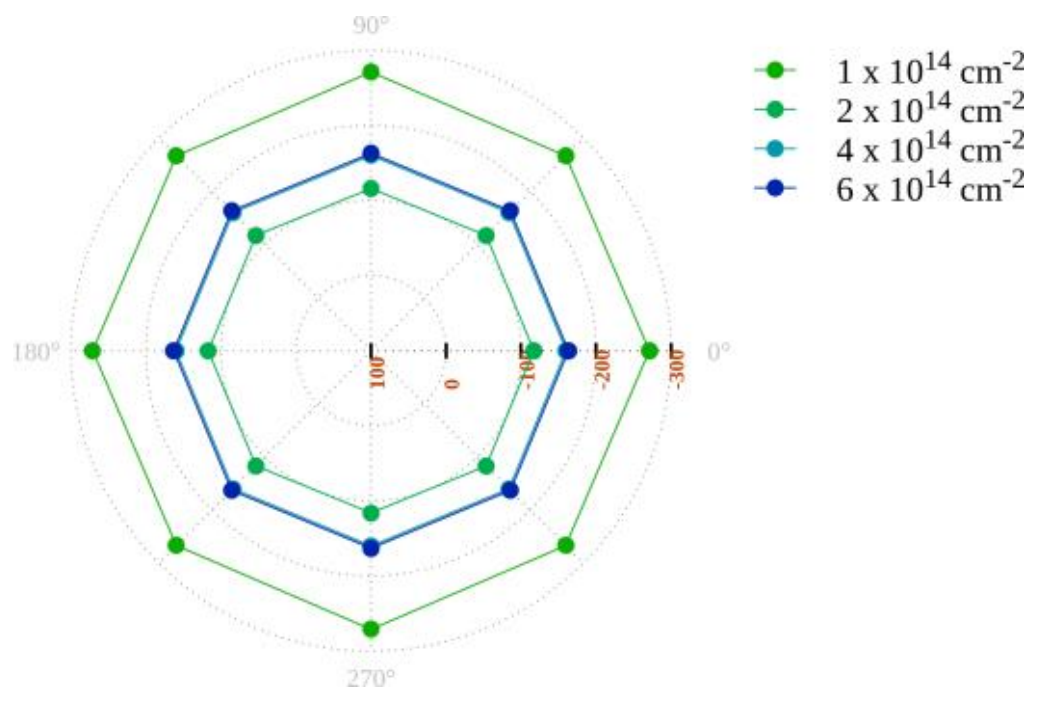

MAE = $E_{\parallel} - E_{\perp}$, a positive (negative) value of MAE indicates the off-plane (in-plane) easy axis.

**Monte Carlo simulations of the normalized magnetization of as a function of temperature**

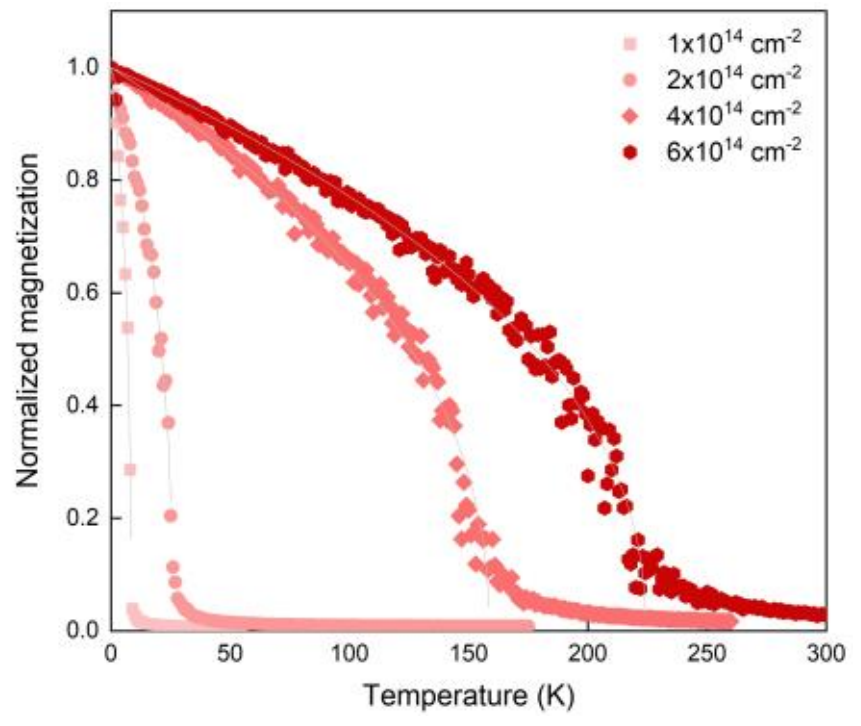

$T_c$: $1\times10^{14}$ cm$^{-2}$: 8 K  $2\times10^{14}$ cm$^{-2}$: 26 K  $4\times10^{14}$ cm$^{-2}$: 158 K  $6\times10^{14}$ cm$^{-2}$: 224 K

# 4. $CaF_2$

| MC2D-ID | C2DB | 2dmat-ID | USPEX | Space group | Band gap (eV) |
| --- | --- | --- | --- | --- | --- |
| - | - | 2dm-253 | - | P3m1 | 7.05 |

**Convex hull** | **Atomic structure** | **Atomic coordinates** | **Phonon dispersion curve**

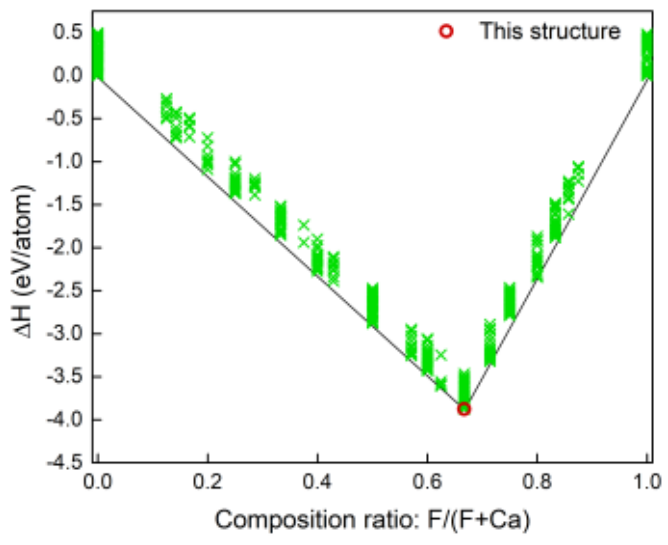

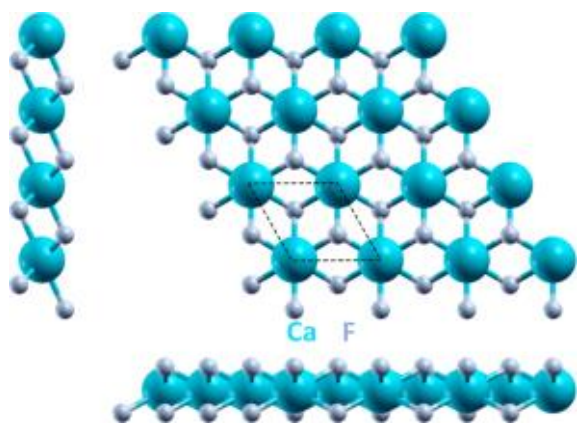

```
Ca1F2
1.00000000000000
3.6146954836474197 0.0000000000000000 0.0000000000000000
-1.8073574282525919 3.1304124666103319 0.0000000000000000
0.0000000000000000 0.0000000000000000 23.0162860000000009
Ca F
1 2
Direct
0.9999999944597491 0.9999999764067269 0.0000000056559131
0.6666661430052940 0.3333340805736000 0.0416578000119898
0.3333338625349569 0.6666659430196731 0.9583421943320971
```

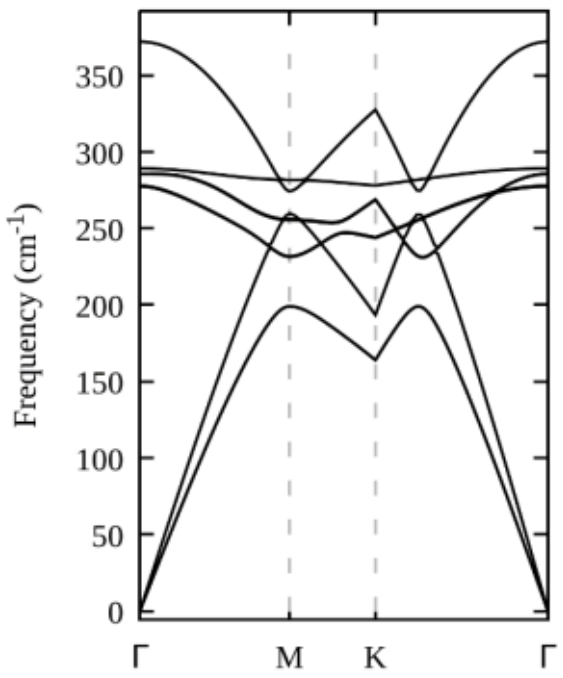

**Projected band structure and density of states** | **Magnetic moment and spin polarization energy as a function of hole doping concentration**

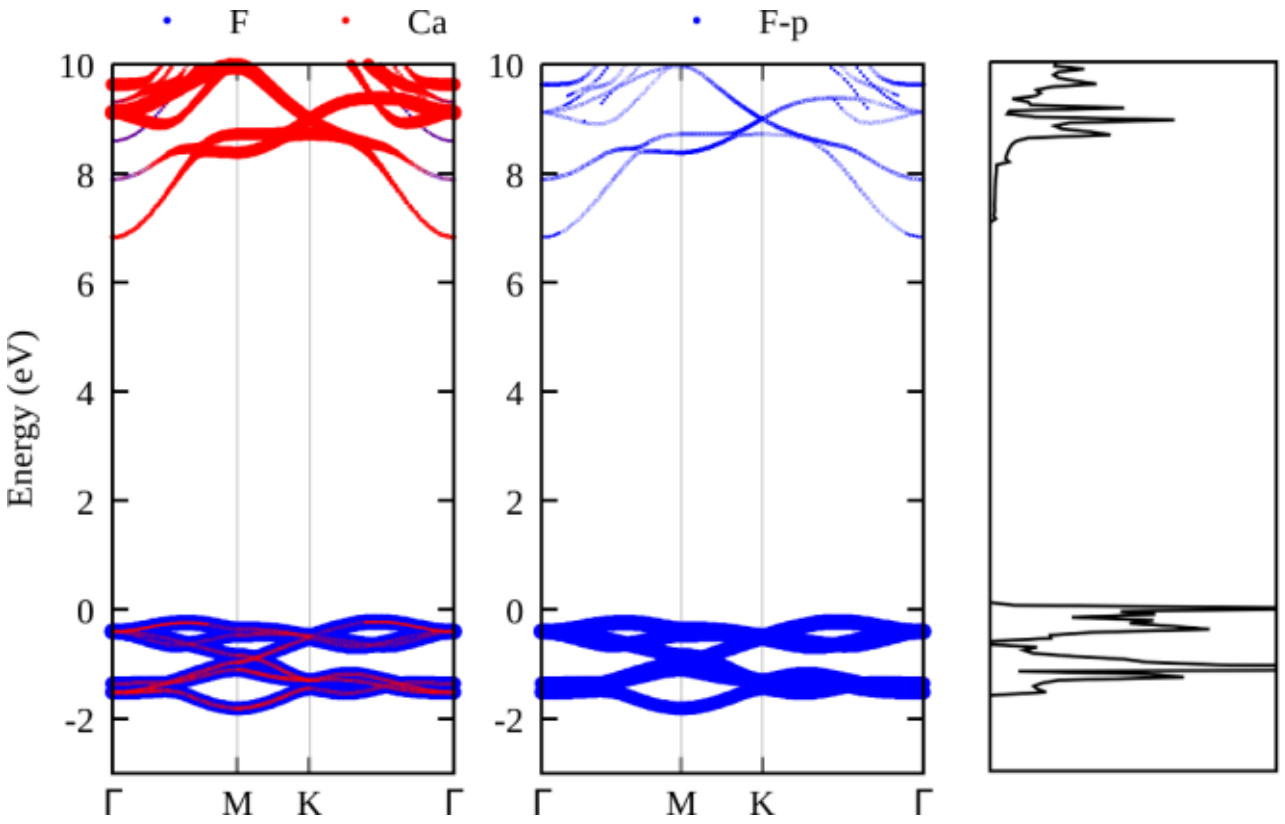

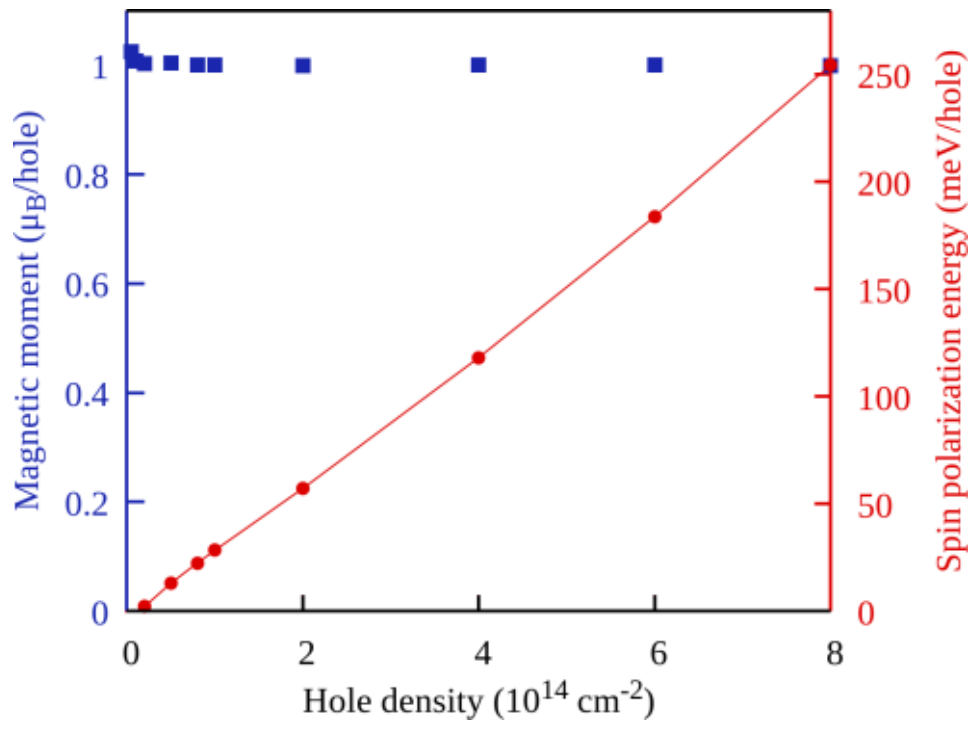

**Magnetic configurations and spin Hamiltonian** | **Magnetic exchange coupling parameters**

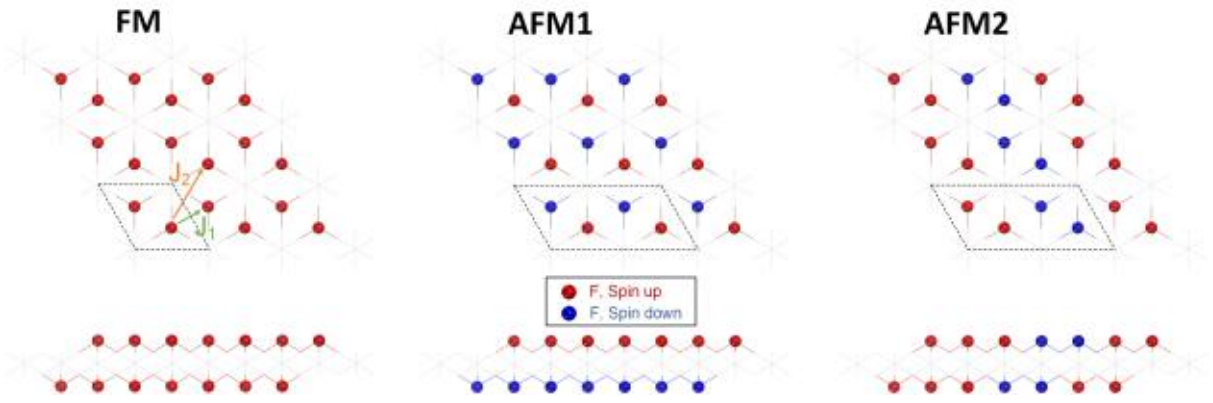

$$E_{FM} = E_0 - 3J_1S^2 - 6J_2S^2$$
$$E_{AFM1} = E_0 + 3J_1S^2 - 6J_2S^2$$
$$E_{AFM2} = E_0 - J_1S^2 + 2J_2S^2$$

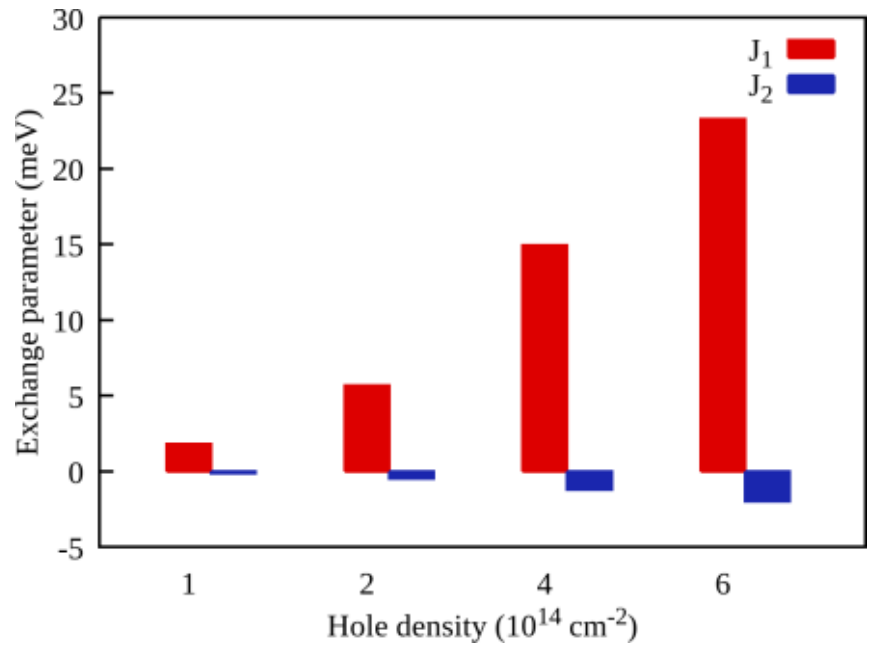

**Magnetic anisotropy energy (MAE, μeV) per magnetic atom** | **Monte Carlo simulations of the normalized magnetization of as a function of temperature**

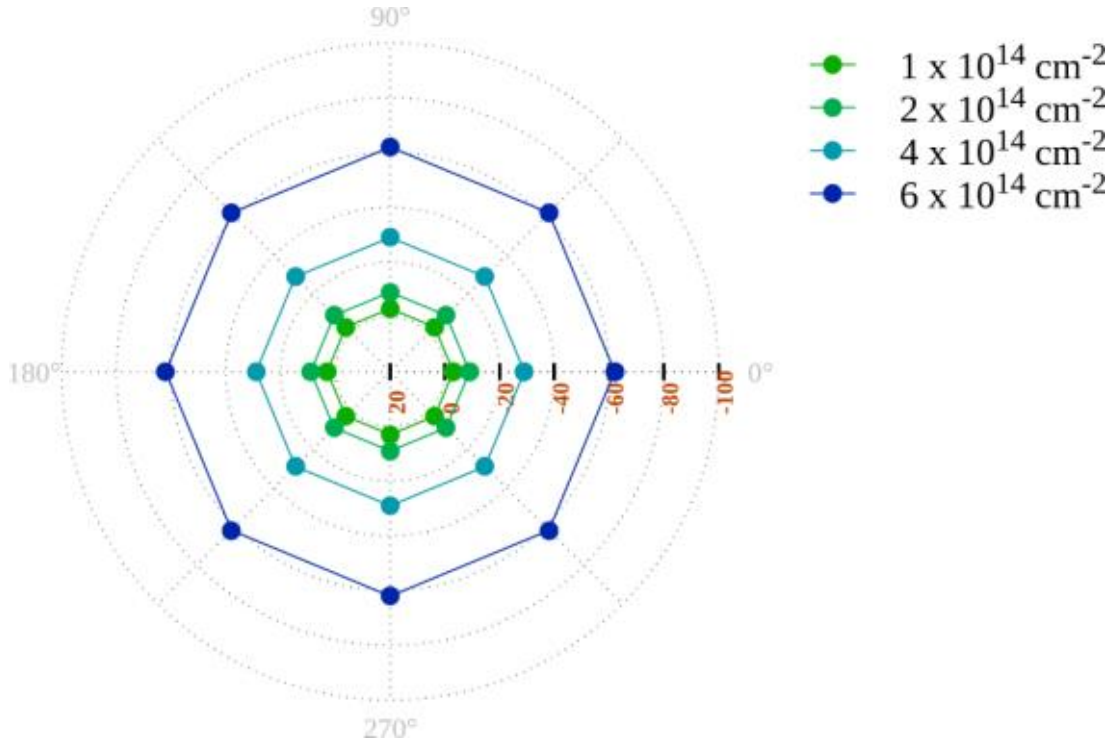

MAE = $E_\parallel - E_\perp$, a positive (negative) value of MAE indicates the off-plane (in-plane) easy axis.

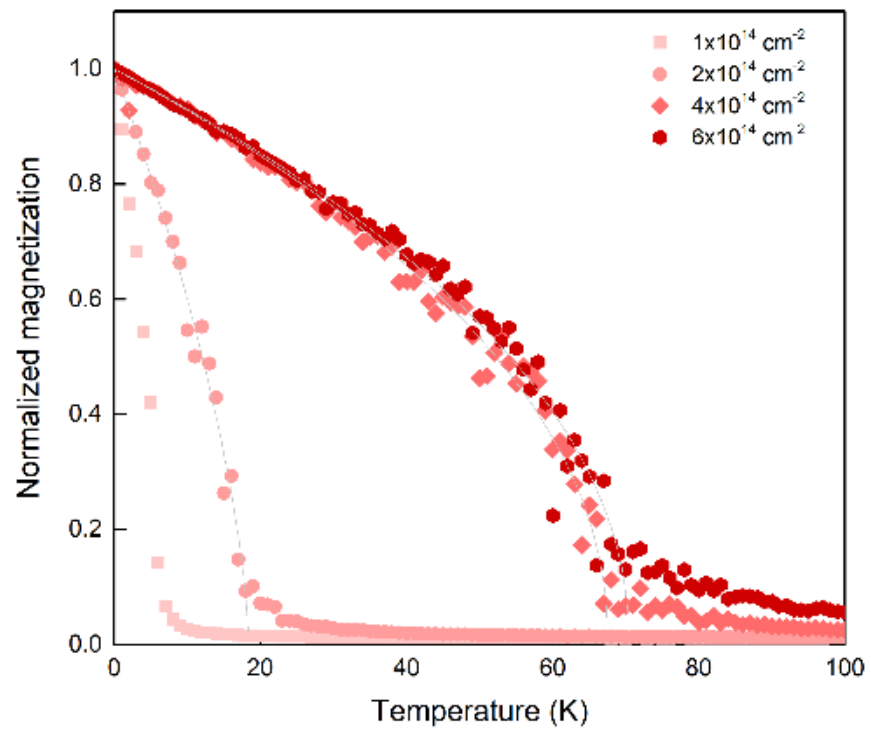

$T_c$: $1\times10^{14}$ cm$^{-2}$: 7 K  $2\times10^{14}$ cm$^{-2}$: 18 K  $4\times10^{14}$ cm$^{-2}$: 67 K  $6\times10^{14}$ cm$^{-2}$: 70 K

# 5. $CaCl_2$

| MC2D-ID | C2DB | 2dmat-ID | USPEX | Space group | Band gap (eV) |
|---|---|---|---|---|---|
| - | - | 2dm-290 | - | P3m1 | 5.80 |

**Convex hull** | **Atomic structure** | **Atomic coordinates** | **Phonon dispersion curve**

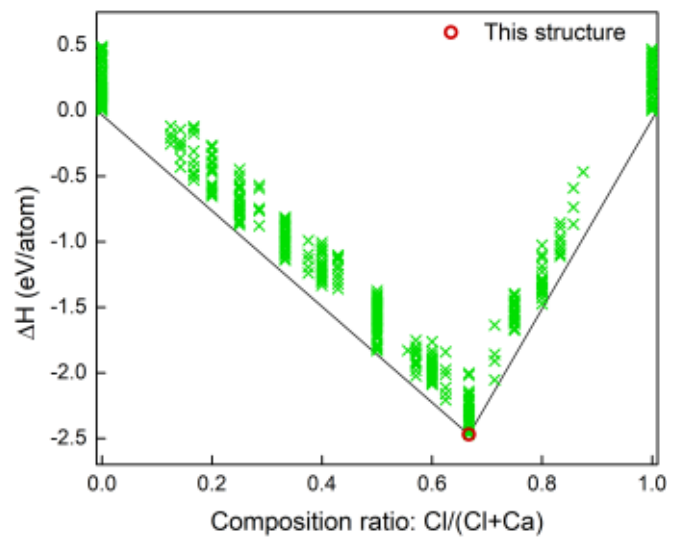

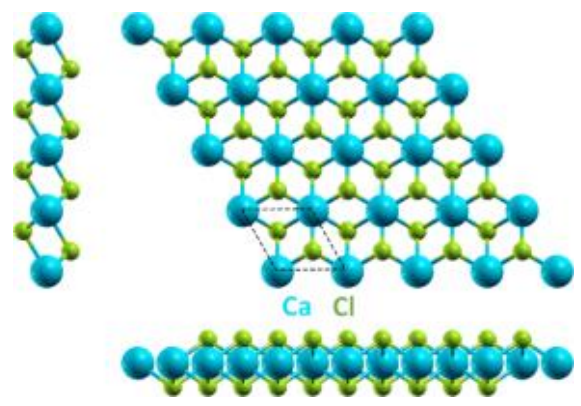

```
Ca1Cl2
   1.0000000000000
   4.1328322659408245   0.0000000000000000   0.0000000000000000
  -2.0664142106727197   3.5791387990563859   0.0000000000000000
   0.0000000000000000   0.0000000000000000   23.0162860000000009
   Ca   Cl
   1    2
Direct
  0.0000000041059991   0.0000000017533566   0.9999999944880287
  0.6666666435974022   0.3333333674675671   0.0609587123737469
  0.3333333522965987   0.6666666307790834   0.9390412931382244
```

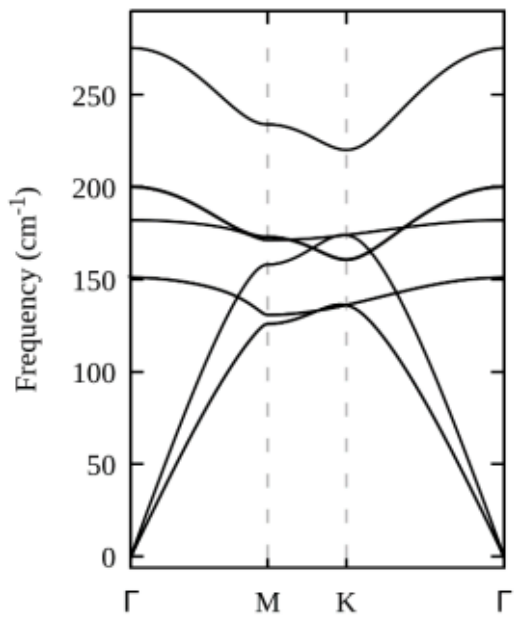

**Projected band structure and density of states** | **Magnetic moment and spin polarization energy as a function of hole doping concentration**

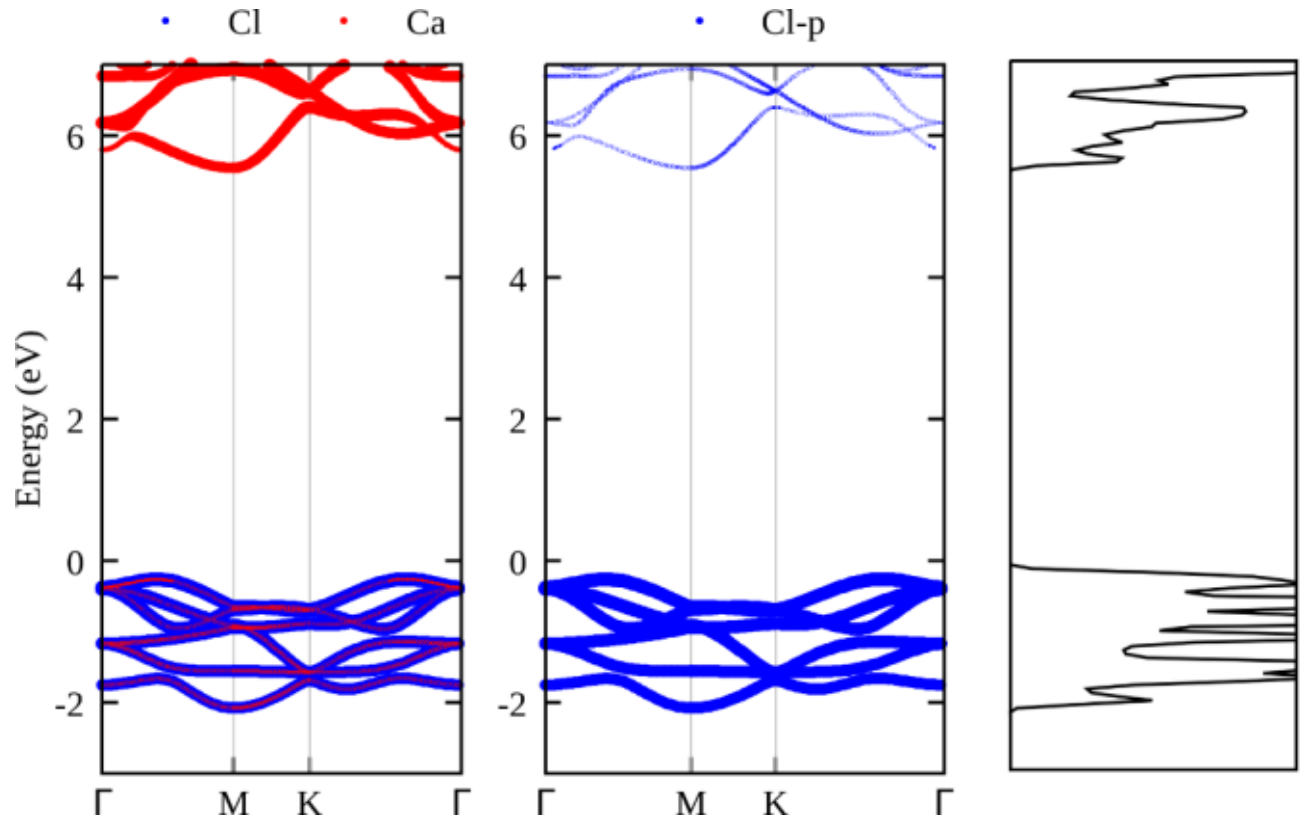

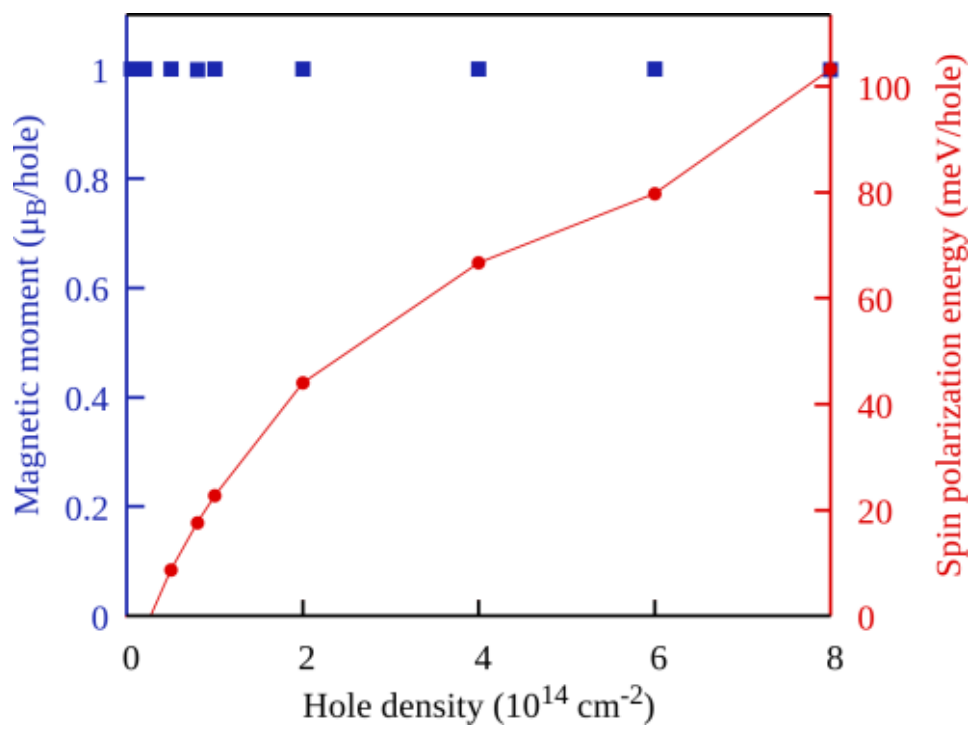

**Magnetic configurations and spin Hamiltonian** | **Magnetic exchange coupling parameters**

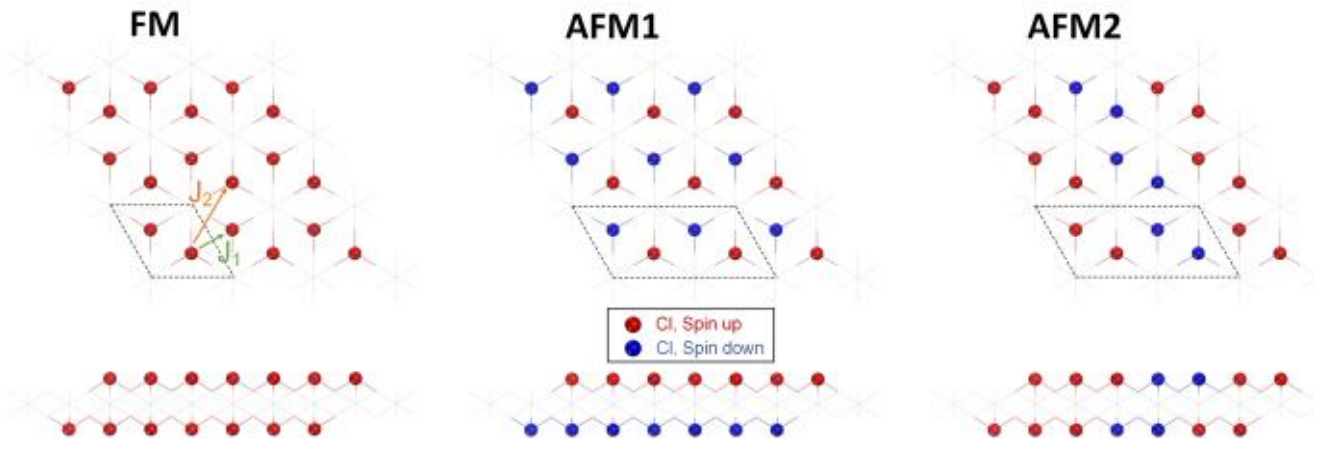

$$E_{FM} = E_0 - 3J_1S^2 - 6J_2S^2$$
$$E_{AFM1} = E_0 + 3J_1S^2 - 6J_2S^2$$
$$E_{AFM2} = E_0 - J_1S^2 + 2J_2S^2$$

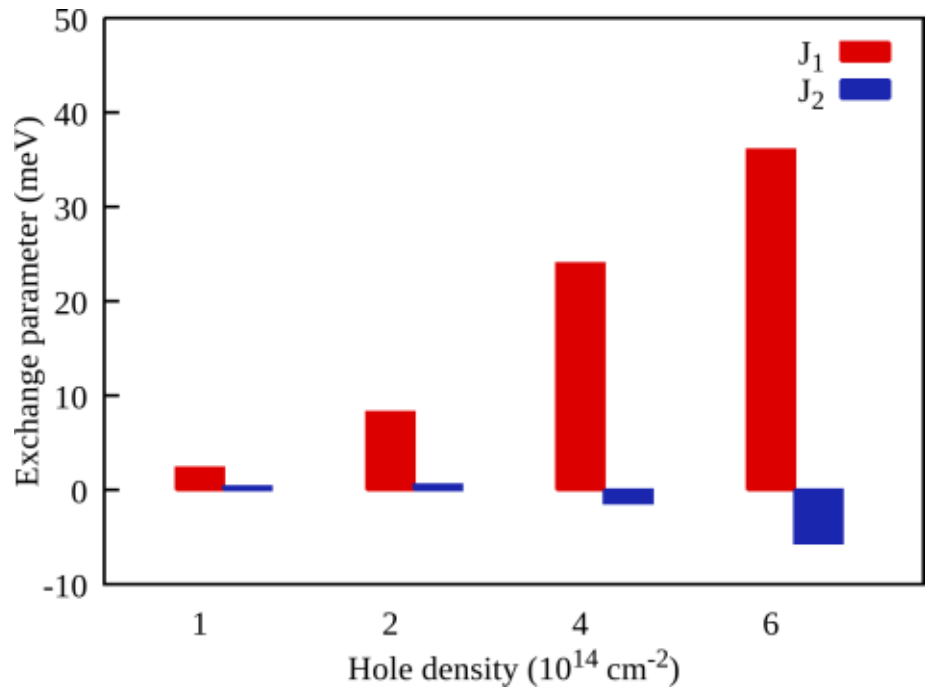

**Magnetic anisotropy energy (MAE, μeV) per magnetic atom** | **Monte Carlo simulations of the normalized magnetization of as a function of temperature**

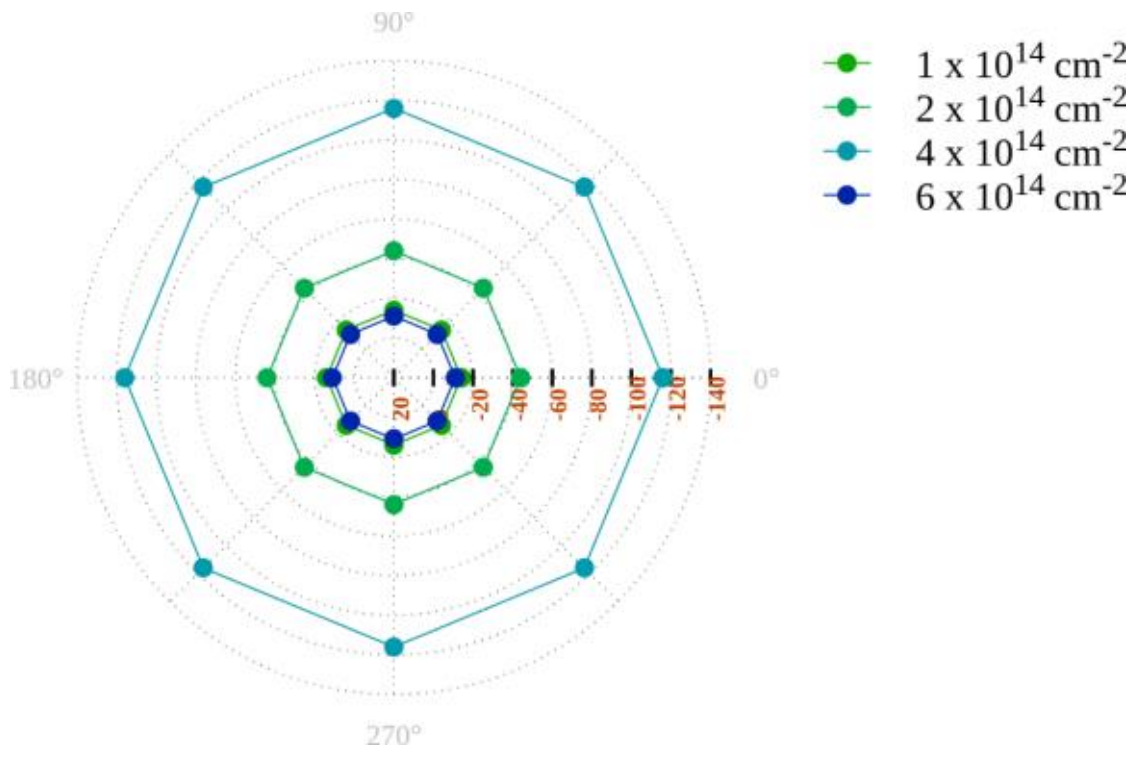

MAE = $E_{\parallel} - E_{\perp}$, a positive (negative) value of MAE indicates the off-plane (in-plane) easy axis.

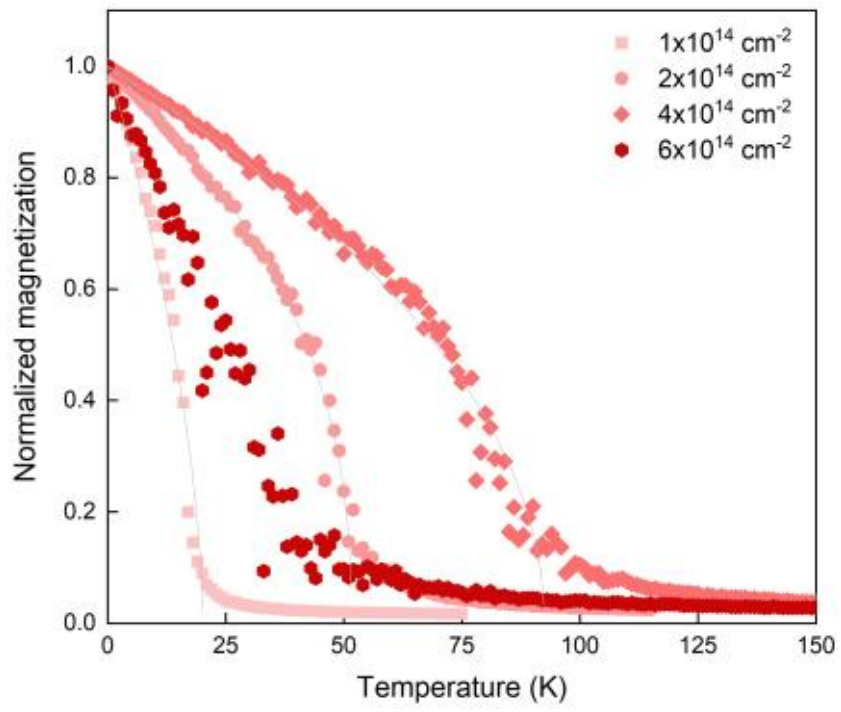

$T_c$: $1\times10^{14}$ cm$^{-2}$: 20 K  $2\times10^{14}$ cm$^{-2}$: 52 K  $4\times10^{14}$ cm$^{-2}$: 92 K  $6\times10^{14}$ cm$^{-2}$: 49 K

# 6. $CaBr_2$

| MC2D-ID | C2DB | 2dmat-ID | USPEX | Space group | Band gap (eV) |
|---|---|---|---|---|---|
| - | - | 2dm-618 | - | P3m1 | 5.02 |

**Convex hull** | **Atomic structure** | **Atomic coordinates** | **Phonon dispersion curve**

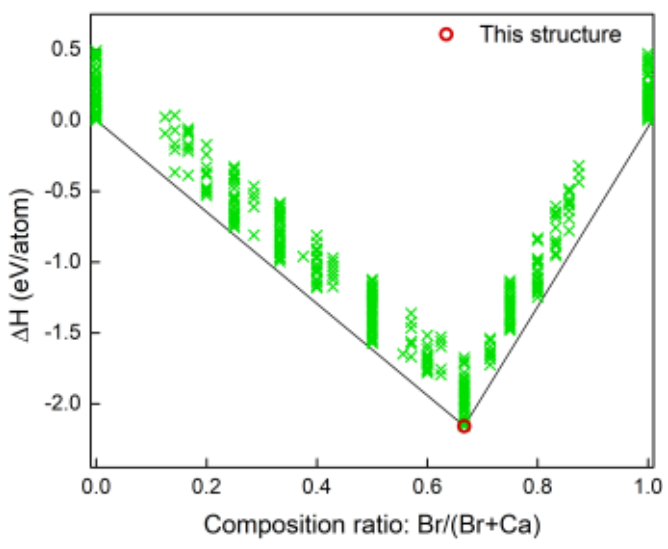

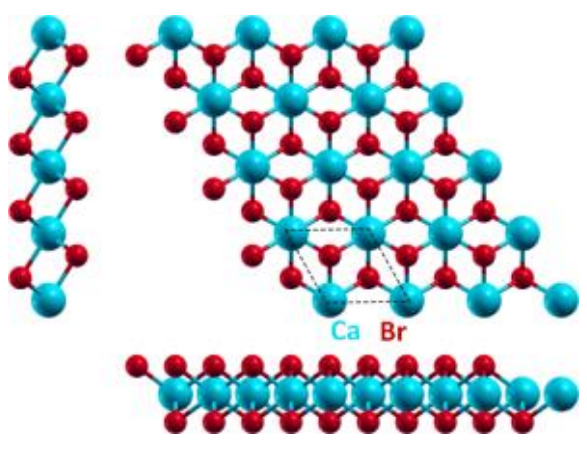

```
Ca1Br2
   1.00000000000000
   4.2897231099517441   0.0000000000000000   0.0000000000000000
  -2.1448649768165291   3.7150068475578428   0.0000000000000000
   0.0000000000000000   0.0000000000000000  22.7427169999999990
   Ca   Br
    1    2
Direct
  0.9999999673048805  0.0000000562025804  0.9999999967490183
  0.6666668435005079  0.3333332155573601  0.9312910311879961
  0.3333331891946187  0.6666667282400596  0.0687089720629928
```

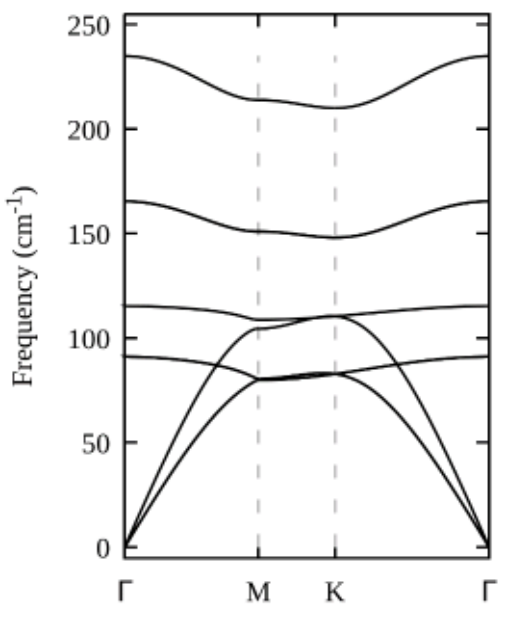

## Projected band structure and density of states

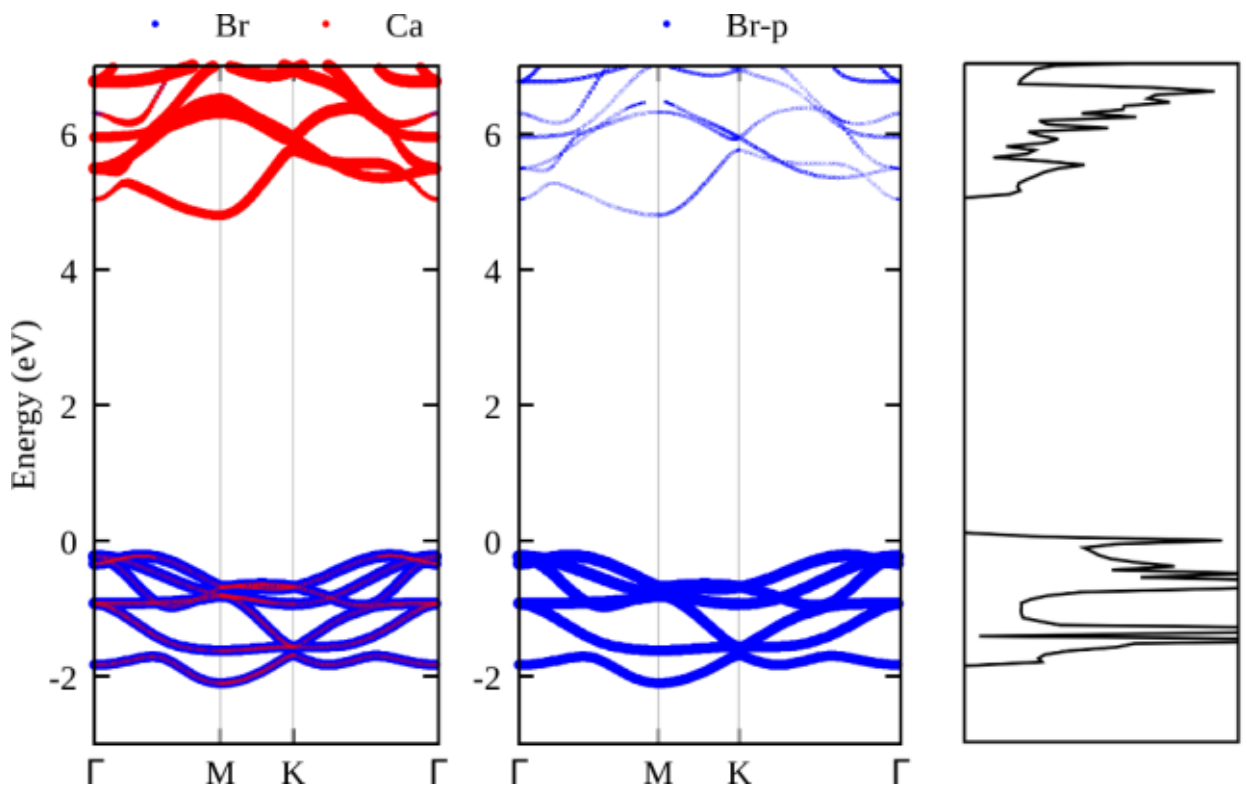

## Magnetic moment and spin polarization energy as a function of hole doping concentration

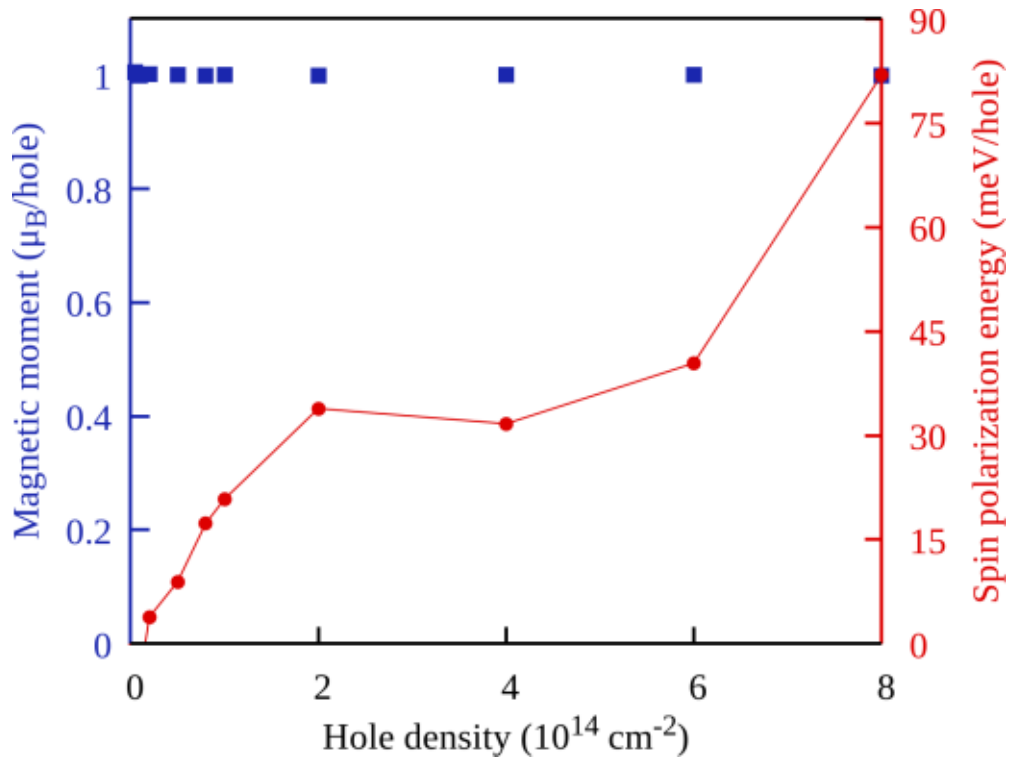

## Magnetic configurations and spin Hamiltonian

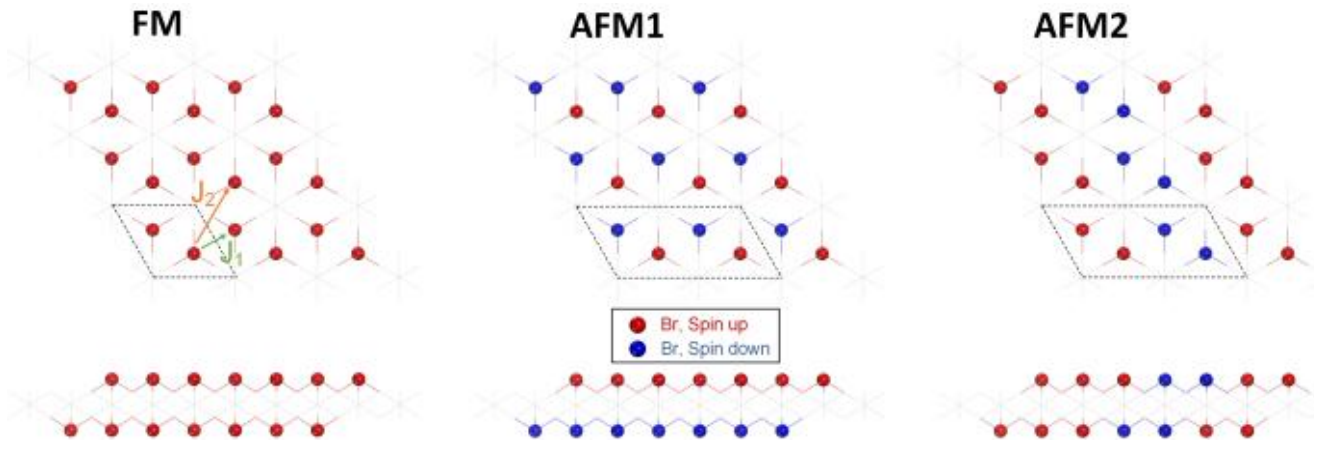

$$E_{FM} = E_0 - 3J_1S^2 - 6J_2S^2$$
$$E_{AFM1} = E_0 + 3J_1S^2 - 6J_2S^2$$
$$E_{AFM2} = E_0 - J_1S^2 + 2J_2S^2$$

## Magnetic exchange coupling parameters

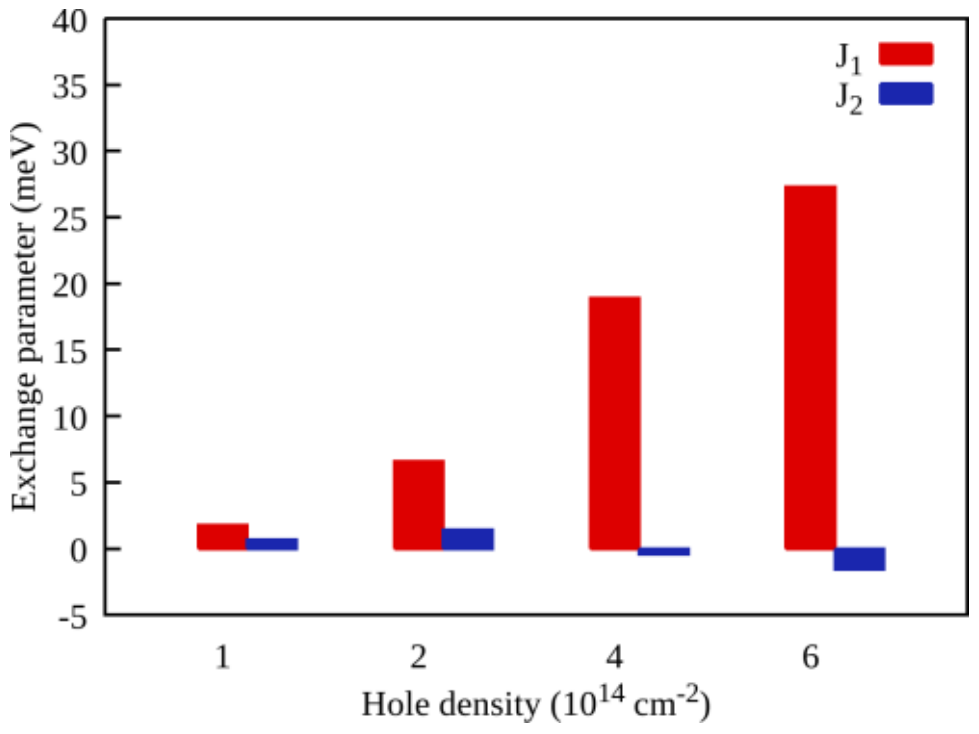

## Magnetic anisotropy energy (MAE, μeV) per magnetic atom

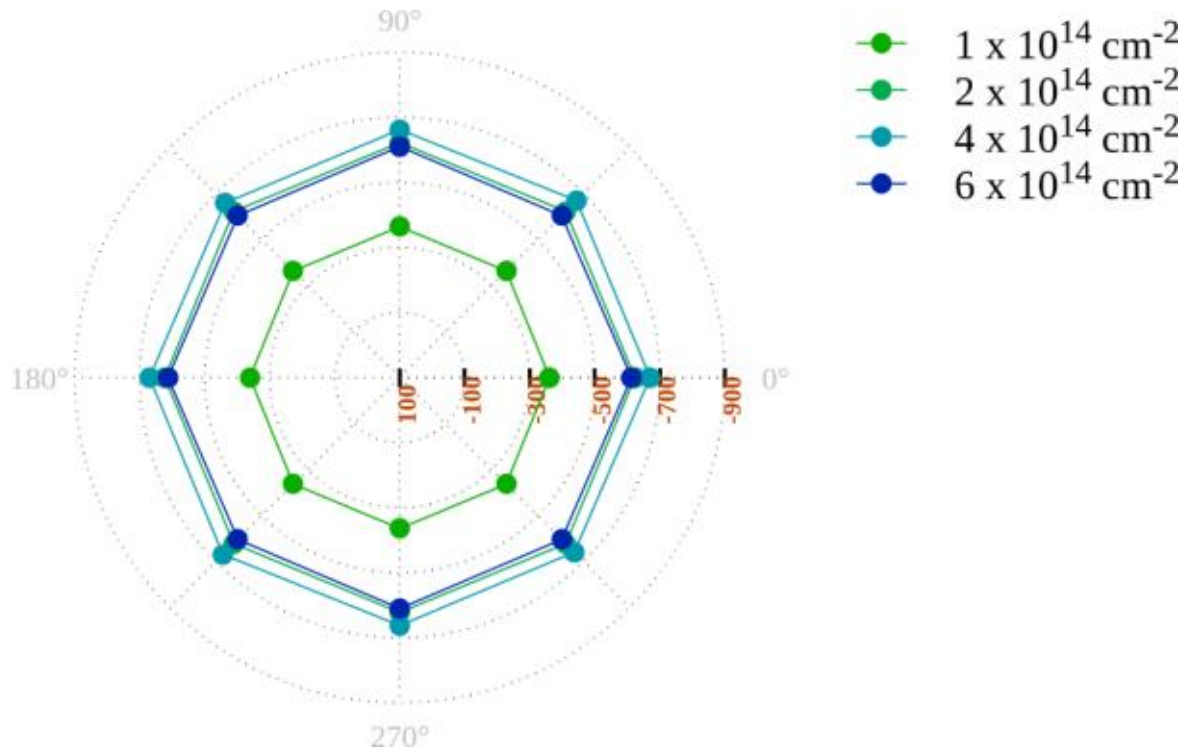

MAE = $E_{\parallel} - E_{\perp}$, a positive (negative) value of MAE indicates the off-plane (in-plane) easy axis.

## Monte Carlo simulations of the normalized magnetization of as a function of temperature

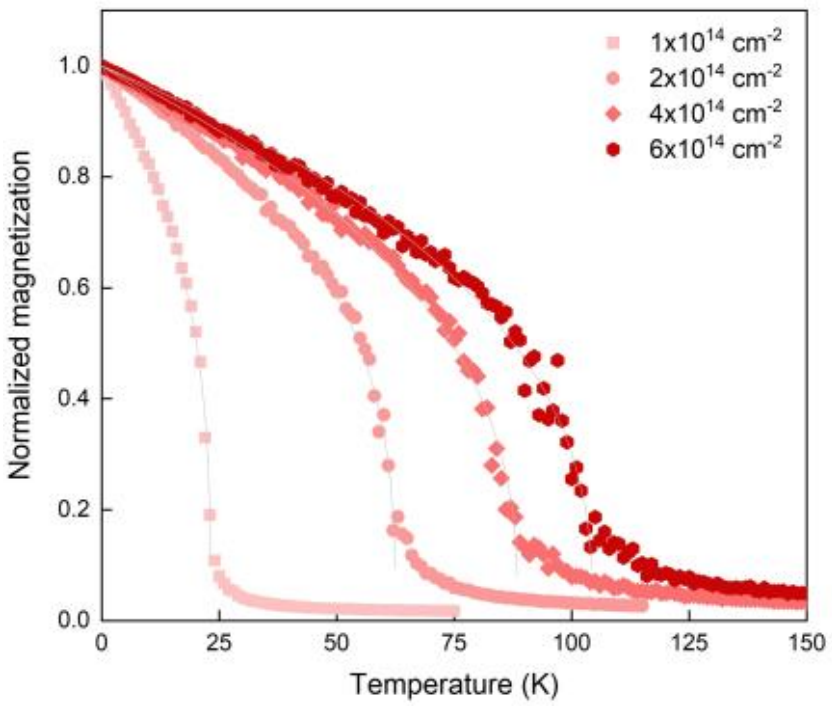

$T_c$: $1\times10^{14}$ cm$^{-2}$: 23 K  $2\times10^{14}$ cm$^{-2}$: 62 K  $4\times10^{14}$ cm$^{-2}$: 89 K  $6\times10^{14}$ cm$^{-2}$: 104 K

# 7. $CaI_2$

| MC2D-ID | C2DB | 2dmat-ID | USPEX | Space group | Band gap (eV) |
|---|---|---|---|---|---|
| 29 | ✓ | - | - | P3m1 | 3.91 |

**Convex hull** | **Atomic structure** | **Atomic coordinates** | **Phonon dispersion curve**

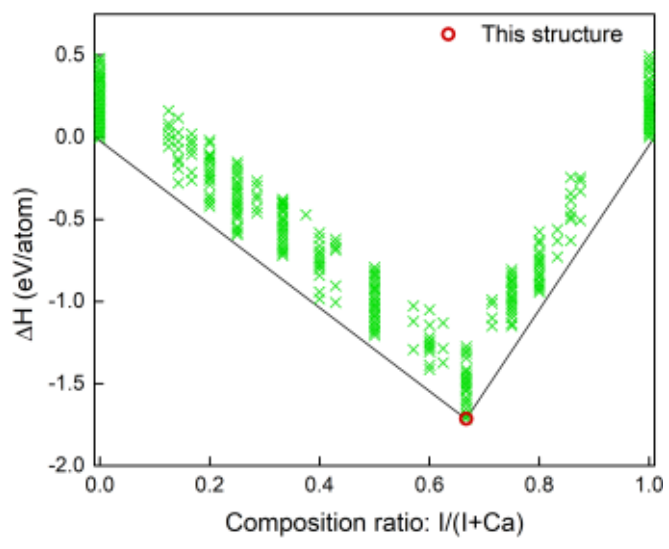

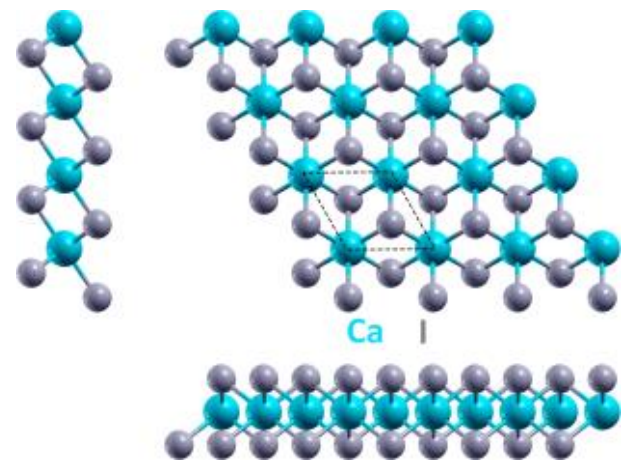

Ca1I2
1.00000000000000
4.5415685628997693 0.0000000000000000 0.0000000000000000
-2.2707073727749525 3.9331250512874485 0.0000000000000000
0.0000000000000000 0.0000000000000000 18.5370006561279332
Ca I
1 2
Direct
0.9999770747201140 0.9999822193259575 0.5000122657880297
0.6665529080537027 0.3332915086209738 0.5953312750419357
0.3332904005901298 0.6665868203645928 0.4046833791565660

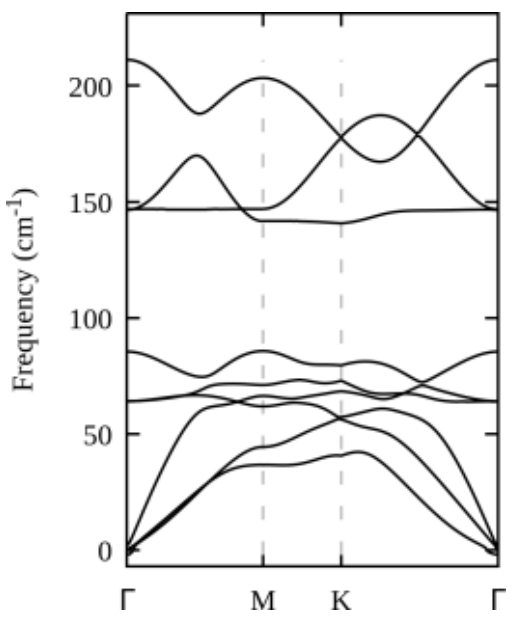

## Projected band structure and density of states

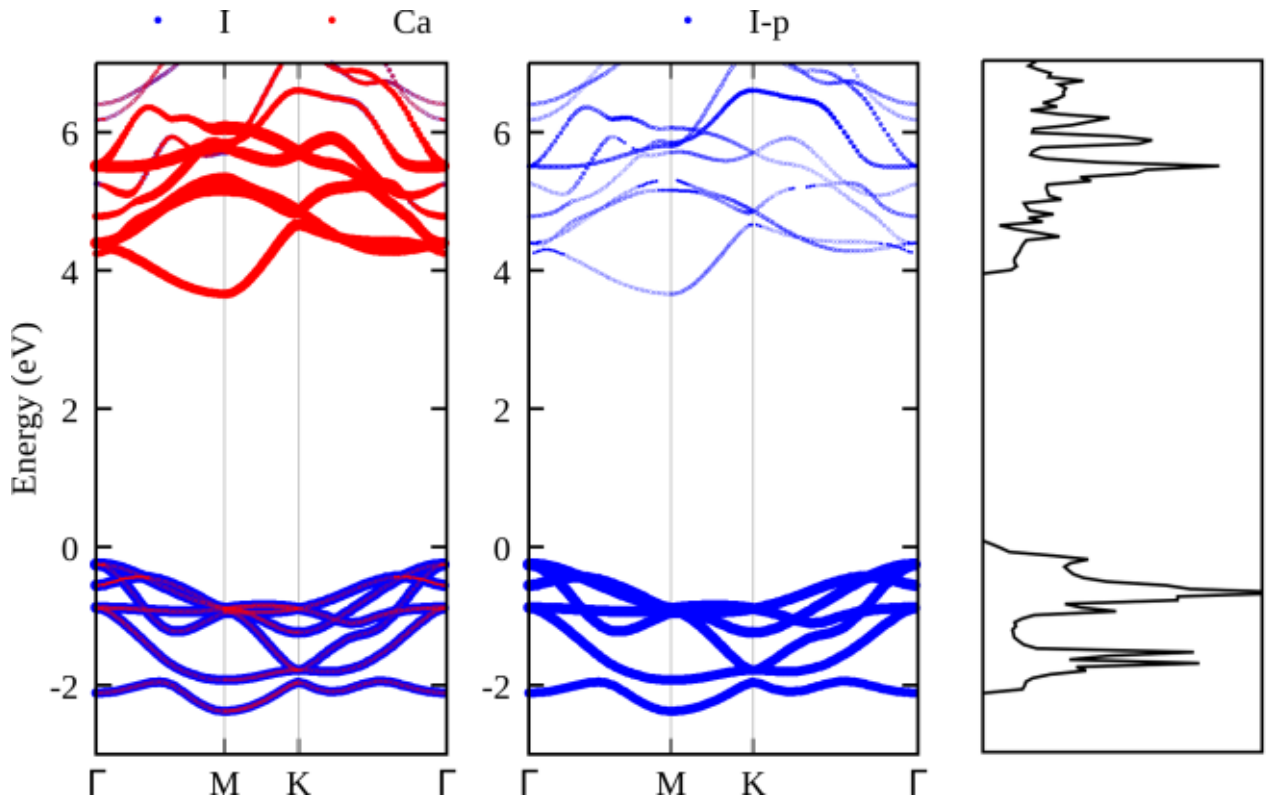

## Magnetic moment and spin polarization energy as a function of hole doping concentration

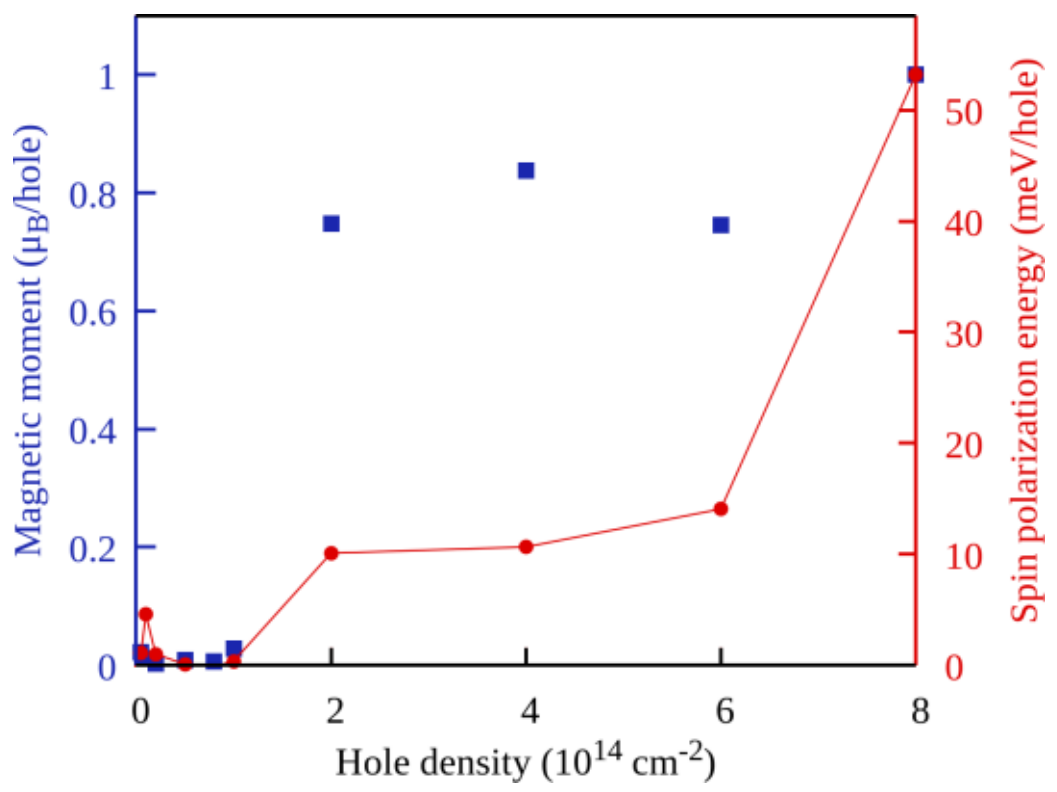

## Magnetic configurations and spin Hamiltonian

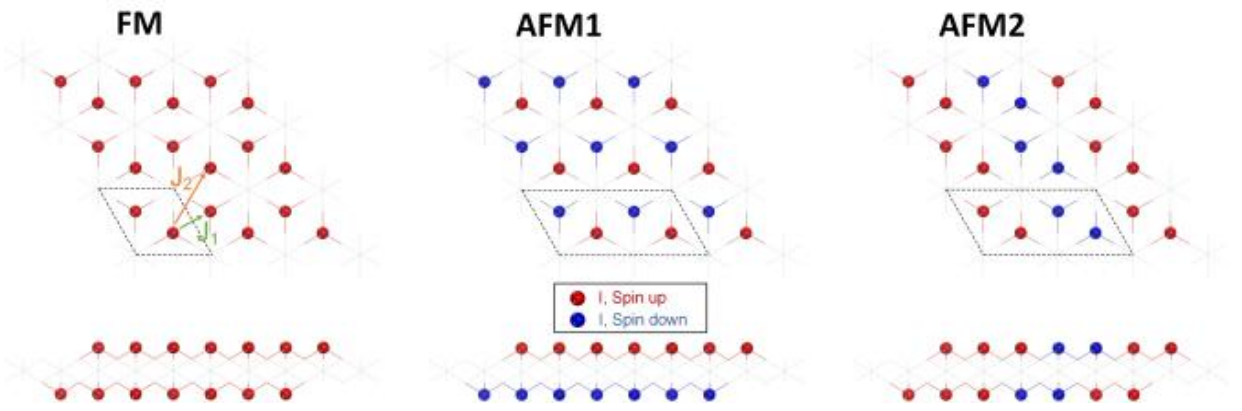

$$E_{FM} = E_0 - 3J_1S^2 - 6J_2S^2$$
$$E_{AFM1} = E_0 + 3J_1S^2 - 6J_2S^2$$
$$E_{AFM2} = E_0 - J_1S^2 + 2J_2S^2$$

## Magnetic exchange coupling parameters

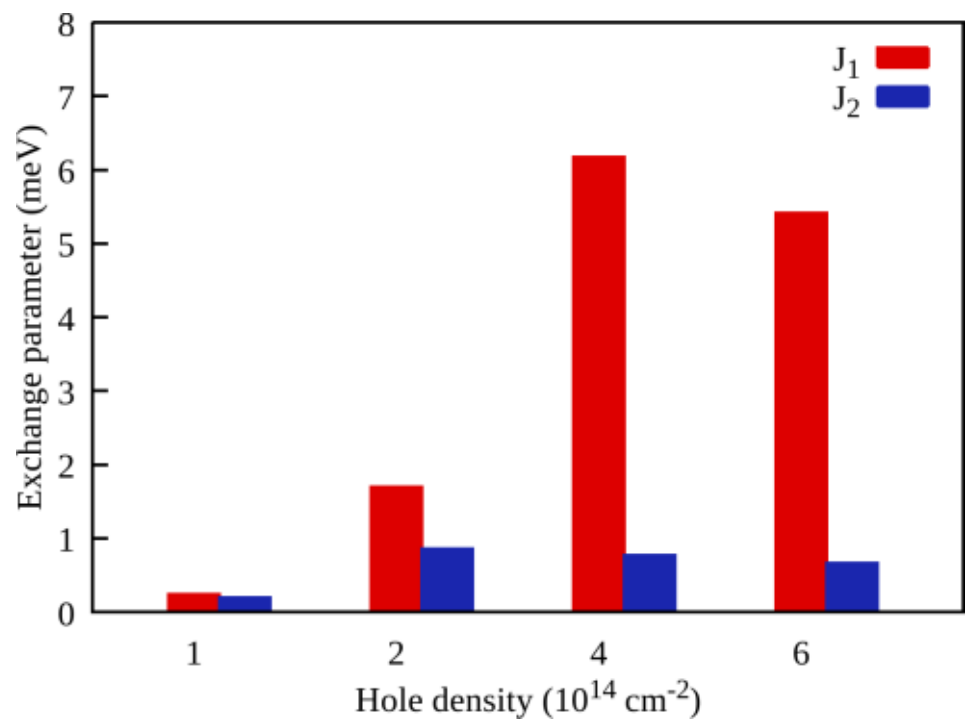

## Magnetic anisotropy energy (MAE, μeV) per magnetic atom

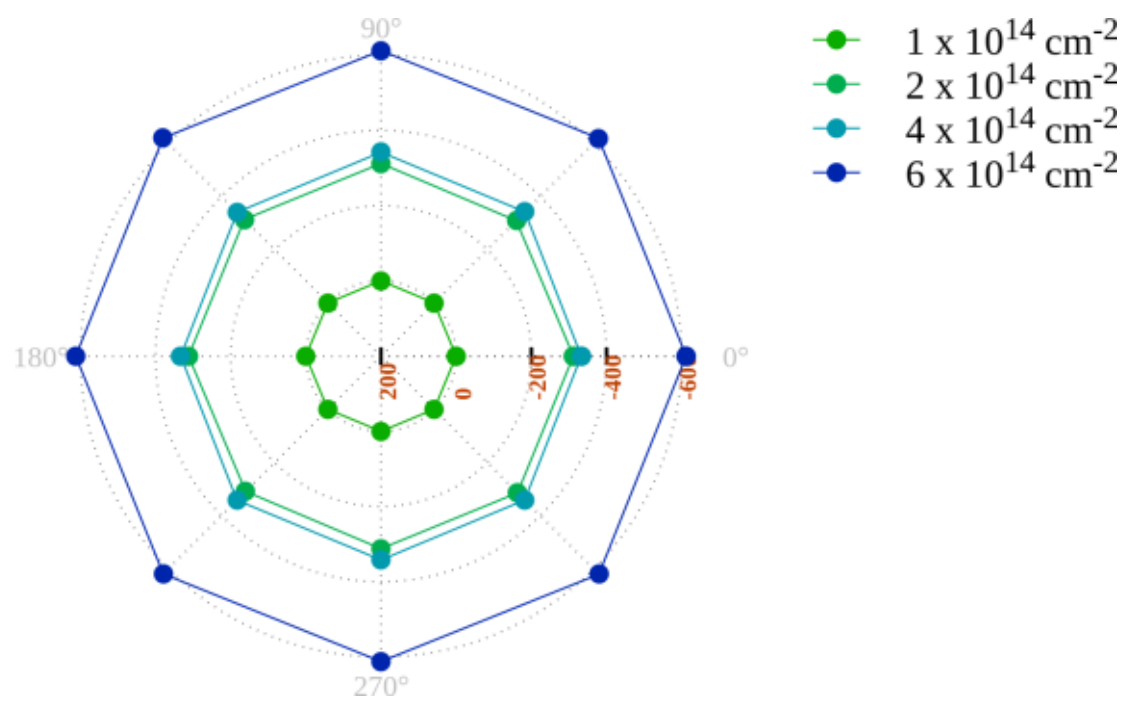

MAE = $E_{\parallel} - E_{\perp}$, a positive (negative) value of MAE indicates the off-plane (in-plane) easy axis.

## Monte Carlo simulations of the normalized magnetization of as a function of temperature

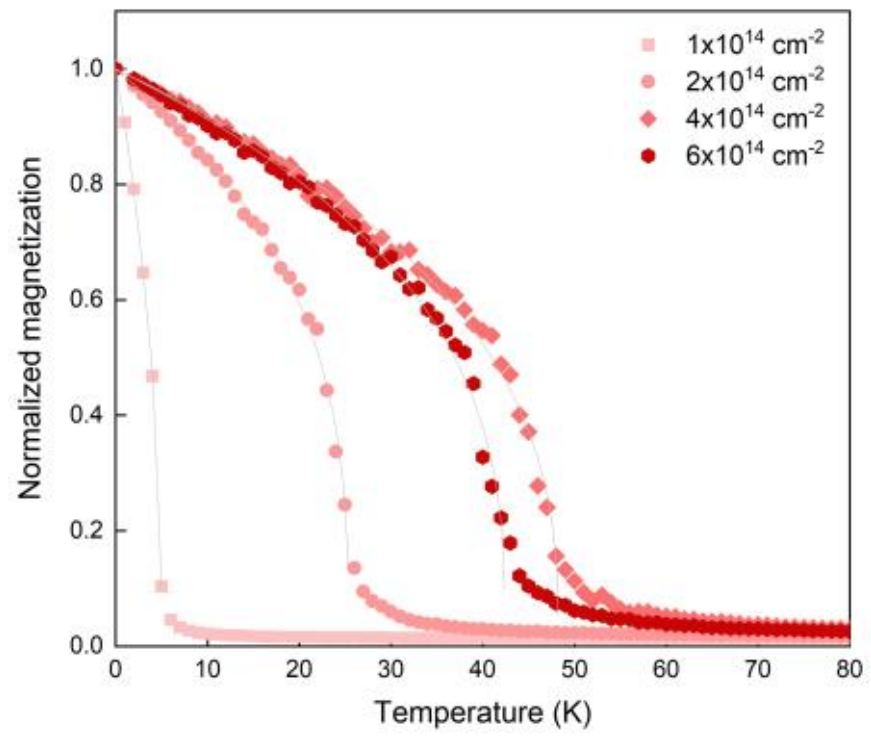

$T_c$: $1\times10^{14}$ cm$^{-2}$: 5 K  $2\times10^{14}$ cm$^{-2}$: 25 K  $4\times10^{14}$ cm$^{-2}$:48 K  $6\times10^{14}$ cm$^{-2}$: 42 K

# 8. $SrF_2$

| MC2D-ID | C2DB | 2dmat-ID | USPEX | Space group | Band gap (eV) |
|---|---|---|---|---|---|
| - | - | 2dm-325 | - | P3m1 | 6.54 |

**Convex hull** | **Atomic structure** | **Atomic coordinates** | **Phonon dispersion curve**

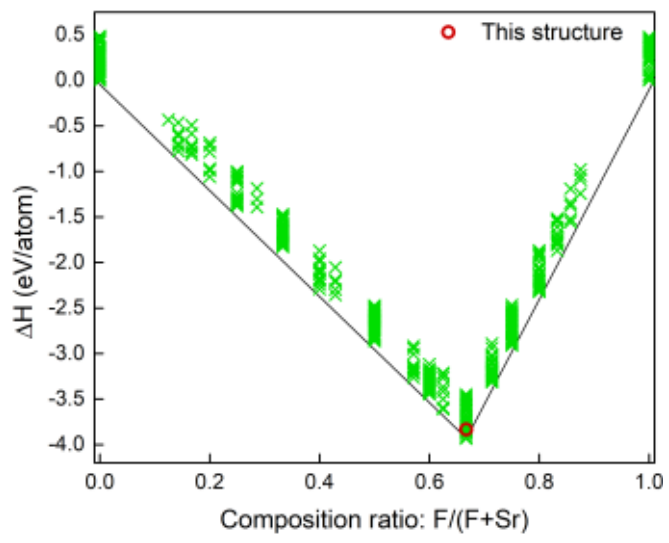

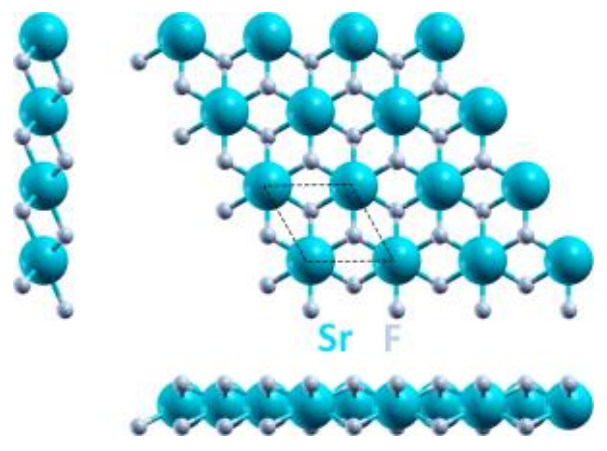

```
Sr1F2
   1.00000000000000
   3.9100925209269715    0.0000000000000000    0.0000000000000000
  -1.9550419704333364    3.3862411587467638    0.0000000000000000
   0.0000000000000000    0.0000000000000000   23.5298450000000017
   Sr    F
   1     2
Direct
  0.9999999919565923  0.9999999886292770  0.0000000036421426
  0.6666661064332828  0.3333339471161452  0.0412259079451118
  0.3333339016101320  0.6666660642545779  0.9587740884127456
```

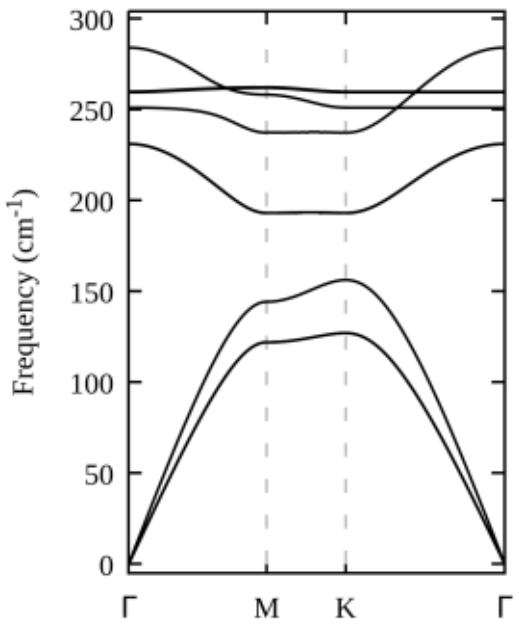

**Projected band structure and density of states** | **Magnetic moment and spin polarization energy as a function of hole doping concentration**

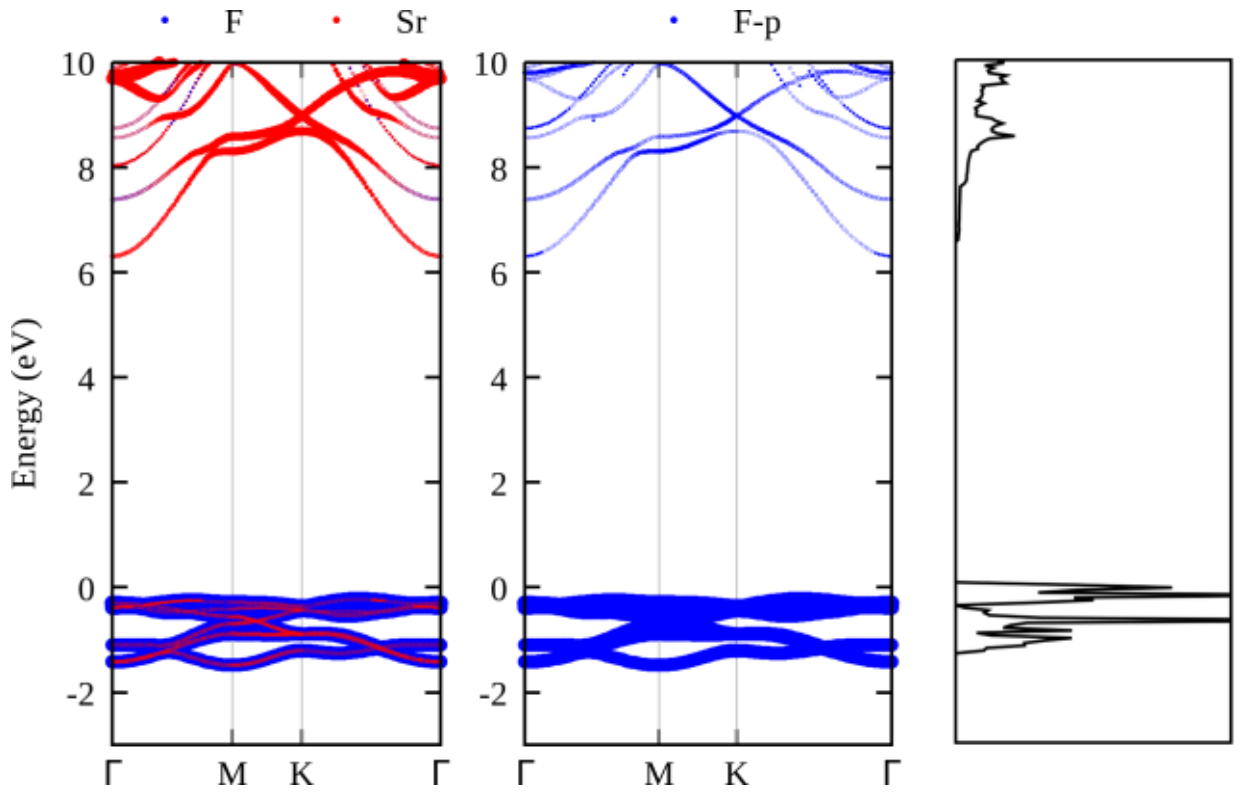

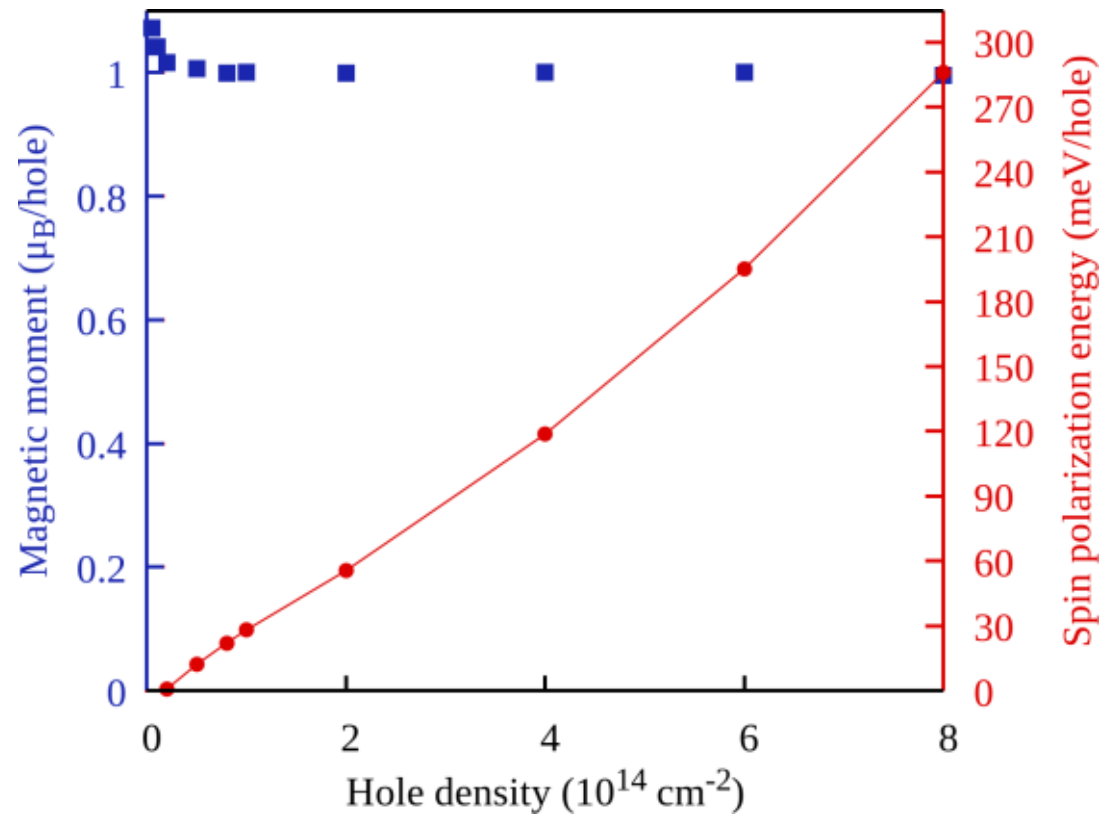

**Magnetic configurations and spin Hamiltonian** | **Magnetic exchange coupling parameters**

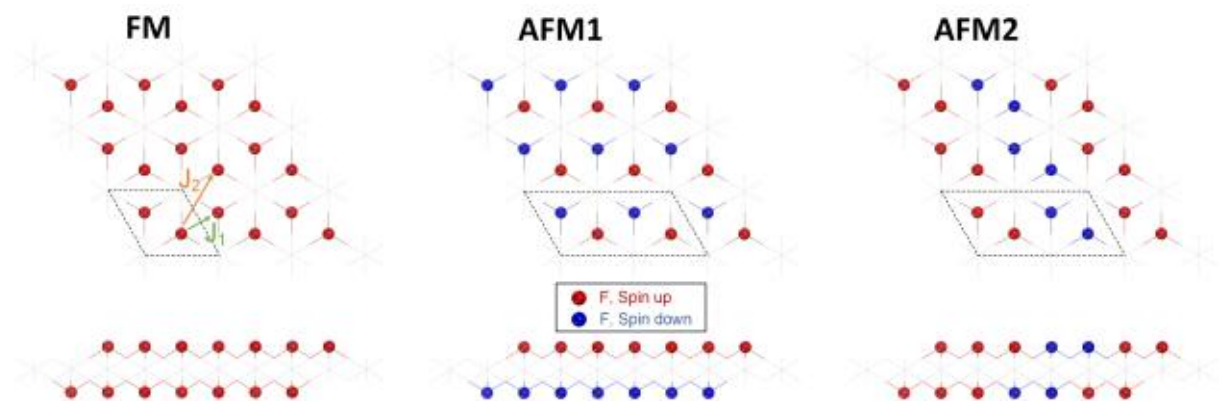

$$E_{FM} = E_0 - 3J_1S^2 - 6J_2S^2$$
$$E_{AFM1} = E_0 + 3J_1S^2 - 6J_2S^2$$
$$E_{AFM2} = E_0 - J_1S^2 + 2J_2S^2$$

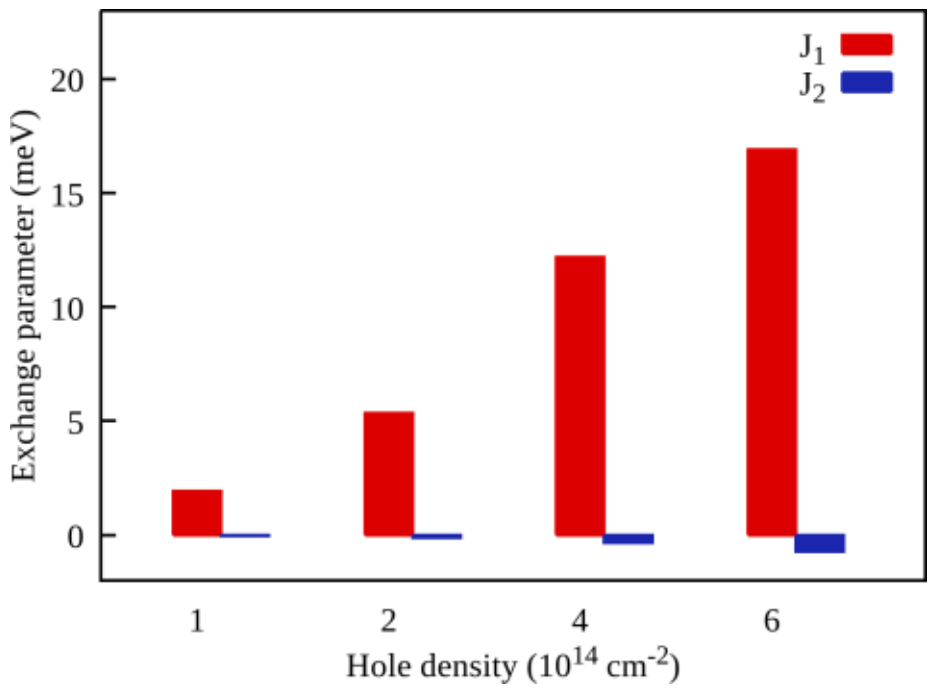

**Magnetic anisotropy energy (MAE, μeV) per magnetic atom** | **Monte Carlo simulations of the normalized magnetization of as a function of temperature**

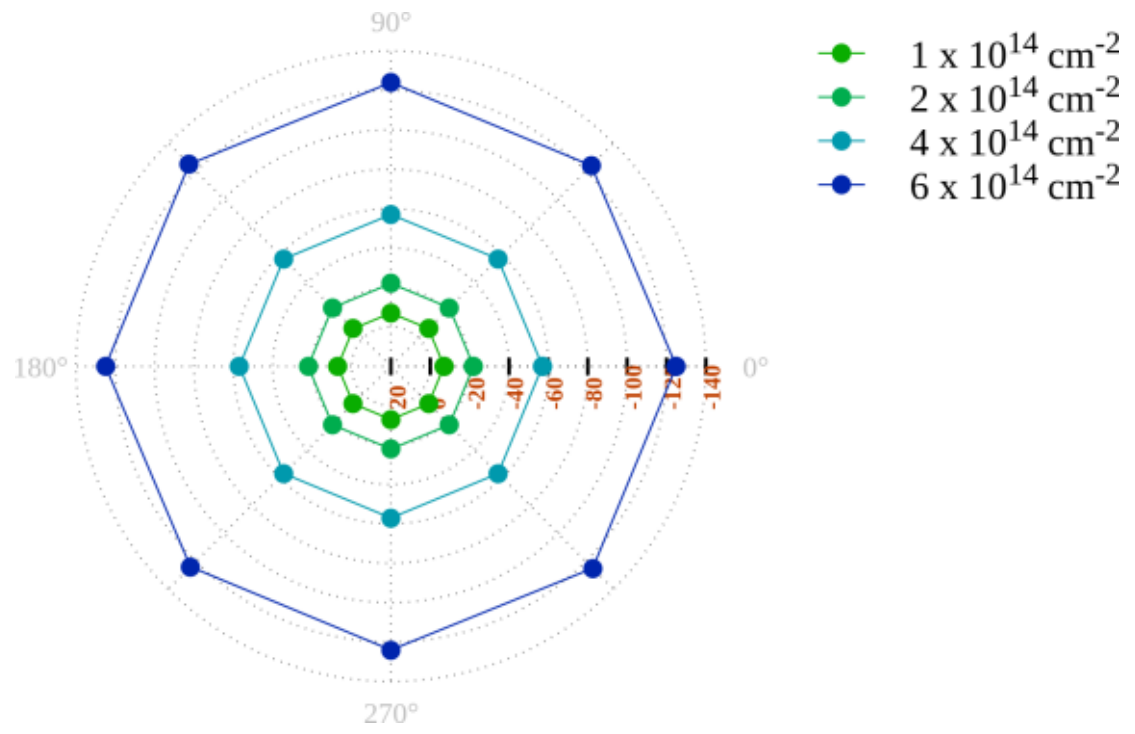

MAE = $E_{\parallel} - E_{\perp}$, a positive (negative) value of MAE indicates the off-plane (in-plane) easy axis.

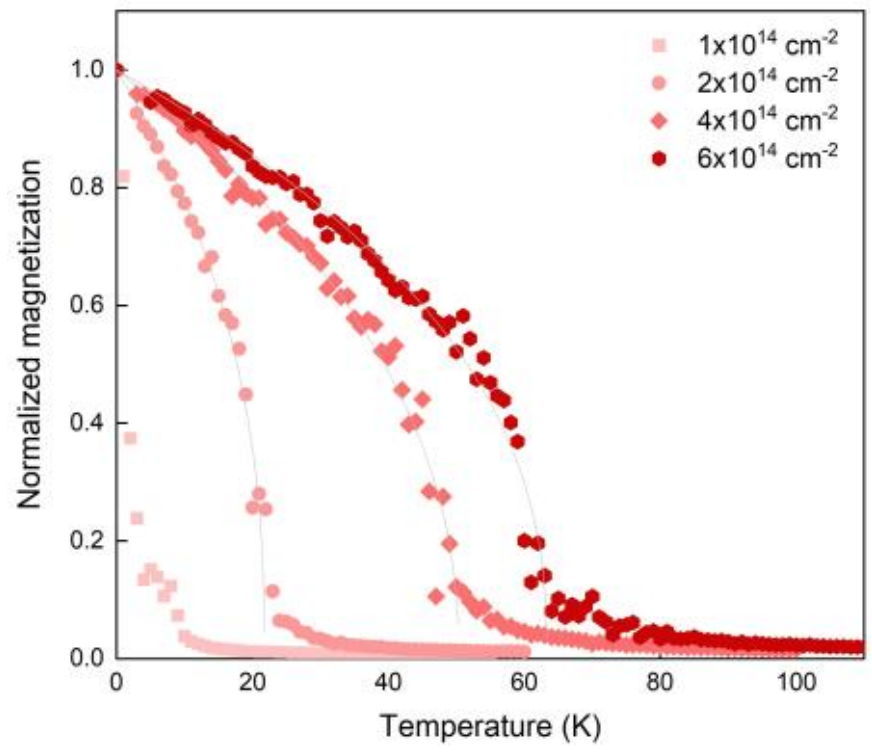

$T_c$: $1\times10^{14}$ cm$^{-2}$: 9 K  $2\times10^{14}$ cm$^{-2}$: 22 K  $4\times10^{14}$ cm$^{-2}$: 50 K  $6\times10^{14}$ cm$^{-2}$: 63 K

# 9. $SrCl_2$

| MC2D-ID | C2DB | 2dmat-ID | USPEX | Space group | Band gap (eV) |
|---|---|---|---|---|---|
| - | ✓ | 2dm-491 | - | P3m1 | 5.76 |

**Convex hull** | **Atomic structure** | **Atomic coordinates** | **Phonon dispersion curve**

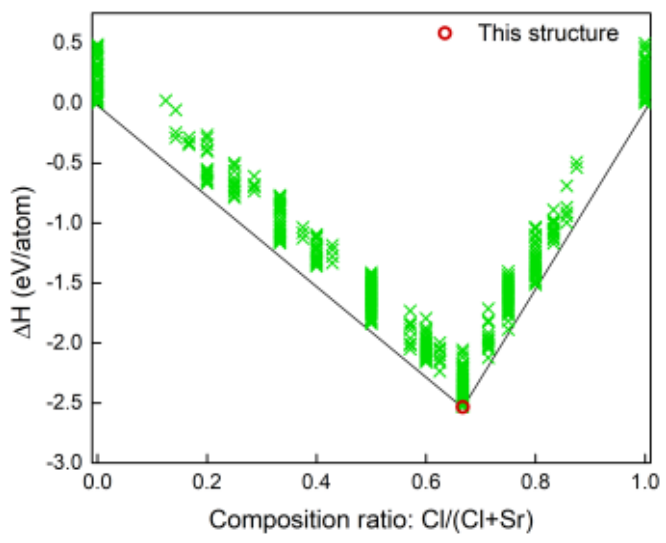

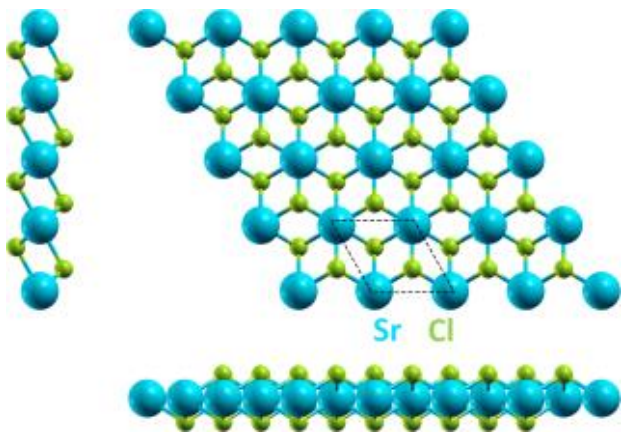

```
Sr1Cl2
   1.00000000000000
   4.4521103983943764    0.0000000000000000    0.0000000000000000
  -2.2260537991918081    3.8556412623226648    0.0000000000000000
   0.0000000000000000    0.0000000000000000   23.5298450000000017
   Sr   Cl
   1     2
Direct
  0.0000000015717418  0.0000000177476167  0.0000000092065093
  0.6666665225228101  0.3333334107469312  0.0596025144725019
  0.3333334759054551  0.6666665715054592  0.9403974763209888
```

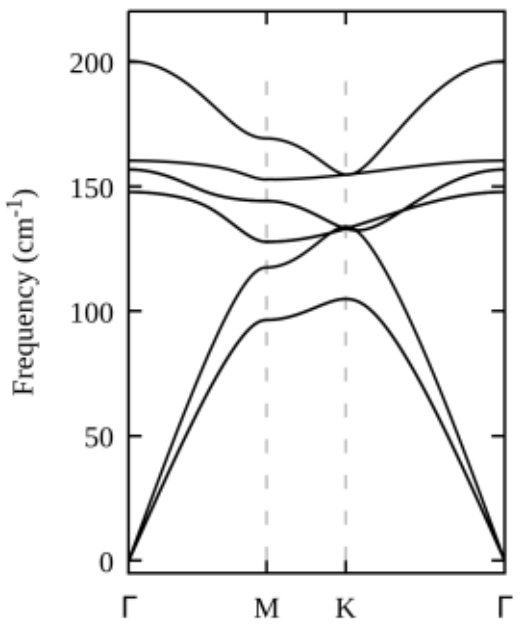

## Projected band structure and density of states

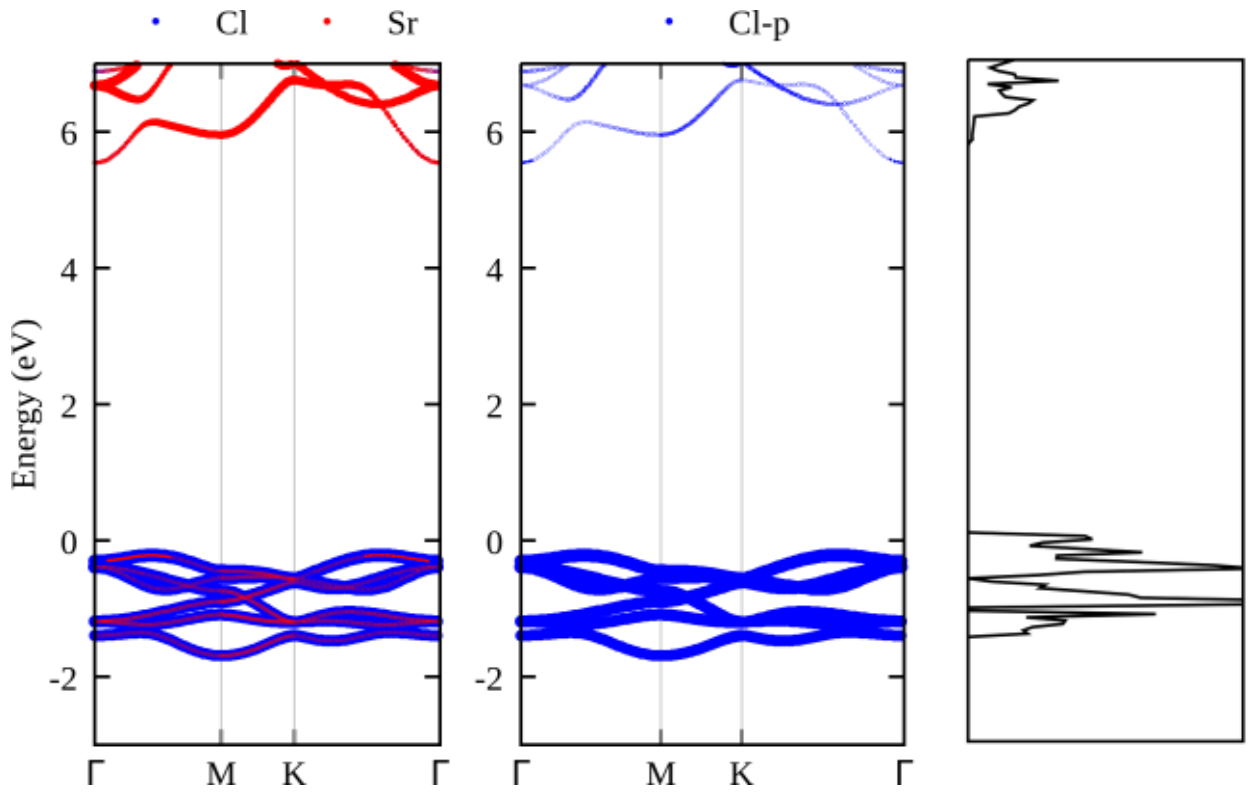

## Magnetic moment and spin polarization energy as a function of hole doping concentration

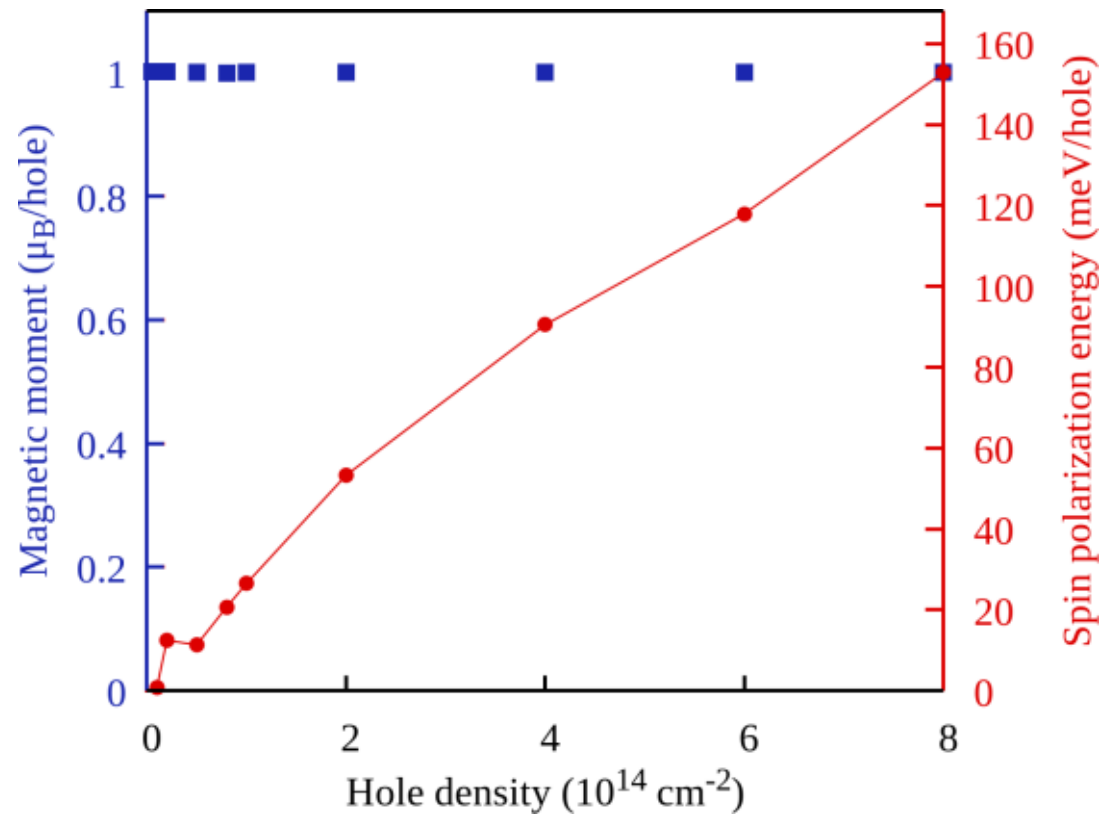

## Magnetic configurations and spin Hamiltonian

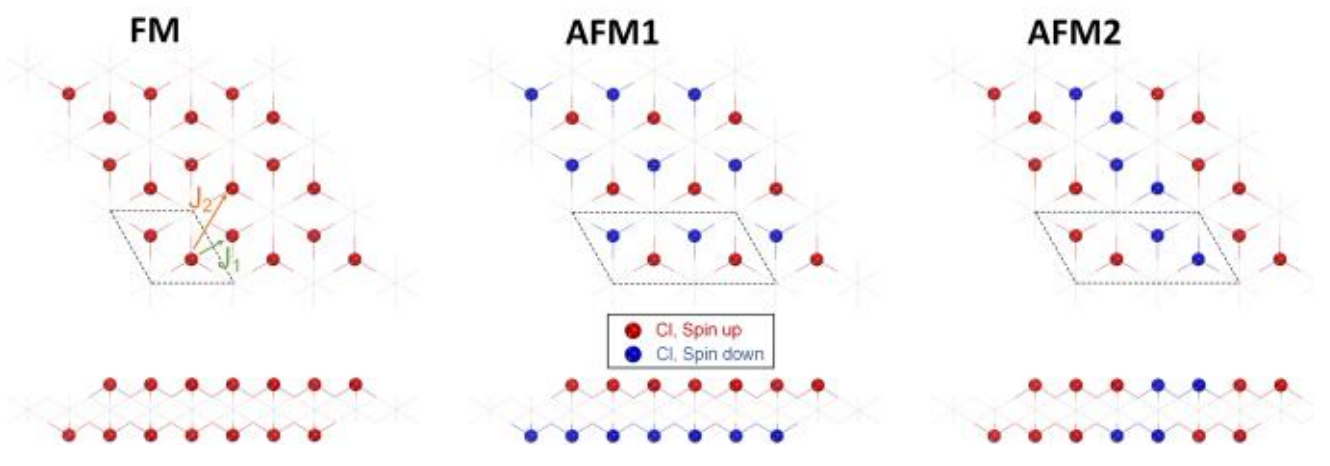

$$E_{FM} = E_0 - 3J_1S^2 - 6J_2S^2$$
$$E_{AFM1} = E_0 + 3J_1S^2 - 6J_2S^2$$
$$E_{AFM2} = E_0 - J_1S^2 + 2J_2S^2$$

## Magnetic exchange coupling parameters

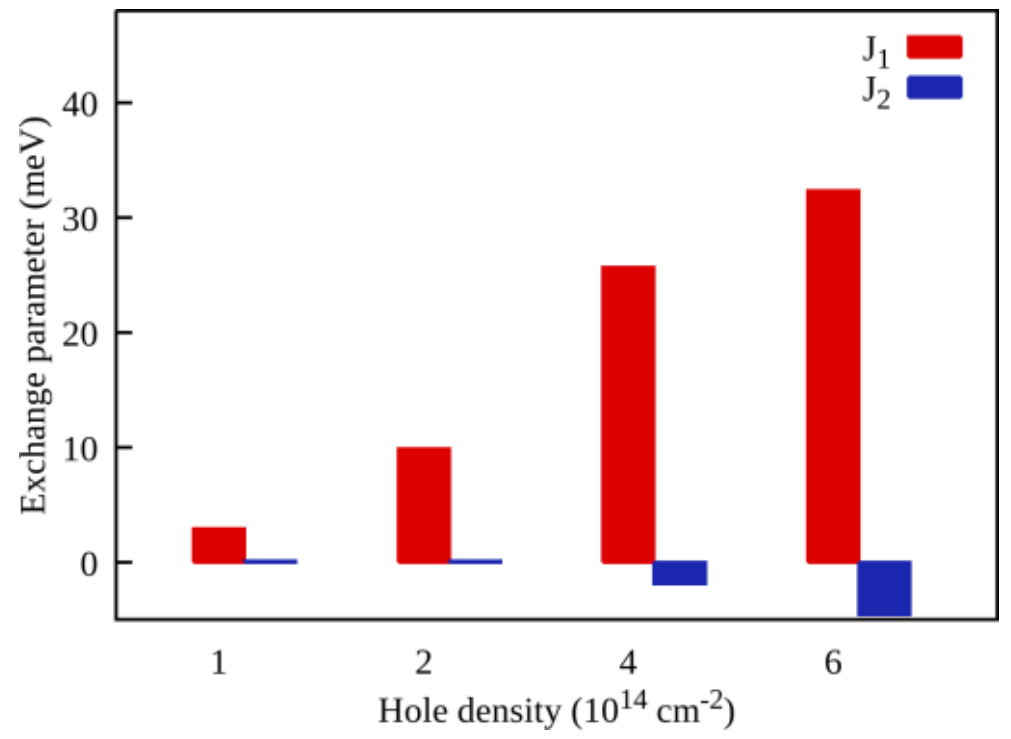

## Magnetic anisotropy energy (MAE, μeV) per magnetic atom

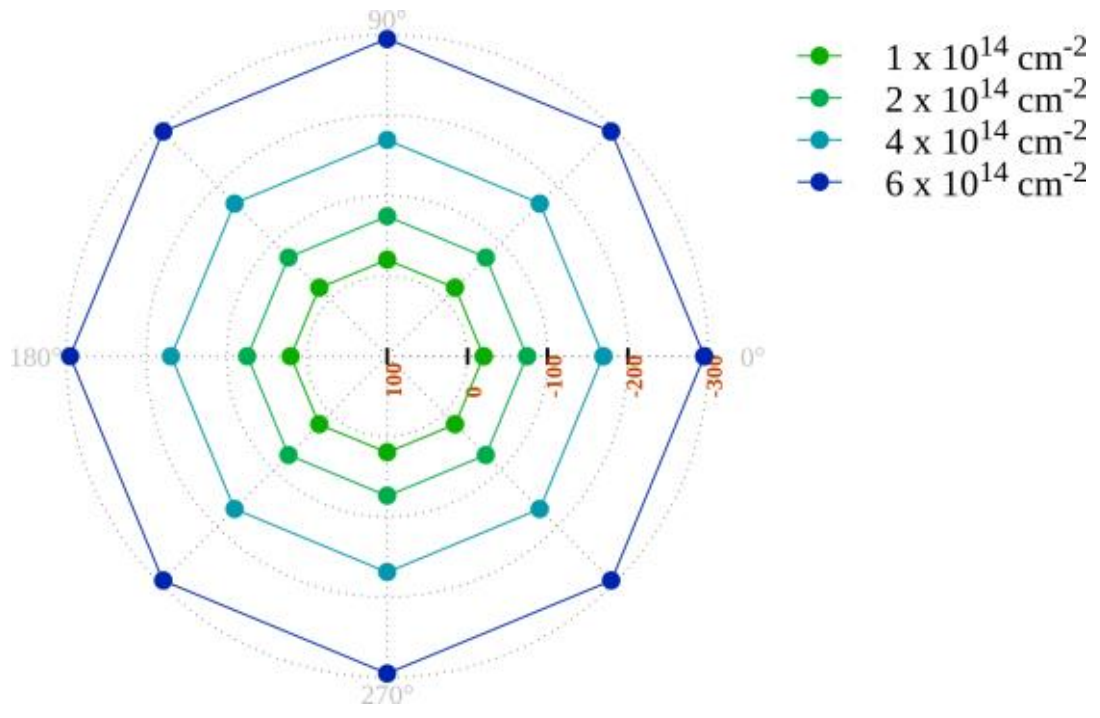

MAE = $E_{\parallel} - E_{\perp}$, a positive (negative) value of MAE indicates the off-plane (in-plane) easy axis.

## Monte Carlo simulations of the normalized magnetization of as a function of temperature

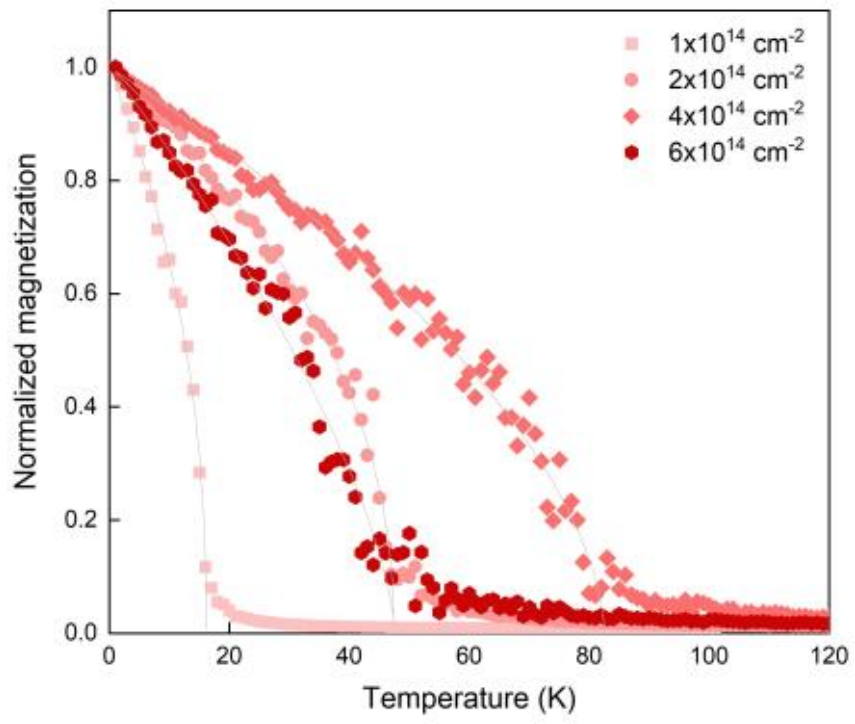

$T_c$: $1\times10^{14}$ cm$^{-2}$: 16 K  $2\times10^{14}$ cm$^{-2}$: 47 K
$4\times10^{14}$ cm$^{-2}$: 83 K  $6\times10^{14}$ cm$^{-2}$: 48 K

# 10.$SrBr_2$

| MC2D-ID | C2DB | 2dmat-ID | USPEX | Space group | Band gap (eV) |
|---|---|---|---|---|---|
| - | ✓ | 2dm-349 | - | P3m1 | 5.00 |

**Convex hull** | **Atomic structure** | **Atomic coordinates** | **Phonon dispersion curve**

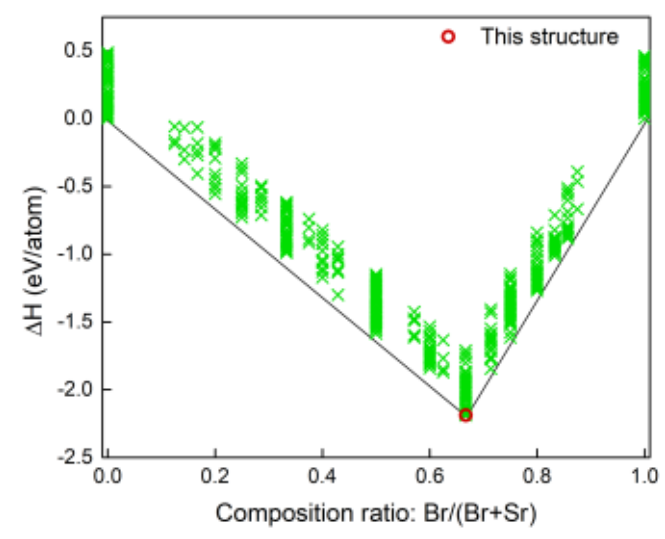

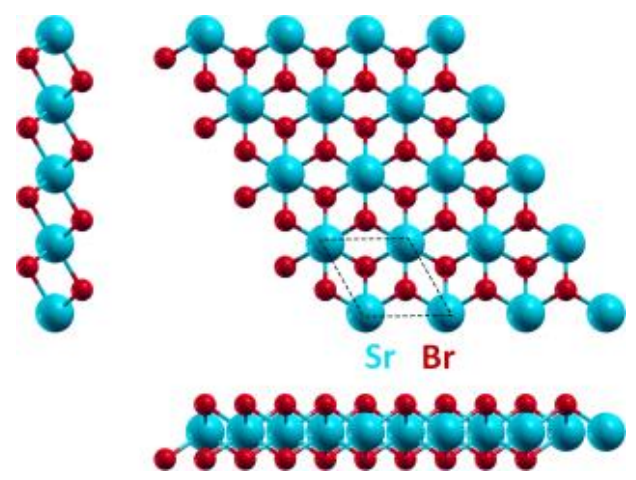

```
Sn1Te1
   1.00000000000000
   4.1801374371344826    0.0000000000000000    0.0000000000000000
  -2.0900848226707285    3.6201327930791019    0.0000000000000000
   0.0000000000000000    0.0000000000000000   22.3597700000000010
   Sn    Te
   1     1
Direct
  0.9999798074185620  0.9999593082243052  0.5382850831818260
  0.3333351925814441  0.6666706917756997  0.4626419168181712
```

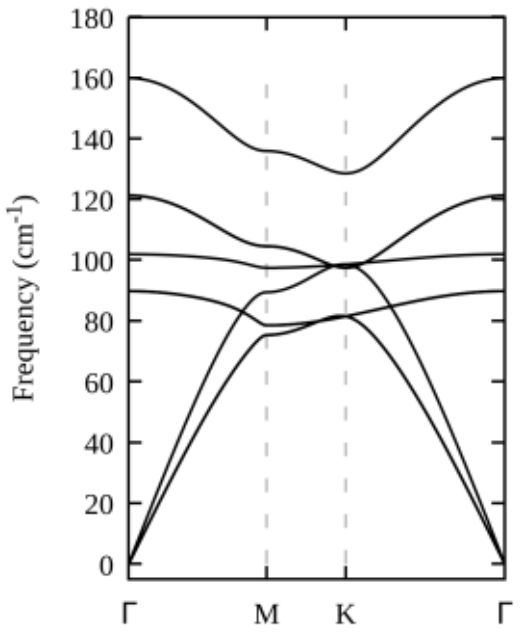

**Projected band structure and density of states** | **Magnetic moment and spin polarization energy as a function of hole doping concentration**

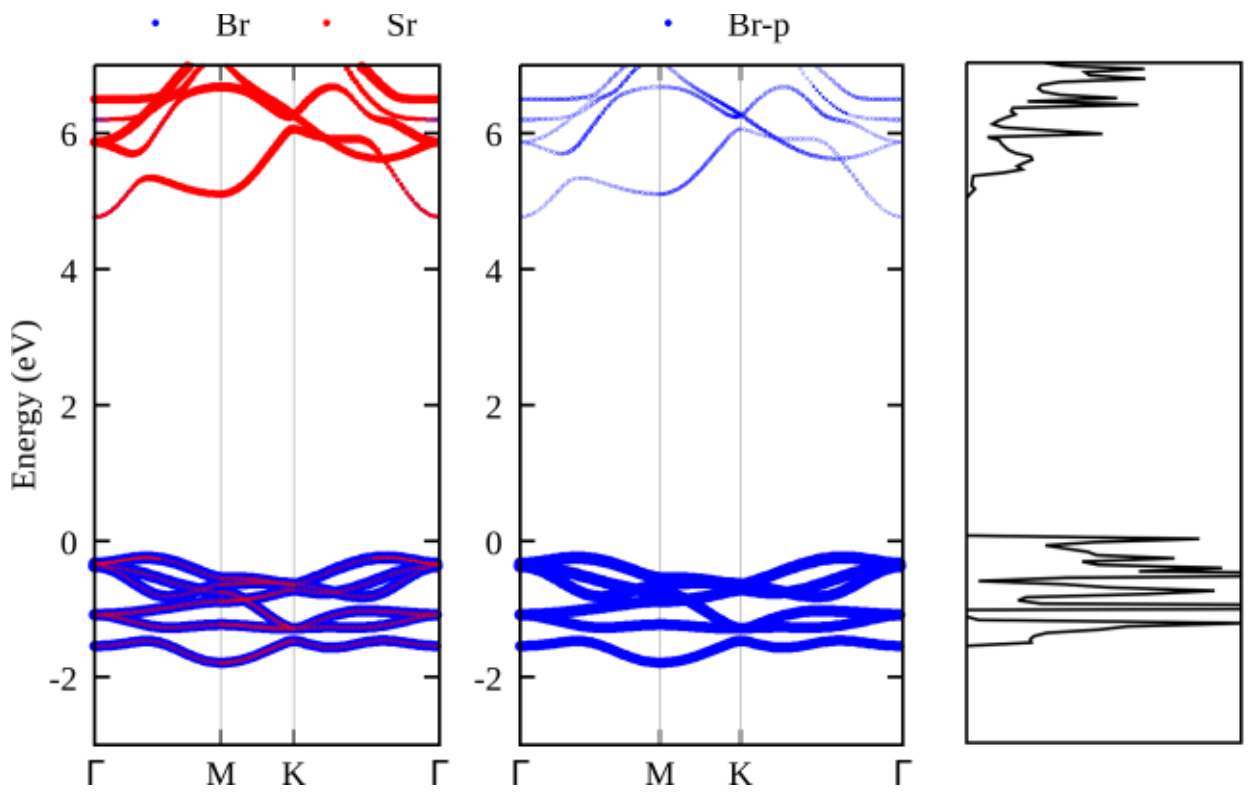

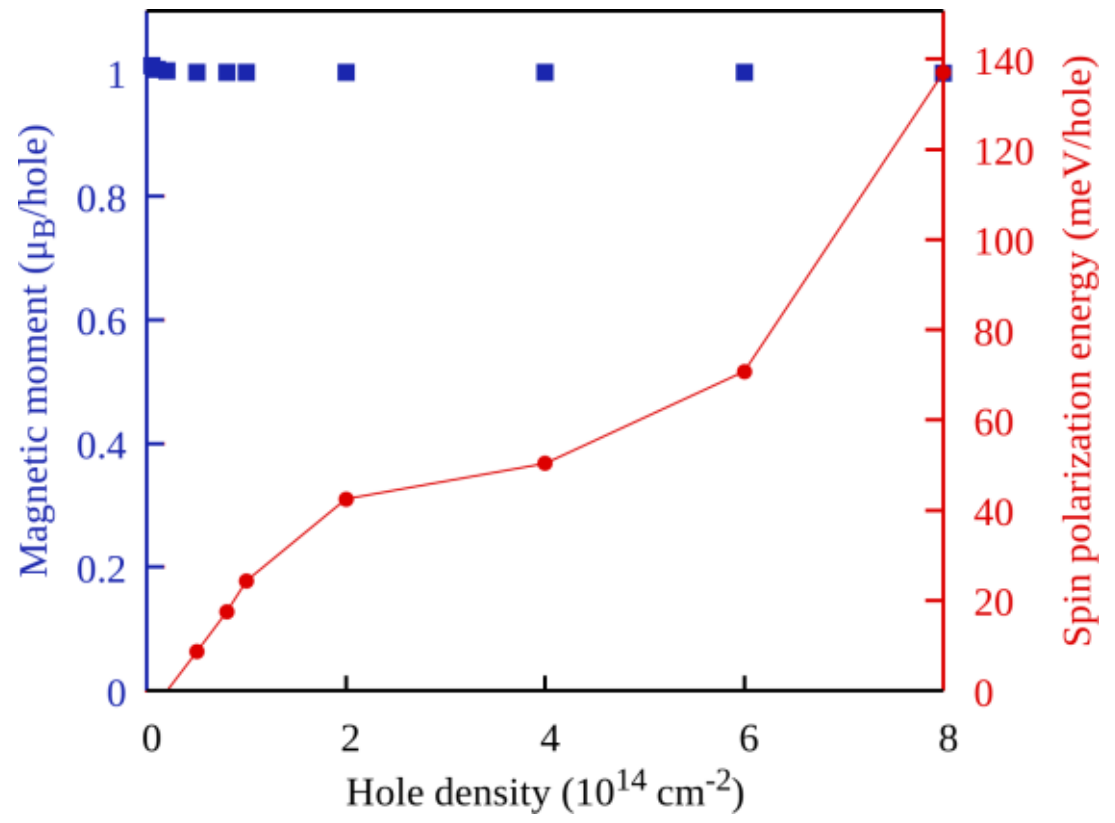

**Magnetic configurations and spin Hamiltonian** | **Magnetic exchange coupling parameters**

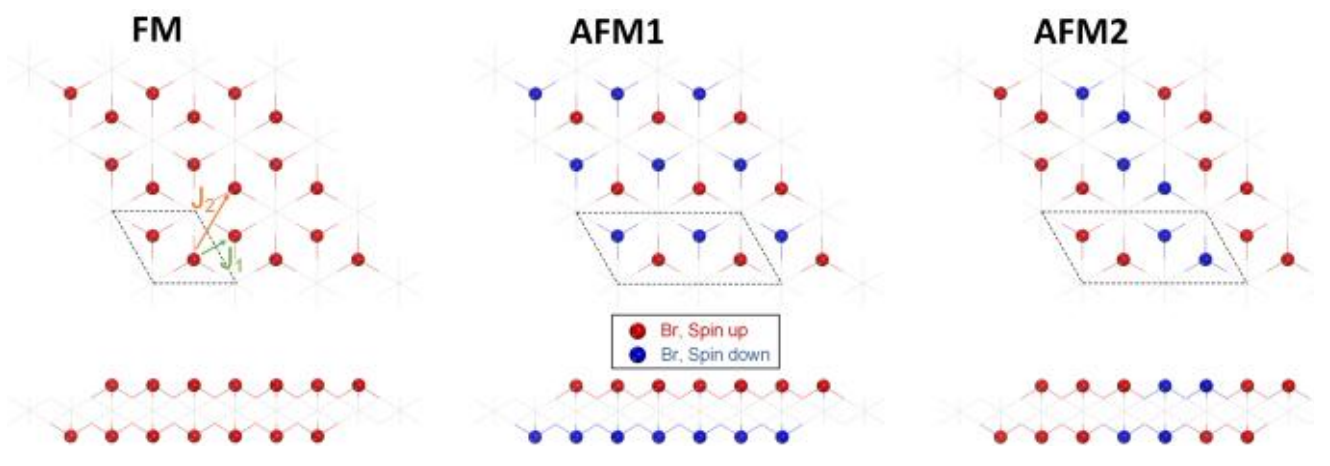

$$E_{FM} = E_0 - 3J_1S^2 - 6J_2S^2$$
$$E_{AFM1} = E_0 + 3J_1S^2 - 6J_2S^2$$
$$E_{AFM2} = E_0 - J_1S^2 + 2J_2S^2$$

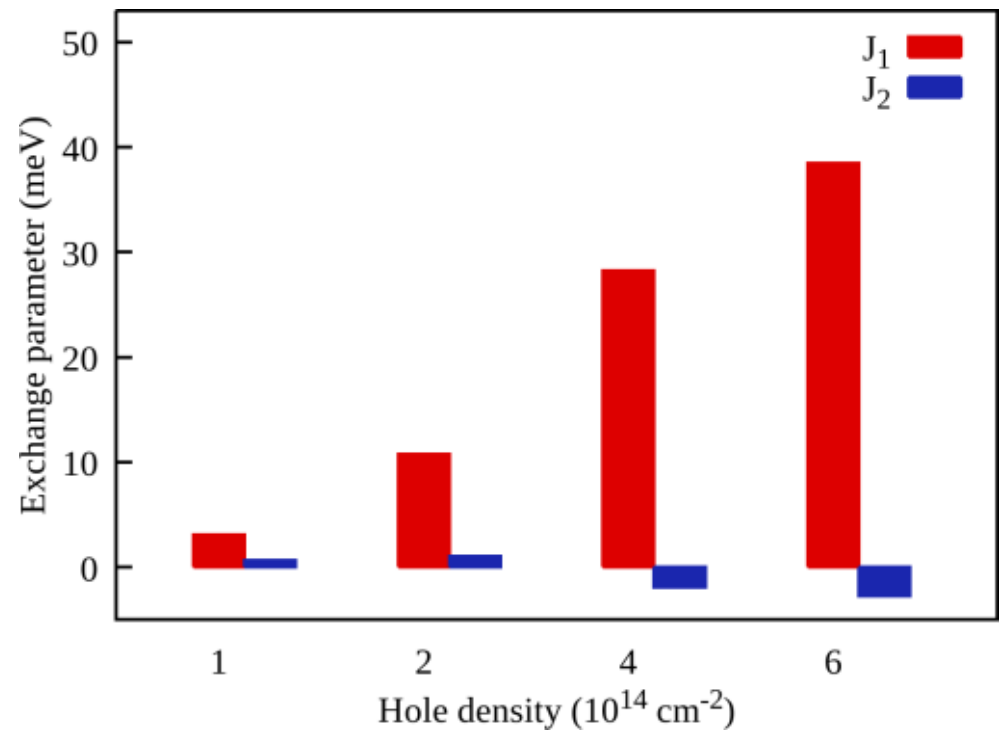

**Magnetic anisotropy energy (MAE, μeV) per magnetic atom** | **Monte Carlo simulations of the normalized magnetization of as a function of temperature**

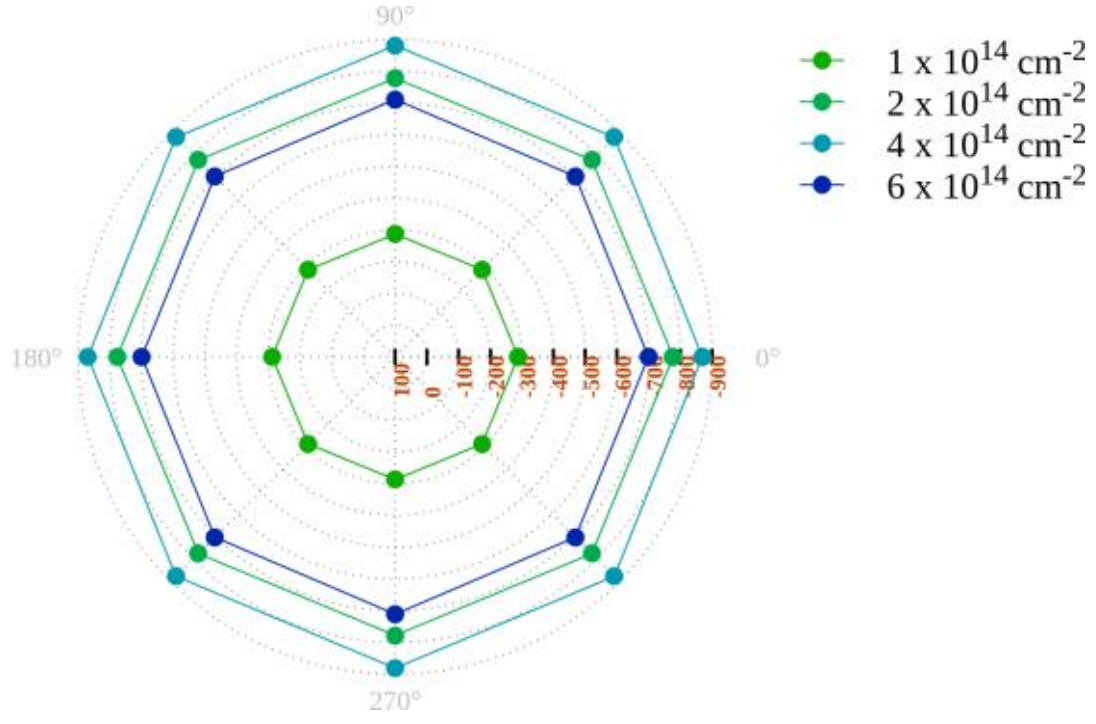

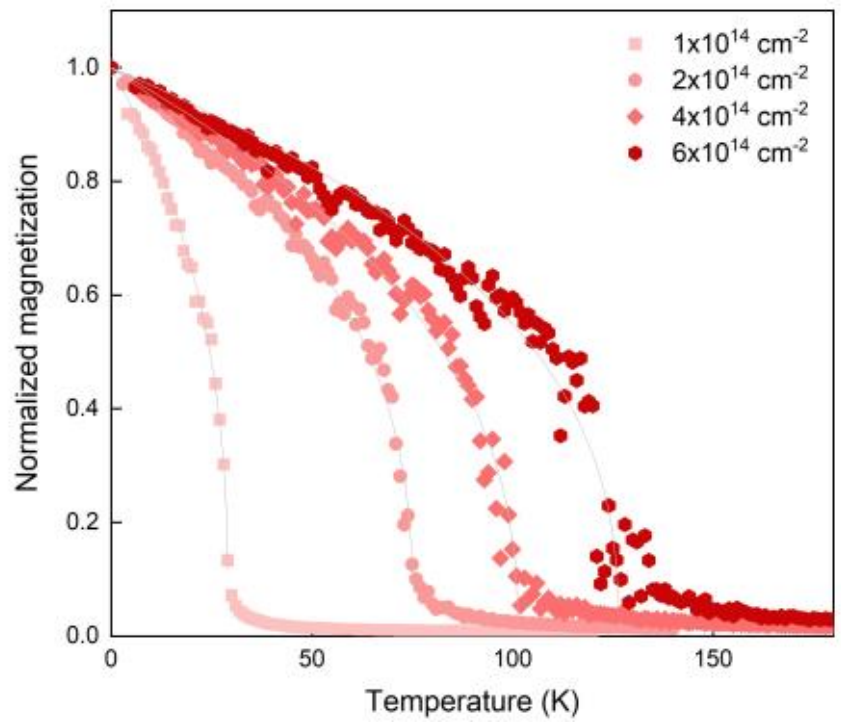

MAE = $E_{\parallel} - E_{\perp}$, a positive (negative) value of MAE indicates the off-plane (in-plane) easy axis.

$T_c$: $1\times10^{14}$ cm$^{-2}$: 29 K  $2\times10^{14}$ cm$^{-2}$: 75 K  $4\times10^{14}$ cm$^{-2}$: 101 K  $6\times10^{14}$ cm$^{-2}$: 126 K

# 11.$SrI_2$

| MC2D-ID | C2DB | 2dmat-ID | USPEX | Space group | Band gap (eV) |
|---|---|---|---|---|---|
| - | ✓ | 2dm-980 | - | P3m1 | 4.33 |

**Convex hull** | **Atomic structure** | **Atomic coordinates** | **Phonon dispersion curve**

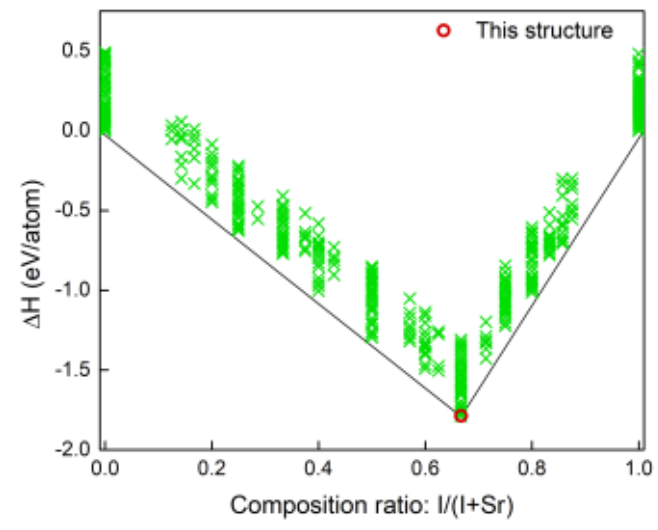

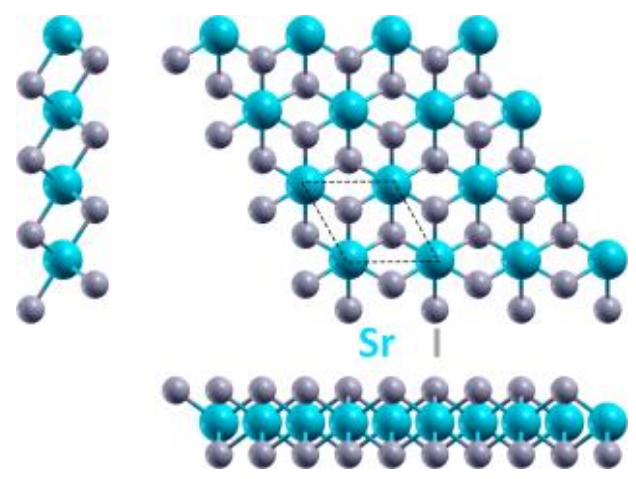

```
Sr1I2
   1.00000000000000
   4.8374045014613749    0.0000000000000000    0.0000000000000000
  -2.4187021135054843    4.1893150489074769    0.0000000000000000
   0.0000000000000000    0.0000000000000000   22.7427169999999990
   Sr    I
   1     2
Direct
  0.9999999763343652  0.9999999463252820  0.0000000200908161
  0.6666667330449343  0.3333332401574580  0.9211748513637446
  0.3333332906207005  0.6666668135172600  0.0788251285454393
```

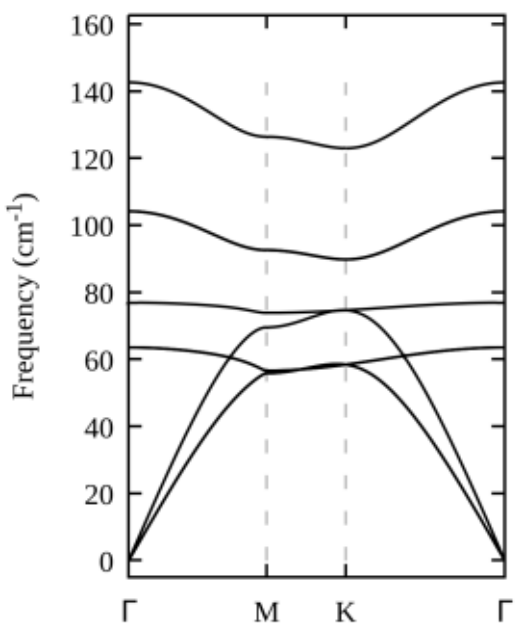

## Projected band structure and density of states

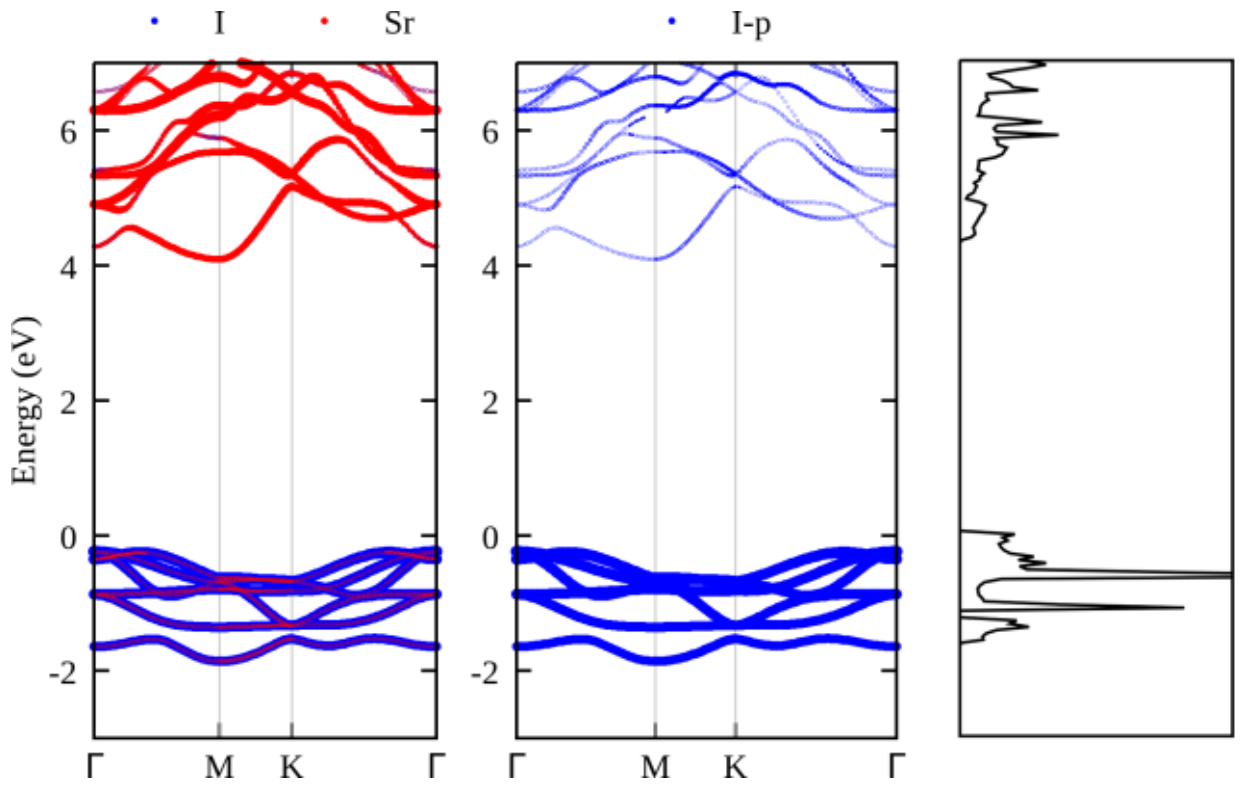

## Magnetic moment and spin polarization energy as a function of hole doping concentration

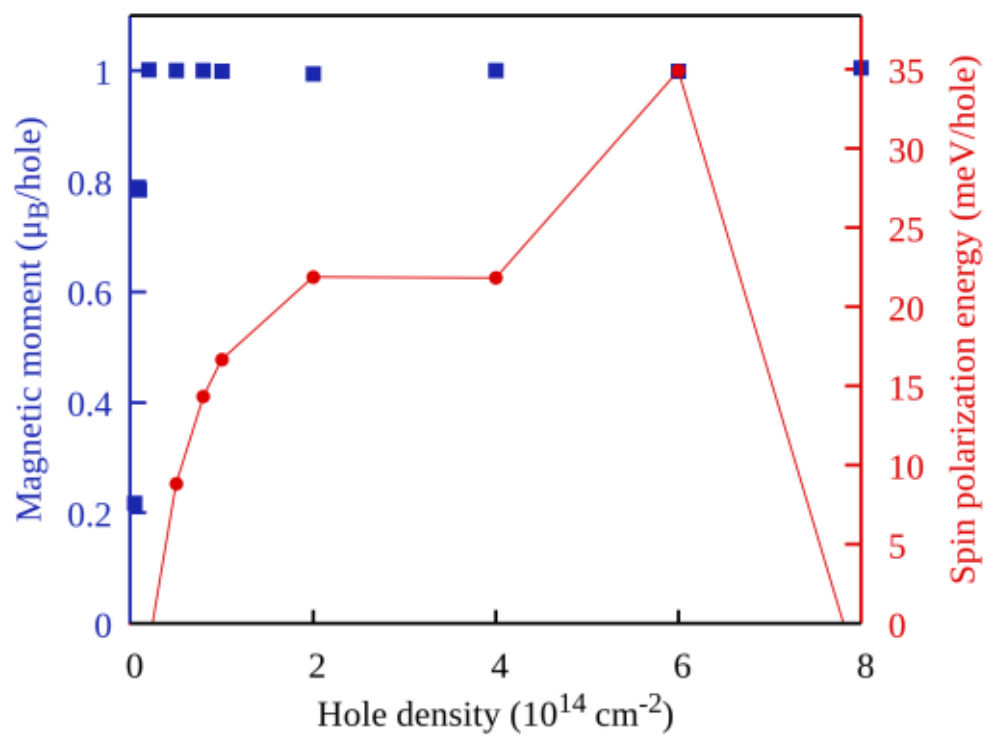

## Magnetic configurations and spin Hamiltonian

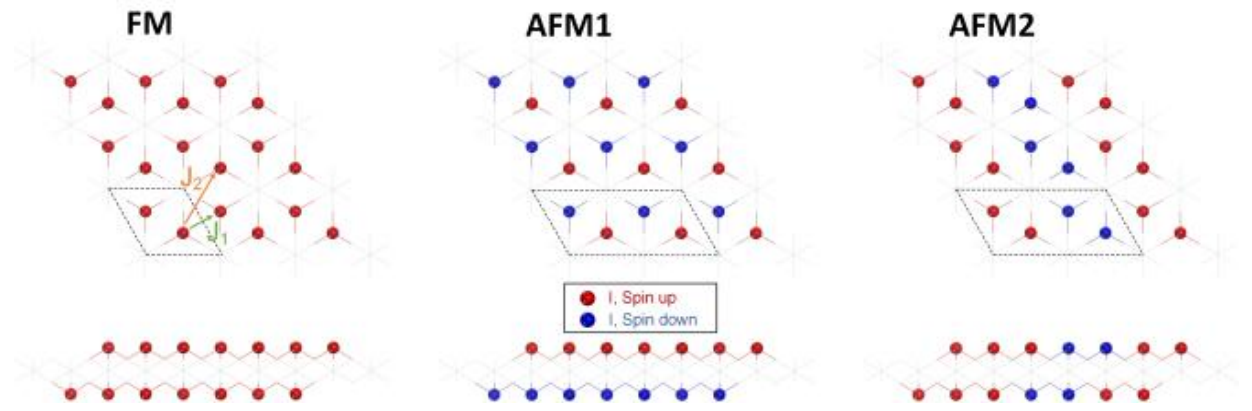

$$E_{FM} = E_0 - 3J_1S^2 - 6J_2S^2$$
$$E_{AFM1} = E_0 + 3J_1S^2 - 6J_2S^2$$
$$E_{AFM2} = E_0 - J_1S^2 + 2J_2S^2$$

## Magnetic exchange coupling parameters

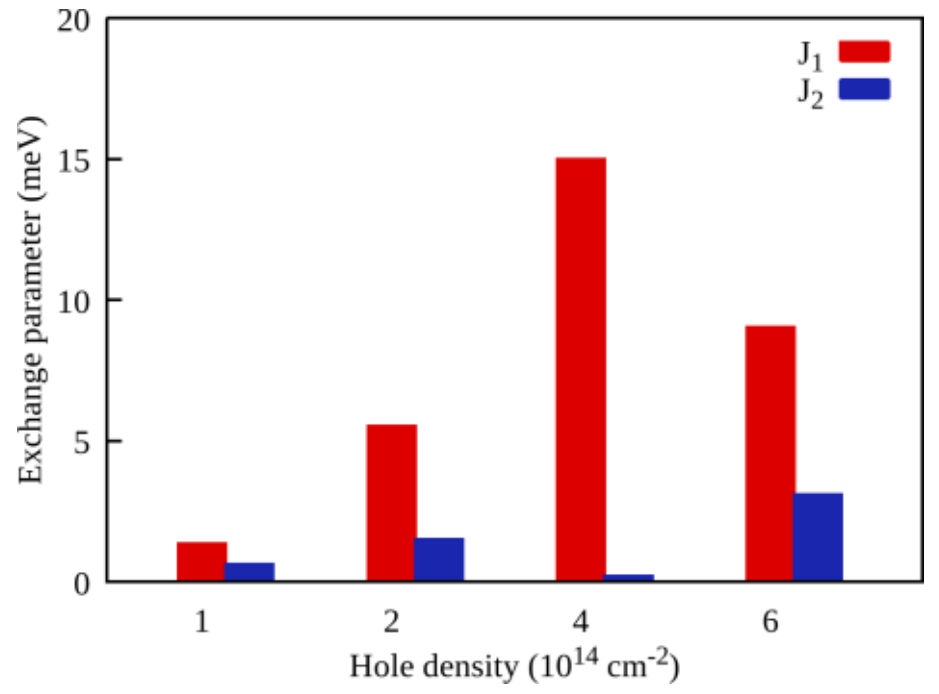

## Magnetic anisotropy energy (MAE, μeV) per magnetic atom

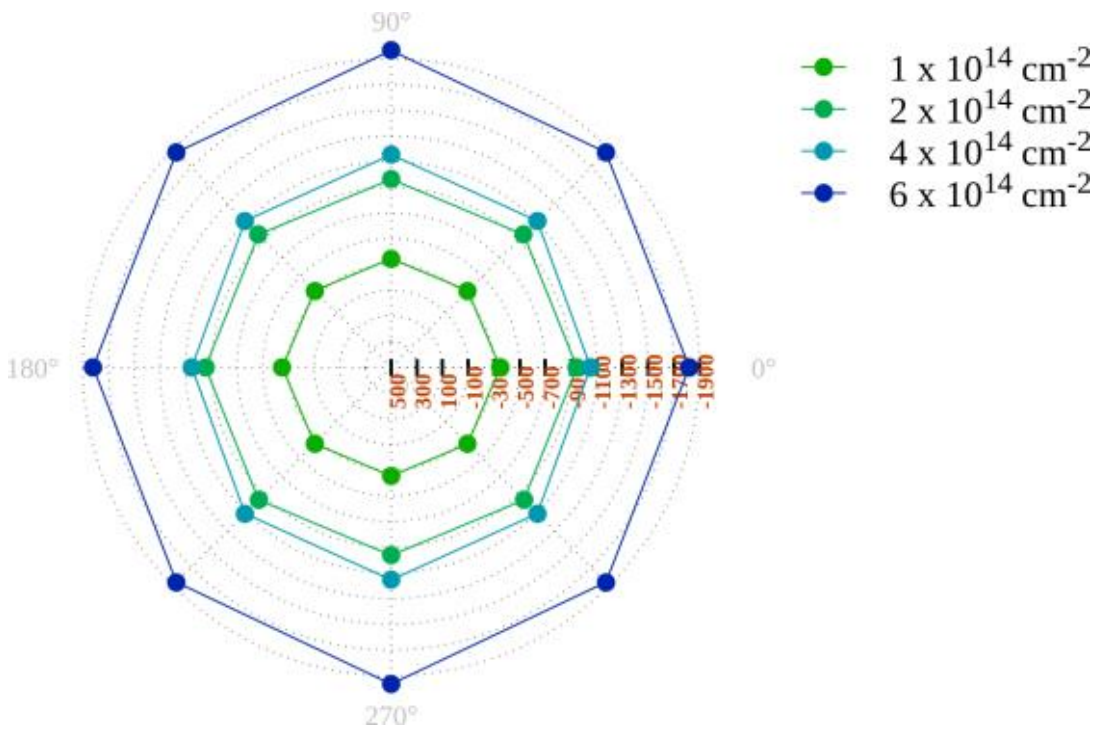

MAE = $E_{\parallel} - E_{\perp}$, a positive (negative) value of MAE indicates the off-plane (in-plane) easy axis.

## Monte Carlo simulations of the normalized magnetization of as a function of temperature

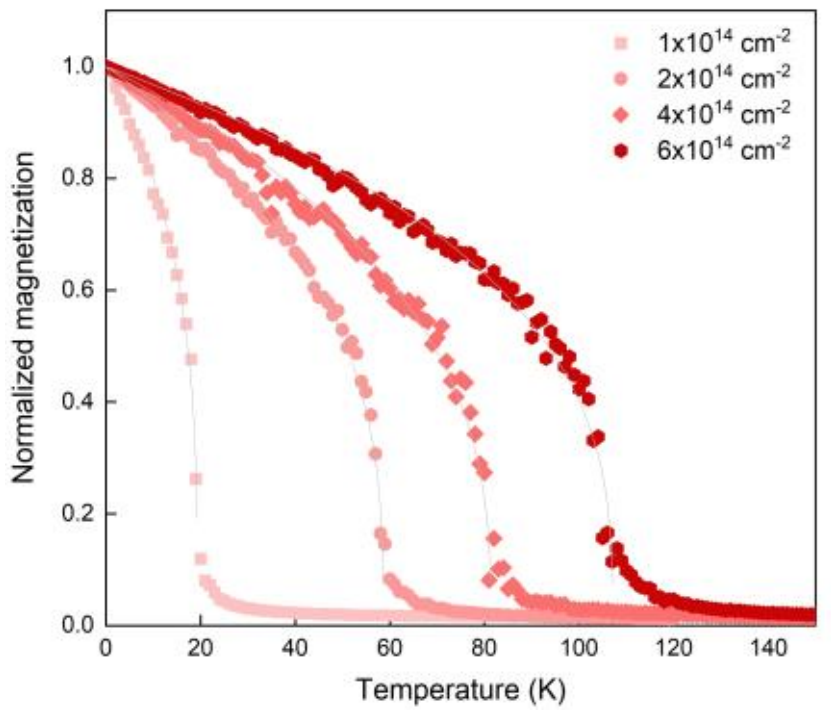

$T_c$: $1\times10^{14}$ cm$^{-2}$: 19 K  $2\times10^{14}$ cm$^{-2}$: 59 K
$4\times10^{14}$ cm$^{-2}$: 81 K  $6\times10^{14}$ cm$^{-2}$: 107 K

# 12. $BaF_2$

| MC2D-ID | C2DB | 2dmat-ID | USPEX | Space group | Band gap (eV) |
|---|---|---|---|---|---|
| - | - | 2dm-519 | - | P3m1 | 6.75 |

**Convex hull** | **Atomic structure** | **Atomic coordinates** | **Phonon dispersion curve**

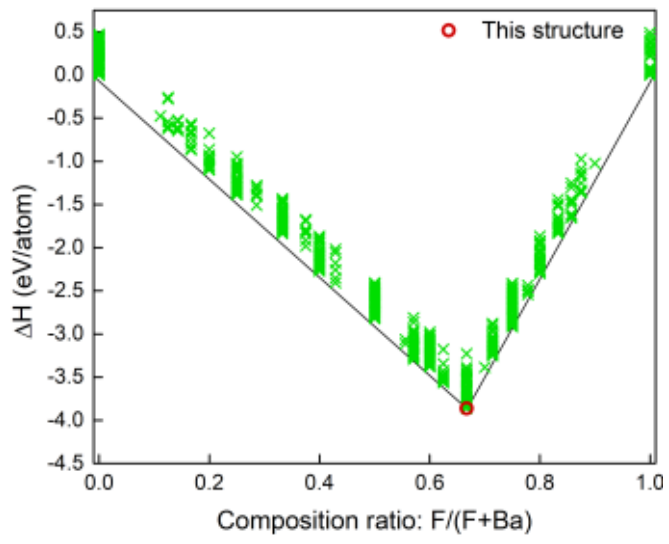

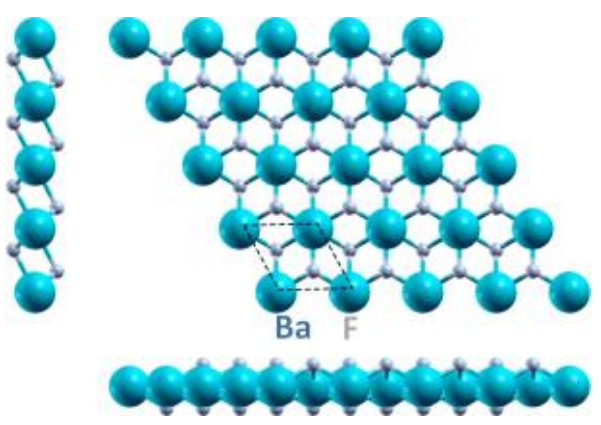

```
Ba1F2
   1.00000000000000
     3.9612593115589059    0.0000000000000000    0.0000000000000000
    -1.9806442192289506    3.4306998803184459    0.0000000000000000
     0.0000000000000000    0.0000000000000000   23.5298450000000017
   Ba    F
     1     2
Direct
  0.0000012602392090  0.0000002237310426  0.9999981429803526
  0.6666713792017092  0.3333413211465981  0.0500095127199103
  0.3333273605590890  0.6666584551223664  0.9499923442997513
```

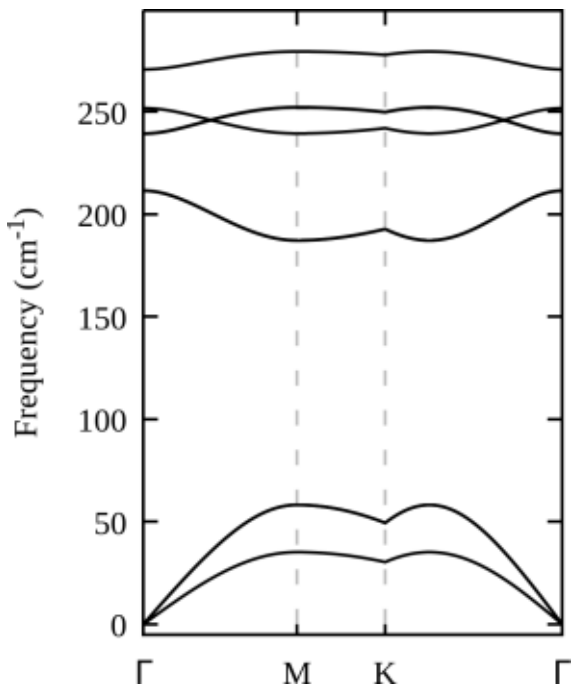

## Projected band structure and density of states

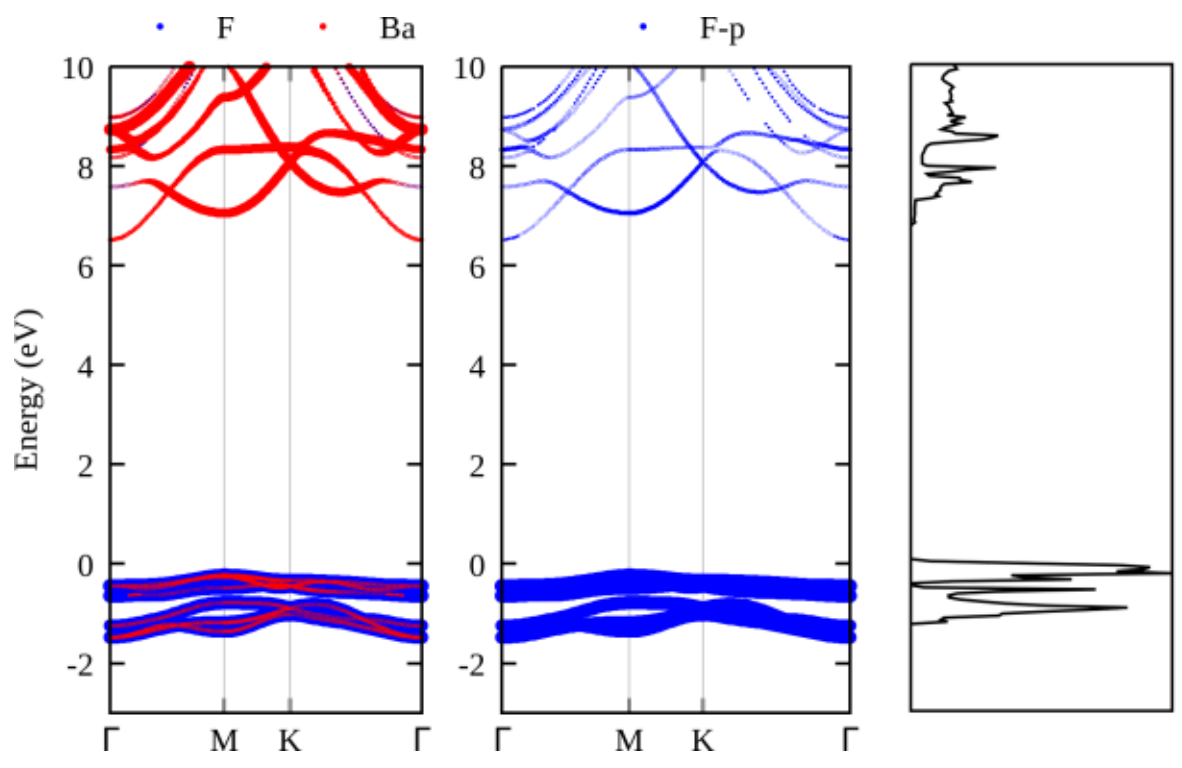

## Magnetic moment and spin polarization energy as a function of hole doping concentration

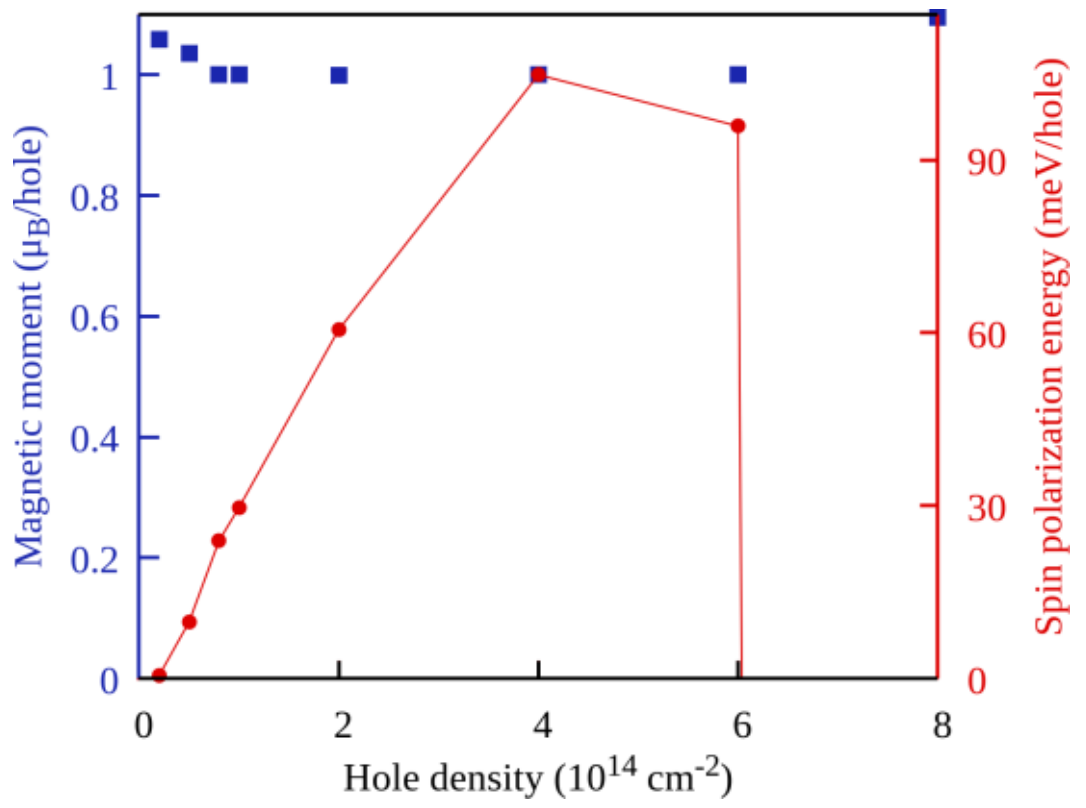

## Magnetic configurations and spin Hamiltonian

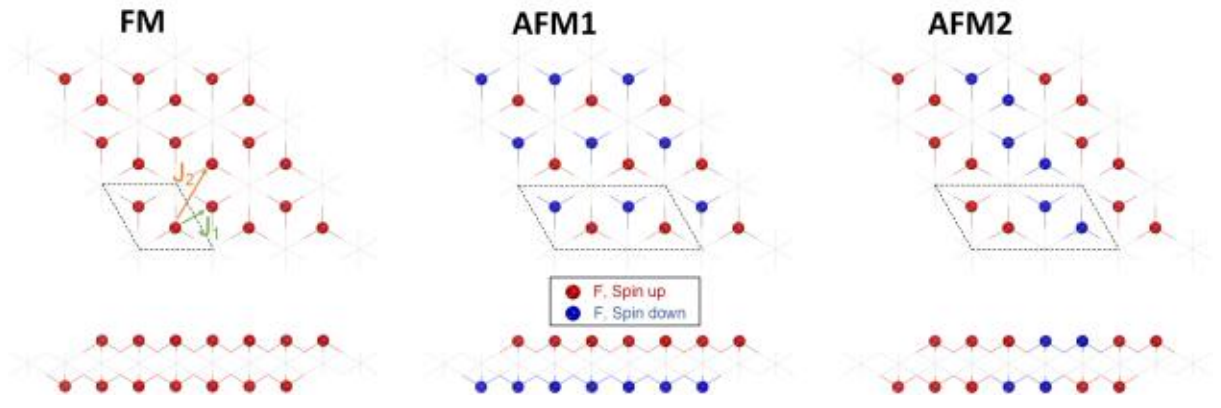

$$E_{FM} = E_0 - 3J_1S^2 - 6J_2S^2$$
$$E_{AFM1} = E_0 + 3J_1S^2 - 6J_2S^2$$
$$E_{AFM2} = E_0 - J_1S^2 + 2J_2S^2$$

## Magnetic exchange coupling parameters

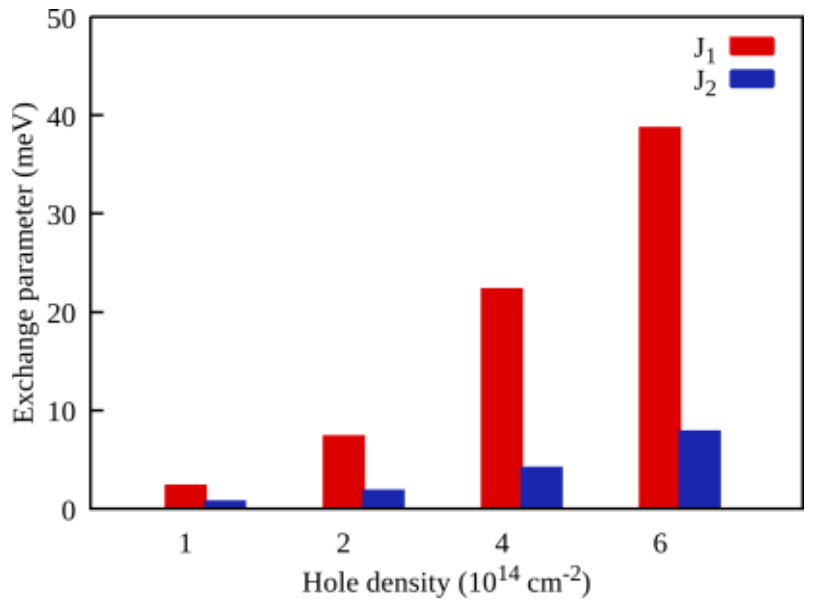

## Magnetic anisotropy energy (MAE, μeV) per magnetic atom

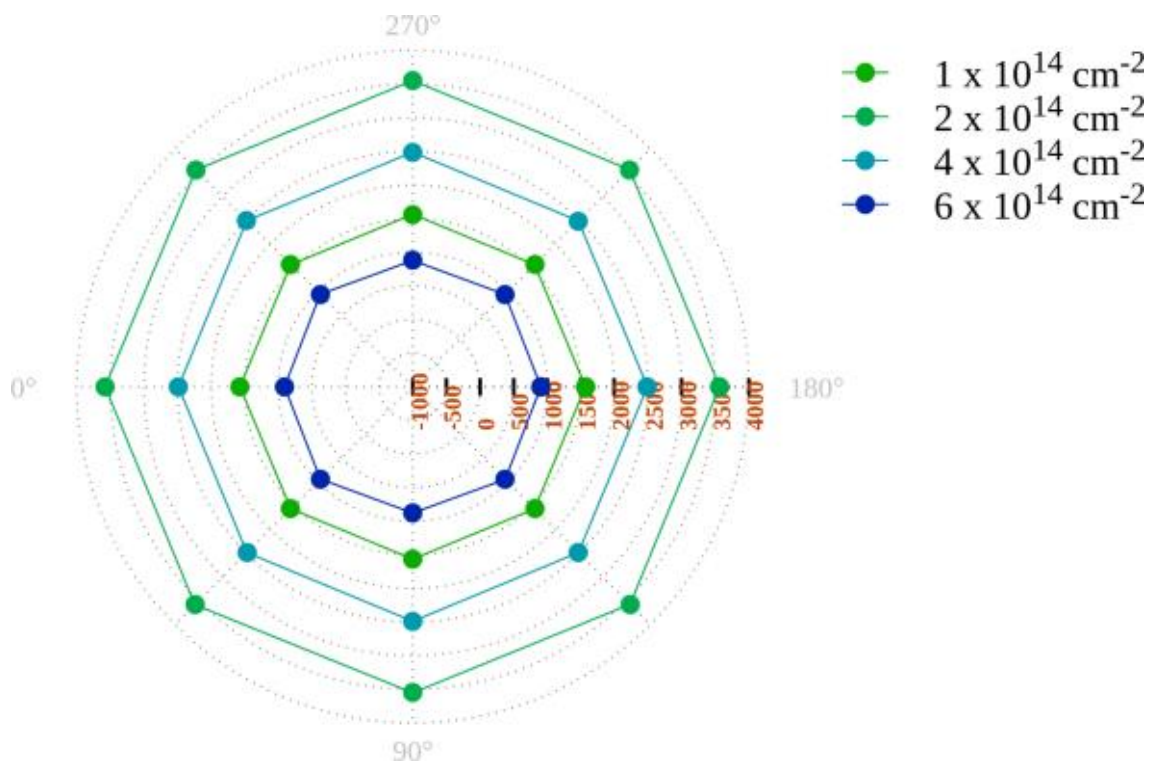

MAE = $E_{\parallel} - E_{\perp}$, a positive (negative) value of MAE indicates the off-plane (in-plane) easy axis.

## Monte Carlo simulations of the normalized magnetization of as a function of temperature

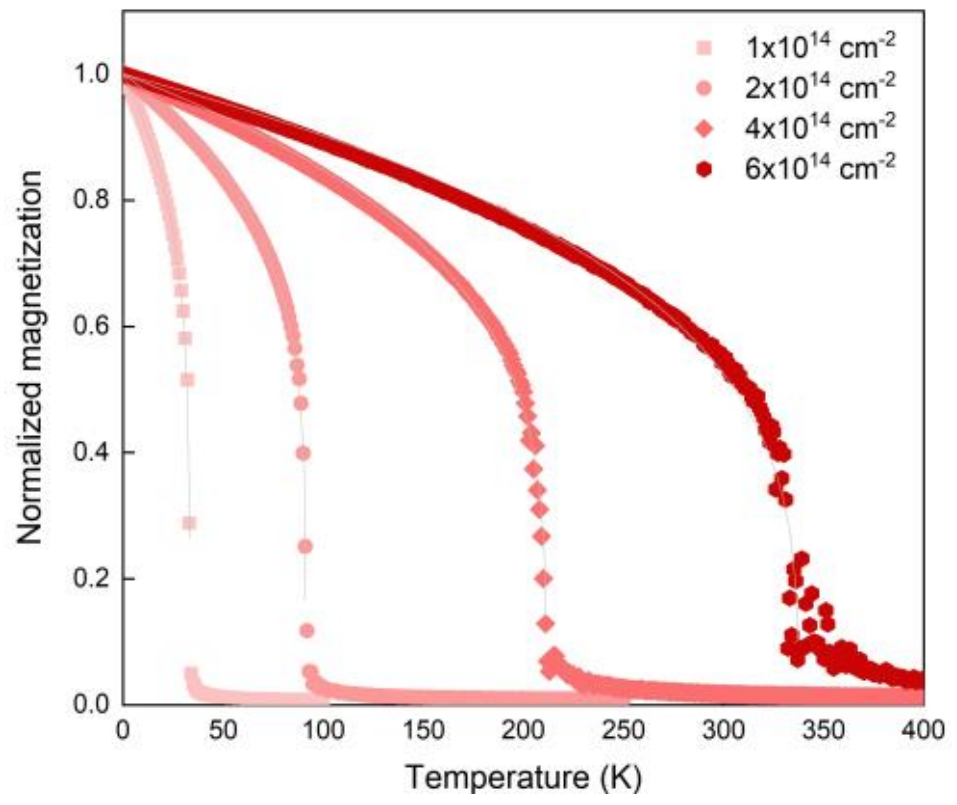

$T_c$: $1\times10^{14}$ cm$^{-2}$: 33 K  $2\times10^{14}$ cm$^{-2}$: 91 K

$4\times10^{14}$ cm$^{-2}$: 211 K  $6\times10^{14}$ cm$^{-2}$: 337

# 13. $BaCl_2$

| MC2D-ID | C2DB | 2dmat-ID | USPEX | Space group | Band gap (eV) |
|---|---|---|---|---|---|
| - | - | 2dm-375 | - | P3m1 | 5.62 |

**Convex hull**

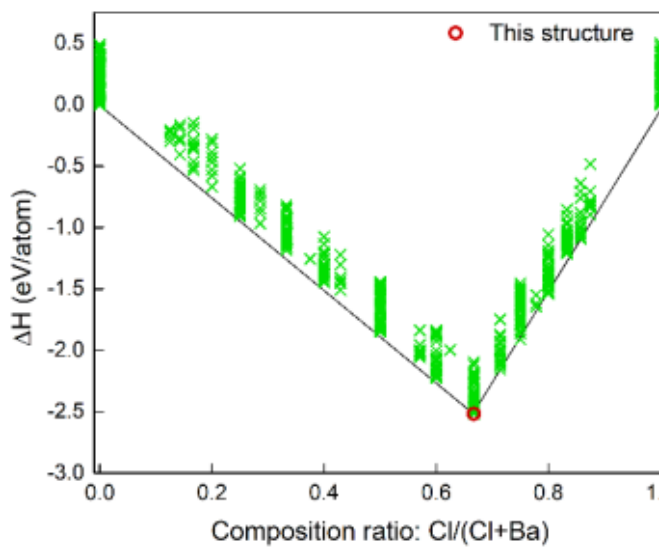

**Atomic structure**

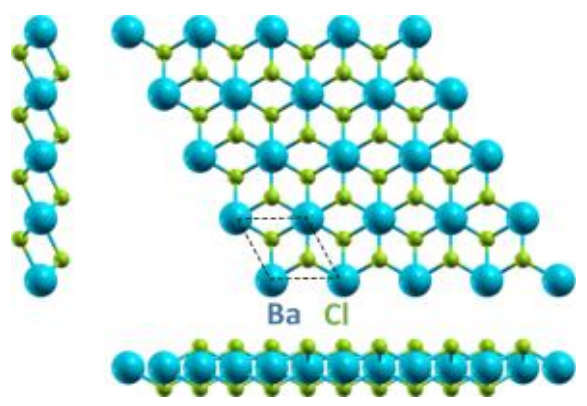

**Atomic coordinates**

```
Ba Cl2
   1.00000000000000
   4.8143921693898761    0.0000000000000000    0.0000000000000000
  -2.4071905294981946    4.1693892796060368    0.0000000000000000
   0.0000000000000000    0.0000000000000000   23.5298450000000017
   Ba    Cl
   1     2
Direct
  0.0000000216941487   0.0000000156677160   0.0000000113663816
  0.6666662872184119   0.3333337859237773   0.0591696875755261
  0.3333336910874394   0.6666661984085067   0.9408303010580994
```

**Phonon dispersion curve**

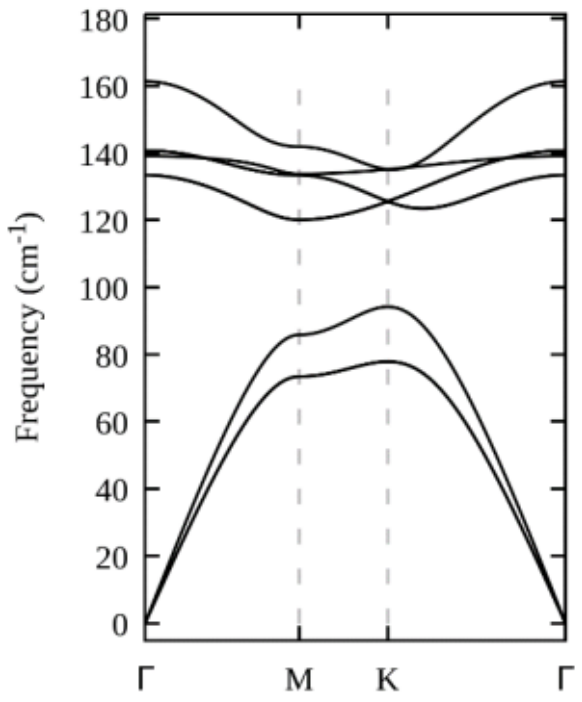

**Projected band structure and density of states**

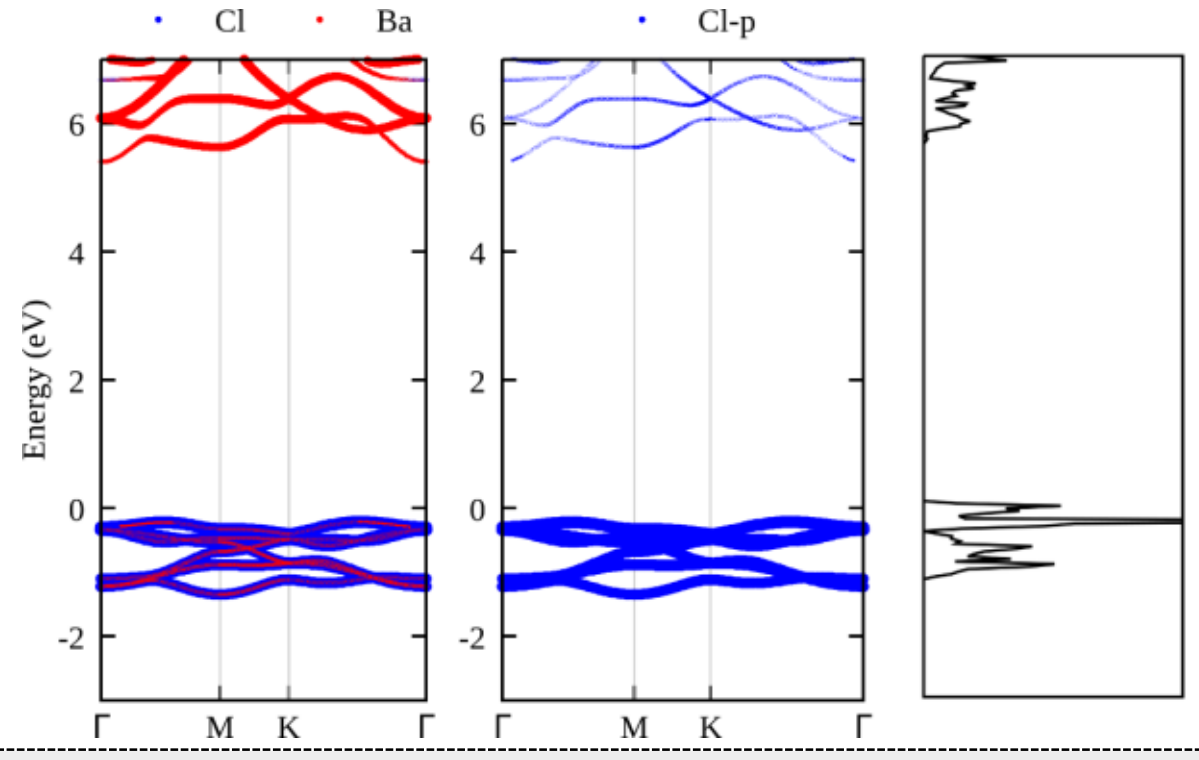

**Magnetic moment and spin polarization energy as a function of hole doping concentration**

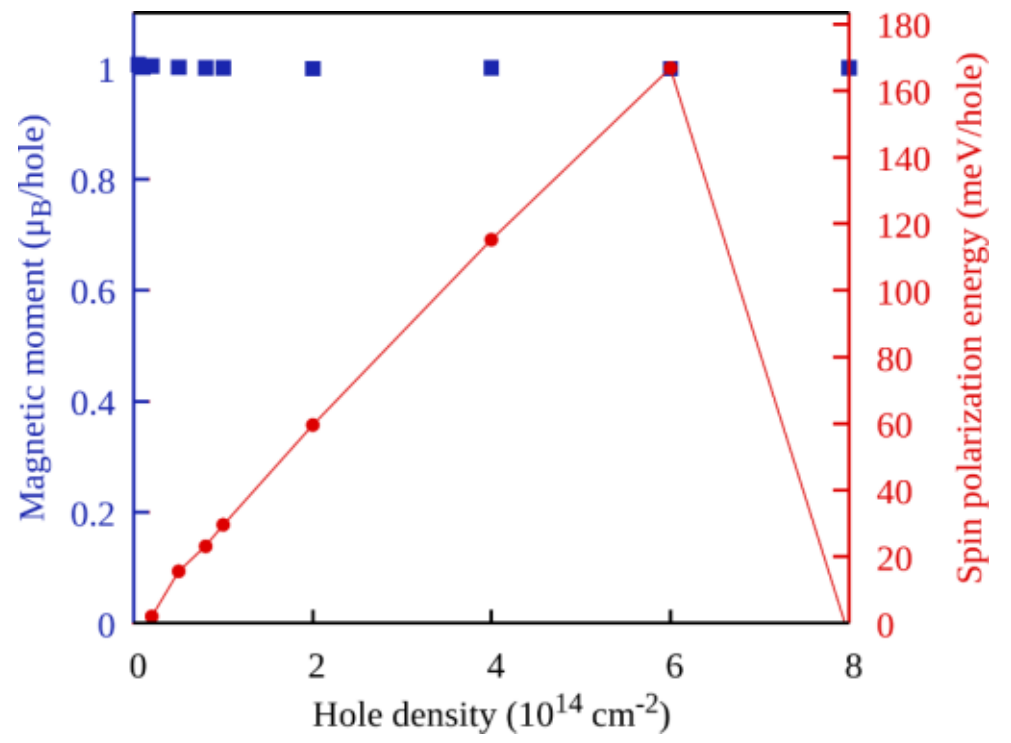

**Magnetic configurations and spin Hamiltonian**

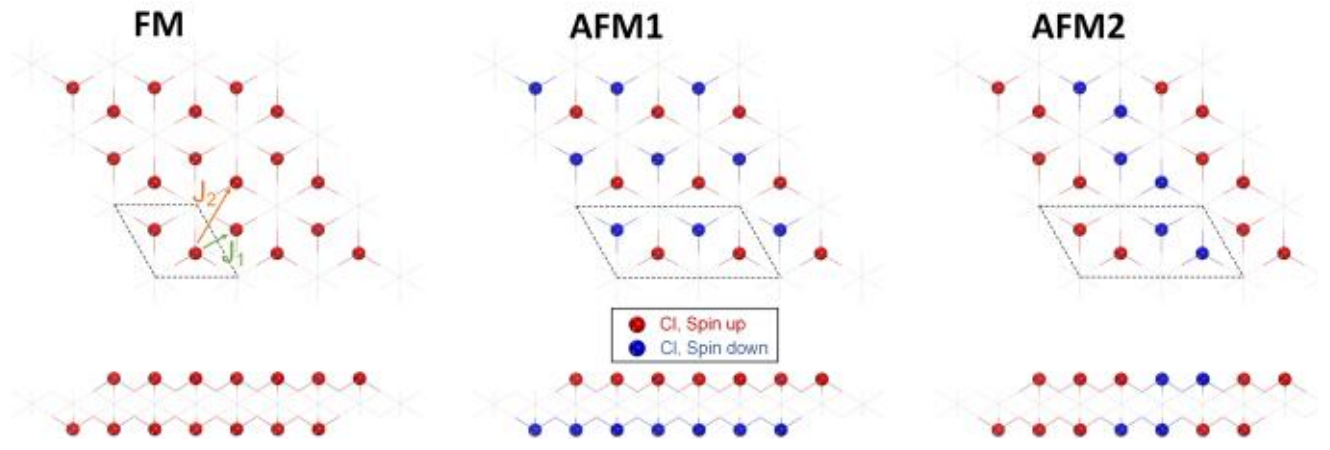

$$E_{FM} = E_0 - 3J_1S^2 - 6J_2S^2$$
$$E_{AFM1} = E_0 + 3J_1S^2 - 6J_2S^2$$
$$E_{AFM2} = E_0 - J_1S^2 + 2J_2S^2$$

**Magnetic exchange coupling parameters**

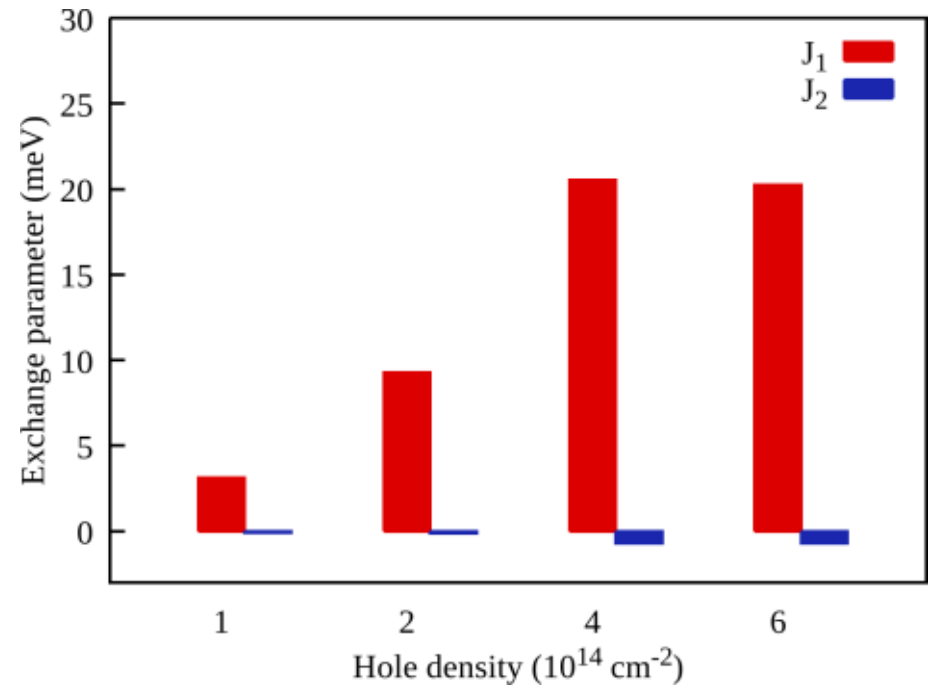

**Magnetic anisotropy energy (MAE, μeV) per magnetic atom**

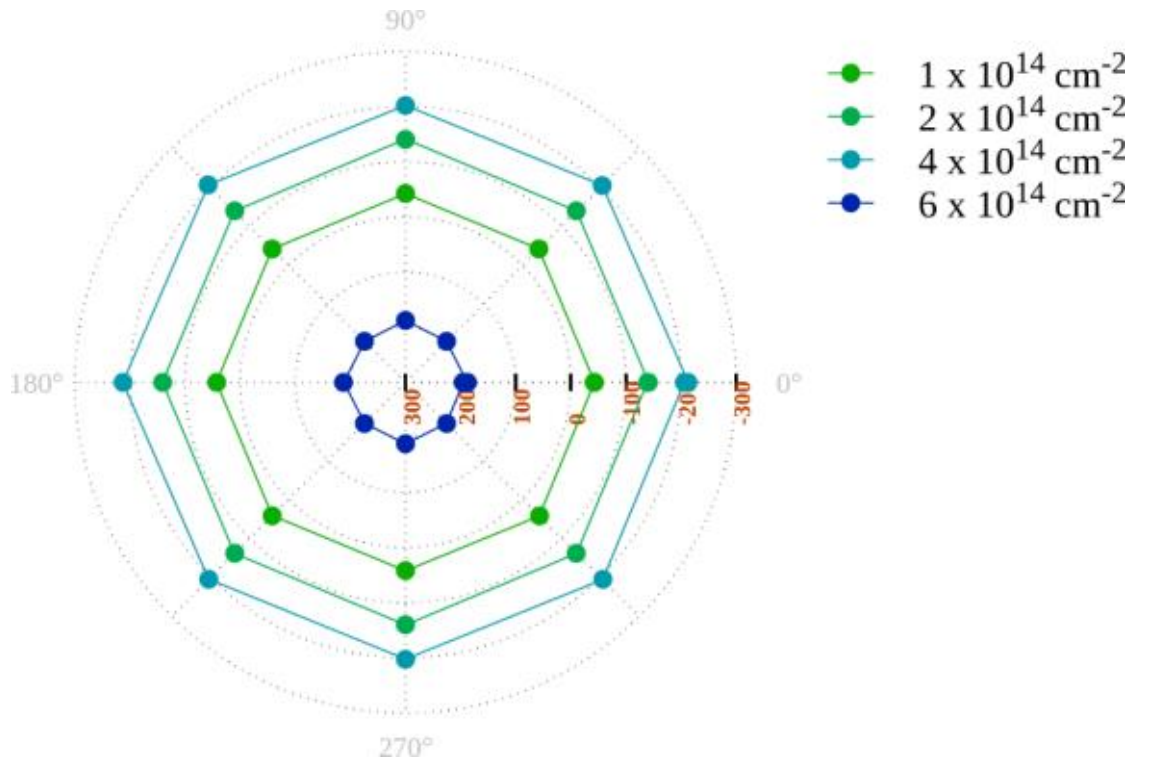

MAE = $E_{\parallel} - E_{\perp}$, a positive (negative) value of MAE indicates the off-plane (in-plane) easy axis.

**Monte Carlo simulations of the normalized magnetization of as a function of temperature**

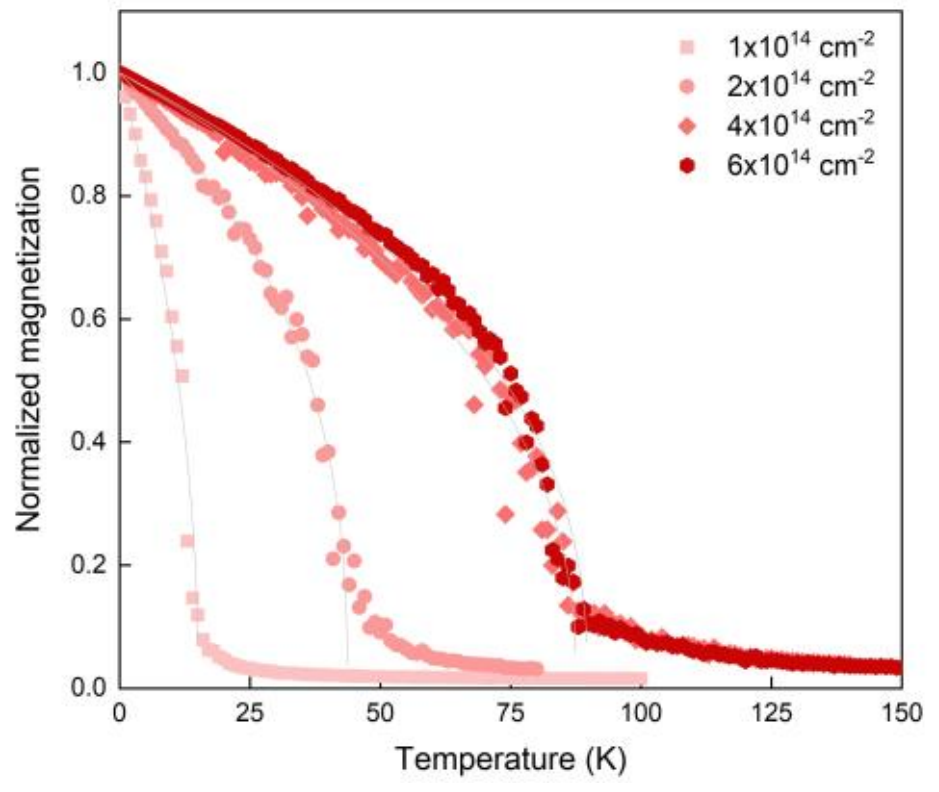

+

$T_c$: $1\times10^{14}$ cm$^{-2}$: 15 K  $2\times10^{14}$ cm$^{-2}$: 44 K

$4\times10^{14}$ cm$^{-2}$: 88 K  $6\times10^{14}$ cm$^{-2}$: 90 K

# 14. $BaBr_2$

| MC2D-ID | C2DB | 2dmat-ID | USPEX | Space group | Band gap (eV) |
|---|---|---|---|---|---|
| - | - | 2dm-606 | - | P3m1 | 4.95 |

**Convex hull**

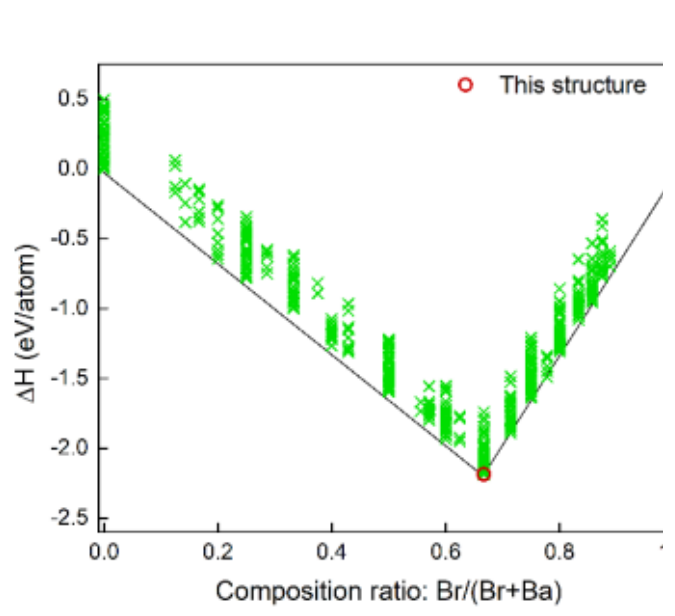

**Atomic structure**

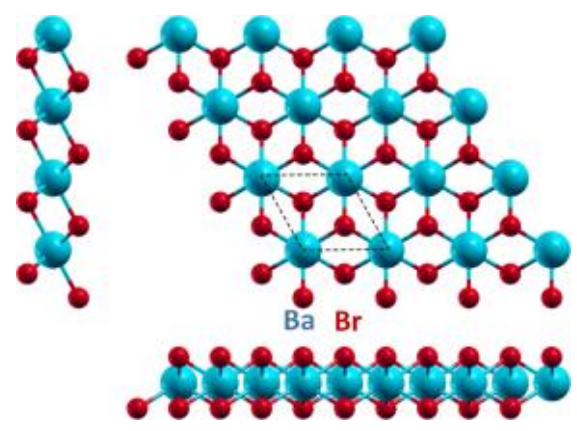

**Atomic coordinates**

```
Ba1Br2
   1.00000000000000
   4.9850549887801128   0.0000000000000000   0.0000000000000000
  -2.4925186364639837   4.3171932506893418   0.0000000000000000
   0.0000000000000000   0.0000000000000000  23.5298450000000017
   Ba   Br
   1    2
Direct
  0.9999999932224100  0.9999999868052072  0.0000000047005386
  0.6666665621886381  0.3333334368866261  0.0655372747501346
  0.3333334445889591  0.6666665763081667  0.9344627205493268
```

**Phonon dispersion curve**

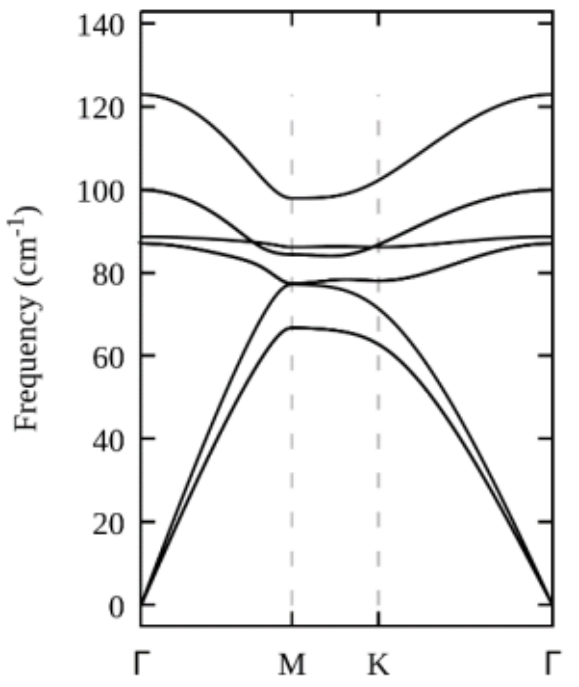

**Projected band structure and density of states**

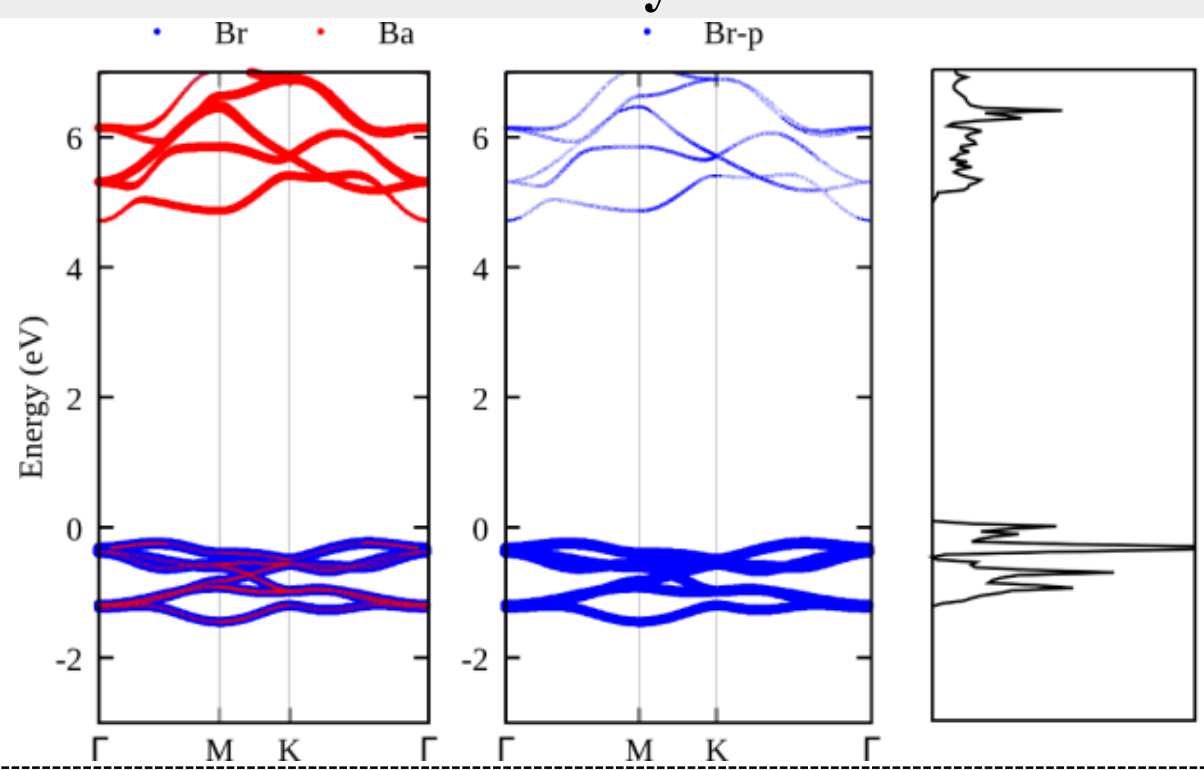

**Magnetic moment and spin polarization energy as a function of hole doping concentration**

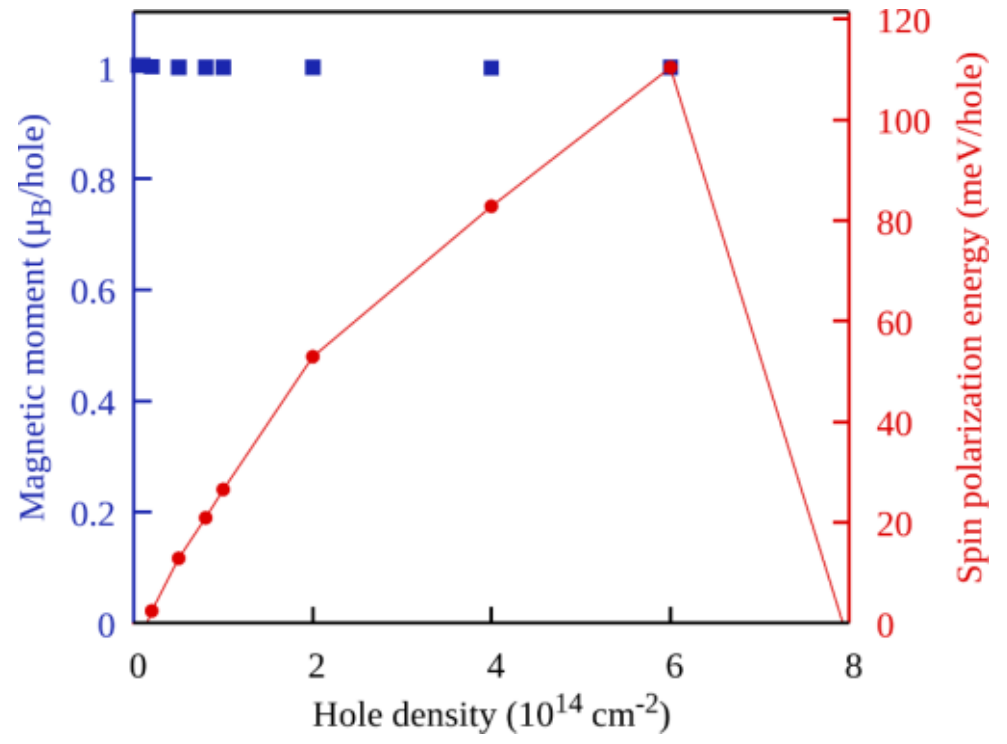

**Magnetic configurations and spin Hamiltonian**

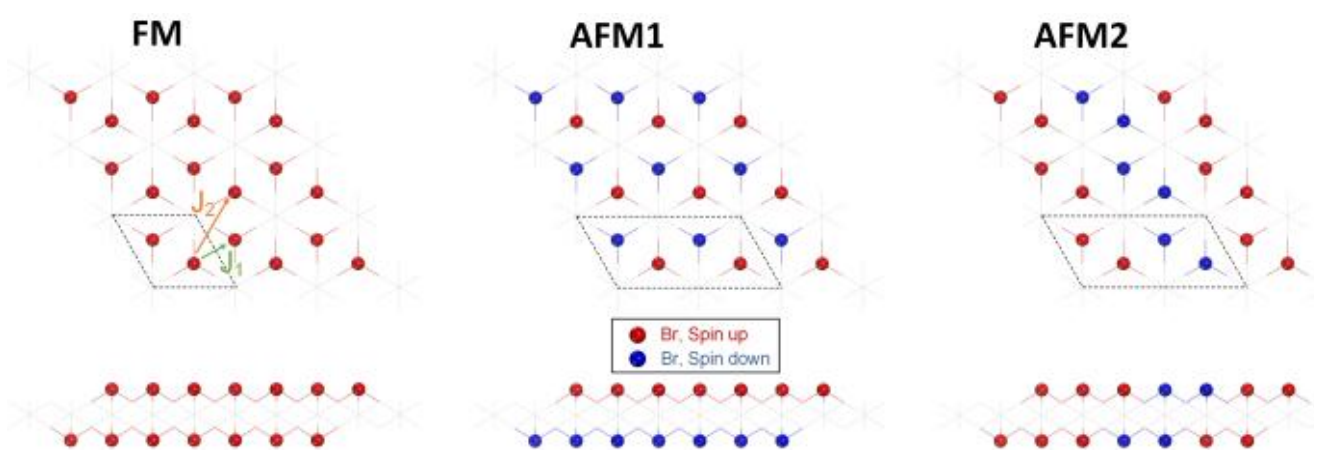

$$E_{FM} = E_0 - 3J_1S^2 - 6J_2S^2$$

$$E_{AFM1} = E_0 + 3J_1S^2 - 6J_2S^2$$

$$E_{AFM2} = E_0 - J_1S^2 + 2J_2S^2$$

**Magnetic exchange coupling parameters**

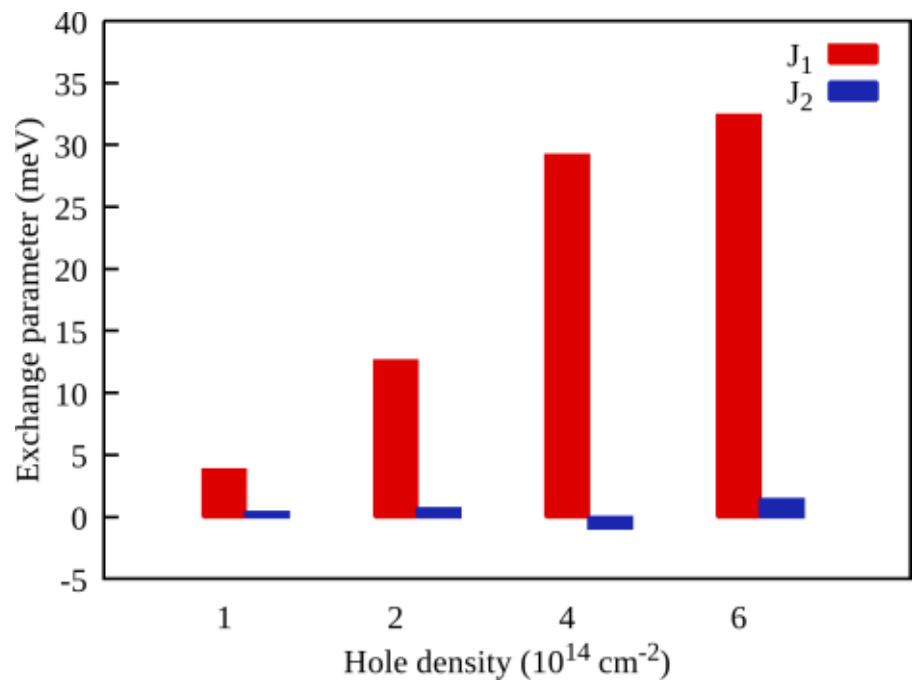

**Magnetic anisotropy energy (MAE, μeV) per magnetic atom**

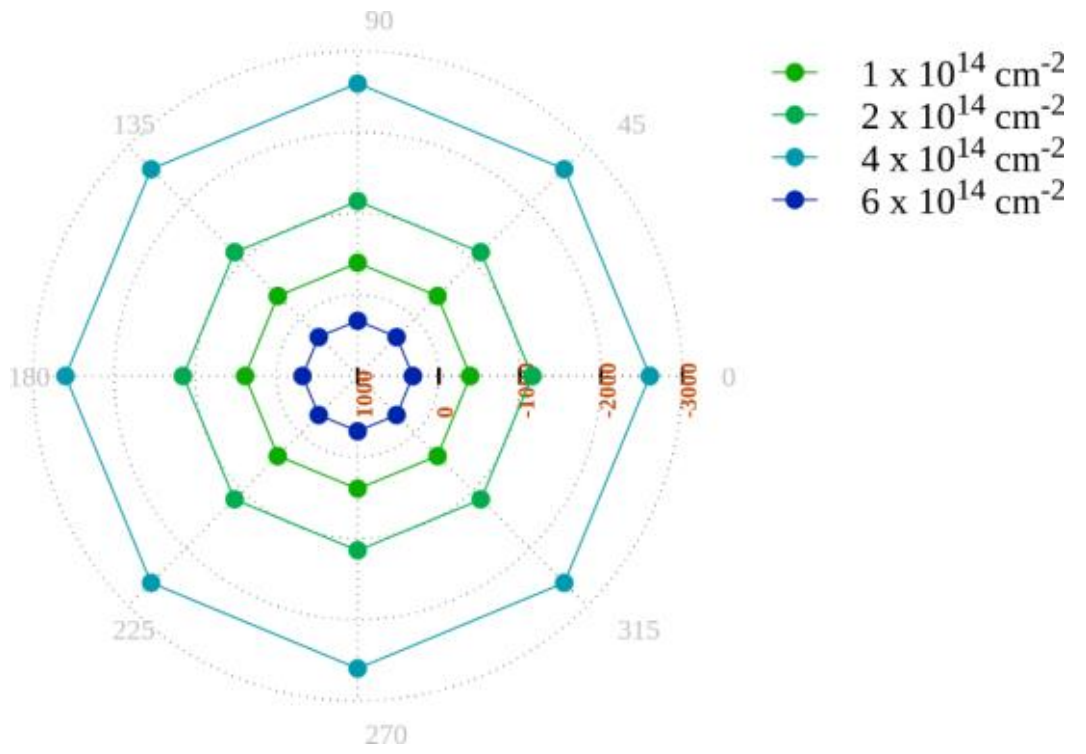

MAE = $E_{\parallel} - E_{\perp}$, a positive (negative) value of MAE indicates the off-plane (in-plane) easy axis.

**Monte Carlo simulations of the normalized magnetization of as a function of temperature**

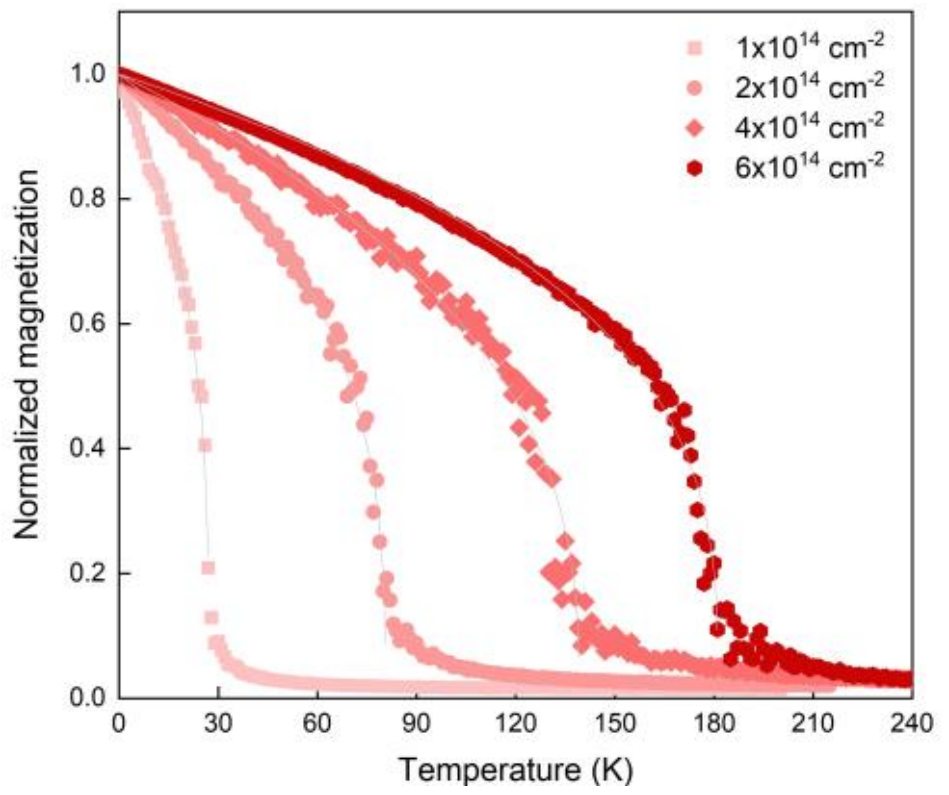

$T_c$: $1\times10^{14}$ cm$^{-2}$: 27 K  $2\times10^{14}$ cm$^{-2}$: 81 K

$4\times10^{14}$ cm$^{-2}$: 139 K  $6\times10^{14}$ cm$^{-2}$: 181

# 15. $BaI_2$

| MC2D-ID | C2DB | 2dmat-ID | USPEX | Space group | Band gap (eV) |
|---|---|---|---|---|---|
| - | - | 2dm-804 | - | P3m1 | 4.15 |

**Convex hull** | **Atomic structure** | **Atomic coordinates** | **Phonon dispersion curve**

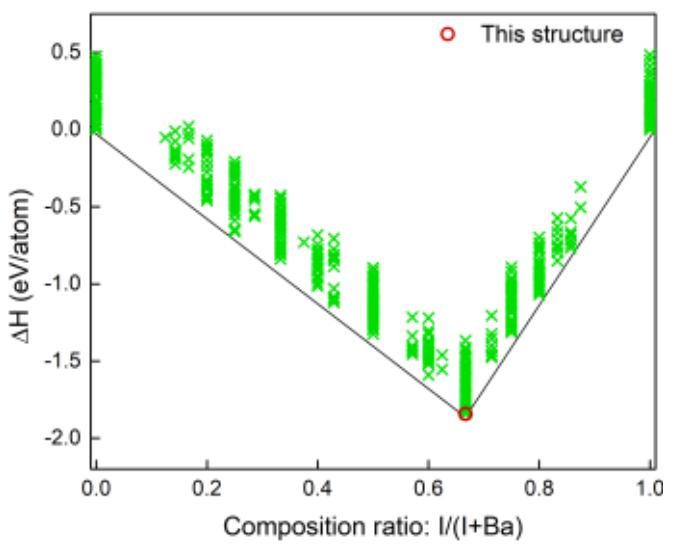

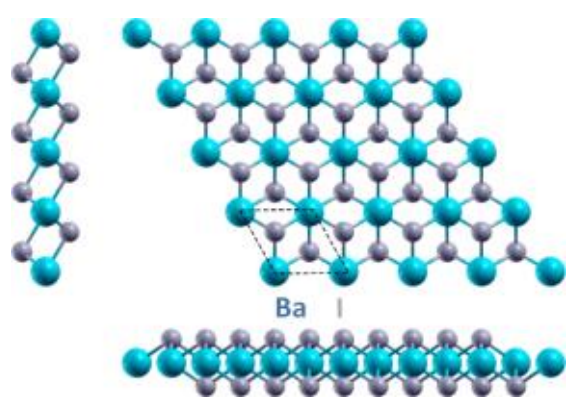

```
Ba1I2
   1.00000000000000
   5.2073983008725655   0.0000000000000000   0.0000000000000000
  -2.6037054740318659    .5097379194447873   0.0000000000000000
   0.0000000000000000   0.0000000000000000  22.7427169999999990
   Ba   I
   1     2
Direct
  0.0000000031733549   0.0000000081148528   0.9999999910277495
  0.6666664477720730   0.3333334854018304   0.9217489752268904
  0.3333335490545721   0.6666665064833168   0.0782510337453601
```

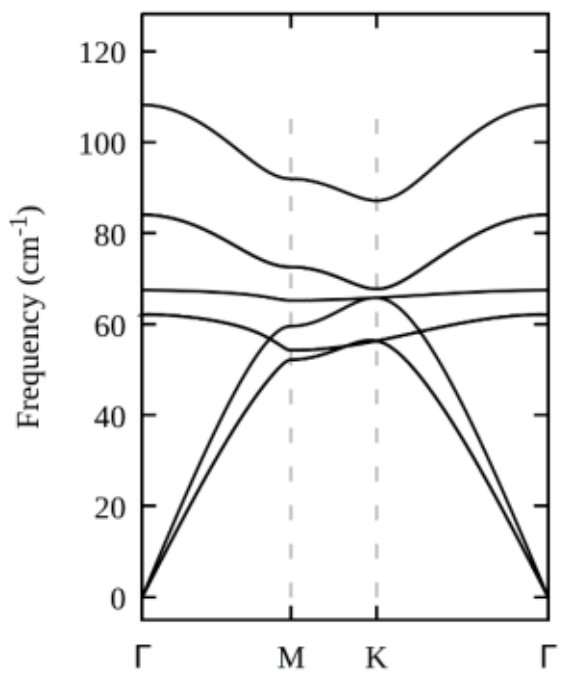

**Projected band structure and density of states** | **Magnetic moment and spin polarization energy as a function of hole doping concentration**

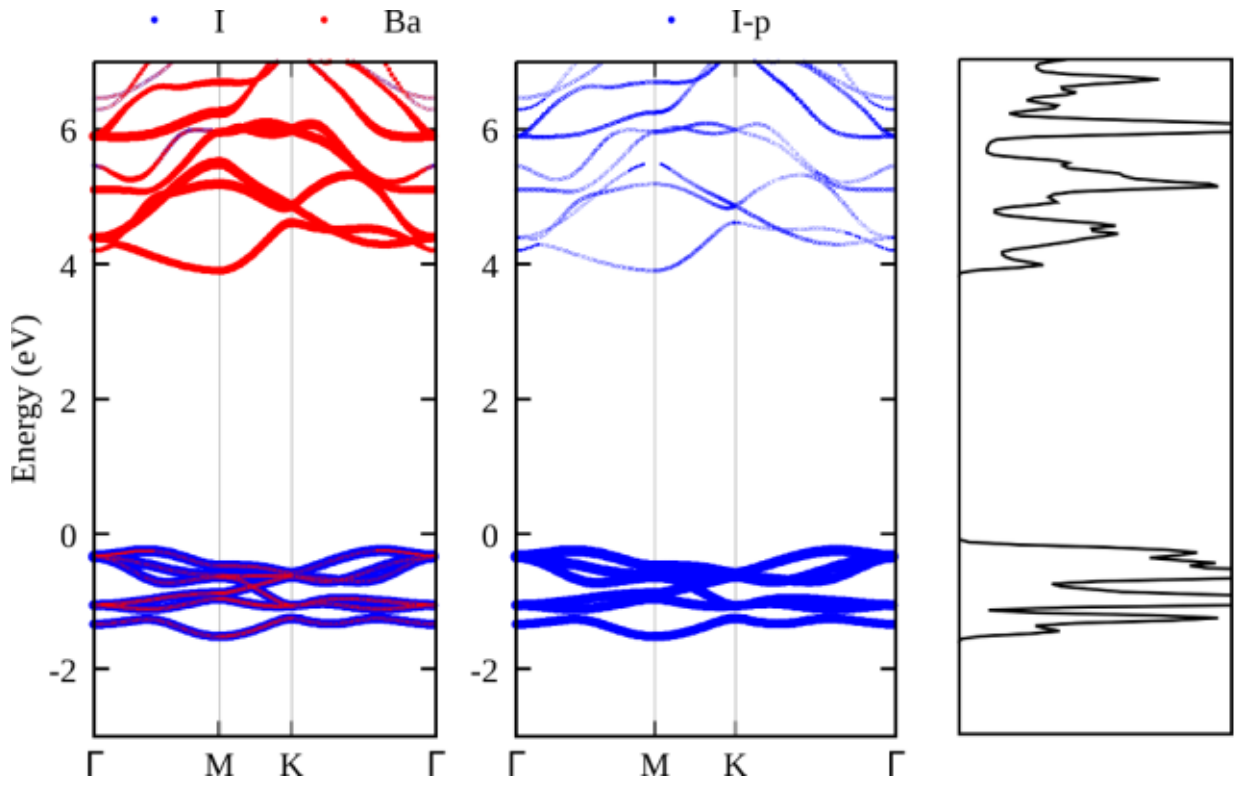

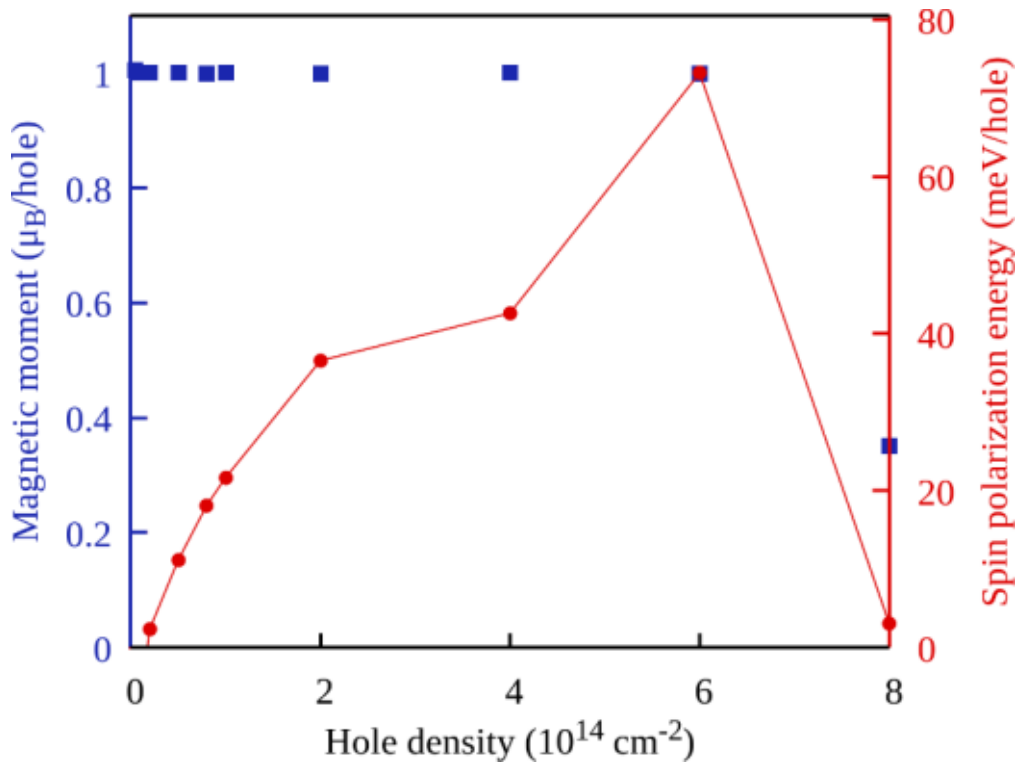

**Magnetic configurations and spin Hamiltonian** | **Magnetic exchange coupling parameters**

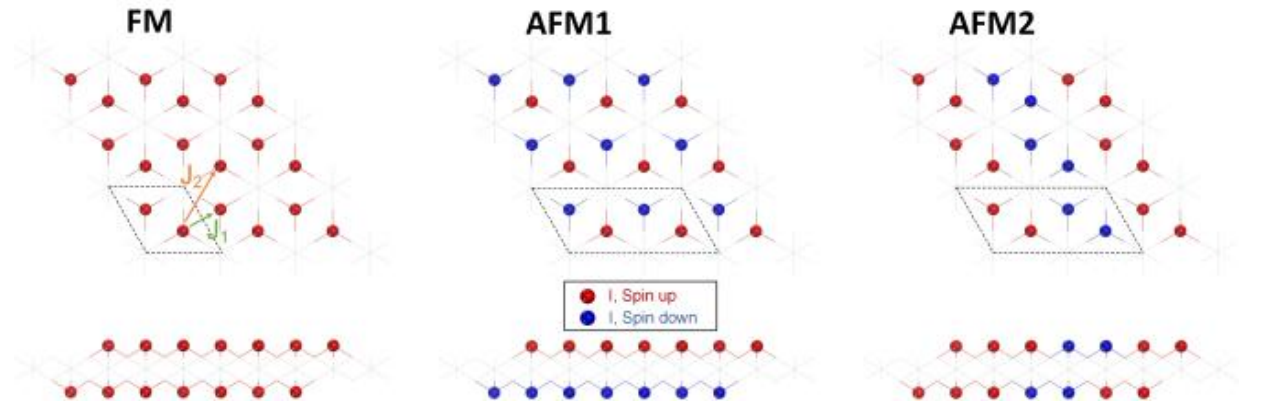

$$E_{FM} = E_0 - 3J_1S^2 - 6J_2S^2$$
$$E_{AFM1} = E_0 + 3J_1S^2 - 6J_2S^2$$
$$E_{AFM2} = E_0 - J_1S^2 + 2J_2S^2$$

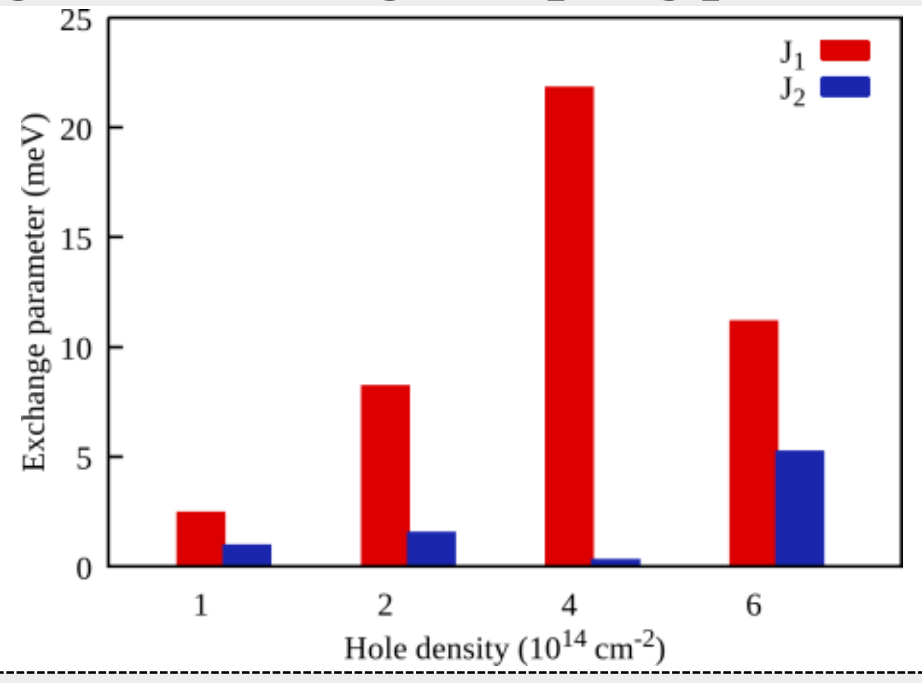

**Magnetic anisotropy energy (MAE, μeV) per magnetic atom** | **Monte Carlo simulations of the normalized magnetization of as a function of temperature**

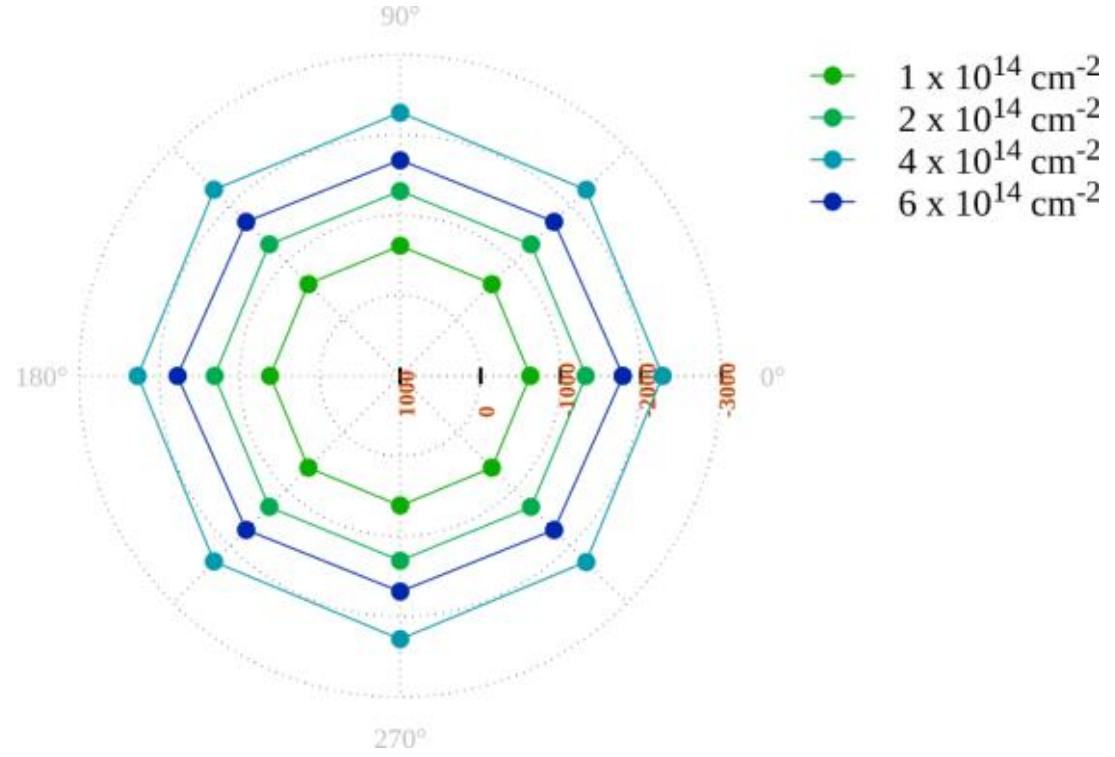

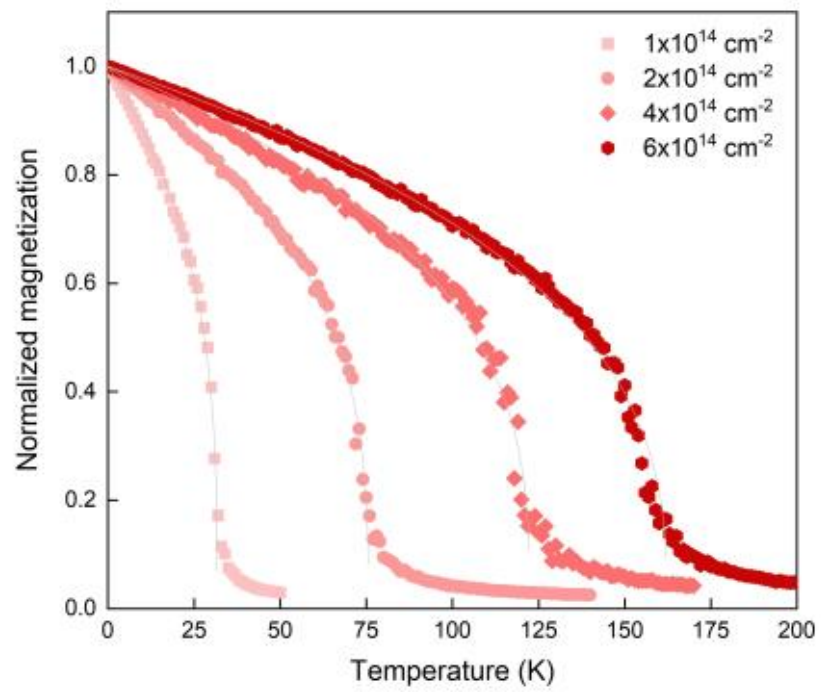

MAE = $E_{\parallel} - E_{\perp}$, a positive (negative) value of MAE indicates the off-plane (in-plane) easy axis.

$T_c$: $1\times10^{14}$ cm$^{-2}$: 32 K  $2\times10^{14}$ cm$^{-2}$: 76 K  $4\times10^{14}$ cm$^{-2}$: 122 K  $6\times10^{14}$ cm$^{-2}$: 162

# 16.ZnF$_2$

| MC2D-ID | C2DB | 2dmat-ID | USPEX | Space group | Band gap (eV) |
|---|---|---|---|---|---|
| - | - | - | ✓ | P3m1 | 4.47 |

## Convex hull

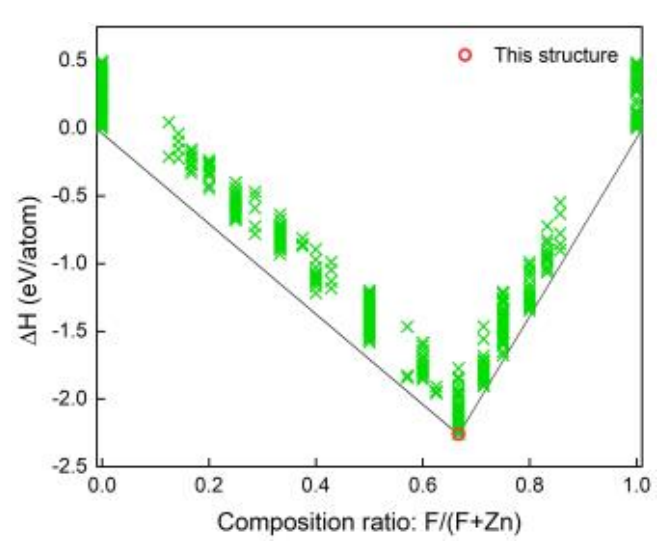

## Atomic structure

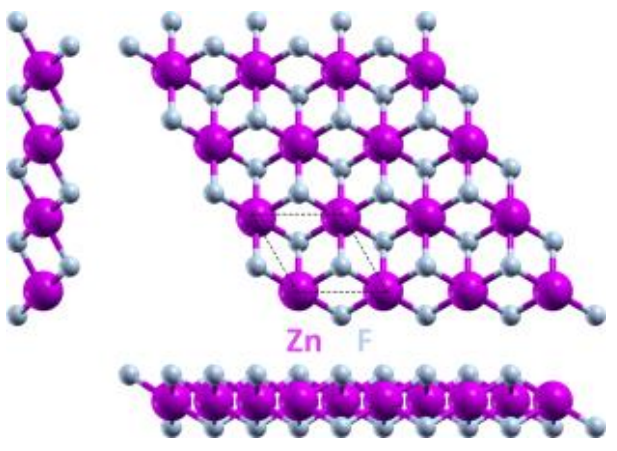

## Atomic coordinates

```
Zn1F2
1.00000000000000
3.1800763879302671    0.0000000000000000    0.0000000000000000
-1.5901079974308439    2.7539876665380705    0.0000000000000000
0.0000000000000014    0.0000000000000024    22.7725935562999986
Zn    F
1    2
Direct
0.6666707971307986    0.3333293903718726    0.5000079107596136
0.3333324786651346    0.6666674978137763    0.5431399079587464
0.9999967242521777    0.0000031119105515    0.4568600920412536
```

## Phonon dispersion curve

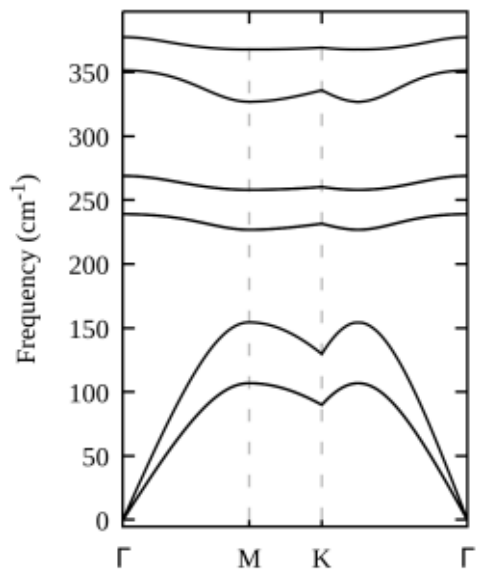

## Projected band structure and density of states

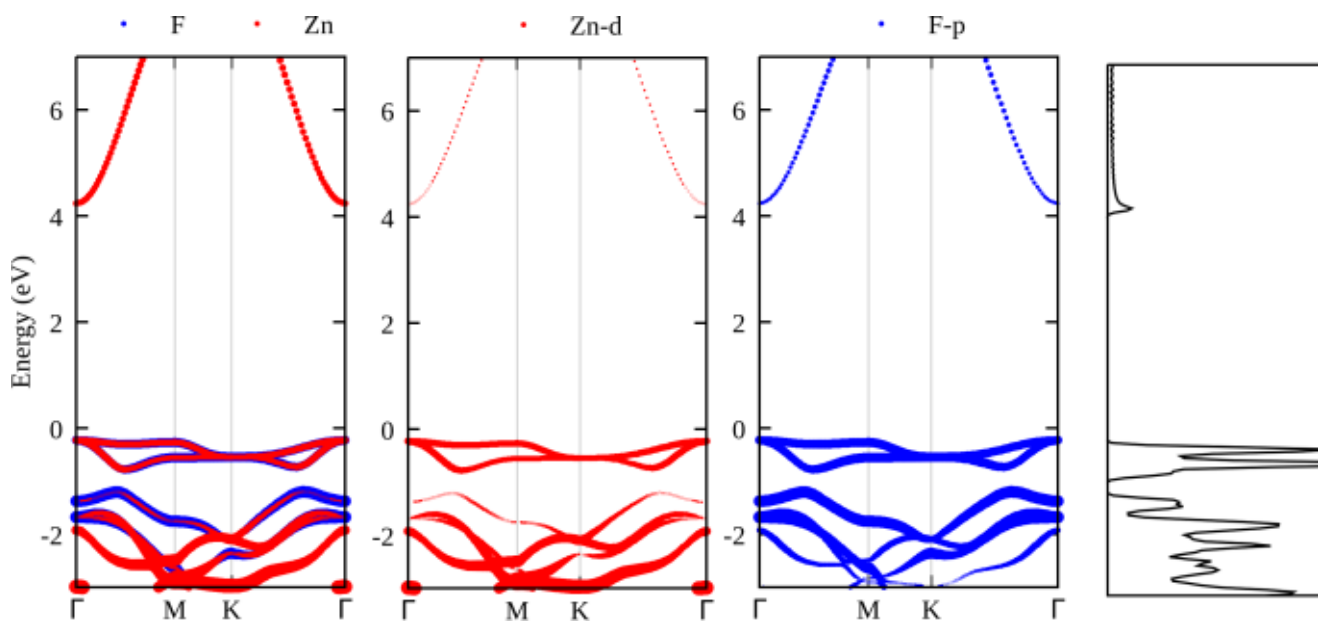

## Magnetic moment and spin polarization energy as a function of hole doping concentration

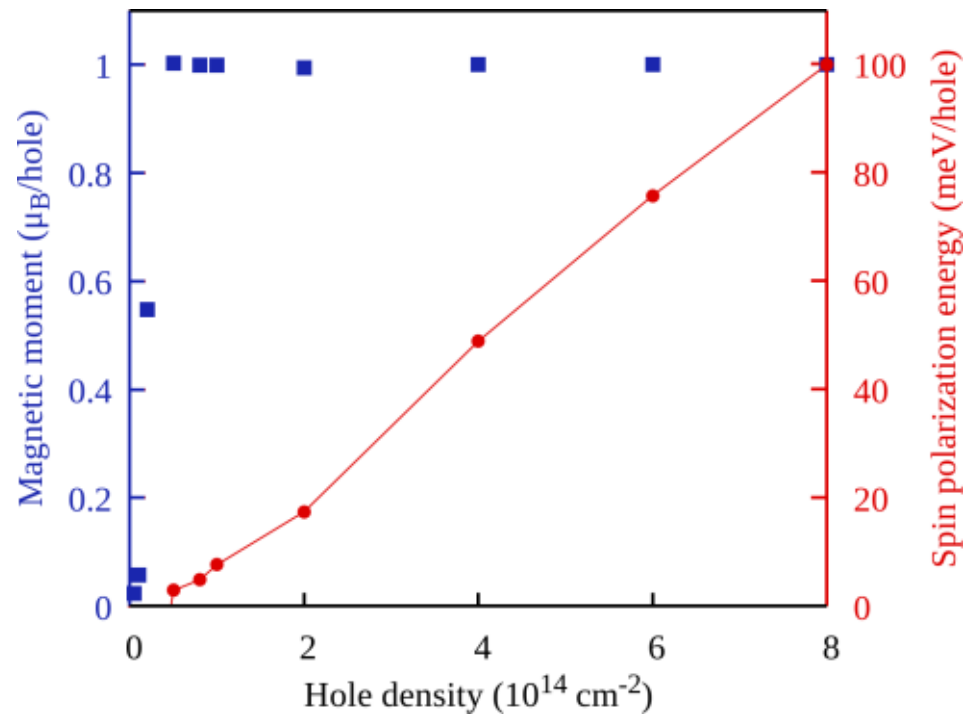

## Magnetic configurations and spin Hamiltonian

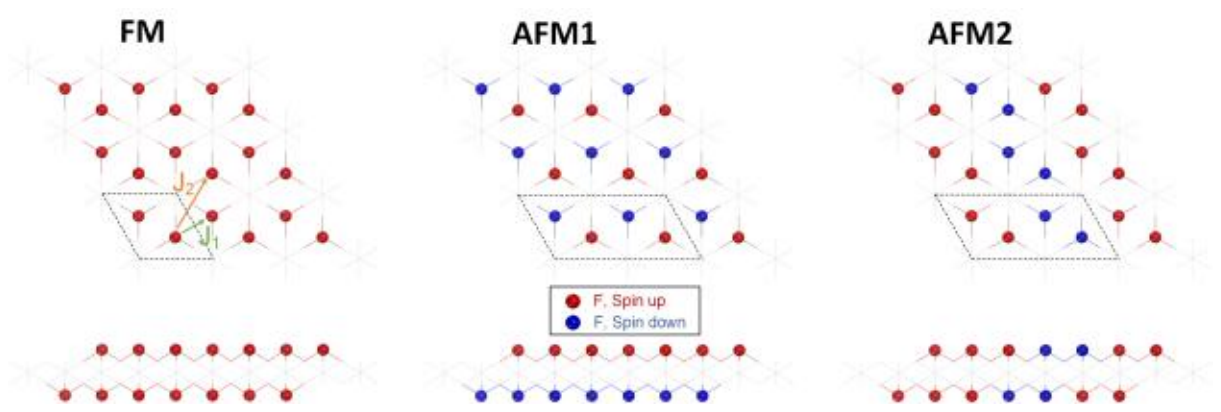

$$E_{FM} = E_0 - 3J_1S^2 - 6J_2S^2$$
$$E_{AFM1} = E_0 + 3J_1S^2 - 6J_2S^2$$
$$E_{AFM2} = E_0 - J_1S^2 + 2J_2S^2$$

## Magnetic exchange coupling parameters

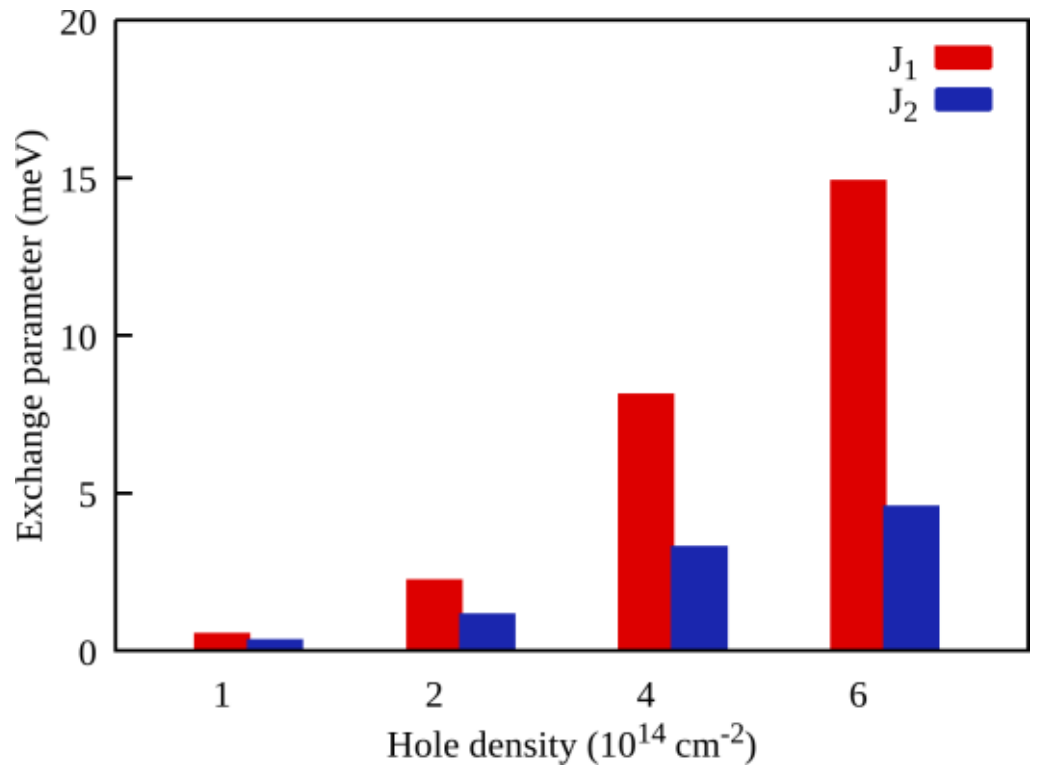

## Magnetic anisotropy energy (MAE, μeV) per magnetic atom

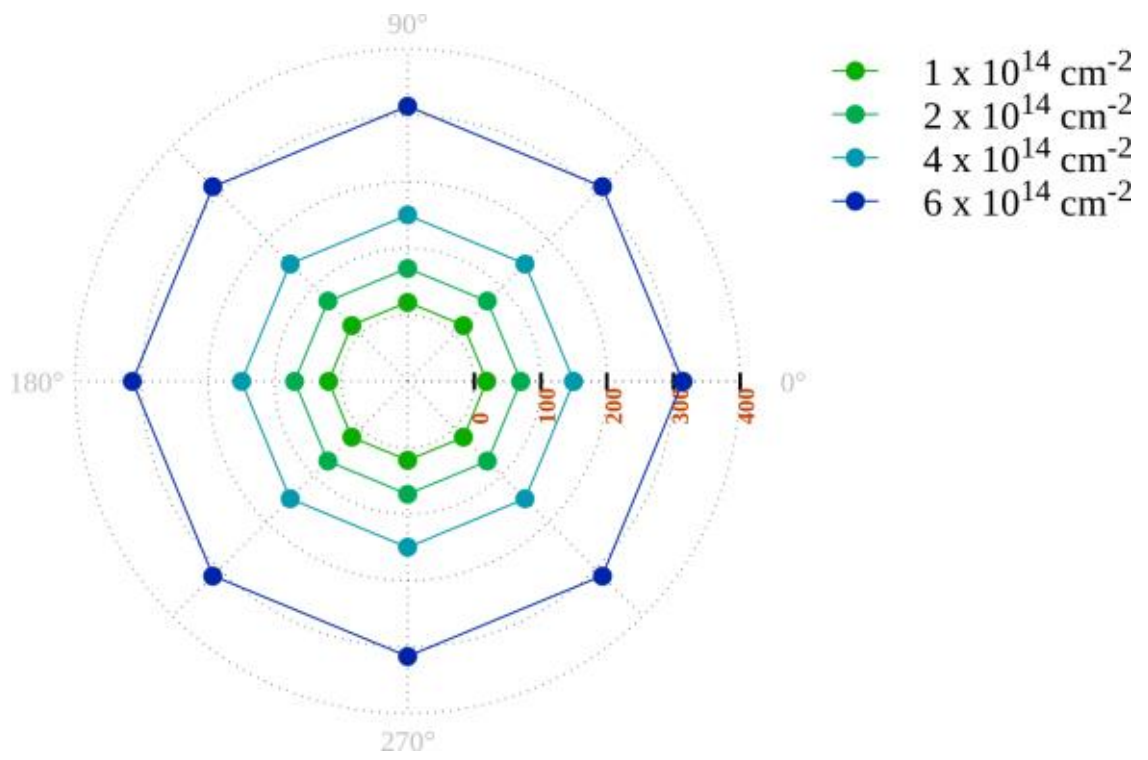

MAE = $E_{\parallel} - E_{\perp}$, a positive (negative) value of MAE indicates the off-plane (in-plane) easy axis.

## Monte Carlo simulations of the normalized magnetization of as a function of temperature

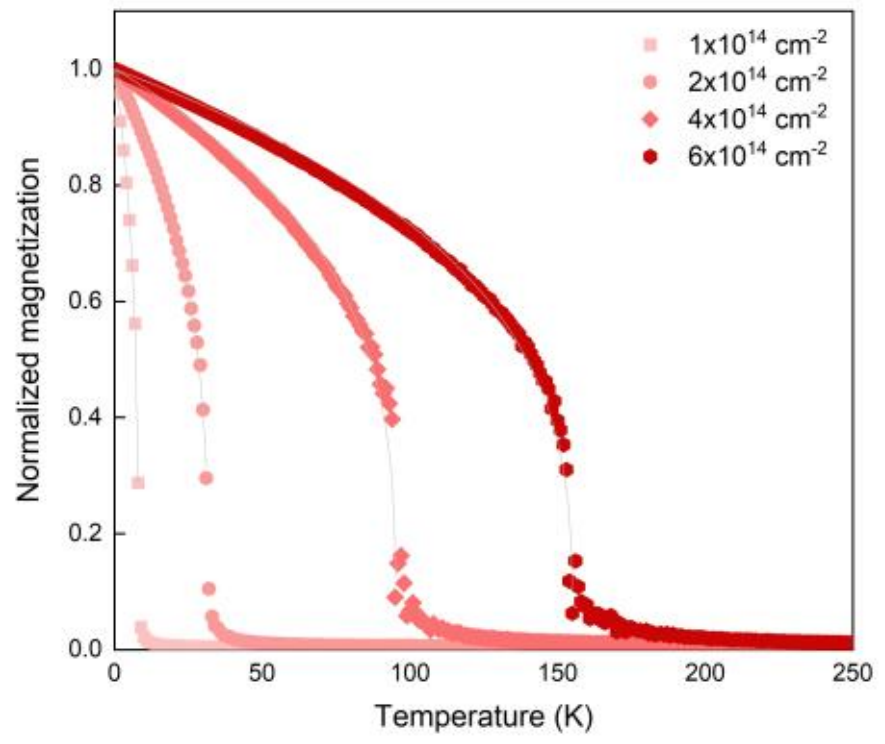

$T_c$: $1\times10^{14}$ cm$^{-2}$: 8 K    $2\times10^{14}$ cm$^{-2}$: 32 K
$4\times10^{14}$ cm$^{-2}$: 95 K    $6\times10^{14}$ cm$^{-2}$: 155 K

# 17.$ZnCl_2$

| MC2D-ID | C2DB | 2dmat-ID | USPEX | Space group | Band gap (eV) |
|---|---|---|---|---|---|
| 238 | ✓ | 2dm-4402 | - | P3m1 | 4.46 |

**Convex hull** | **Atomic structure** | **Atomic coordinates** | **Phonon dispersion curve**

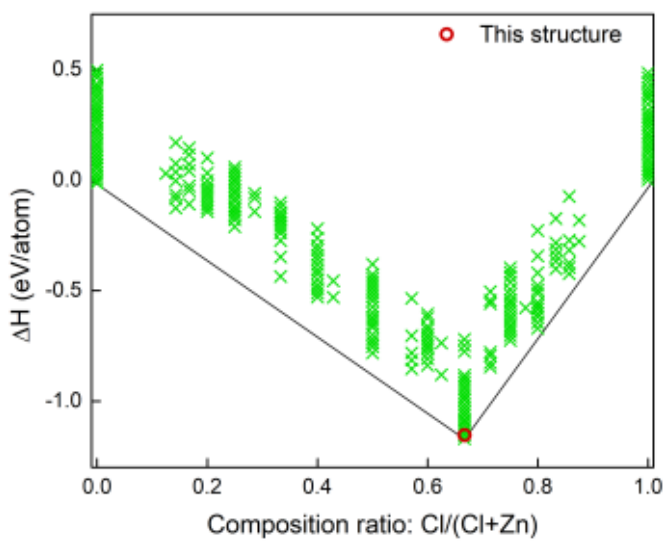

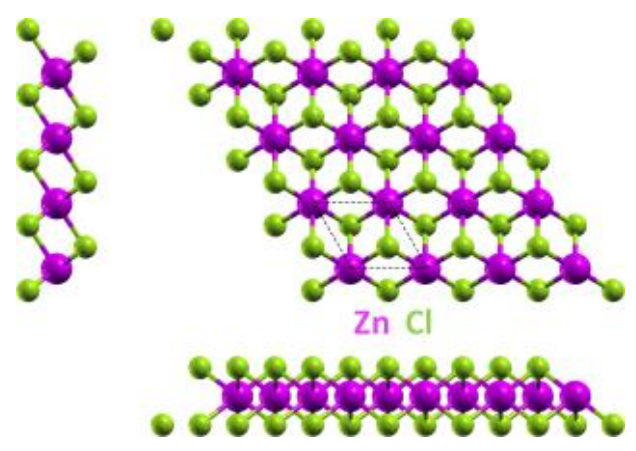

```
Zn1Cl2
   1.00000000000000
     3.6006943636000002    0.0000000000000000    0.0000000000000000
    -1.8003471818000001    3.1182927901999999    0.0000000000000000
     0.0000000000000000    0.0000000000000000   22.7725935562999986
   Zn   Cl
     1     2
Direct
  0.6666666666880445  0.3333333333760891  0.4999999818575773
  0.3333333333440223  0.6666666666880445  0.5601819124713145
  0.0000000000160369  0.0000000000320668  0.4398181056733037
```

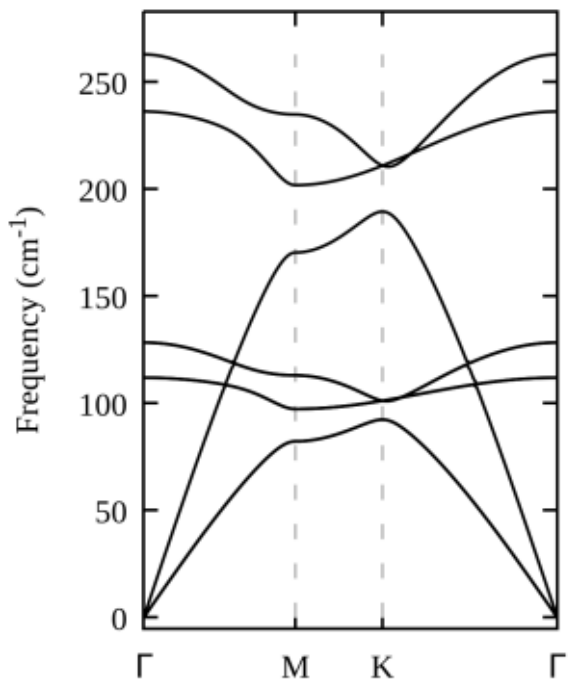

## Projected band structure and density of states

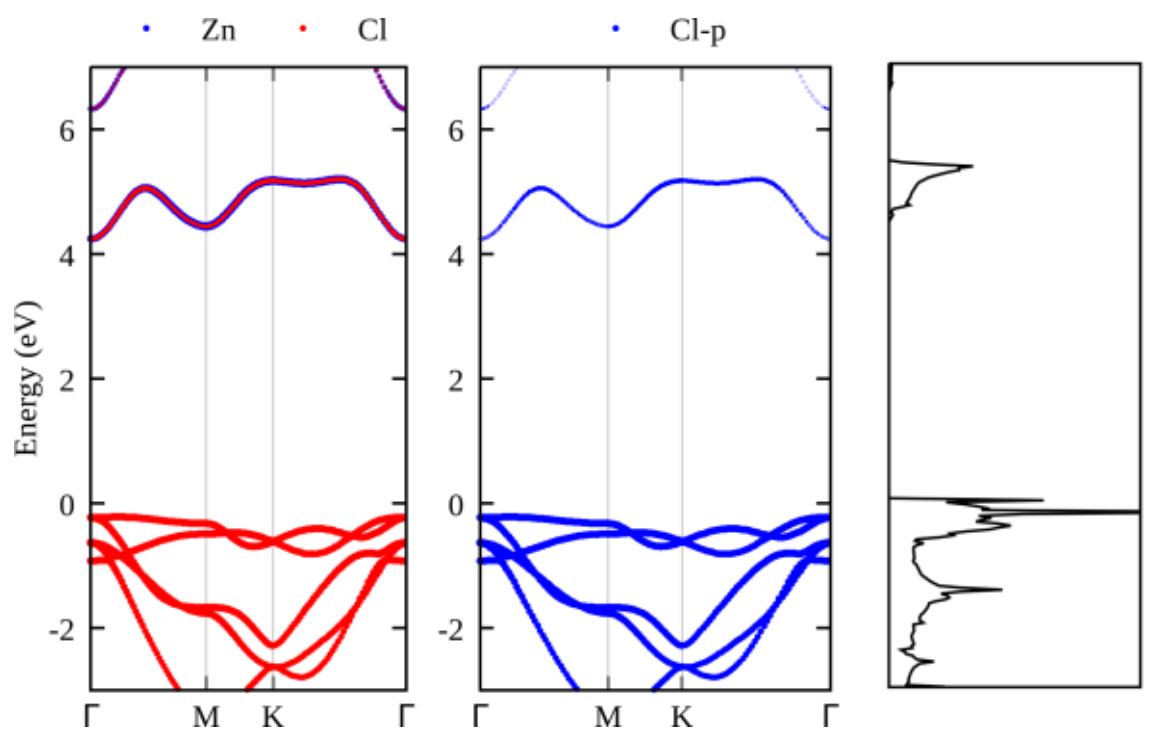

## Magnetic moment and spin polarization energy as a function of hole doping concentration

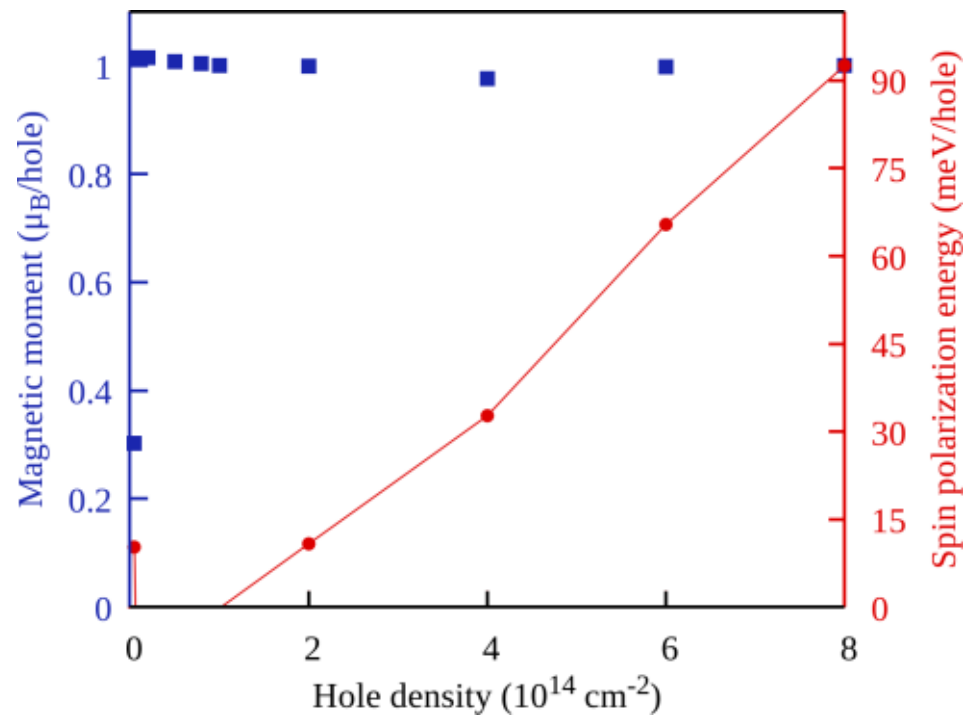

## Magnetic configurations and spin Hamiltonian

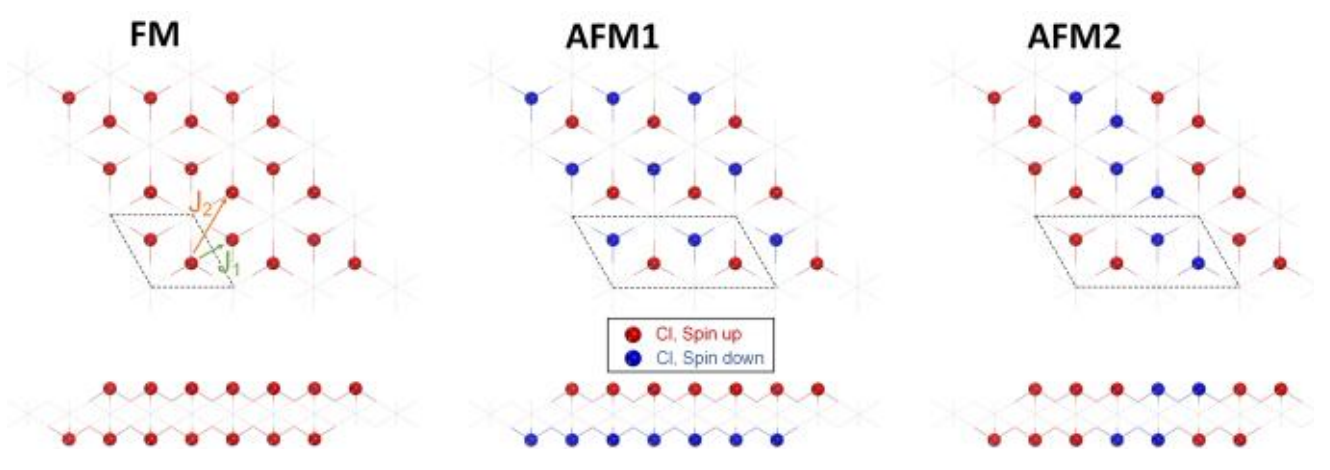

$$E_{FM} = E_0 - 3J_1S^2 - 6J_2S^2$$
$$E_{AFM1} = E_0 + 3J_1S^2 - 6J_2S^2$$
$$E_{AFM2} = E_0 - J_1S^2 + 2J_2S^2$$

## Magnetic exchange coupling parameters

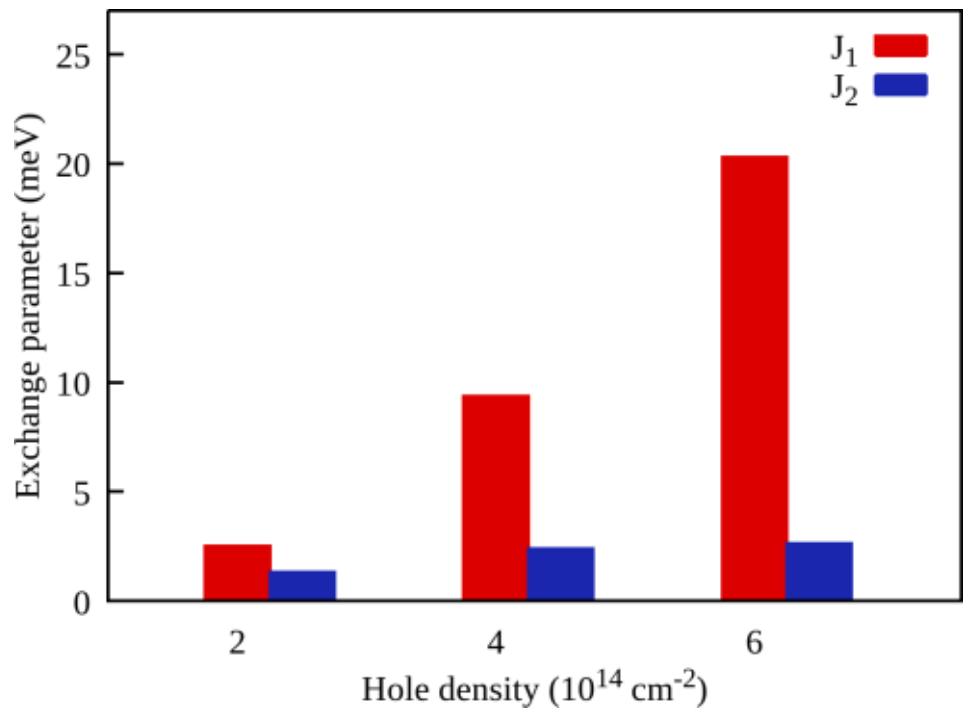

## Magnetic anisotropy energy (MAE, μeV) per magnetic atom

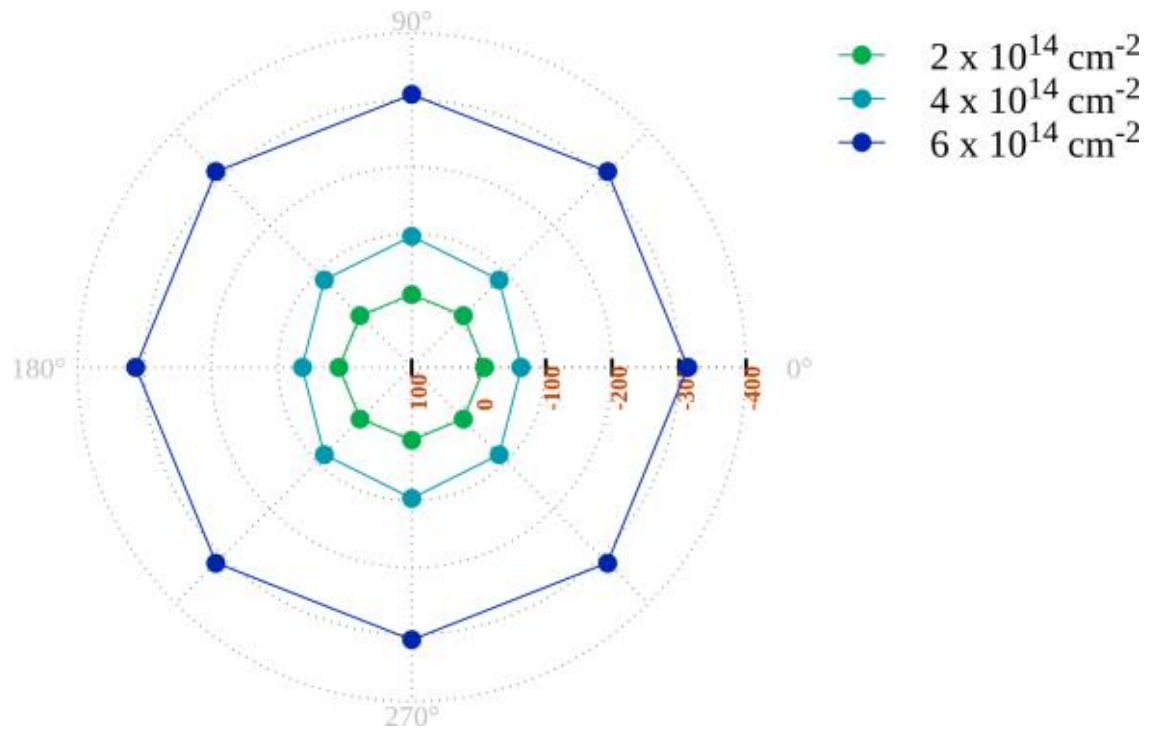

MAE = $E_{\parallel} - E_{\perp}$, a positive (negative) value of MAE indicates the off-plane (in-plane) easy axis.

## Monte Carlo simulations of the normalized magnetization of as a function of temperature

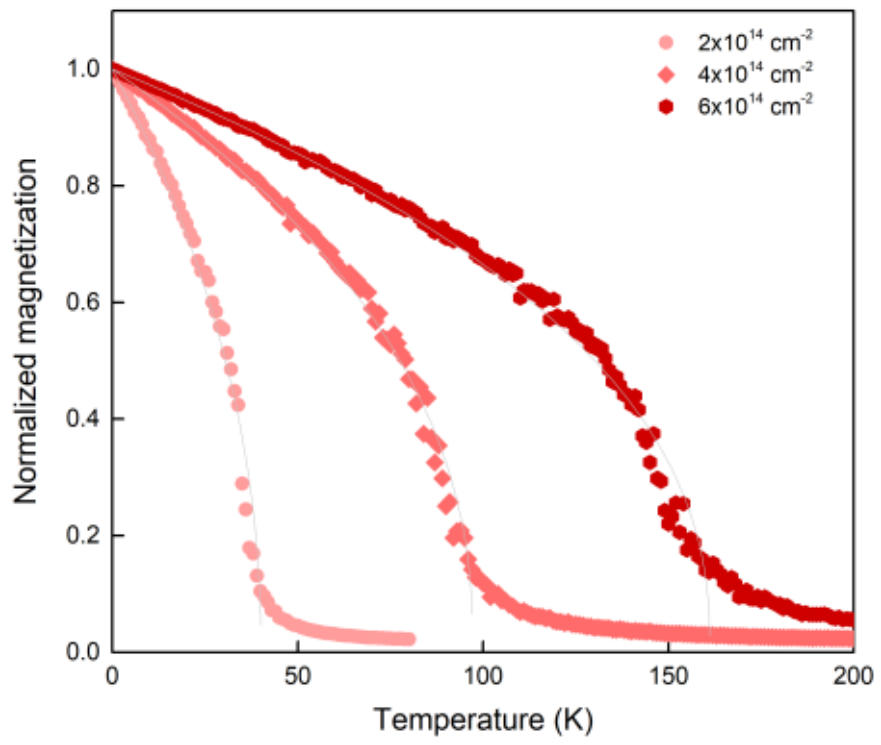

$T_c$: $1\times10^{14}$ cm$^{-2}$: - K  $2\times10^{14}$ cm$^{-2}$:40 K

$4\times10^{14}$ cm$^{-2}$: 97 K  $6\times10^{14}$ cm$^{-2}$: 161 K

# 18.ZnBr2`

| MC2D-ID | C2DB | 2dmat-ID | USPEX | Space group | Band gap (eV) |
|---|---|---|---|---|---|
| 237 | ✓ | 2dm-5204 | - | P3m1 | 3.47 |

**Convex hull** | **Atomic structure** | **Atomic coordinates** | **Phonon dispersion curve**

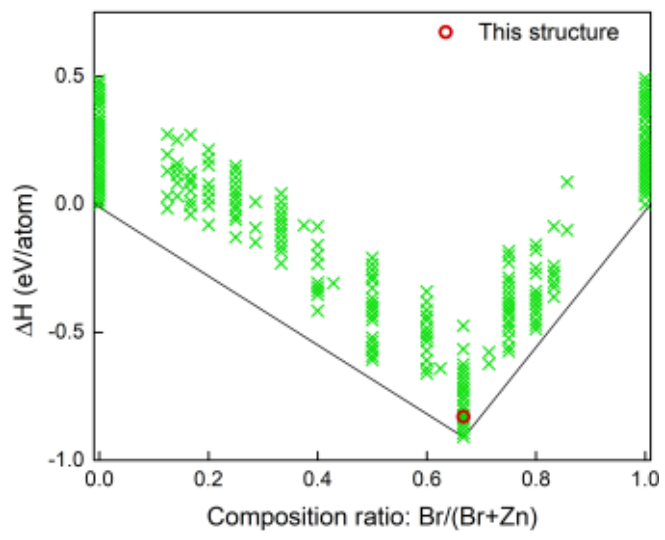

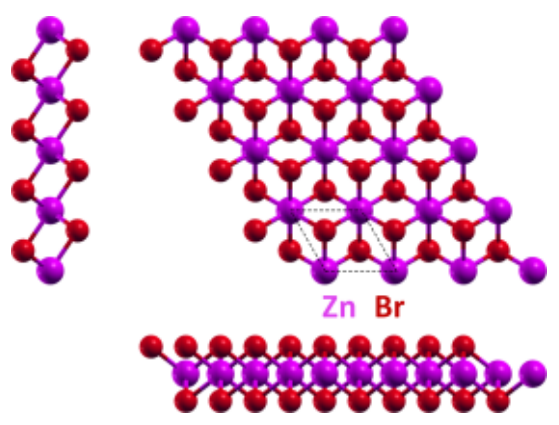

```
Zn1Br2
   1.00000000000000
   3.8002466363168996   0.0000000000000000   0.0000000000000000
   1.9001385057259335   3.2911013689901698   0.0000000000000000
   0.0000000000000000   0.0000000000000000   23.0561270000000000
   Zn   Br
   1    2
Direct
  0.0000000025429330  0.9999999899984786  0.4999999992549746
  0.6666660428341586  0.6666660416888561  0.4355832714836581
  0.3333339546229013  0.3333339683126582  0.5644167292613744
```

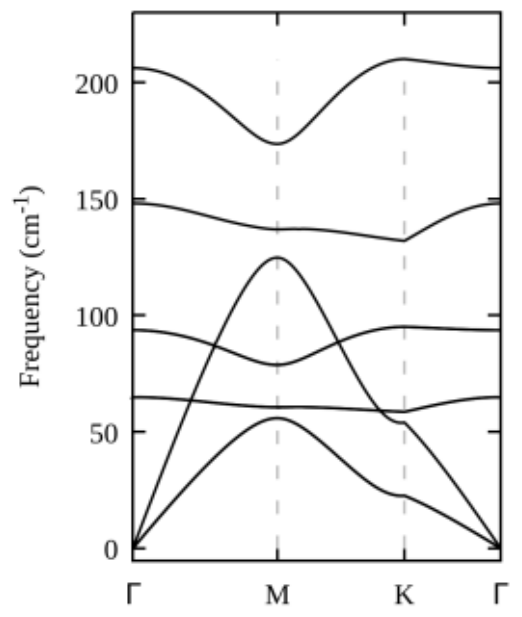

## Projected band structure and density of states

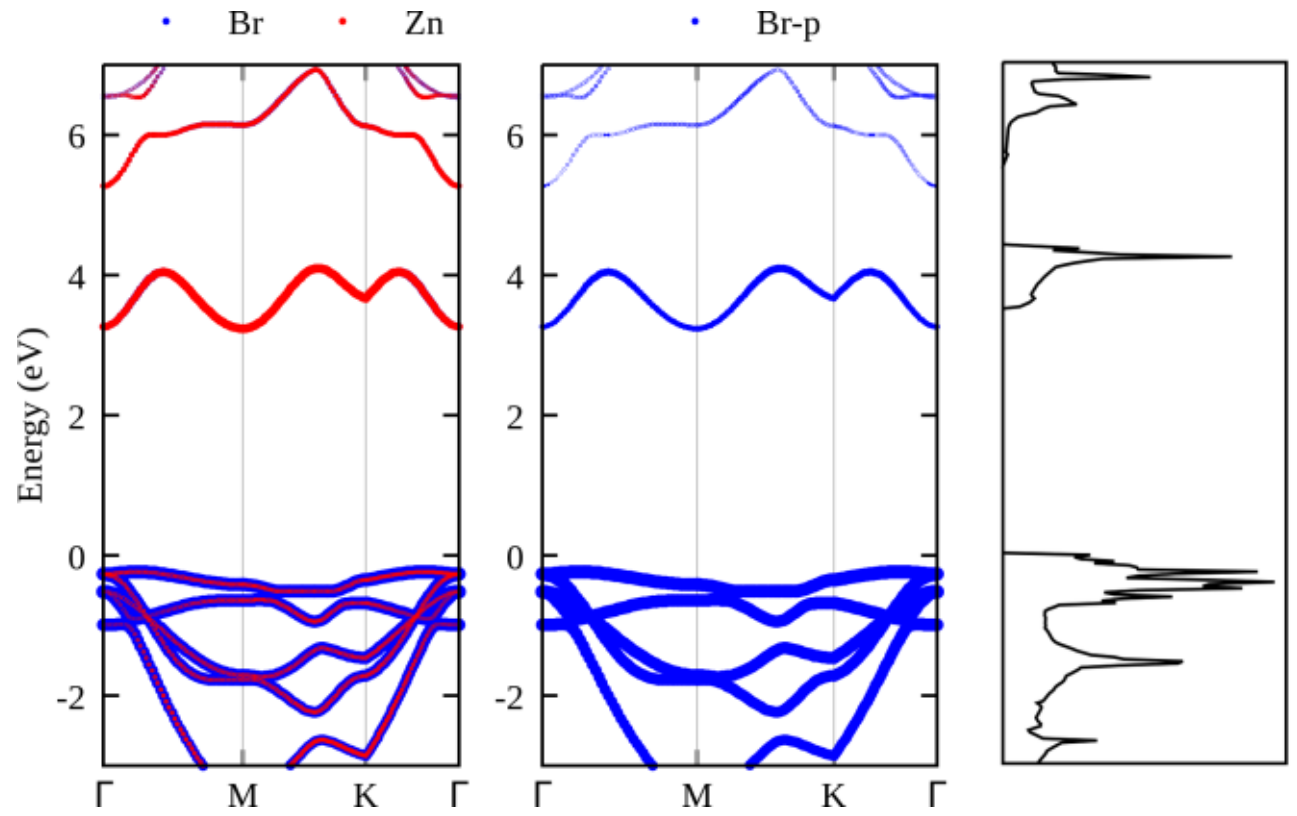

## Magnetic moment and spin polarization energy as a function of hole doping concentration

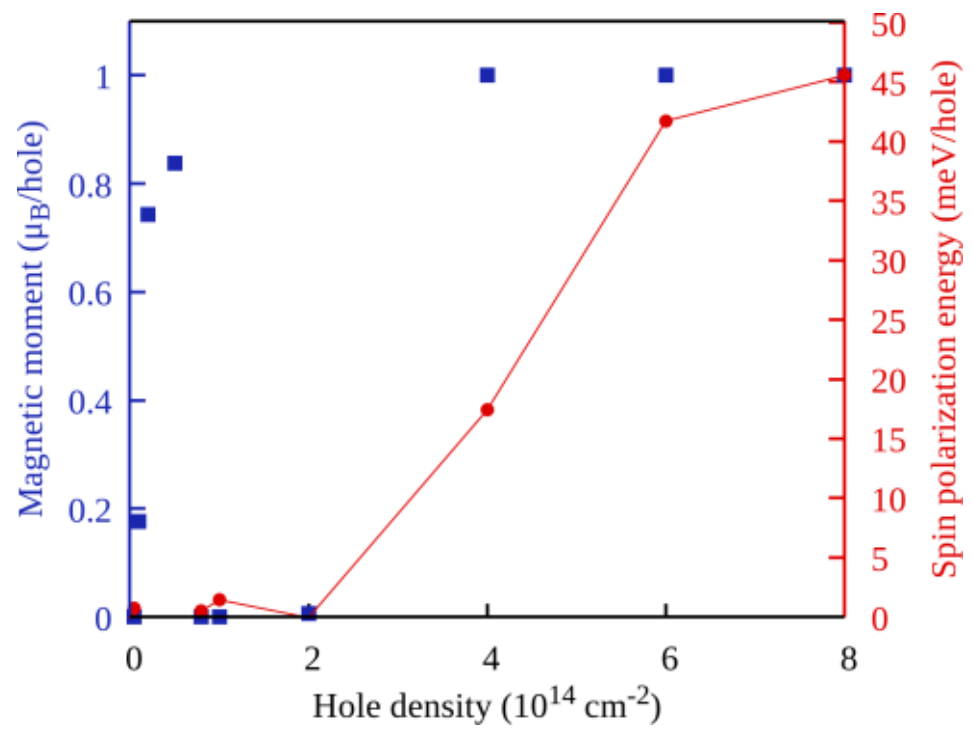

## Magnetic configurations and spin Hamiltonian

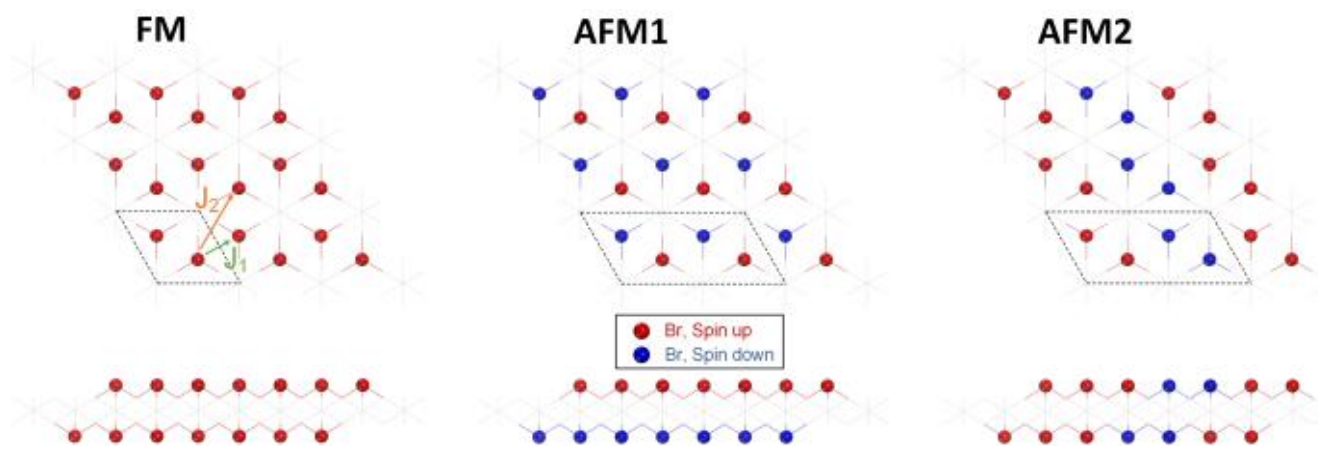

$$E_{FM} = E_0 - 3J_1S^2 - 6J_2S^2$$
$$E_{AFM1} = E_0 + 3J_1S^2 - 6J_2S^2$$
$$E_{AFM2} = E_0 - J_1S^2 + 2J_2S^2$$

## Magnetic exchange coupling parameters

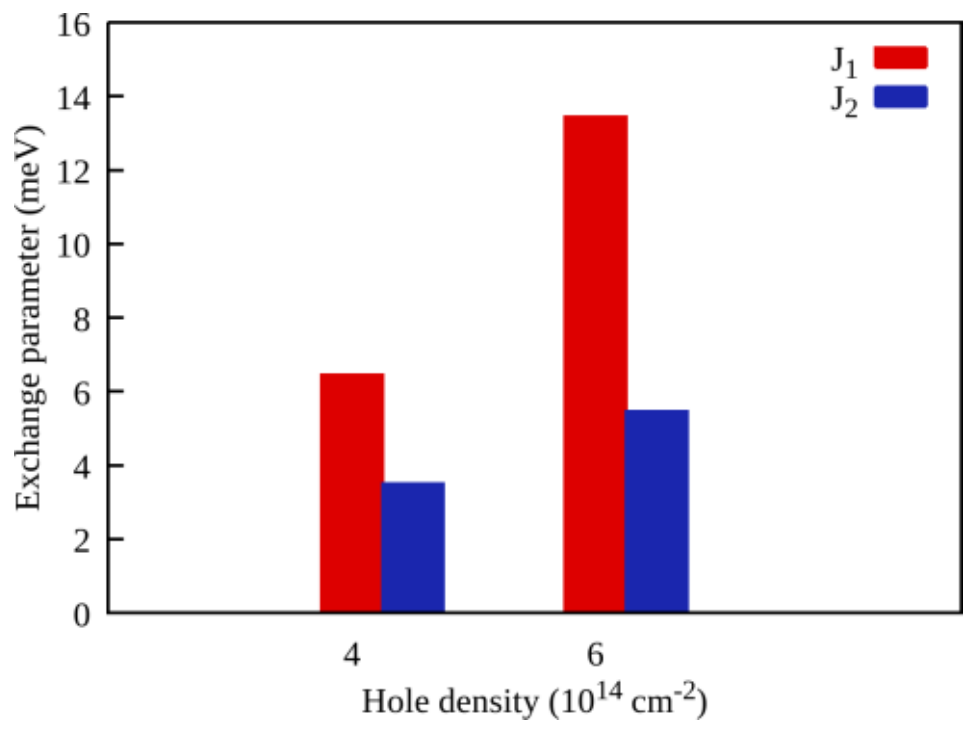

## Magnetic anisotropy energy (MAE, μeV) per magnetic atom

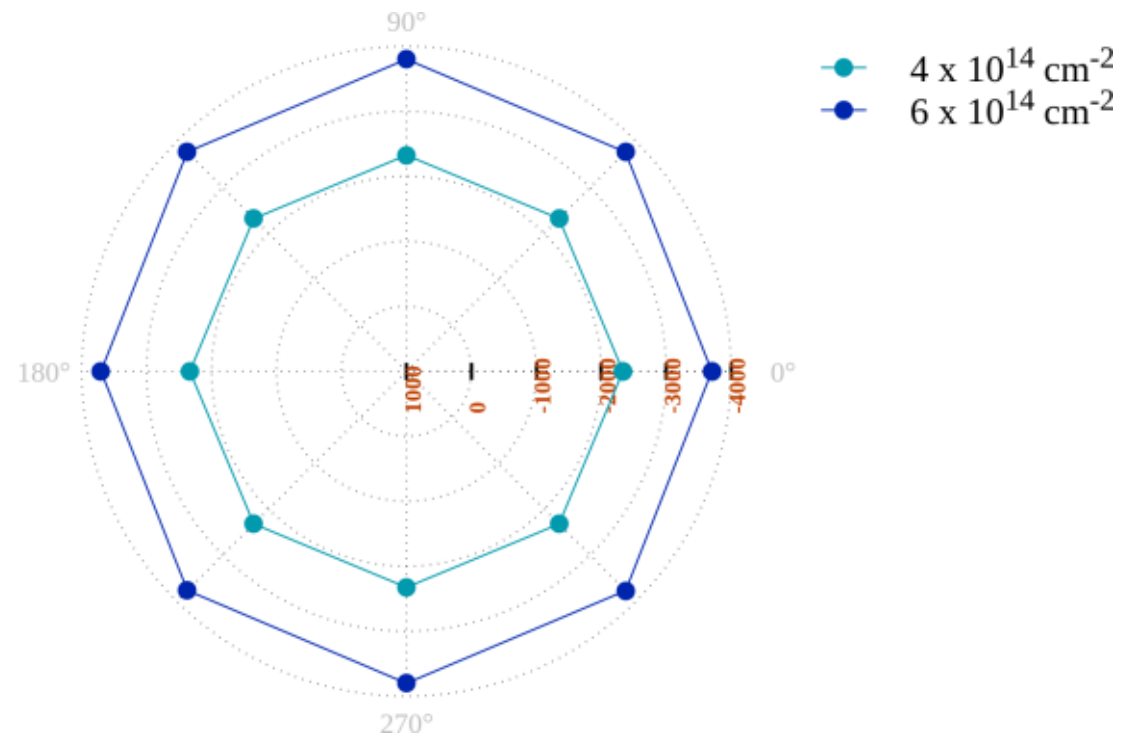

MAE = $E_{\parallel} - E_{\perp}$, a positive (negative) value of MAE indicates the off-plane (in-plane) easy axis.

## Monte Carlo simulations of the normalized magnetization of as a function of temperature

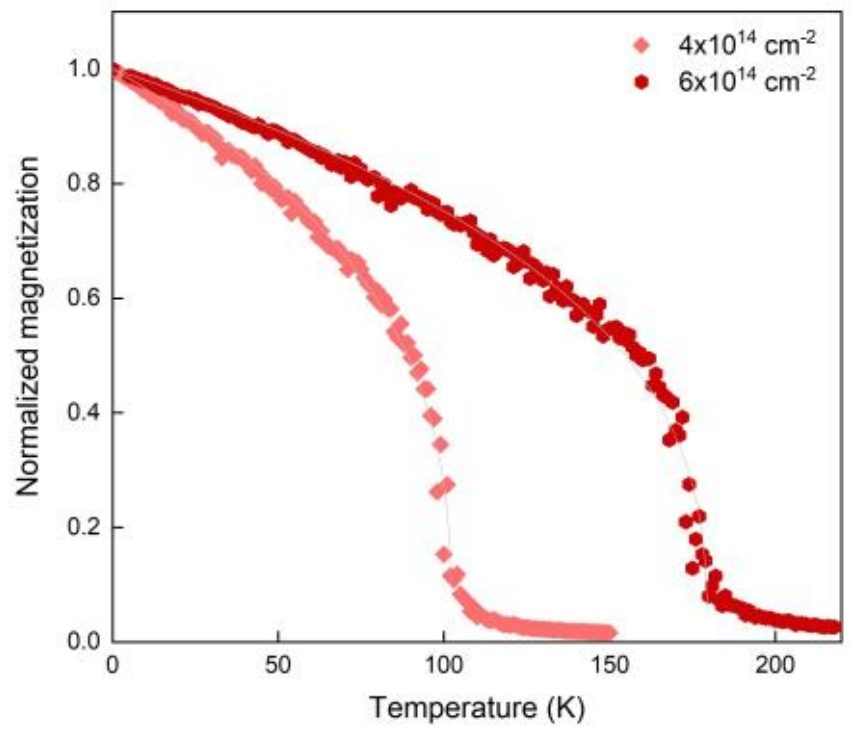

$T_c$: $1\times10^{14}$ cm$^{-2}$: - K  $2\times10^{14}$ cm$^{-2}$: - K

$4\times10^{14}$ cm$^{-2}$: 102 K  $6\times10^{14}$ cm$^{-2}$: 179 K

# 19. $CdF_2$

| MC2D-ID | C2DB | 2dmat-ID | USPEX | Space group | Band gap (eV) |
|---|---|---|---|---|---|
| - | - | 2dm-766 | - | P3m1 | 3.83 |

**Convex hull** | **Atomic structure** | **Atomic coordinates** | **Phonon dispersion curve**

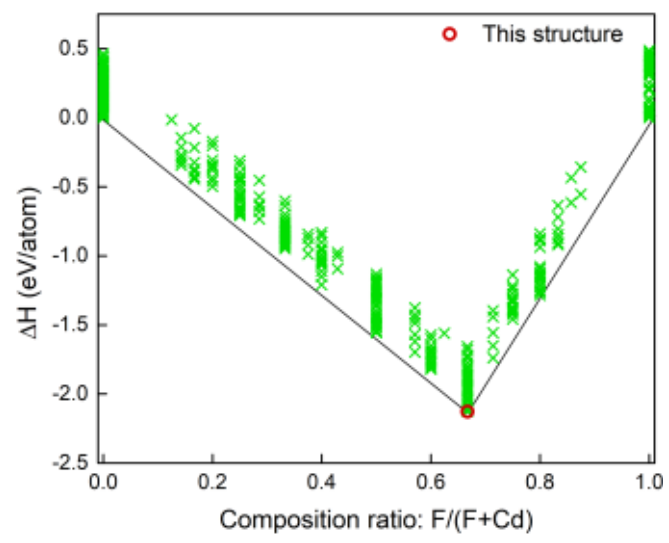

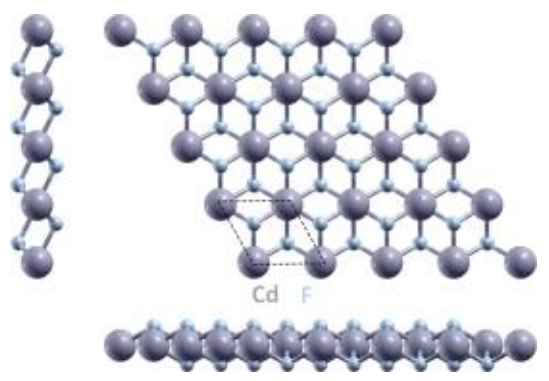

```
Cd1F2
   1.00000000000000
   3.5610764138459019    0.0000000000000000    0.0000000000000000
  -1.7808839336124016    3.0837829331036968    0.0000000000000000
   0.0000000000000000    0.0000000000000000   22.9560519999999997
   Cd    F
   1     2
Direct
  0.0000000010281257  0.0000000007603228  0.9999999996078505
  0.3333344642687417  0.6666654662189303  0.0436518055167469
  0.6666655347031326  0.3333345330207468  0.9563481948754026
```

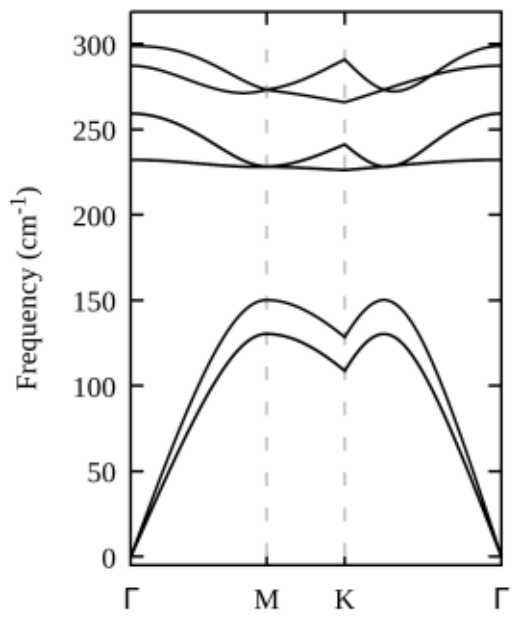

**Projected band structure and density of states**

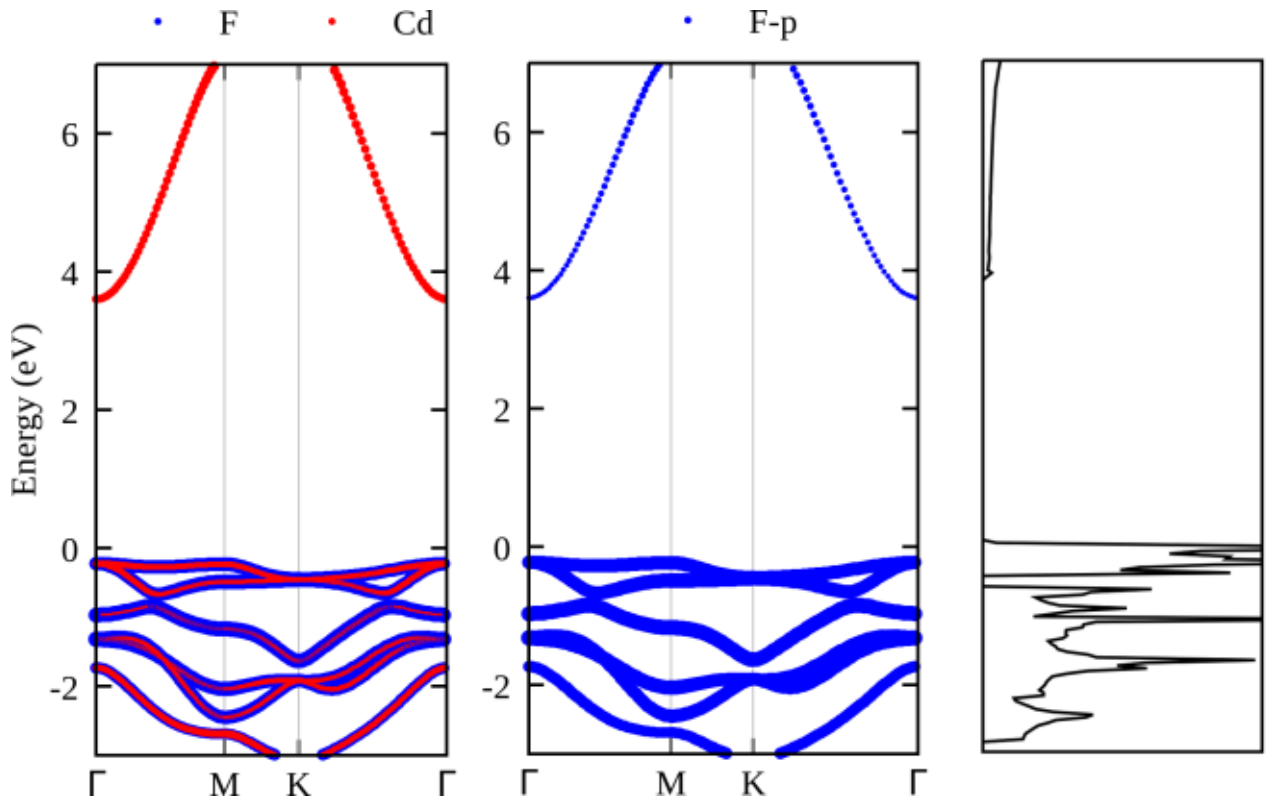

**Magnetic moment and spin polarization energy as a function of hole doping concentration**

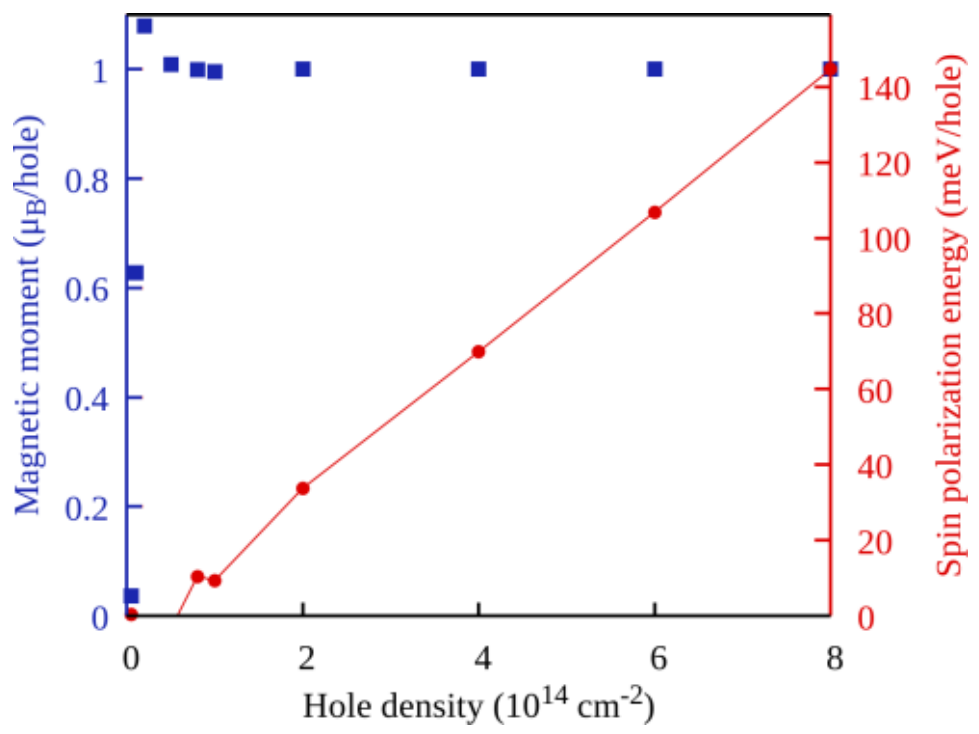

**Magnetic configurations and spin Hamiltonian**

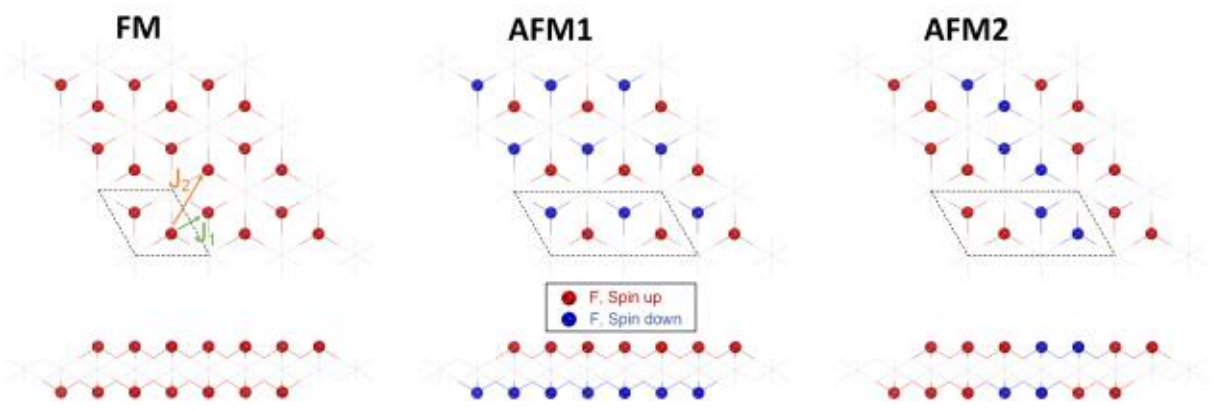

$$E_{FM} = E_0 - 3J_1S^2 - 6J_2S^2$$
$$E_{AFM1} = E_0 + 3J_1S^2 - 6J_2S^2$$
$$E_{AFM2} = E_0 - J_1S^2 + 2J_2S^2$$

**Magnetic exchange coupling parameters**

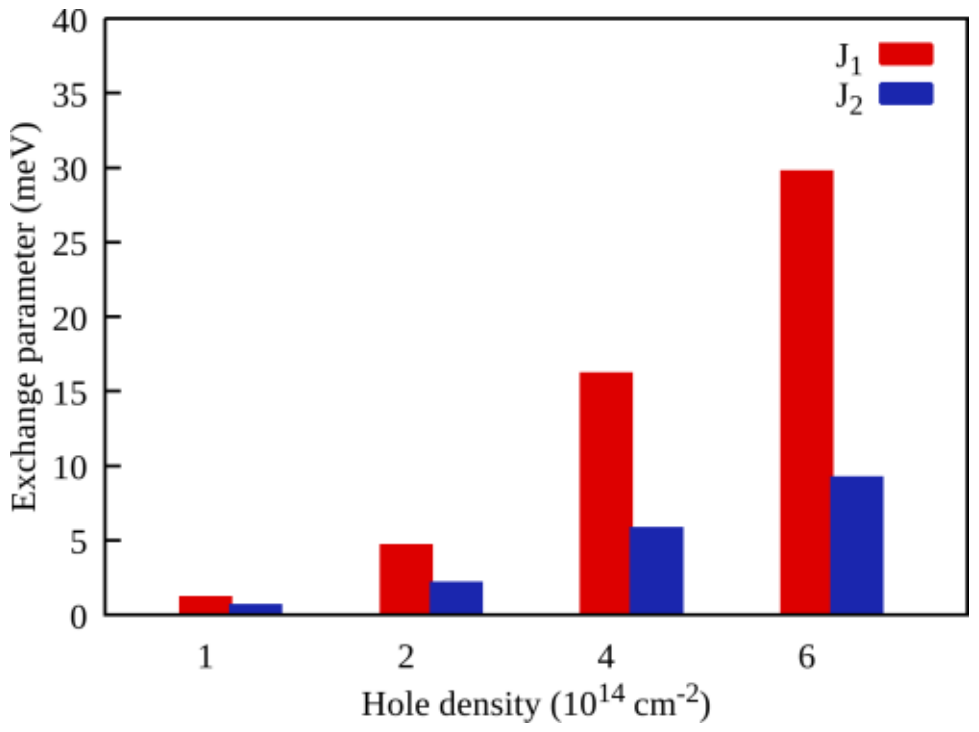

**Magnetic anisotropy energy (MAE, μeV) per magnetic atom**

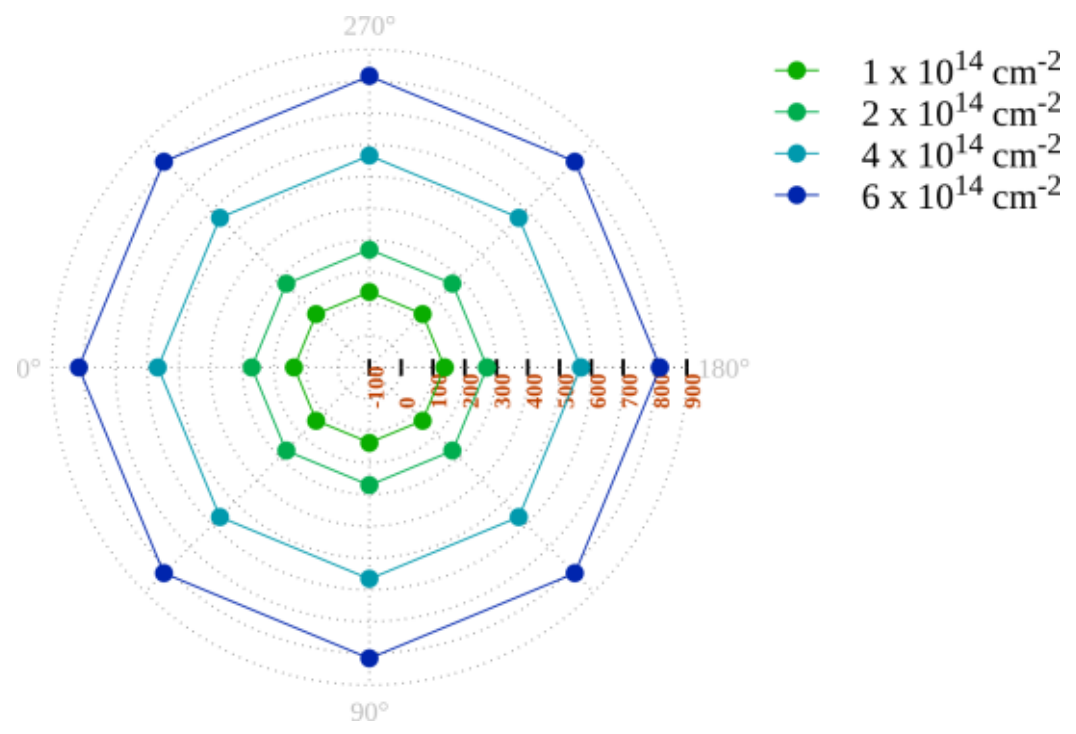

MAE = $E_{\parallel} - E_{\perp}$, a positive (negative) value of MAE indicates the off-plane (in-plane) easy axis.

**Monte Carlo simulations of the normalized magnetization of as a function of temperature**

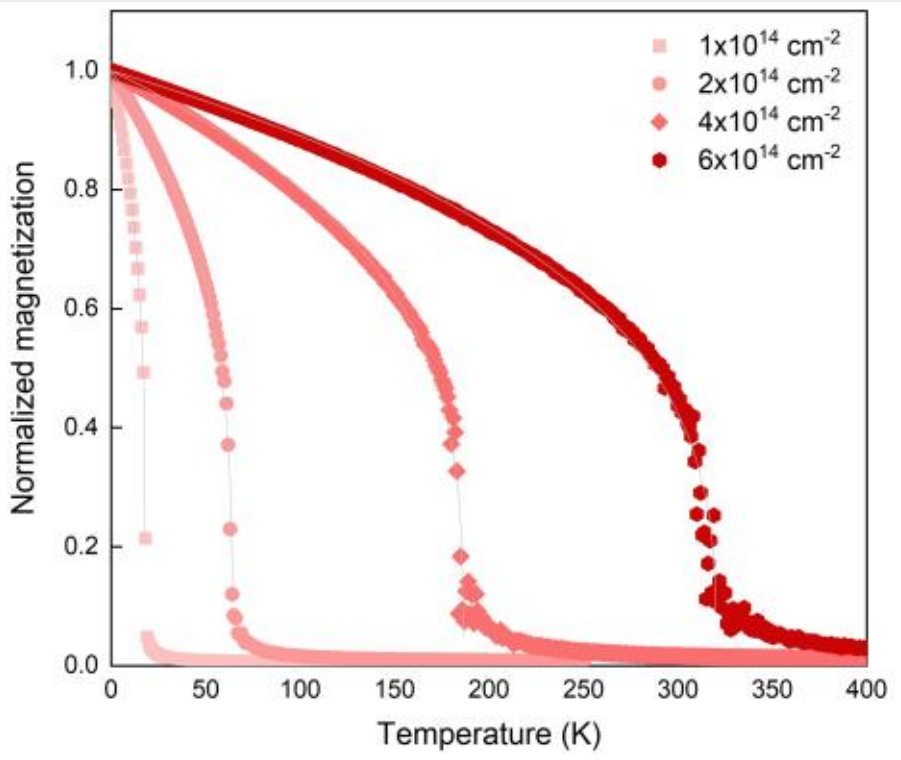

$T_c$: $1\times10^{14}$ cm$^{-2}$: 18 K  $2\times10^{14}$ cm$^{-2}$: 64 K  $4\times10^{14}$ cm$^{-2}$: 186 K  $6\times10^{14}$ cm$^{-2}$: 320 K

# 20. $CdCl_2$

| MC2D-ID | C2DB | 2dmat-ID | USPEX | Space group | Band gap (eV) |
|---|---|---|---|---|---|
| 32 | - | 2dm-3485 | - | P3m1 | 3.89 |

## Convex hull

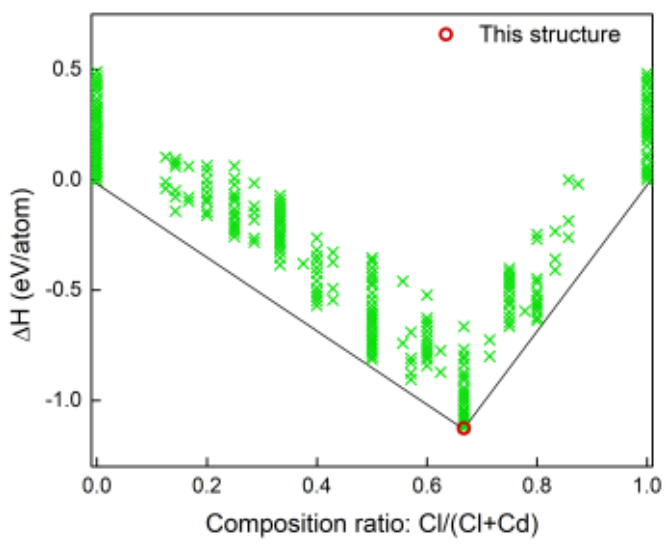

## Atomic structure

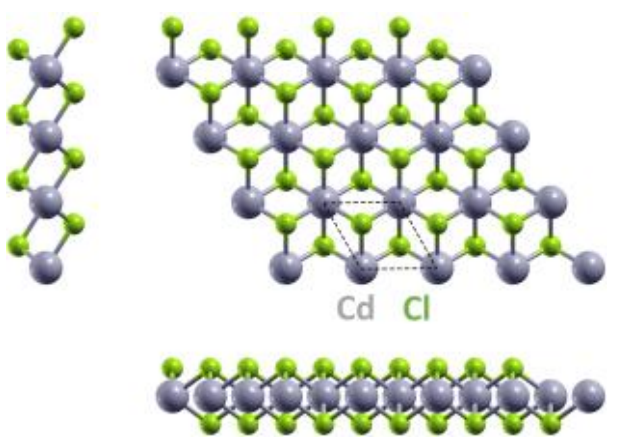

## Atomic coordinates

```
Cd1Cl2
   1.00000000000000
   3.9076173896460085   0.0000000000000000   0.0000000000000000
   1.9538089758997854   3.3840958059197010   0.0000000000000000
   0.0000000000000000   0.0000000000000000  22.8652279999999983
   Cd   Cl
   1    2
Direct
  0.9999999951674496  0.0000000068514439  0.4999999994666027
  0.6666667349110043  0.6666666788783928  0.4370233396324821
  0.3333332699215532  0.3333333142701633  0.5629766609009152
```

## Phonon dispersion curve

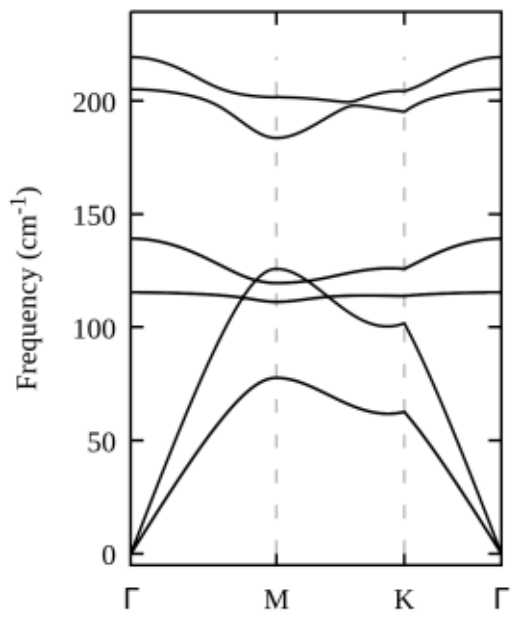

## Projected band structure and density of states

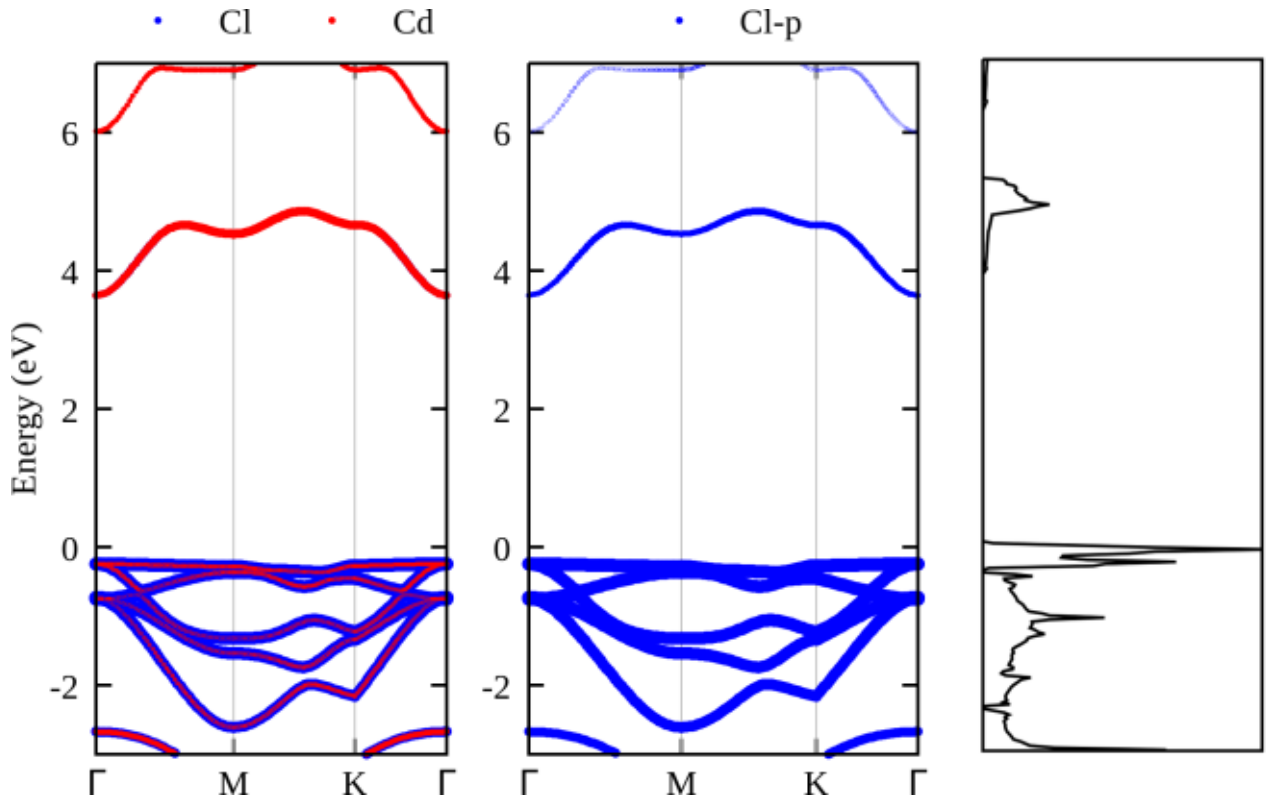

## Magnetic moment and spin polarization energy as a function of hole doping concentration

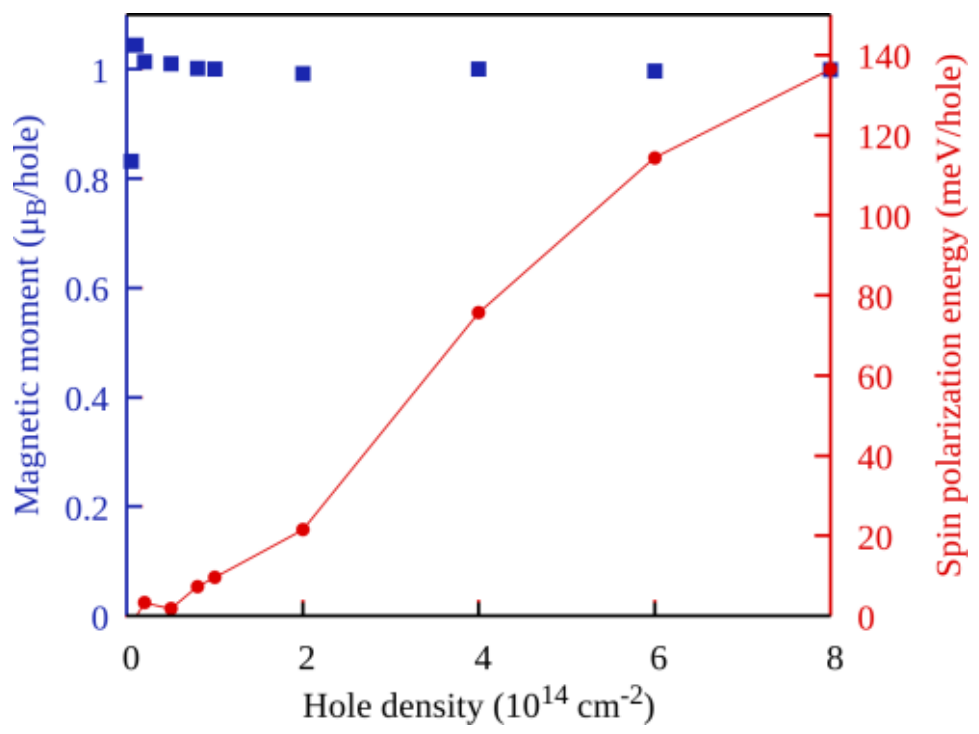

## Magnetic configurations and spin Hamiltonian

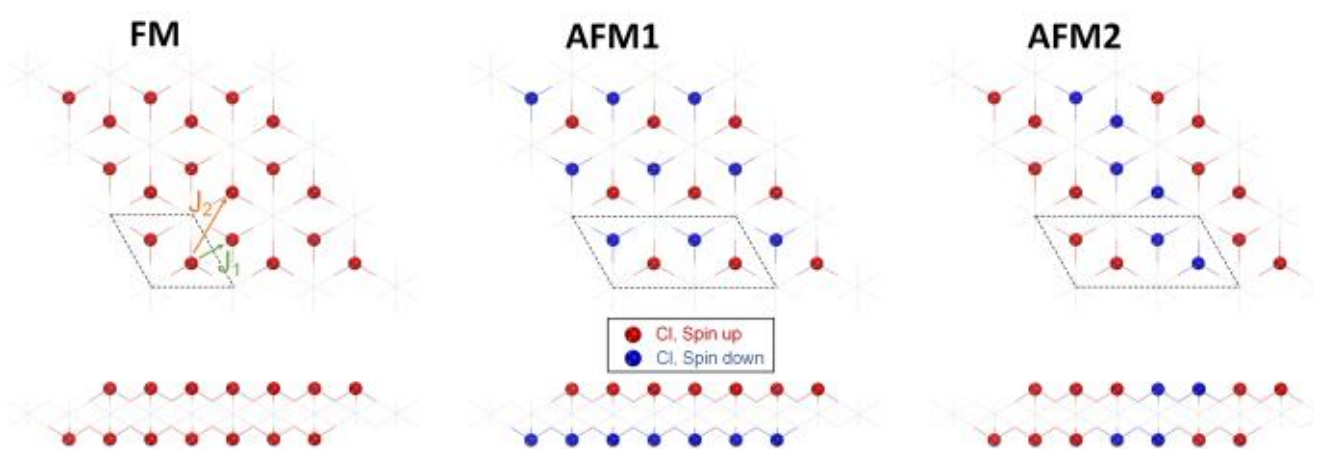

$$E_{FM} = E_0 - 3J_1S^2 - 6J_2S^2$$
$$E_{AFM1} = E_0 + 3J_1S^2 - 6J_2S^2$$
$$E_{AFM2} = E_0 - J_1S^2 + 2J_2S^2$$

## Magnetic exchange coupling parameters

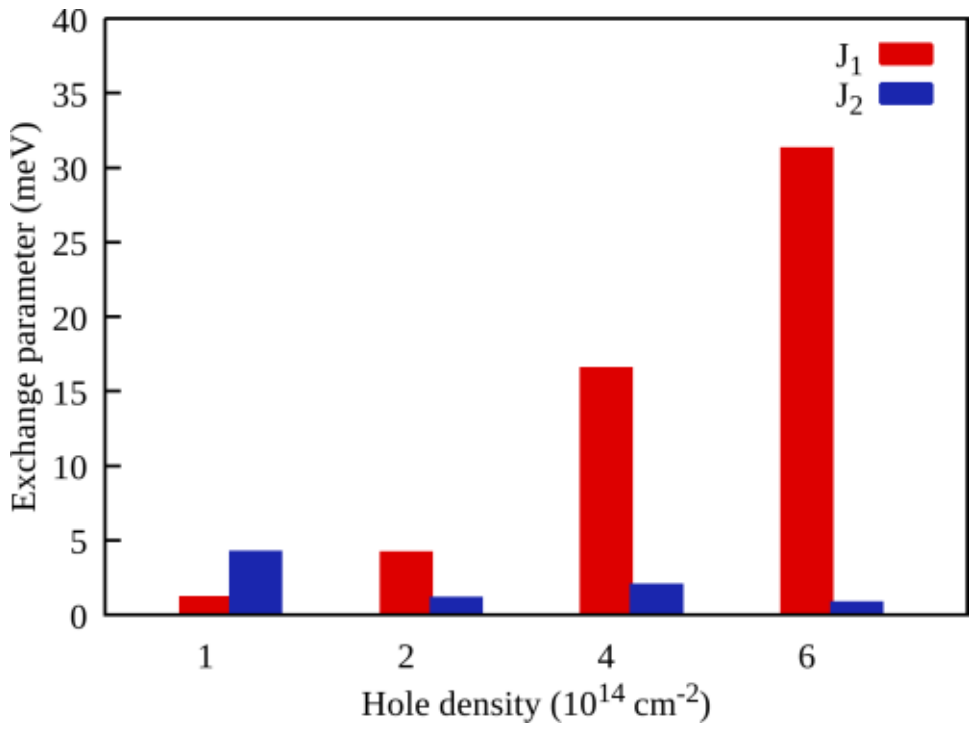

## Magnetic anisotropy energy (MAE, μeV) per magnetic atom

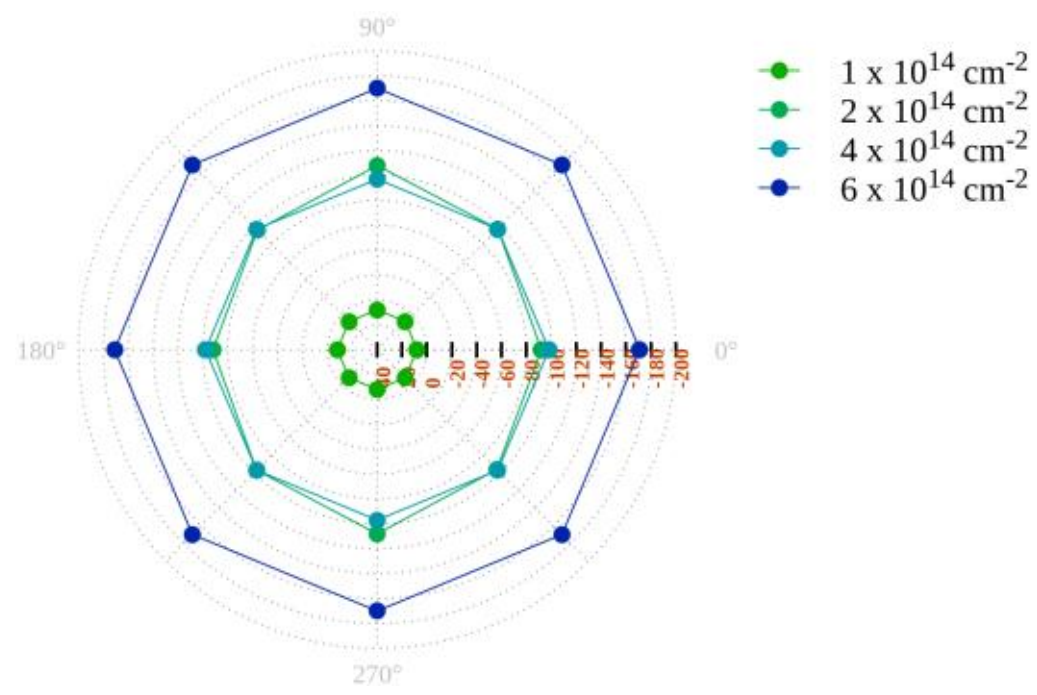

MAE = $E_{\parallel} - E_{\perp}$, a positive (negative) value of MAE indicates the off-plane (in-plane) easy axis.

## Monte Carlo simulations of the normalized magnetization of as a function of temperature

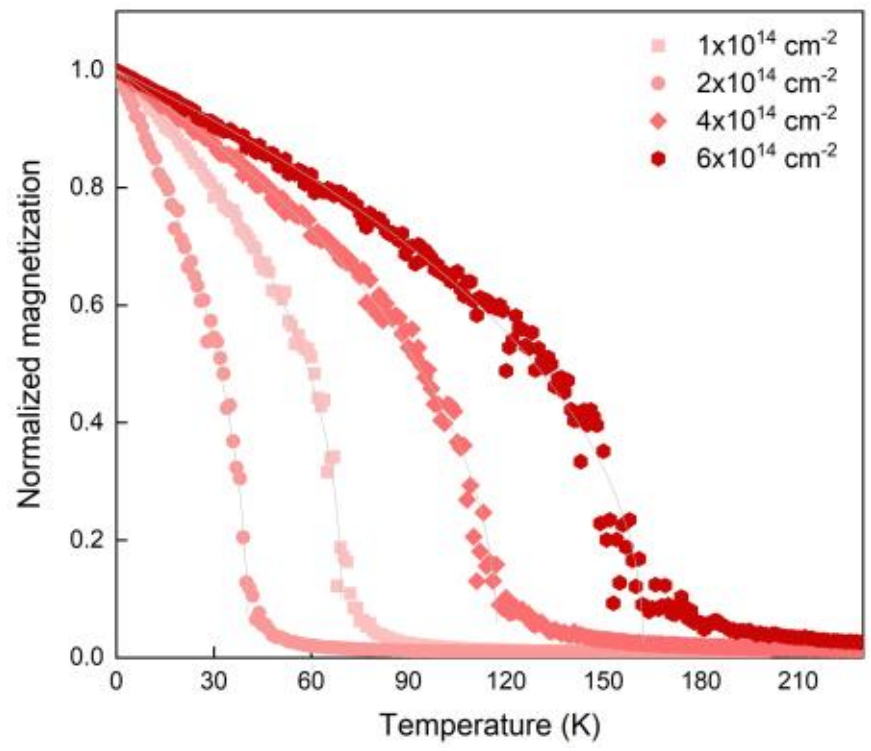

$T_c$: $1\times10^{14}$ cm$^{-2}$: 70 K  $2\times10^{14}$ cm$^{-2}$: 41 K  $4\times10^{14}$ cm$^{-2}$: 117 K  $6\times10^{14}$ cm$^{-2}$: 162 K

# 21.CdBr2

| MC2D-ID | C2DB | 2dmat-ID | USPEX | Space group | Band gap (eV) |
|---|---|---|---|---|---|
| 31 | - | 2dm-3696 | - | P3m1 | 3.22 |

**Convex hull**

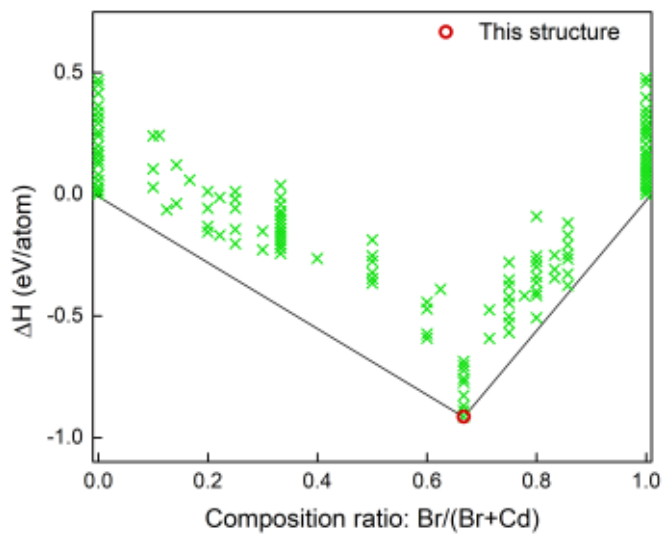

**Atomic structure**

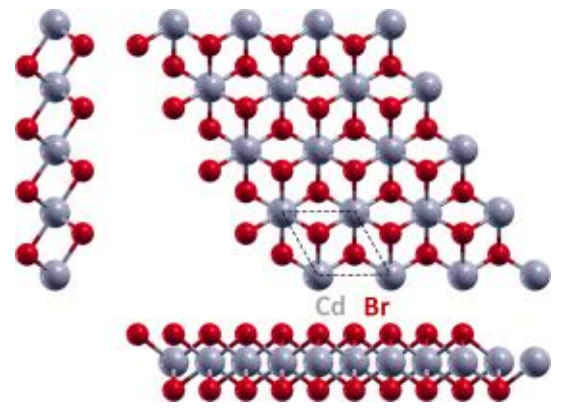

**Atomic coordinates**

```
Cd1Br2
   1.00000000000000
   4.0706621156352378    0.0000000000000000    0.0000000000000000
  -2.0353408016766656    3.5252915802523042    0.0000000000000000
   0.0000000000000000    0.0000000000000000   23.2415160000000007
   Cd    Br
   1      2
Direct
  0.9999999992555857  0.0000000008387815  0.5000121988331898
  0.3333332777723470  0.6666666985006202  0.5674896742607487
  0.6666667229720673  0.3333333006605983  0.4324991269060590
```

**Phonon dispersion curve**

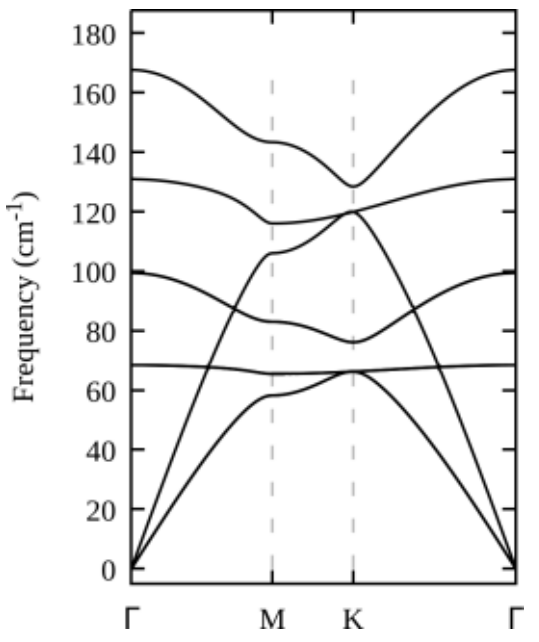

**Projected band structure and density of states**

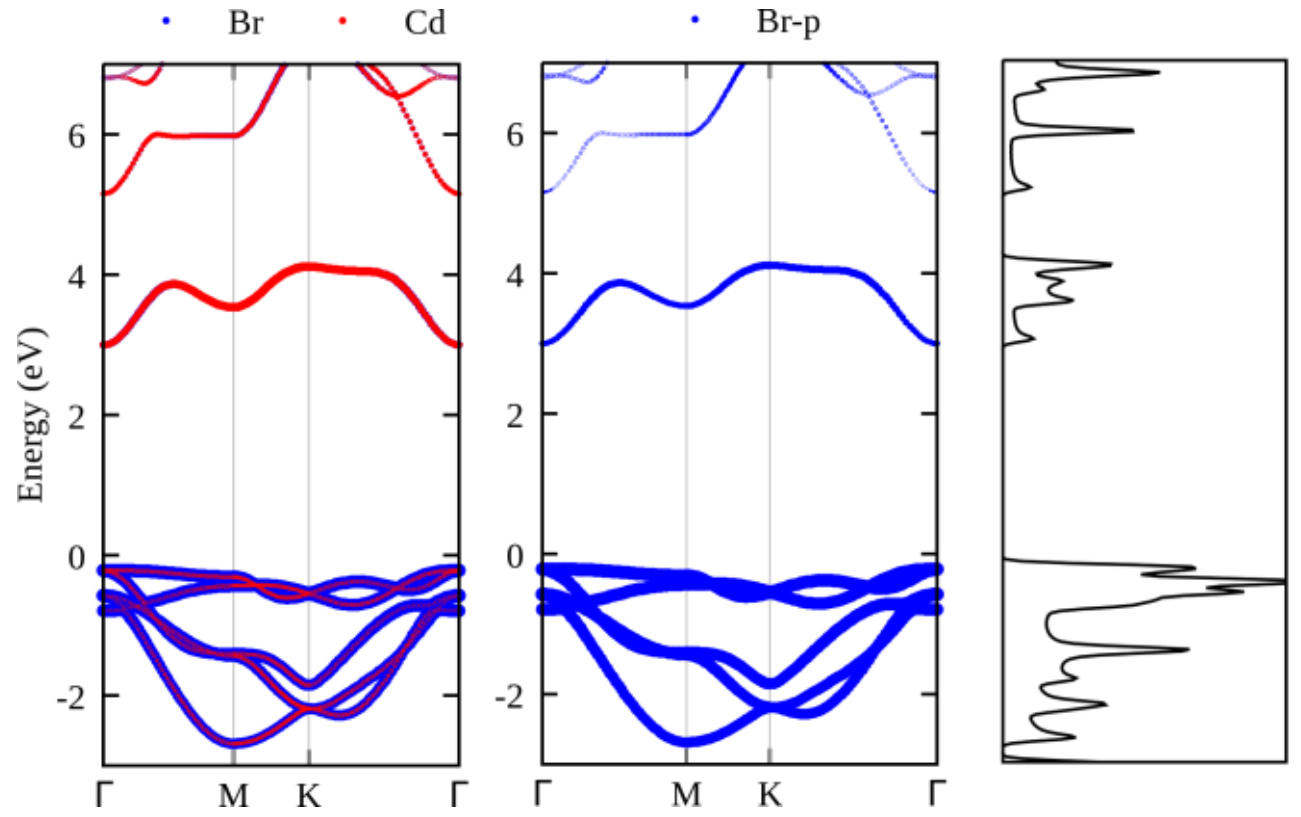

**Magnetic moment and spin polarization energy as a function of hole doping concentration**

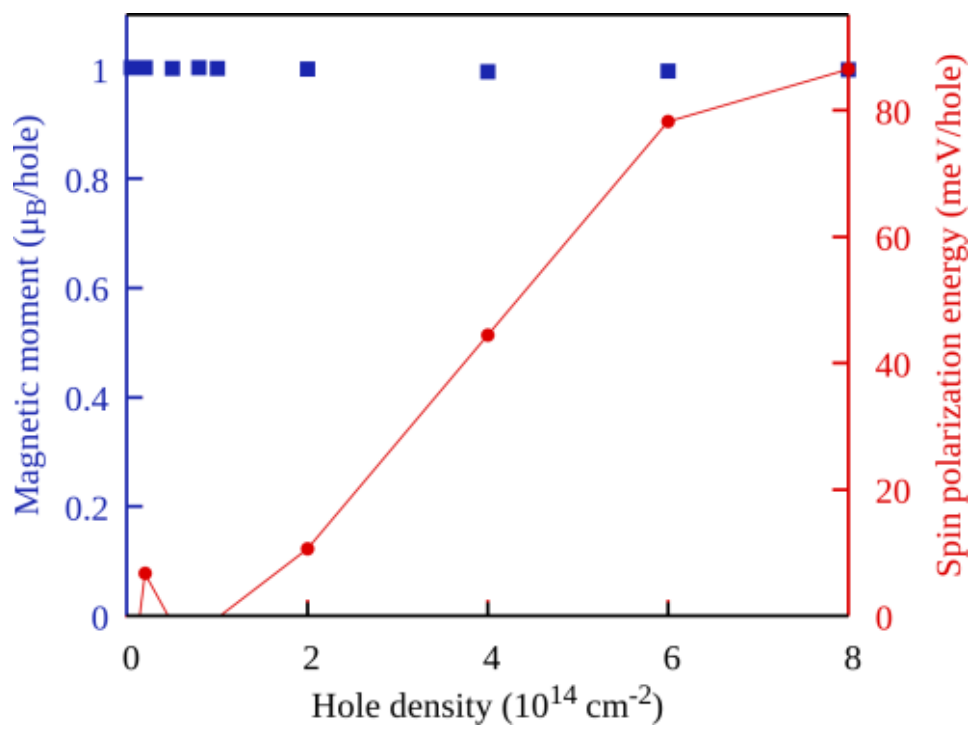

**Magnetic configurations and spin Hamiltonian**

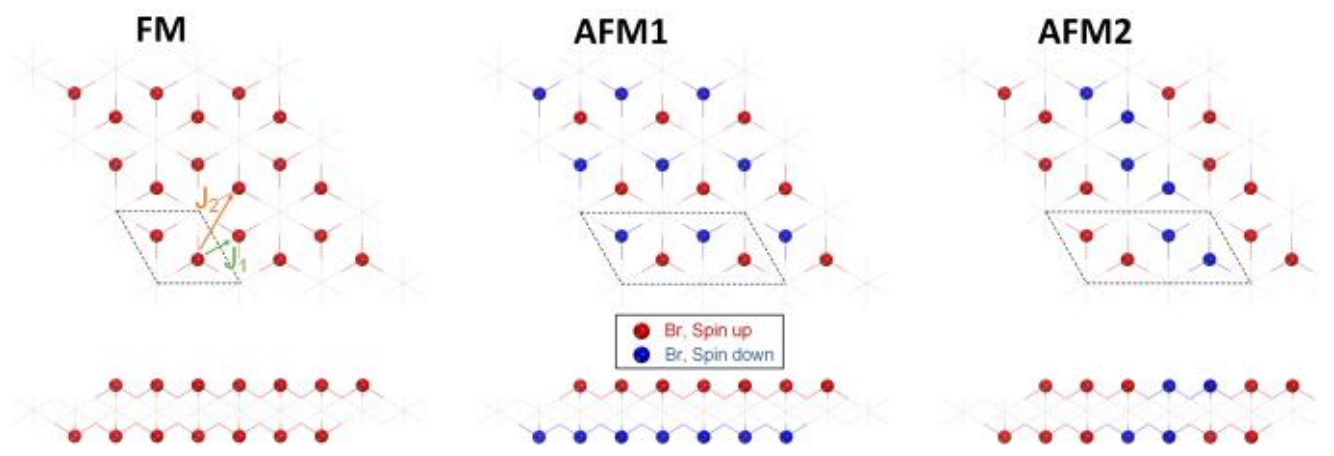

$$E_{FM} = E_0 - 3J_1S^2 - 6J_2S^2$$
$$E_{AFM1} = E_0 + 3J_1S^2 - 6J_2S^2$$
$$E_{AFM2} = E_0 - J_1S^2 + 2J_2S^2$$

**Magnetic exchange coupling parameters**

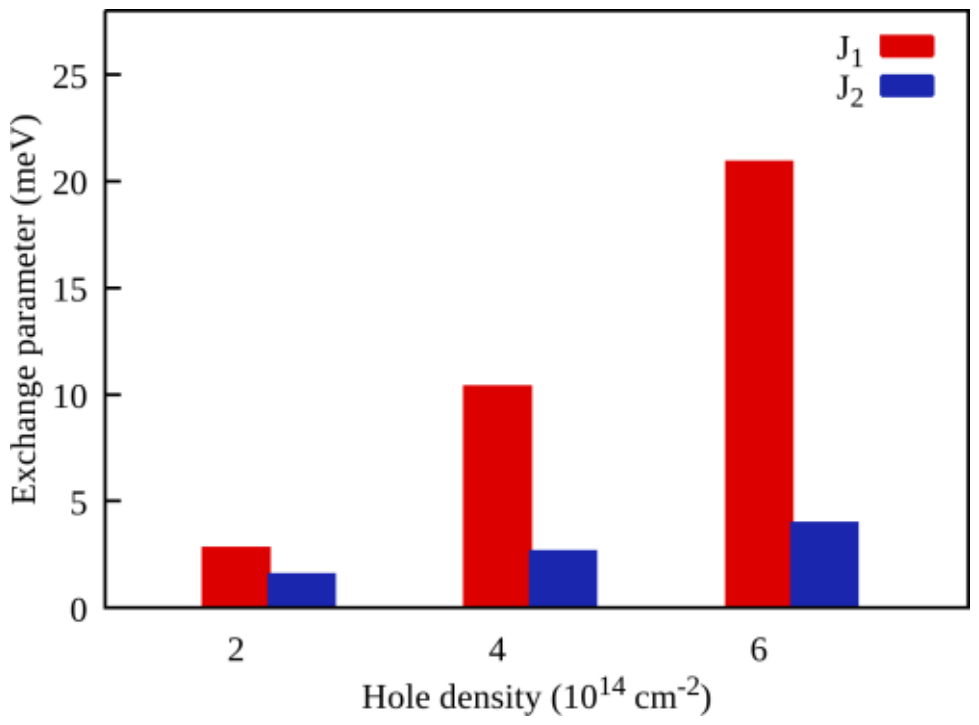

**Magnetic anisotropy energy (MAE, μeV) per magnetic atom**

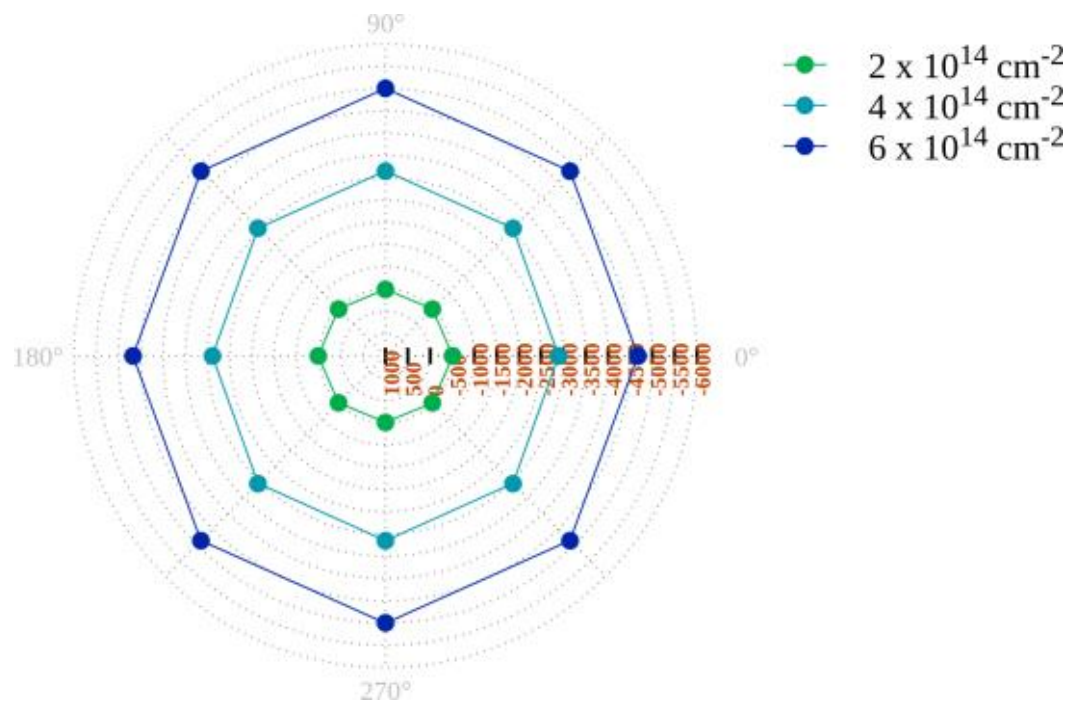

MAE = $E_{\parallel} - E_{\perp}$, a positive (negative) value of MAE indicates the off-plane (in-plane) easy axis.

**Monte Carlo simulations of the normalized magnetization of as a function of temperature**

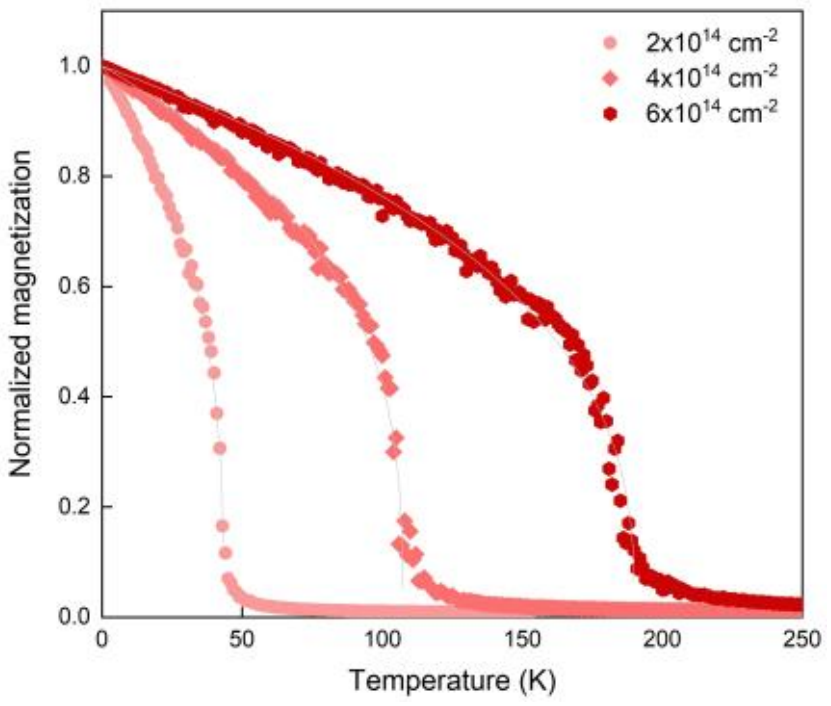

$T_c$: 1×$10^{14}$ cm$^{-2}$: - K  2×$10^{14}$ cm$^{-2}$: 43 K

4×$10^{14}$ cm$^{-2}$: 107 K  6×$10^{14}$ cm$^{-2}$: 190 K

# 22.$CdI_2$

| MC2D-ID | C2DB | 2dmat-ID | USPEX | Space group | Band gap (eV) |
|---|---|---|---|---|---|
| 33 | - | 2dm-4402 | - | P3m1 | 2.49 |

**Convex hull** | **Atomic structure** | **Atomic coordinates** | **Phonon dispersion curve**

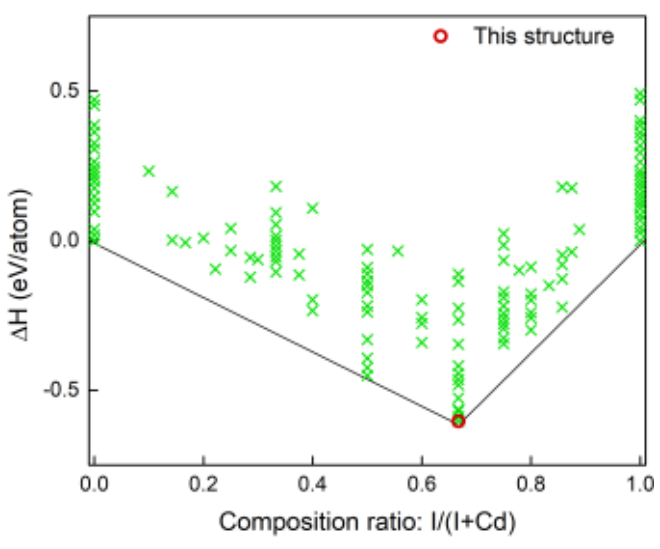

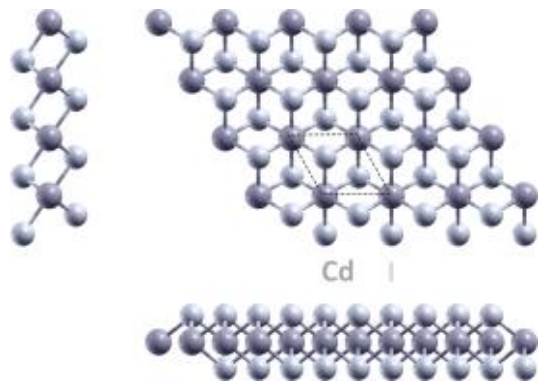

```
CdI2
1.0000000000000000
4.3316988530728873 0.0000000000000000 0.0000000000000000
-2.1658519942253878 3.7513592953485837 0.0000000000000000
0.0000000000000000 0.0000000000000000 23.4358480000000000
Cd I
1 2
Direct
0.0000000000086544 0.9999999991882902 0.5000000011422969
0.3333332763331356 0.6666666941107380 0.5732266571983189
0.6666667236582100 0.3333333067009789 0.4267733416593913
```

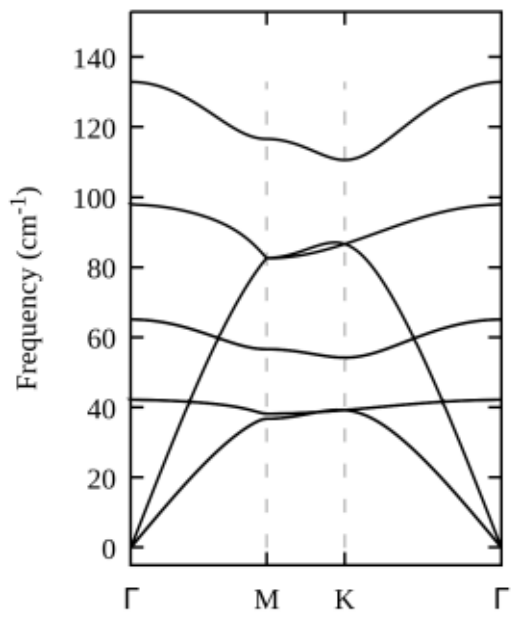

**Projected band structure and density of states**

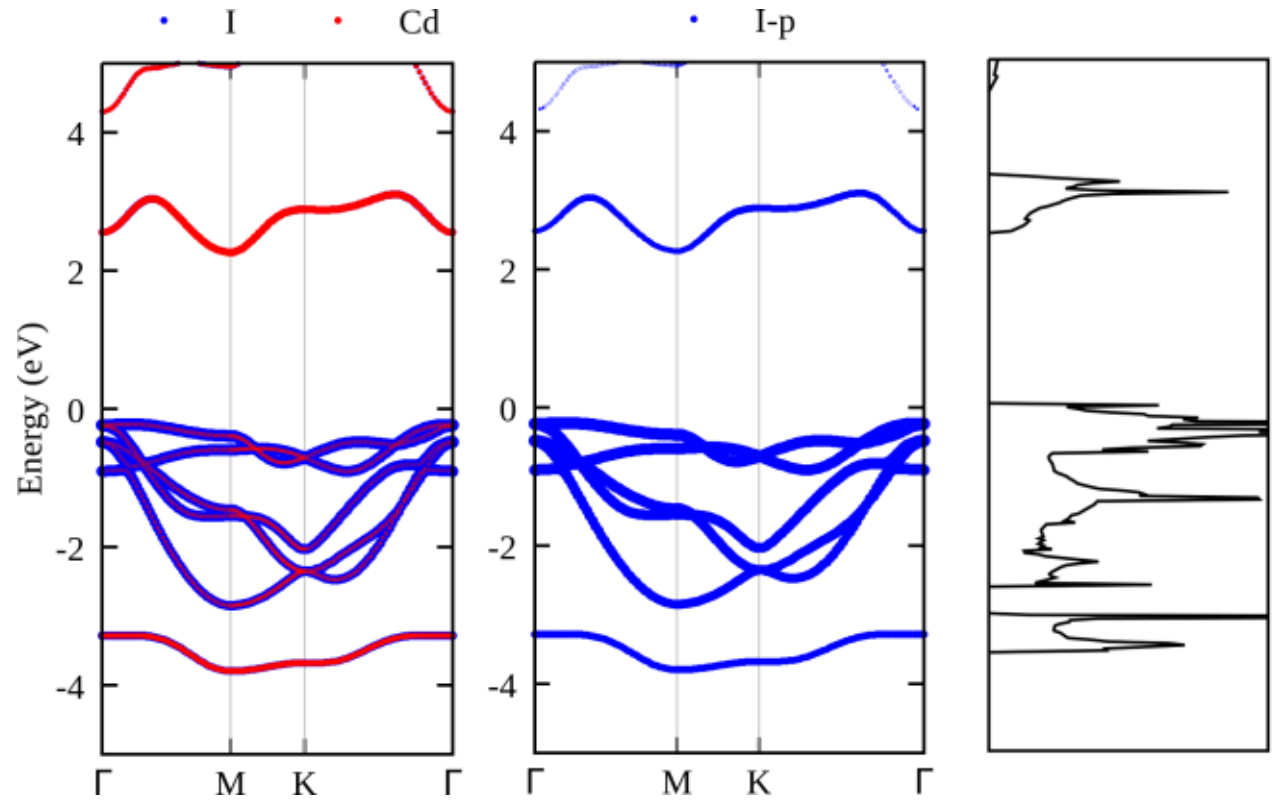

**Magnetic moment and spin polarization energy as a function of hole doping concentration**

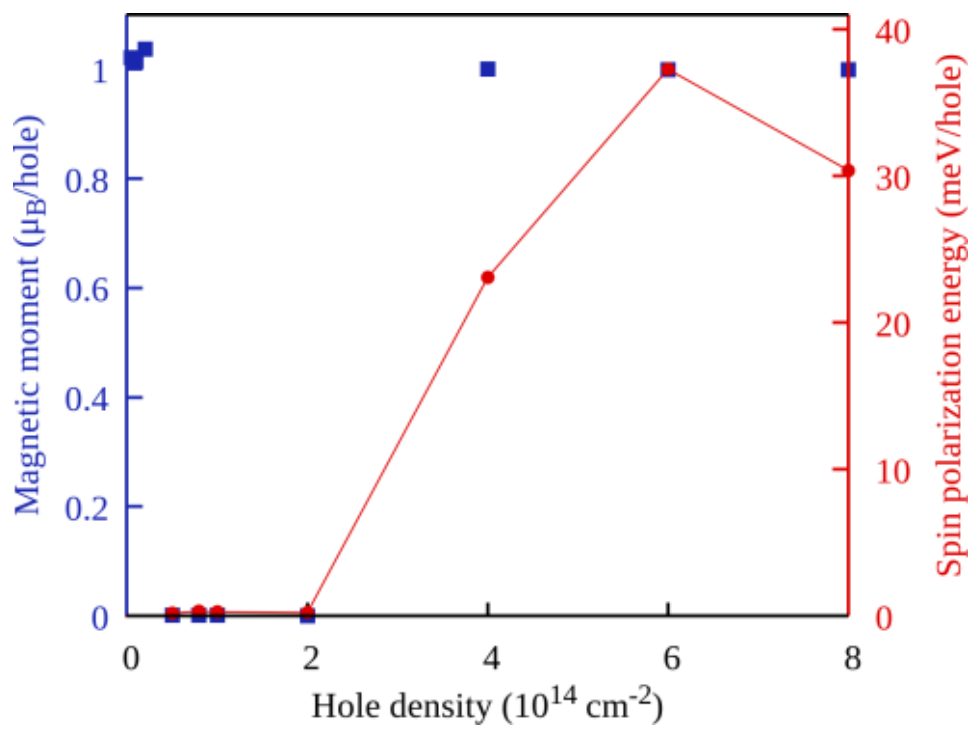

**Magnetic configurations and spin Hamiltonian**

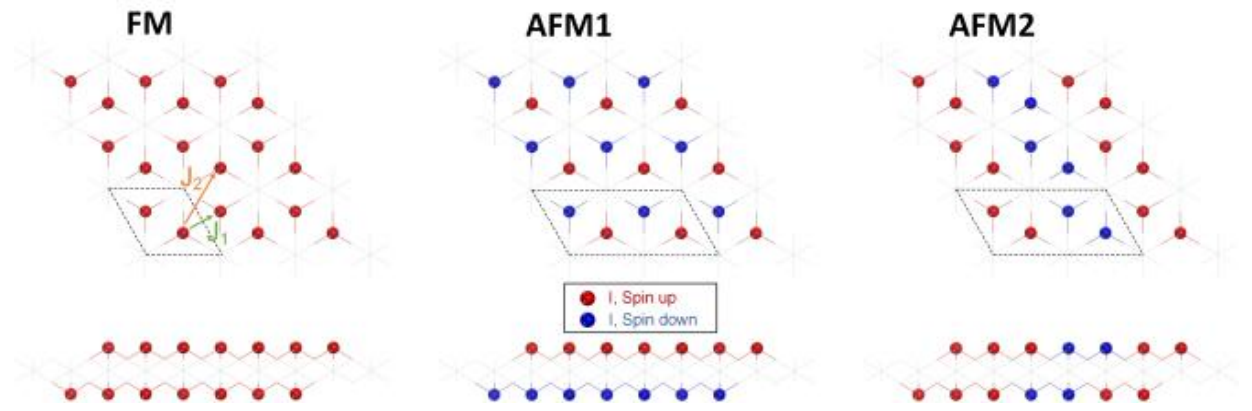

$$E_{FM} = E_0 - 3J_1S^2 - 6J_2S^2$$
$$E_{AFM1} = E_0 + 3J_1S^2 - 6J_2S^2$$
$$E_{AFM2} = E_0 - J_1S^2 + 2J_2S^2$$

**Magnetic exchange coupling parameters**

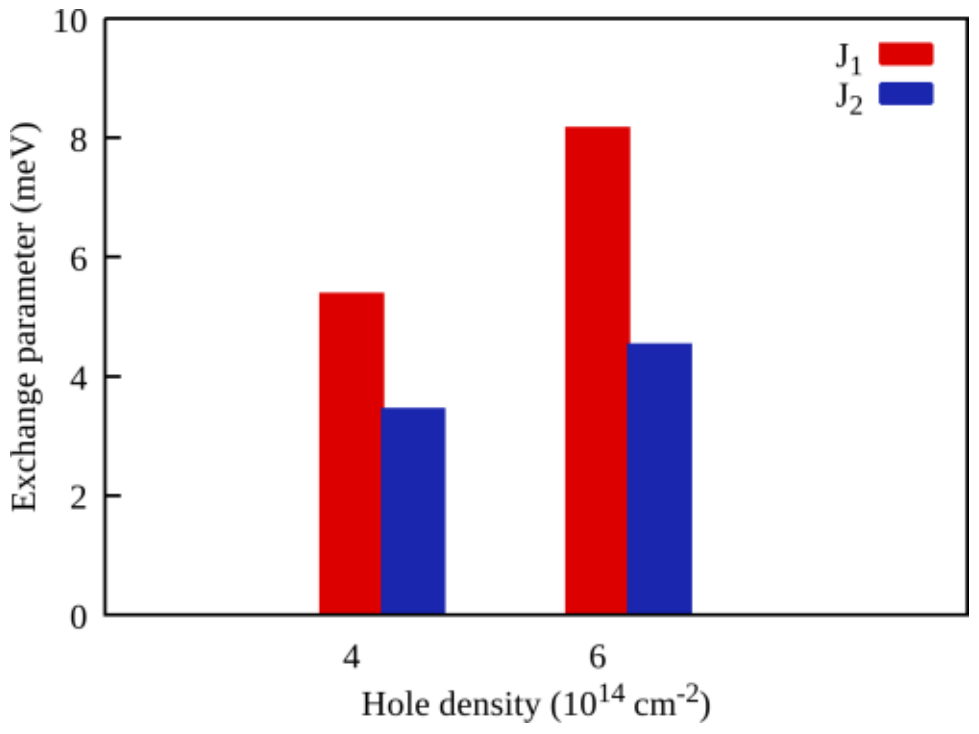

**Magnetic anisotropy energy (MAE, μeV) per magnetic atom**

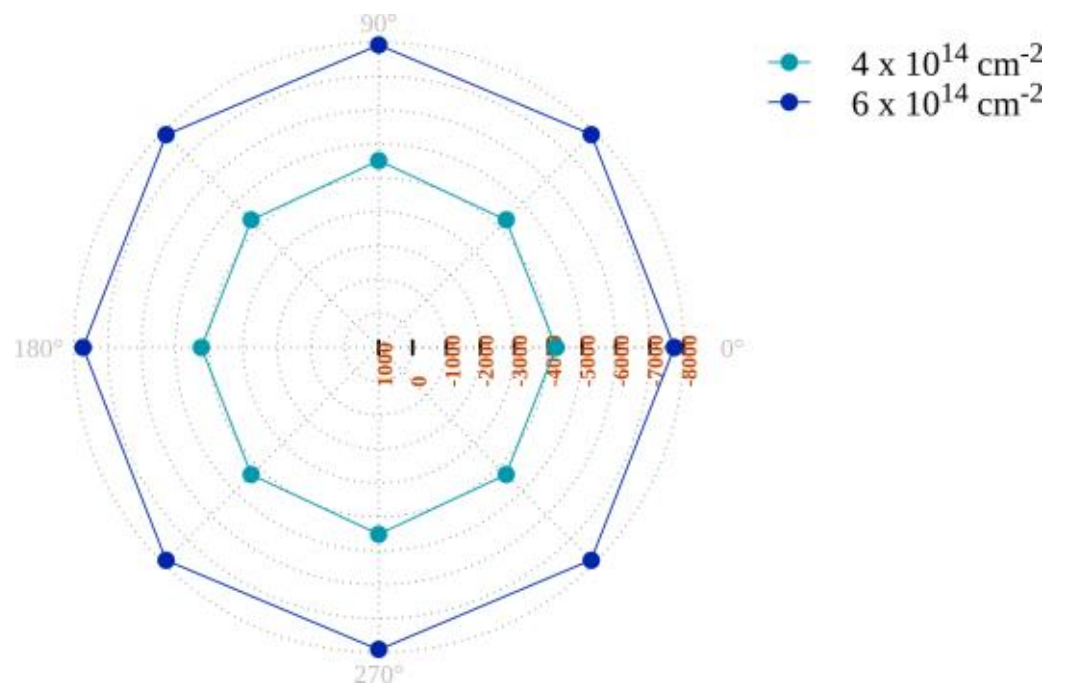

MAE = $E_\parallel - E_\perp$, a positive (negative) value of MAE indicates the off-plane (in-plane) easy axis.

**Monte Carlo simulations of the normalized magnetization of as a function of temperature**

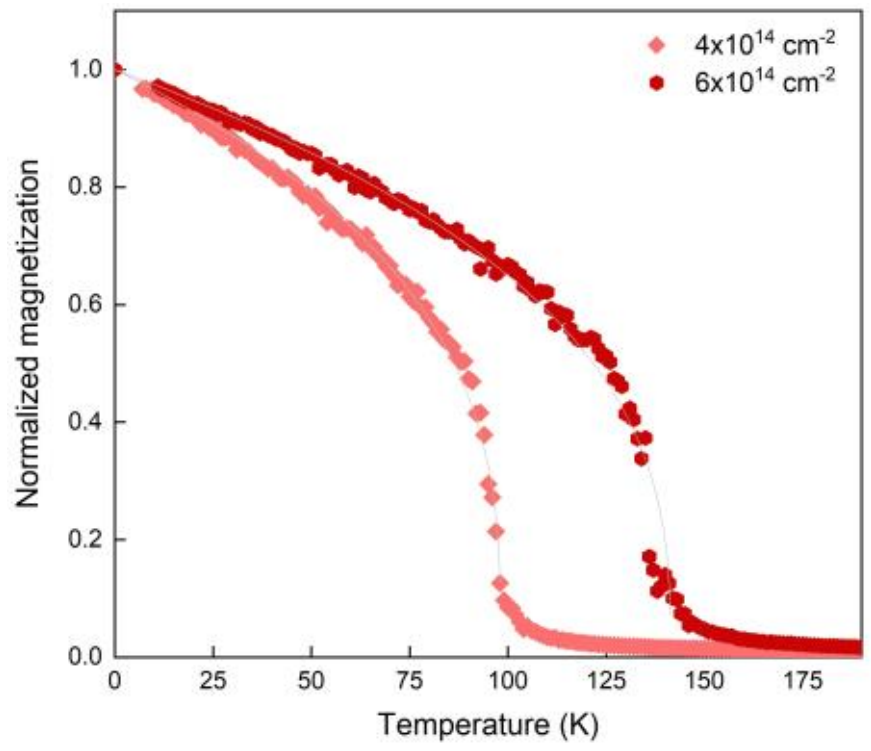

$T_c$: $1\times10^{14}$ cm$^{-2}$: - K $2\times10^{14}$ cm$^{-2}$: - K

$4\times10^{14}$ cm$^{-2}$: 98 K $6\times10^{14}$ cm$^{-2}$: 141 K

# 23. $HgF_2$

| MC2D-ID | C2DB | 2dmat-ID | USPEX | Space group | Band gap (eV) |
|---|---|---|---|---|---|
| - | - | 2dm-1267 | - | P3m1 | 1.83 |

**Convex hull** | **Atomic structure** | **Atomic coordinates** | **Phonon dispersion curve**

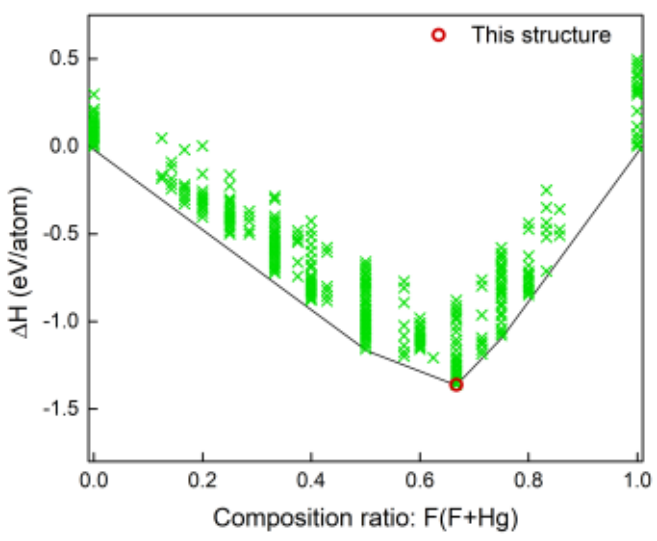

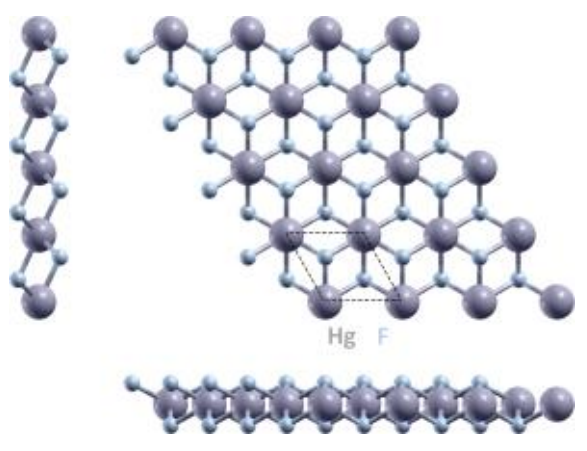

```
Hg1F2
   1.00000000000000
   3.6961332543963730    0.0000000000000000    0.0000000000000000
  -1.8480904290674907    3.2009317551727383    0.0000000000000000
   0.0000000000000000    0.0000000000000000   22.9560519999999997
   Hg    F
    1     2
Direct
  0.9999999999731912  0.0000000000039009  0.0000000000325286
  0.3333340419202813  0.6666660076113686  0.0438198432261103
  0.6666659581065275  0.3333339923847305  0.9561801567413610
```

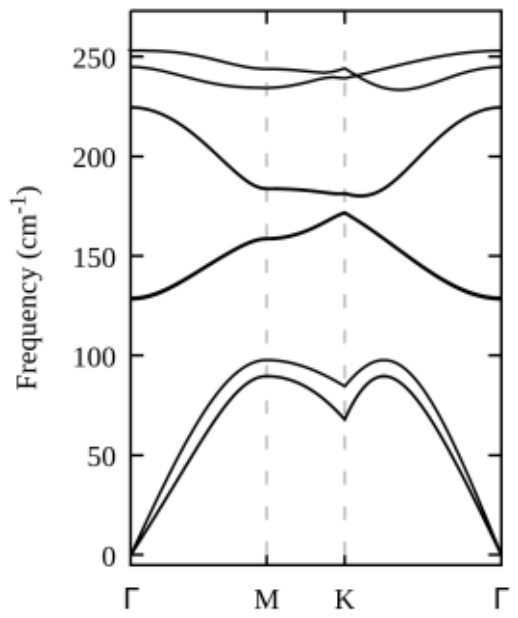

**Projected band structure and density of states**

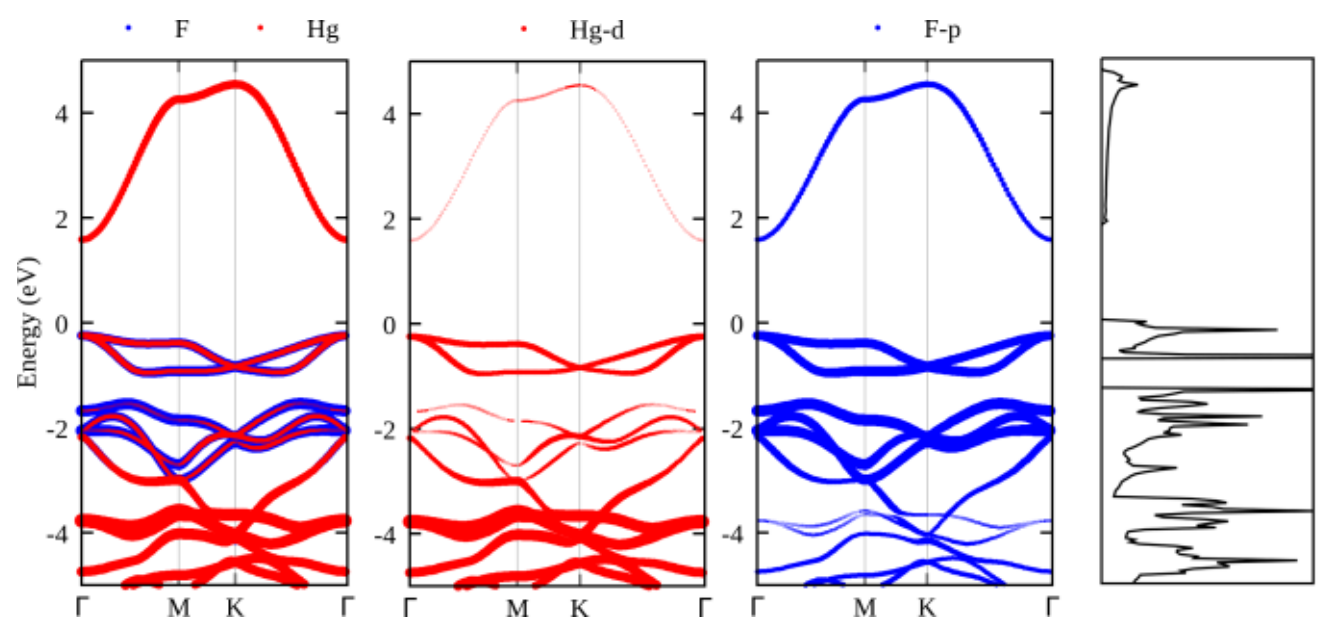

**Magnetic moment and spin polarization energy as a function of hole doping concentration**

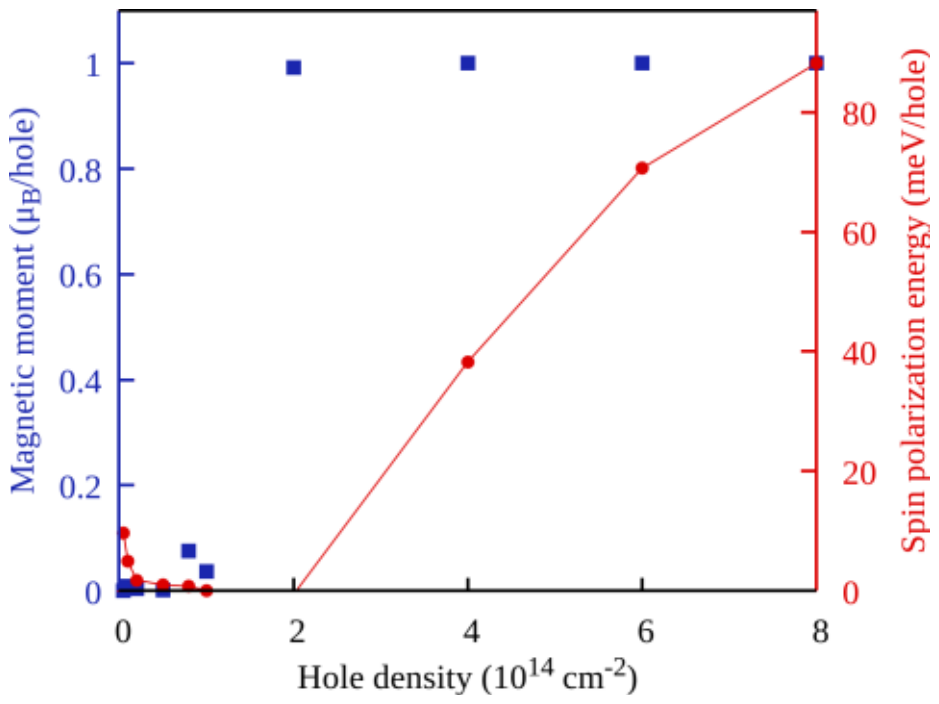

**Magnetic configurations and spin Hamiltonian**

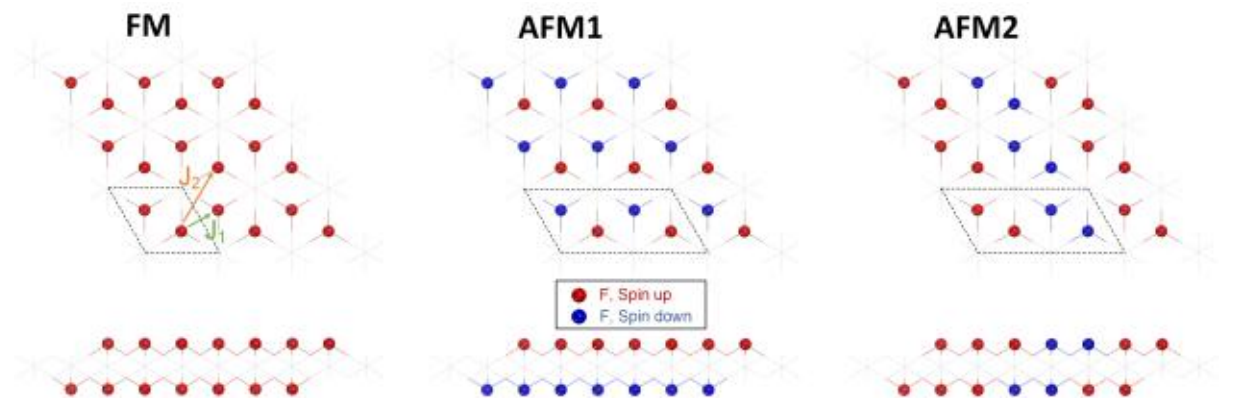

$$E_{FM} = E_0 - 3J_1S^2 - 6J_2S^2$$
$$E_{AFM1} = E_0 + 3J_1S^2 - 6J_2S^2$$
$$E_{AFM2} = E_0 - J_1S^2 + 2J_2S^2$$

**Magnetic exchange coupling parameters**

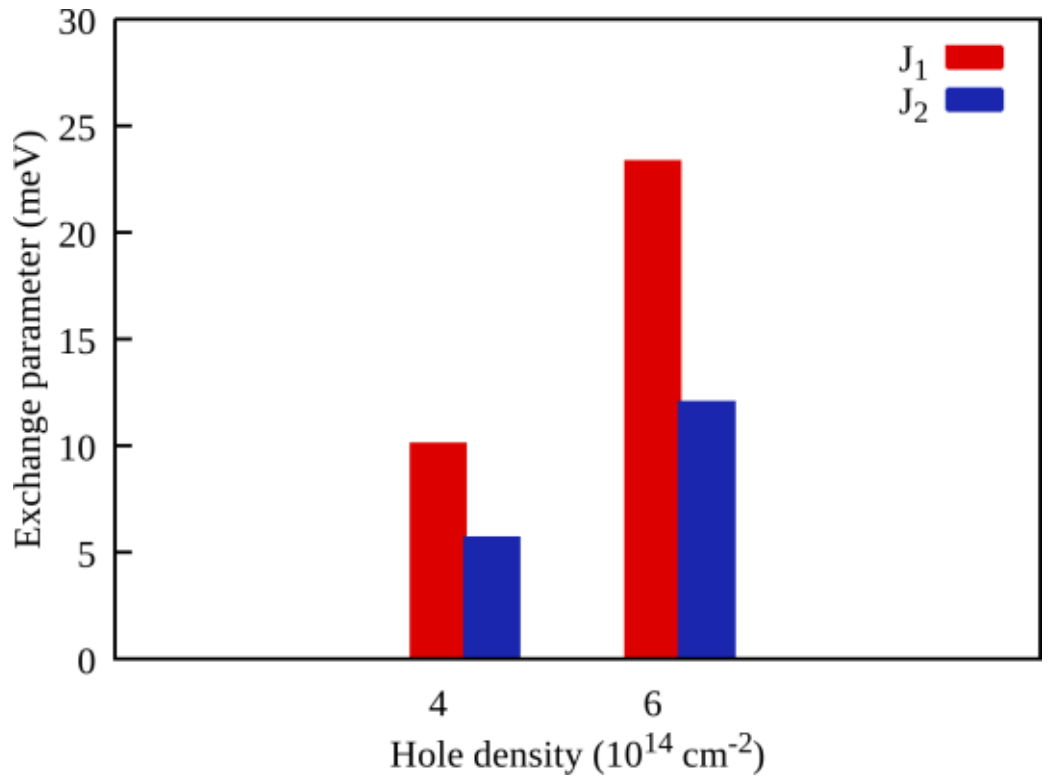

**Magnetic anisotropy energy (MAE, μeV) per magnetic atom**

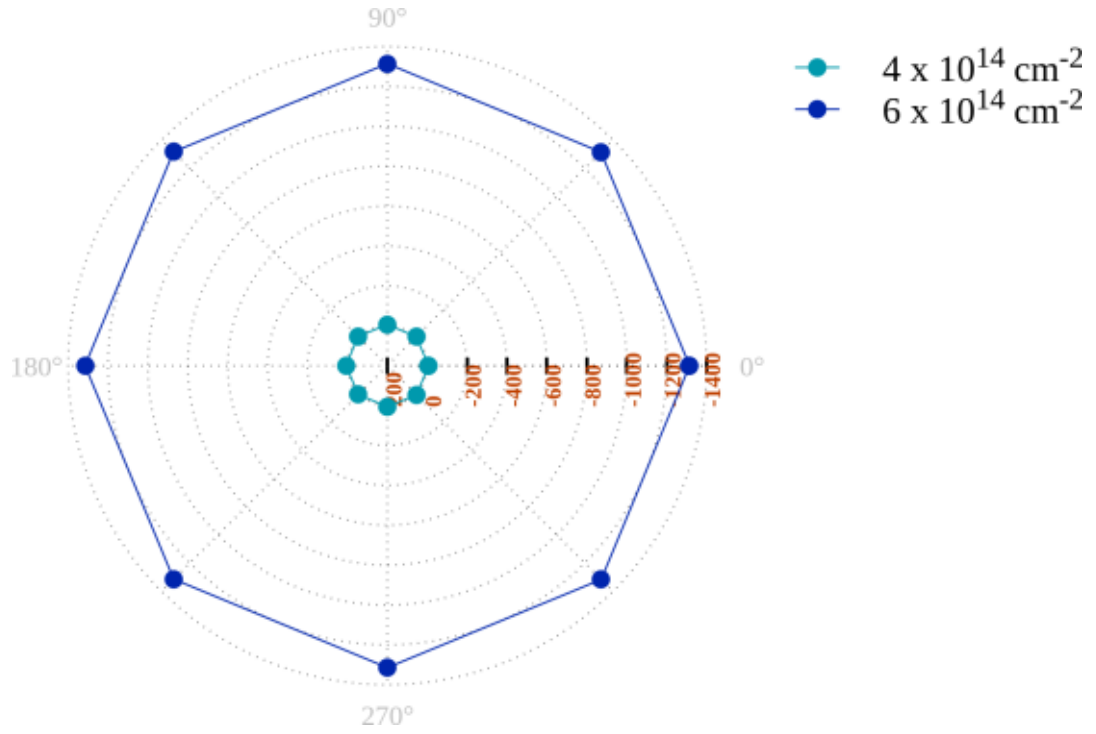

MAE = $E_{\parallel} - E_{\perp}$, a positive (negative) value of MAE indicates the off-plane (in-plane) easy axis.

**Monte Carlo simulations of the normalized magnetization of as a function of temperature**

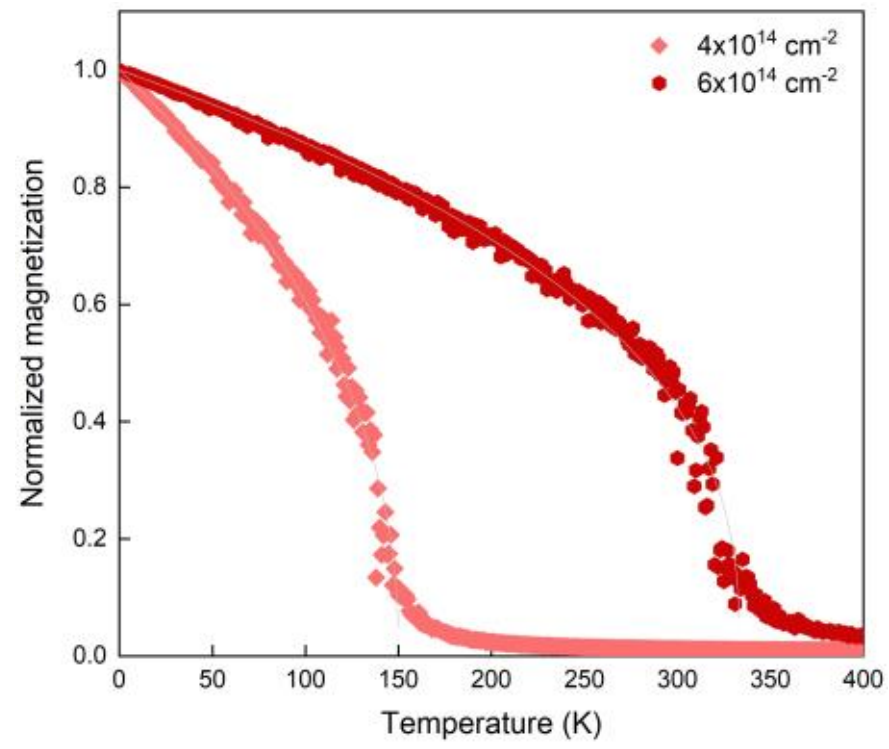

$T_c$: $1\times10^{14}$ cm$^{-2}$: - K  $2\times10^{14}$ cm$^{-2}$: - K

$4\times10^{14}$ cm$^{-2}$: 150 K  $6\times10^{14}$ cm$^{-2}$: 333 K

# 24. $HgCl_2$

| MC2D-ID | C2DB | 2dmat-ID | USPEX | Space group | Band gap (eV) |
|---|---|---|---|---|---|
| - | - | 2dm-704 | - | P3m1 | 2.43 |

**Convex hull** | **Atomic structure** | **Atomic coordinates** | **Phonon dispersion curve**

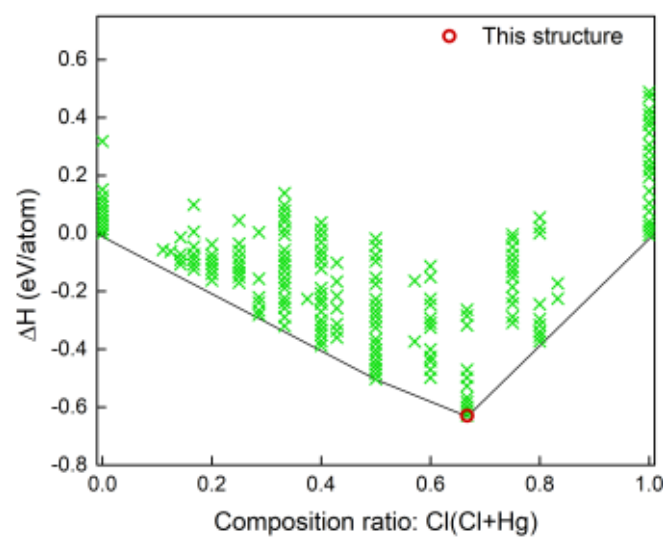

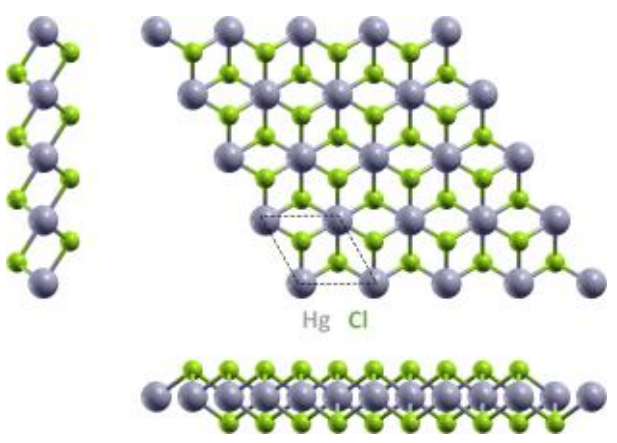

```
Hg1Cl2
   1.00000000000000
   3.9910470260351625   0.0000000000000000   0.0000000000000000
  -1.9955138134886130   3.4563539216977830   0.0000000000000000
   0.0000000000000000   0.0000000000000000  22.9560519999999997
   Hg   Cl
   1    2
Direct
  0.0000000003411529  0.0000000009434515  0.0000000002844232
  0.3333329915070777  0.6666669812547212  0.0636426794320073
  0.6666670081517623  0.3333330178018272  0.9363573202835695
```

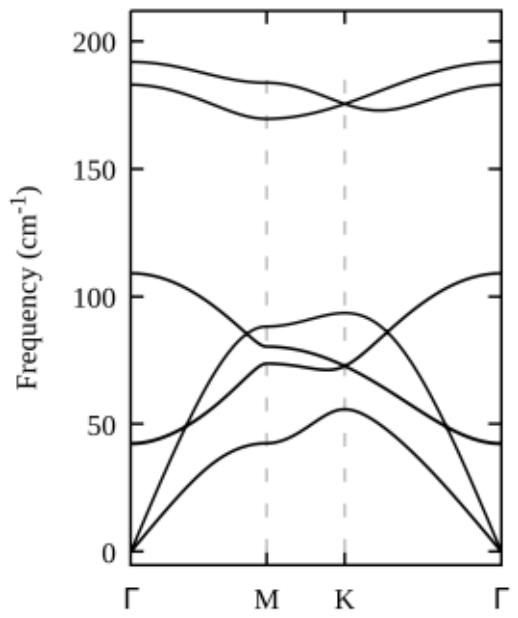

**Projected band structure and density of states** | **Magnetic moment and spin polarization energy as a function of hole doping concentration**

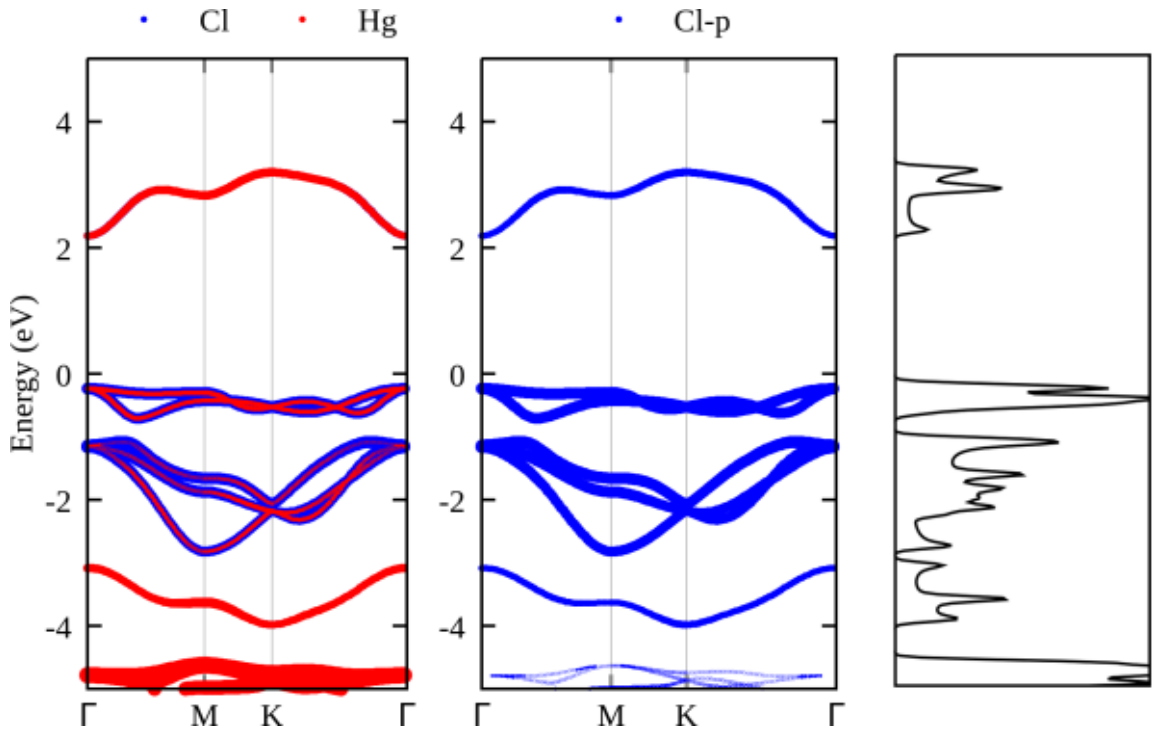

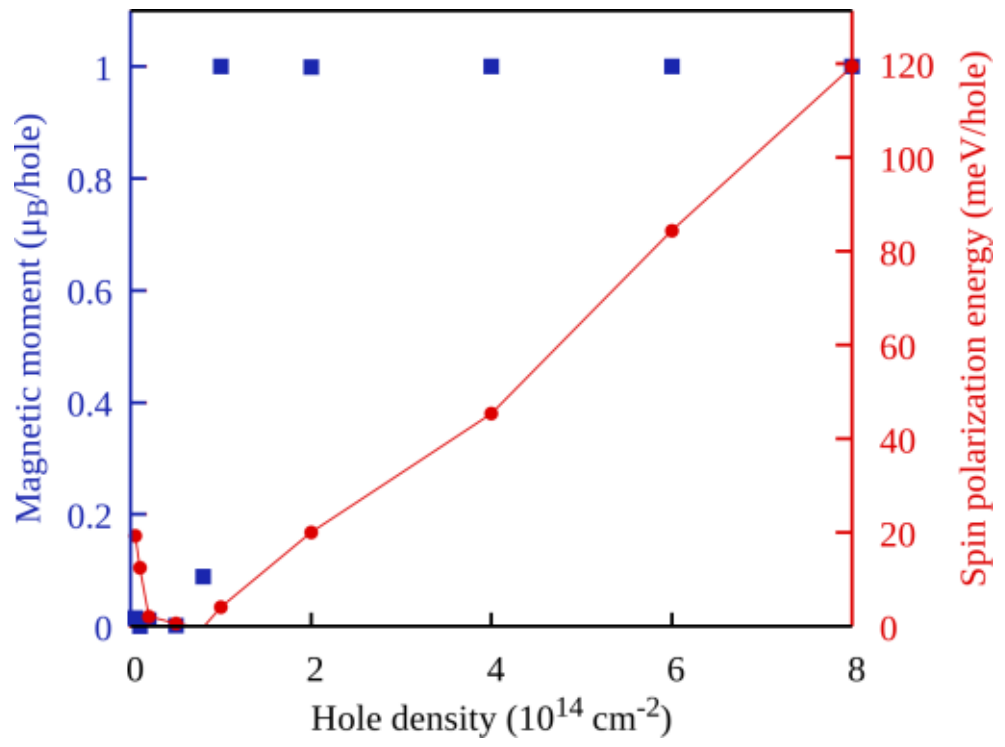

**Magnetic configurations and spin Hamiltonian** | **Magnetic exchange coupling parameters**

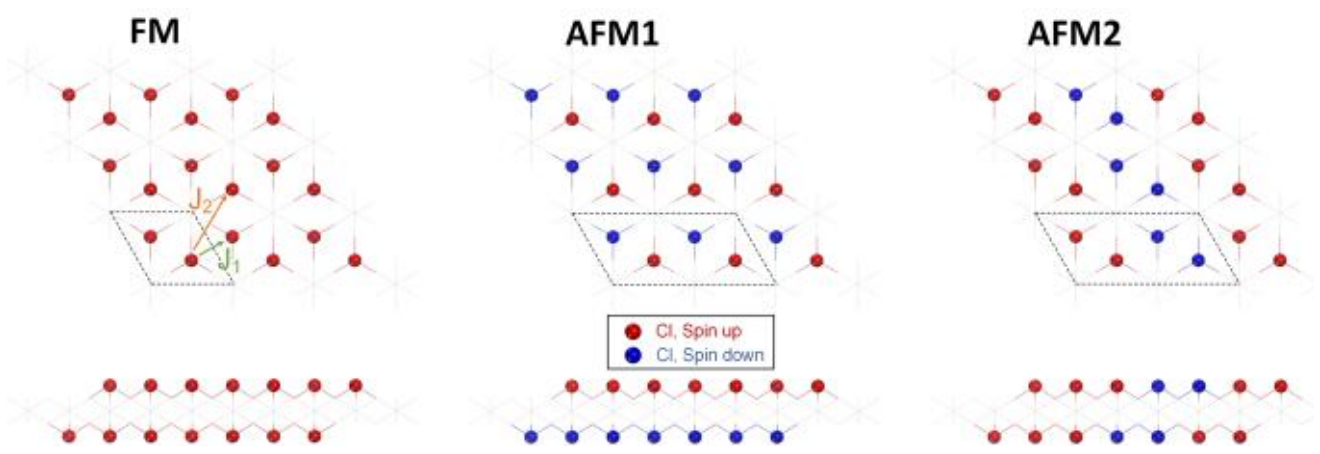

$$E_{FM} = E_0 - 3J_1S^2 - 6J_2S^2$$
$$E_{AFM1} = E_0 + 3J_1S^2 - 6J_2S^2$$
$$E_{AFM2} = E_0 - J_1S^2 + 2J_2S^2$$

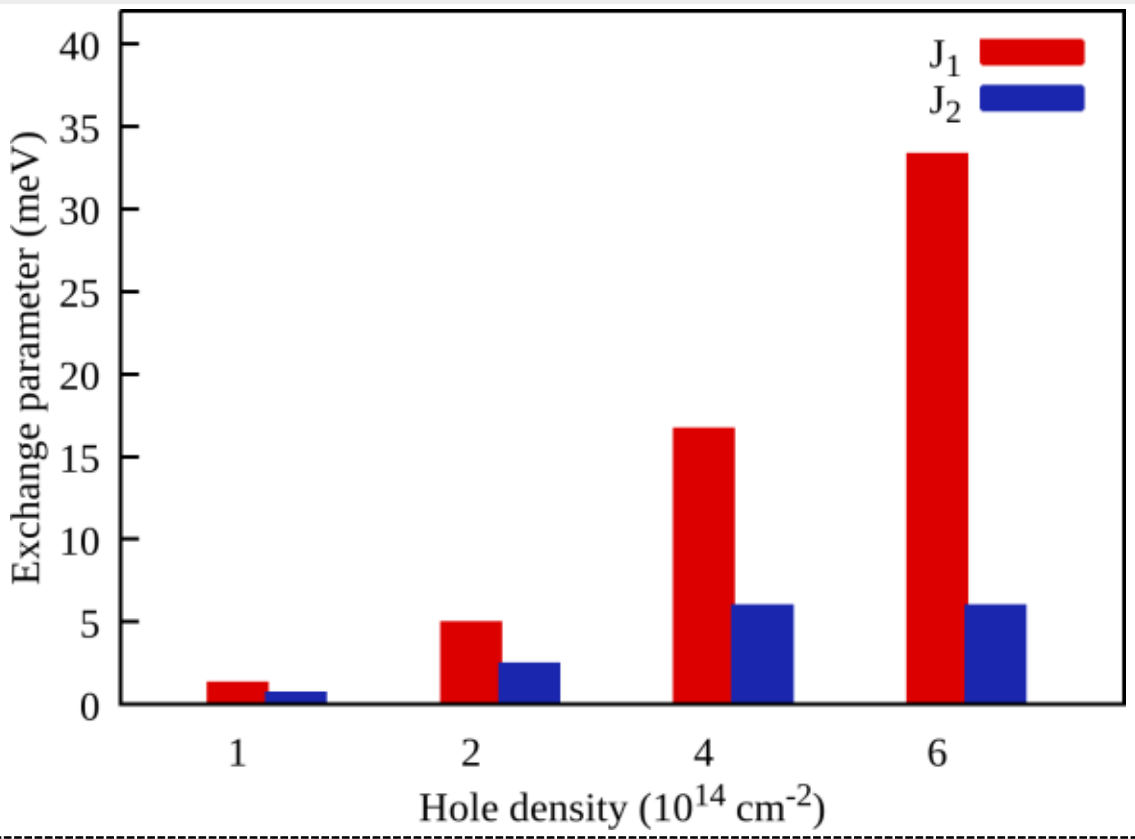

**Magnetic anisotropy energy (MAE, μeV) per magnetic atom** | **Monte Carlo simulations of the normalized magnetization of as a function of temperature**

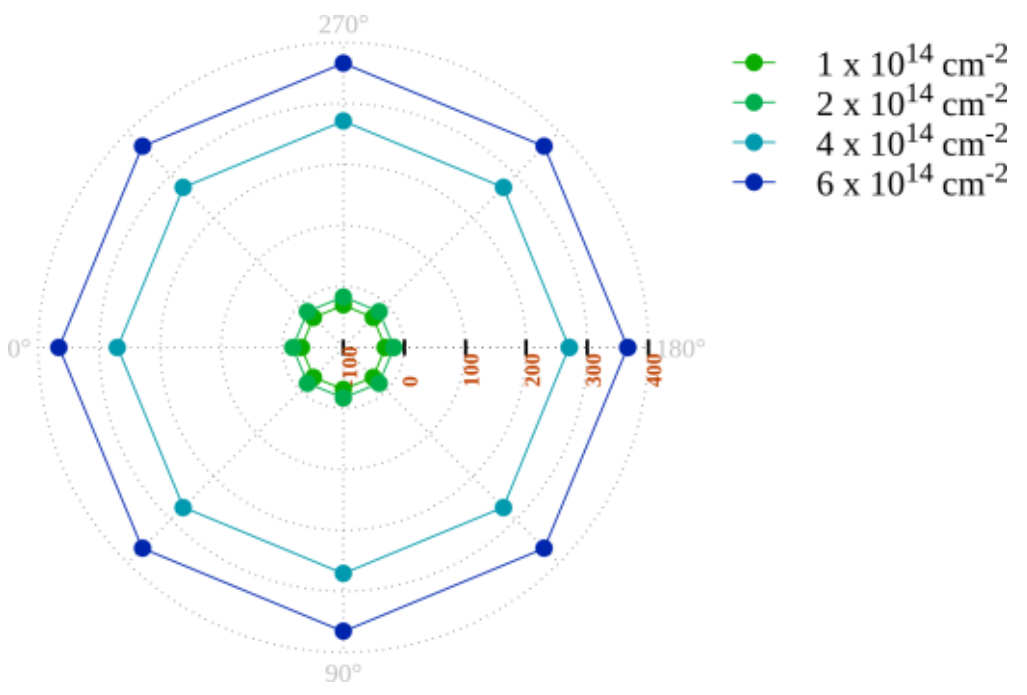

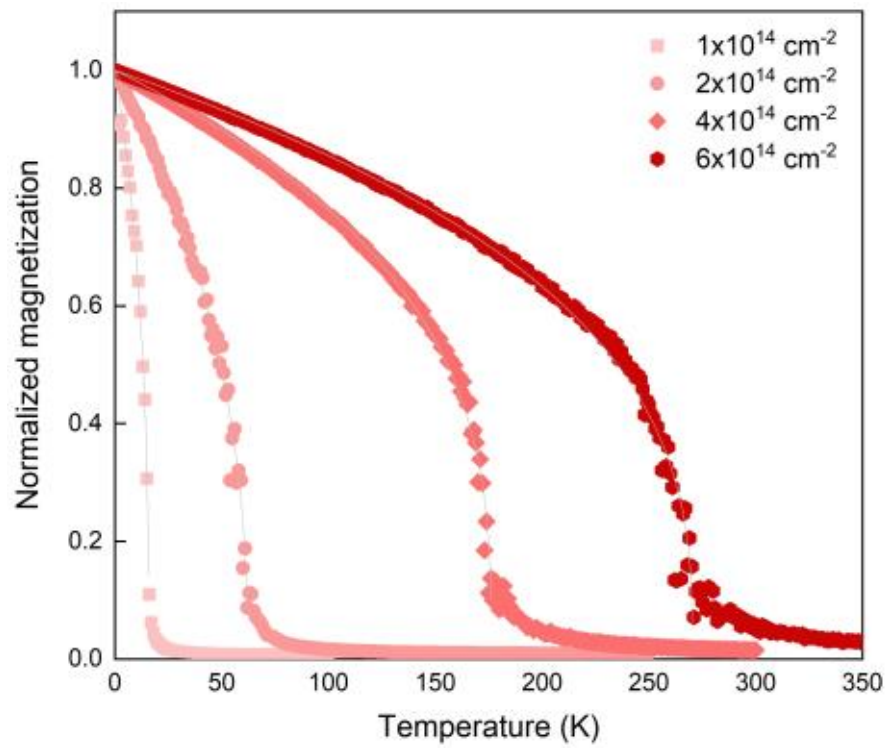

MAE = $E_{\parallel} - E_{\perp}$, a positive (negative) value of MAE indicates the off-plane (in-plane) easy axis.

$T_c$: $1\times10^{14}$ cm$^{-2}$: 16 K  $2\times10^{14}$ cm$^{-2}$: 62 K  $4\times10^{14}$ cm$^{-2}$: 176 K  $6\times10^{14}$ cm$^{-2}$: 270 K

# 25.HgBr$_2$

| MC2D-ID | C2DB | 2dmat-ID | USPEX | Space group | Band gap (eV) |
|---|---|---|---|---|---|
| - | - | 2dm-5582 | - | P3m1 | 2.03 |

**Convex hull**

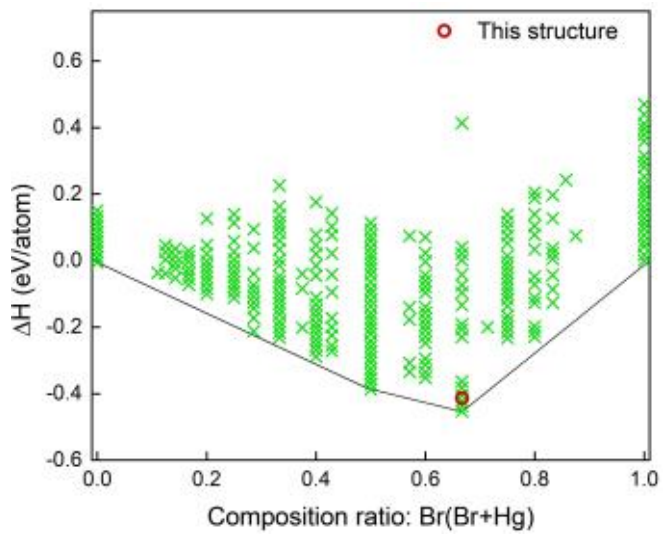

**Atomic structure**

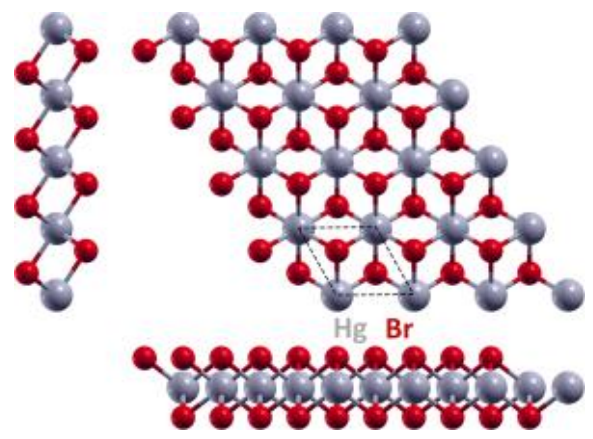

**Atomic coordinates**

```
Hg1Br2
1.00000000000000
4.1414124754368009    0.0000000000000000    0.0000000000000000
-2.0707091579513657    3.5865667601677167    0.0000000000000000
0.0000000000000014    0.0000000000000024    22.9560519999999997
Hg    Br
1    2
Direct
0.0000000013484254    0.0000000062100511    0.4999999727847353
0.3333329961901370    0.6666670063158548    0.5691642222798095
0.6666670024614234    0.3333329874740940    0.4308357777201905
```

**Phonon dispersion curve**

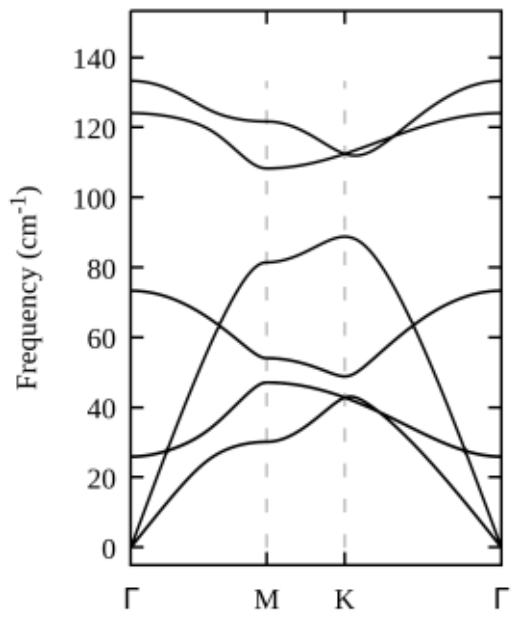

**Projected band structure and density of states**

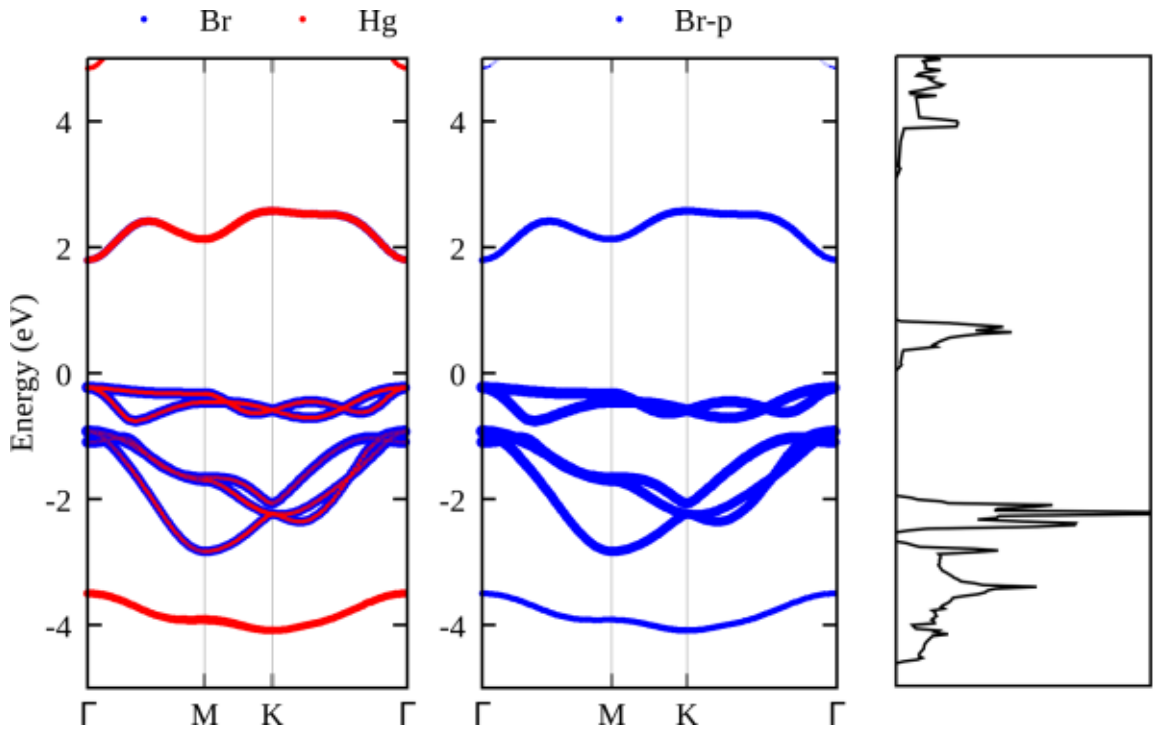

**Magnetic moment and spin polarization energy as a function of hole doping concentration**

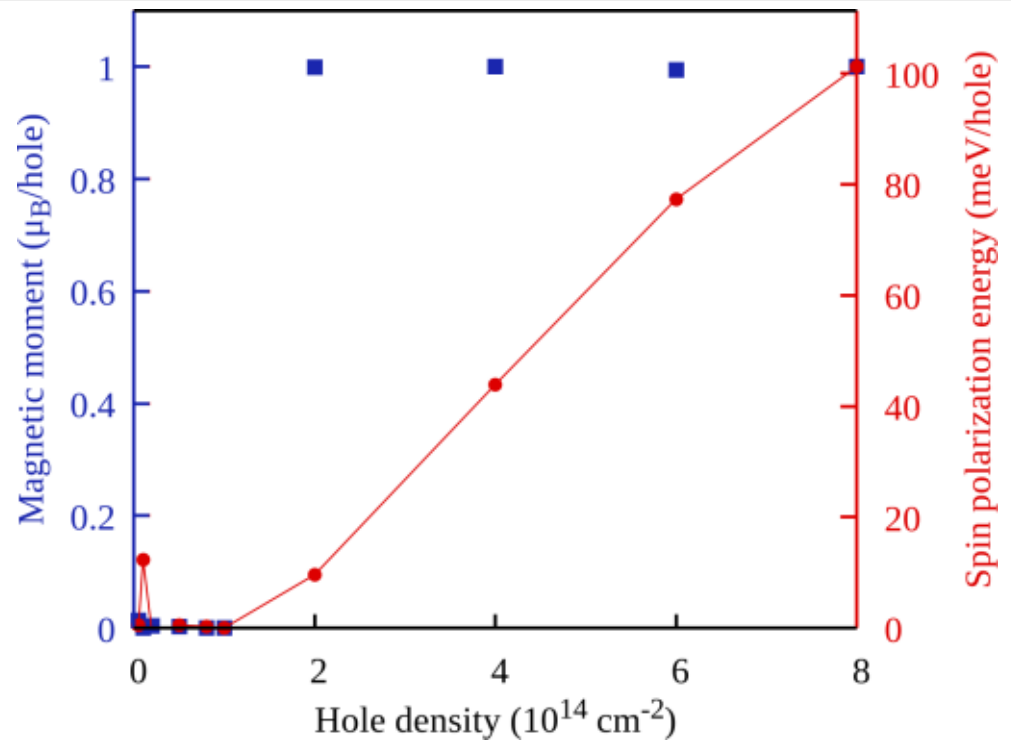

**Magnetic configurations and spin Hamiltonian**

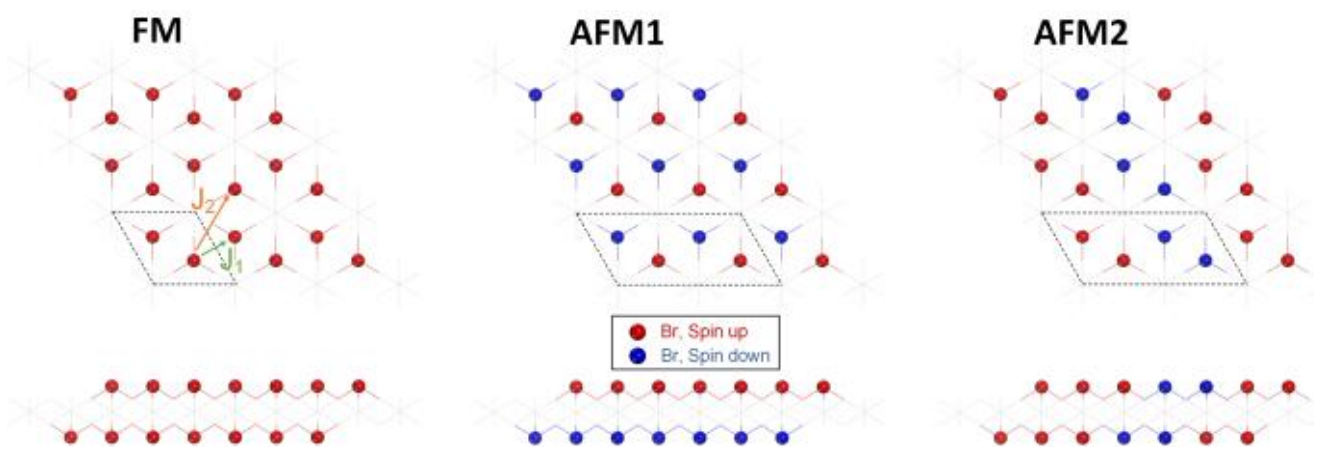

$$E_{FM} = E_0 - 3J_1S^2 - 6J_2S^2$$

$$E_{AFM1} = E_0 + 3J_1S^2 - 6J_2S^2$$

$$E_{AFM2} = E_0 - J_1S^2 + 2J_2S^2$$

**Magnetic exchange coupling parameters**

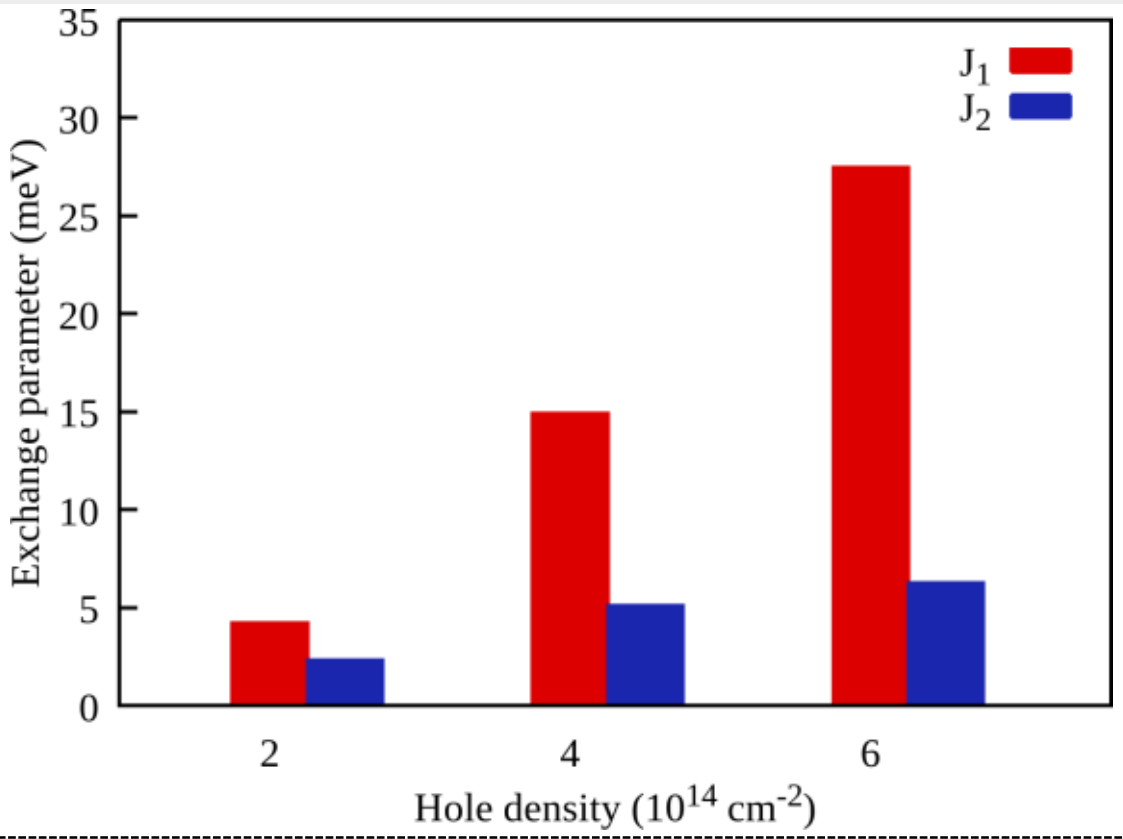

**Magnetic anisotropy energy (MAE, μeV) per magnetic atom**

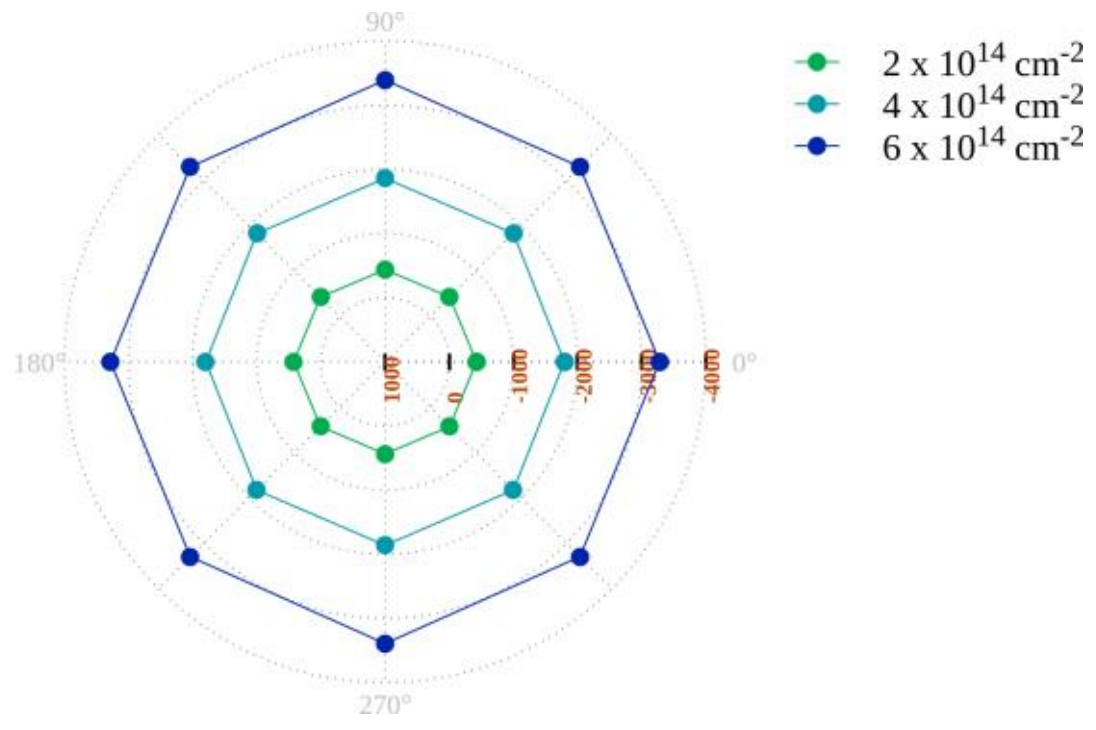

MAE = $E_{\parallel} - E_{\perp}$, a positive (negative) value of MAE indicates the off-plane (in-plane) easy axis.

**Monte Carlo simulations of the normalized magnetization of as a function of temperature**

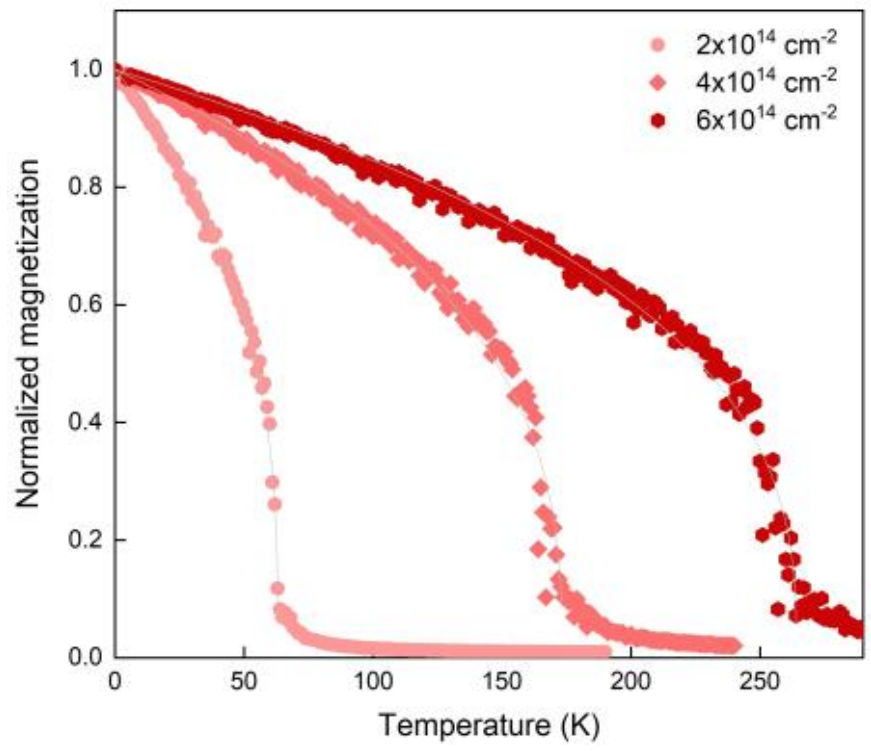

$T_c$: $1\times10^{14}$ cm$^{-2}$: - K  $2\times10^{14}$ cm$^{-2}$: 63 K

$4\times10^{14}$ cm$^{-2}$: 173 K  $6\times10^{14}$ cm$^{-2}$: 264 K

# 26. $HgI_2$

| MC2D-ID | C2DB | 2dmat-ID | USPEX | Space group | Band gap (eV) |
|---|---|---|---|---|---|
| - | - | 2dm-1787 | - | P3m1 | 1.42 |

**Convex hull**

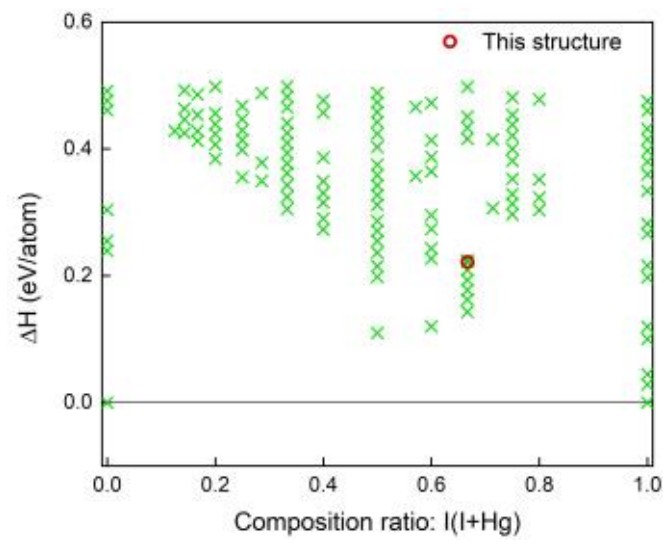

**Atomic structure**

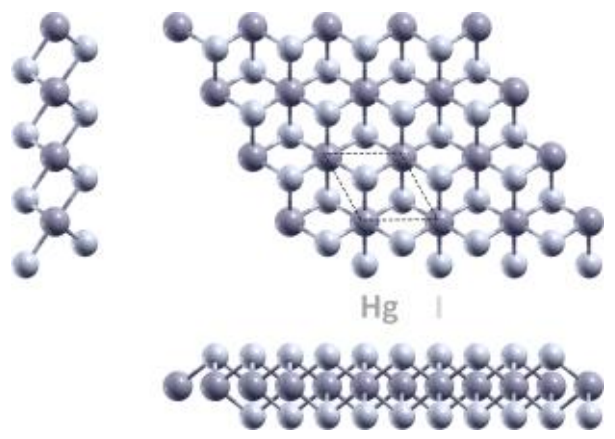

**Atomic coordinates**

```
Hg1I2
1.00000000000000
4.3801552151490224    0.0000000000000000    0.0000000000000000
-2.1900890609255428    3.7933179985396399    0.0000000000000000
0.0000000000000014    0.0000000000000024    22.9560519999999997
Hg    I
1    2
Direct
0.9999999734892455    0.9999999764134557    0.4999999925911638
0.3333333609336862    0.6666664630857753    0.5758827326418228
0.6666666655770612    0.3333335605007690    0.4241172673581772
```

**Phonon dispersion curve**

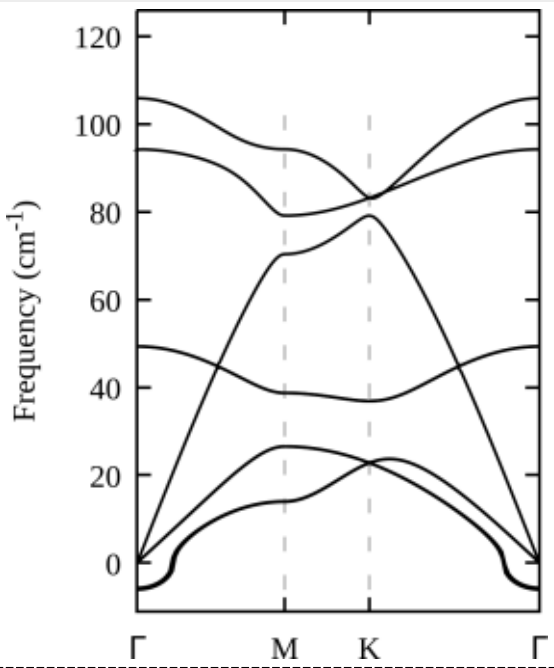

**Projected band structure and density of states**

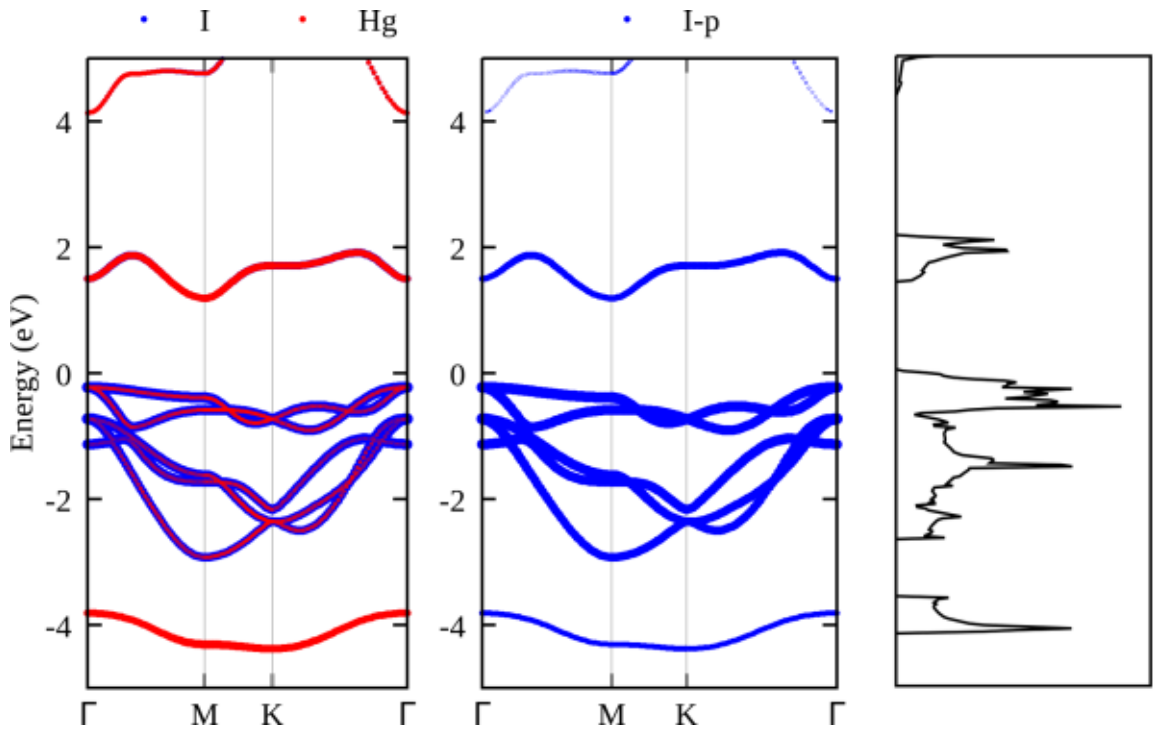

**Magnetic moment and spin polarization energy as a function of hole doping concentration**

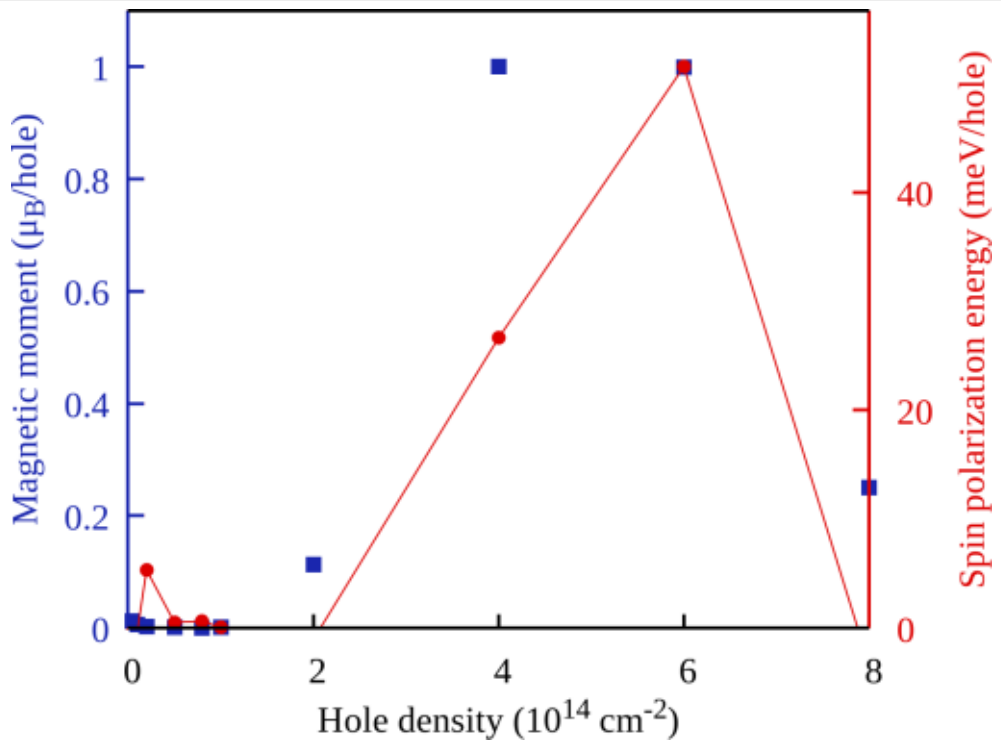

**Magnetic configurations and spin Hamiltonian**

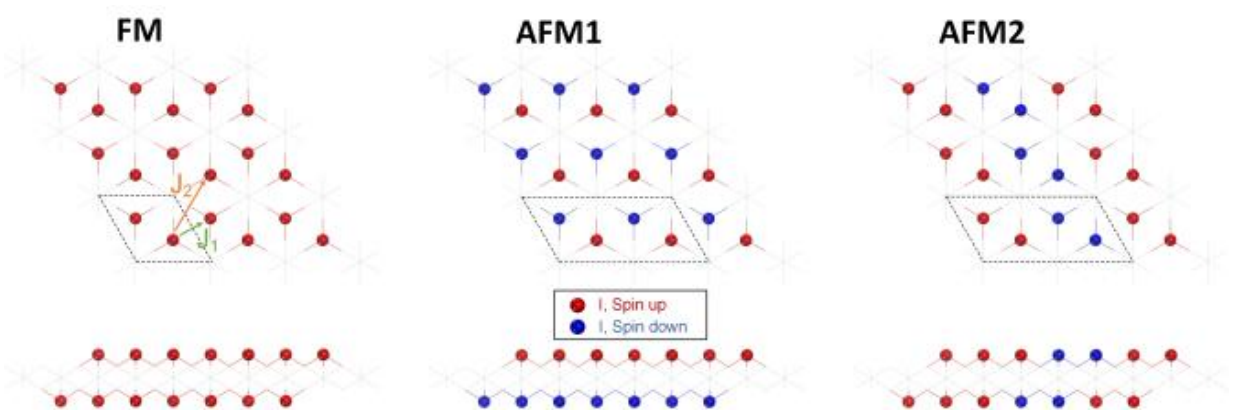

$$E_{FM} = E_0 - 3J_1S^2 - 6J_2S^2$$
$$E_{AFM1} = E_0 + 3J_1S^2 - 6J_2S^2$$
$$E_{AFM2} = E_0 - J_1S^2 + 2J_2S^2$$

**Magnetic exchange coupling parameters**

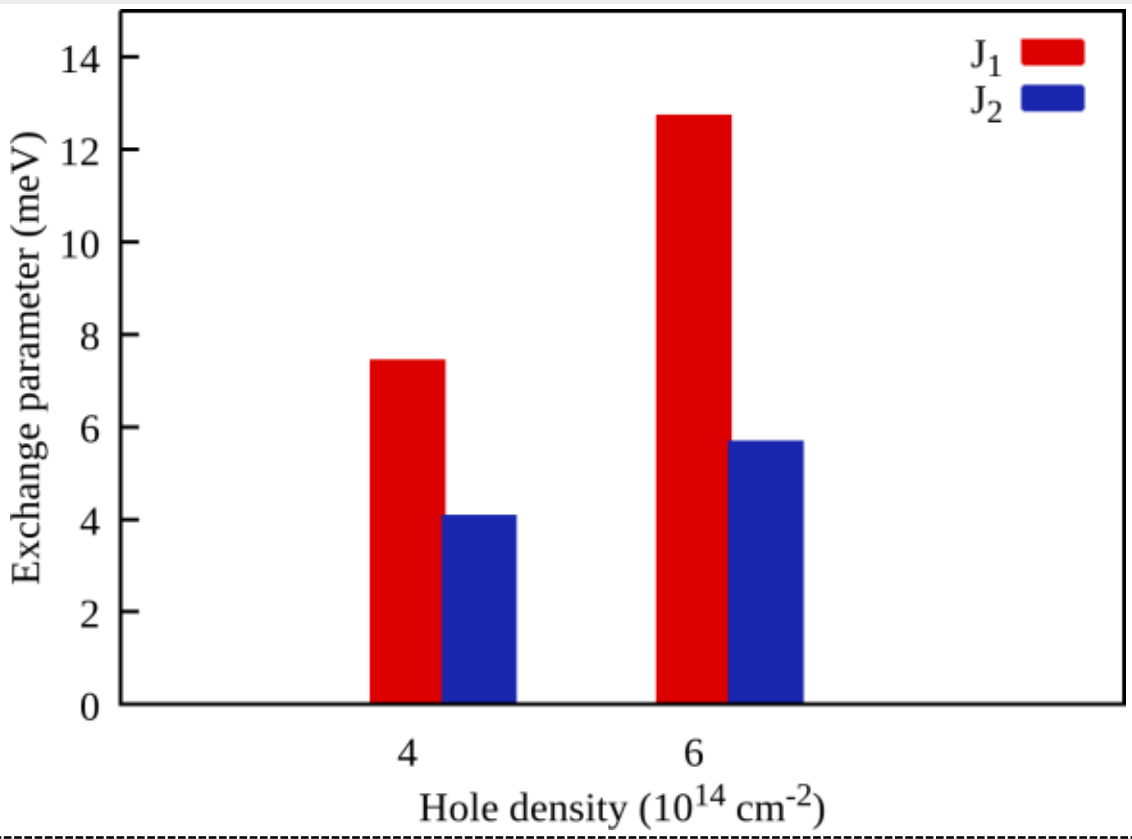

**Magnetic anisotropy energy (MAE, μeV) per magnetic atom**

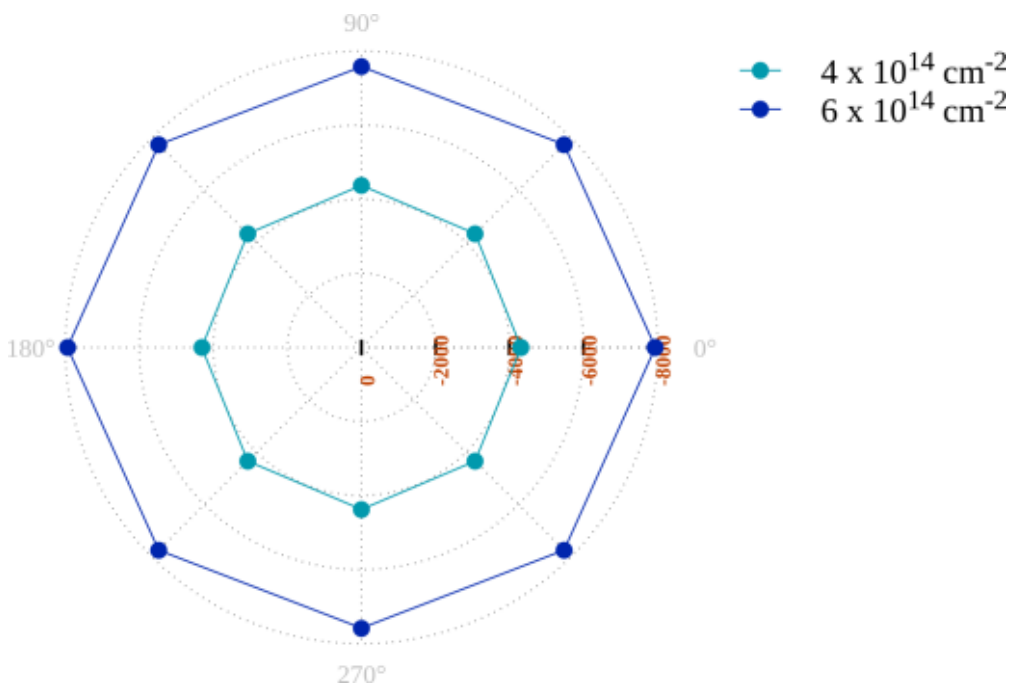

MAE = $E_{\parallel} - E_{\perp}$, a positive (negative) value of MAE indicates the off-plane (in-plane) easy axis.

**Monte Carlo simulations of the normalized magnetization of as a function of temperature**

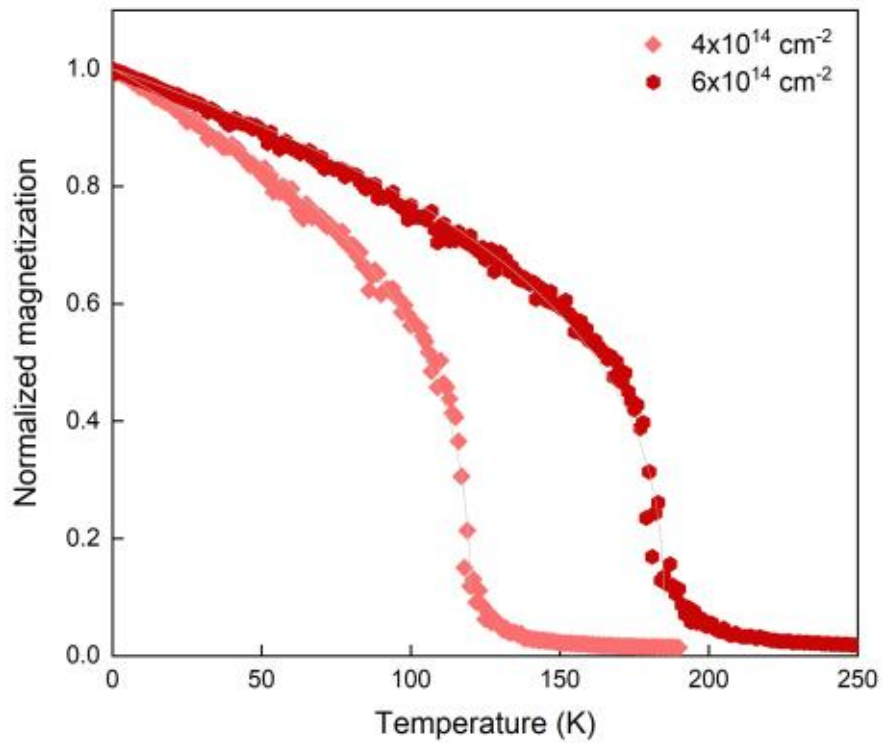

$T_c$: $1\times10^{14}$ cm$^{-2}$: - K    $2\times10^{14}$ cm$^{-2}$: - K

$4\times10^{14}$ cm$^{-2}$: 120 K    $6\times10^{14}$ cm$^{-2}$: 185 K

# 27. $PbCl_2$

| MC2D-ID | C2DB | 2dmat-ID | USPEX | Space group | Band gap (eV) |
|---|---|---|---|---|---|
| - | ✓ | 2dm-1488 | - | P3m1 | 3.07 |

**Convex hull** | **Atomic structure** | **Atomic coordinates** | **Phonon dispersion curve**

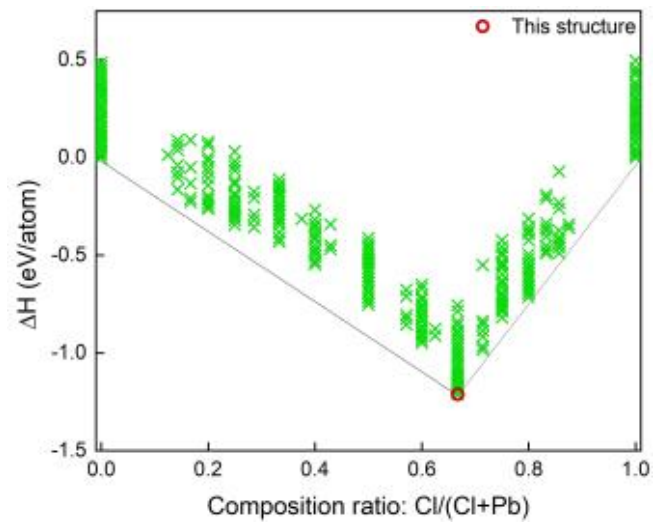

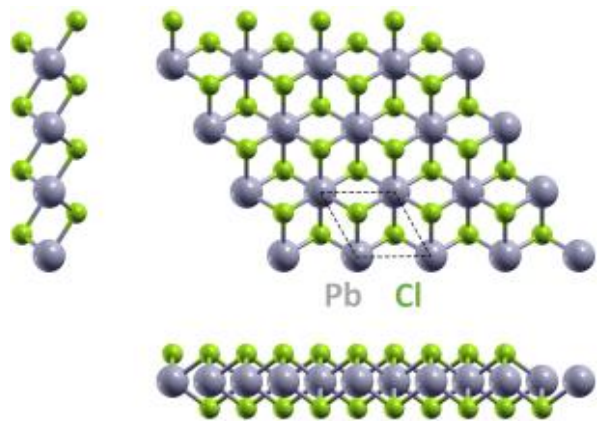

```
PbCl2
1.00000000000000
4.3970613879130758    0.0000000000000000    0.0000000000000000
-2.1985156727696702    3.8079873908668000    0.0000000000000000
0.0000000000000014    0.0000000000000025    23.1145000457999998
Pb    Cl
1    2
Direct
0.9999999958593691    0.9999999535866948    0.5000068738244536
0.3333332074778212    0.6666674573040368    0.4357256432128835
0.6666667966628239    0.3333325891092471    0.5642743567871165
```

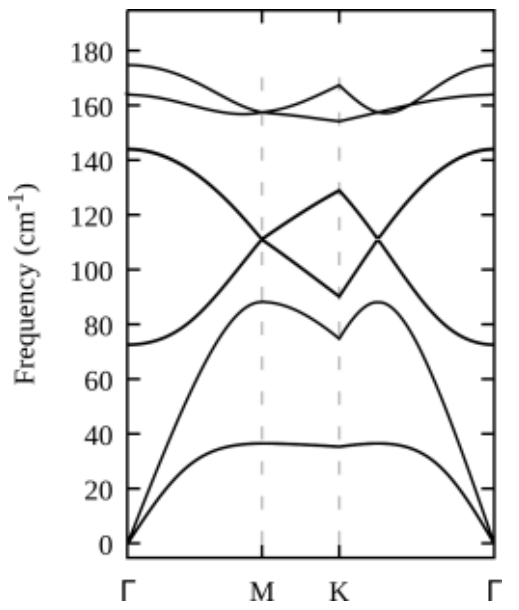

**Projected band structure and density of states**

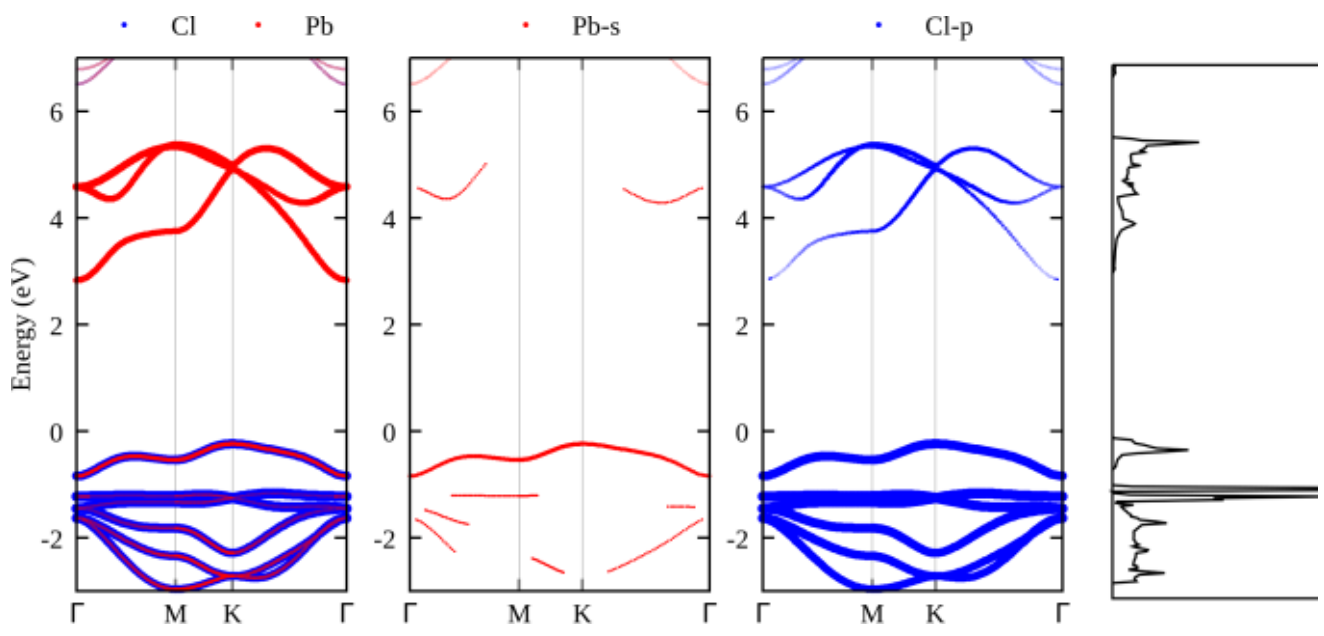

**Magnetic moment and spin polarization energy as a function of hole doping concentration**

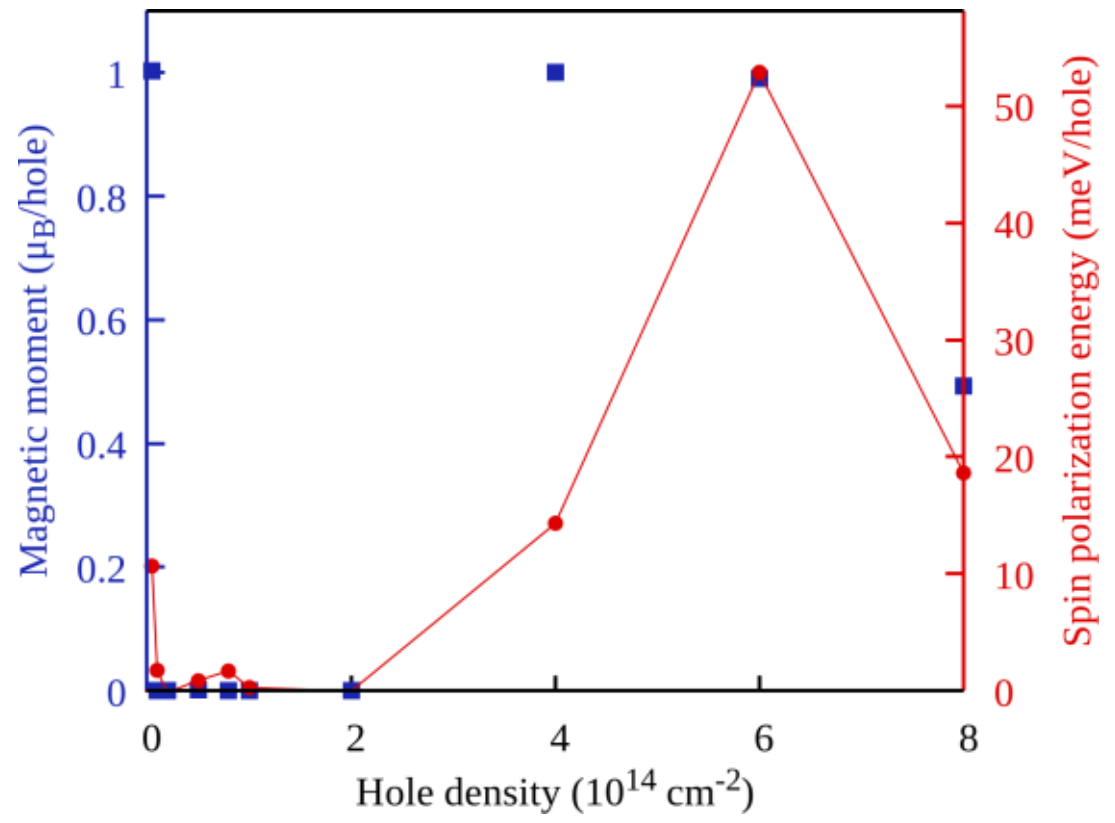

**Magnetic configurations and spin Hamiltonian**

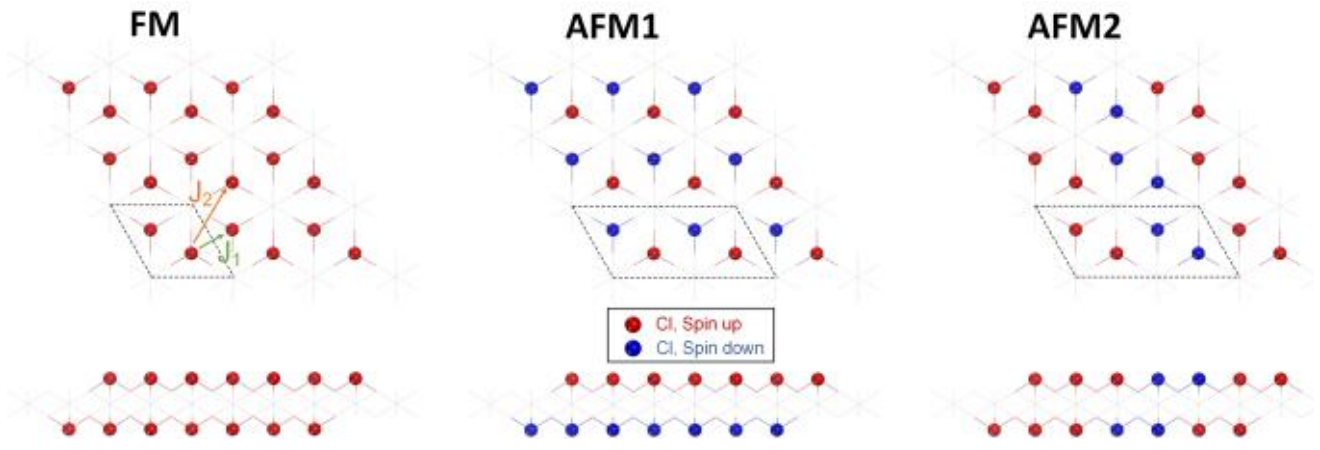

$$E_{FM} = E_0 - 3J_1S^2 - 6J_2S^2$$
$$E_{AFM1} = E_0 + 3J_1S^2 - 6J_2S^2$$
$$E_{AFM2} = E_0 - J_1S^2 + 2J_2S^2$$

**Magnetic exchange coupling parameters**

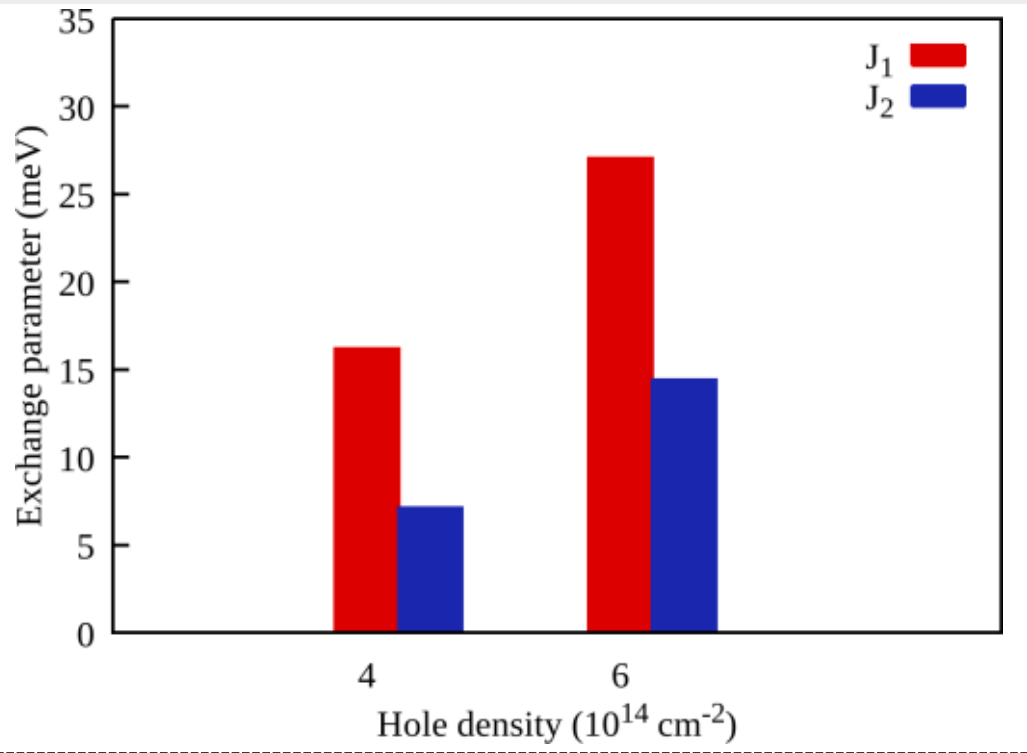

**Magnetic anisotropy energy (MAE, μeV) per magnetic atom**

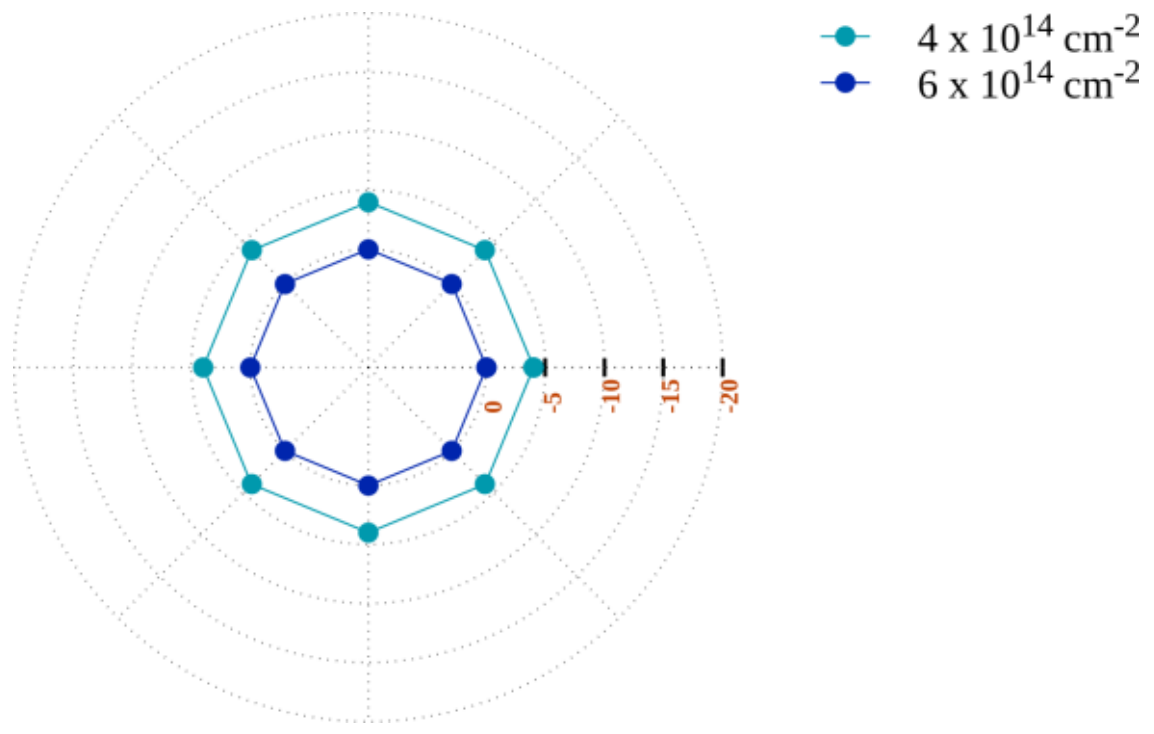

MAE = $E_{\parallel} - E_{\perp}$, a positive (negative) value of MAE indicates the off-plane (in-plane) easy axis.

**Monte Carlo simulations of the normalized magnetization of as a function of temperature**

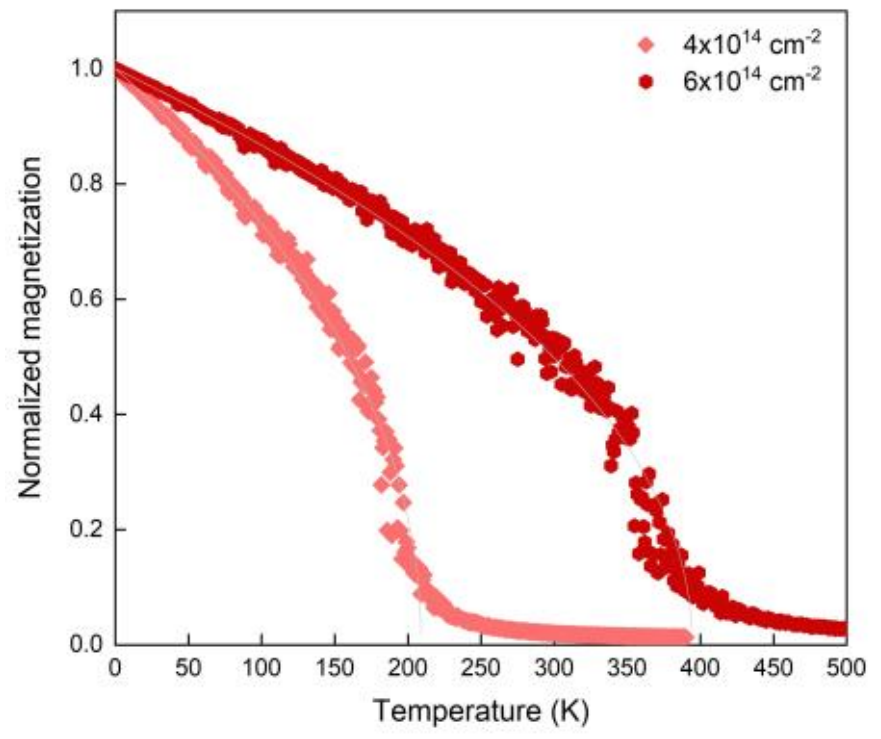

$T_c$: $1\times10^{14}$ cm$^{-2}$: - K  $2\times10^{14}$ cm$^{-2}$: - K

$4\times10^{14}$ cm$^{-2}$: 209 K  $6\times10^{14}$ cm$^{-2}$: 394 K

# 28. $PbBr_2$

| MC2D-ID | C2DB | 2dmat-ID | USPEX | Space group | Band gap (eV) |
|---|---|---|---|---|---|
| - | ✓ | - | ✓ | P3m1 | 2.75 |

**Convex hull**

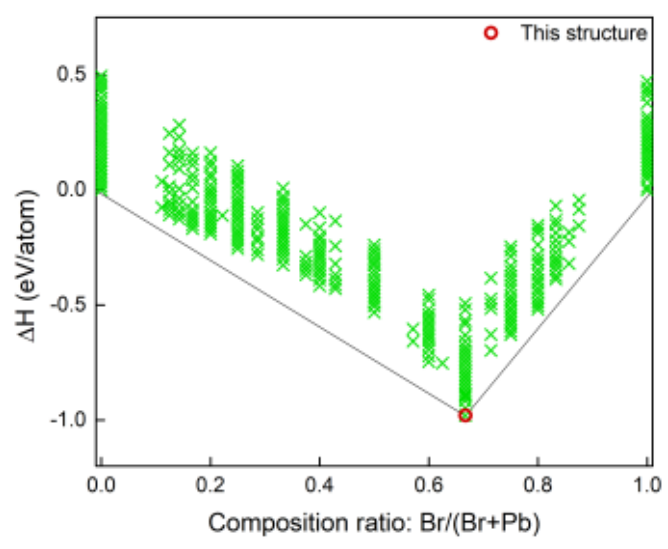

**Atomic structure**

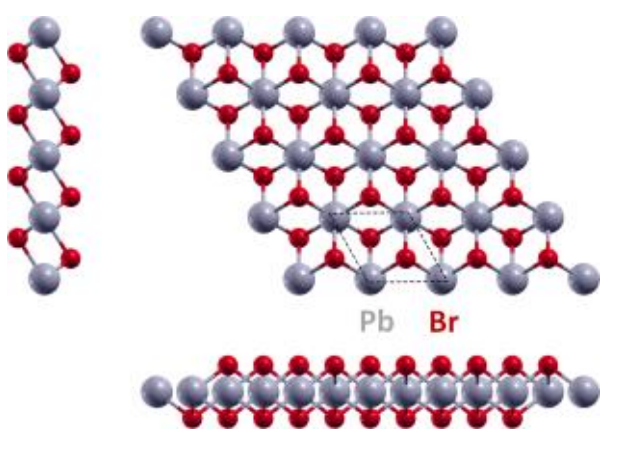

**Atomic coordinates**

```
Pb1Br2
   1.00000000000000
   4.4853000640999996    0.0000000000000000    0.0000000000000000
  -2.2426500319999998    3.8843837991000001    0.0000000000000000
   0.0000000000000000    0.0000000000000000   23.1145000457999998
   Pb    Br
   1     2
Direct
  0.0000000000000000  0.0000000000000000  0.5000000000000000
  0.3333333730000021  0.6666666870000029  0.4278699759999967
  0.6666666269999979  0.3333333129999971  0.5721300240000033
```

**Phonon dispersion curve**

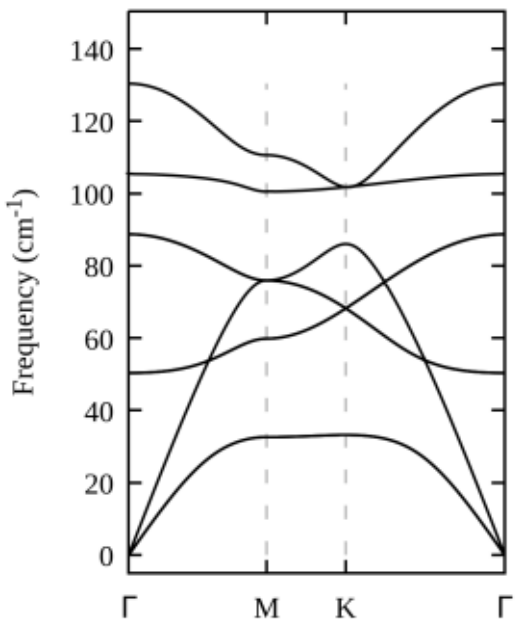

**Projected band structure and density of states**

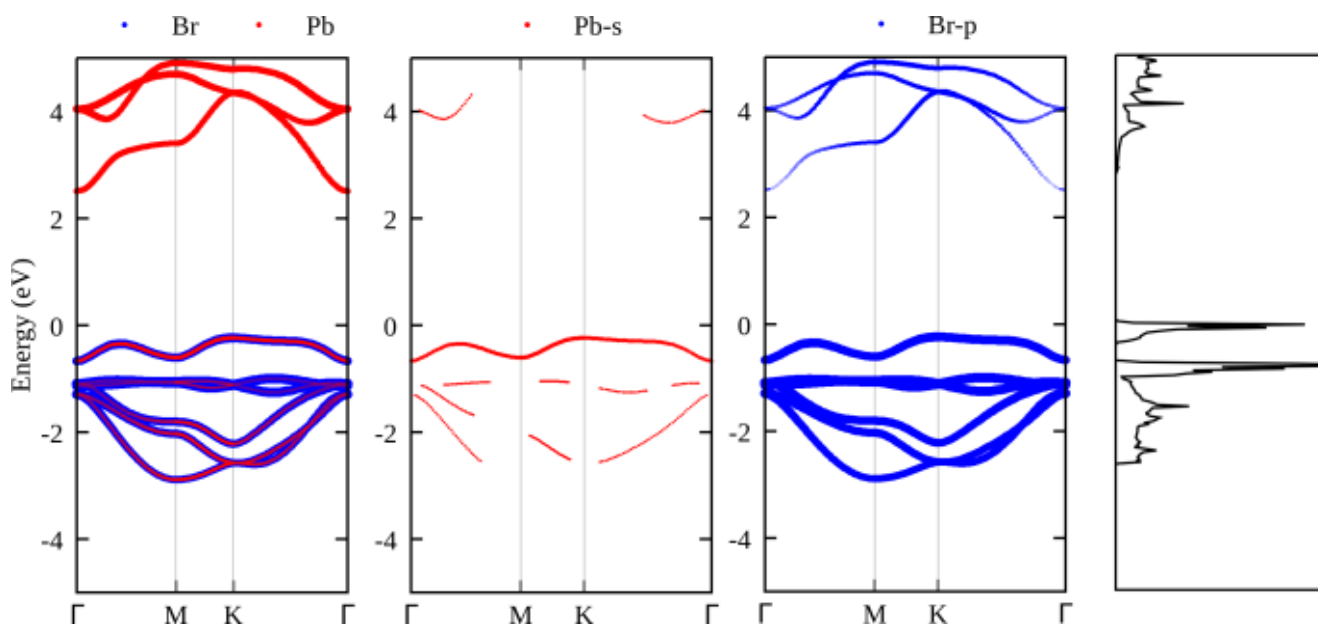

**Magnetic moment and spin polarization energy as a function of hole doping concentration**

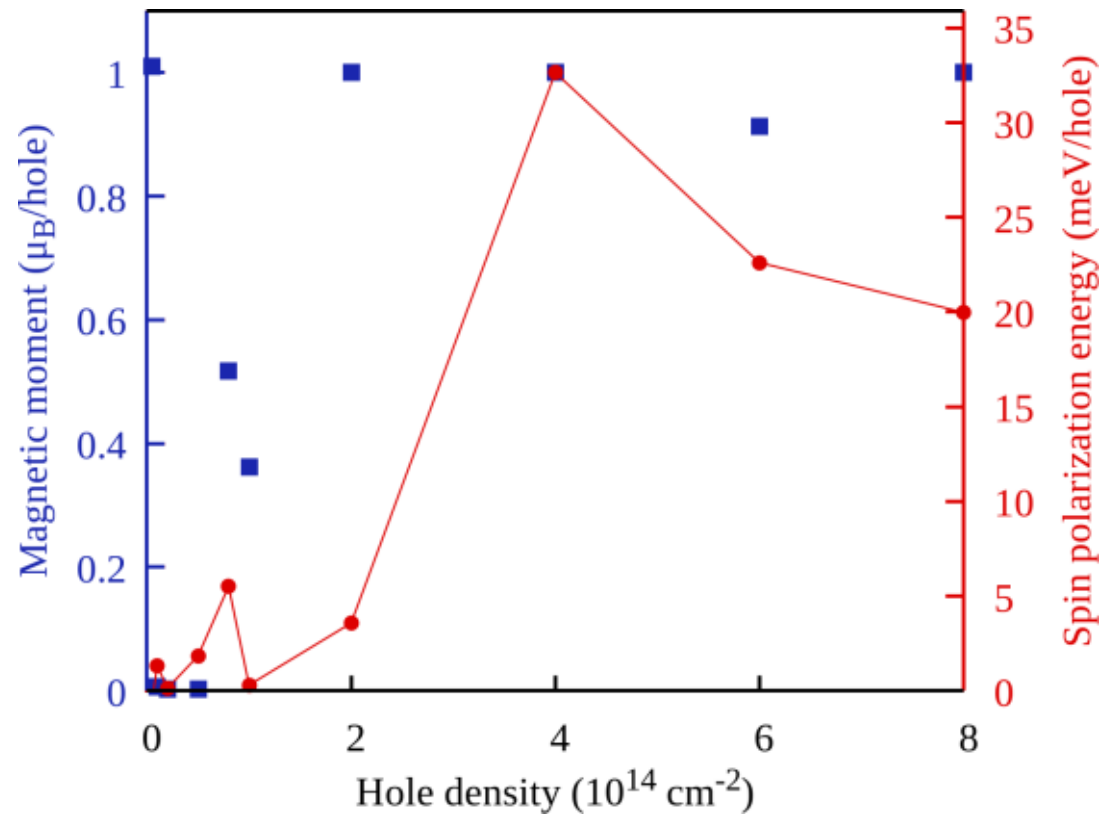

**Magnetic configurations and spin Hamiltonian**

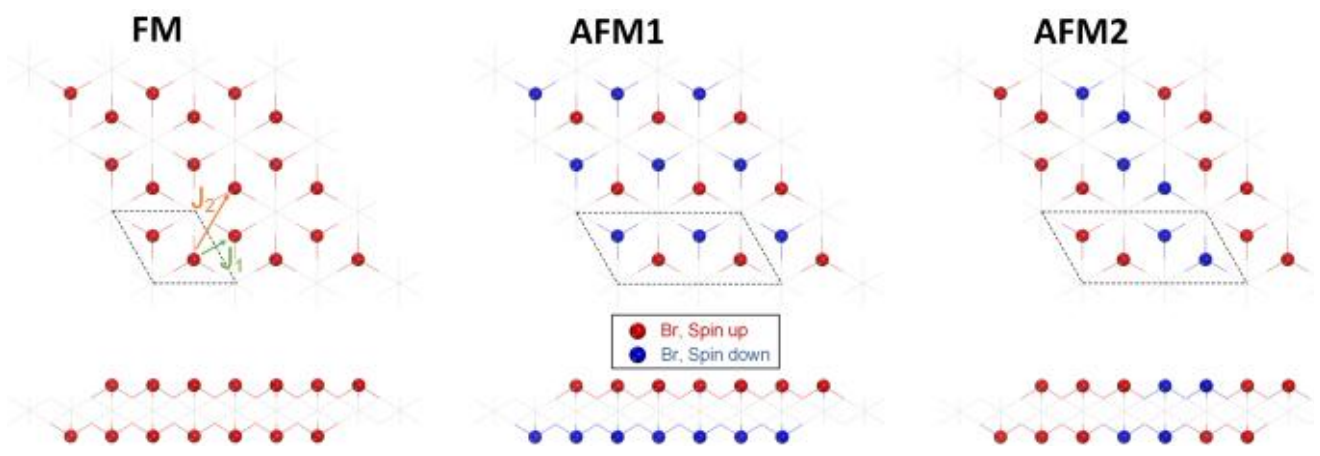

$$E_{FM} = E_0 - 3J_1S^2 - 6J_2S^2$$

$$E_{AFM1} = E_0 + 3J_1S^2 - 6J_2S^2$$

$$E_{AFM2} = E_0 - J_1S^2 + 2J_2S^2$$

**Magnetic exchange coupling parameters**

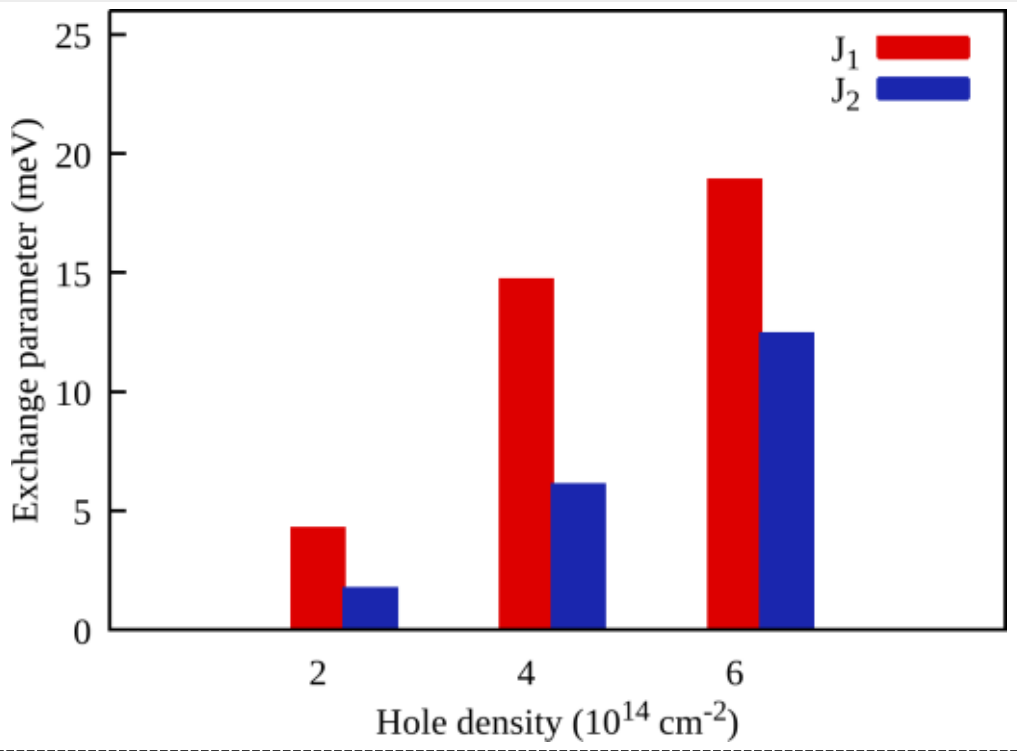

**Magnetic anisotropy energy (MAE, μeV) per magnetic atom**

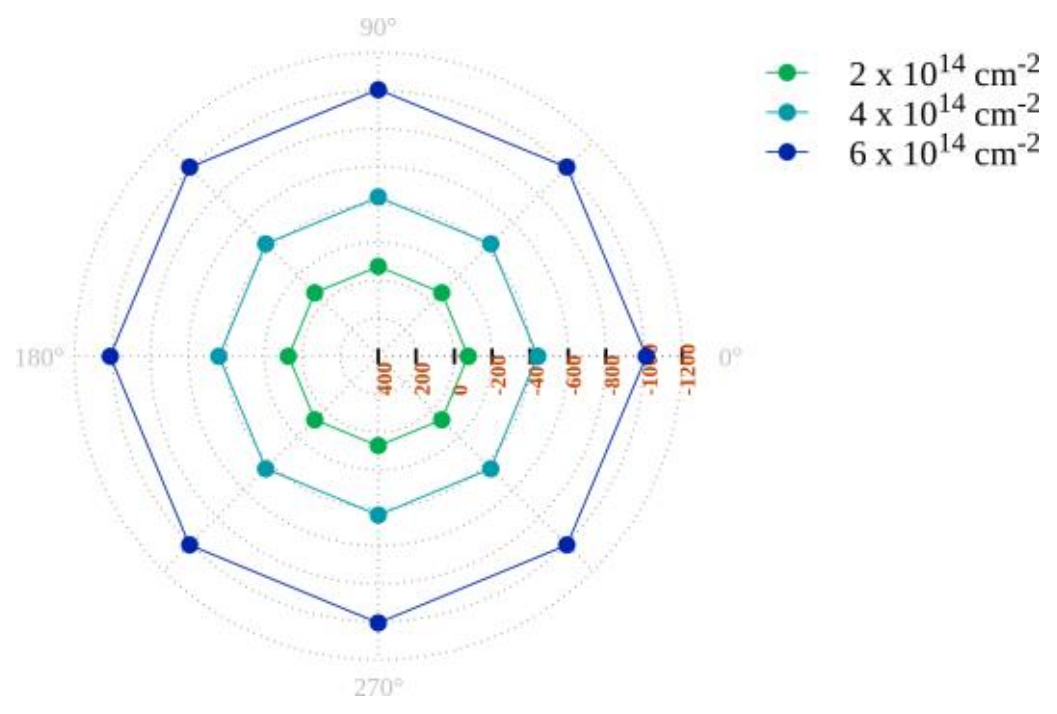

MAE = $E_{\parallel} - E_{\perp}$, a positive (negative) value of MAE indicates the off-plane (in-plane) easy axis.

**Monte Carlo simulations of the normalized magnetization of as a function of temperature**

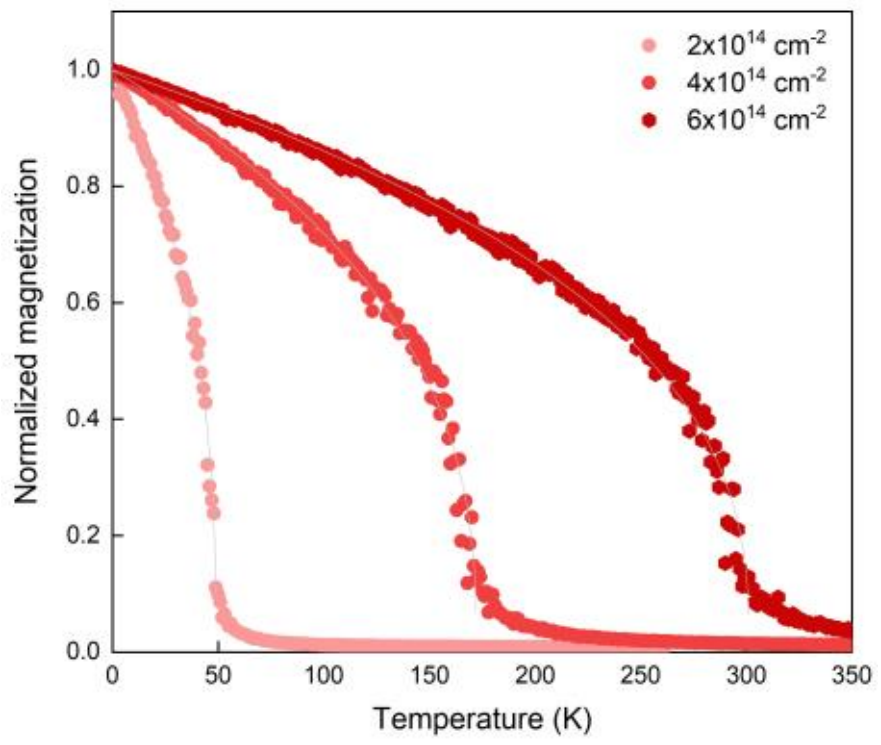

$T_c$: $1\times10^{14}$ cm$^{-2}$: - K  $2\times10^{14}$ cm$^{-2}$: 49 K

$4\times10^{14}$ cm$^{-2}$: 172 K  $6\times10^{14}$ cm$^{-2}$: 301 K

# 29. $Al_2S_2$

| MC2D-ID | C2DB | 2dmat-ID | USPEX | Space group | Band gap (eV) |
|---|---|---|---|---|---|
| - | ✓ | 2dm-1791 | - | P3m1 | 2.15 |

**Convex hull** | **Atomic structure** | **Atomic coordinates** | **Phonon dispersion curve**

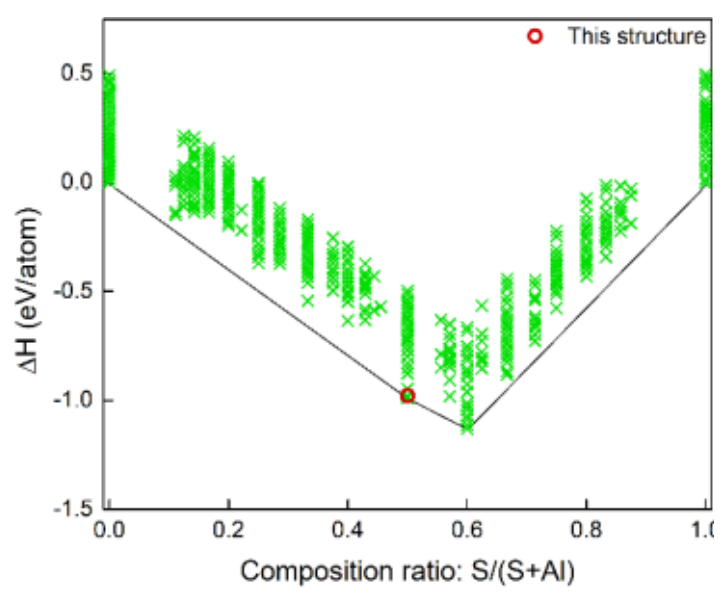

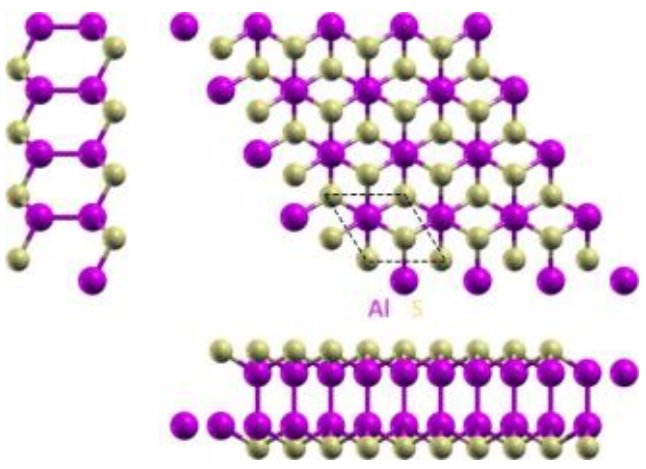

Al2S2
1.00000000000000
3.5921342265381568 0.0000000000000000 0.0000000000000000
-1.7960664300799600 3.1108798811072038 0.0000000000000000
0.0000000000000000 0.0000000000000000 24.6231550000000006
Al S
2 2
Direct
0.0000001739833166 0.9999997859377459 0.0073385258748786
0.9999998212522385 0.0000002233968743 0.1125265962129731
0.6666667216689461 0.3333331687641419 0.9642231695601424
0.3333332830954916 0.6666668219012379 0.1556407083519957

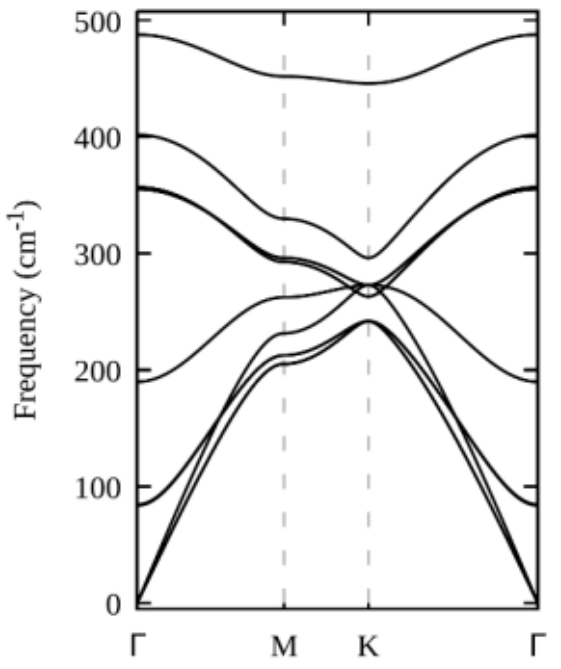

**Projected band structure and density of states** | **Magnetic moment and spin polarization energy as a function of hole doping concentration**

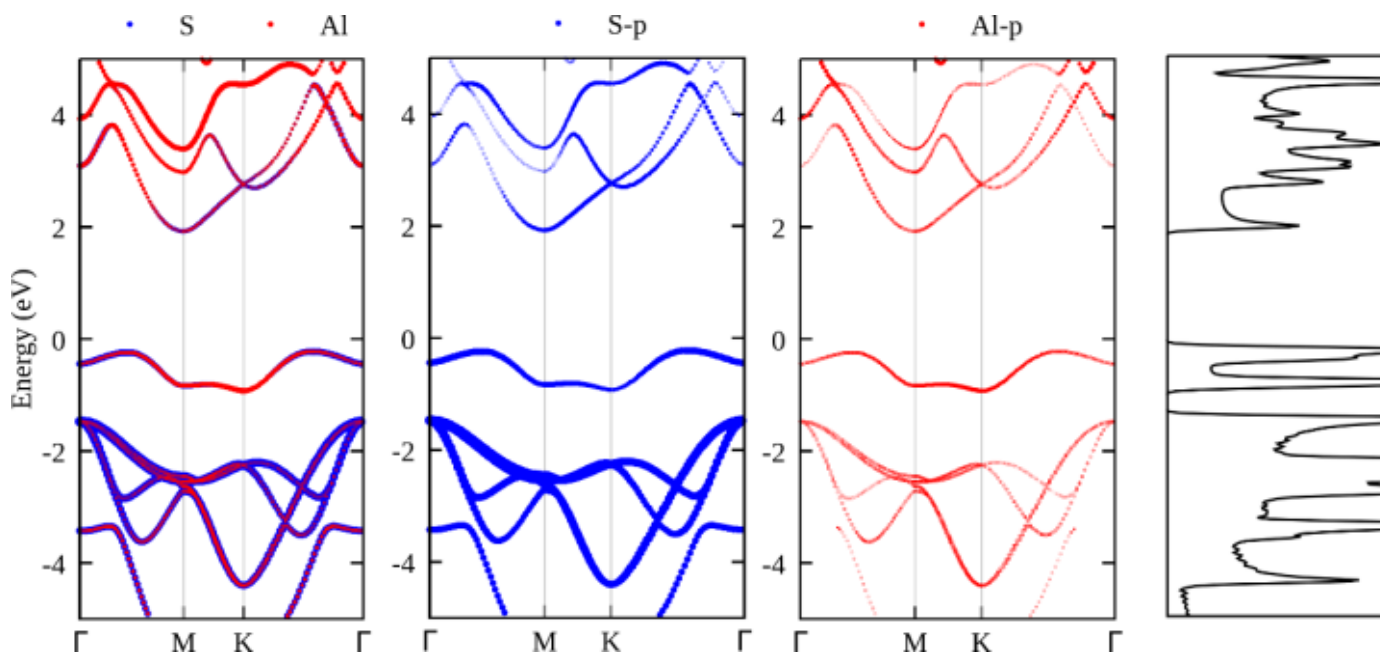

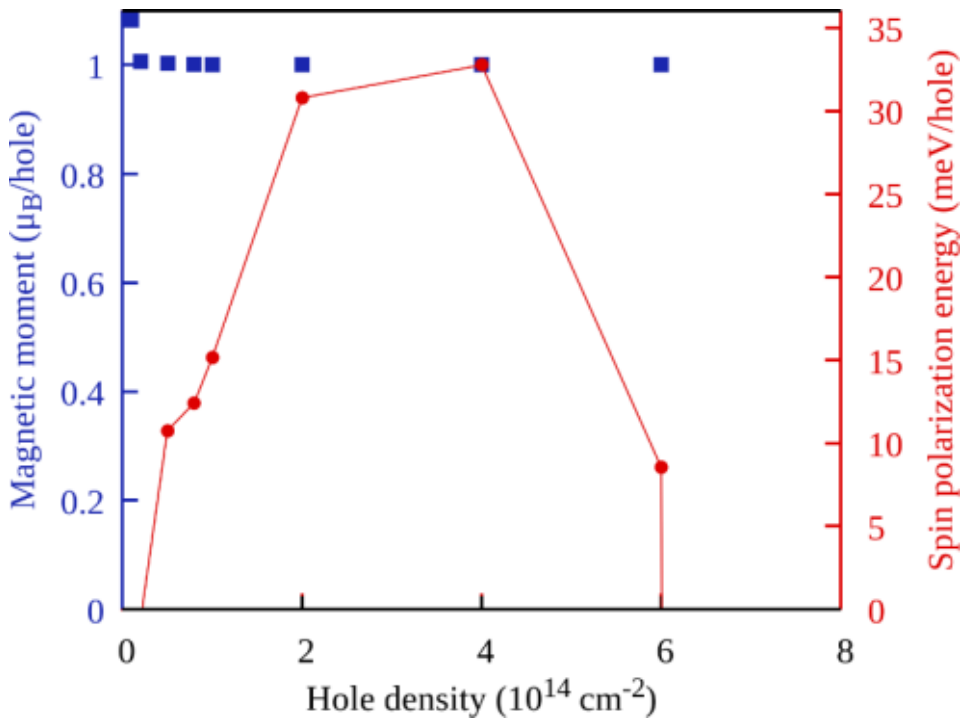

**Magnetic configurations and spin Hamiltonian** | **Magnetic exchange coupling parameters**

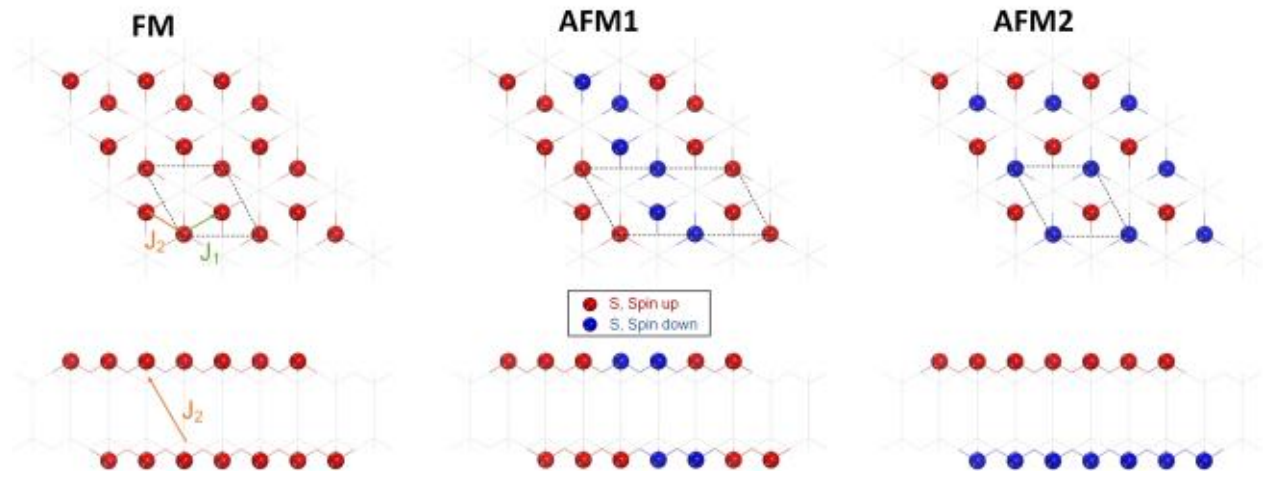

$$E_{FM} = E_0 - 6J_1S^2 - J_2S^2$$
$$E_{AFM1} = E_0 + 2J_1S^2 - J_2S^2$$
$$E_{AFM2} = E_0 - 6J_1S^2 + 3J_2S^2$$

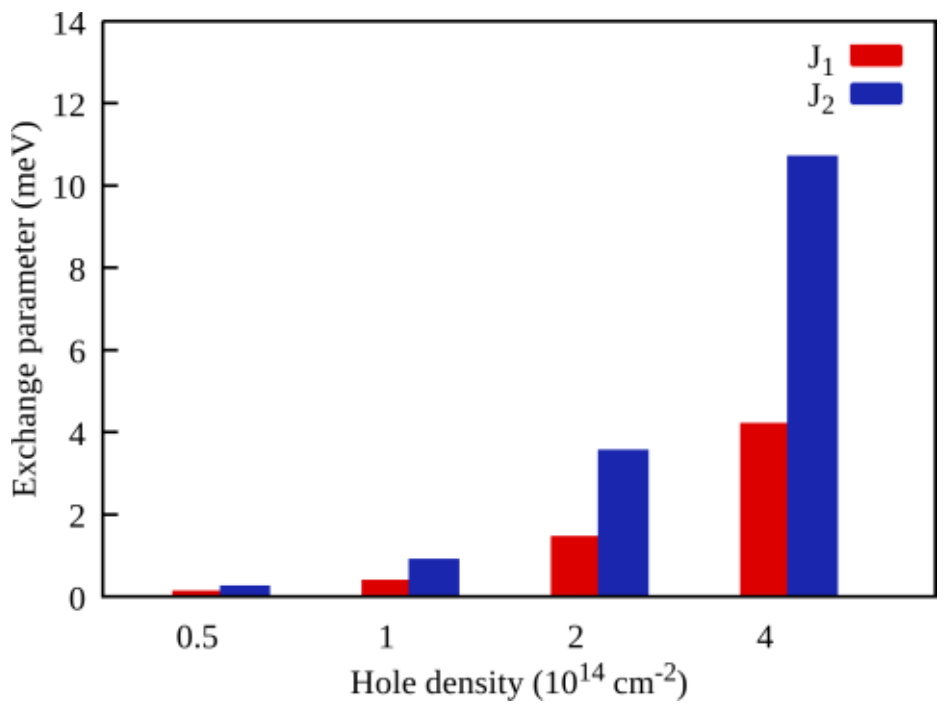

**Magnetic anisotropy energy (MAE, μeV) per magnetic atom** | **Monte Carlo simulations of the normalized magnetization of as a function of temperature**

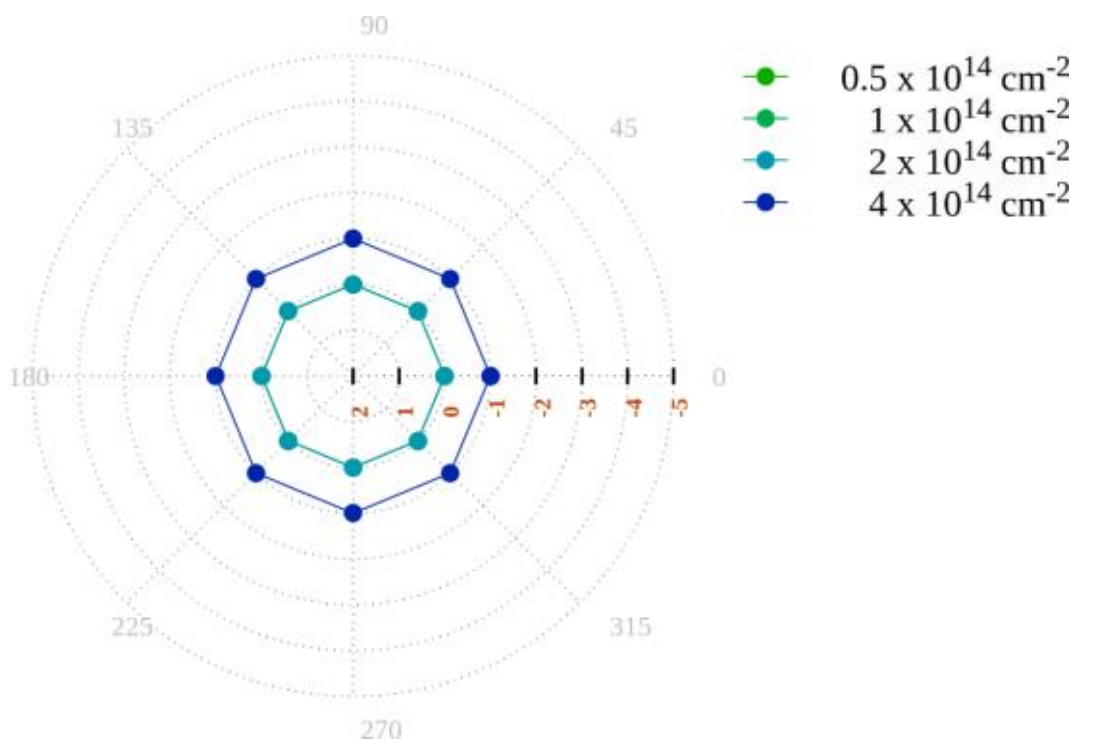

MAE = $E_{\parallel} - E_{\perp}$, a positive (negative) value of MAE indicates the off-plane (in-plane) easy axis.

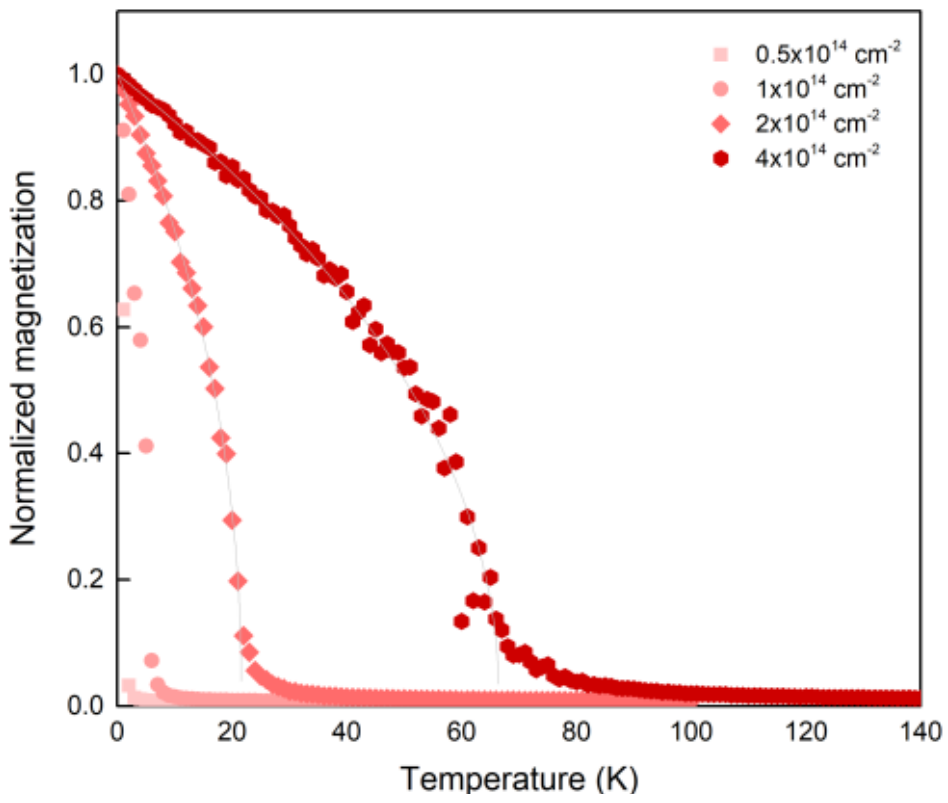

$T_c$: $0.5\times10^{14}$ cm$^{-2}$: 2 K  $1\times10^{14}$ cm$^{-2}$: 6 K  $2\times10^{14}$ cm$^{-2}$: 21 K  $4\times10^{14}$ cm$^{-2}$: 66 K

# 30. $Al_2Se_2$

| MC2D-ID | C2DB | 2dmat-ID | USPEX | Space group | Band gap (eV) |
|---|---|---|---|---|---|
| - | ✓ | 2dm-835 | - | P3m1 | 2.14 |

**Convex hull** | **Atomic structure** | **Atomic coordinates** | **Phonon dispersion curve**

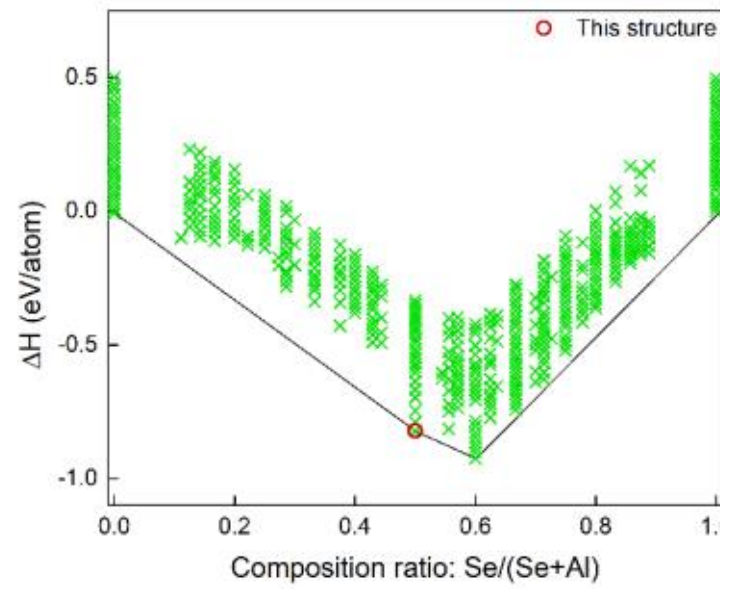

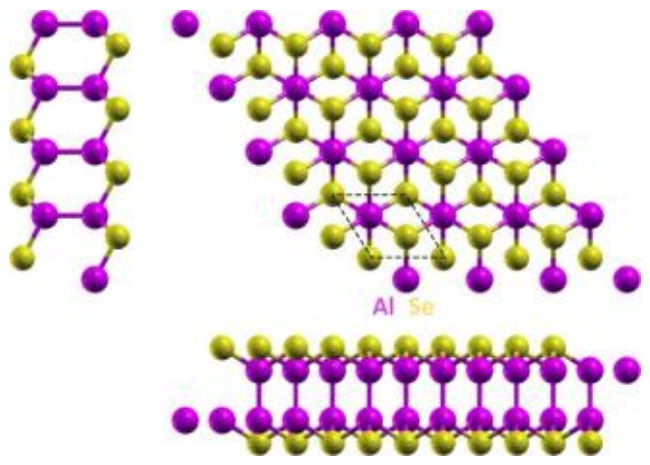

```
Al2Se2
   1.00000000000000
   3.7890287407636372   0.0000000000000000   0.0000000000000000
  -1.8945134127802470   3.2813958126517888   0.0000000000000000
   0.0000000000000000   0.0000000000000000  24.6231550000000006
   Al   Se
   2    2
Direct
  0.0000001475127291   0.9999998252788131   0.0076291995589983
  0.9999998536617909   0.0000001758314951   0.1122351700220108
  0.6666668171954555   0.3333331551883560   0.9605922183525735
  0.3333331816300174   0.6666668437013357   0.1592724120664144
```

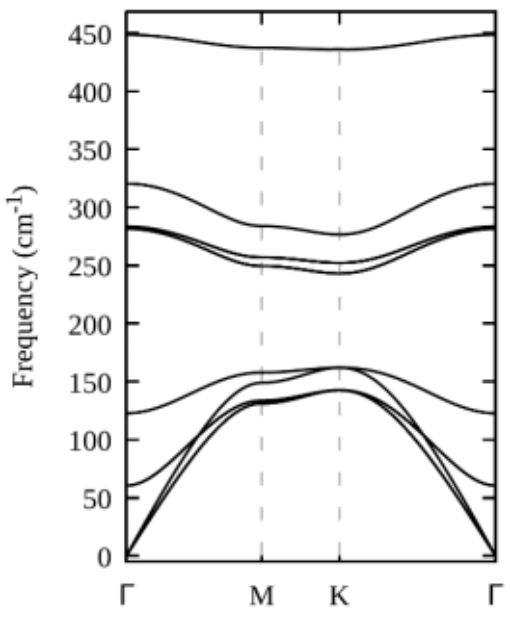

**Projected band structure and density of states**

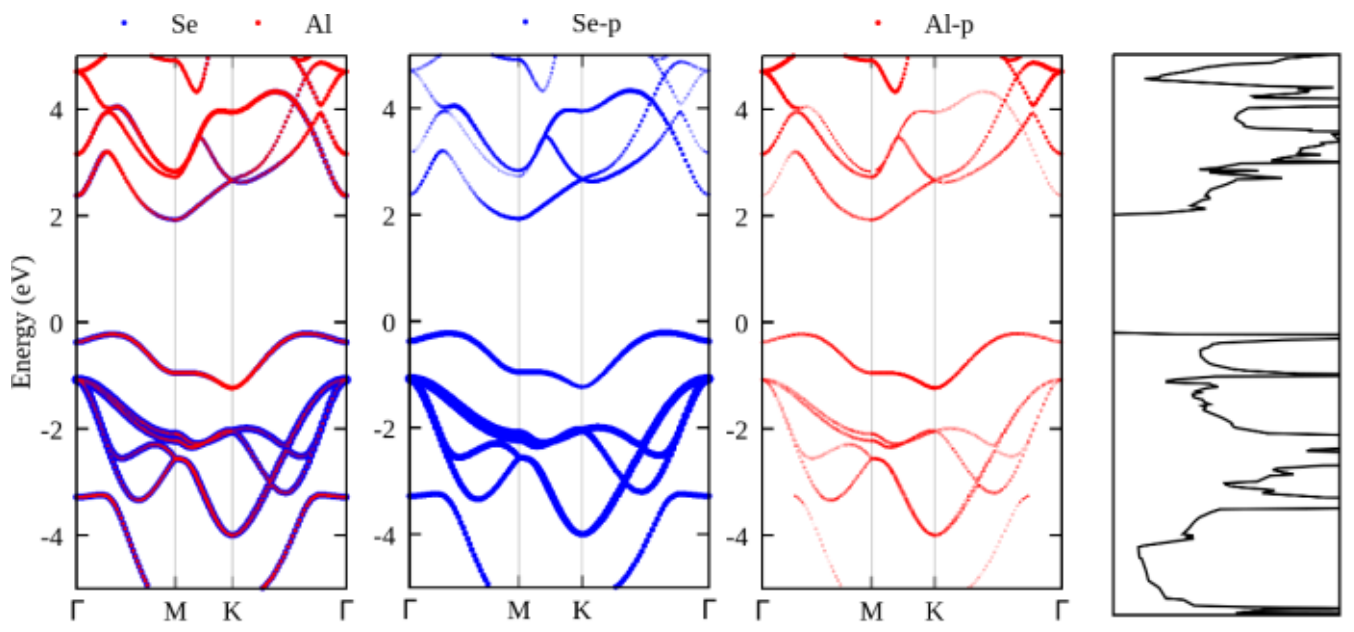

**Magnetic moment and spin polarization energy as a function of hole doping concentration**

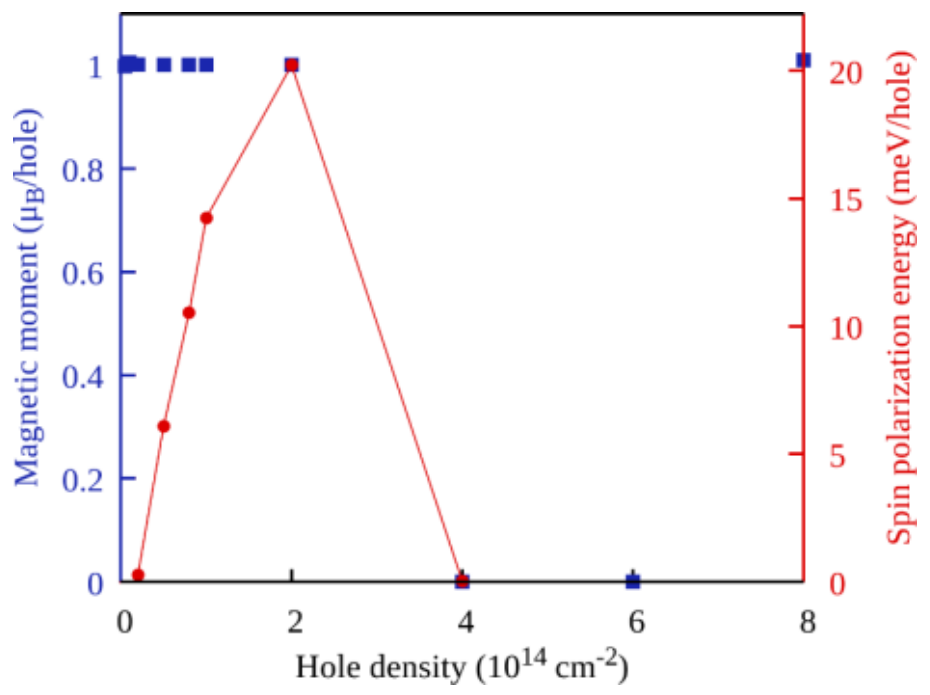

**Magnetic configurations and spin Hamiltonian**

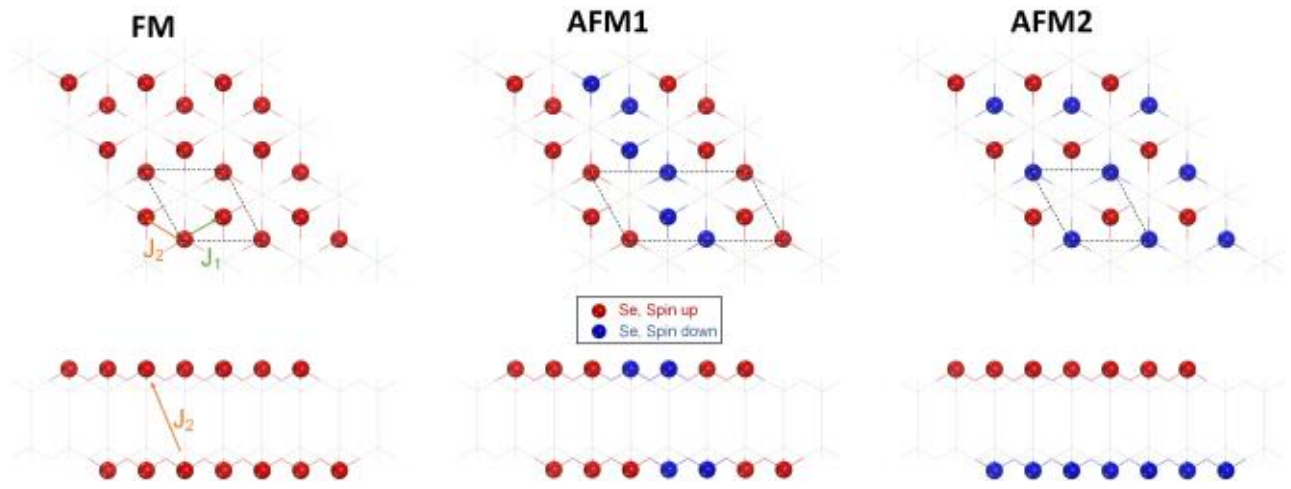

$$E_{FM} = E_0 - 6J_1S^2 - 3J_2S^2$$
$$E_{AFM1} = E_0 + 2J_1S^2 - J_2S^2$$
$$E_{AFM2} = E_0 - 6J_1S^2 + 3J_2S^2$$

**Magnetic exchange coupling parameters**

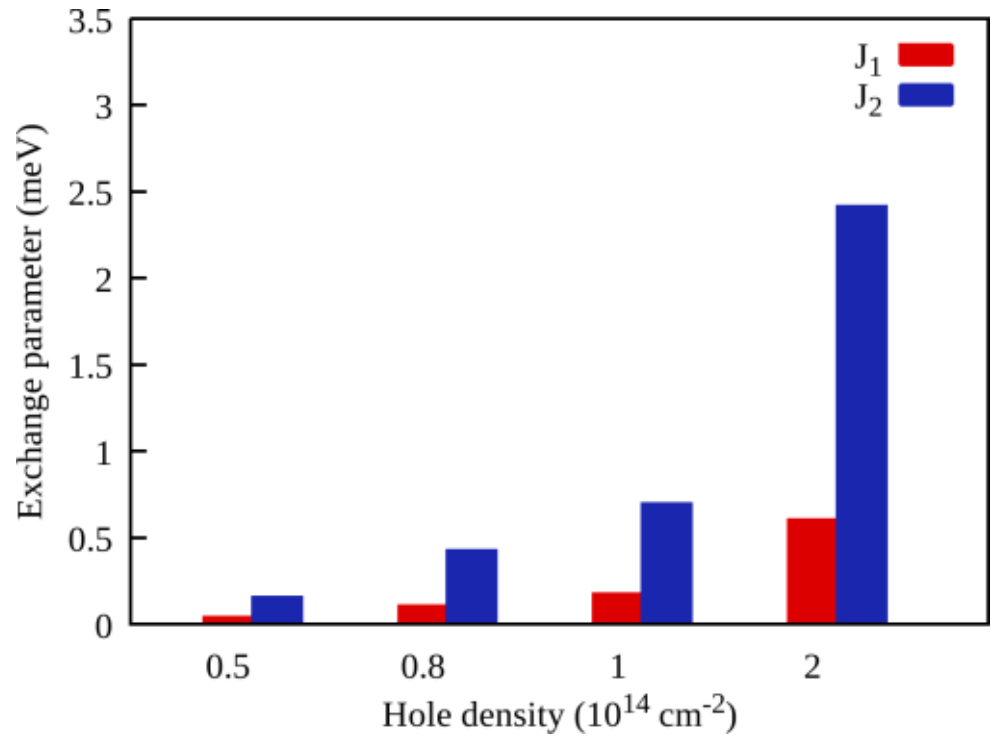

**Magnetic anisotropy energy (MAE, μeV) per magnetic atom**

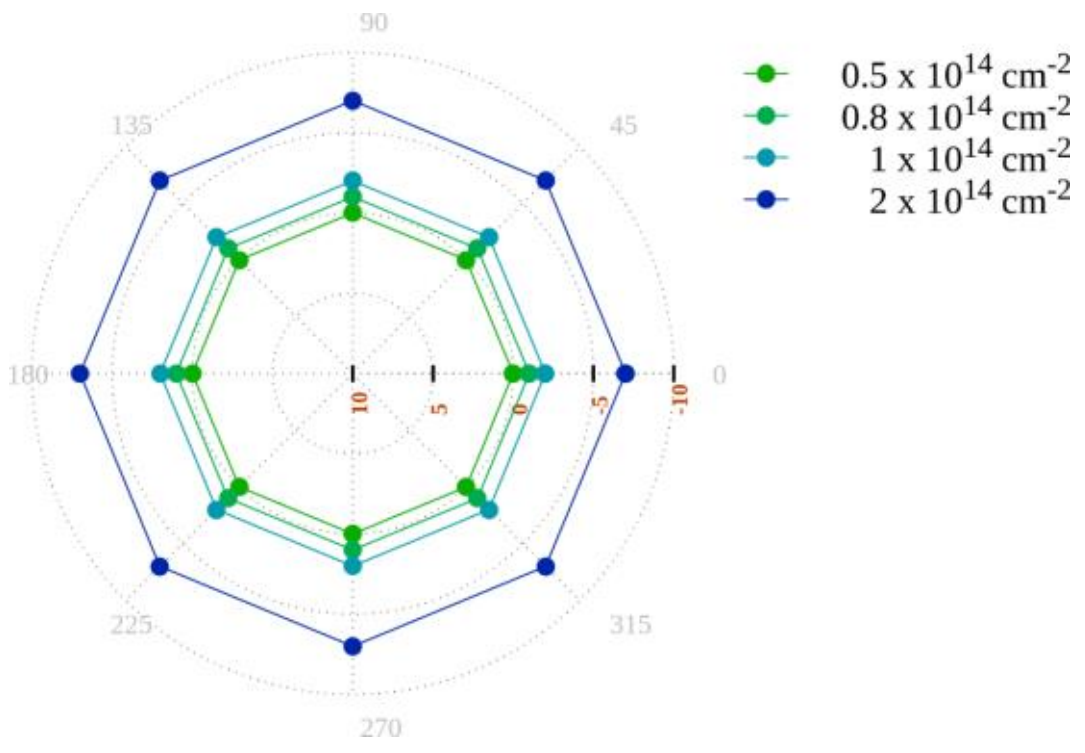

MAE = $E_{\parallel} - E_{\perp}$, a positive (negative) value of MAE indicates the off-plane (in-plane) easy axis.

**Monte Carlo simulations of the normalized magnetization of as a function of temperature**

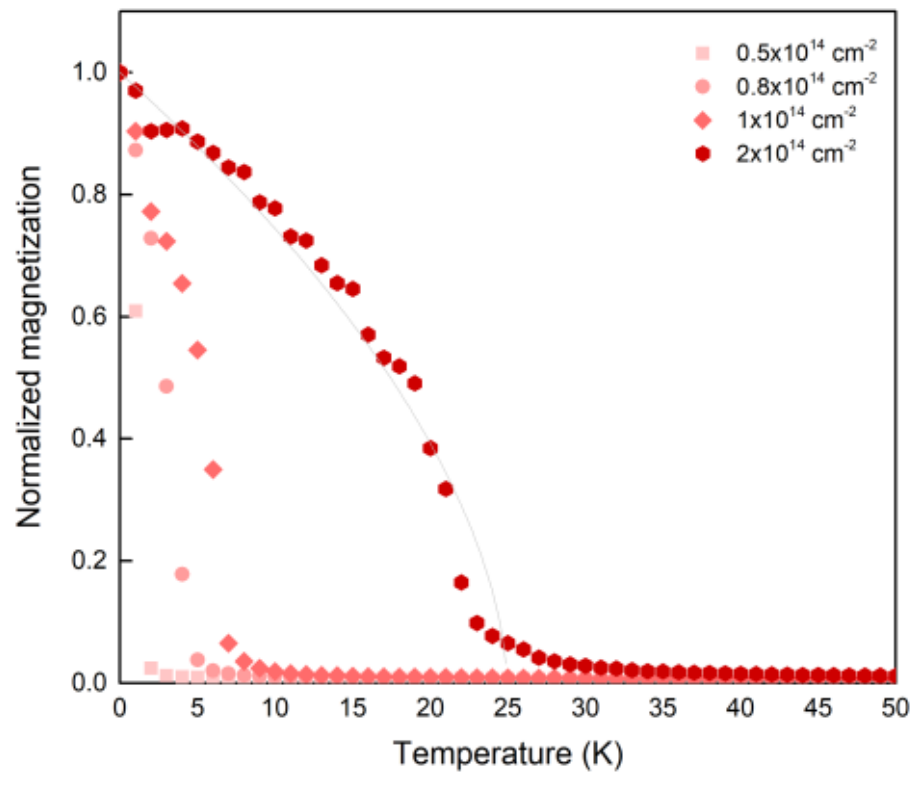

$T_c$: $0.5\times10^{14}$ cm$^{-2}$: 2 K  $0.8\times10^{14}$ cm$^{-2}$: 5 K  $1\times10^{14}$ cm$^{-2}$: 7 K  $2\times10^{14}$ cm$^{-2}$: 25 K

# 31. $In_2S_2$

| MC2D-ID | C2DB | 2dmat-ID | USPEX | Space group | Band gap (eV) |
|---|---|---|---|---|---|
| - | ✓ | 2dm-1655 | - | P3m1 | 1.59 |

**Convex hull** | **Atomic structure** | **Atomic coordinates** | **Phonon dispersion curve**

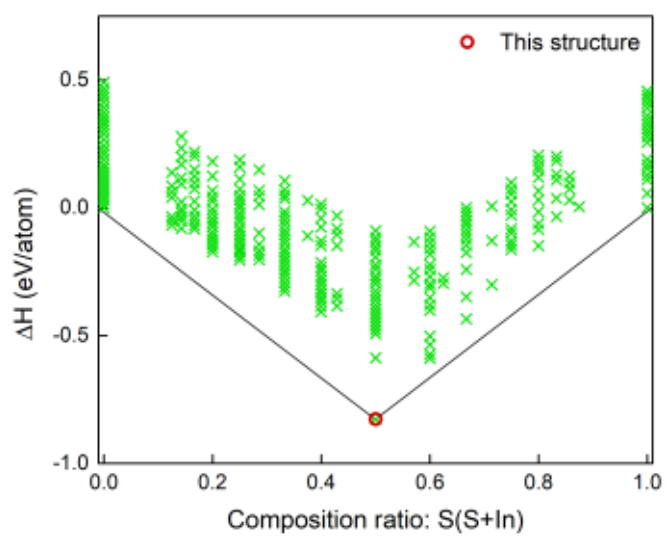

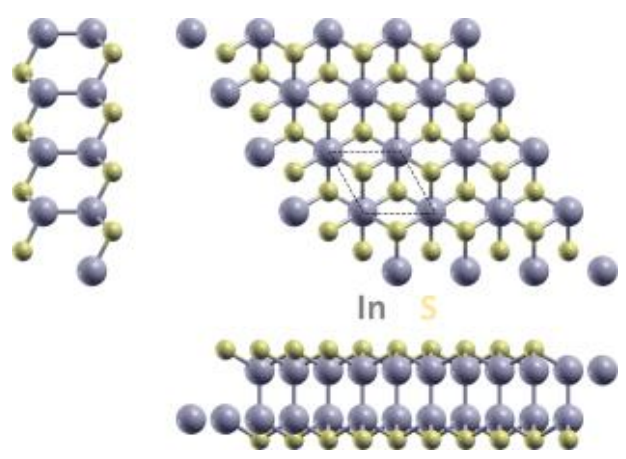

```
In2S2
  1.00000000000000
  3.9255056553423078   0.0000000000000000   0.0000000000000000
 -1.9627590902974188   3.3995840234451506   0.0000000000000000
  0.0000000000000000   0.0000000000000000  24.6231550000000006
  In   S
  2    2
Direct
 0.0000002887760360  0.9999998500440341  0.0030620583553400
 0.9999997100388427  0.0000001837059500  0.1167939975452299
 0.6666671347126112  0.3333330130271150  0.9556137015571053
 0.3333328664725101  0.6666669532229008  0.1642592425423217
```

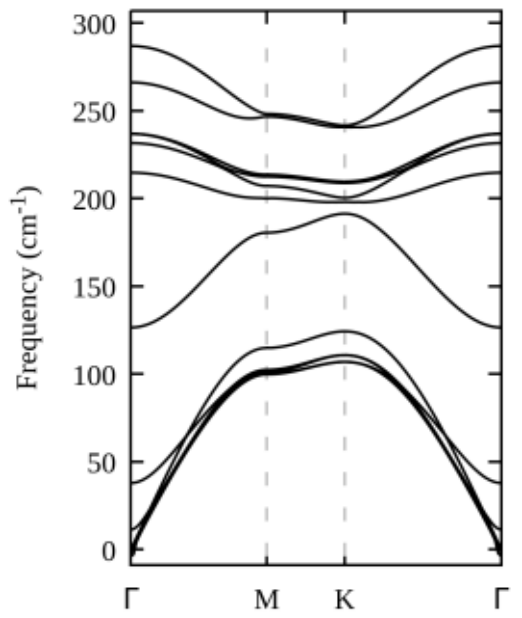

## Projected band structure and density of states

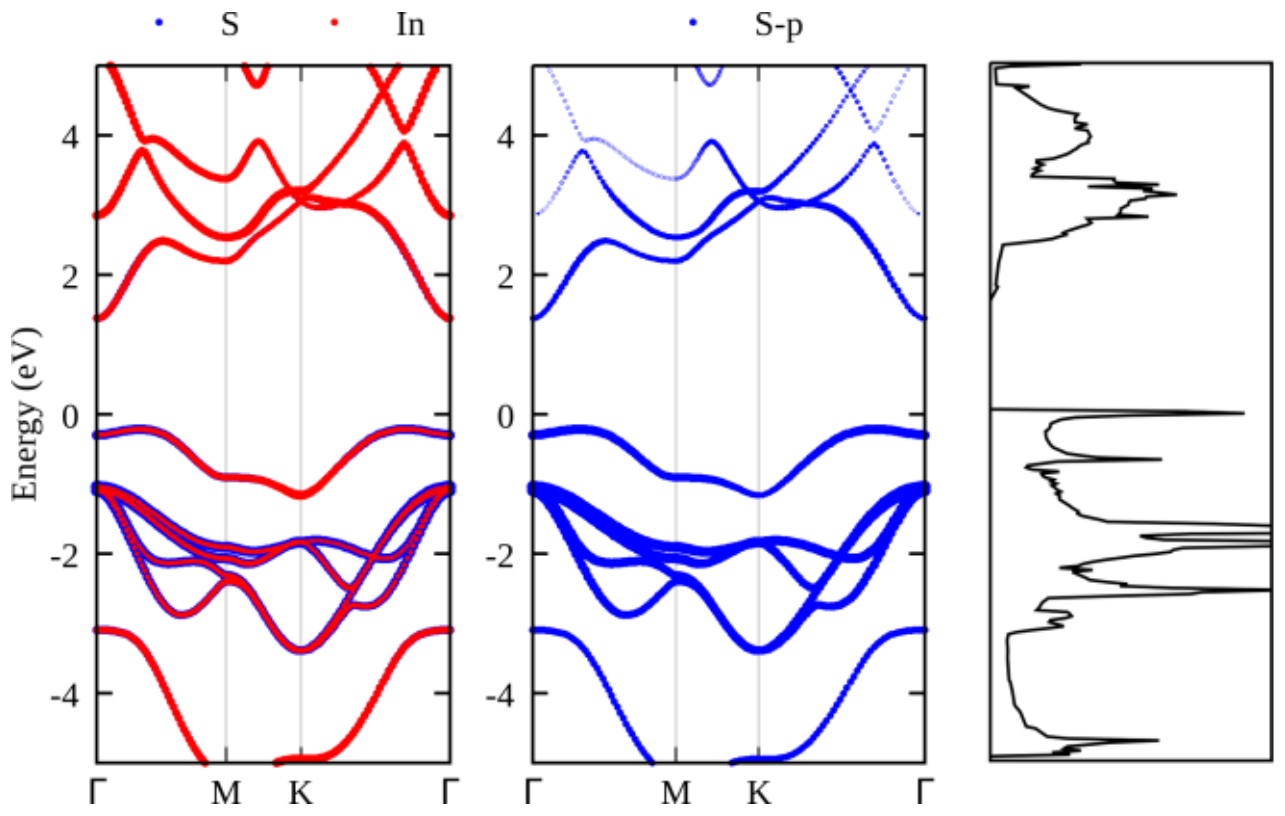

## Magnetic moment and spin polarization energy as a function of hole doping concentration

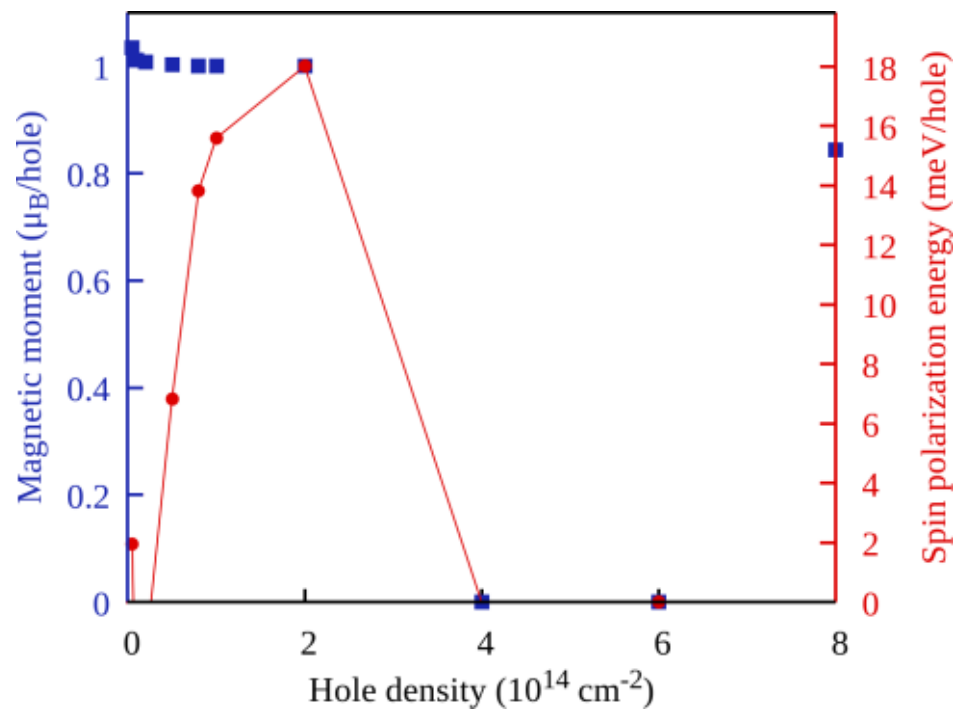

## Magnetic configurations and spin Hamiltonian

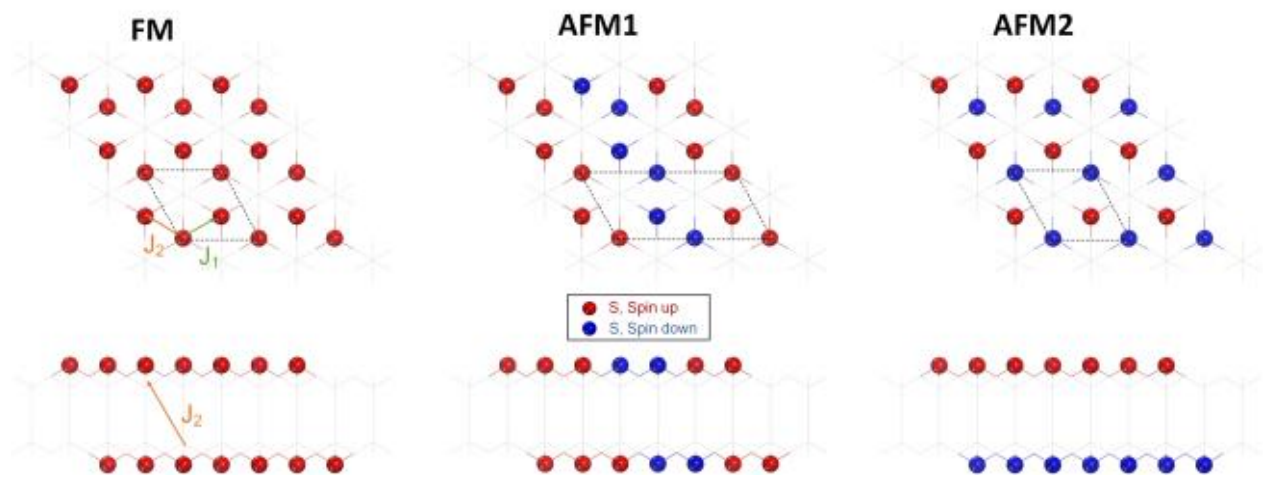

$$E_{FM} = E_0 - 6J_1S^2 - 3J_2S^2$$

$$E_{AFM1} = E_0 + 2J_1S^2 - J_2S^2$$

$$E_{AFM2} = E_0 - 6J_1S^2 + 3J_2S^2$$

## Magnetic exchange coupling parameters

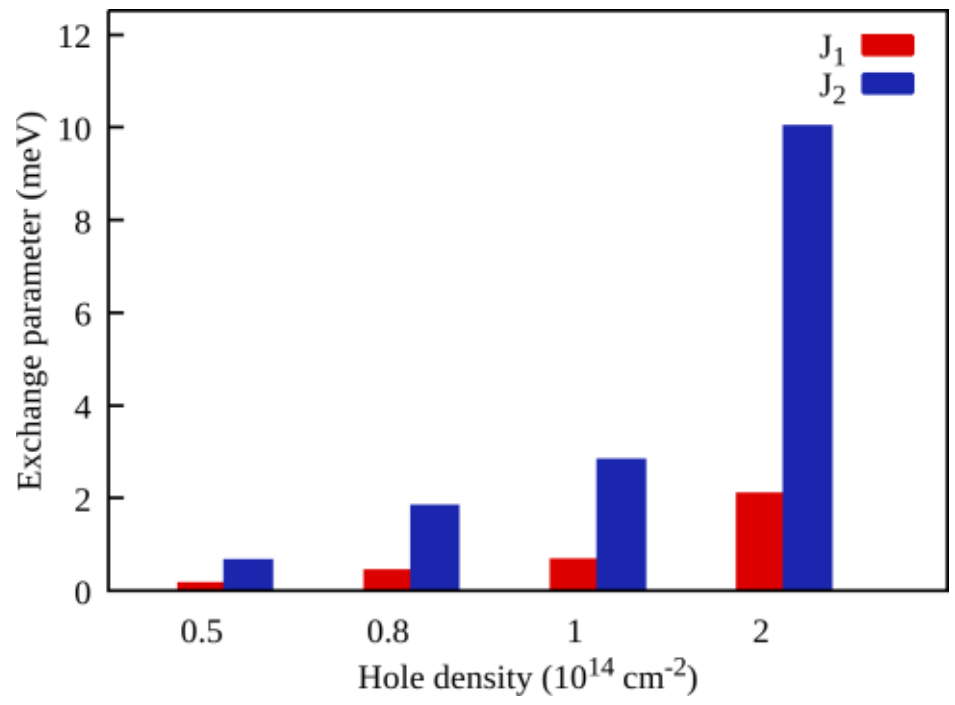

## Magnetic anisotropy energy (MAE, μeV) per magnetic atom

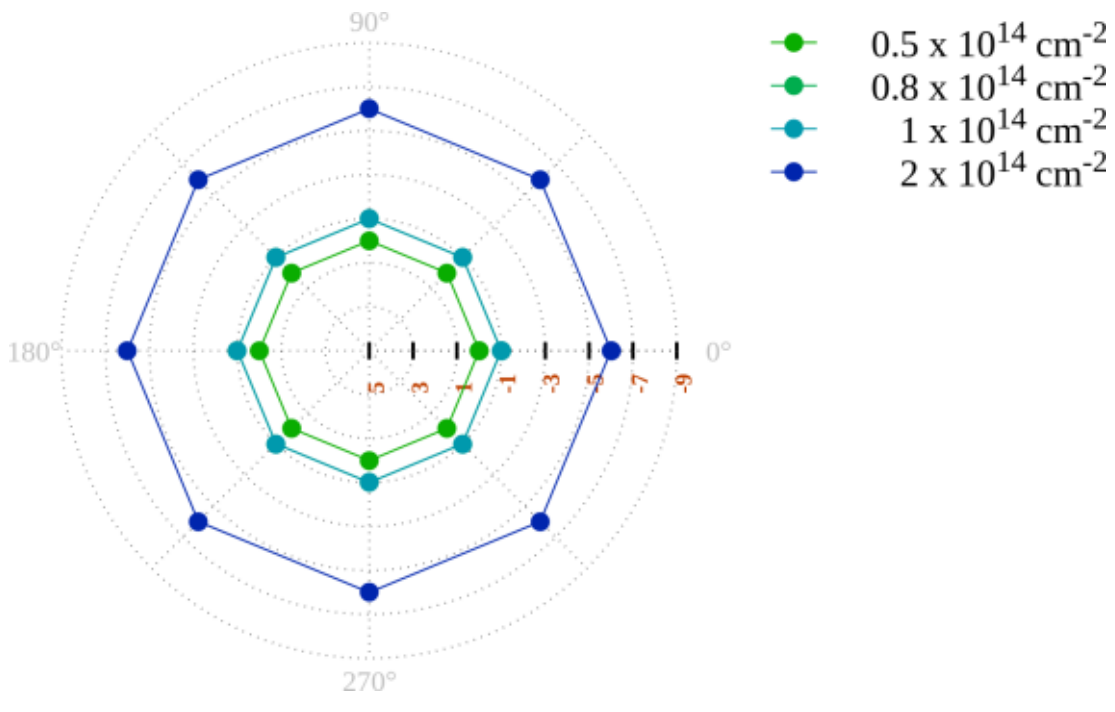

MAE = $E_{\parallel} - E_{\perp}$, a positive (negative) value of MAE indicates the off-plane (in-plane) easy axis.

## Monte Carlo simulations of the normalized magnetization of as a function of temperature

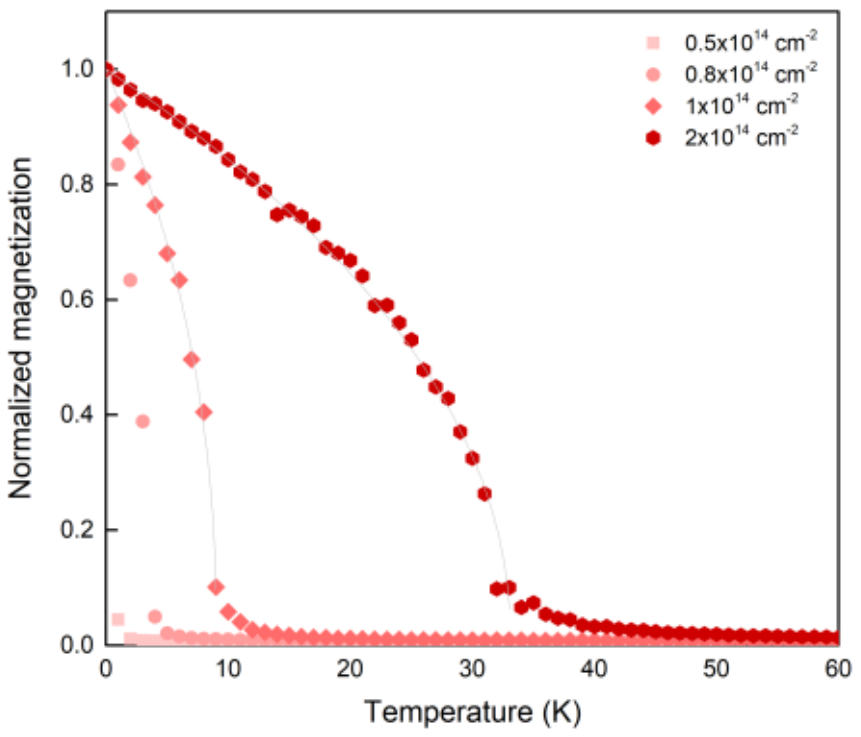

$T_c$: $0.5\times10^{14}$ cm$^{-2}$: 2 K　$0.8\times10^{14}$ cm$^{-2}$: 4 K

$1\times10^{14}$ cm$^{-2}$: 9 K　$2\times10^{14}$ cm$^{-2}$: 30 K

# 32. SiS

| MC2D-ID | C2DB | 2dmat-ID | USPEX | Space group | Band gap (eV) |
|---|---|---|---|---|---|
| - | - | 2dm-776 | - | P3m1 | 2.20 |

**Convex hull** | **Atomic structure** | **Atomic coordinates** | **Phonon dispersion curve**

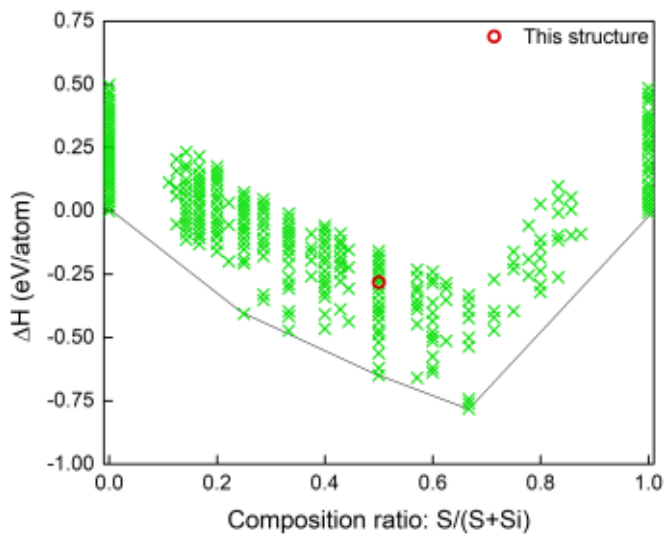

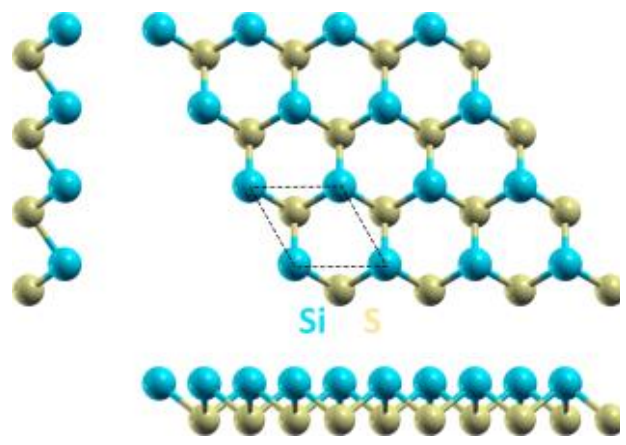

Si1S1
1.00000000000000
3.3033376893731692 0.0000000000000000 0.0000000000000000
-1.6516683655338755 2.8607747078457786 0.0000000000000000
0.0000000000000000 0.0000000000000000 21.4786740000000016
Si S
1 1
Direct
0.3333332618974936 0.6666667344429342 0.1629379298674891
0.6666667381025064 0.3333332655570658 0.1012060701325126

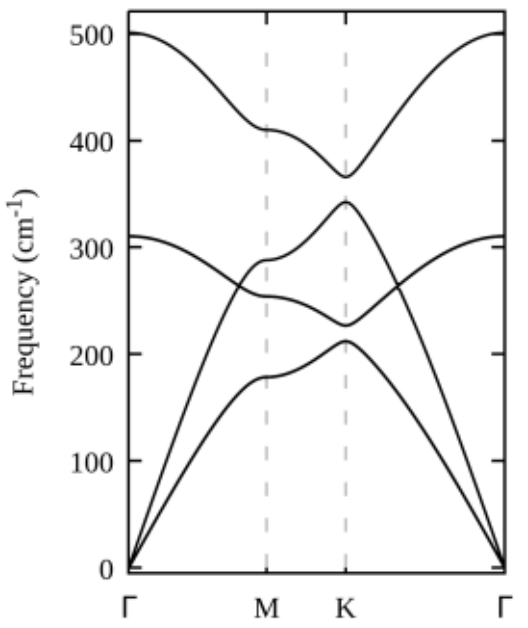

**Projected band structure and density of states** | **Magnetic moment and spin polarization energy as a function of hole doping concentration**

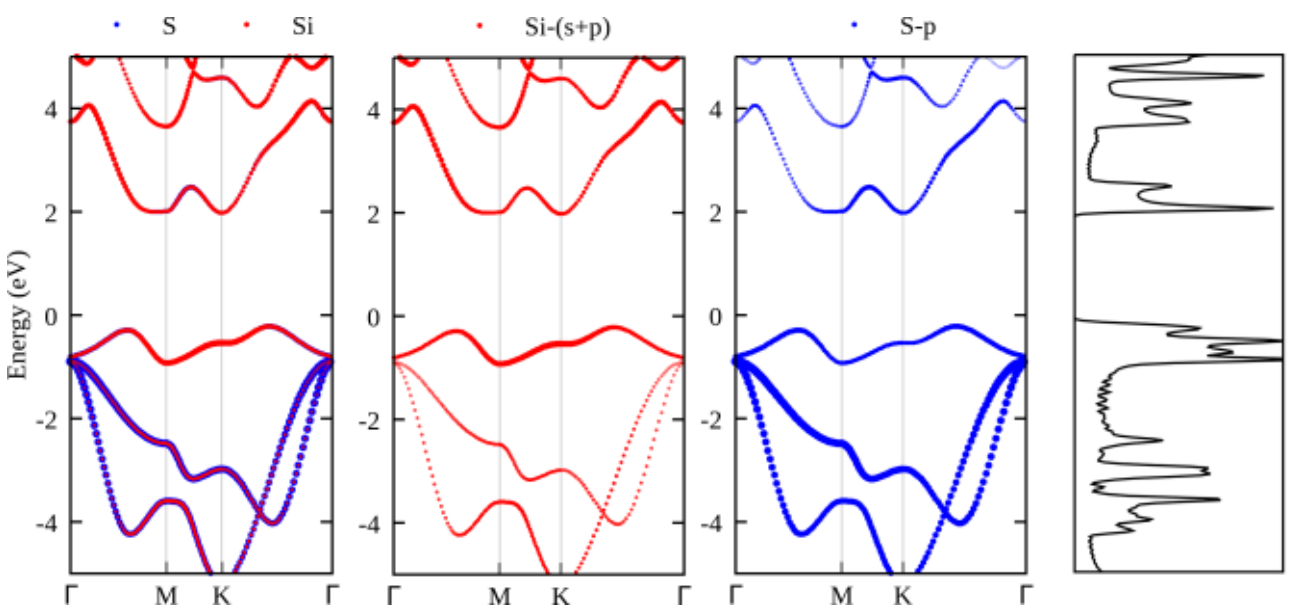

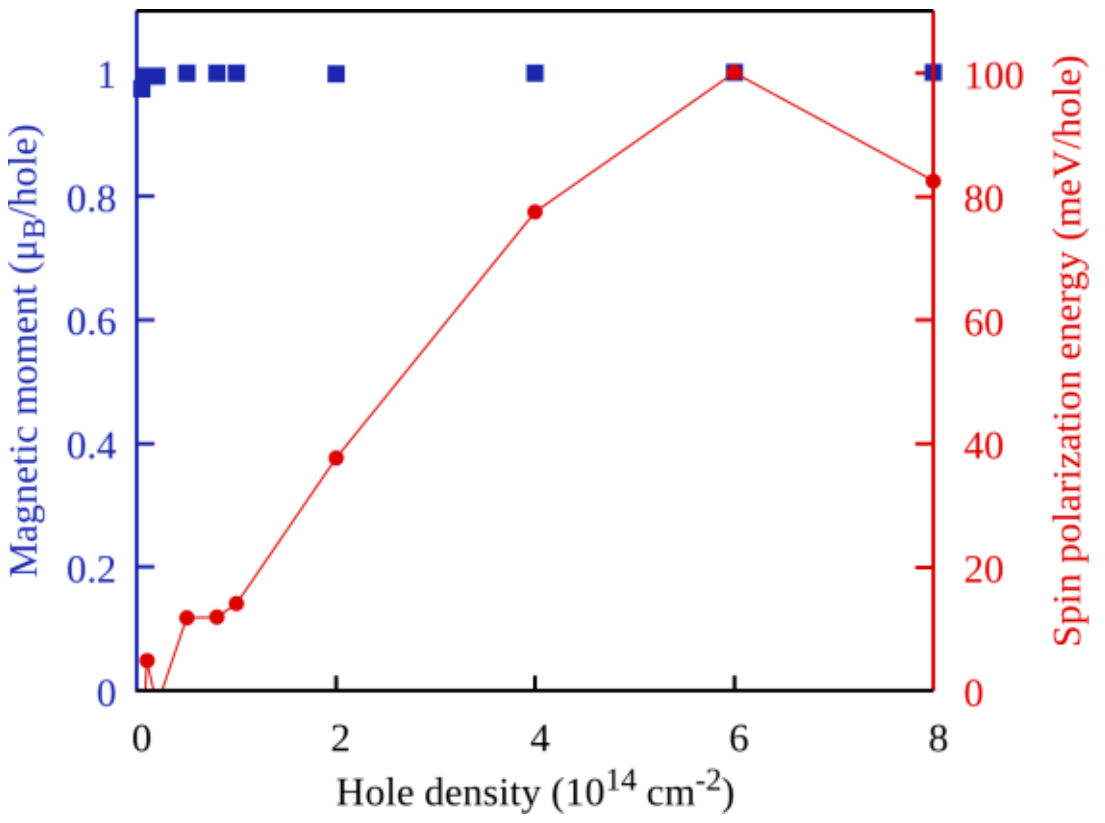

**Magnetic configurations and spin Hamiltonian** | **Magnetic exchange coupling parameters**

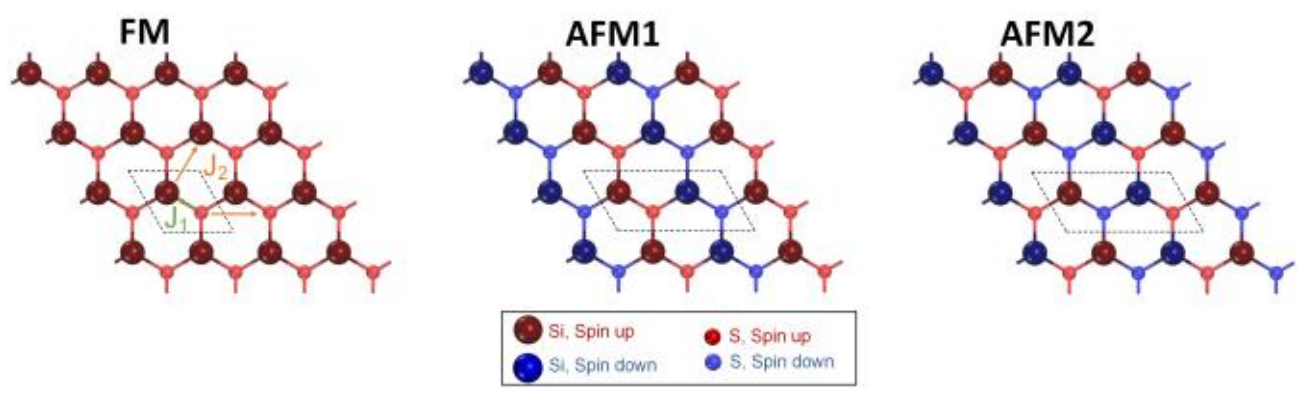

$$E_{FM} = E_0 - 3J_1S^2 - 6J_2S^2$$
$$E_{AFM1} = E_0 - J_1S^2 + 2J_2S^2$$
$$E_{AFM2} = E_0 + J_1S^2 + 2J_2S^2$$

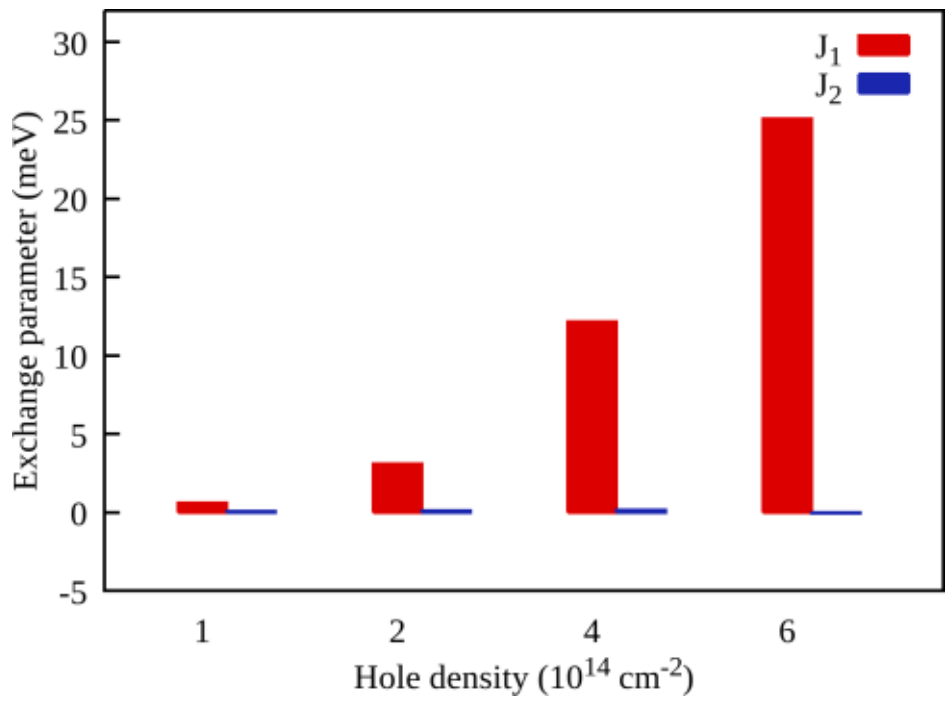

**Magnetic anisotropy energy (MAE, μeV) per magnetic atom** | **Monte Carlo simulations of the normalized magnetization of as a function of temperature**

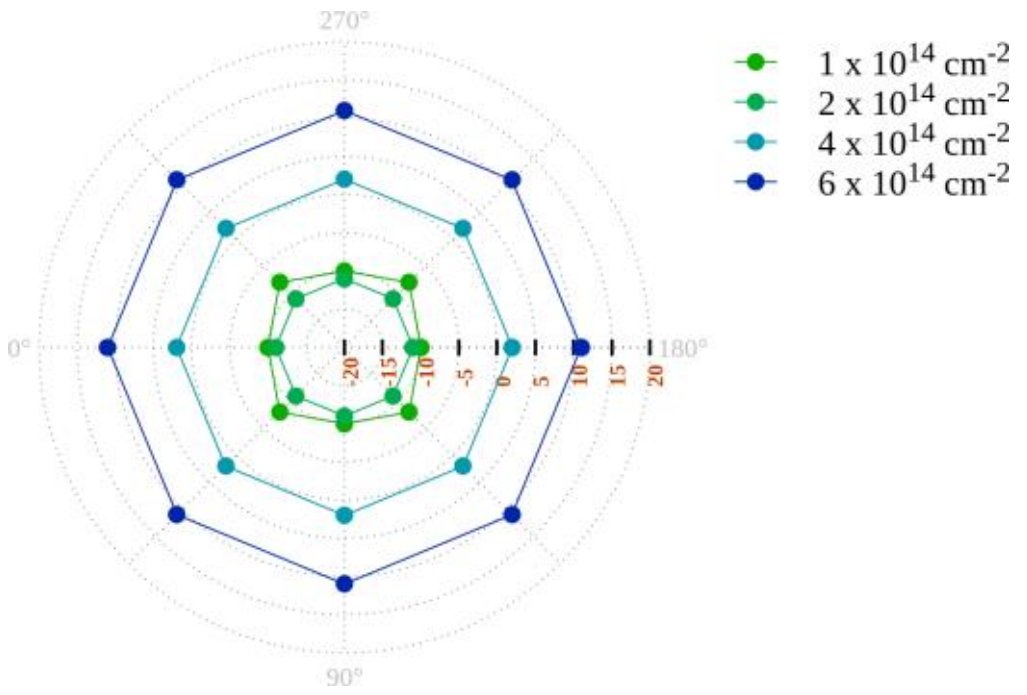

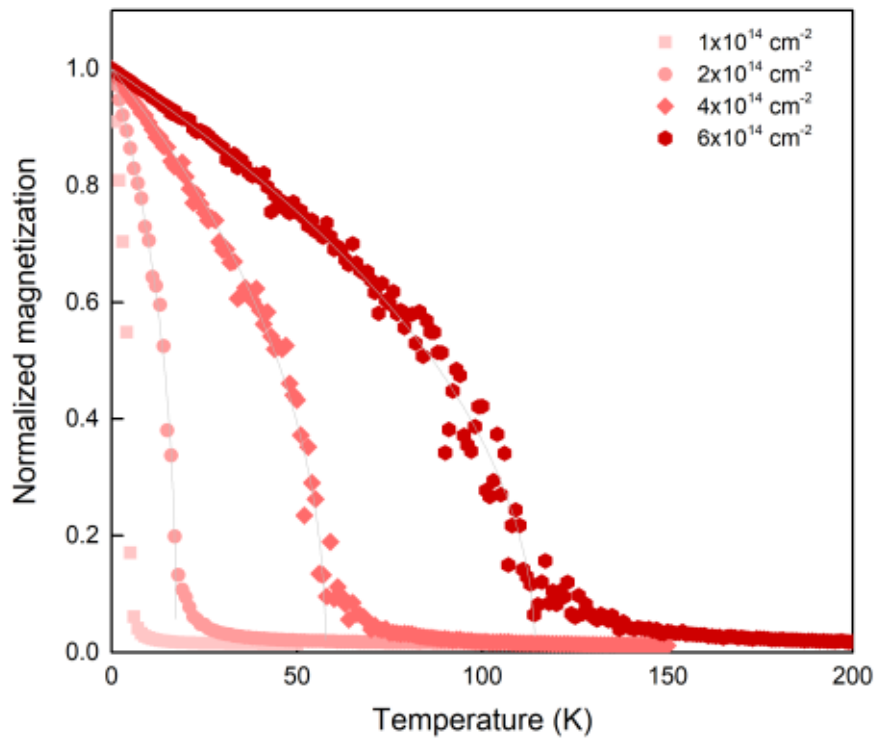

MAE = $E_{\parallel} - E_{\perp}$, a positive (negative) value of MAE indicates the off-plane (in-plane) easy axis.

$T_c$: $1\times10^{14}$ cm$^{-2}$: 6 K  $2\times10^{14}$ cm$^{-2}$: 17 K  $4\times10^{14}$ cm$^{-2}$: 58 K  $6\times10^{14}$ cm$^{-2}$: 114 K

# 33. SnS

| MC2D-ID | C2DB | 2dmat-ID | USPEX | Space group | Band gap (eV) |
|---|---|---|---|---|---|
| - | ✓ | 2dm-280 | - | P3m1 | 2.32 |

**Convex hull** | **Atomic structure** | **Atomic coordinates** | **Phonon dispersion curve**

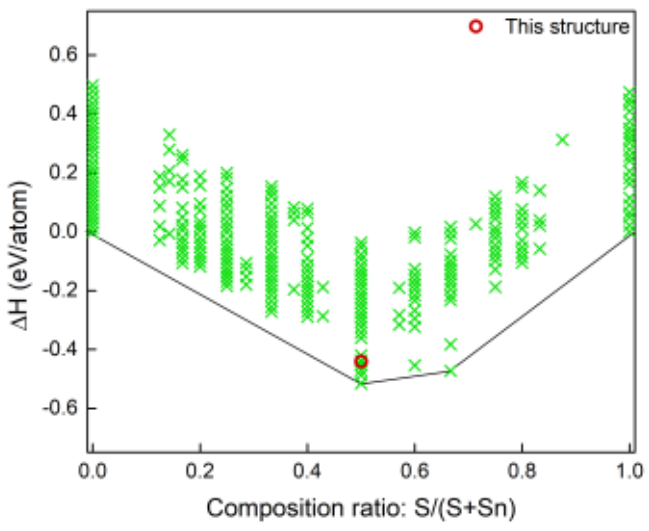

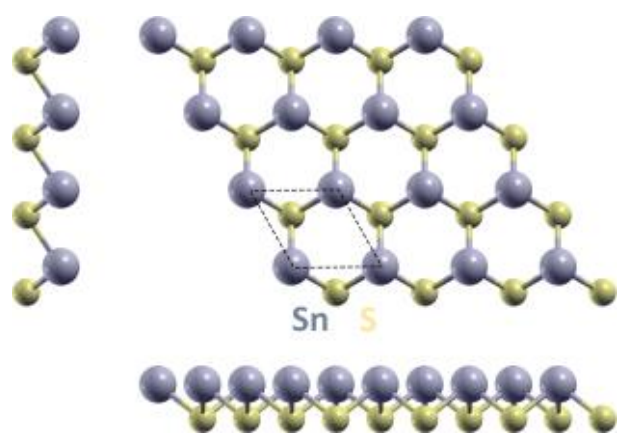

```
Sn1S1
   1.00000000000000
   3.7550081475776218   0.0000000000000000   0.0000000000000000
  -1.8775079252228946   3.2519302232273097   0.0000000000000000
   0.0000000000000000   0.0000000000000000  21.4786740000000016
   Sn   S
   1    1
Direct
  0.3333332934743112  0.6666667044025658  0.1660752026681962
  0.6666667065256888  0.3333332955974342  0.0980687973318055
```

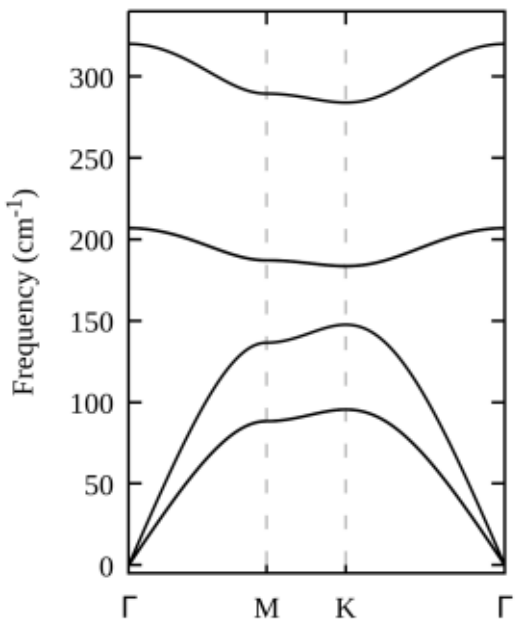

## Projected band structure and density of states

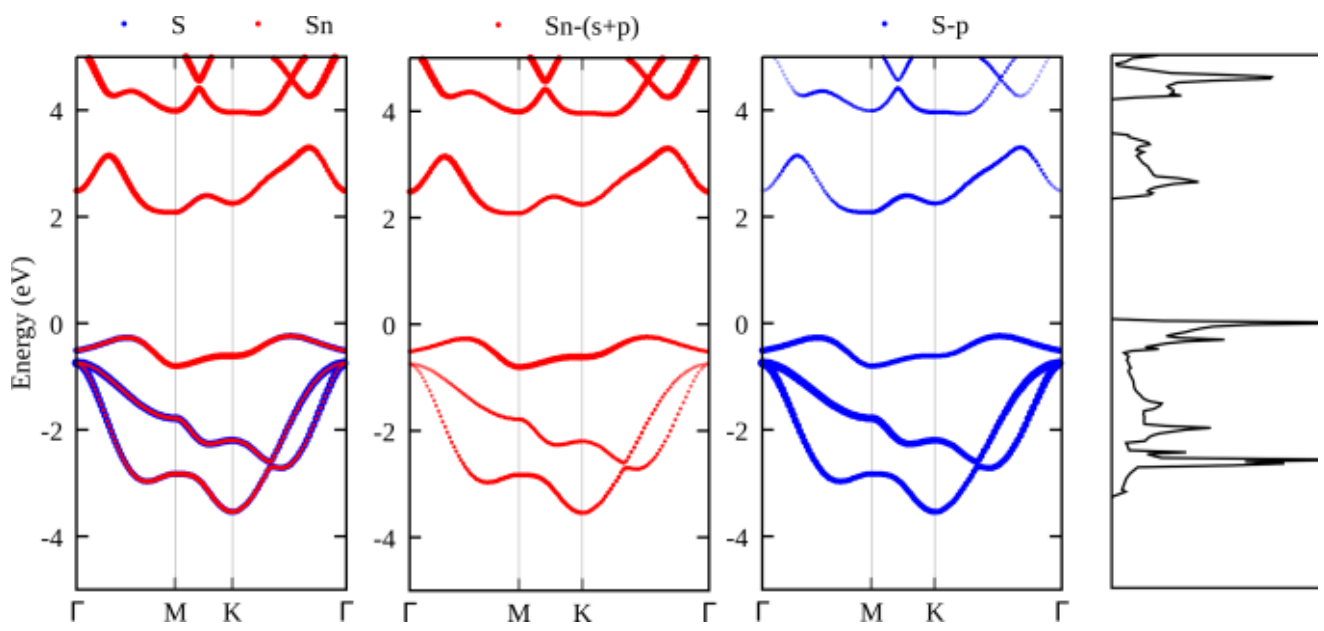

## Magnetic moment and spin polarization energy as a function of hole doping concentration

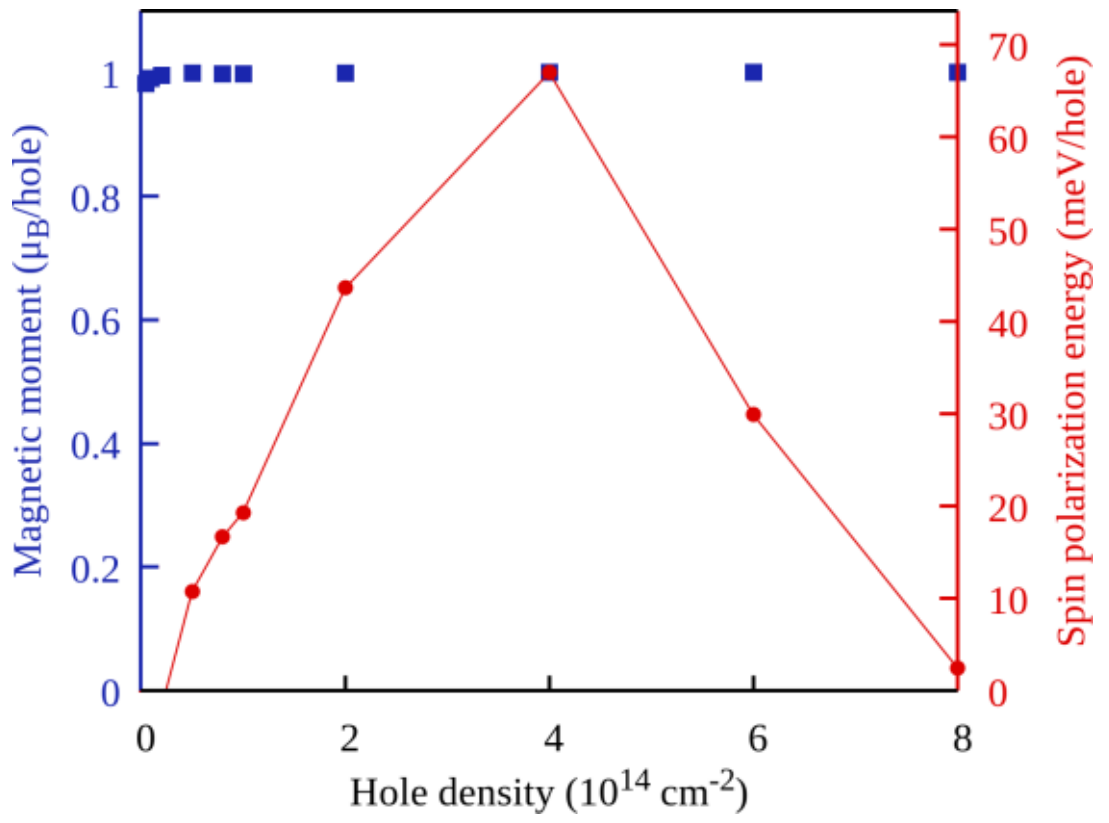

## Magnetic configurations and spin Hamiltonian

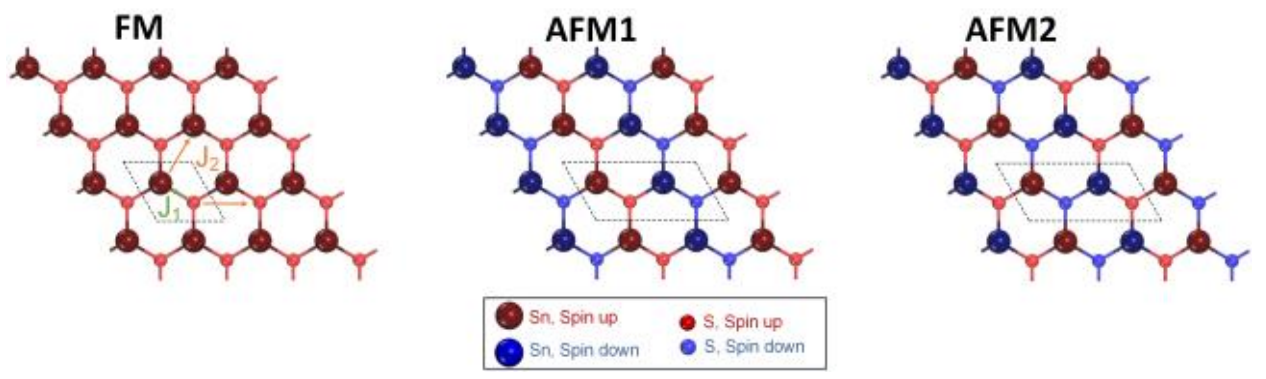

$$E_{FM} = E_0 - 3J_1S^2 - 6J_2S^2$$

$$E_{AFM1} = E_0 - J_1S^2 + 2J_2S^2$$

$$E_{AFM2} = E_0 + J_1S^2 + 2J_2S^2$$

## Magnetic exchange coupling parameters

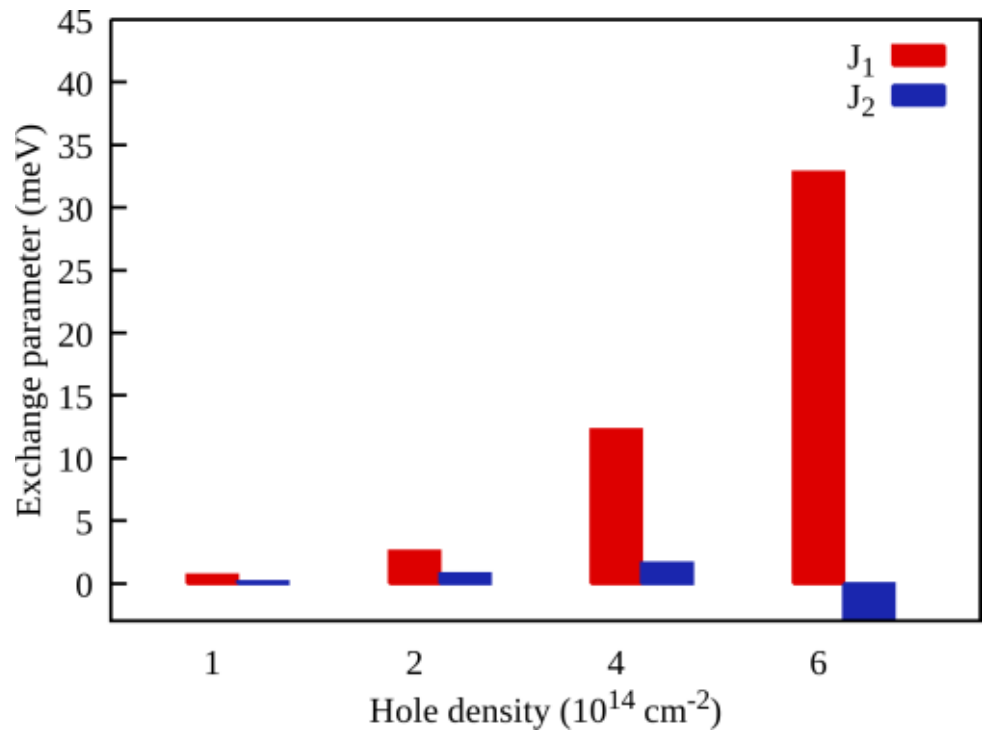

## Magnetic anisotropy energy (MAE, μeV) per magnetic atom

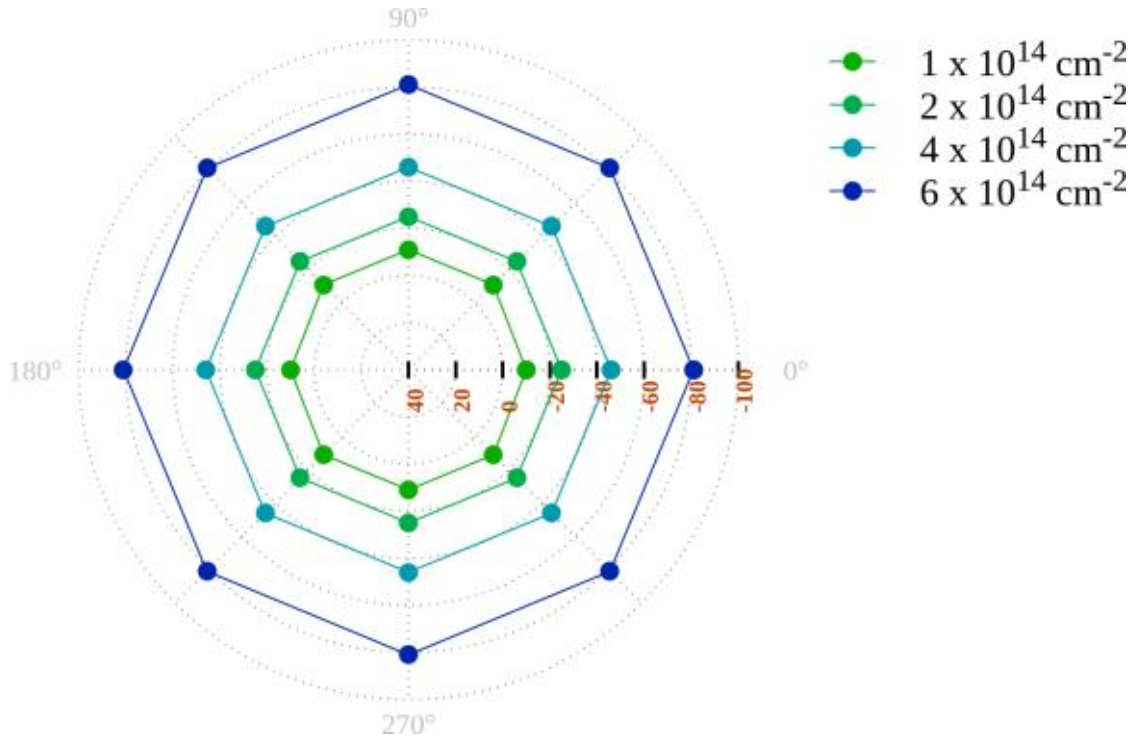

MAE = $E_{\parallel} - E_{\perp}$, a positive (negative) value of MAE indicates the off-plane (in-plane) easy axis.

## Monte Carlo simulations of the normalized magnetization of as a function of temperature

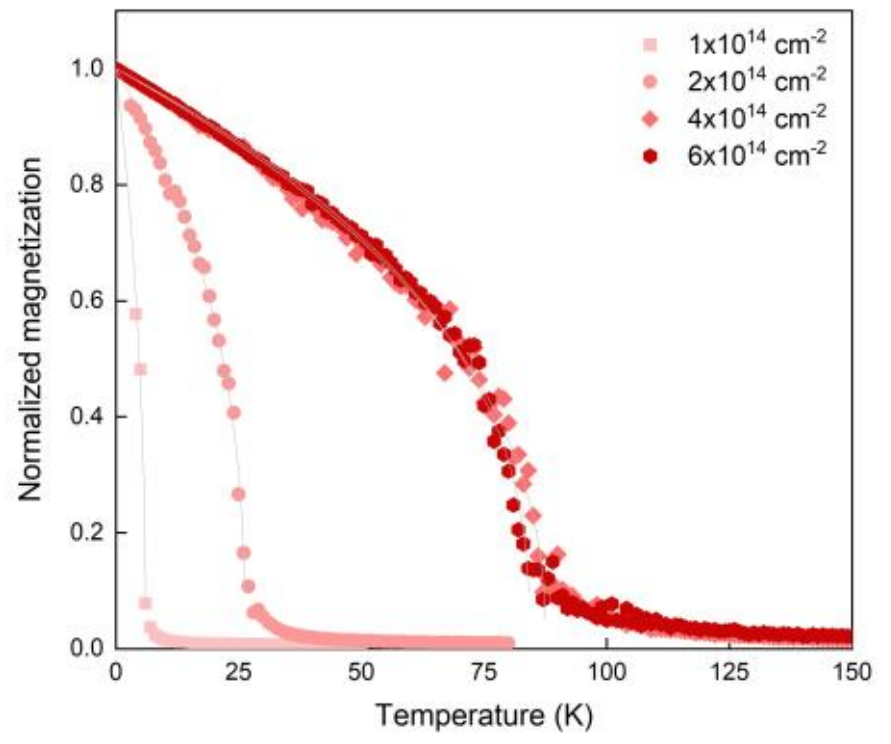

$T_c$: $1\times10^{14}$ cm$^{-2}$: 6 K  $2\times10^{14}$ cm$^{-2}$: 26 K

$4\times10^{14}$ cm$^{-2}$: 88 K  $6\times10^{14}$ cm$^{-2}$: 84 K

# 34. PbS

| MC2D-ID | C2DB | 2dmat-ID | USPEX | Space group | Band gap (eV) |
|---|---|---|---|---|---|
| - | ✓ | 2dm-3013 | - | P3m1 | 2.05 |

**Convex hull**

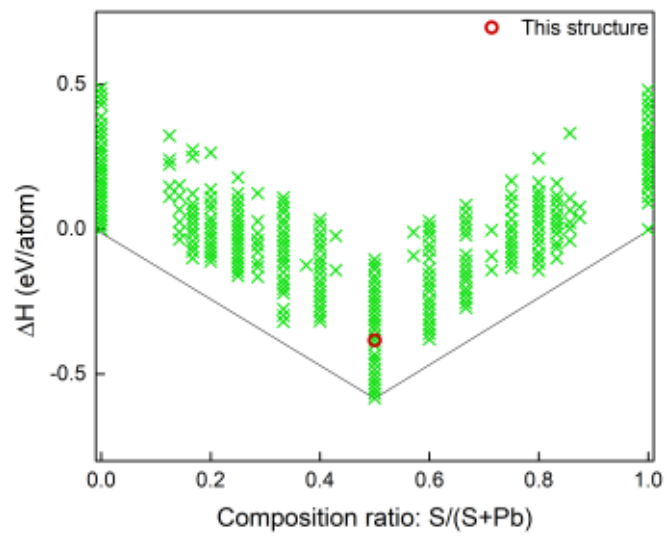

**Atomic structure**

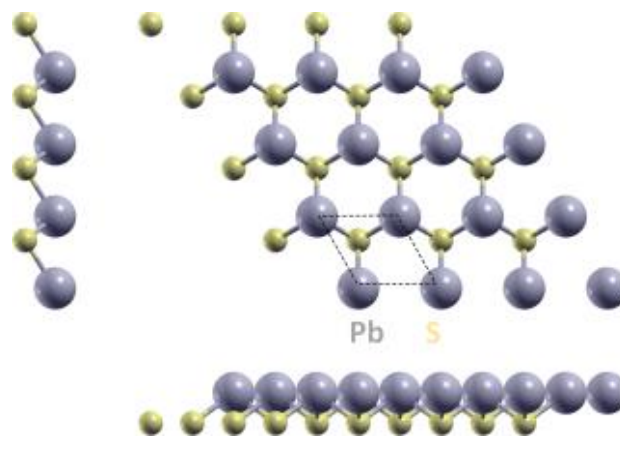

**Atomic coordinates**

```
Pb1S1
  1.00000000000000
  3.9574464751196028   0.0000000000000000   0.0000000000000000
  1.9787250178403992   3.4272481510099406   0.0000000000000000
  0.0000000000000000   0.0000000000000000  21.5987730000000013
  Pb   S
  1    1
Direct
  0.9999998641125032  0.9999998528727545  0.5332669634183915
  0.3333331358875000  0.3333331471272487  0.4667330365816085
```

**Phonon dispersion curve**

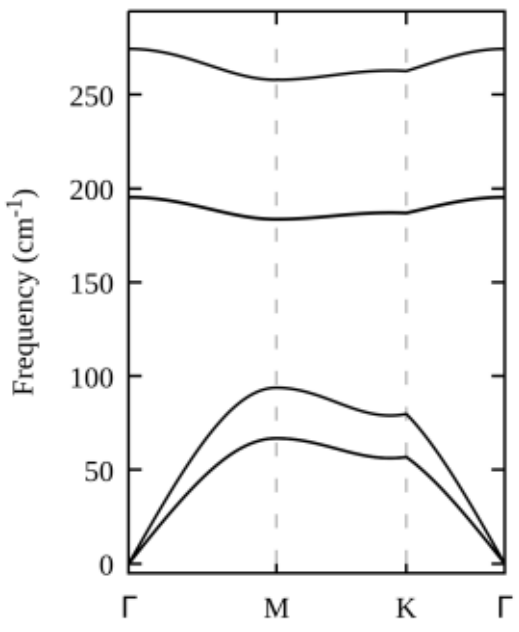

**Projected band structure and density of states**

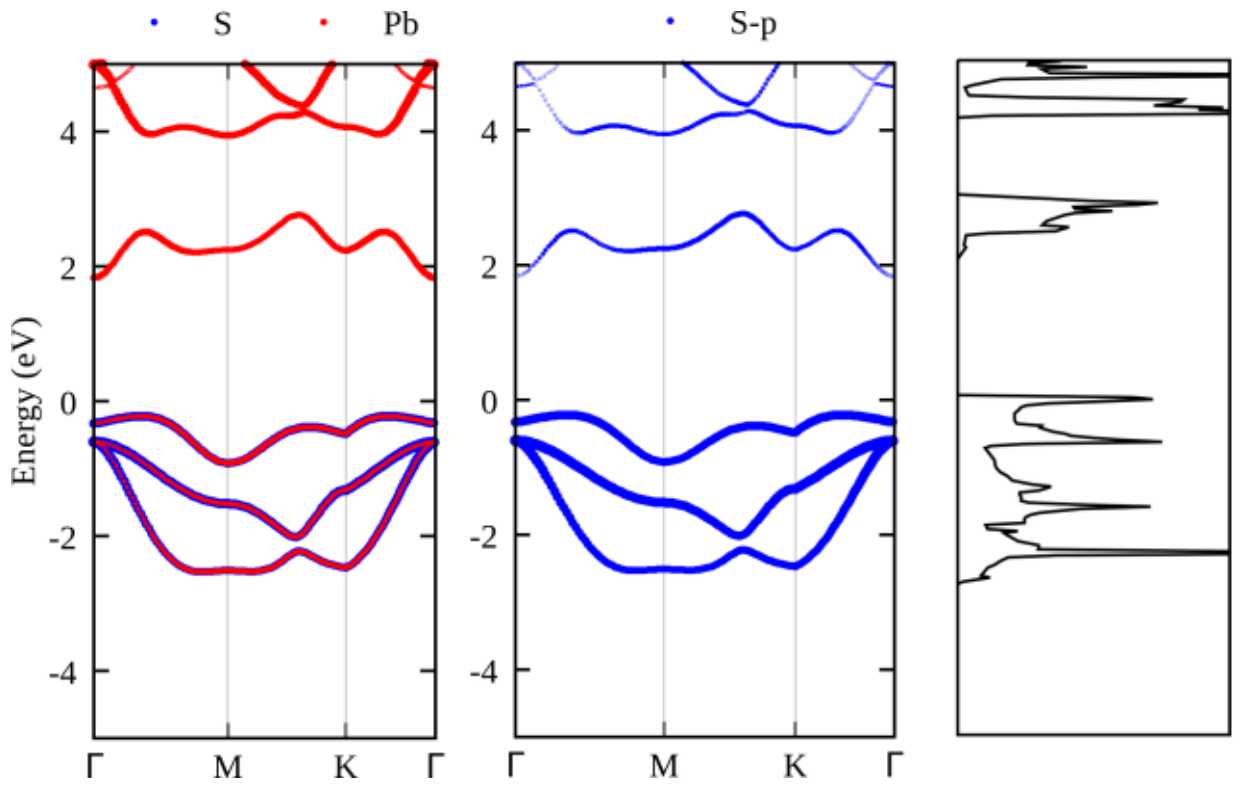

**Magnetic moment and spin polarization energy as a function of hole doping concentration**

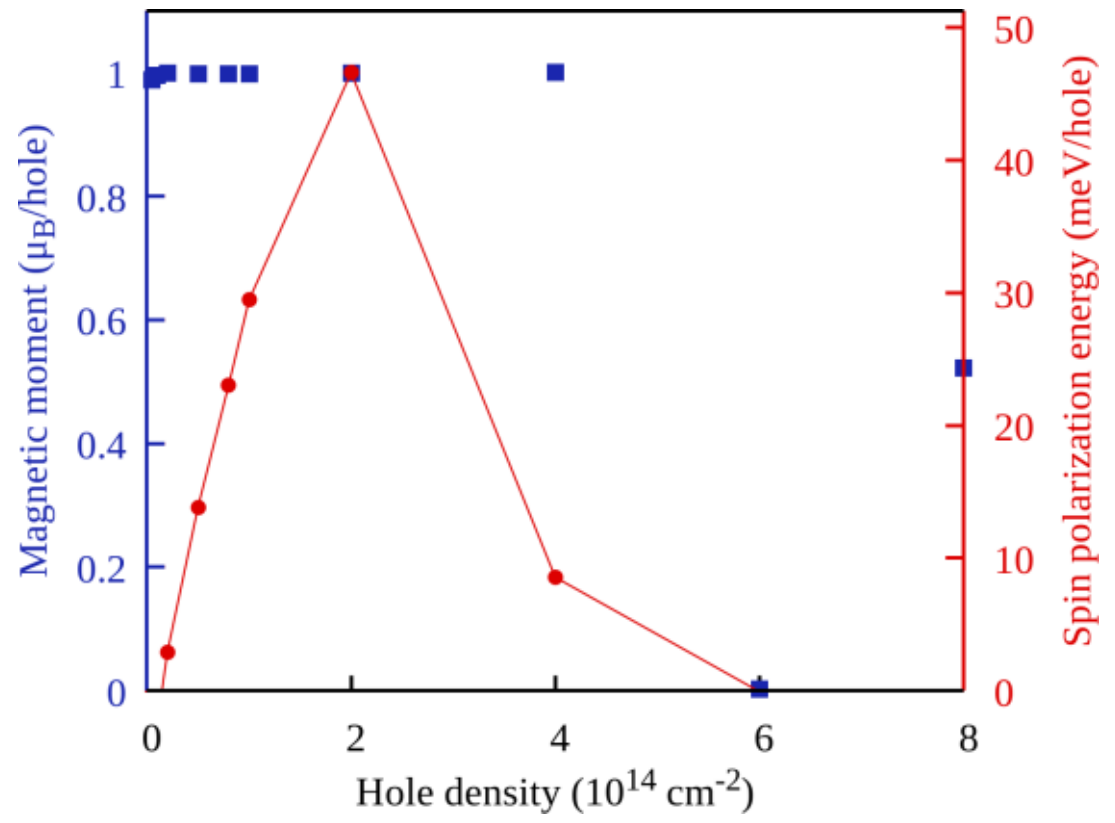

**Magnetic configurations and spin Hamiltonian**

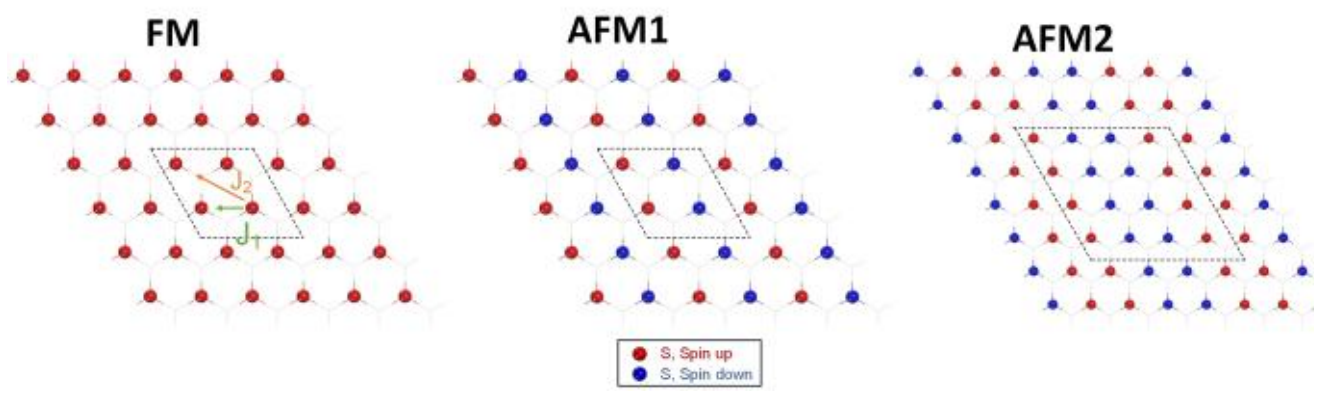

$$E_{FM} = E_0 - 12J_1S^2 - 12J_2S^2$$
$$E_{AFM1} = E_0 + 4J_1S^2 + 4J_2S^2$$
$$E_{AFM2} = E_0 - 4J_1S^2 + 4J_2S^2$$

**Magnetic exchange coupling parameters**

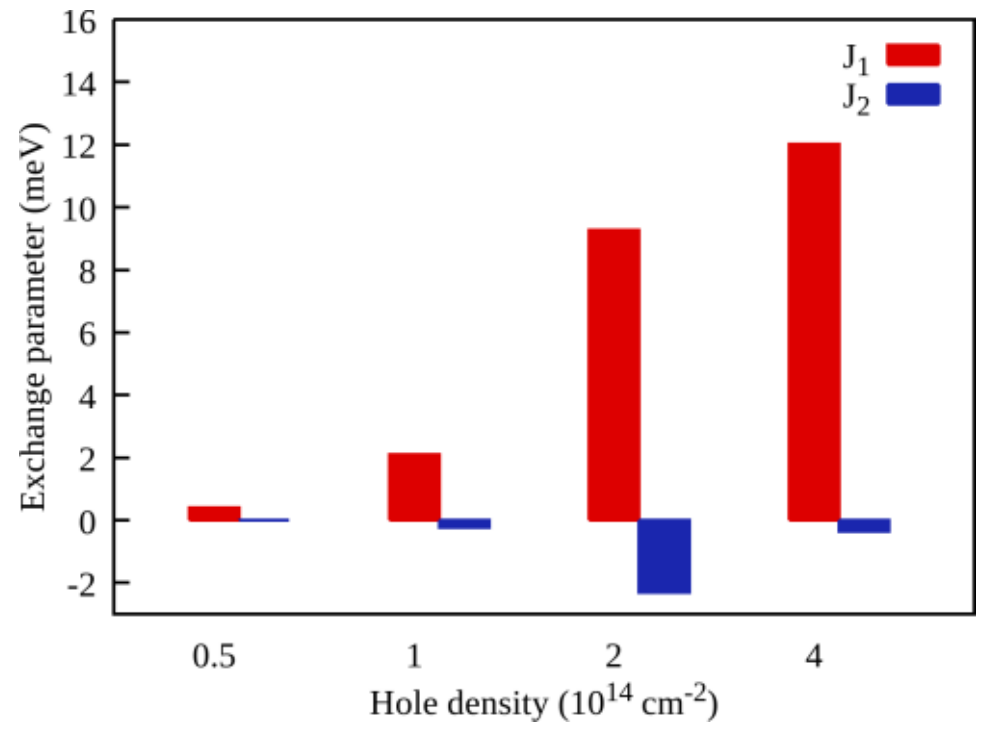

**Magnetic anisotropy energy (MAE, μeV) per magnetic atom**

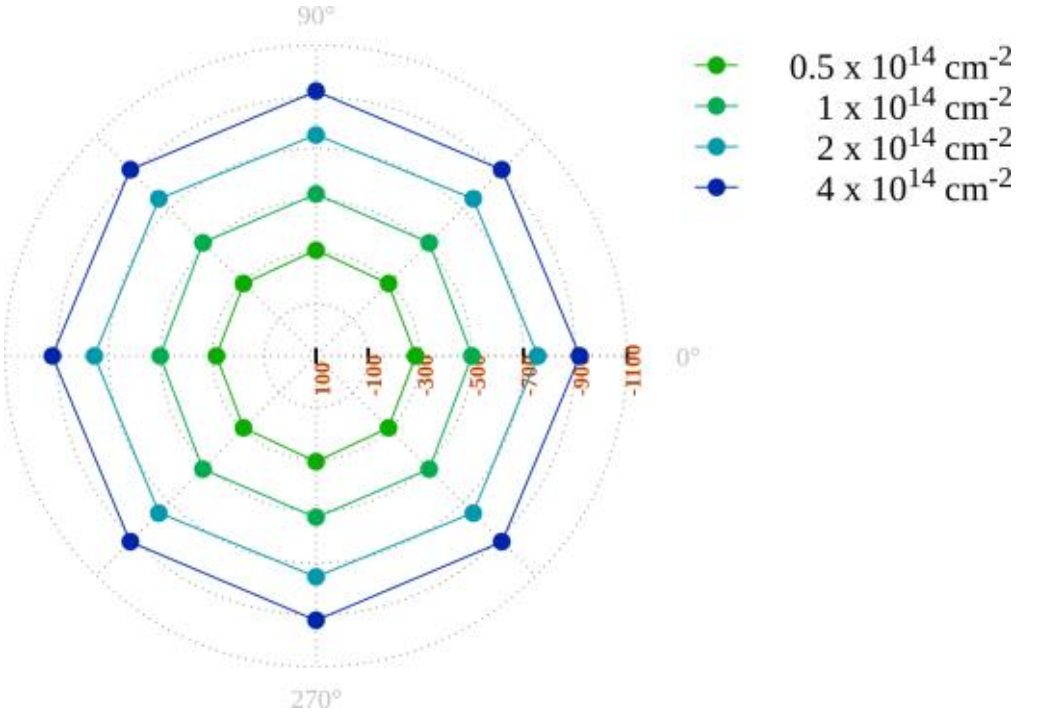

MAE = $E_{\parallel} - E_{\perp}$, a positive (negative) value of MAE indicates the off-plane (in-plane) easy axis.

**Monte Carlo simulations of the normalized magnetization of as a function of temperature**

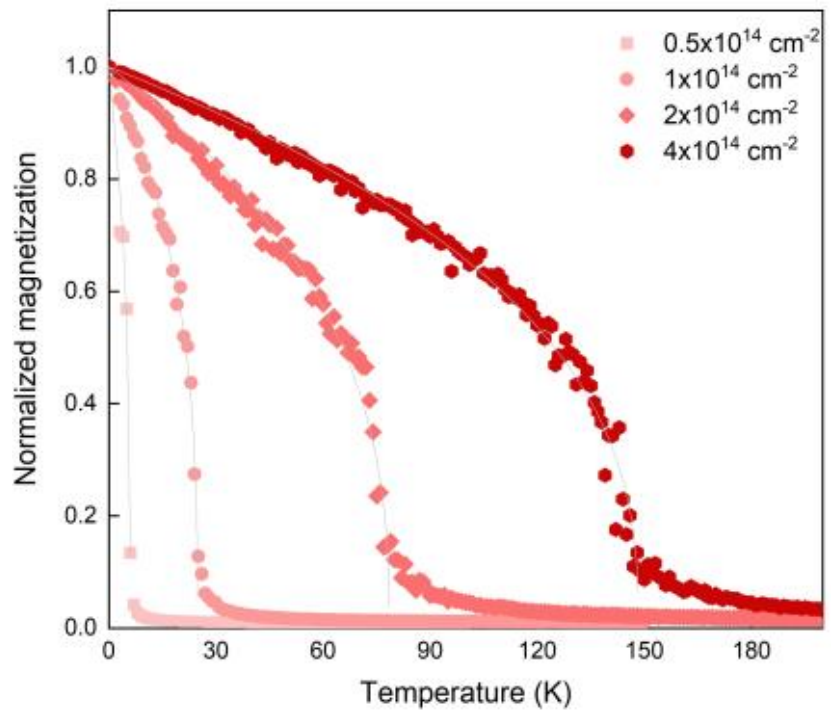

$T_c$: $0.5\times10^{14}$ cm$^{-2}$: 6 K  $1\times10^{14}$ cm$^{-2}$: 25 K  $2\times10^{14}$ cm$^{-2}$: 78 K  $4\times10^{14}$ cm$^{-2}$: 148 K

# 35. PbSe

| MC2D-ID | C2DB | 2dmat-ID | USPEX | Space group | Band gap (eV) |
|---|---|---|---|---|---|
| - | ✓ | 2dm-458 | - | P3m1 | 1.87 |

**Convex hull** | **Atomic structure** | **Atomic coordinates** | **Phonon dispersion curve**

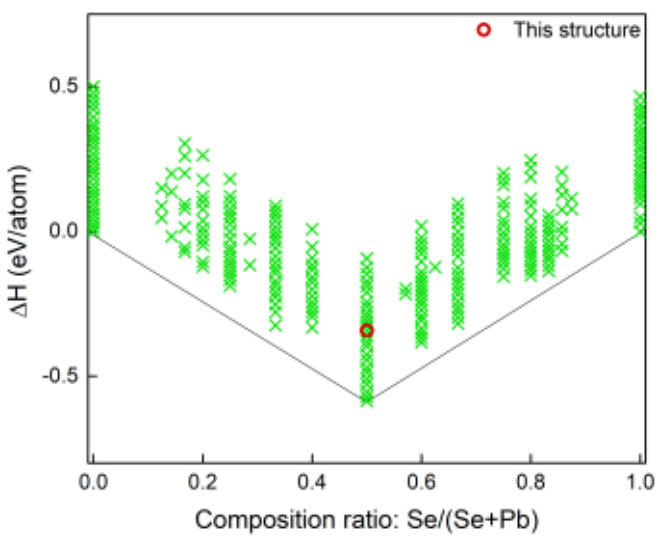

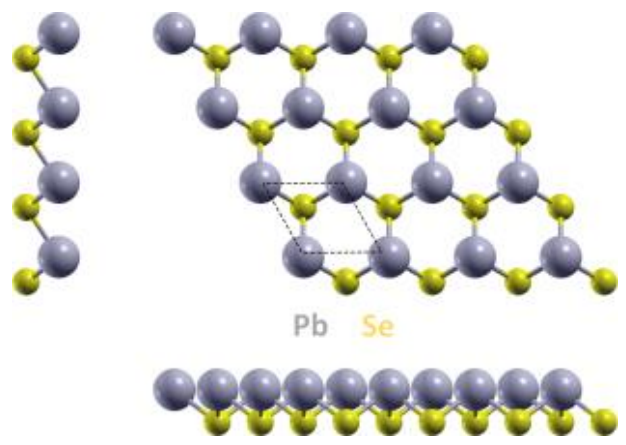

```
Pb1Se1
   1.00000000000000
   4.0889834380812813    0.0000000000000000    0.0000000000000000
  -2.0444941902647757    3.5411621784365397    0.0000000000000000
   0.0000000000000000    0.0000000000000000   21.4786740000000016
   Pb    Se
   1     1
Direct
  0.3333333170308492   0.6666666697625487   0.1680887920243137
  0.6666666829691508   0.3333333302374513   0.0960552079756880
```

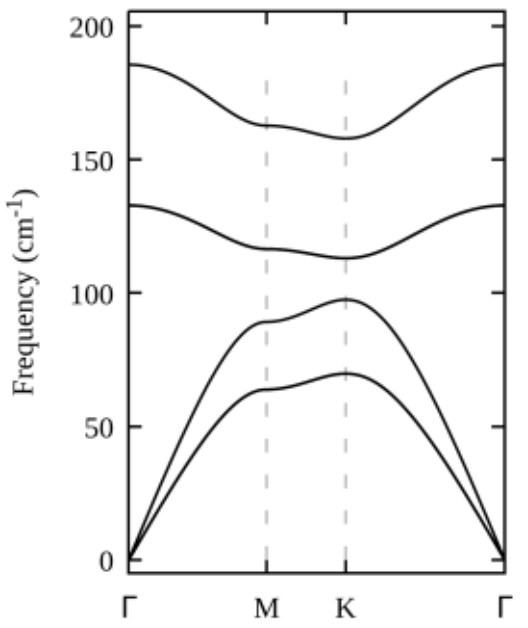

**Projected band structure and density of states** | **Magnetic moment and spin polarization energy as a function of hole doping concentration**

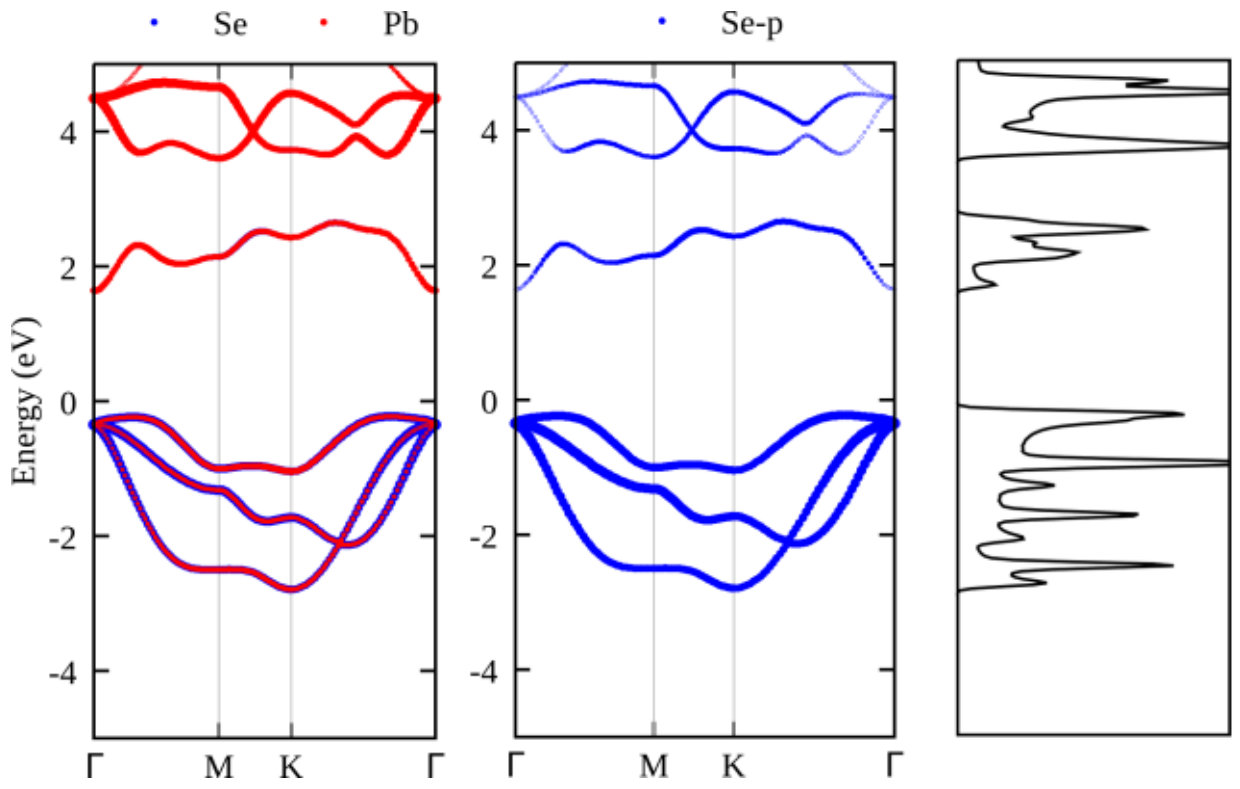

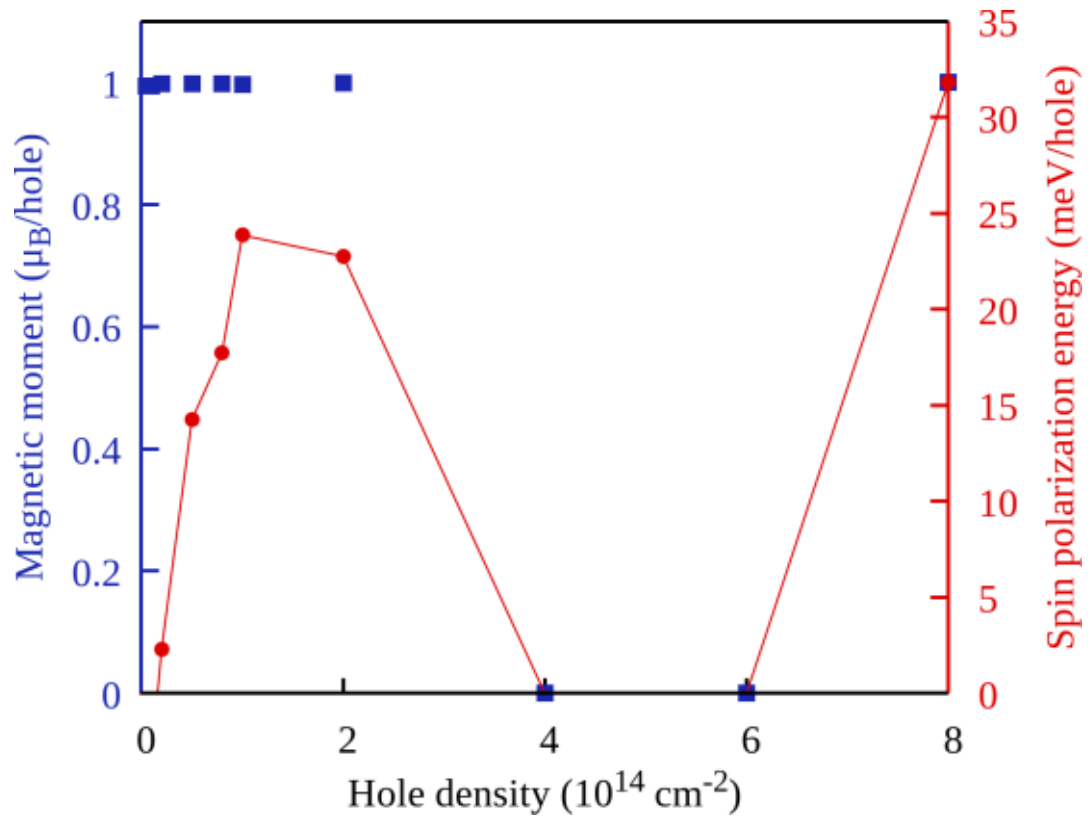

**Magnetic configurations and spin Hamiltonian** | **Magnetic exchange coupling parameters**

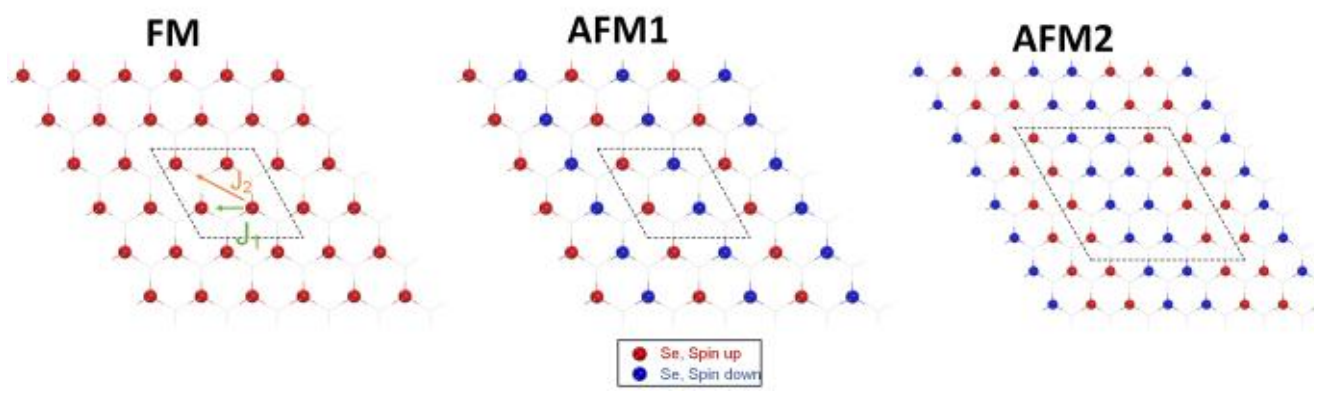

$$E_{FM} = E_0 - 12J_1S^2 - 12J_2S^2$$
$$E_{AFM1} = E_0 + 4J_1S^2 + 4J_2S^2$$
$$E_{AFM2} = E_0 - 4J_1S^2 + 4J_2S^2$$

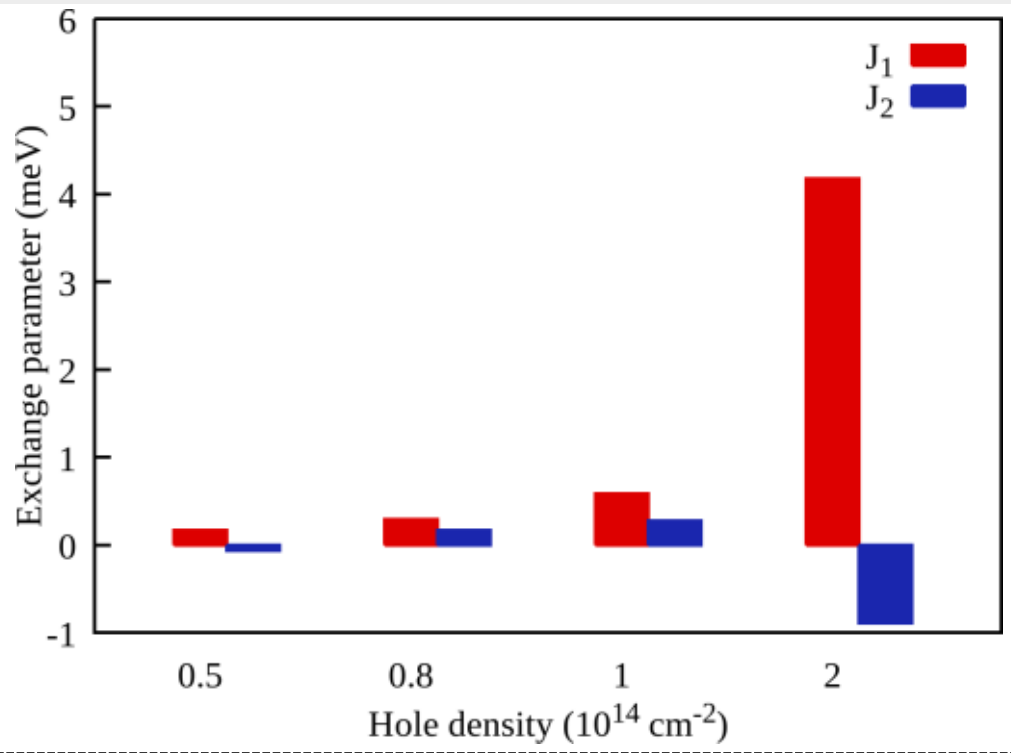

**Magnetic anisotropy energy (MAE, μeV) per magnetic atom** | **Monte Carlo simulations of the normalized magnetization of as a function of temperature**

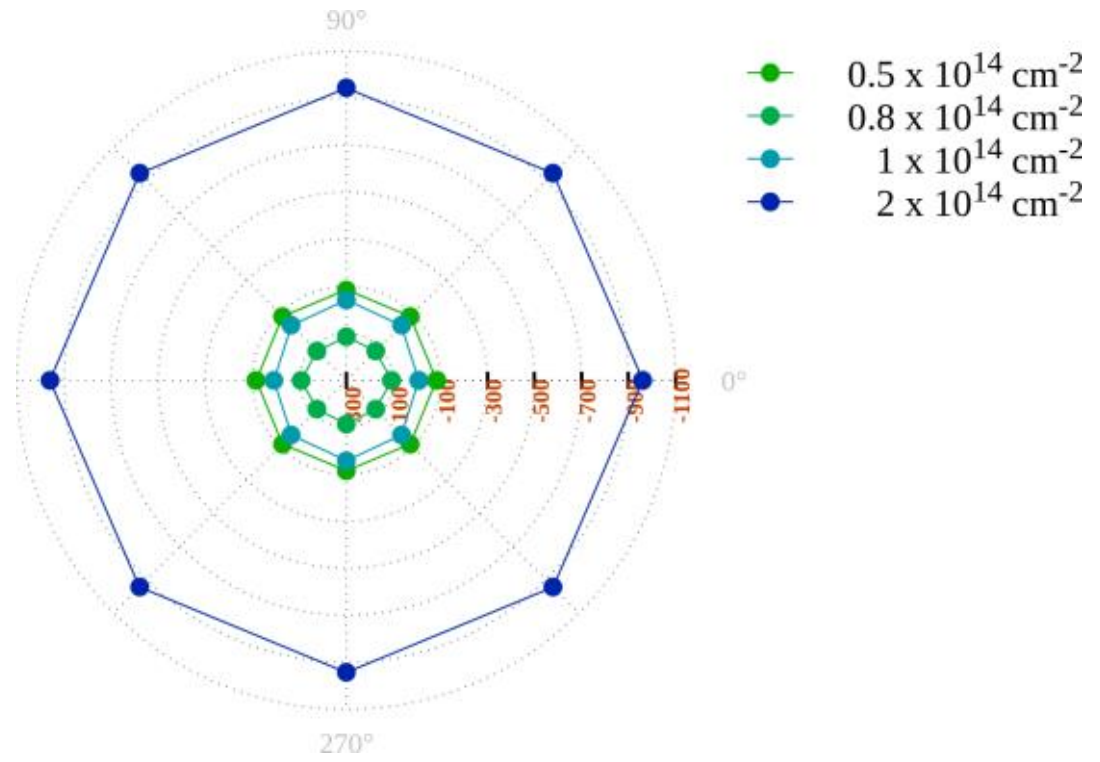

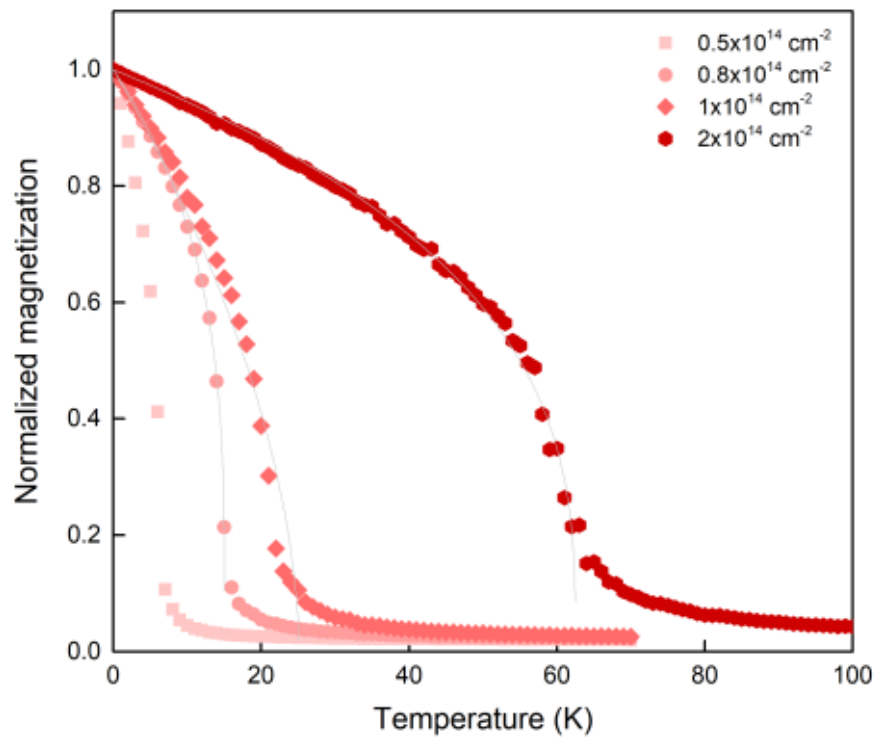

MAE = $E_{\parallel} - E_{\perp}$, a positive (negative) value of MAE indicates the off-plane (in-plane) easy axis.

$T_c$: $0.5\times10^{14}$ cm$^{-2}$: - K  $0.8\times10^{14}$ cm$^{-2}$: 7 K  $1\times10^{14}$ cm$^{-2}$: 11 K  $2\times10^{14}$ cm$^{-2}$: 41 K

# 36.ZnSe

| MC2D-ID | C2DB | 2dmat-ID | USPEX | Space group | Band gap (eV) |
|---|---|---|---|---|---|
| - | - | 2dm-6047 | - | P3m1 | 1.75 |

**Convex hull** | **Atomic structure** | **Atomic coordinates** | **Phonon dispersion curve**

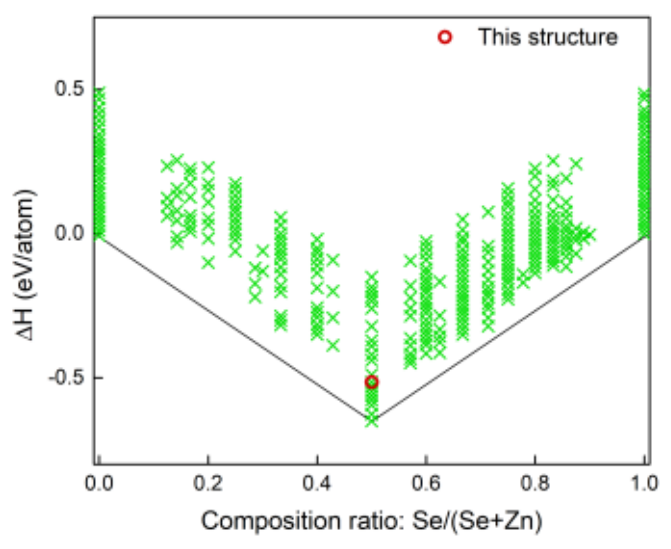

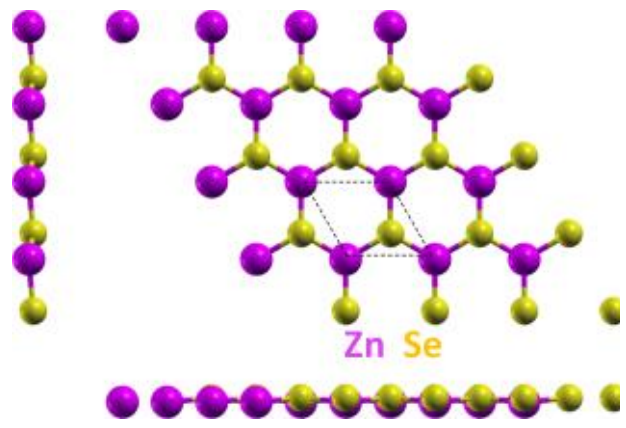

Zn1Se1
1.00000000000000
4.0978712846507142 0.0000000000000000 0.0000000000000000
-2.0488732777045082 3.5488967196092260 0.0000000000000000
0.0000000000000000 0.0000000000000000 20.1669300000000007
Zn Se
1 1
Direct
0.0000053148122987 0.9999947018611266 0.4953029702306182
0.6666616851876981 0.3333382981388766 0.5046970297693818

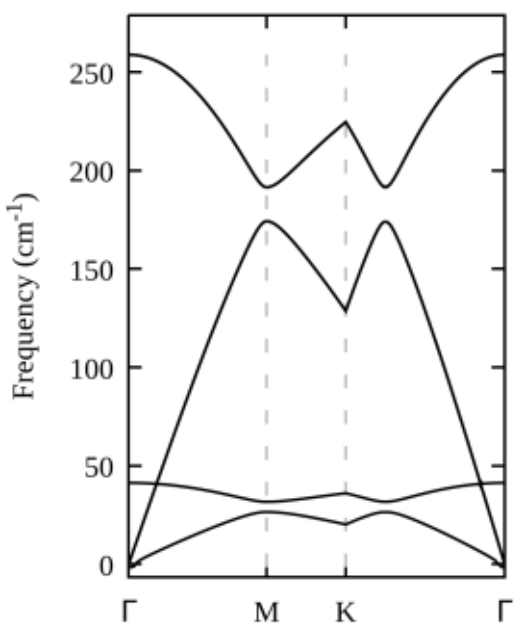

## Projected band structure and density of states

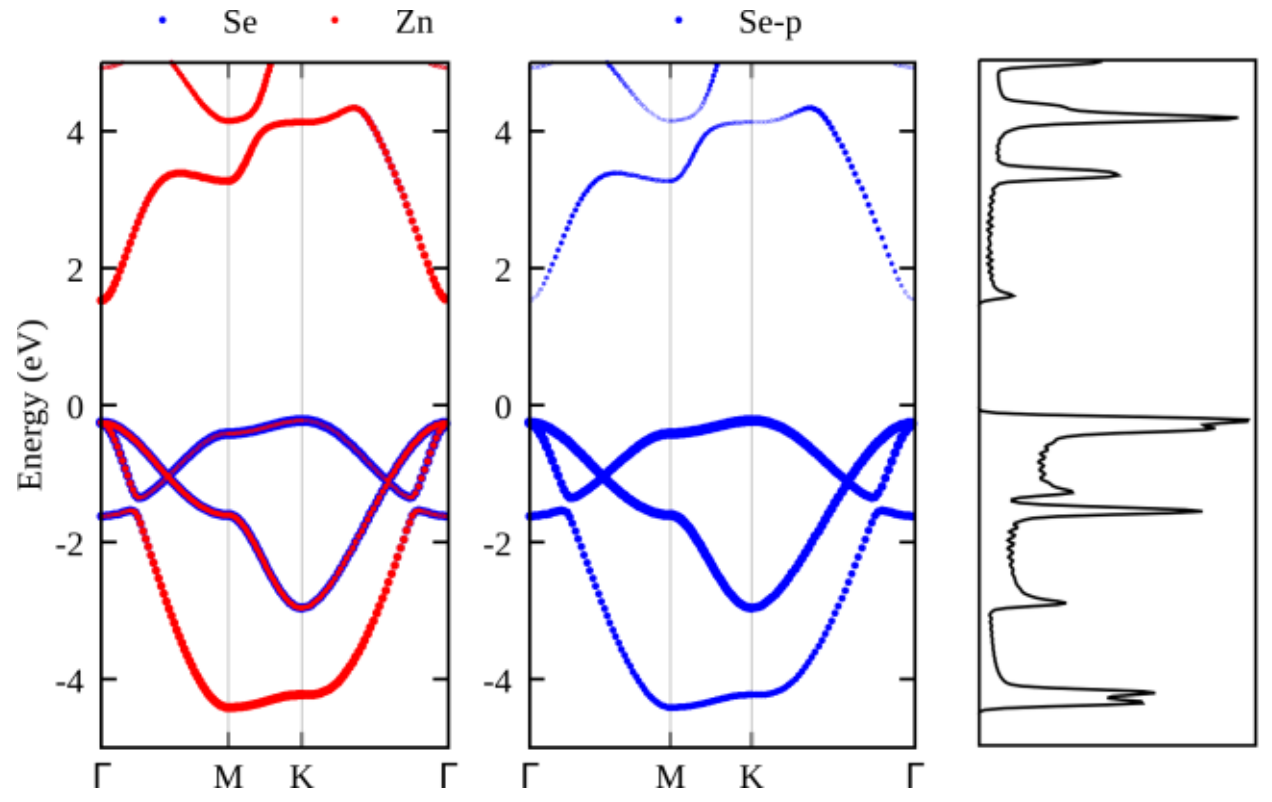

## Magnetic moment and spin polarization energy as a function of hole doping concentration

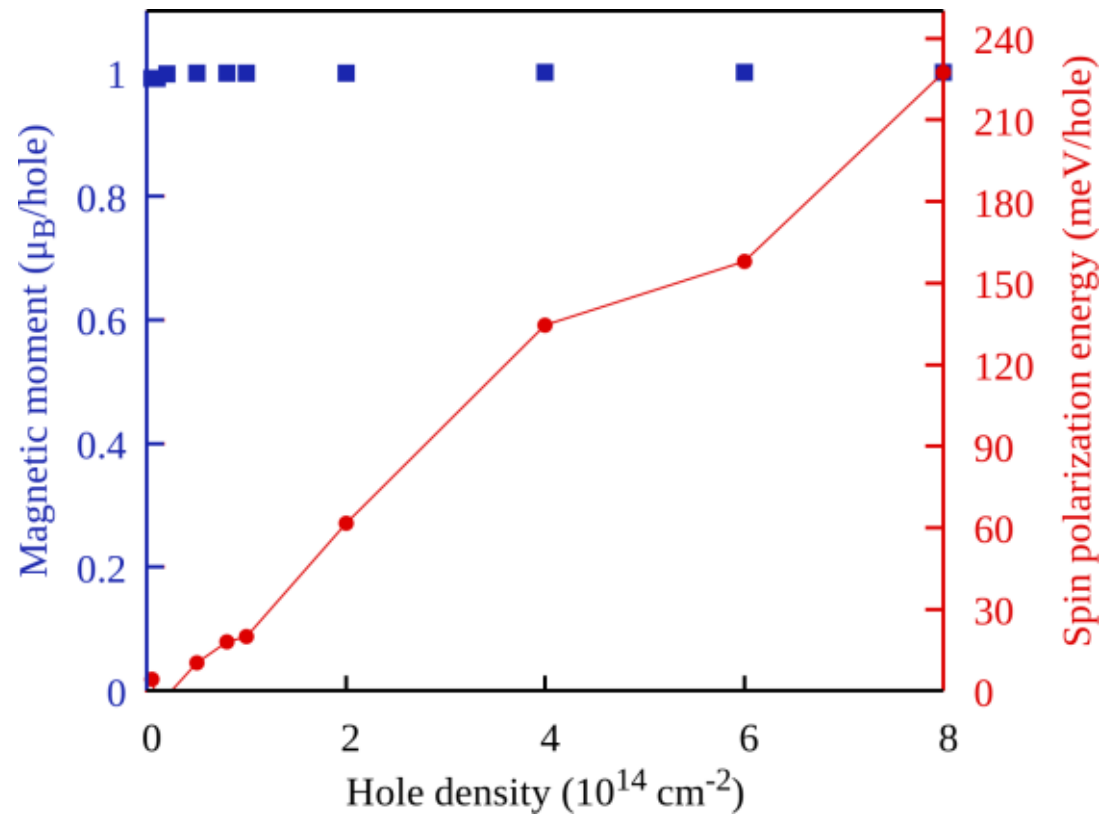

## Magnetic configurations and spin Hamiltonian

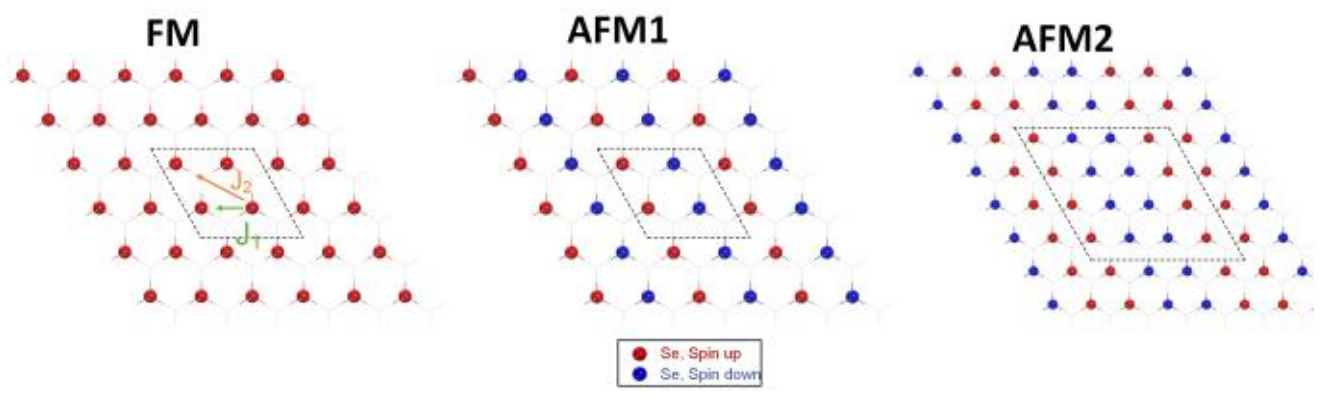

$$E_{FM} = E_0 - 12J_1S^2 - 12J_2S^2$$
$$E_{AFM1} = E_0 + 4J_1S^2 + 4J_2S^2$$
$$E_{AFM2} = E_0 - 4J_1S^2 + 4J_2S^2$$

## Magnetic exchange coupling parameters

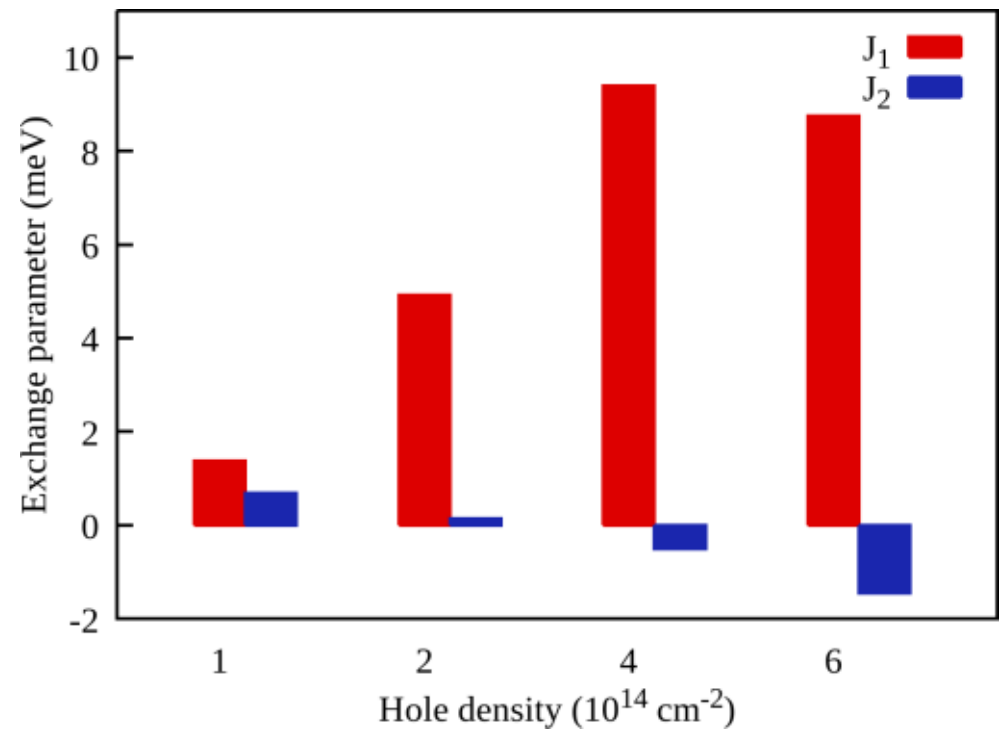

## Magnetic anisotropy energy (MAE, μeV) per magnetic atom

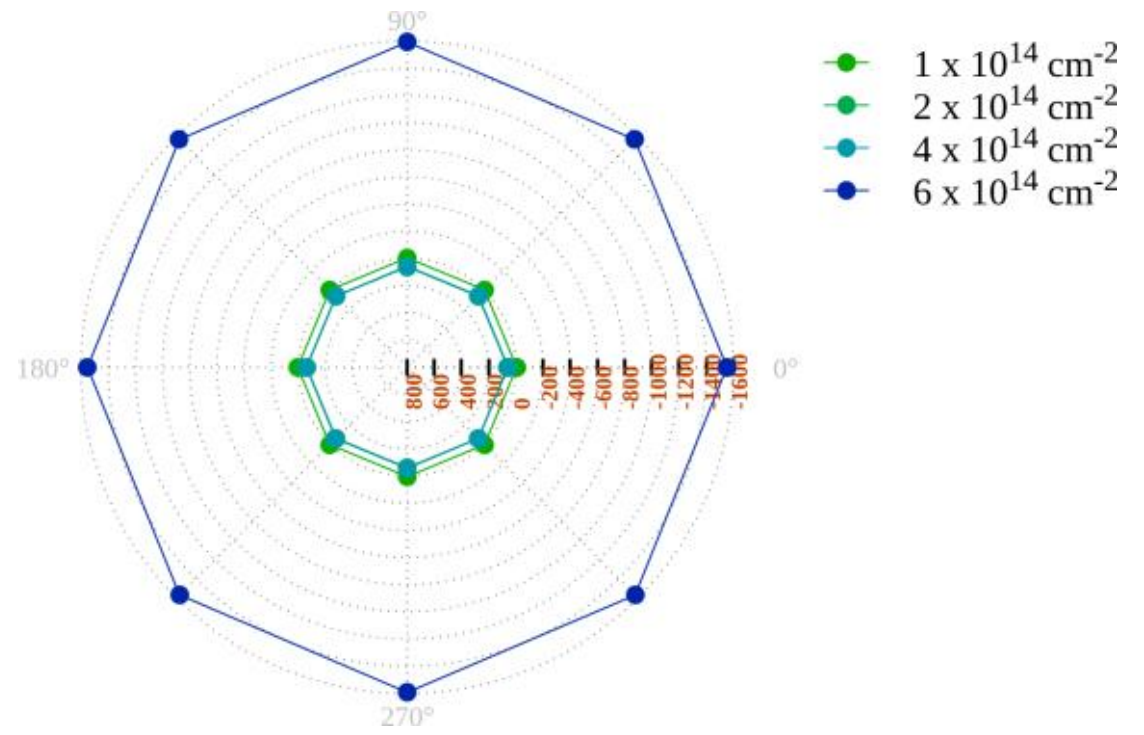

MAE = $E_{\parallel} - E_{\perp}$, a positive (negative) value of MAE indicates the off-plane (in-plane) easy axis.

## Monte Carlo simulations of the normalized magnetization of as a function of temperature

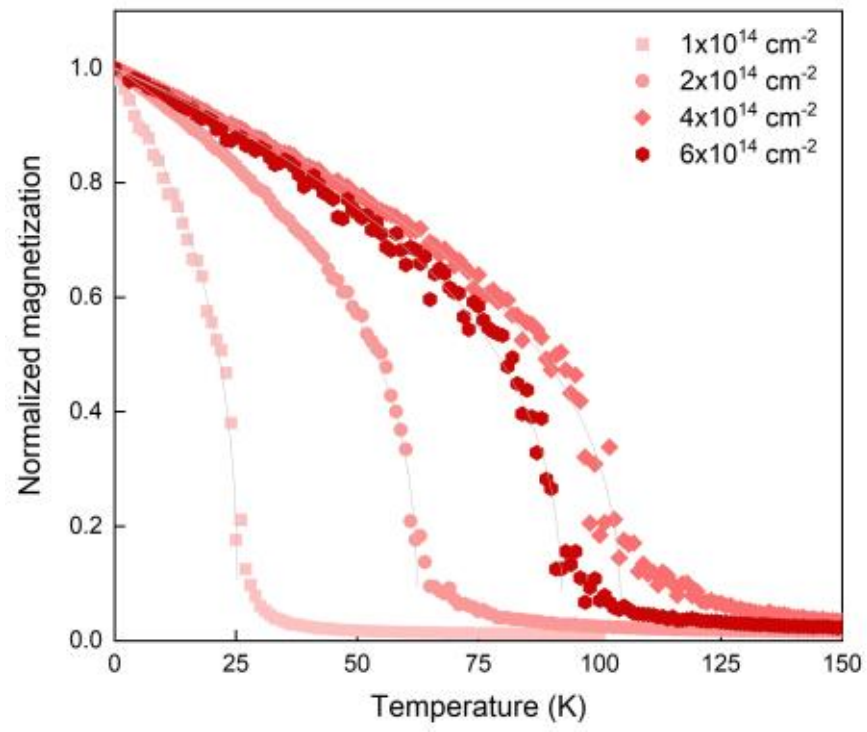

$T_c$: $1\times10^{14}$ cm$^{-2}$: 25 K  $2\times10^{14}$ cm$^{-2}$: 62 K  $4\times10^{14}$ cm$^{-2}$: 104 K  $6\times10^{14}$ cm$^{-2}$: 92 K

# 37. $PdS_2$

| MC2D-ID | C2DB | 2dmat-ID | USPEX | Space group | Band gap (eV) |
|---|---|---|---|---|---|
| - | ✓ | 2dm-71 | - | P3m1 | 1.26 |

## Convex hull

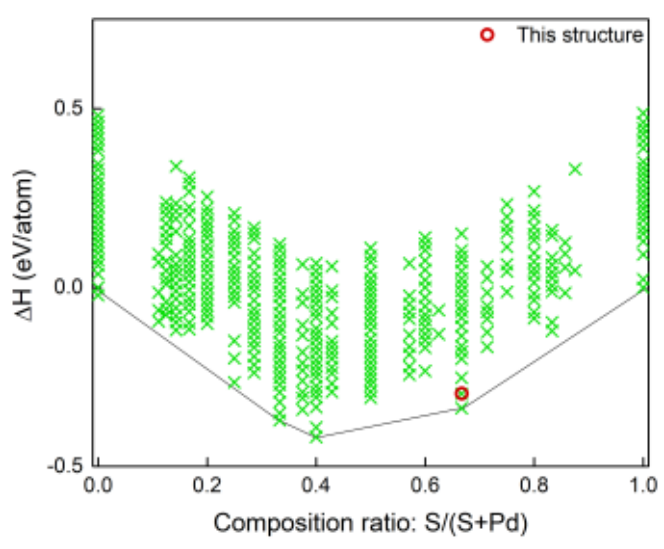

## Atomic structure

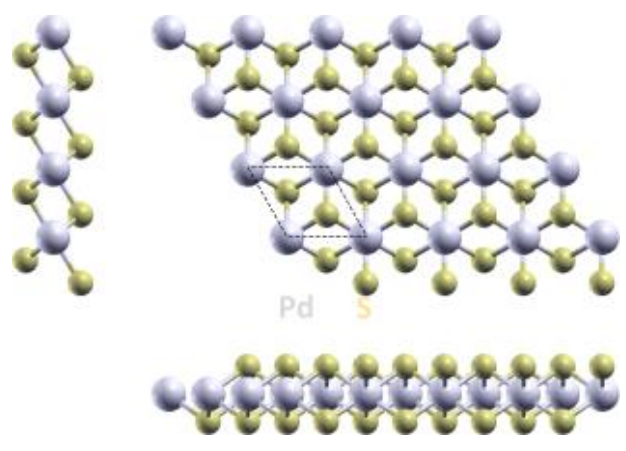

## Atomic coordinates

```
Pd1S2
   1.00000000000000
   3.5495563932969119    0.0000000000000000    0.0000000000000000
  -1.7747775327287703    3.0740067111284404    0.0000000000000000
   0.0000000000000000    0.0000000000000000   22.6134009999999996
   Pd    S
   1     2
Direct
  0.0000000094028891  0.9999999939163047  0.0000001727369536
  0.3333334999086475  0.6666665007981649  0.9451683186251998
  0.6666664906884634  0.3333335052855233  0.0548315086378466
```

## Phonon dispersion curve

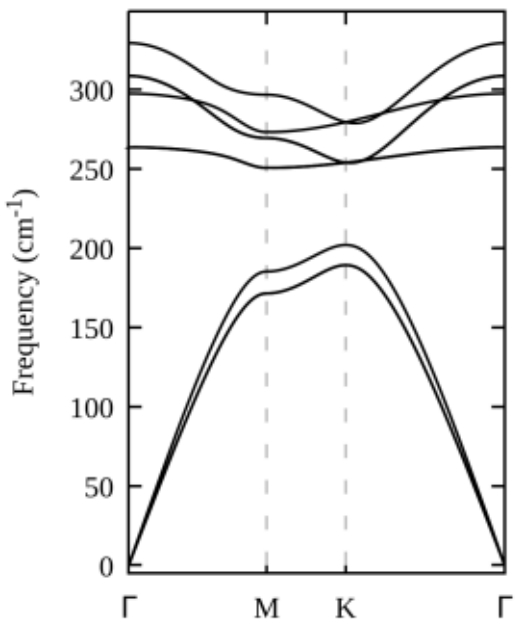

## Projected band structure and density of states

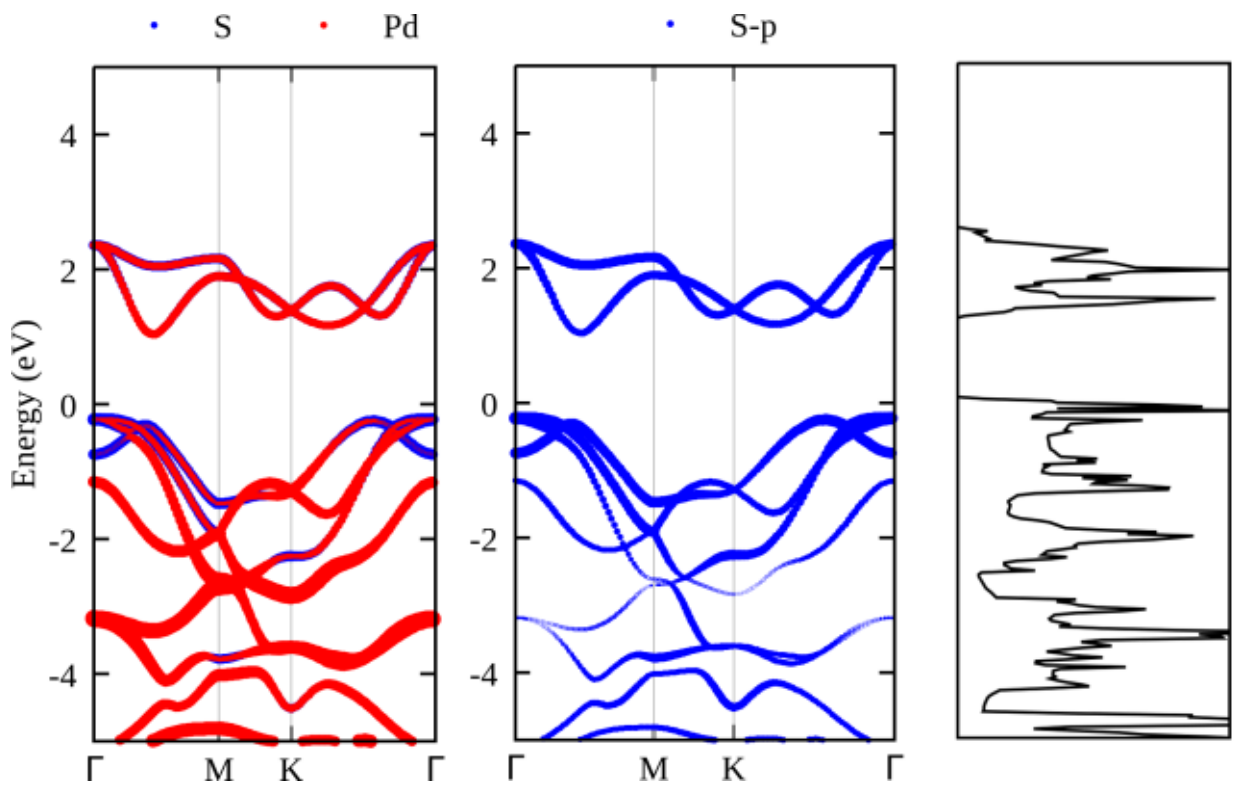

## Magnetic moment and spin polarization energy as a function of hole doping concentration

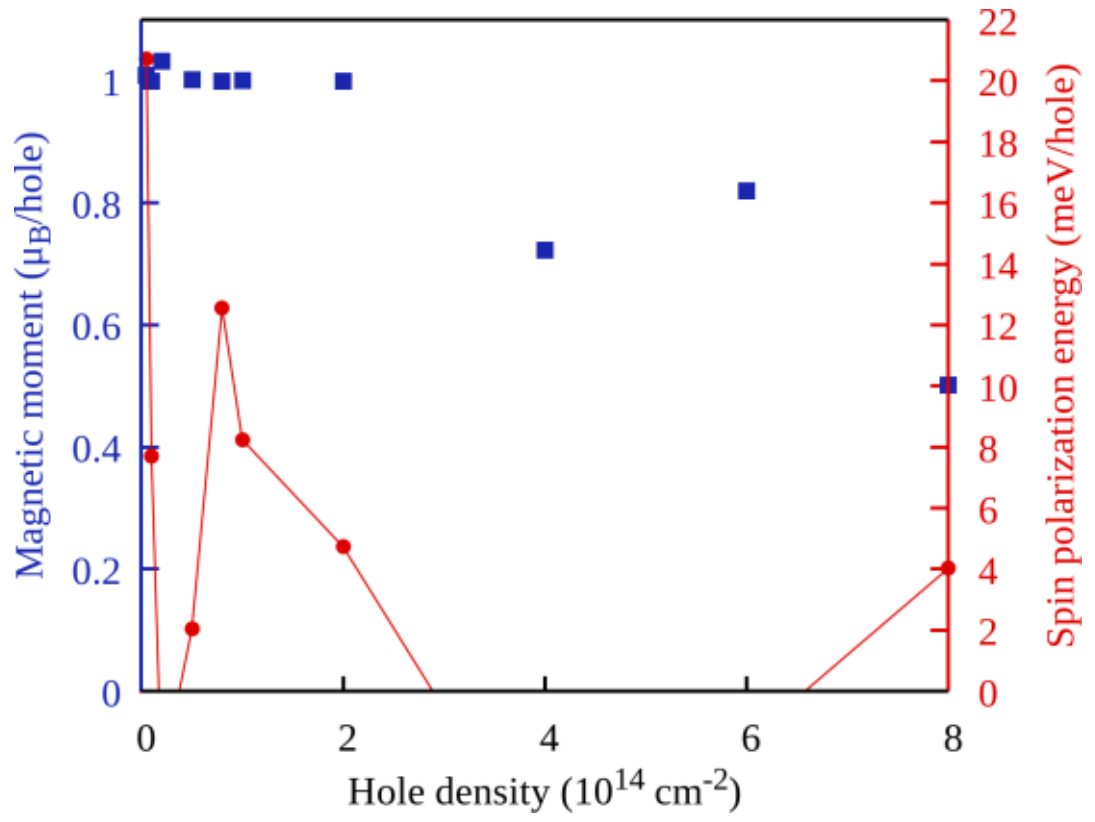

## Magnetic configurations and spin Hamiltonian

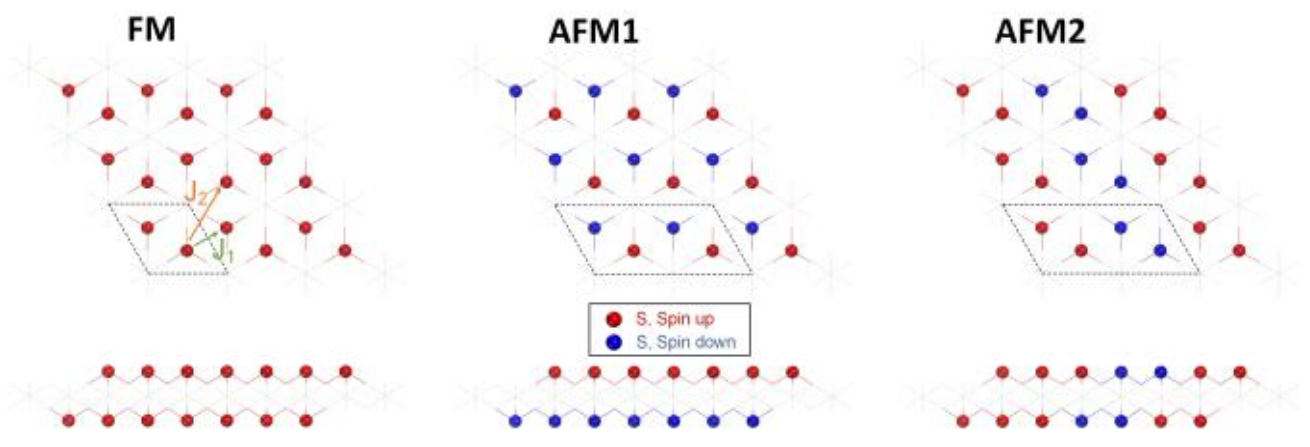

$$E_{FM} = E_0 - 3J_1S^2 - 6J_2S^2$$
$$E_{AFM1} = E_0 + 3J_1S^2 - 6J_2S^2$$
$$E_{AFM2} = E_0 - J_1S^2 + 2J_2S^2$$

## Magnetic exchange coupling parameters

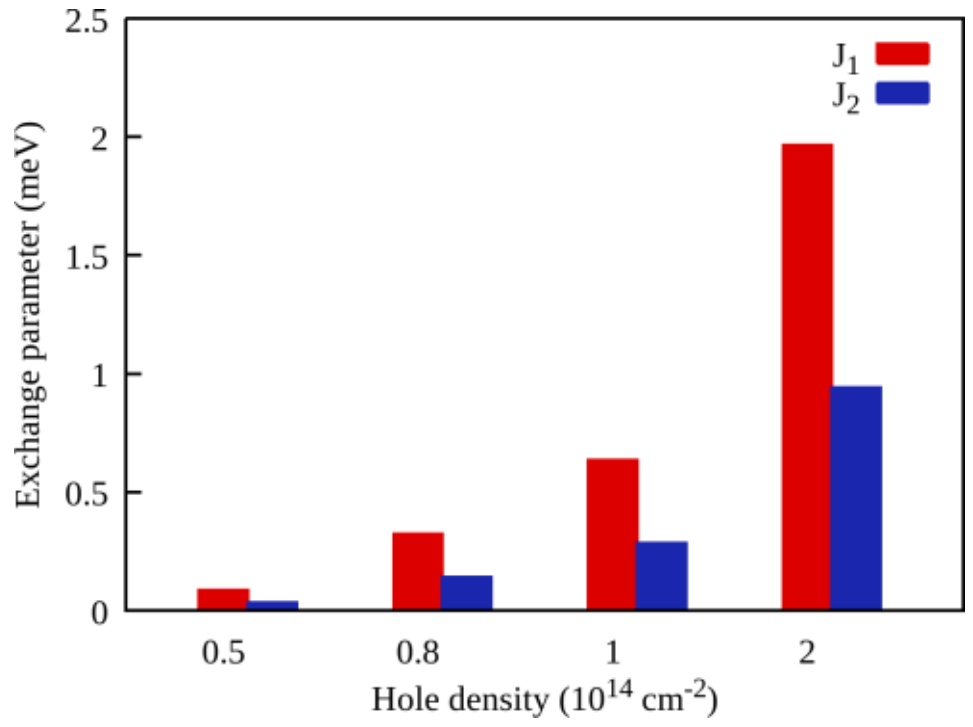

## Magnetic anisotropy energy (MAE, μeV) per magnetic atom

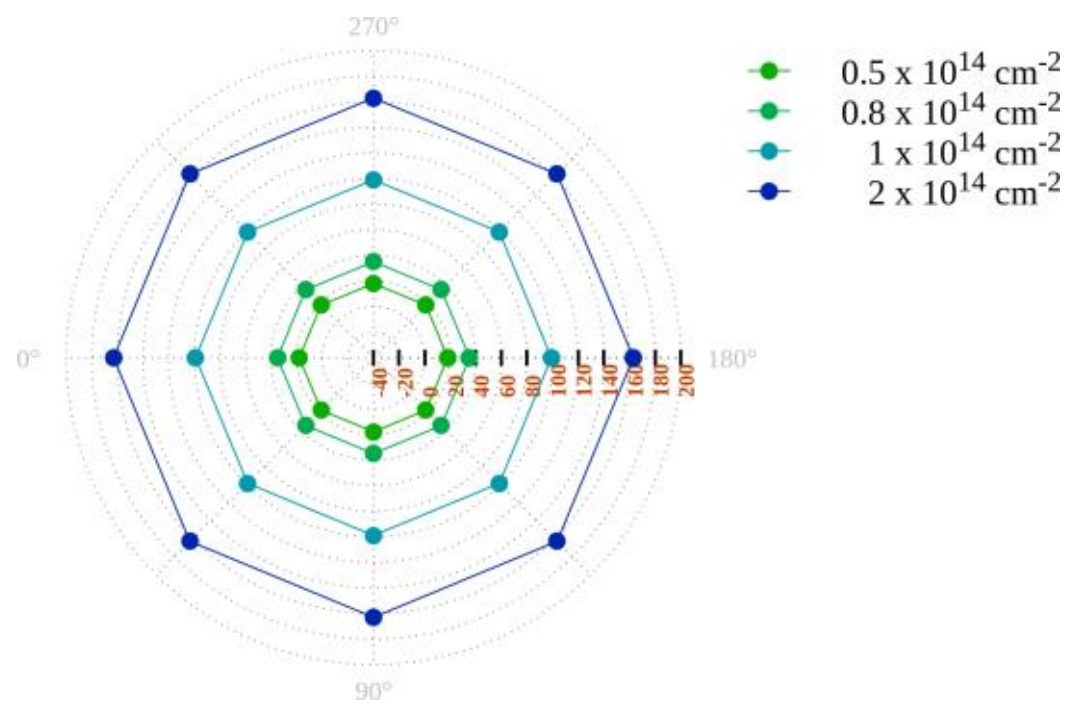

MAE = $E_{\parallel} - E_{\perp}$, a positive (negative) value of MAE indicates the off-plane (in-plane) easy axis.

## Monte Carlo simulations of the normalized magnetization of as a function of temperature

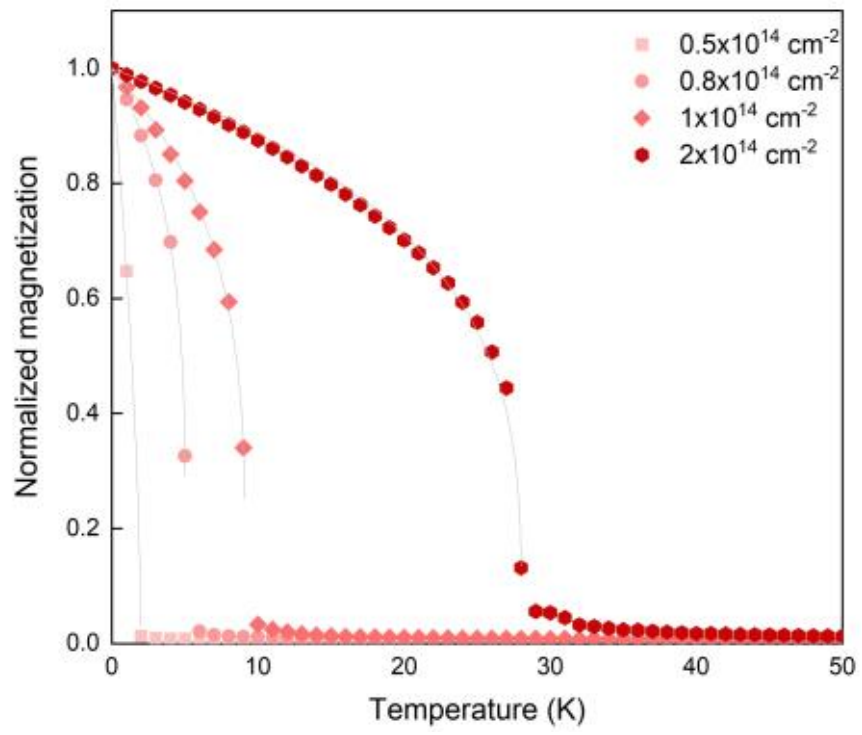

$T_c$: $1\times10^{14}$ cm$^{-2}$: 2 K  $2\times10^{14}$ cm$^{-2}$: 5 K

$4\times10^{14}$ cm$^{-2}$: 9 K  $6\times10^{14}$ cm$^{-2}$: 28 K

# 38. $SnS_2$

| MC2D-ID | C2DB | 2dmat-ID | USPEX | Space group | Band gap (eV) |
|---|---|---|---|---|---|
| 184 | ✓ | 2dm-3203 | - | P3m1 | 1.58 |

## Convex hull

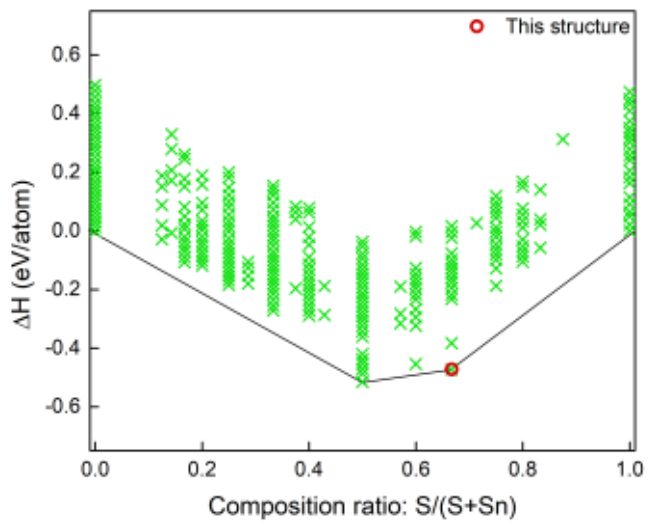

## Atomic structure

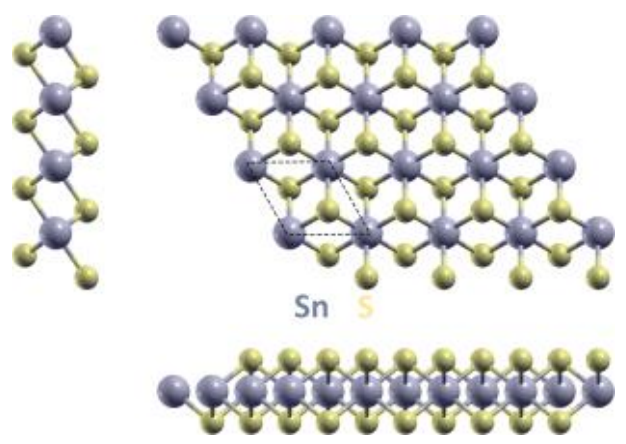

## Atomic coordinates

```
Sn1S2
   1.0000000000000000
   3.6986775778313943    0.0000000000000000    0.0000000000000000
  -1.8493394205076867    3.2031503712734675    0.0000000000000000
   0.0000000000000000    0.0000000000000000   22.9654742180999989
   Sn    S
   1     2
Direct
  0.0000000020600694  0.9999999885573345  0.5000000109287441
  0.3333333536954370  0.6666666786313584  0.4357795055875542
  0.6666666442172726  0.3333333328113000  0.5642204834815274
```

## Phonon dispersion curve

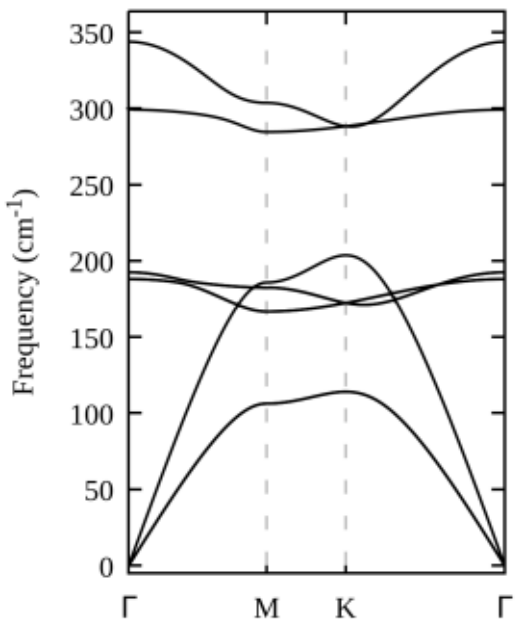

## Projected band structure and density of states

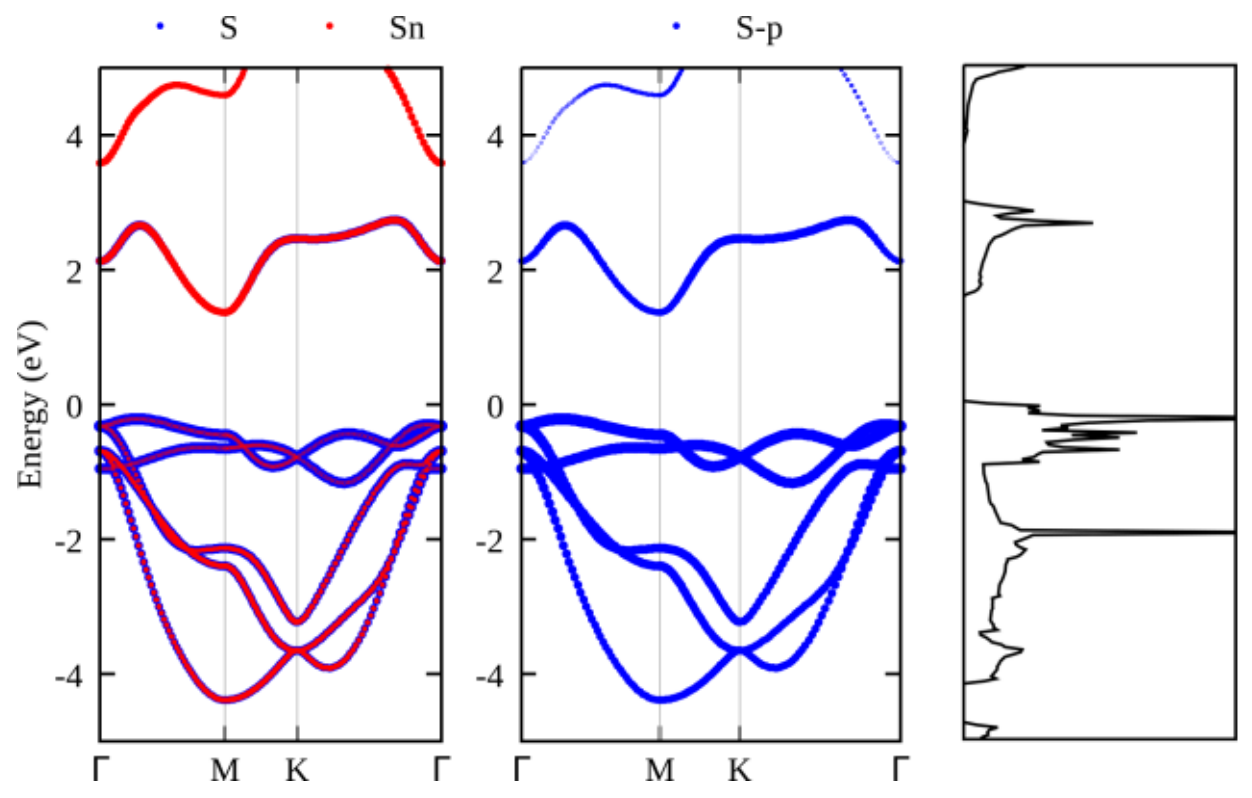

## Magnetic moment and spin polarization energy as a function of hole doping concentration

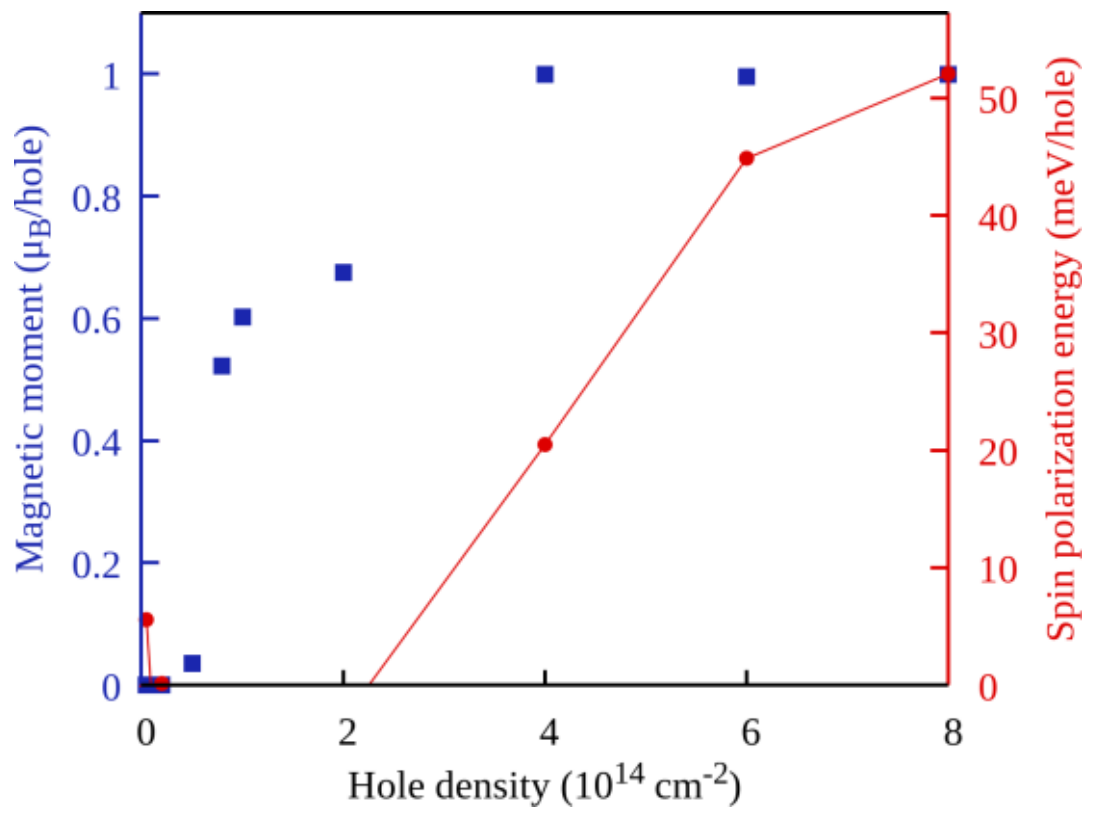

## Magnetic configurations and spin Hamiltonian

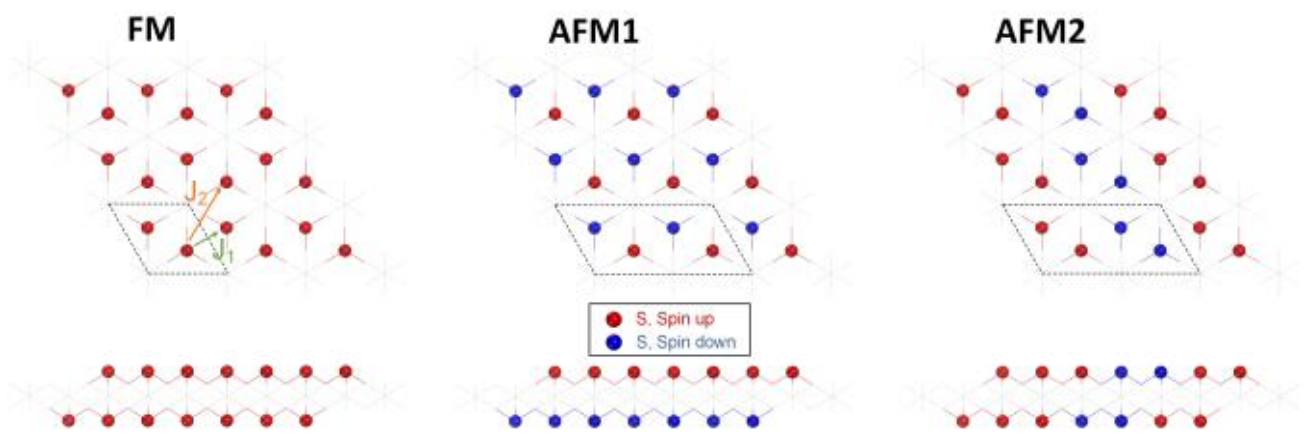

$$E_{FM} = E_0 - 3J_1S^2 - 6J_2S^2$$
$$E_{AFM1} = E_0 + 3J_1S^2 - 6J_2S^2$$
$$E_{AFM2} = E_0 - J_1S^2 + 2J_2S^2$$

## Magnetic exchange coupling parameters

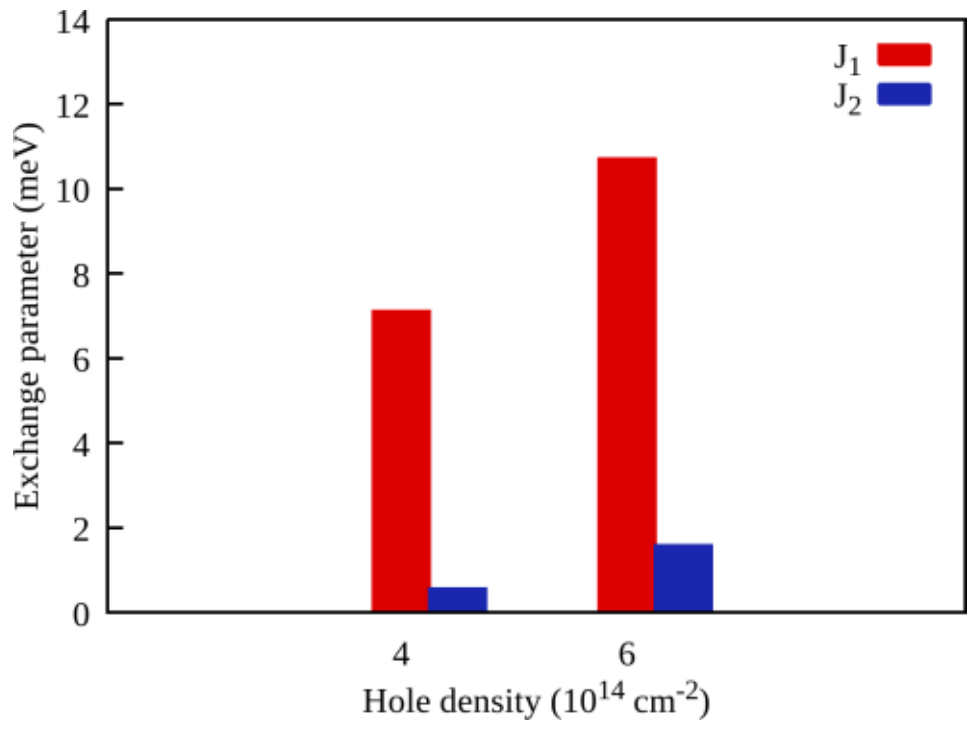

## Magnetic anisotropy energy (MAE, μeV) per magnetic atom

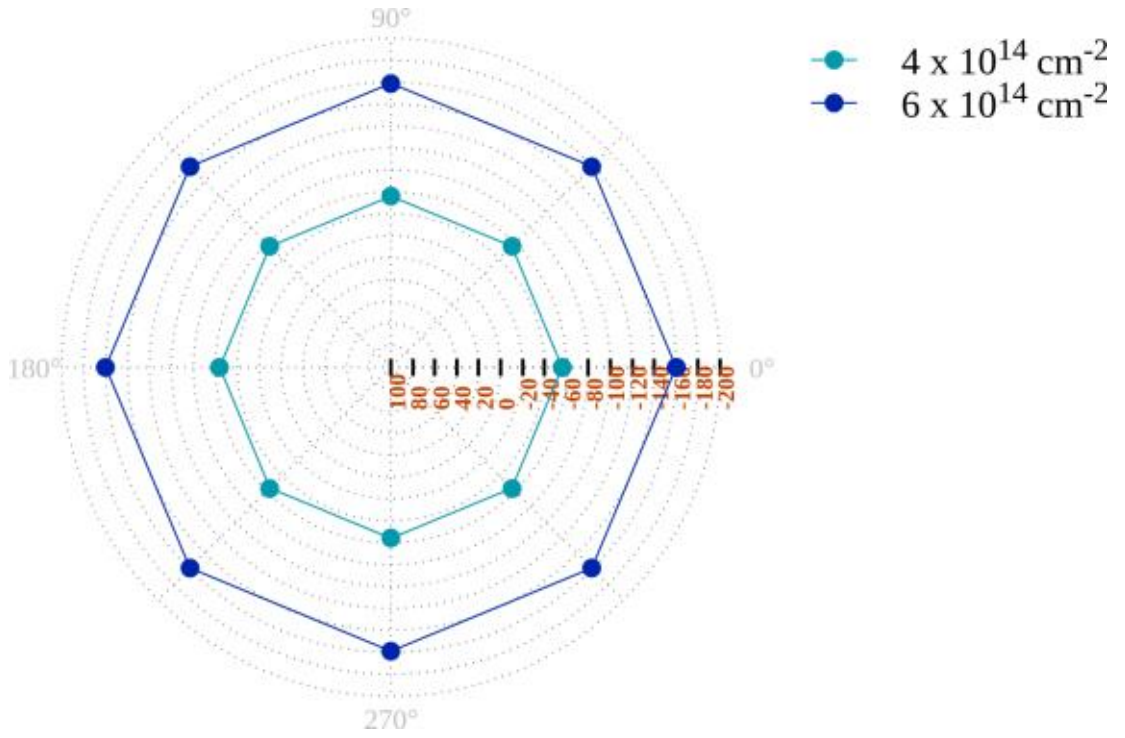

MAE = $E_{\parallel} - E_{\perp}$, a positive (negative) value of MAE indicates the off-plane (in-plane) easy axis.

## Monte Carlo simulations of the normalized magnetization of as a function of temperature

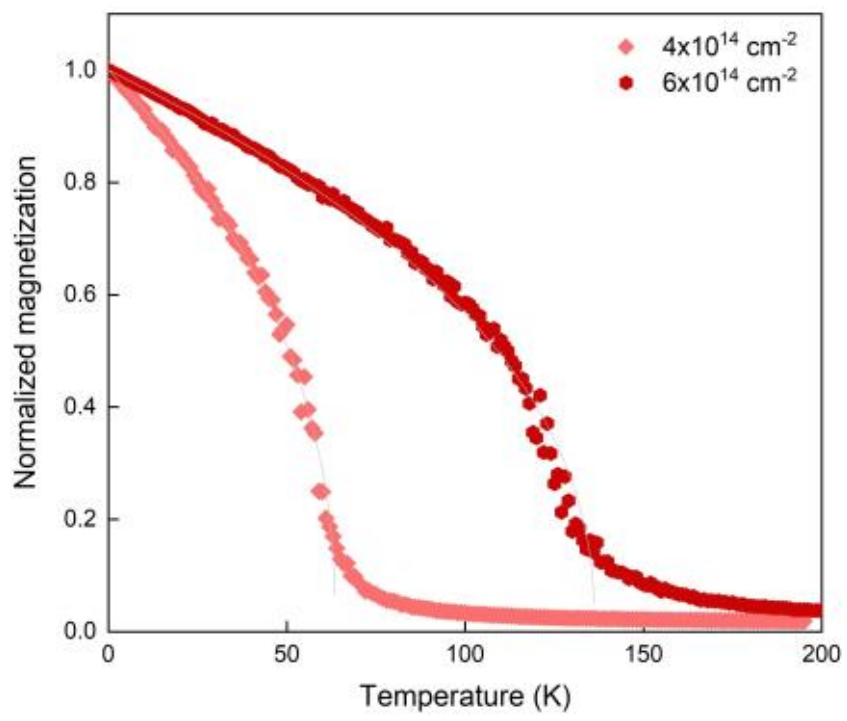

$T_c$: $1\times10^{14}$ cm$^{-2}$: - K  $2\times10^{14}$ cm$^{-2}$: - K

$4\times10^{14}$ cm$^{-2}$: 64 K  $6\times10^{14}$ cm$^{-2}$: 136 K

# 39. $PbS_2$

| MC2D-ID | C2DB | 2dmat-ID | USPEX | Space group | Band gap (eV) |
|---|---|---|---|---|---|
| - | ✓ | 2dm-773 | - | P3m1 | 0.72 |

**Convex hull** | **Atomic structure** | **Atomic coordinates** | **Phonon dispersion curve**

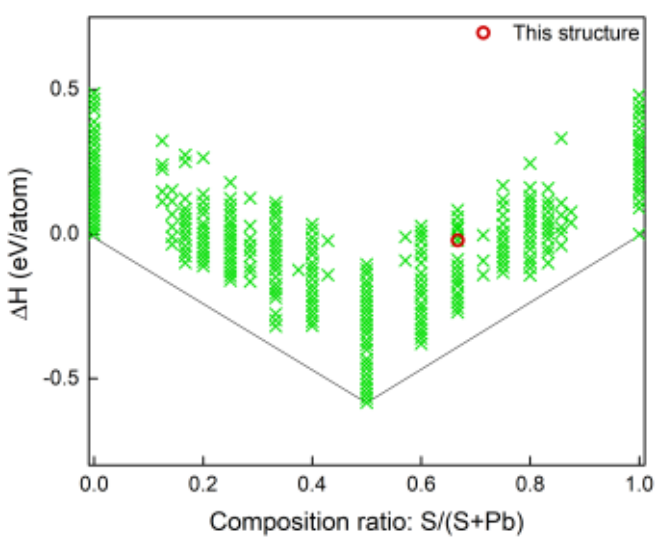

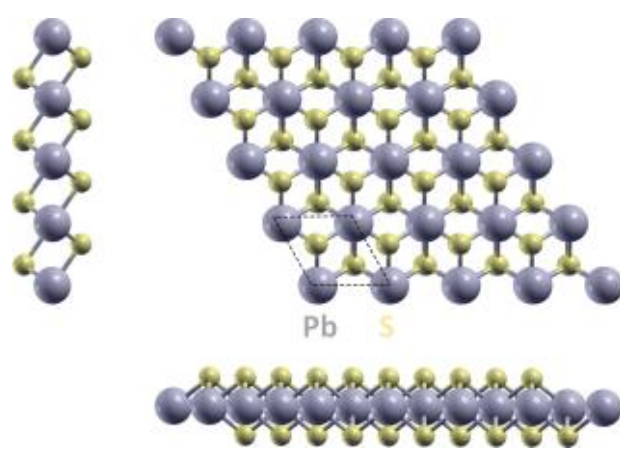

```
Pb1S2
   1.00000000000000
   3.8539737416739341    0.0000000000000000    0.0000000000000000
  -1.9269871390013520    3.3376389717959429    0.0000000000000000
   0.0000000000000000    0.0000000000000000   23.1928920000000005
   Pb    S
   1     2
Direct
  0.0000000094714636  0.0000000083801268  0.9999999986081747
  0.3333331807494631  0.6666667919214007  0.0665891499451021
  0.6666668097790733  0.3333331996984725  0.9334108514467232
```

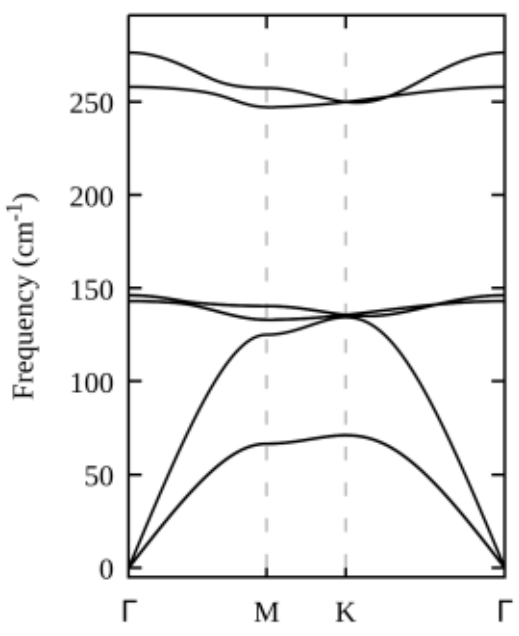

**Projected band structure and density of states** | **Magnetic moment and spin polarization energy as a function of hole doping concentration**

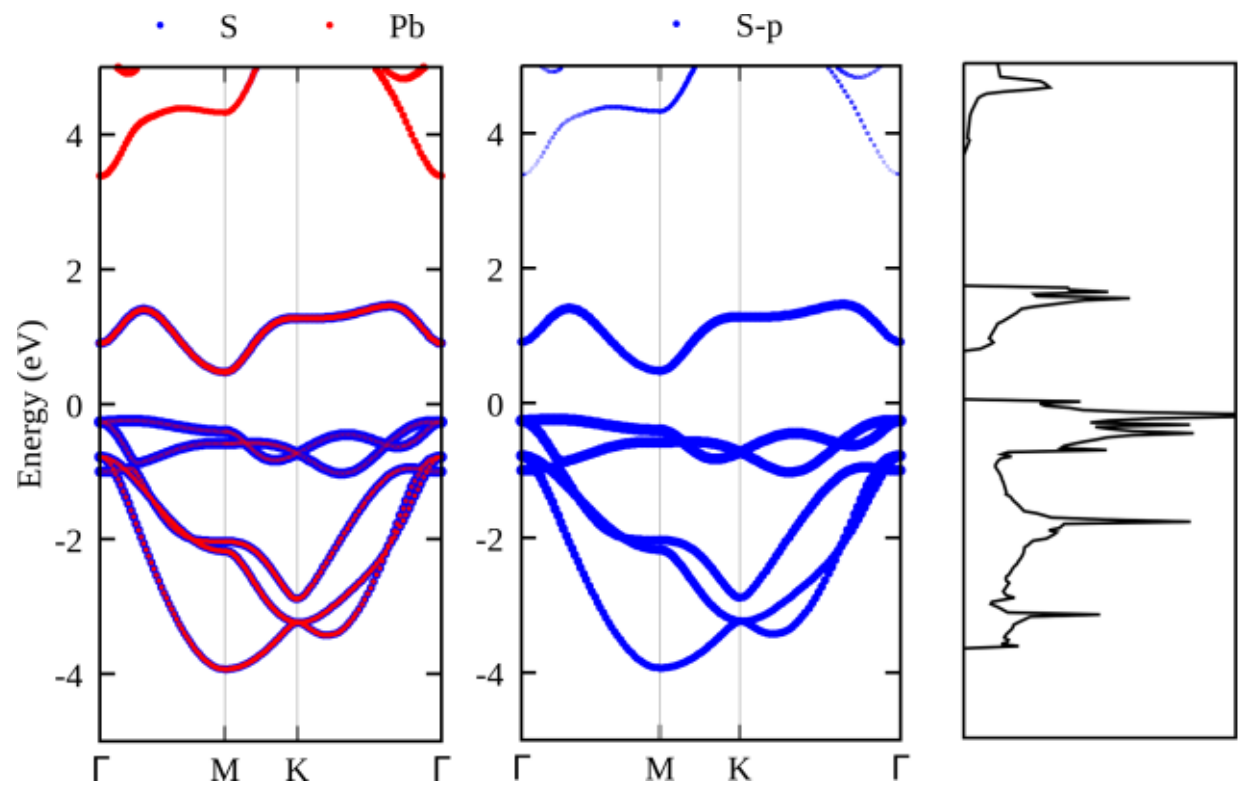

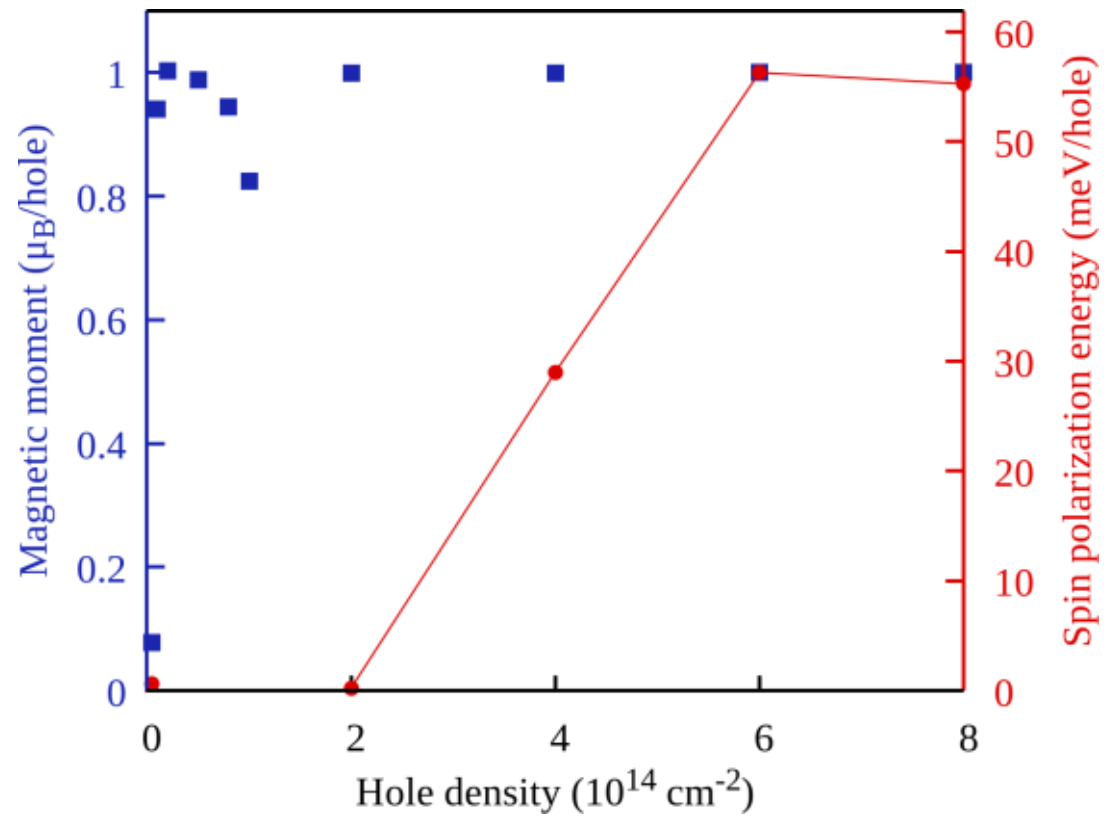

**Magnetic configurations and spin Hamiltonian** | **Magnetic exchange coupling parameters**

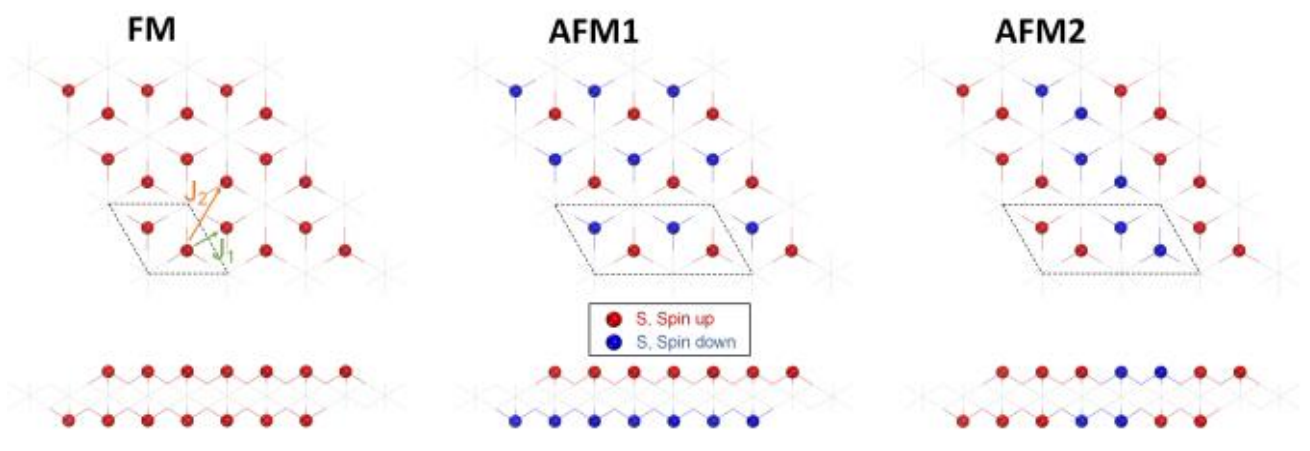

$$E_{FM} = E_0 - 3J_1S^2 - 6J_2S^2$$
$$E_{AFM1} = E_0 + 3J_1S^2 - 6J_2S^2$$
$$E_{AFM2} = E_0 - J_1S^2 + 2J_2S^2$$

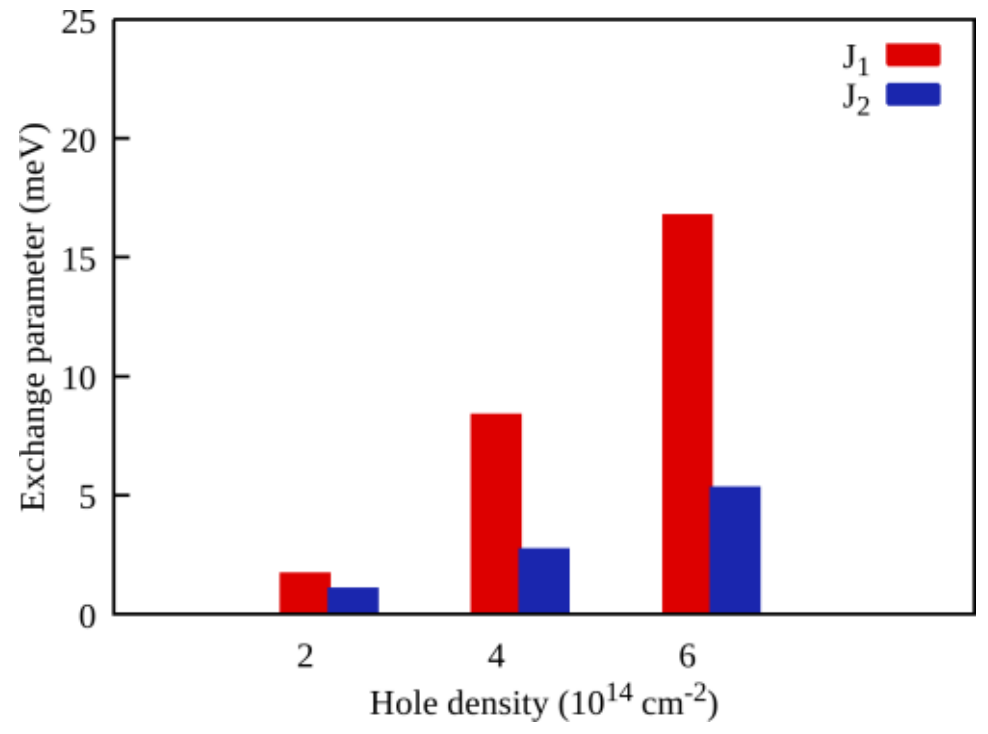

**Magnetic anisotropy energy (MAE, μeV) per magnetic atom** | **Monte Carlo simulations of the normalized magnetization of as a function of temperature**

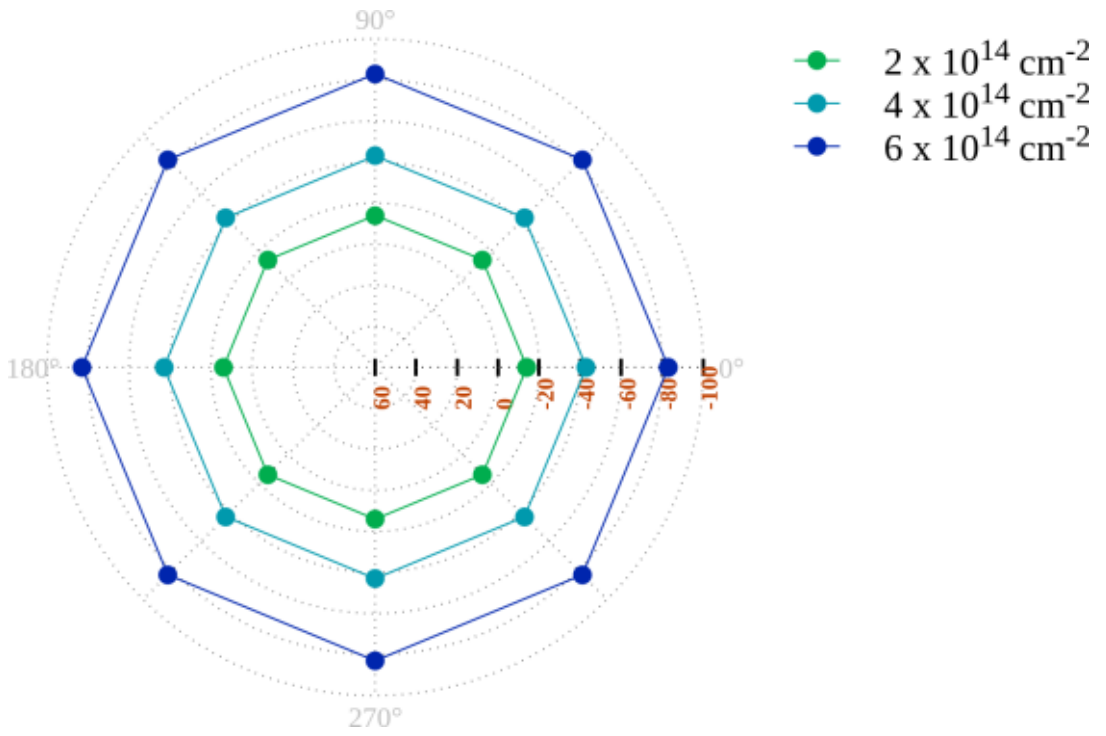

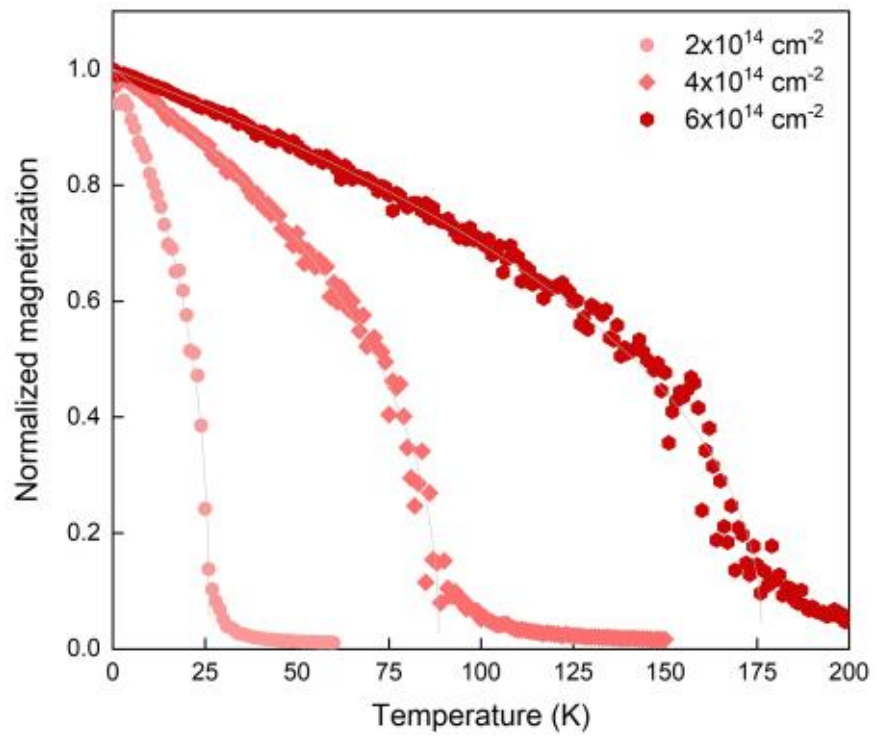

MAE = $E_{\parallel} - E_{\perp}$, a positive (negative) value of MAE indicates the off-plane (in-plane) easy axis.

$T_c$: $1\times10^{14}$ $cm^{-2}$: - K  $2\times10^{14}$ $cm^{-2}$: 26 K  $4\times10^{14}$ $cm^{-2}$: 88 K  $6\times10^{14}$ $cm^{-2}$: 176 K

# 40. $Al_2O_3$

| MC2D-ID | C2DB | 2dmat-ID | USPEX | Space group | Band gap (eV) |
|---|---|---|---|---|---|
| - | - | 2dm-2031 | - | P3m1 | 5.06 |

**Convex hull**

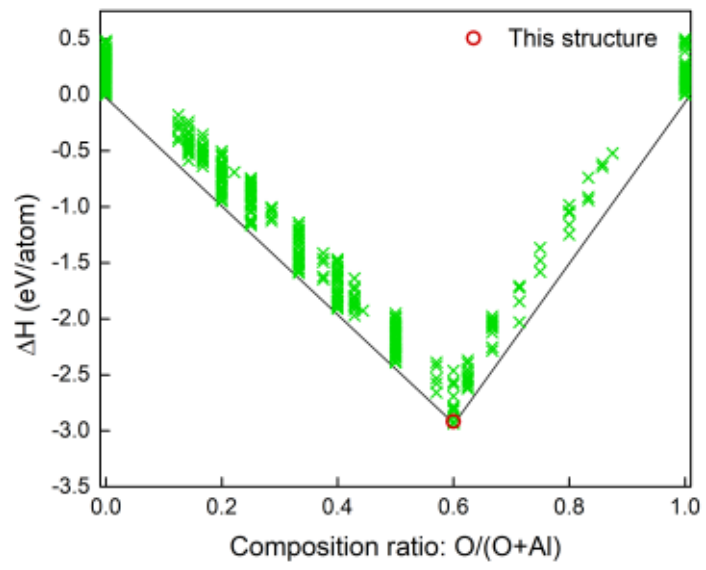

**Atomic structure**

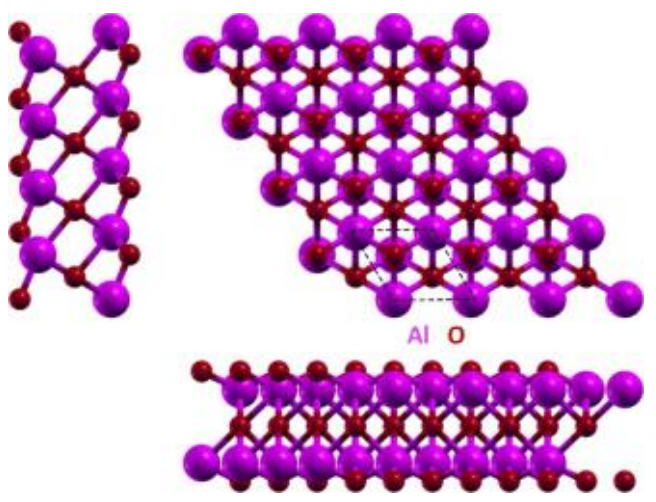

**Atomic coordinates**

```
Al2 O3
   1.00000000000000
   2.8880760942522561   0.0000000000000000   0.0000000000000000
  -1.4440323919937297   2.5011505816539961   0.0000000000000000
   0.0000000000000000   0.0000000000000000  24.6292340000000003
   Al   O
   2    3
Direct
  0.9999991221962148  0.0000009400556067  0.6275054727579317
  0.3333338851025545  0.6666660406841700  0.5156122354868558
  0.3333333154995515  0.6666667389266365  0.6550782369124093
  0.9999997577686770  0.0000001839094494  0.4880375566537793
  0.6666669194329984  0.3333330964241341  0.5715634981890148
```

**Phonon dispersion curve**

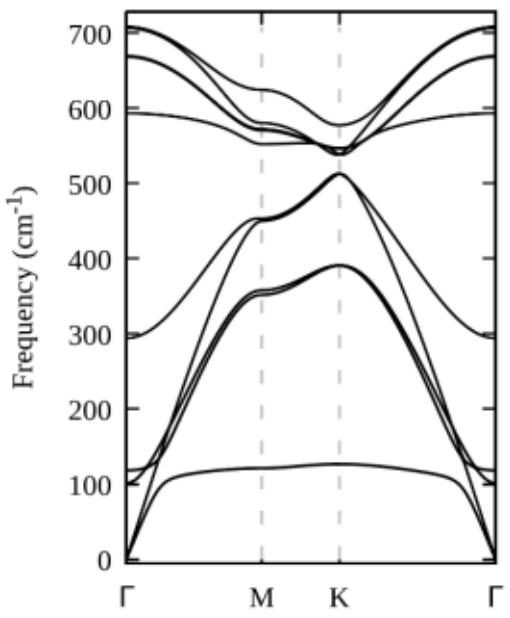

**Projected band structure and density of states**

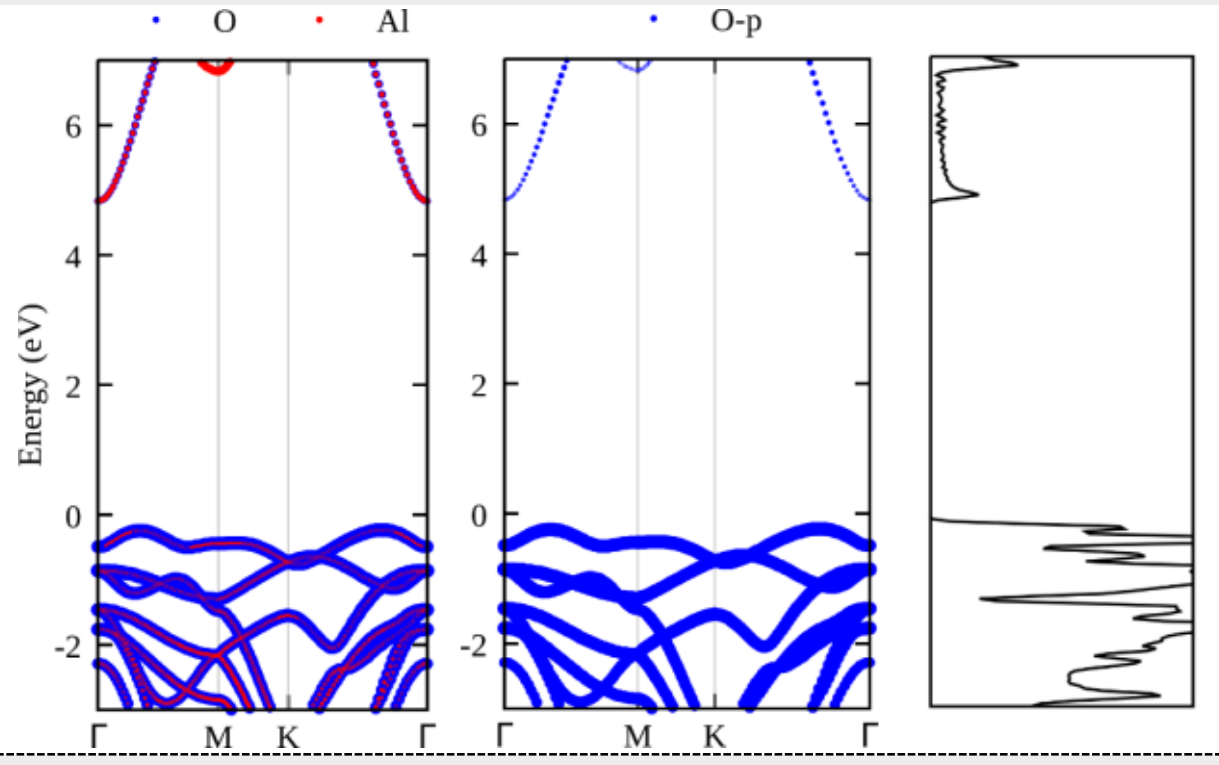

**Magnetic moment and spin polarization energy as a function of hole doping concentration**

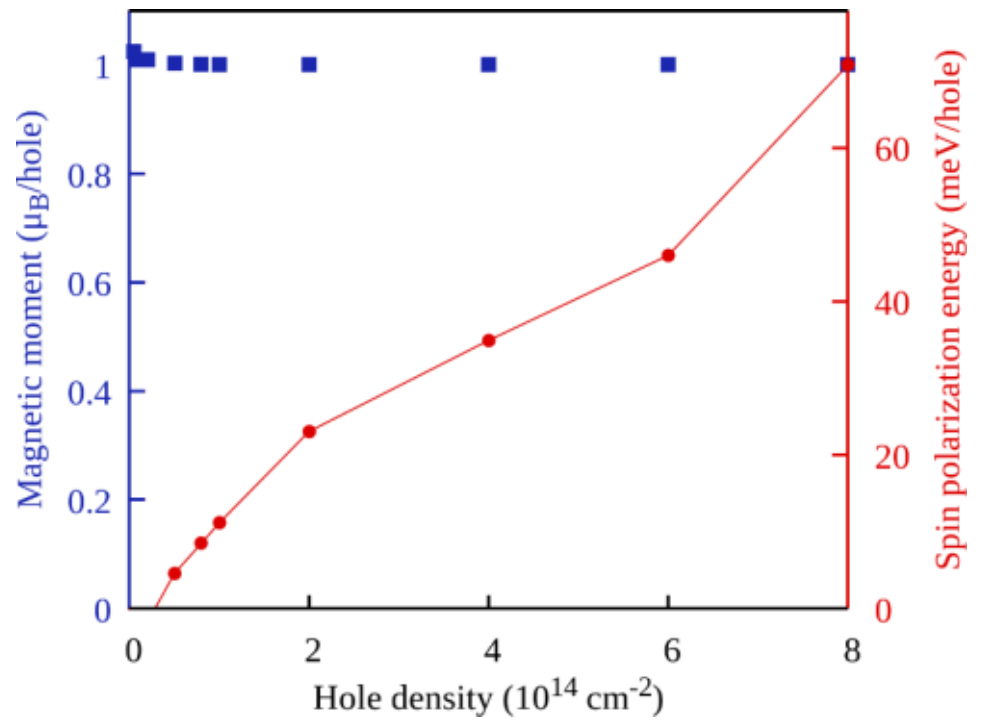

**Magnetic configurations and spin Hamiltonian**

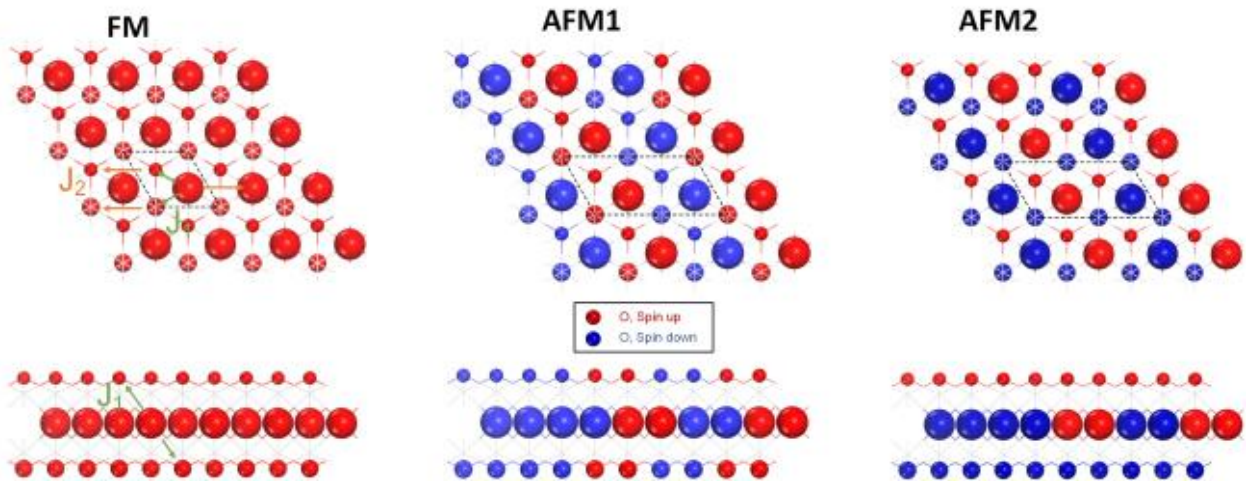

$$E_{FM} = E_0 - 6J_1S^2 - 9J_2S^2$$

$$E_{AFM1} = E_0 + 3J_2S^2$$

$$E_{AFM2} = E_0 - J_1S^2 - 5J_2S^2$$

**Magnetic exchange coupling parameters**

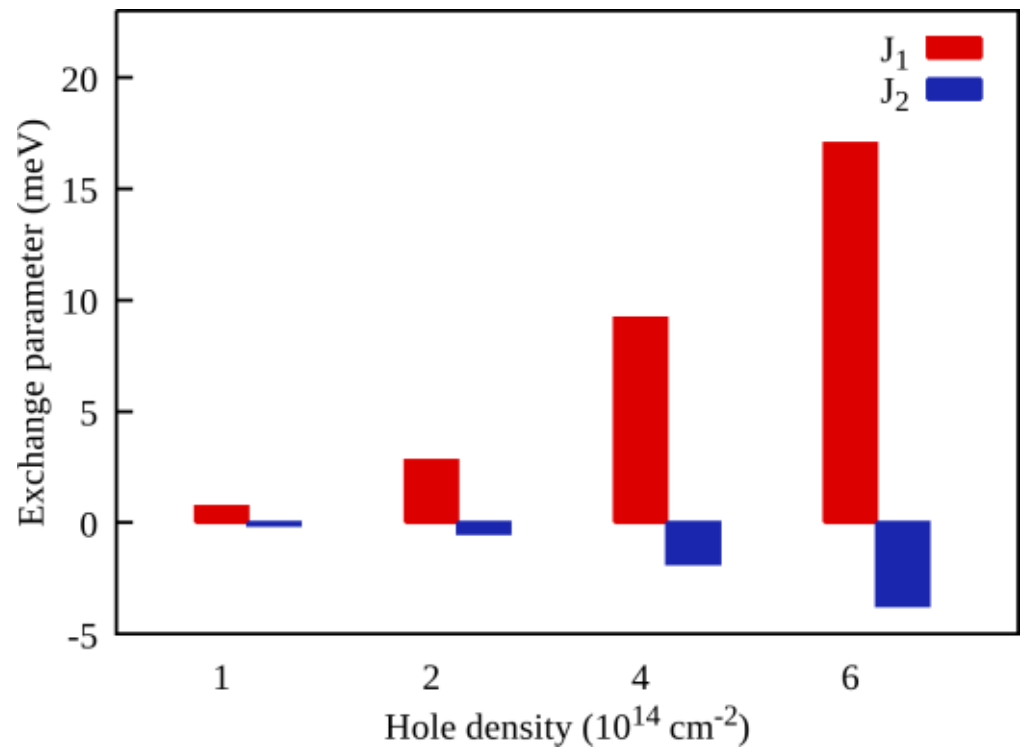

**Magnetic anisotropy energy (MAE, μeV) per magnetic atom**

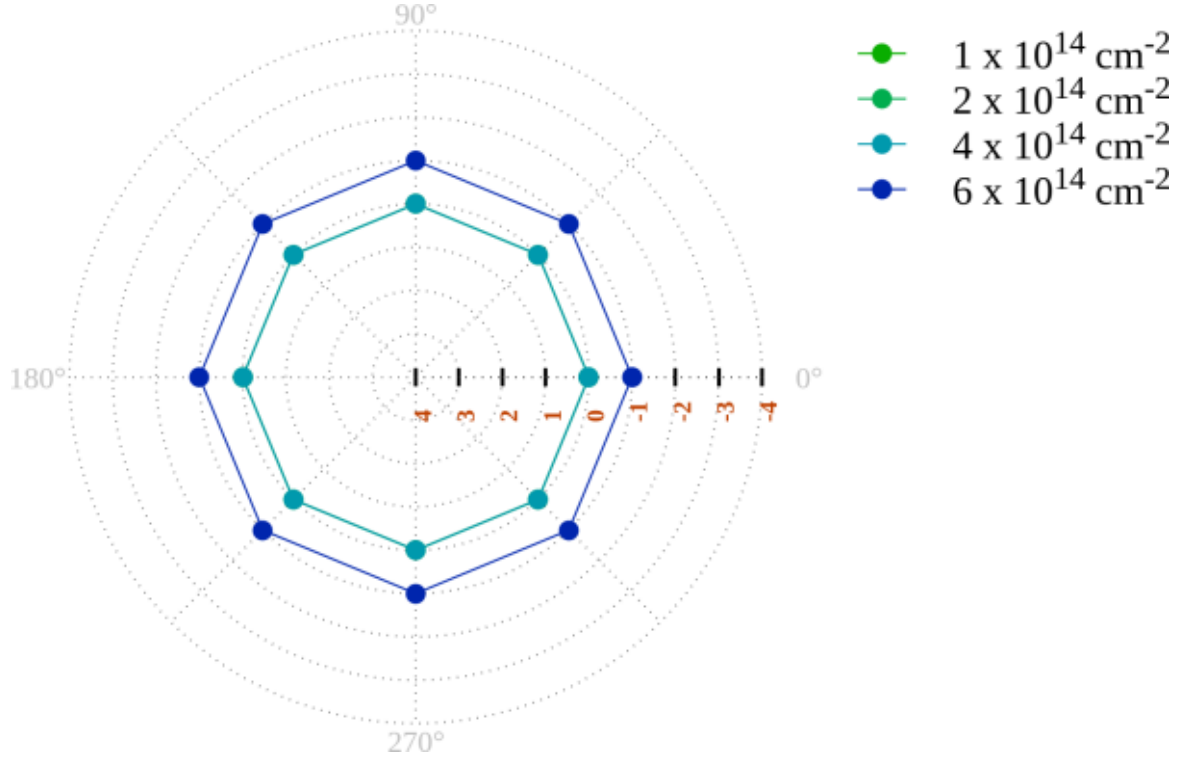

MAE = $E_{\parallel} - E_{\perp}$, a positive (negative) value of MAE indicates the off-plane (in-plane) easy axis.

**Monte Carlo simulations of the normalized magnetization of as a function of temperature**

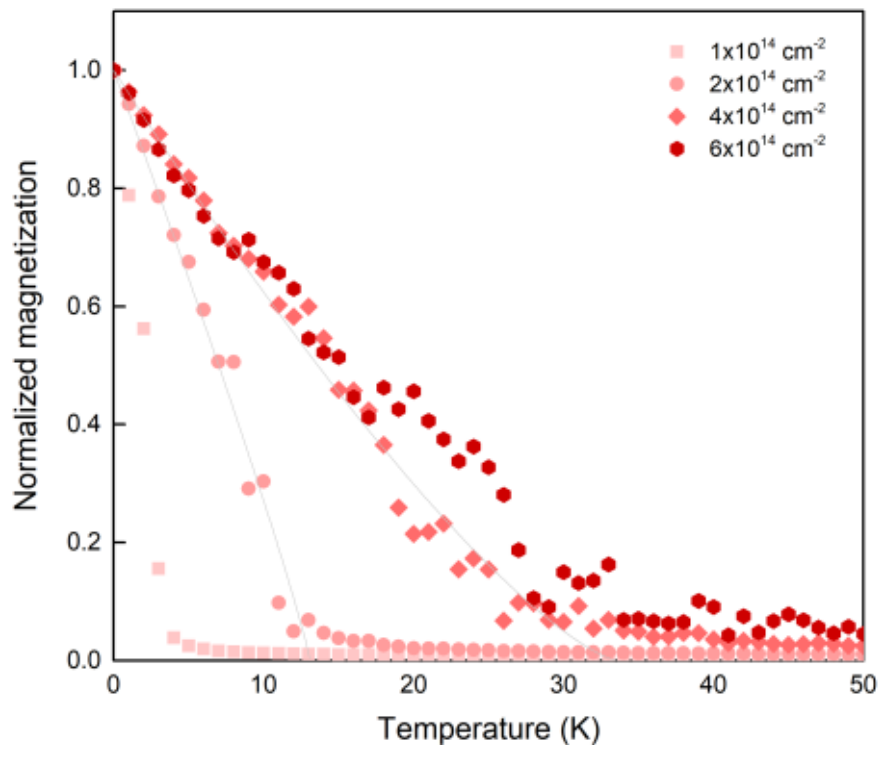

$T_c$: $1\times10^{14}$ cm$^{-2}$: 5 K  $2\times10^{14}$ cm$^{-2}$: 11 K

$4\times10^{14}$ cm$^{-2}$: 26 K  $6\times10^{14}$ cm$^{-2}$: 34 K

# 41. $GeO_2$

| MC2D-ID | C2DB | 2dmat-ID | USPEX | Space group | Band gap (eV) |
|---|---|---|---|---|---|
| - | ✓ | 2dm-6440 | ✓ | P3m1 | 3.59 |

**Convex hull** | **Atomic structure** | **Atomic coordinates** | **Phonon dispersion curve**

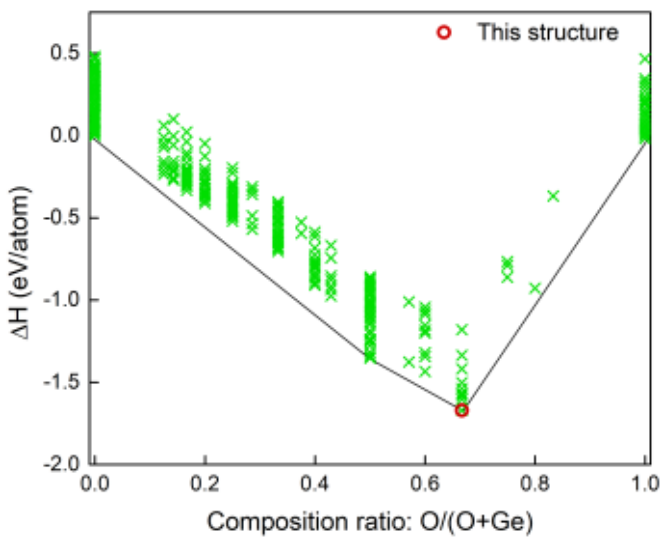

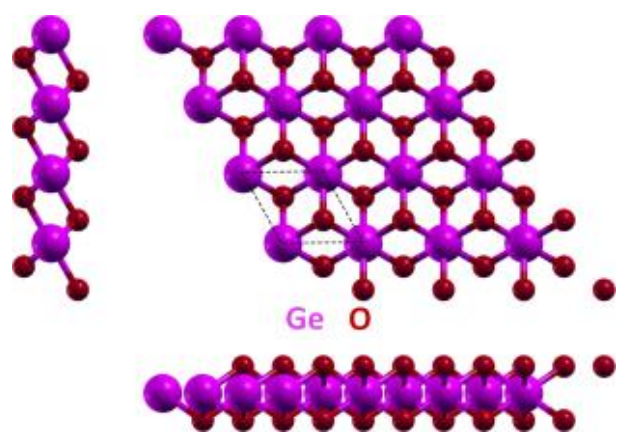

```
Ge1O2
   1.00000000000000
   1.4533940999104427    2.5173465541177111    0.0000000000000000
  -1.4533932615547911    2.5173466481221562    0.0000000000000000
   0.0000000000000000    0.0000000000000000   20.0000000000000000
   Ge    O
   1     2
Direct
  0.9999999915906770  0.9999999726070286  0.4999999723726987
  0.3333330089489692  0.3333331709755214  0.5488004338288164
  0.6666669994603538  0.6666668564174500  0.4511995937984778
```

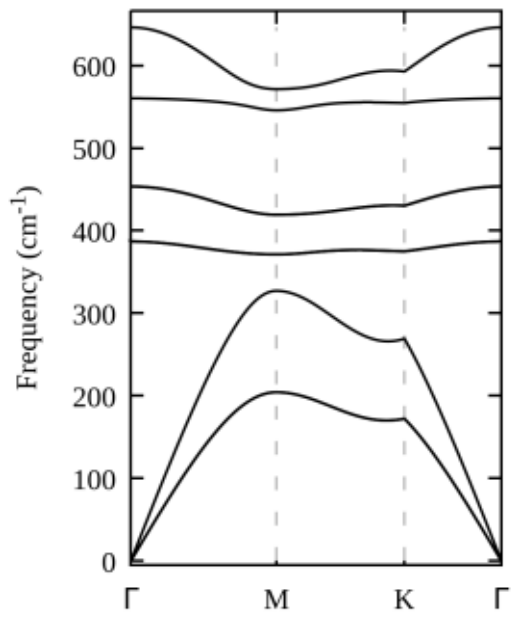

**Projected band structure and density of states** | **Magnetic moment and spin polarization energy as a function of hole doping concentration**

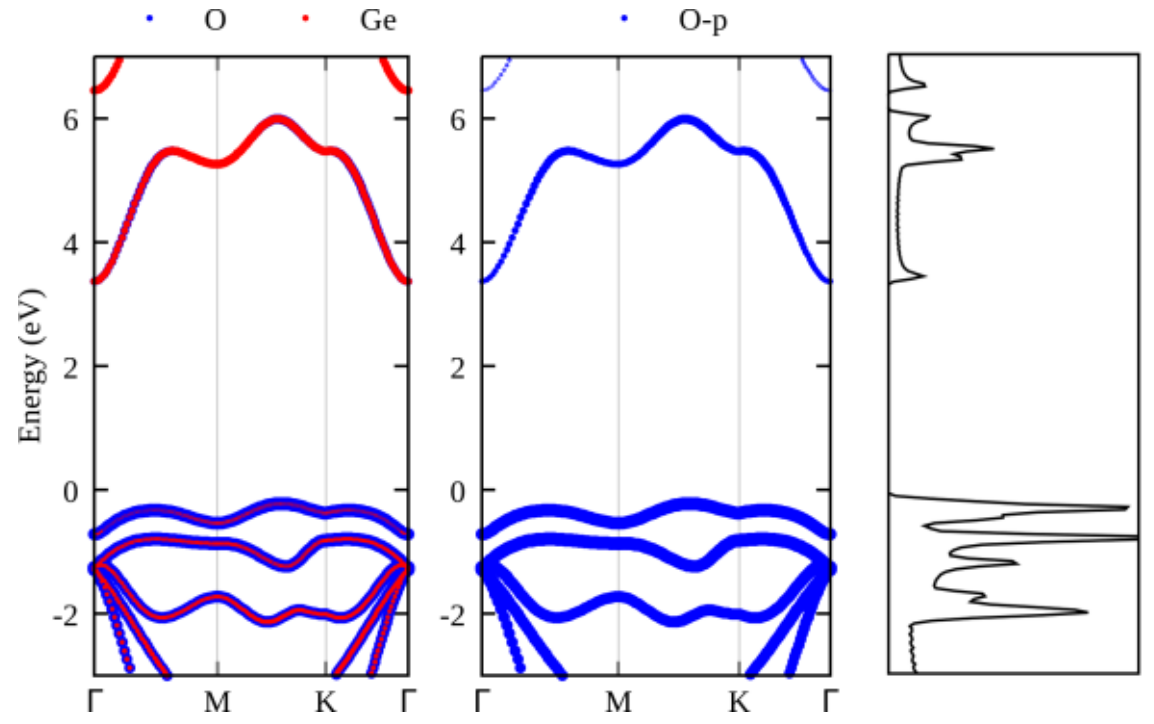

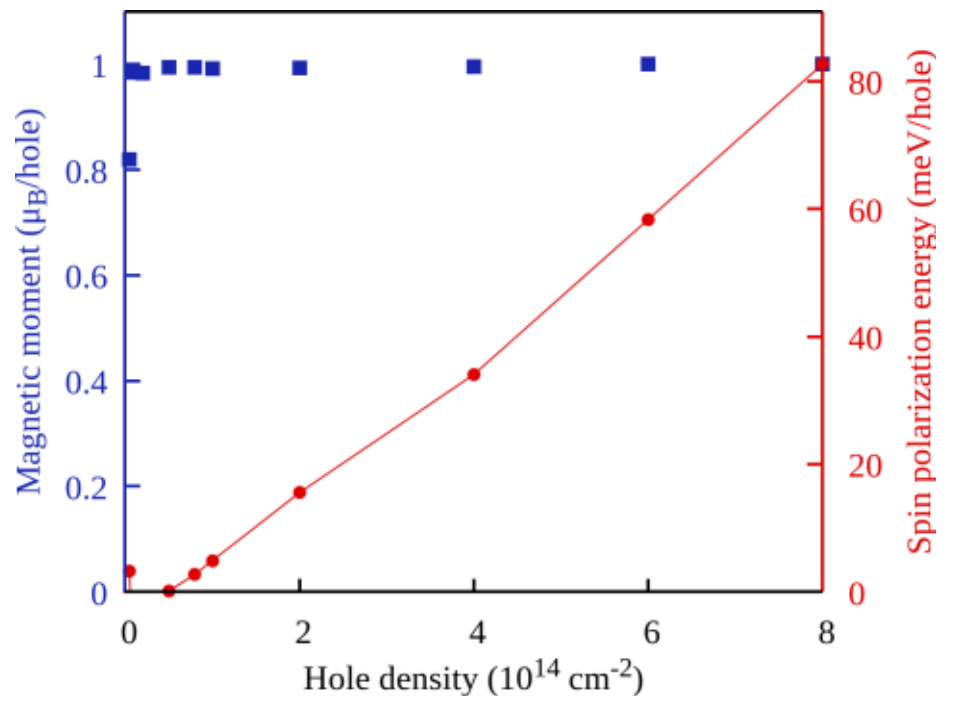

**Magnetic configurations and spin Hamiltonian** | **Magnetic exchange coupling parameters**

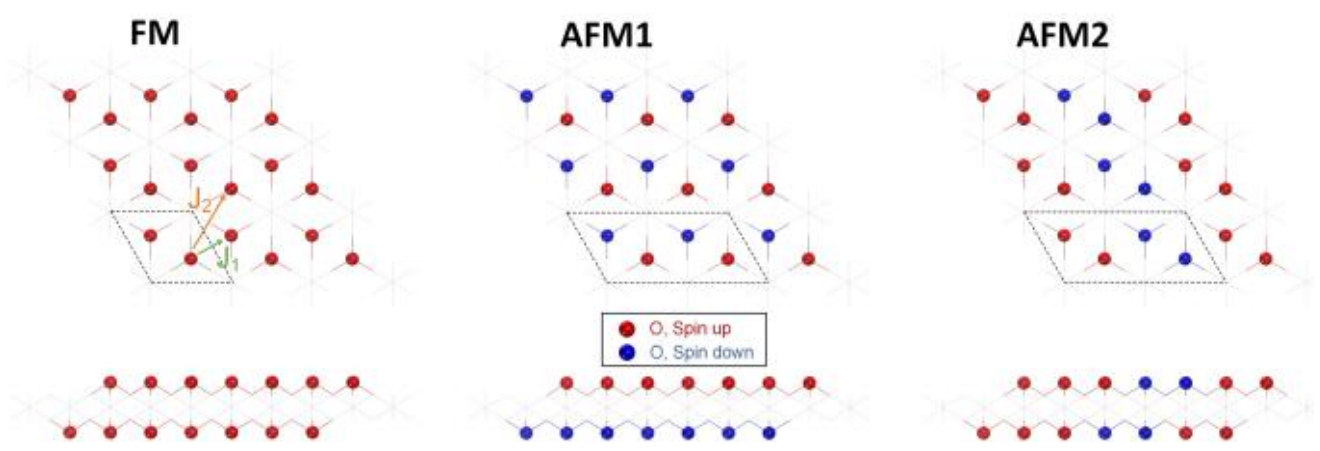

$$E_{FM} = E_0 - 3J_1S^2 - 6J_2S^2$$
$$E_{AFM1} = E_0 + 3J_1S^2 - 6J_2S^2$$
$$E_{AFM2} = E_0 - J_1S^2 + 2J_2S^2$$

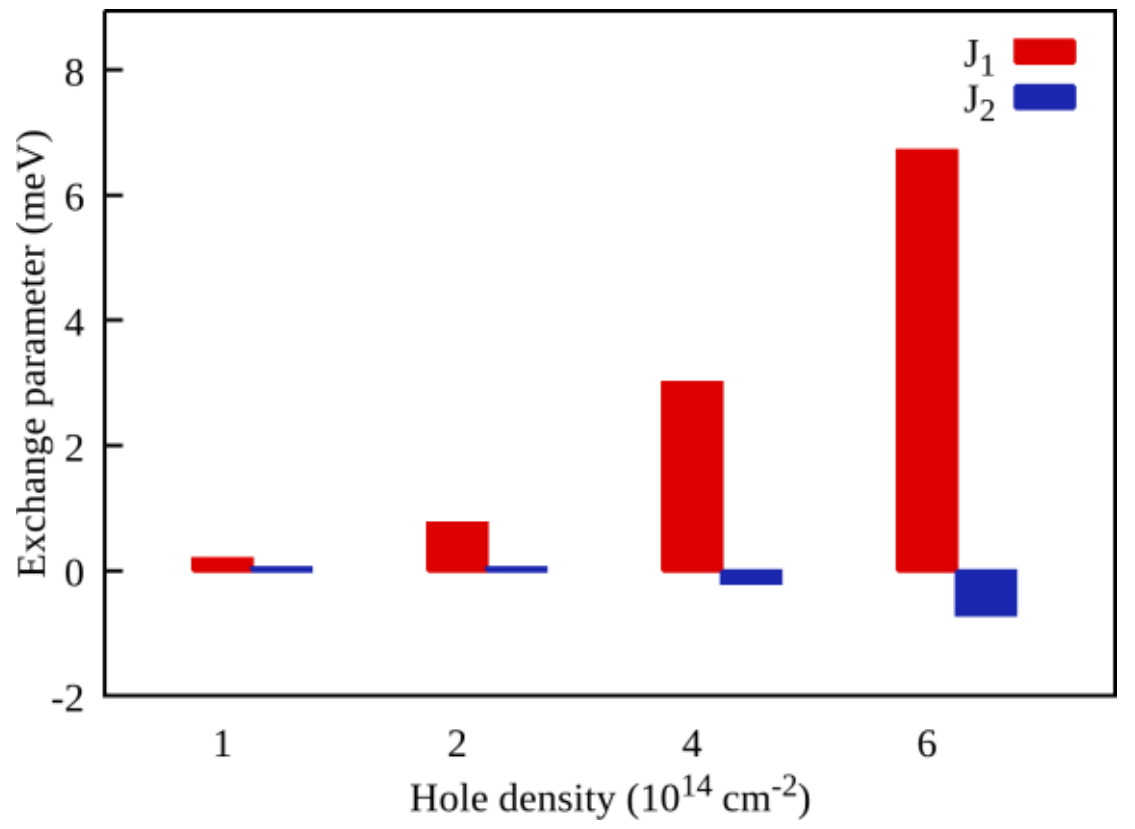

**Magnetic anisotropy energy (MAE, μeV) per magnetic atom** | **Monte Carlo simulations of the normalized magnetization of as a function of temperature**

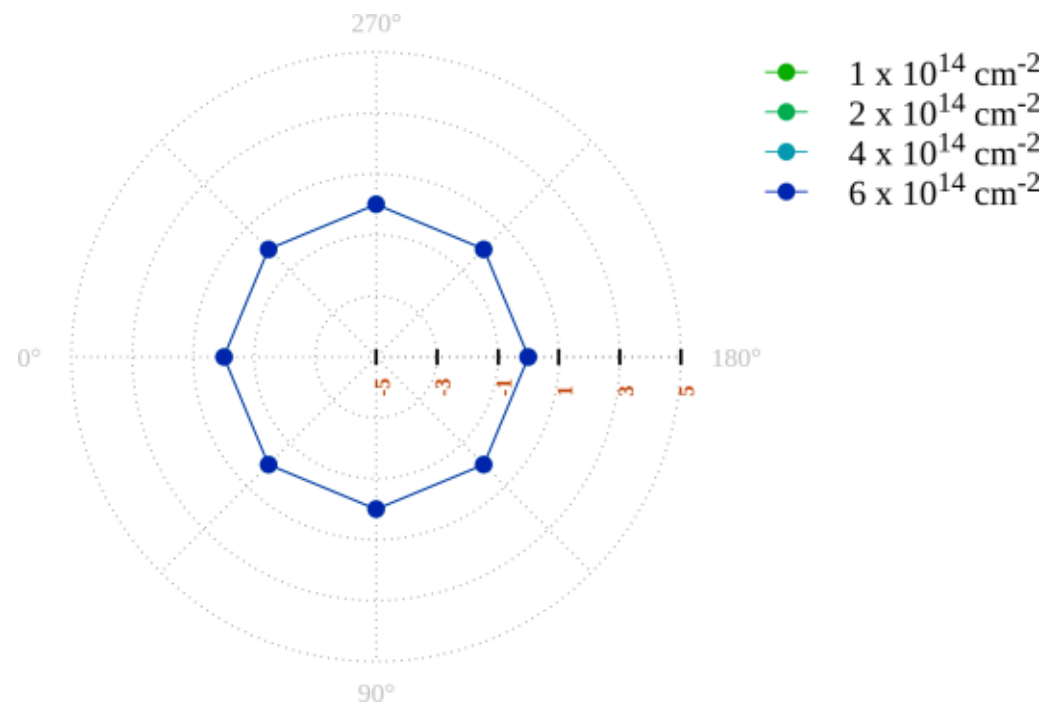

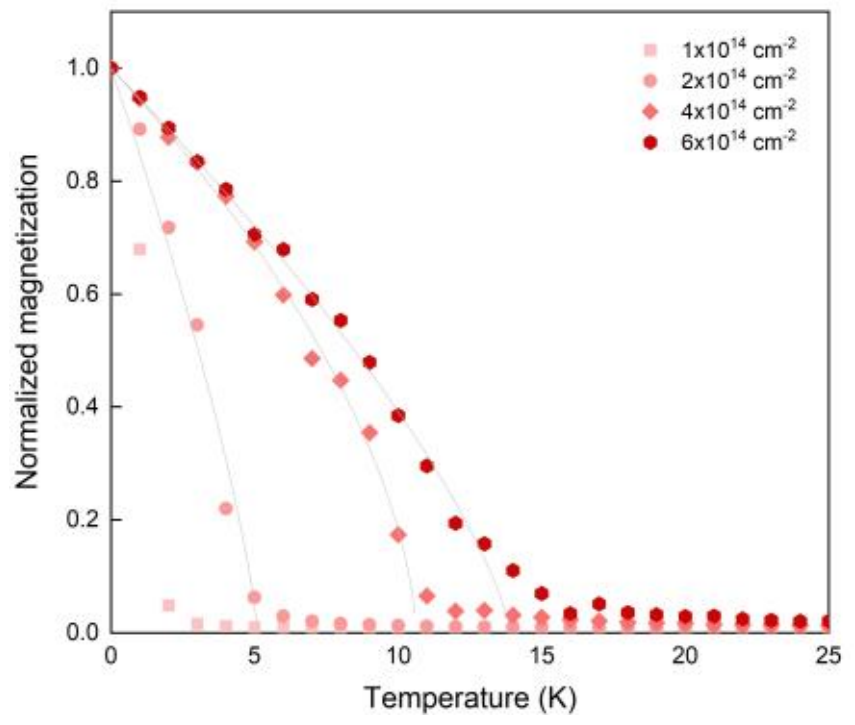

MAE = $E_{\parallel} - E_{\perp}$, a positive (negative) value of MAE indicates the off-plane (in-plane) easy axis.

$T_c$: $1\times10^{14}$ cm$^{-2}$: 2 K  $2\times10^{14}$ cm$^{-2}$: 5 K  $4\times10^{14}$ cm$^{-2}$: 11 K  $6\times10^{14}$ cm$^{-2}$: 14 K

# 42. $SnO_2$

| MC2D-ID | C2DB | 2dmat-ID | USPEX | Space group | Band gap (eV) |
|---|---|---|---|---|---|
| - | ✓ | 2dm-5126 | - | P3m1 | 2.58 |

**Convex hull** | **Atomic structure** | **Atomic coordinates** | **Phonon dispersion curve**

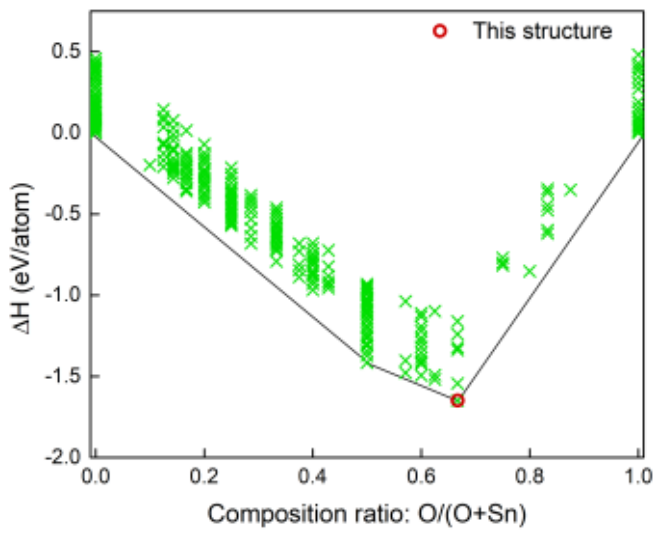

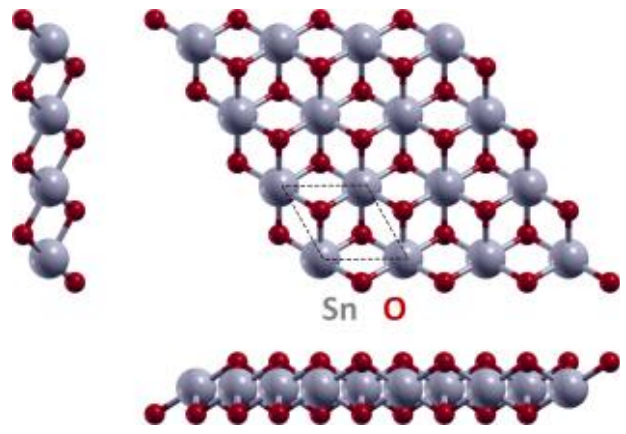

```
Sn1O2
1.00000000000000
3.2314916001454064 0.0000000000000000 0.0000000000000000
1.6157408761073357 2.7985563250660106 0.0000000000000000
0.0000000000000000 0.0000000000000000 22.0306889999999989
Sn O
1 2
Direct
0.6666666176169969 0.6666665950884152 0.4999997058801355
0.9999999018715400 0.9999999166747600 0.5460274559632978
0.3333334805114632 0.3333334882368177 0.4539728381565666
```

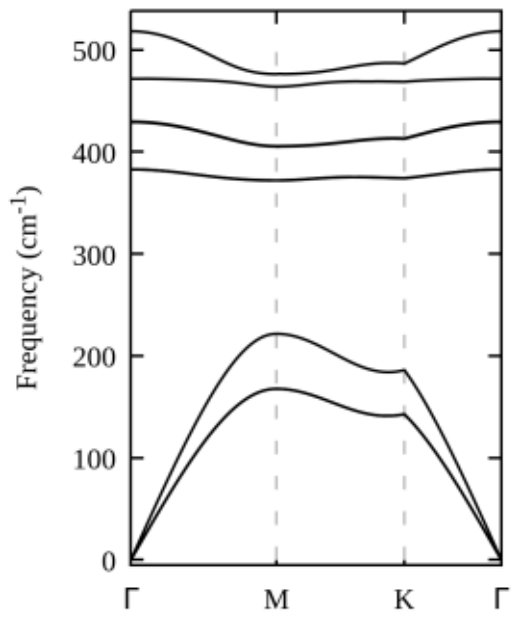

**Projected band structure and density of states** | **Magnetic moment and spin polarization energy as a function of hole doping concentration**

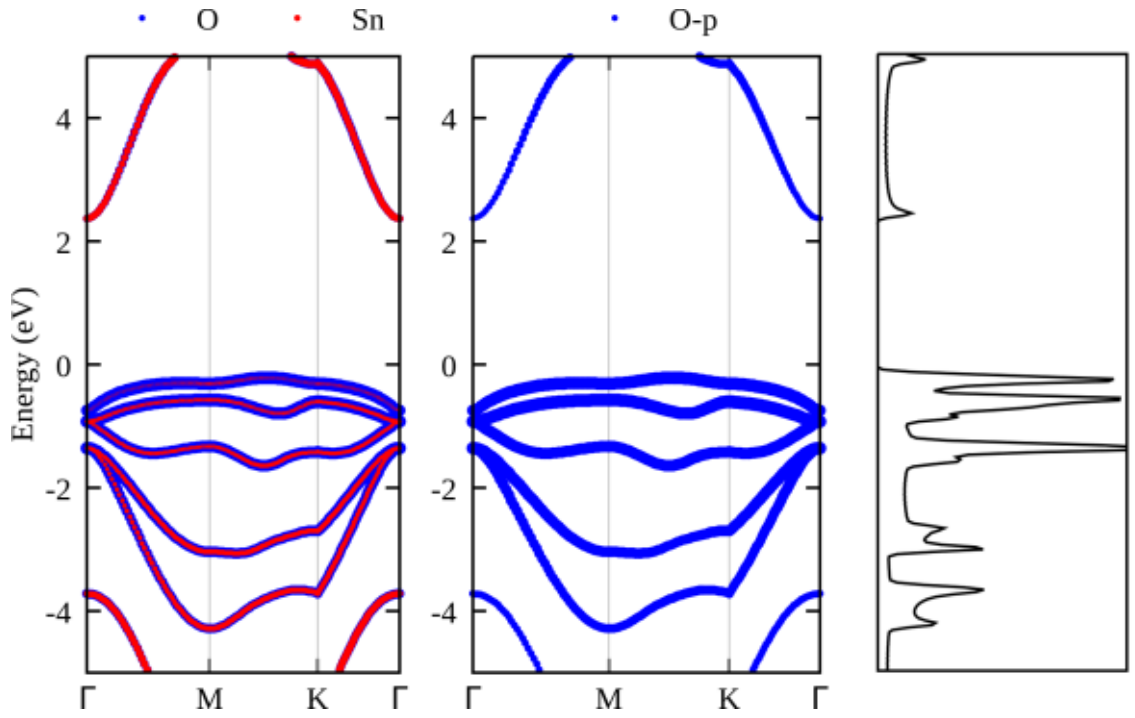

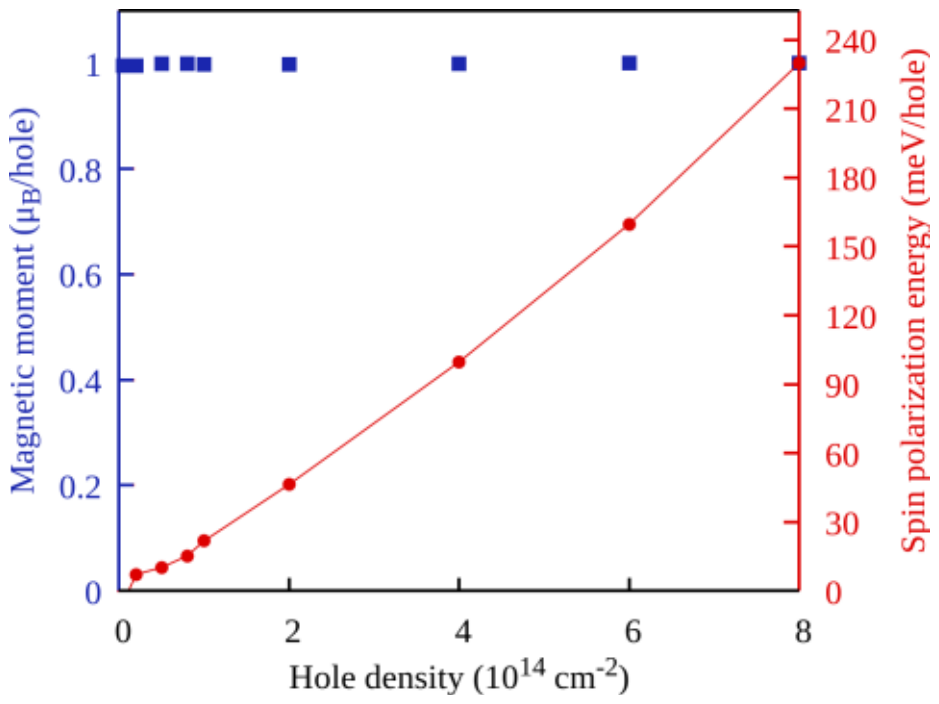

**Magnetic configurations and spin Hamiltonian** | **Magnetic exchange coupling parameters**

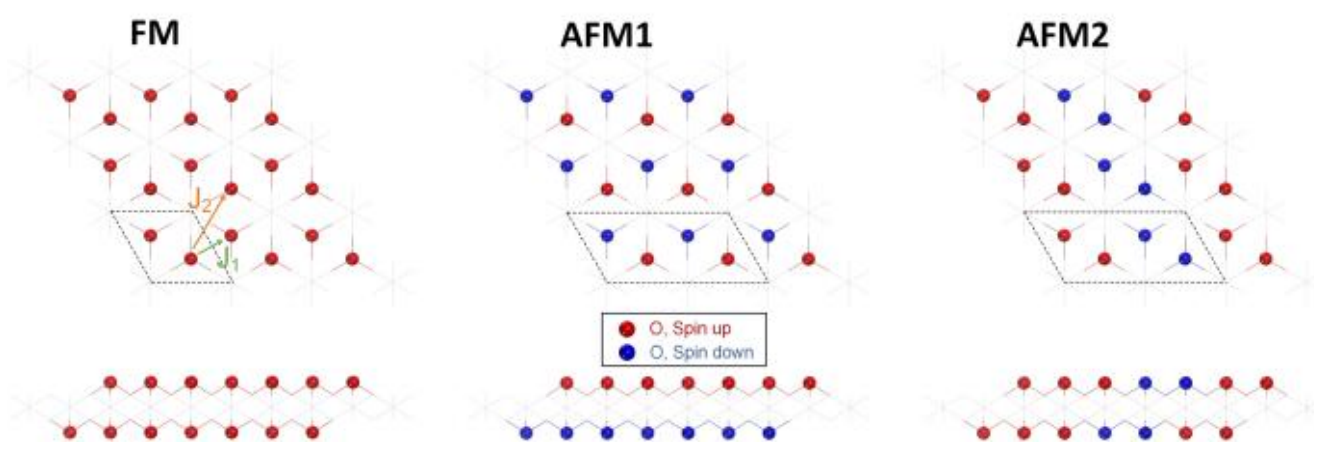

$$E_{FM} = E_0 - 3J_1S^2 - 6J_2S^2$$
$$E_{AFM1} = E_0 + 3J_1S^2 - 6J_2S^2$$
$$E_{AFM2} = E_0 - J_1S^2 + 2J_2S^2$$

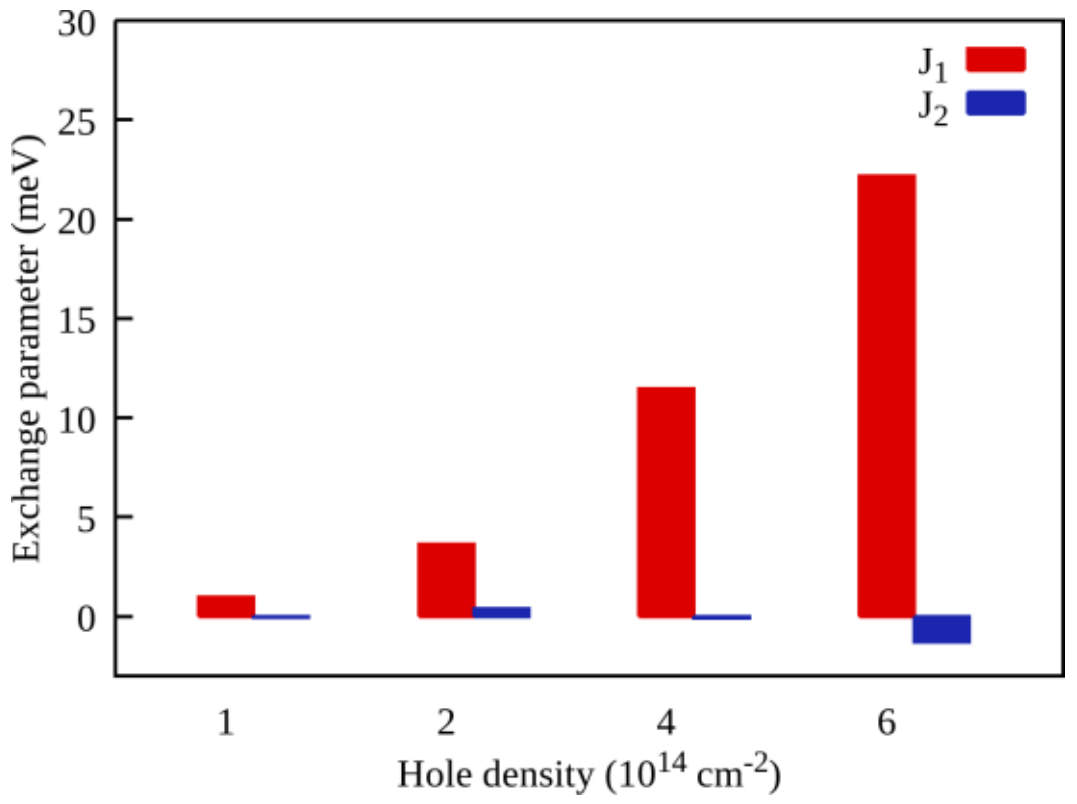

**Magnetic anisotropy energy (MAE, μeV) per magnetic atom** | **Monte Carlo simulations of the normalized magnetization of as a function of temperature**

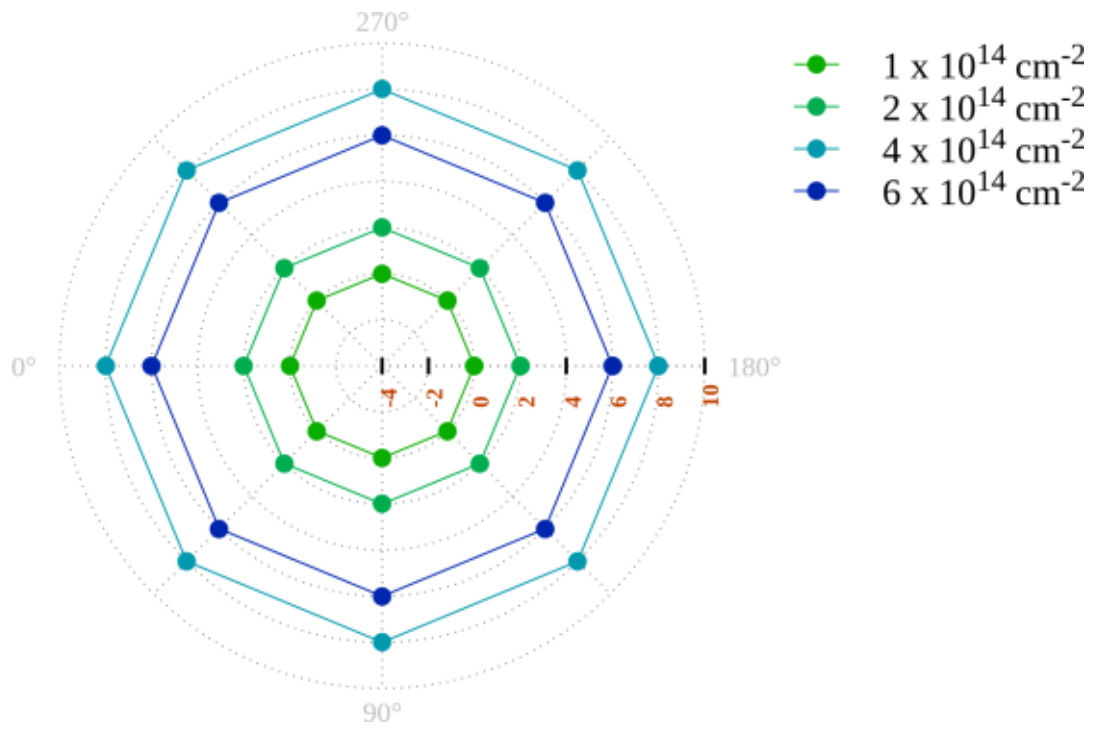

MAE = $E_{\parallel} - E_{\perp}$, a positive (negative) value of MAE indicates the off-plane (in-plane) easy axis.

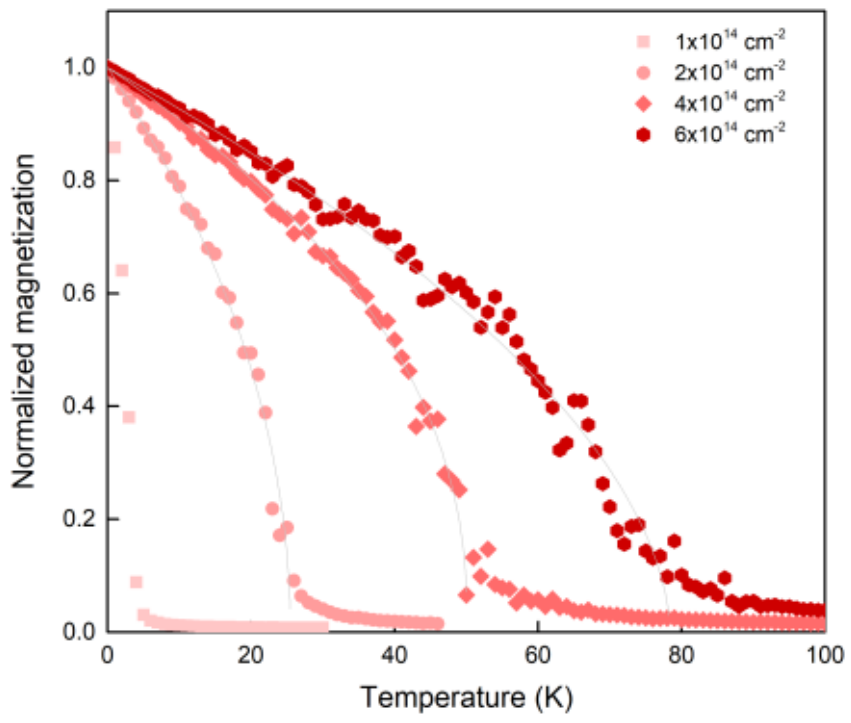

$T_c$: $1\times10^{14}$ cm$^{-2}$: 4 K  $2\times10^{14}$ cm$^{-2}$: 25 K  $4\times10^{14}$ cm$^{-2}$: 50 K  $6\times10^{14}$ cm$^{-2}$: 78 K

# 43. $TiO_2$

| MC2D-ID | C2DB | 2dmat-ID | USPEX | Space group | Band gap (eV) |
|---|---|---|---|---|---|
| - |  | 2dm-3816 | - | P3m1 | 2.72 |

**Convex hull**

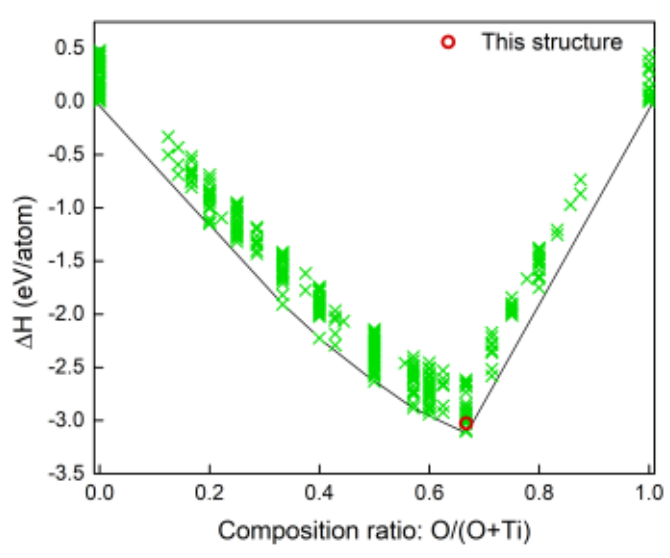

**Atomic structure**

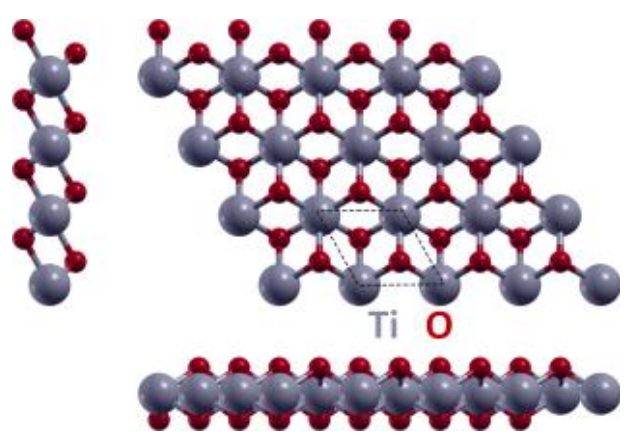

**Atomic coordinates**

```
Ti1O2
1.0000000000000000
2.9924450441581008 0.0000000000000000 0.0000000000000000
1.4962176577629793 2.5915362450782649 0.0000000000000000
0.0000000000000000 0.0000000000000000 21.9352699999999992
Ti O
1 2
Direct
0.9999999976287199 0.0000000025867664 0.5000000167018683
0.6666668161742066 0.6666668110108134 0.5440614346086221
0.3333331861970734 0.3333331864024203 0.4559385486895025
```

**Phonon dispersion curve**

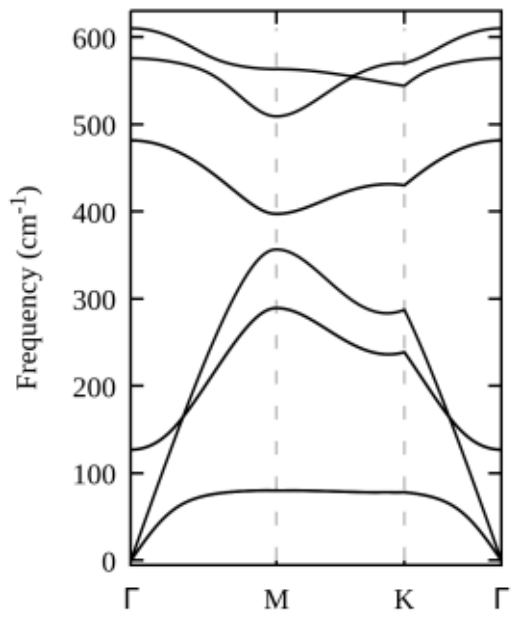

**Projected band structure and density of states**

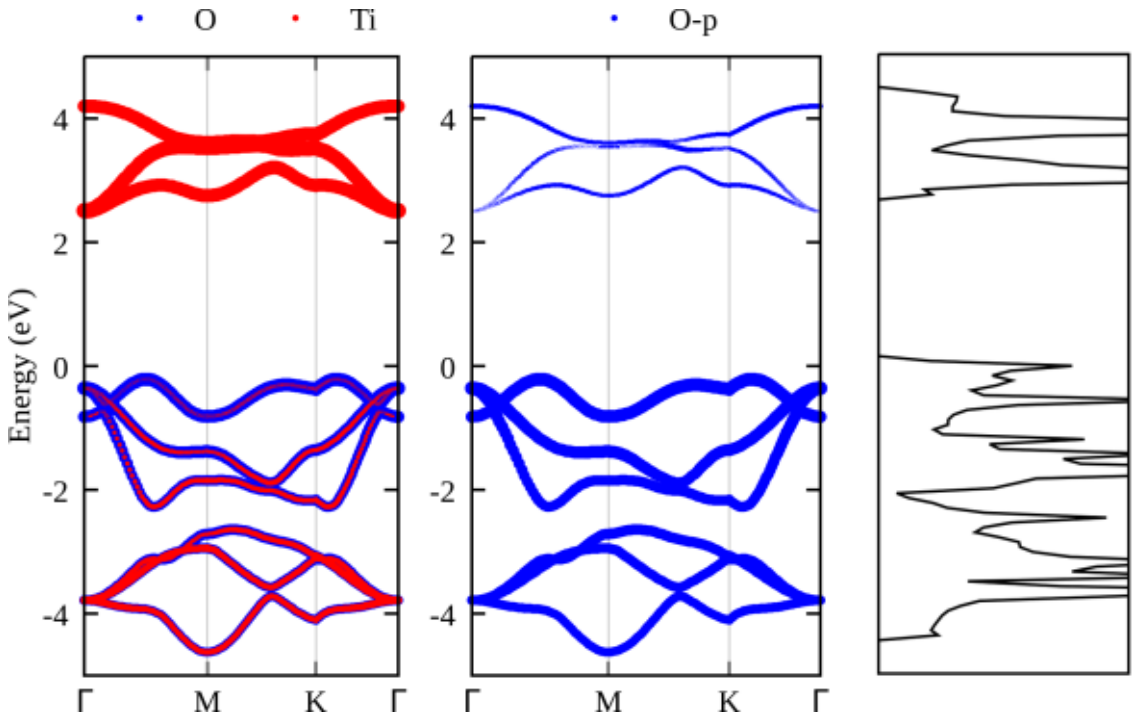

**Magnetic moment and spin polarization energy as a function of hole doping concentration**

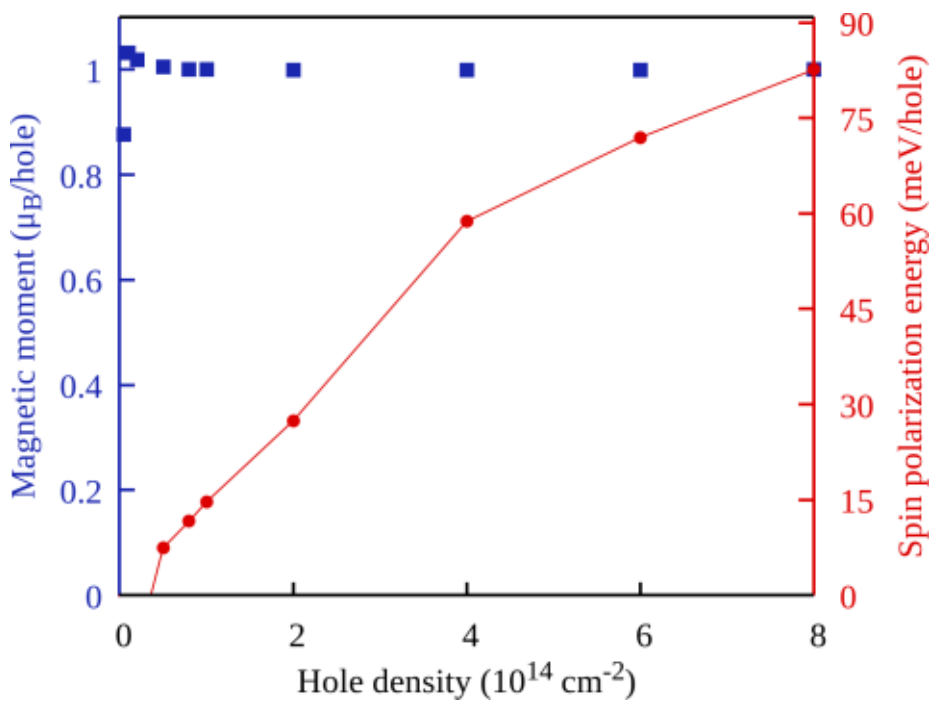

**Magnetic configurations and spin Hamiltonian**

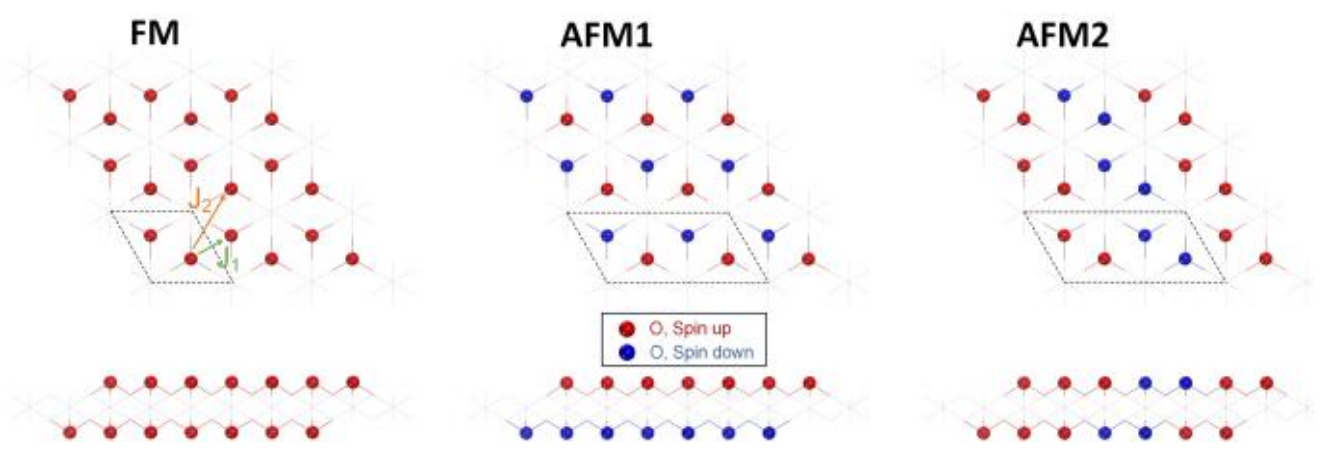

$$E_{FM} = E_0 - 3J_1S^2 - 6J_2S^2$$
$$E_{AFM1} = E_0 + 3J_1S^2 - 6J_2S^2$$
$$E_{AFM2} = E_0 - J_1S^2 + 2J_2S^2$$

**Magnetic exchange coupling parameters**

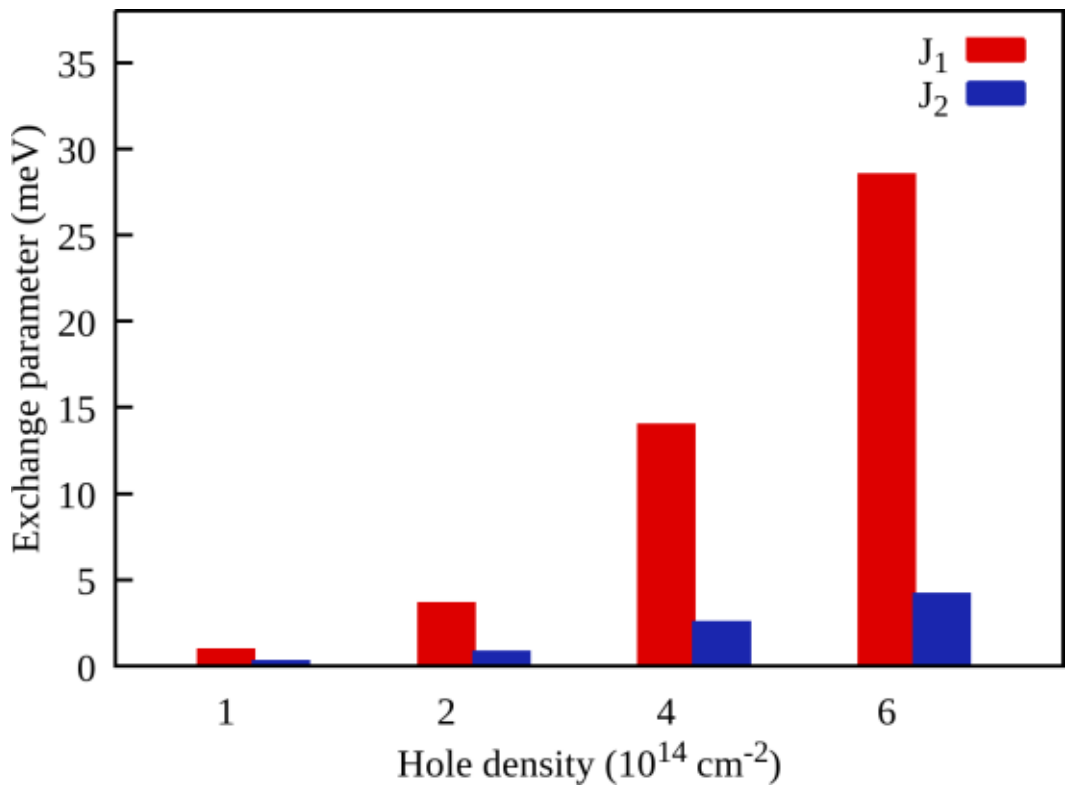

**Magnetic anisotropy energy (MAE, μeV) per magnetic atom**

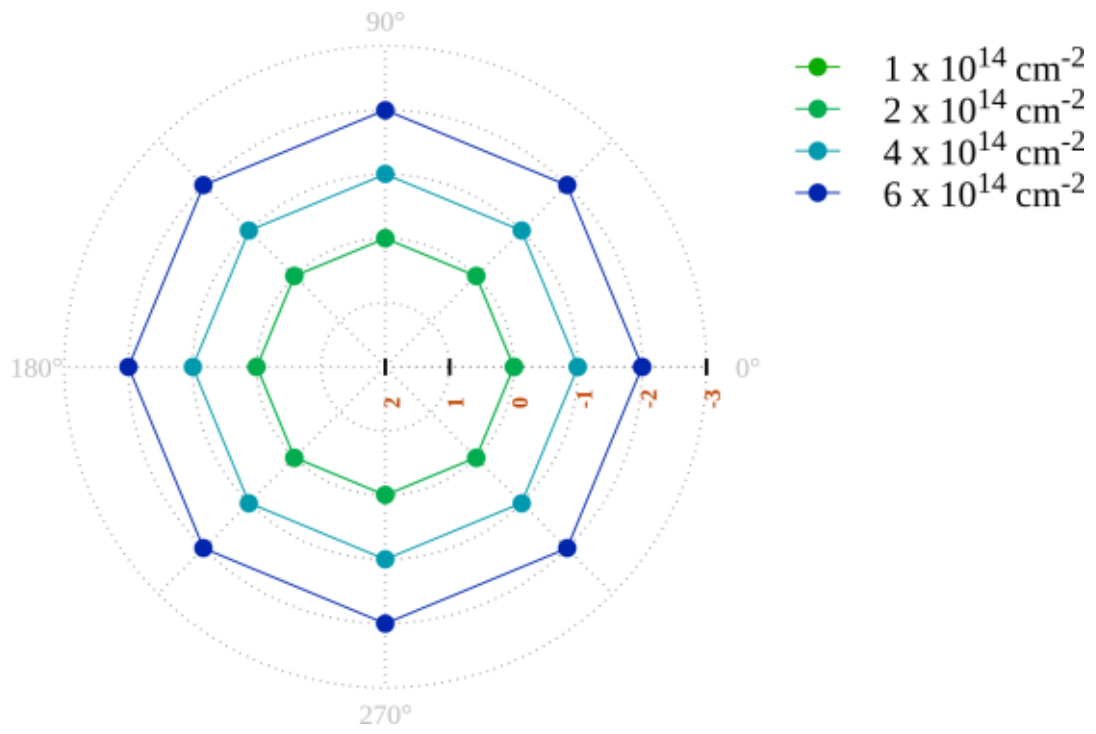

MAE = $E_{\parallel} - E_{\perp}$, a positive (negative) value of MAE indicates the off-plane (in-plane) easy axis.

**Monte Carlo simulations of the normalized magnetization of as a function of temperature**

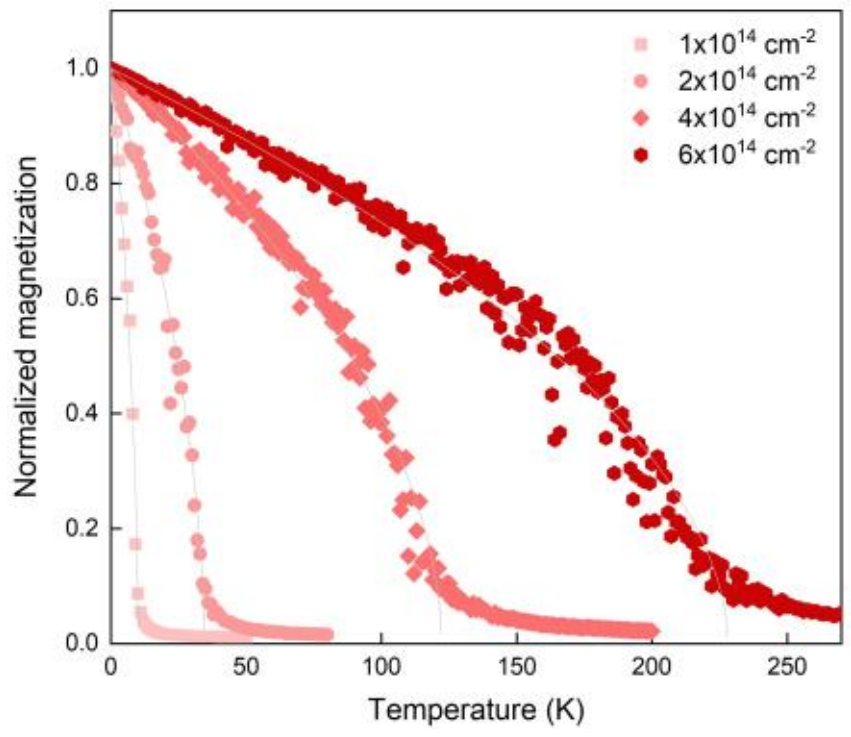

$T_c$: $1\times10^{14}$ cm$^{-2}$: 10 K  $2\times10^{14}$ cm$^{-2}$: 34 K  $4\times10^{14}$ cm$^{-2}$: 122 K  $6\times10^{14}$ cm$^{-2}$: 228 K

# 44. $ZrO_2$

| MC2D-ID | C2DB | 2dmat-ID | USPEX | Space group | Band gap (eV) |
|---|---|---|---|---|---|
| - | ✓ | 2dm-2760 | - | P3m1 | 4.41 |

**Convex hull**

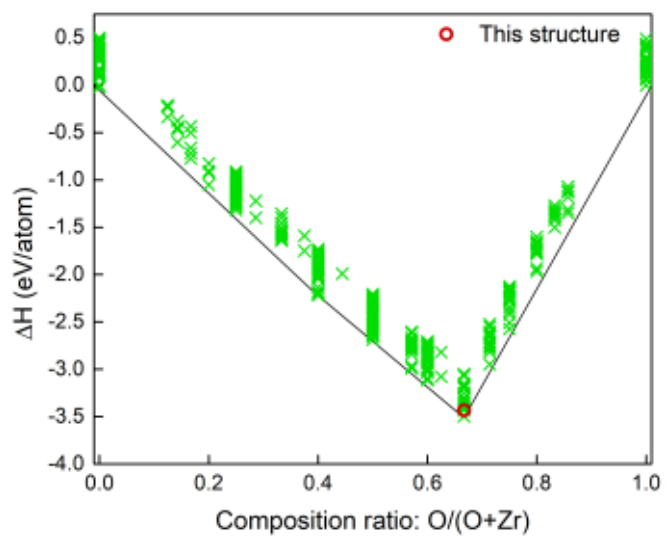

**Atomic structure**

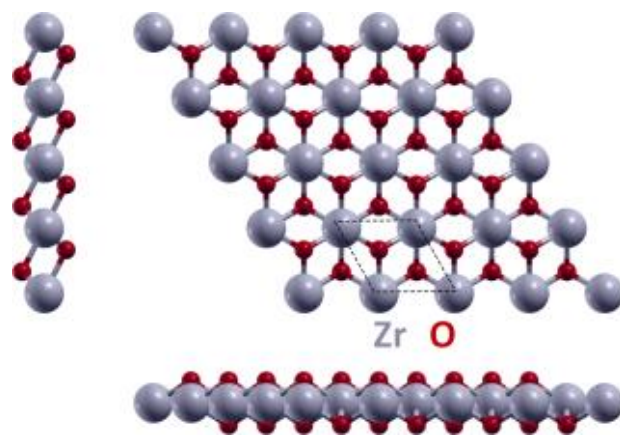

**Atomic coordinates**

```
Zr1O2
   1.00000000000000
   3.2782740240424788    0.0000000000000000    0.0000000000000000
  -1.6391299774822352    2.8390725995268471    0.0000000000000000
   0.0000000000000000    0.0000000000000000   21.9372839999999982
   Zr    O
   1     2
Direct
  0.0000000003210161  0.0000000002877698  0.0000000040734349
  0.3333335748378232  0.6666664039804999  0.0439843153940842
  0.6666664248411678  0.3333335957317303  0.9560156805324809
```

**Phonon dispersion curve**

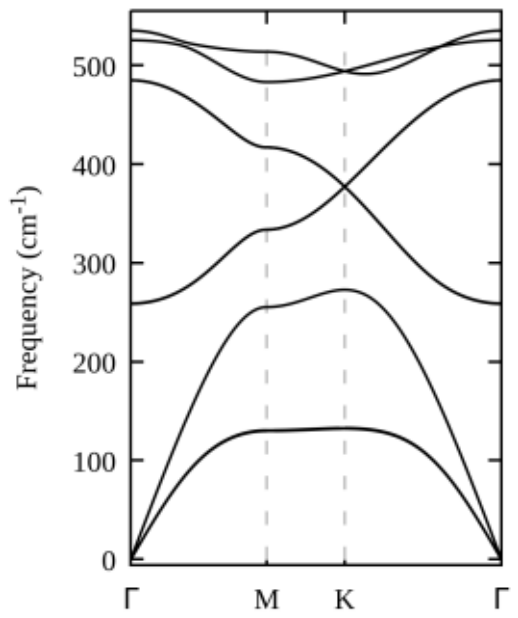

**Projected band structure and density of states**

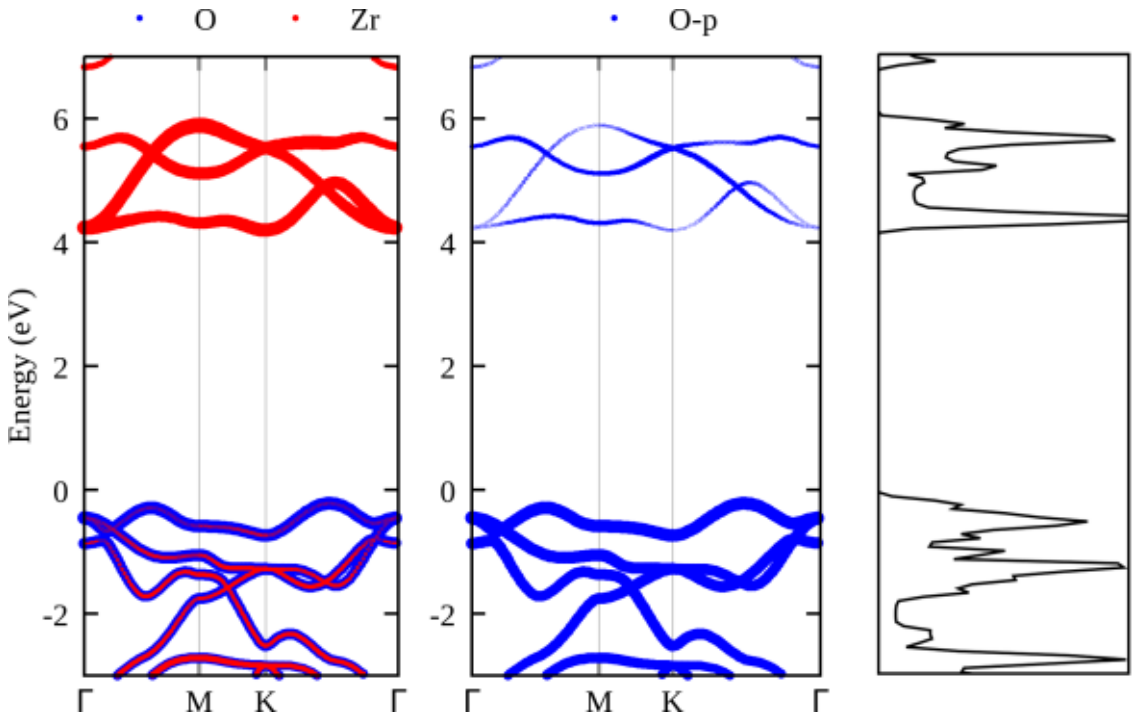

**Magnetic moment and spin polarization energy as a function of hole doping concentration**

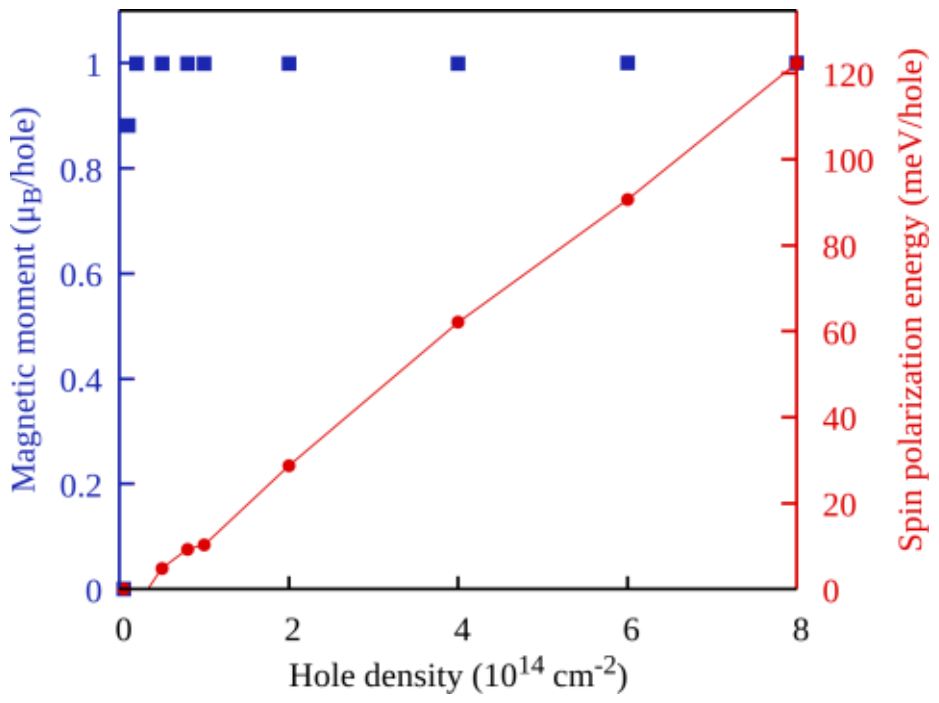

**Magnetic configurations and spin Hamiltonian**

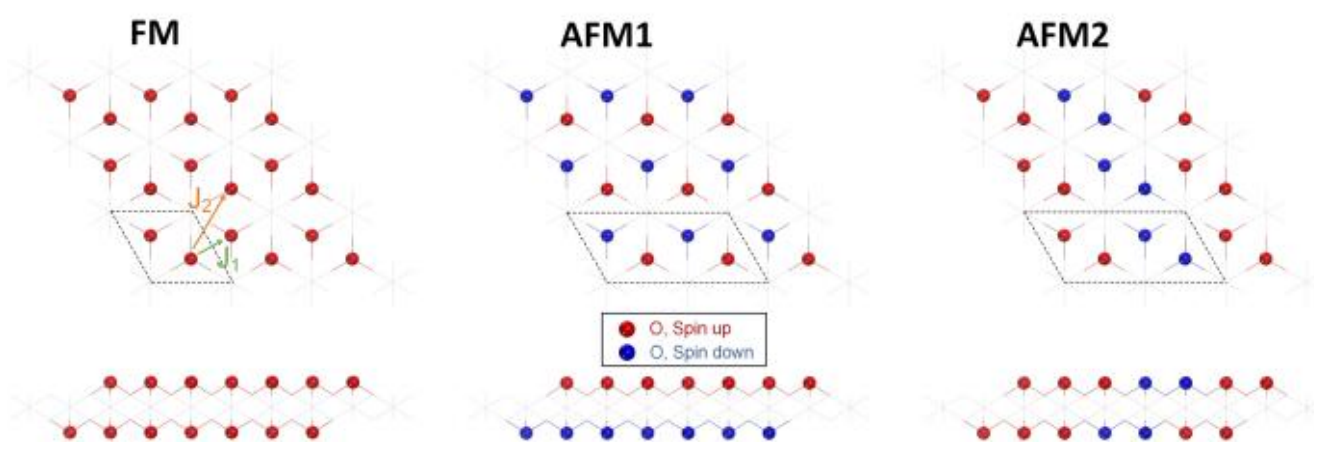

$$E_{FM} = E_0 - 3J_1S^2 - 6J_2S^2$$
$$E_{AFM1} = E_0 + 3J_1S^2 - 6J_2S^2$$
$$E_{AFM2} = E_0 - J_1S^2 + 2J_2S^2$$

**Magnetic exchange coupling parameters**

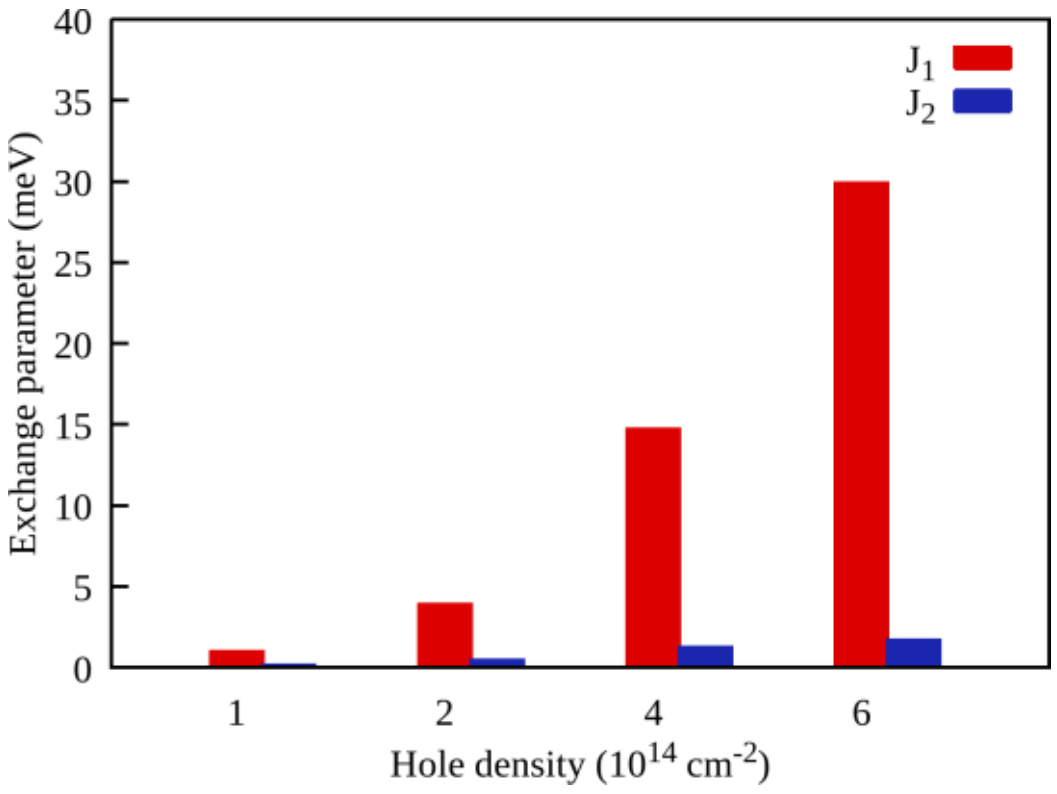

**Magnetic anisotropy energy (MAE, μeV) per magnetic atom**

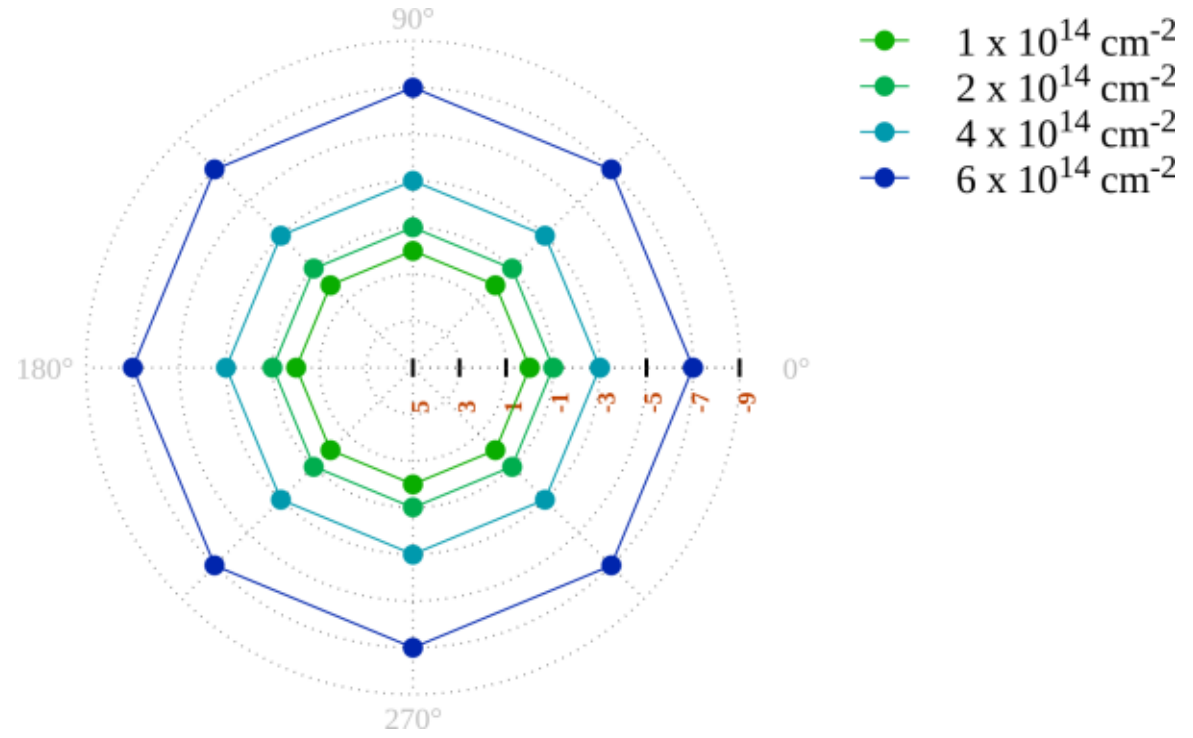

MAE = $E_{\parallel} - E_{\perp}$, a positive (negative) value of MAE indicates the off-plane (in-plane) easy axis.

**Monte Carlo simulations of the normalized magnetization of as a function of temperature**

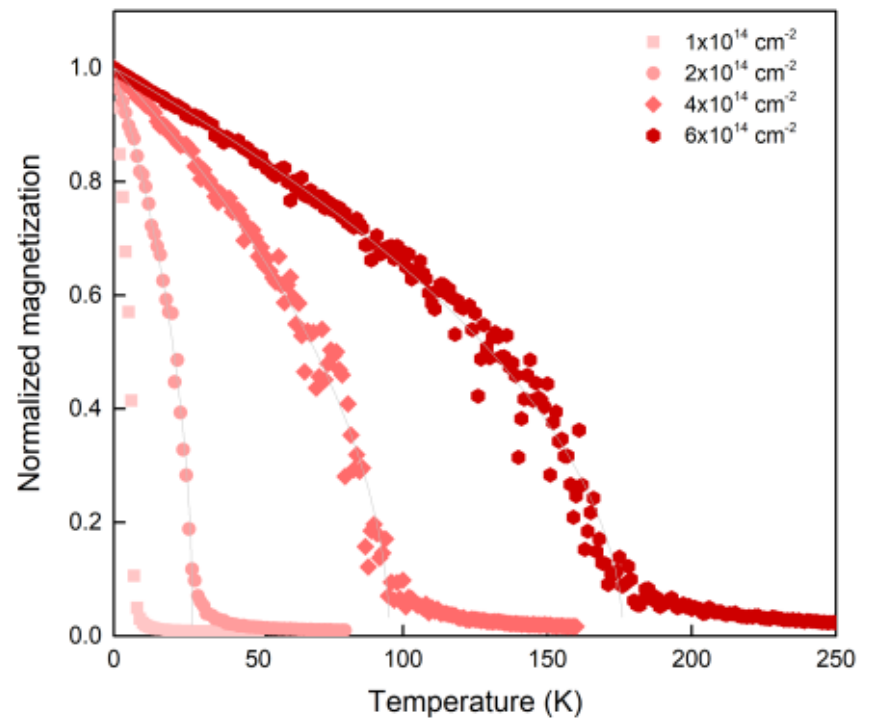

$T_c$: $1\times10^{14}$ cm$^{-2}$: 8 K  $2\times10^{14}$ cm$^{-2}$: 27 K  $4\times10^{14}$ cm$^{-2}$: 95 K  $6\times10^{14}$ cm$^{-2}$: 176 K

# 45. $Al_2N_2$

| MC2D-ID | C2DB | 2dmat-ID | USPEX | Space group | Band gap (eV) |
|---|---|---|---|---|---|
| - | - | - | ✓ | P3m1 | 3.50 |

**Convex hull** | **Atomic structure** | **Atomic coordinates** | **Phonon dispersion curve**

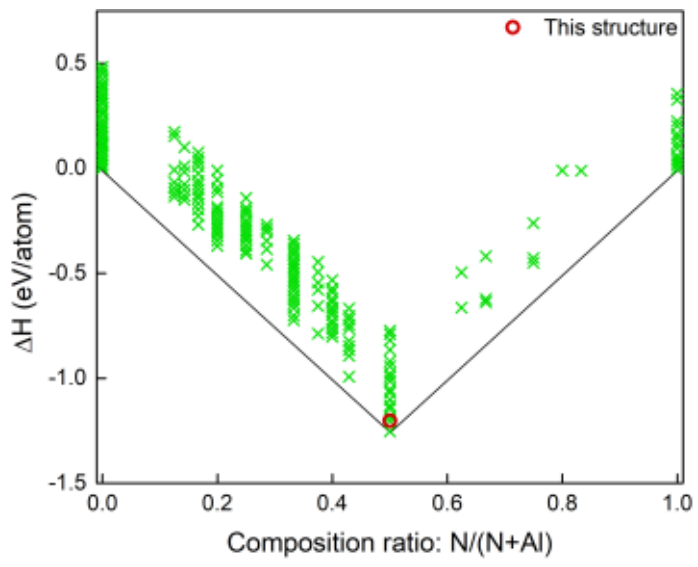

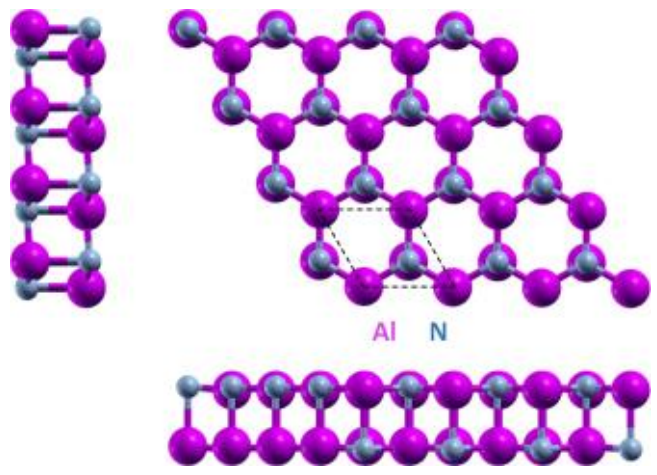

```
Al2N2
1.0000000000000000
3.1569069837959960 0.5321837153485067 0.0000000000000000
1.1181249120853491 3.0003695011230347 0.0000000000000000
0.0000000000000000 0.0000000000000000 20.0000000000000000
Al N
2 2
Direct
0.3676633334244457 0.4537661869794078 0.4494394326520350
0.0349973019316394 0.1202368071170525 0.5505605699356337
0.3682612689215006 0.4535880144369600 0.5563400399759786
0.0343940957224120 0.1204139914665703 0.4436599574363527
```

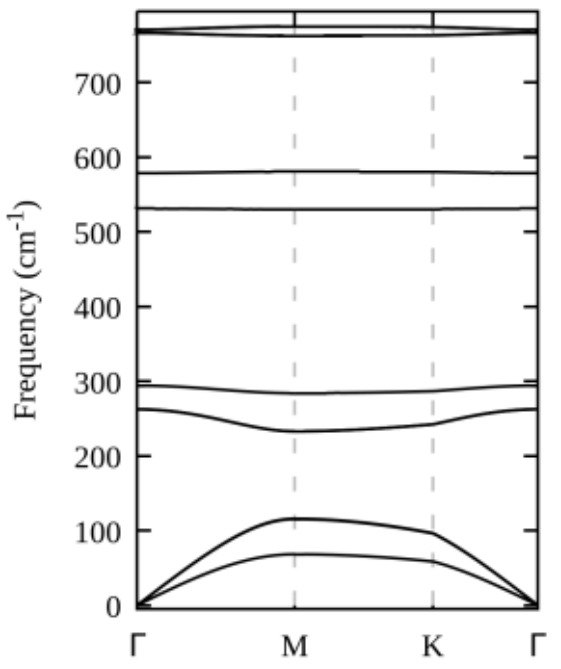

**Projected band structure and density of states**

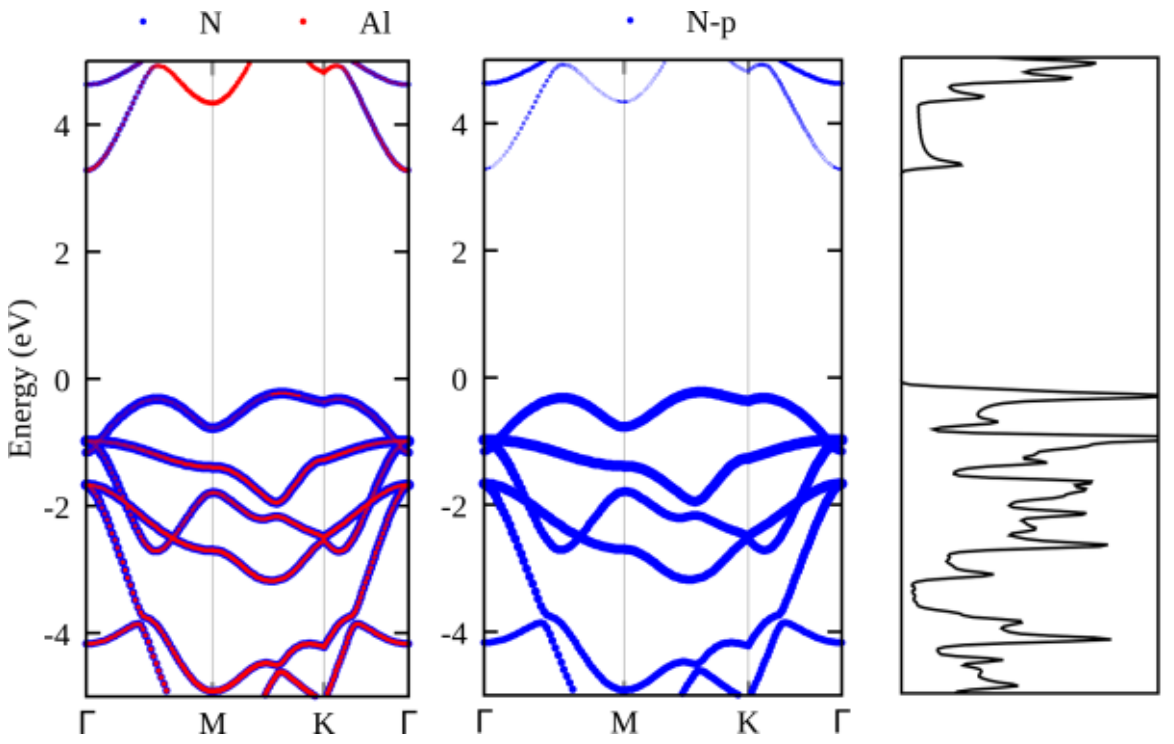

**Magnetic moment and spin polarization energy as a function of hole doping concentration**

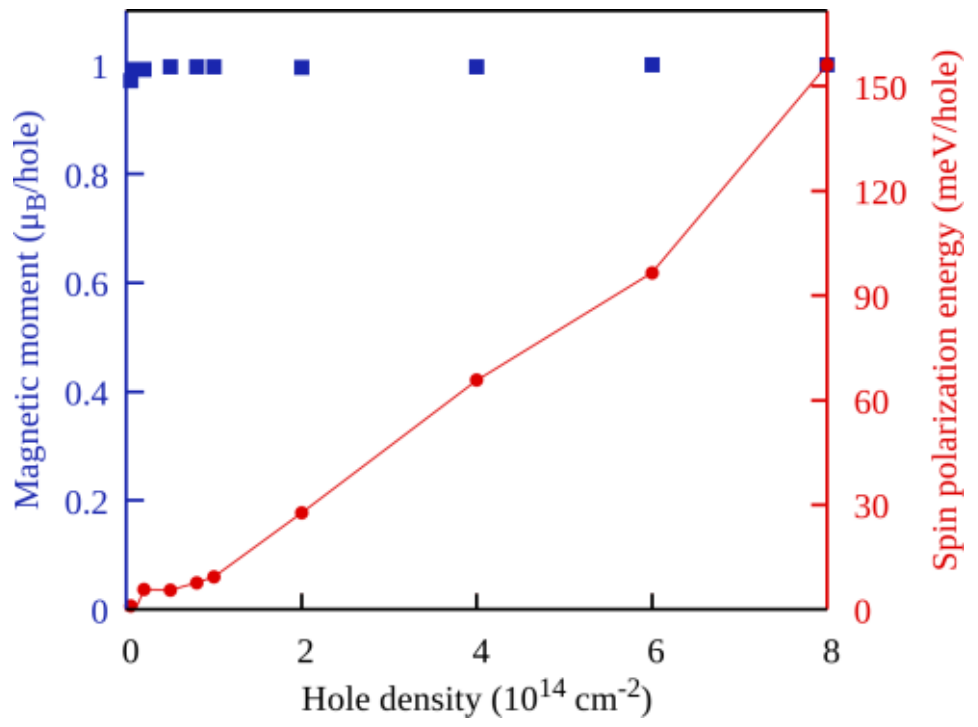

**Magnetic configurations and spin Hamiltonian**

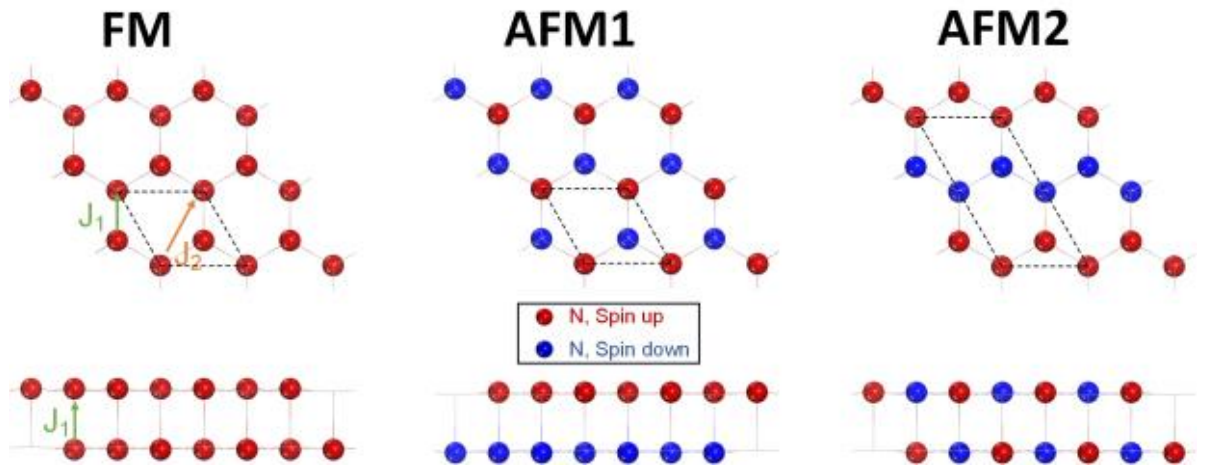

$$E_{FM} = E_0 - 3J_1S^2 - 6J_2S^2$$

$$E_{AFM1} = E_0 + 3J_1S^2 - 6J_2S^2$$

$$E_{AFM2} = E_0 - J_1S^2 + 2J_2S^2$$

**Magnetic exchange coupling parameters**

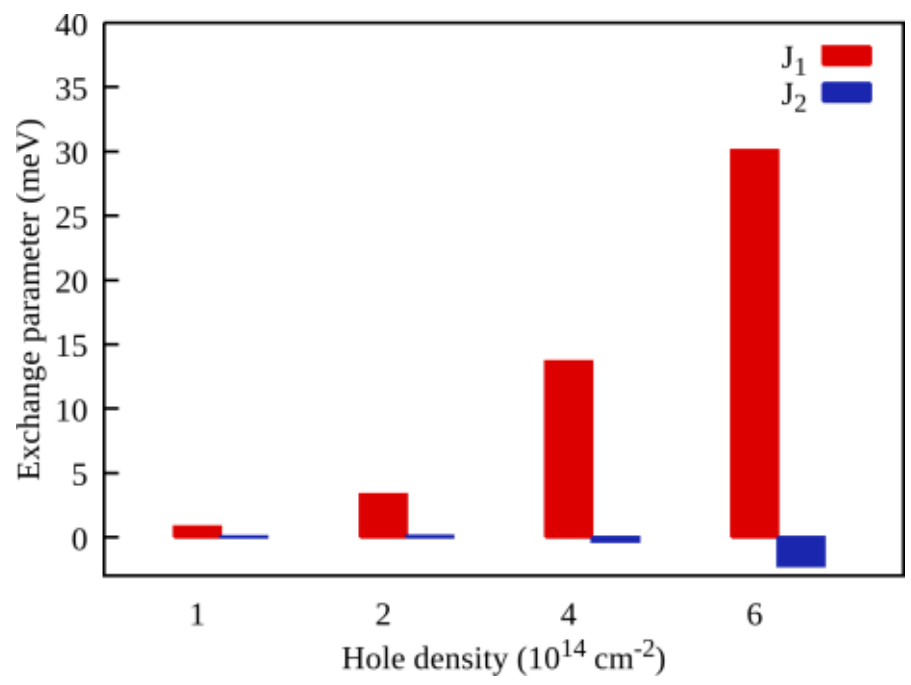

**Magnetic anisotropy energy (MAE, μeV) per magnetic atom**

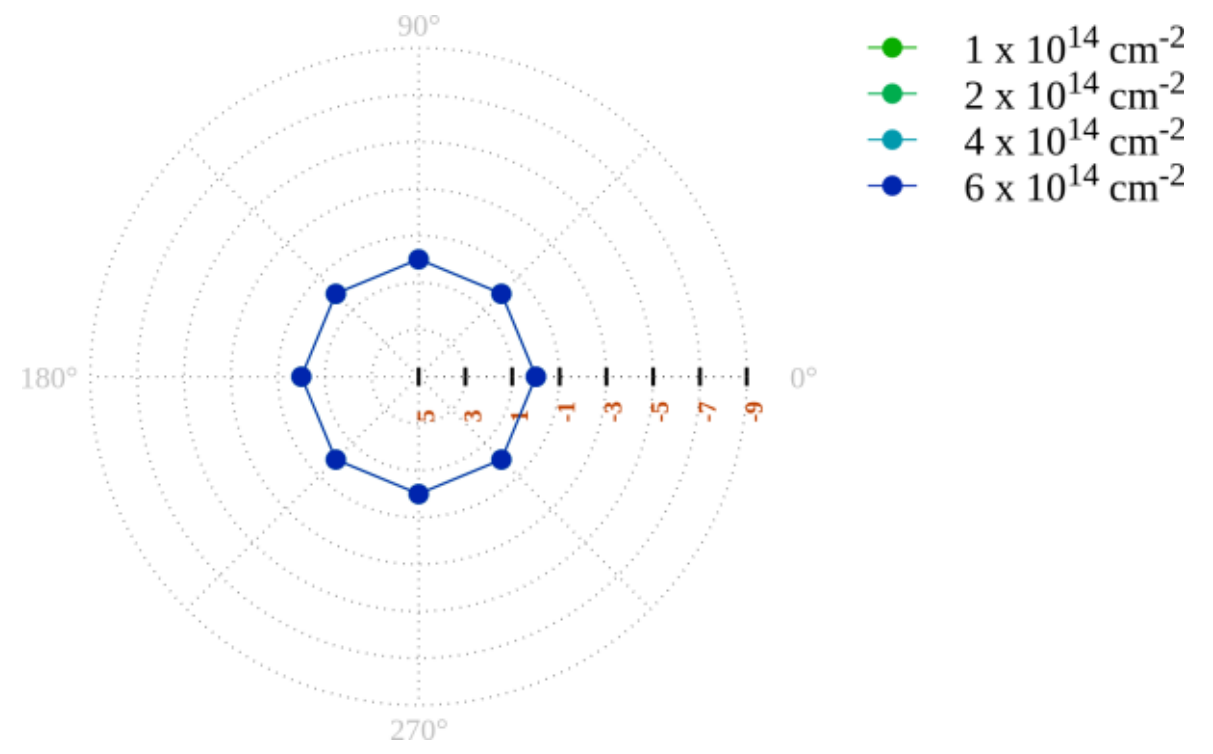

MAE = $E_{\parallel} - E_{\perp}$, a positive (negative) value of MAE indicates the off-plane (in-plane) easy axis.

**Monte Carlo simulations of the normalized magnetization of as a function of temperature**

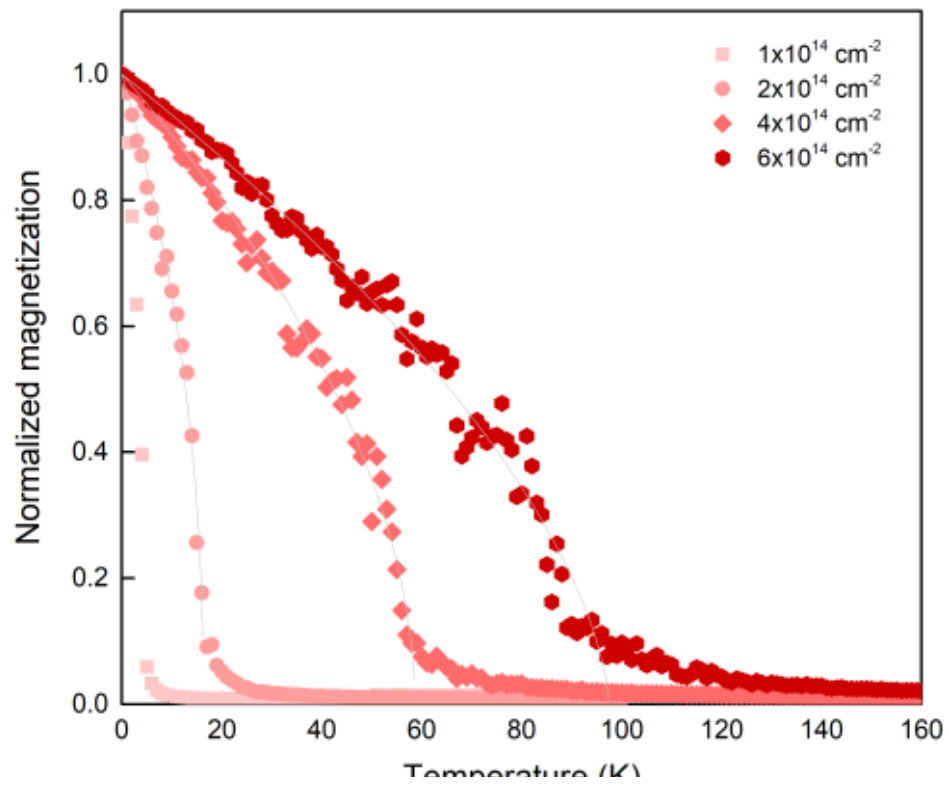

$T_c$: $1\times10^{14}$ cm$^{-2}$: 5 K  $2\times10^{14}$ cm$^{-2}$: 16 K

$4\times10^{14}$ cm$^{-2}$: 59K  $6\times10^{14}$ cm$^{-2}$: 97 K

# 46. $Mg(OH)_2$

| MC2D-ID | C2DB | 2dmat-ID | USPEX | Space group | Band gap (eV) |
|---|---|---|---|---|---|
| 116 | ✓ | 2dm-5739 | - | P3m1 | 3.30 |

**Convex hull**

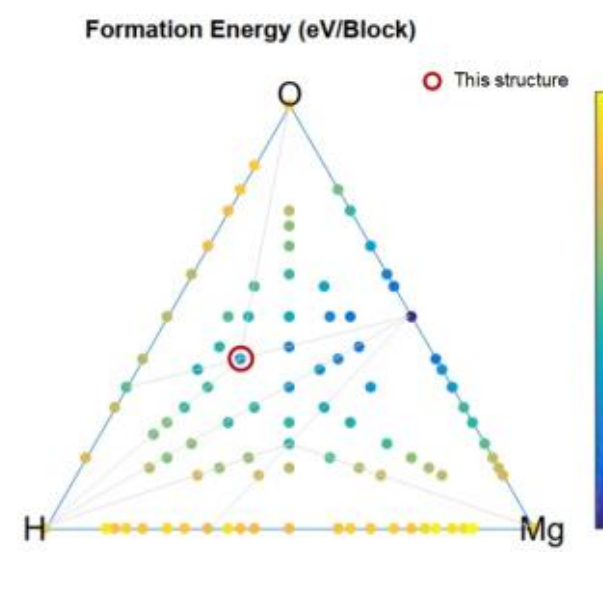

**Atomic structure**

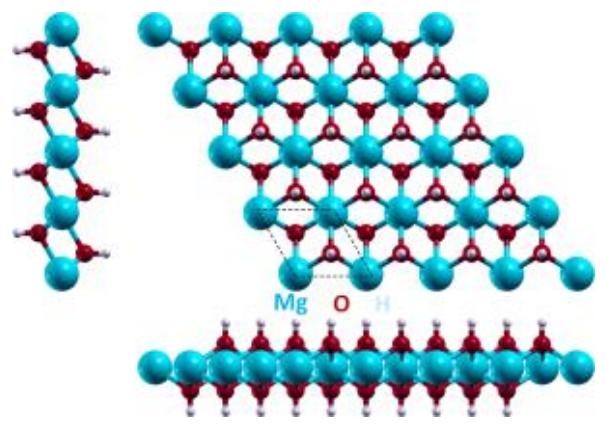

**Atomic coordinates**

```
Mg1H2O2
   1.00000000000000
   3.1751312257357447    0.0000000000000000    0.0000000000000000
  -1.5876568617590670    2.7496917417189422    0.0000000000000000
   0.0000000000000000    0.0000000000000000   24.0290010000000009
   Mg   H   O
   1    2    2
Direct
  0.0000000000649081  0.0000000000122924  0.4999999989443822
  0.6666665528492572  0.3333333508677612  0.5833277350975337
  0.3333334469813565  0.6666666491729387  0.4166722634303568
  0.6666689772502252  0.3333311958666556  0.5431118051139165
  0.3333310228542530  0.6666688040803521  0.4568881974138108
```

**Phonon dispersion curve**

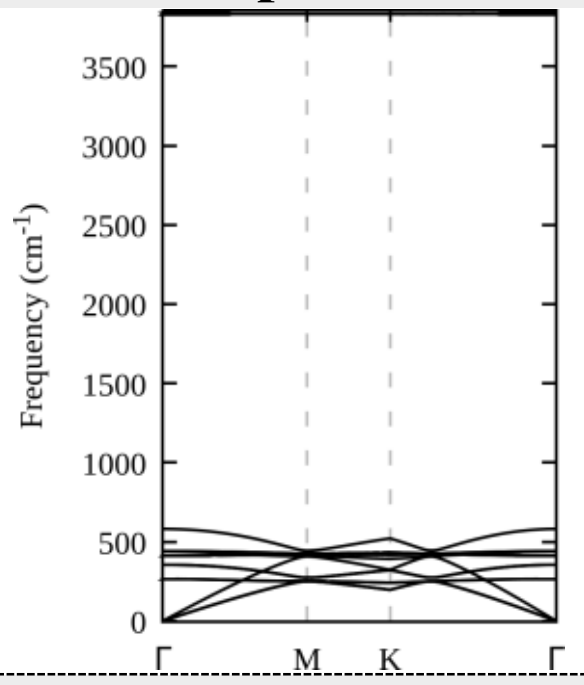

**Projected band structure and density of states**

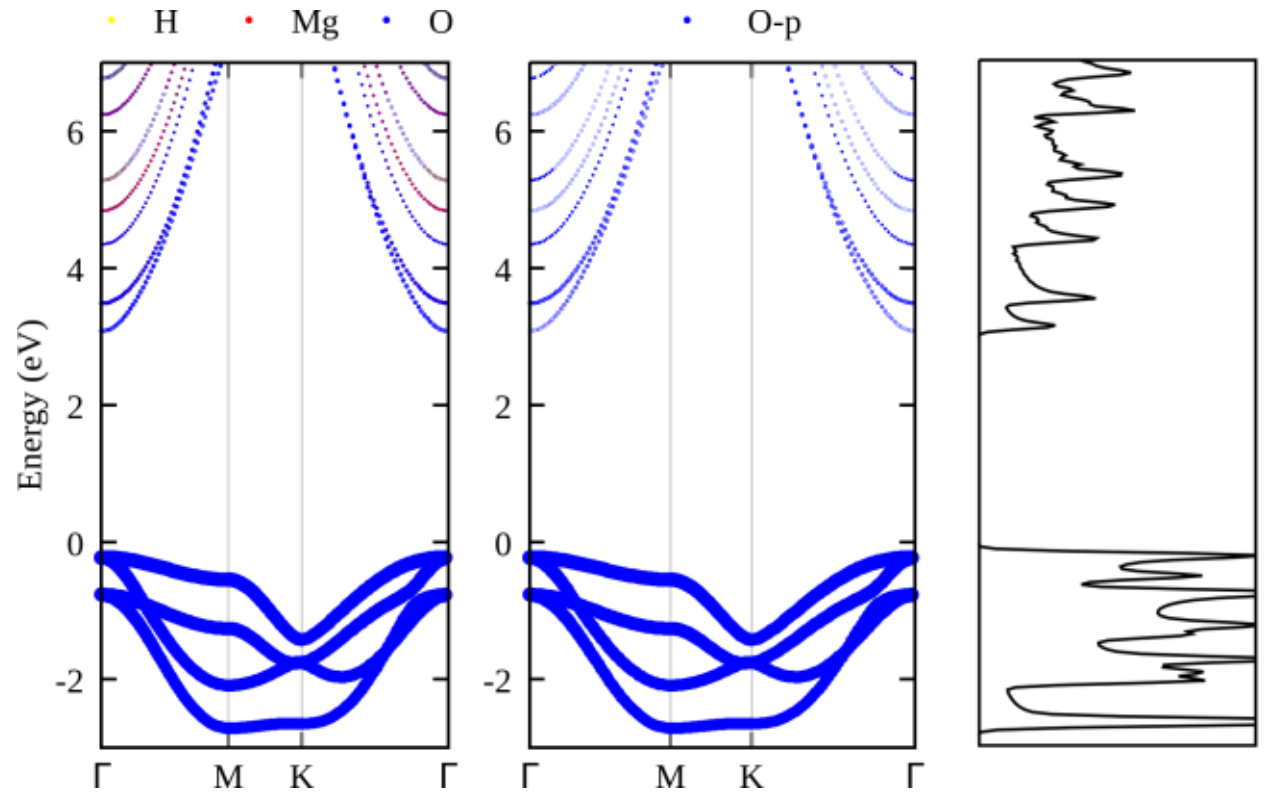

**Magnetic moment and spin polarization energy as a function of hole doping concentration**

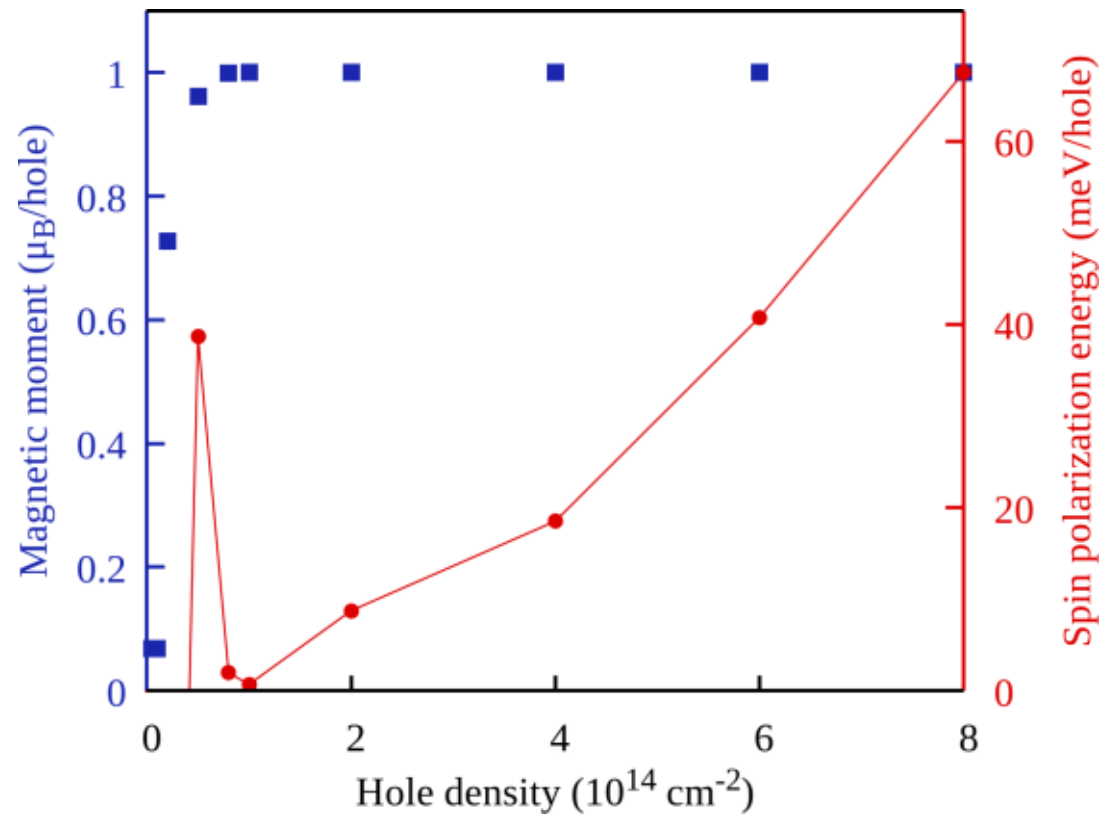

**Magnetic configurations and spin Hamiltonian**

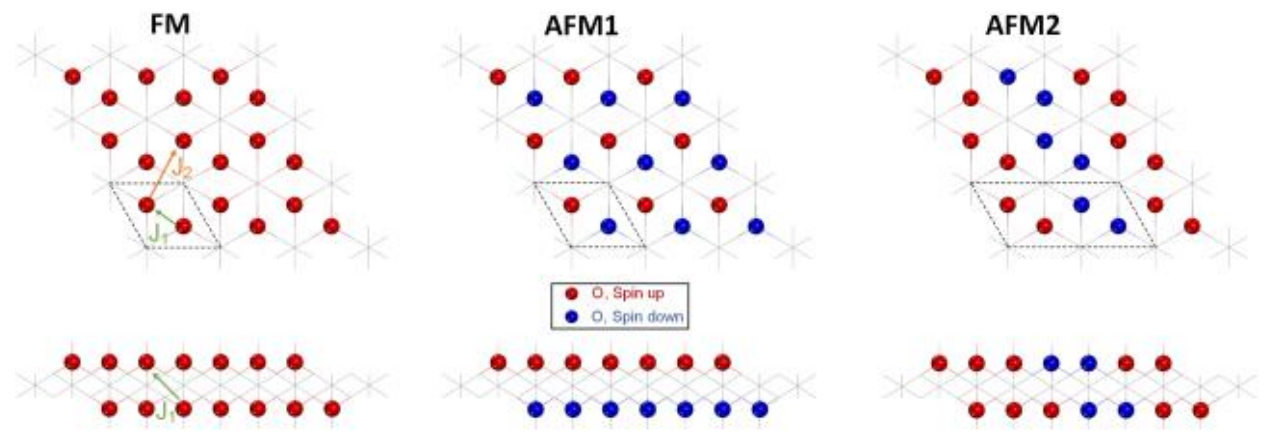

$$E_{FM} = E_0 - 3J_1S^2 - 6J_2S^2$$
$$E_{AFM1} = E_0 + 3J_1S^2 - 6J_2S^2$$
$$E_{AFM2} = E_0 - J_1S^2 + 2J_2S^2$$

**Magnetic exchange coupling parameters**

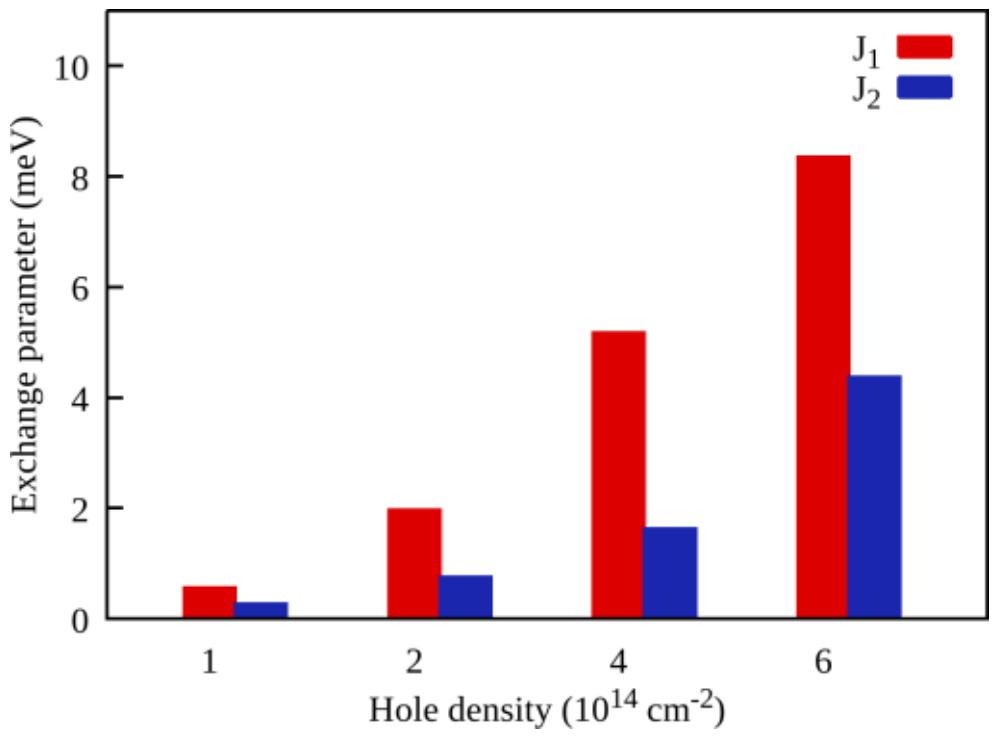

**Magnetic anisotropy energy (MAE, μeV) per magnetic atom**

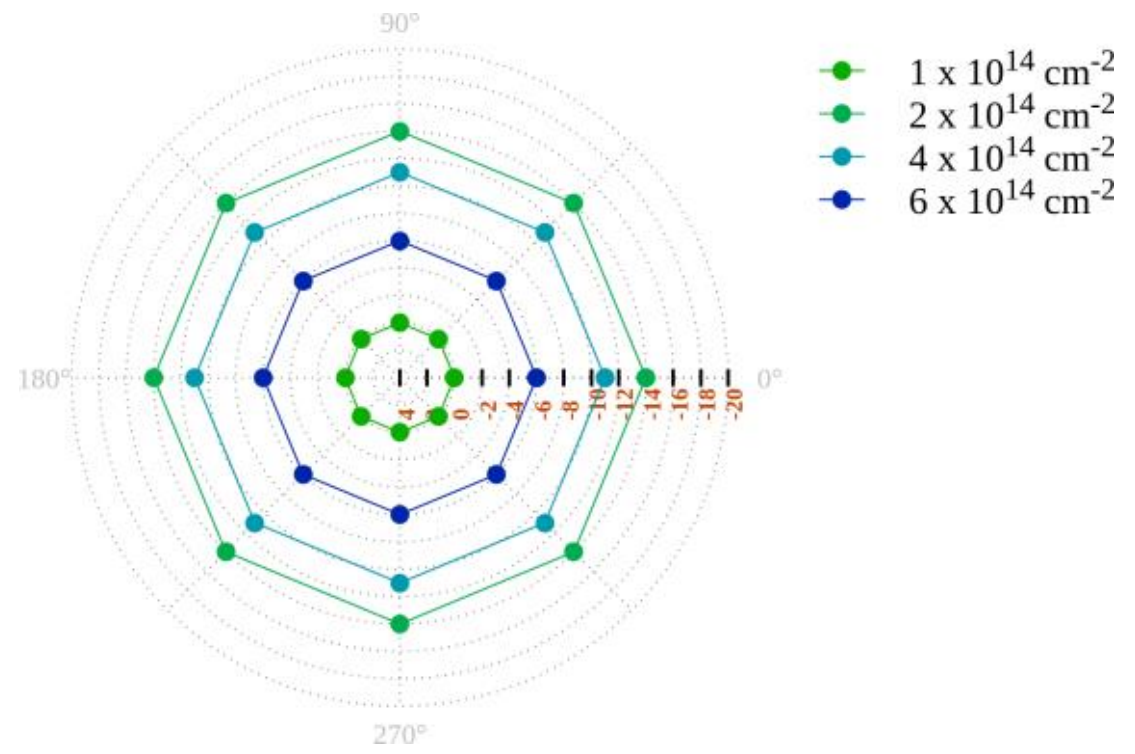

MAE = $E_{\parallel} - E_{\perp}$, a positive (negative) value of MAE indicates the off-plane (in-plane) easy axis.

**Monte Carlo simulations of the normalized magnetization of as a function of temperature**

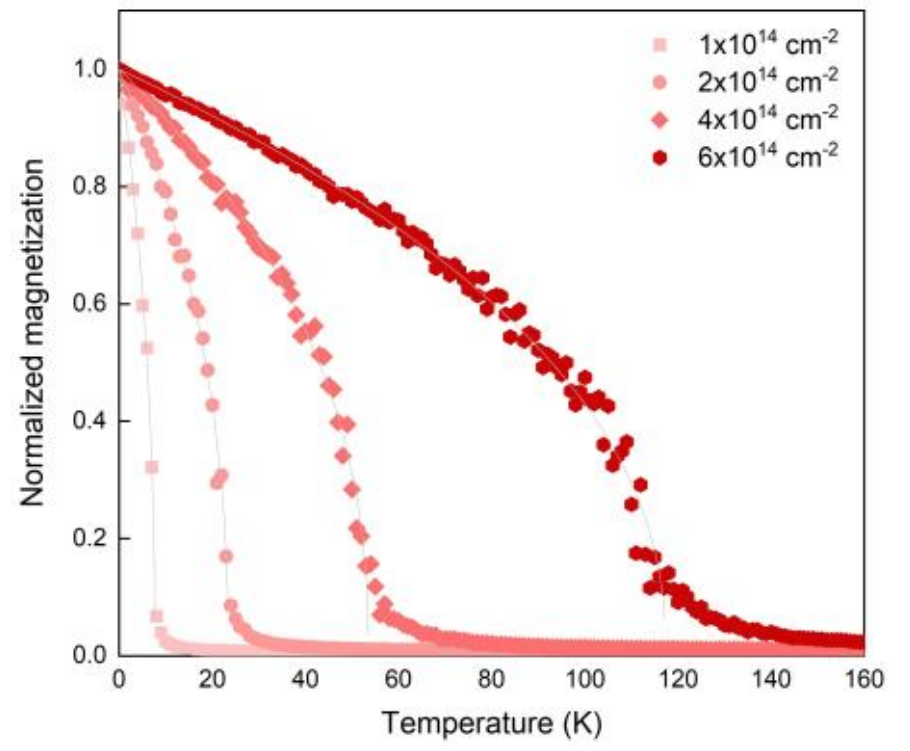

$T_c$: $1\times10^{14}$ cm$^{-2}$: 8 K $2\times10^{14}$ cm$^{-2}$: 23 K

$4\times10^{14}$ cm$^{-2}$: 53 K $6\times10^{14}$ cm$^{-2}$: 117 K

# 47. $Ca(OH)_2$

| MC2D-ID | C2DB | 2dmat-ID | USPEX | Space group | Band gap (eV) |
|---|---|---|---|---|---|
| 26 | - | 2dm-3595 | - | P3m1 | 3.68 |

**Convex hull** **Atomic structure** **Atomic coordinates** **Phonon dispersion curve**

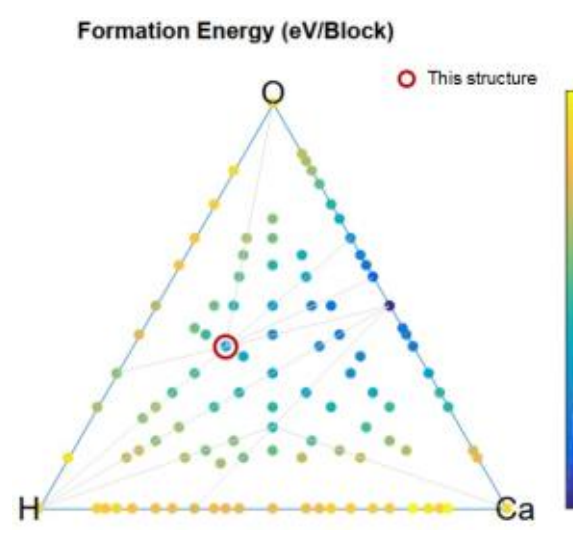

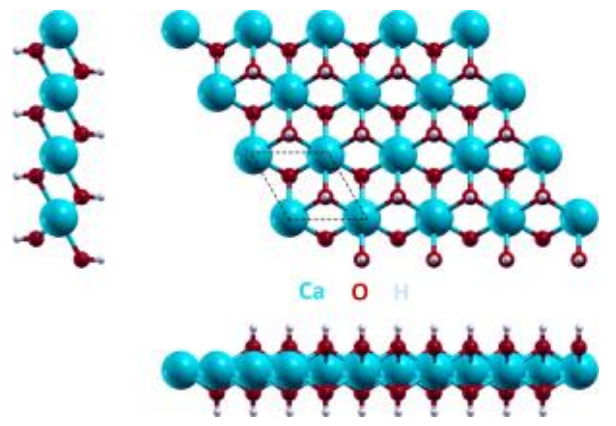

```
Ca1H2O2
   1.00000000000000
   3.6323417965460472    0.0000000000000000    0.0000000000000000
  -1.8161768156685321    3.1456965616529682    0.0000000000000000
   0.0000000000000000    0.0000000000000000   24.2520099999999985
   Ca   H   O
   1    2    2
Direct
  0.0000000001817000  0.9999999996367777  0.4999999993959605
  0.6666669284561024  0.3333330562961052  0.5869570112582636
  0.3333330701477379  0.6666669444442306  0.4130429870705470
  0.6666667311249554  0.3333334392343161  0.5469845406018976
  0.3333332700895042  0.6666665603885633  0.4530154616733242
```

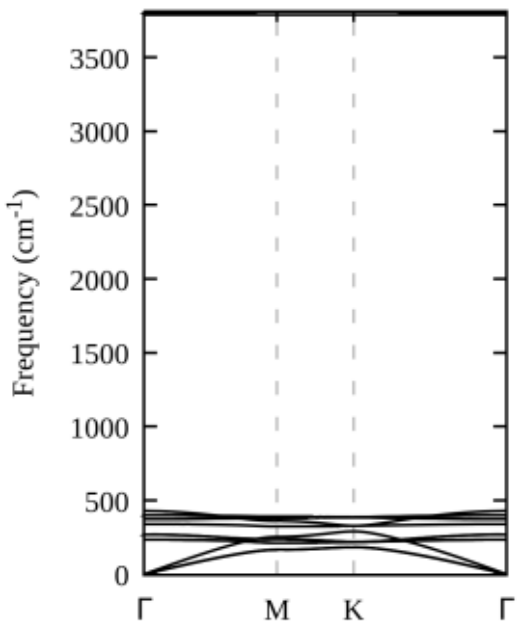

**Projected band structure and density of states**

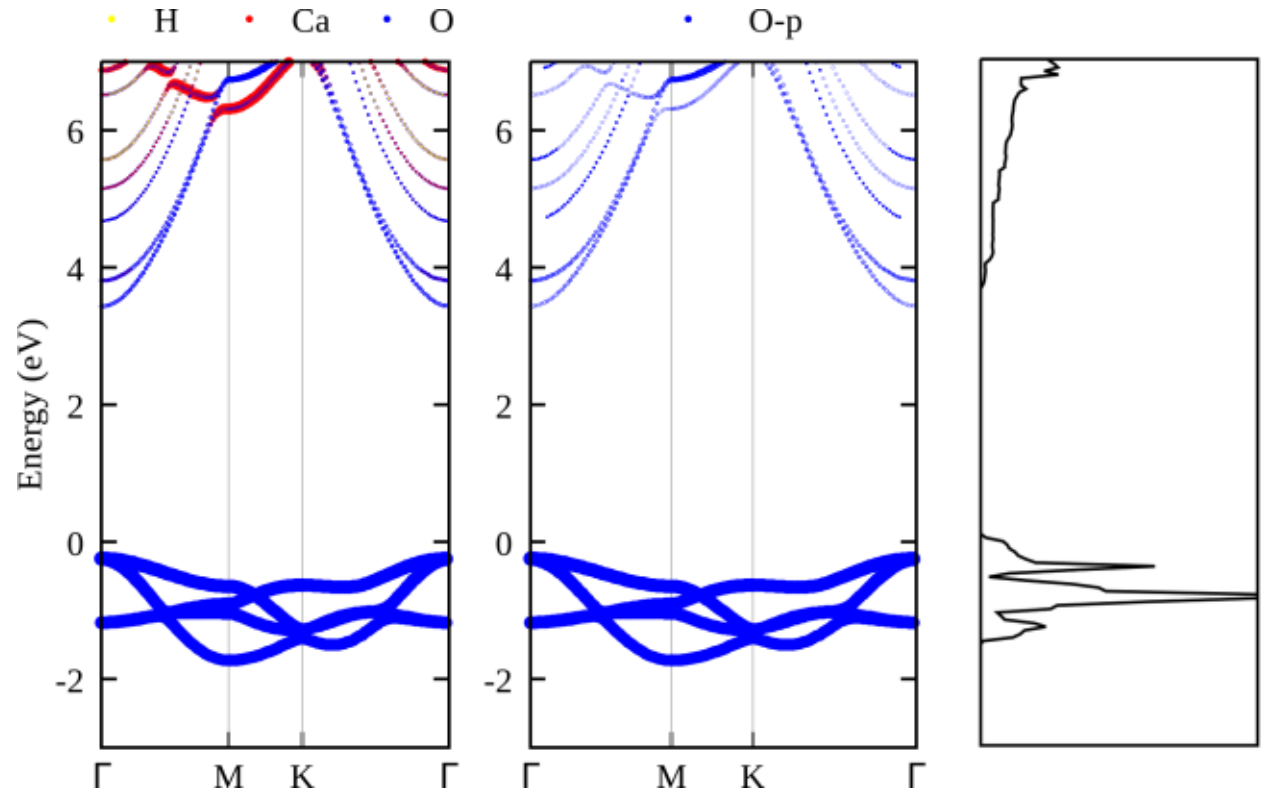

**Magnetic moment and spin polarization energy as a function of hole doping concentration**

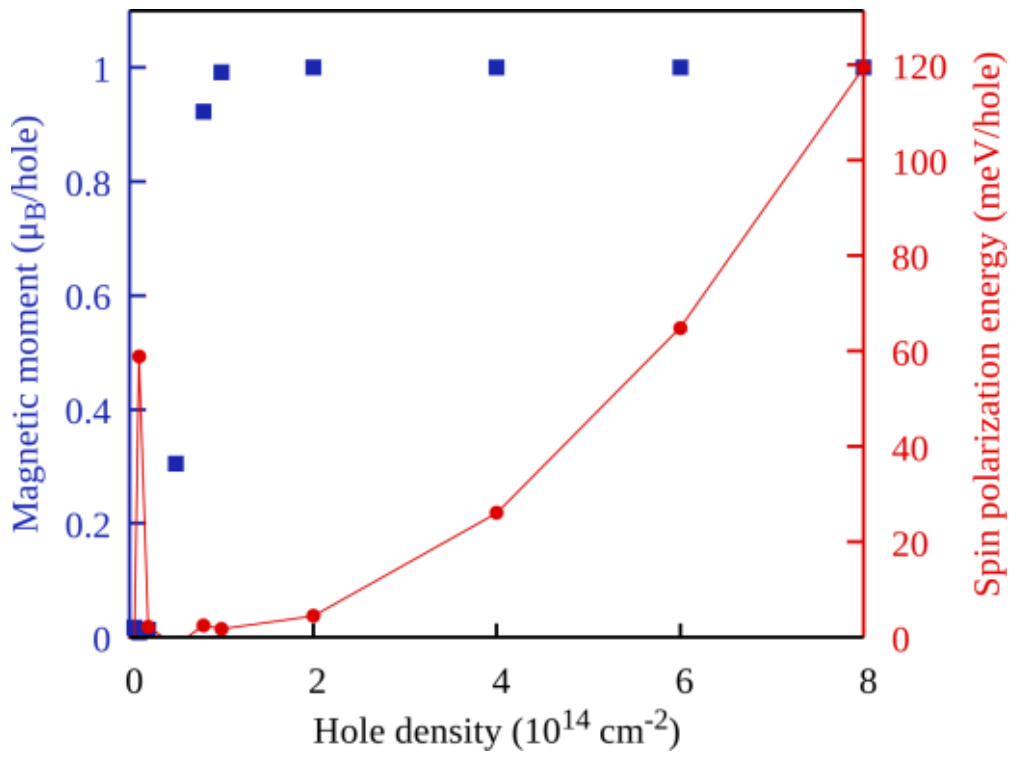

**Magnetic configurations and spin Hamiltonian**

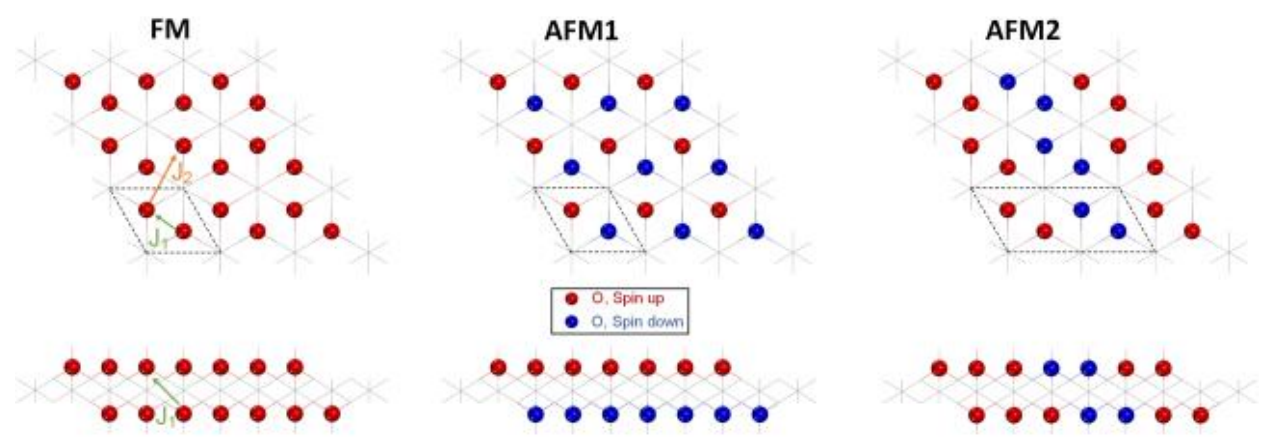

$$E_{FM} = E_0 - 3J_1S^2 - 6J_2S^2$$
$$E_{AFM1} = E_0 + 3J_1S^2 - 6J_2S^2$$
$$E_{AFM2} = E_0 - J_1S^2 + 2J_2S^2$$

**Magnetic exchange coupling parameters**

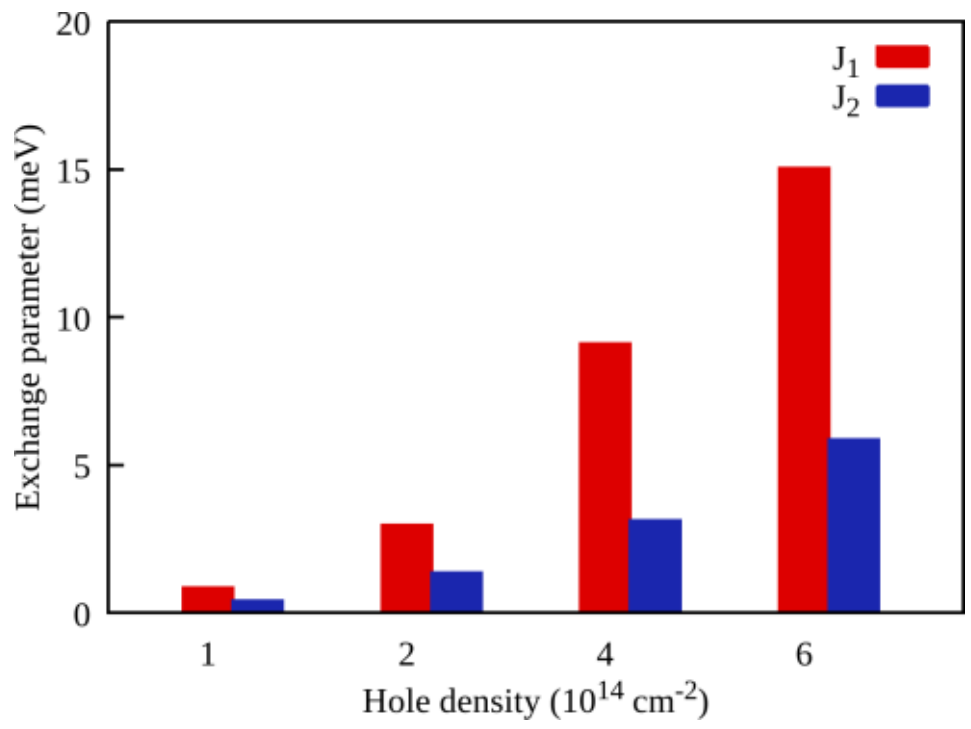

**Magnetic anisotropy energy (MAE, μeV) per magnetic atom**

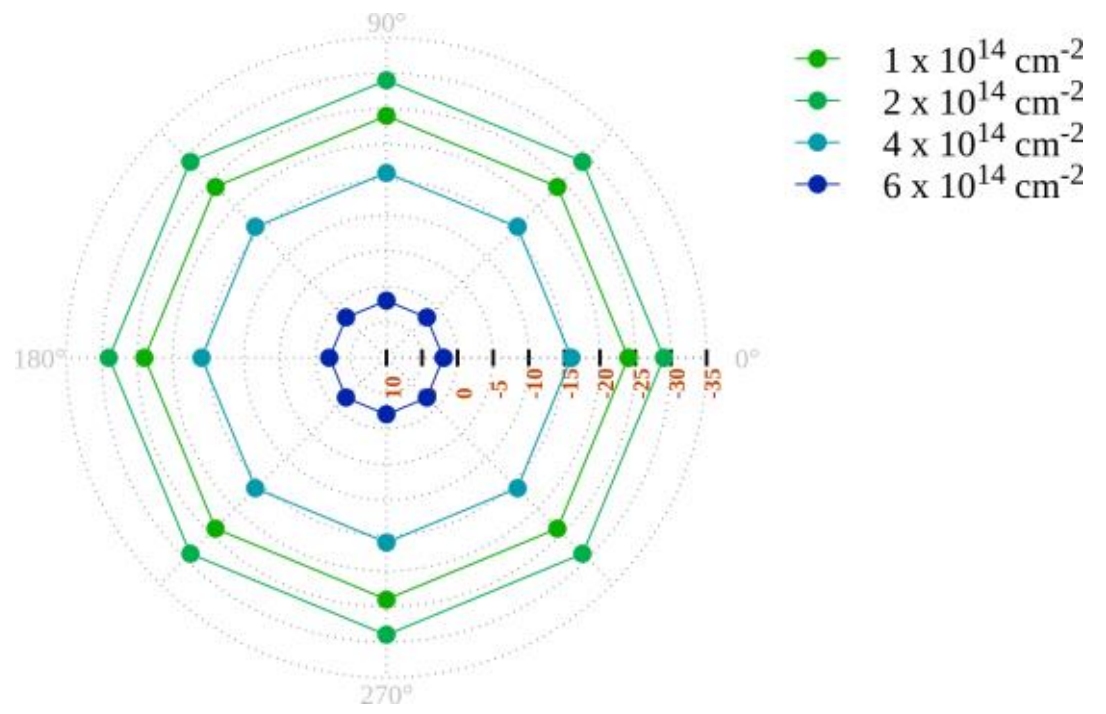

MAE = $E_{\parallel} - E_{\perp}$, a positive (negative) value of MAE indicates the off-plane (in-plane) easy axis.

**Monte Carlo simulations of the normalized magnetization of as a function of temperature**

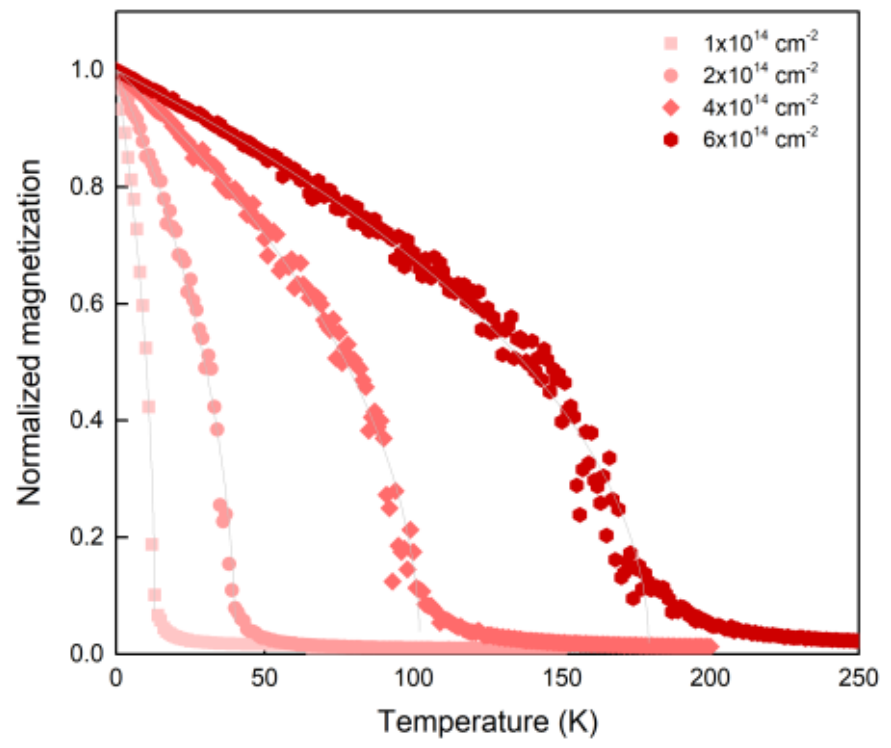

$T_c$: $1\times10^{14}$ cm$^{-2}$: 13 K  $2\times10^{14}$ cm$^{-2}$: 40 K

$4\times10^{14}$ cm$^{-2}$: 102 K  $6\times10^{14}$ cm$^{-2}$: 179 K

# 48. $Zn(OH)_2$

| MC2D-ID | C2DB | 2dmat-ID | USPEX | Space group | Band gap (eV) |
|---|---|---|---|---|---|
| - | ✓ | 2dm-3835 | - | P3m1 | 2.31 |

**Convex hull**

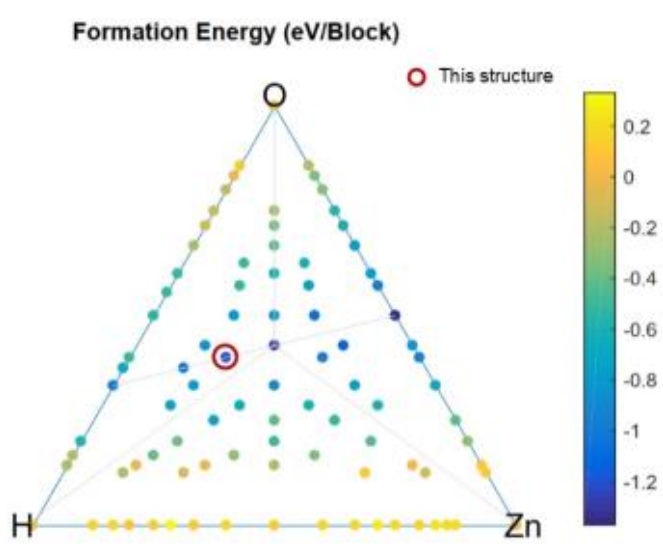

**Atomic structure**

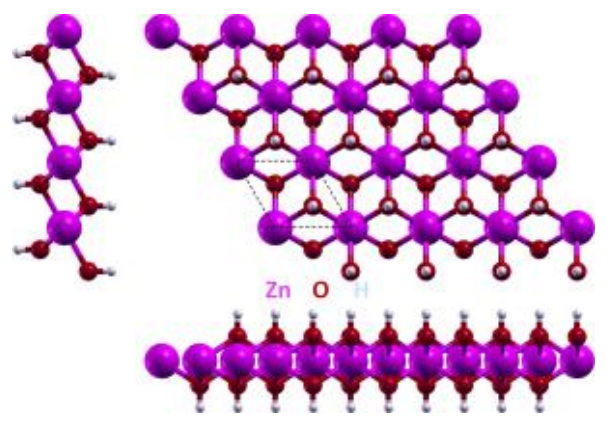

**Atomic coordinates**

```
Zn1H2O2
   1.00000000000000
   3.2353808825138515    0.0000000000000000    0.0000000000000000
  -1.6180078068454582    2.8017390107423221    0.0000000000000000
   0.0000000000000000    0.0000000000000000   24.0492699999999999
   Zn   H   O
   1    2   2
Direct
  0.0000000029575205  0.9999999973844496  0.4999999581771419
  0.6666625554332768  0.3333374095621267  0.5832637052899159
  0.3333374438495511  0.6666625913620479  0.4167362088128357
  0.6666674842100448  0.3333328930466024  0.5429472369259187
  0.3333325135496068  0.6666671086447735  0.4570528907941878
```

**Phonon dispersion curve**

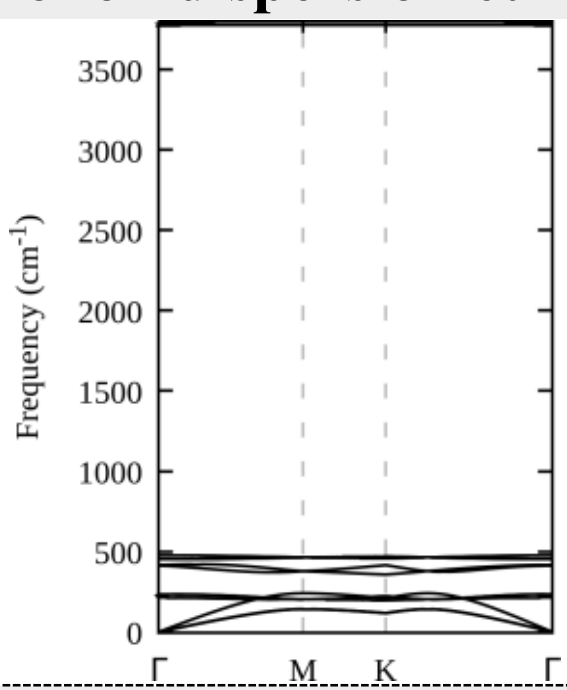

**Projected band structure and density of states**

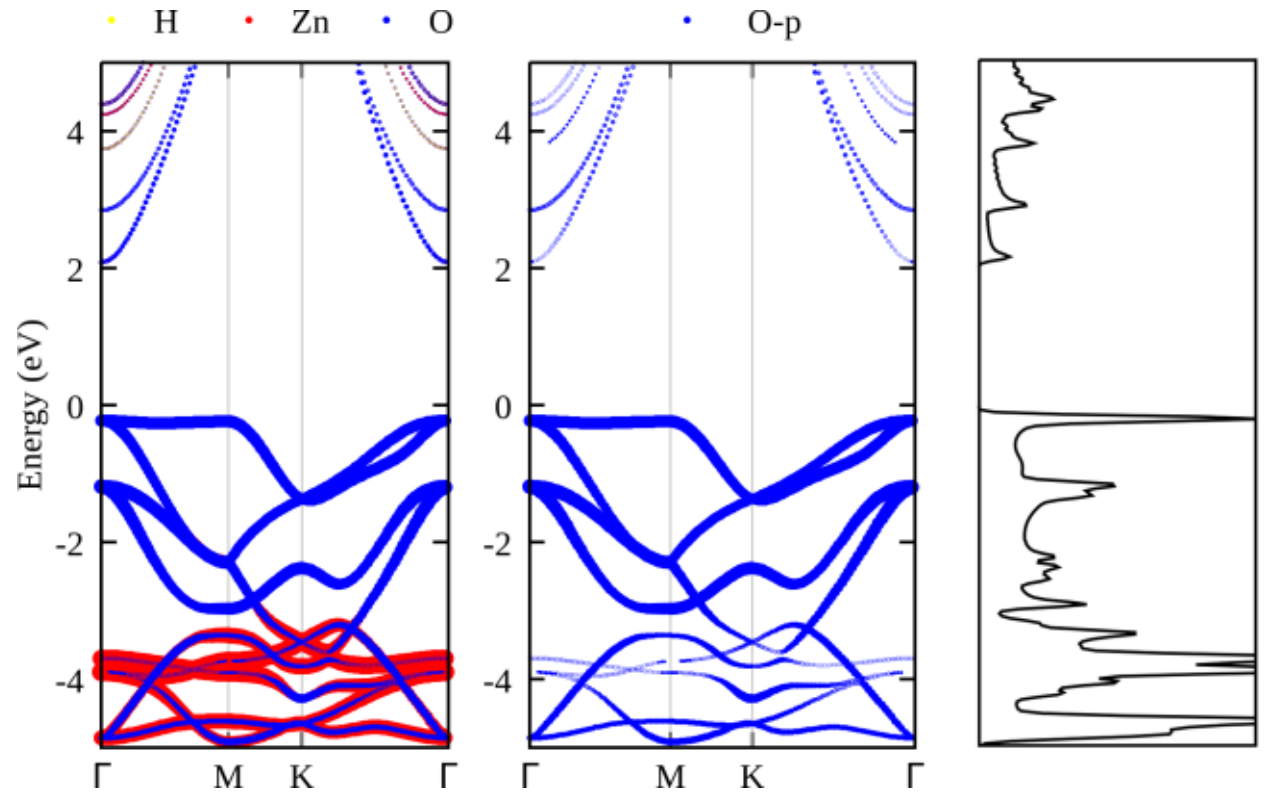

**Magnetic moment and spin polarization energy as a function of hole doping concentration**

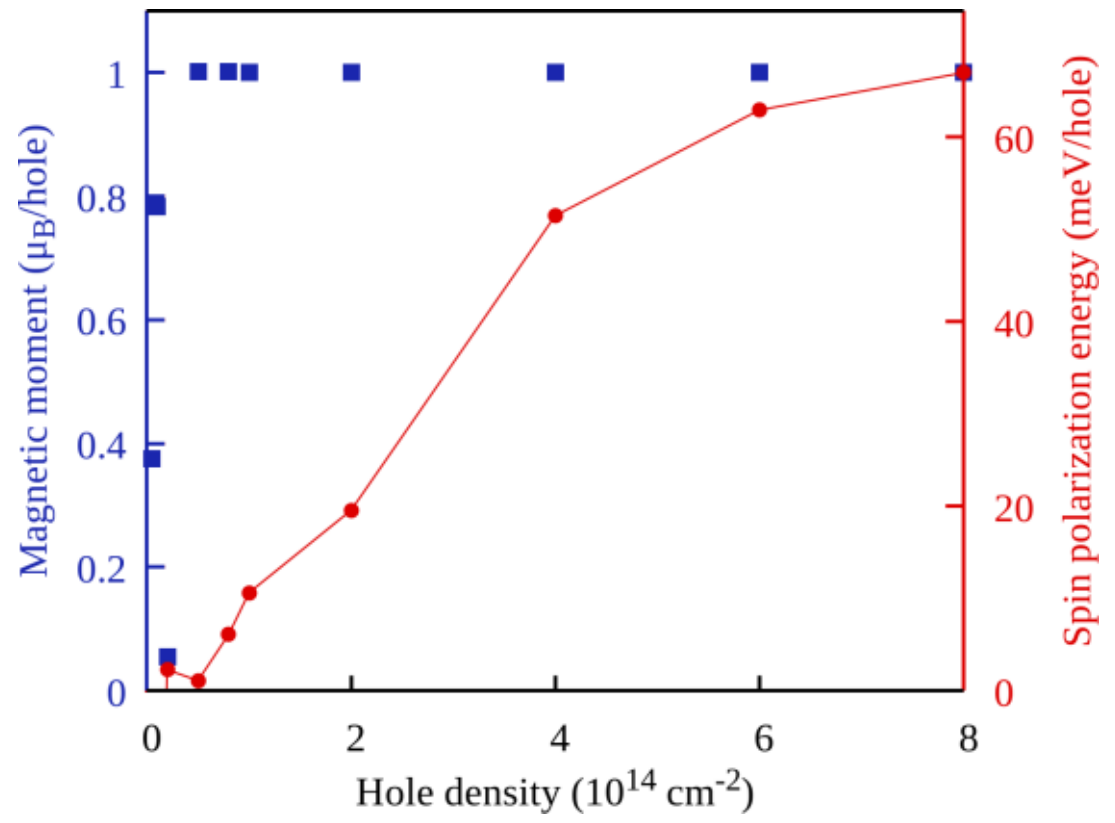

**Magnetic configurations and spin Hamiltonian**

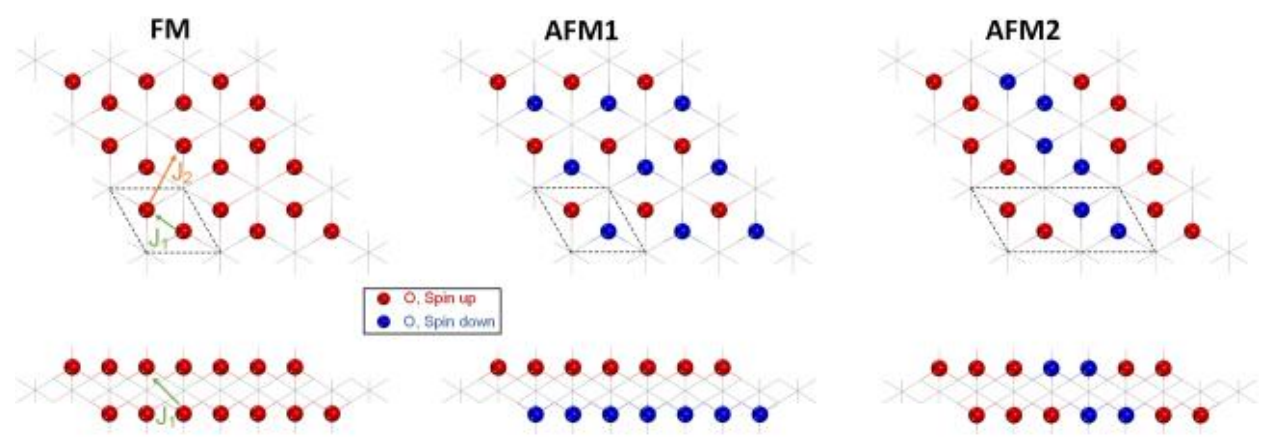

$$E_{FM} = E_0 - 3J_1S^2 - 6J_2S^2$$
$$E_{AFM1} = E_0 + 3J_1S^2 - 6J_2S^2$$
$$E_{AFM2} = E_0 - J_1S^2 + 2J_2S^2$$

**Magnetic exchange coupling parameters**

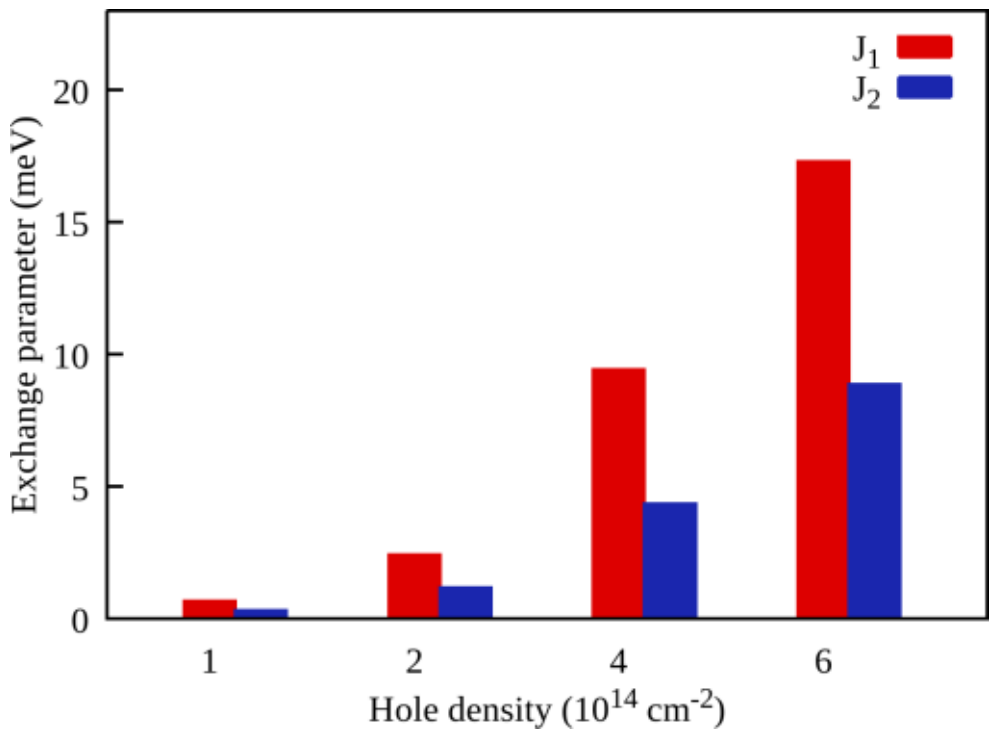

**Magnetic anisotropy energy (MAE, μeV) per magnetic atom**

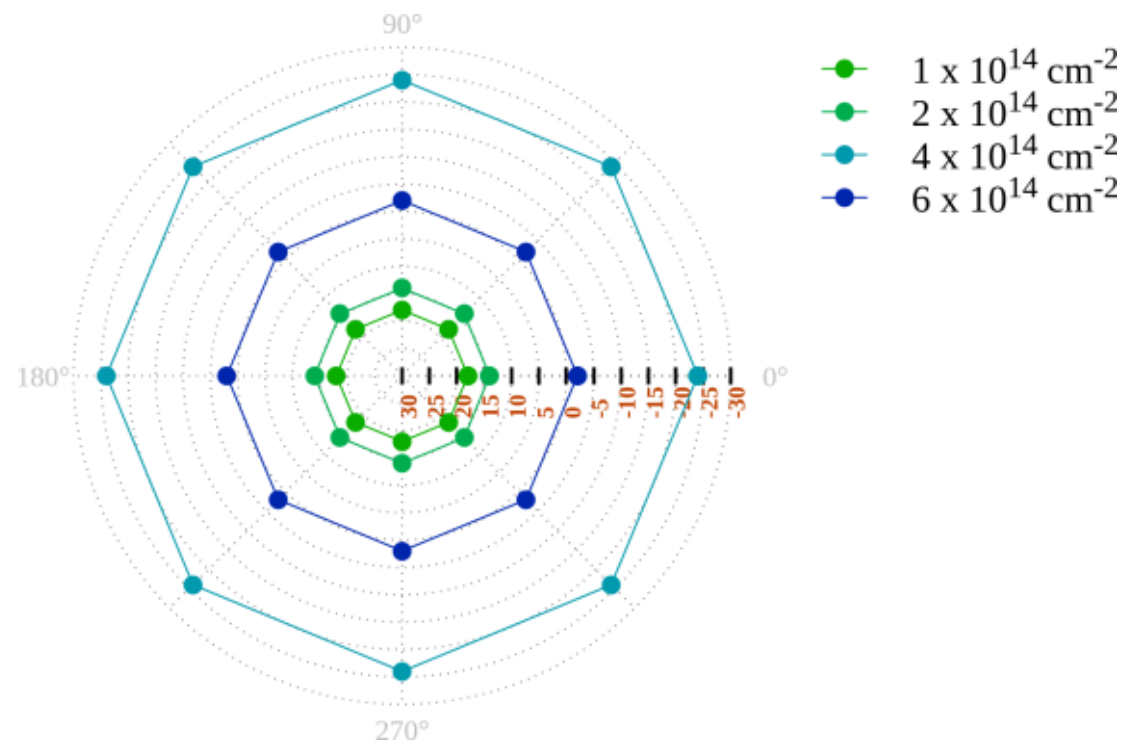

MAE = $E_{\parallel} - E_{\perp}$, a positive (negative) value of MAE indicates the off-plane (in-plane) easy axis.

**Monte Carlo simulations of the normalized magnetization of as a function of temperature**

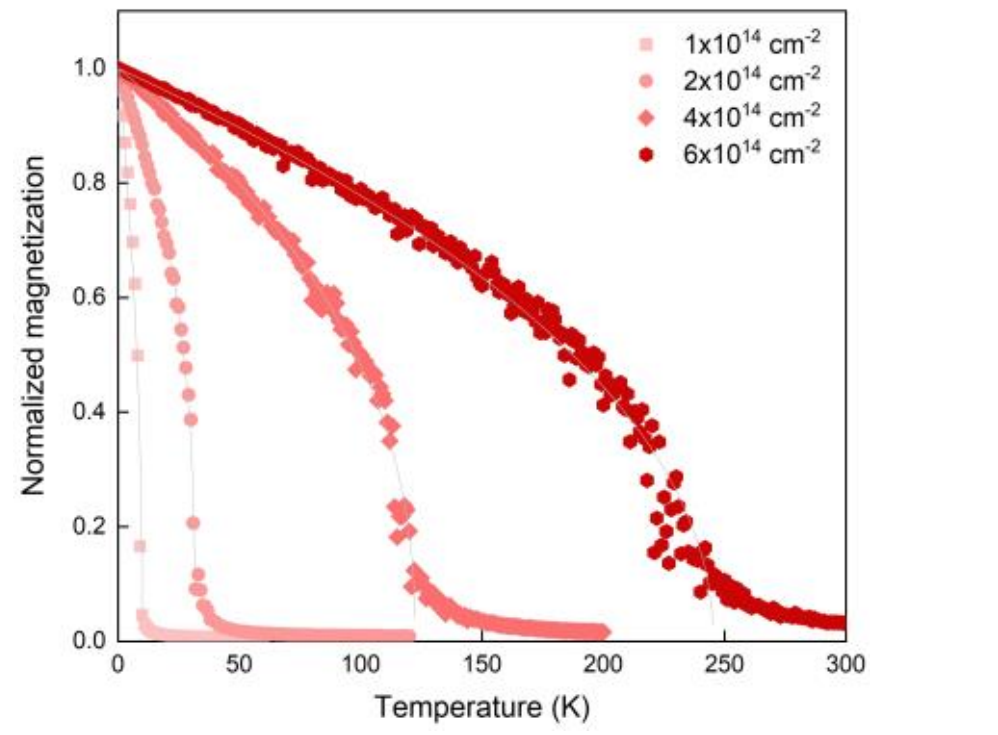

$T_c$: $1\times10^{14}$ cm$^{-2}$: 10 K  $2\times10^{14}$ cm$^{-2}$: 32 K  $4\times10^{14}$ cm$^{-2}$: 122 K  $6\times10^{14}$ cm$^{-2}$: 245 K

# 49. $Cd(OH)_2$

| MC2D-ID | C2DB | 2dmat-ID | USPEX | Space group | Band gap (eV) |
|---|---|---|---|---|---|
| 30 | - | 2dm-4273 | - | P3m1 | 2.27 |

**Convex hull** | **Atomic structure** | **Atomic coordinates** | **Phonon dispersion curve**

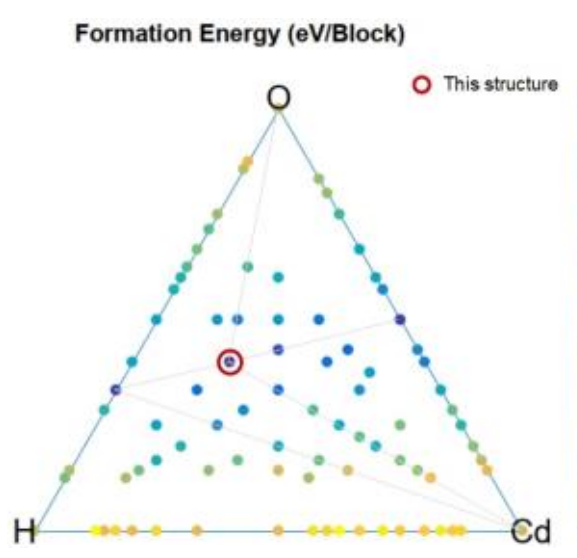

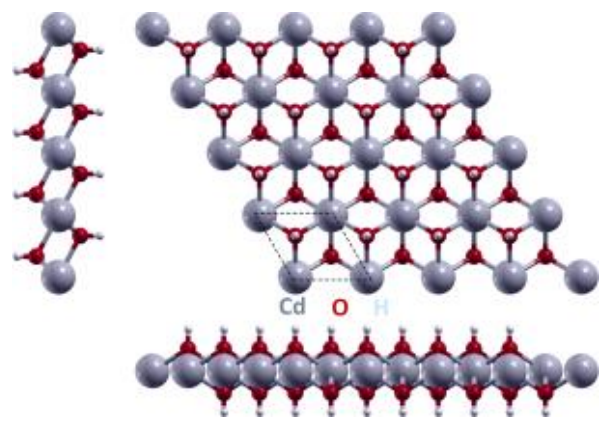

```
Cd1H2O2
   1.00000000000000
   3.5799420594205853    0.0000000000000000    0.0000000000000000
  -1.7900444418957295    3.1002784089441624    0.0000000000000000
   0.0000000000000000    0.0000000000000000   24.1669234368999994
     Cd     H      O
      1      2      2
Direct
  0.9999999998872511  0.0000000000548539  0.5000000004885621
  0.3333334950215061  0.6666664932644011  0.5857309909998492
  0.6666665049997746  0.3333335067164072  0.4142690100591722
  0.3333333550734849  0.6666666401951815  0.5455607873114161
  0.6666666449619711  0.3333333597691635  0.4544392111389257
```

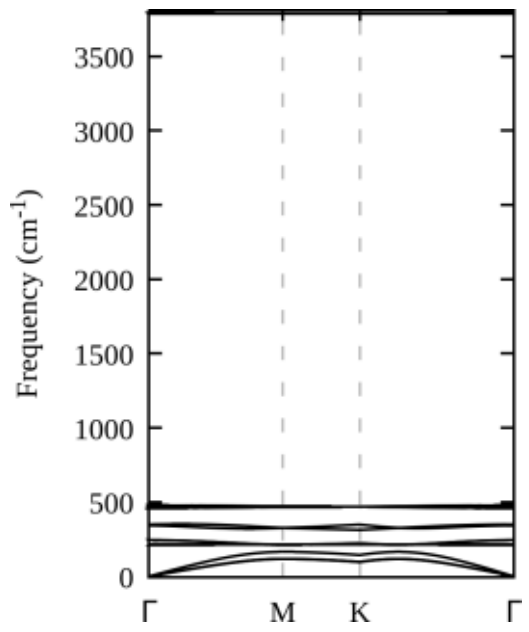

## Projected band structure and density of states

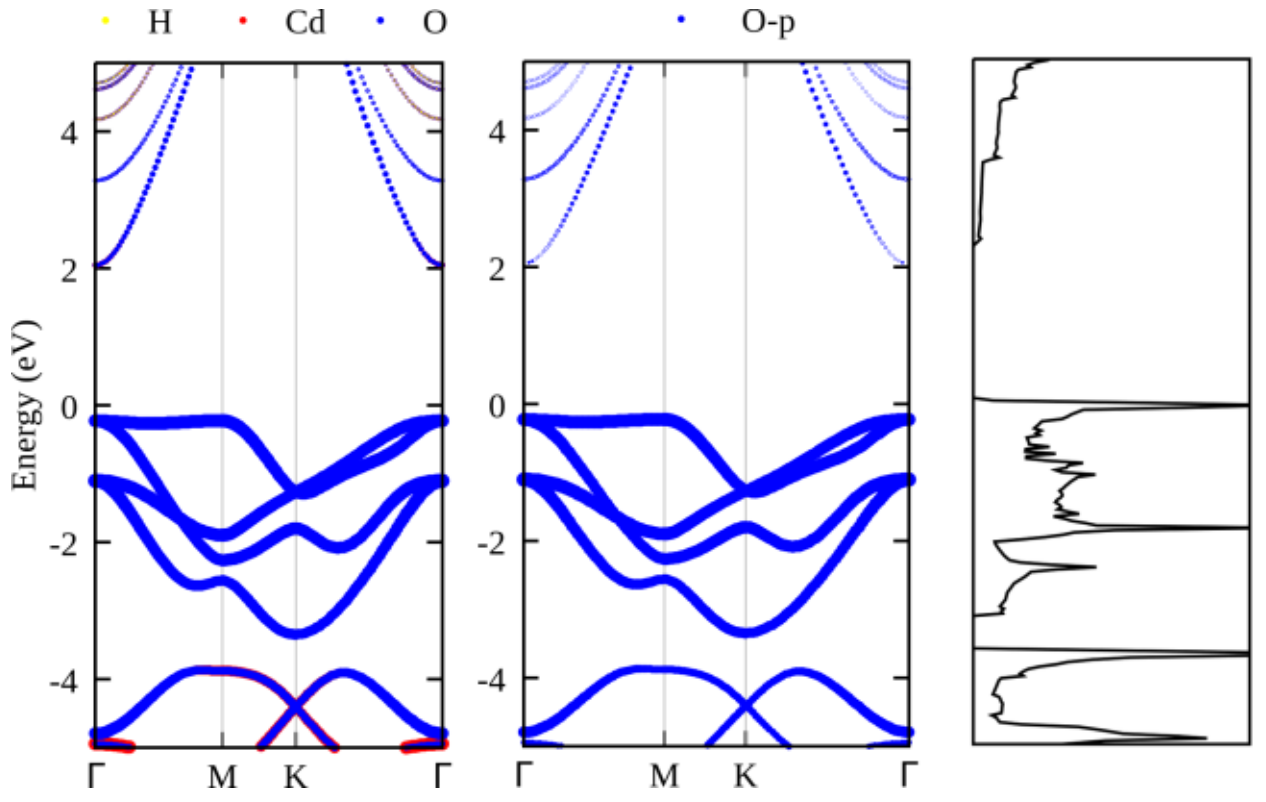

## Magnetic moment and spin polarization energy as a function of hole doping concentration

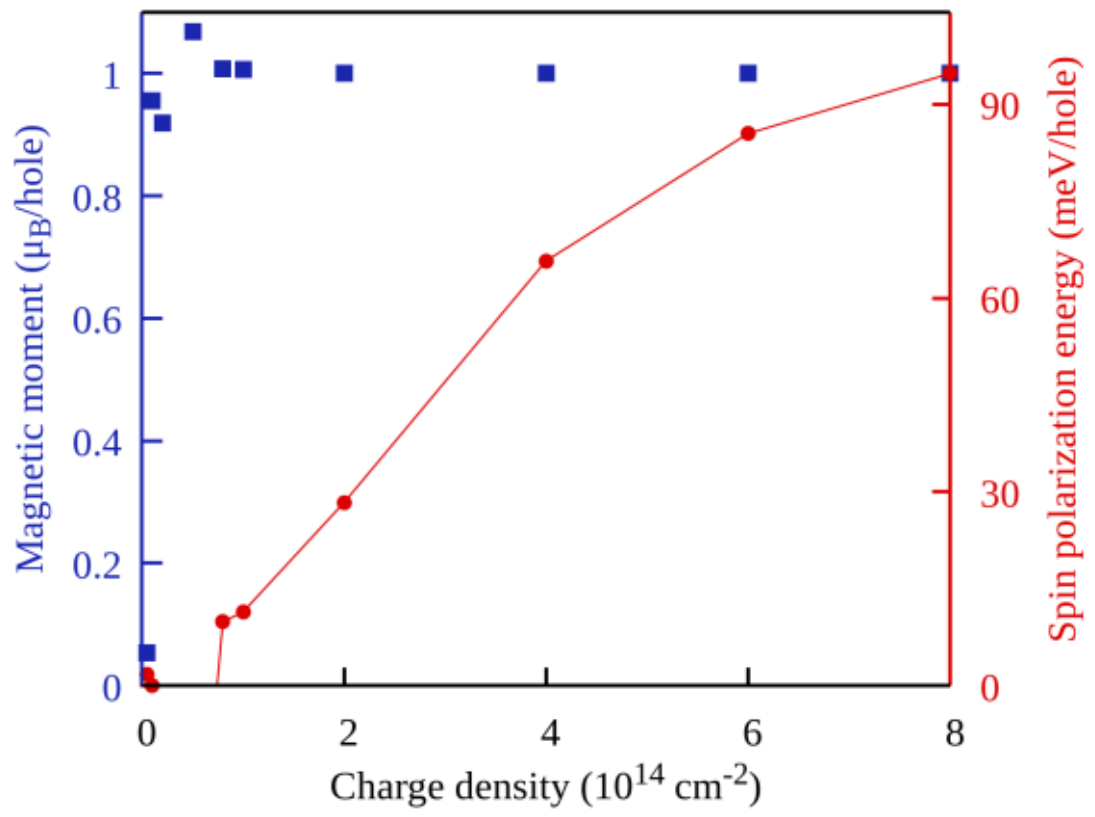

## Magnetic configurations and spin Hamiltonian

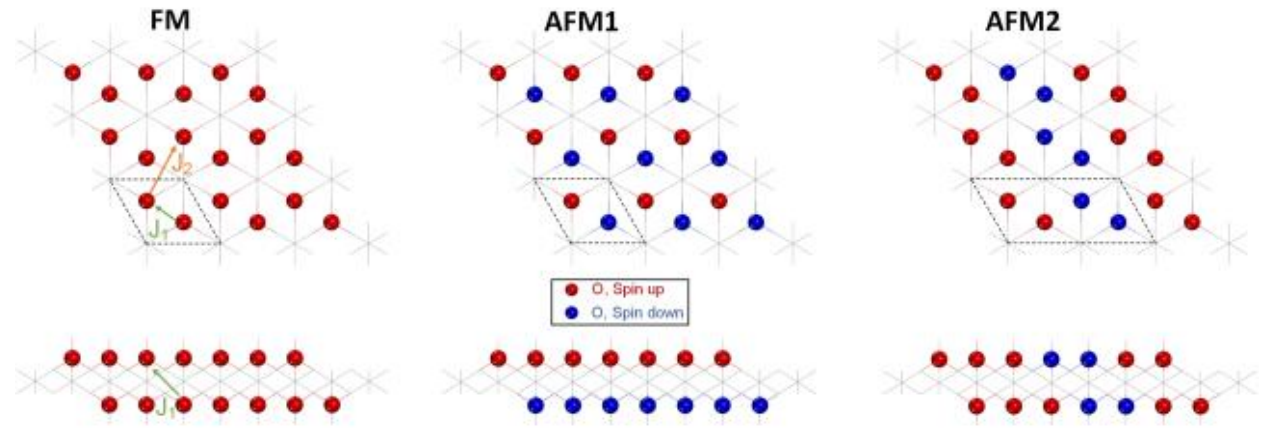

$$E_{FM} = E_0 - 3J_1S^2 - 6J_2S^2$$
$$E_{AFM1} = E_0 + 3J_1S^2 - 6J_2S^2$$
$$E_{AFM2} = E_0 - J_1S^2 + 2J_2S^2$$

## Magnetic exchange coupling parameters

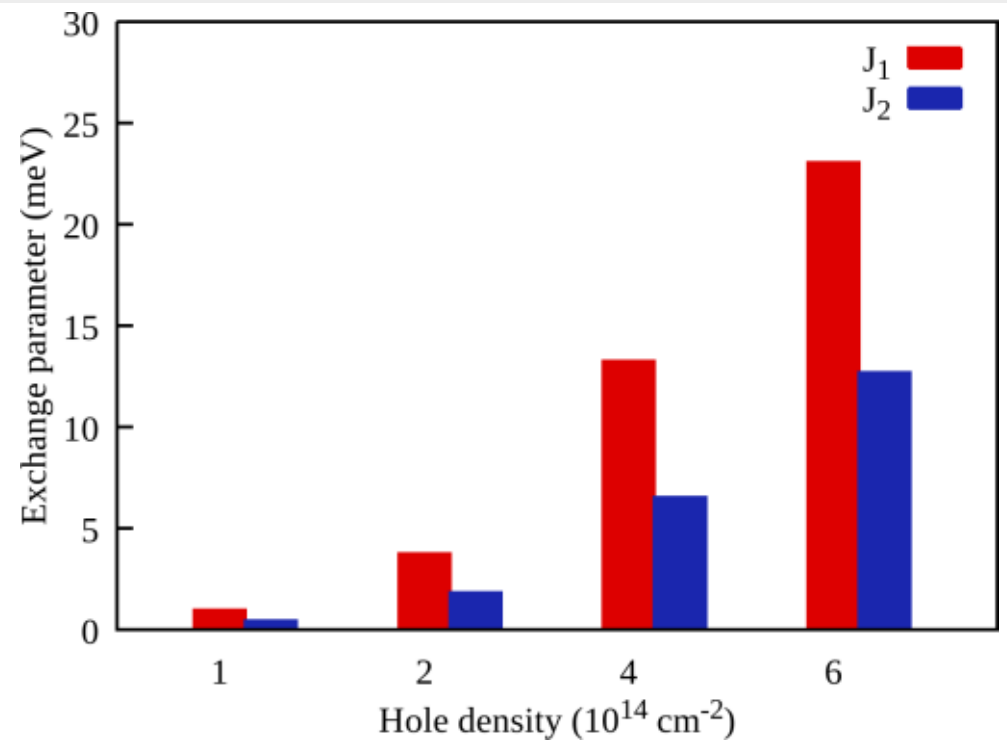

## Magnetic anisotropy energy (MAE, μeV) per magnetic atom

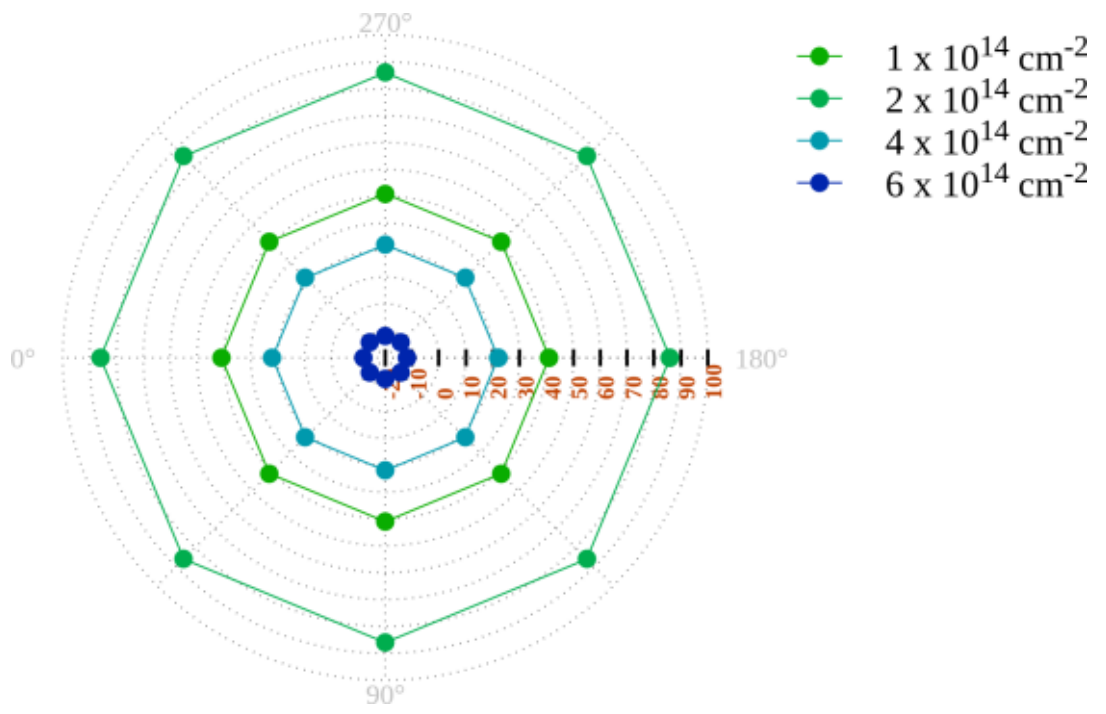

MAE = $E_{\parallel} - E_{\perp}$, a positive (negative) value of MAE indicates the off-plane (in-plane) easy axis.

## Monte Carlo simulations of the normalized magnetization of as a function of temperature

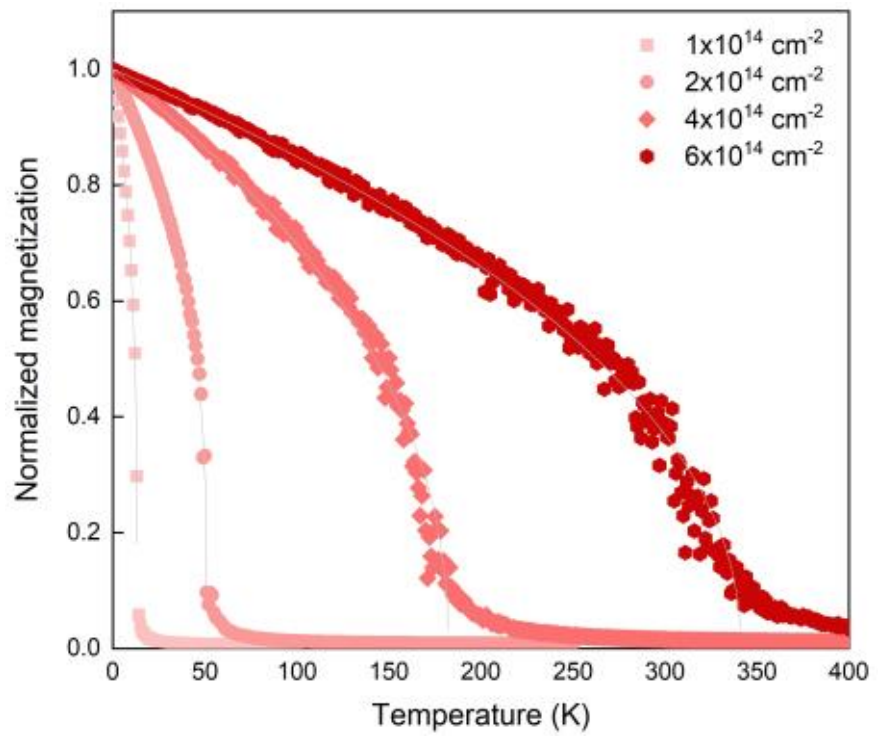

$T_c$: $1\times10^{14}$ cm$^{-2}$: 13 K  $2\times10^{14}$ cm$^{-2}$: 51 K  $4\times10^{14}$ cm$^{-2}$: 182 K  $6\times10^{14}$ cm$^{-2}$: 341 K

# 50. TeC

| MC2D-ID | C2DB | 2dmat-ID | USPEX | Space group | Band gap (eV) |
|---|---|---|---|---|---|
| - | - | 2dm-6385 | - | P3m1 | 1.29 |

**Convex hull**

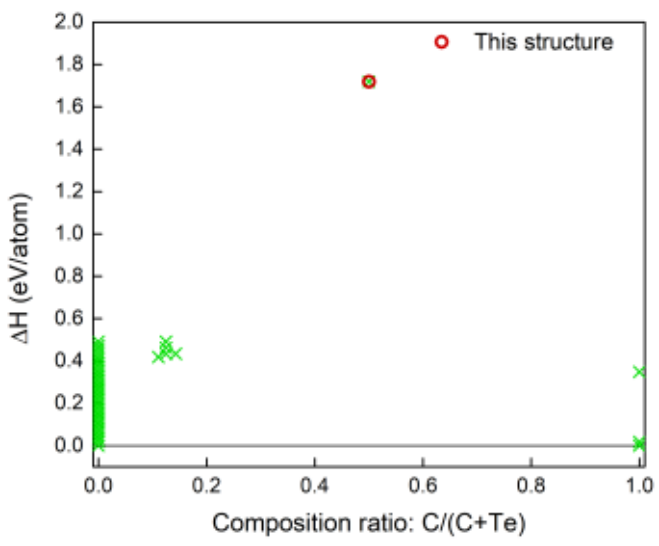

**Atomic structure**

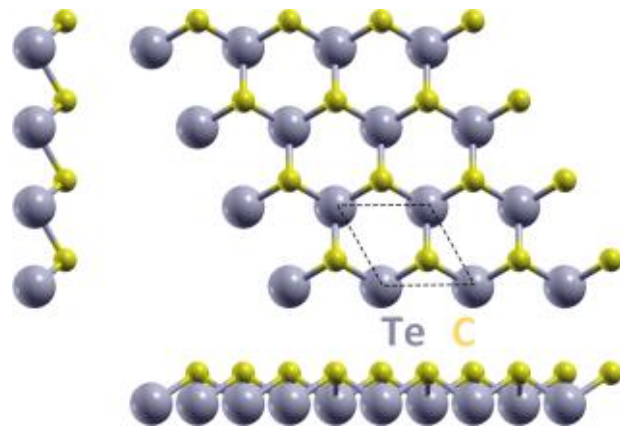

**Atomic coordinates**

```
Te1C1
   1.00000000000000
   3.3479315019084788    0.0000000000000000    0.0000000000000000
  -1.6739654406742093    2.8993885712899390    0.0000000000000000
   0.0000000000000000    0.0000000000000000   19.0773769999999985
   Te    C
   1     1
Direct
  0.3333397911702534   0.6666799631394653   0.4712258947687573
  0.9999752088297456   0.9999500368605325   0.5297011052312470
```

**Phonon dispersion curve**

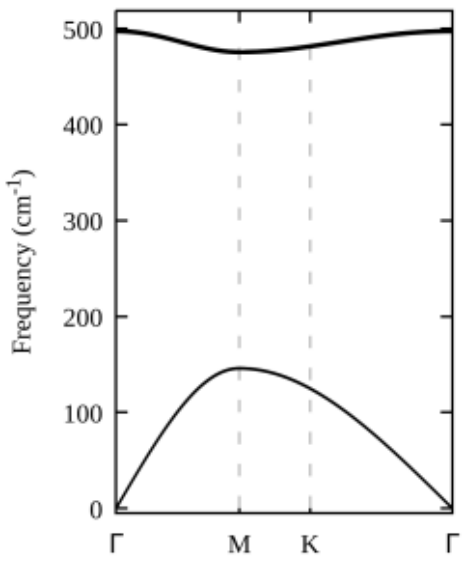

**Projected band structure and density of states**

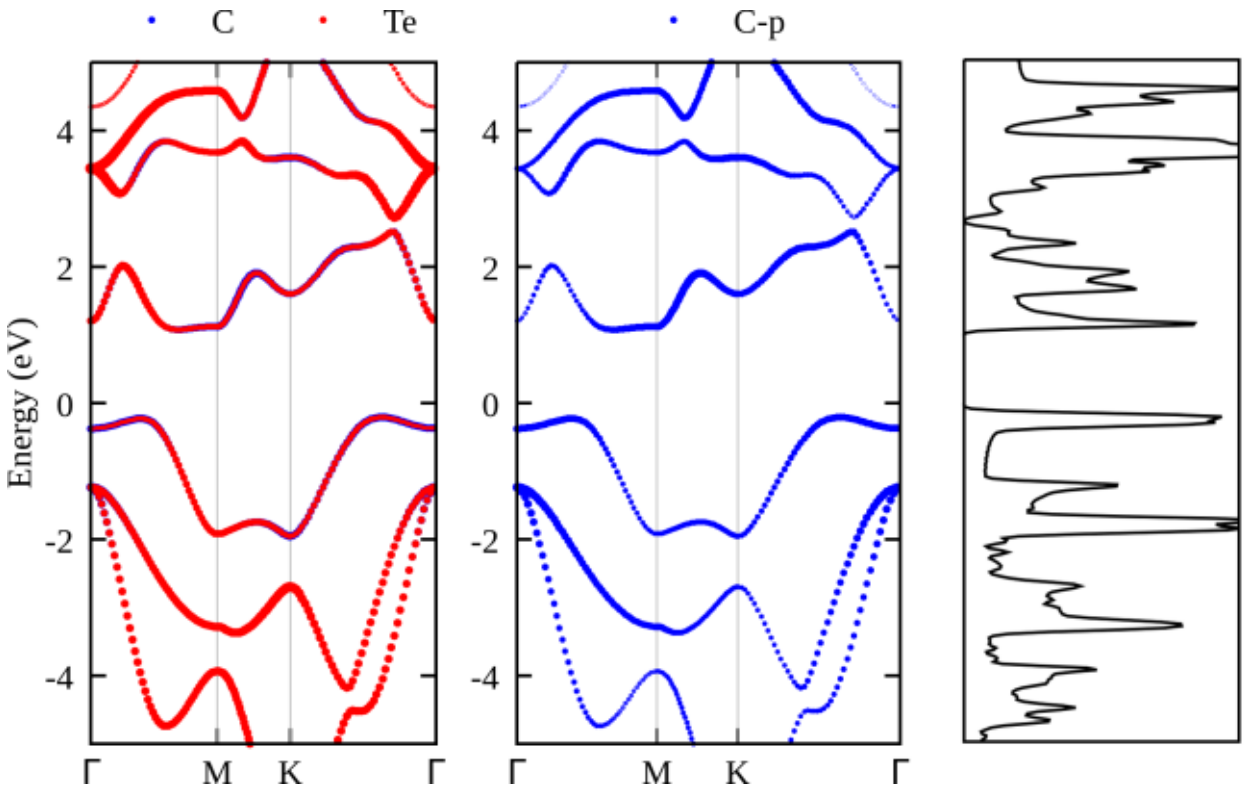

**Magnetic moment and spin polarization energy as a function of hole doping concentration**

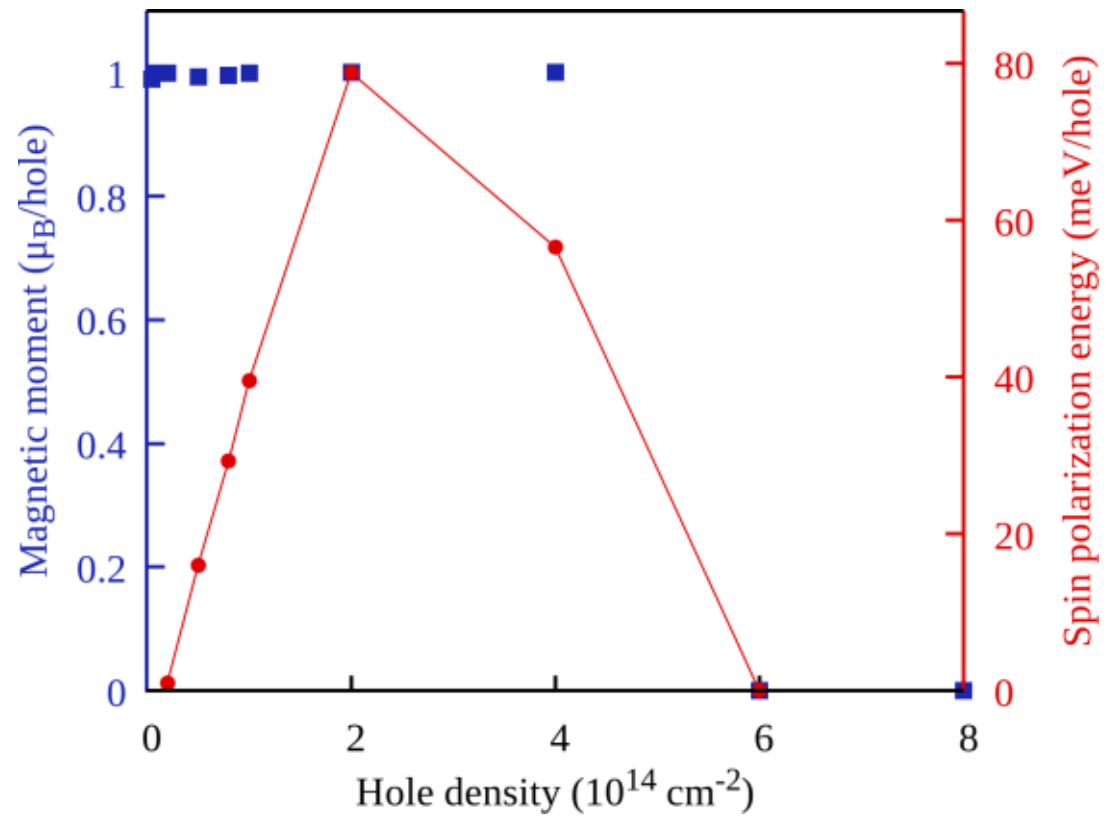

**Magnetic configurations and spin Hamiltonian**

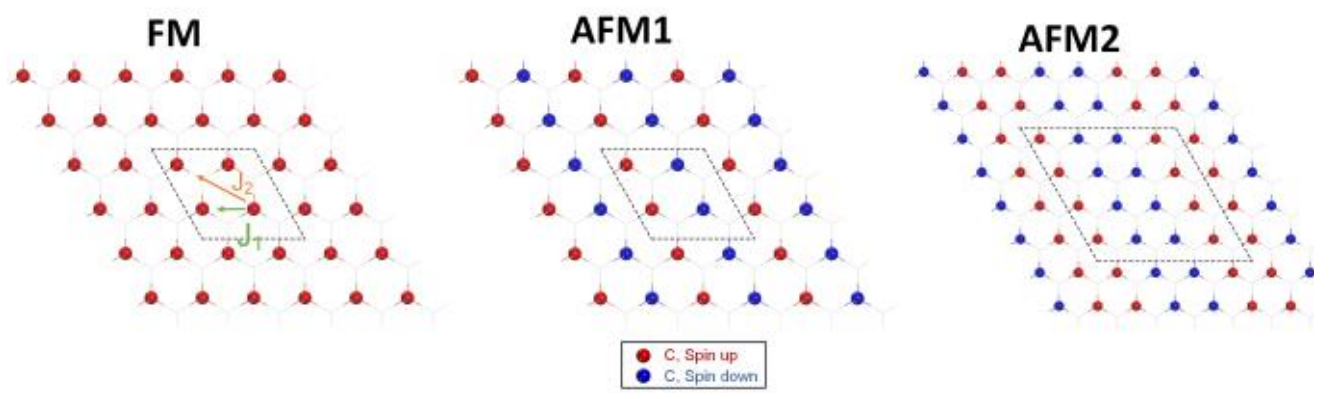

$$E_{FM} = E_0 - 12J_1S^2 - 12J_2S^2$$
$$E_{AFM1} = E_0 + 4J_1S^2 + 4J_2S^2$$
$$E_{AFM2} = E_0 - 4J_1S^2 + 4J_2S^2$$

**Magnetic exchange coupling parameters**

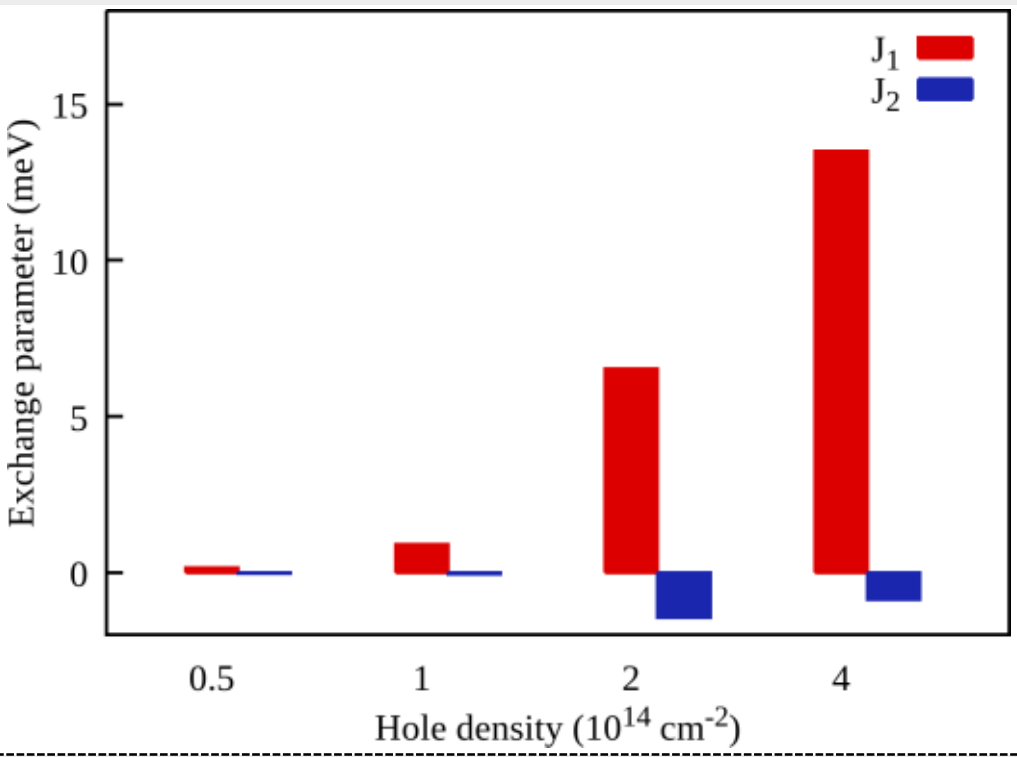

**Magnetic anisotropy energy (MAE, μeV) per magnetic atom**

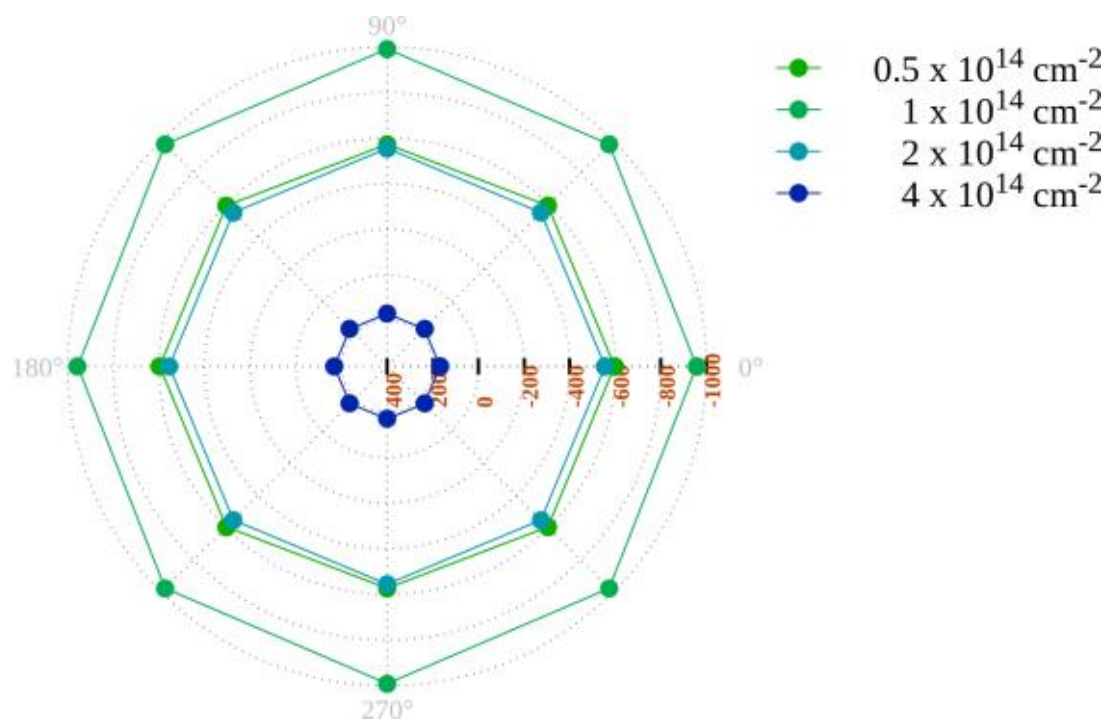

MAE = $E_{\parallel} - E_{\perp}$, a positive (negative) value of MAE indicates the off-plane (in-plane) easy axis.

**Monte Carlo simulations of the normalized magnetization of as a function of temperature**

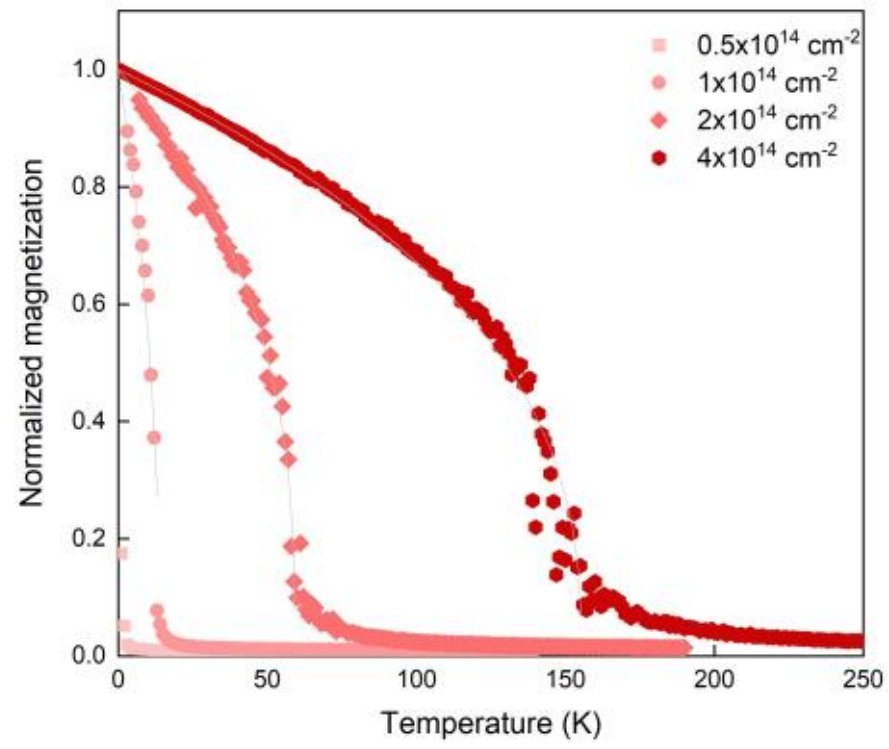

$T_c$: $0.5\times10^{14}$ cm$^{-2}$: 2 K  $1\times10^{14}$ cm$^{-2}$: 13 K

$2\times10^{14}$ cm$^{-2}$: 59 K  $4\times10^{14}$ cm$^{-2}$: 155 K

# 51. $PbCl_2$

| MC2D-ID | C2DB | 2dmat-ID | USPEX | Space group | Band gap (eV) |
|---|---|---|---|---|---|
| - | ✓ | 2dm-2389 | - | P6m2 | 3.51 |

**Convex hull** | **Atomic structure** | **Atomic coordinates** | **Phonon dispersion curve**

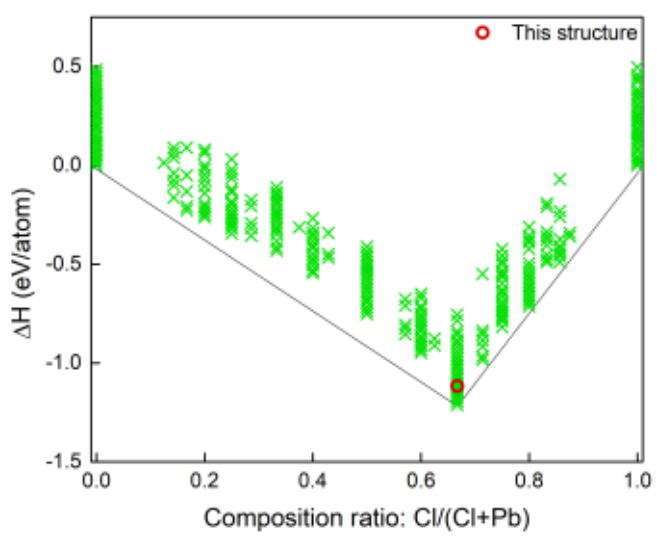

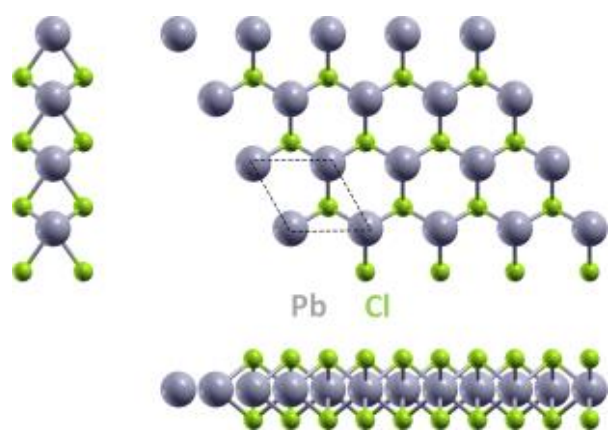

```
Pb1Cl2
   1.00000000000000
   4.1792382900134513    0.0000000000000000    0.0000000000000000
  -2.0896228122357243    3.6193252552410371    0.0000000000000000
   0.0000000000000000    0.0000000000000000   23.8380000000000010
   Pb    Cl
   1     2
Direct
  0.0000000478759219  0.9999999537932069  0.0000000242073526
  0.6666669767475994  0.3333329992222076  0.0713952306129286
  0.6666669753764722  0.3333330469845919  0.9286047451797188
```

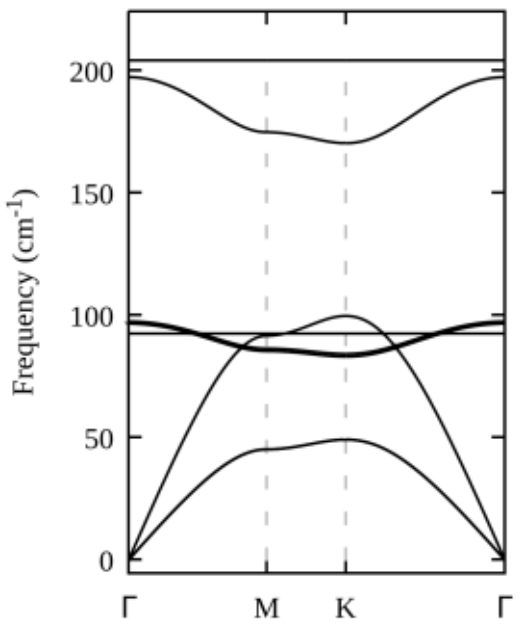

**Projected band structure and density of states**

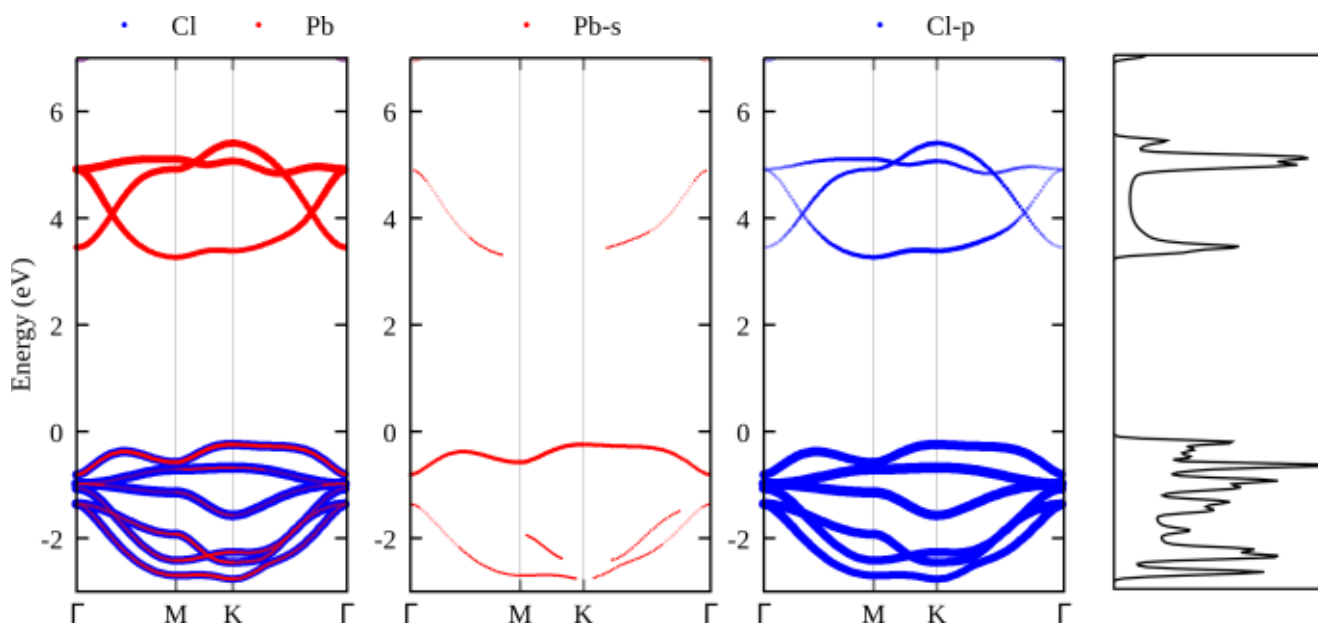

**Magnetic moment and spin polarization energy as a function of hole doping concentration**

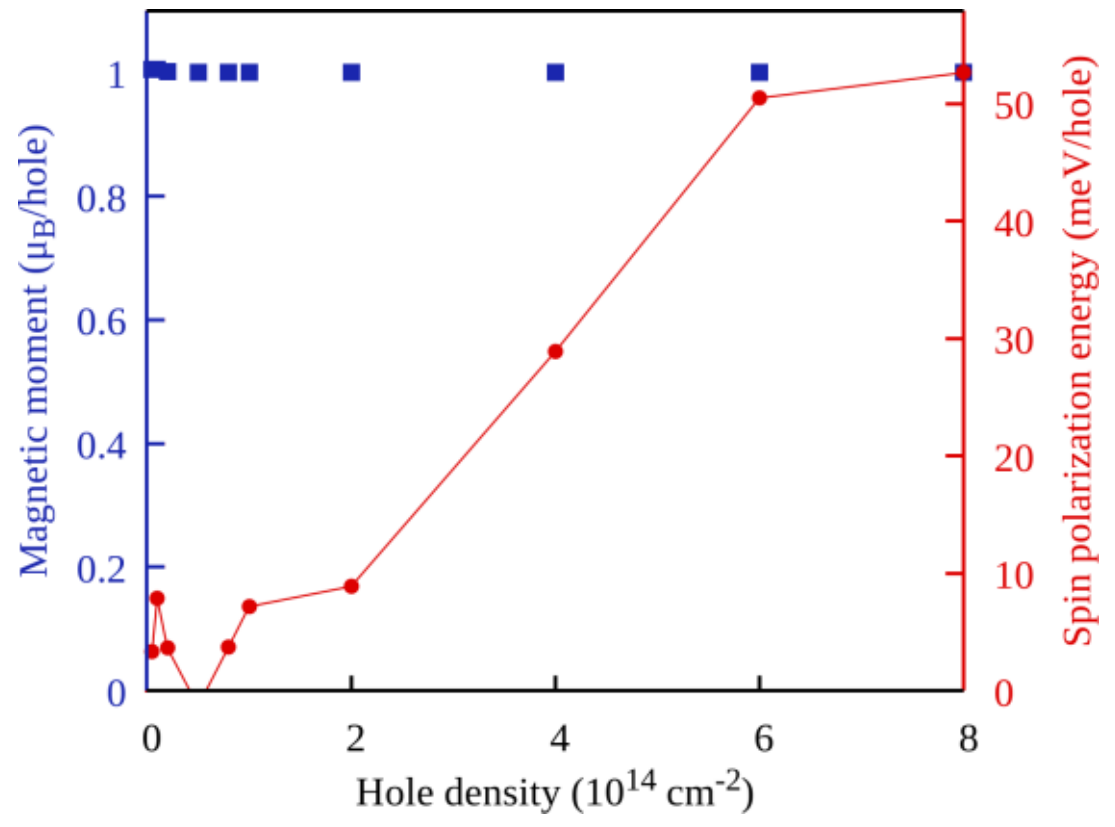

**Magnetic configurations and spin Hamiltonian**

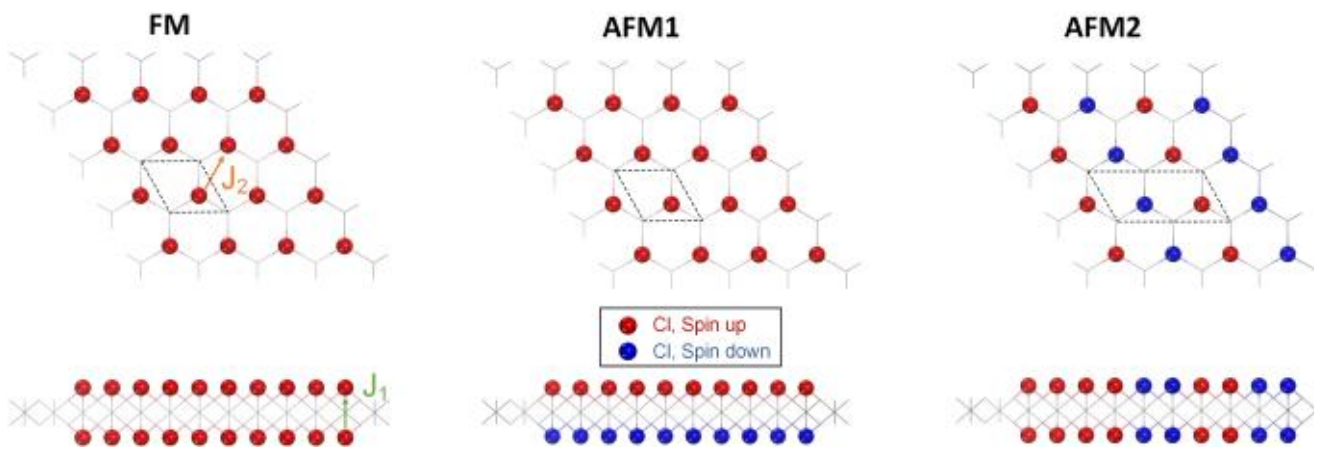

$$E_{FM} = E_0 - J_1S^2 - 6J_2S^2$$
$$E_{AFM1} = E_0 + J_1S^2 - 6J_2S^2$$
$$E_{AFM2} = E_0 - J_1S^2 + 2J_2S^2$$

**Magnetic exchange coupling parameters**

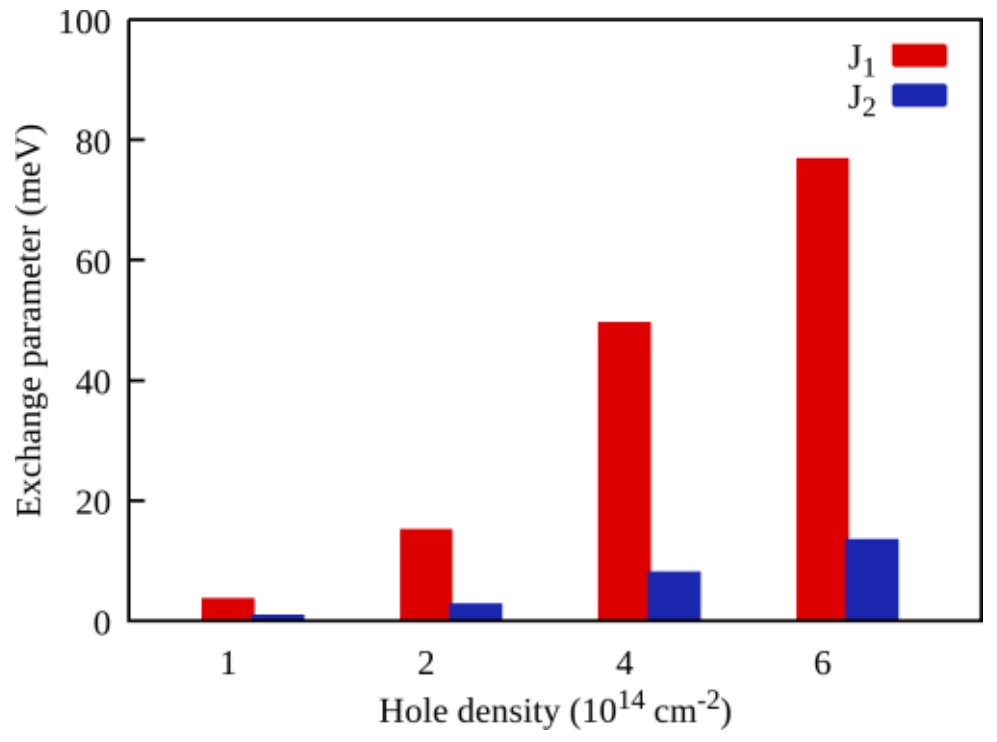

**Magnetic anisotropy energy (MAE, μeV) per magnetic atom**

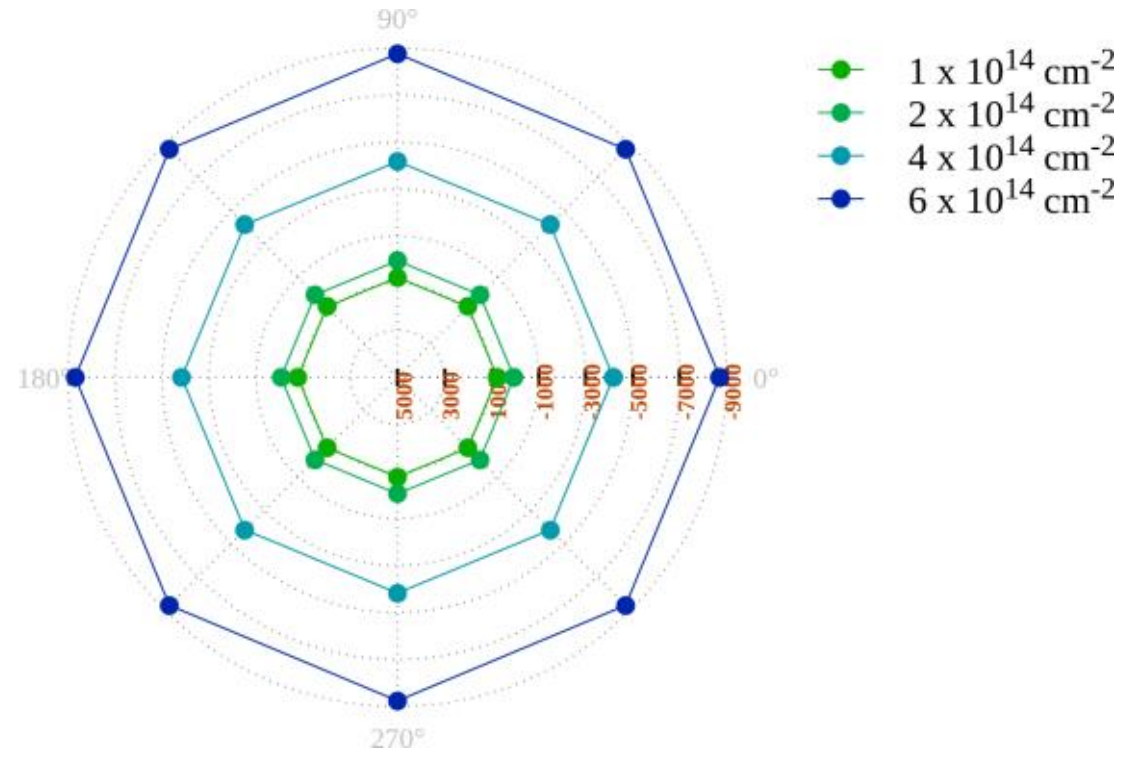

MAE = $E_{\parallel} - E_{\perp}$, a positive (negative) value of MAE indicates the off-plane (in-plane) easy axis.

**Monte Carlo simulations of the normalized magnetization of as a function of temperature**

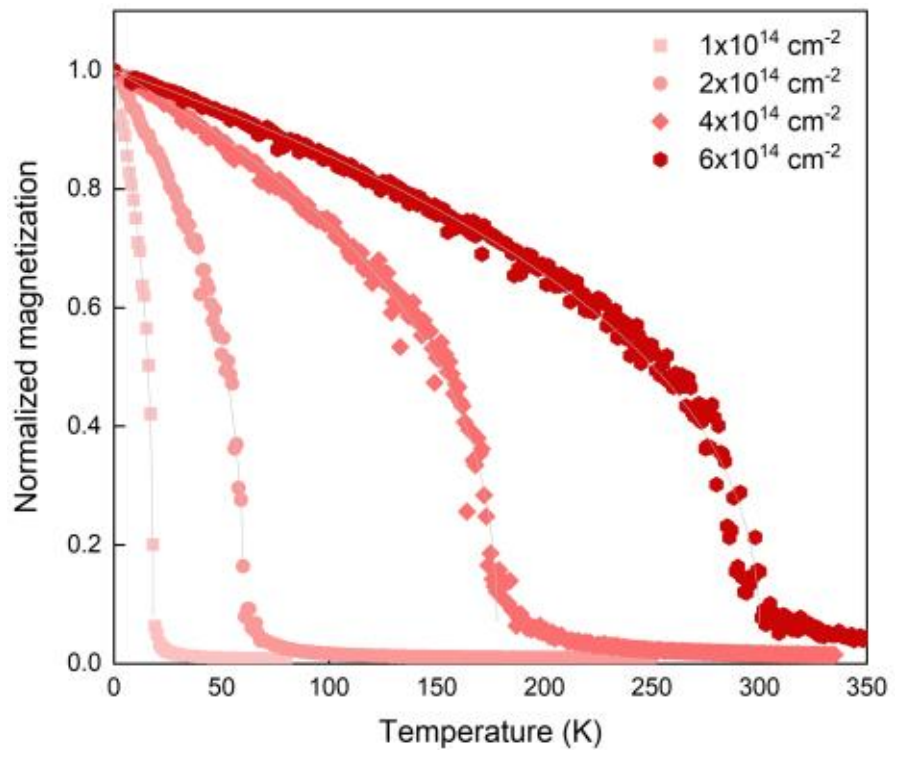

$T_c$: $1\times10^{14}$ cm$^{-2}$: 18 K  $2\times10^{14}$ cm$^{-2}$: 60 K  $4\times10^{14}$ cm$^{-2}$: 178 K  $6\times10^{14}$ cm$^{-2}$: 300 K

# 52.$PbBr_2$

| MC2D-ID | C2DB | 2dmat-ID | USPEX | Space group | Band gap (eV) |
|---|---|---|---|---|---|
| - | ✓ | 2dm-2251 | - | P6m2 | 3.20 |

**Convex hull** | **Atomic structure** | **Atomic coordinates** | **Phonon dispersion curve**

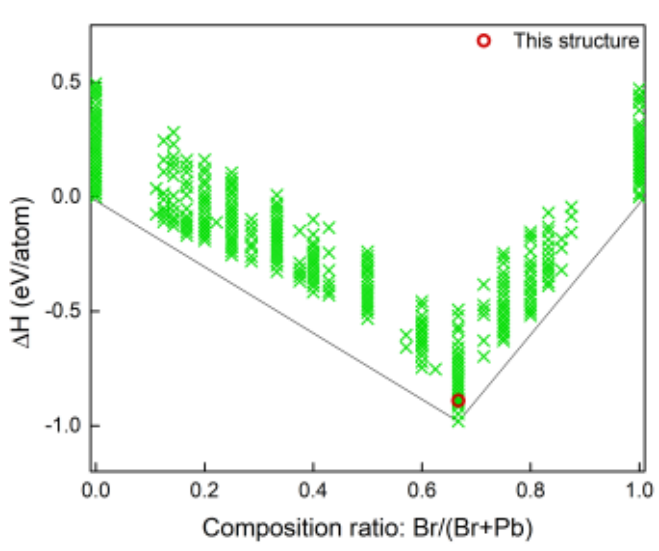

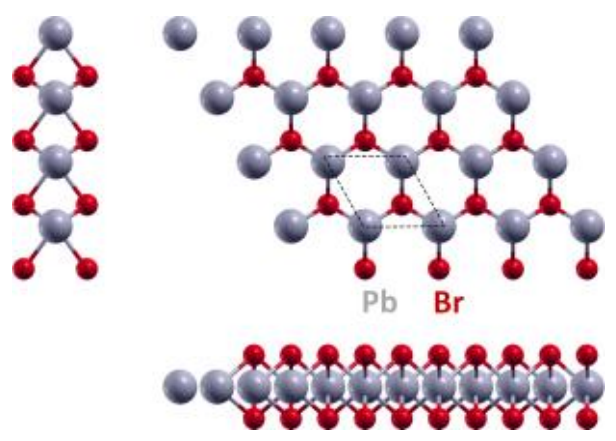

```
Pb1Br2
   1.00000000000000
   4.3075062621630185    0.0000000000000000    0.0000000000000000
  -2.1537530794790190    3.7304106452124888    0.0000000000000000
   0.0000000000000000    0.0000000000000000   23.8380000000000010
   Pb    Br
   1     2
Direct
  0.0000003932458341  0.9999995832829853  0.0000000056039582
  0.6666668147712116  0.3333332242805724  0.0773418257317360
  0.6666667919829479  0.3333331924364487  0.9226581686643058
```

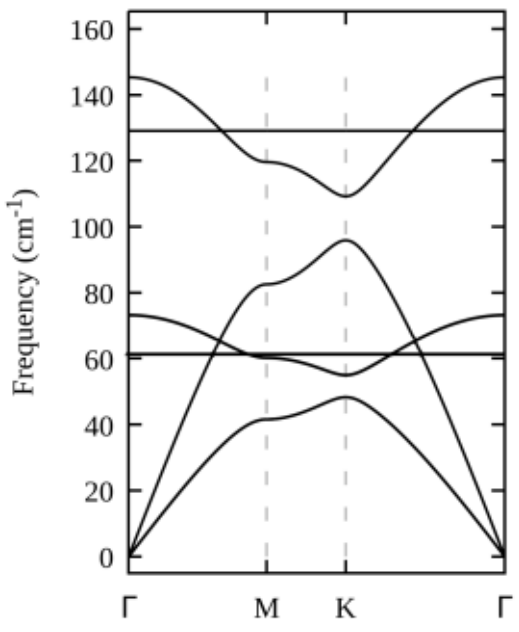

**Projected band structure and density of states** | **Magnetic moment and spin polarization energy as a function of hole doping concentration**

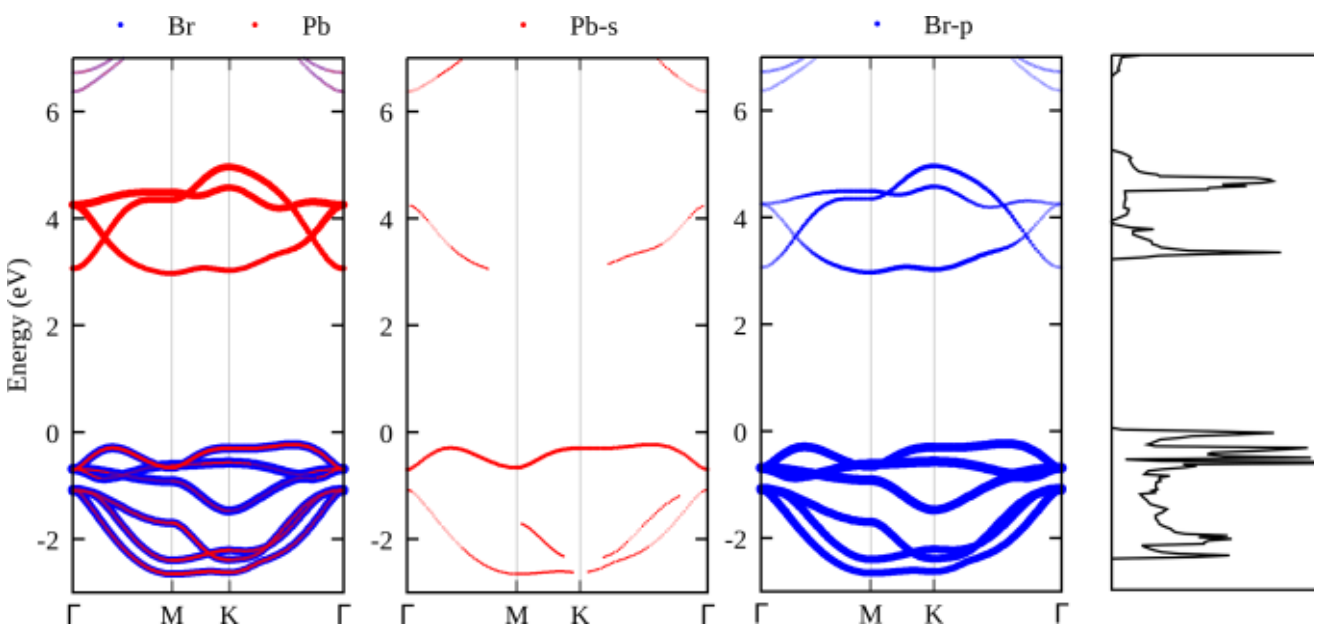

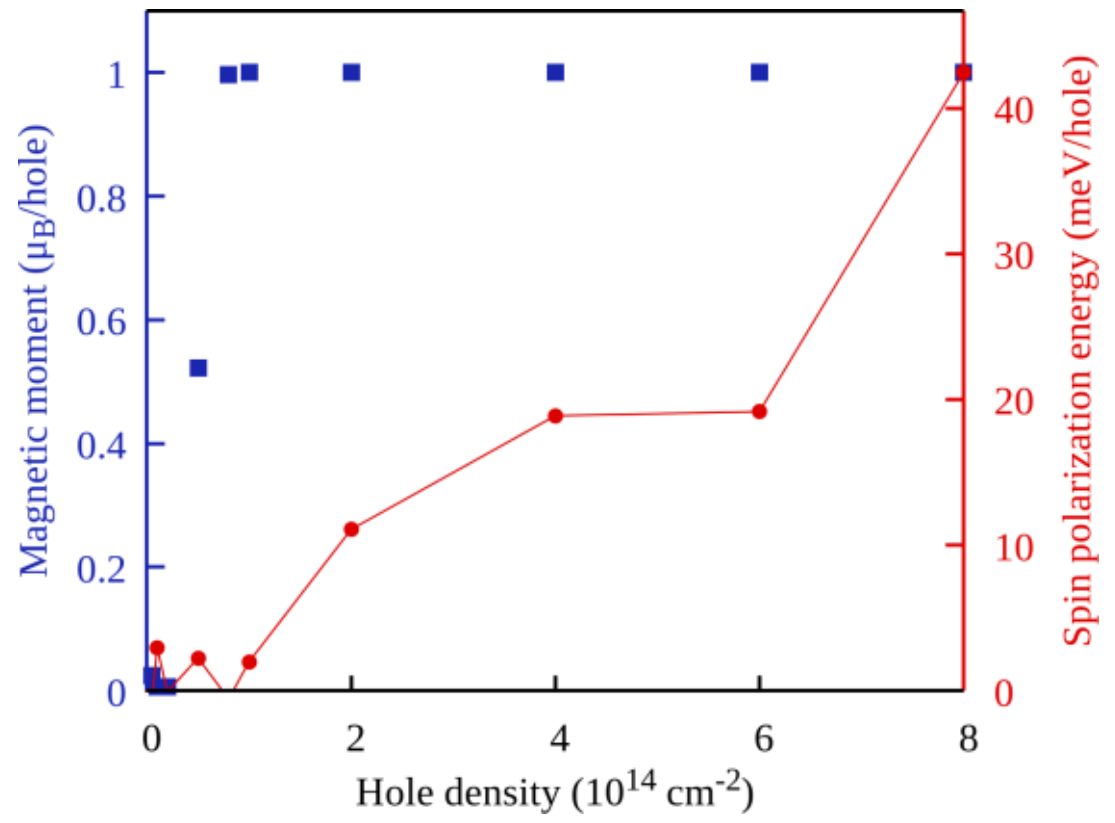

**Magnetic configurations and spin Hamiltonian** | **Magnetic exchange coupling parameters**

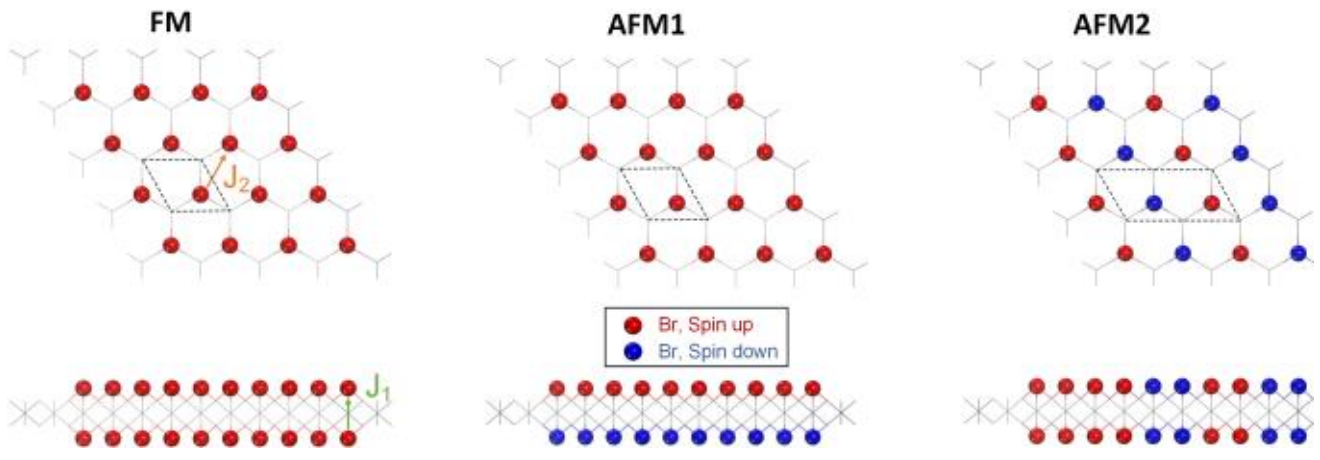

$$E_{FM} = E_0 - J_1S^2 - 6J_2S^2$$
$$E_{AFM1} = E_0 + J_1S^2 - 6J_2S^2$$
$$E_{AFM2} = E_0 - J_1S^2 + 2J_2S^2$$

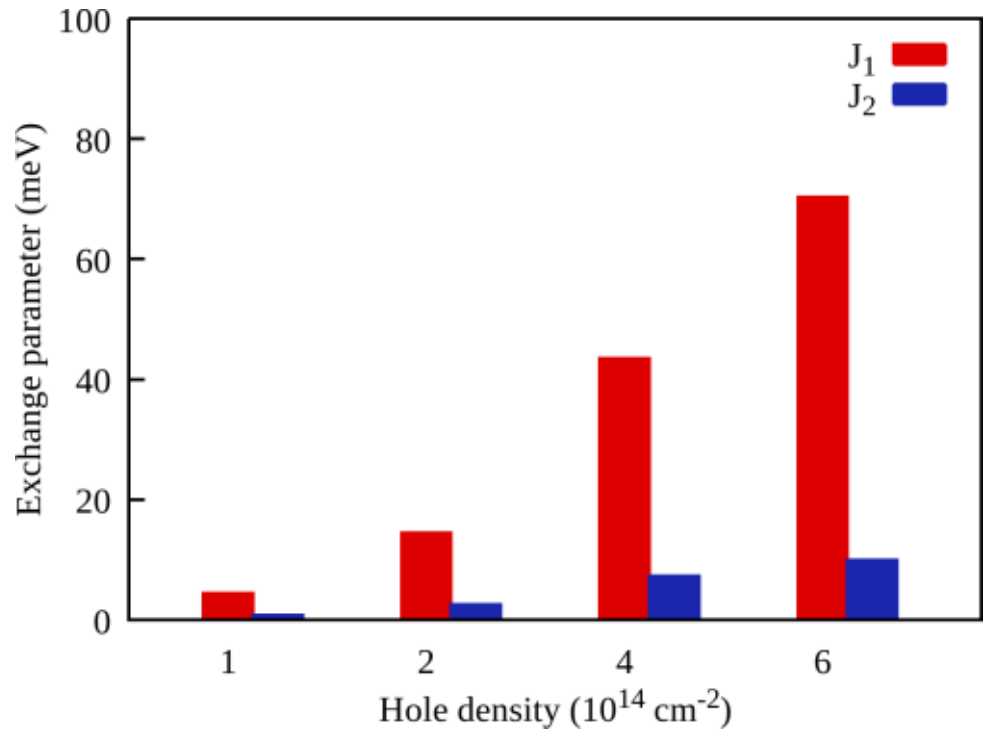

**Magnetic anisotropy energy (MAE, μeV) per magnetic atom** | **Monte Carlo simulations of the normalized magnetization of as a function of temperature**

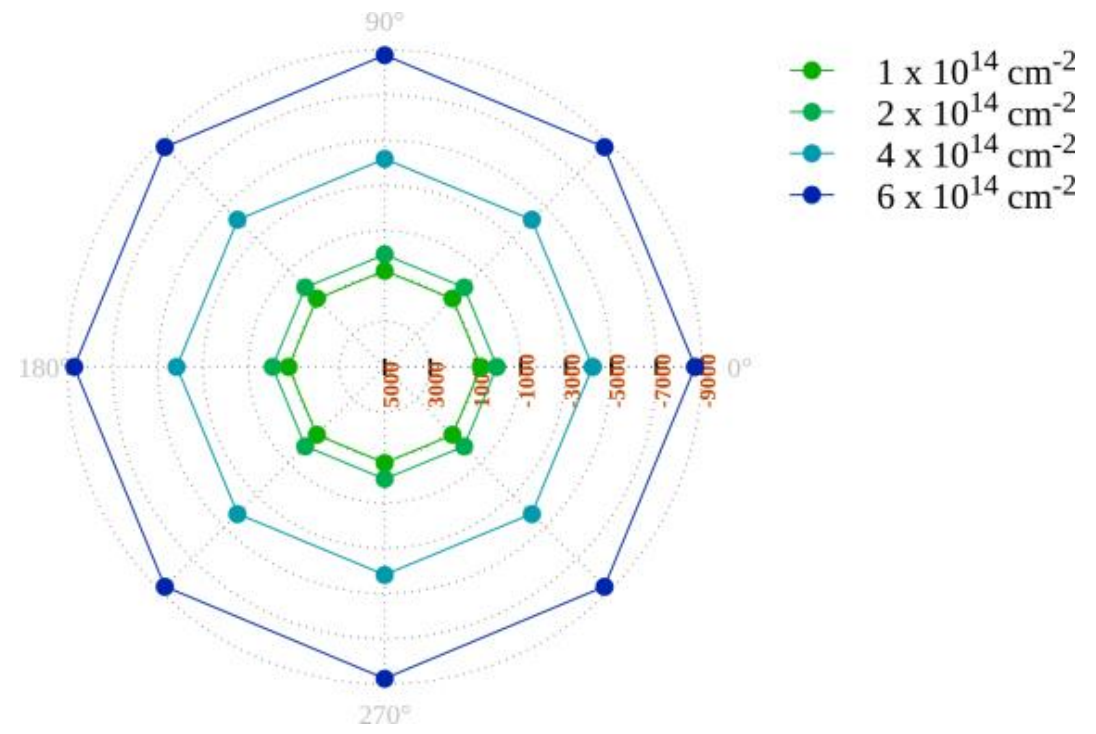

MAE = $E_{\parallel} - E_{\perp}$, a positive (negative) value of MAE indicates the off-plane (in-plane) easy axis.

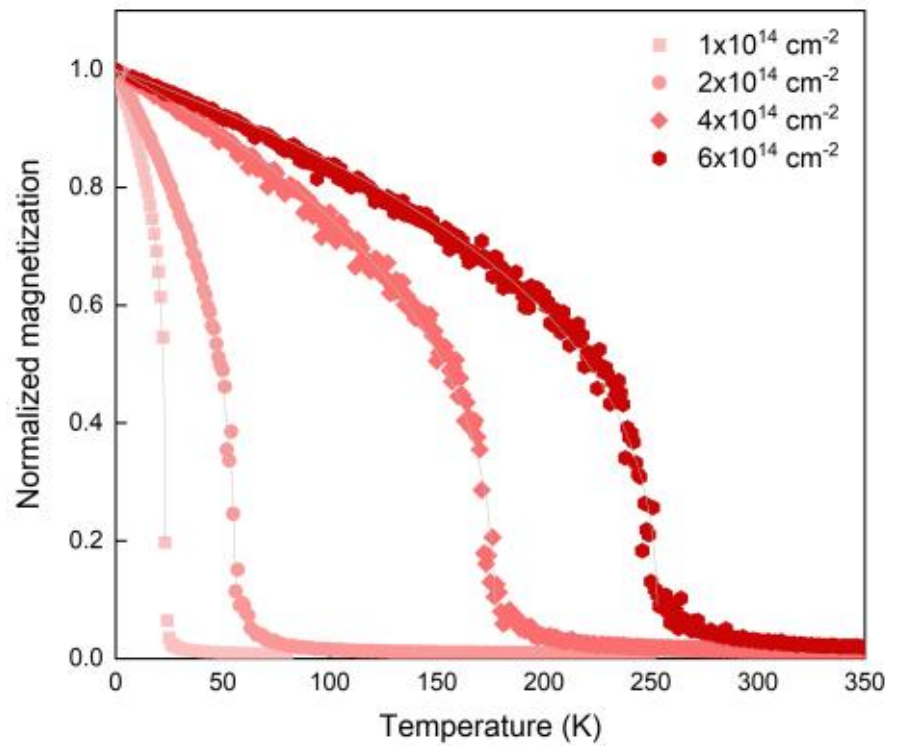

$T_c$: $1\times10^{14}$ cm$^{-2}$: 23 K  $2\times10^{14}$ cm$^{-2}$: 56 K  $4\times10^{14}$ cm$^{-2}$: 177 K  $6\times10^{14}$ cm$^{-2}$: 253 K

# 53. $Al_2S_2$

| MC2D-ID | C2DB | 2dmat-ID | USPEX | Space group | Band gap (eV) |
|---|---|---|---|---|---|
| - | ✓ | 2dm-289 | - | P6m2 | 2.10 |

**Convex hull** | **Atomic structure** | **Atomic coordinates** | **Phonon dispersion curve**

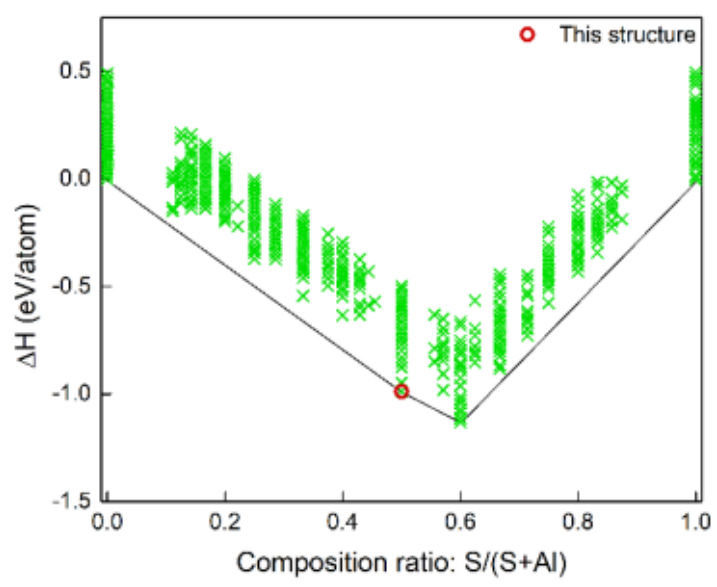

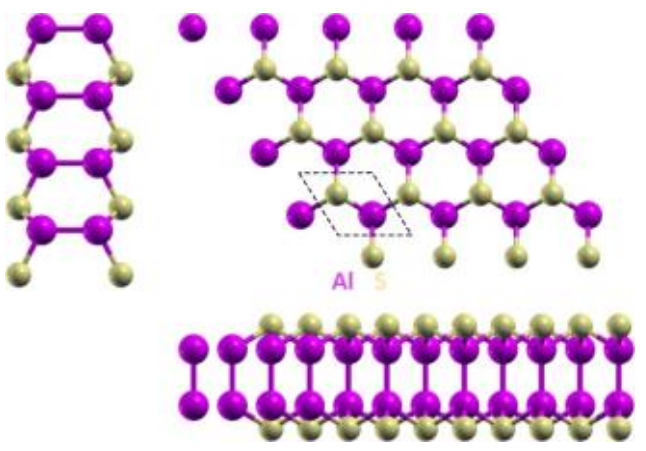

```
Al2S2
   1.00000000000000
   3.5798922574179861   0.0000000000000000   0.0000000000000000
  -1.7899470606356469   3.1002771729430374   0.0000000000000000
   0.0000000000000000   0.0000000000000000  25.0192729999999983
   Al   S
   2    2
Direct
  0.0000001465943669  0.9999998397079537  0.6845740298811052
  0.0000001467123454  0.9999998475117806  0.7883356698716995
  0.6666668464147989  0.3333331535408490  0.8309130204768067
  0.6666668602784895  0.3333331592394160  0.6419972797703934
```

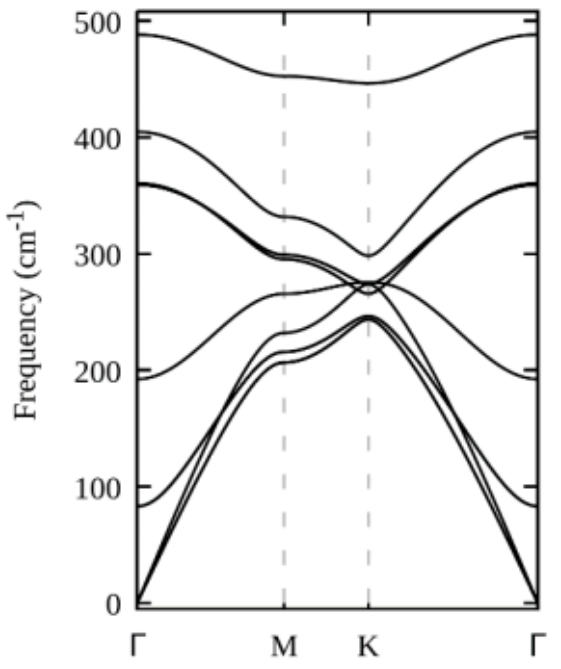

**Projected band structure and density of states** | **Magnetic moment and spin polarization energy as a function of hole doping concentration**

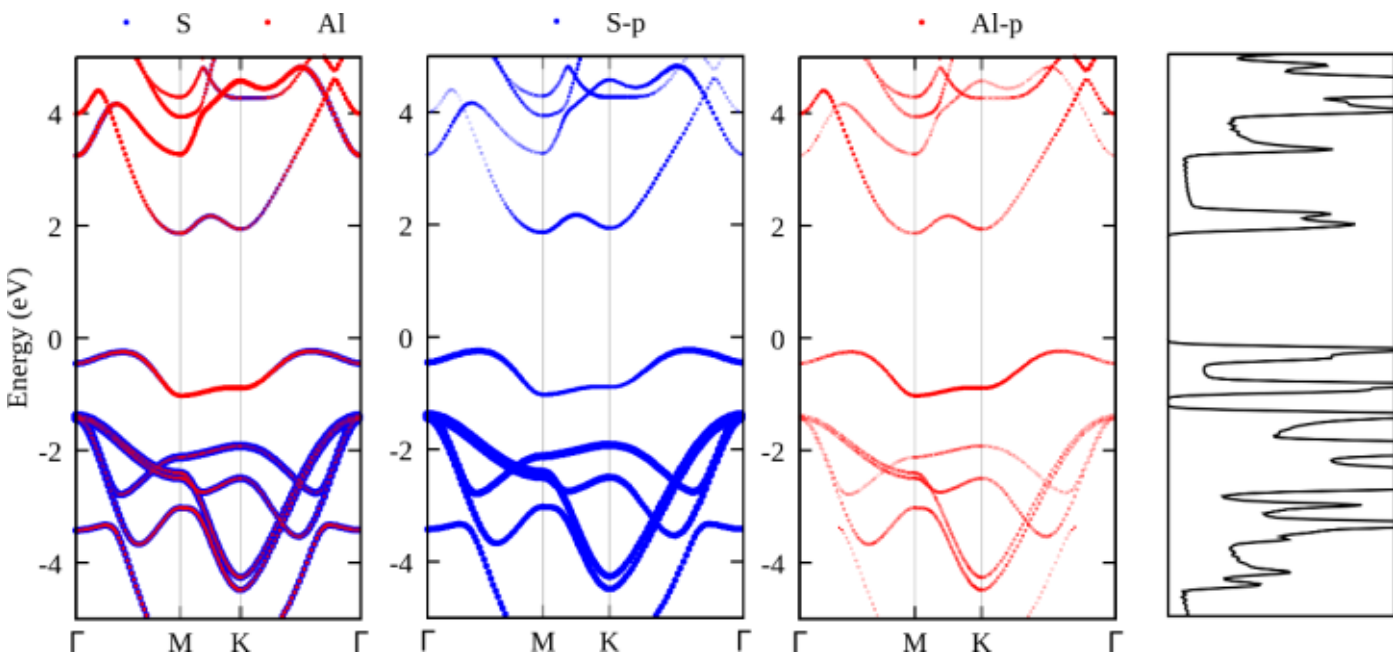

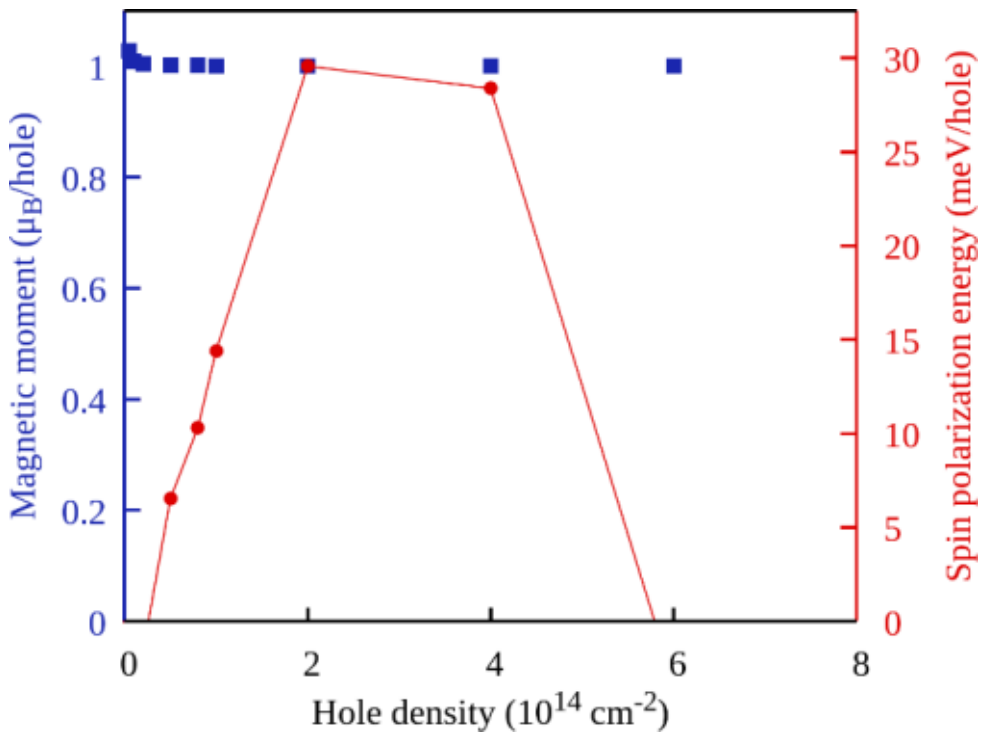

**Magnetic configurations and spin Hamiltonian** | **Magnetic exchange coupling parameters**

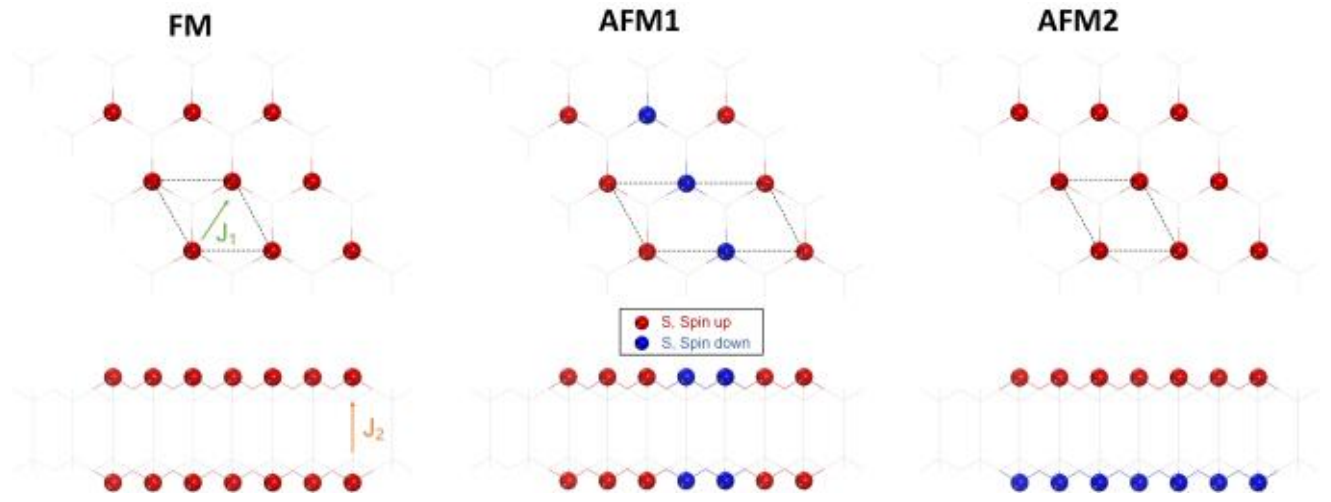

$$E_{FM} = E_0 - 6J_1S^2 - J_2S^2$$

$$E_{AFM1} = E_0 + 2J_1S^2 - J_2S^2$$

$$E_{AFM2} = E_0 - 6J_1S^2 + J_2S^2$$

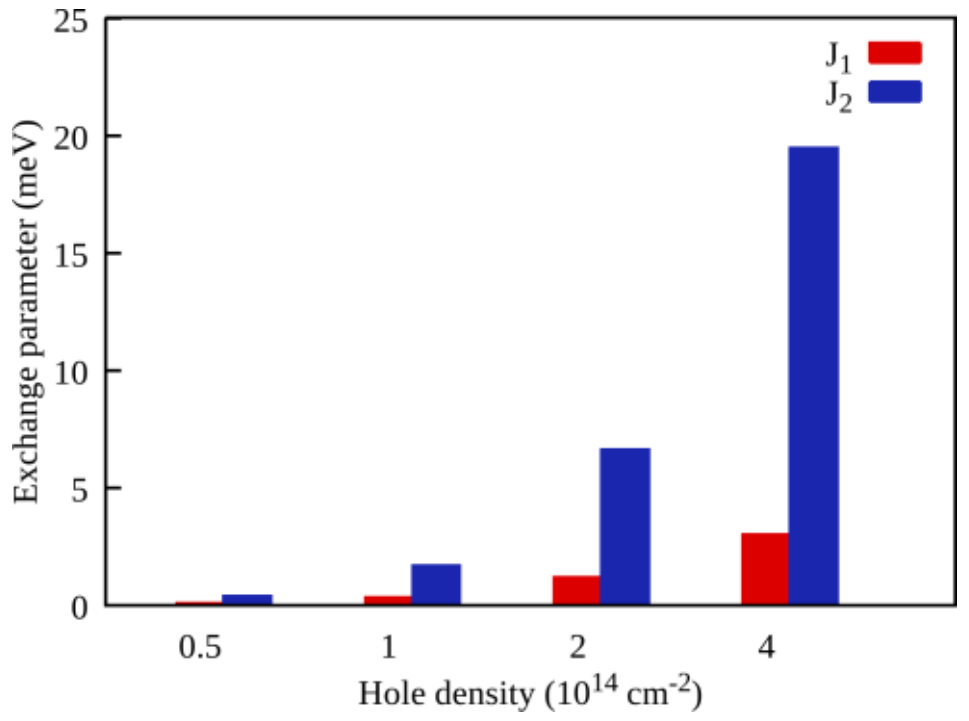

**Magnetic anisotropy energy (MAE, μeV) per magnetic atom** | **Monte Carlo simulations of the normalized magnetization of as a function of temperature**

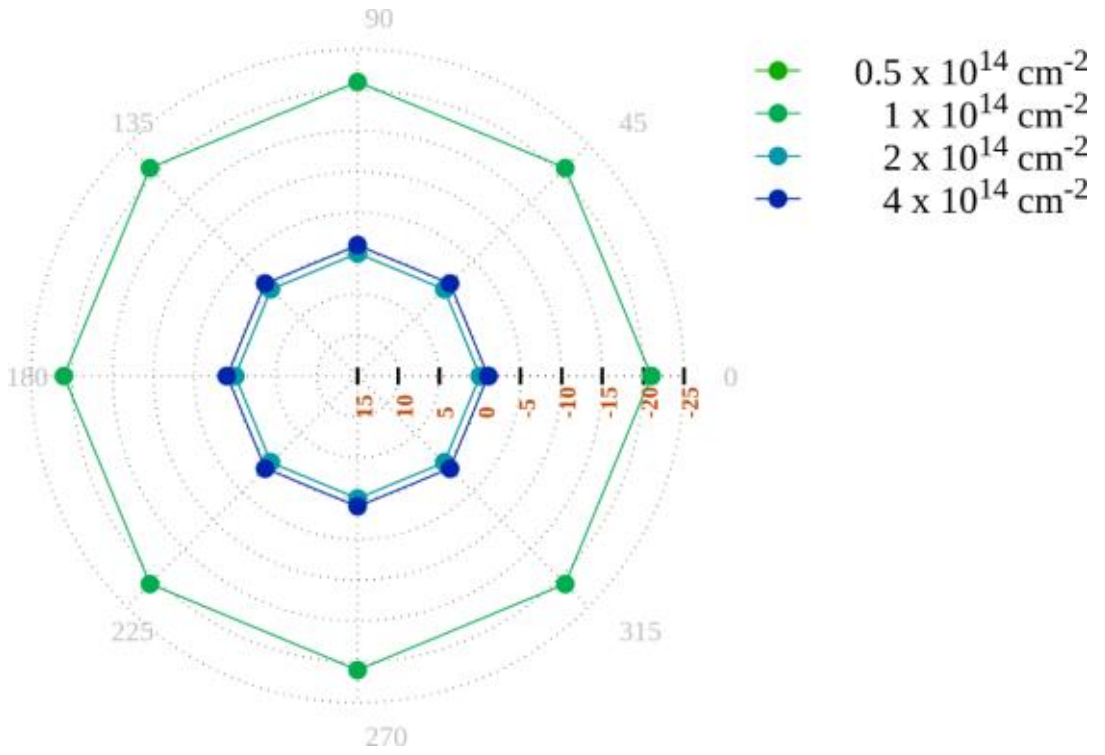

MAE = $E_{\parallel} - E_{\perp}$, a positive (negative) value of MAE indicates the off-plane (in-plane) easy axis.

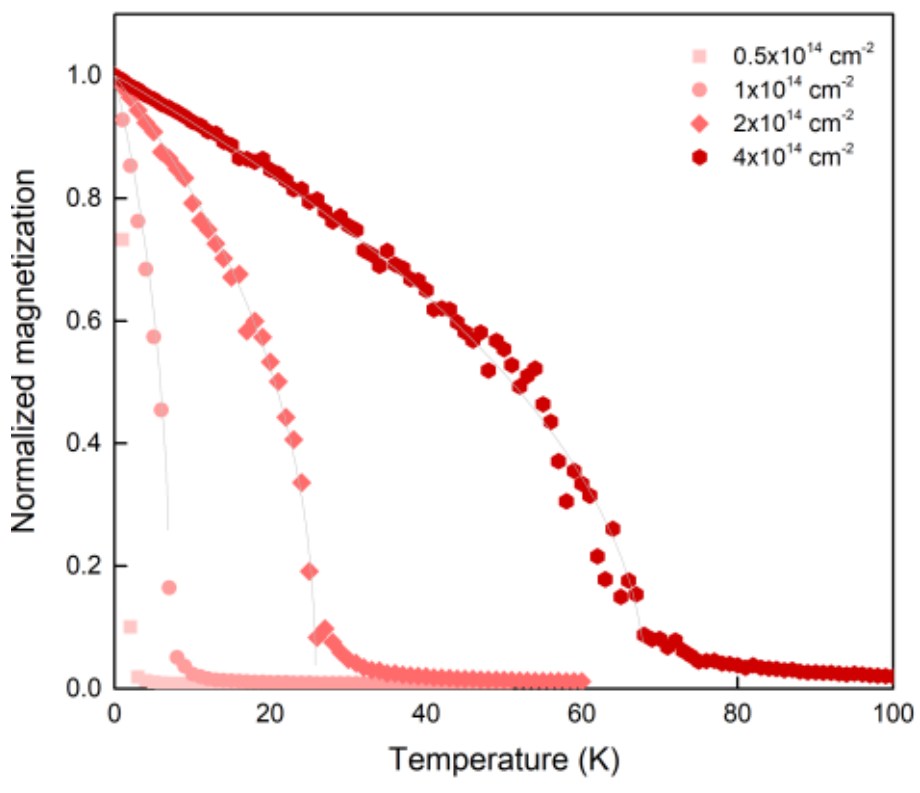

$T_c$: $0.5\times10^{14}$ cm$^{-2}$: 3 K  $1\times10^{14}$ cm$^{-2}$: 7 K

$2\times10^{14}$ cm$^{-2}$: 26 K  $4\times10^{14}$ cm$^{-2}$: 36 K

# 54. $Al_2Se_2$

| MC2D-ID | C2DB | 2dmat-ID | USPEX | Space group | Band gap (eV) |
|---|---|---|---|---|---|
| - | ✓ | 2dm-868 | - | P6m2 | 2.00 |

**Convex hull**

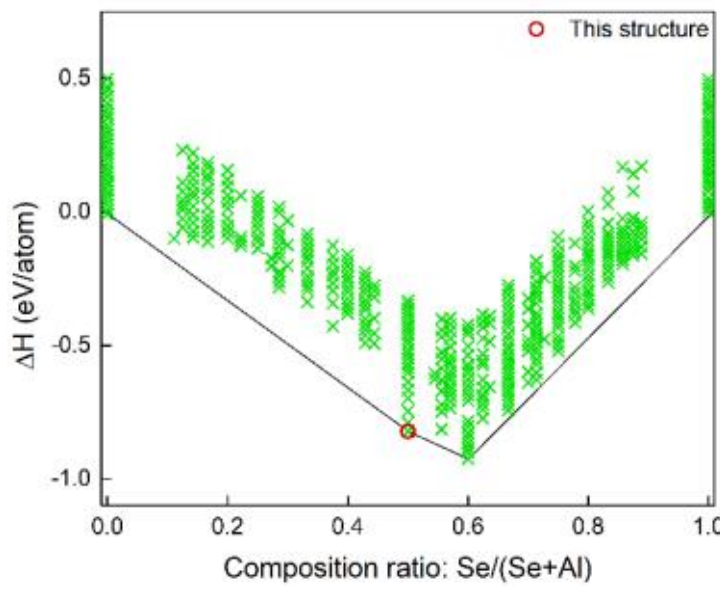

**Atomic structure**

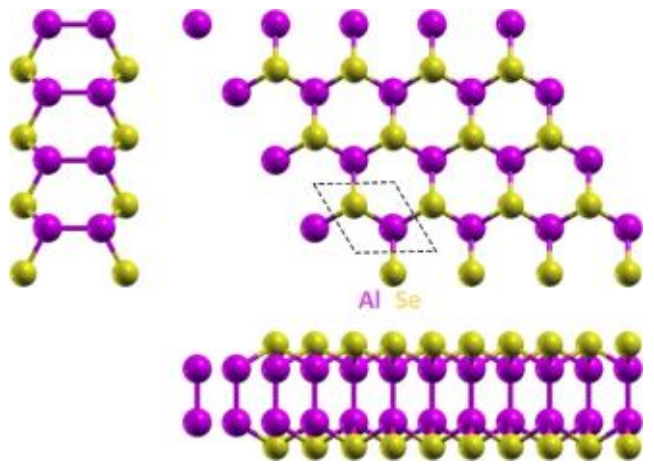

**Atomic coordinates**

```
Al2Se2
1.0000000000000
3.7778825182436613    0.0000000000000000    0.0000000000000000
-1.8889414578327082    3.2717419256532256    0.0000000000000000
0.0000000000000000    0.0000000000000000    25.0192729999999983
Al    Se
2    2
Direct
0.0000001429421630    0.9999998645452521    0.6848207534099870
0.0000001493541149    0.9999998696059151    0.7880892077043526
0.6666668562835198    0.3333331334181153    0.8346121006108405
0.6666668514202030    0.3333331324307238    0.6382979382748175
```

**Phonon dispersion curve**

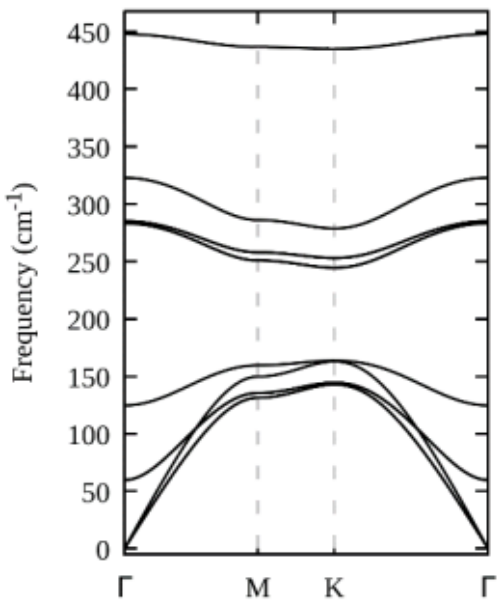

**Projected band structure and density of states**

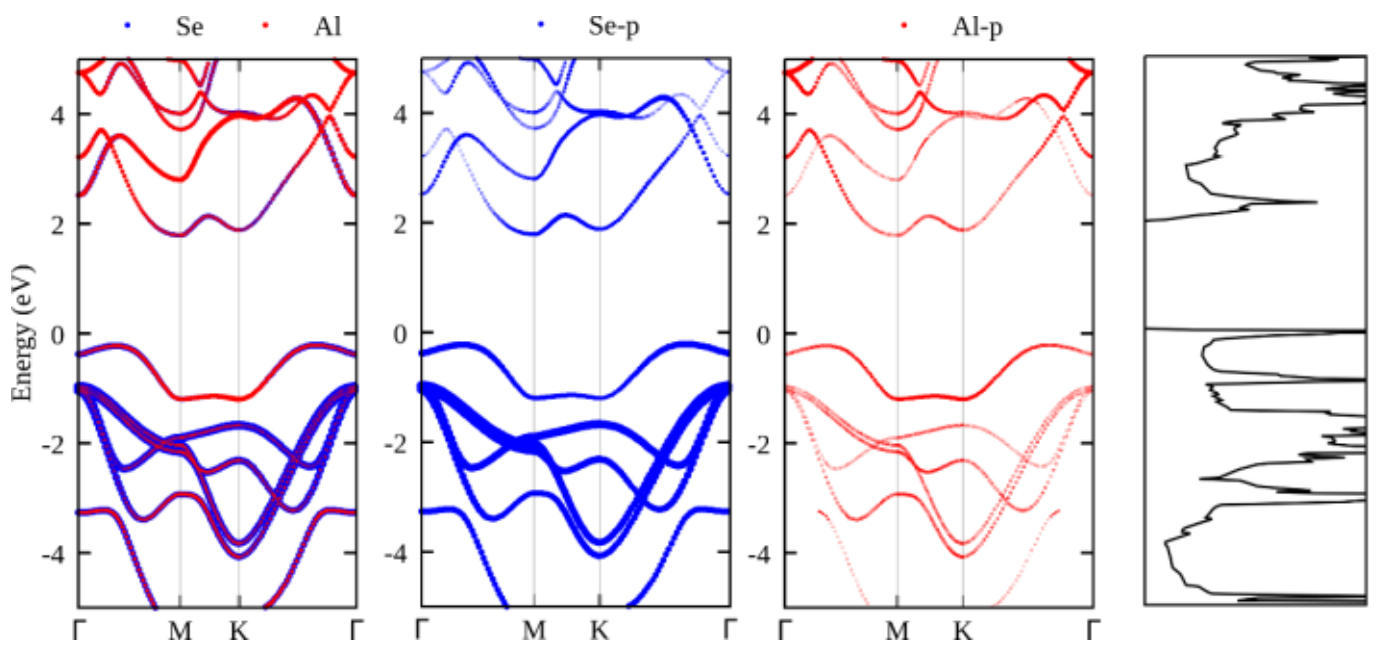

**Magnetic moment and spin polarization energy as a function of hole doping concentration**

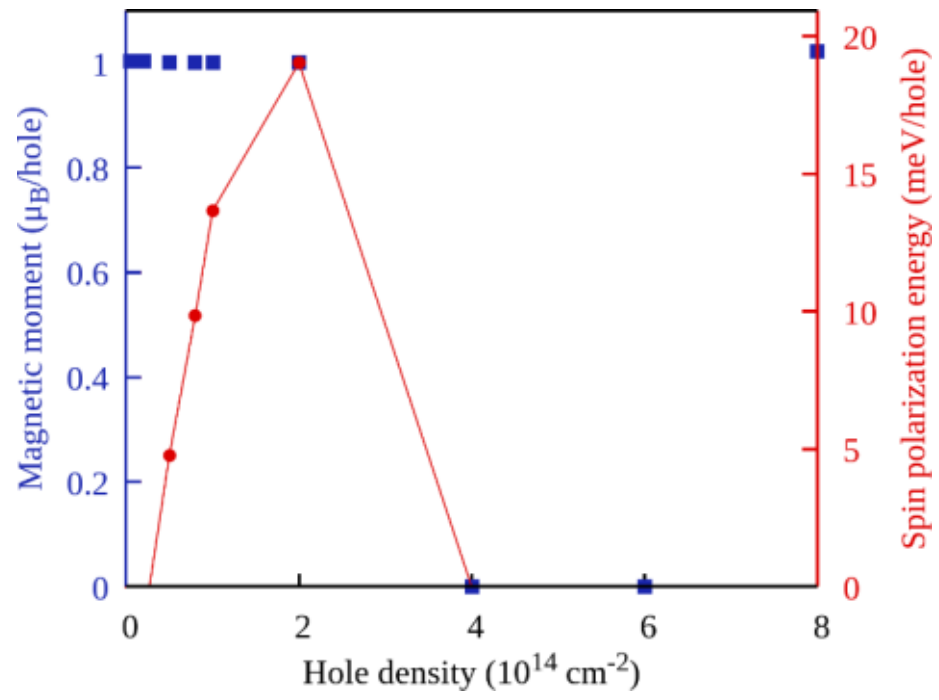

**Magnetic configurations and spin Hamiltonian**

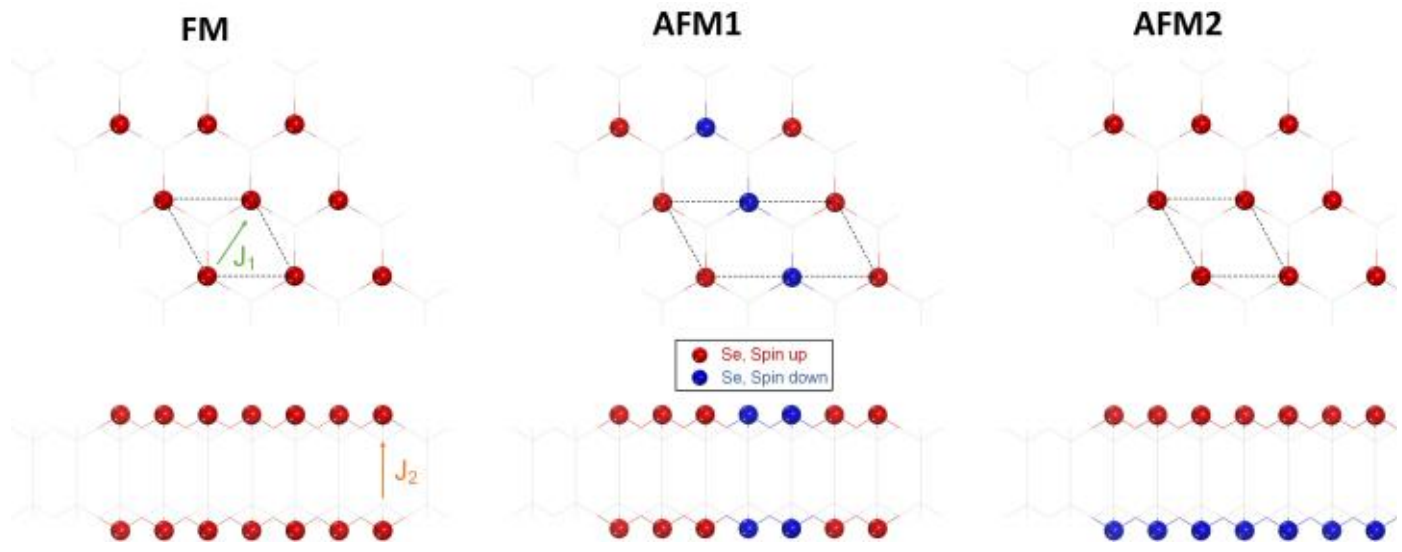

$$E_{FM} = E_0 - 6J_1S^2 - J_2S^2$$
$$E_{AFM1} = E_0 + 2J_1S^2 - J_2S^2$$
$$E_{AFM2} = E_0 - 6J_1S^2 + J_2S^2$$

**Magnetic exchange coupling parameters**

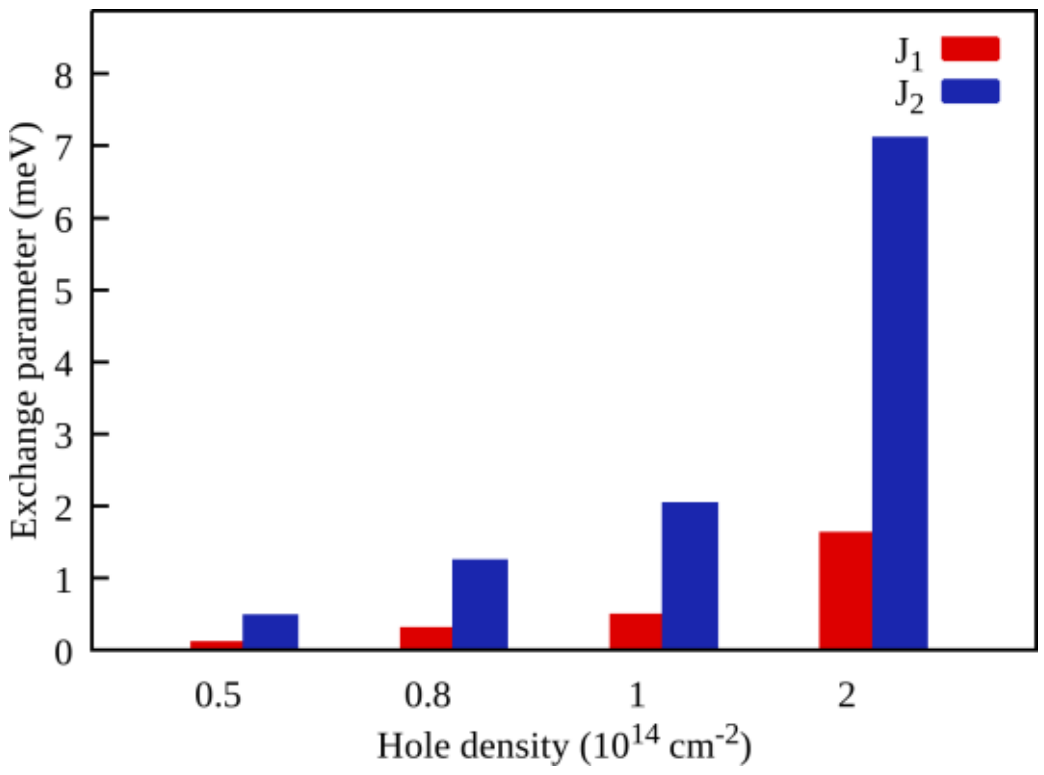

**Magnetic anisotropy energy (MAE, μeV) per magnetic atom**

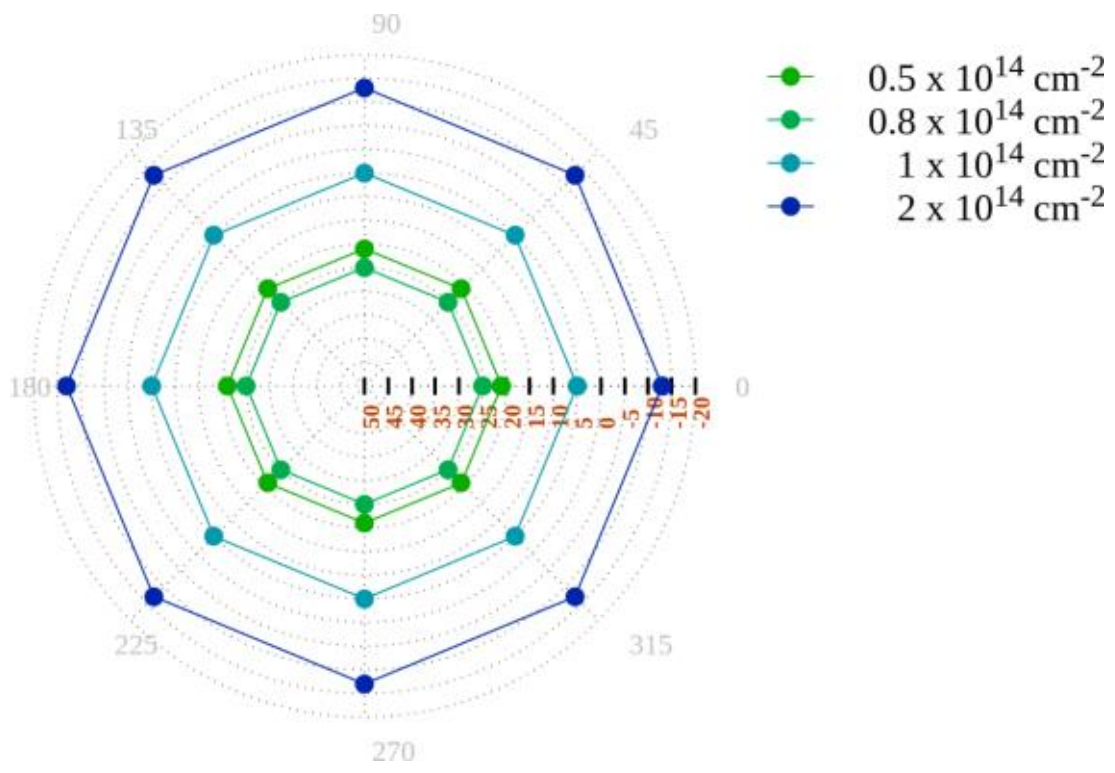

MAE = $E_{\parallel} - E_{\perp}$, a positive (negative) value of MAE indicates the off-plane (in-plane) easy axis.

**Monte Carlo simulations of the normalized magnetization of as a function of temperature**

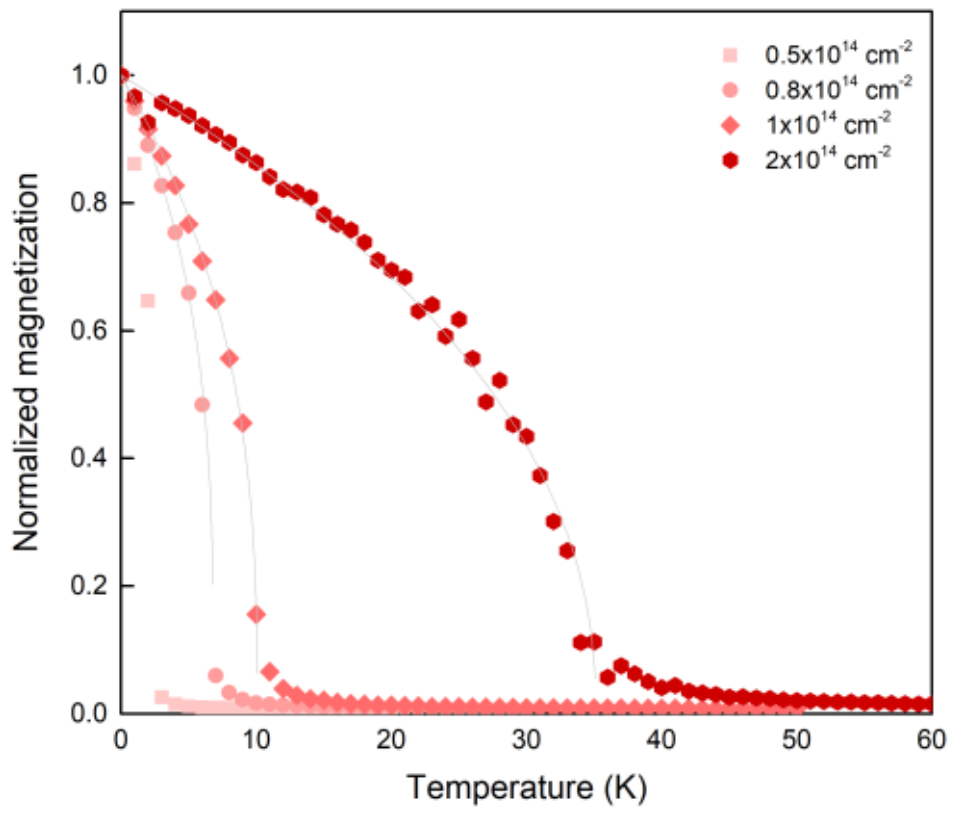

$T_c$: $0.5\times10^{14}$ cm$^{-2}$: 2 K  $0.8\times10^{14}$ cm$^{-2}$: 6K
$1\times10^{14}$ cm$^{-2}$: 10 K  $2\times10^{14}$ cm$^{-2}$: 35K

# 55. $Ga_2S_2$

| MC2D-ID | C2DB | 2dmat-ID | USPEX | Space group | Band gap (eV) |
|---|---|---|---|---|---|
| - | ✓ | 2dm-3608 | - | P6m2 | 2.37 |

## Convex hull

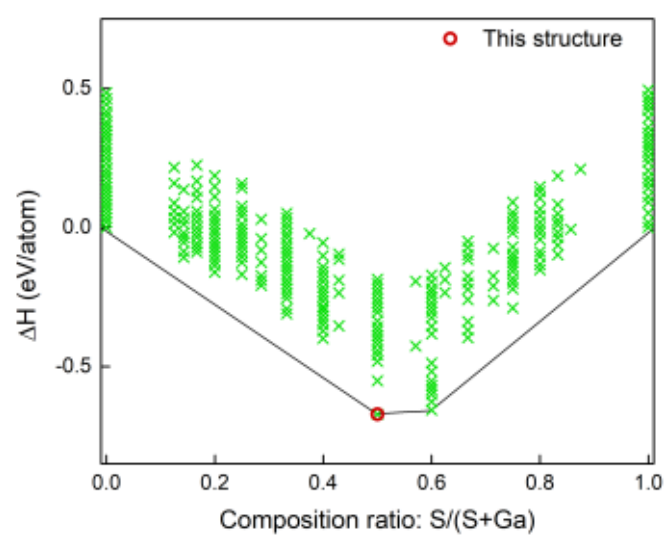

## Atomic structure

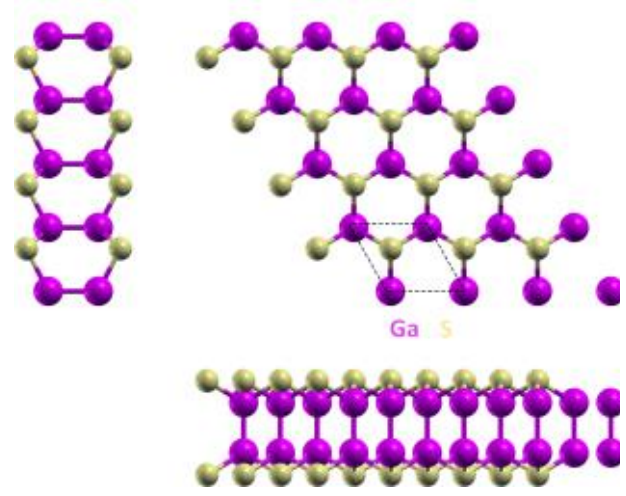

## Atomic coordinates

```
Ga2S2
   1.00000000000000
   3.6284444130234483    0.0000000000000000    0.0000000000000000
  -1.8142235231982928    3.1423242251638892    0.0000000000000000
   0.0000000000000000    0.0000000000000000   24.6403500000000015
   Ga    S
   2     2
Direct
  0.9999998311713725  0.0000003277878093  0.4498288151669172
  0.9999996742776460  0.0000001562129910  0.5501698700550293
  0.3333333133584020  0.6666668247670984  0.4058138022918598
  0.3333331811925930  0.6666666912320949  0.5941875124862008
```

## Phonon dispersion curve

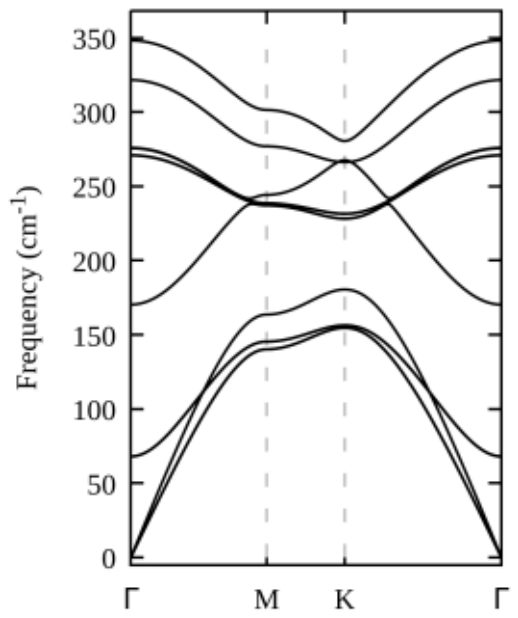

## Projected band structure and density of states

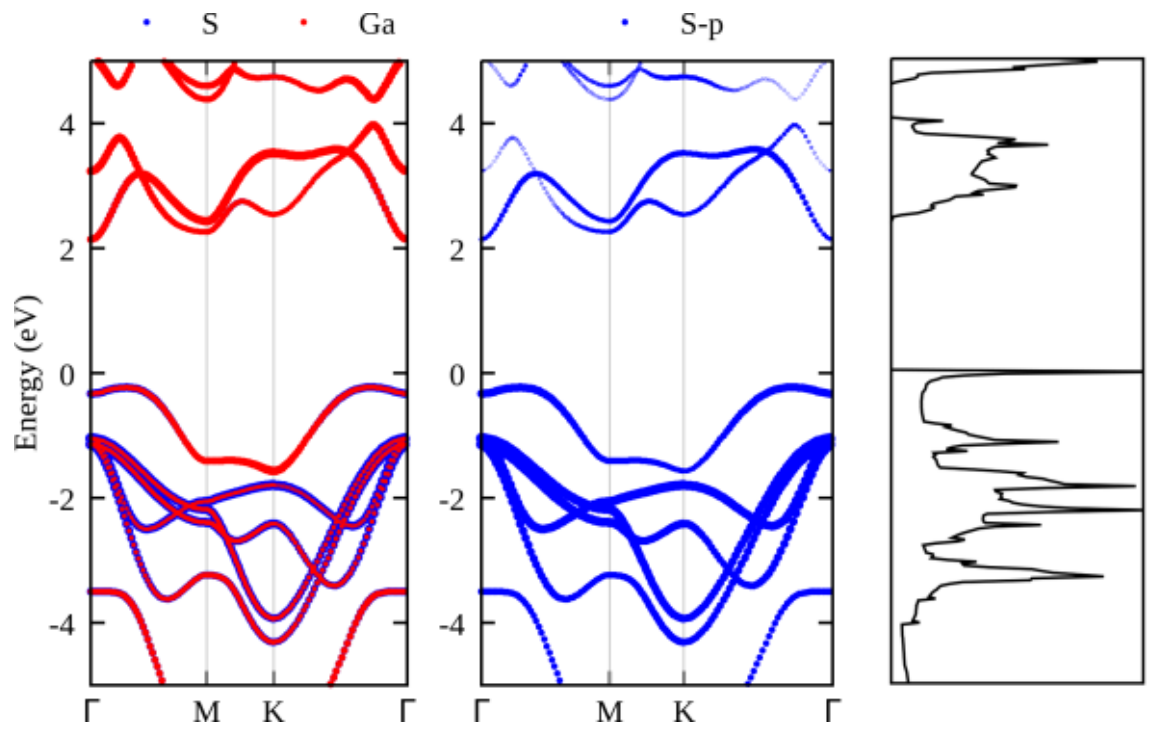

## Magnetic moment and spin polarization energy as a function of hole doping concentration

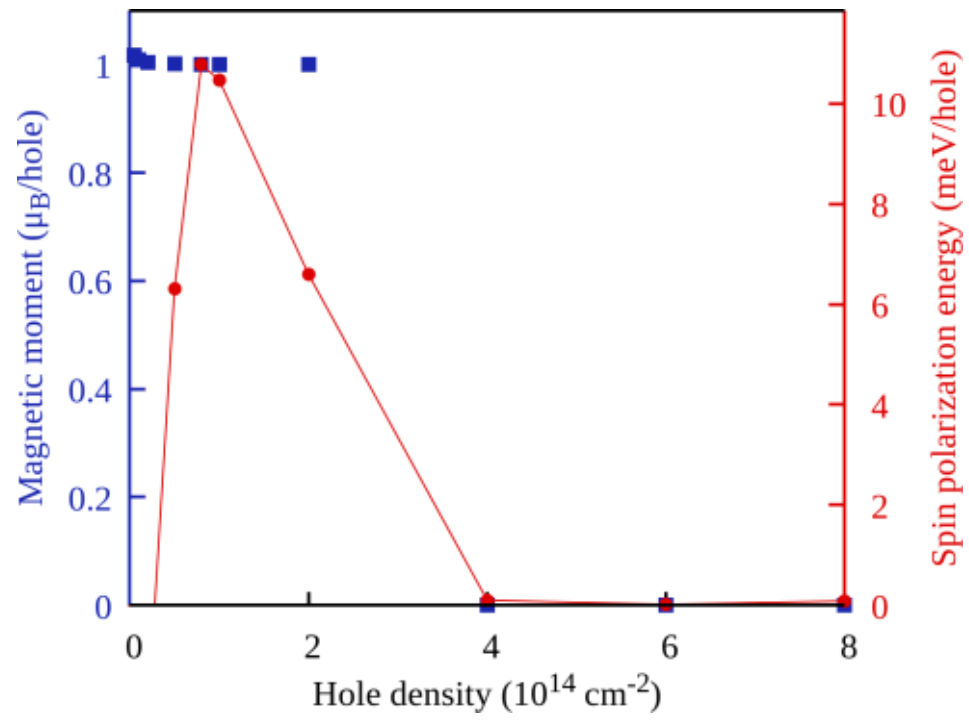

## Magnetic configurations and spin Hamiltonian

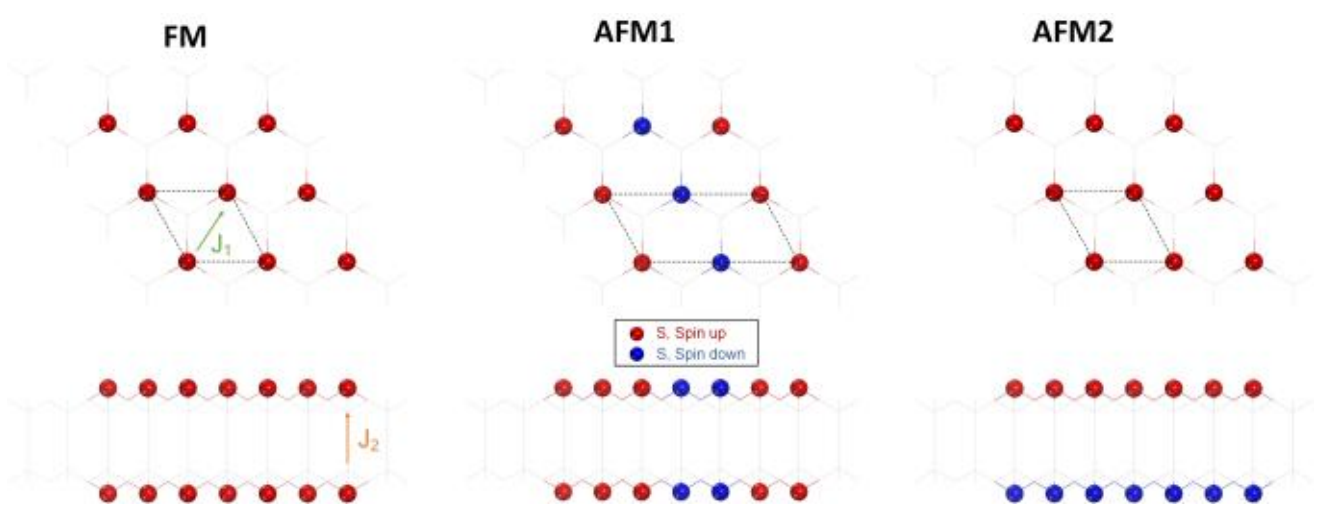

$$E_{FM} = E_0 - 6J_1S^2 - J_2S^2$$

$$E_{AFM1} = E_0 + 2J_1S^2 - J_2S^2$$

$$E_{AFM2} = E_0 - 6J_1S^2 + J_2S^2$$

## Magnetic exchange coupling parameters

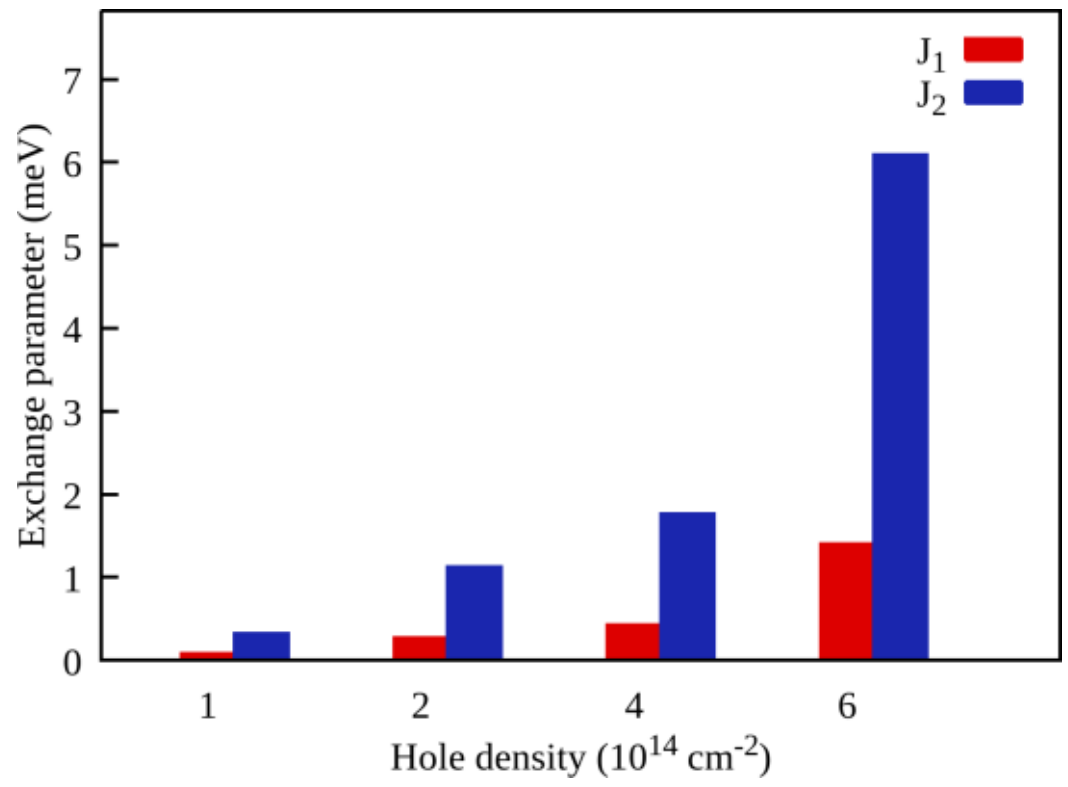

## Magnetic anisotropy energy (MAE, μeV) per magnetic atom

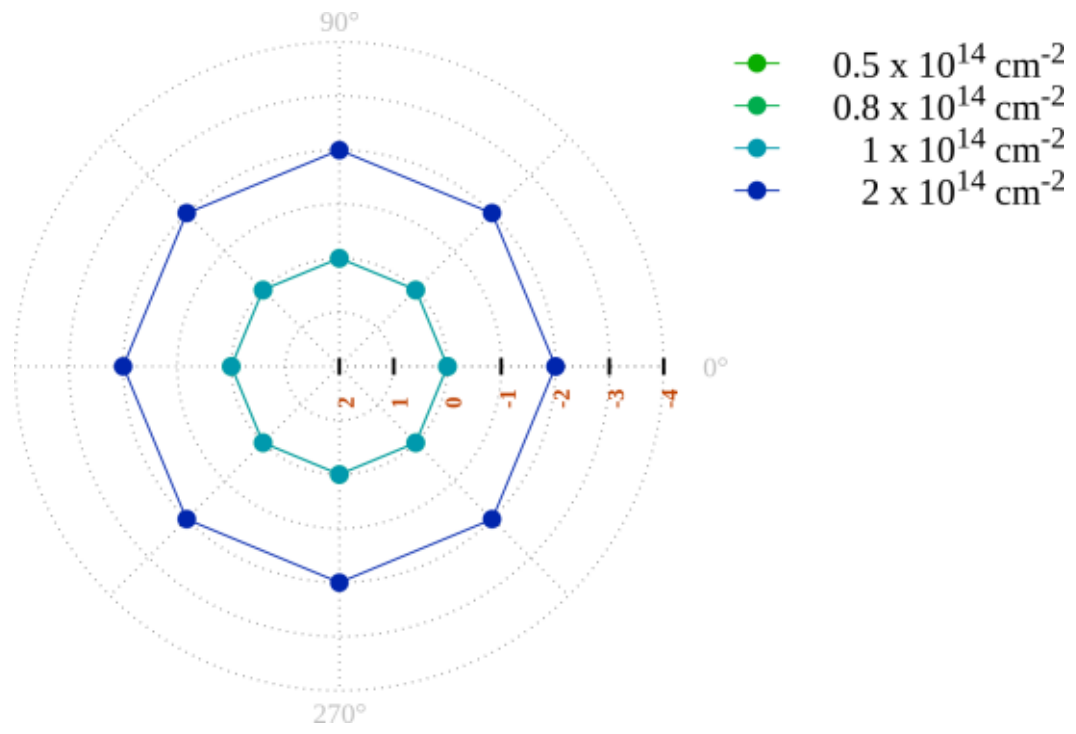

MAE = $E_{\parallel} - E_{\perp}$, a positive (negative) value of MAE indicates the off-plane (in-plane) easy axis.

## Monte Carlo simulations of the normalized magnetization of as a function of temperature

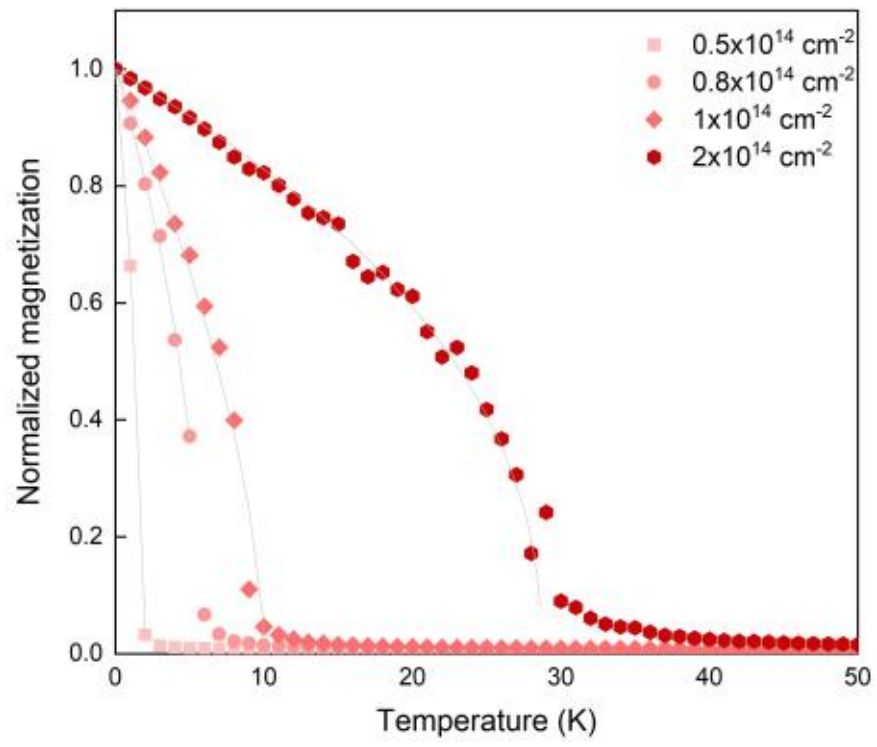

$T_c$: $0.5\times10^{14}$ cm$^{-2}$: 2 K  $0.8\times10^{14}$ cm$^{-2}$: 6 K

$1\times10^{14}$ cm$^{-2}$: 10 K  $2\times10^{14}$ cm$^{-2}$: 29 K

# 56.$In_2S_2$

| MC2D-ID | C2DB | 2dmat-ID | USPEX | Space group | Band gap (eV) |
|---|---|---|---|---|---|
| 92 | ✓ | 2dm-341 | - | P6m2 | 1.67 |

**Convex hull** | **Atomic structure** | **Atomic coordinates** | **Phonon dispersion curve**

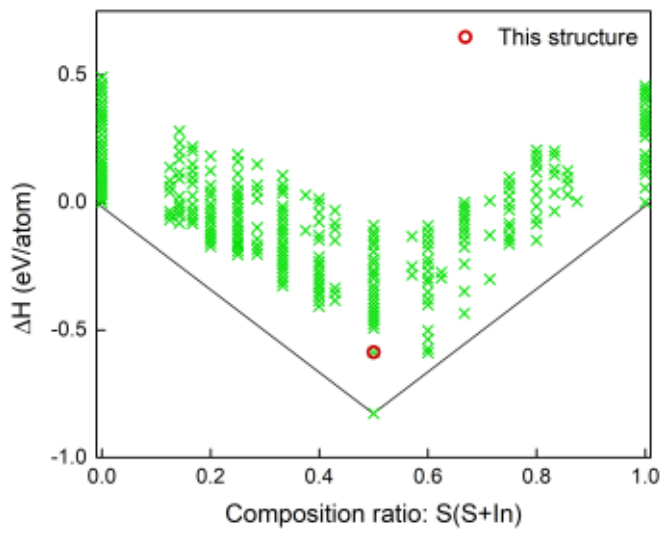

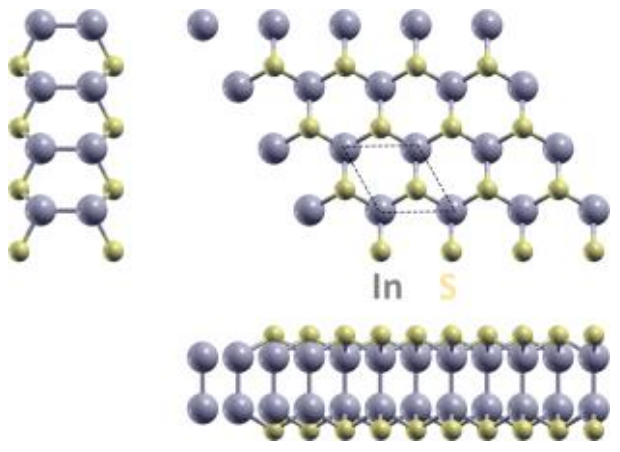

```
In2S2
   1.00000000000000
   3.9185666080547219    0.0000000000000000    0.0000000000000000
  -1.9592897715604429    3.3935744799206717    0.0000000000000000
   0.0000000000000000    0.0000000000000000   25.0192729999999983
   In    S
    2     2
Direct
  0.0000001819039639  0.9999998993744370  0.6799257115131994
  0.0000000934144708  0.9999998335445142  0.7929862264336265
  0.6666668458848051  0.3333331219688347  0.8401302385281824
  0.6666668787967538  0.3333331451122206  0.6327778235249895
```

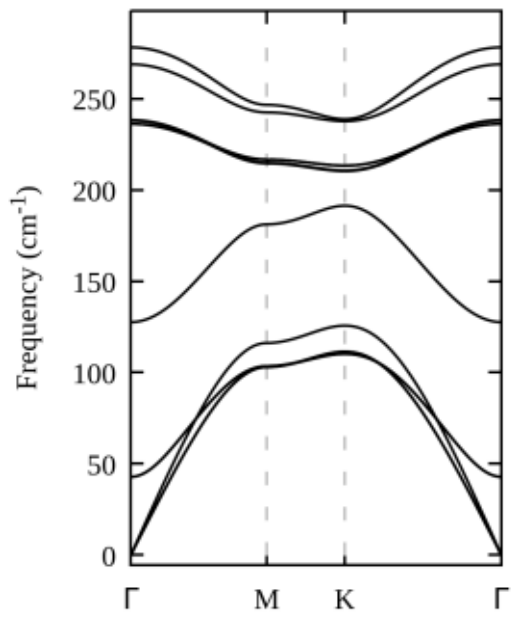

**Projected band structure and density of states**

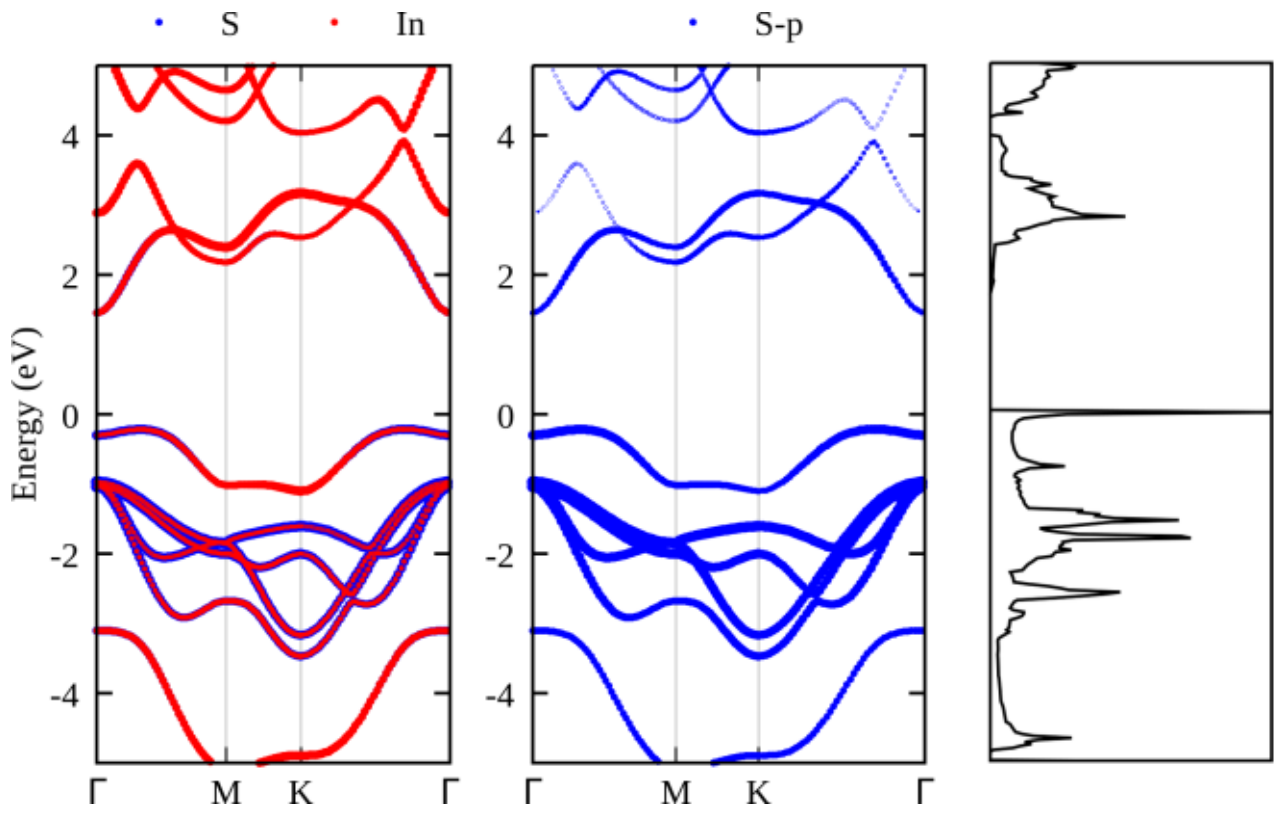

**Magnetic moment and spin polarization energy as a function of hole doping concentration**

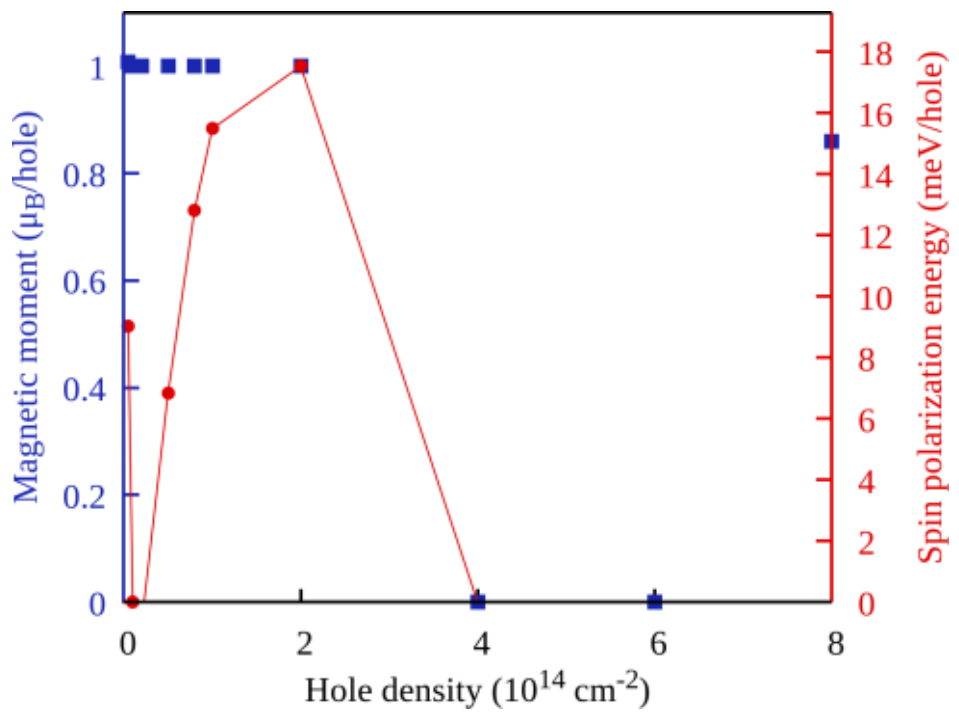

**Magnetic configurations and spin Hamiltonian**

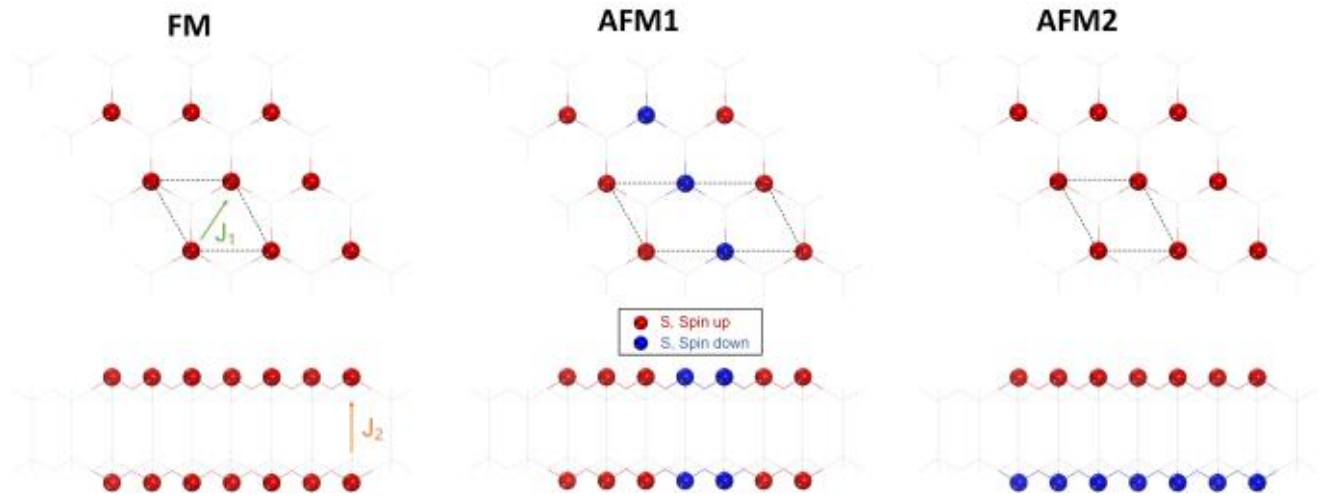

$$E_{FM} = E_0 - 6J_1S^2 - J_2S^2$$
$$E_{AFM1} = E_0 + 2J_1S^2 - J_2S^2$$
$$E_{AFM2} = E_0 - 6J_1S^2 + J_2S^2$$

**Magnetic exchange coupling parameters**

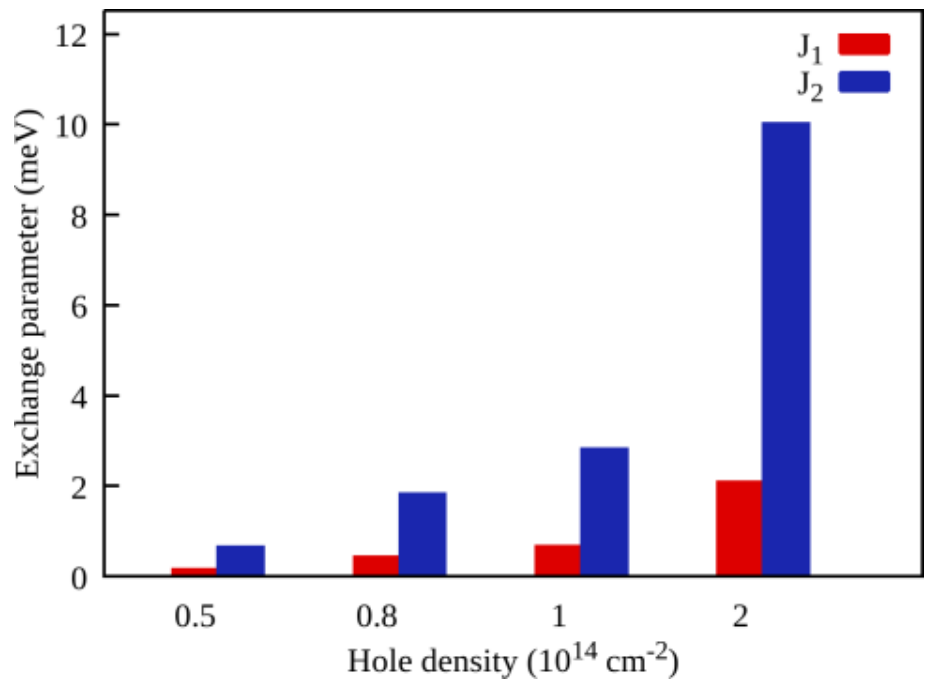

**Magnetic anisotropy energy (MAE, μeV) per magnetic atom**

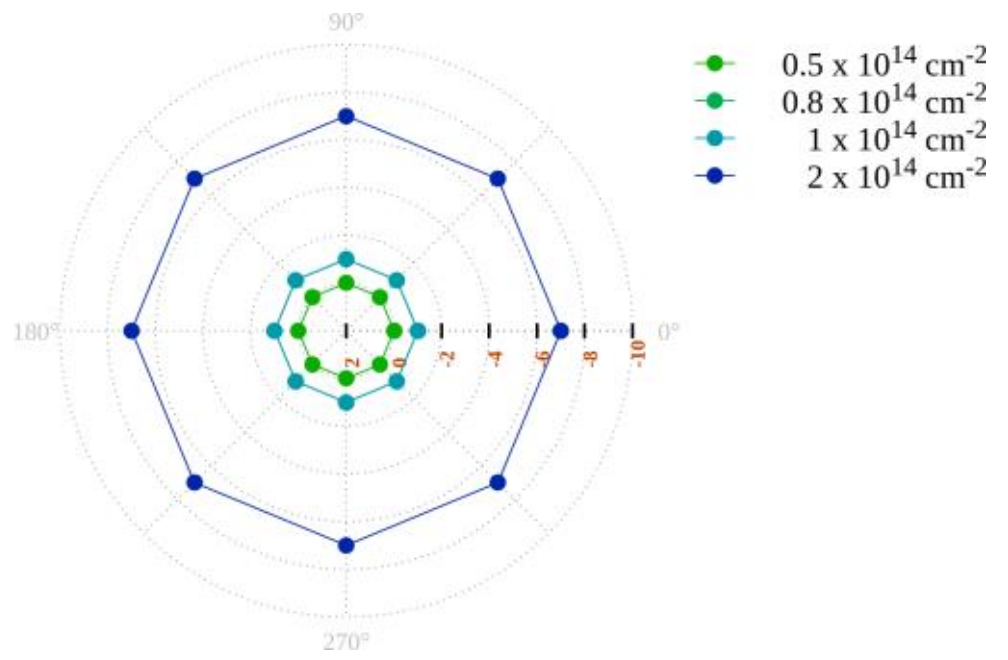

MAE = $E_{\parallel} - E_{\perp}$, a positive (negative) value of MAE indicates the off-plane (in-plane) easy axis.

**Monte Carlo simulations of the normalized magnetization of as a function of temperature**

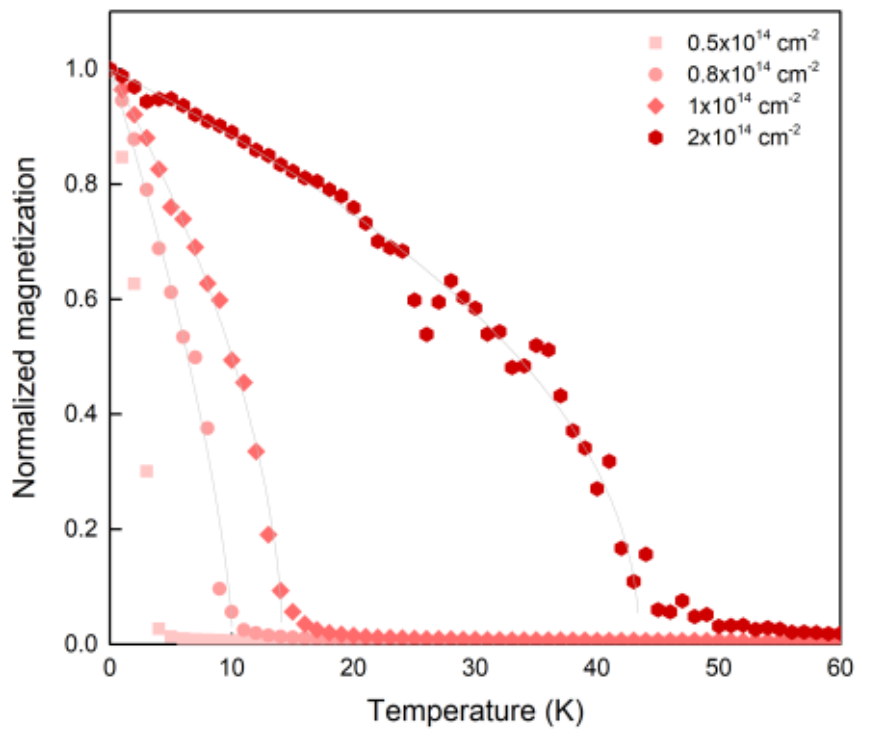

$T_c$: $0.5\times10^{14}$ cm$^{-2}$: 4 K  $0.8\times10^{14}$ cm$^{-2}$: 10 K

$1\times10^{14}$ cm$^{-2}$: 14 K  $2\times10^{14}$ cm$^{-2}$: 43 K

# 57. $Tl_2S_2$

| MC2D-ID | C2DB | 2dmat-ID | USPEX | Space group | Band gap (eV) |
|---|---|---|---|---|---|
| - | ✓ | 2dm-59 | - | P6m2 | 0.64 |

## Convex hull

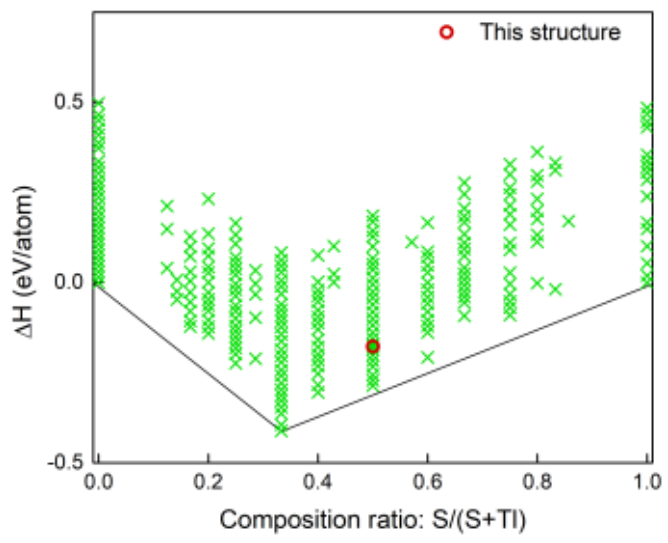

## Atomic structure

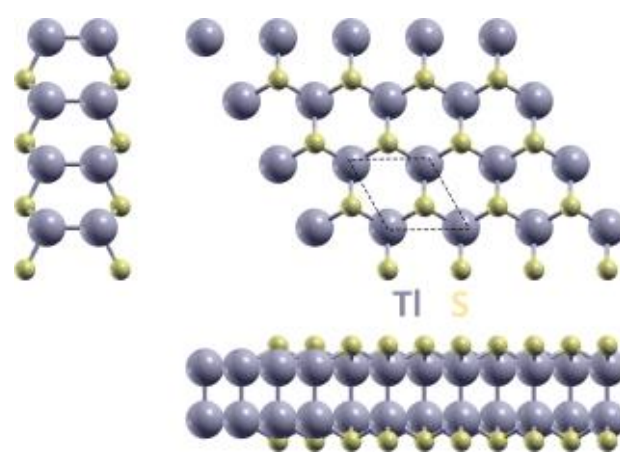

## Atomic coordinates

```
Tl2S2
   1.00000000000000
   4.0686518437142709    0.0000000000000000    0.0000000000000000
  -2.0343271845599413    3.5235546946074092    0.0000000000000000
   0.0000000000000000    0.0000000000000000   25.0192729999999983
   Tl    S
   2     2
Direct
  0.0000002362939000   0.9999998127287384   0.6785093255514667
  0.0000001279460093   0.9999998324554156   0.7944036021698508
  0.6666667720654900   0.3333331170876477   0.8423251117015980
  0.6666668636945943   0.3333332377282048   0.6305819605770822
```

## Phonon dispersion curve

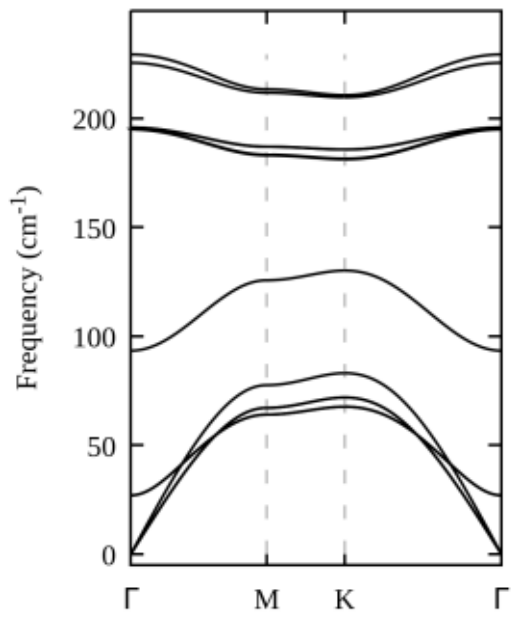

## Projected band structure and density of states

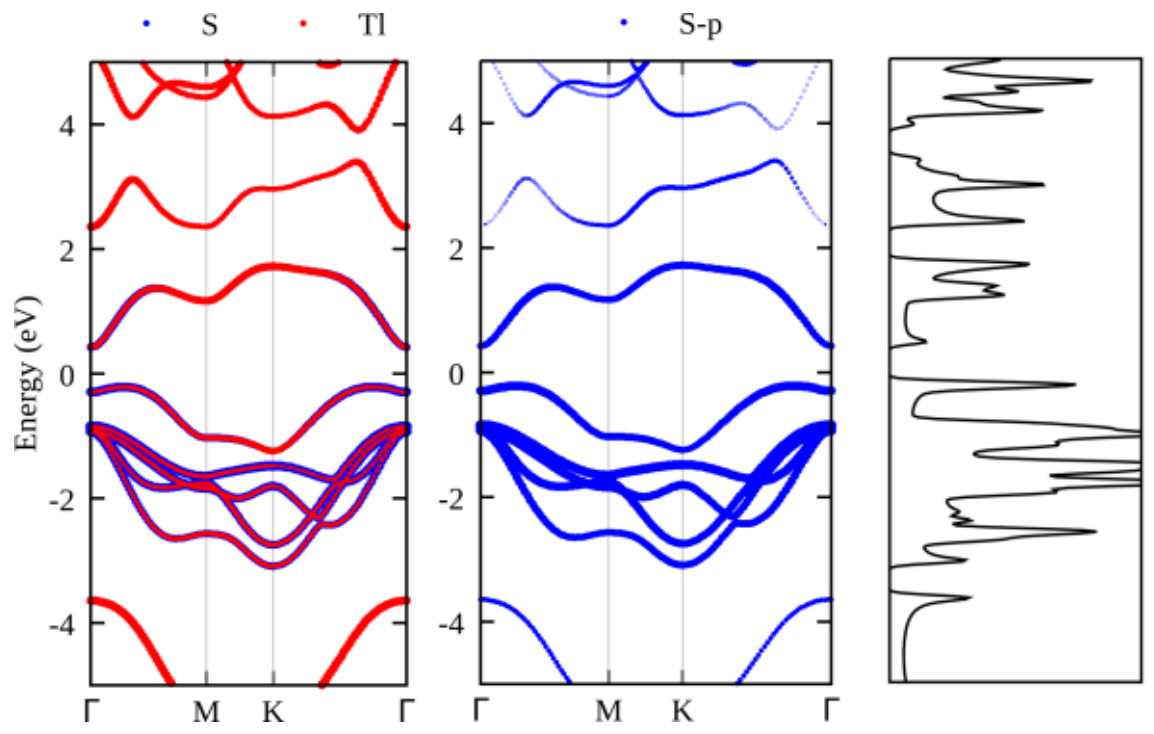

## Magnetic moment and spin polarization energy as a function of hole doping concentration

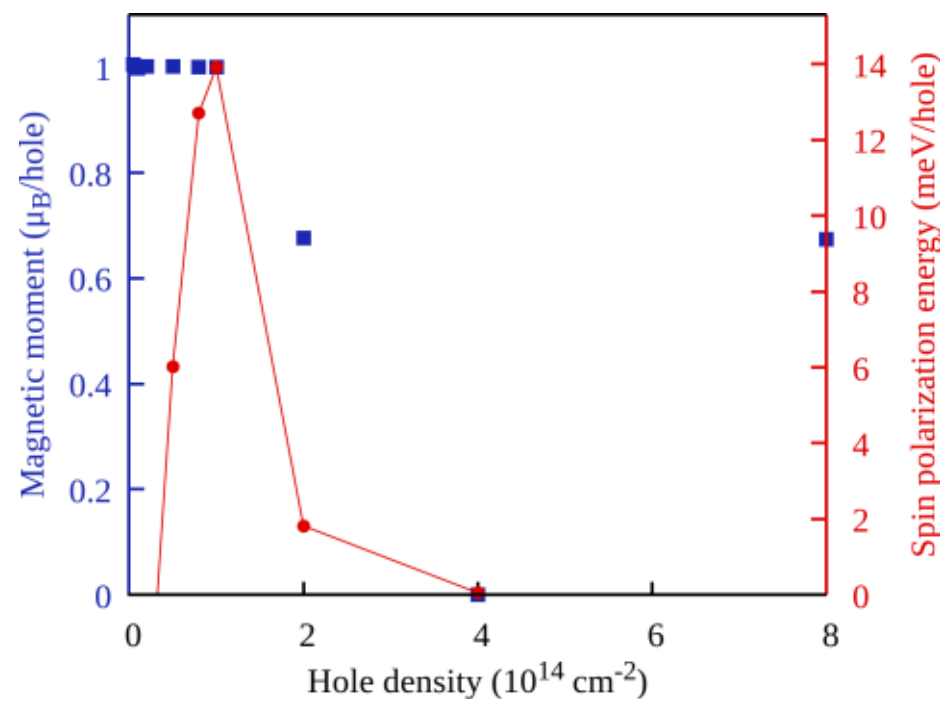

## Magnetic configurations and spin Hamiltonian

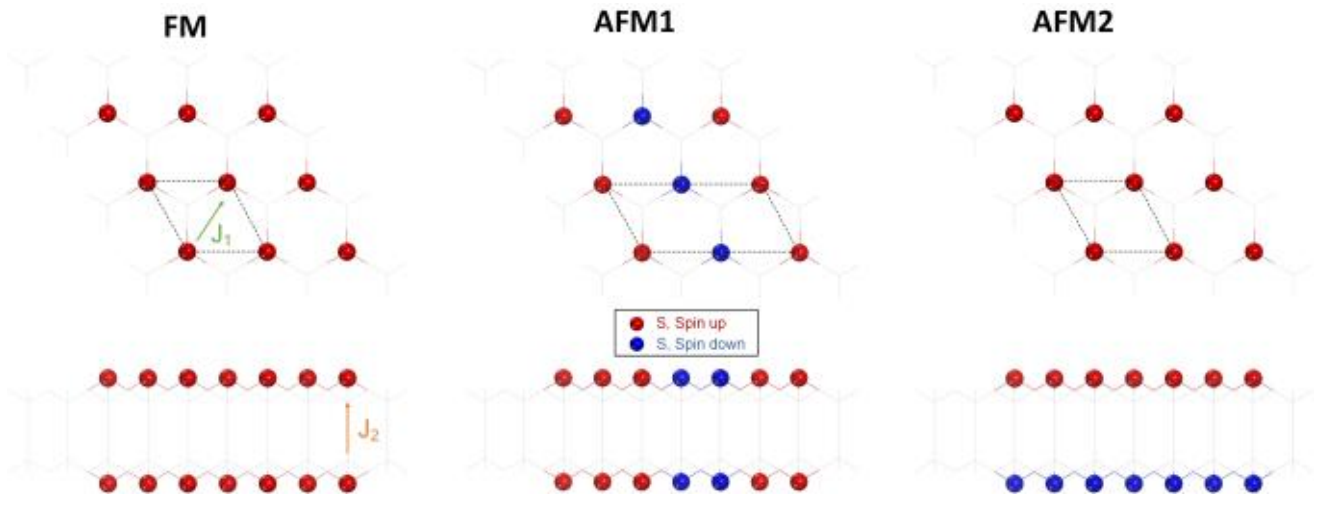

$$E_{FM} = E_0 - 6J_1S^2 - J_2S^2$$
$$E_{AFM1} = E_0 + 2J_1S^2 - J_2S^2$$
$$E_{AFM2} = E_0 - 6J_1S^2 + J_2S^2$$

## Magnetic exchange coupling parameters

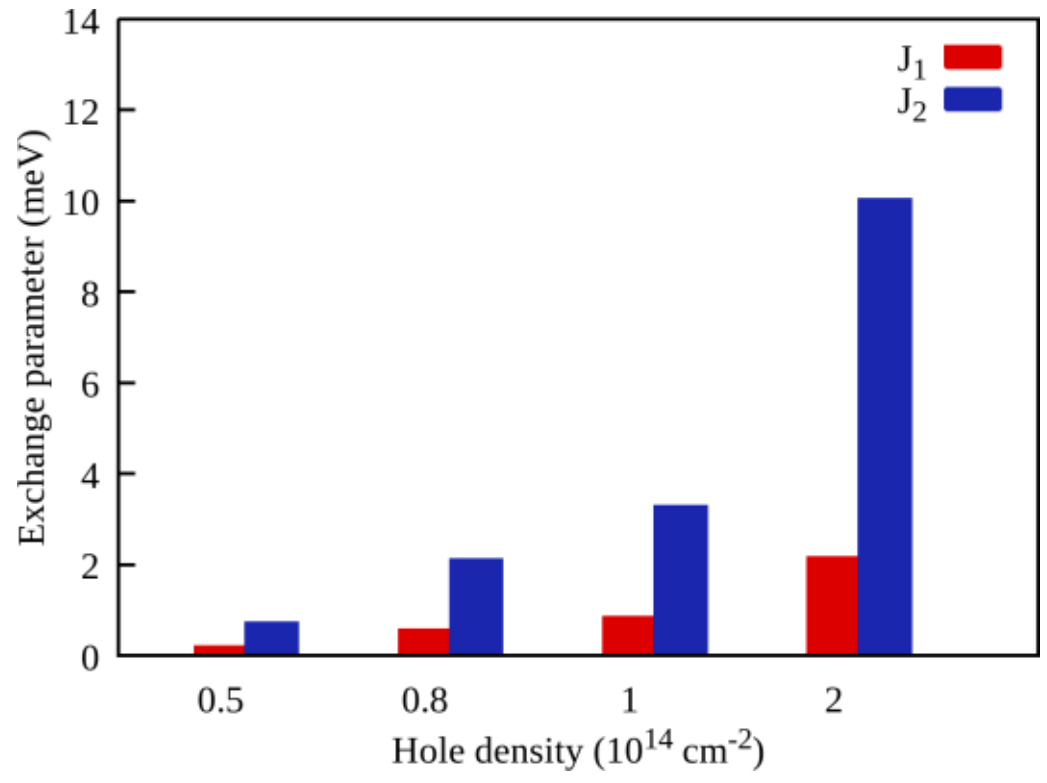

## Magnetic anisotropy energy (MAE, μeV) per magnetic atom

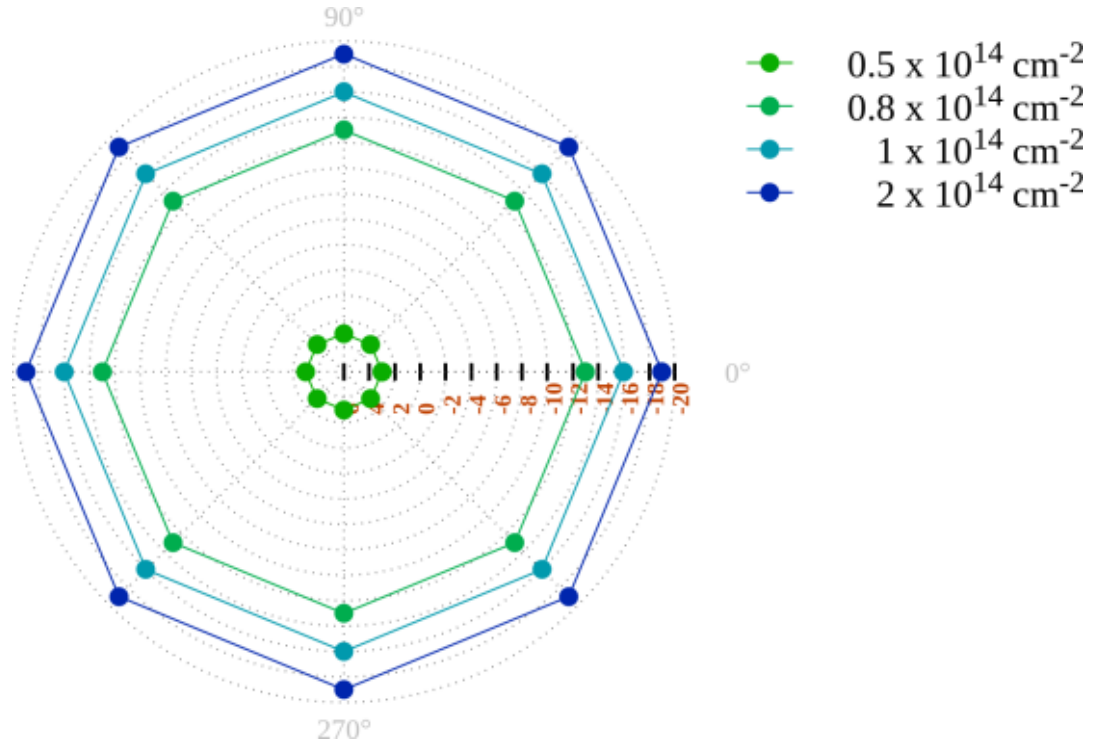

MAE = $E_{\parallel} - E_{\perp}$, a positive (negative) value of MAE indicates the off-plane (in-plane) easy axis.

## Monte Carlo simulations of the normalized magnetization of as a function of temperature

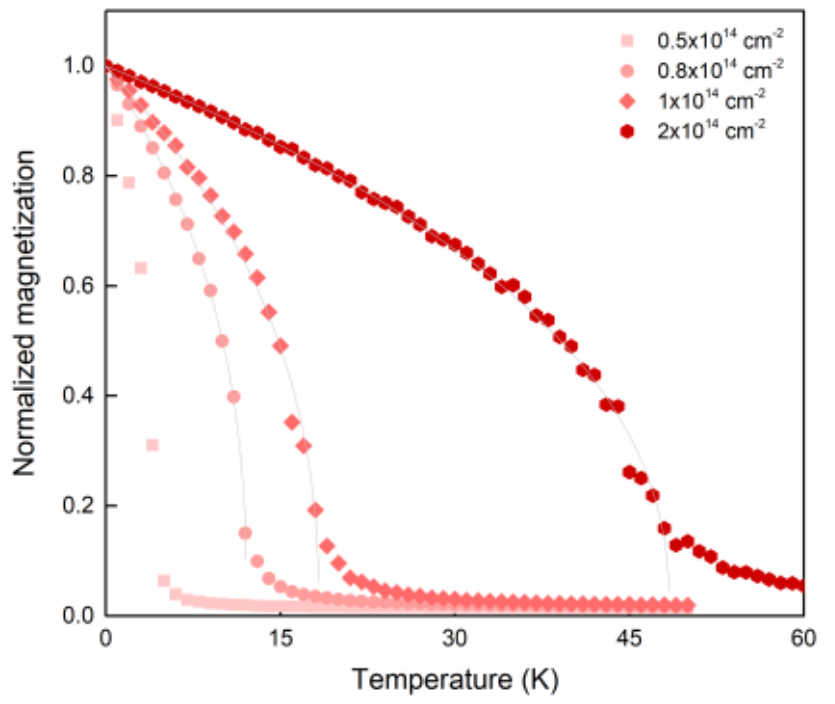

$T_c$: $0.5\times10^{14}$ cm$^{-2}$: 5 K  $0.8\times10^{14}$ cm$^{-2}$: 12 K  $1\times10^{14}$ cm$^{-2}$: 18 K  $2\times10^{14}$ cm$^{-2}$: 48 K

# 58.CdS

| MC2D-ID | C2DB | 2dmat-ID | USPEX | Space group | Band gap (eV) |
|---|---|---|---|---|---|
| - | ✓ | 2dm-3242 | - | P6m2 | 1.63 |

**Convex hull**

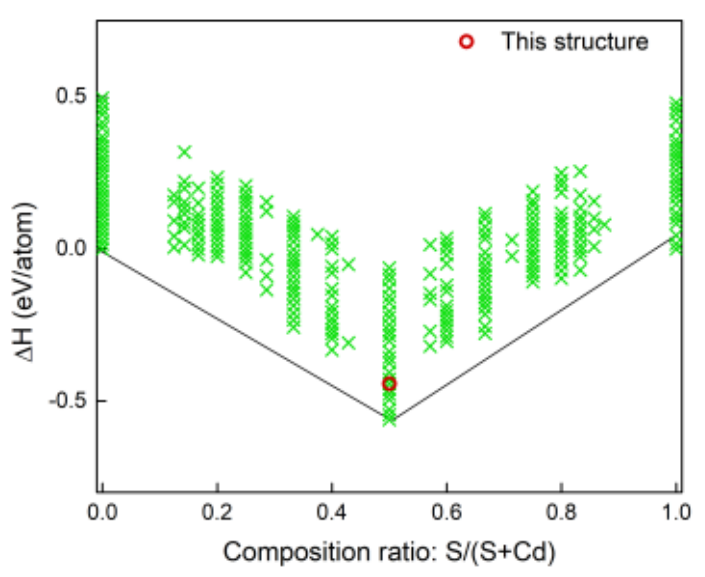

**Atomic structure**

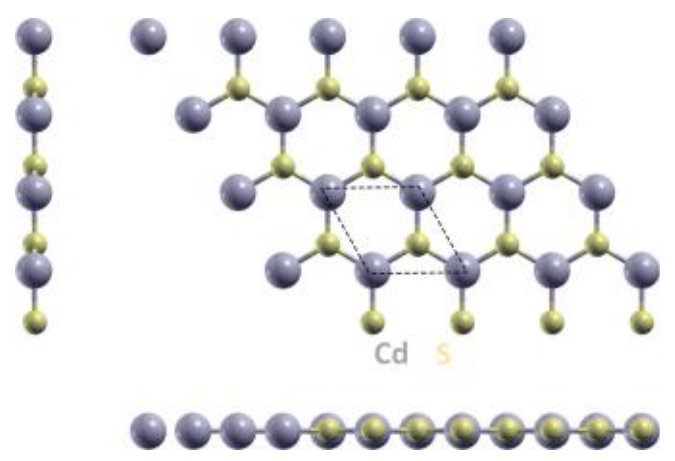

**Atomic coordinates**

```
Cd1S1
  1.00000000000000
  4.2580819629675677   0.0000000000000000   0.0000000000000000
 -2.1290382627071618   3.6876071496348870   0.0000000000000000
  0.0000000000000000   0.0000000000000000  20.1669300000000007
  Cd   S
  1    1
Direct
  0.0000005926146045   0.9999996012260013   0.4990068239281555
  0.6666664073853923   0.3333333987740019   0.5009931760718445
```

**Phonon dispersion curve**

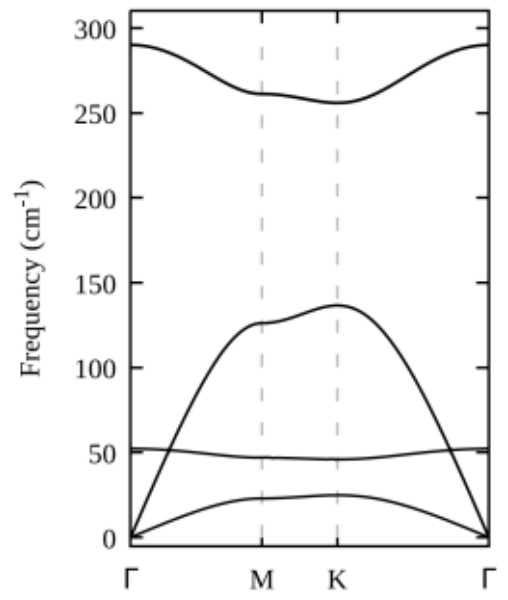

**Projected band structure and density of states**

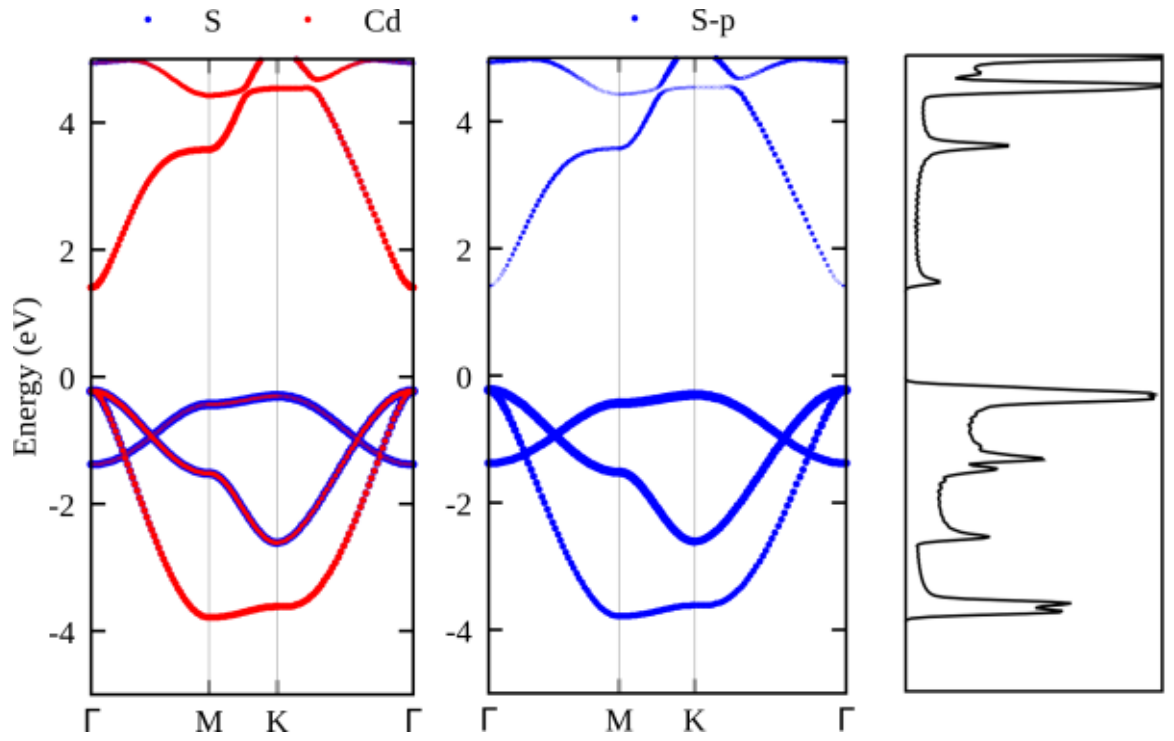

**Magnetic moment and spin polarization energy as a function of hole doping concentration**

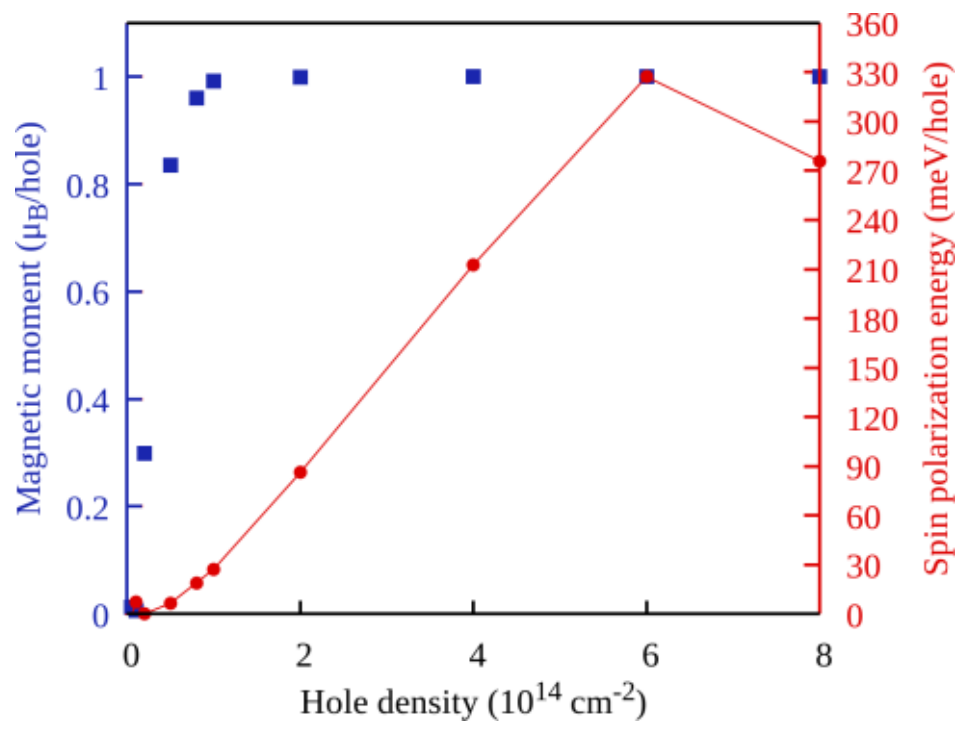

**Magnetic configurations and spin Hamiltonian**

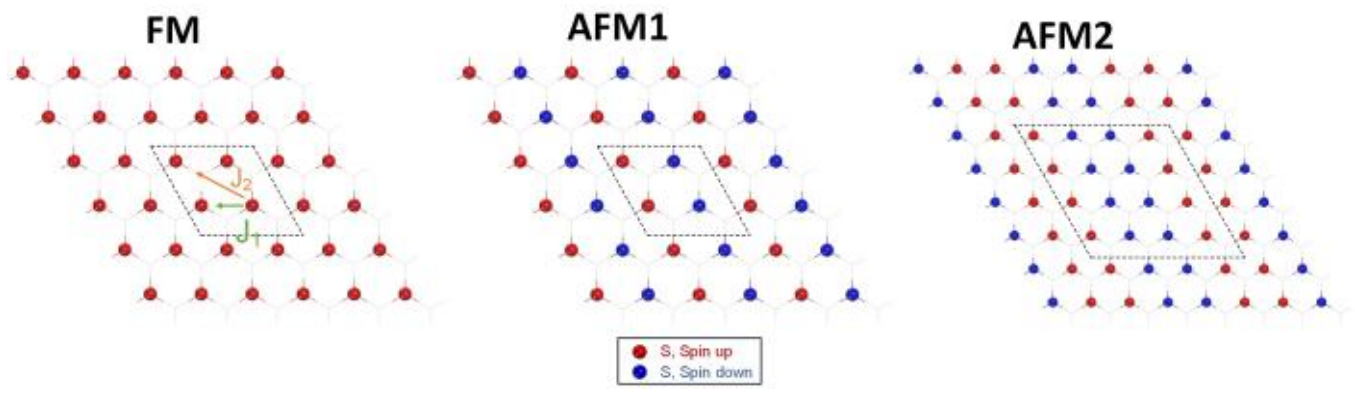

$$E_{FM} = E_0 - 12J_1S^2 - 12J_2S^2$$
$$E_{AFM1} = E_0 + 4J_1S^2 + 4J_2S^2$$
$$E_{AFM2} = E_0 - 4J_1S^2 + 4J_2S^2$$

**Magnetic exchange coupling parameters**

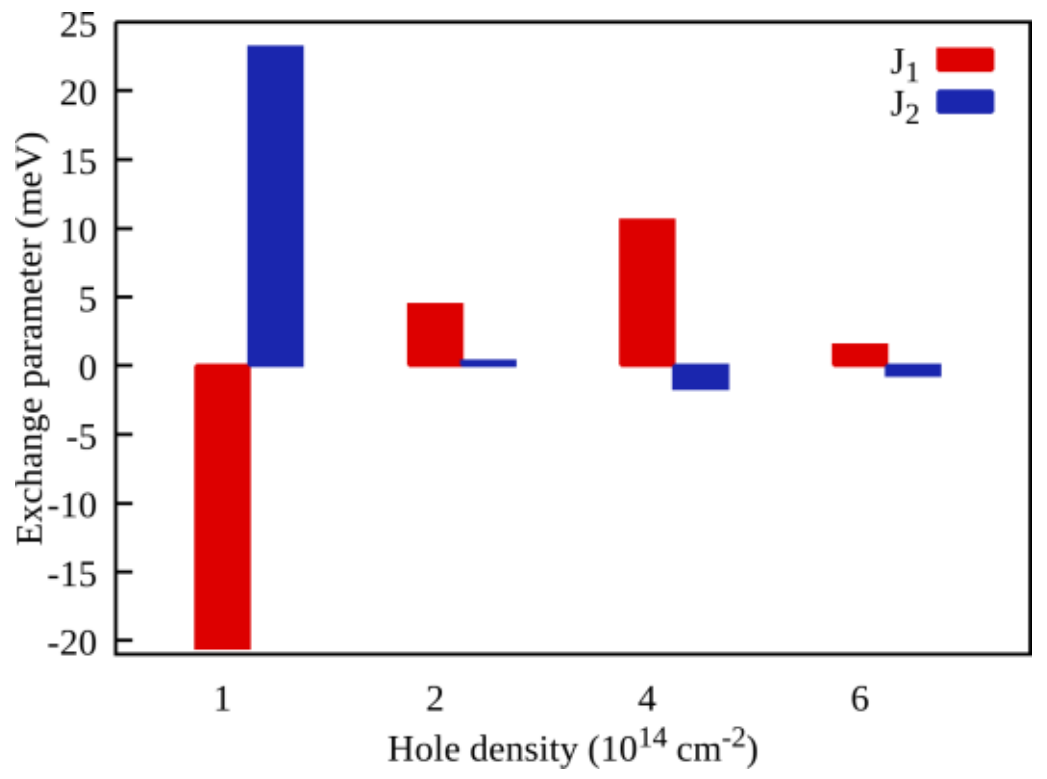

**Magnetic anisotropy energy (MAE, μeV) per magnetic atom**

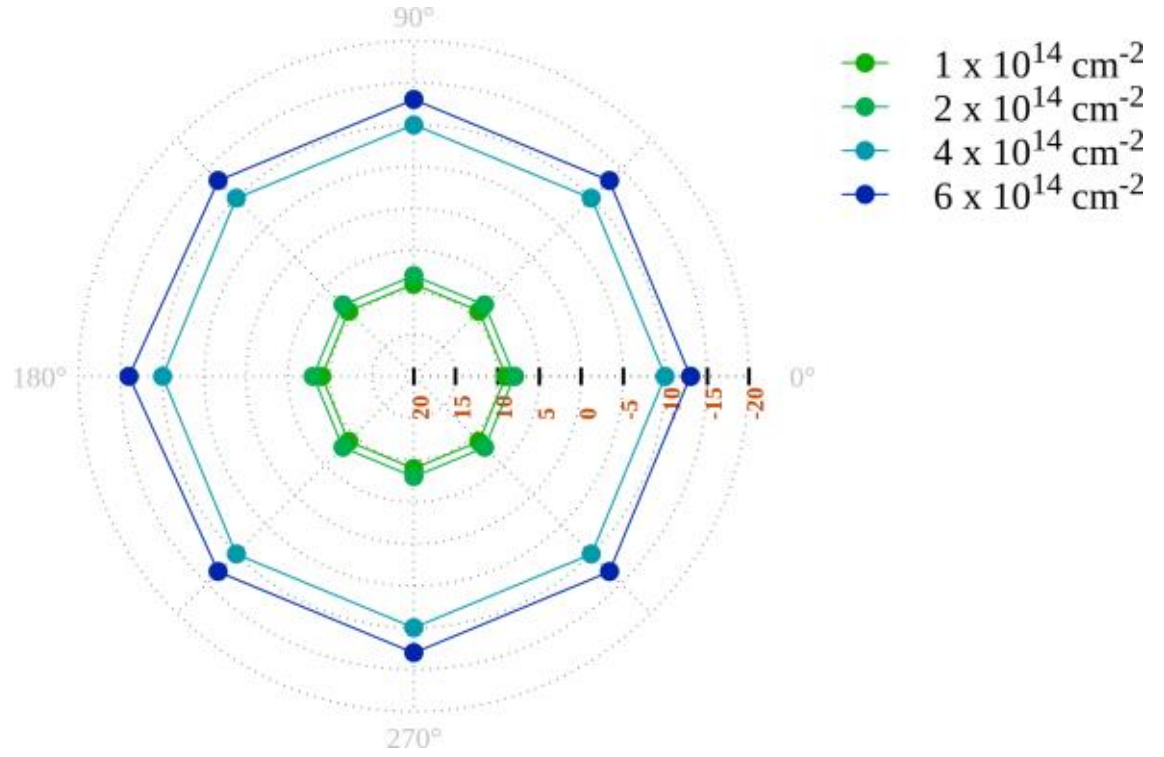

MAE = $E_{\parallel} - E_{\perp}$, a positive (negative) value of MAE indicates the off-plane (in-plane) easy axis.

**Monte Carlo simulations of the normalized magnetization of as a function of temperature**

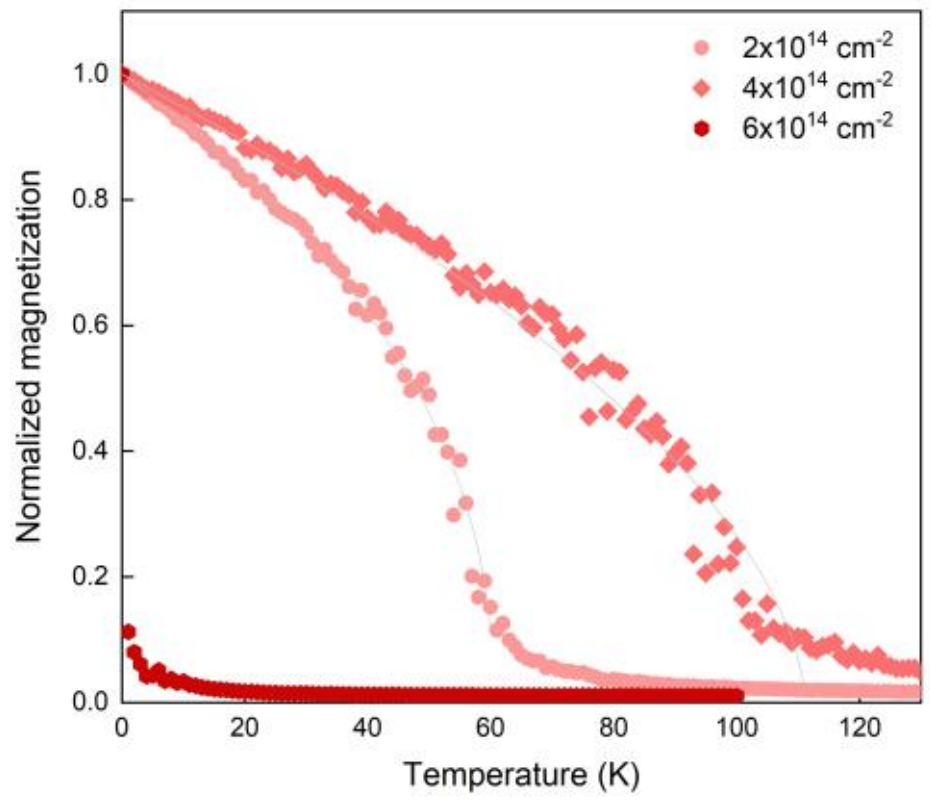

$T_c$: $0.5\times10^{14}$ cm$^{-2}$: - K  $0.8\times10^{14}$ cm$^{-2}$: 60
$1\times10^{14}$ cm$^{-2}$: 111 K  $2\times10^{14}$ cm$^{-2}$: - K

# 59.ZnS

| MC2D-ID | C2DB | 2dmat-ID | USPEX | Space group | Band gap (eV) |
|---|---|---|---|---|---|
| - | - | 2dm-6003 | - | P6m2 | 2.54 |

**Convex hull** | **Atomic structure** | **Atomic coordinates** | **Phonon dispersion curve**

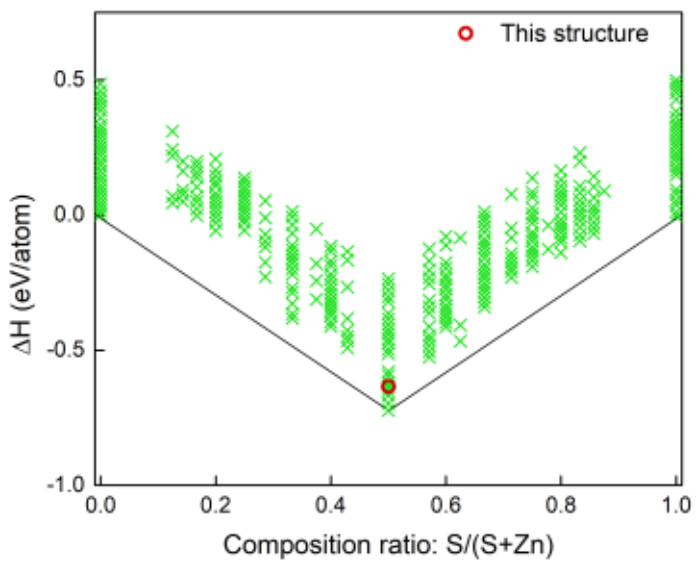

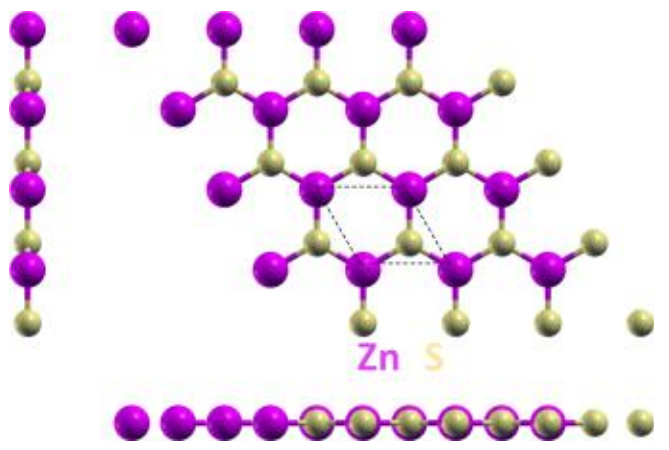

```
Zn1S1
   1.00000000000000
   3.8920917828964909   0.0000000000000000   0.0000000000000000
  -1.9460098501003000   3.3706717941884605   0.0000000000000000
   0.0000000000000000   0.0000000000000000  20.1669300000000007
   Zn   S
   1    1
Direct
  0.0000055430588475  0.9999947329815484  0.5000067577056129
  0.6666614569411493  0.3333382670184548  0.4999932422943871
```

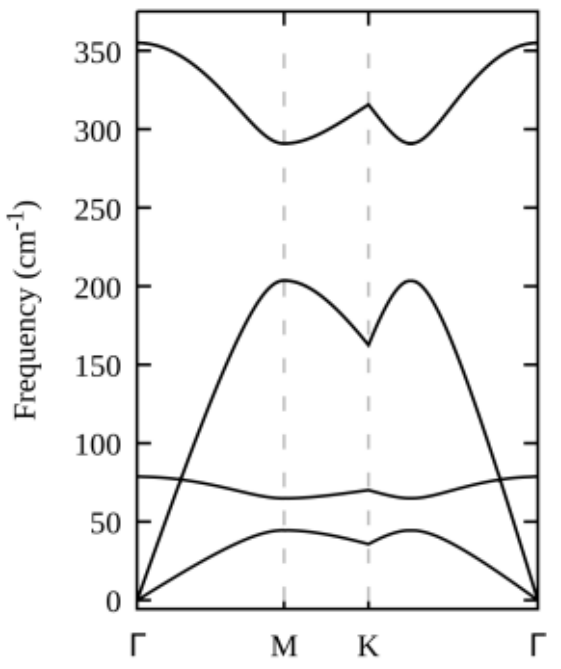

## Projected band structure and density of states

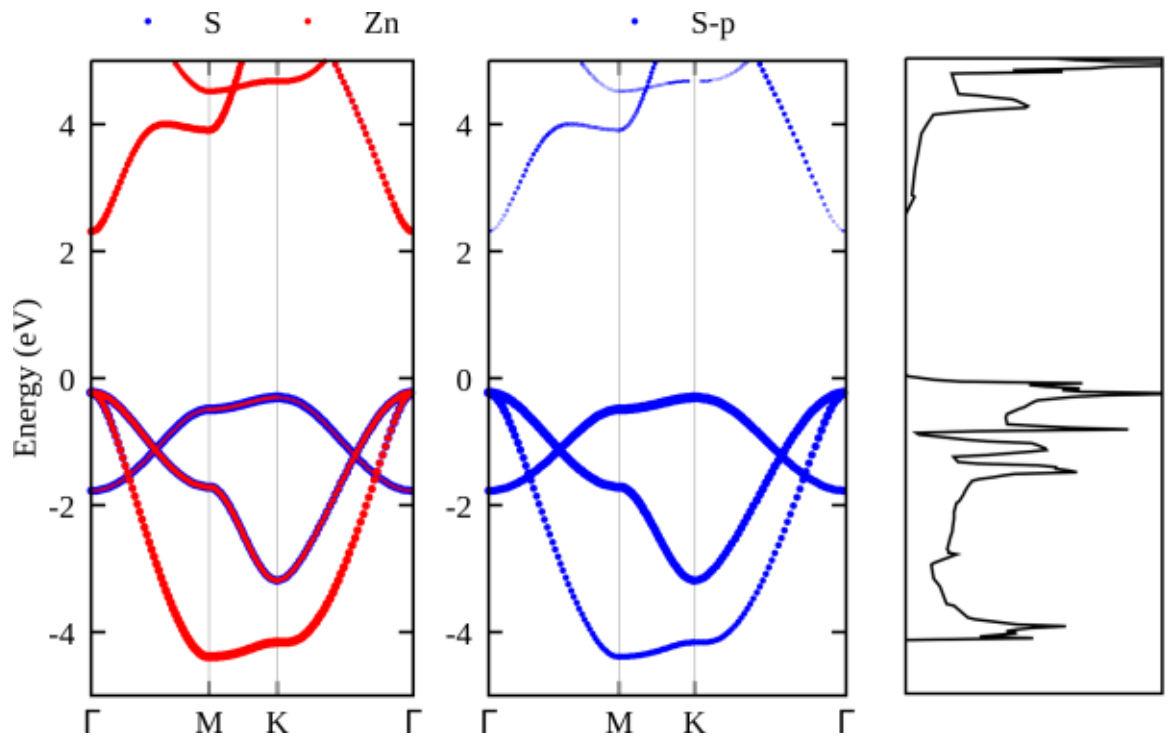

## Magnetic moment and spin polarization energy as a function of hole doping concentration

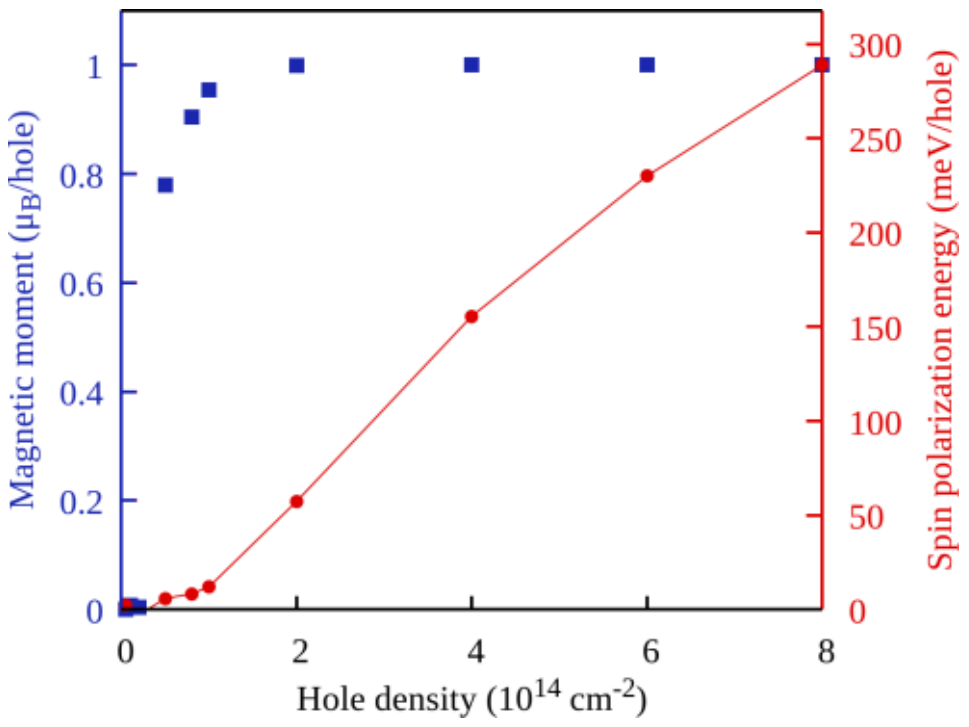

## Magnetic configurations and spin Hamiltonian

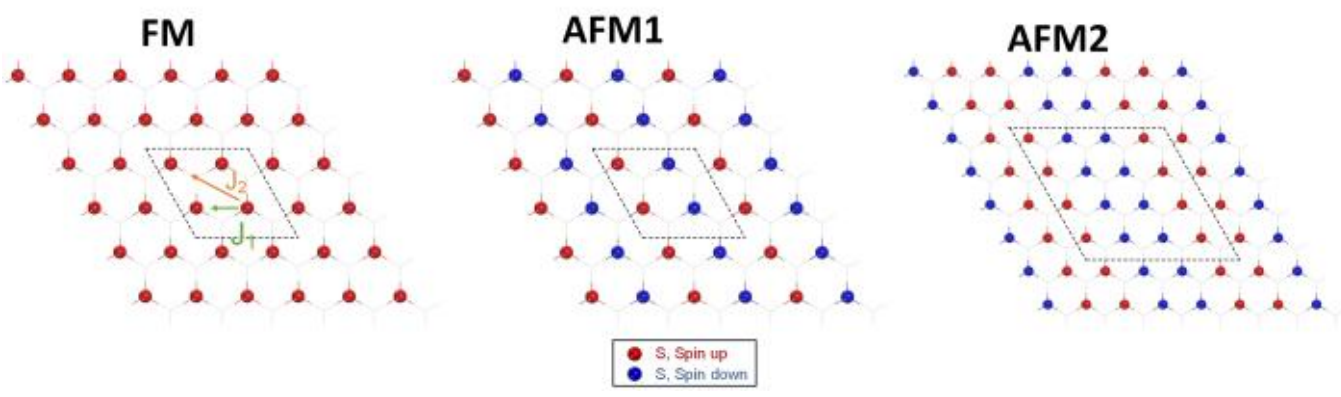

$$E_{FM} = E_0 - 12J_1S^2 - 12J_2S^2$$
$$E_{AFM1} = E_0 + 4J_1S^2 + 4J_2S^2$$
$$E_{AFM2} = E_0 - 4J_1S^2 + 4J_2S^2$$

## Magnetic exchange coupling parameters

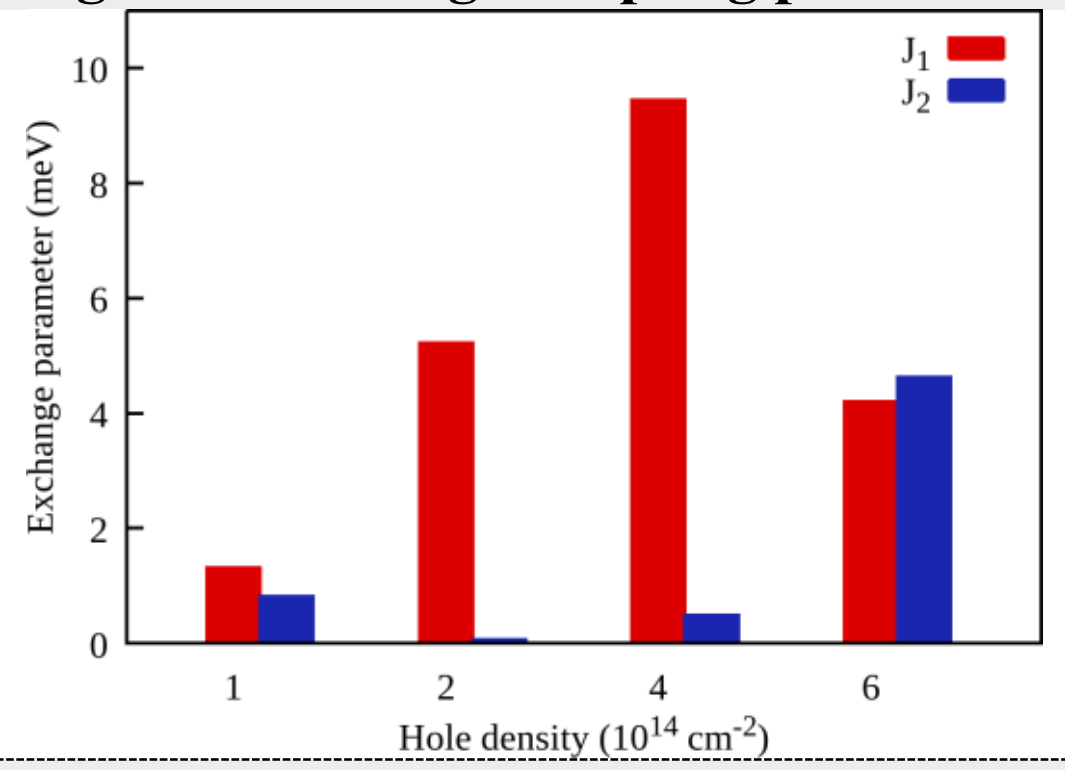

## Magnetic anisotropy energy (MAE, μeV) per magnetic atom

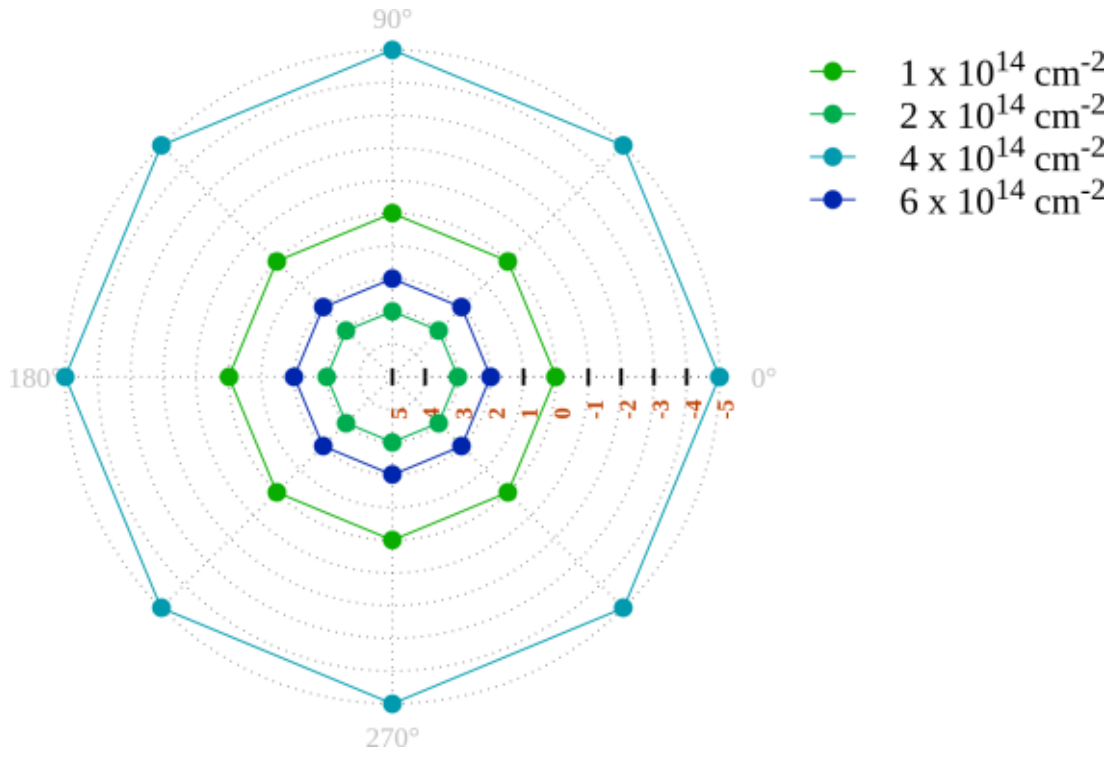

MAE = $E_{\parallel} - E_{\perp}$, a positive (negative) value of MAE indicates the off-plane (in-plane) easy axis.

## Monte Carlo simulations of the normalized magnetization of as a function of temperature

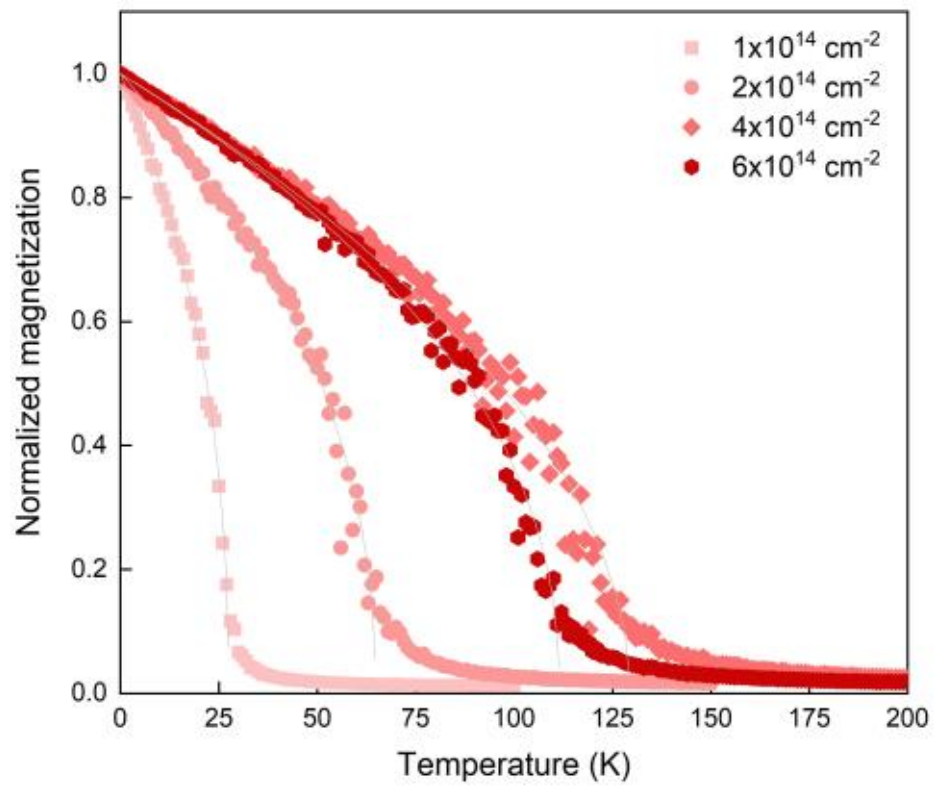

$T_c$: $0.5\times10^{14}$ cm$^{-2}$: 27 $0.8\times10^{14}$ cm$^{-2}$: 65

$1\times10^{14}$ cm$^{-2}$: 129 K $2\times10^{14}$ cm$^{-2}$: 111 K

# 60. BeO

| MC2D-ID | C2DB | 2dmat-ID | USPEX | Space group | Band gap (eV) |
|---|---|---|---|---|---|
| - | - | - | ✓ | P6m2 | 5.64 |

## Convex hull

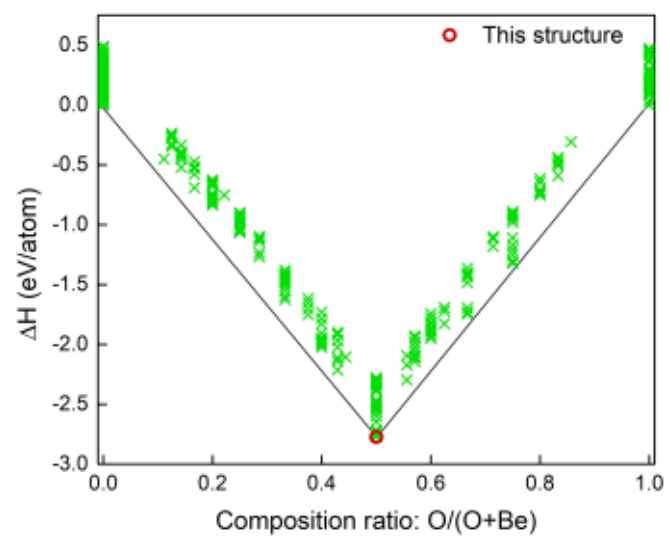

## Atomic structure

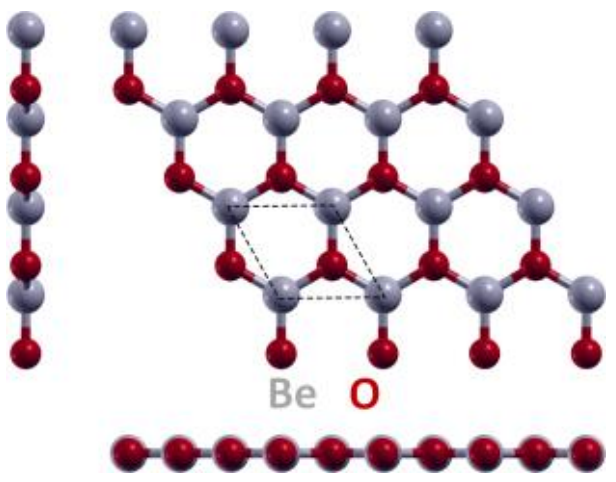

## Atomic coordinates

```
Be1O1
   1.00000000000000
   2.6794513013437462   0.0000000000000000   0.0000000000000000
   1.3385932197899217   2.3222466752701587   0.0000000000000000
   0.0000000000000000   0.0000000000000000  20.0000000000000000
   Be   O
   1    1
Direct
  0.2477499428836225  0.9866502521319376  0.4999999576193730
  0.9144080571163826  0.6531267478680647  0.5000000423806270
```

## Phonon dispersion curve

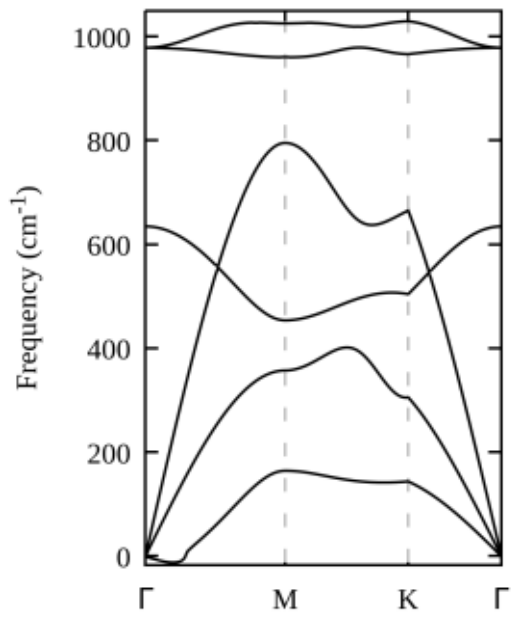

## Projected band structure and density of states

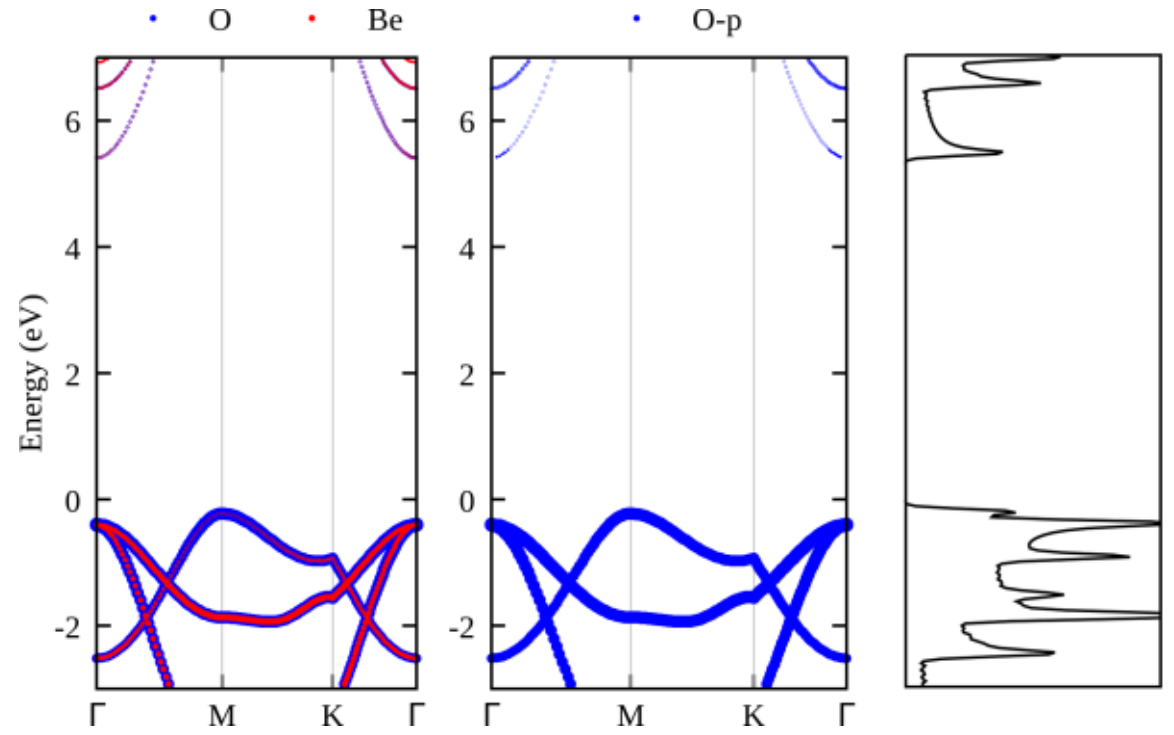

## Magnetic moment and spin polarization energy as a function of hole doping concentration

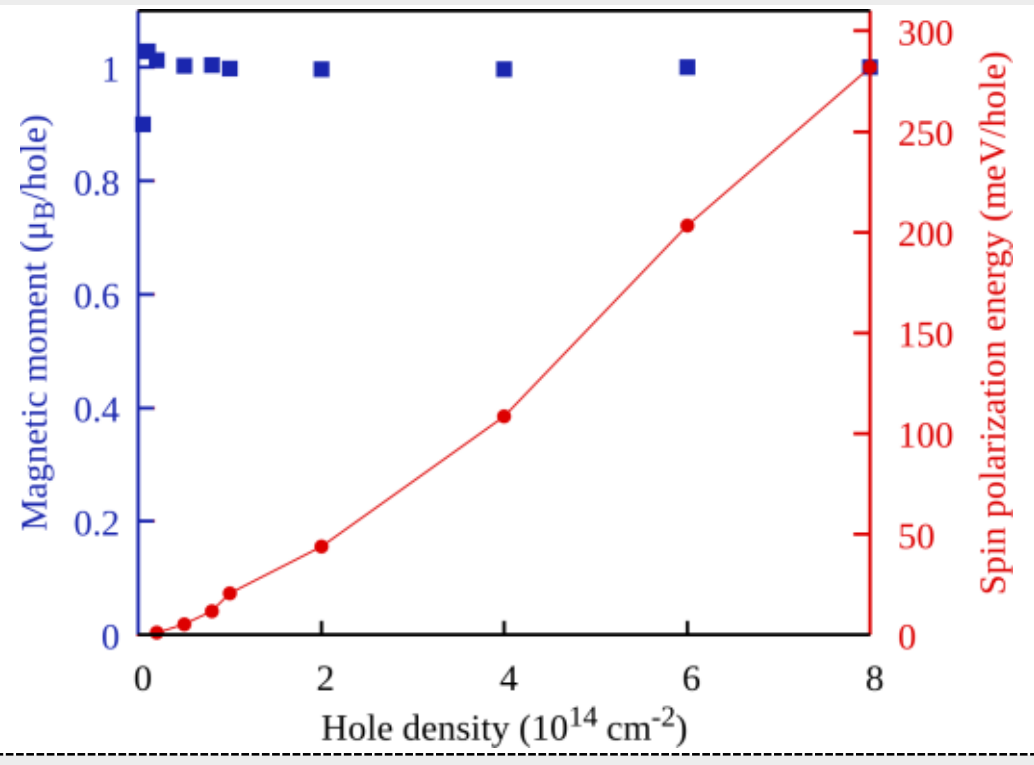

## Magnetic configurations and spin Hamiltonian

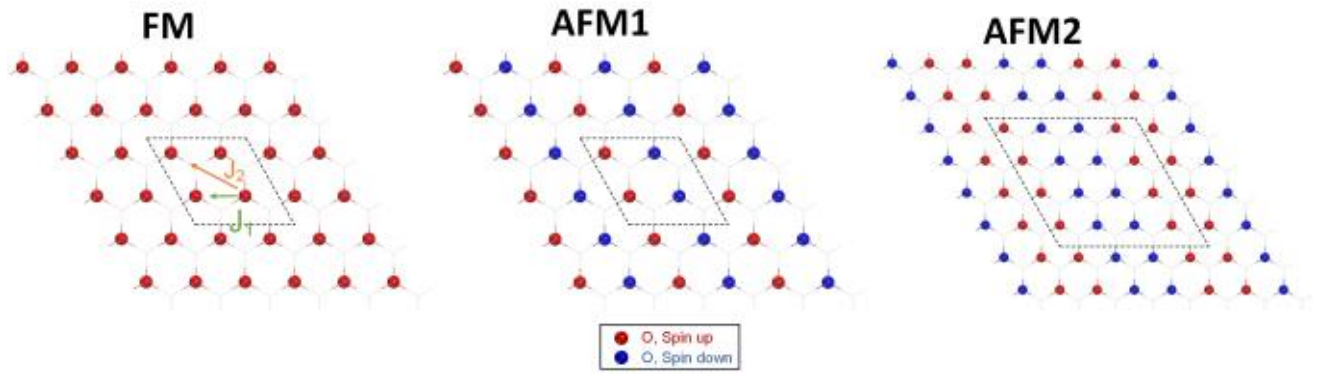

$$E_{FM} = E_0 - 12J_1S^2 - 12J_2S^2$$

$$E_{AFM1} = E_0 + 4J_1S^2 + 4J_2S^2$$

$$E_{AFM2} = E_0 - 4J_1S^2 + 4J_2S^2$$

## Magnetic exchange coupling parameters

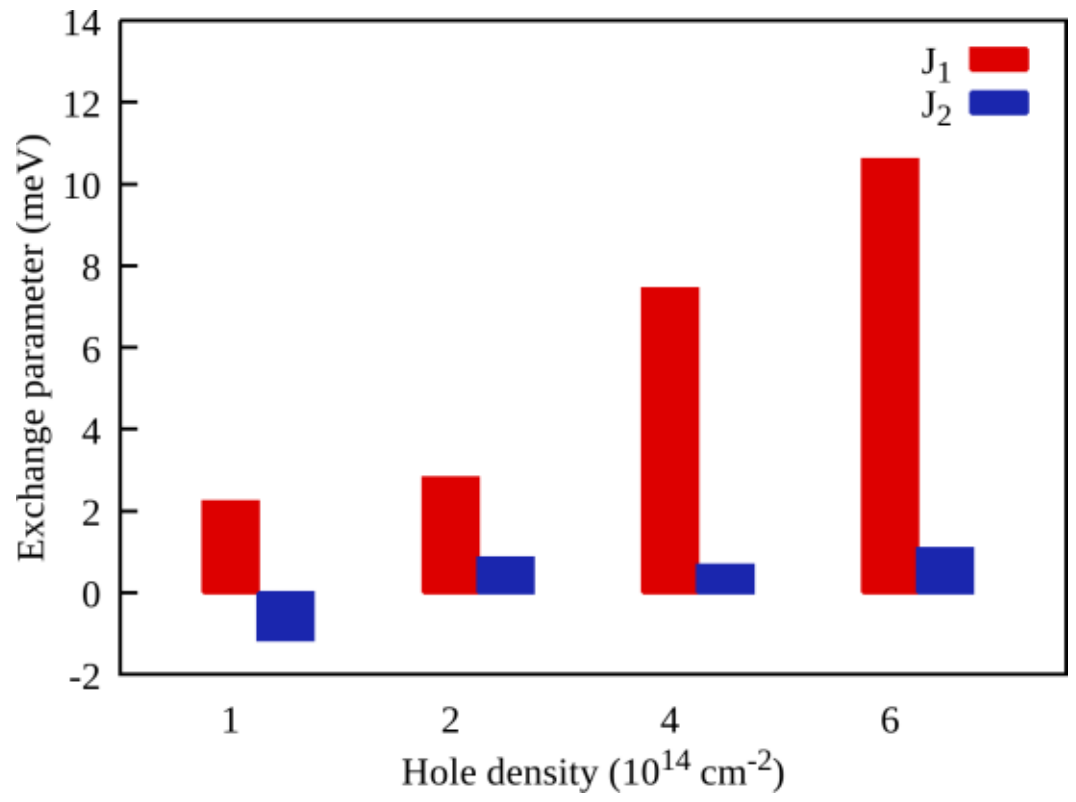

## Magnetic anisotropy energy (MAE, μeV) per magnetic atom

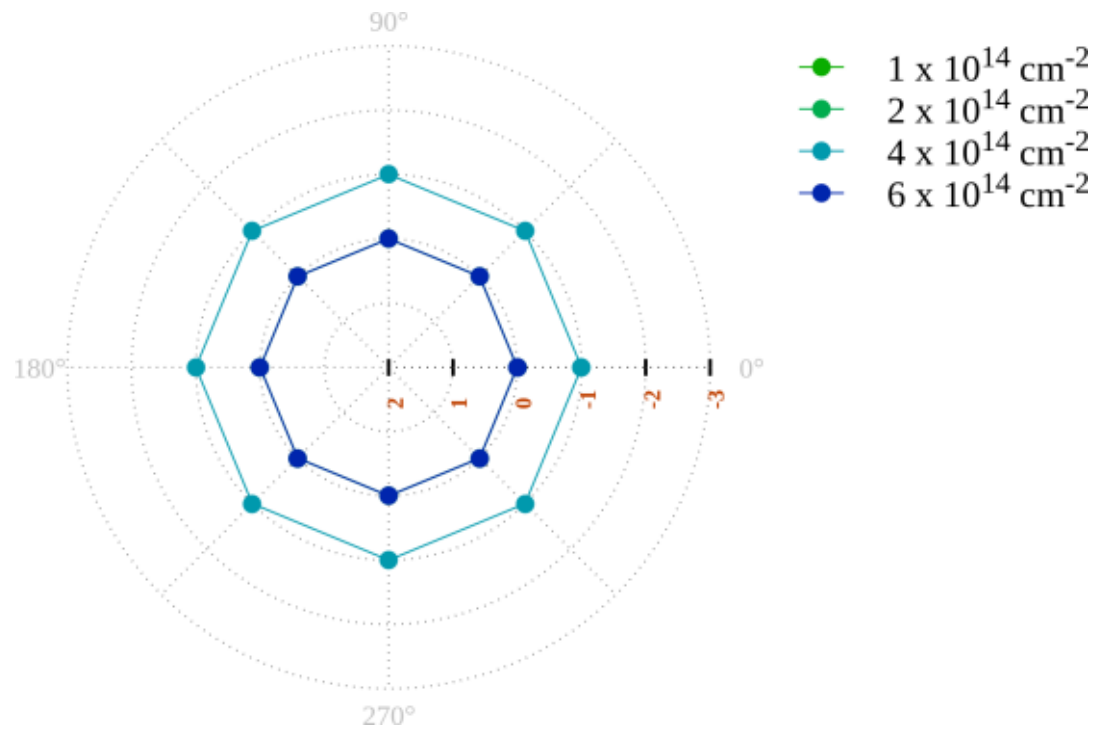

MAE = $E_{\parallel} - E_{\perp}$, a positive (negative) value of MAE indicates the off-plane (in-plane) easy axis.

## Monte Carlo simulations of the normalized magnetization of as a function of temperature

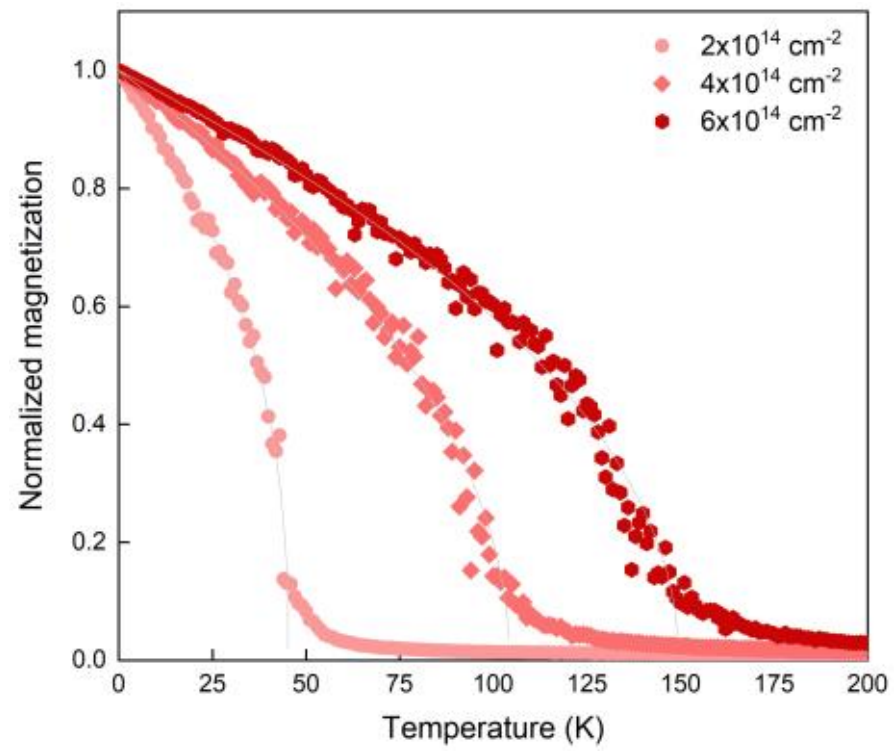

$T_c$: $1\times10^{14}$ cm$^{-2}$: - K   $2\times10^{14}$ cm$^{-2}$: 45 K

$4\times10^{14}$ cm$^{-2}$: 104 K   $6\times10^{14}$ cm$^{-2}$: 149 K

# 61.CdO

| MC2D-ID | C2DB | 2dmat-ID | USPEX | Space group | Band gap (eV) |
|---|---|---|---|---|---|
| - | ✓ | 2dm-6249 | - | P6m2 | 0.82 |

**Convex hull** | **Atomic structure** | **Atomic coordinates** | **Phonon dispersion curve**

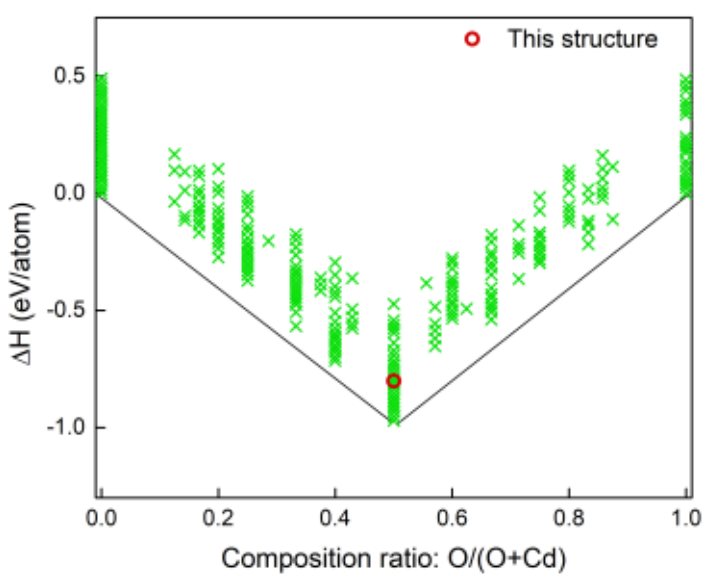

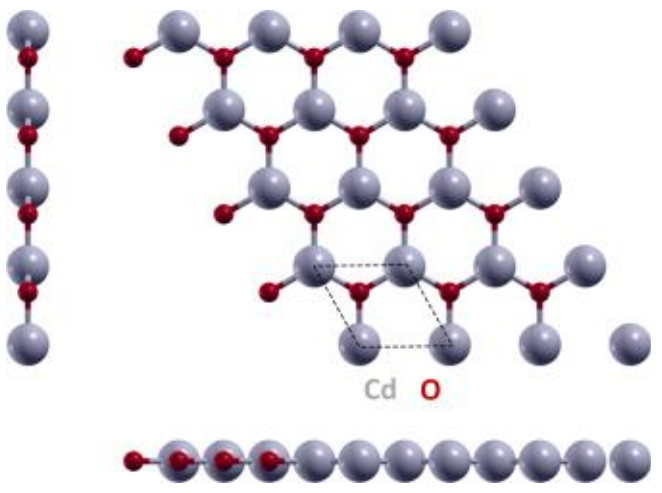

```
Cd1O1
   1.00000000000000
 3.6825886440488960   0.0000000000000000   0.0000000000000000
 -1.8412959146795791   3.1892143383270488   0.0000000000000000
 0.0000000000000000   0.0000000000000000   20.6349209999999985
   Cd   O
   1    1
Direct
  0.9999998982562843   0.0000000992136862   0.5000065496344277
  0.3333331017437189   0.6666669007863106   0.4999934503655723
```

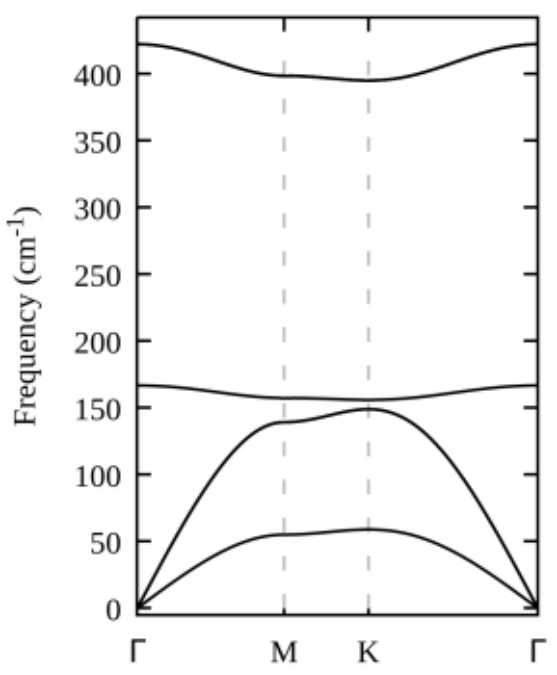

## Projected band structure and density of states

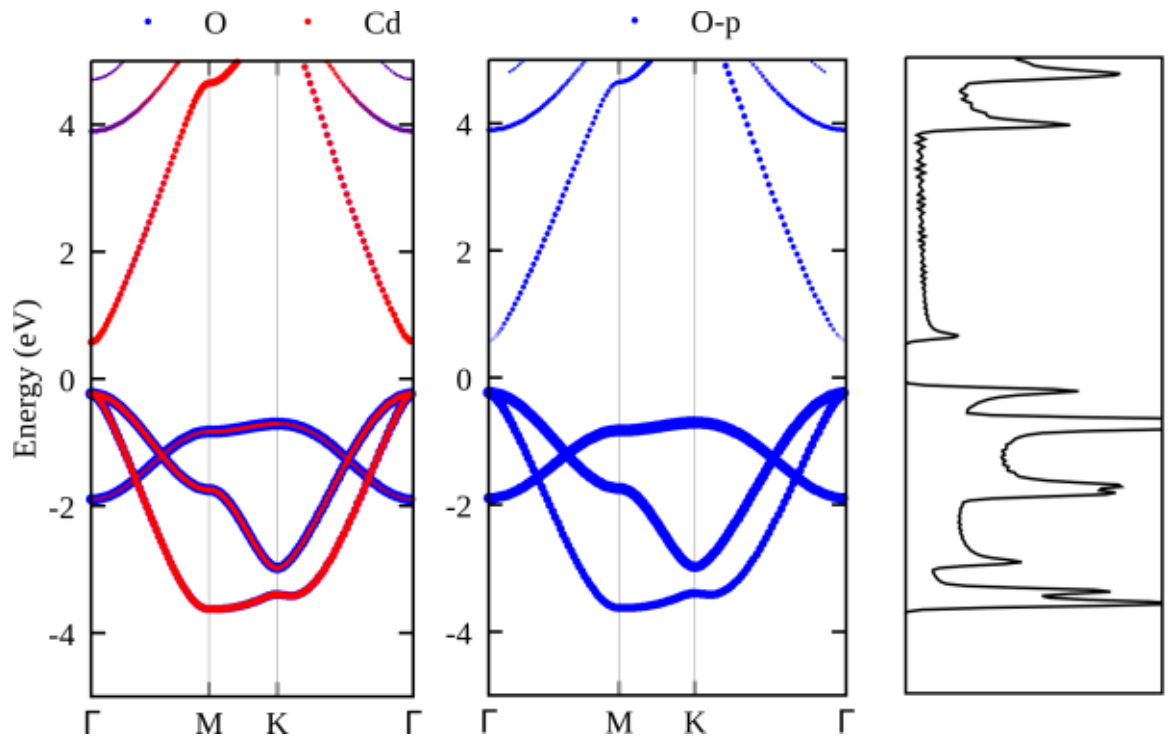

## Magnetic moment and spin polarization energy as a function of hole doping concentration

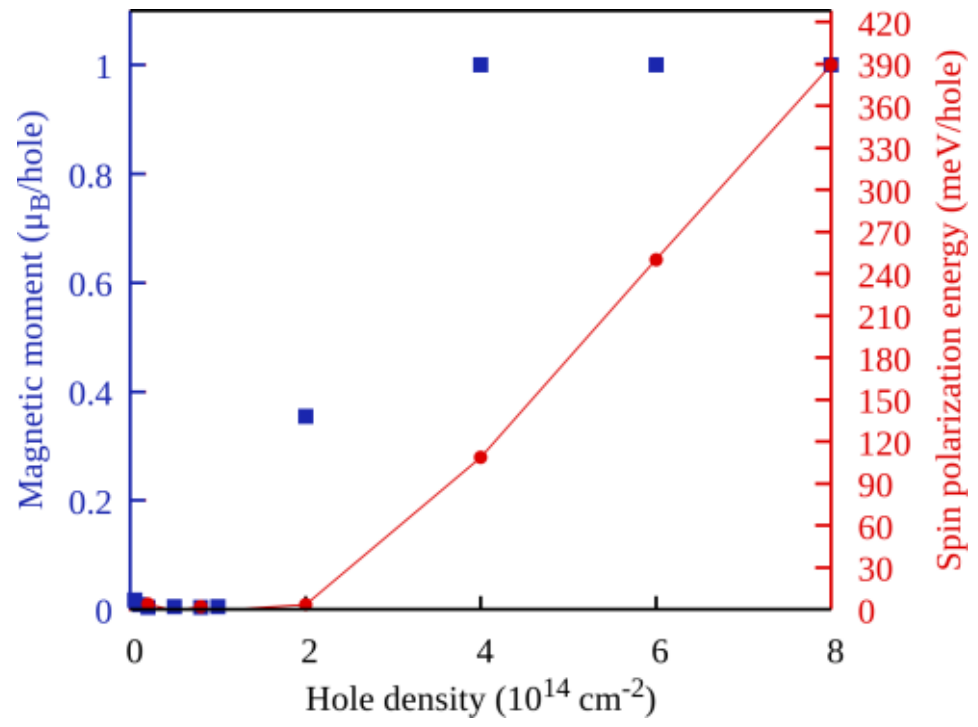

## Magnetic configurations and spin Hamiltonian

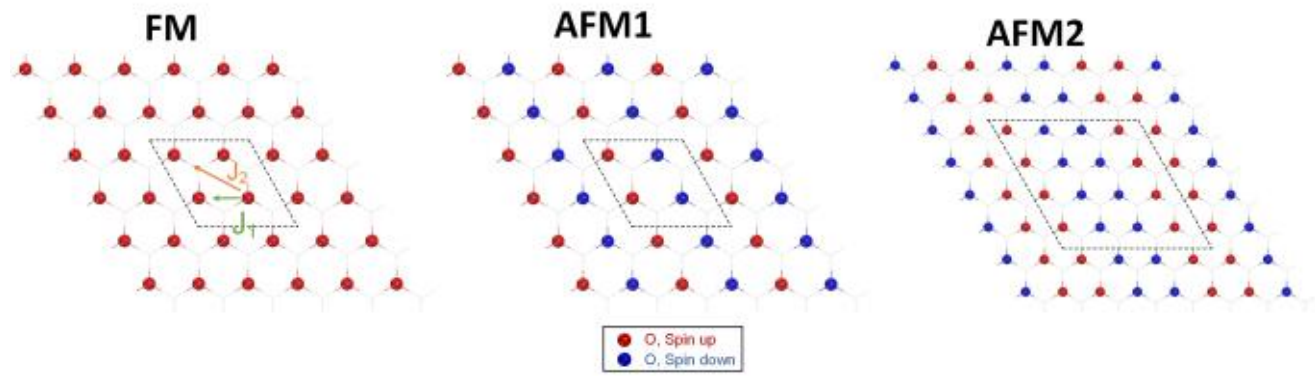

$$E_{FM} = E_0 - 12J_1S^2 - 12J_2S^2$$
$$E_{AFM1} = E_0 + 4J_1S^2 + 4J_2S^2$$
$$E_{AFM2} = E_0 - 4J_1S^2 + 4J_2S^2$$

## Magnetic exchange coupling parameters

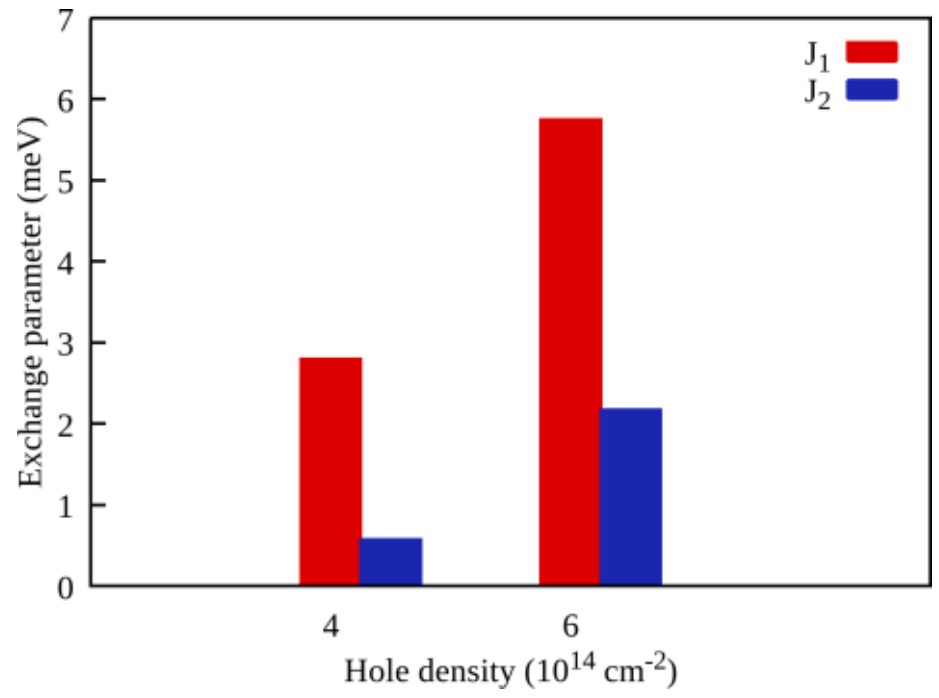

## Magnetic anisotropy energy (MAE, μeV) per magnetic atom

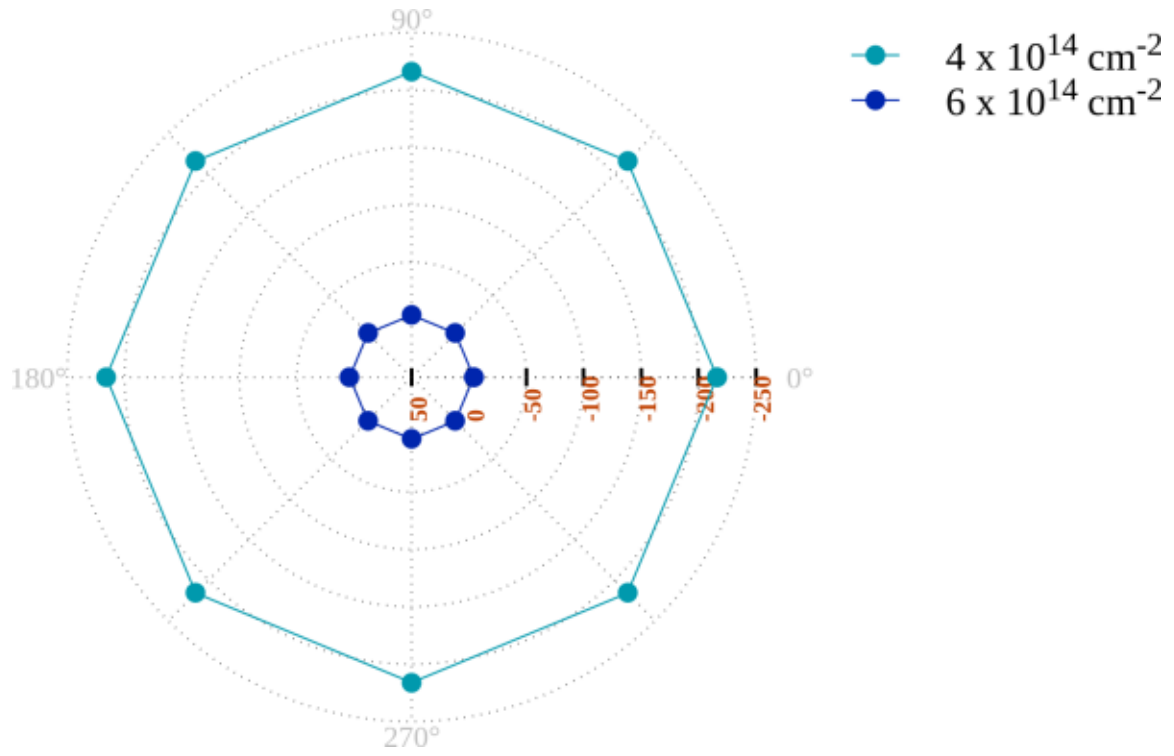

MAE = $E_{\parallel} - E_{\perp}$, a positive (negative) value of MAE indicates the off-plane (in-plane) easy axis.

## Monte Carlo simulations of the normalized magnetization of as a function of temperature

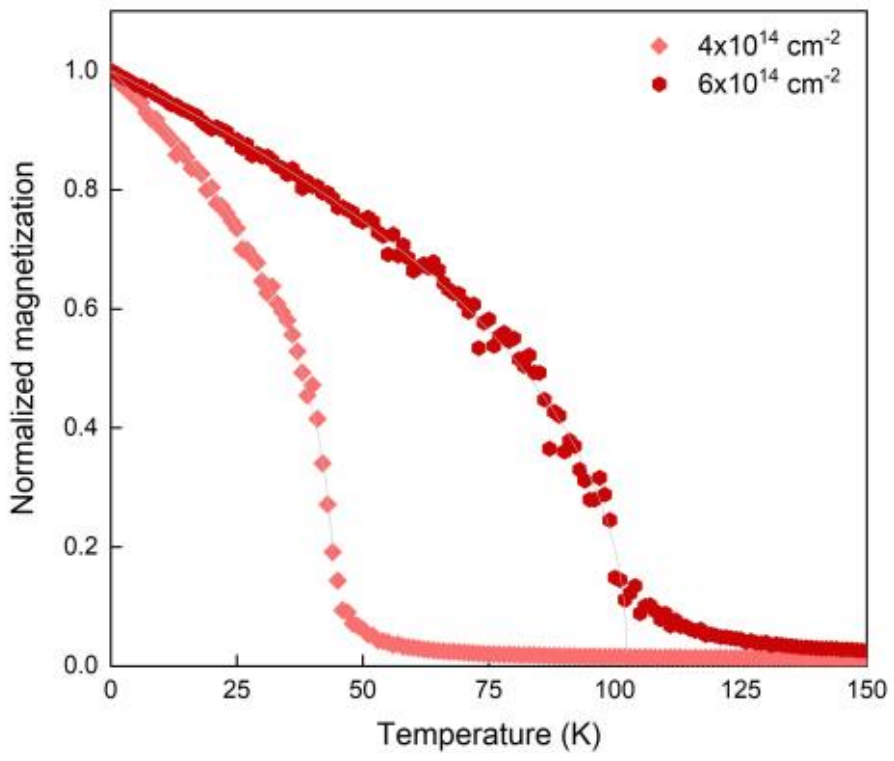

$T_c$: $1\times10^{14}$ cm$^{-2}$: - K   $2\times10^{14}$ cm$^{-2}$: - K

$4\times10^{14}$ cm$^{-2}$: 45 K   $6\times10^{14}$ cm$^{-2}$: 102 K

# 62. $TeO_3$

| MC2D-ID | C2DB | 2dmat-ID | USPEX | Space group | Band gap (eV) |
|---|---|---|---|---|---|
| - | - | 2dm-4882 | - | P6m2 | 0.86 |

## Convex hull

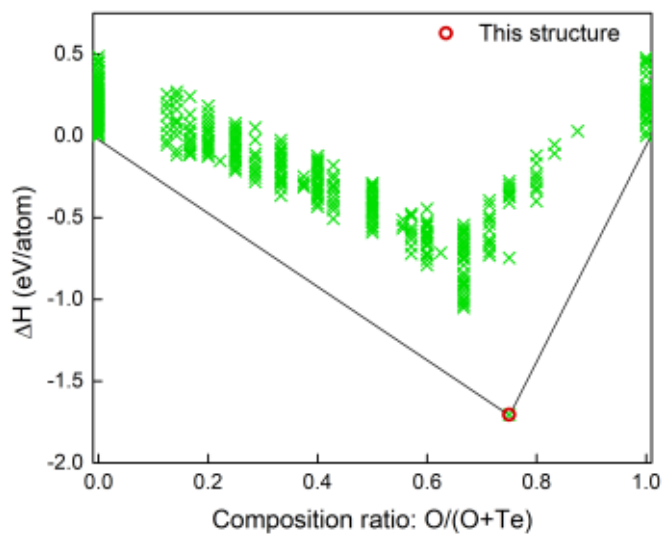

## Atomic structure

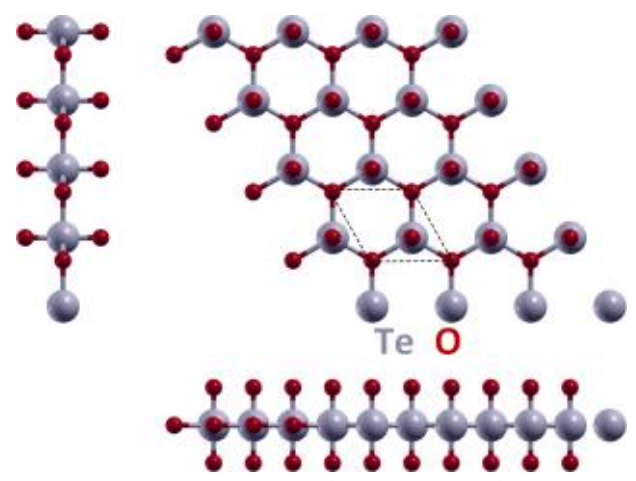

## Atomic coordinates

```
Te1O3
   1.00000000000000
   3.7689421325705479    0.0000000000000000    0.0000000000000000
  -1.8844707233302402    3.2639992105142546    0.0000000000000000
   0.0000000000000000    0.0000000000000000   23.5916649999999990
   Te   O
   1    3
Direct
  0.9999999967672437  0.0000000340517587  0.5000001130785705
  0.9999999778634034  0.9999999944631028  0.4238697578018247
  0.9999999783810836  0.9999999934514747  0.5761301269029389
  0.3333330469882725  0.6666669780336605  0.5000000022166660
```

## Phonon dispersion curve

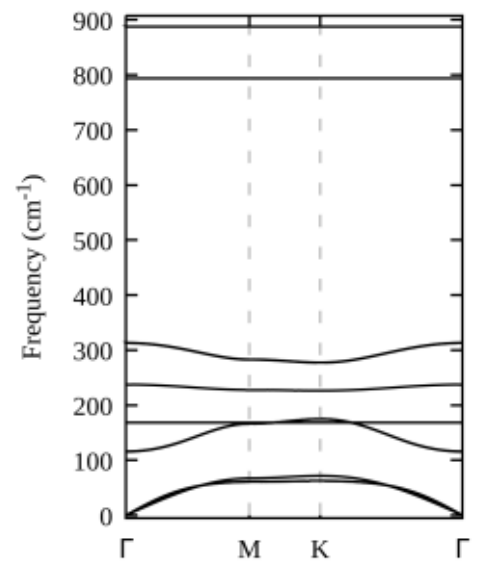

## Projected band structure and density of states

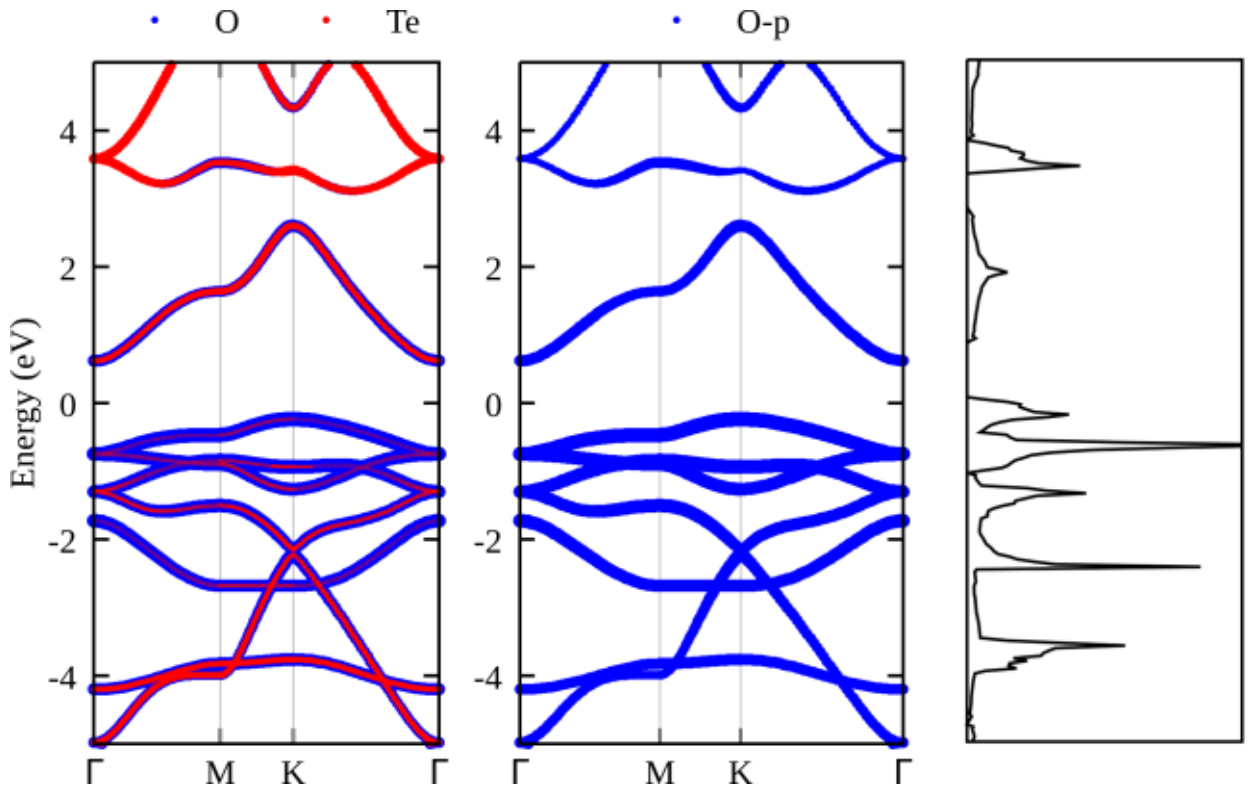

## Magnetic moment and spin polarization energy as a function of hole doping concentration

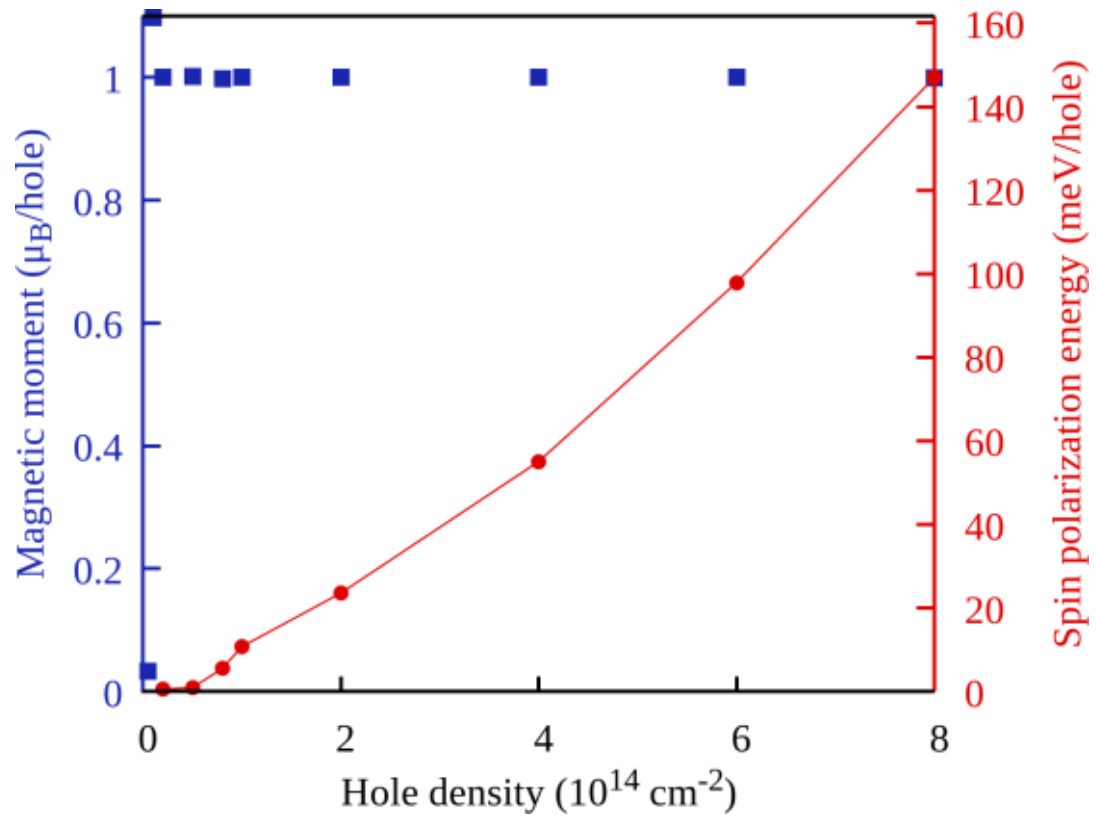

## Magnetic configurations and spin Hamiltonian

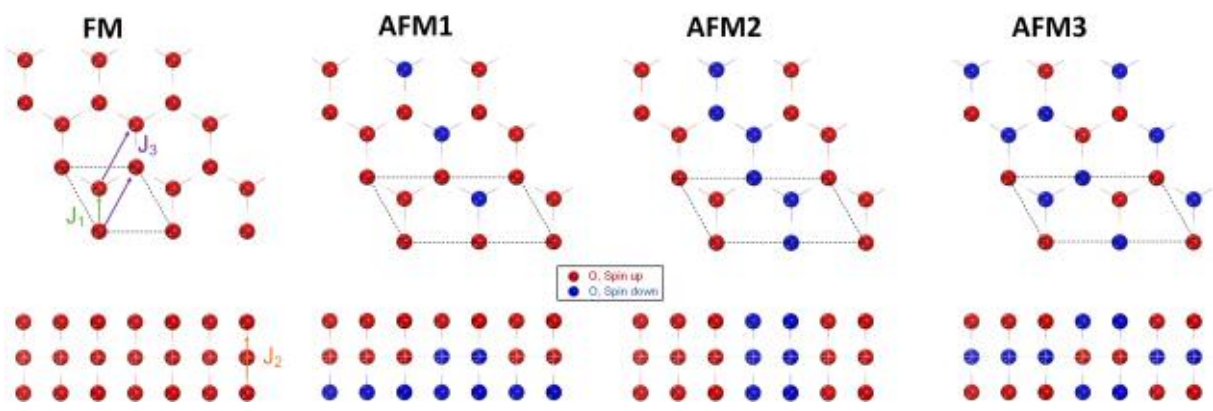

$$E_{FM} = E_0 - 6J_1S^2 - J_2S^2 - 9J_3S^2$$

$$E_{AFM1} = E_0 + J_2S^2 - 5J_3S^2$$

$$E_{AFM2} = E_0 - 2J_1S^2 - J_2S^2 + 3J_3S^2$$

$$E_{AFM3} = E_0 + 2J_1S^2 - J_2S^2 + 3J_3S^2$$

## Magnetic exchange coupling parameters

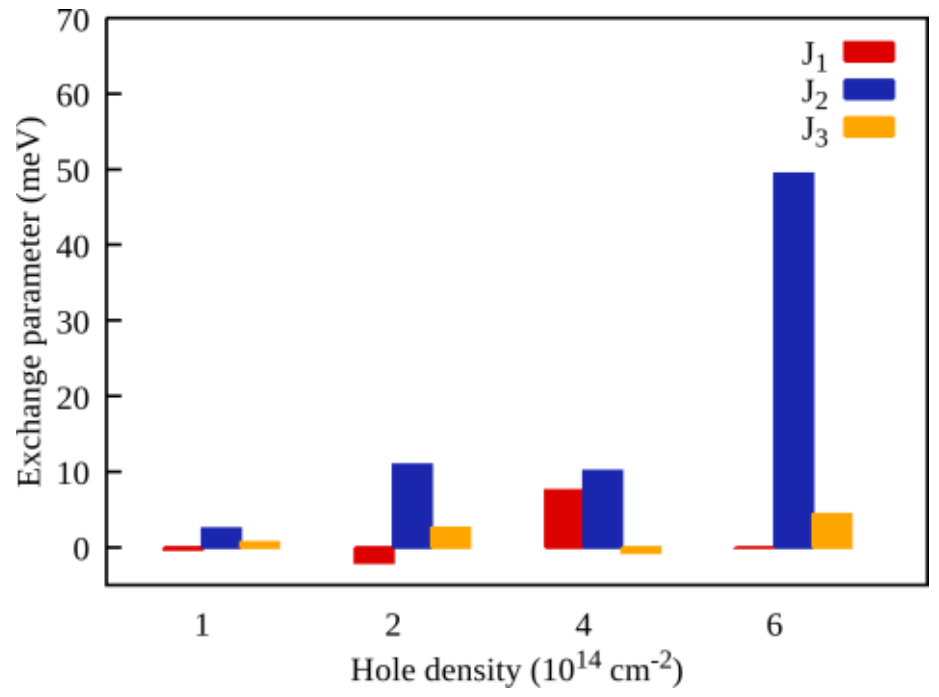

## Magnetic anisotropy energy (MAE, μeV) per magnetic atom

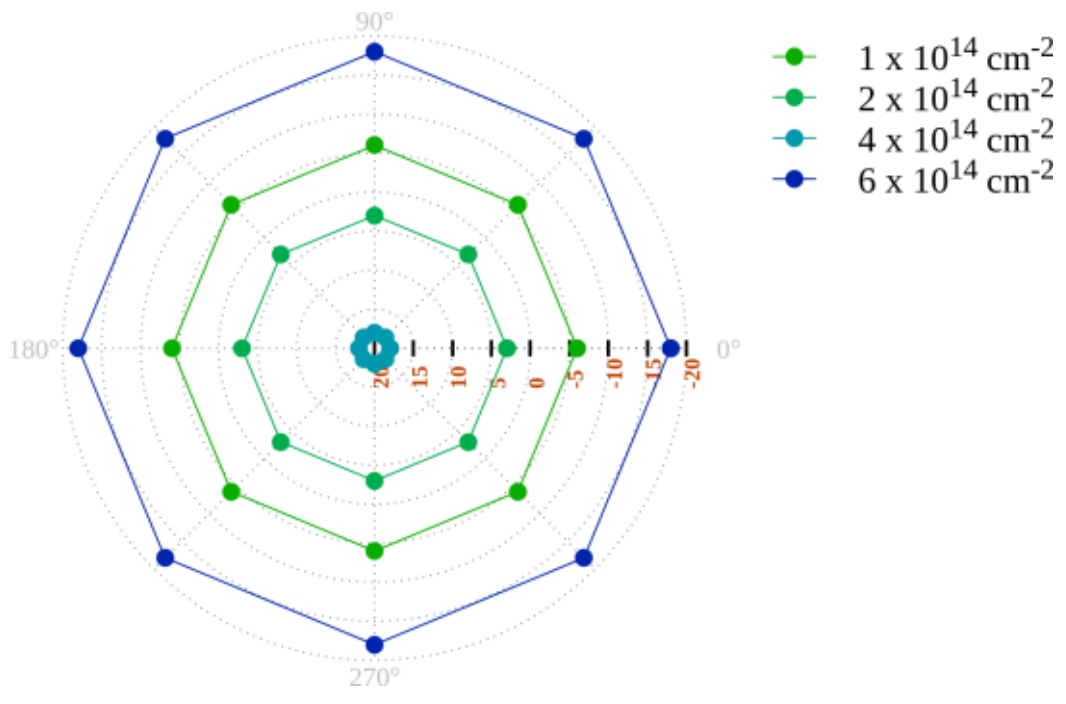

MAE = $E_{\parallel} - E_{\perp}$, a positive (negative) value of MAE indicates the off-plane (in-plane) easy axis.

## Monte Carlo simulations of the normalized magnetization of as a function of temperature

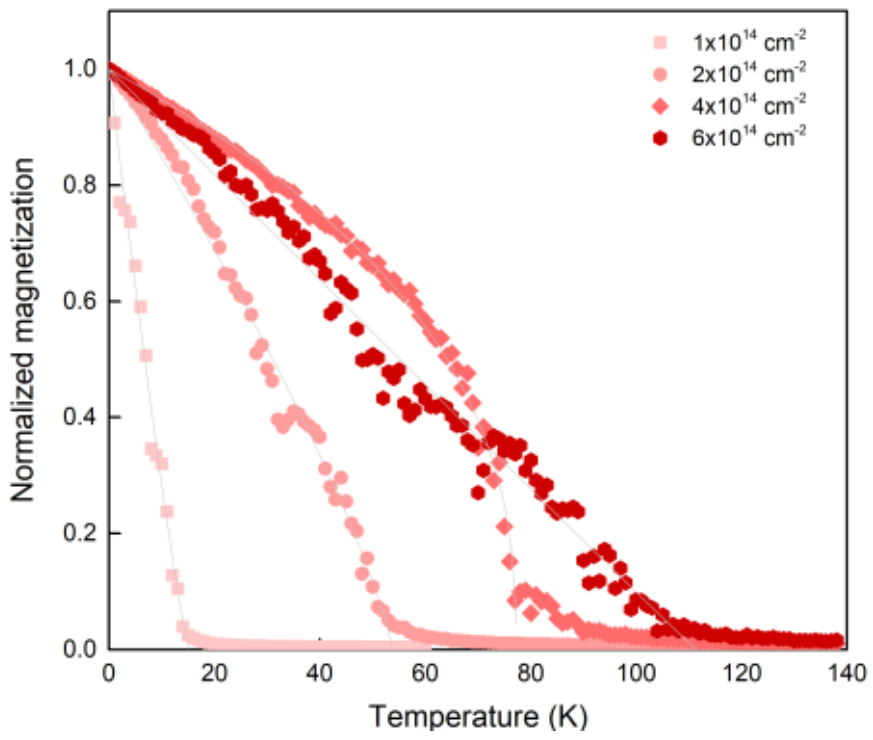

$T_c$: $1\times10^{14}$ cm$^{-2}$: 15 K  $2\times10^{14}$ cm$^{-2}$: 54 K

$4\times10^{14}$ cm$^{-2}$: 77 K  $6\times10^{14}$ cm$^{-2}$: 111 K

# 63. AlN

| MC2D-ID | C2DB | 2dmat-ID | USPEX | Space group | Band gap (eV) |
| --- | --- | --- | --- | --- | --- |
| - | ✓ | 2dm-3085 | - | P6m2 | 2.91 |

## Convex hull

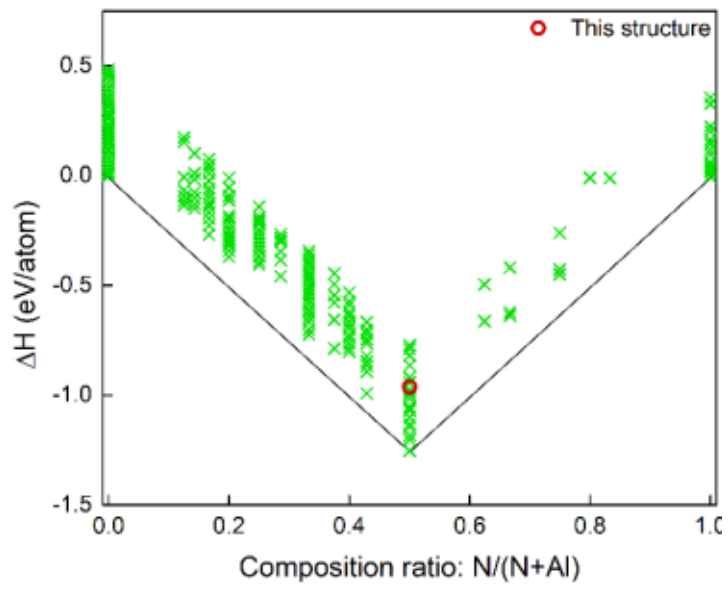

## Atomic structure

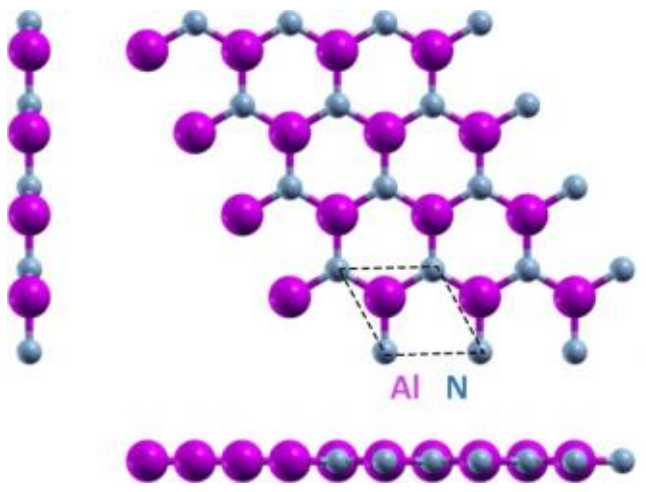

## Atomic coordinates

```
Al N
1.0000000000000000
3.1261529833522546   0.0000000000000000   0.0000000000000000
-1.5630816990343190  2.7073252116660003   0.0000000000000000
0.0000000000000000   0.0000000000000000   20.0000000000000000
Al N
1 1
Direct
0.3333330967632762   0.6666668749795690   0.5000000011677059
0.9999999032367271   0.0000001250204278   0.4999999988322941
```

## Phonon dispersion curve

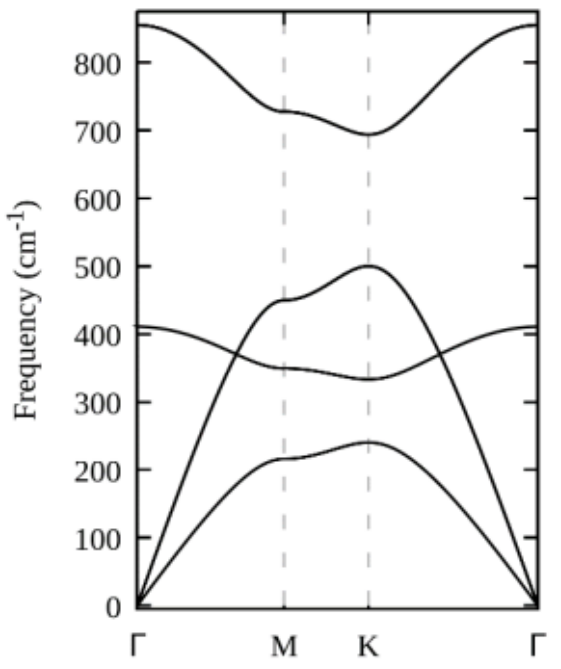

## Projected band structure and density of states

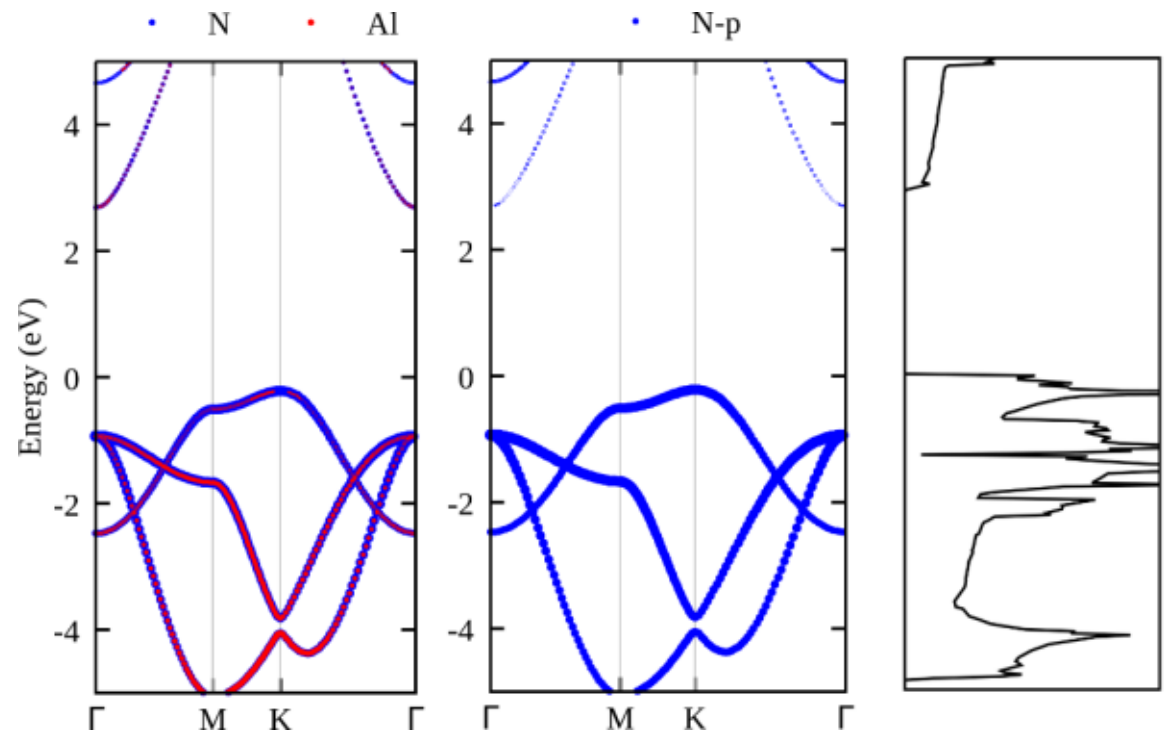

## Magnetic moment and spin polarization energy as a function of hole doping concentration

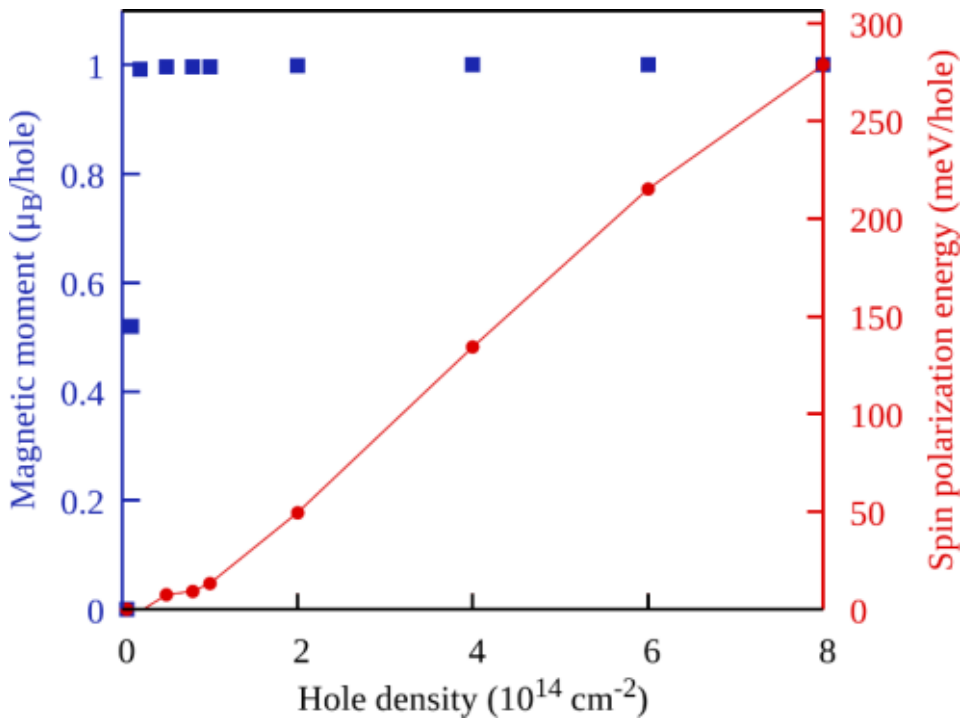

## Magnetic configurations and spin Hamiltonian

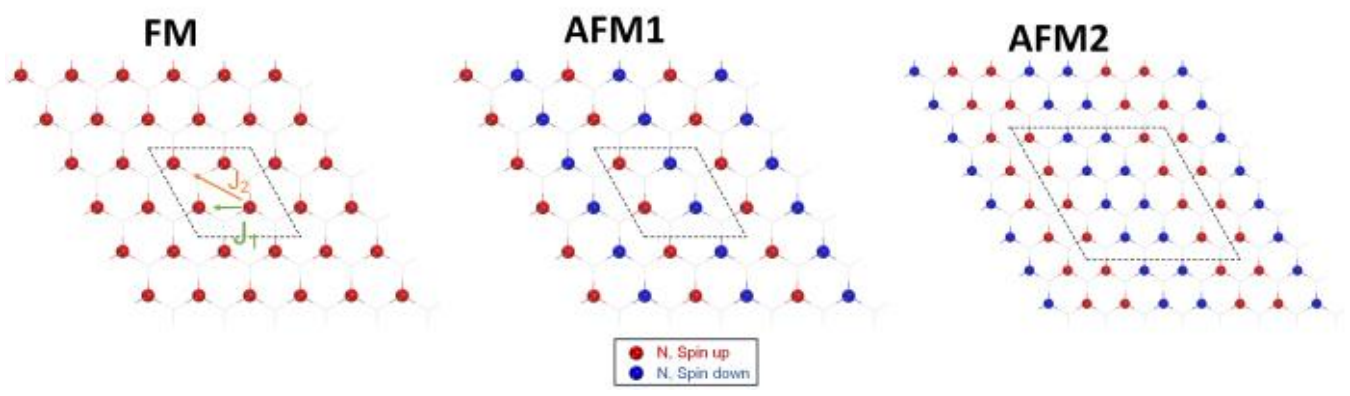

$$E_{FM} = E_0 - 12J_1S^2 - 12J_2S^2$$
$$E_{AFM1} = E_0 + 4J_1S^2 + 4J_2S^2$$
$$E_{AFM2} = E_0 - 4J_1S^2 + 4J_2S^2$$

## Magnetic exchange coupling parameters

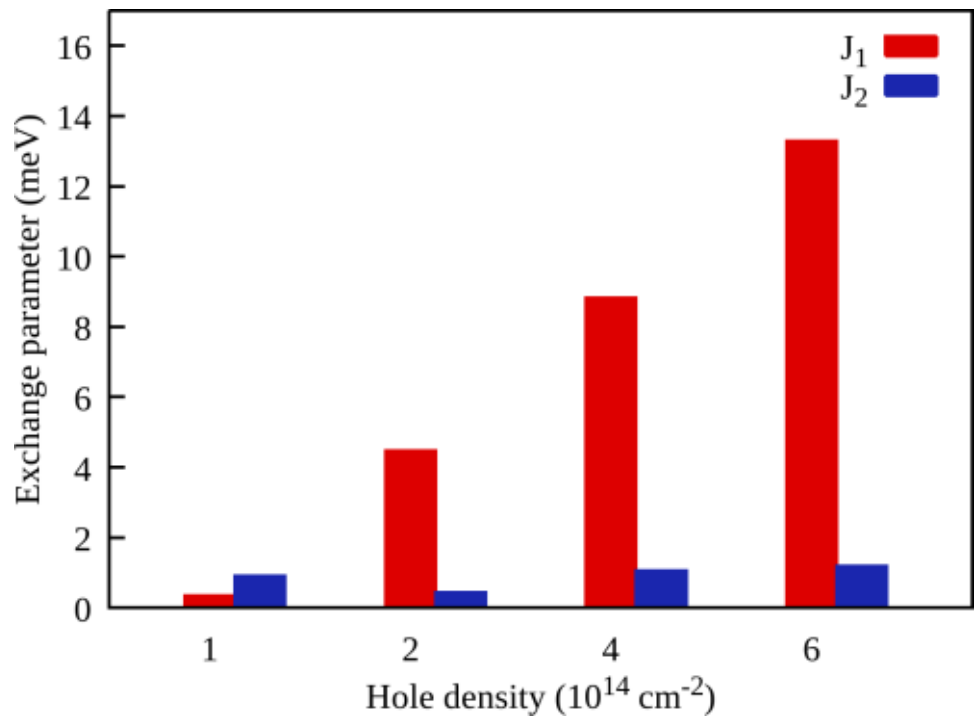

## Magnetic anisotropy energy (MAE, μeV) per magnetic atom

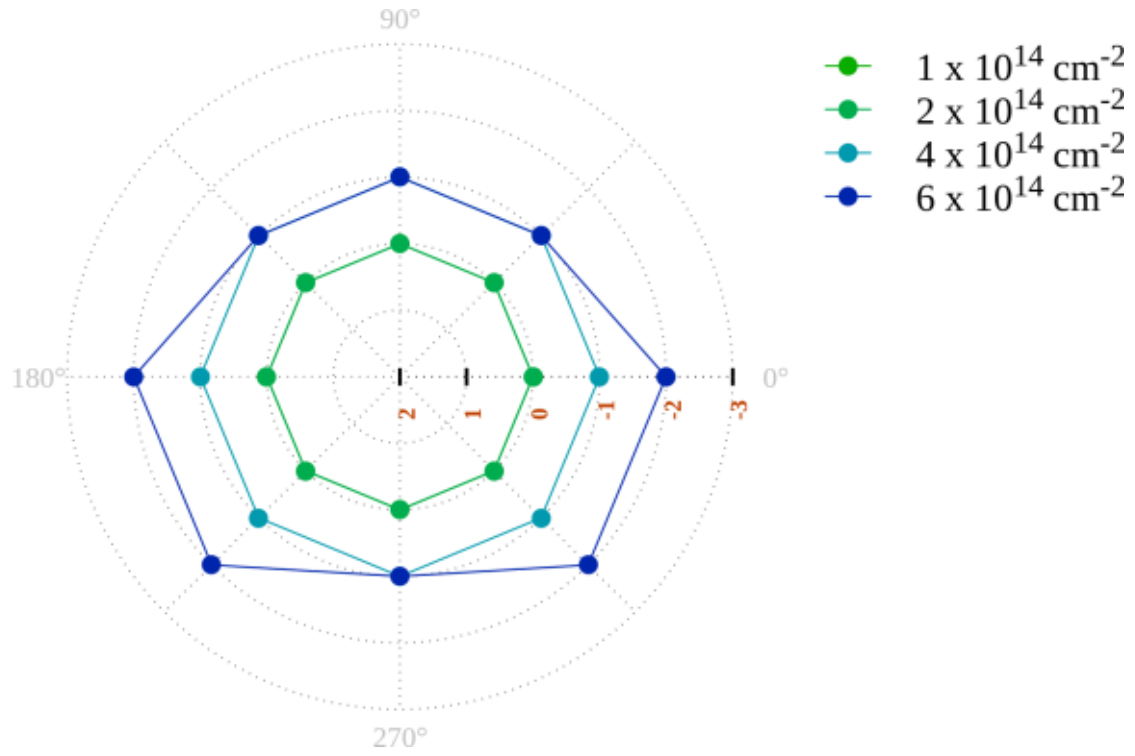

MAE = $E_{\parallel} - E_{\perp}$, a positive (negative) value of MAE indicates the off-plane (in-plane) easy axis.

## Monte Carlo simulations of the normalized magnetization of as a function of temperature

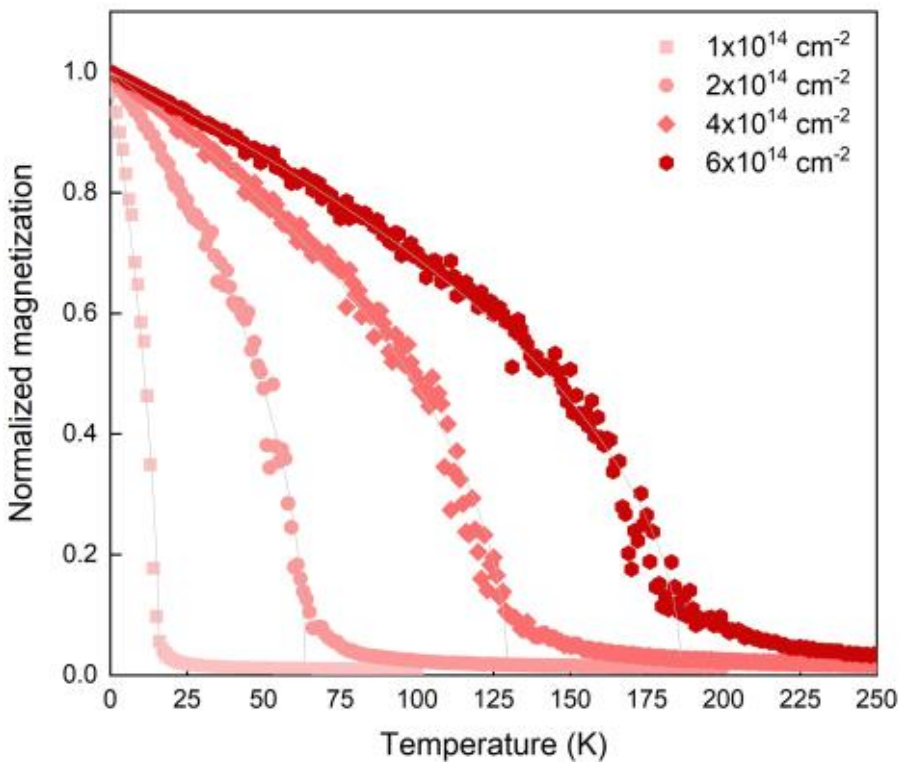

$T_c$: $1\times10^{14}$ cm$^{-2}$: 16 K  $2\times10^{14}$ cm$^{-2}$: 63 K

$4\times10^{14}$ cm$^{-2}$: 129 K  $6\times10^{14}$ cm$^{-2}$: 186 K

# 64.GaN

| MC2D-ID | C2DB | 2dmat-ID | USPEX | Space group | Band gap (eV) |
|---|---|---|---|---|---|
| - | ✓ | 2dm-2992 | - | P6m2 | 2.16 |

**Convex hull**

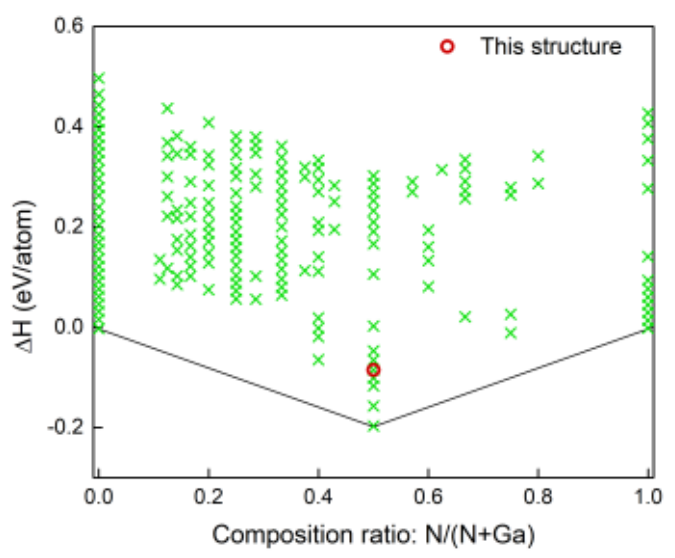

**Atomic structure**

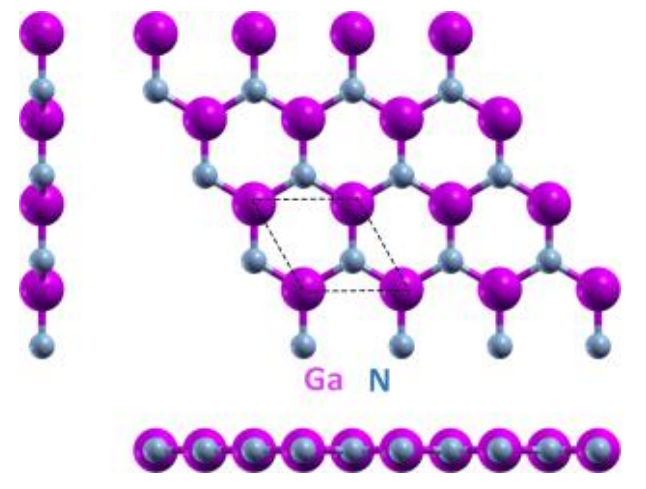

**Atomic coordinates**

```
Ga1N1
   1.00000000000000
   3.2095416635794085    0.0000000000000000    0.0000000000000000
  -1.6047941441788967    2.7795964982737638    0.0000000000000000
   0.0000000000000000    0.0000000000000000   20.0053009999999993
   Ga   N
   1    1
Direct
  0.3333558881396570  0.6667181924672363  0.4999606694951524
  0.0000061118603440  0.0000058075327587  0.5000393305048476
```

**Phonon dispersion curve**

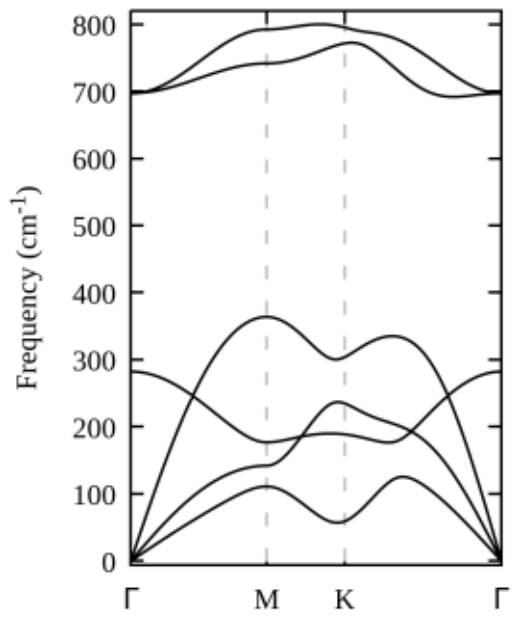

**Projected band structure and density of states**

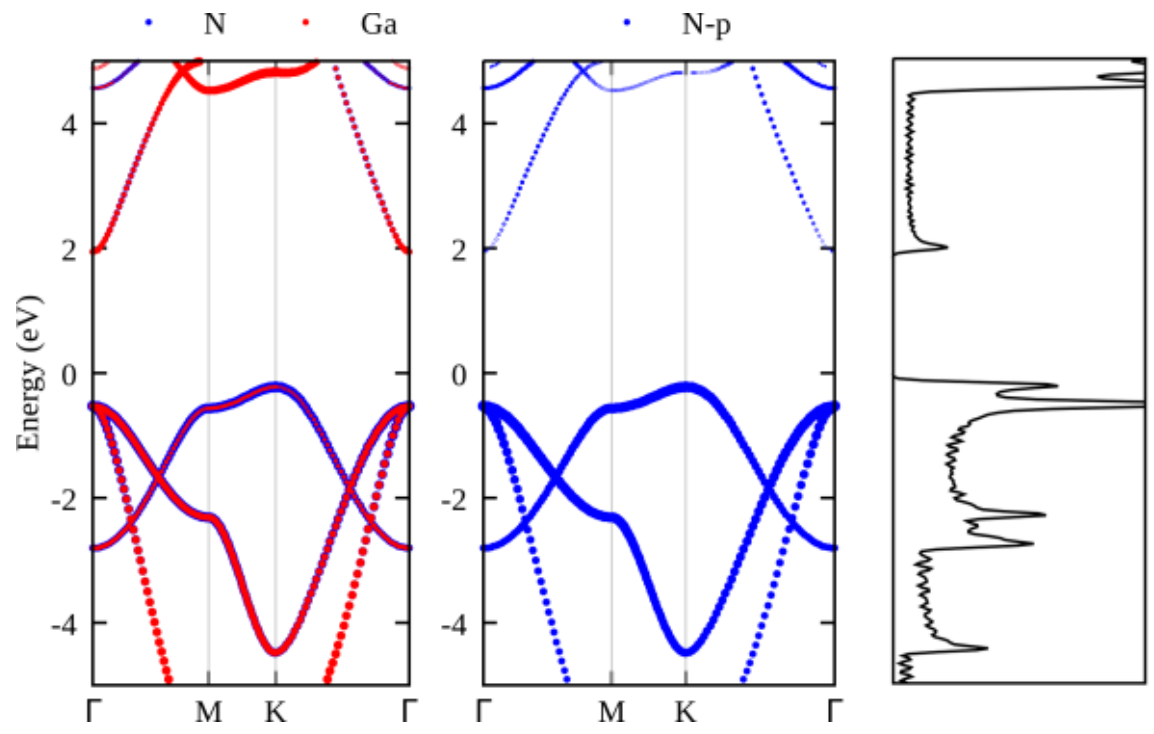

**Magnetic moment and spin polarization energy as a function of hole doping concentration**

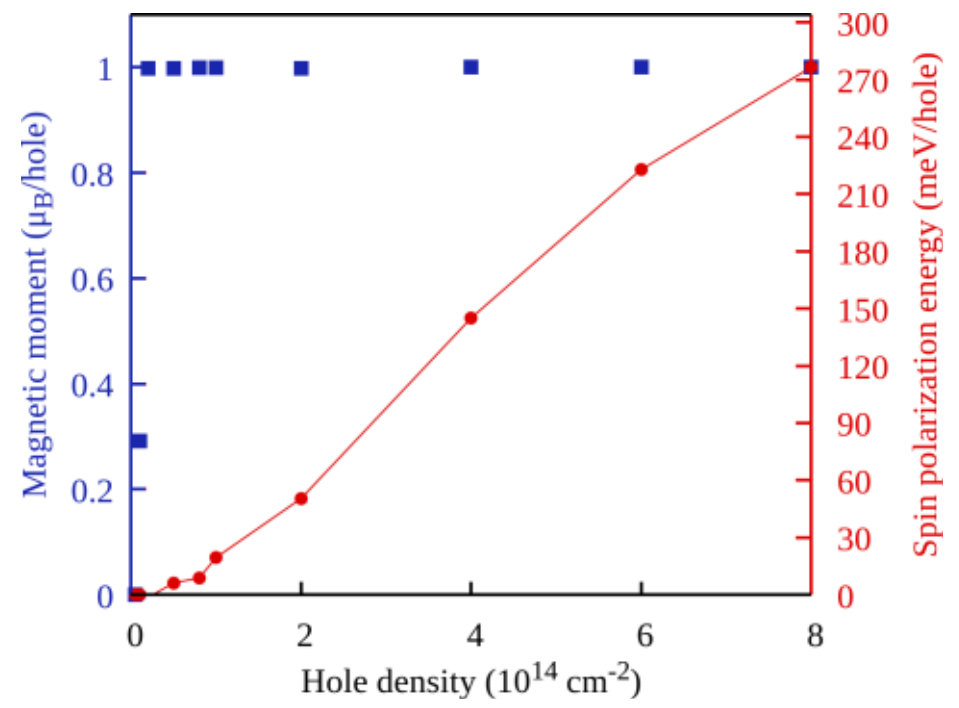

**Magnetic configurations and spin Hamiltonian**

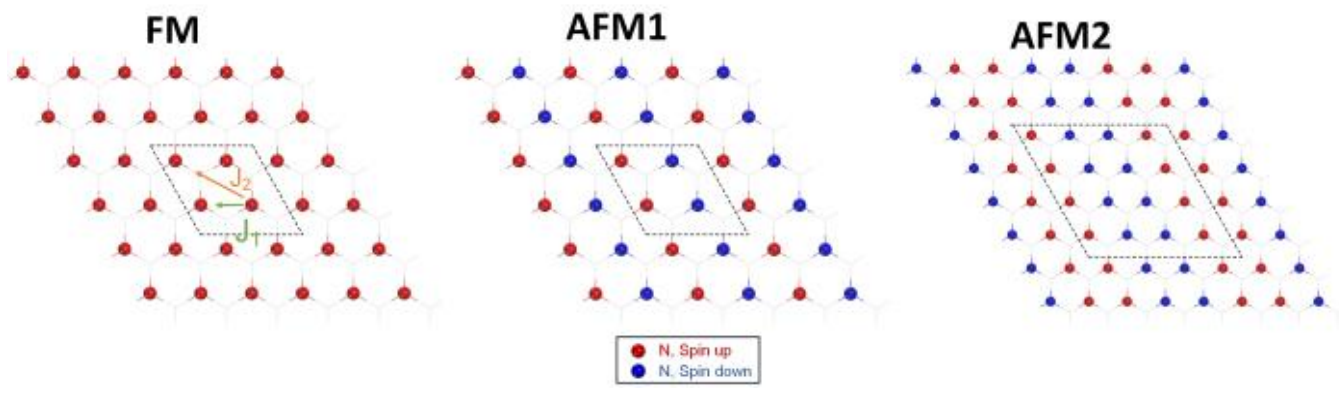

$$E_{FM} = E_0 - 12J_1S^2 - 12J_2S^2$$
$$E_{AFM1} = E_0 + 4J_1S^2 + 4J_2S^2$$
$$E_{AFM2} = E_0 - 4J_1S^2 + 4J_2S^2$$

**Magnetic exchange coupling parameters**

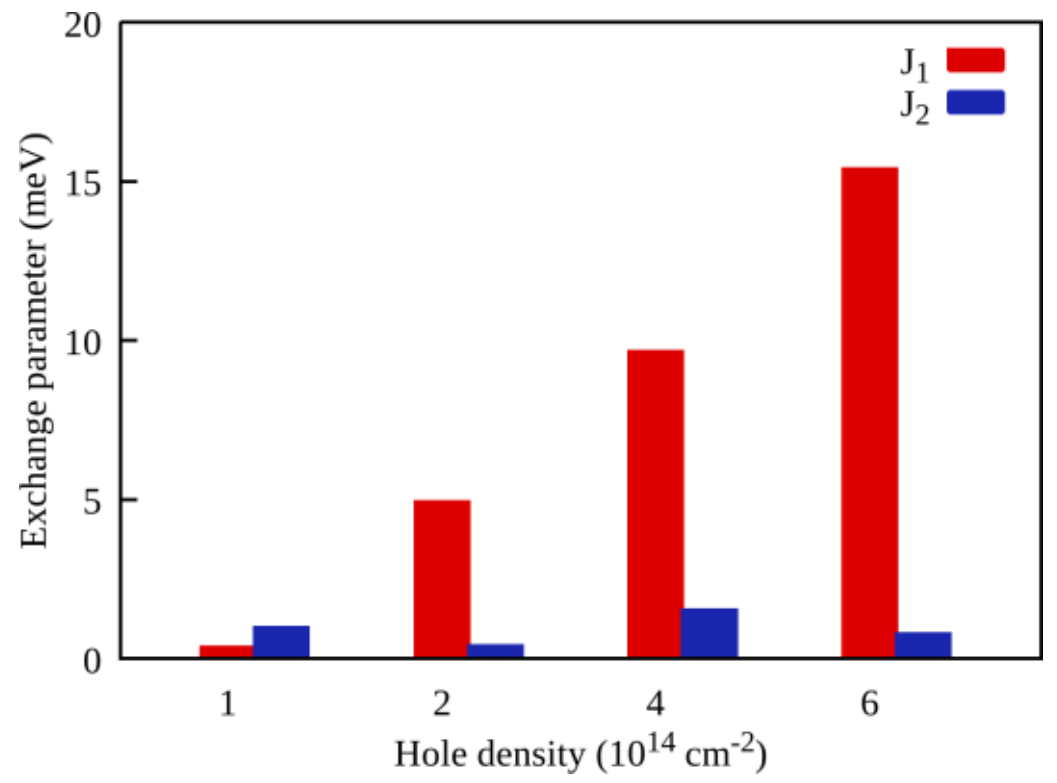

**Magnetic anisotropy energy (MAE, μeV) per magnetic atom**

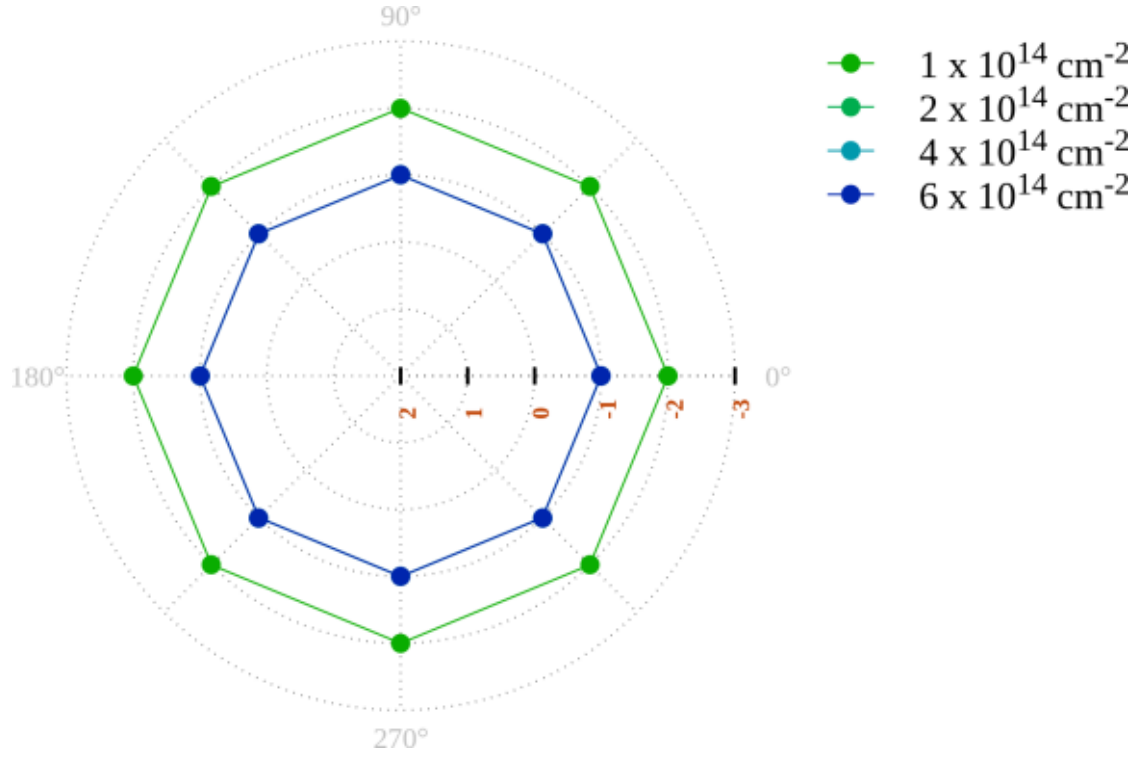

MAE = $E_{\parallel} - E_{\perp}$, a positive (negative) value of MAE indicates the off-plane (in-plane) easy axis.

**Monte Carlo simulations of the normalized magnetization of as a function of temperature**

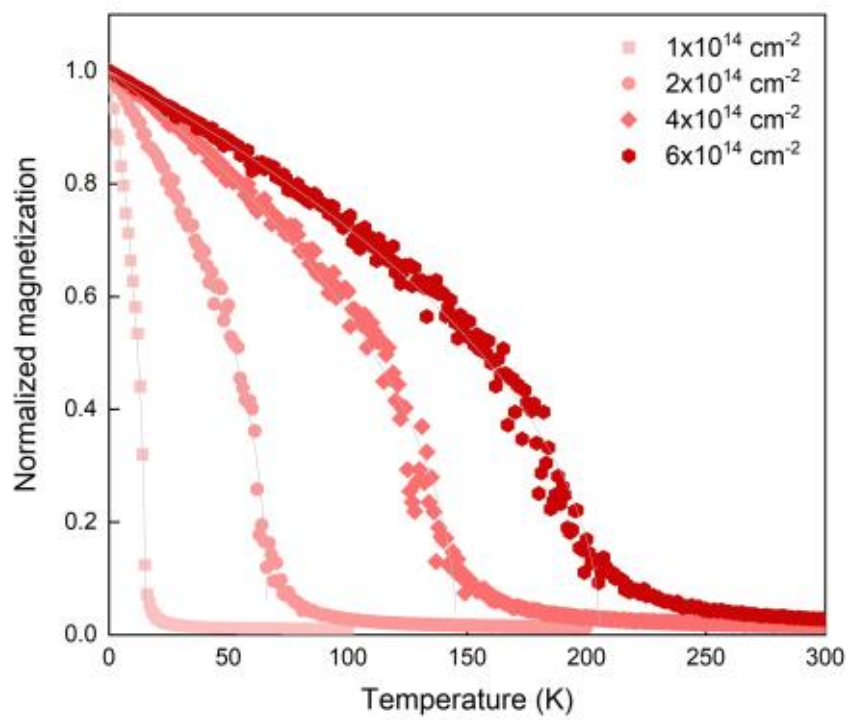

$T_c$: $1\times10^{14}$ cm$^{-2}$: 15 K  $2\times10^{14}$ cm$^{-2}$: 66 K

$4\times10^{14}$ cm$^{-2}$: 145 K  $6\times10^{14}$ cm$^{-2}$: 205 K

# 65.InN

| MC2D-ID | C2DB | 2dmat-ID | USPEX | Space group | Band gap (eV) |
|---|---|---|---|---|---|
| - | - | 2dm-6424 | - | P6m2 | 0.57 |

**Convex hull** | **Atomic structure** | **Atomic coordinates** | **Phonon dispersion curve**

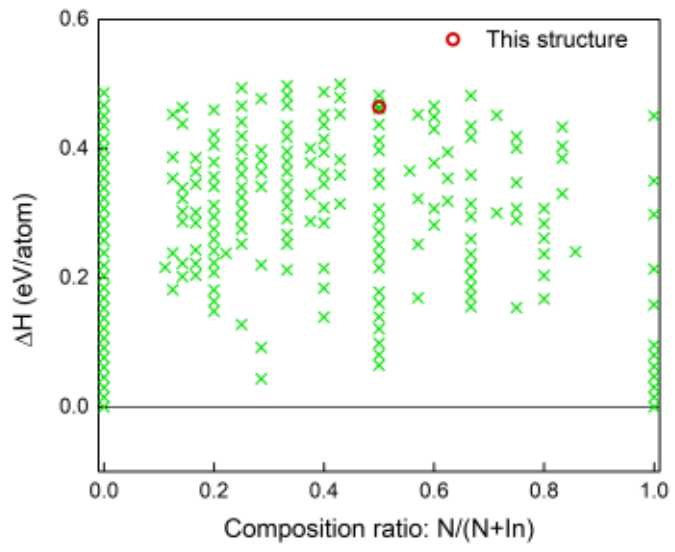

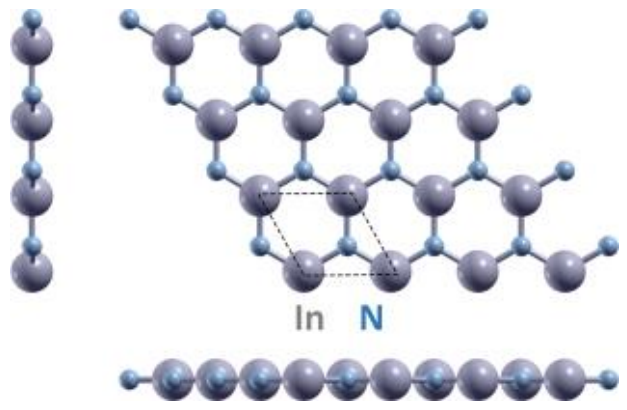

```
In1N1
   1.00000000000000
   3.5856211591877236    0.0000000000000000    0.0000000000000000
  -1.7928138495758739    3.1052374082318037    0.0000000000000000
   0.0000000000000000    0.0000000000000000   20.0000000000000000
   In    N
   1     1
Direct
  0.3333329143395574  0.6666670480444168  0.3853517022048223
  0.0000000856604458  0.9999999519555800  0.3853742977951811
```

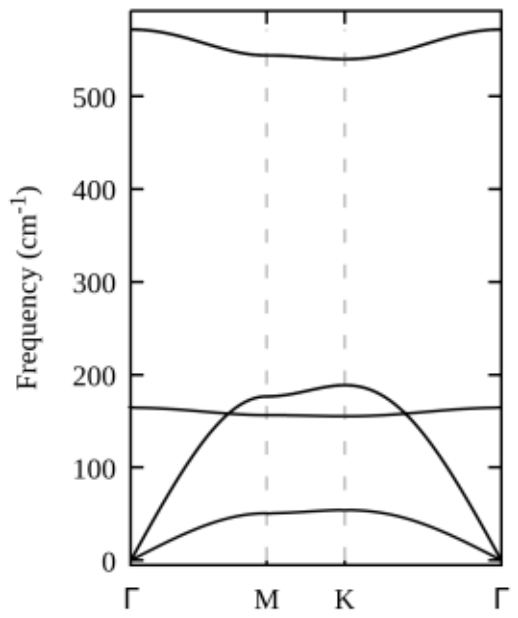

**Projected band structure and density of states**

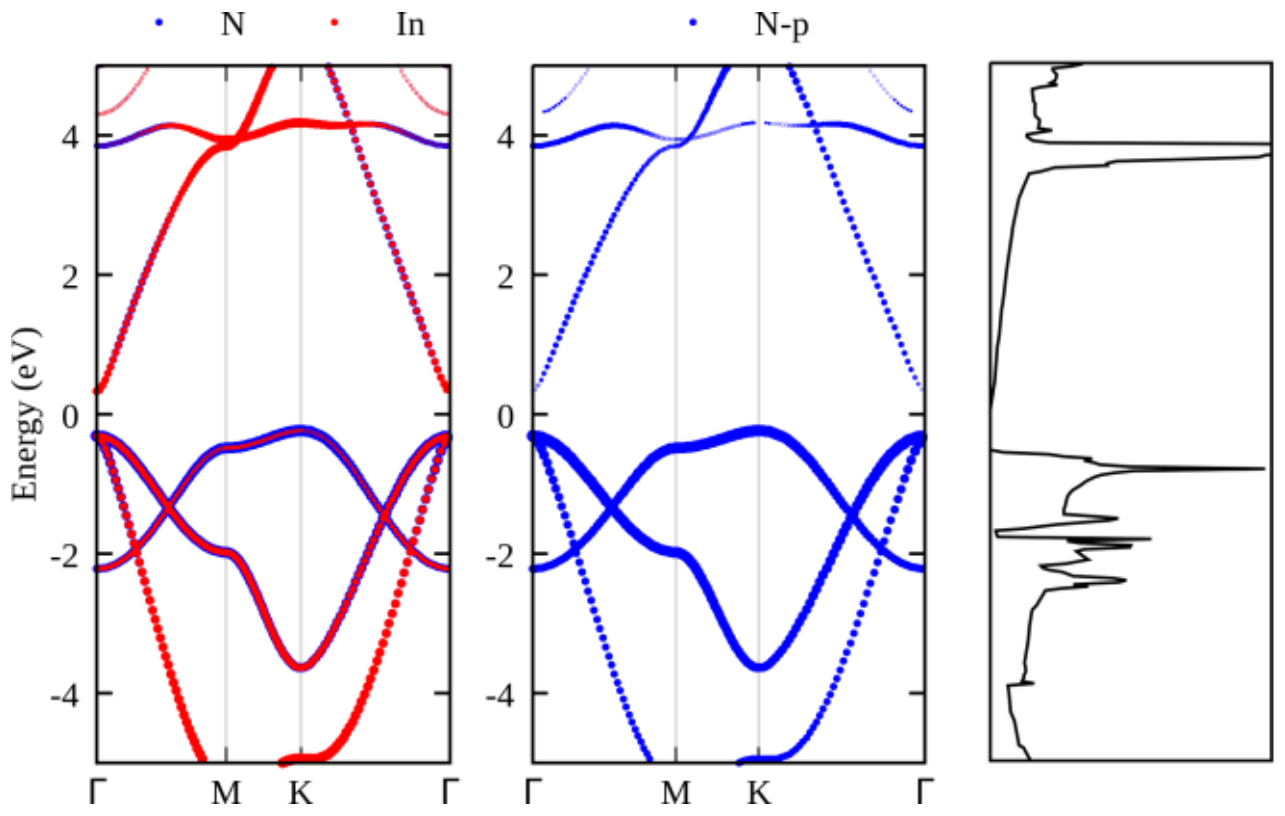

**Magnetic moment and spin polarization energy as a function of hole doping concentration**

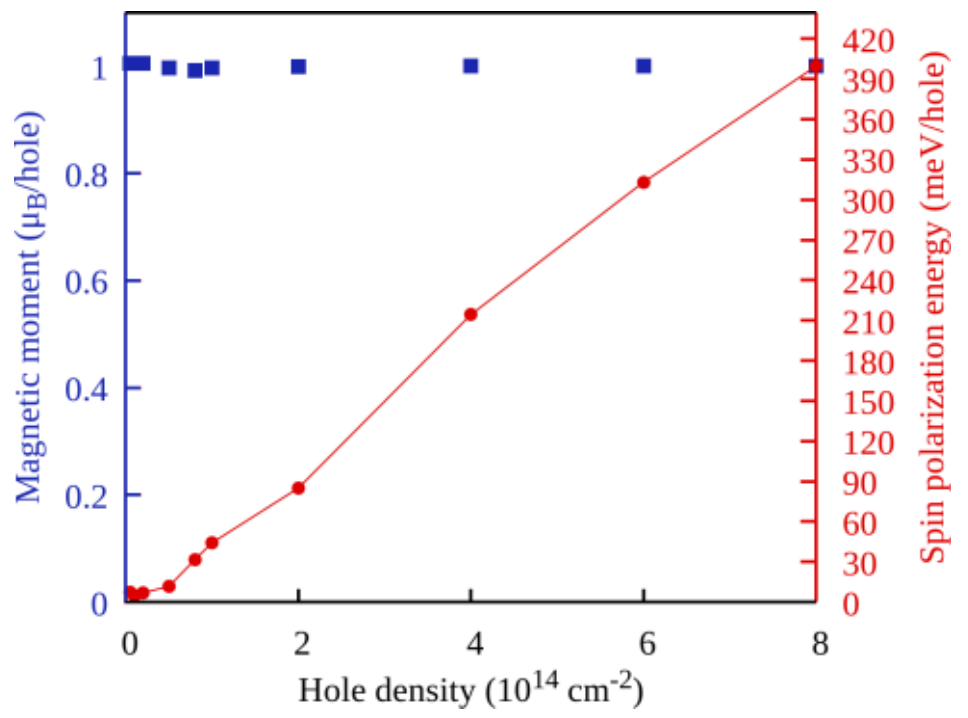

**Magnetic configurations and spin Hamiltonian**

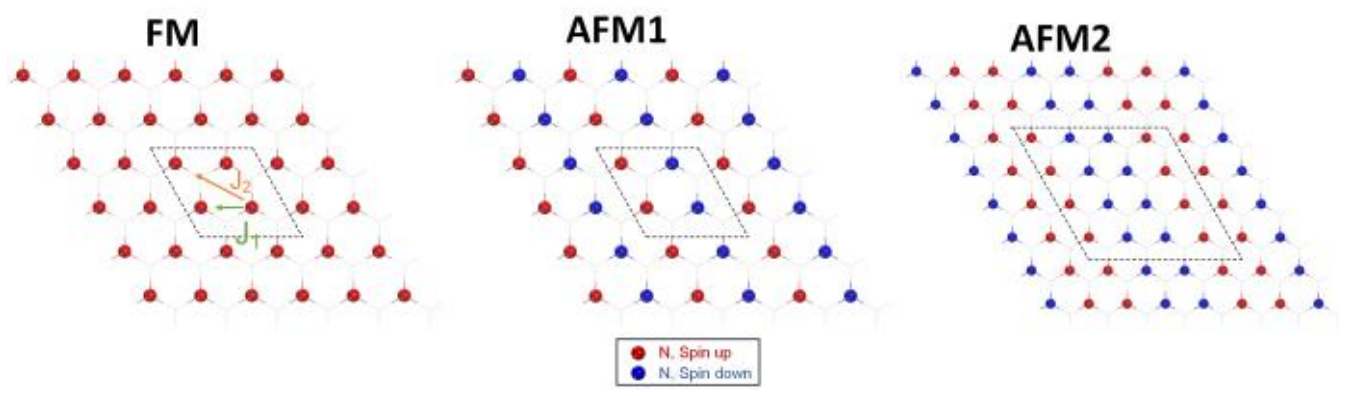

$$E_{FM} = E_0 - 12J_1S^2 - 12J_2S^2$$
$$E_{AFM1} = E_0 + 4J_1S^2 + 4J_2S^2$$
$$E_{AFM2} = E_0 - 4J_1S^2 + 4J_2S^2$$

**Magnetic exchange coupling parameters**

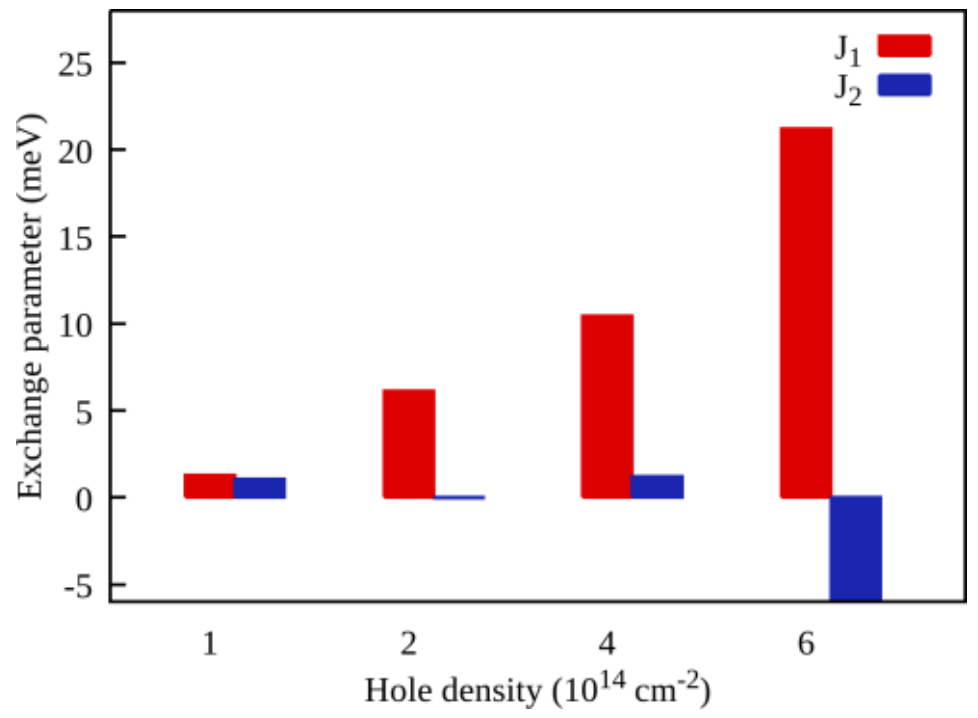

**Magnetic anisotropy energy (MAE, μeV) per magnetic atom**

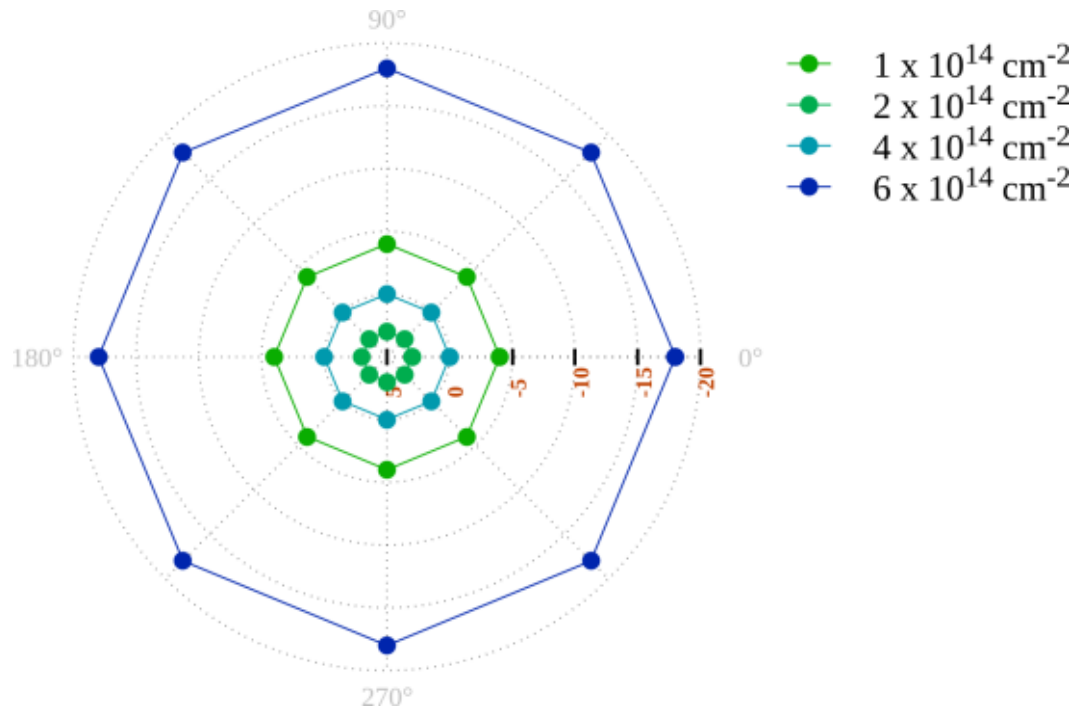

MAE = $E_{\parallel} - E_{\perp}$, a positive (negative) value of MAE indicates the off-plane (in-plane) easy axis.

**Monte Carlo simulations of the normalized magnetization of as a function of temperature**

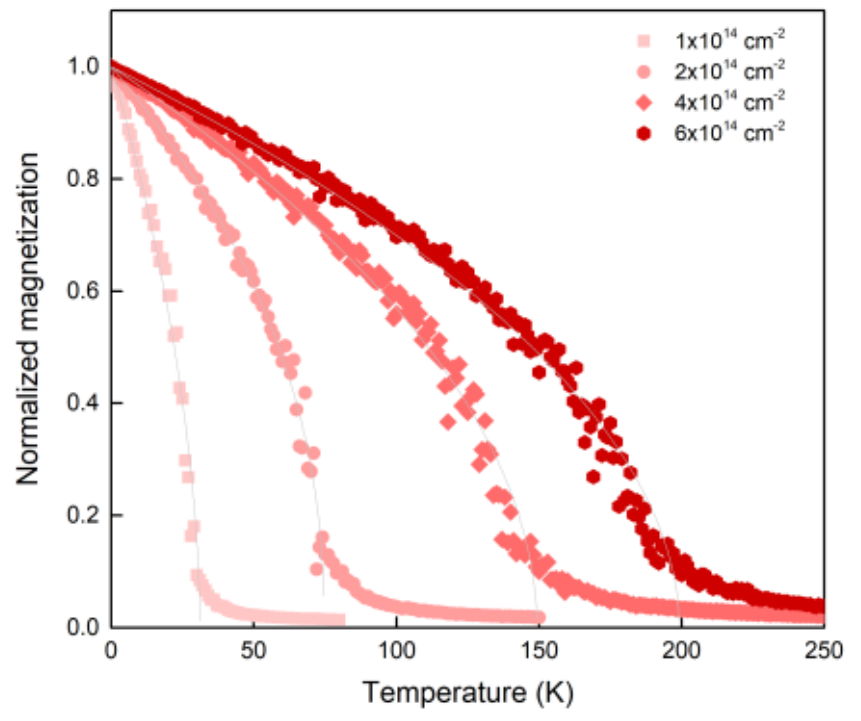

$T_c$: $1\times10^{14}$ cm$^{-2}$: 29 K  $2\times10^{14}$ cm$^{-2}$: 75 K
$4\times10^{14}$ cm$^{-2}$: 151 K  $6\times10^{14}$ cm$^{-2}$: 173 K

# 66. TlN

| MC2D-ID | C2DB | 2dmat-ID | USPEX | Space group | Band gap (eV) |
|---|---|---|---|---|---|
| - | | 2dm-6412 | - | P6m2 | 0.002 |

## Convex hull

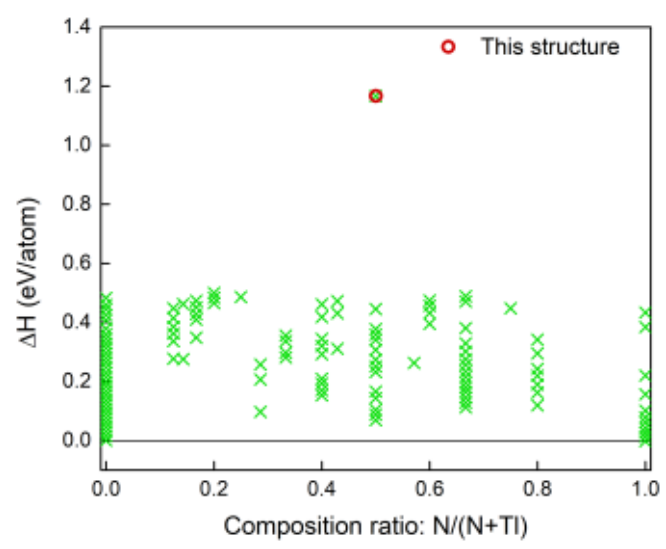

## Atomic structure

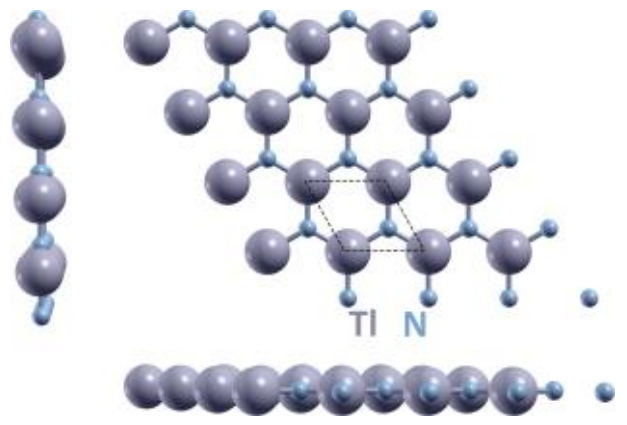

## Atomic coordinates

```
Tl1N1
   1.00000000000000
   3.7270386939767035    0.0000000000000000    0.0000000000000000
  -1.8635196930302853    3.2277107372986182    0.0000000000000000
   0.0000000000000000    0.0000000000000000   20.0000000000000000
   Tl    N
   1     1
Direct
  0.3333330275056099  0.6666669015254953  0.3853632018469071
  0.9999999724943933  0.0000000984745014  0.3853627981530892
```

## Phonon dispersion curve

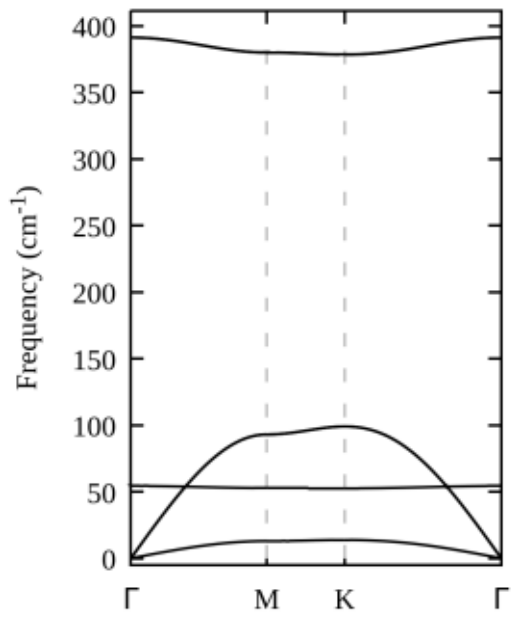

## Projected band structure and density of states

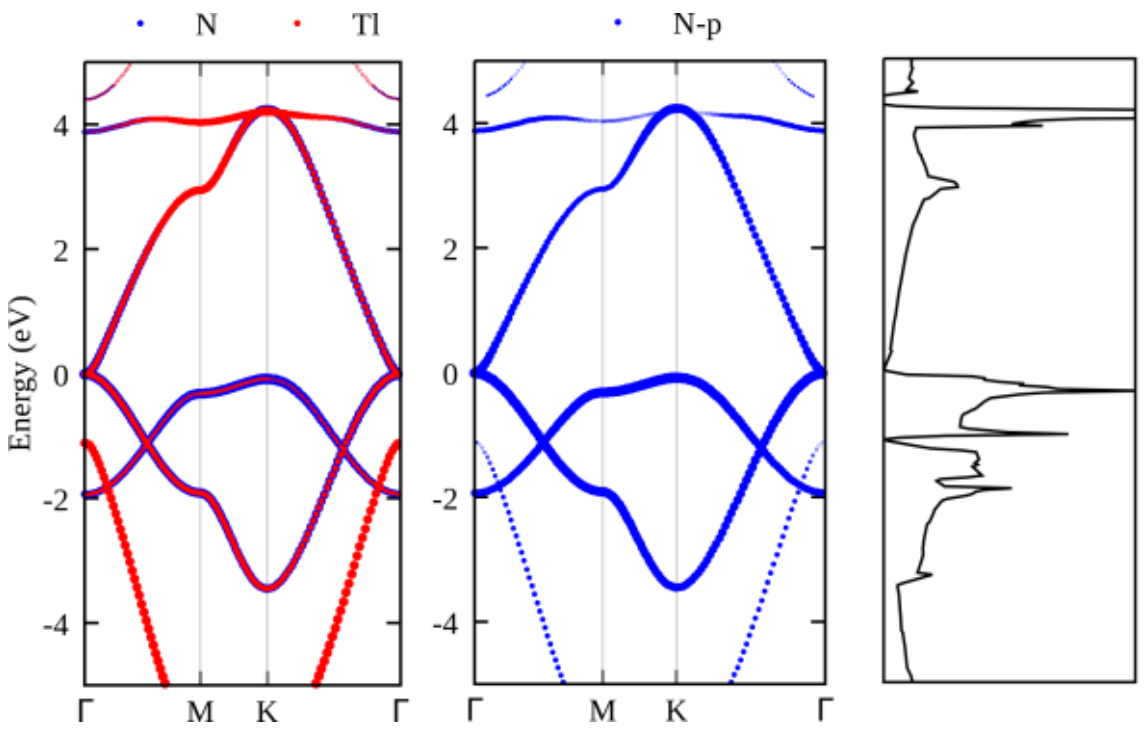

## Magnetic moment and spin polarization energy as a function of hole doping concentration

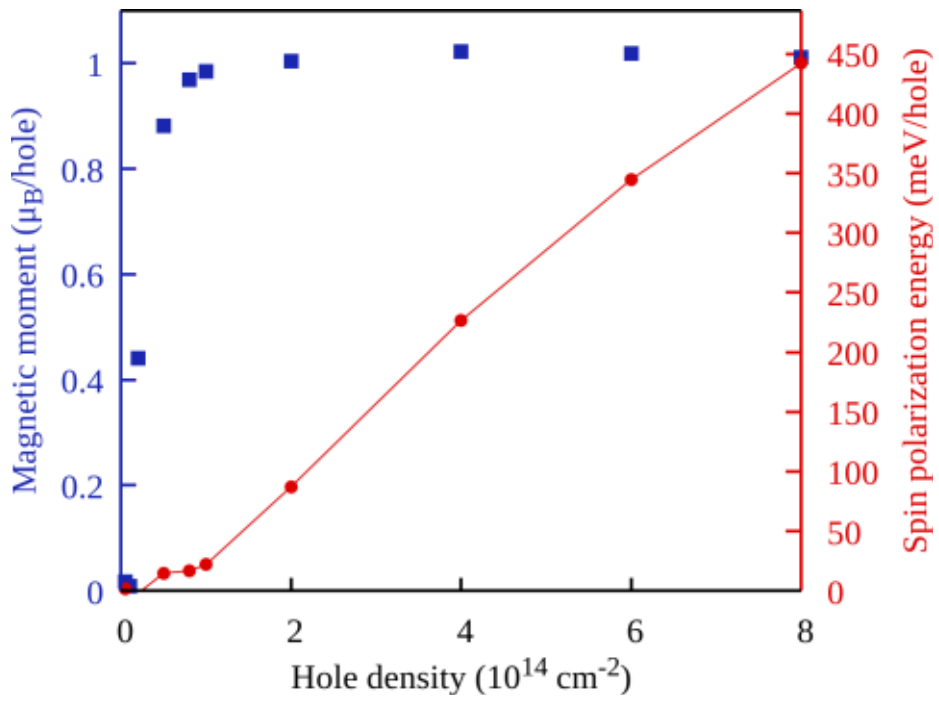

## Magnetic configurations and spin Hamiltonian

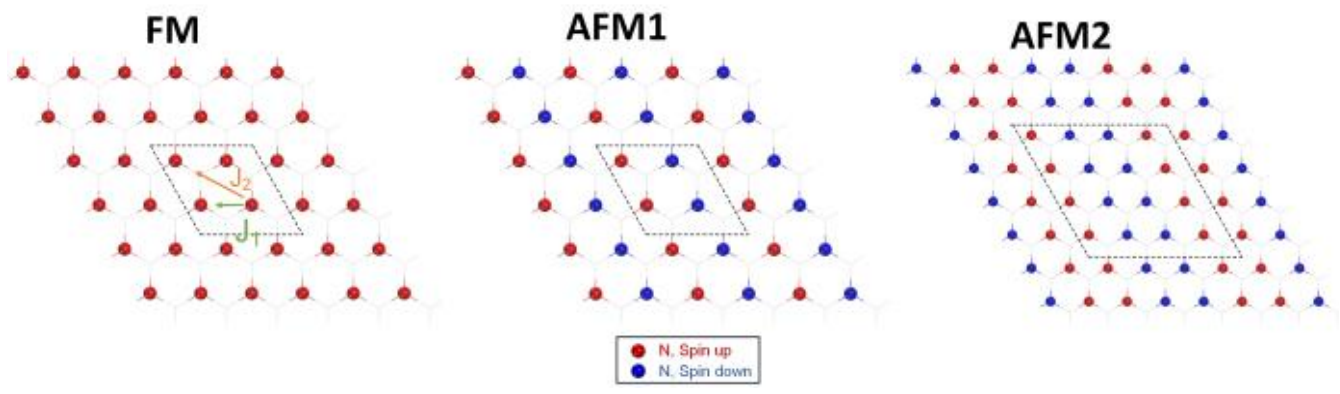

$$E_{FM} = E_0 - 12J_1S^2 - 12J_2S^2$$
$$E_{AFM1} = E_0 + 4J_1S^2 + 4J_2S^2$$
$$E_{AFM2} = E_0 - 4J_1S^2 + 4J_2S^2$$

## Magnetic exchange coupling parameters

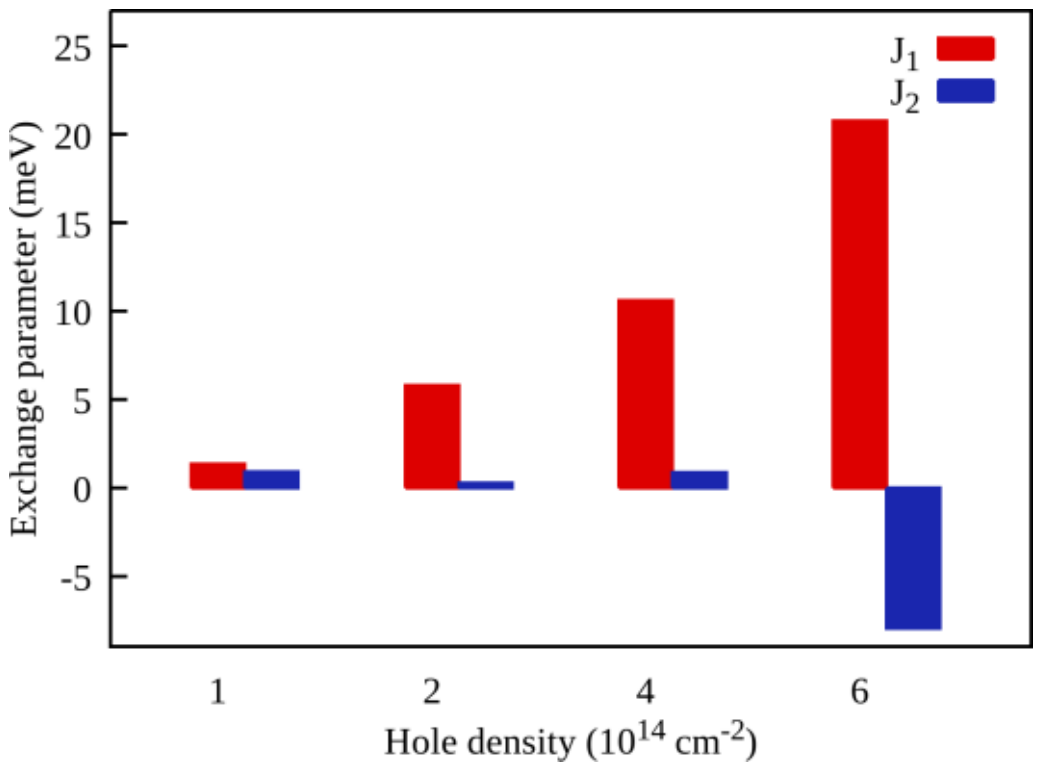

## Magnetic anisotropy energy (MAE, μeV) per magnetic atom

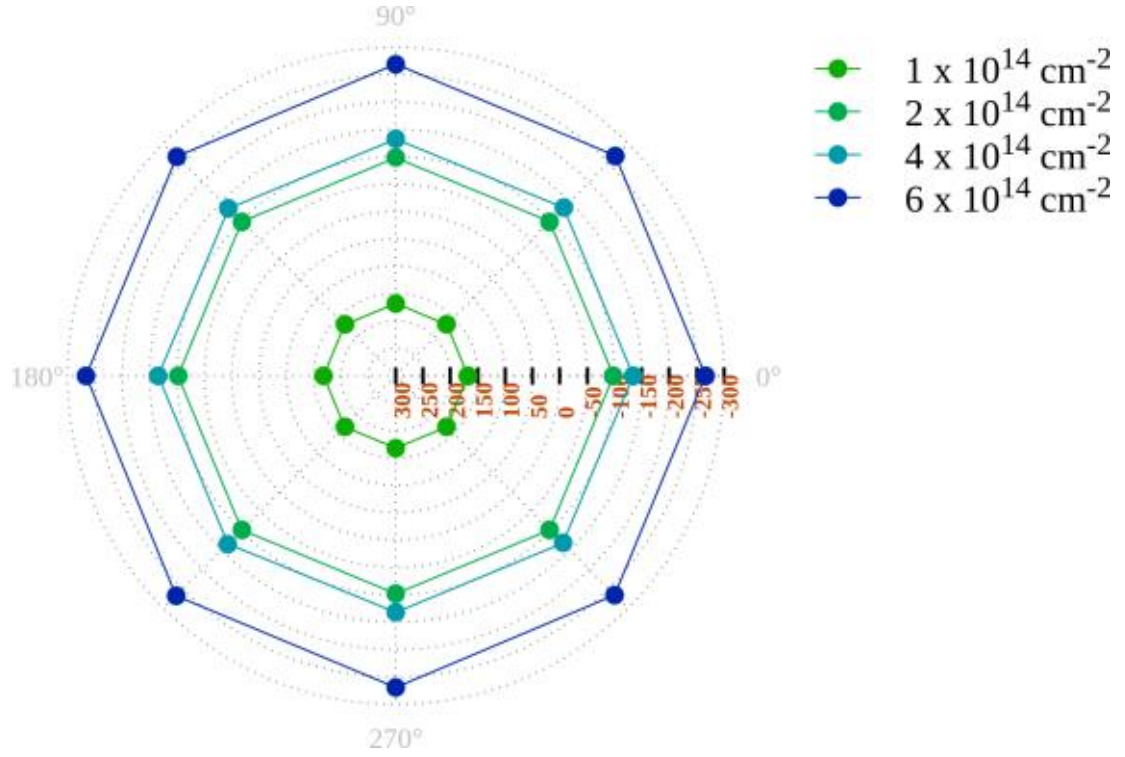

MAE = $E_{\parallel} - E_{\perp}$, a positive (negative) value of MAE indicates the off-plane (in-plane) easy axis.

## Monte Carlo simulations of the normalized magnetization of as a function of temperature

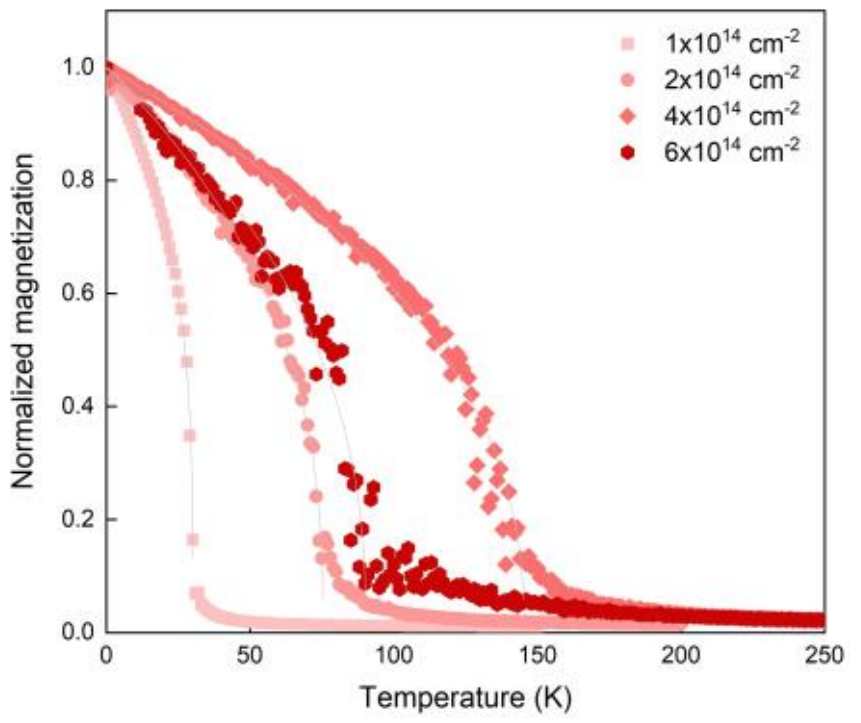

$T_c$: $1\times10^{14}$ cm$^{-2}$: 30 K  $2\times10^{14}$ cm$^{-2}$: 75 K

$4\times10^{14}$ cm$^{-2}$: 145 K  $6\times10^{14}$ cm$^{-2}$: 90 K

# 67. $BeF_2$

| MC2D-ID | C2DB | 2dmat-ID | USPEX | Space group | Band gap (eV) |
|---|---|---|---|---|---|
| - | - | 2dm-951 | - | P4m2 | 8.57 |

## Convex hull

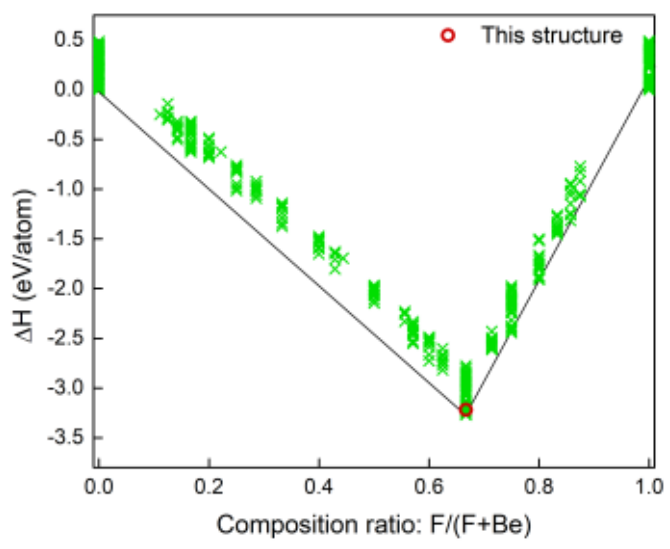

## Atomic structure

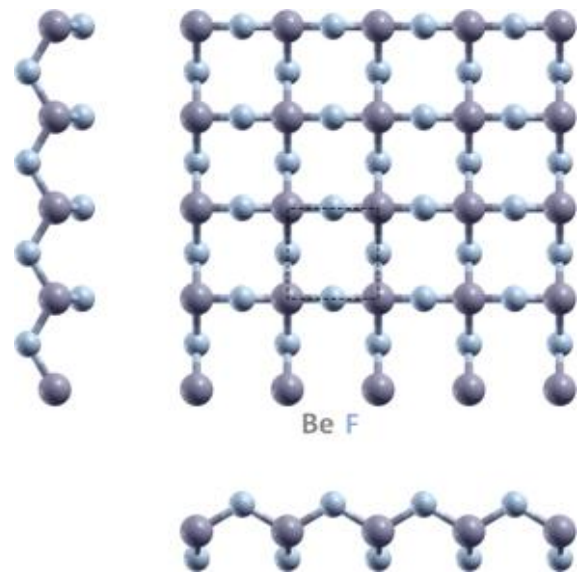

## Atomic coordinates

```
Be1 F2
   1.00000000000000
   2.7287100377127125   0.0000000000000000   0.0000000000000000
   0.0000000000000000   2.7287099143497966   0.0000000000000000
   0.0000000000000000   0.0000000000000000   22.6561909999999997
   Be   F
   1    2
Direct
  0.9999999995816324   0.0000000004229719   0.0000000105880460
  0.0000000005111787   0.4999999998487610   0.9641986322388192
  0.4999999999071889   0.9999999997282671   0.0358013571731419
```

## Phonon dispersion curve

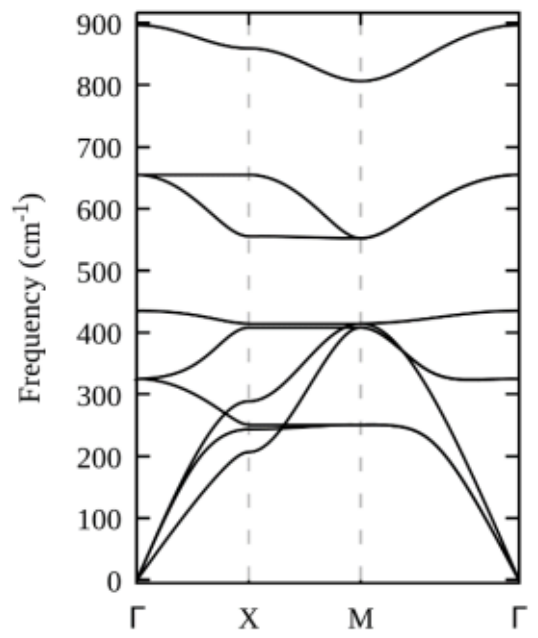

## Projected band structure and density of states

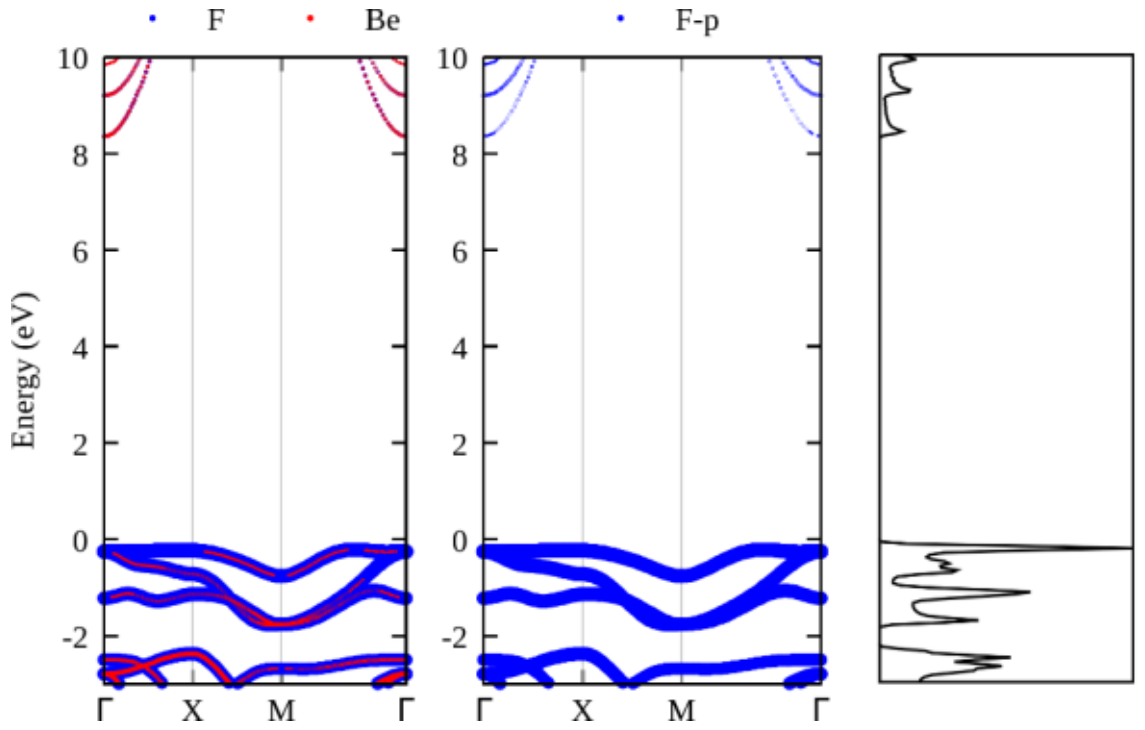

## Magnetic moment and spin polarization energy as a function of hole doping concentration

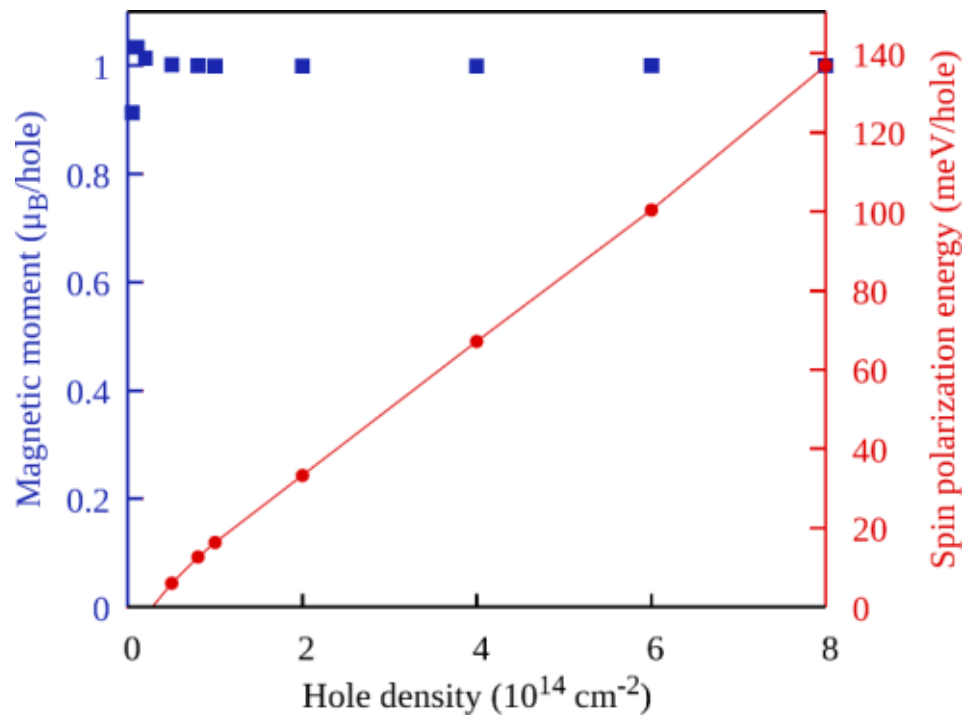

## Magnetic configurations and spin Hamiltonian

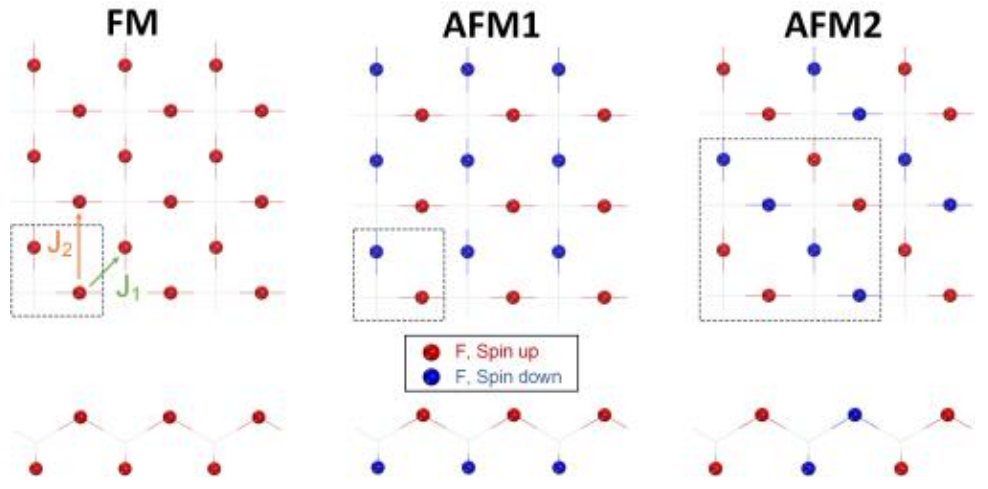

$$E_{FM} = E_0 - 4J_1S^2 - 4J_2S^2$$

$$E_{AFM1} = E_0 + 4J_1S^2 - 4J_2S^2$$

$$E_{AFM2} = E_0 + 4J_2S^2$$

## Magnetic exchange coupling parameters

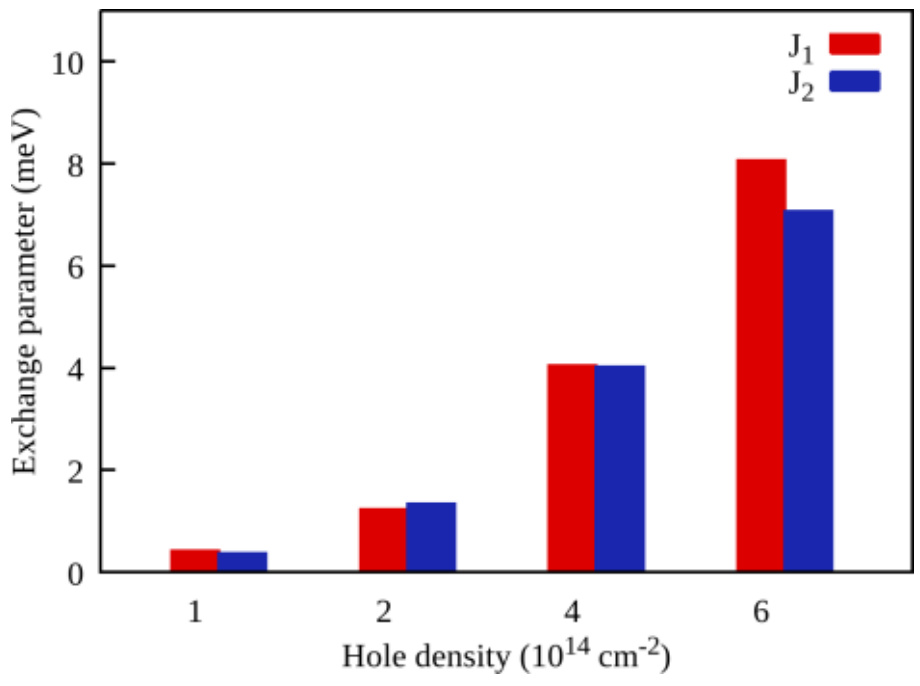

## Magnetic anisotropy energy (MAE, μeV) per magnetic atom

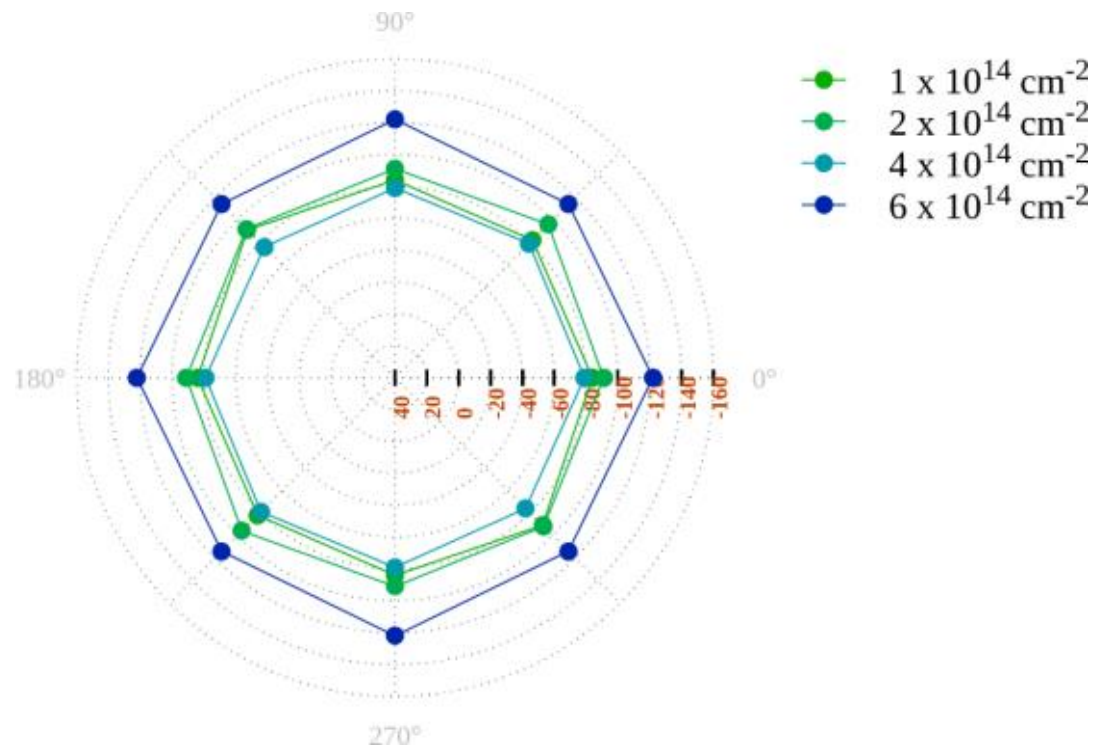

MAE = $E_{\parallel} - E_{\perp}$, a positive (negative) value of MAE indicates the off-plane (in-plane) easy axis.

## Monte Carlo simulations of the normalized magnetization of as a function of temperature

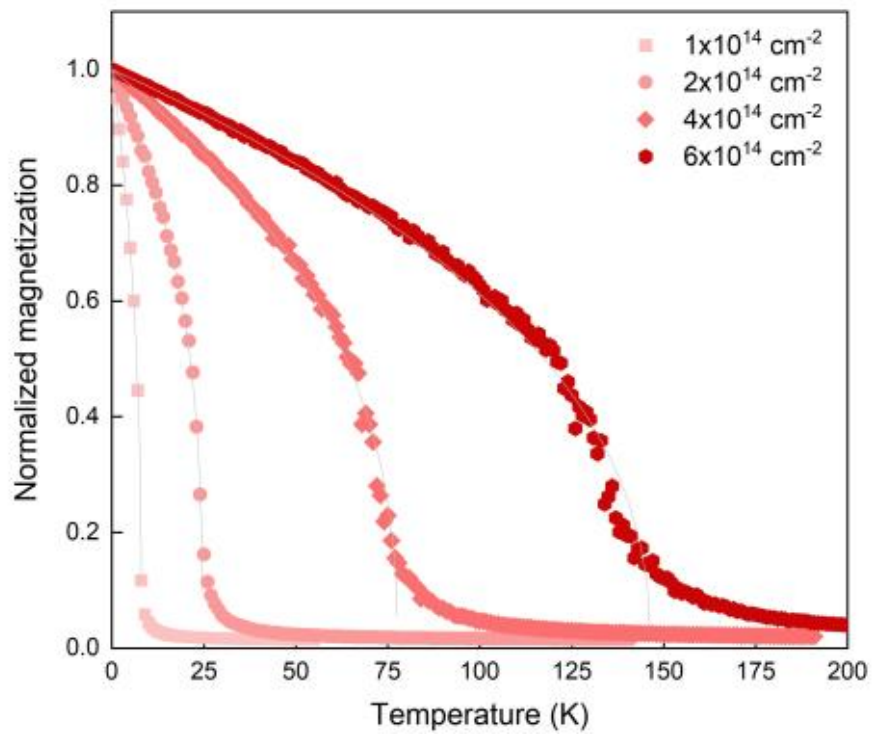

$T_c$: $1\times10^{14}$ cm$^{-2}$: 8 K  $2\times10^{14}$ cm$^{-2}$: 25 K

$4\times10^{14}$ cm$^{-2}$: 77 K  $6\times10^{14}$ cm$^{-2}$: 146 K

# 68. $BeCl_2$

| MC2D-ID | C2DB | 2dmat-ID | USPEX | Space group | Band gap (eV) |
|---|---|---|---|---|---|
| - | - | 2dm-1681 | - | P4m2 | 5.40 |

| Convex hull | Atomic structure | Atomic coordinates | Phonon dispersion curve |
|---|---|---|---|

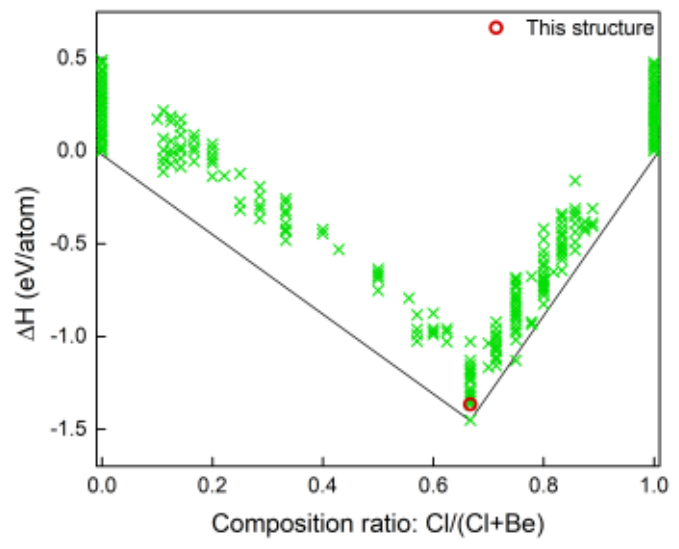

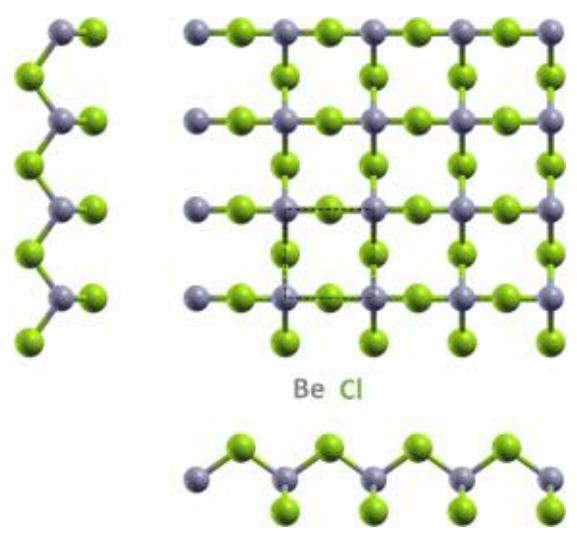

```
Be1Cl2
   1.00000000000000
   3.3430584071816685   0.0000000000000000   0.0000000000000000
   0.0000000000000000   3.3430589003723514   0.0000000000000000
   0.0000000000000000   0.0000000000000000   22.6561909999999997
   Be   Cl
   1    2
Direct
  0.0000000210738023   0.9999999953296026   0.9999999744550081
  0.9999999542964204   0.5000000206131787   0.9471076325499652
  0.5000000246297702   0.9999999840572187   0.0528923929950338
```

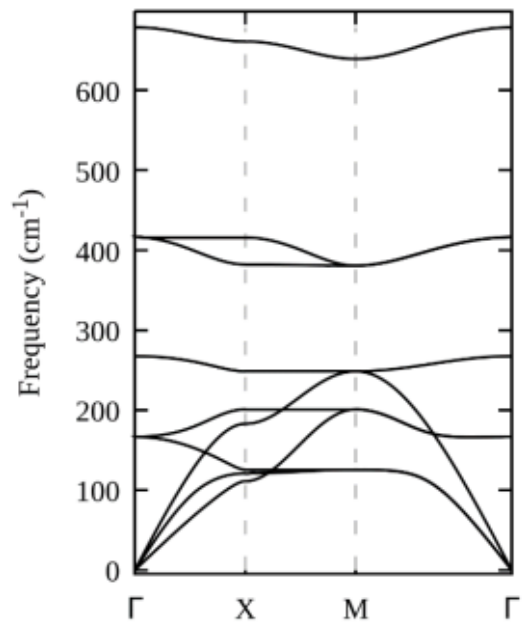

## Projected band structure and density of states

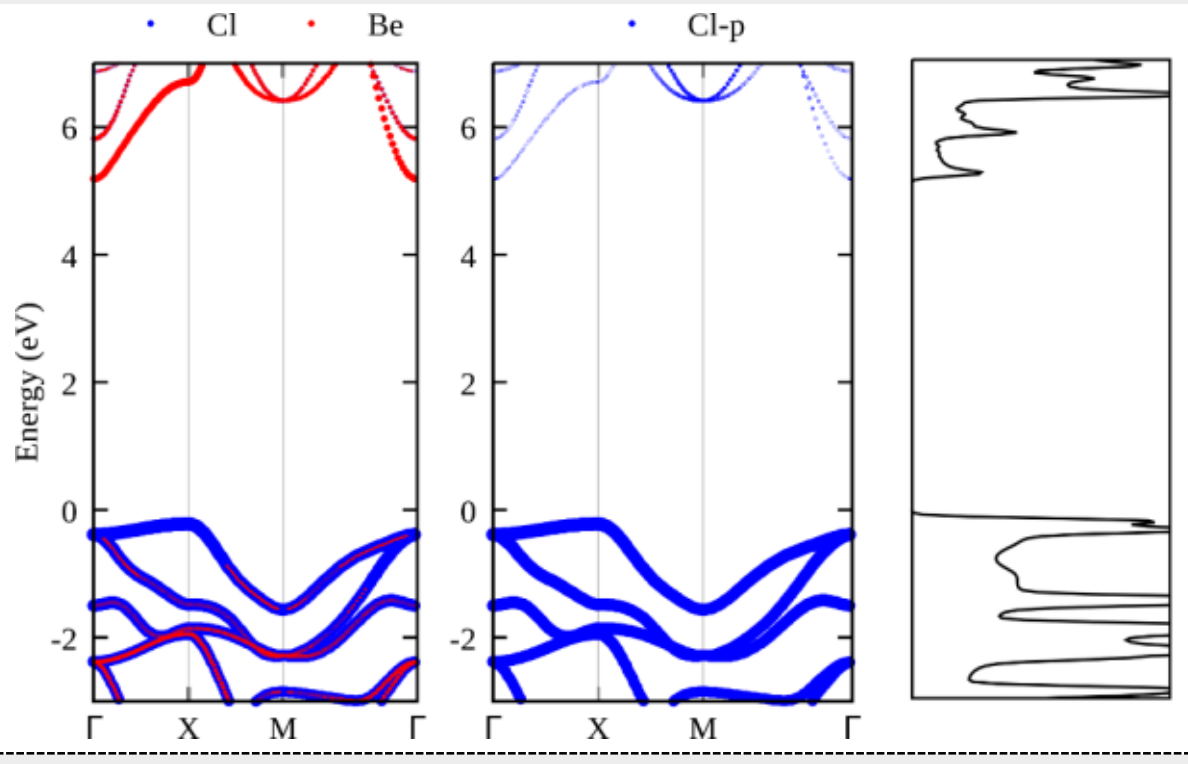

## Magnetic moment and spin polarization energy as a function of hole doping concentration

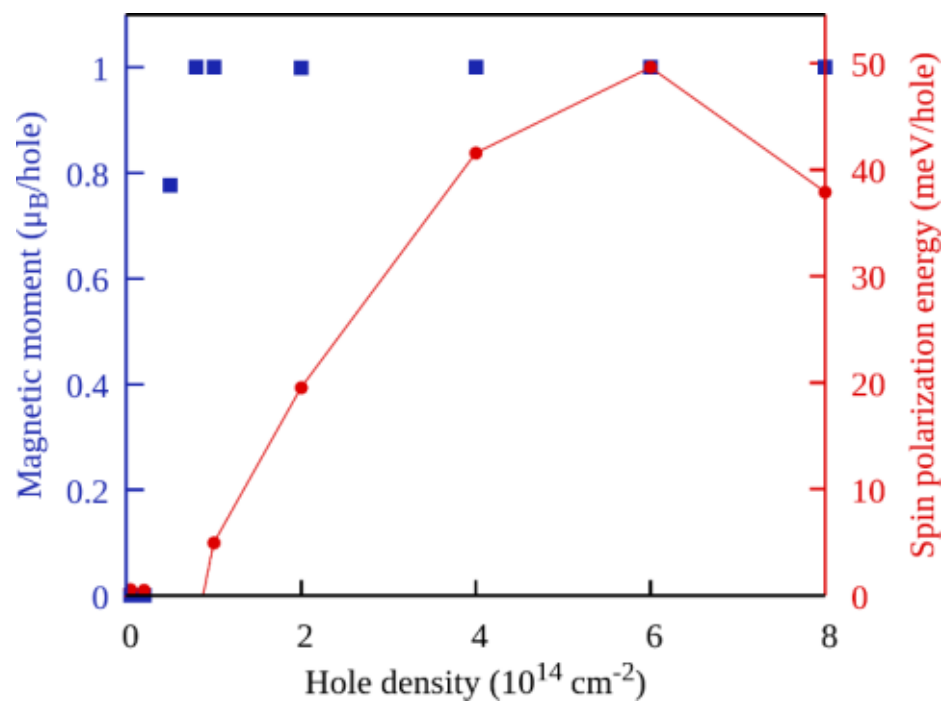

## Magnetic configurations and spin Hamiltonian

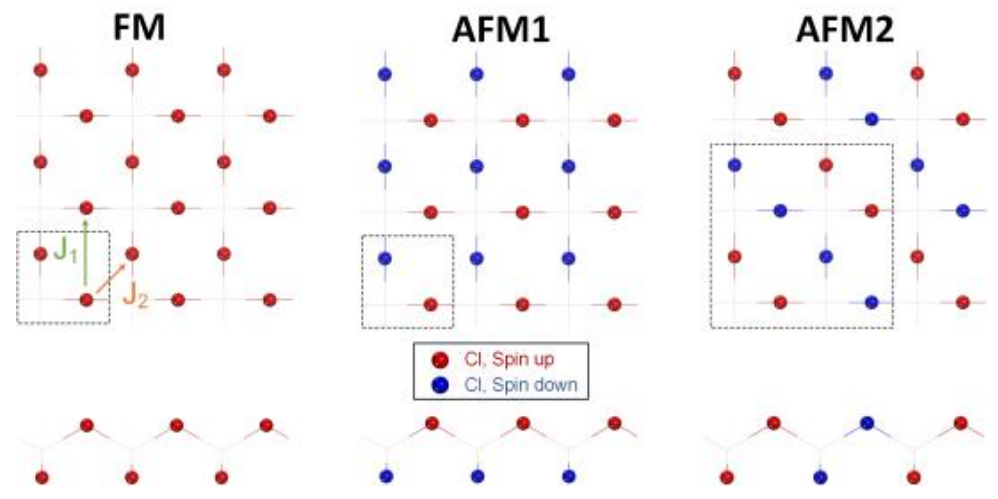

$$E_{FM} = E_0 - 4J_1S^2 - 4J_2S^2$$

$$E_{AFM1} = E_0 - 4J_1S^2 + 4J_2S^2$$

$$E_{AFM2} = E_0 + 4J_1S^2$$

## Magnetic exchange coupling parameters

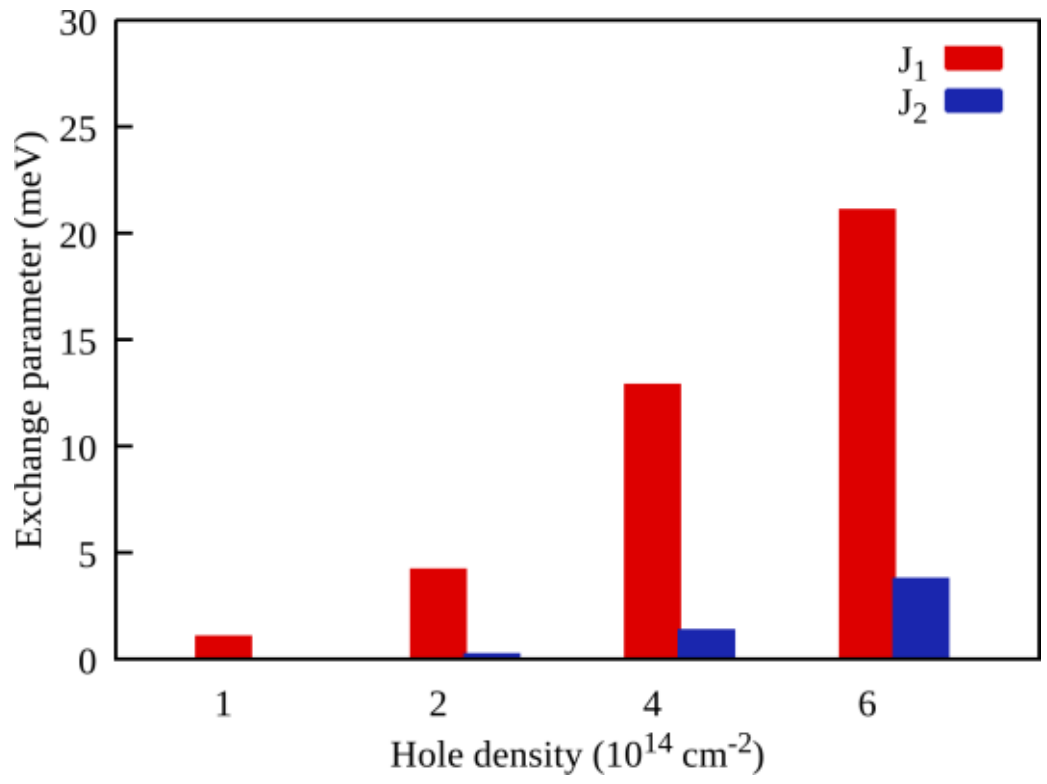

## Magnetic anisotropy energy (MAE, μeV) per magnetic atom

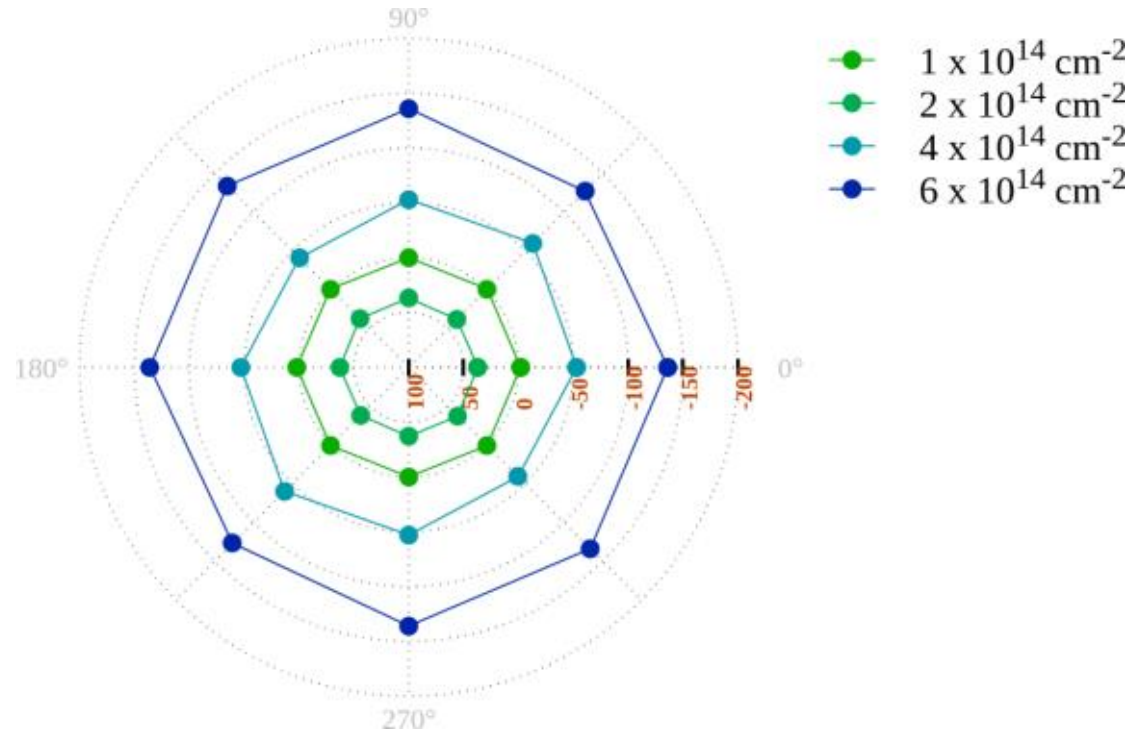

MAE = $E_{\parallel} - E_{\perp}$, a positive (negative) value of MAE indicates the off-plane (in-plane) easy axis.

## Monte Carlo simulations of the normalized magnetization of as a function of temperature

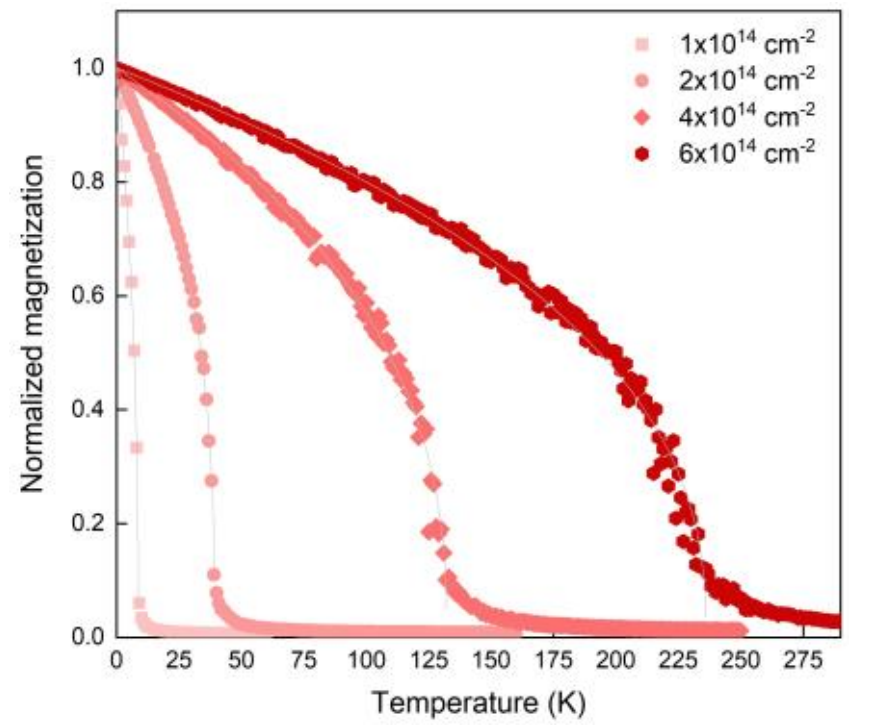

$T_c$: $1\times10^{14}$ cm$^{-2}$: 9 K   $2\times10^{14}$ cm$^{-2}$: 39 K   $4\times10^{14}$ cm$^{-2}$: 132 K   $6\times10^{14}$ cm$^{-2}$: 236 K

# 69. $MgF_2$

| MC2D-ID | C2DB | 2dmat-ID | USPEX | Space group | Band gap (eV) |
|---|---|---|---|---|---|
| - | - | 2dm-1799 | - | P4m2 | 6.63 |

**Convex hull**

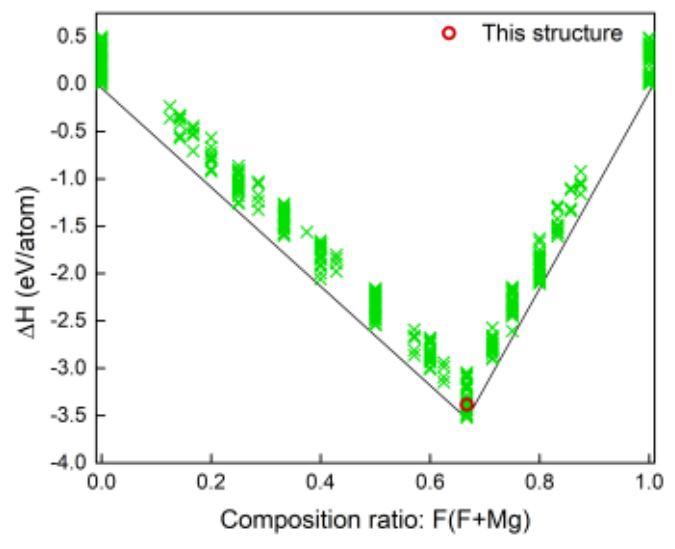

**Atomic structure**

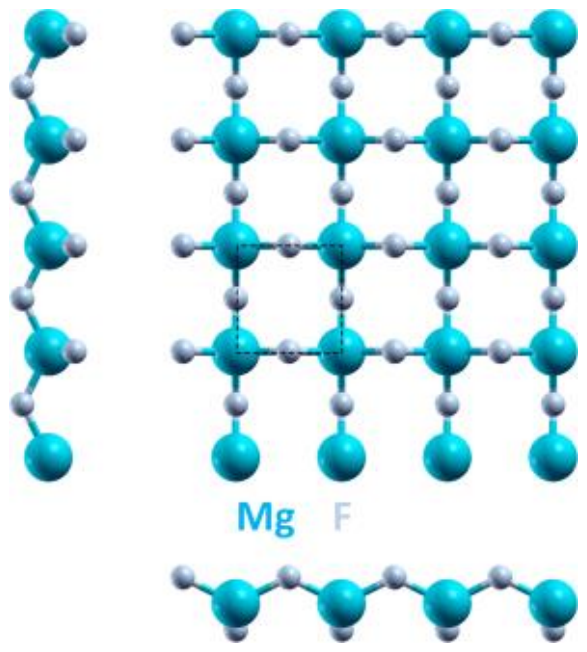

**Atomic coordinates**

```
Mg1F2
1.00000000000000
3.3964268531687636 0.0000000000000000 0.0000000000000000
0.0000000000000000 3.3966978452804462 0.0000000000000000
0.0000000000000000 0.0000000000000000 22.6561909999999997
Mg F
1 2
Direct
0.9999950786037601 0.0000018575789156 0.0000100712674111
0.9999979803661745 0.5000084214424518 0.9628688135455903
0.5000069410300654 0.9999907209786372 0.0371211151869986
```

**Phonon dispersion curve**

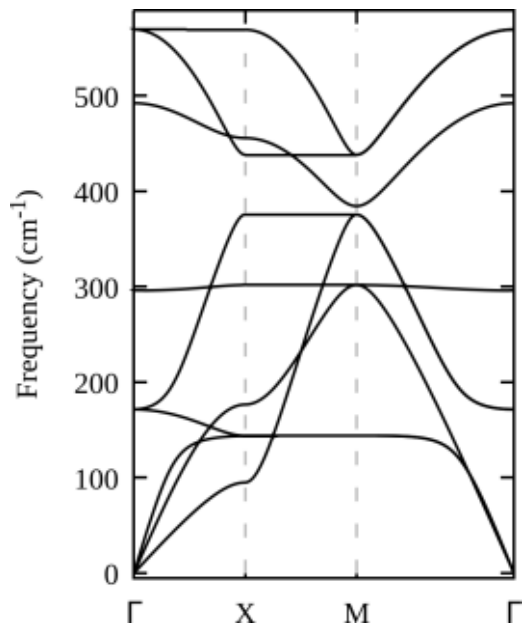

**Projected band structure and density of states**

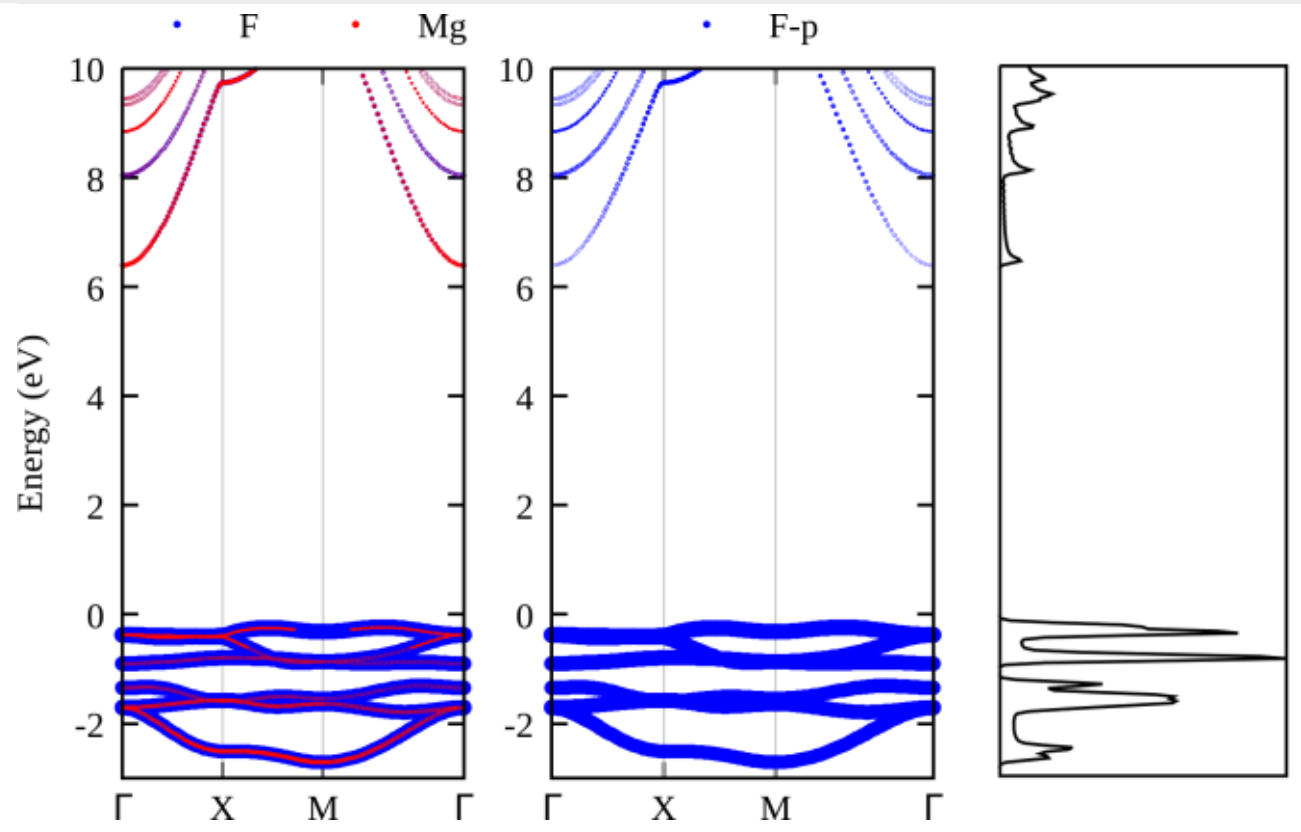

**Magnetic moment and spin polarization energy as a function of hole doping concentration**

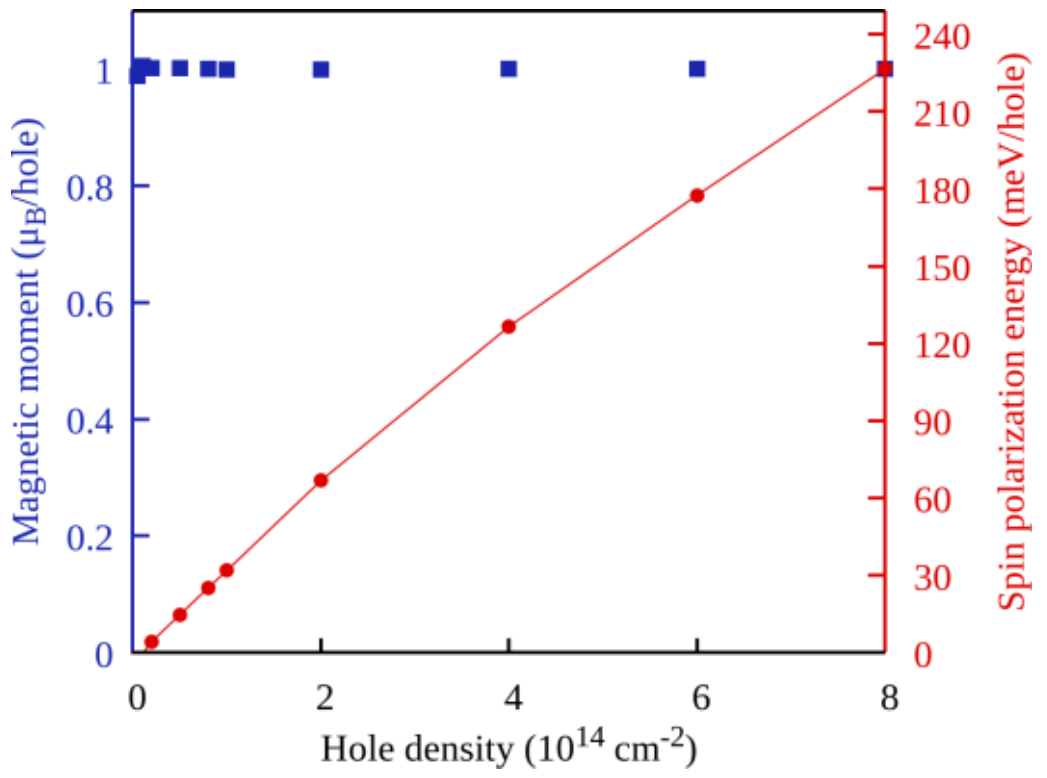

**Magnetic configurations and spin Hamiltonian**

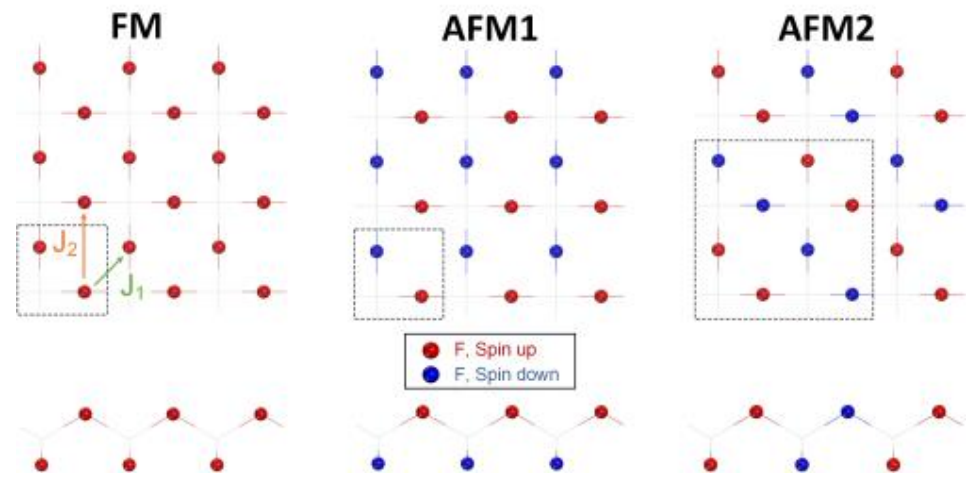

$$E_{FM} = E_0 - 4J_1S^2 - 4J_2S^2$$
$$E_{AFM1} = E_0 + 4J_1S^2 - 4J_2S^2$$
$$E_{AFM2} = E_0 + 4J_2S^2$$

**Magnetic exchange coupling parameters**

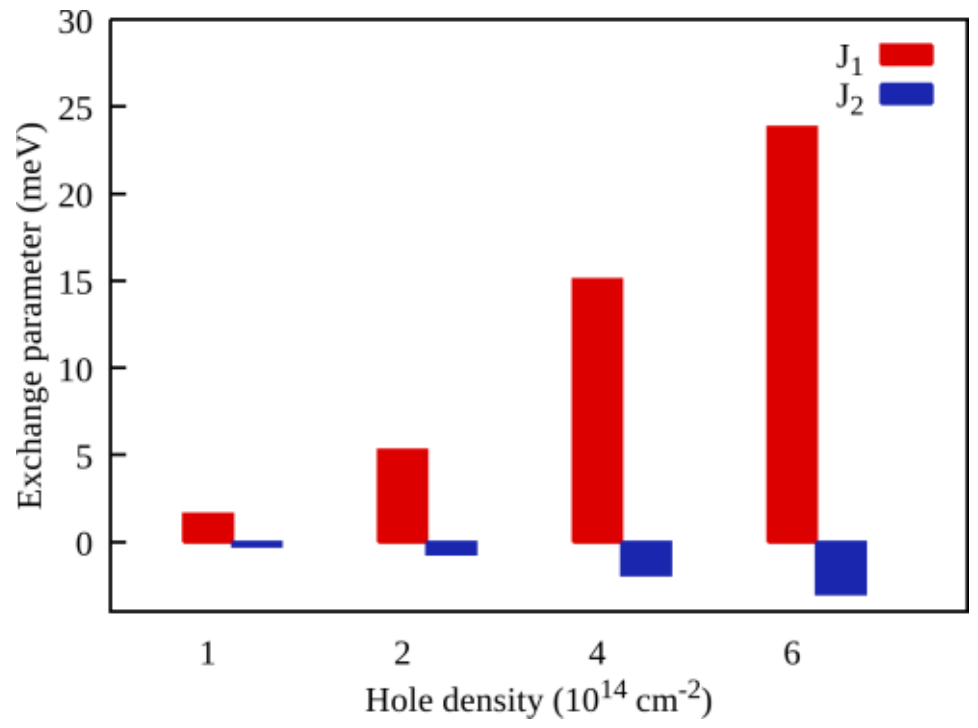

**Magnetic anisotropy energy (MAE, μeV) per magnetic atom**

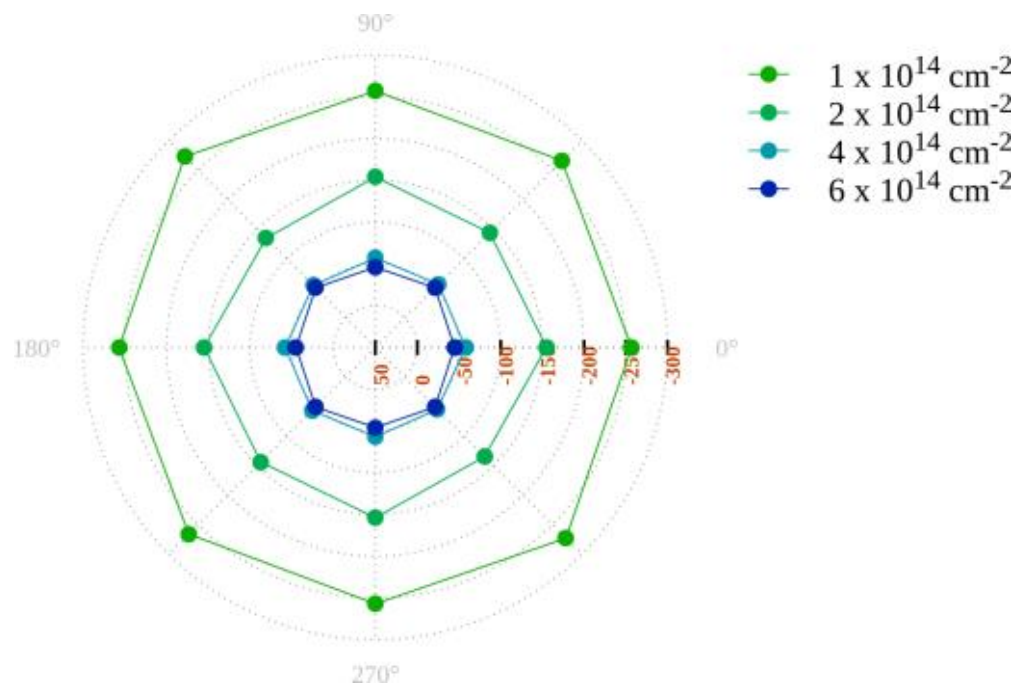

MAE = $E_{\parallel} - E_{\perp}$, a positive (negative) value of MAE indicates the off-plane (in-plane) easy axis.

**Monte Carlo simulations of the normalized magnetization of as a function of temperature**

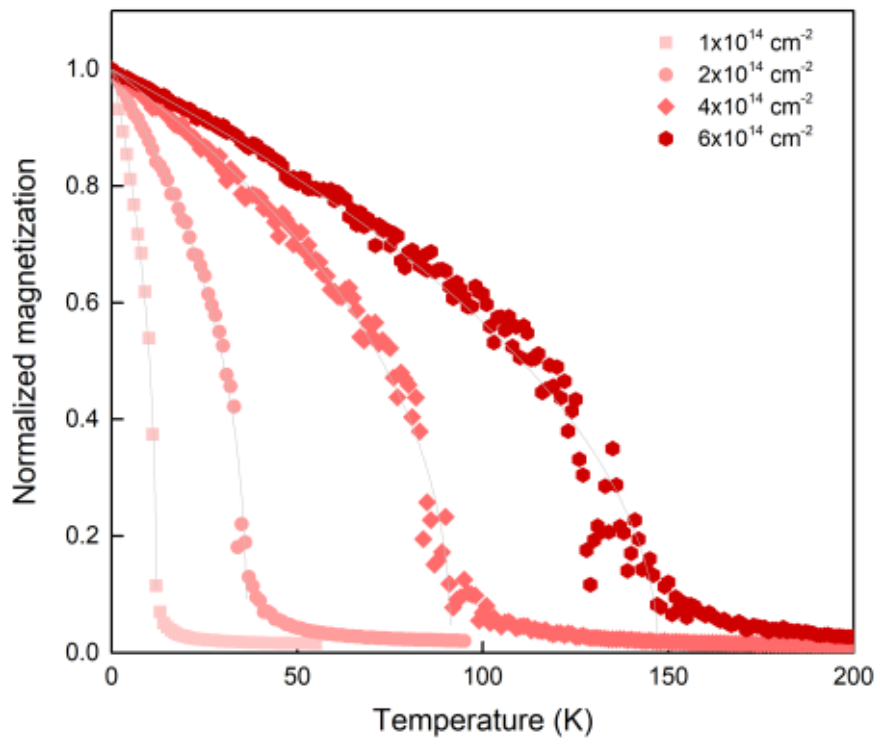

$T_c$: $1\times10^{14}$ cm$^{-2}$: 12 K  $2\times10^{14}$ cm$^{-2}$: 36 K

$4\times10^{14}$ cm$^{-2}$: 91 K  $6\times10^{14}$ cm$^{-2}$: 147 K

# 70.$MgCl_2$

| MC2D-ID | C2DB | 2dmat-ID | USPEX | Space group | Band gap (eV) |
|---|---|---|---|---|---|
| - | ✓ | 2dm-4009 | - | P4m2 | 5.30 |

**Convex hull**

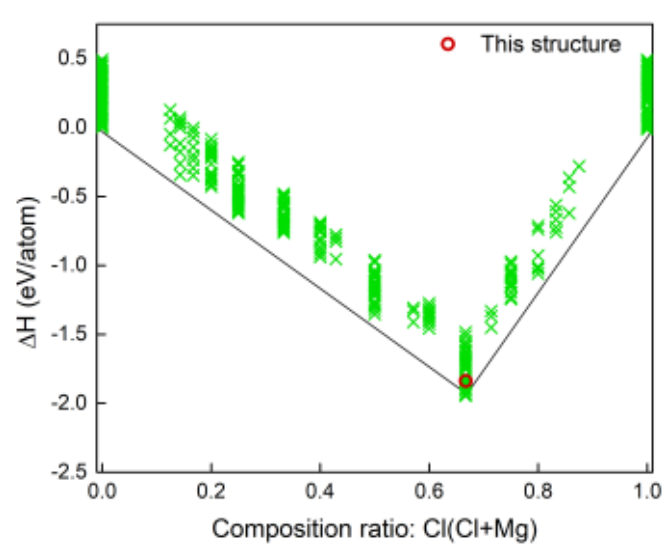

**Atomic structure**

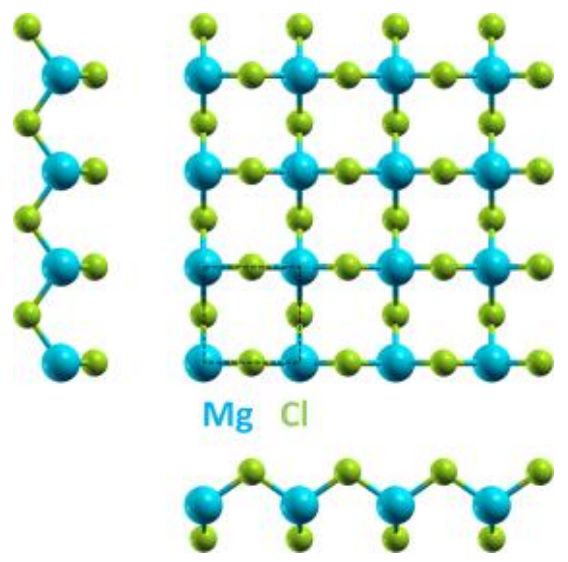

**Atomic coordinates**

```
Mg1Cl2
   1.00000000000000
   3.8860546408558880   0.0000000000000000   0.0000000000000000
   0.0000000000000000   3.8860549545644787   0.0000000000000000
   0.0000000000000000   0.0000000000000000   22.6561909999999997
   Mg   Cl
   1    2
Direct
  0.4999999972785929  0.5000000026059581  0.4999999975733260
  0.5000000116682841  0.9999999847521011  0.4402447806052976
  0.9999999910531159  0.5000000126419408  0.5597552218213693
```

**Phonon dispersion curve**

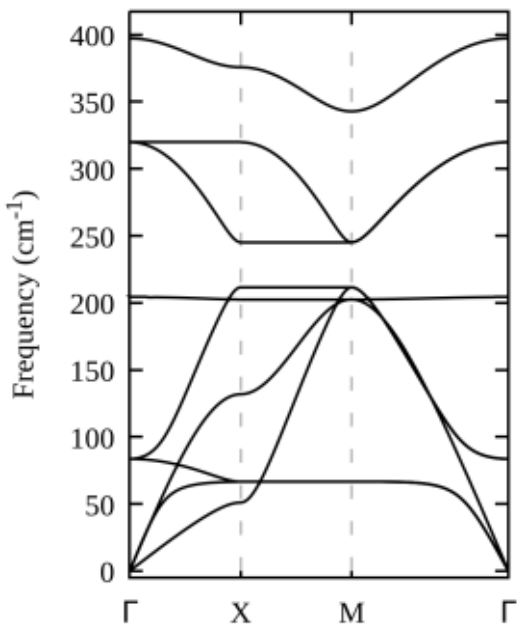

**Projected band structure and density of states**

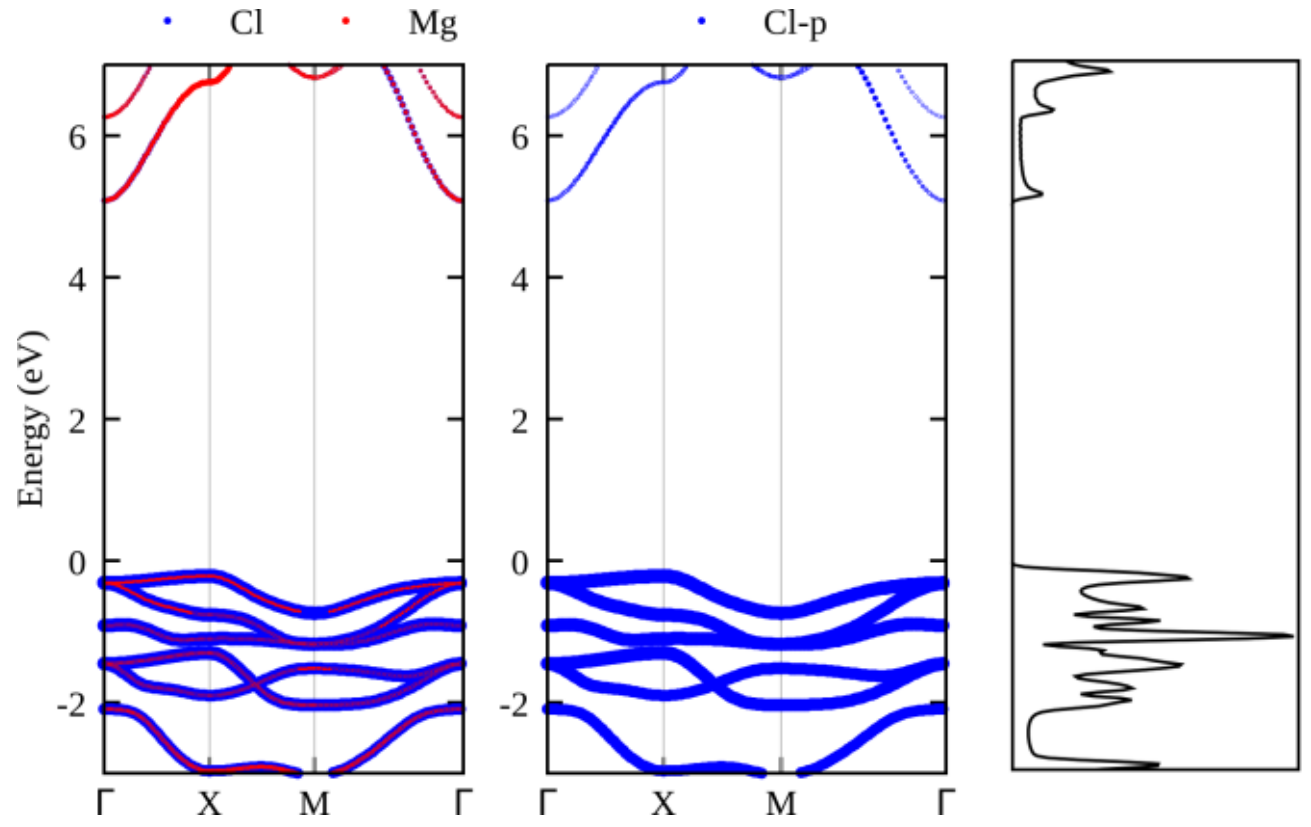

**Magnetic moment and spin polarization energy as a function of hole doping concentration**

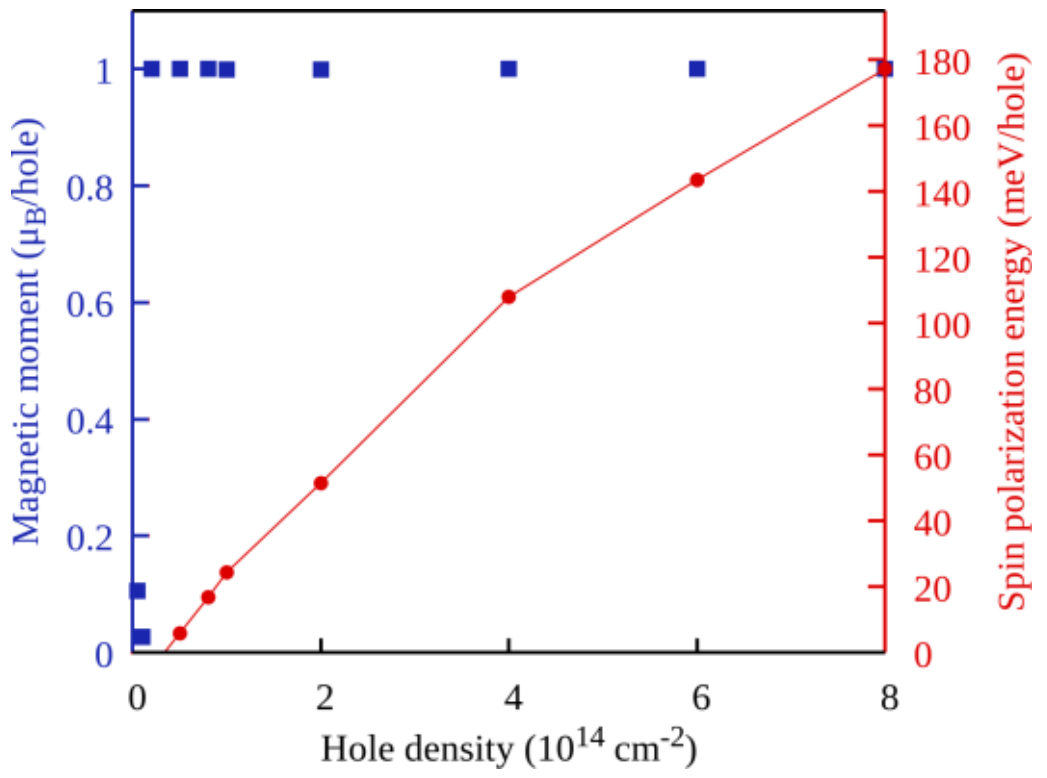

**Magnetic configurations and spin Hamiltonian**

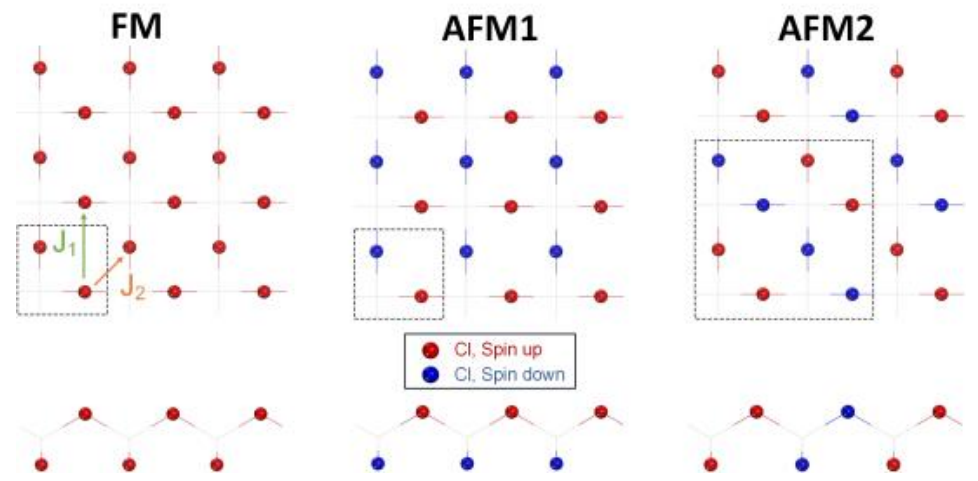

$$E_{FM} = E_0 - 4J_1S^2 - 4J_2S^2$$

$$E_{AFM1} = E_0 - 4J_1S^2 + 4J_2S^2$$

$$E_{AFM2} = E_0 + 4J_1S^2$$

**Magnetic exchange coupling parameters**

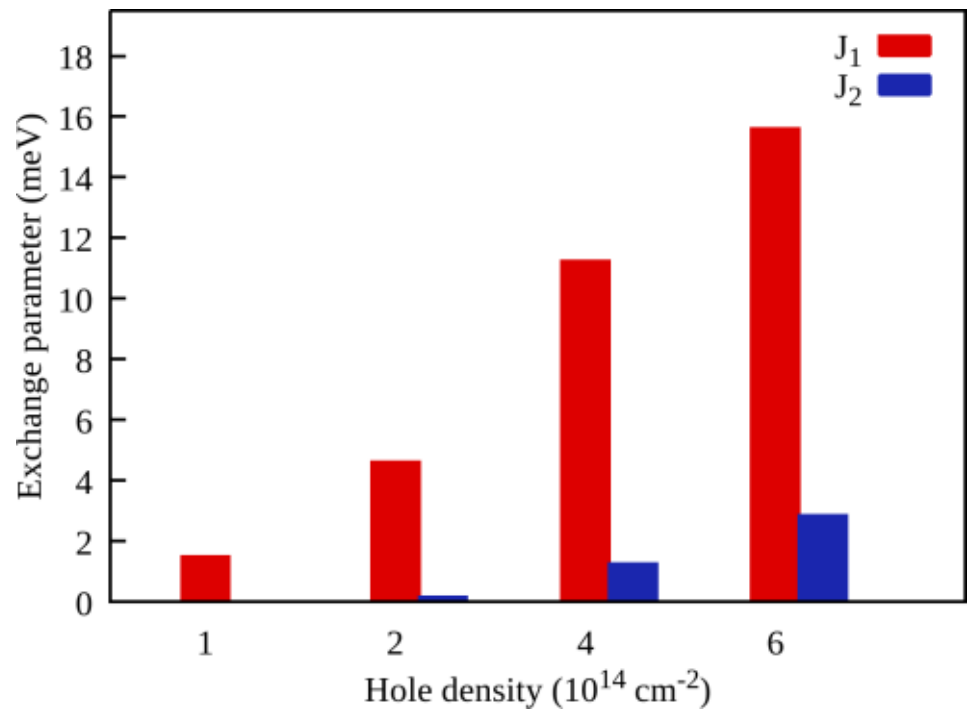

**Magnetic anisotropy energy (MAE, μeV) per magnetic atom**

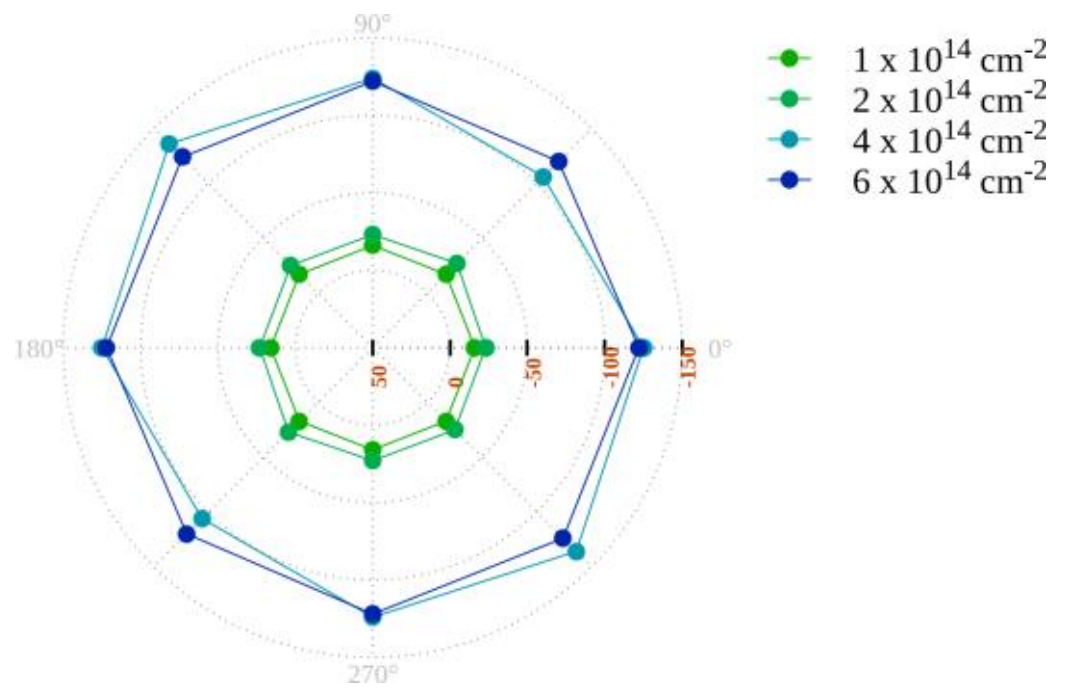

MAE = $E_{\parallel} - E_{\perp}$, a positive (negative) value of MAE indicates the off-plane (in-plane) easy axis.

**Monte Carlo simulations of the normalized magnetization of as a function of temperature**

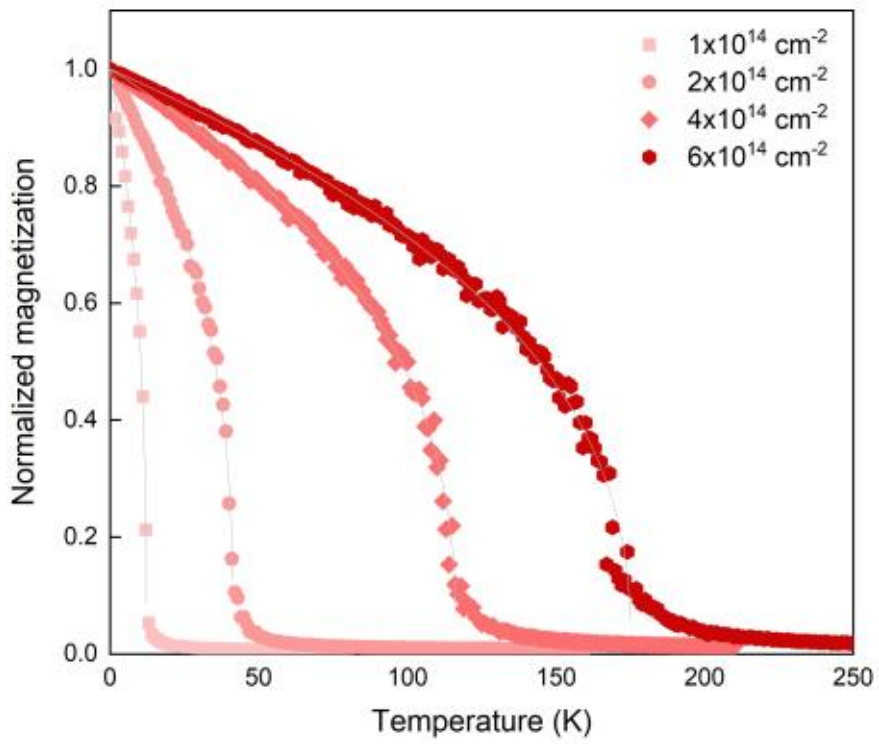

$T_c$: $1\times10^{14}$ cm$^{-2}$: 12 K   $2\times10^{14}$ cm$^{-2}$: 41 K

$4\times10^{14}$ cm$^{-2}$: 116 K   $6\times10^{14}$ cm$^{-2}$: 175 K

# 71. $MgBr_2$

| MC2D-ID | C2DB | 2dmat-ID | USPEX | Space group | Band gap (eV) |
|---|---|---|---|---|---|
| - | - | 2dm-1719 | - | P4m2 | 4.29 |

## Convex hull

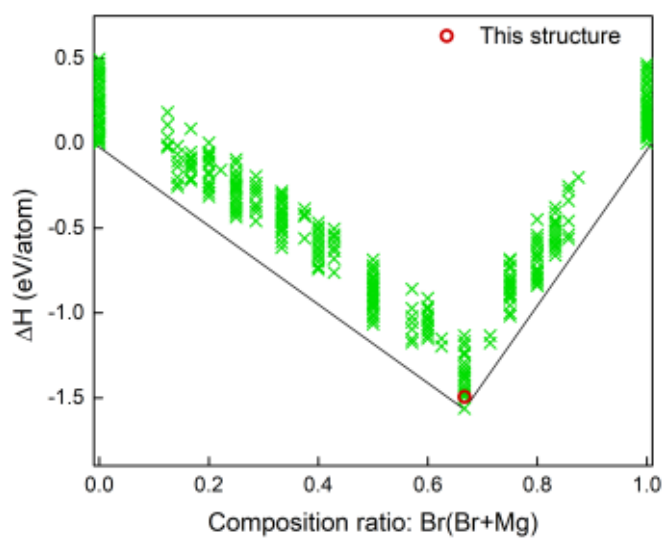

## Atomic structure

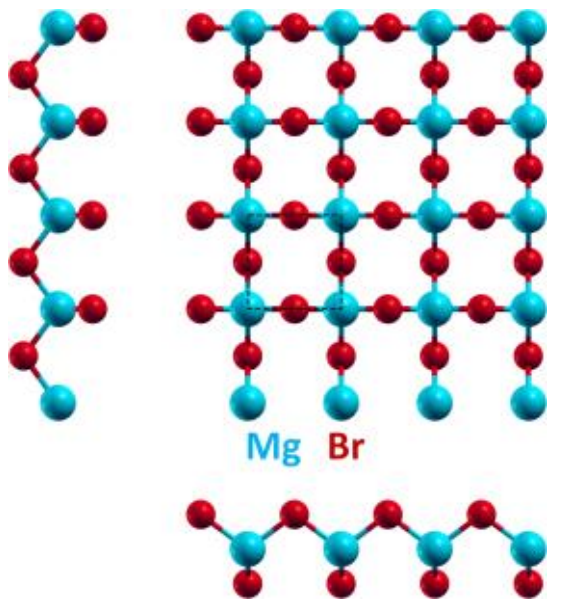

## Atomic coordinates

```
Mg1Br2
   1.00000000000000
   4.0734703651003379    0.0000000000000000    0.0000000000000000
   0.0000000000000000    4.0734704136992779    0.0000000000000000
   0.0000000000000000    0.0000000000000000   22.6561909999999997
   Mg    Br
   1     2
Direct
  0.9999999865177287  0.0000000583975179  0.9999999995217266
  0.9999999528752852  0.5000000356119969  0.9335674807916021
  0.5000000606069790  0.9999999059904923  0.0664325196866713
```

## Phonon dispersion curve

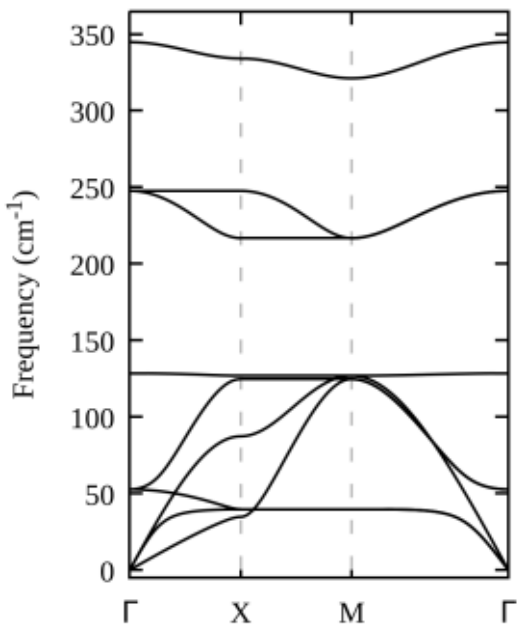

## Projected band structure and density of states

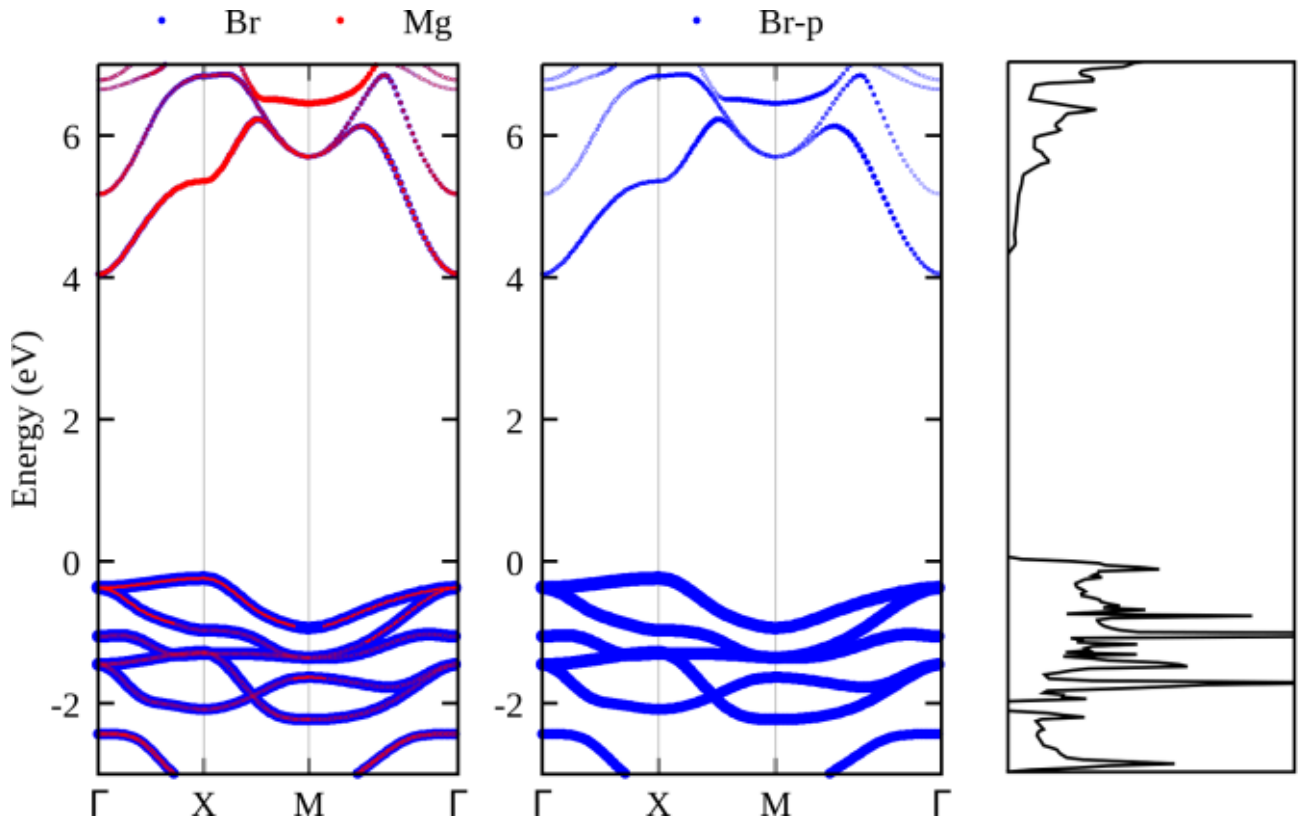

## Magnetic moment and spin polarization energy as a function of hole doping concentration

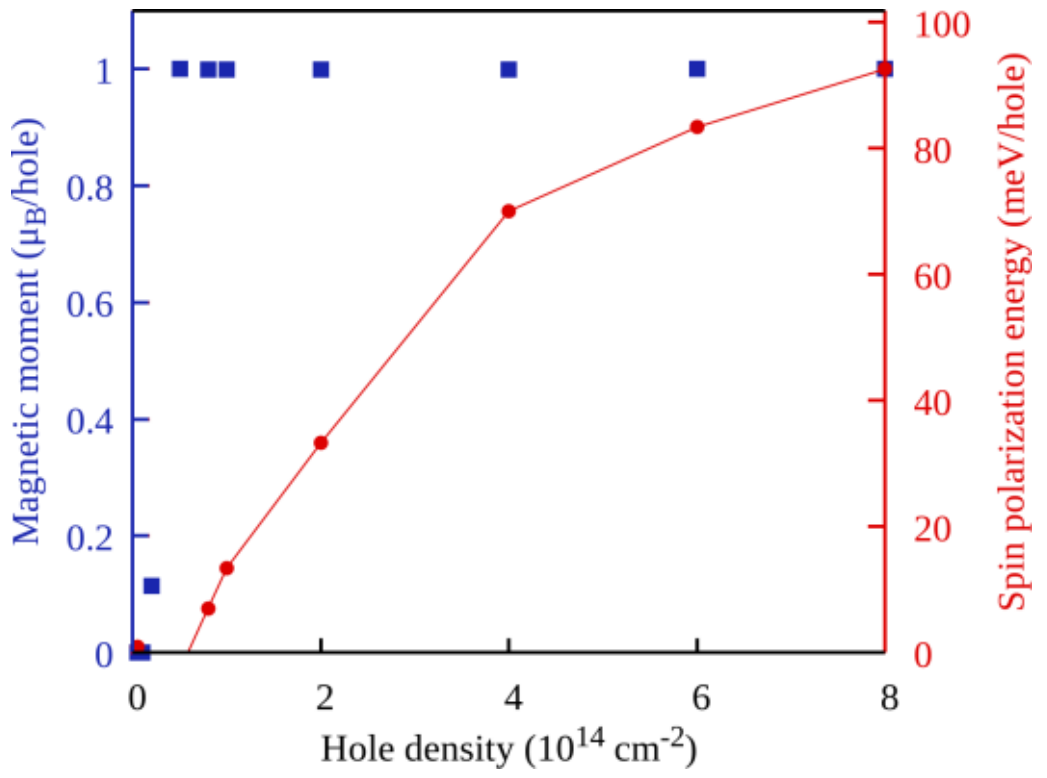

## Magnetic configurations and spin Hamiltonian

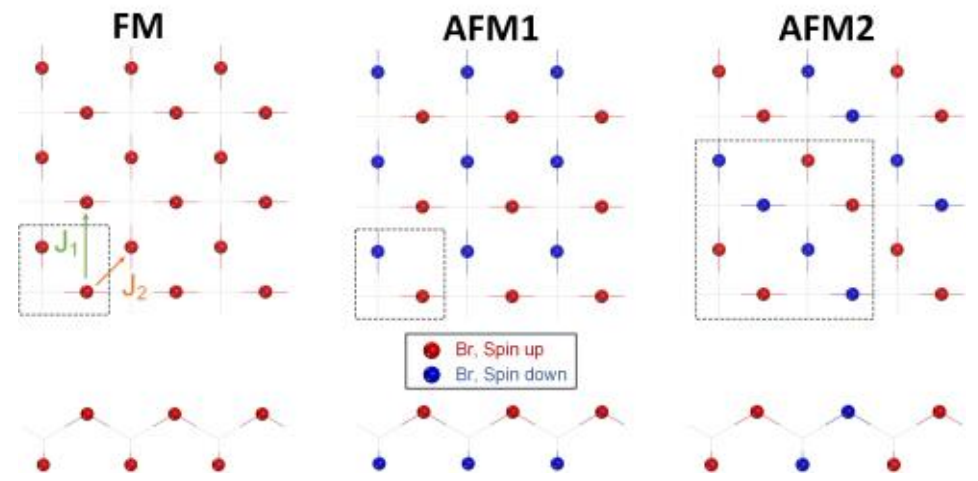

$$E_{FM} = E_0 - 4J_1S^2 - 4J_2S^2$$

$$E_{AFM1} = E_0 - 4J_1S^2 + 4J_2S^2$$

$$E_{AFM2} = E_0 + 4J_1S^2$$

## Magnetic exchange coupling parameters

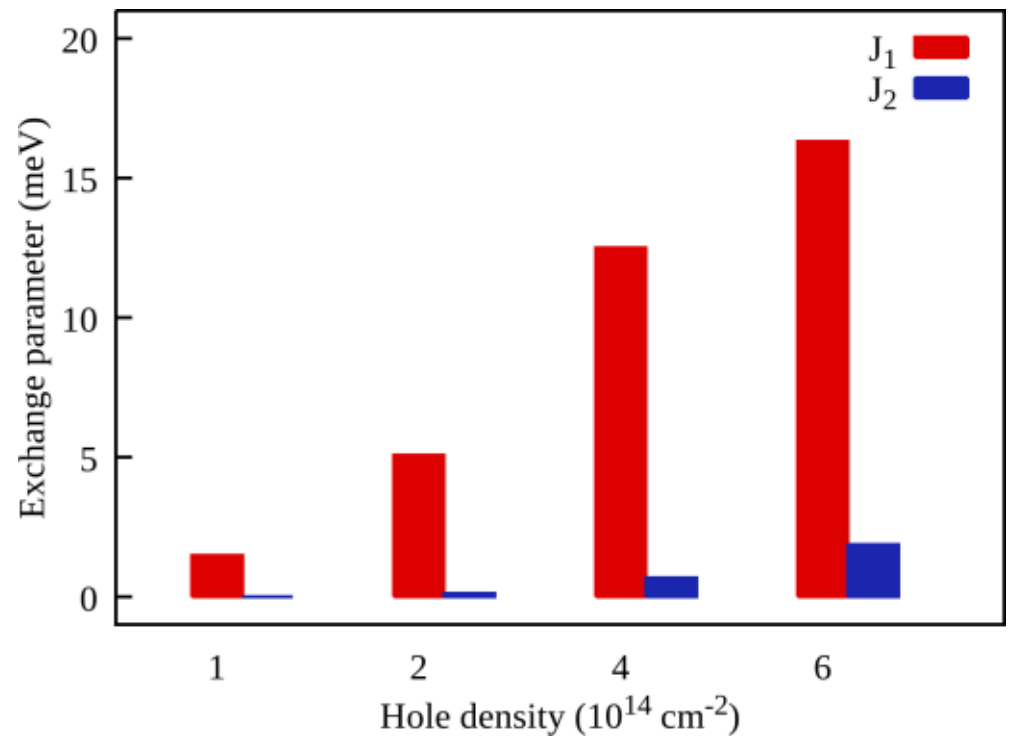

## Magnetic anisotropy energy (MAE, μeV) per magnetic atom

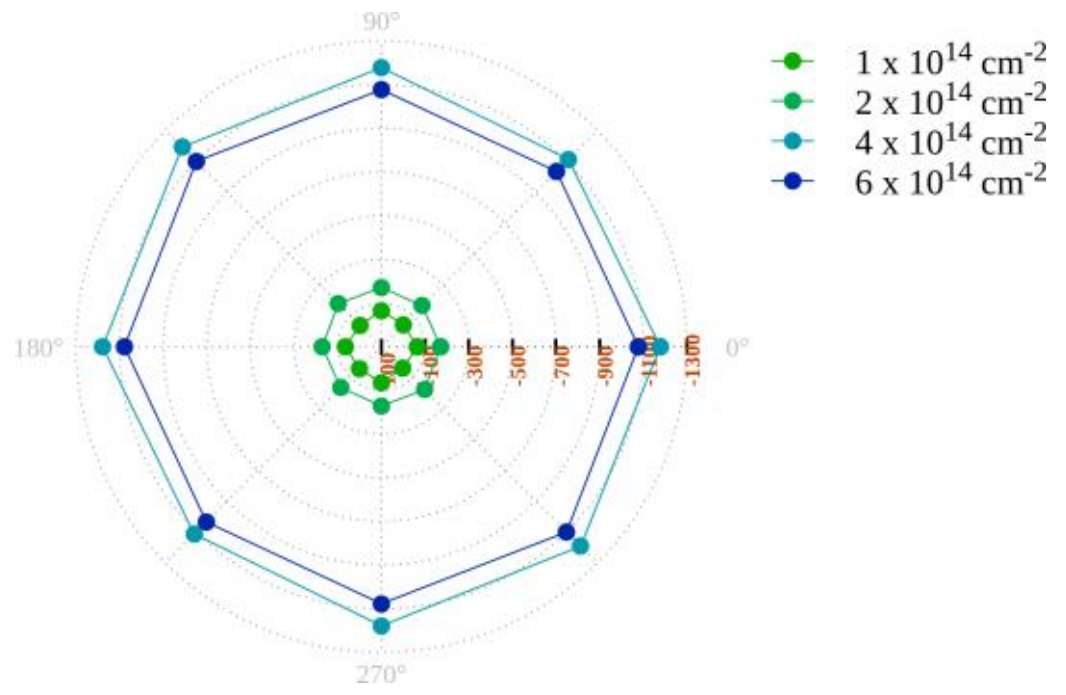

MAE = $E_{\parallel} - E_{\perp}$, a positive (negative) value of MAE indicates the off-plane (in-plane) easy axis.

## Monte Carlo simulations of the normalized magnetization of as a function of temperature

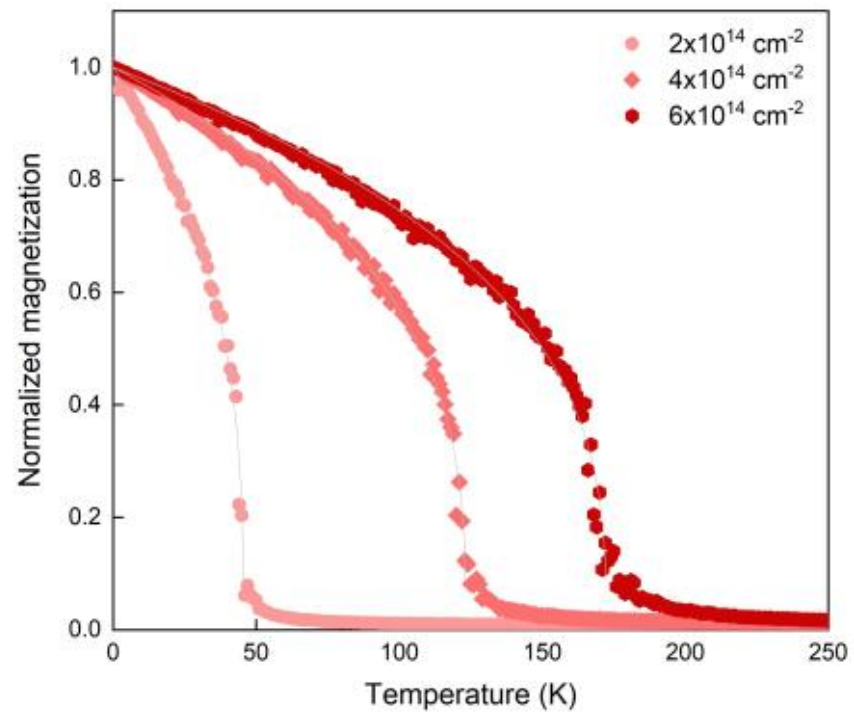

$T_c$: $1\times10^{14}$ cm$^{-2}$: - K   $2\times10^{14}$ cm$^{-2}$: 45 K

$4\times10^{14}$ cm$^{-2}$: 123 K   $6\times10^{14}$ cm$^{-2}$: 172 K

# 72. $CaF_2$

| MC2D-ID | C2DB | 2dmat-ID | USPEX | Space group | Band gap (eV) |
|---|---|---|---|---|---|
| - | - | 2dm-690 | - | P4m2 | 6.35 |

## Convex hull

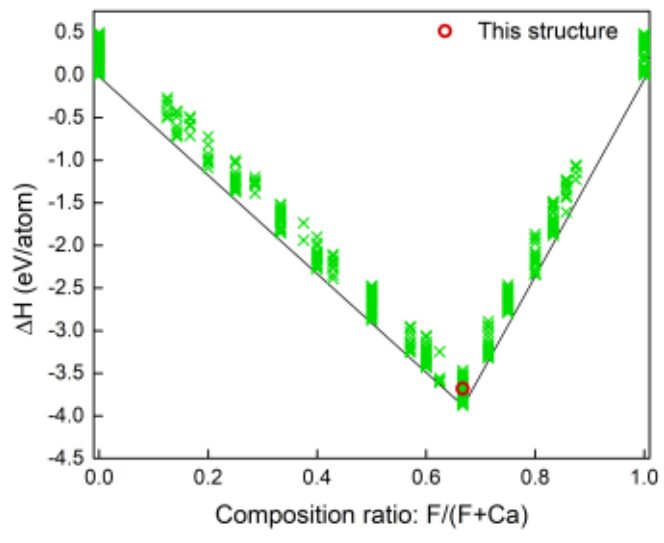

## Atomic structure

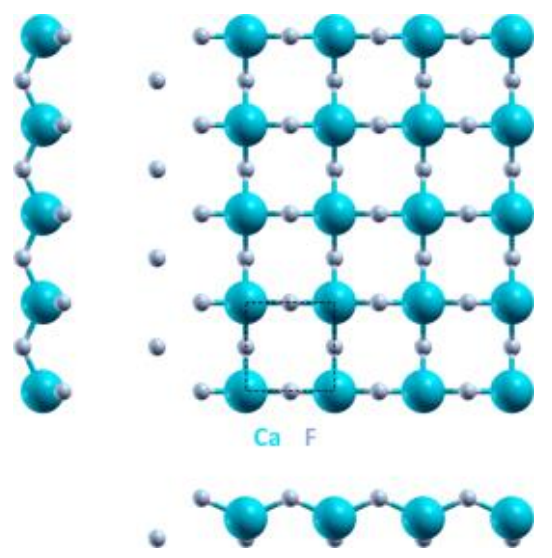

## Atomic coordinates

```
Ca1F2
   1.00000000000000
   3.9545539778349492   0.0000000000000000   0.0000000000000000
   0.0000000000000000   3.9545558625799551   0.0000000000000000
   0.0000000000000000   0.0000000000000000   22.6561909999999997
   Ca    F
   1     2
Direct
  0.9999999378009363   0.0000000158025415   0.9999999761766531
  0.0000001156571017   0.4999999831472124   0.9603490458055433
  0.4999999465419620   0.0000000010502461   0.0396509780178036
```

## Phonon dispersion curve

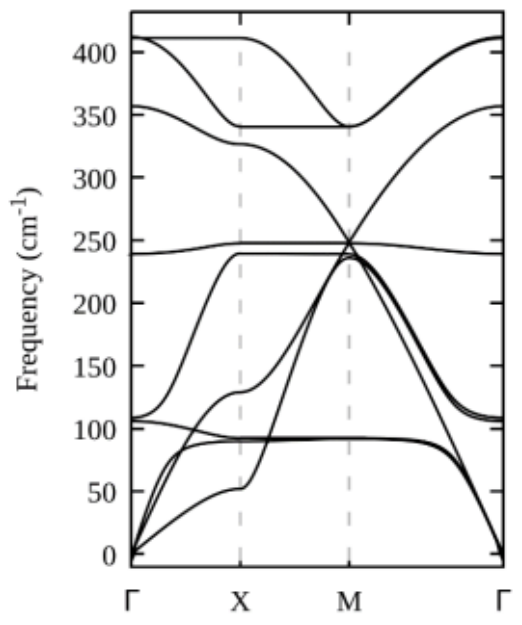

## Projected band structure and density of states

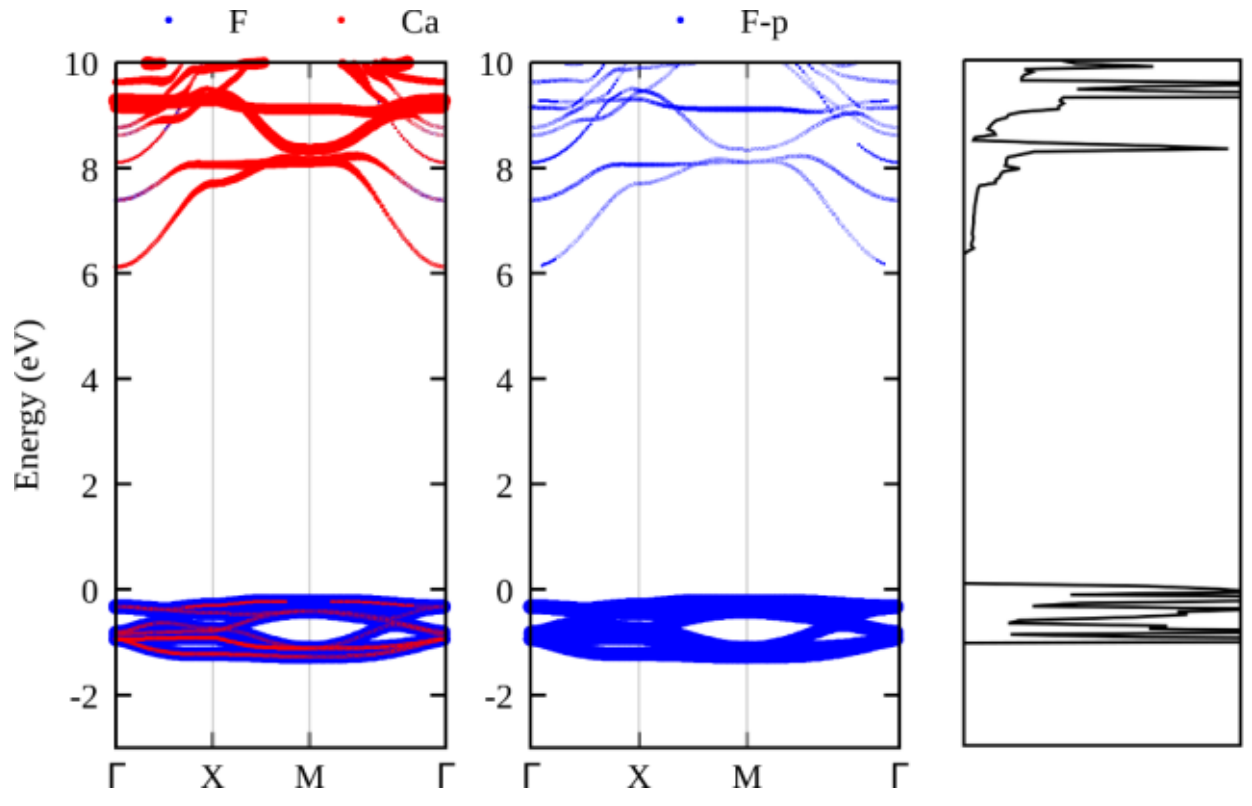

## Magnetic moment and spin polarization energy as a function of hole doping concentration

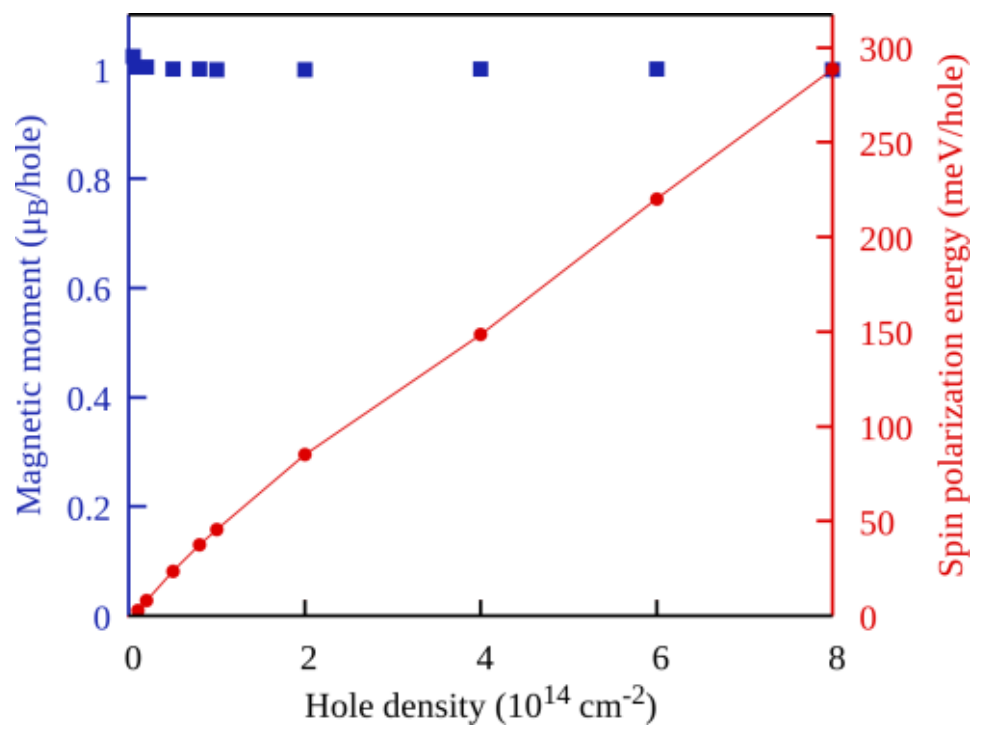

## Magnetic configurations and spin Hamiltonian

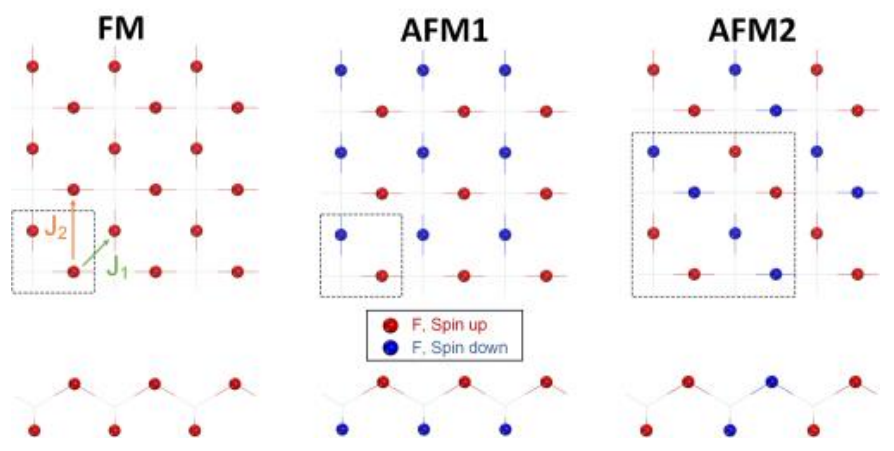

$$E_{FM} = E_0 - 4J_1S^2 - 4J_2S^2$$
$$E_{AFM1} = E_0 + 4J_1S^2 - 4J_2S^2$$
$$E_{AFM2} = E_0 + 4J_2S^2$$

## Magnetic exchange coupling parameters

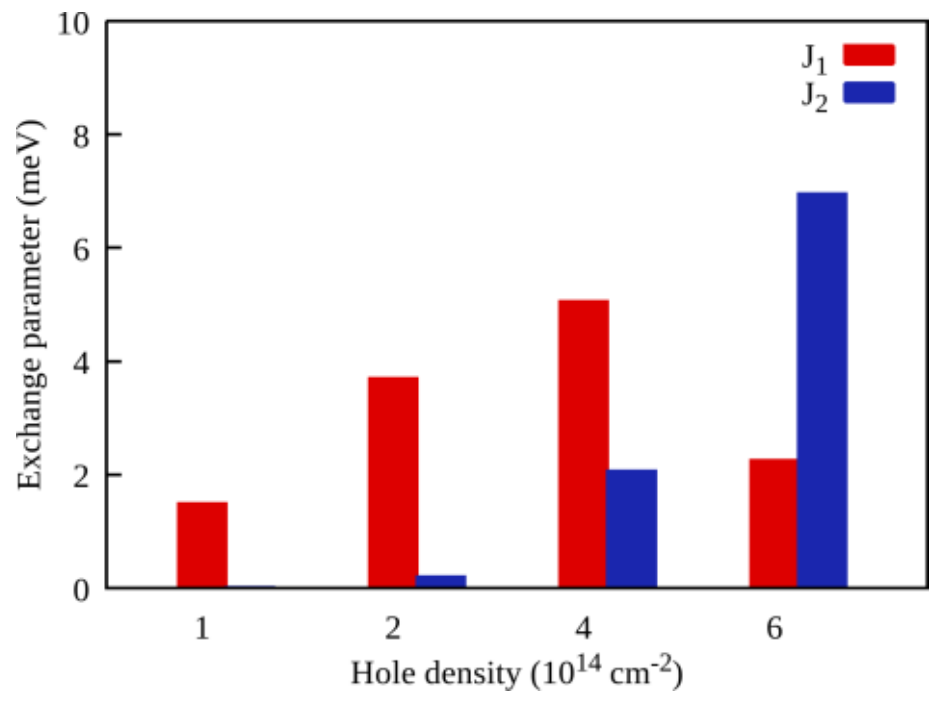

## Magnetic anisotropy energy (MAE, μeV) per magnetic atom

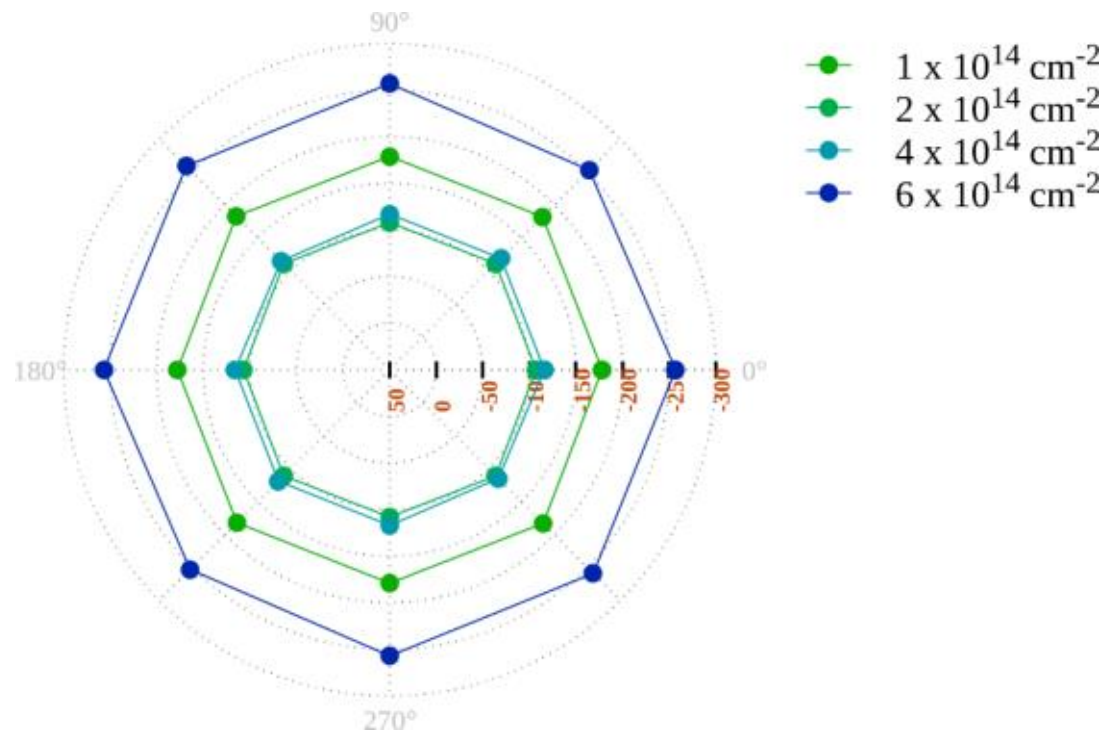

MAE = $E_{\parallel} - E_{\perp}$, a positive (negative) value of MAE indicates the off-plane (in-plane) easy axis.

## Monte Carlo simulations of the normalized magnetization of as a function of temperature

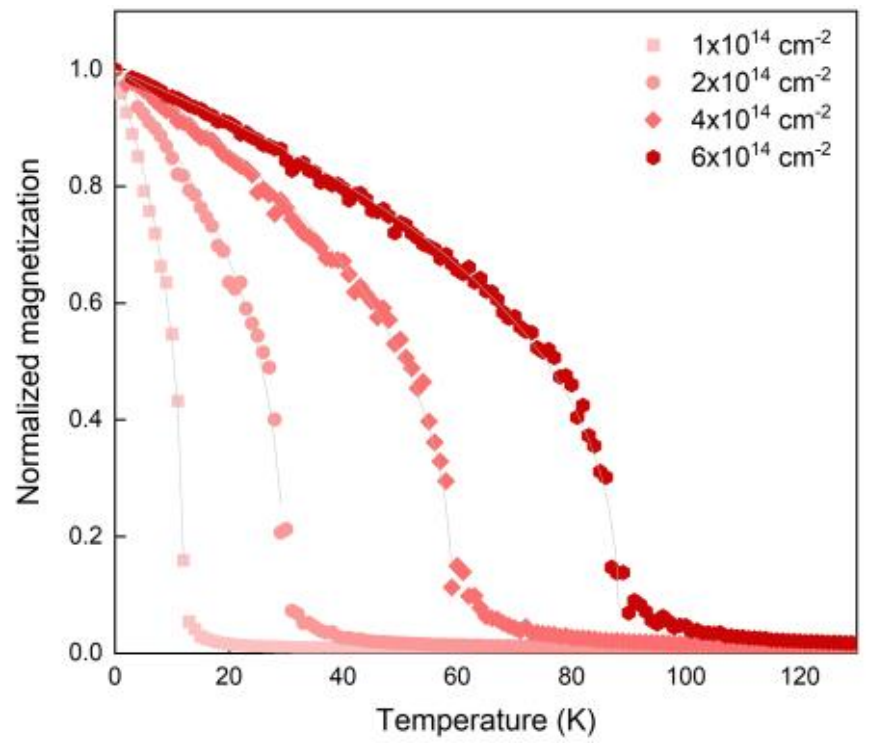

$T_c$: $1\times10^{14}$ cm$^{-2}$: 12 K   $2\times10^{14}$ cm$^{-2}$: 30 K

$4\times10^{14}$ cm$^{-2}$: 59 K   $6\times10^{14}$ cm$^{-2}$: 88 K

# 73.$CaCl_2$

| MC2D-ID | C2DB | 2dmat-ID | USPEX | Space group | Band gap (eV) |
| --- | --- | --- | --- | --- | --- |
| - | - | 2dm-1617 | - | P4m2 | 5.57 |

**Convex hull** | **Atomic structure** | **Atomic coordinates** | **Phonon dispersion curve**

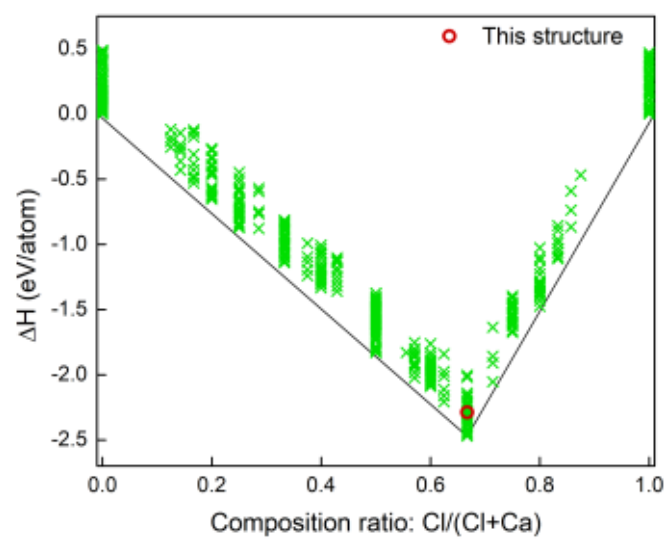

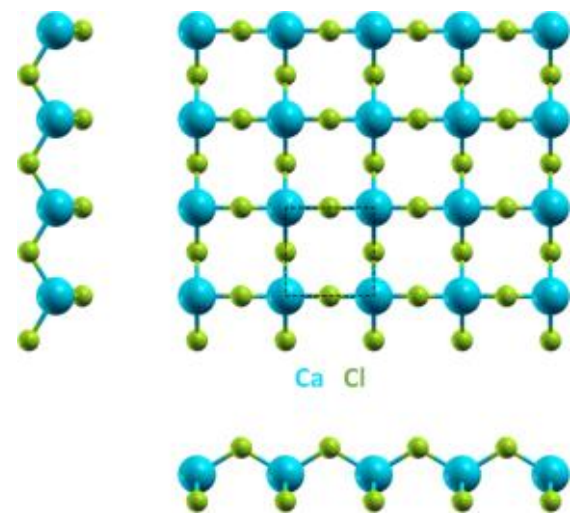

```
Ca1Cl2
   1.00000000000000
   4.5059930501459551   0.0000000000000000   0.0000000000000000
   0.0000000000000000   4.5059939016548638   0.0000000000000000
   0.0000000000000000   0.0000000000000000   22.6561909999999997
   Ca   Cl
   1    2
Direct
  0.0000000006244889   0.9999999674575051   0.9999999883601873
  0.0000000040950212   0.5000000350374876   0.9398917786764045
   0.4999999952804899   0.9999999975050073   0.0601082329634011
```

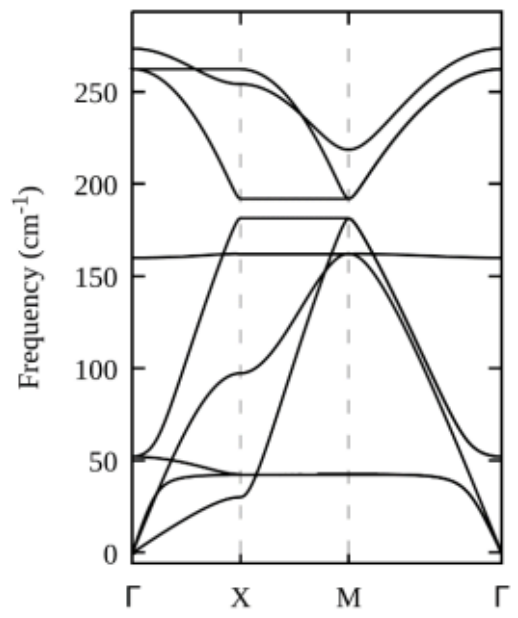

**Projected band structure and density of states**

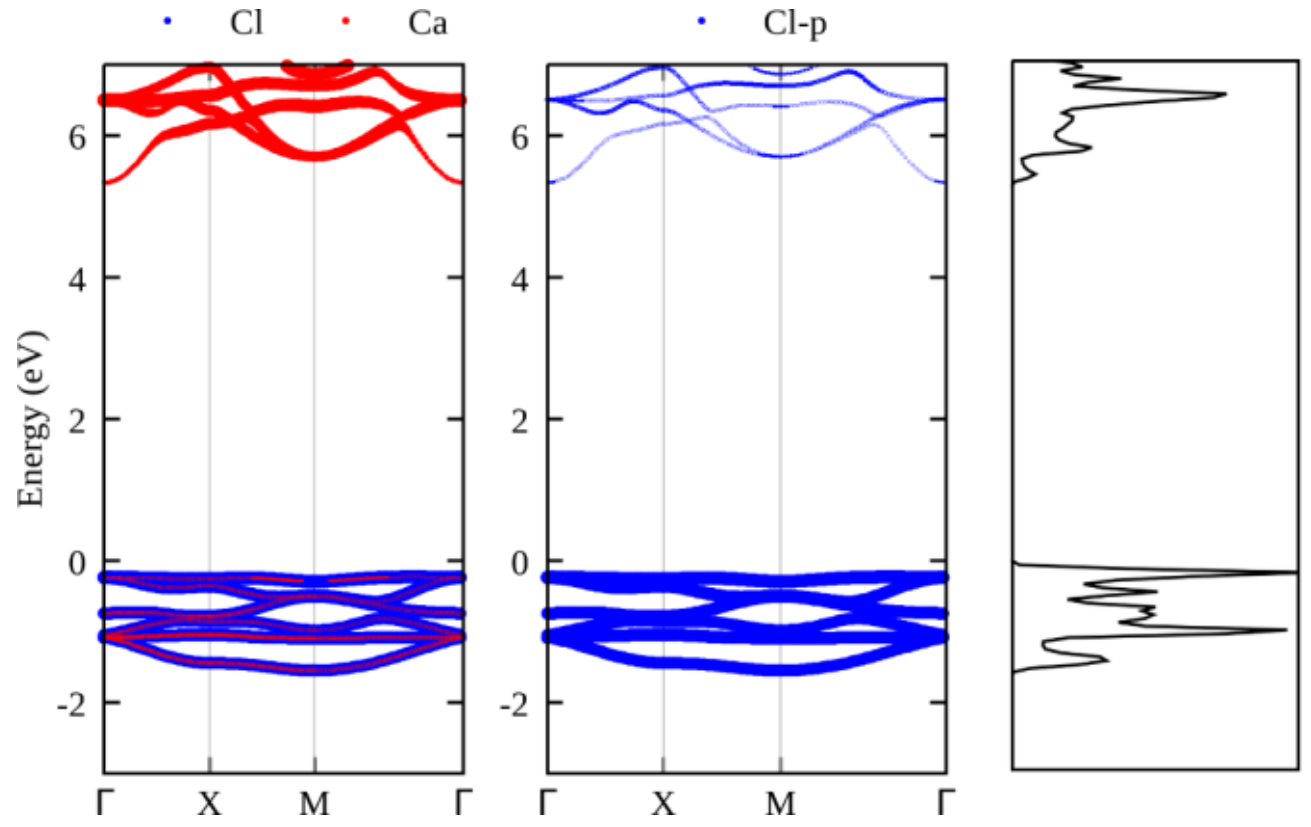

**Magnetic moment and spin polarization energy as a function of hole doping concentration**

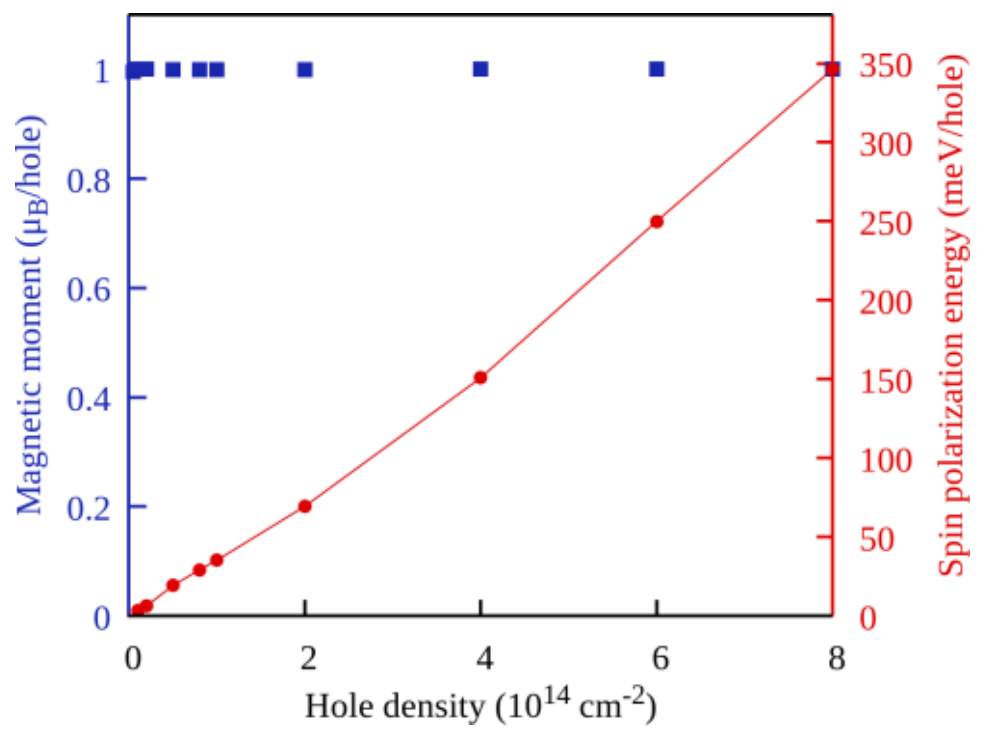

**Magnetic configurations and spin Hamiltonian**

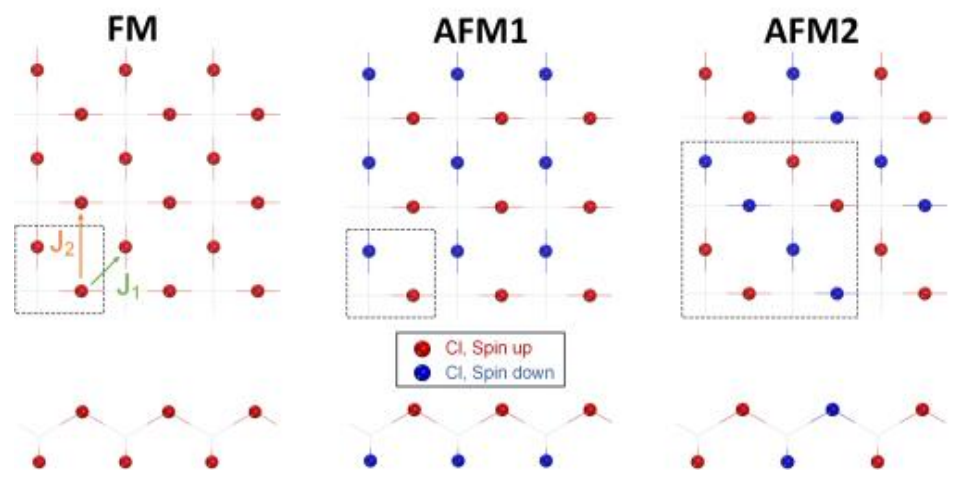

$$E_{FM} = E_0 - 4J_1S^2 - 4J_2S^2$$
$$E_{AFM1} = E_0 + 4J_1S^2 - 4J_2S^2$$
$$E_{AFM2} = E_0 + 4J_2S^2$$

**Magnetic exchange coupling parameters**

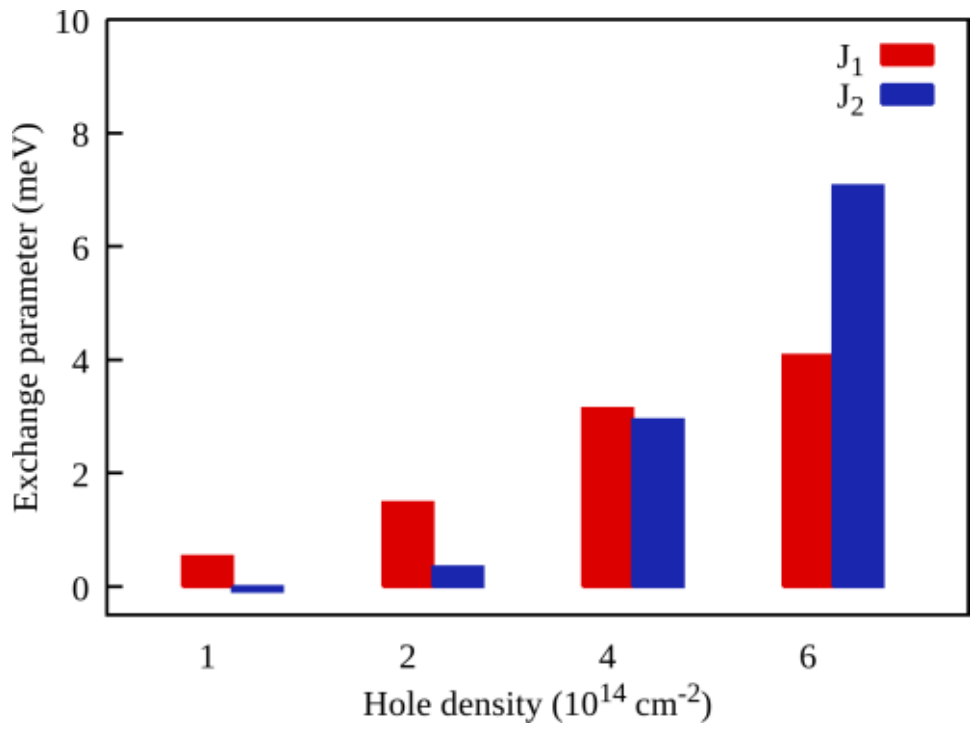

**Magnetic anisotropy energy (MAE, μeV) per magnetic atom**

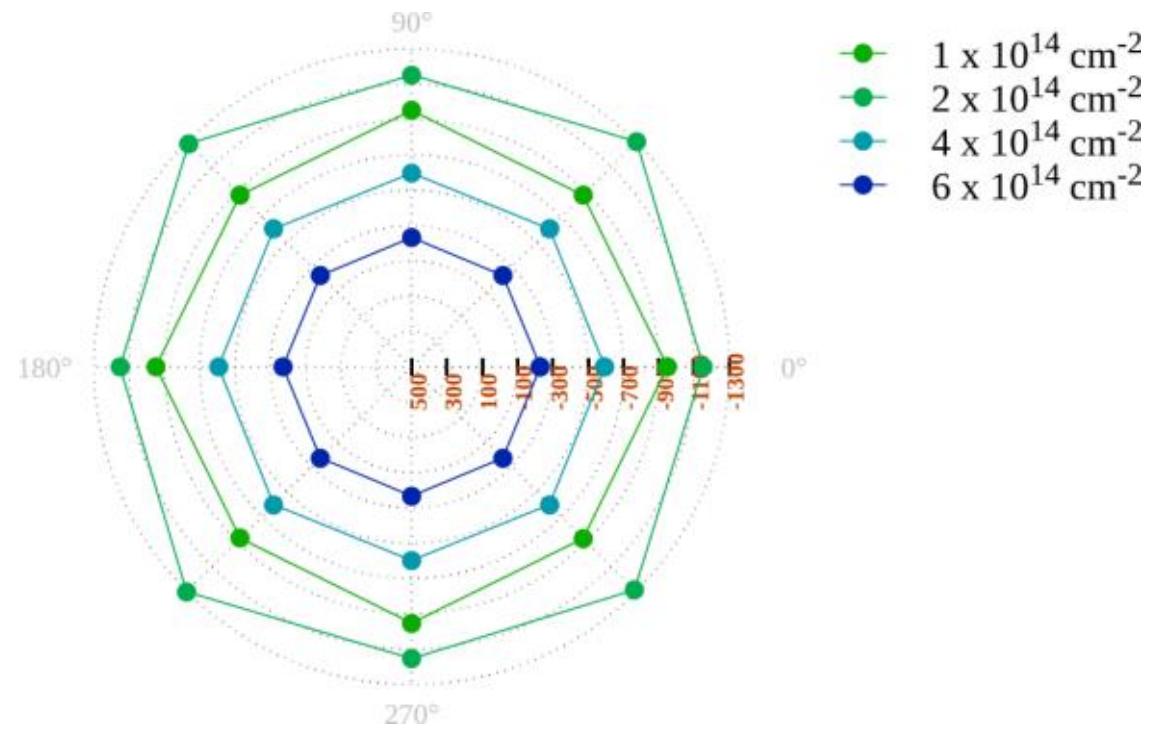

MAE = $E_{\parallel} - E_{\perp}$, a positive (negative) value of MAE indicates the off-plane (in-plane) easy axis.

**Monte Carlo simulations of the normalized magnetization of as a function of temperature**

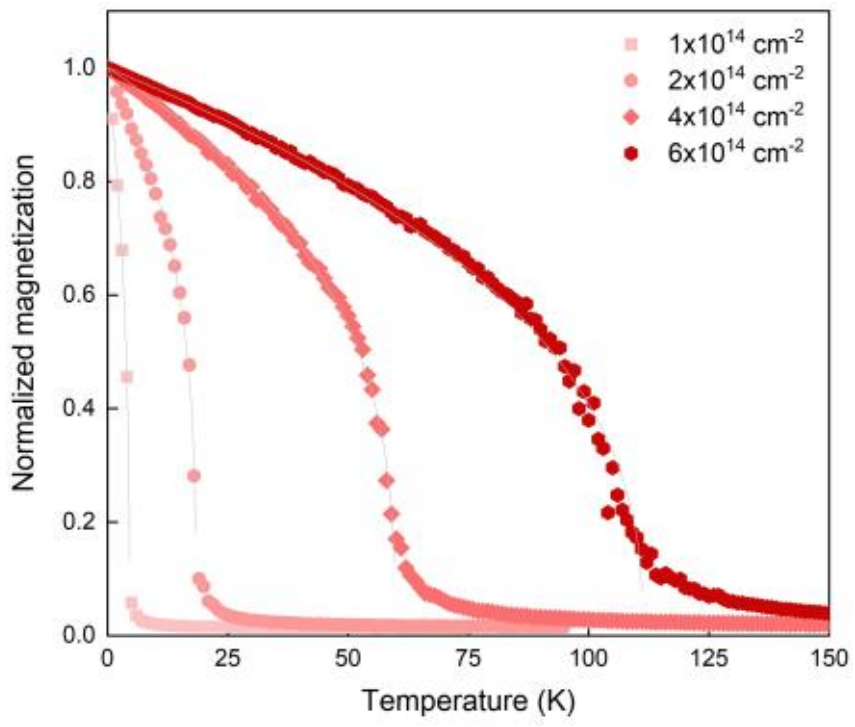

$T_c$: $1\times10^{14}$ cm$^{-2}$: 5 K  $2\times10^{14}$ cm$^{-2}$: 18 K  $4\times10^{14}$ cm$^{-2}$: 60 K  $6\times10^{14}$ cm$^{-2}$: 111 K

# 74. $CaBr_2$

| MC2D-ID | C2DB | 2dmat-ID | USPEX | Space group | Band gap (eV) |
|---|---|---|---|---|---|
| - | - | 2dm-265 | - | P4m2 | 4.83 |

**Convex hull**

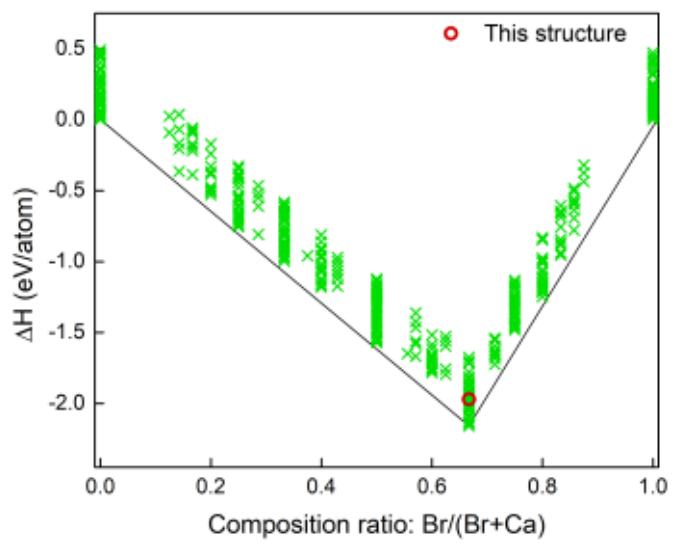

**Atomic structure**

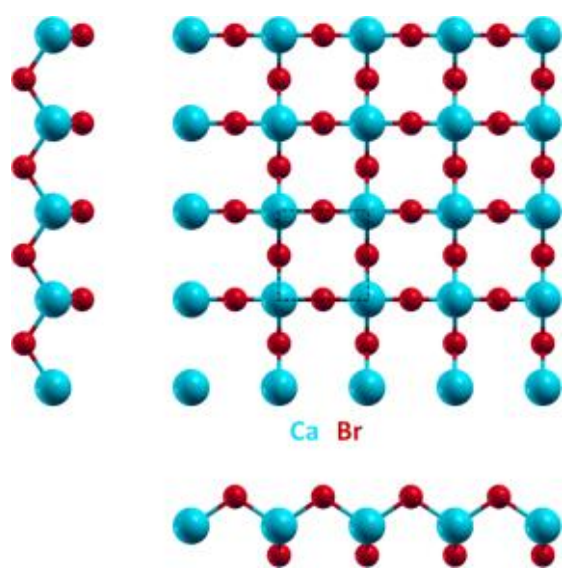

**Atomic coordinates**

```
Ca1 Br2
1.00000000000000
4.6464431803270090 0.0000000000000000 0.0000000000000000
0.0000000000000000 4.6464438158227459 0.0000000000000000
0.0000000000000000 0.0000000000000000 22.6561909999999997
Ca Br
1 2
Direct
0.0000000070004447 0.0000000073439566 0.0000000149294976
0.9999999878355510 0.4999999976954754 0.9320430549211309
0.5000000051640043 0.9999999949605680 0.0679569301493785
```

**Phonon dispersion curve**

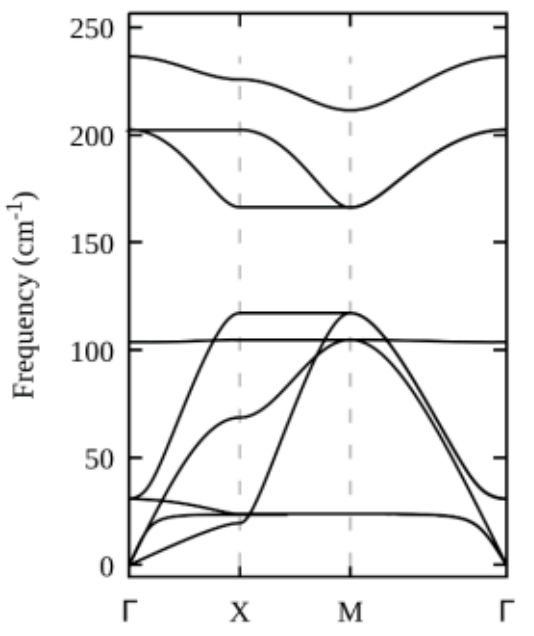

**Projected band structure and density of states**

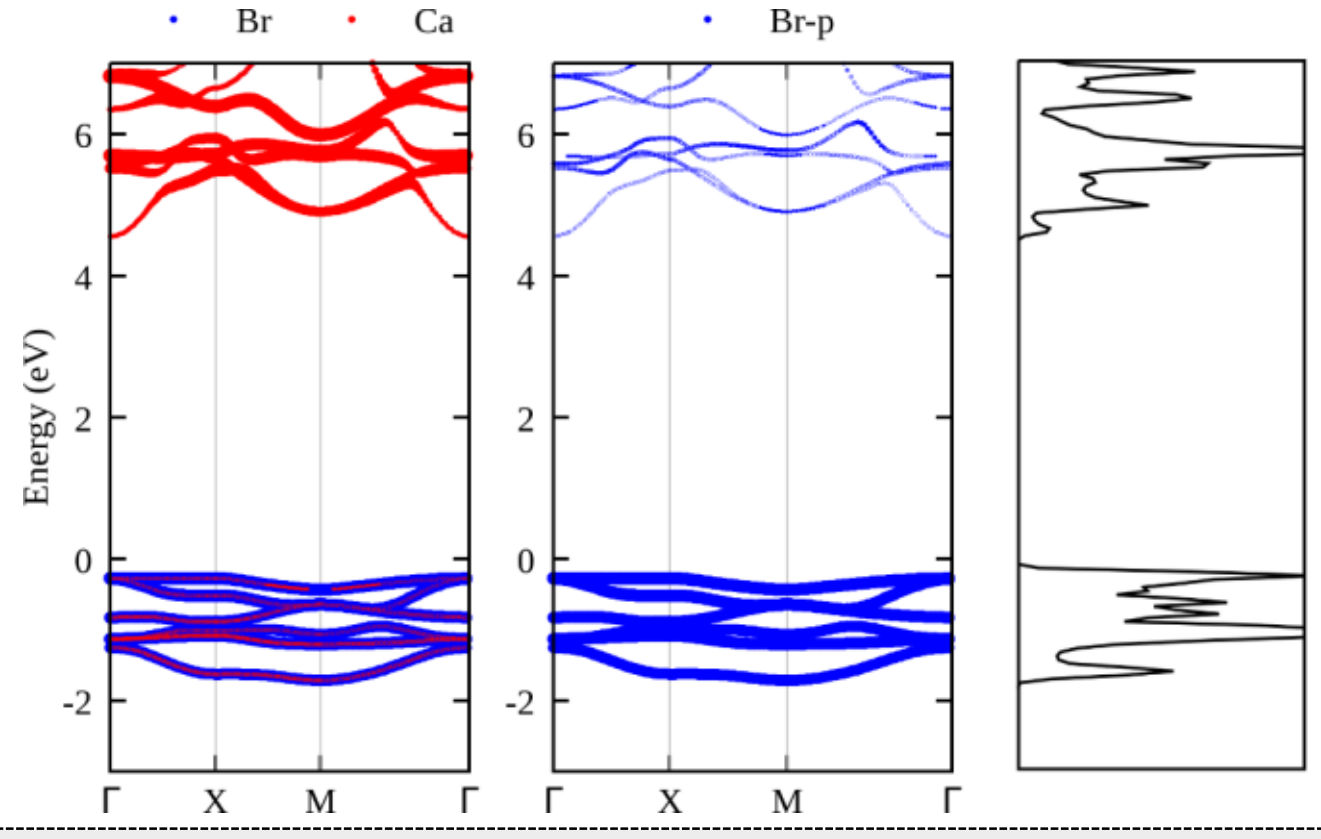

**Magnetic moment and spin polarization energy as a function of hole doping concentration**

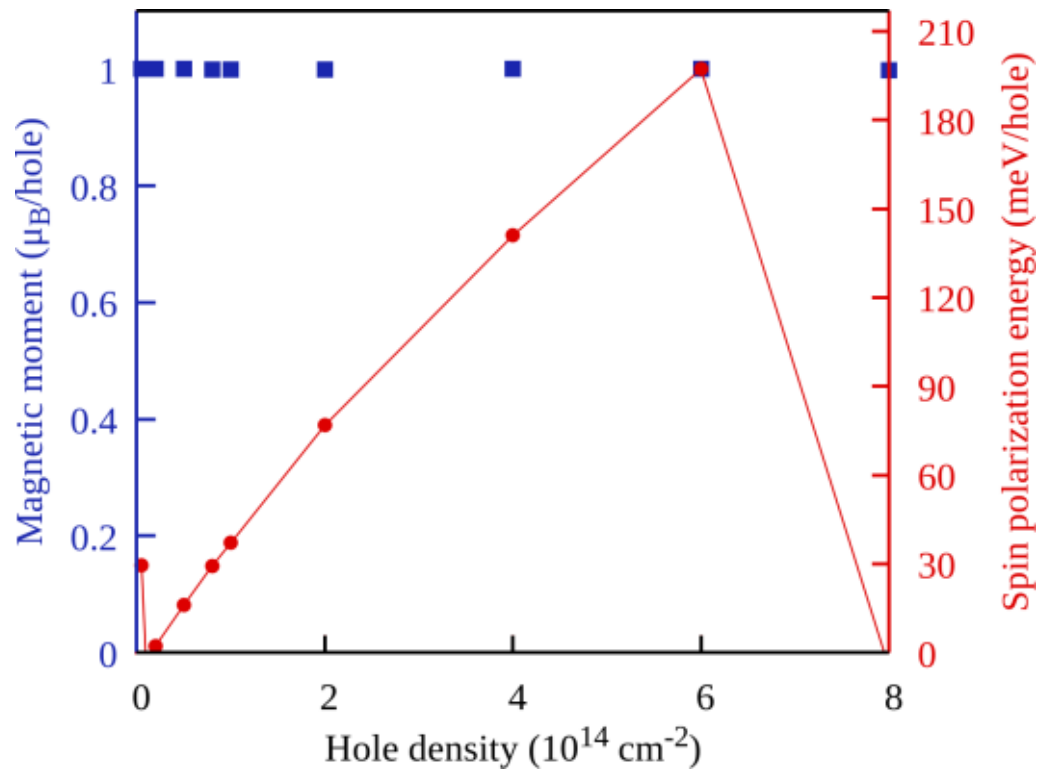

**Magnetic configurations and spin Hamiltonian**

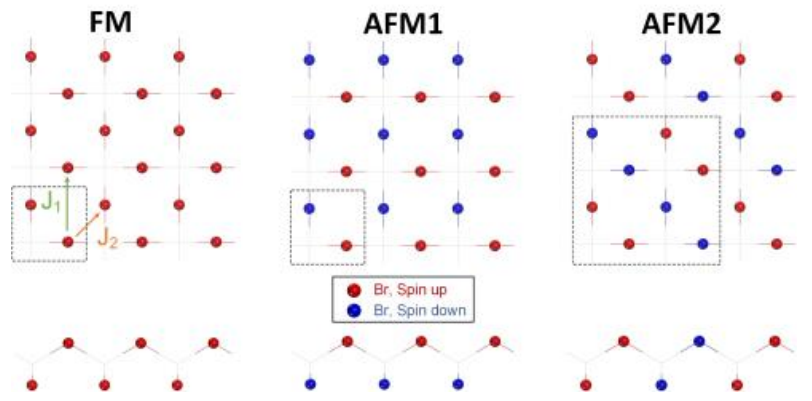

$$E_{FM} = E_0 - 4J_1S^2 - 4J_2S^2$$

$$E_{AFM1} = E_0 - 4J_1S^2 + 4J_2S^2$$

$$E_{AFM2} = E_0 + 4J_1S^2$$

**Magnetic exchange coupling parameters**

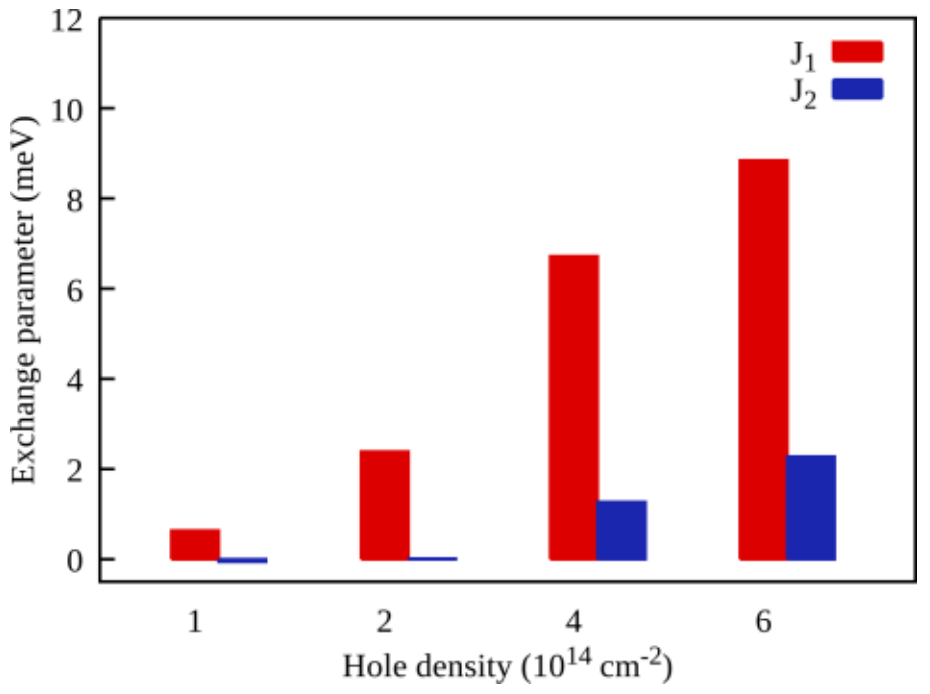

**Magnetic anisotropy energy (MAE, μeV) per magnetic atom**

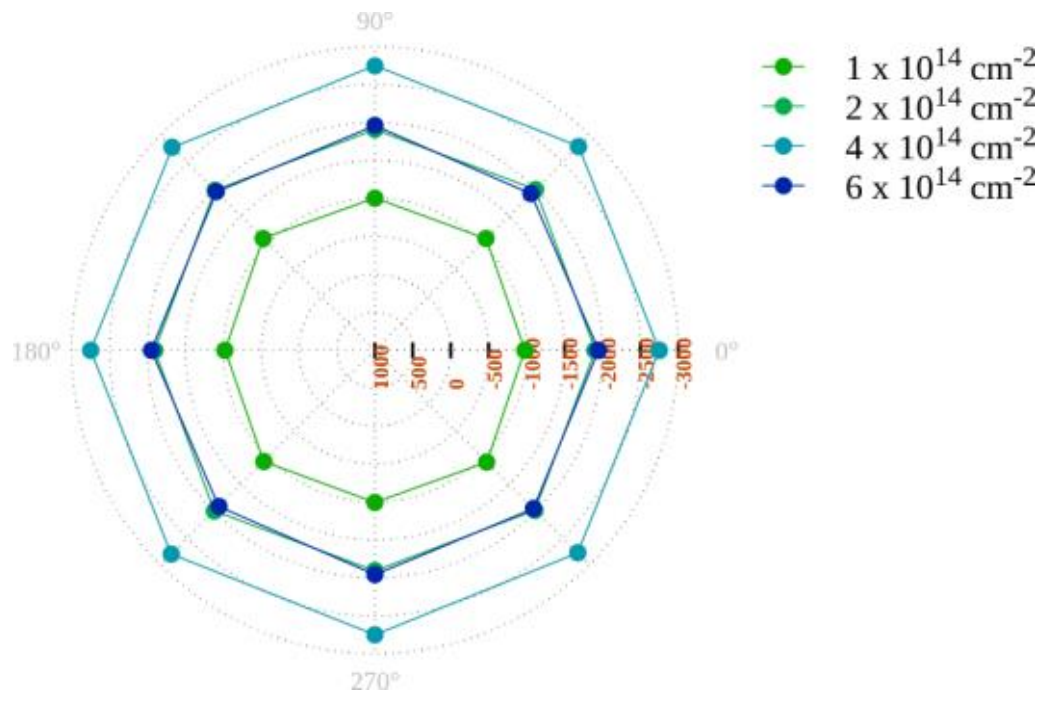

MAE = $E_{\parallel} - E_{\perp}$, a positive (negative) value of MAE indicates the off-plane (in-plane) easy axis.

**Monte Carlo simulations of the normalized magnetization of as a function of temperature**

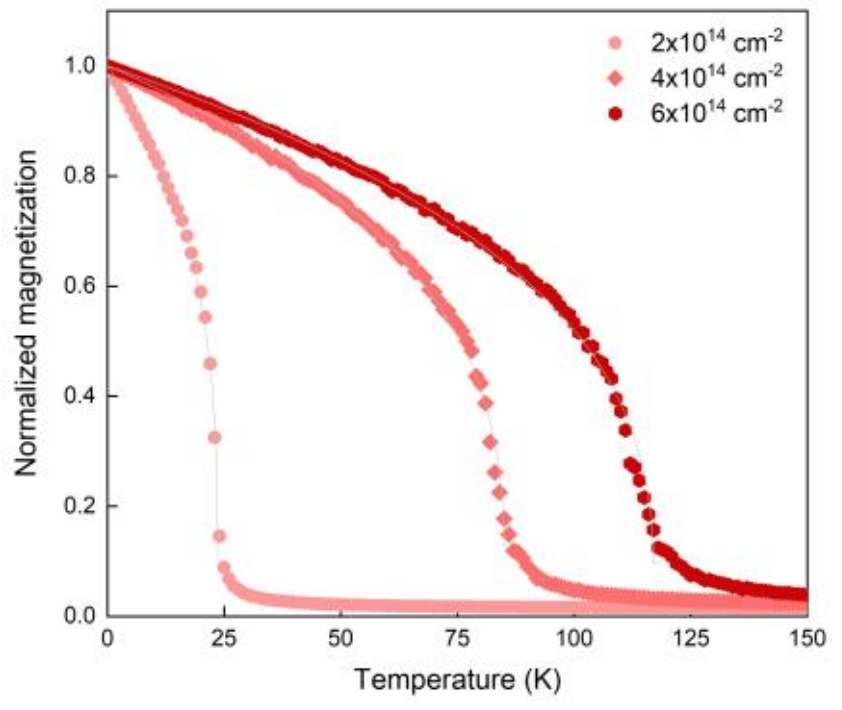

$T_c$: $1\times10^{14}$ cm$^{-2}$: - K $2\times10^{14}$ cm$^{-2}$: 24 K

$4\times10^{14}$ cm$^{-2}$: 85 K $6\times10^{14}$ cm$^{-2}$: 117 K

# 75. $CaI_2$

| MC2D-ID | C2DB | 2dmat-ID | USPEX | Space group | Band gap (eV) |
|---|---|---|---|---|---|
| 29 | ✓ | - | - | P3m1 | 3.91 |

## Convex hull

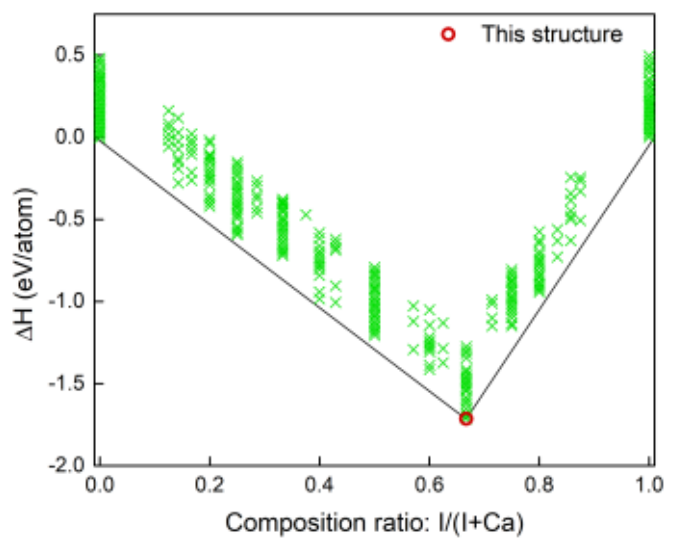

## Atomic structure

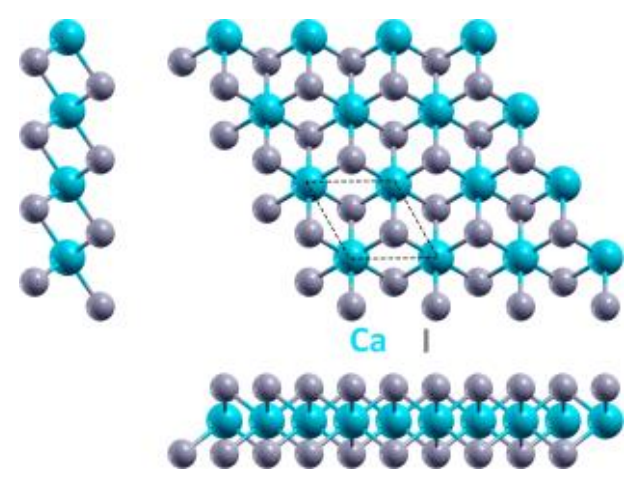

## Atomic coordinates

```
Ca1I2
   1.00000000000000
     4.5415685628997693    0.0000000000000000    0.0000000000000000
    -2.2707073727749525    3.9331250512874485    0.0000000000000000
     0.0000000000000000    0.0000000000000000   18.5370006561279332
   Ca    I
    1     2
Direct
  0.9999770747201140  0.9999822193259575  0.5000122657880297
  0.6665529080537027  0.3332915086209738  0.5953312750419357
  0.3332904005901298  0.6665868203645928  0.4046833791565660
```

## Phonon dispersion curve

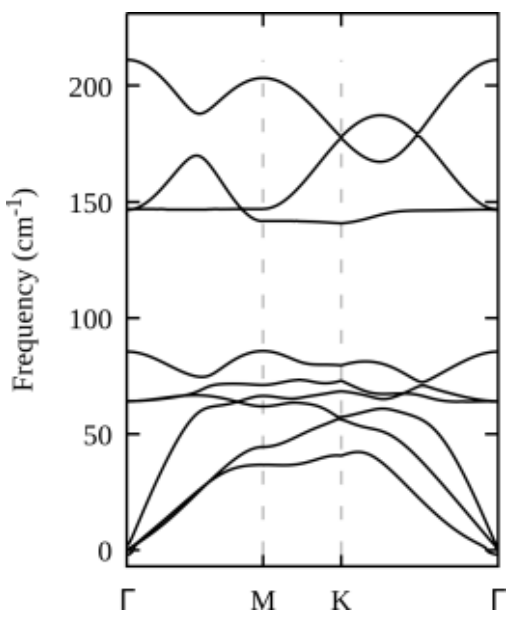

## Projected band structure and density of states

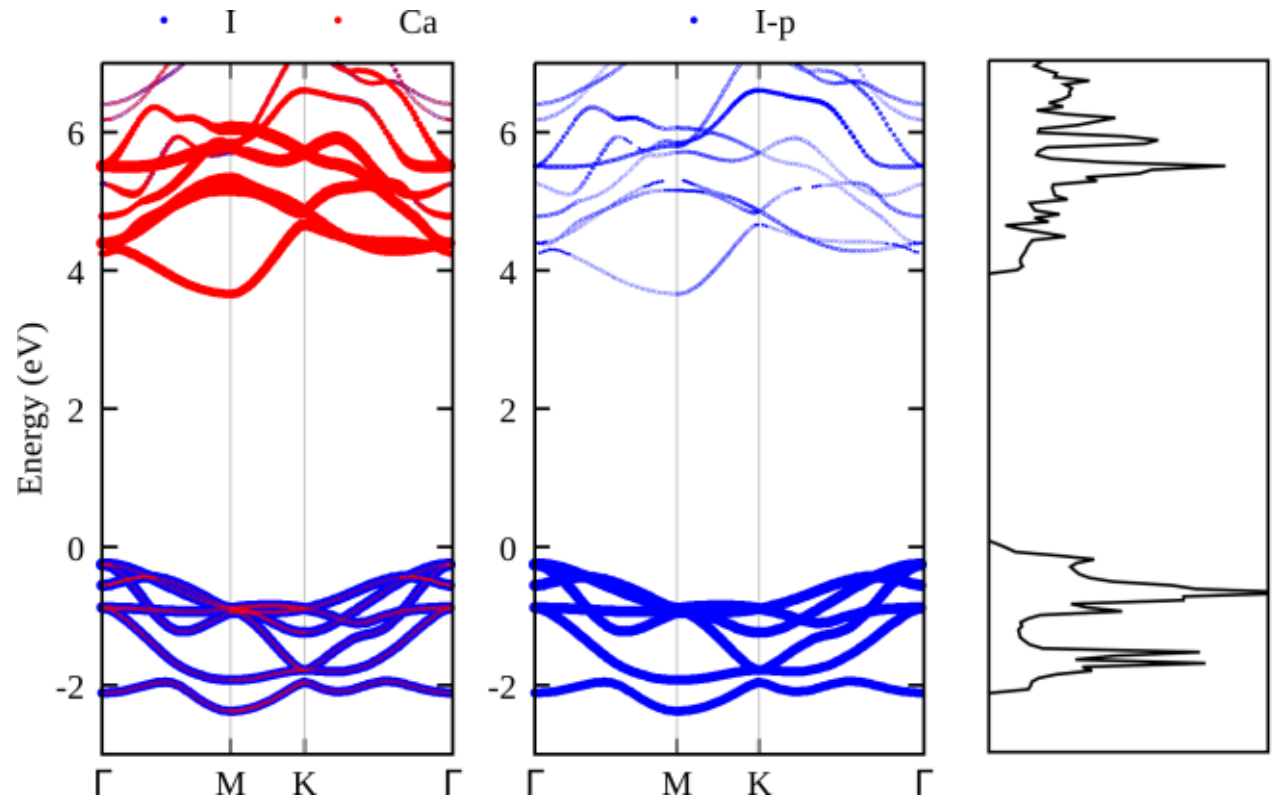

## Magnetic moment and spin polarization energy as a function of hole doping concentration

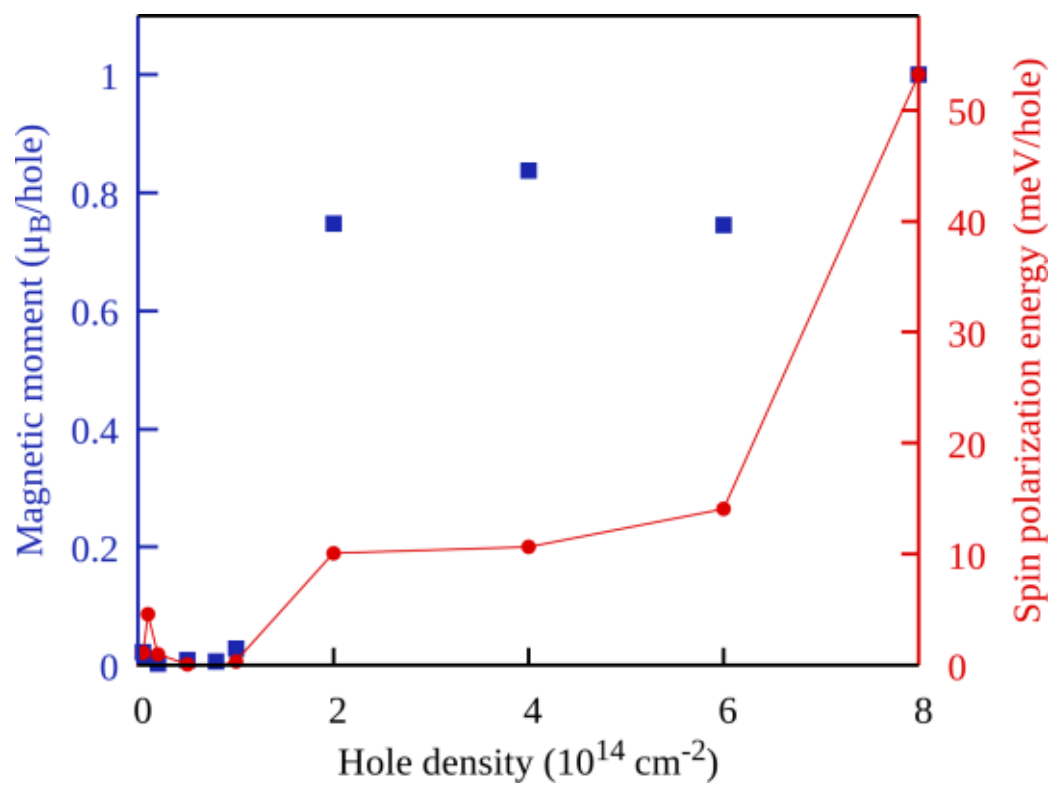

## Magnetic configurations and spin Hamiltonian

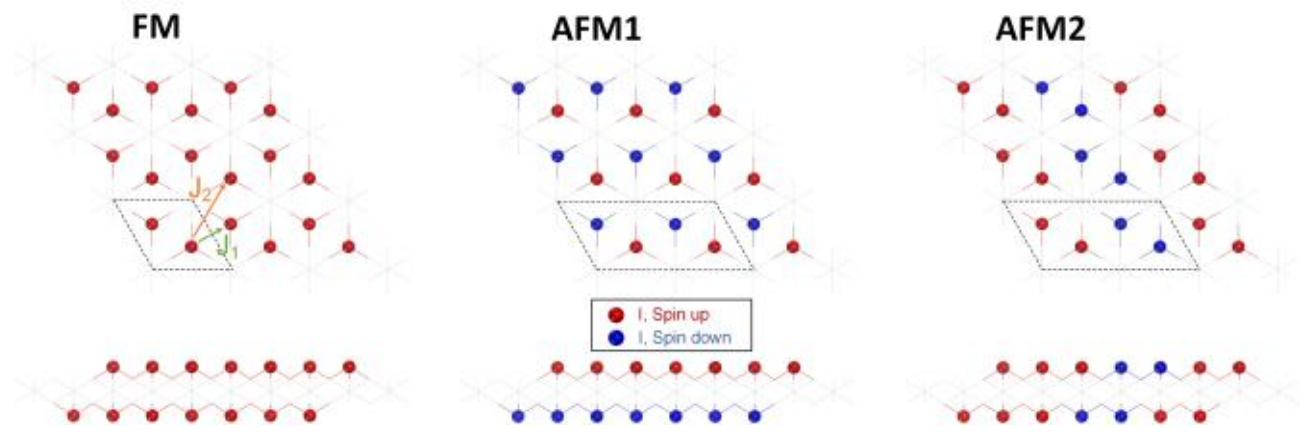

$$E_{FM} = E_0 - 3J_1S^2 - 6J_2S^2$$
$$E_{AFM1} = E_0 + 3J_1S^2 - 6J_2S^2$$
$$E_{AFM2} = E_0 - J_1S^2 + 2J_2S^2$$

## Magnetic exchange coupling parameters

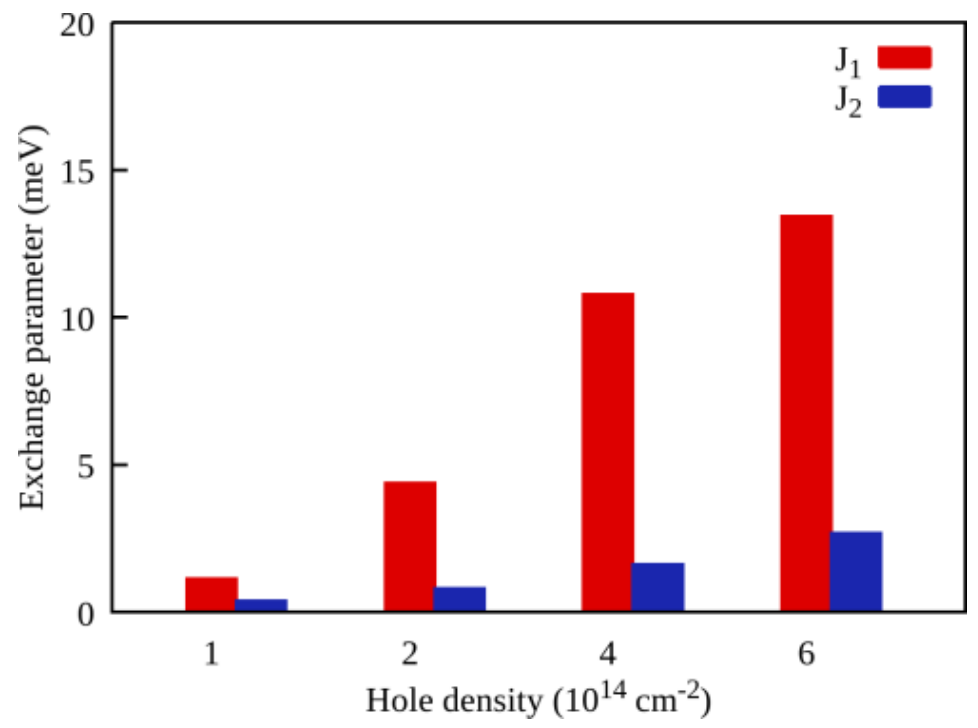

## Magnetic anisotropy energy (MAE, μeV) per magnetic atom

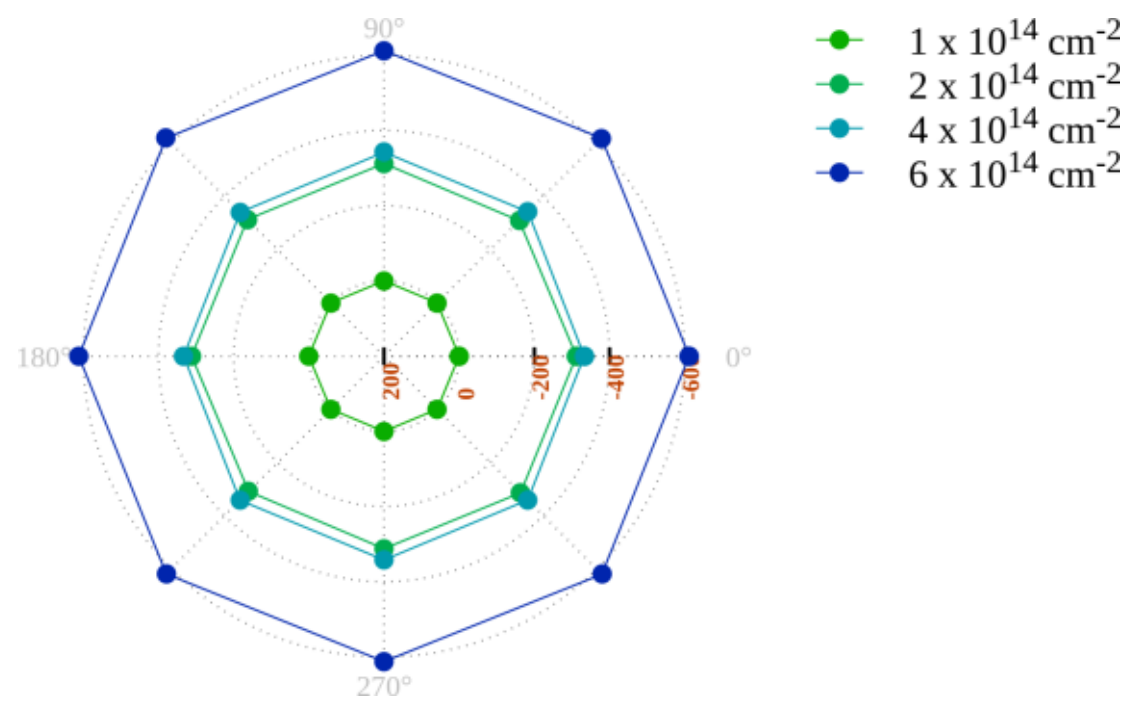

MAE = $E_{\parallel} - E_{\perp}$, a positive (negative) value of MAE indicates the off-plane (in-plane) easy axis.

## Monte Carlo simulations of the normalized magnetization of as a function of temperature

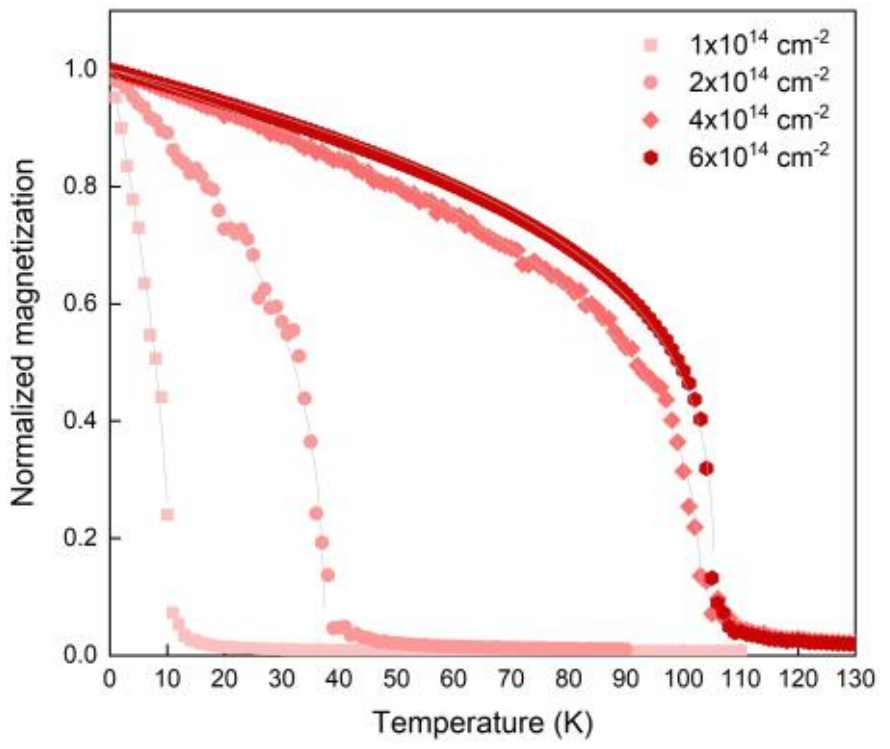

$T_c$: $1\times10^{14}$ cm$^{-2}$: 11 K   $2\times10^{14}$ cm$^{-2}$: 37 K

$4\times10^{14}$ cm$^{-2}$: 103 K   $6\times10^{14}$ cm$^{-2}$: 105 K

# 76.$SrF_2$

| MC2D-ID | C2DB | 2dmat-ID | USPEX | Space group | Band gap (eV) |
|---|---|---|---|---|---|
| - | - | 2dm-736 | - | P4m2 | 5.95 |

**Convex hull**

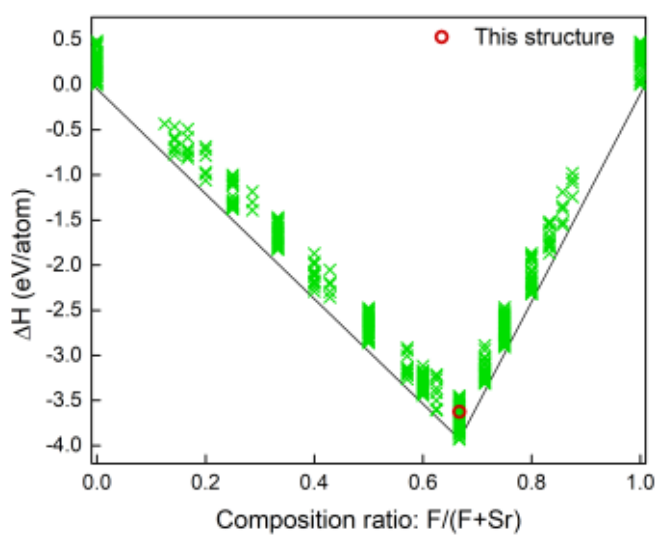

**Atomic structure**

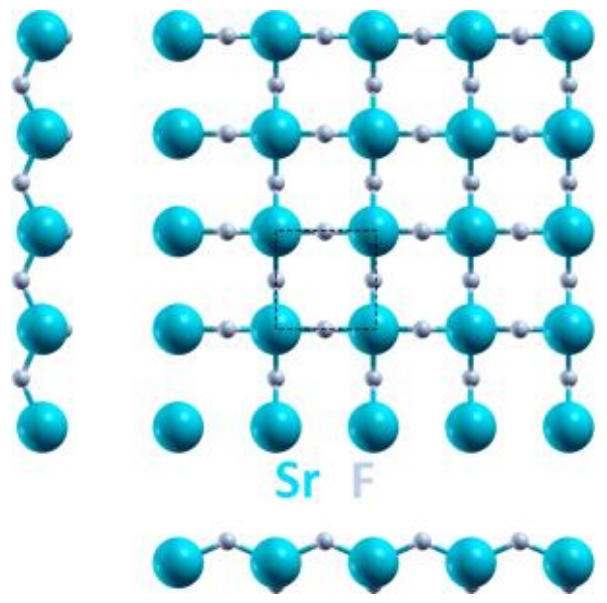

**Atomic coordinates**

```
Sr1F2
   1.00000000000000
   4.2629461998541327    0.0000000000000000    0.0000000000000000
   0.0000000000000000    4.2629736922051258    0.0000000000000000
   0.0000000000000000    0.0000000000000000   22.6561909999999997
   Sr    F
   1     2
Direct
  0.0000003237238957  0.0000003252872460  0.9999991898838232
  0.9999998753638977  0.4999997930998674  0.9587276808113430
  0.4999998009122066  0.9999998816128794  0.0412731293048409
```

**Phonon dispersion curve**

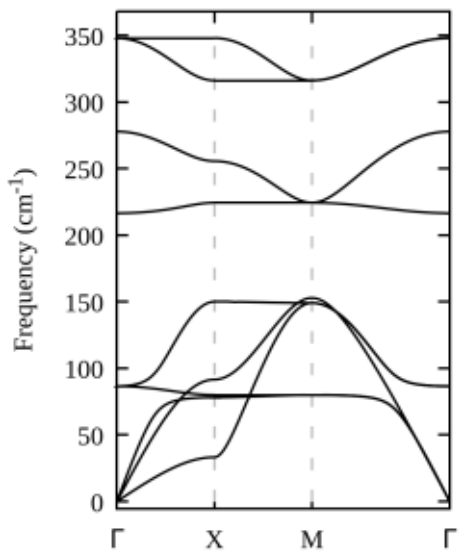

**Projected band structure and density of states**

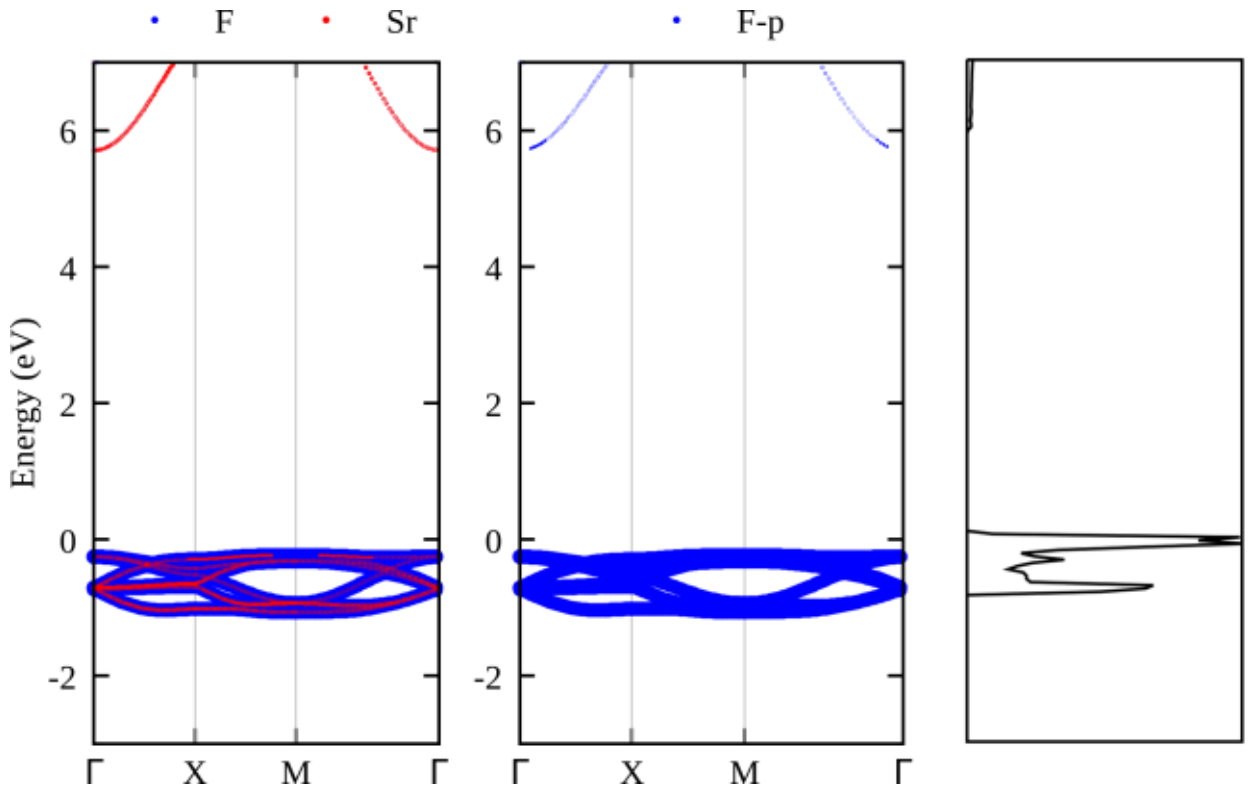

**Magnetic moment and spin polarization energy as a function of hole doping concentration**

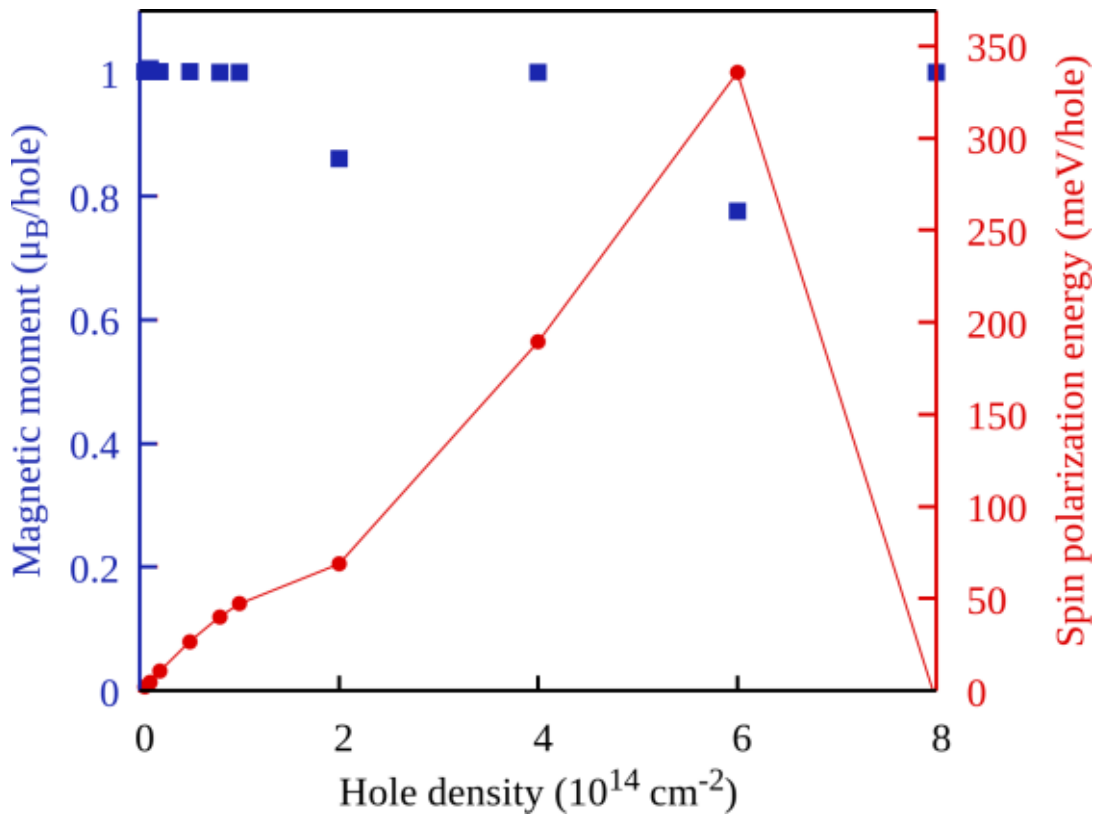

**Magnetic configurations and spin Hamiltonian**

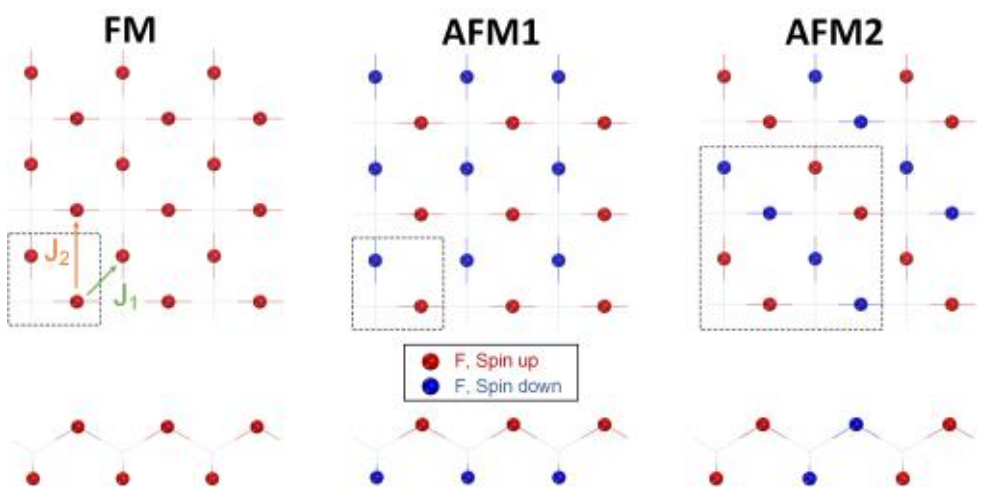

$$E_{FM} = E_0 - 4J_1S^2 - 4J_2S^2$$

$$E_{AFM1} = E_0 + 4J_1S^2 - 4J_2S^2$$

$$E_{AFM2} = E_0 + 4J_2S^2$$

**Magnetic exchange coupling parameters**

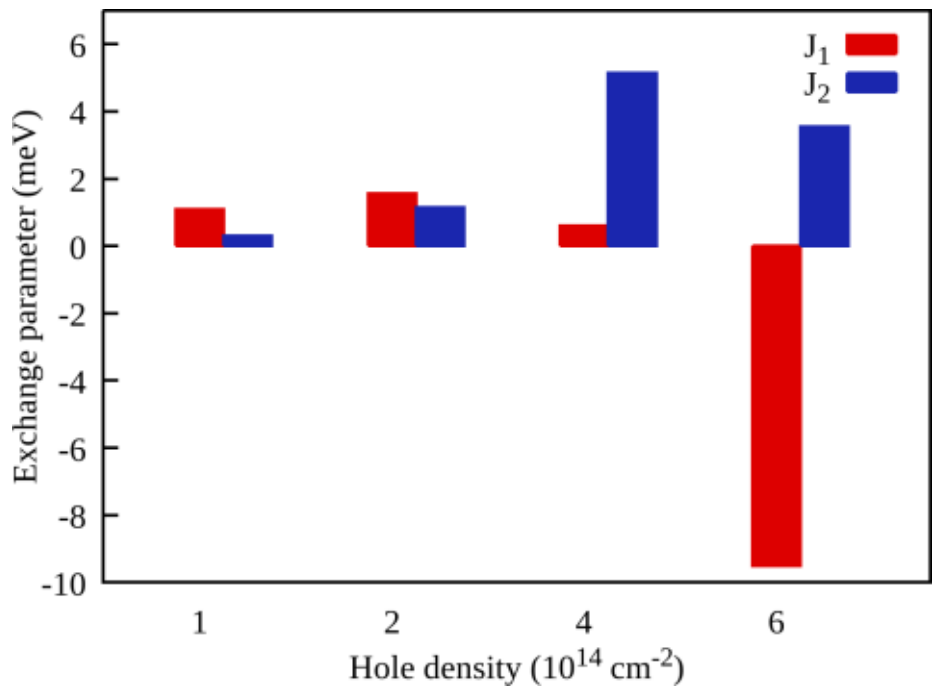

**Magnetic anisotropy energy (MAE, μeV) per magnetic atom**

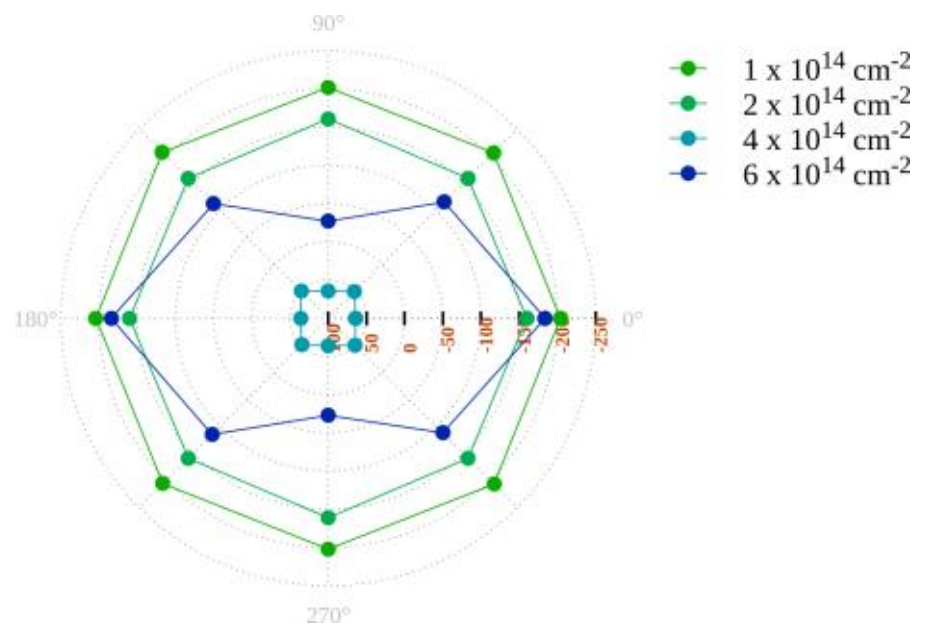

MAE = $E_{\parallel} - E_{\perp}$, a positive (negative) value of MAE indicates the off-plane (in-plane) easy axis.

**Monte Carlo simulations of the normalized magnetization of as a function of temperature**

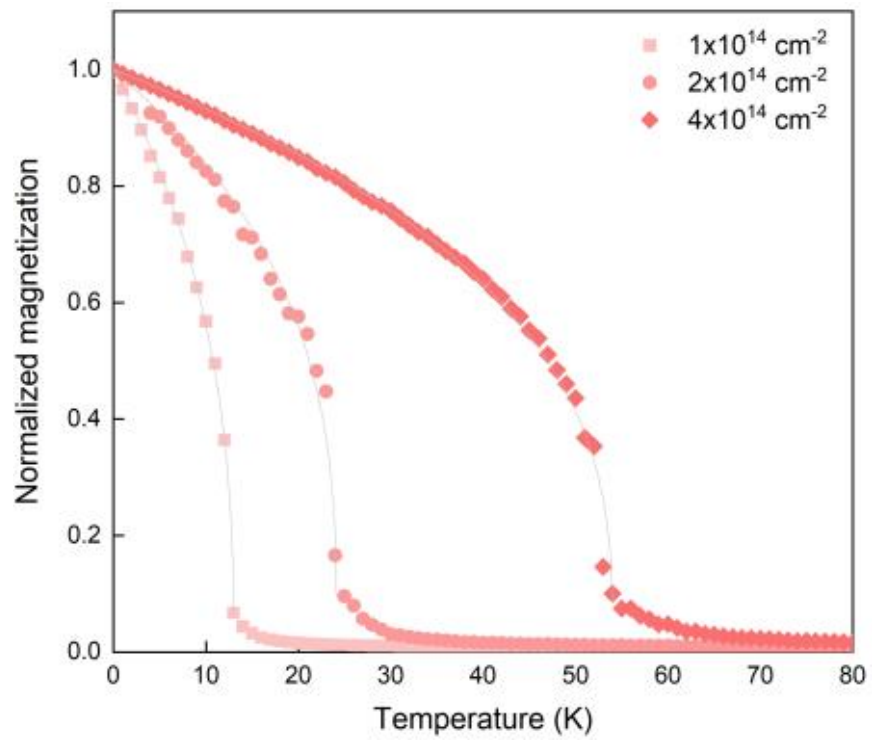

$T_c$: $1\times10^{14}$ cm$^{-2}$: 13 K  $2\times10^{14}$ cm$^{-2}$: 24 K

$4\times10^{14}$ cm$^{-2}$: 54 K  $6\times10^{14}$ cm$^{-2}$: - K

# 77.SrCl₂

| MC2D-ID | C2DB | 2dmat-ID | USPEX | Space group | Band gap (eV) |
|---|---|---|---|---|---|
| - | - | 2dm-1453 | - | P4m2 | 5.31 |

**Convex hull** | **Atomic structure** | **Atomic coordinates** | **Phonon dispersion curve**

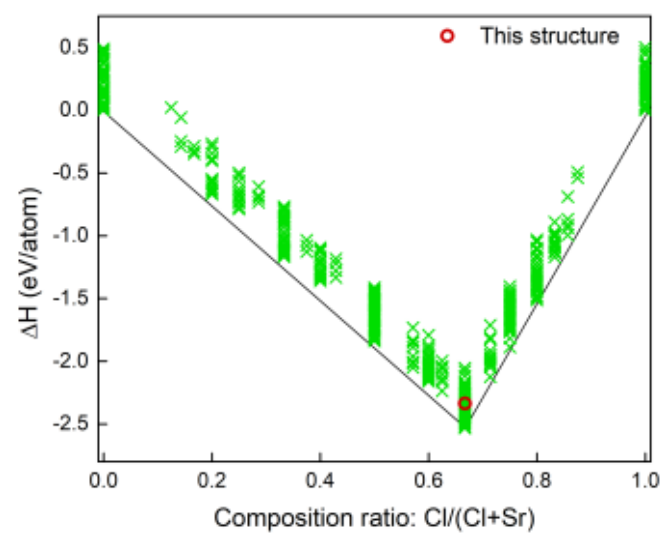

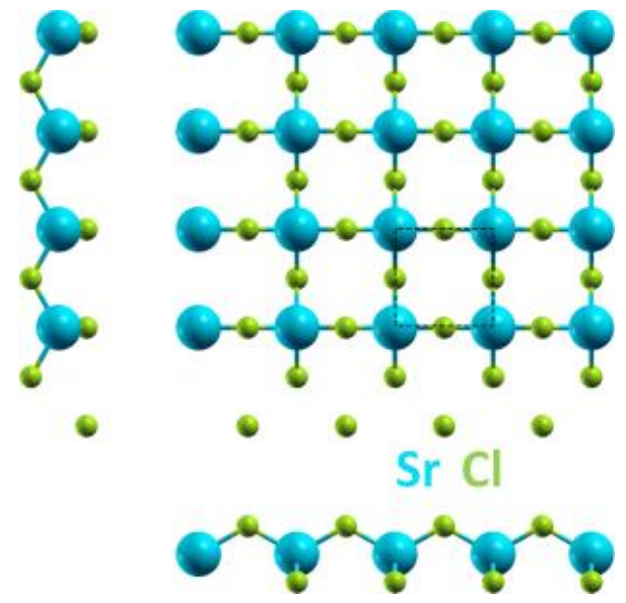

```
Sr1Cl2
   1.00000000000000
   4.8543299045047013   0.0000000000000000   0.0000000000000000
   0.0000000000000000   4.8543300630774118   0.0000000000000000
   0.0000000000000000   0.0000000000000000   22.6561909999999997
   Sr   Cl
   1    2
Direct
  0.0000000025609310  0.9999999744235950  0.9999999912219621
  0.9999999772453876  0.4999999599091751  0.9392024719621190
  0.5000000201936814  0.0000000656672441  0.0607975368159188
```

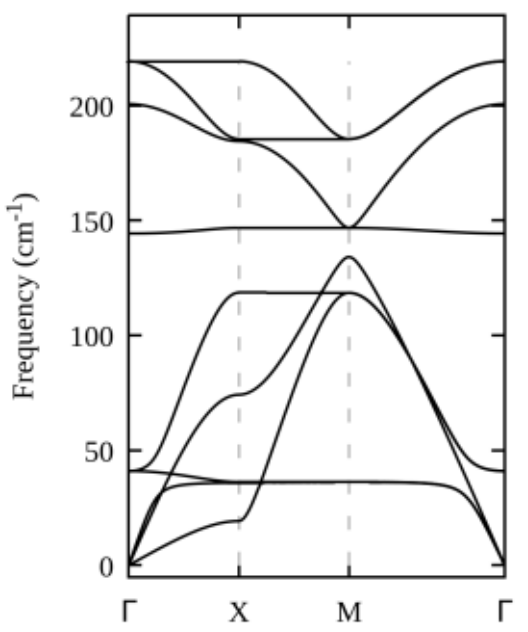

**Projected band structure and density of states** | **Magnetic moment and spin polarization energy as a function of hole doping concentration**

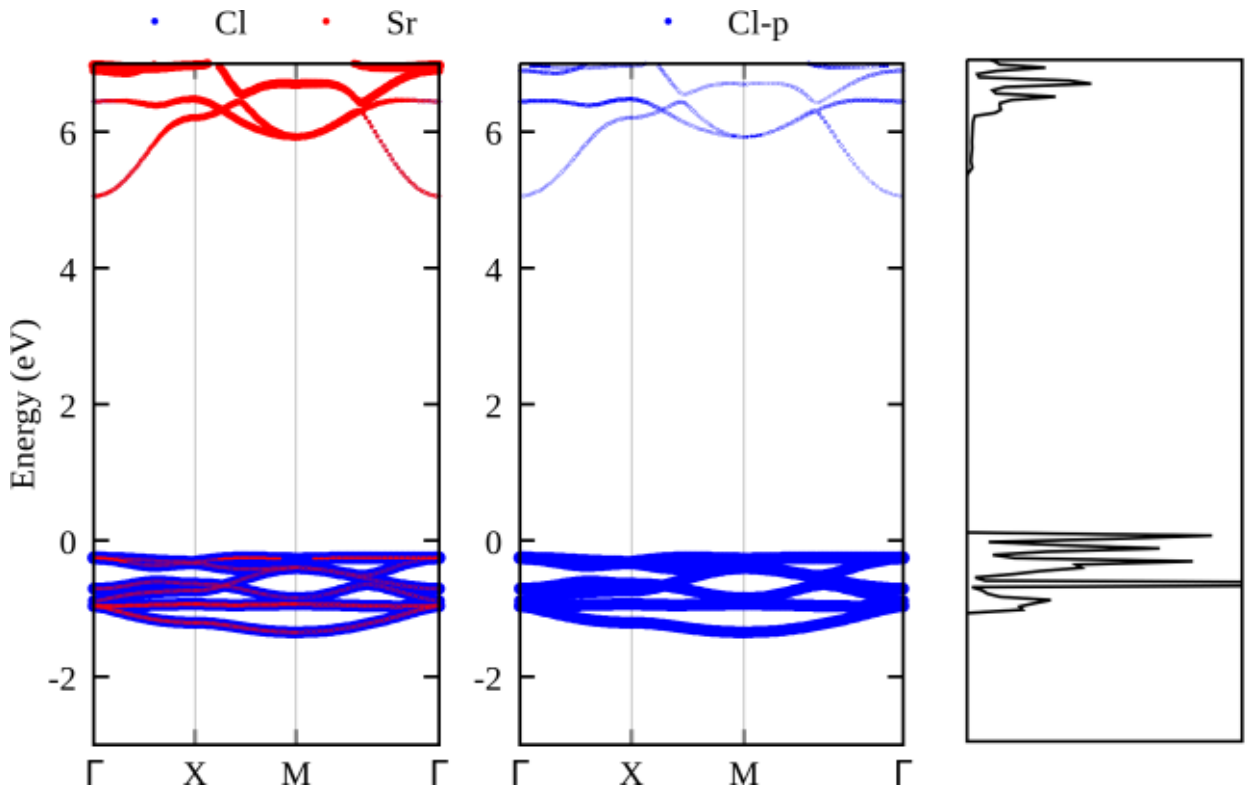

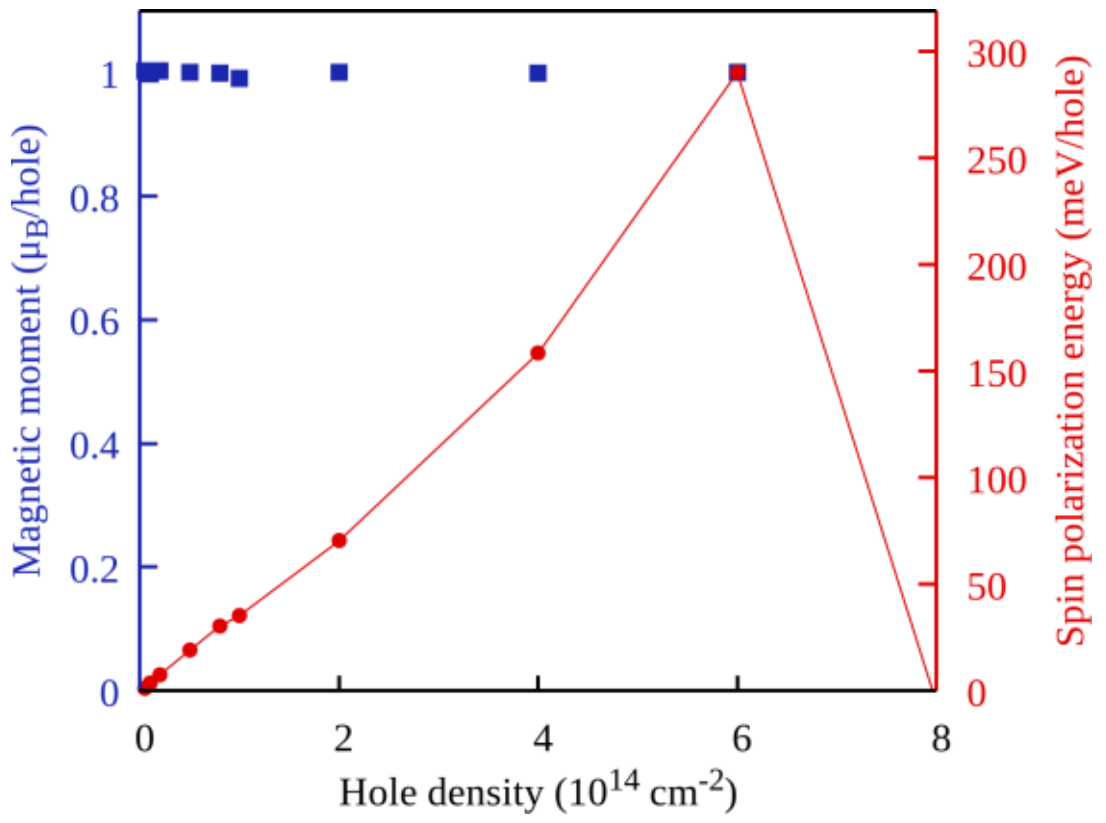

**Magnetic configurations and spin Hamiltonian** | **Magnetic exchange coupling parameters**

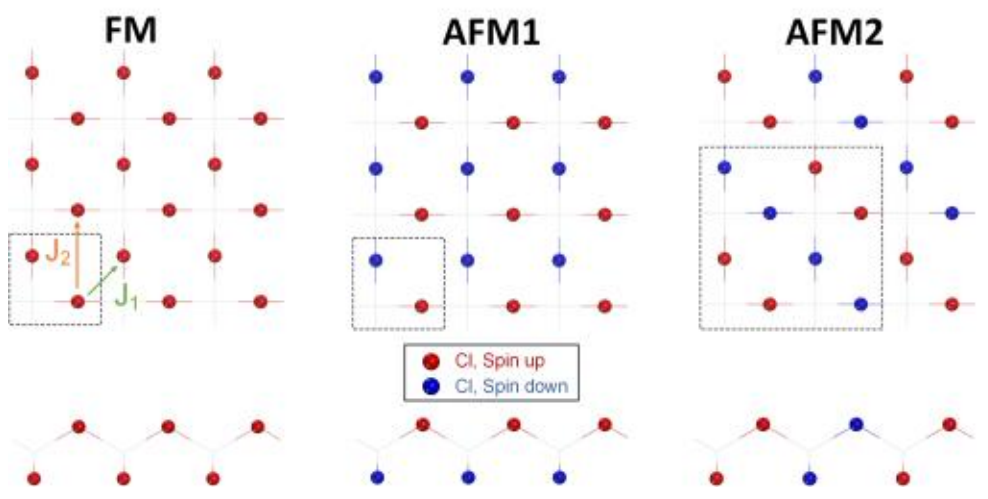

$$E_{FM} = E_0 - 4J_1S^2 - 4J_2S^2$$
$$E_{AFM1} = E_0 + 4J_1S^2 - 4J_2S^2$$
$$E_{AFM2} = E_0 + 4J_2S^2$$

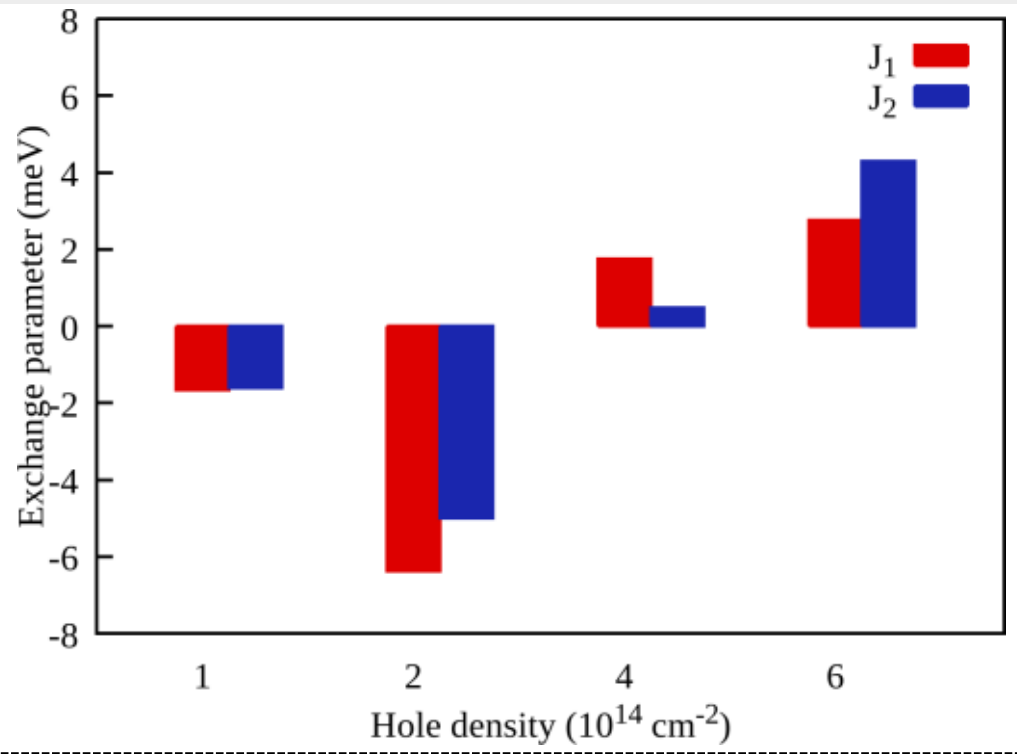

**Magnetic anisotropy energy (MAE, μeV) per magnetic atom** | **Monte Carlo simulations of the normalized magnetization of as a function of temperature**

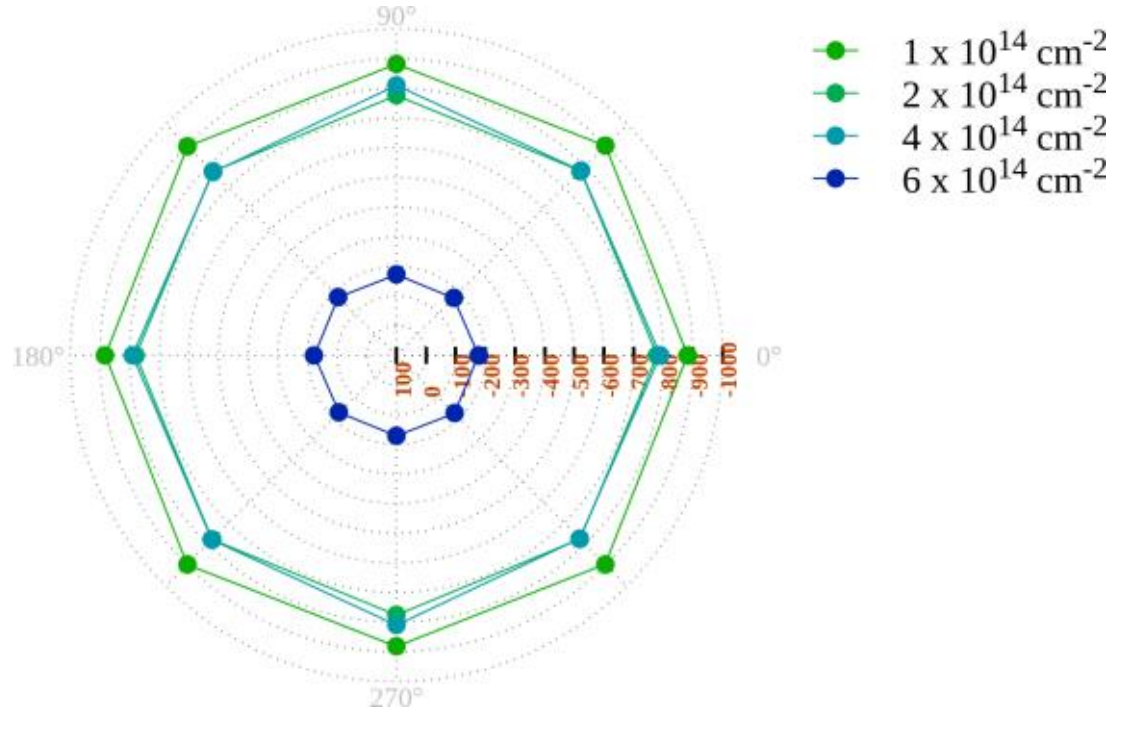

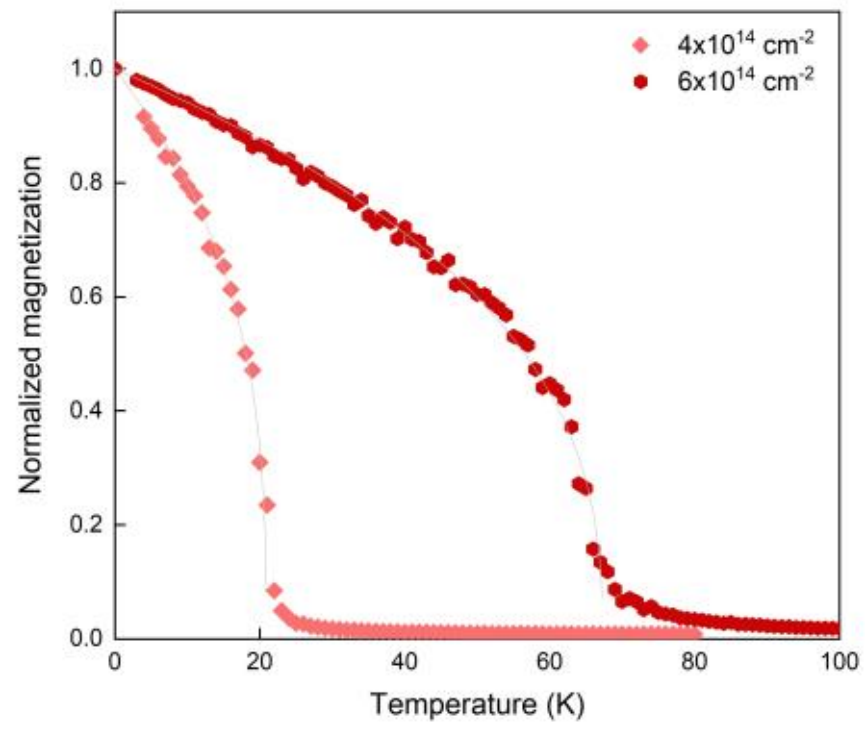

MAE = $E_{\parallel} - E_{\perp}$, a positive (negative) value of MAE indicates the off-plane (in-plane) easy axis.

$T_c$: $1\times10^{14}$ cm$^{-2}$: - K  $2\times10^{14}$ cm$^{-2}$: - K  $4\times10^{14}$ cm$^{-2}$: 21 K  $6\times10^{14}$ cm$^{-2}$: 67 K

# 78.$SrBr_2$

| MC2D-ID | C2DB | 2dmat-ID | USPEX | Space group | Band gap (eV) |
|---|---|---|---|---|---|
| - | - | 2dm-470 | - | P4m2 | 4.67 |

**Convex hull**

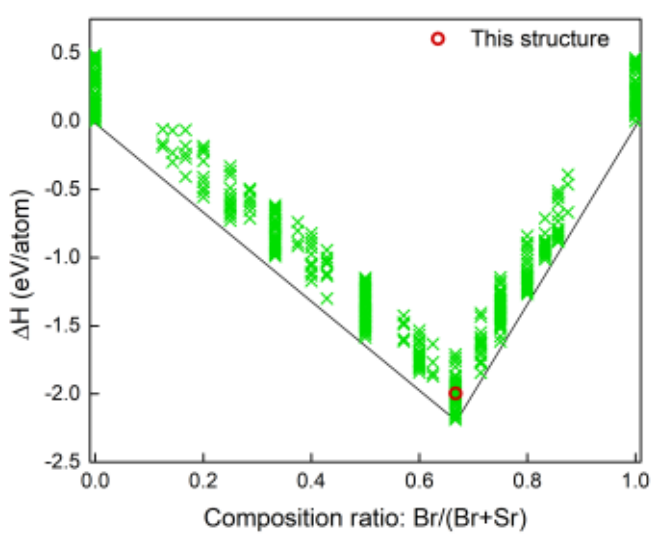

**Atomic structure**

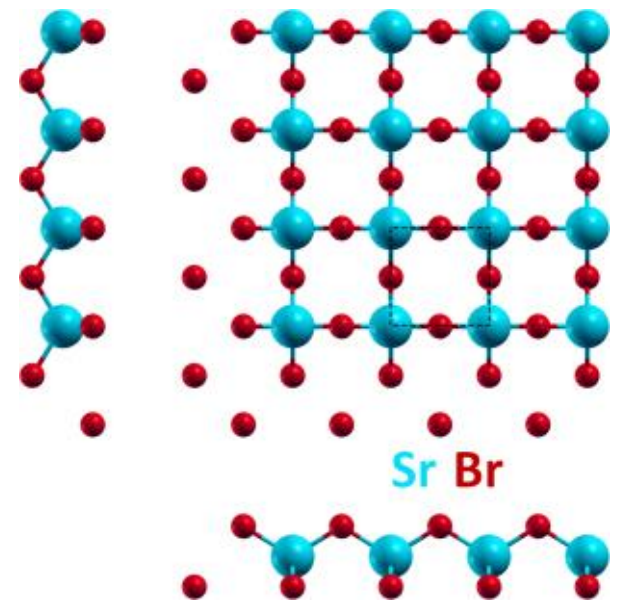

**Atomic coordinates**

```
Sr1Br2
   1.00000000000000
   5.0131425786126185   0.0000000000000000   0.0000000000000000
   0.0000000000000000   5.0131451710533899   0.0000000000000000
   0.0000000000000000   0.0000000000000000   22.6561909999999997
   Sr   Br
   1    2
Direct
  0.9999999712321994  0.9999999511638364  0.9999999691335972
  0.0000000168529155  0.4999999431819333  0.9317467581101084
  0.5000000119148922  0.0000001056542231  0.0682532727562943
```

**Phonon dispersion curve**

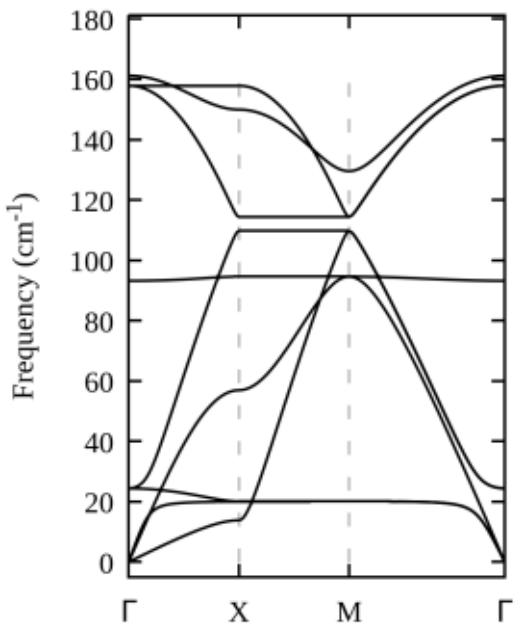

**Projected band structure and density of states**

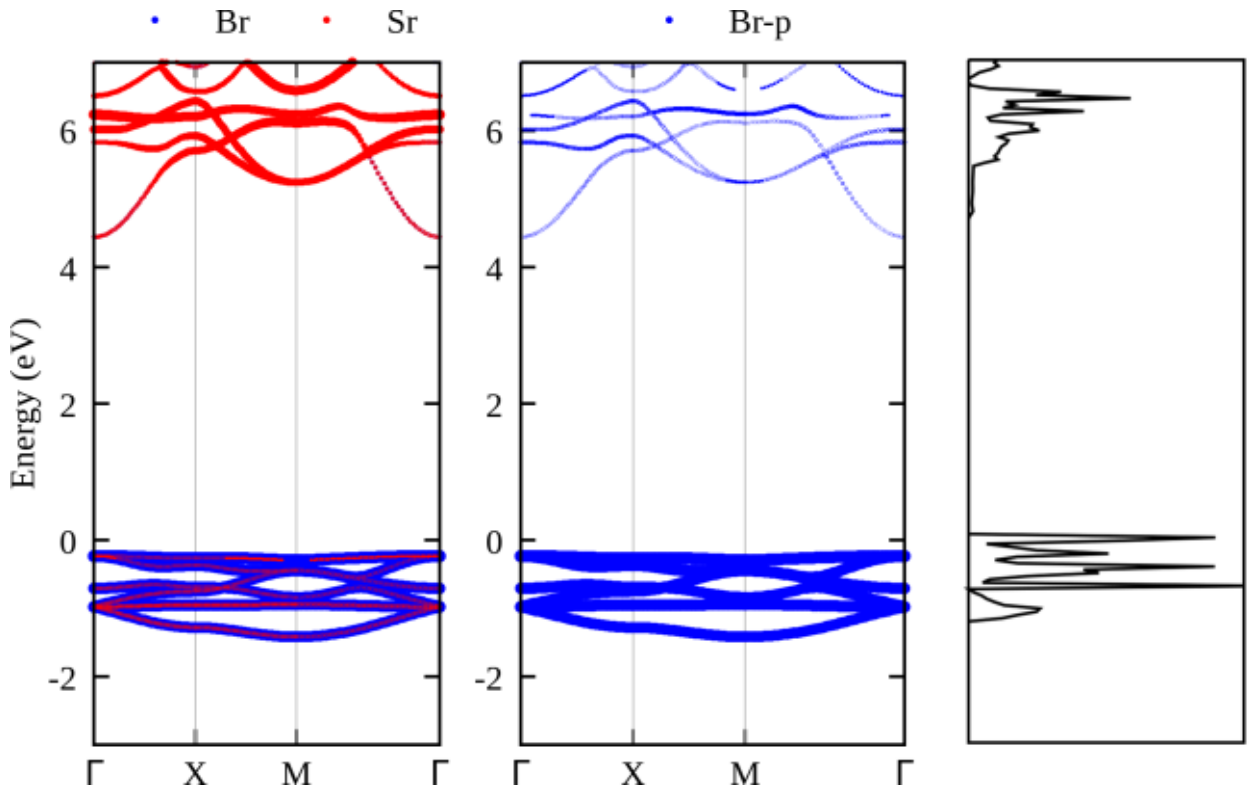

**Magnetic moment and spin polarization energy as a function of hole doping concentration**

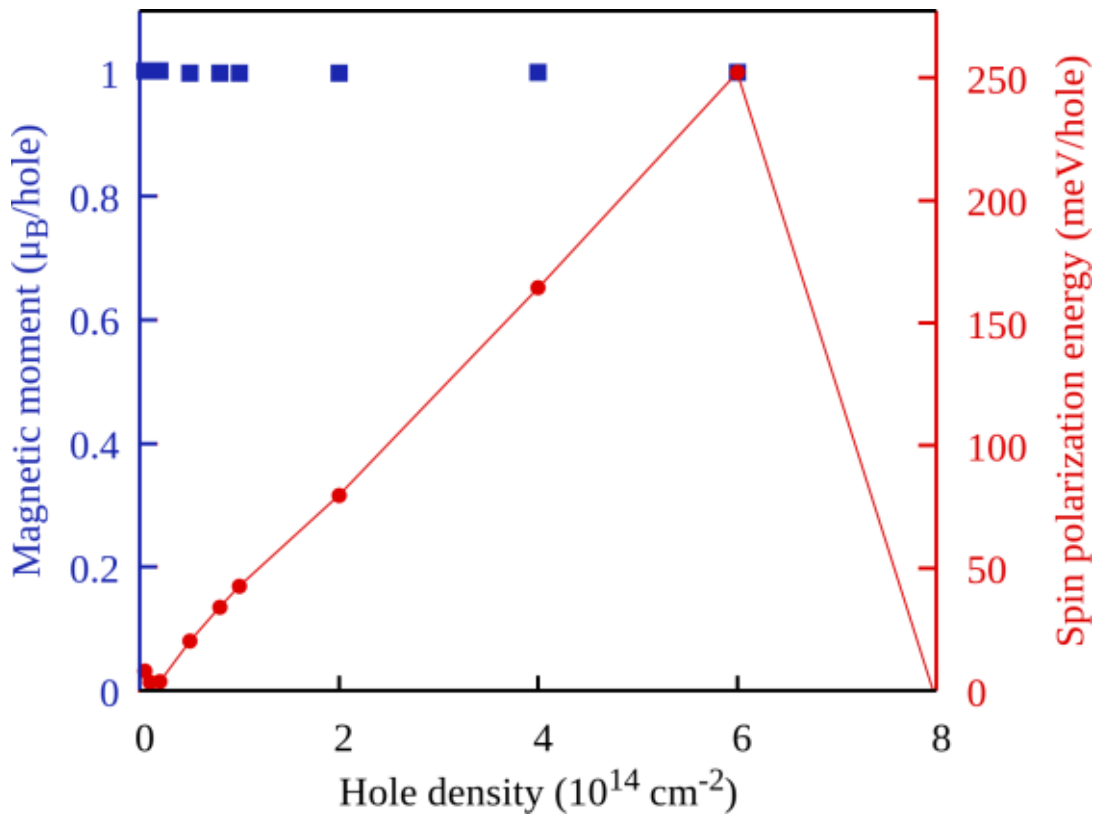

**Magnetic configurations and spin Hamiltonian**

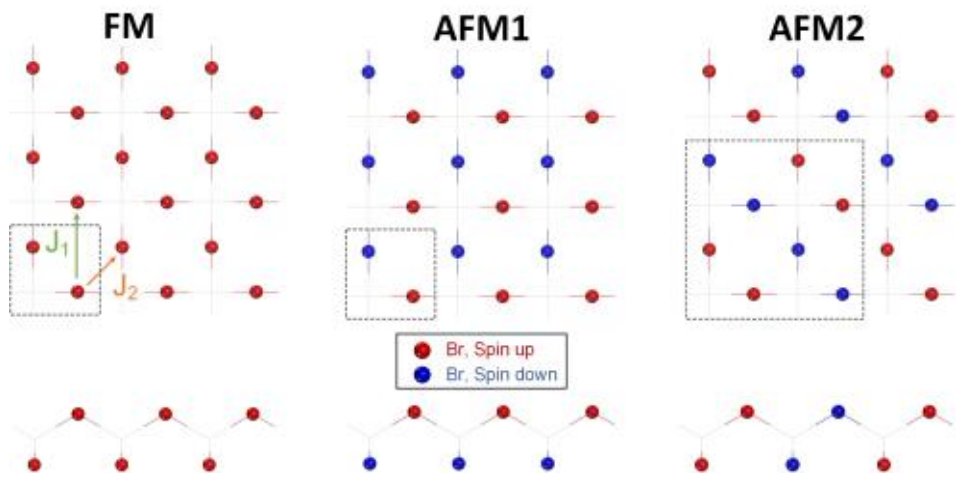

$$E_{FM} = E_0 - 4J_1S^2 - 4J_2S^2$$

$$E_{AFM1} = E_0 - 4J_1S^2 + 4J_2S^2$$

$$E_{AFM2} = E_0 + 4J_1S^2$$

**Magnetic exchange coupling parameters**

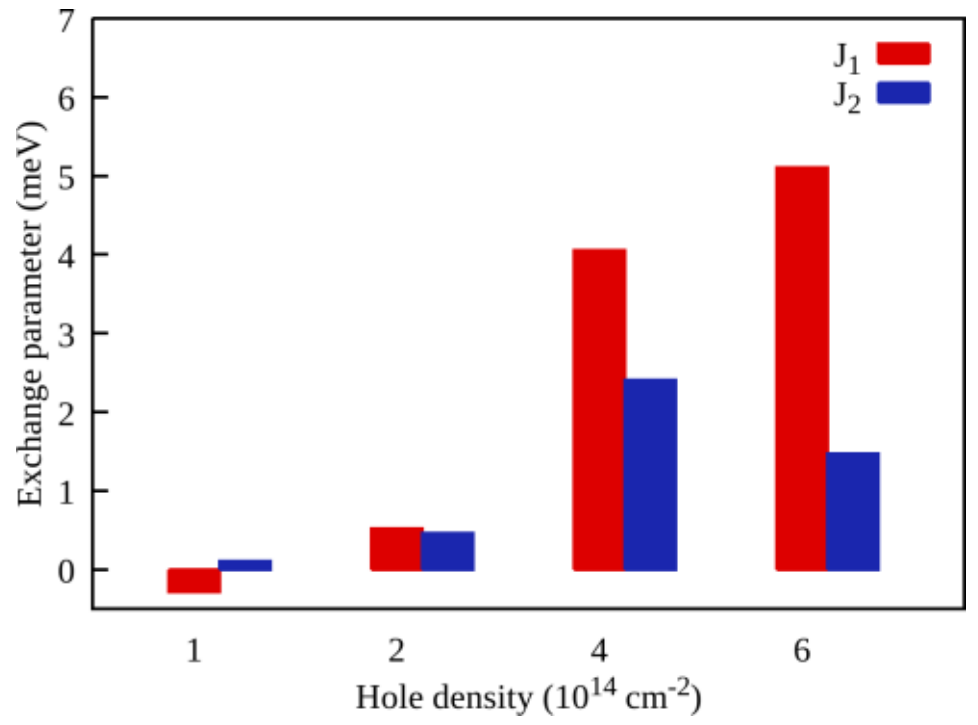

**Magnetic anisotropy energy (MAE, μeV) per magnetic atom**

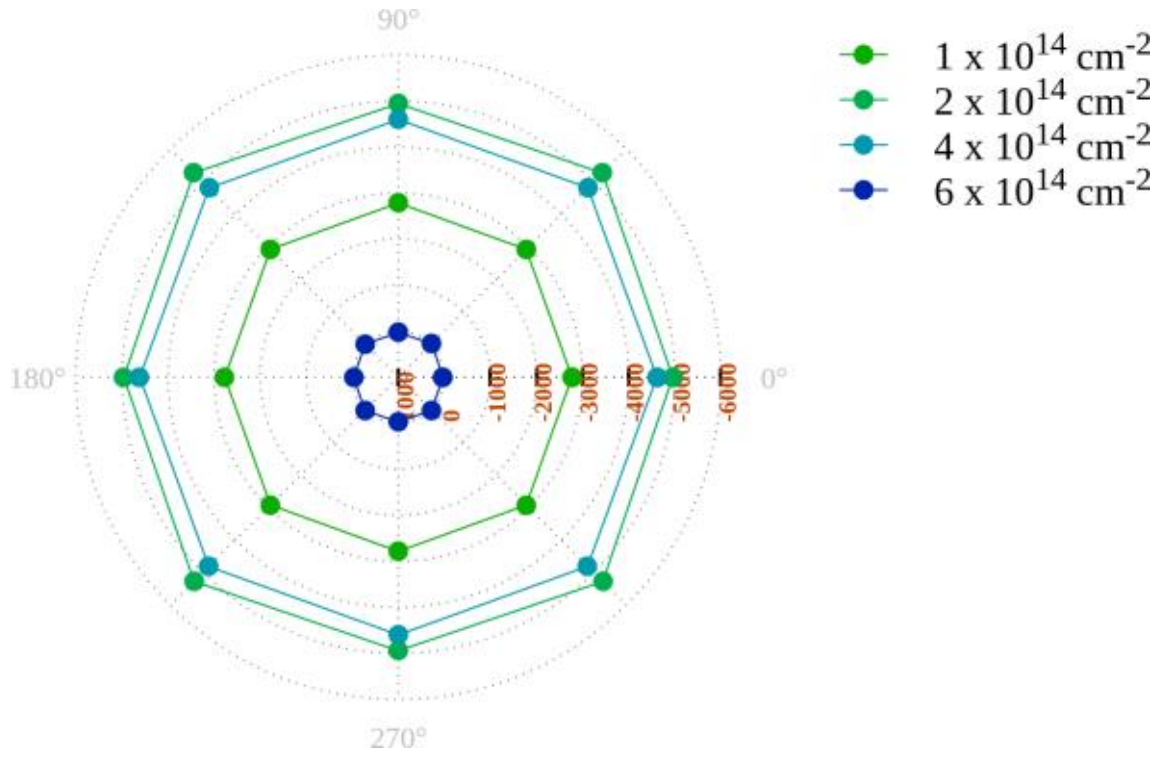

MAE = $E_{\parallel} - E_{\perp}$, a positive (negative) value of MAE indicates the off-plane (in-plane) easy axis.

**Monte Carlo simulations of the normalized magnetization of as a function of temperature**

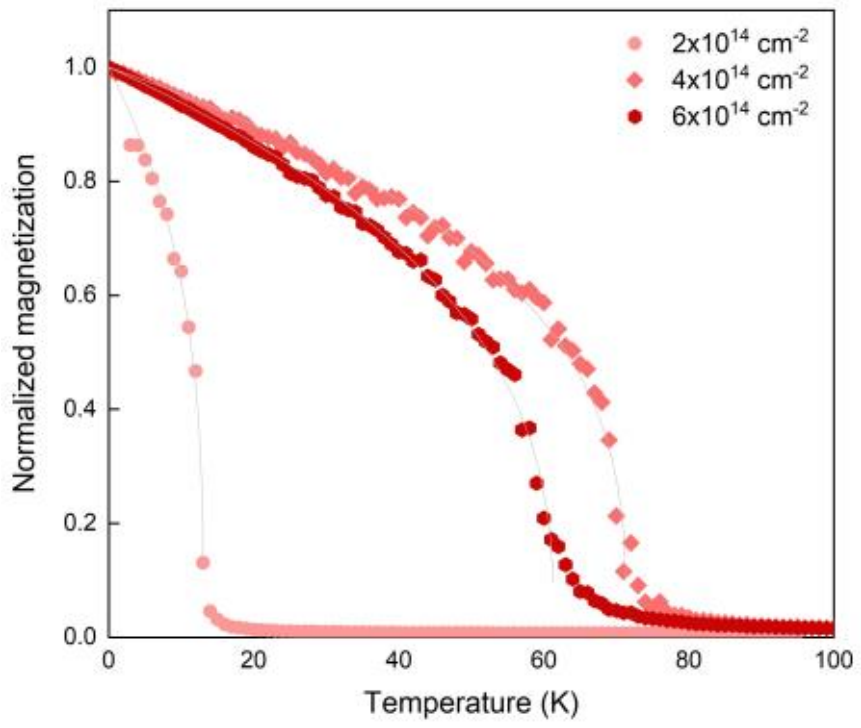

$T_c$: $1\times10^{14}$ cm$^{-2}$: - K  $2\times10^{14}$ cm$^{-2}$: 13 K  $4\times10^{14}$ cm$^{-2}$: 71 K  $6\times10^{14}$ cm$^{-2}$: 61 K

# 79.$SrI_2$

| MC2D-ID | C2DB | 2dmat-ID | USPEX | Space group | Band gap (eV) |
|---|---|---|---|---|---|
| - | - | 2dm-1251 | - | P4m2 | 4.15 |

| Convex hull | Atomic structure | Atomic coordinates | Phonon dispersion curve |
|---|---|---|---|

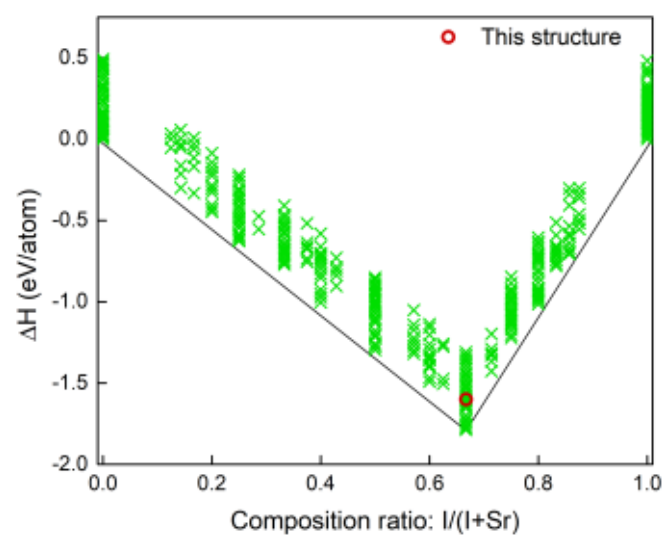

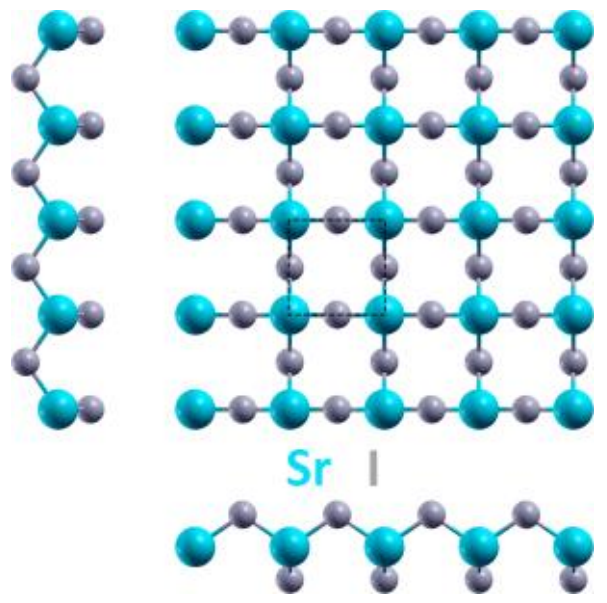

Sr1I2
1.00000000000000
5.2456463437361229 0.0000000000000000 0.0000000000000000
0.0000000000000000 5.2456598323804133 0.0000000000000000
0.0000000000000000 0.0000000000000000 22.6561909999999997
Sr I
1 2
Direct
0.0000000830145623 0.0000000068729662 0.9999998411190703
0.9999998019985270 0.4999999113557791 0.9208822368191676
0.5000001149869178 0.0000000817712547 0.0791179220617622

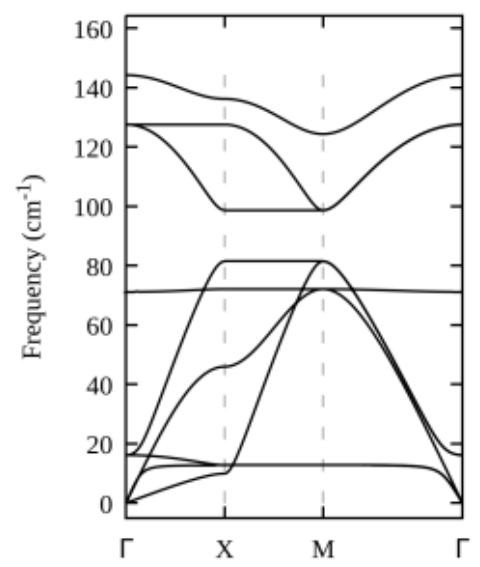

## Projected band structure and density of states

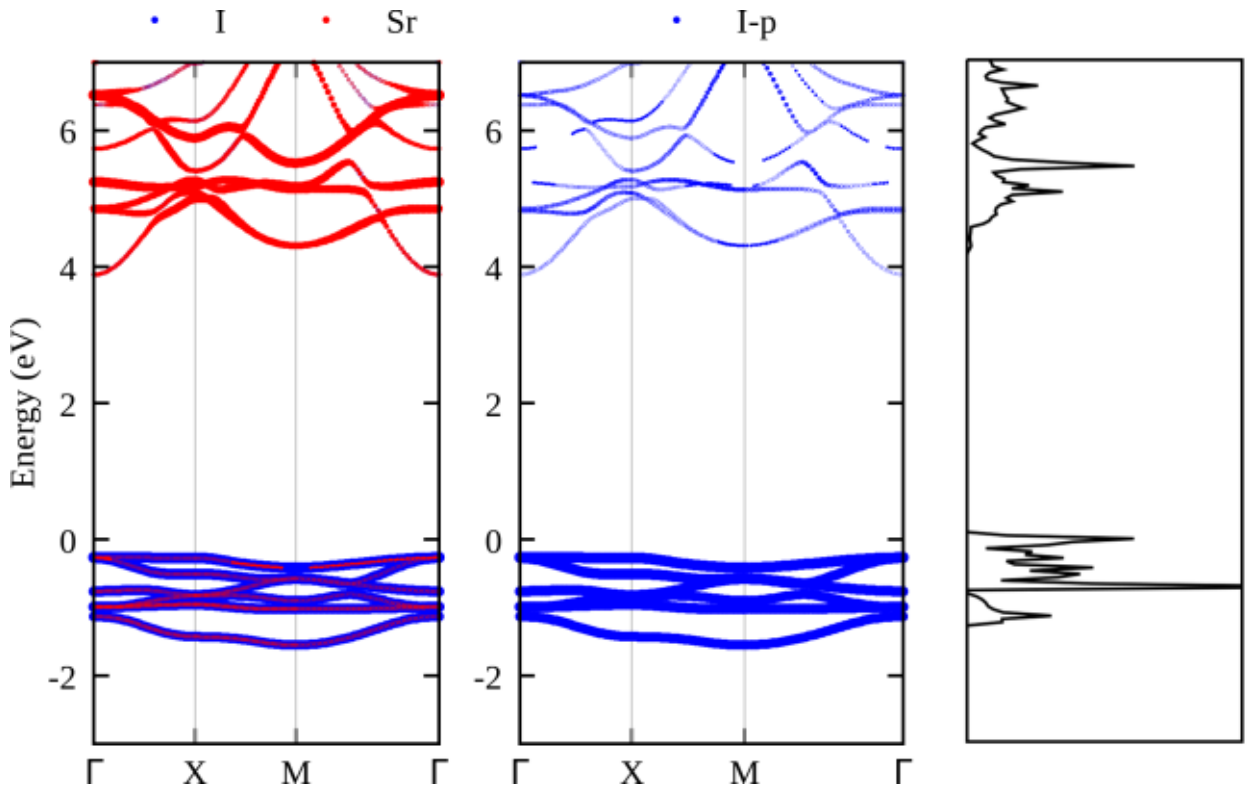

## Magnetic moment and spin polarization energy as a function of hole doping concentration

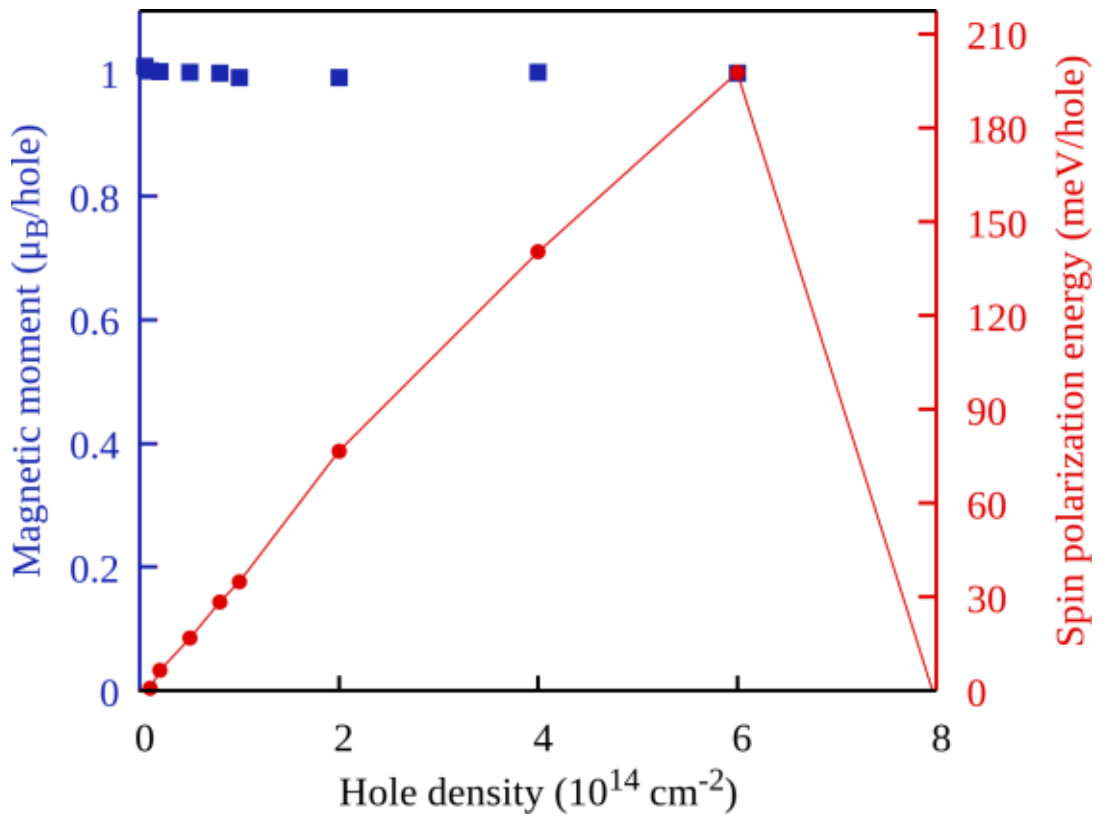

## Magnetic configurations and spin Hamiltonian

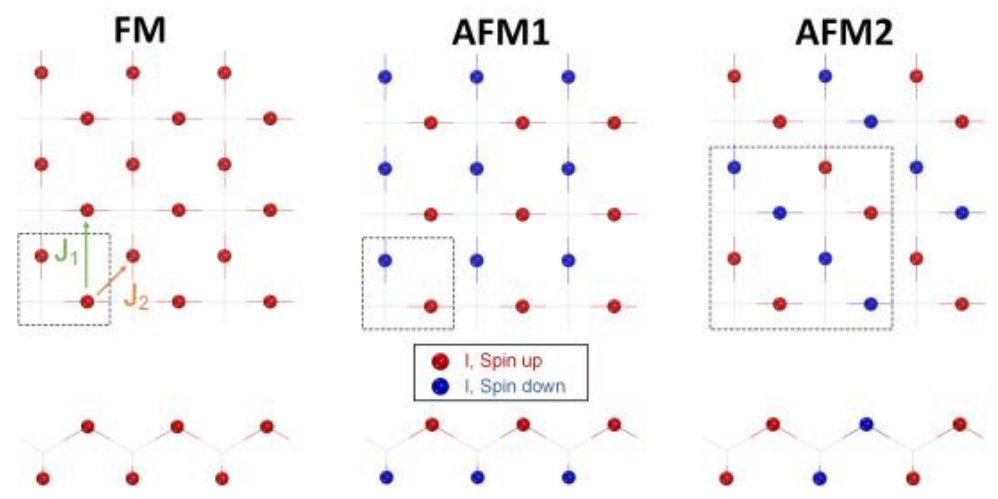

$$E_{FM} = E_0 - 4J_1S^2 - 4J_2S^2$$
$$E_{AFM1} = E_0 - 4J_1S^2 + 4J_2S^2$$
$$E_{AFM2} = E_0 + 4J_1S^2$$

## Magnetic exchange coupling parameters

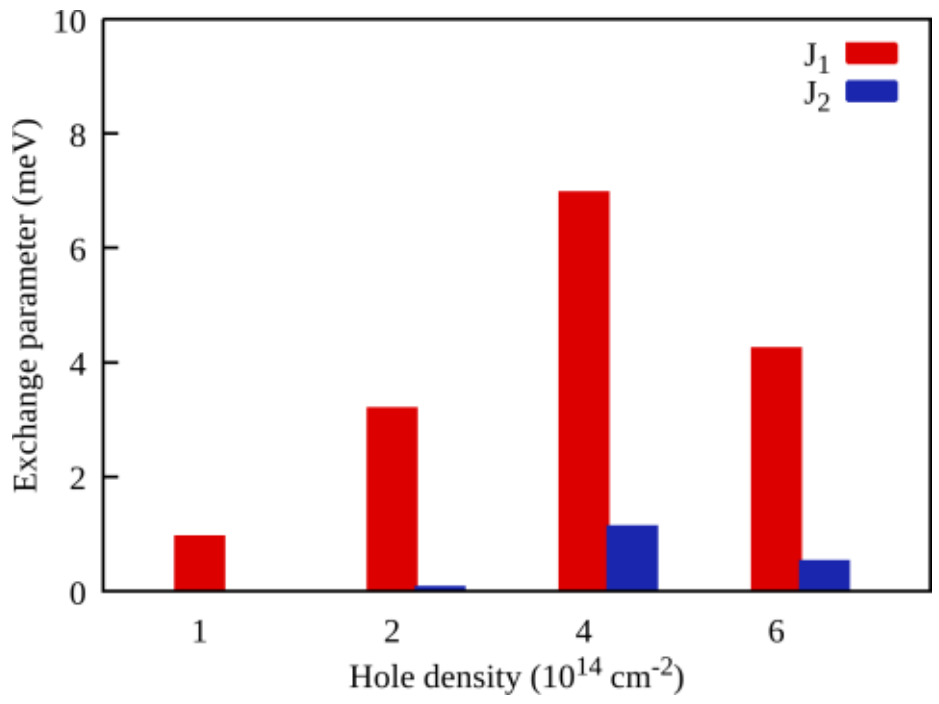

## Magnetic anisotropy energy (MAE, μeV) per magnetic atom

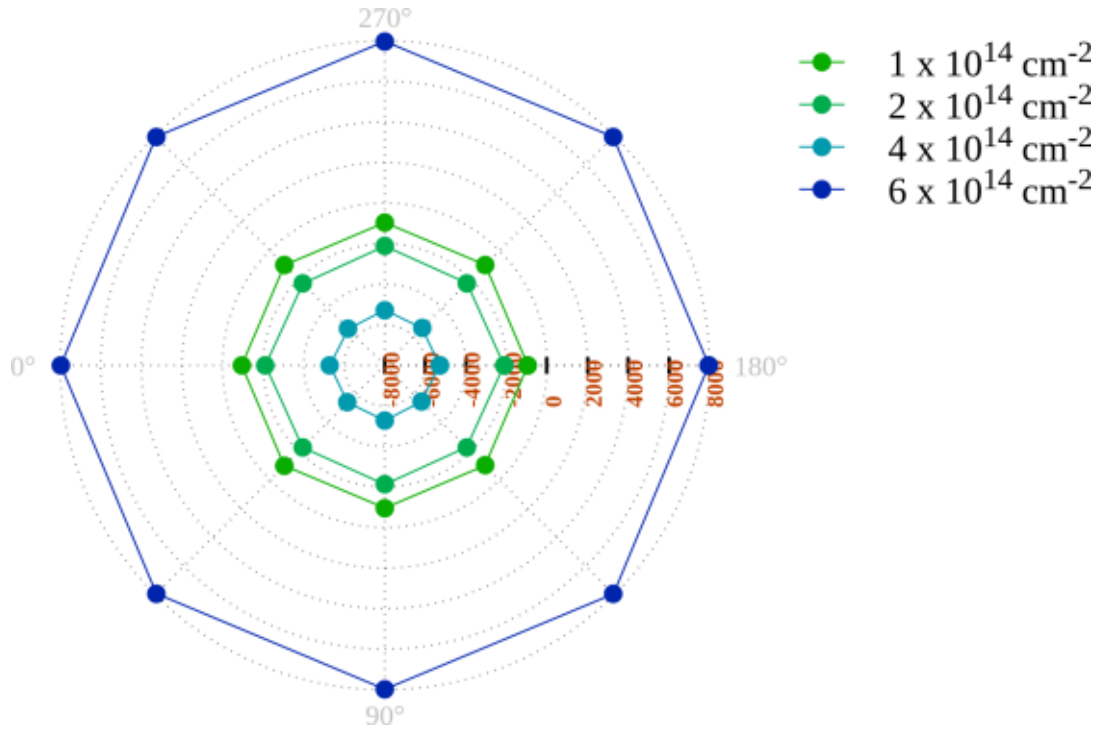

MAE = $E_{\parallel} - E_{\perp}$, a positive (negative) value of MAE indicates the off-plane (in-plane) easy axis.

## Monte Carlo simulations of the normalized magnetization of as a function of temperature

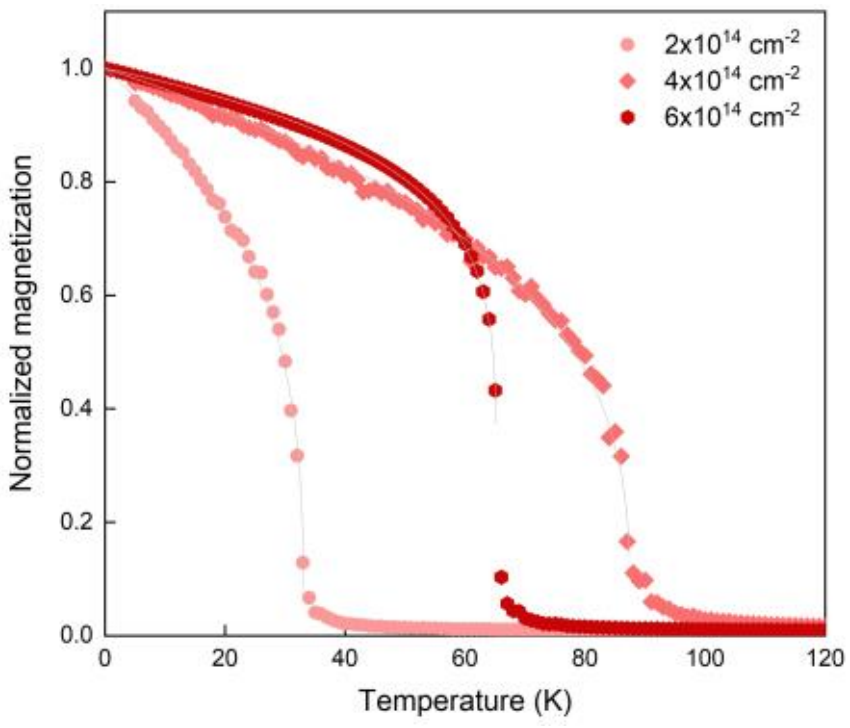

$T_c$: $1\times10^{14}$ cm$^{-2}$: - K　$2\times10^{14}$ cm$^{-2}$: 33 K

$4\times10^{14}$ cm$^{-2}$: 87 K　$6\times10^{14}$ cm$^{-2}$: 65 K

# 80.$BaF_2$

| MC2D-ID | C2DB | 2dmat-ID | USPEX | Space group | Band gap (eV) |
|---|---|---|---|---|---|
| - | - | 2dm-880 | - | P4m2 | 5.65 |

**Convex hull** | **Atomic structure** | **Atomic coordinates** | **Phonon dispersion curve**

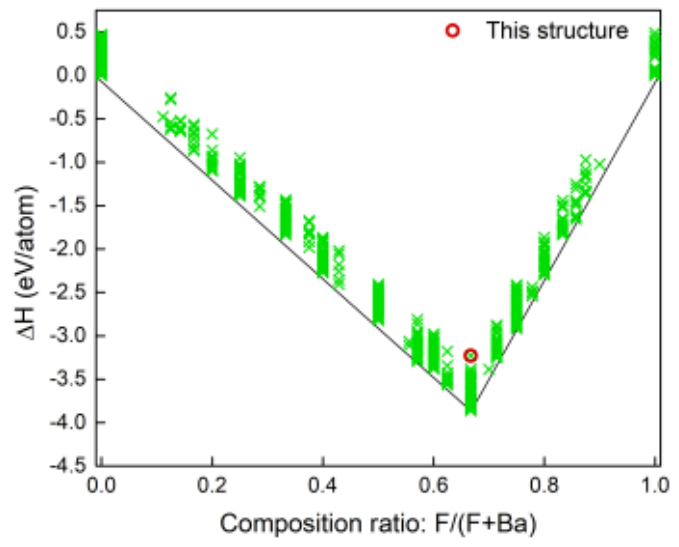

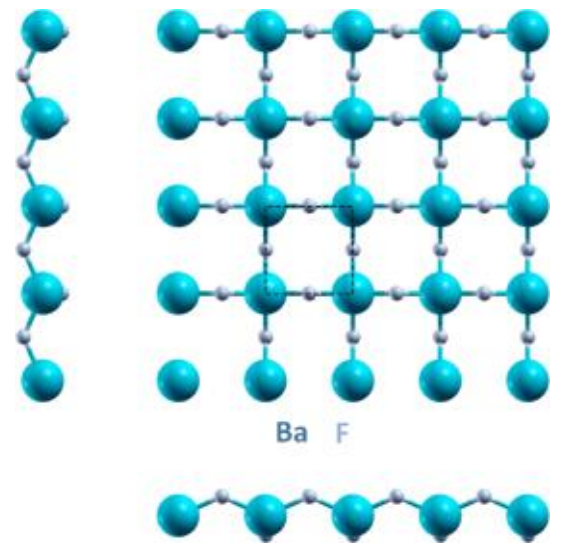

```
Ba1F2
1.00000000000000
4.5690708279870531    0.0000000000000000    0.0000000000000000
0.0000000000000000    4.5690607374883241    0.0000000000000000
\   0.0000000000000000    0.0000000000000000    22.6561909999999997
Ba    F
1     2
Direct
0.0000000045197126    0.0000000149795838    0.0000000784861740
0.9999999917091813    0.5000000175521677    0.9556433456282321
0.5000000037711061    0.9999999674682556    0.0443565758855939
```

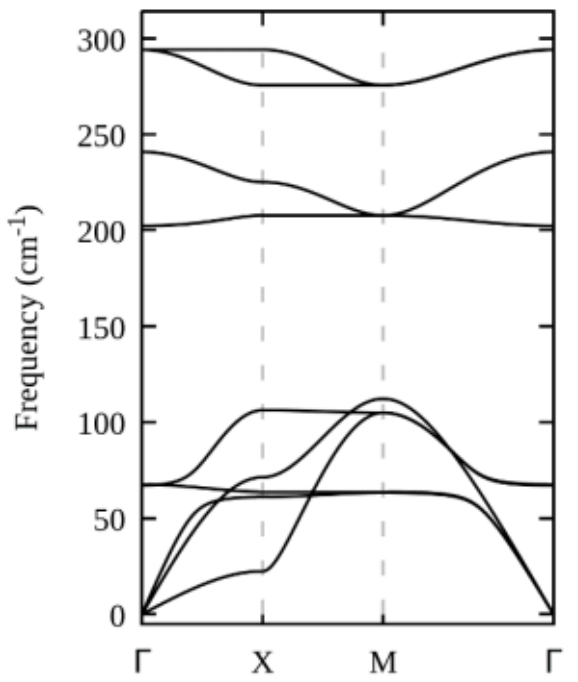

## Projected band structure and density of states

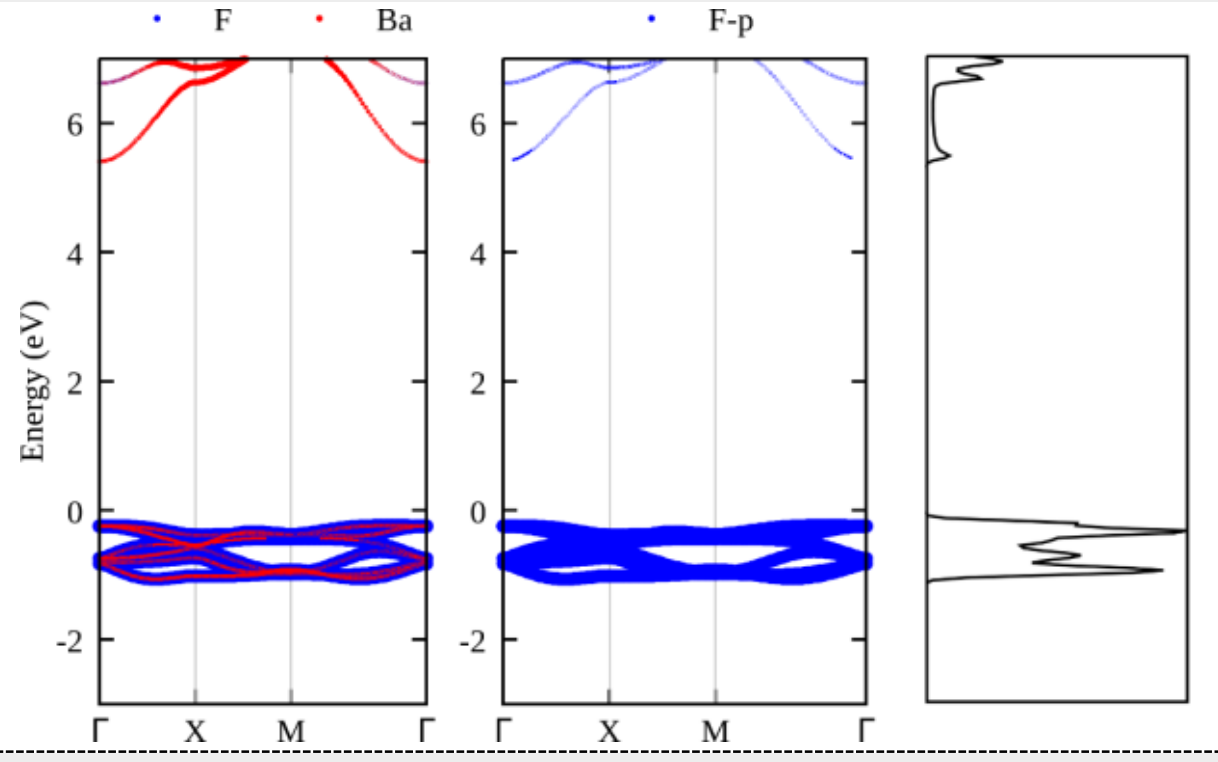

## Magnetic moment and spin polarization energy as a function of hole doping concentration

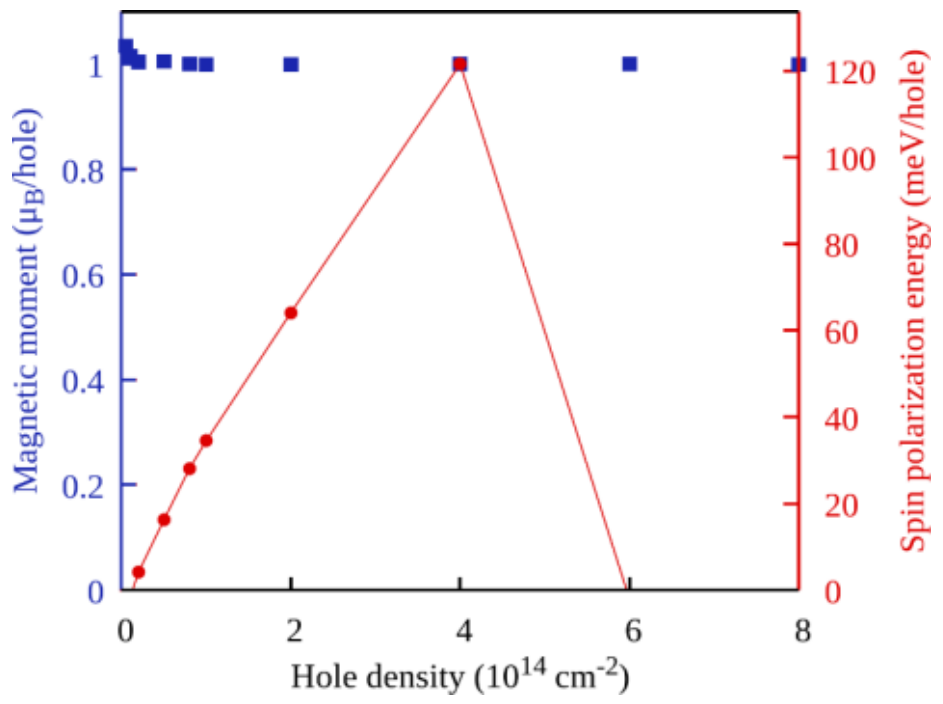

## Magnetic configurations and spin Hamiltonian

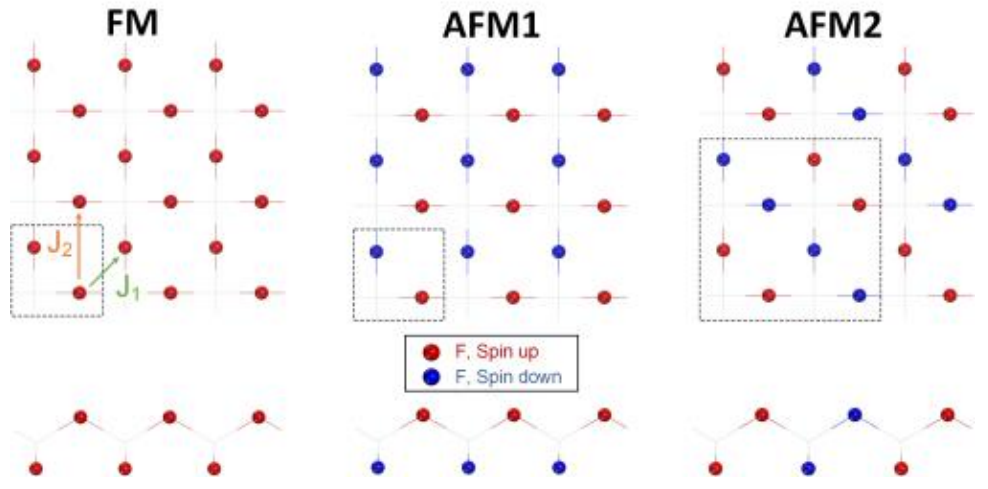

$$E_{FM} = E_0 - 4J_1S^2 - 4J_2S^2$$

$$E_{AFM1} = E_0 + 4J_1S^2 - 4J_2S^2$$

$$E_{AFM2} = E_0 + 4J_2S^2$$

## Magnetic exchange coupling parameters

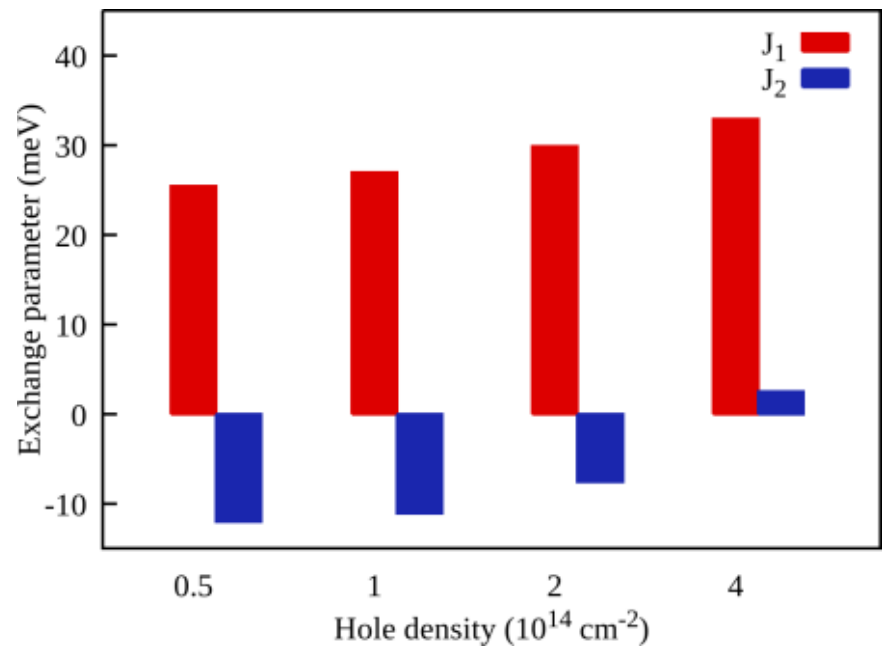

## Magnetic anisotropy energy (MAE, μeV) per magnetic atom

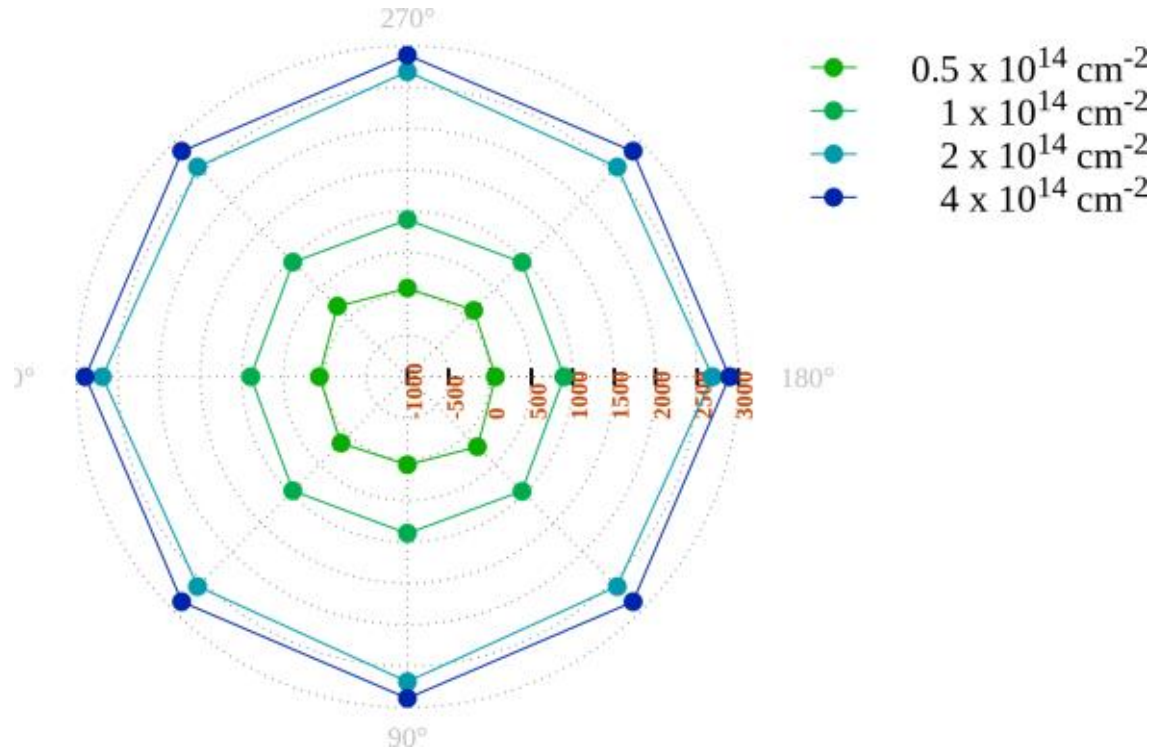

MAE = $E_{\parallel} - E_{\perp}$, a positive (negative) value of MAE indicates the off-plane (in-plane) easy axis.

## Monte Carlo simulations of the normalized magnetization of as a function of temperature

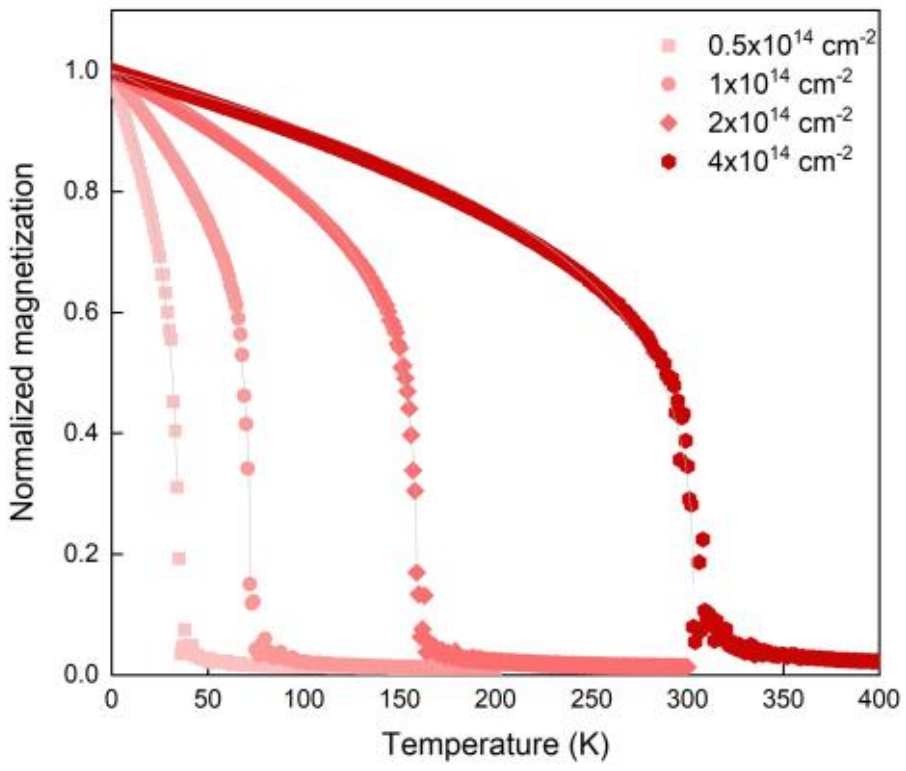

$T_c$: $0.5\times10^{14}$ cm$^{-2}$: 35 K  $1\times10^{14}$ cm$^{-2}$: 72 K

$2\times10^{14}$ cm$^{-2}$: 159 K  $4\times10^{14}$ cm$^{-2}$: 303 K

# 81. $BaCl_2$

| MC2D-ID | C2DB | 2dmat-ID | USPEX | Space group | Band gap (eV) |
|---|---|---|---|---|---|
| - | - | 2dm-1724 | - | P4m2 | 5.19 |

**Convex hull** | **Atomic structure** | **Atomic coordinates** | **Phonon dispersion curve**

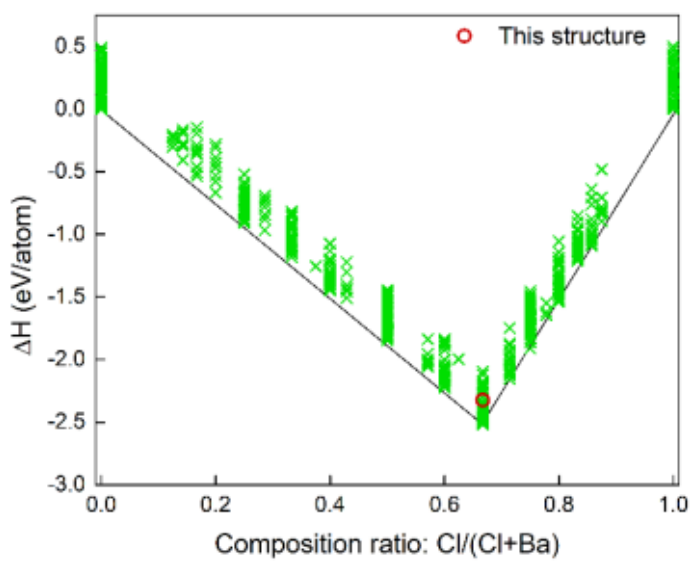

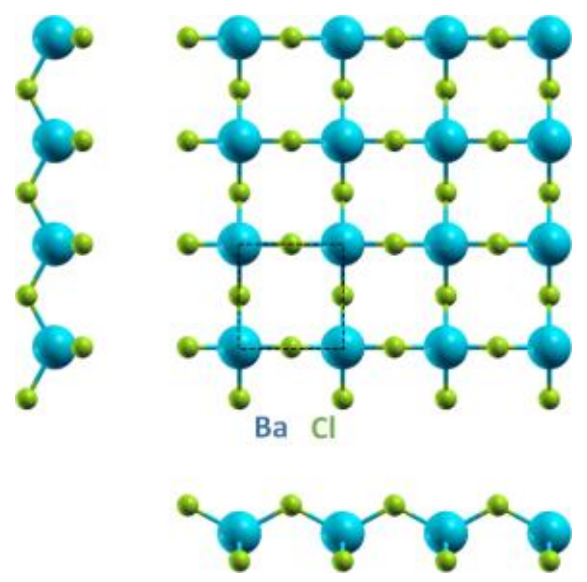

Ba1Cl2
1.00000000000000
5.2204455449080065 0.0000000000000000 0.0000000000000000
0.0000000000000000 5.2204455608024087 0.0000000000000000
0.0000000000000000 0.0000000000000000 22.6561909999999997
Ba Cl
1 2
Direct
0.9999999895210436 0.9999998649340824 0.0000000120736772
0.9999999136415667 0.4999997978403741 0.9379403053198416
0.5000000968373897 0.0000003372255364 0.0620596826064812

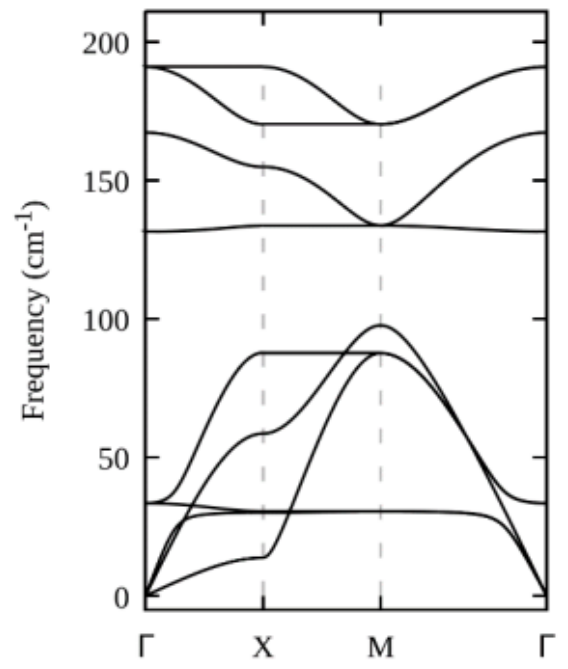

**Projected band structure and density of states**

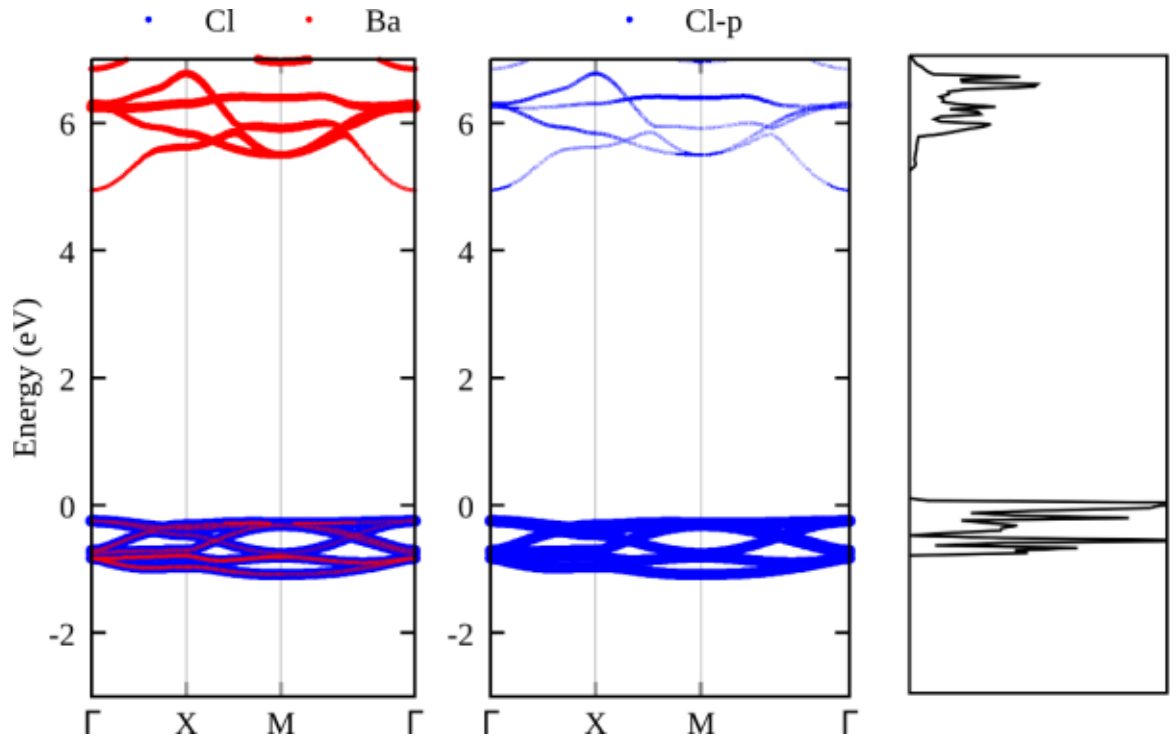

**Magnetic moment and spin polarization energy as a function of hole doping concentration**

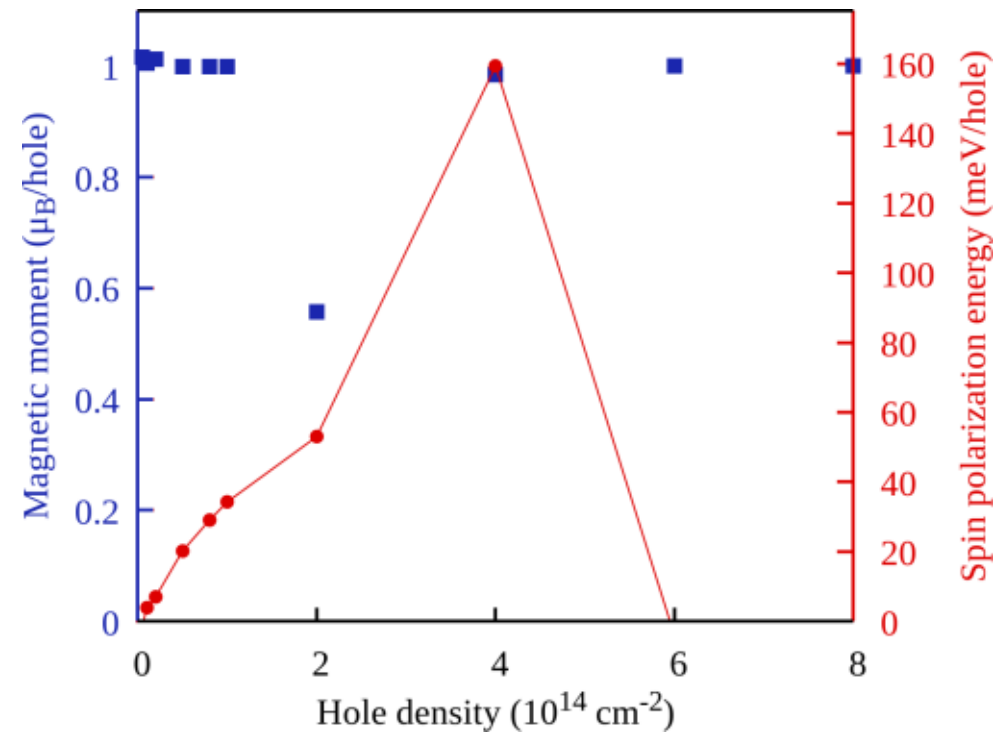

**Magnetic configurations and spin Hamiltonian**

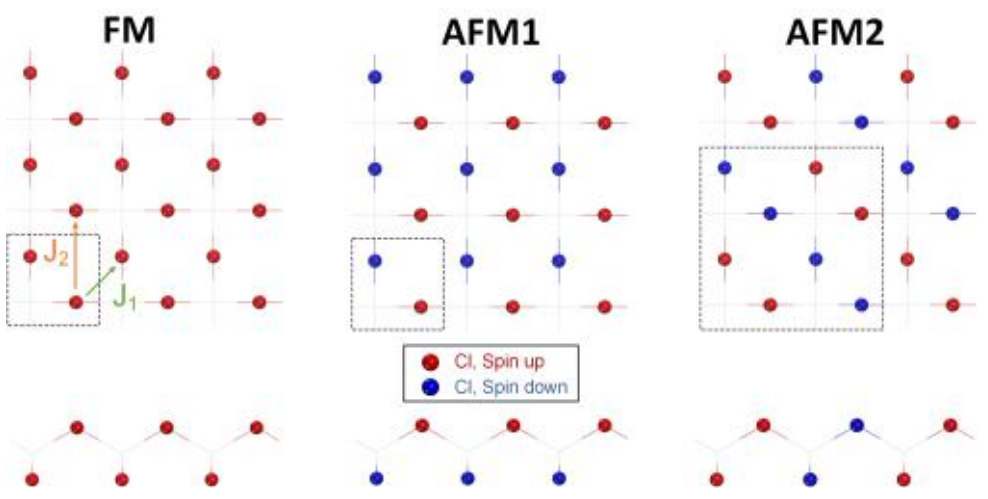

$$E_{FM} = E_0 - 4J_1S^2 - 4J_2S^2$$
$$E_{AFM1} = E_0 + 4J_1S^2 - 4J_2S^2$$
$$E_{AFM2} = E_0 + 4J_2S^2$$

**Magnetic exchange coupling parameters**

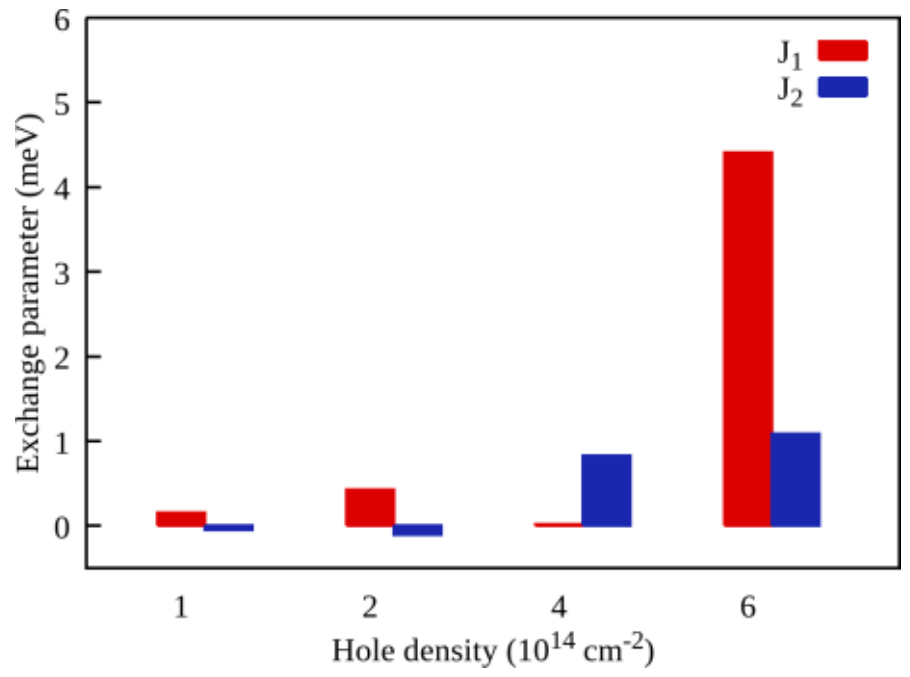

**Magnetic anisotropy energy (MAE, μeV) per magnetic atom**

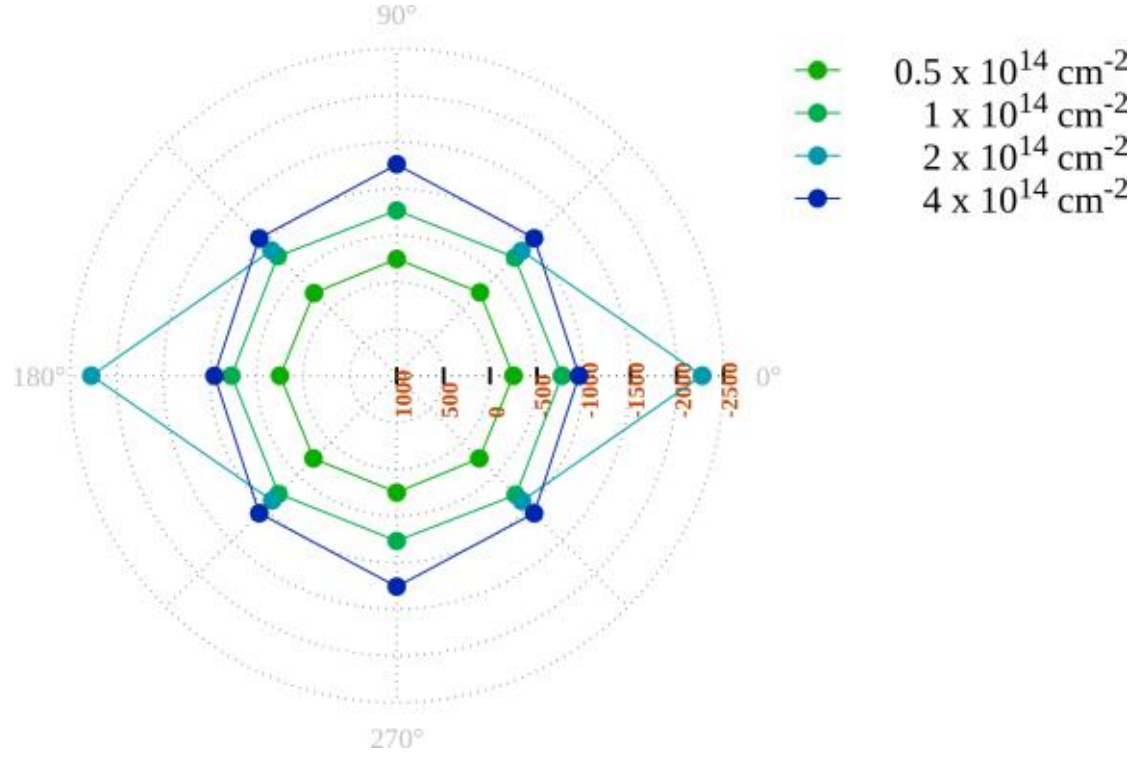

MAE = $E_{\parallel} - E_{\perp}$, a positive (negative) value of MAE indicates the off-plane (in-plane) easy axis.

**Monte Carlo simulations of the normalized magnetization of as a function of temperature**

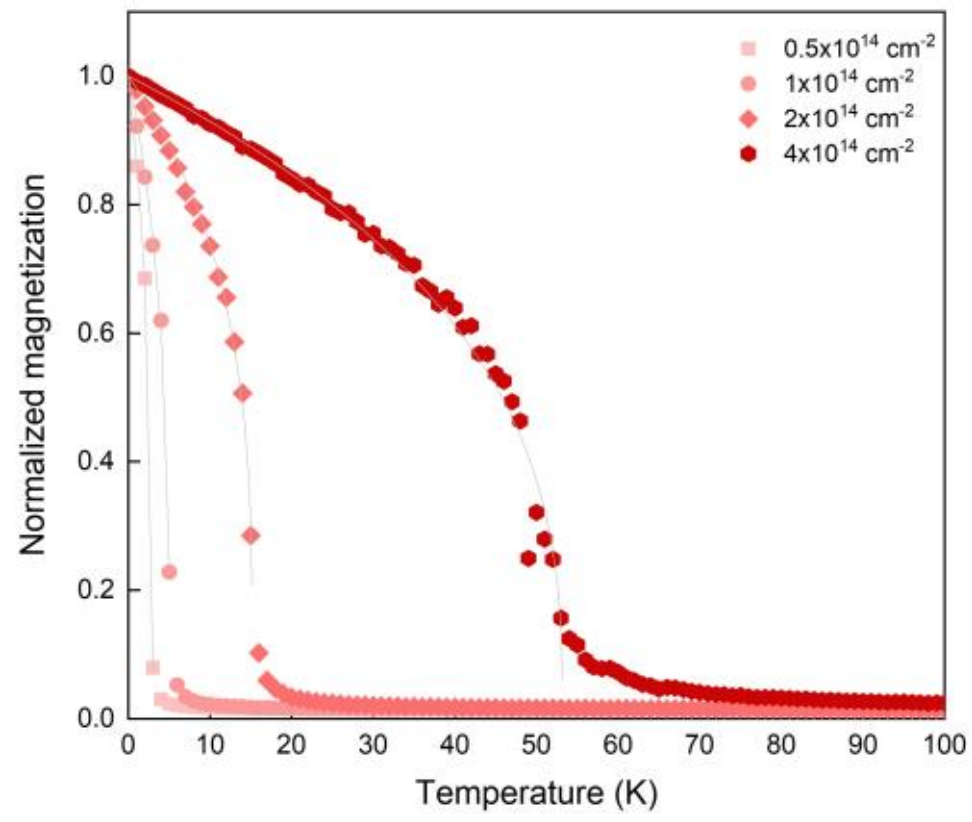

$T_c$: $0.5\times10^{14}$ cm$^{-2}$: 3 K $1\times10^{14}$ cm$^{-2}$: 5 K
$2\times10^{14}$ cm$^{-2}$: 15 K $4\times10^{14}$ cm$^{-2}$: 53 K

# 82.$BaBr_2$

| MC2D-ID | C2DB | 2dmat-ID | USPEX | Space group | Band gap (eV) |
|---|---|---|---|---|---|
| - | - | 2dm-940 | - | P4m2 | 4.65 |

**Convex hull** | **Atomic structure** | **Atomic coordinates** | **Phonon dispersion curve**

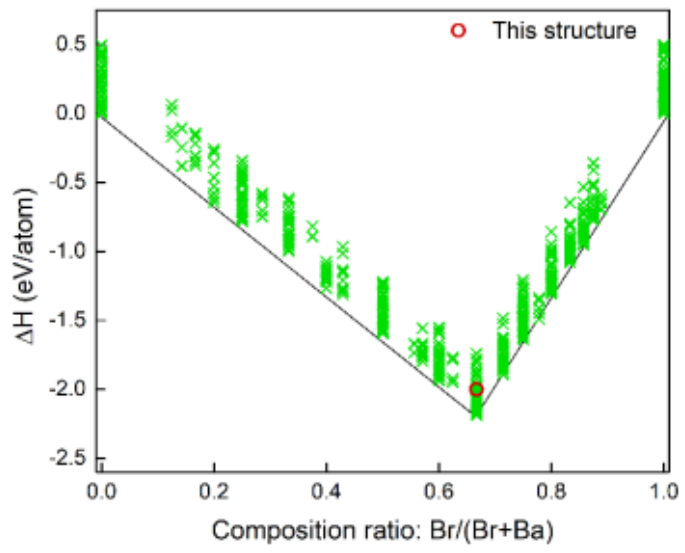

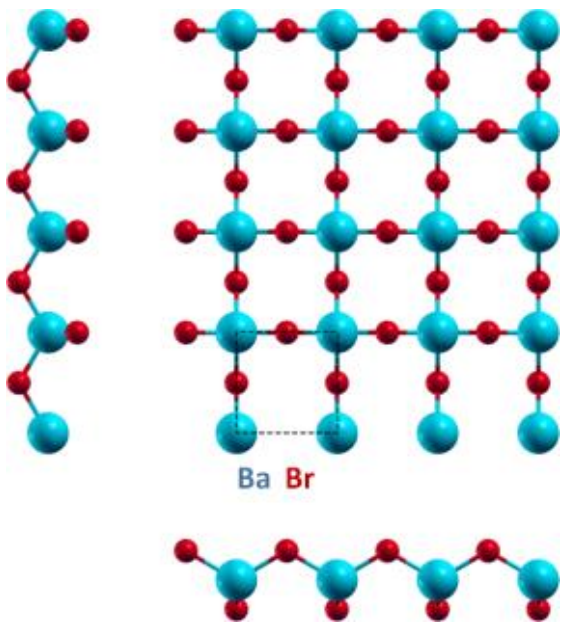

```
Ba1Br2
   1.00000000000000
   5.3962707755781327   0.0000000000000000   0.0000000000000000
   0.0000000000000000   5.3962708320850812   0.0000000000000000
   0.0000000000000000   0.0000000000000000   22.6561909999999997
   Ba    Br
   1     2
Direct
  0.9999999445990824   0.0000000188573068   0.9999999890720801
  0.0000000144325156   0.4999999835966094   0.9309651109781285
  0.5000000409684091   0.9999999975460838   0.0690348999497914
```

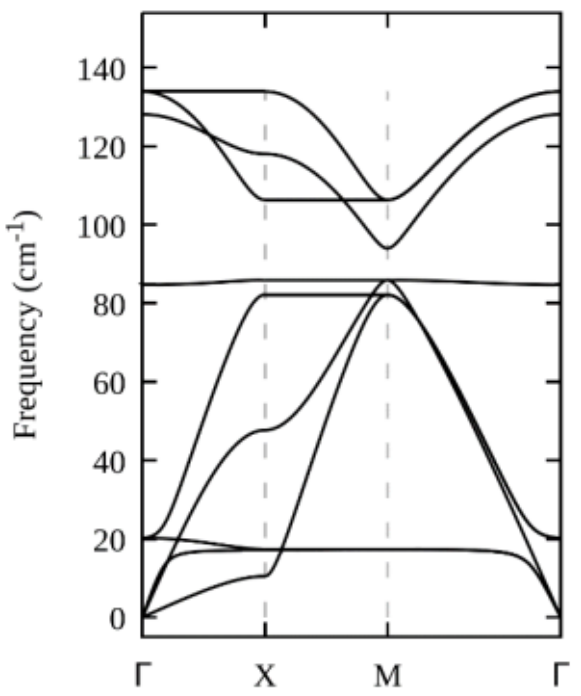

**Projected band structure and density of states**

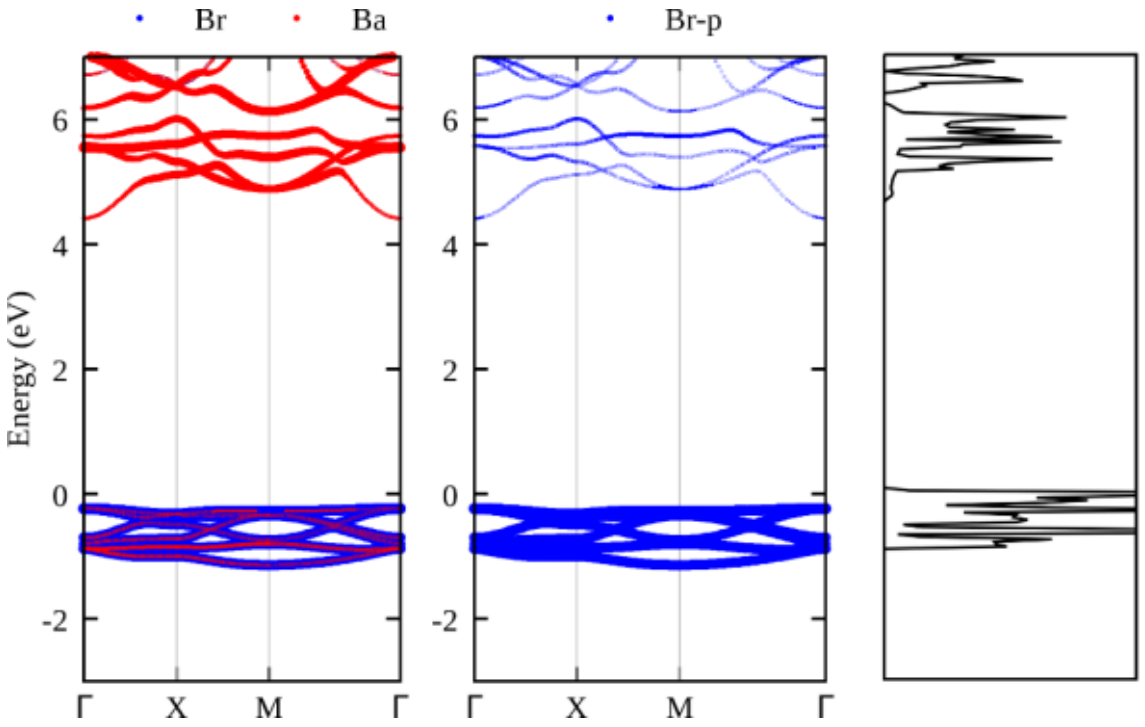

**Magnetic moment and spin polarization energy as a function of hole doping concentration**

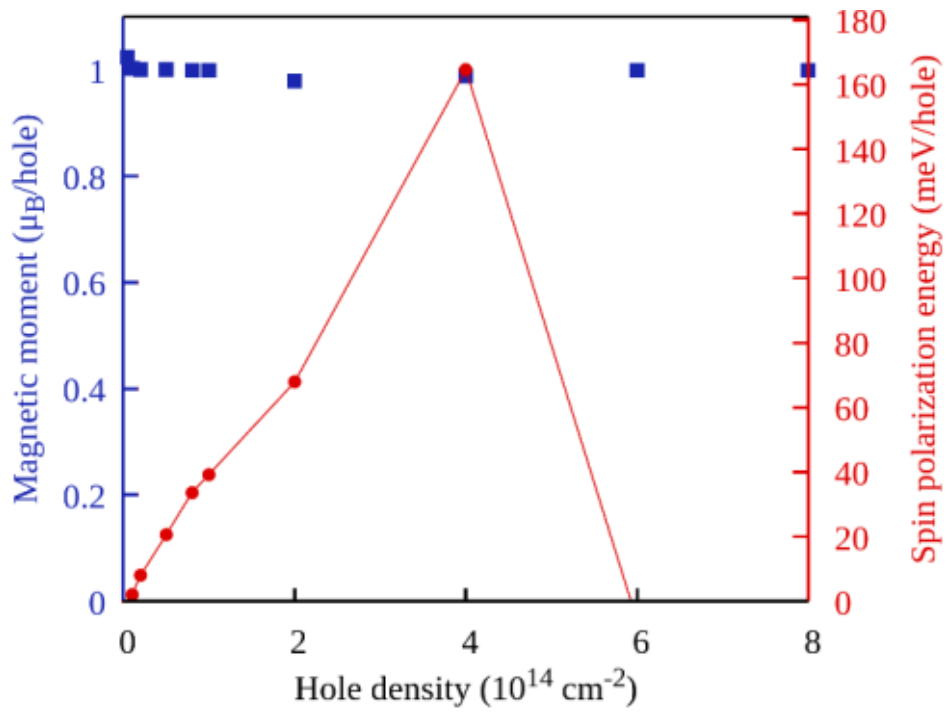

**Magnetic configurations and spin Hamiltonian**

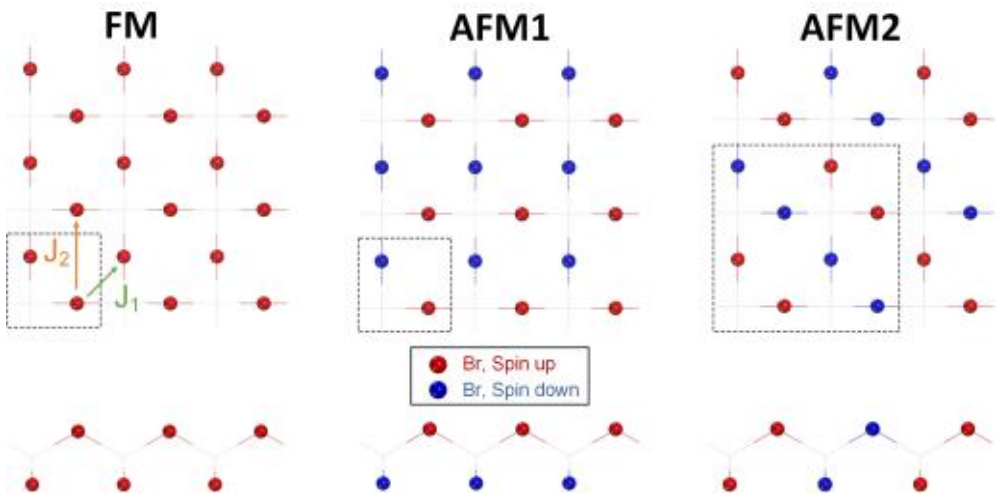

$$E_{FM} = E_0 - 4J_1S^2 - 4J_2S^2$$
$$E_{AFM1} = E_0 + 4J_1S^2 - 4J_2S^2$$
$$E_{AFM2} = E_0 + 4J_2S^2$$

**Magnetic exchange coupling parameters**

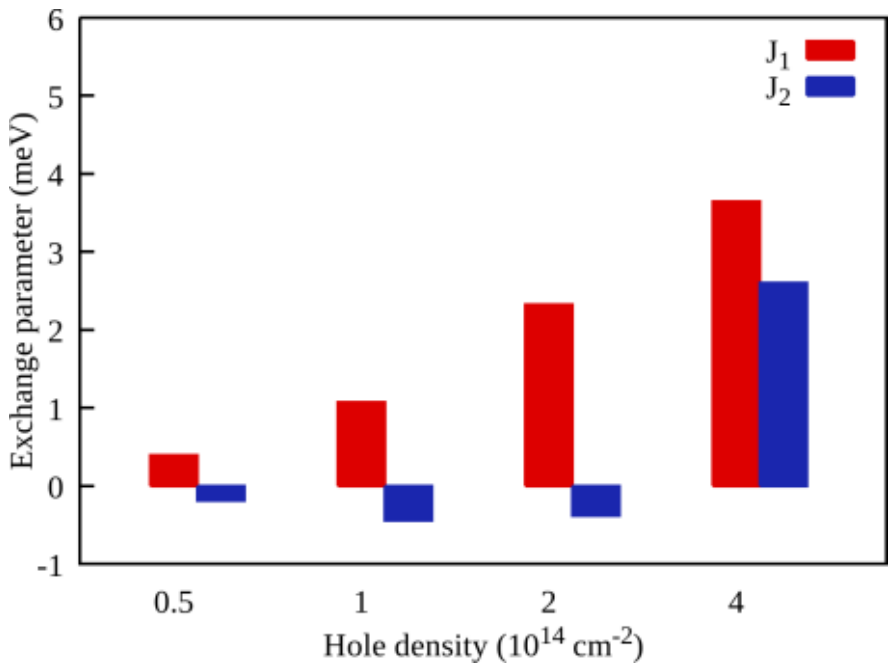

**Magnetic anisotropy energy (MAE, μeV) per magnetic atom**

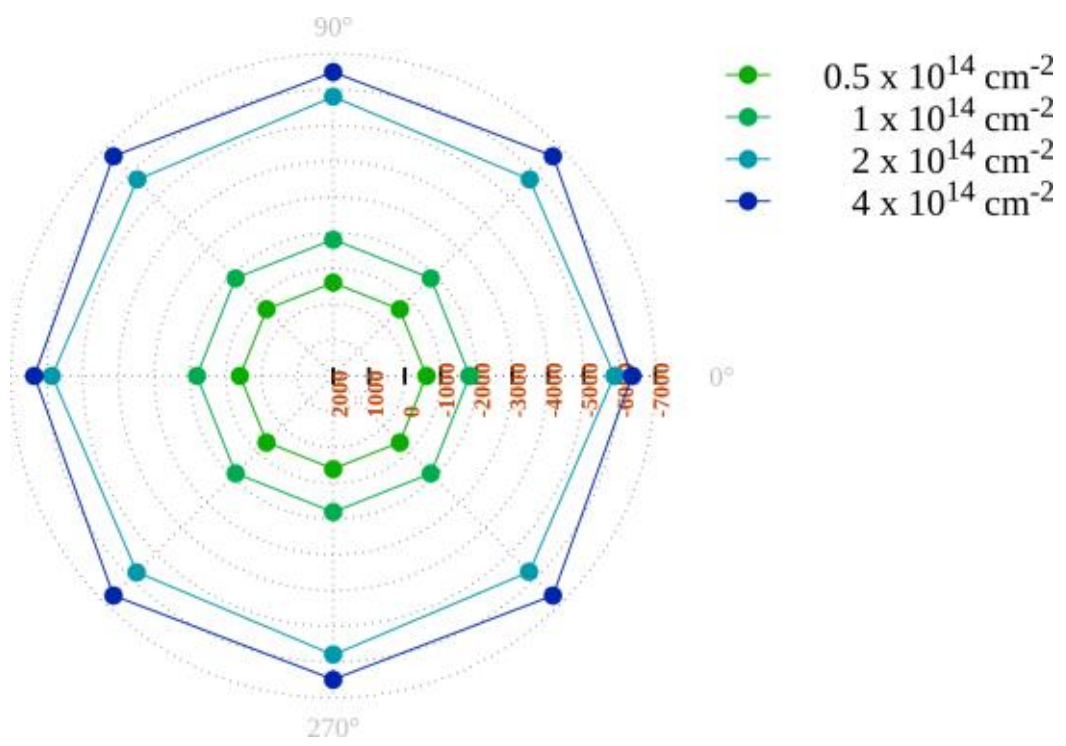

MAE = $E_{\parallel} - E_{\perp}$, a positive (negative) value of MAE indicates the off-plane (in-plane) easy axis.

**Monte Carlo simulations of the normalized magnetization of as a function of temperature**

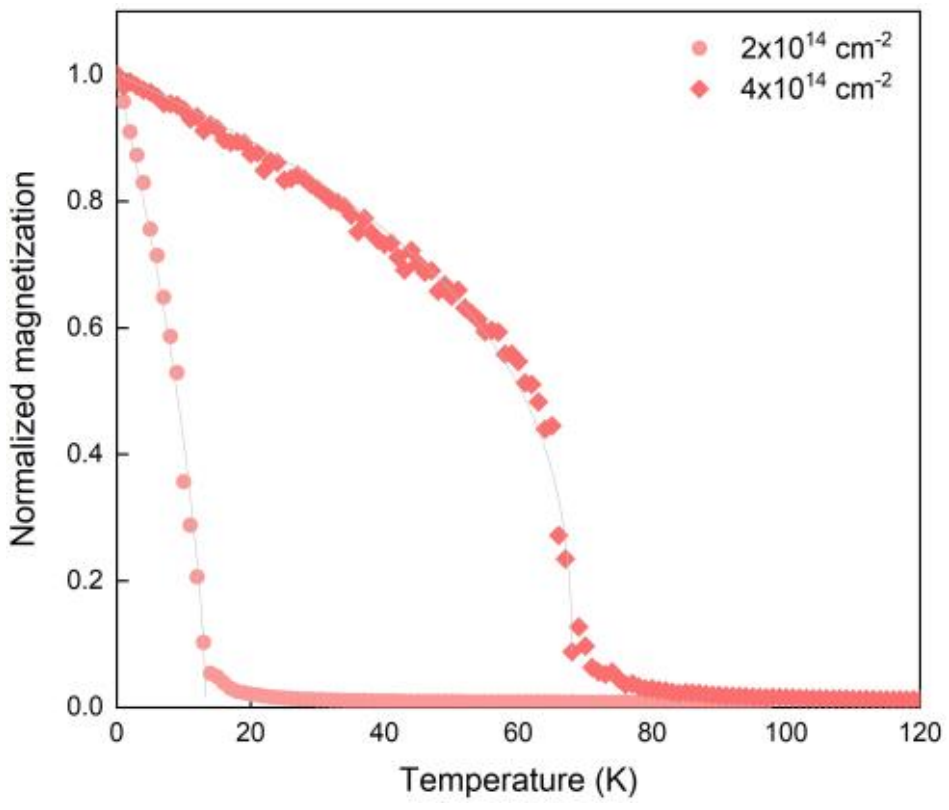

$T_c$: $0.5\times10^{14}$ cm$^{-2}$: - K   $1\times10^{14}$ cm$^{-2}$: - K
$2\times10^{14}$ cm$^{-2}$: 13 K   $4\times10^{14}$ cm$^{-2}$: 68 K

# 83.$BaI_2$

| MC2D-ID | C2DB | 2dmat-ID | USPEX | Space group | Band gap (eV) |
|---|---|---|---|---|---|
| - | - | 2dm-578 | - | P4m2 | 4.17 |

**Convex hull** | **Atomic structure** | **Atomic coordinates** | **Phonon dispersion curve**

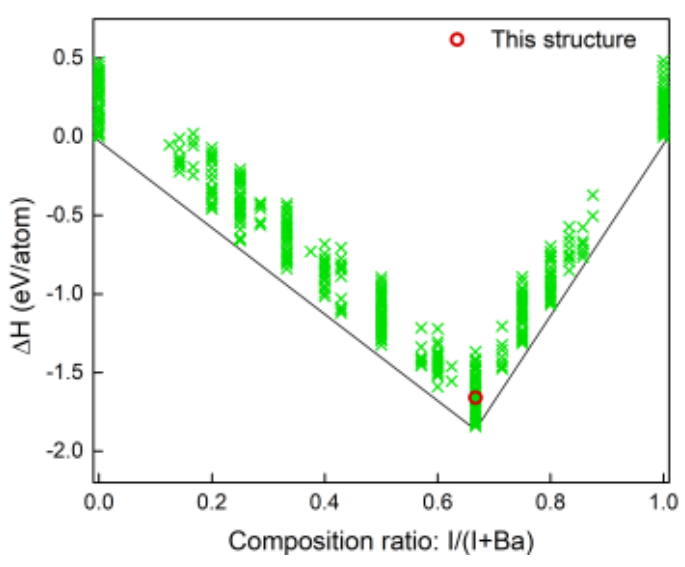

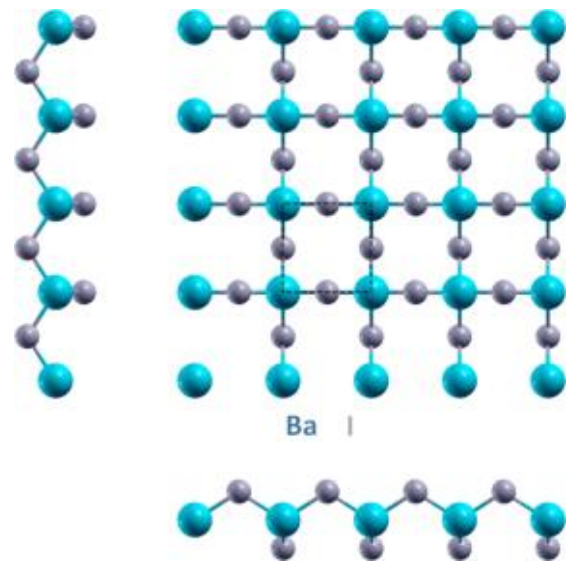

```
Ba1I2
   1.00000000000000
   5.6312877034679518   0.0000000000000000   0.0000000000000000
   0.0000000000000000   5.6312855784550733   0.0000000000000000
   0.0000000000000000   0.0000000000000000   22.6561909999999997
   Ba   I
   1    2
Direct
  0.0000000097016439   0.0000000261087223   0.0000000364624881
  0.9999999760593496   0.5000000066512271   0.9202877037178823
  0.5000000142390064   0.9999999672400506   0.0797122598196296
```

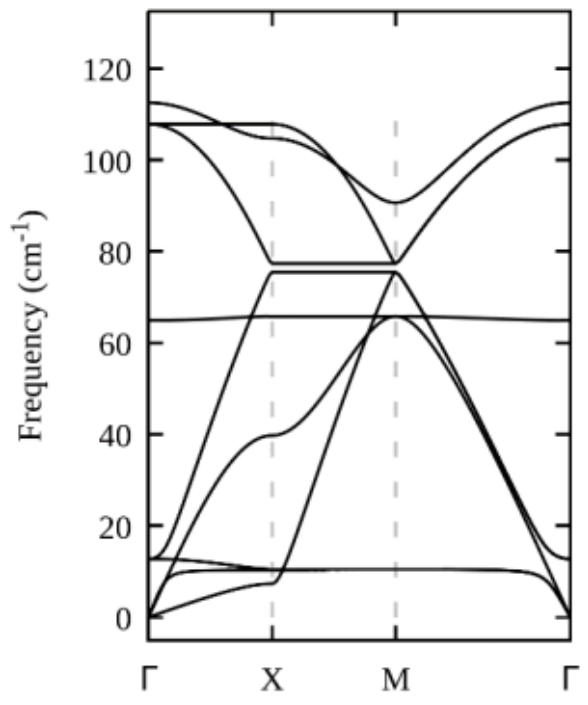

## Projected band structure and density of states

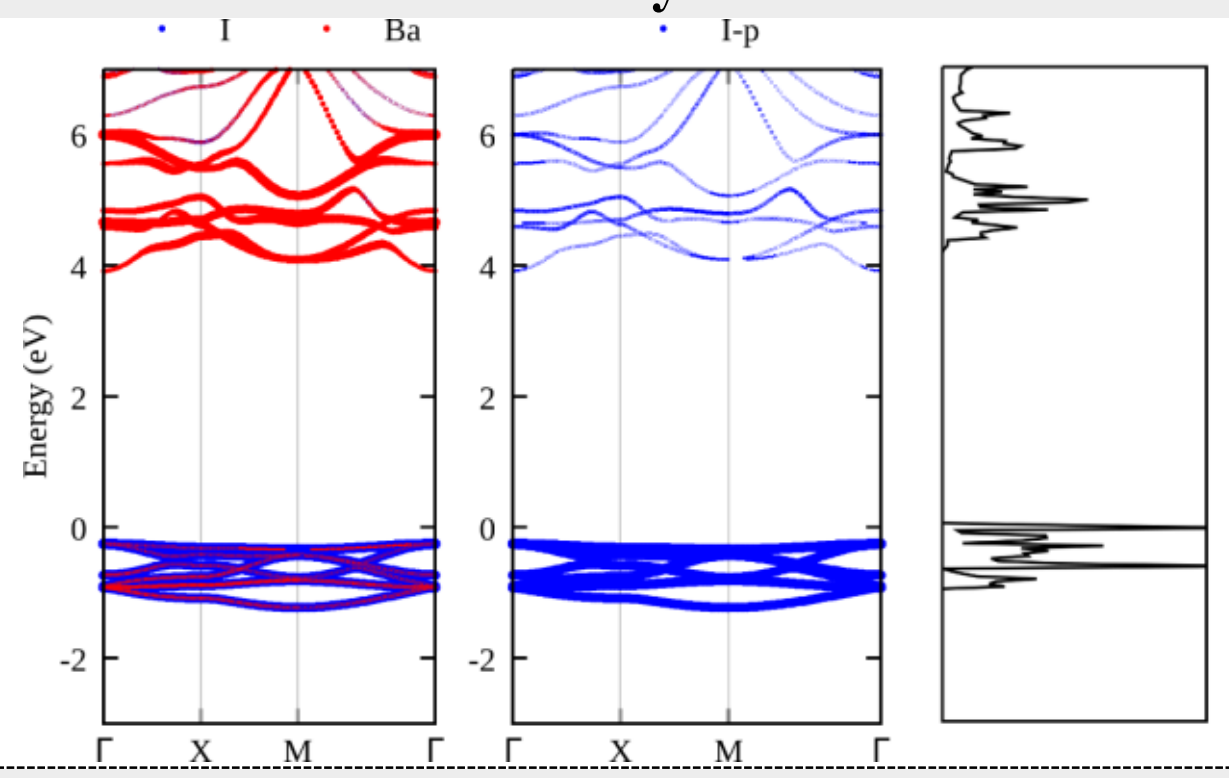

## Magnetic moment and spin polarization energy as a function of hole doping concentration

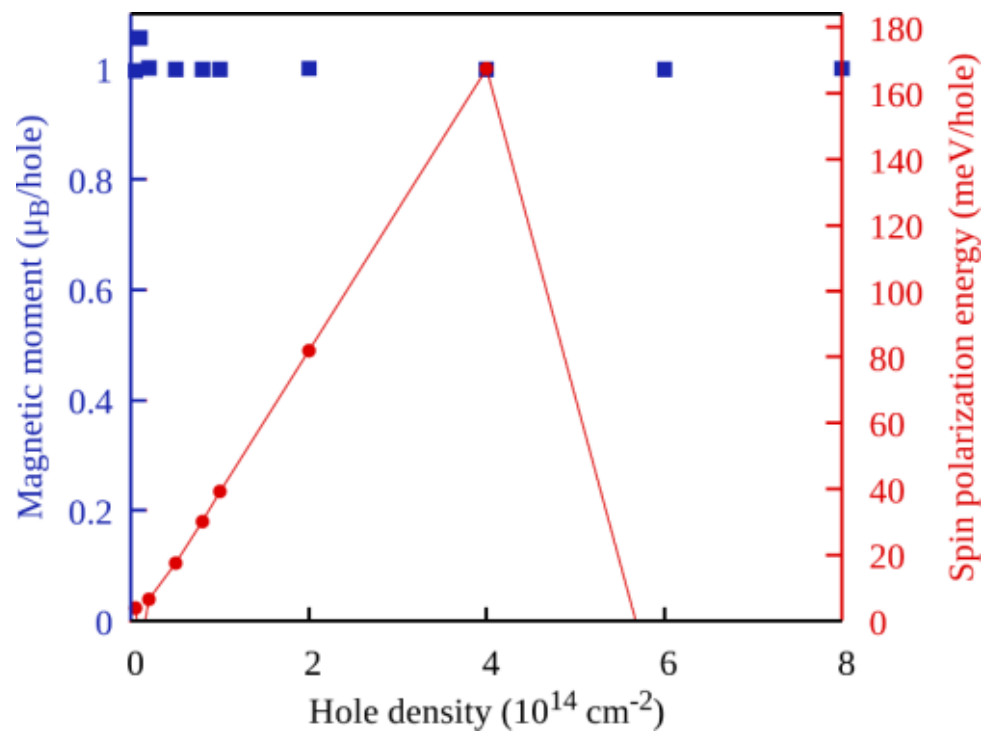

## Magnetic configurations and spin Hamiltonian

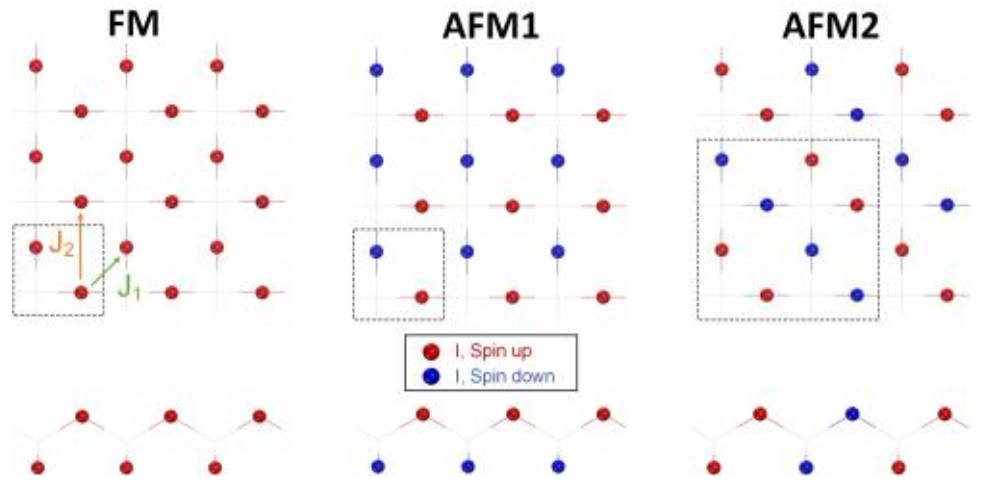

$$E_{FM} = E_0 - 4J_1S^2 - 4J_2S^2$$
$$E_{AFM1} = E_0 + 4J_1S^2 - 4J_2S^2$$
$$E_{AFM2} = E_0 + 4J_2S^2$$

## Magnetic exchange coupling parameters

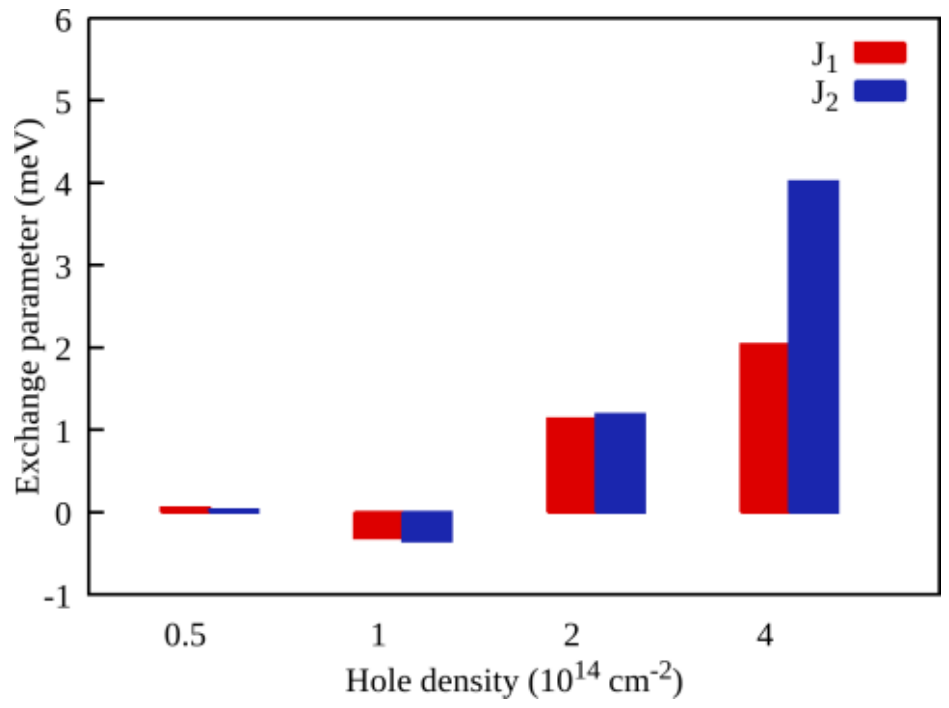

## Magnetic anisotropy energy (MAE, μeV) per magnetic atom

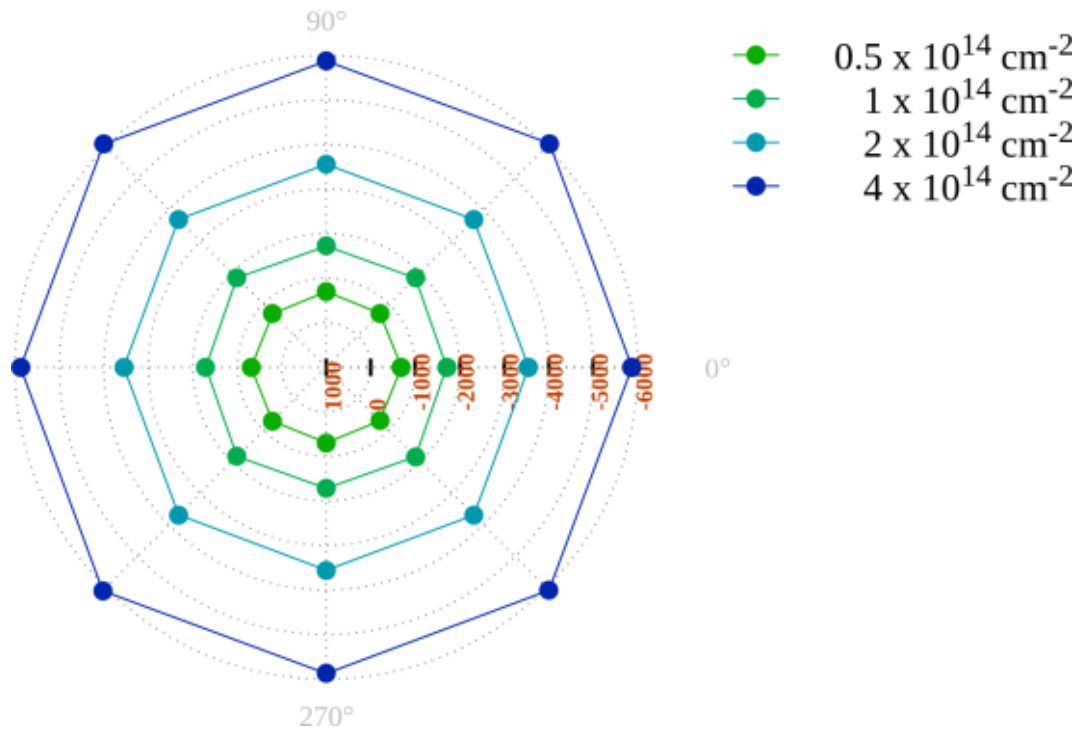

MAE = $E_{\parallel} - E_{\perp}$, a positive (negative) value of MAE indicates the off-plane (in-plane) easy axis.

## Monte Carlo simulations of the normalized magnetization of as a function of temperature

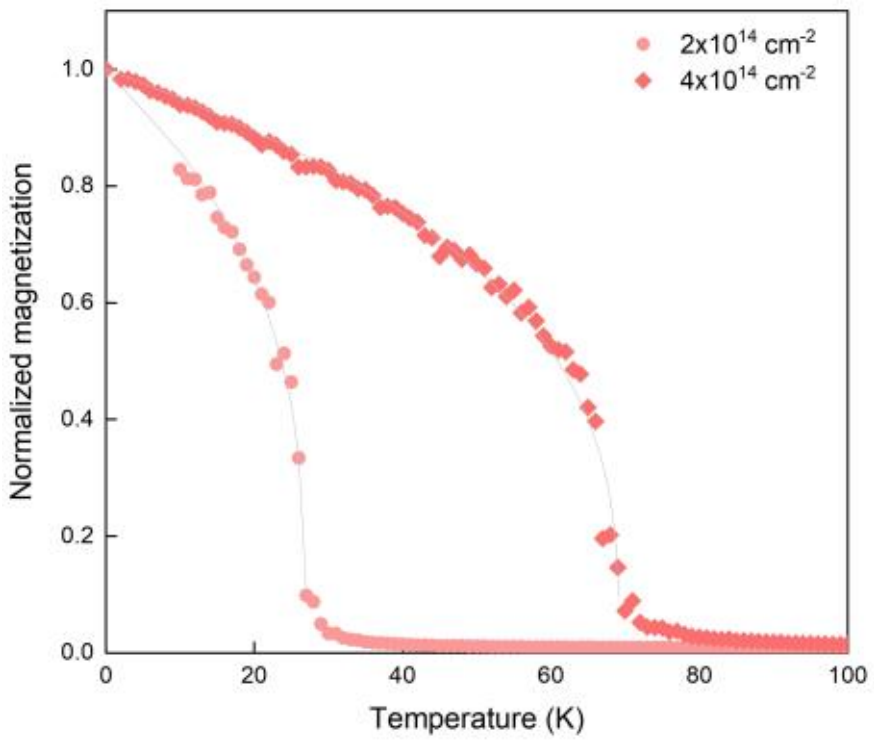

$T_c$: $0.5\times10^{14}$ cm$^{-2}$: - K  $1\times10^{14}$ cm$^{-2}$: 1 K

$2\times10^{14}$ cm$^{-2}$: 27 K  $4\times10^{14}$ cm$^{-2}$: 69 K

# 84.$ZnF_2$

| MC2D-ID | C2DB | 2dmat-ID | USPEX | Space group | Band gap (eV) |
|---|---|---|---|---|---|
| - | ✓ | 2dm-5027 | - | P4m2 | 4.40 |

**Convex hull**

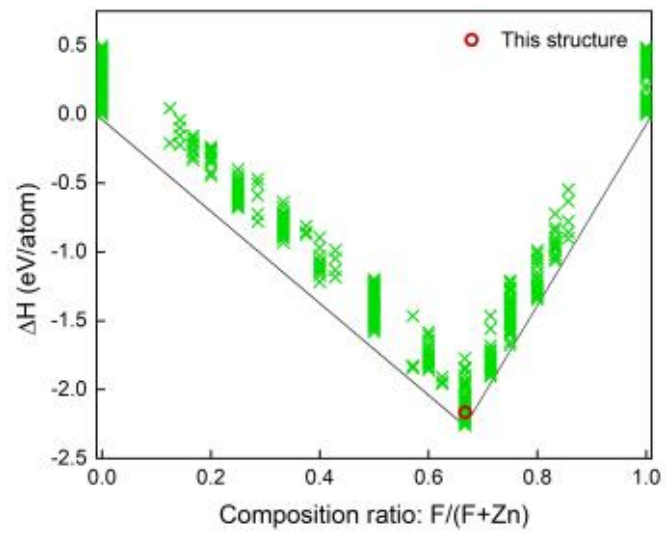

**Atomic structure**

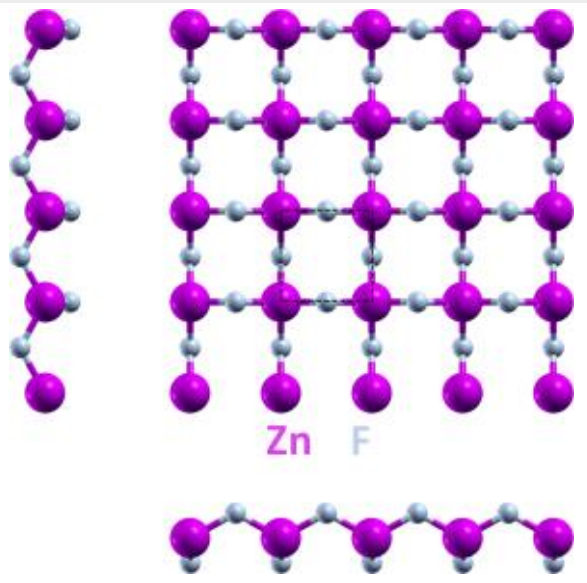

**Atomic coordinates**

```
Zn1F2
1.00000000000000
3.4021277956138194    0.0000000000000000    0.0000000000000000
-0.0000000067903241    3.4021231044657156    0.0000000000000000
0.0000000000000014    0.0000000000000014    22.7593417019000022
Zn    F
1    2
Direct
0.0000000118231895    0.0000000385394259    0.4999999886403614
0.0000000122277015    0.4999999796982166    0.4590367377307842
0.4999999759356513    0.9999999817488643    0.5409632622692158
0.4999999946864051    0.0000000006091483    0.5599368665589637
```

**Phonon dispersion curve**

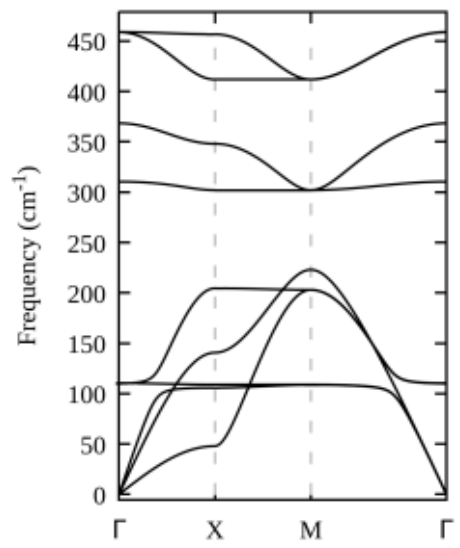

**Projected band structure and density of states**

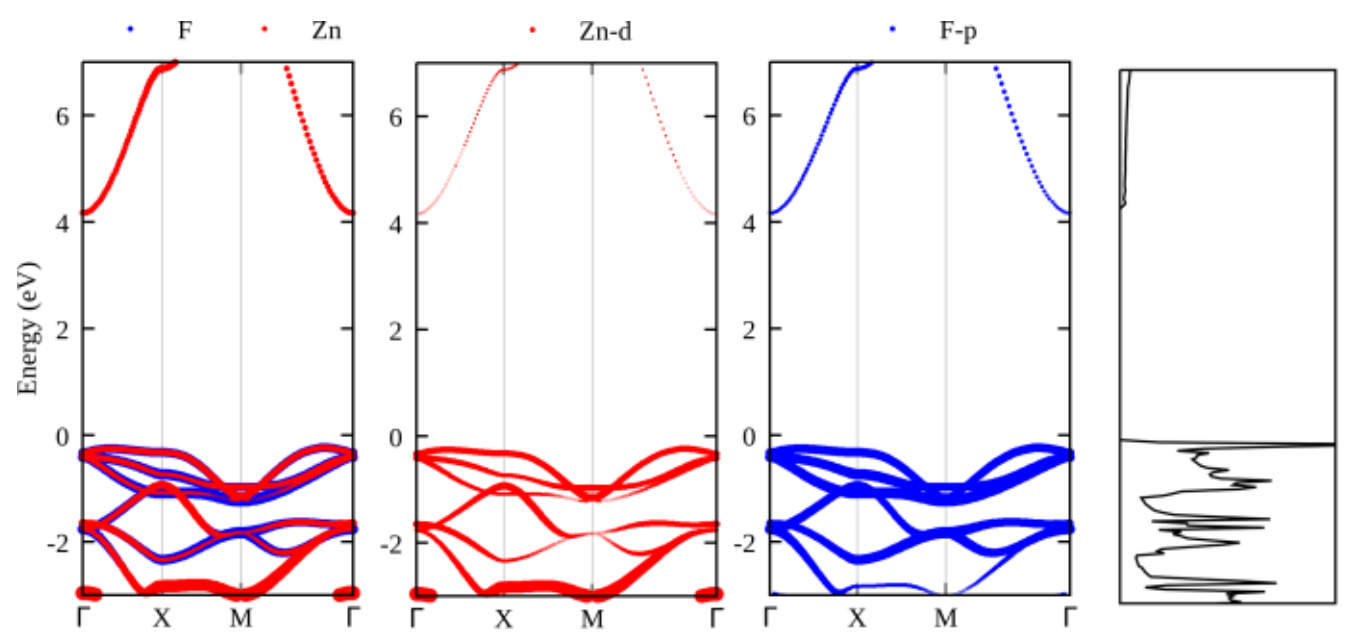

**Magnetic moment and spin polarization energy as a function of hole doping concentration**

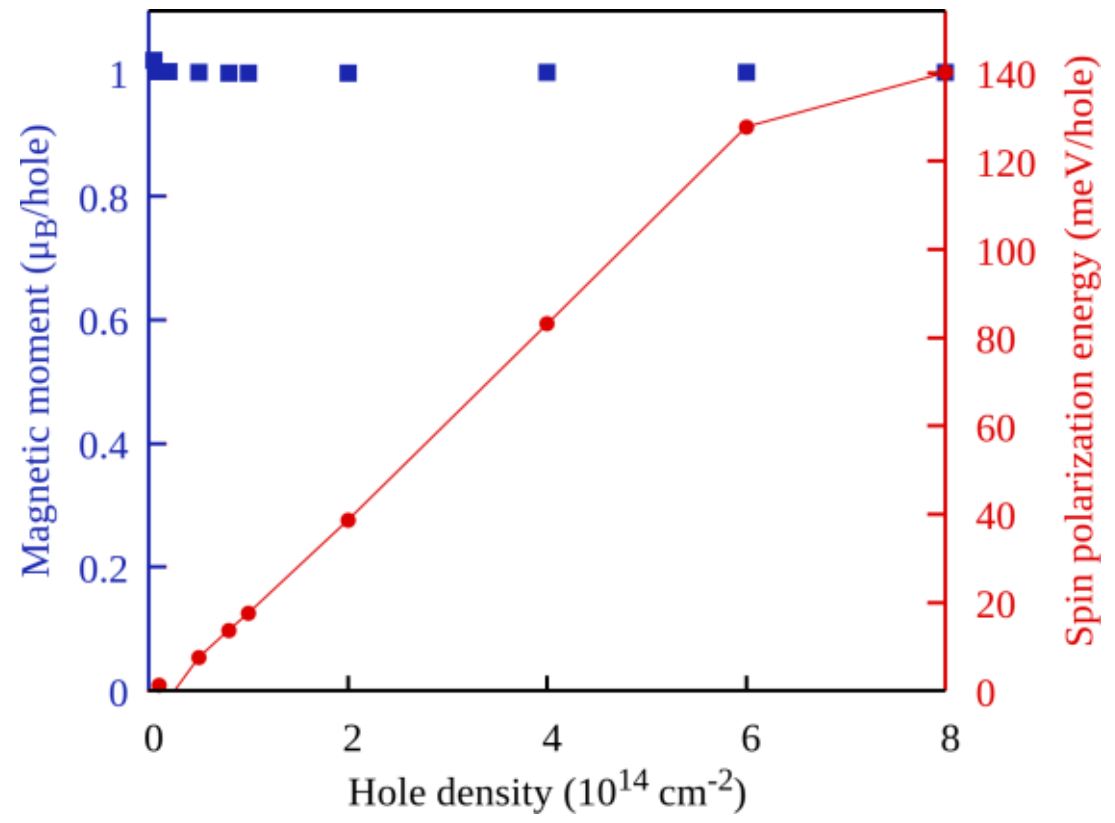

**Magnetic configurations and spin Hamiltonian**

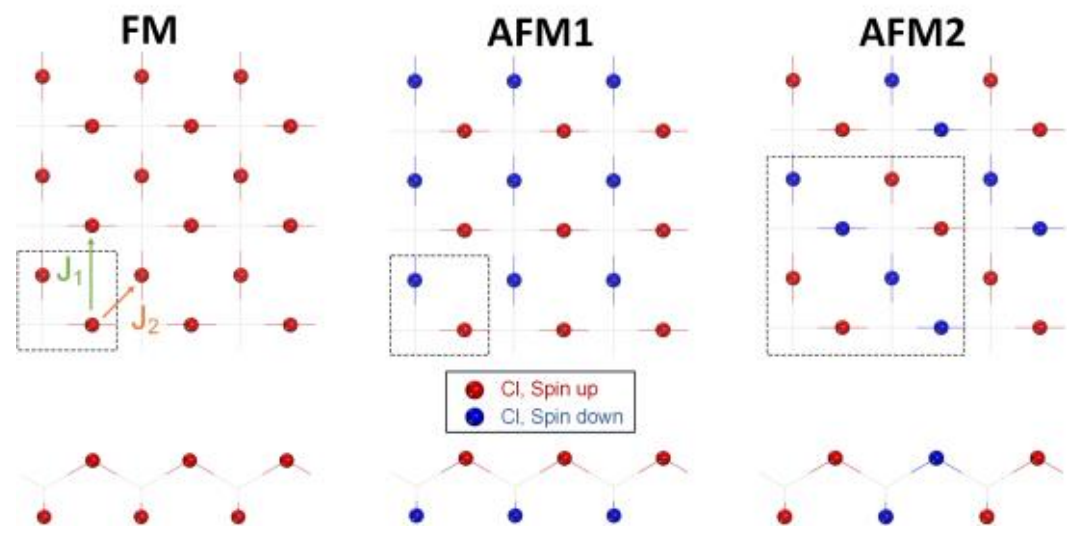

$$E_{FM} = E_0 - 4J_1S^2 - 4J_2S^2$$
$$E_{AFM1} = E_0 - 4J_1S^2 + 4J_2S^2$$
$$E_{AFM2} = E_0 + 4J_1S^2$$

**Magnetic exchange coupling parameters**

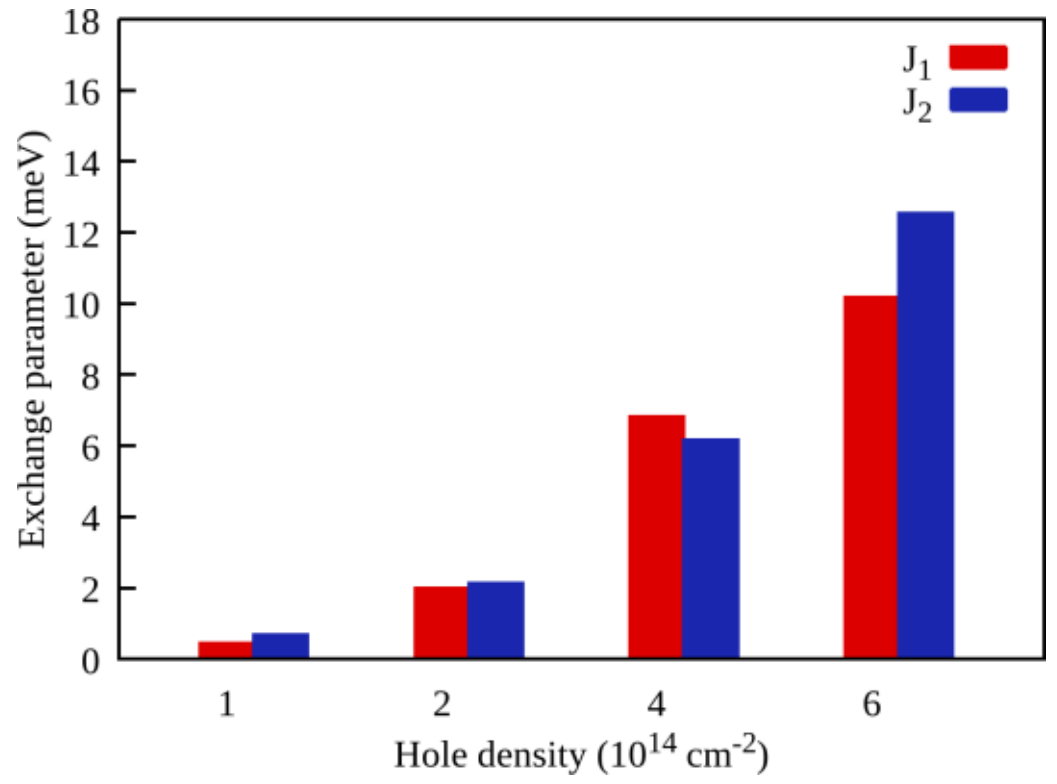

**Magnetic anisotropy energy (MAE, μeV) per magnetic atom**

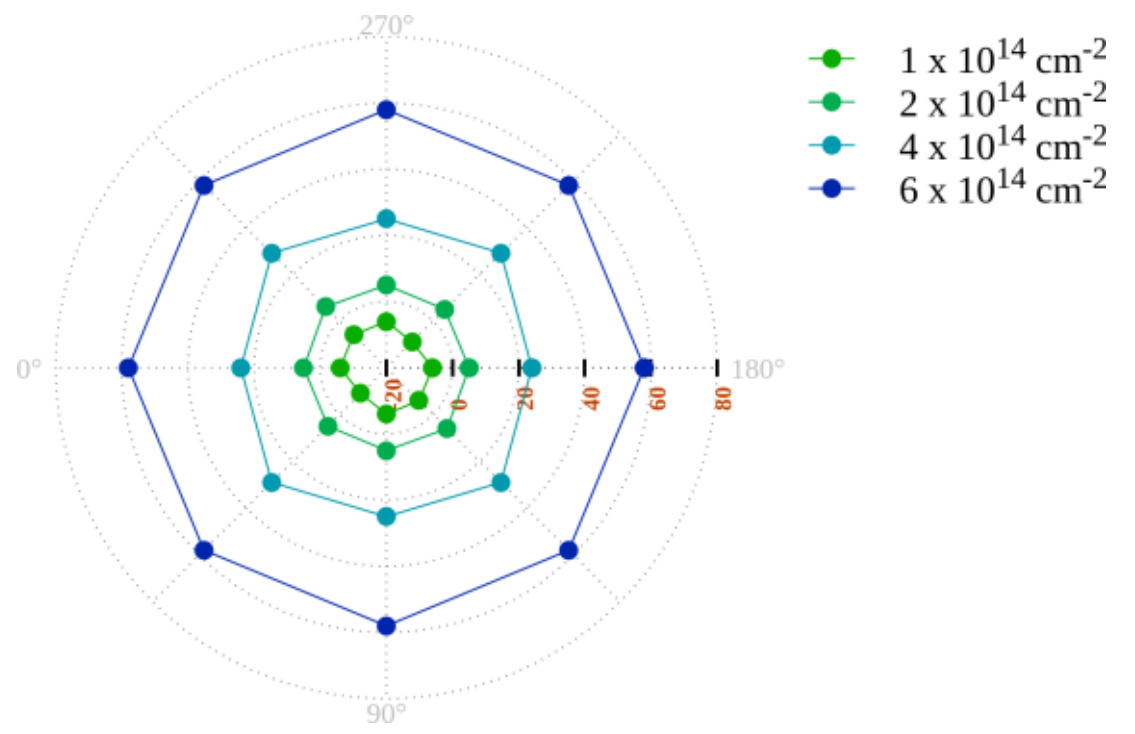

MAE = $E_{\parallel} - E_{\perp}$, a positive (negative) value of MAE indicates the off-plane (in-plane) easy axis.

**Monte Carlo simulations of the normalized magnetization of as a function of temperature**

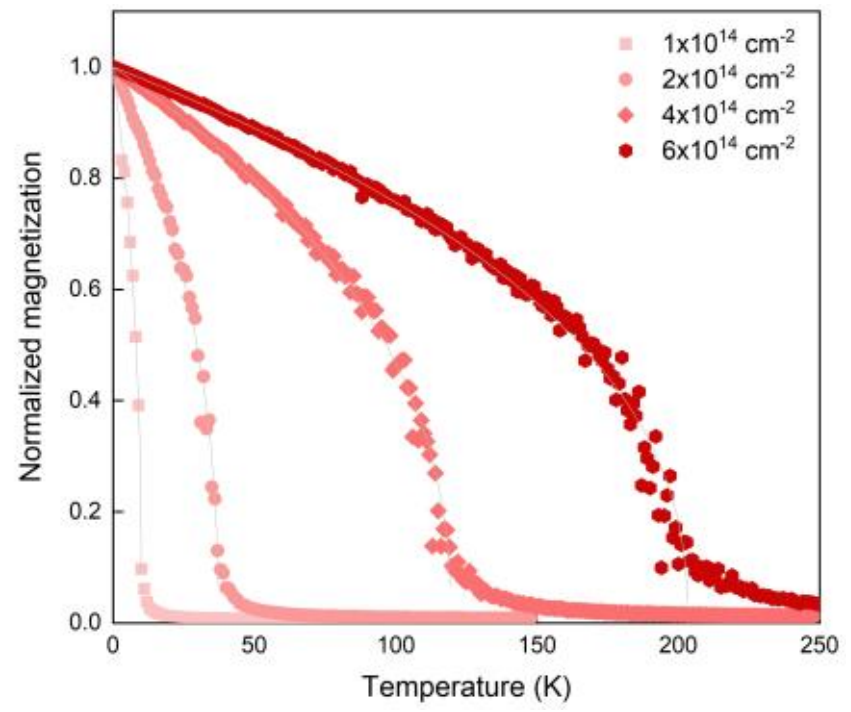

$T_c$: $1\times10^{14}$ cm$^{-2}$: 10 K  $2\times10^{14}$ cm$^{-2}$: 37 K

$4\times10^{14}$ cm$^{-2}$: 119 K  $6\times10^{14}$ cm$^{-2}$: 203 K

# 85.$ZnCl_2$

| MC2D-ID | C2DB | 2dmat-ID | USPEX | Space group | Band gap (eV) |
|---|---|---|---|---|---|
| 239 | ✓ | 2dm-4713 | - | P4m2 | 4.25 |

| Convex hull | Atomic structure | Atomic coordinates | Phonon dispersion curve |
|---|---|---|---|

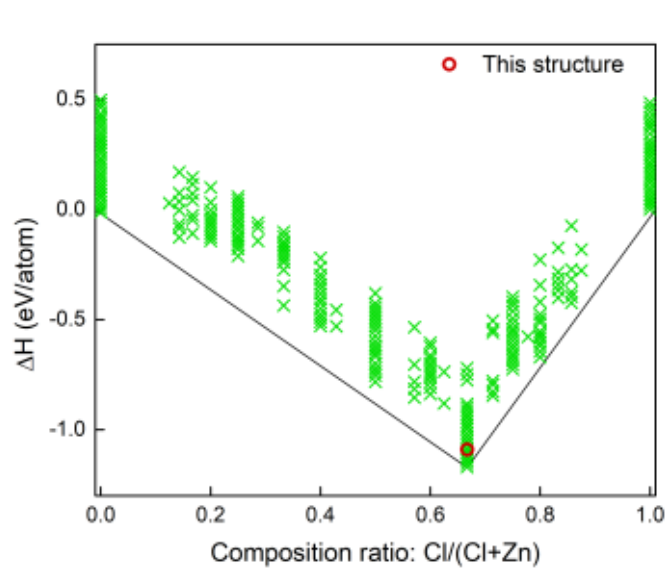

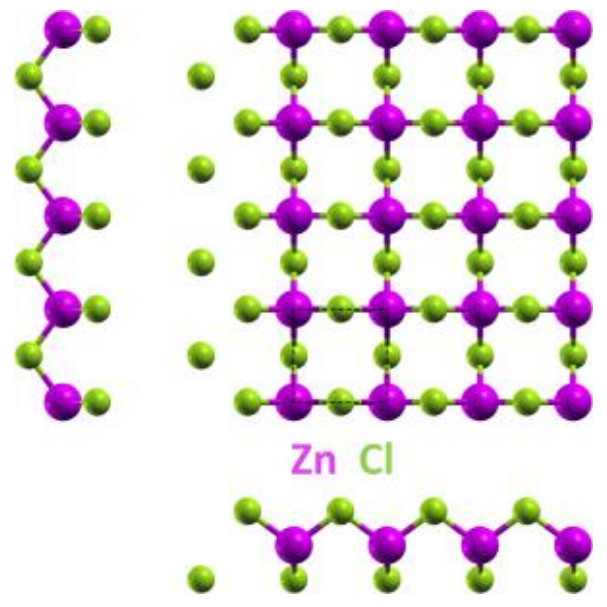

Zn1Cl2
1.00000000000000
3.7221357404532283 0.0000000000000000 0.0000000000000000
0.0000000000000000 3.7221358331008698 0.0000000000000000
0.0000000000000000 0.0000000000000000 22.7501749999999987
Zn Cl
1 2
Direct
0.9999999979183585 0.0000000005683205 0.5000000000157669
0.0000000073952364 0.4999999988225241 0.4400631334252694
0.4999999946864051 0.0000000006091483 0.5599368665589637

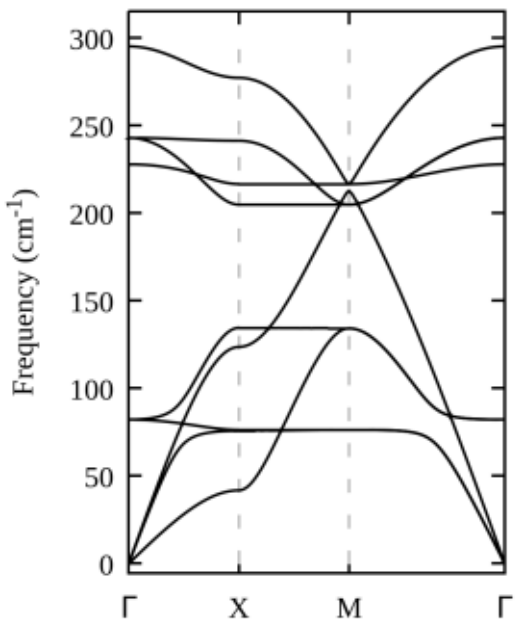

## Projected band structure and density of states

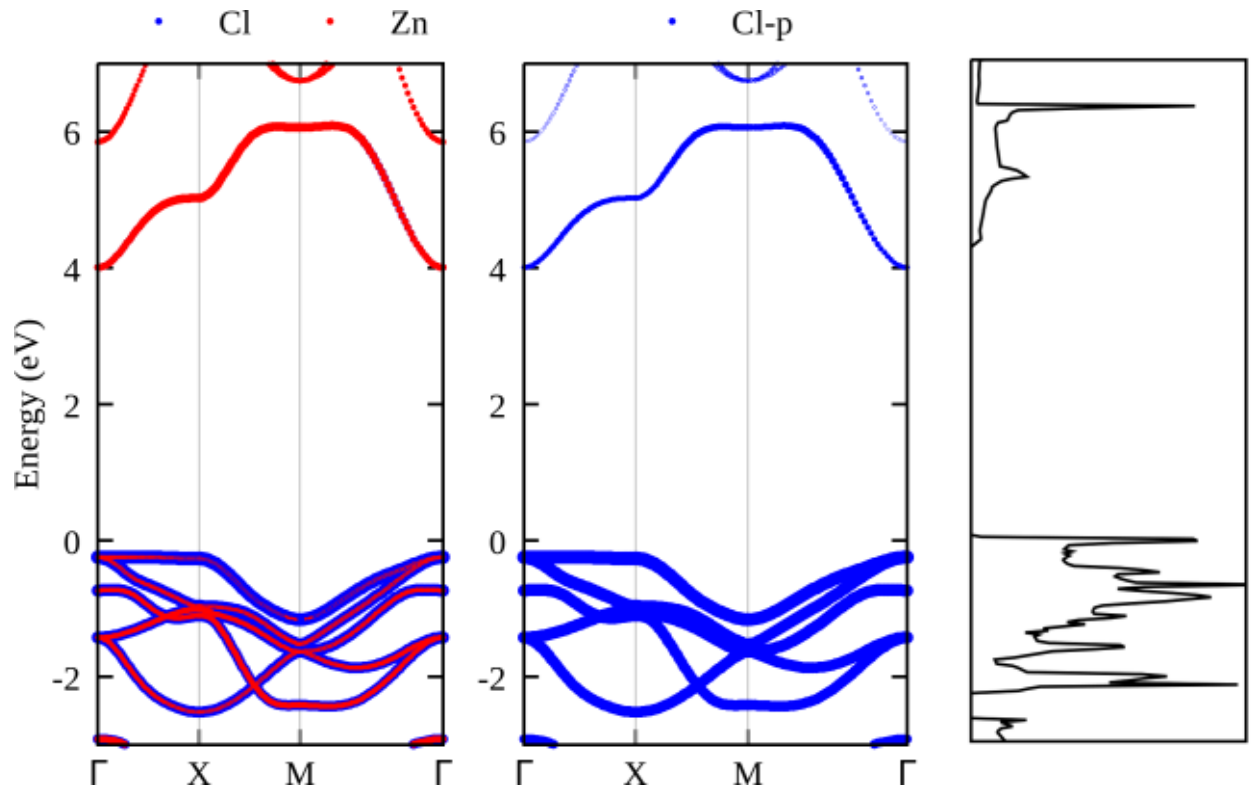

## Magnetic moment and spin polarization energy as a function of hole doping concentration

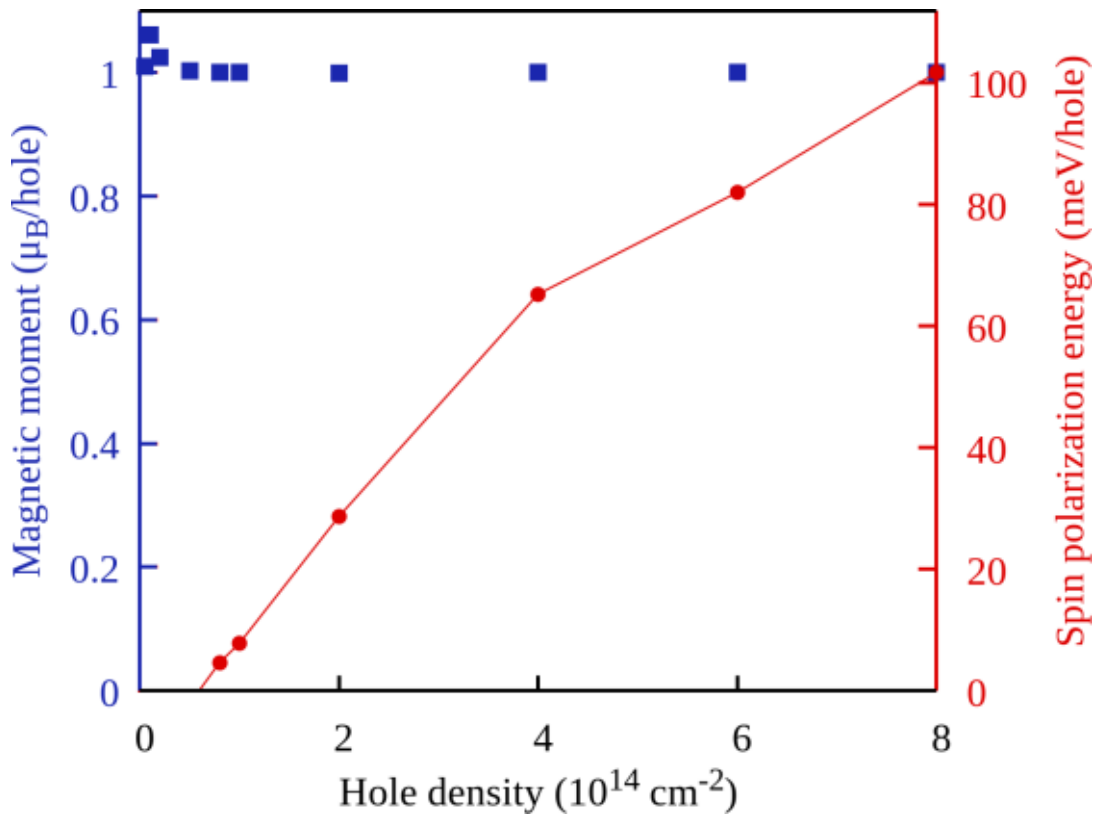

## Magnetic configurations and spin Hamiltonian

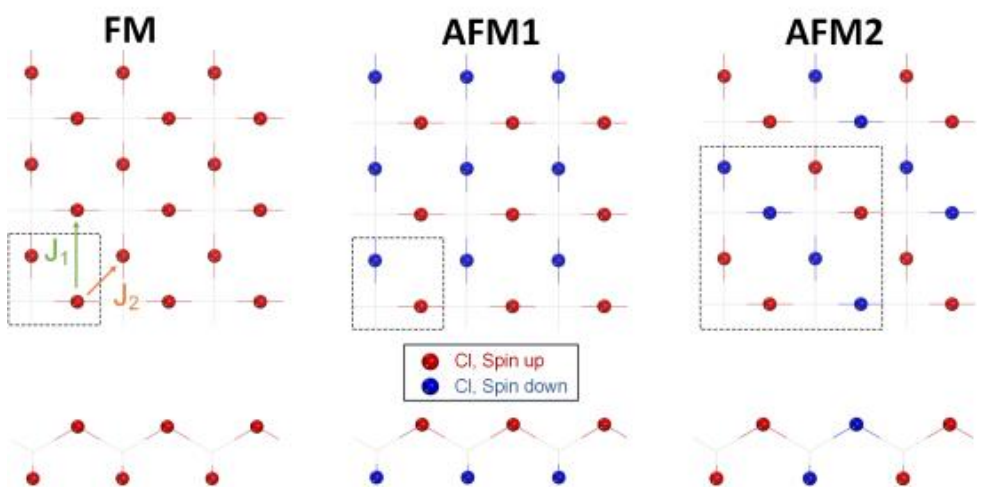

$$E_{FM} = E_0 - 4J_1S^2 - 4J_2S^2$$
$$E_{AFM1} = E_0 - 4J_1S^2 + 4J_2S^2$$
$$E_{AFM2} = E_0 + 4J_1S^2$$

## Magnetic exchange coupling parameters

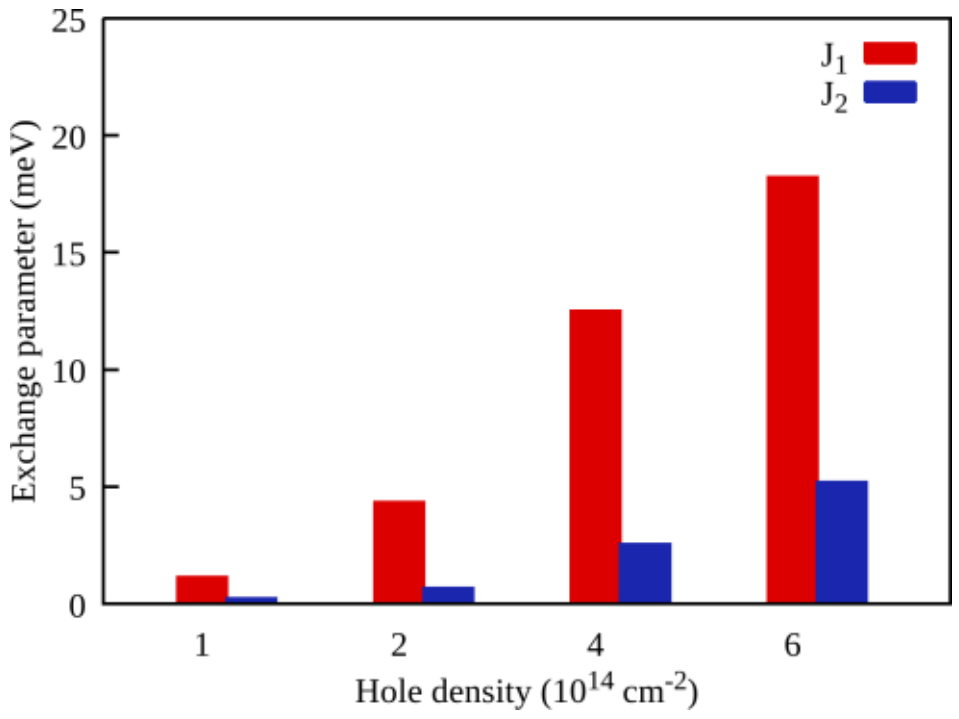

## Magnetic anisotropy energy (MAE, μeV) per magnetic atom

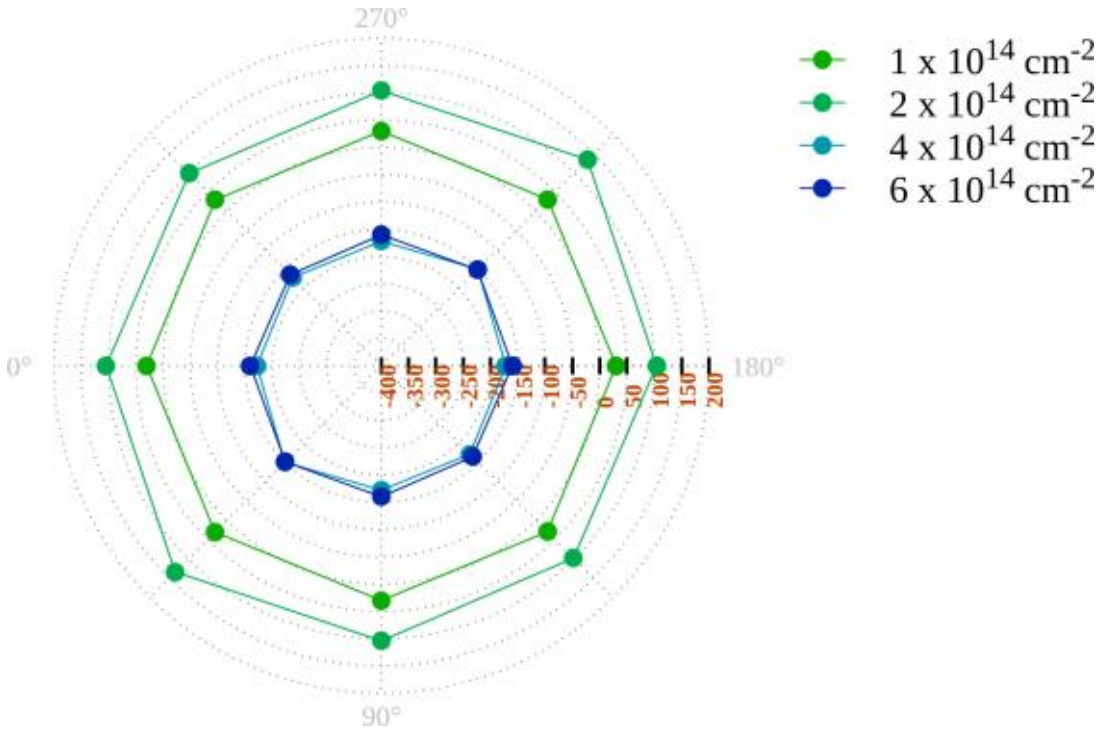

MAE = $E_{\parallel} - E_{\perp}$, a positive (negative) value of MAE indicates the off-plane (in-plane) easy axis.

## Monte Carlo simulations of the normalized magnetization of as a function of temperature

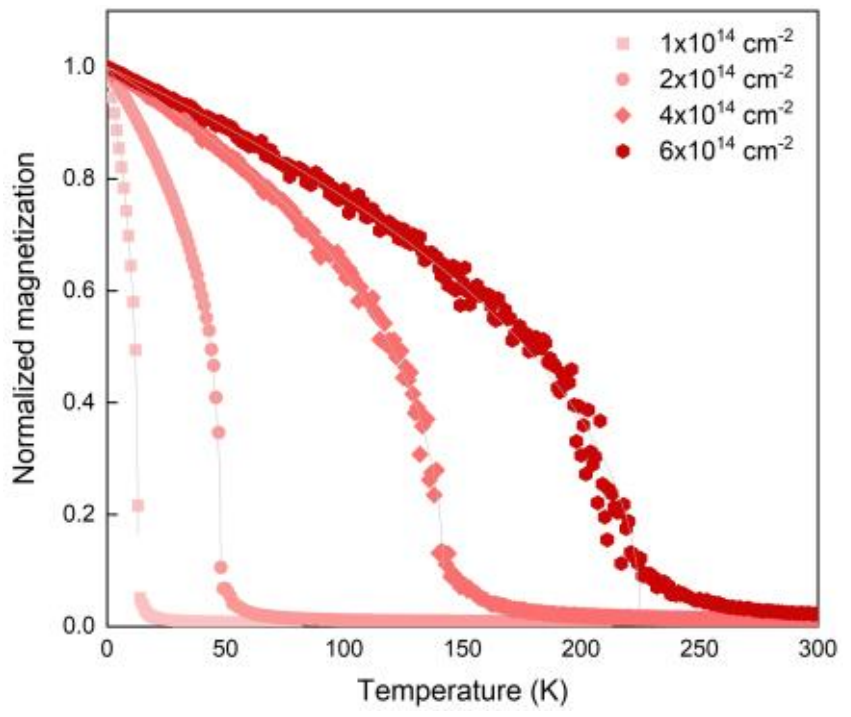

$T_c$: $1\times10^{14}$ cm$^{-2}$: 13 K $2\times10^{14}$ cm$^{-2}$: 48 K $4\times10^{14}$ cm$^{-2}$: 142 K $6\times10^{14}$ cm$^{-2}$: 225 K

# 86.ZnBr2`

| MC2D-ID | C2DB | 2dmat-ID | USPEX | Space group | Band gap (eV) |
|---|---|---|---|---|---|
| - | ✓ | 2dm-186 | - | P4m2 | 3.37 |

**Convex hull**

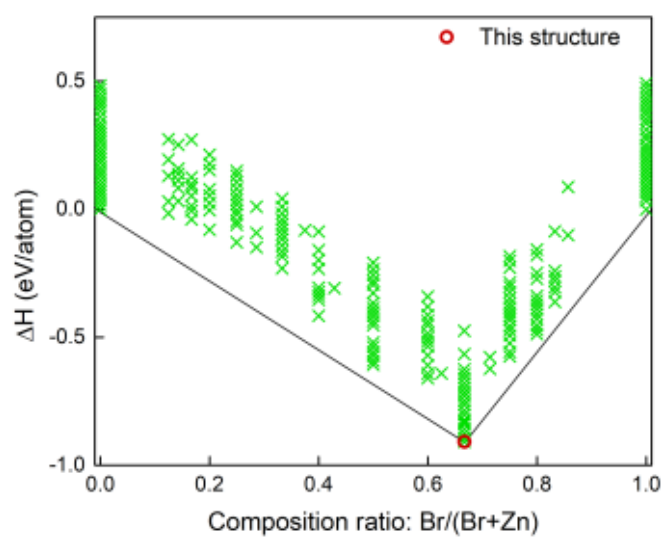

**Atomic structure**

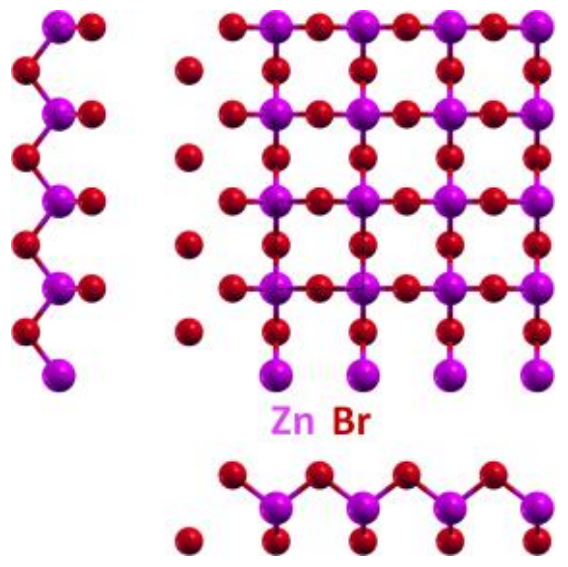

**Atomic coordinates**

```
Zn1Br2
   1.00000000000000
     3.9153705009491917   0.0000000000000000   0.0000000000000000
     0.0000000000000000   3.9153710571021980   0.0000000000000000
     0.0000000000000000   0.0000000000000000  22.7501749999999987
   Zn   Br
    1    2
Direct
  0.9999999799870594  0.0000000040239740  0.4999999955561520
  0.0000000202587032  0.5000000005451213  0.4347980916009320
  0.4999999997542375  0.9999999954308976  0.5652019128429160
```

**Phonon dispersion curve**

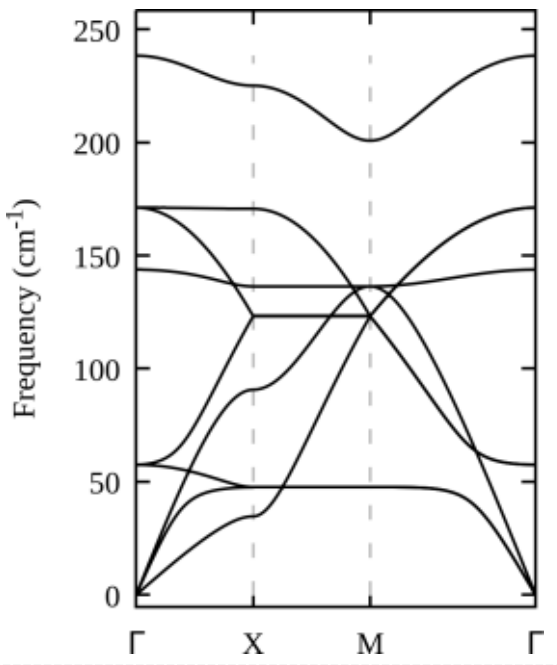

**Projected band structure and density of states**

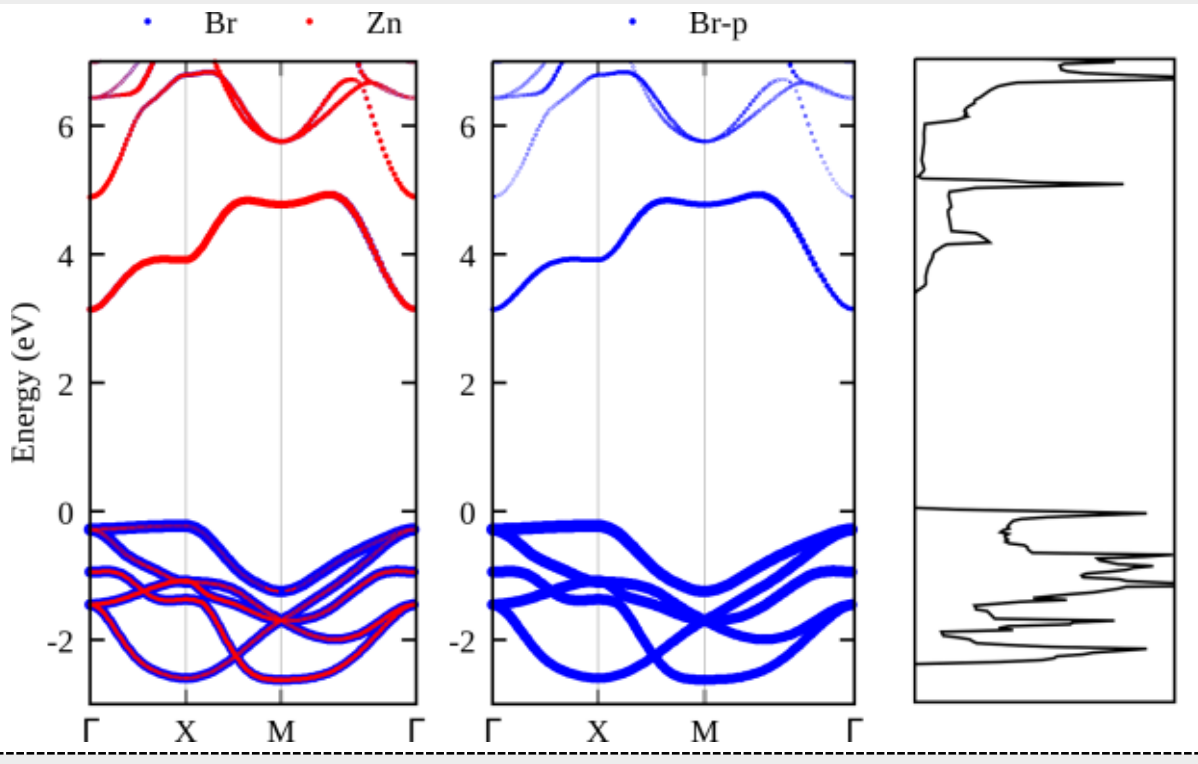

**Magnetic moment and spin polarization energy as a function of hole doping concentration**

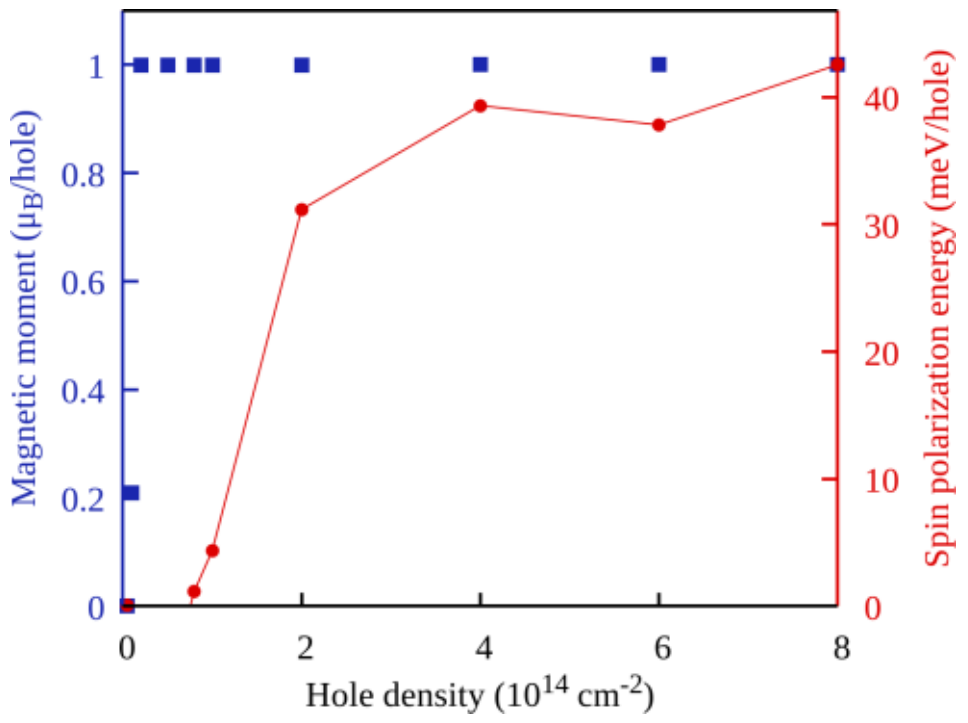

**Magnetic configurations and spin Hamiltonian**

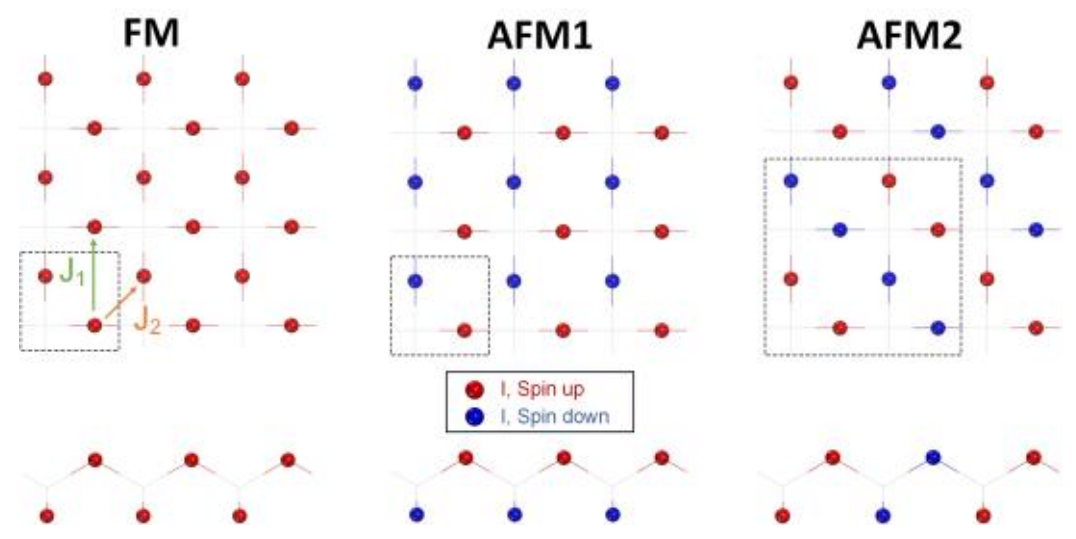

$$E_{FM} = E_0 - 4J_1S^2 - 4J_2S^2$$
$$E_{AFM1} = E_0 - 4J_1S^2 + 4J_2S^2$$
$$E_{AFM2} = E_0 + 4J_1S^2$$

**Magnetic exchange coupling parameters**

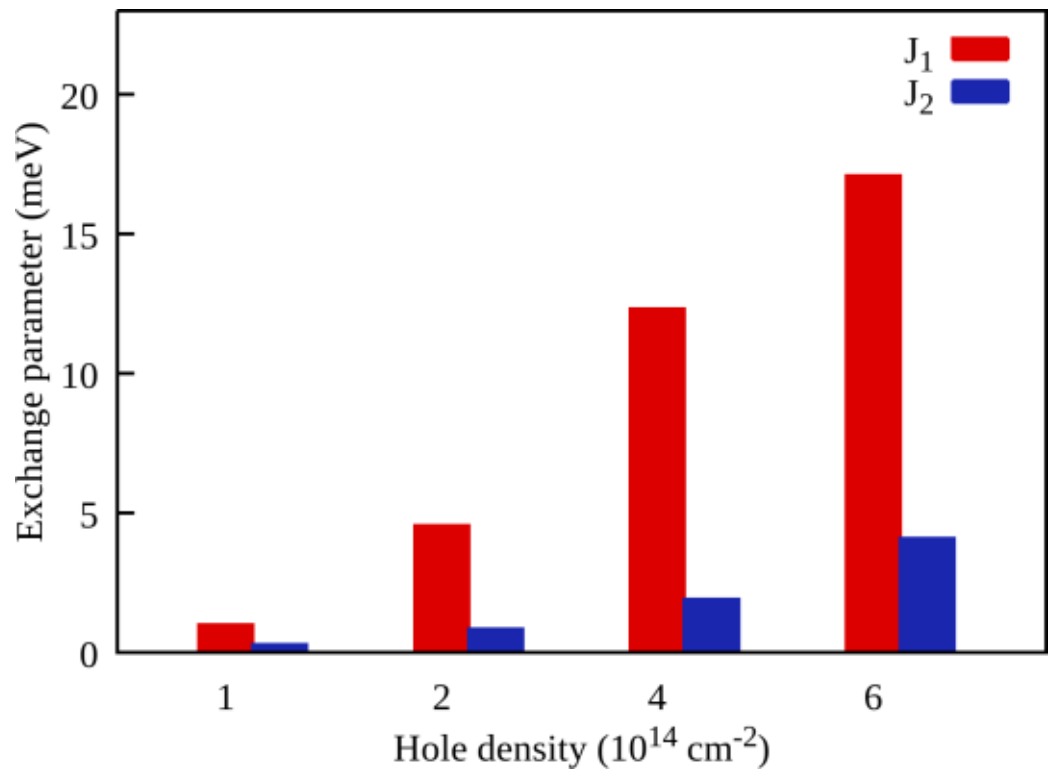

**Magnetic anisotropy energy (MAE, μeV) per magnetic atom**

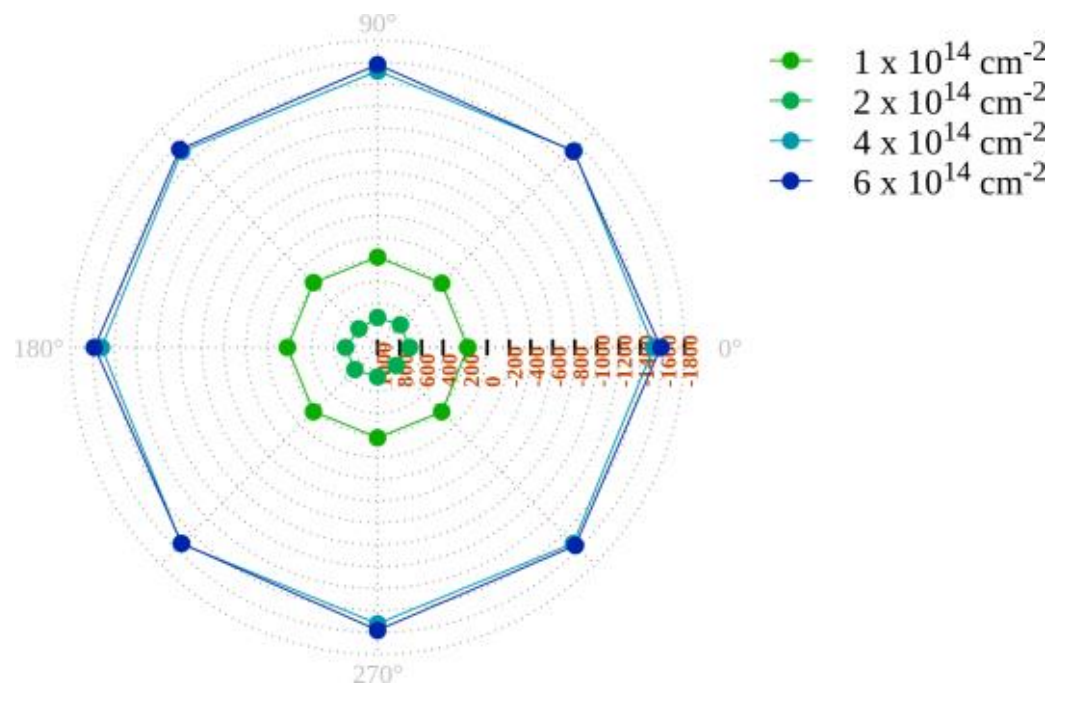

MAE = $E_{\parallel} - E_{\perp}$, a positive (negative) value of MAE indicates the off-plane (in-plane) easy axis.

**Monte Carlo simulations of the normalized magnetization of as a function of temperature**

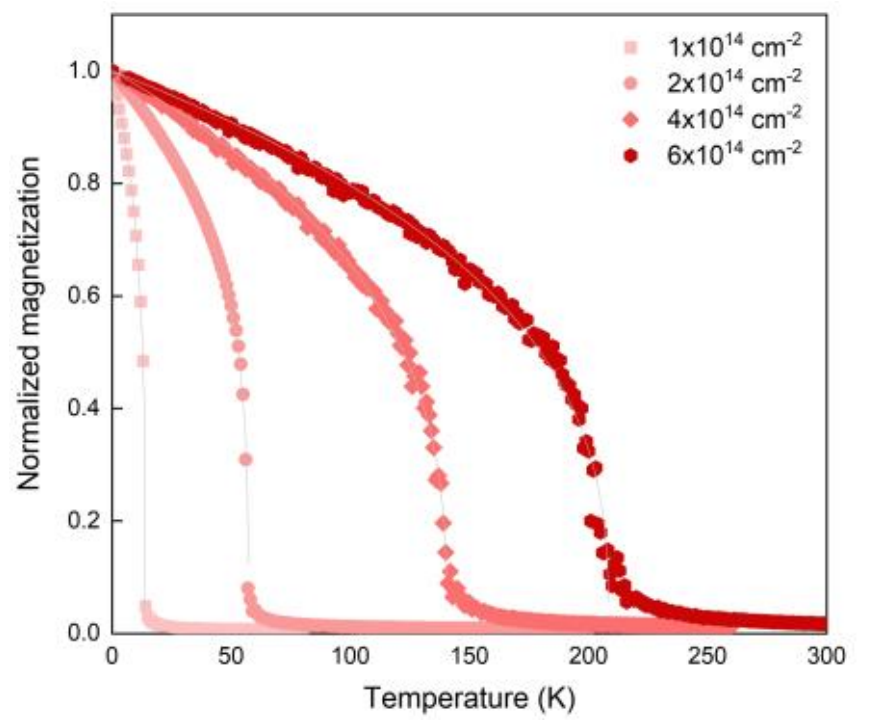

$T_c$: $1\times10^{14}$ cm$^{-2}$: 14 K  $2\times10^{14}$ cm$^{-2}$: 57 K

$4\times10^{14}$ cm$^{-2}$: 140 K  $6\times10^{14}$ cm$^{-2}$: 208 K

# 87. $CdF_2$

| MC2D-ID | C2DB | 2dmat-ID | USPEX | Space group | Band gap (eV) |
|---|---|---|---|---|---|
| - | - | 2dm-1124 | - | P4m2 | 3.78 |

**Convex hull**

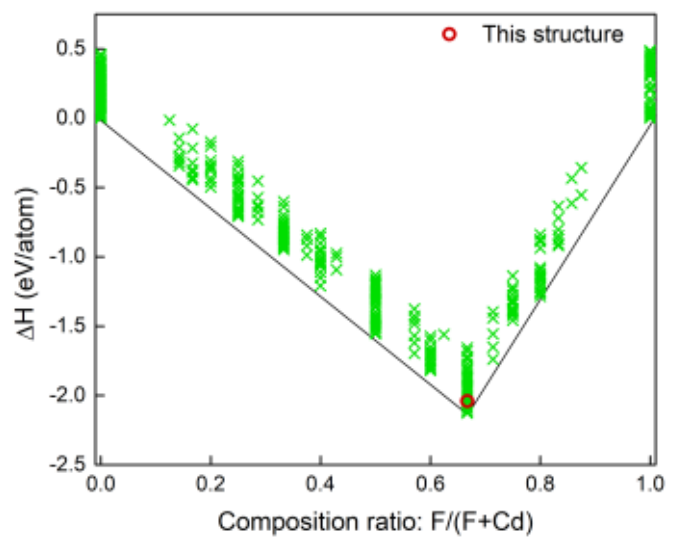

**Atomic structure**

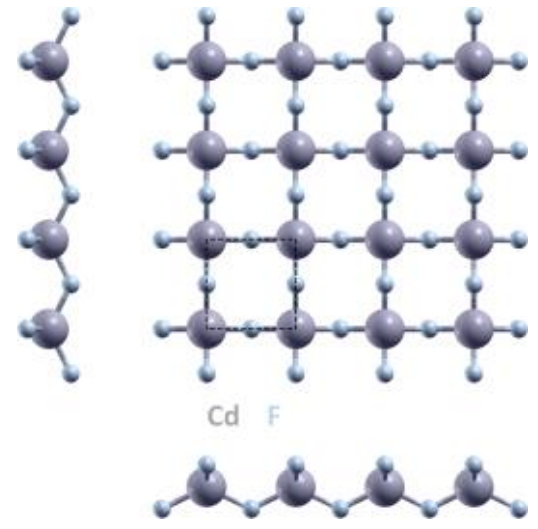

**Atomic coordinates**

```
Cd1F2
1.00000000000000
3.8660617979077605 0.0000000000000000 0.0000000000000000
0.0000000000000000 3.8660617644304405 0.0000000000000000
0.0000000000000000 0.0000000000000000 22.7413689999999988
Cd F
1 2
Direct
0.5000000002372786 0.4999999990947259 0.0000000089976382
0.0000000011284911 0.5000000001739124 0.9582099986260815
0.4999999986342374 0.0000000007313616 0.0417899923762732
```

**Phonon dispersion curve**

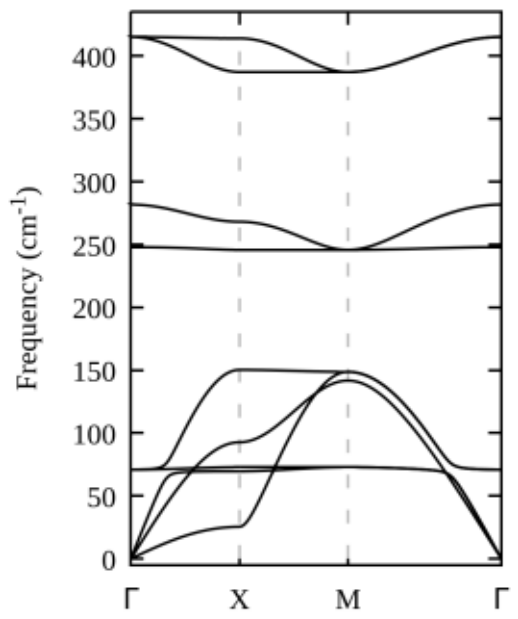

**Projected band structure and density of states**

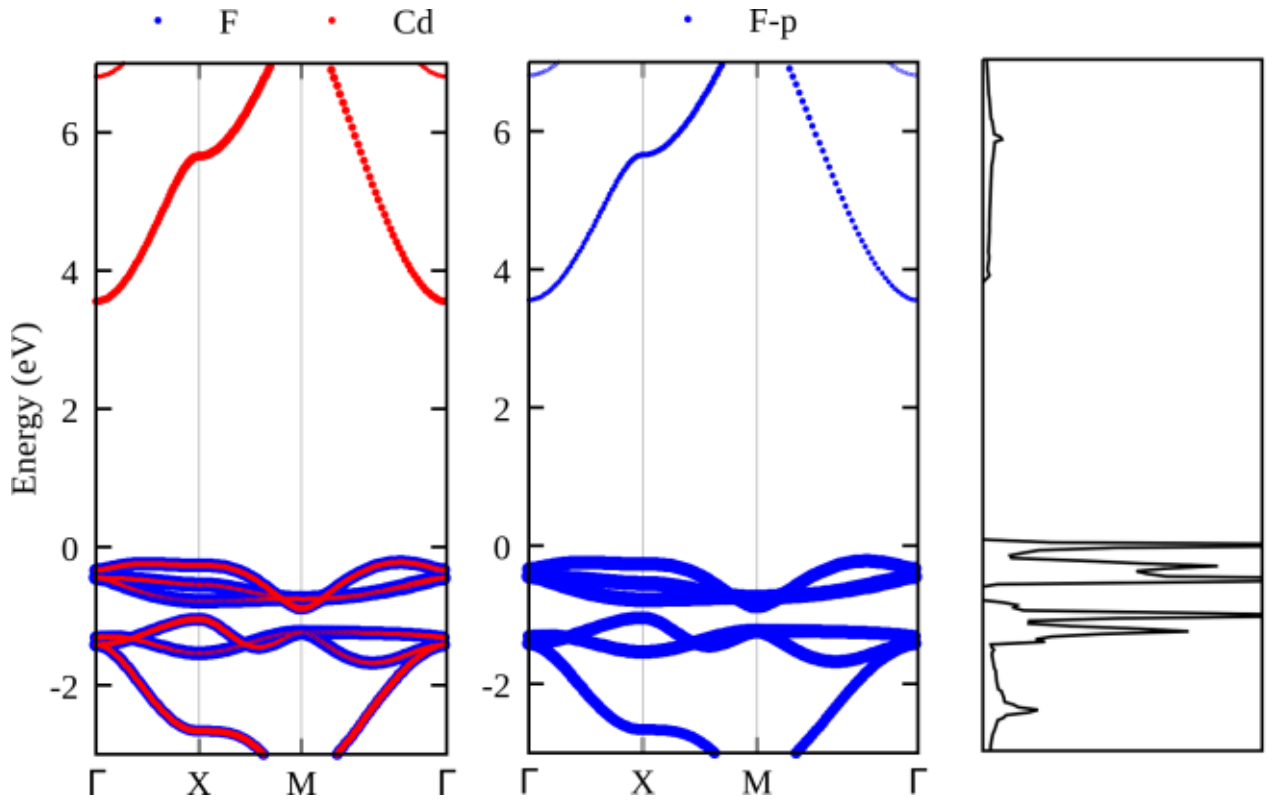

**Magnetic moment and spin polarization energy as a function of hole doping concentration**

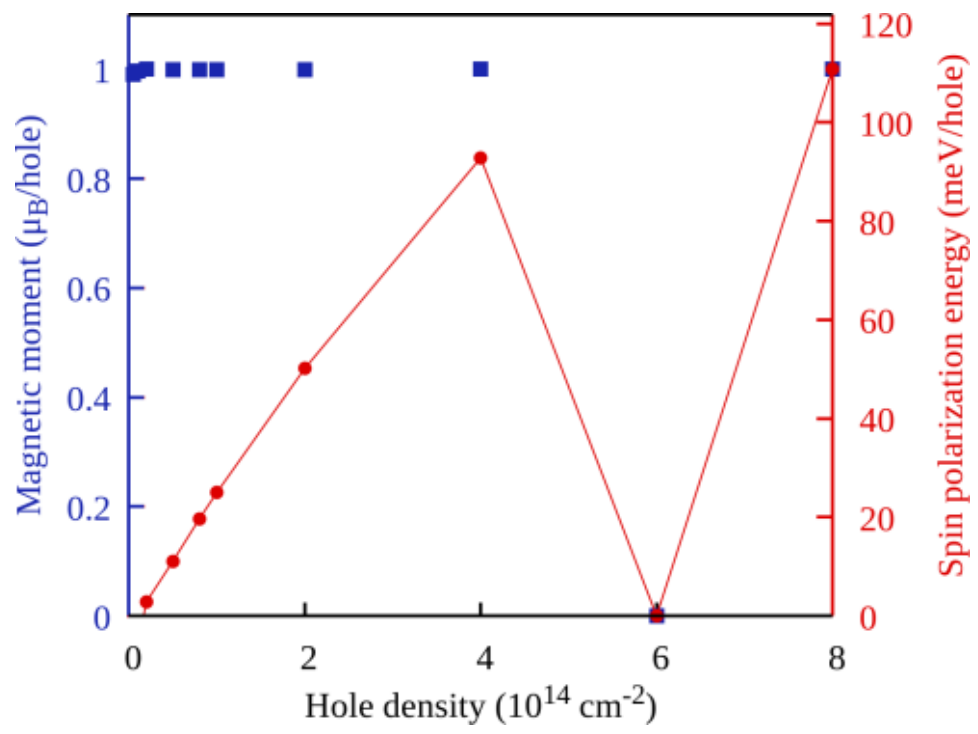

**Magnetic configurations and spin Hamiltonian**

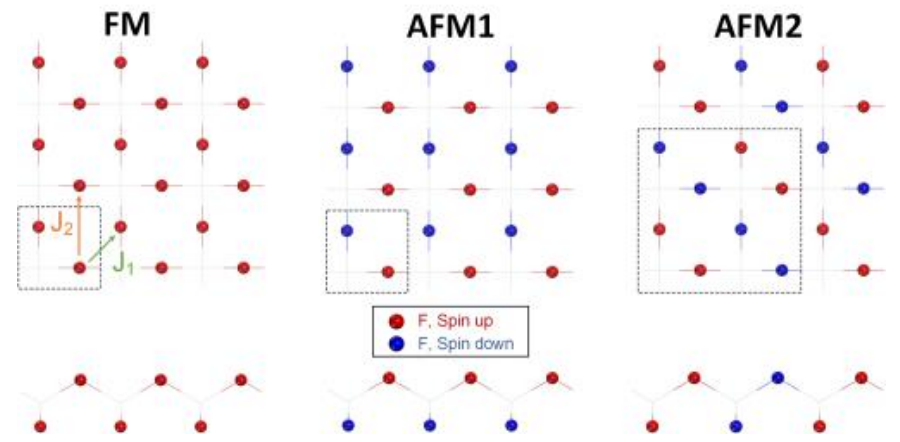

$$E_{FM} = E_0 - 4J_1S^2 - 4J_2S^2$$

$$E_{AFM1} = E_0 + 4J_1S^2 - 4J_2S^2$$

$$E_{AFM2} = E_0 + 4J_2S^2$$

**Magnetic exchange coupling parameters**

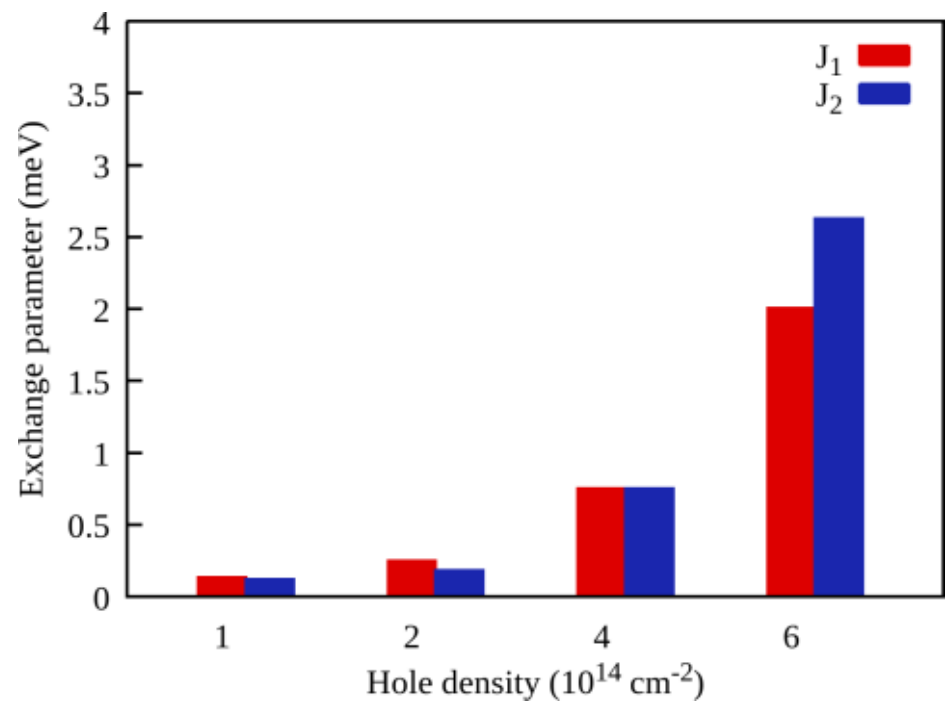

**Magnetic anisotropy energy (MAE, μeV) per magnetic atom**

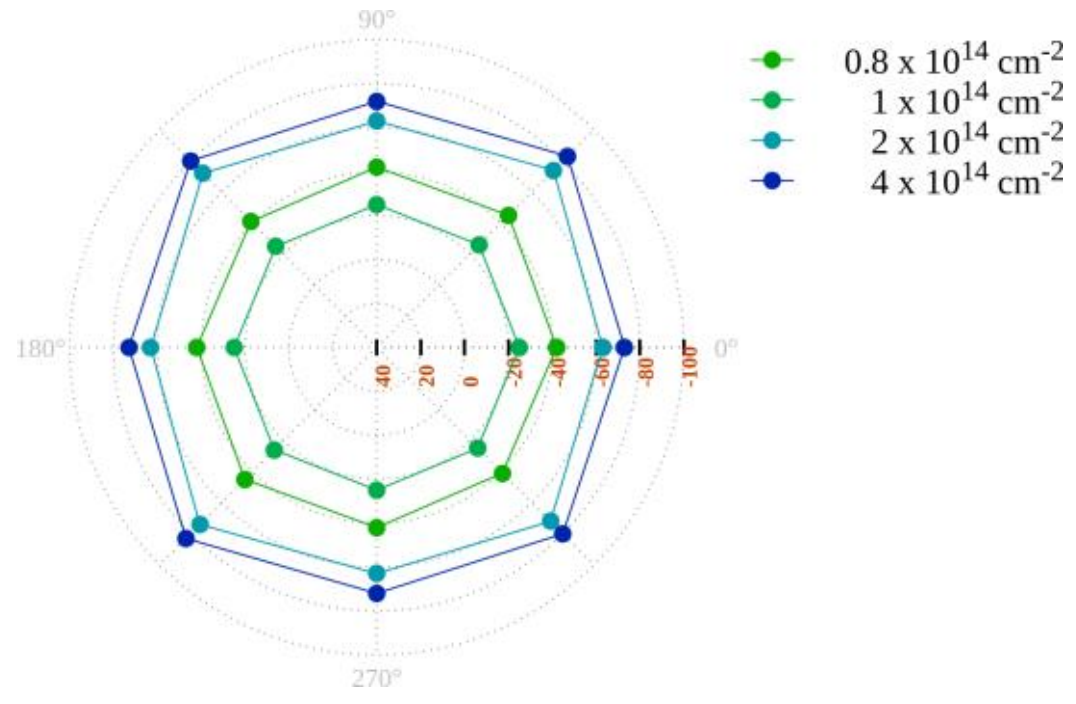

MAE = $E_{\parallel} - E_{\perp}$, a positive (negative) value of MAE indicates the off-plane (in-plane) easy axis.

**Monte Carlo simulations of the normalized magnetization of as a function of temperature**

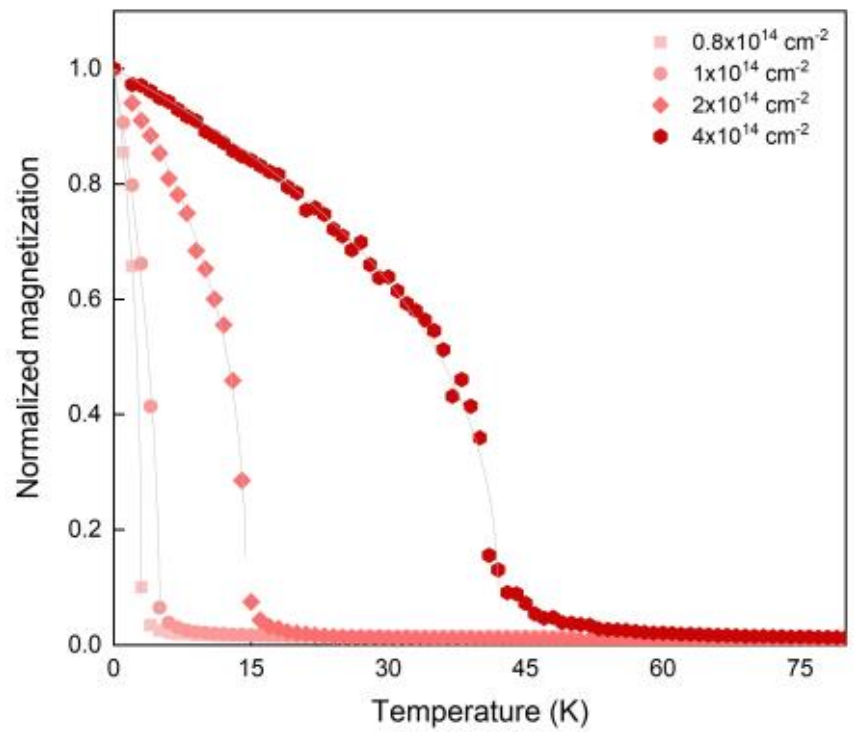

$T_c$: $1\times10^{14}$ cm$^{-2}$: 3 K $2\times10^{14}$ cm$^{-2}$: 5 K

$4\times10^{14}$ cm$^{-2}$: 14 K $6\times10^{14}$ cm$^{-2}$: 42 K

# 88.CdCl₂

| MC2D-ID | C2DB | 2dmat-ID | USPEX | Space group | Band gap (eV) |
|---|---|---|---|---|---|
| - | - | 2dm-5027 | - | P4m2 | 3.63 |

**Convex hull** | **Atomic structure** | **Atomic coordinates** | **Phonon dispersion curve**

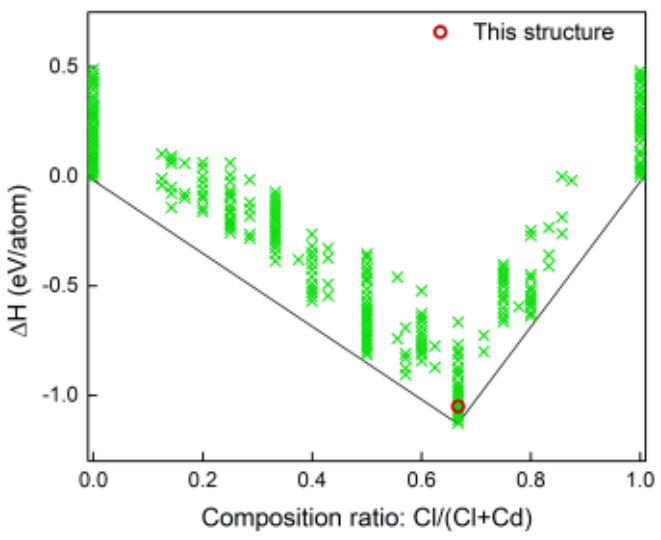

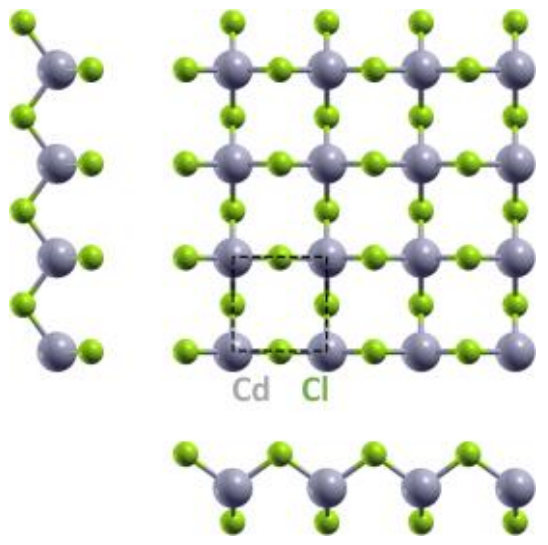

```
Cd1Cl2
  1.00000000000000
  4.0781723446932565   0.0000000000000000   0.0000000000000000
  0.0000000000000000   4.0724475101274082   0.0000000000000000
  0.0000000000000000   0.0000000000000000   22.7505990000000011
  Cd   Cl
  1    2
Direct
 0.8381951176880946  0.1710117550868731  0.5000431008569990
 0.8369366399360061  0.6711498327356509  0.4352580040546670
 0.3380302423758934  0.1725564121774710  0.5646998950883315
```

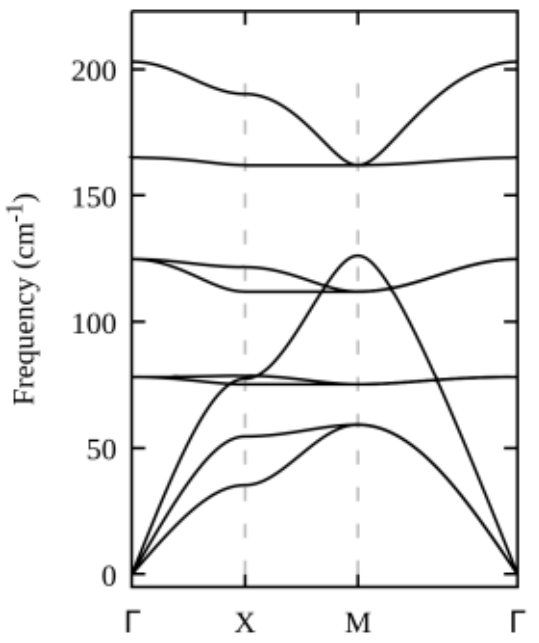

**Projected band structure and density of states** | **Magnetic moment and spin polarization energy as a function of hole doping concentration**

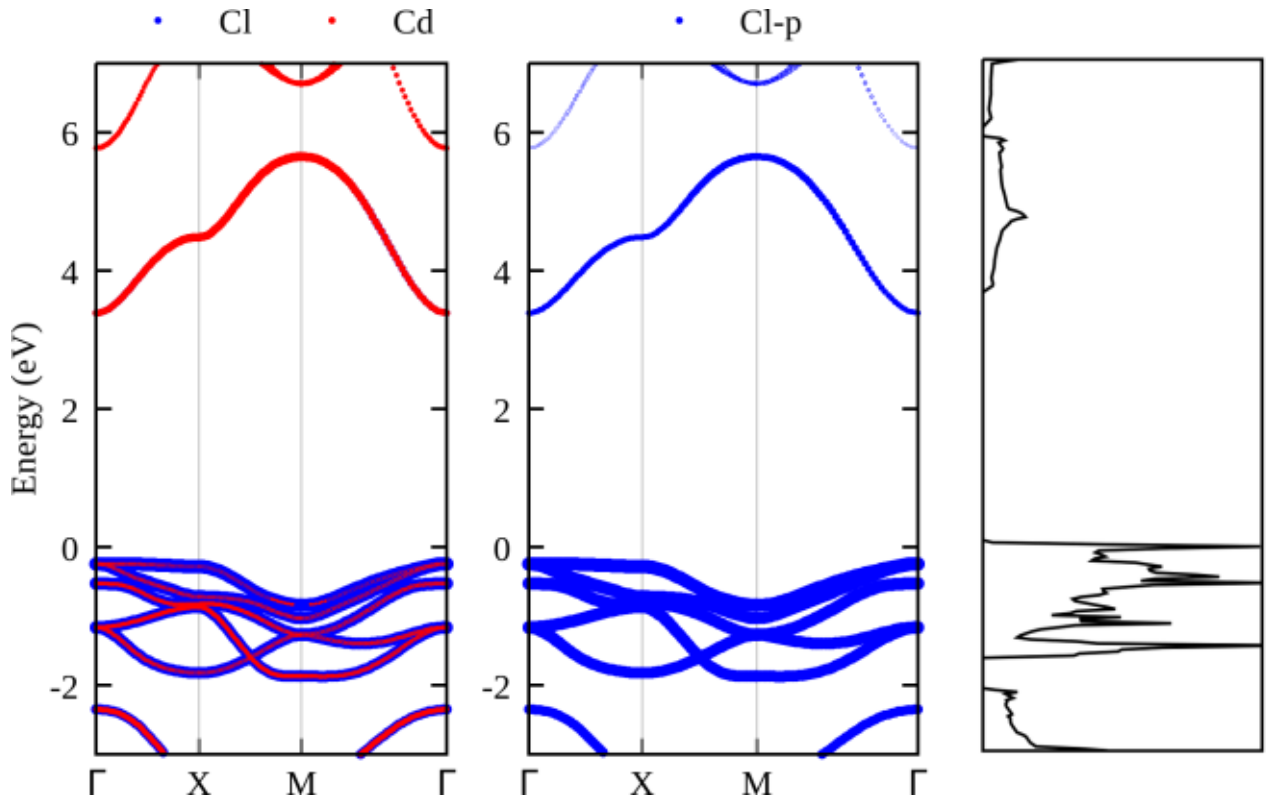

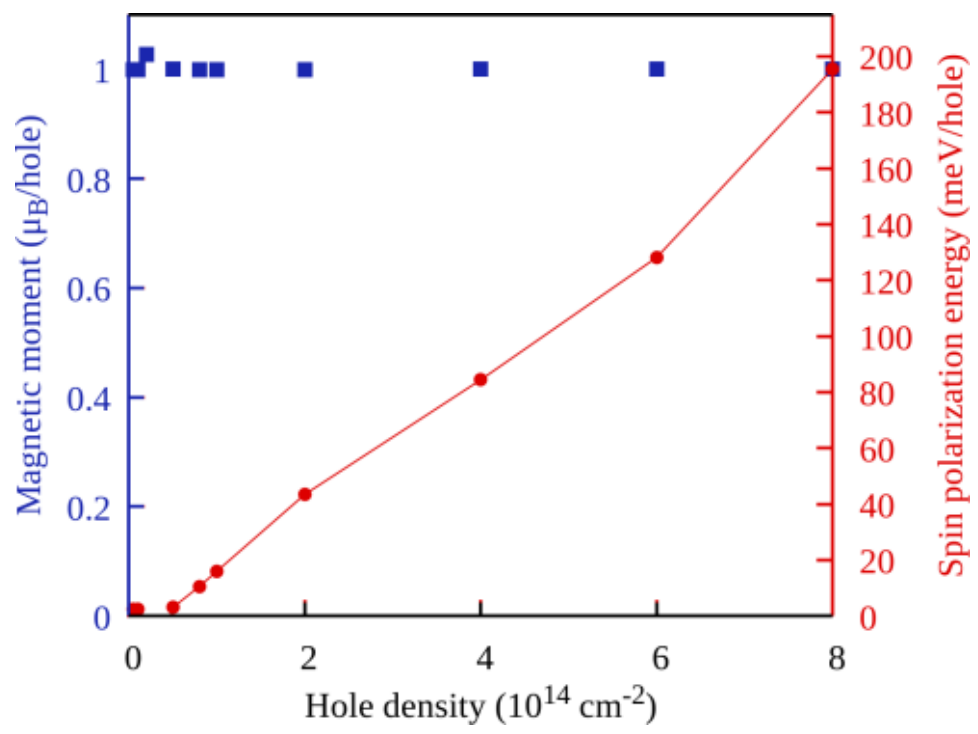

**Magnetic configurations and spin Hamiltonian** | **Magnetic exchange coupling parameters**

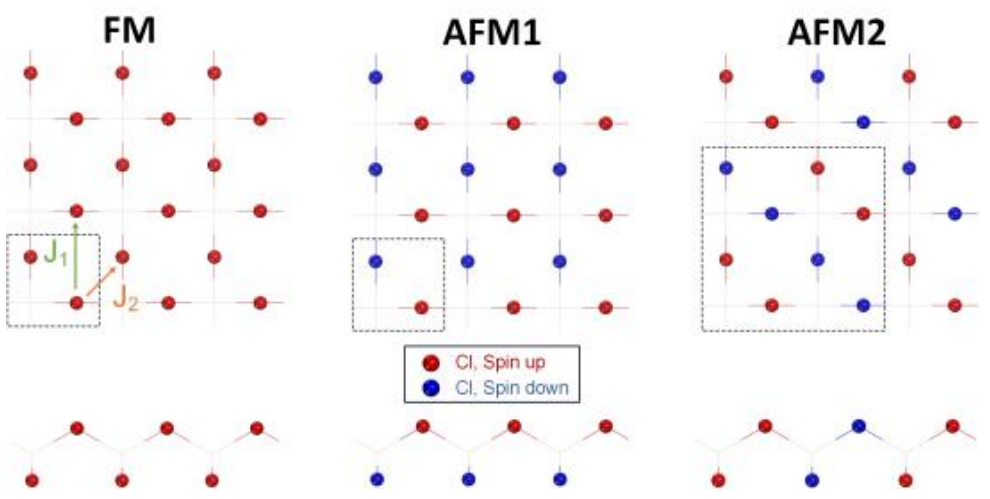

$$E_{FM} = E_0 - 4J_1S^2 - 4J_2S^2$$
$$E_{AFM1} = E_0 - 4J_1S^2 + 4J_2S^2$$
$$E_{AFM2} = E_0 + 4J_1S^2$$

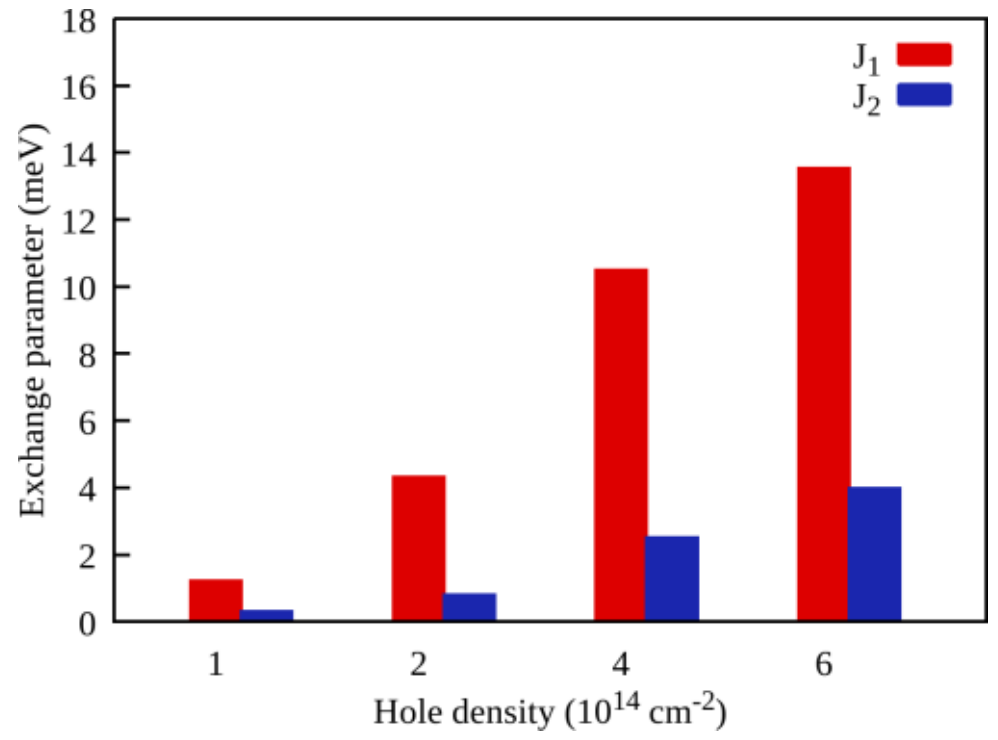

**Magnetic anisotropy energy (MAE, μeV) per magnetic atom** | **Monte Carlo simulations of the normalized magnetization of as a function of temperature**

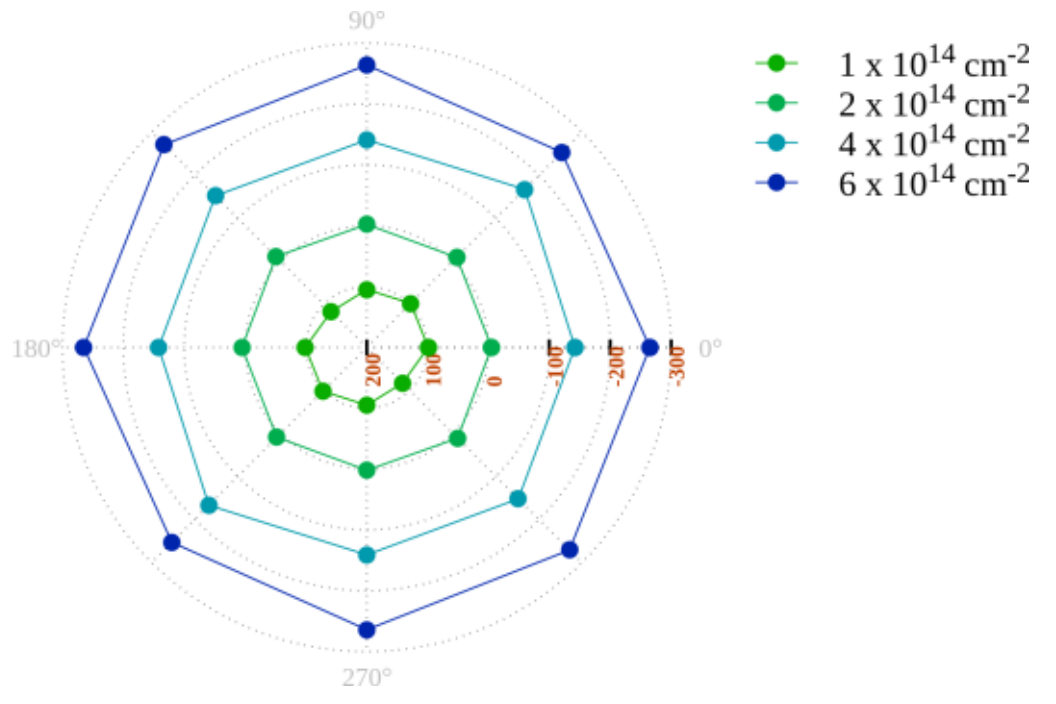

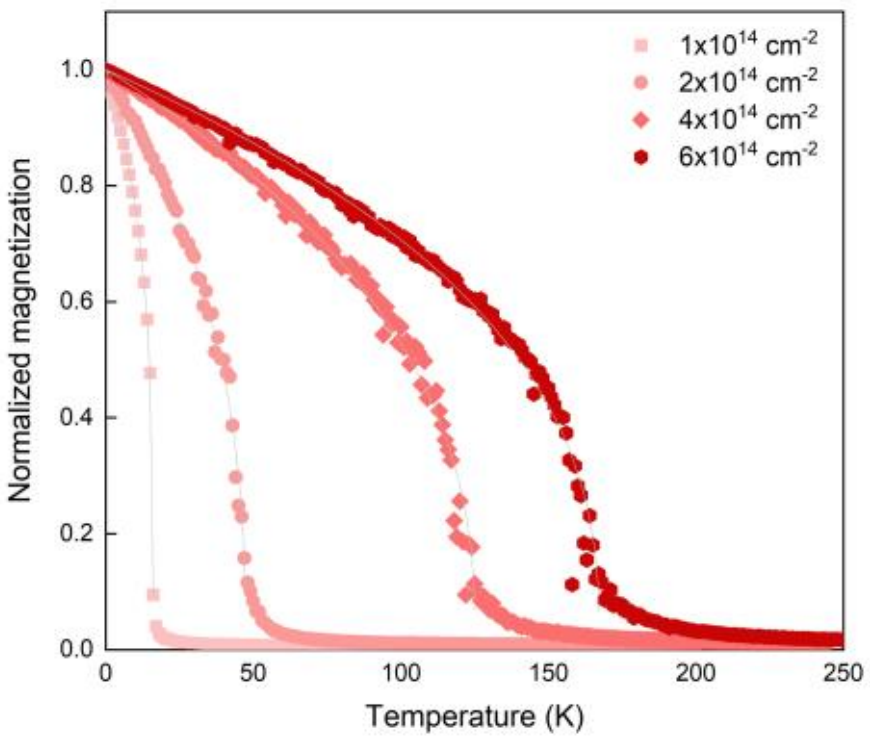

MAE = $E_{\parallel} - E_{\perp}$, a positive (negative) value of MAE indicates the off-plane (in-plane) easy axis.

$T_c$: $1\times10^{14}$ cm$^{-2}$: 16 K  $2\times10^{14}$ cm$^{-2}$: 47 K  $4\times10^{14}$ cm$^{-2}$: 125 K  $6\times10^{14}$ cm$^{-2}$: 167 K

# 89. $CdBr_2$

| MC2D-ID | C2DB | 2dmat-ID | USPEX | Space group | Band gap (eV) |
|---|---|---|---|---|---|
| - | - | 2dm-24 | - | P4m2 | 3.04 |

**Convex hull** | **Atomic structure** | **Atomic coordinates** | **Phonon dispersion curve**

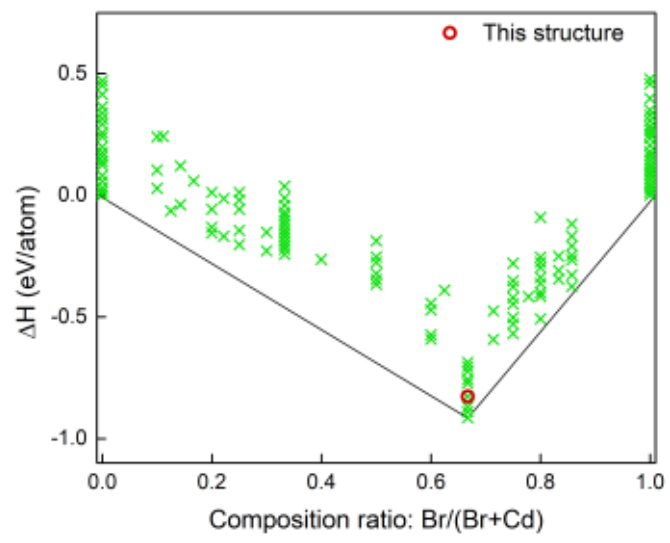

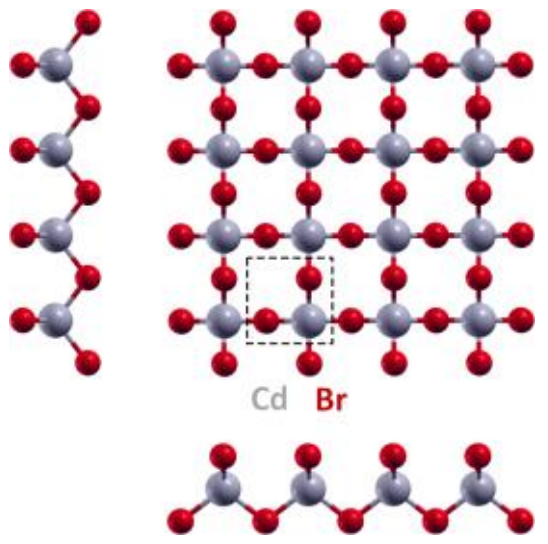

```
Cd1Br2
   1.00000000000000
     4.2145244047588637   0.0000000000000000   0.0000000000000000
     0.0000000000000000   4.2145256921027006   0.0000000000000000
     0.0000000000000000   0.0000000000000000  22.7413689999999988
   Cd   Br
     1     2
Direct
  0.5000000009435723  0.4999999953282668  0.9999999850019776
  0.0000000012902532  0.4999999993134665  0.9291315382588436
  0.4999999977661815  0.0000000053582667  0.0708684767391787
```

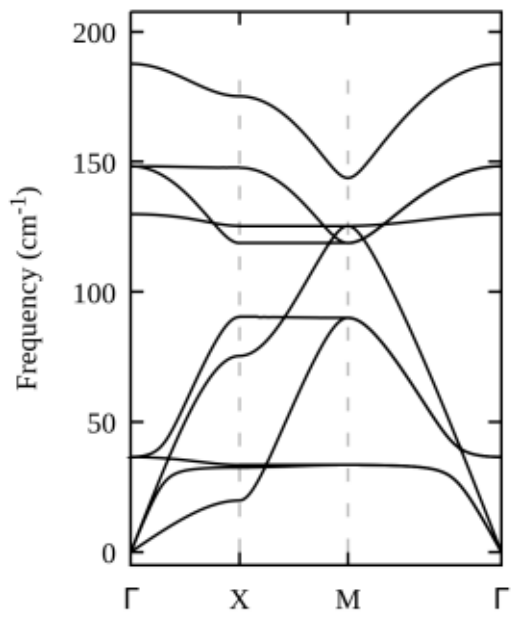

**Projected band structure and density of states** | **Magnetic moment and spin polarization energy as a function of hole doping concentration**

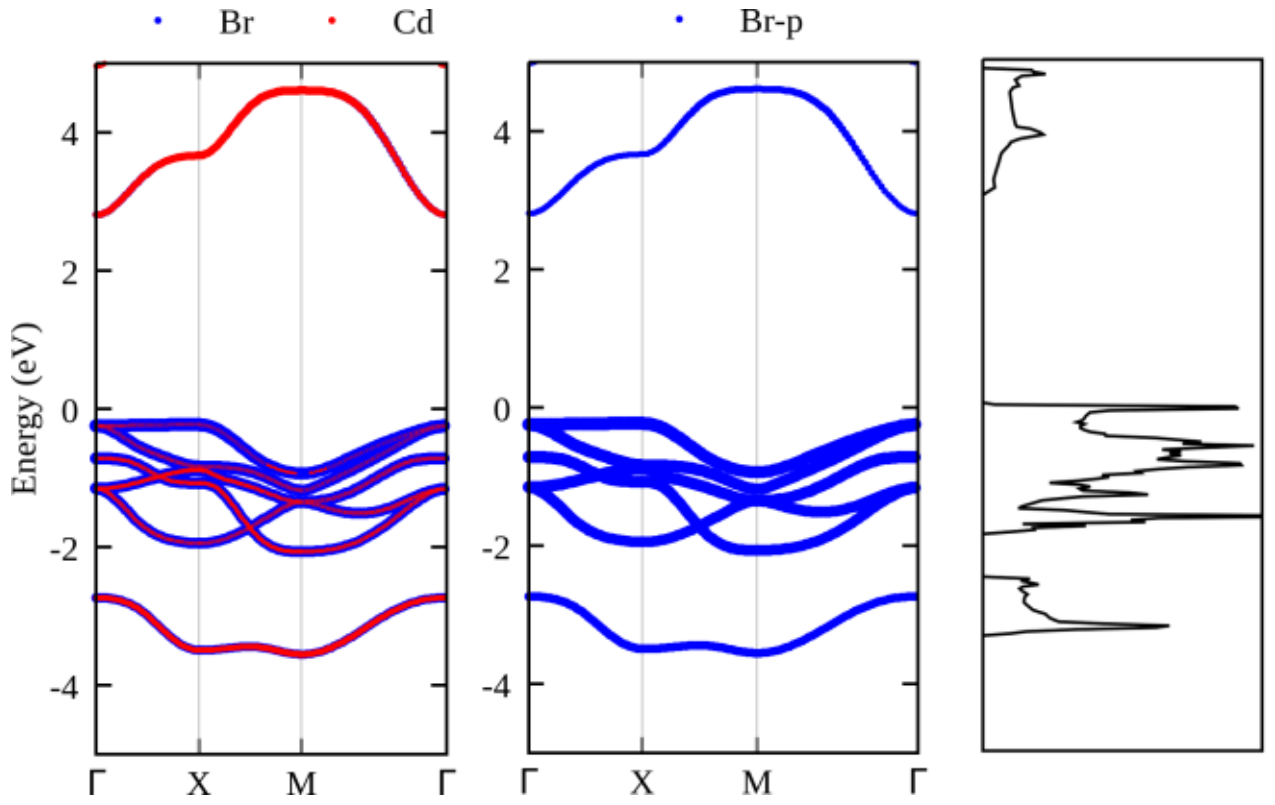

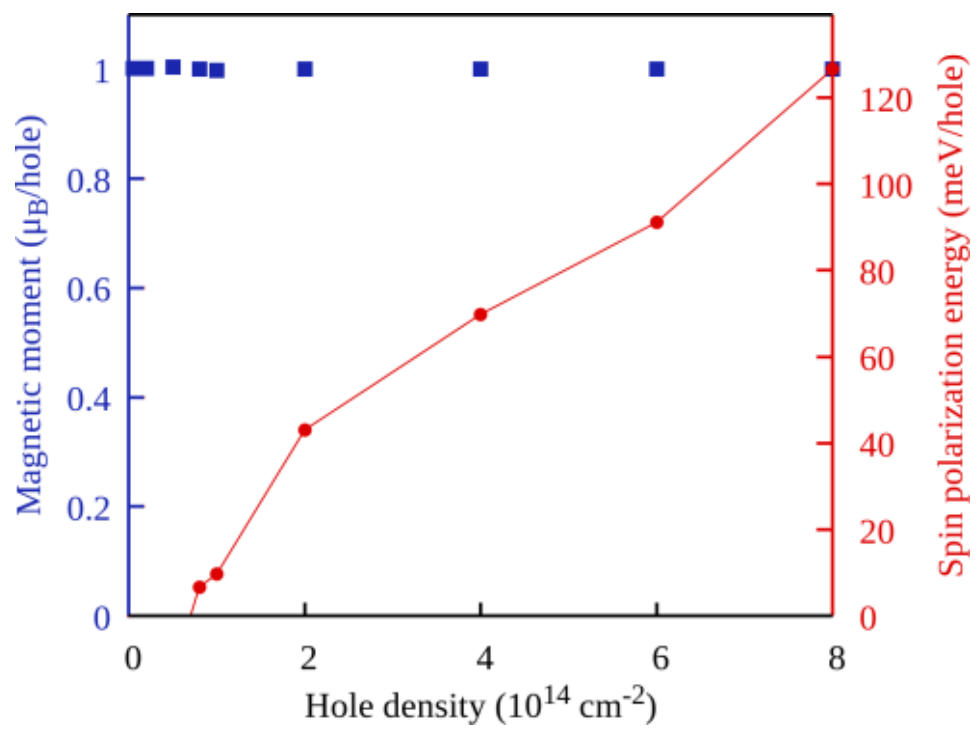

**Magnetic configurations and spin Hamiltonian** | **Magnetic exchange coupling parameters**

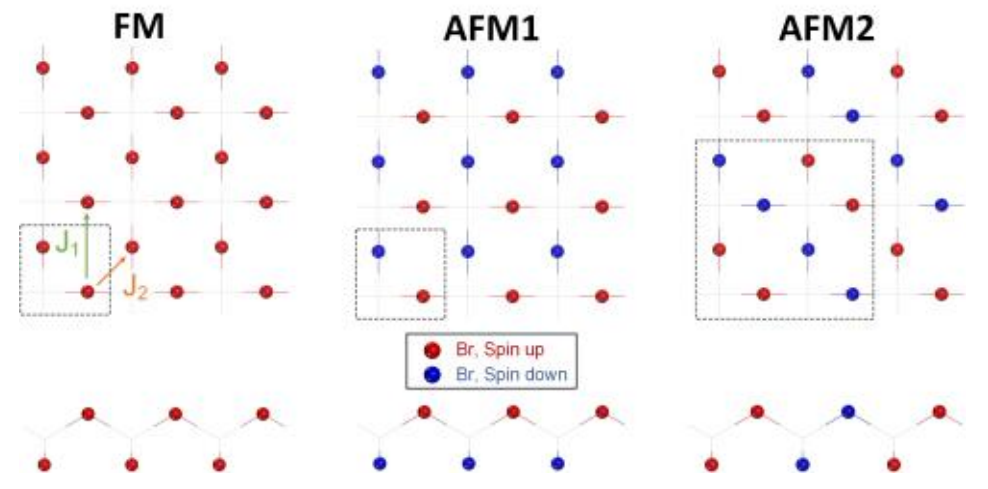

$$E_{FM} = E_0 - 4J_1S^2 - 4J_2S^2$$
$$E_{AFM1} = E_0 - 4J_1S^2 + 4J_2S^2$$
$$E_{AFM2} = E_0 + 4J_1S^2$$

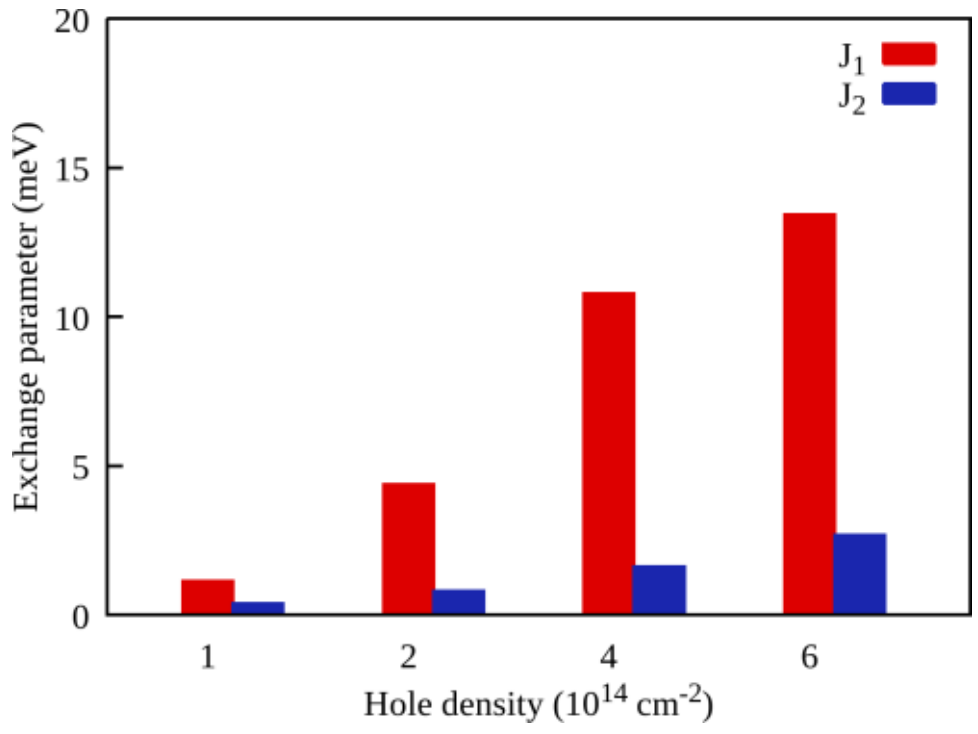

**Magnetic anisotropy energy (MAE, μeV) per magnetic atom** | **Monte Carlo simulations of the normalized magnetization of as a function of temperature**

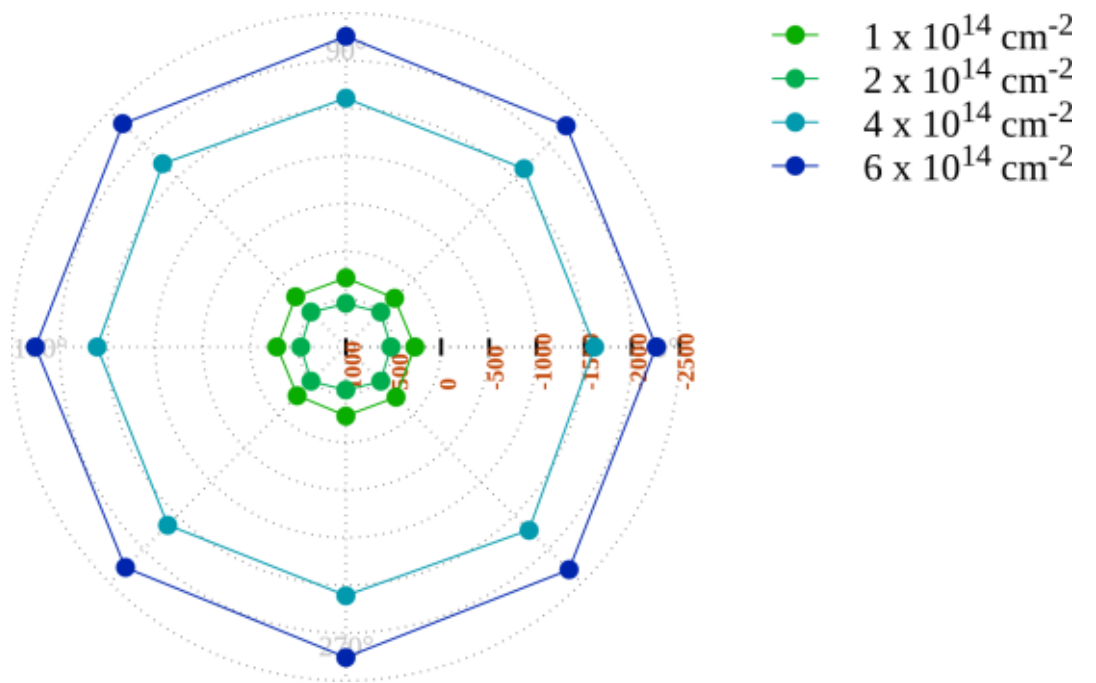

MAE = $E_{\parallel} - E_{\perp}$, a positive (negative) value of MAE indicates the off-plane (in-plane) easy axis.

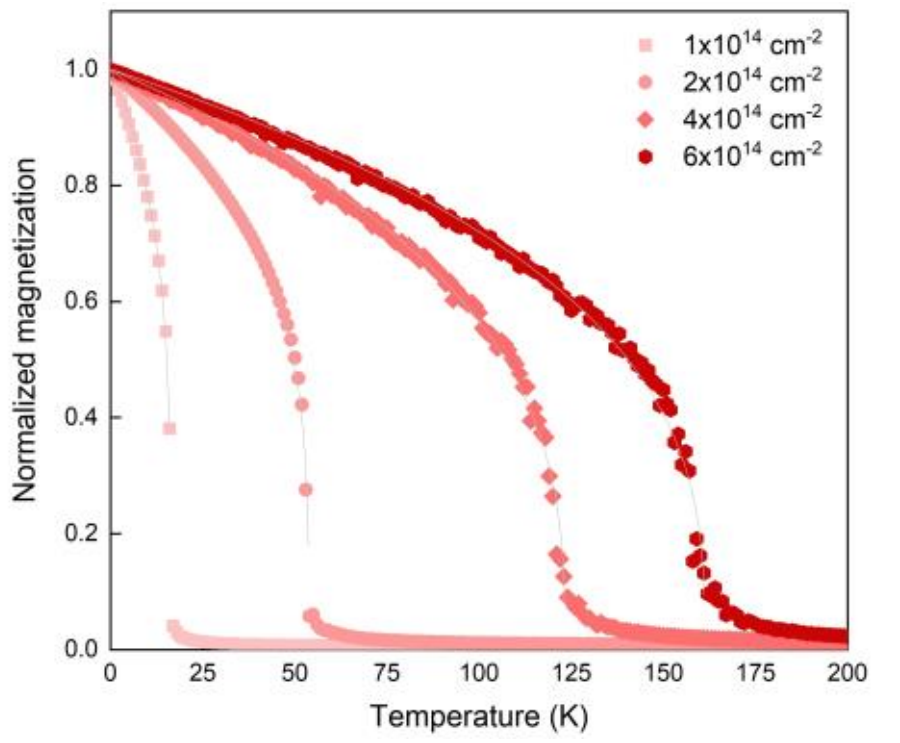

$T_c$: $1\times10^{14}$ cm$^{-2}$: 16 K  $2\times10^{14}$ cm$^{-2}$: 53 K

$4\times10^{14}$ cm$^{-2}$: 123 K  $6\times10^{14}$ cm$^{-2}$: 161 K

# 90. $CdI_2$

| MC2D-ID | C2DB | 2dmat-ID | USPEX | Space group | Band gap (eV) |
|---|---|---|---|---|---|
| - | - | 2dm-498 | - | P4m2 | 2.59 |

**Convex hull**

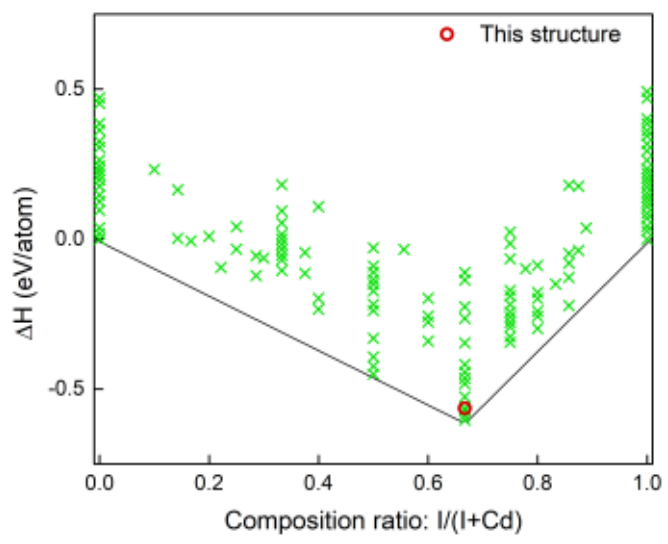

**Atomic structure**

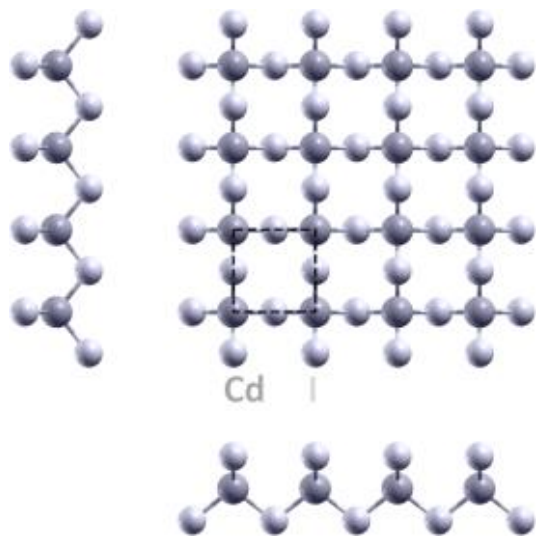

**Atomic coordinates**

```
Cd1I2
   1.00000000000000
   4.4830774580651989   0.0000000000000000   0.0000000000000000
   0.0000000000000000   4.4830733220360397   0.0000000000000000
   0.0000000000000000   0.0000000000000000  22.7413689999999988
   Cd   I
   1    2
Direct
  0.4999999740505032  0.5000000072614341  0.0000000437352909
  0.0000000222610765  0.4999999900661791  0.9230432821464731
  0.5000000036884202  0.0000000026723868  0.0769566741182359
```

**Phonon dispersion curve**

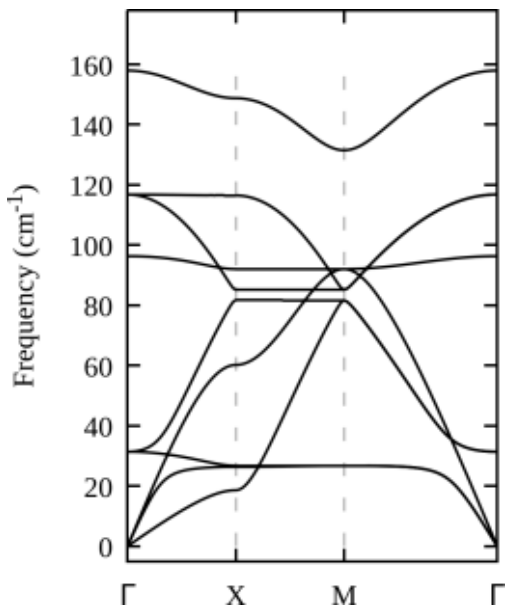

**Projected band structure and density of states**

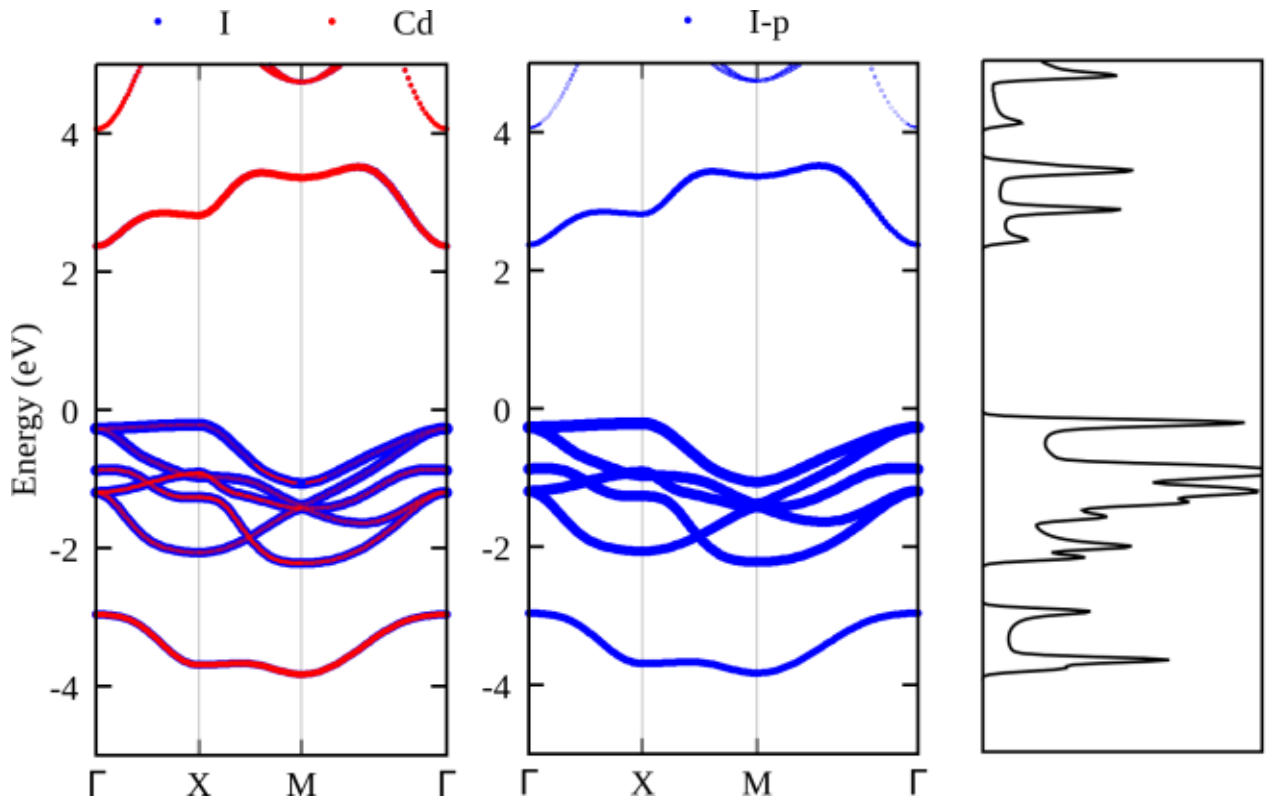

**Magnetic moment and spin polarization energy as a function of hole doping concentration**

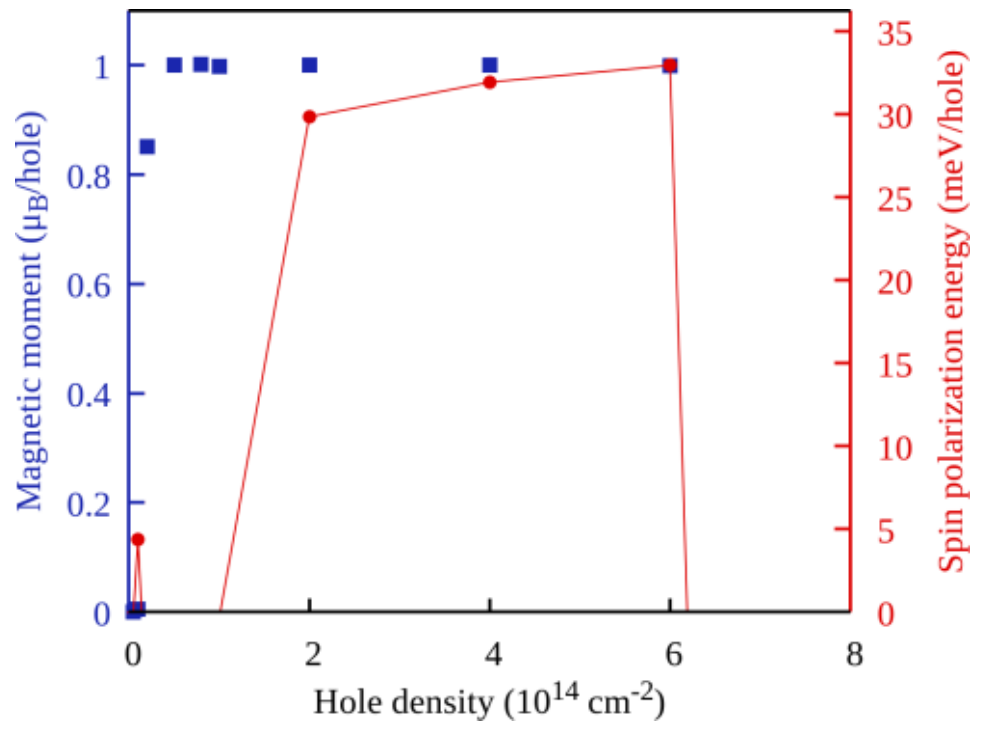

**Magnetic configurations and spin Hamiltonian**

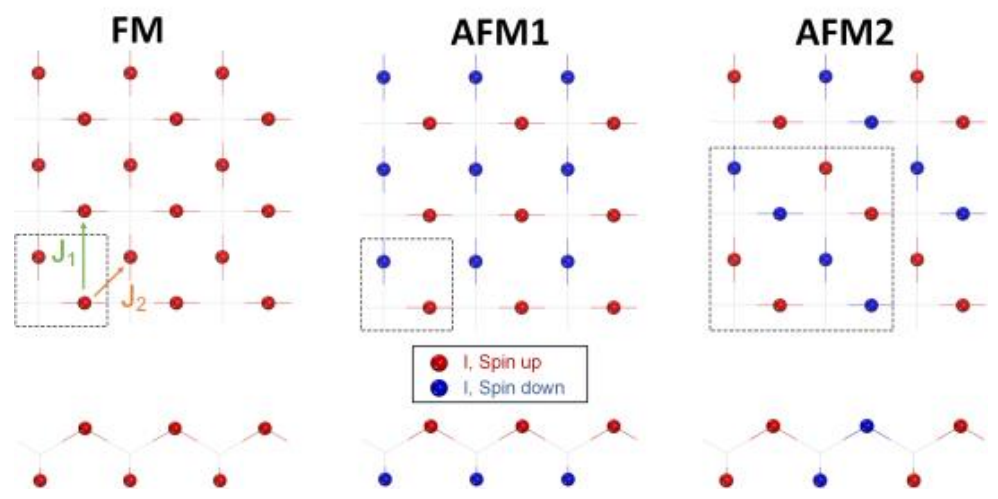

$$E_{FM} = E_0 - 4J_1S^2 - 4J_2S^2$$

$$E_{AFM1} = E_0 - 4J_1S^2 + 4J_2S^2$$

$$E_{AFM2} = E_0 + 4J_1S^2$$

**Magnetic exchange coupling parameters**

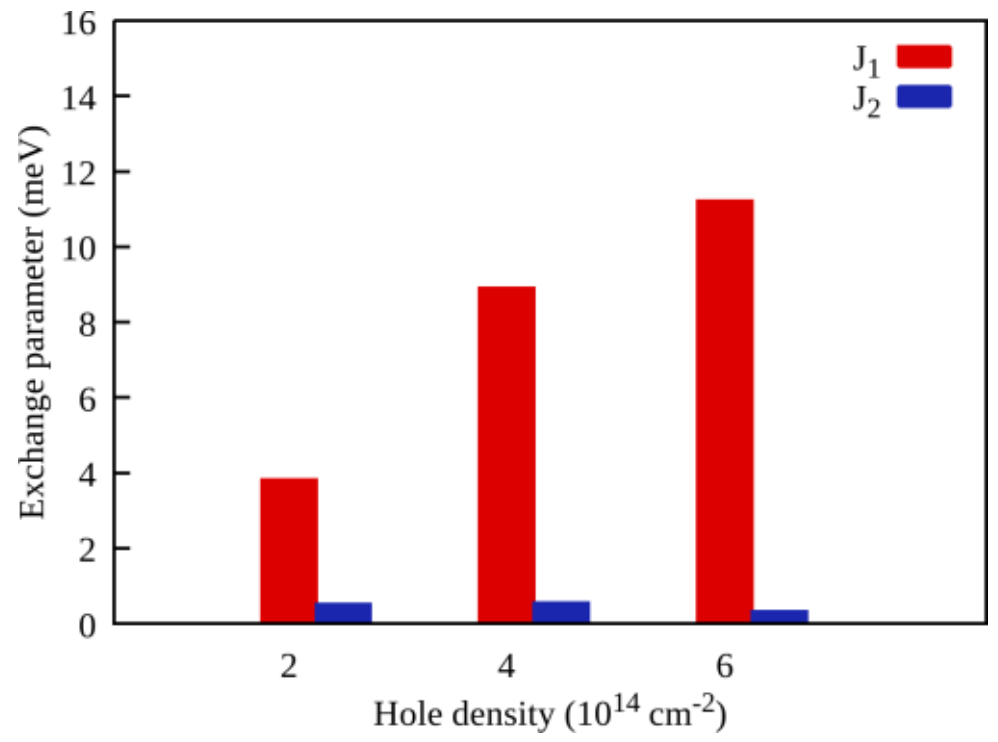

**Magnetic anisotropy energy (MAE, μeV) per magnetic atom**

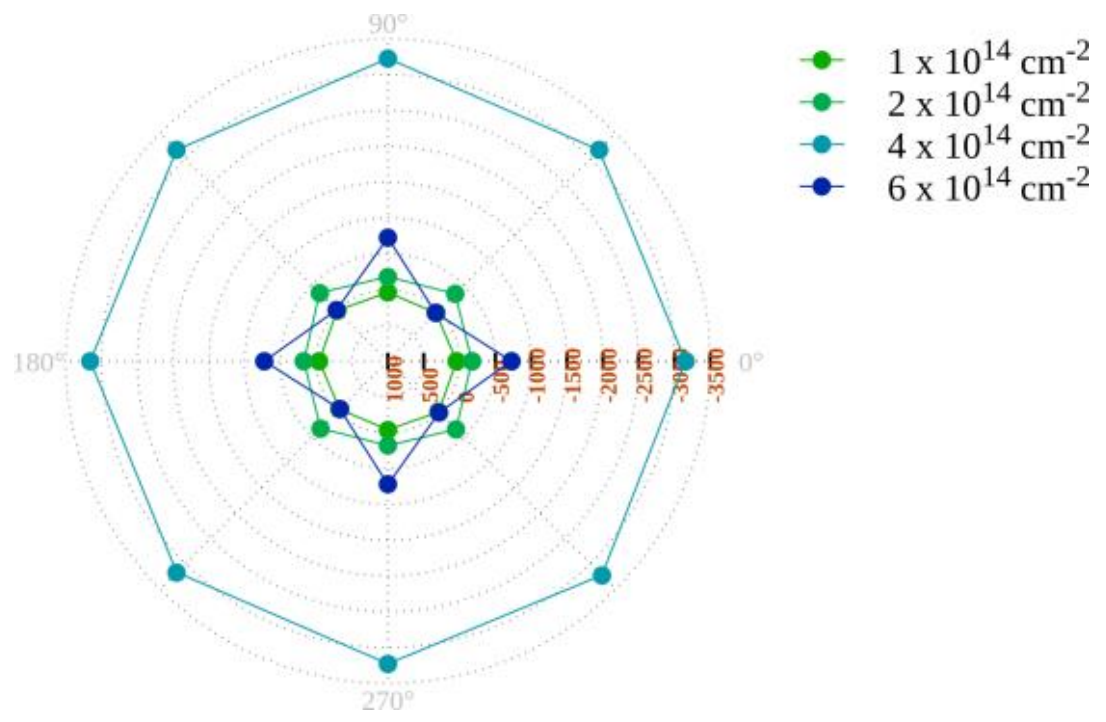

MAE = $E_{\parallel} - E_{\perp}$, a positive (negative) value of MAE indicates the off-plane (in-plane) easy axis.

**Monte Carlo simulations of the normalized magnetization of as a function of temperature**

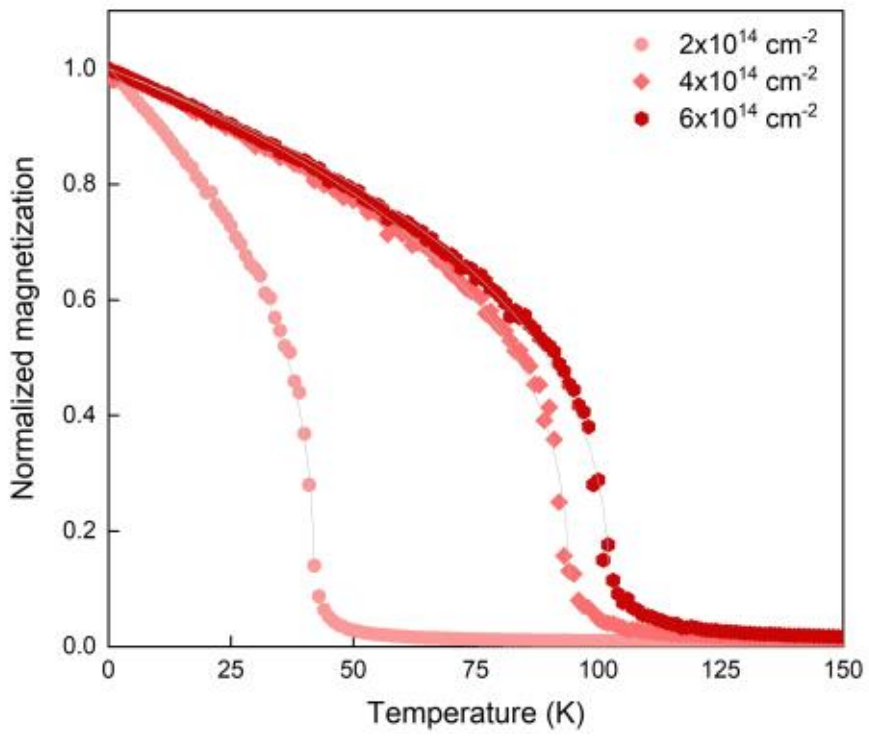

$T_c$: $1\times10^{14}$ cm$^{-2}$: - K  $2\times10^{14}$ cm$^{-2}$: 42 K

$4\times10^{14}$ cm$^{-2}$: 94 K  $6\times10^{14}$ cm$^{-2}$: 102 K

# 91.$SnS_2$

| MC2D-ID | C2DB | 2dmat-ID | USPEX | Space group | Band gap (eV) |
|---|---|---|---|---|---|
| - | ✓ | 2dm-6028 | - | P4m2 | 1.45 |

## Convex hull

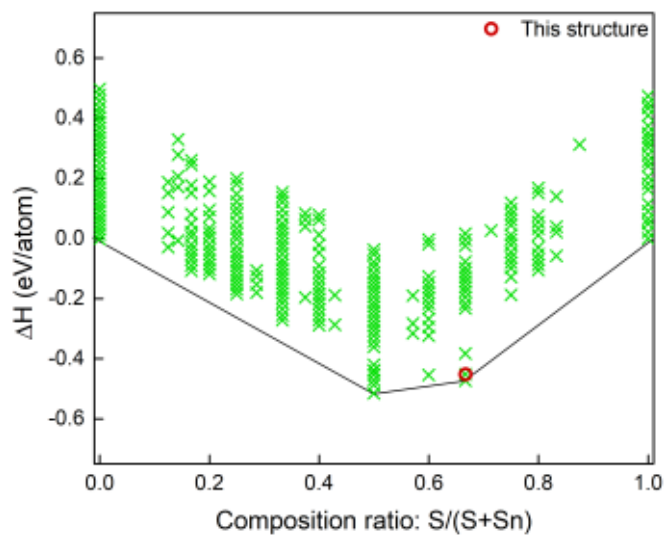

## Atomic structure

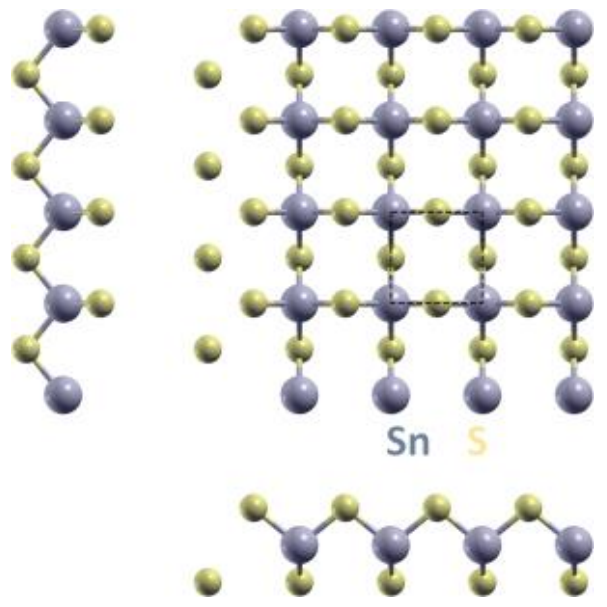

## Atomic coordinates

```
Sn1S2
   1.00000000000000
   3.7843209366167074    0.0000000000000000    0.0000000000000000
   0.0000000000000000    3.7843210308975257    0.0000000000000000
   0.0000000000000000    0.0000000000000000   22.8554570000000012
   Sn   S
   1    2
Direct
  0.9999999970905904  0.0000000007089014  0.5000000052448712
  0.5000000031336782  0.9999999969897146  0.5672176036748198
  0.9999999997757243  0.5000000023013840  0.4327823910803161
```

## Phonon dispersion curve

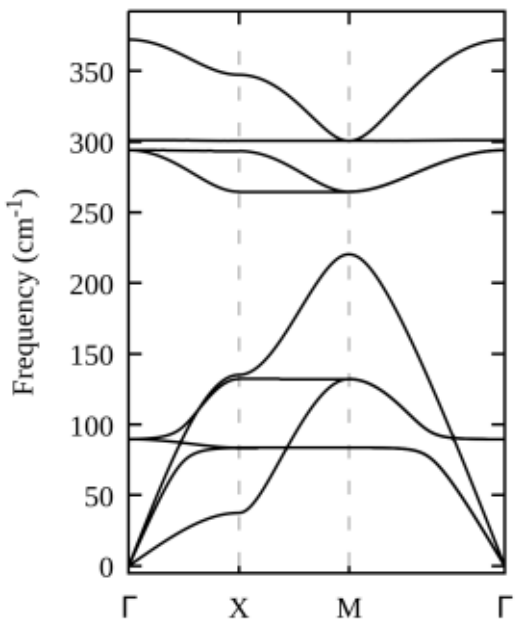

## Projected band structure and density of states

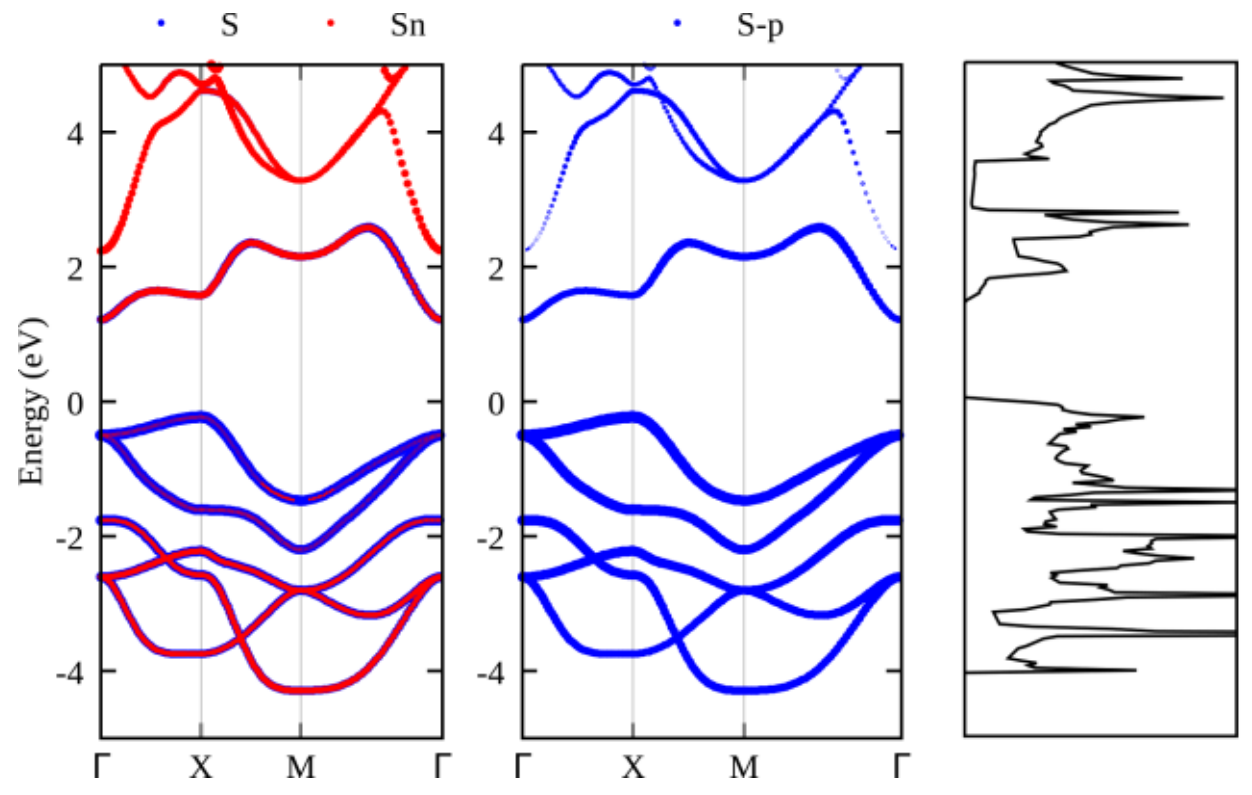

## Magnetic moment and spin polarization energy as a function of hole doping concentration

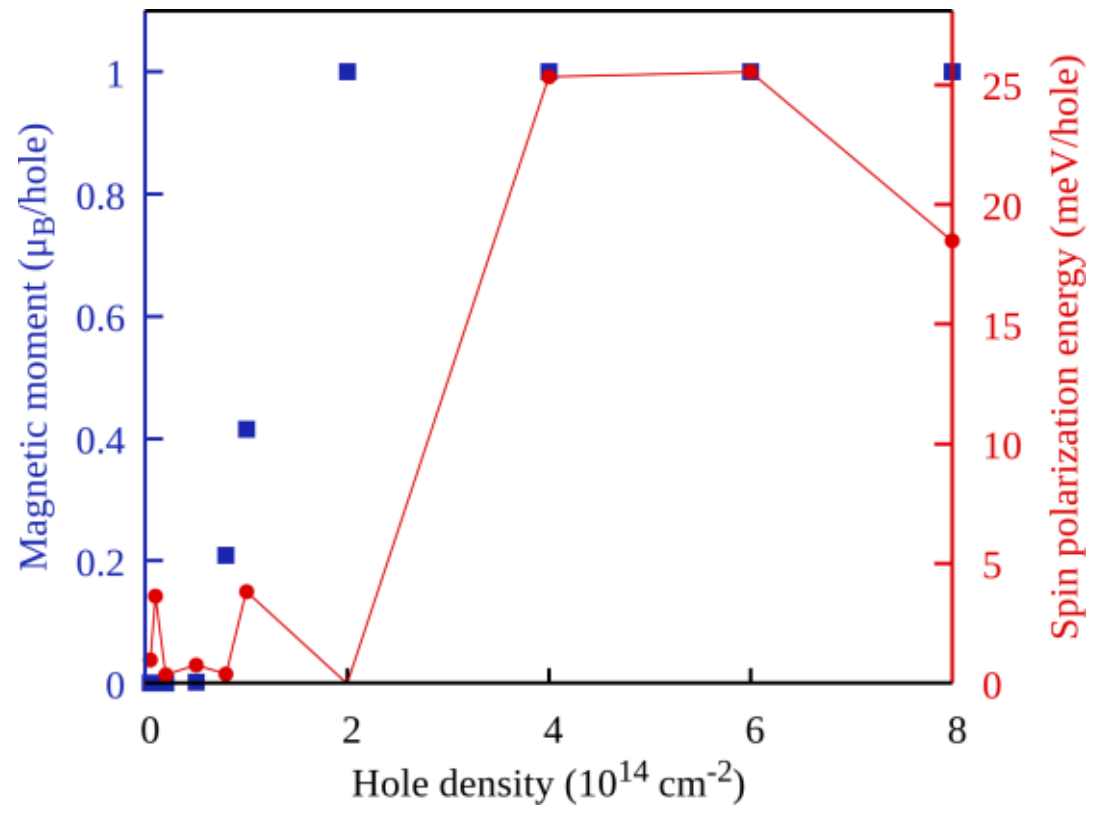

## Magnetic configurations and spin Hamiltonian

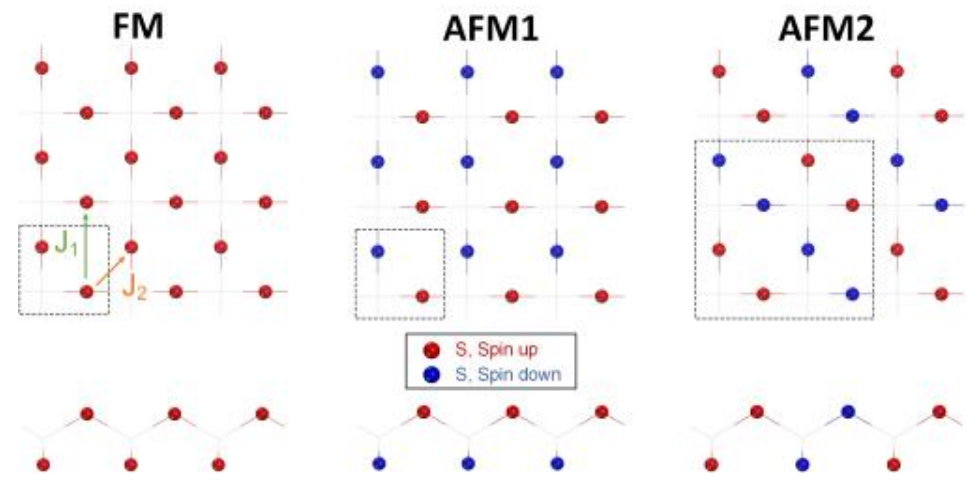

$$E_{FM} = E_0 - 4J_1S^2 - 4J_2S^2$$

$$E_{AFM1} = E_0 - 4J_1S^2 + 4J_2S^2$$

$$E_{AFM2} = E_0 + 4J_1S^2$$

## Magnetic exchange coupling parameters

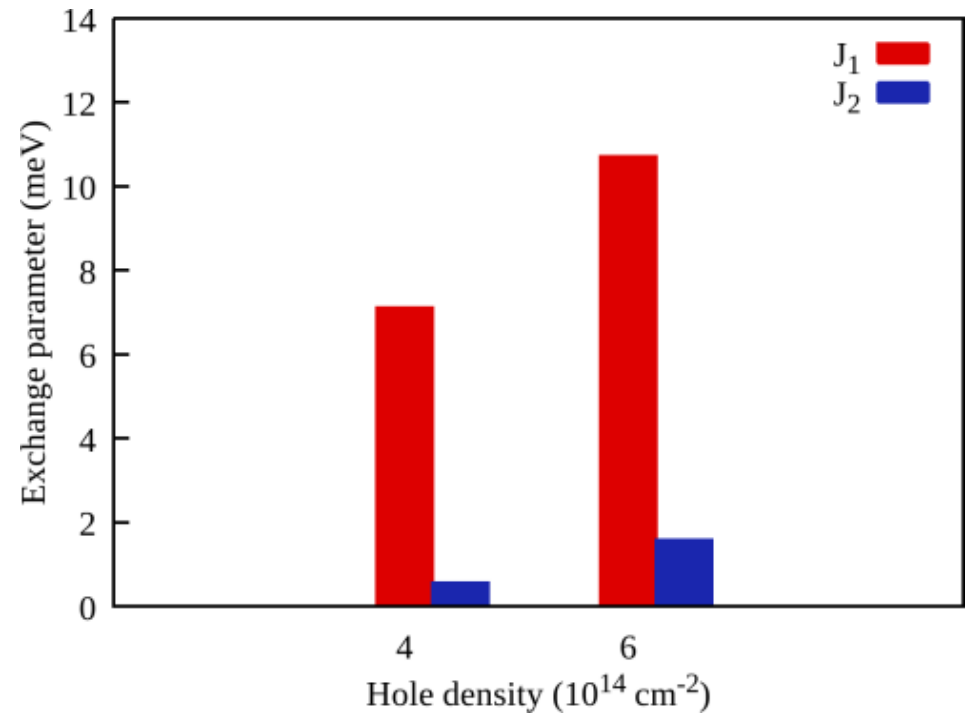

## Magnetic anisotropy energy (MAE, μeV) per magnetic atom

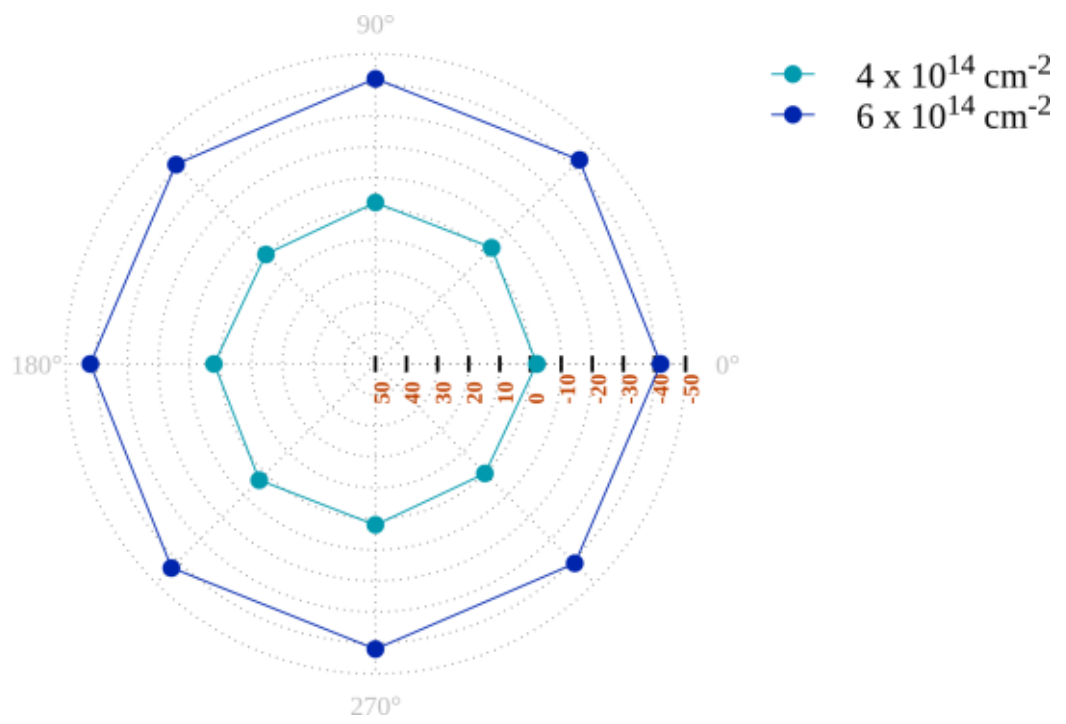

MAE = $E_{\parallel} - E_{\perp}$, a positive (negative) value of MAE indicates the off-plane (in-plane) easy axis.

## Monte Carlo simulations of the normalized magnetization of as a function of temperature

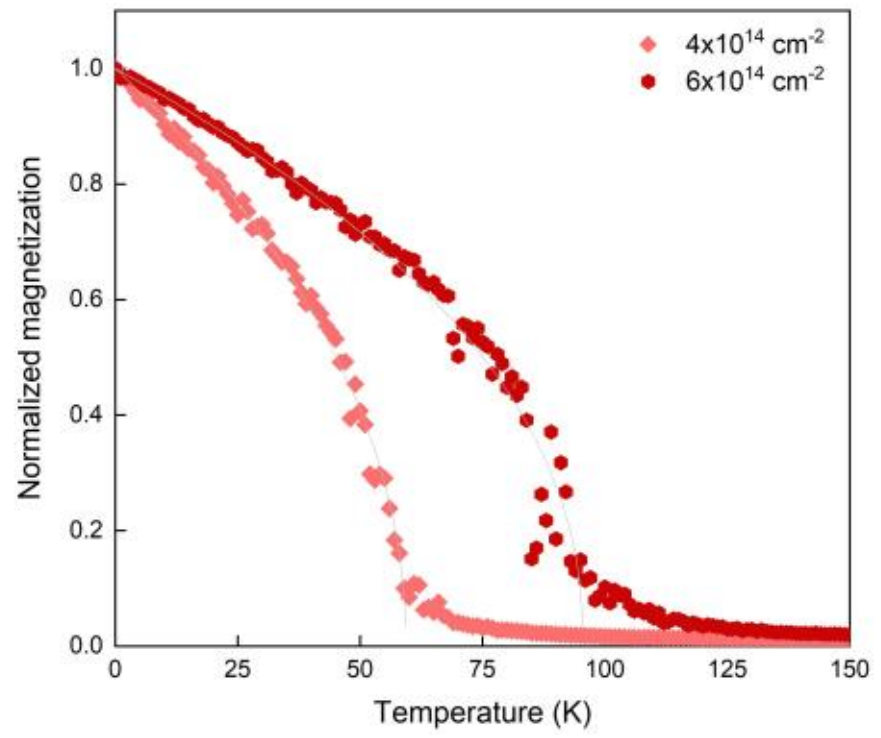

$T_c$: $1\times10^{14}$ cm$^{-2}$: - K  $2\times10^{14}$ cm$^{-2}$: - K

$4\times10^{14}$ cm$^{-2}$: 59 K  $6\times10^{14}$ cm$^{-2}$: 96 K

# 92.$PbS_2$

| MC2D-ID | C2DB | 2dmat-ID | USPEX | Space group | Band gap (eV) |
|---|---|---|---|---|---|
| - | - | 2dm-6220 | - | P4m2 | 0.65 |

**Convex hull**

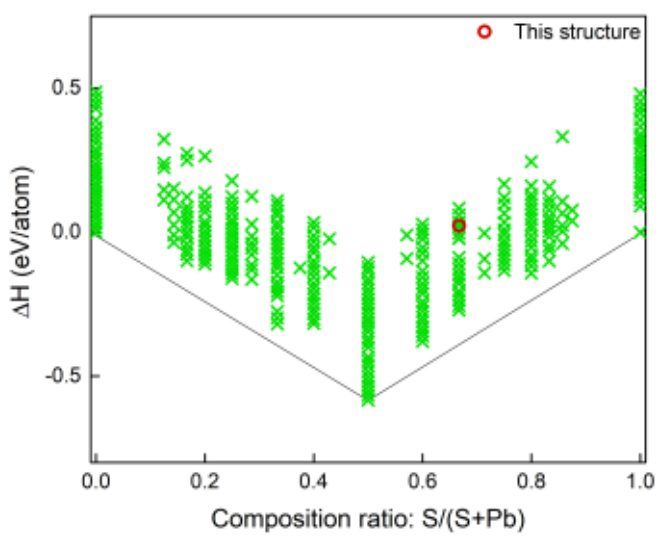

**Atomic structure**

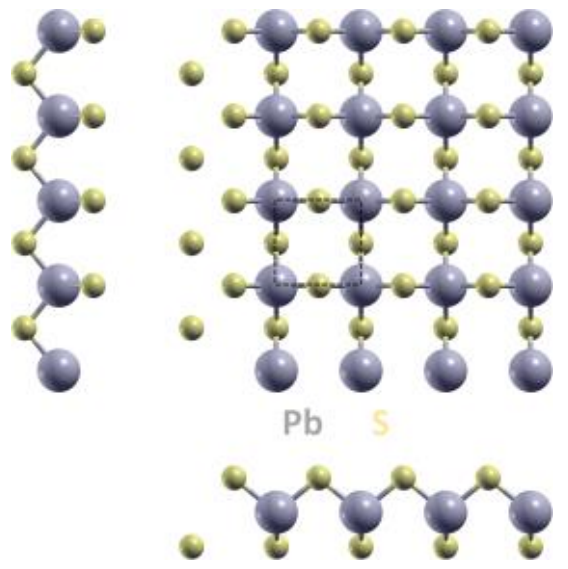

**Atomic coordinates**

```
Pb1S2
    1.00000000000000
    3.9362004097085279    0.0000000000000000    0.0000000000000000
    0.0000000000000000    3.9362002933386218    0.0000000000000000
    0.0000000000000000    0.0000000000000000    22.8554570000000012
    Pb    S
    1     2
Direct
   0.9999999987829113   0.0000000019596129   0.4999999970302511
   0.4999999983130792   0.9999999968400317   0.5699955865671171
   0.0000000029040095   0.5000000012003554   0.4300044164026389
```

**Phonon dispersion curve**

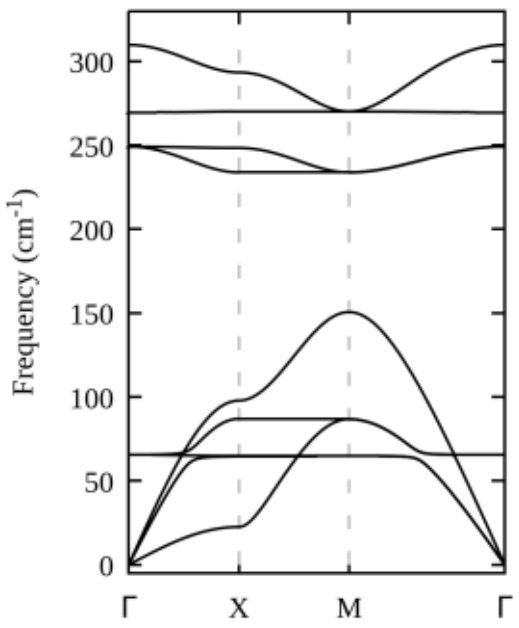

**Projected band structure and density of states**

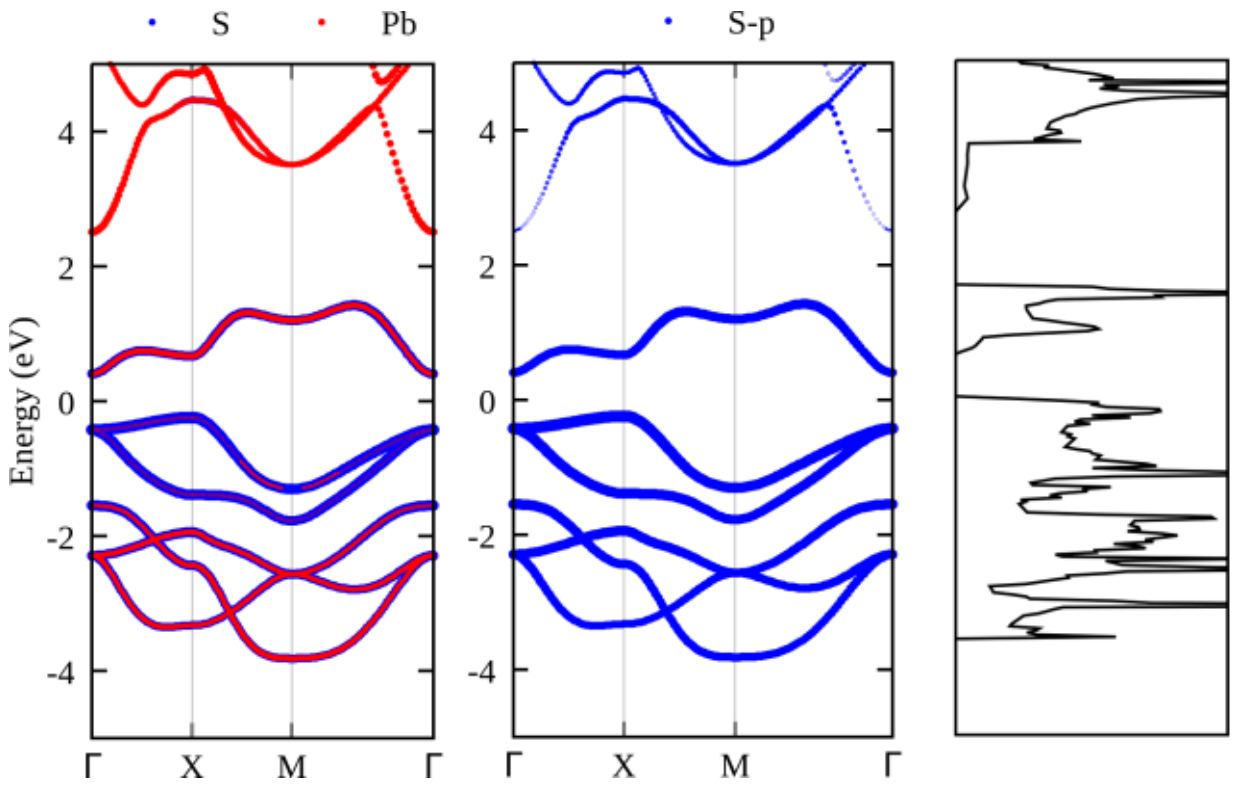

**Magnetic moment and spin polarization energy as a function of hole doping concentration**

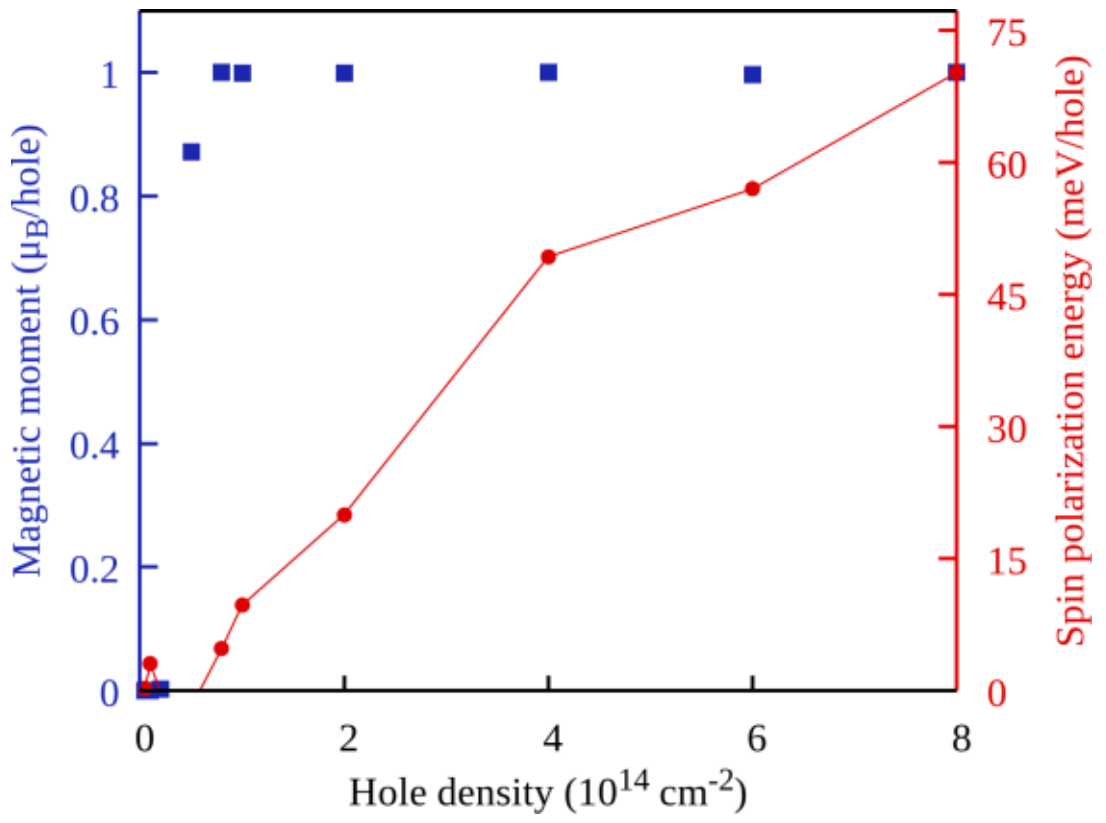

**Magnetic configurations and spin Hamiltonian**

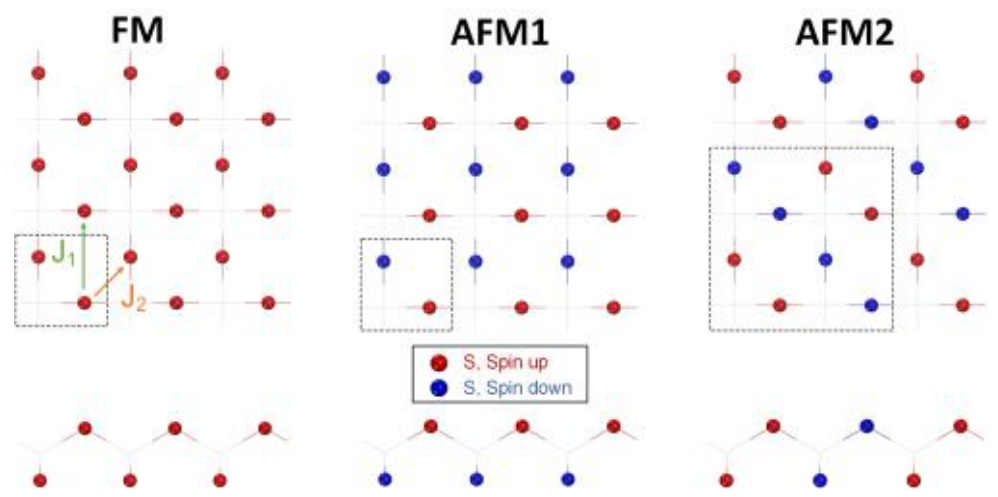

$$E_{FM} = E_0 - 4J_1S^2 - 4J_2S^2$$

$$E_{AFM1} = E_0 - 4J_1S^2 + 4J_2S^2$$

$$E_{AFM2} = E_0 + 4J_1S^2$$

**Magnetic exchange coupling parameters**

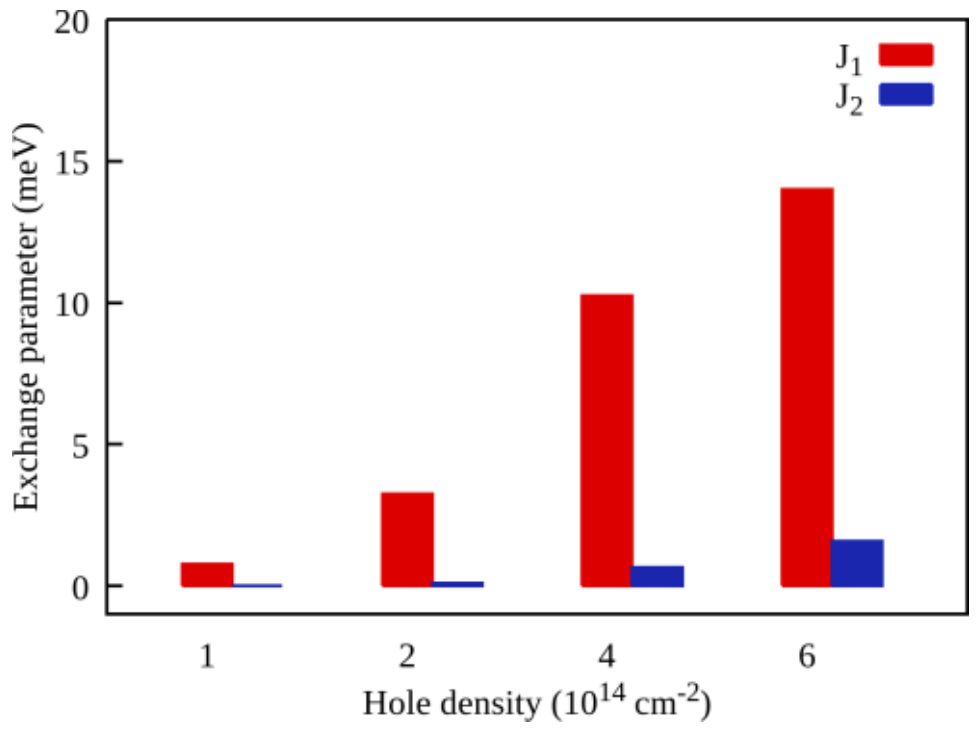

**Magnetic anisotropy energy (MAE, μeV) per magnetic atom**

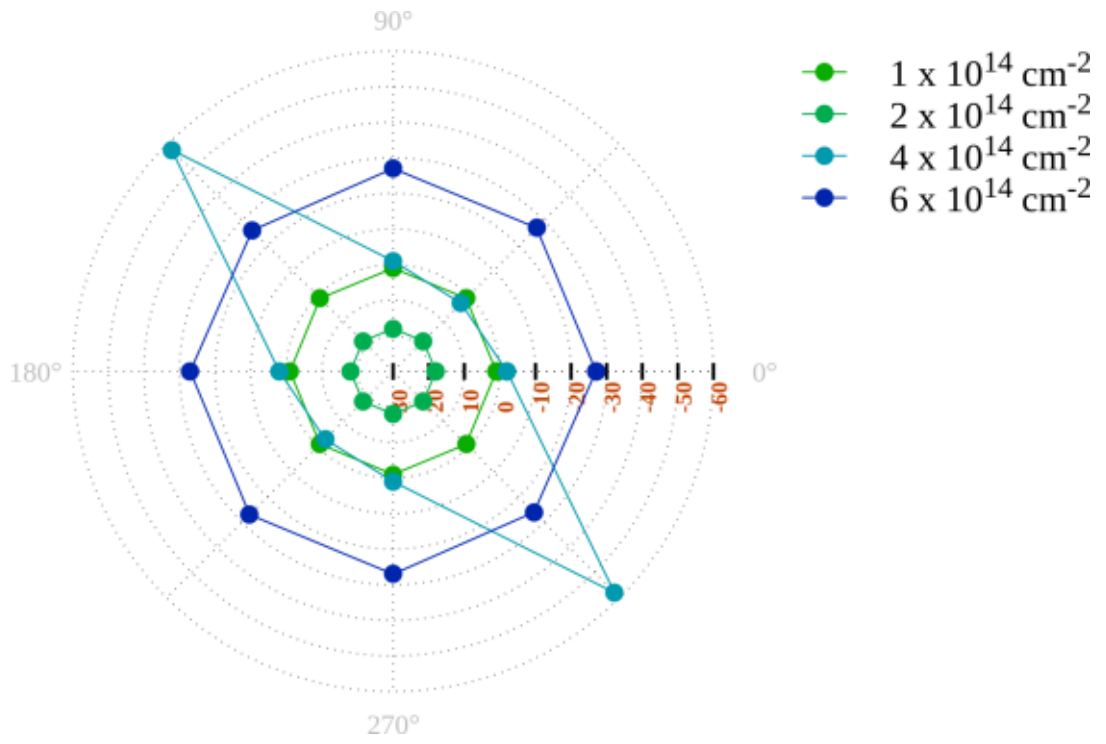

MAE = $E_{\parallel} - E_{\perp}$, a positive (negative) value of MAE indicates the off-plane (in-plane) easy axis.

**Monte Carlo simulations of the normalized magnetization of as a function of temperature**

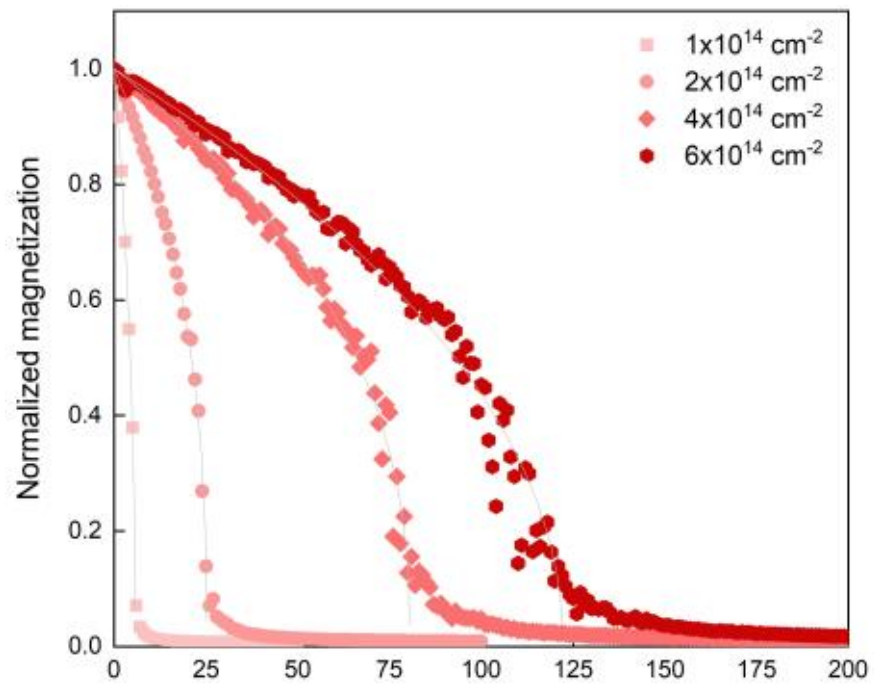

$T_c$: $1\times10^{14}$ cm$^{-2}$: 6 K  $2\times10^{14}$ cm$^{-2}$: 25 K

$4\times10^{14}$ cm$^{-2}$: 81 K  $6\times10^{14}$ cm$^{-2}$: 122 K

# 93. $SiF_4$

| MC2D-ID | C2DB | 2dmat-ID | USPEX | Space group | Band gap (eV) |
|---|---|---|---|---|---|
| - | - | 2dm-1452 | - | P4/mmm | 6.66 |

**Convex hull** | **Atomic structure** | **Atomic coordinates** | **Phonon dispersion curve**

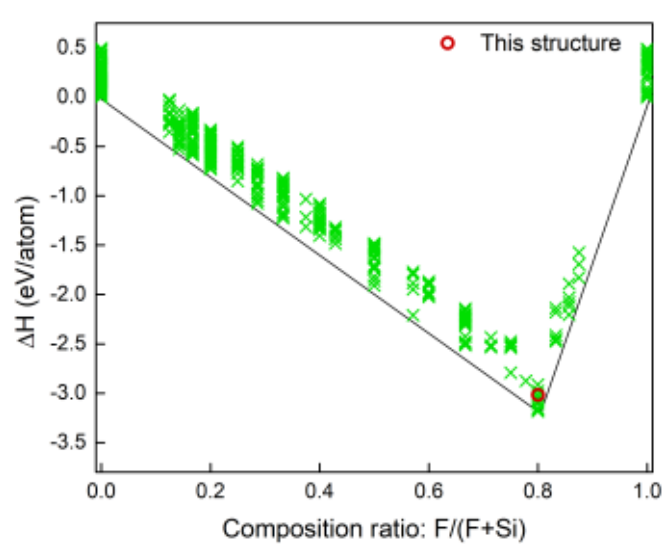

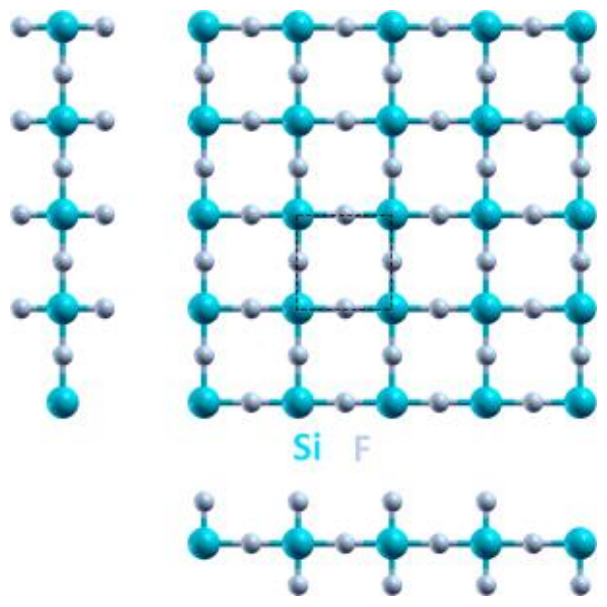

```
SiF4
1.00000000000000
3.6037410880278817 0.0000000000000000 0.0000000000000000
0.0000000000000000 3.6037413636842635 0.0000000000000000
0.0000000000000000 0.0000000000000000 23.8182620000000007
Si F
1 4
Direct
0.0000000001529230 0.0000000001174740 0.0000000008728449
0.9996137853834597 0.9999999996778186 0.9327723585694088
0.4999999997505071 0.0000000005697984 0.9999999998743334
0.0000000006462031 0.4999999997206146 0.9999999998793285
0.0003862140669071 0.9999999999142943 0.0672276408040844
```

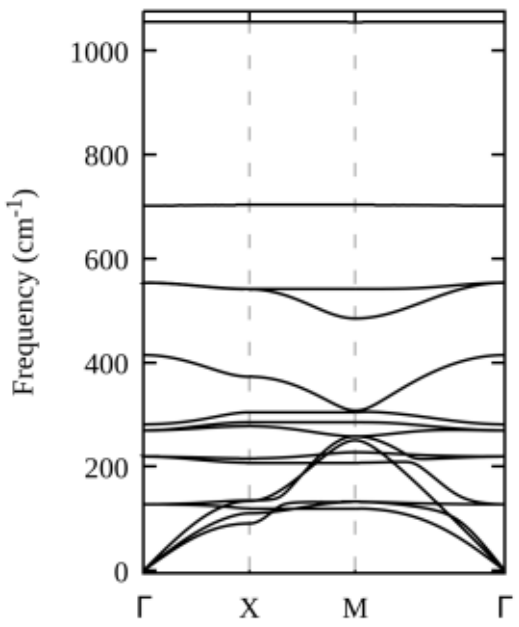

**Projected band structure and density of states**

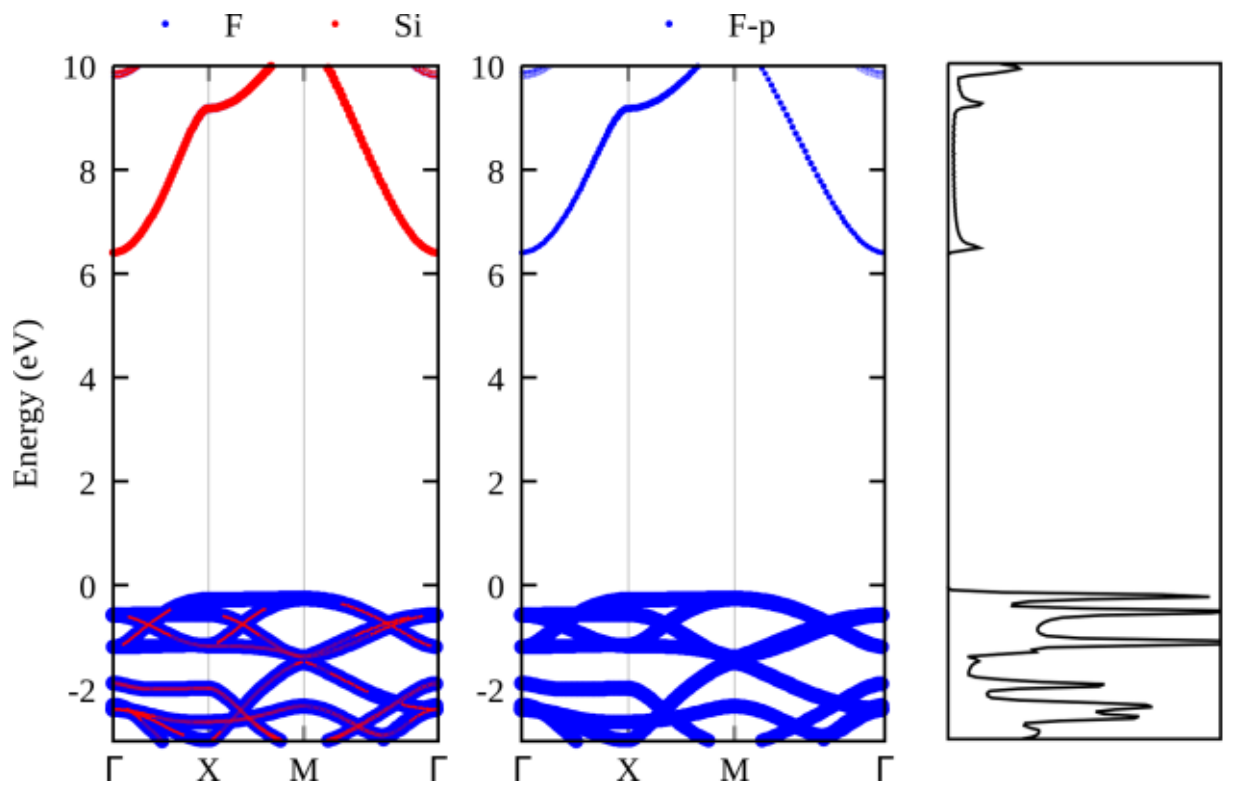

**Magnetic moment and spin polarization energy as a function of hole doping concentration**

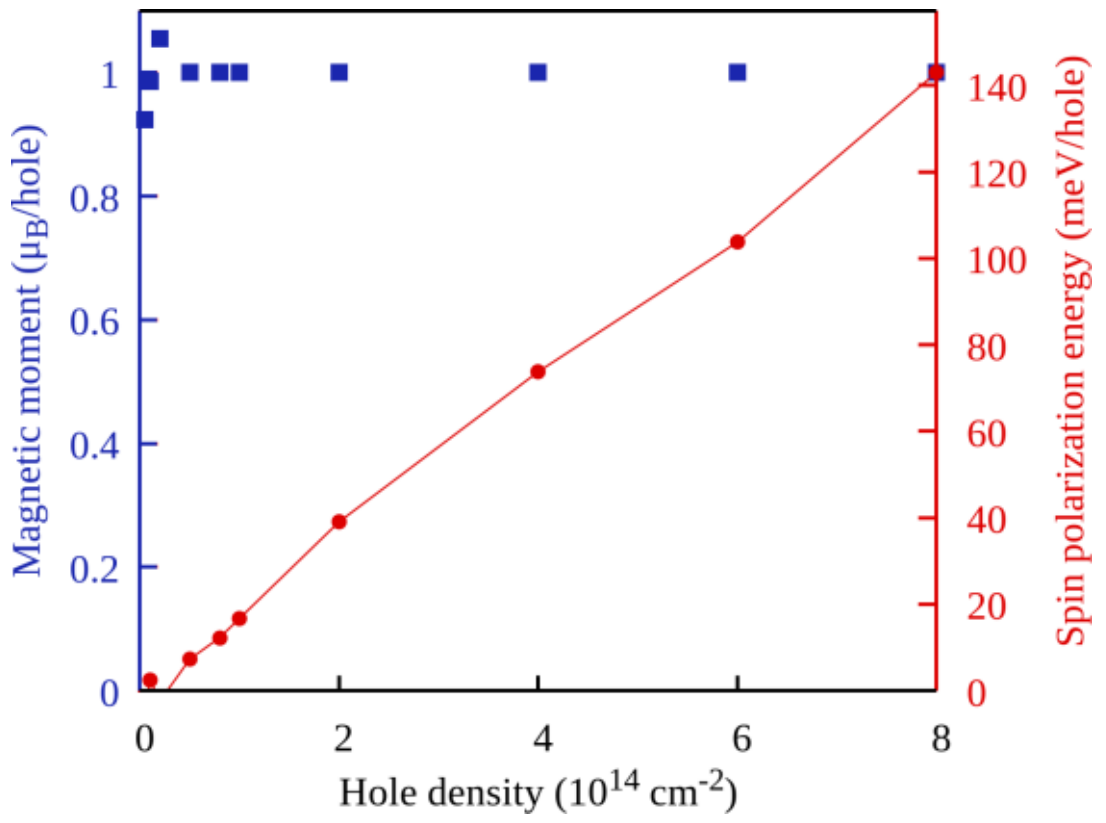

**Magnetic configurations and spin Hamiltonian**

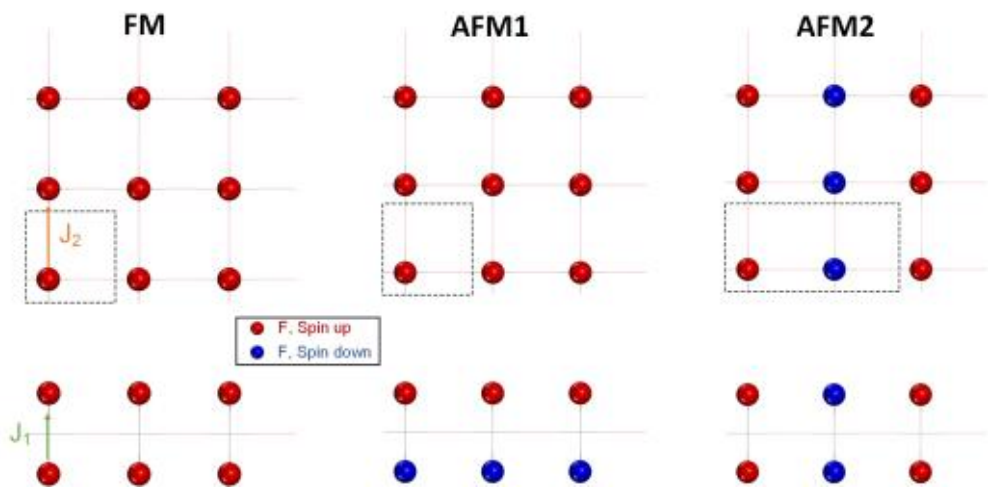

$$E_{FM} = E_0 - J_1S^2 - 4J_2S^2$$
$$E_{AFM1} = E_0 + J_1S^2 - 4J_2S^2$$
$$E_{AFM2} = E_0 - J_1S^2$$

**Magnetic exchange coupling parameters**

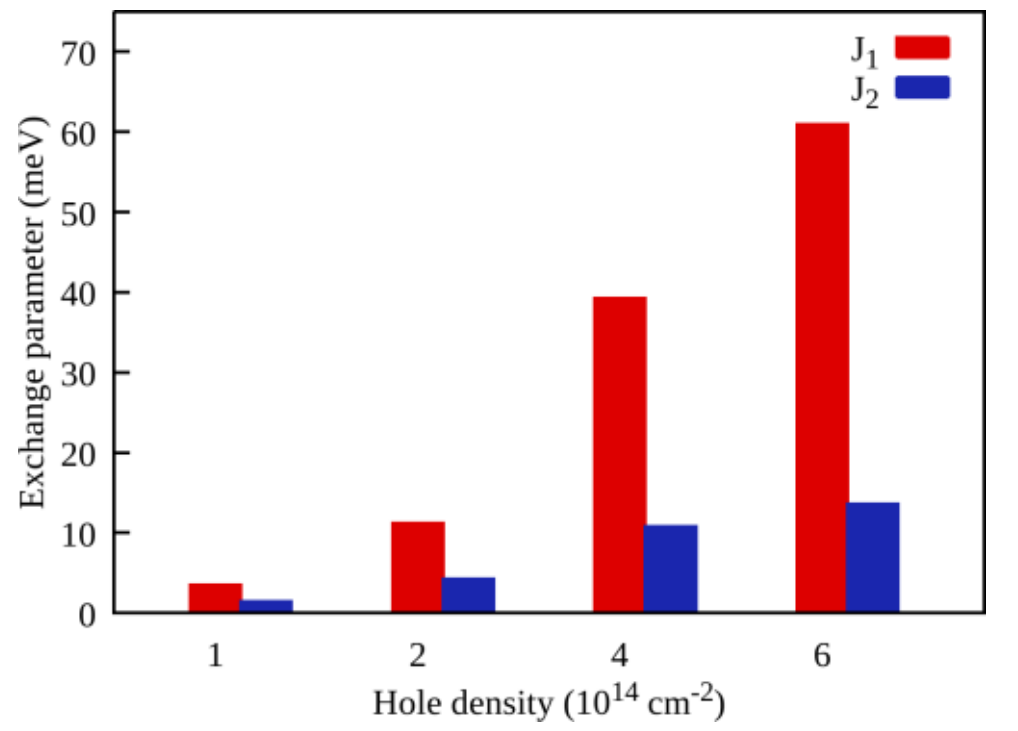

**Magnetic anisotropy energy (MAE, μeV) per magnetic atom**

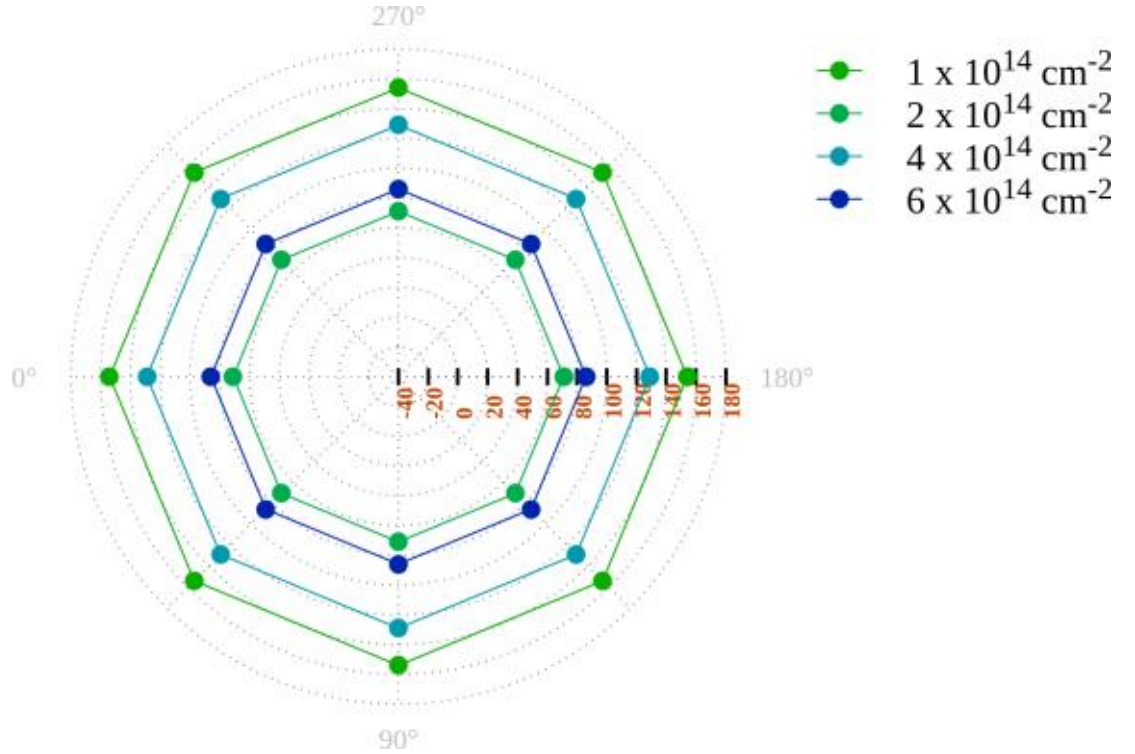

MAE = $E_{\parallel} - E_{\perp}$, a positive (negative) value of MAE indicates the off-plane (in-plane) easy axis.

**Monte Carlo simulations of the normalized magnetization of as a function of temperature**

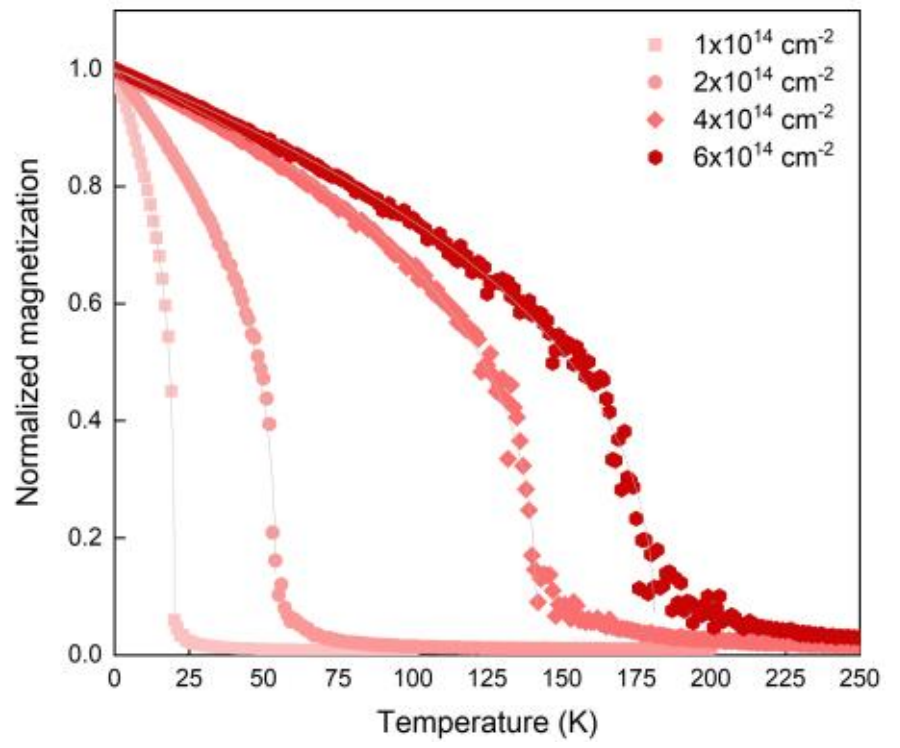

$T_c$: $1\times10^{14}$ cm$^{-2}$: 20 K   $2\times10^{14}$ cm$^{-2}$: 54 K

$4\times10^{14}$ cm$^{-2}$: 142 K   $6\times10^{14}$ cm$^{-2}$: 181 K

# 94. $GeF_4$

| MC2D-ID | C2DB | 2dmat-ID | USPEX | Space group | Band gap (eV) |
|---|---|---|---|---|---|
| - | - | 2dm-1570 | - | P4/mmm | 4.26 |

**Convex hull** | **Atomic structure** | **Atomic coordinates** | **Phonon dispersion curve**

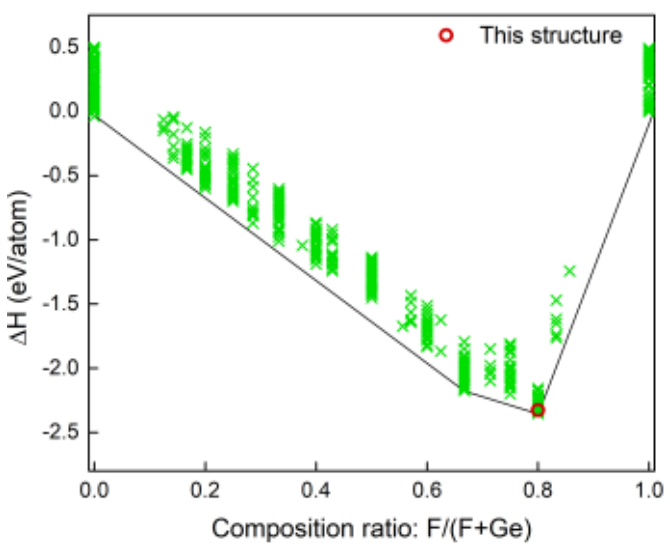

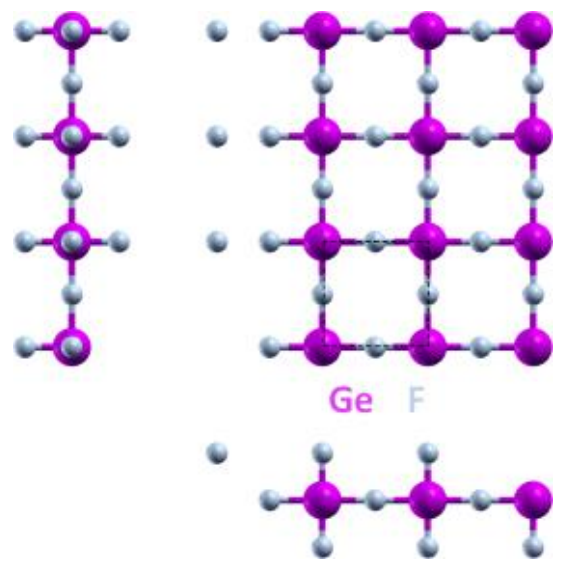

```
Ge1F4
  1.00000000000000
  3.8411487860685023   0.0000000000000000   0.0000000000000000
  0.0000000000000000   3.8411460687975425   0.0000000000000000
  0.0000000000000000   0.0000000000000000   23.8182620000000007
  Ge   F
  1    4
Direct
 0.9999999963869257  0.0000000020198172  0.0000000002135891
 0.9995969344907110  0.0000000015022934  0.9279121530585428
 0.5000000035436329  0.0000000000330260  0.0000000006181509
 0.9999999995331805  0.4999999991252650  0.0000000006424727
 0.0004030660455498  0.9999999973195983  0.0720878454672373
```

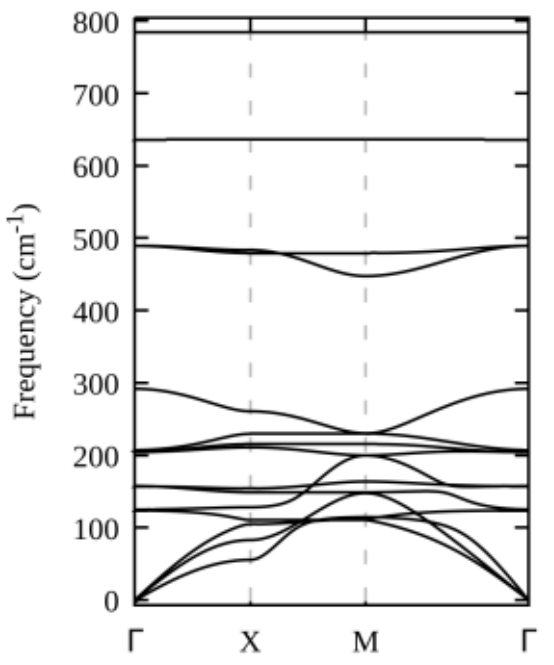

**Projected band structure and density of states**

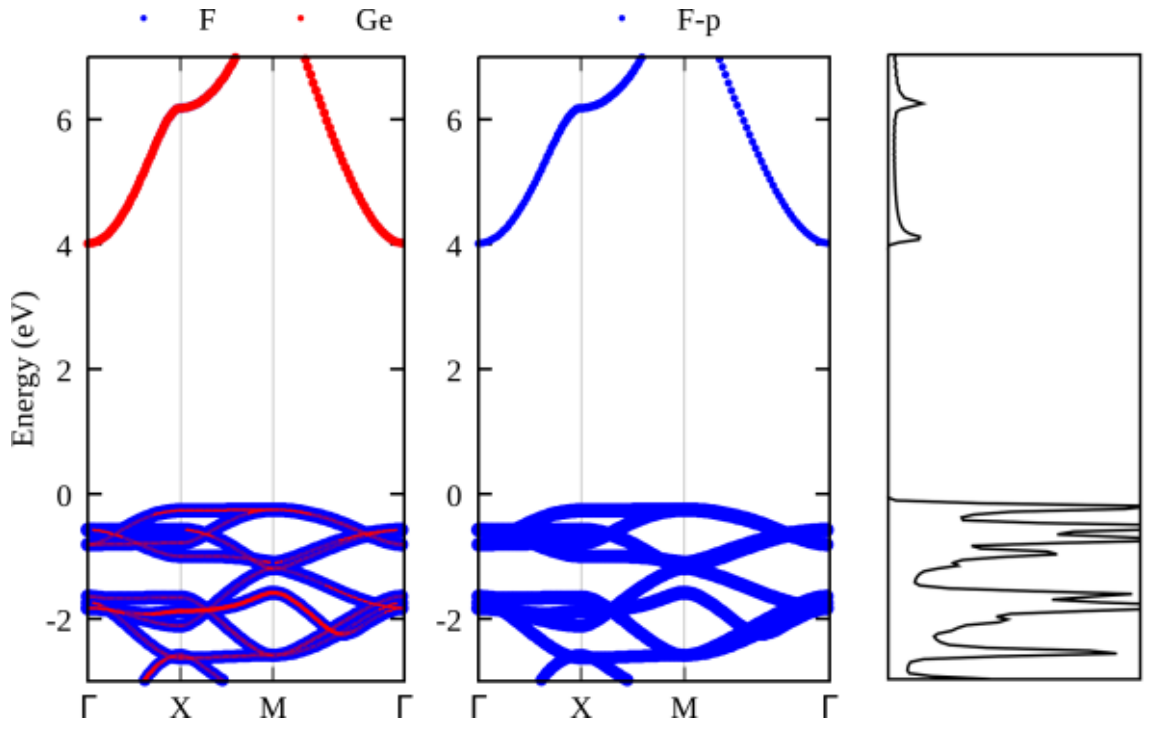

**Magnetic moment and spin polarization energy as a function of hole doping concentration**

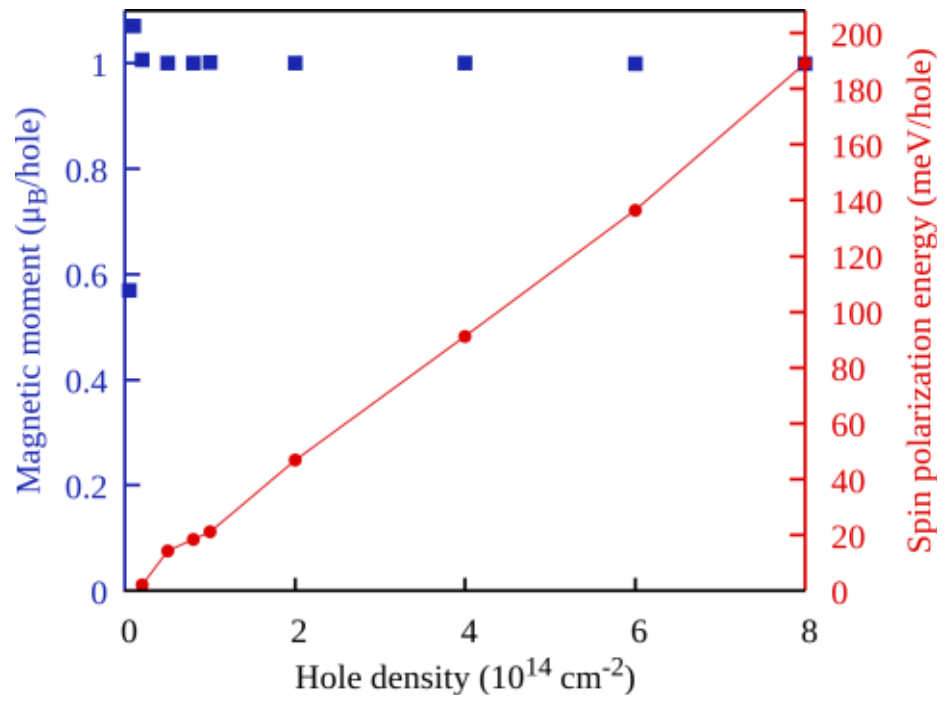

**Magnetic configurations and spin Hamiltonian**

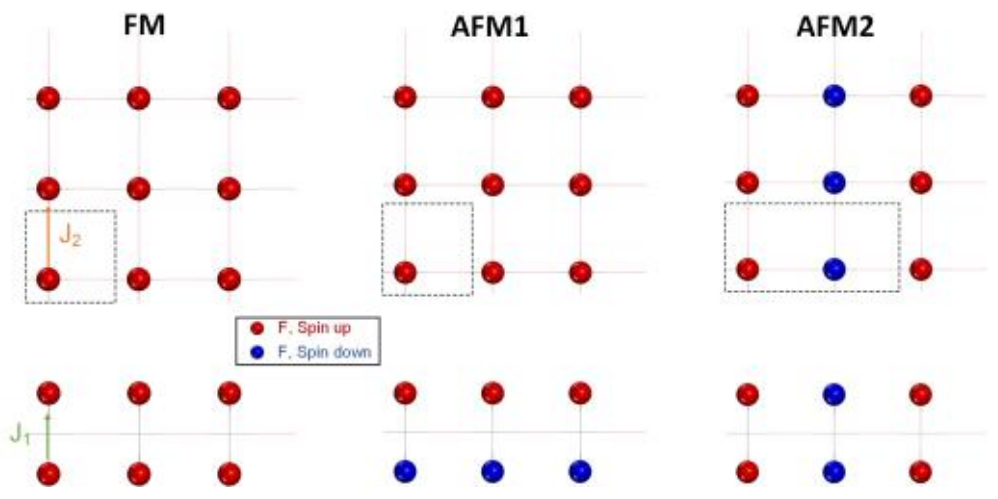

$$E_{FM} = E_0 - J_1S^2 - 4J_2S^2$$
$$E_{AFM1} = E_0 + J_1S^2 - 4J_2S^2$$
$$E_{AFM2} = E_0 - J_1S^2$$

**Magnetic exchange coupling parameters**

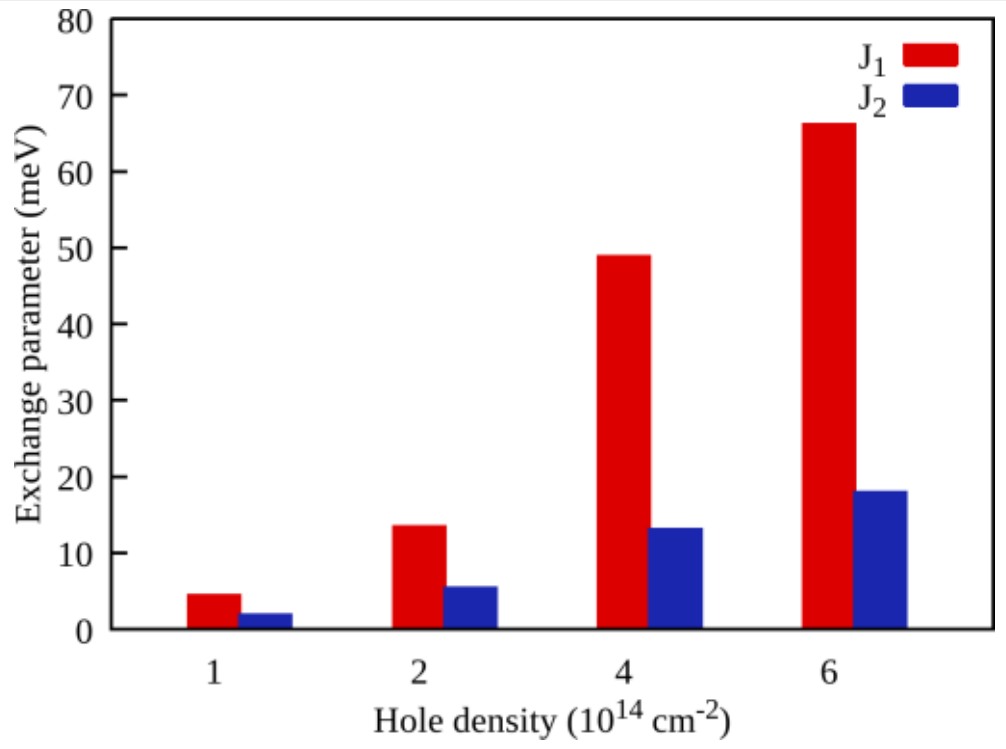

**Magnetic anisotropy energy (MAE, μeV) per magnetic atom**

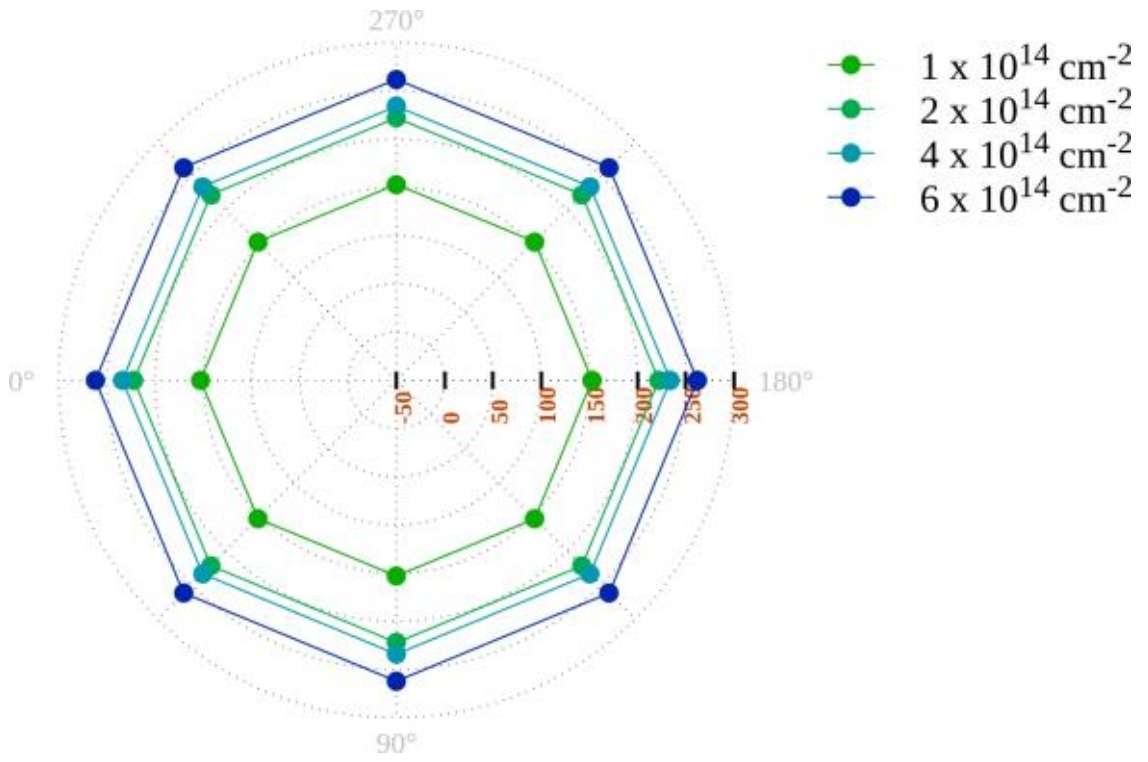

MAE = $E_{\parallel} - E_{\perp}$, a positive (negative) value of MAE indicates the off-plane (in-plane) easy axis.

**Monte Carlo simulations of the normalized magnetization of as a function of temperature**

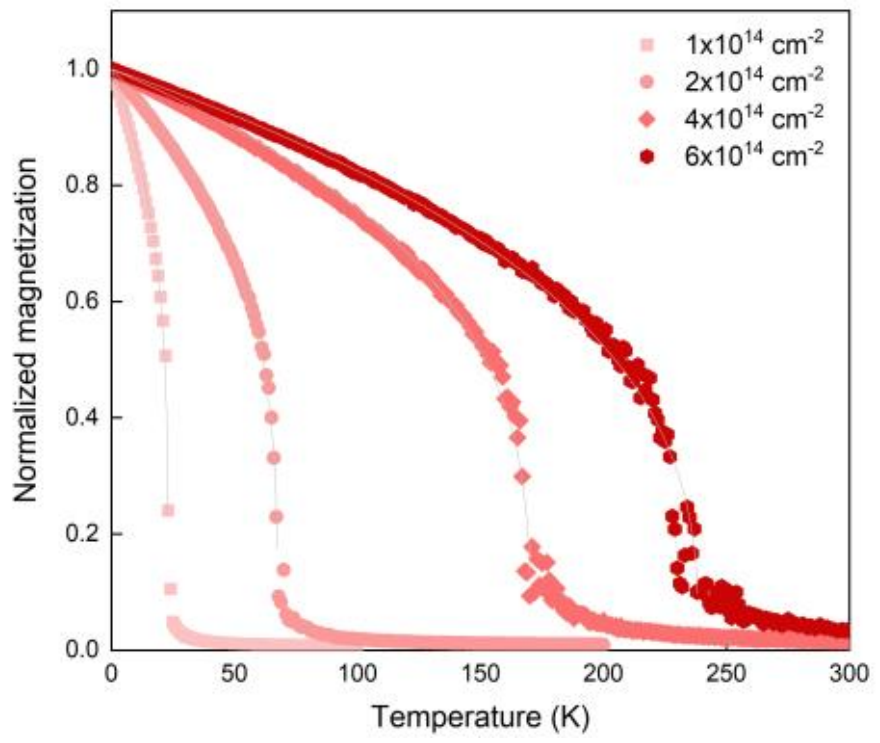

$T_c$: $1\times10^{14}$ cm$^{-2}$: 23 K  $2\times10^{14}$ cm$^{-2}$: 68 K
$4\times10^{14}$ cm$^{-2}$: 170 K  $6\times10^{14}$ cm$^{-2}$: 239 K

# 95. $SnF_4$

| MC2D-ID | C2DB | 2dmat-ID | USPEX | Space group | Band gap (eV) |
|---|---|---|---|---|---|
| 181 | - | 2dm-3719 | - | P4/mmm | 3.85 |

**Convex hull** | **Atomic structure** | **Atomic coordinates** | **Phonon dispersion curve**

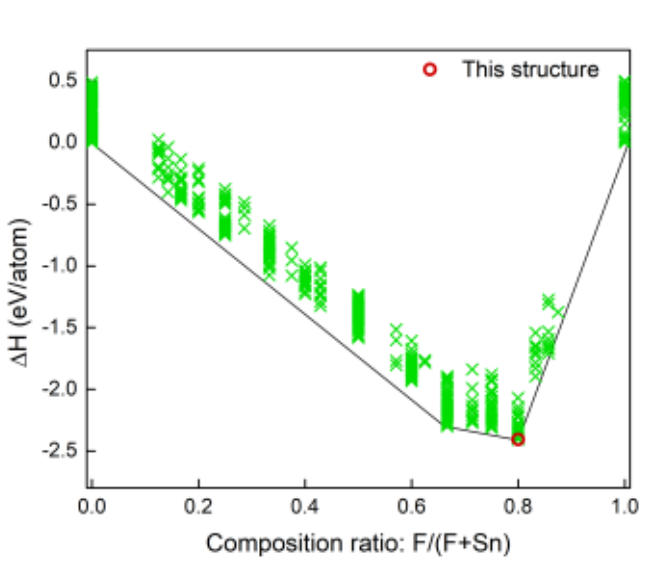

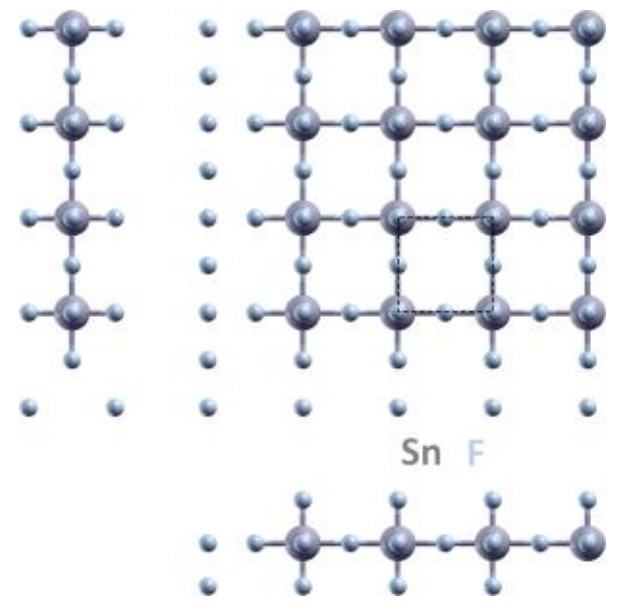

```
Sn1F4
  1.00000000000000
  4.1703085515655678  0.0000000000000000  0.0000000000000000
  0.0000000000000000  4.1703085104161621  0.0000000000000000
  0.0000000000000000  0.0000000000000000  23.8182429999999989
  Sn  F
  1   4
Direct
  0.9999999943846589  0.9999999915730413  0.4999999996232276
  0.9999544901801372  0.0000000014527046  0.5797577264576859
  0.5000000055319589  0.9999999858996134  0.4999999991057820
  0.0000000061119465  0.5000000211374811  0.4999999987110613
  0.0000455037912985  0.9999999999371596  0.4202422761022433
```

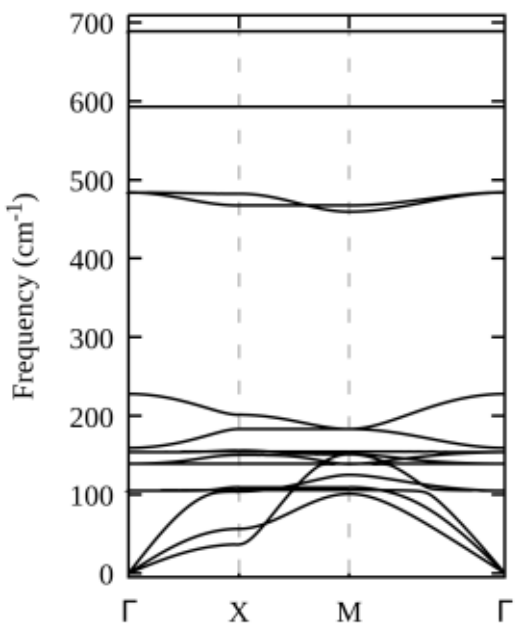

**Projected band structure and density of states**

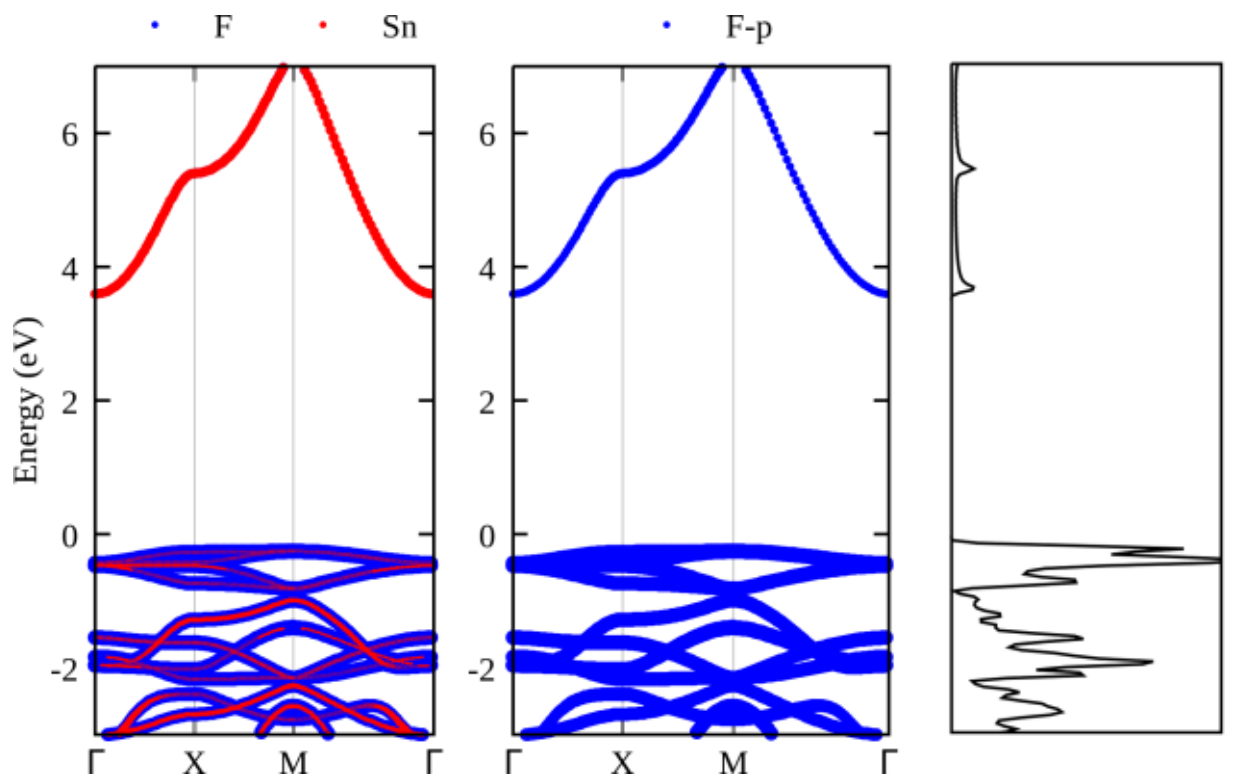

**Magnetic moment and spin polarization energy as a function of hole doping concentration**

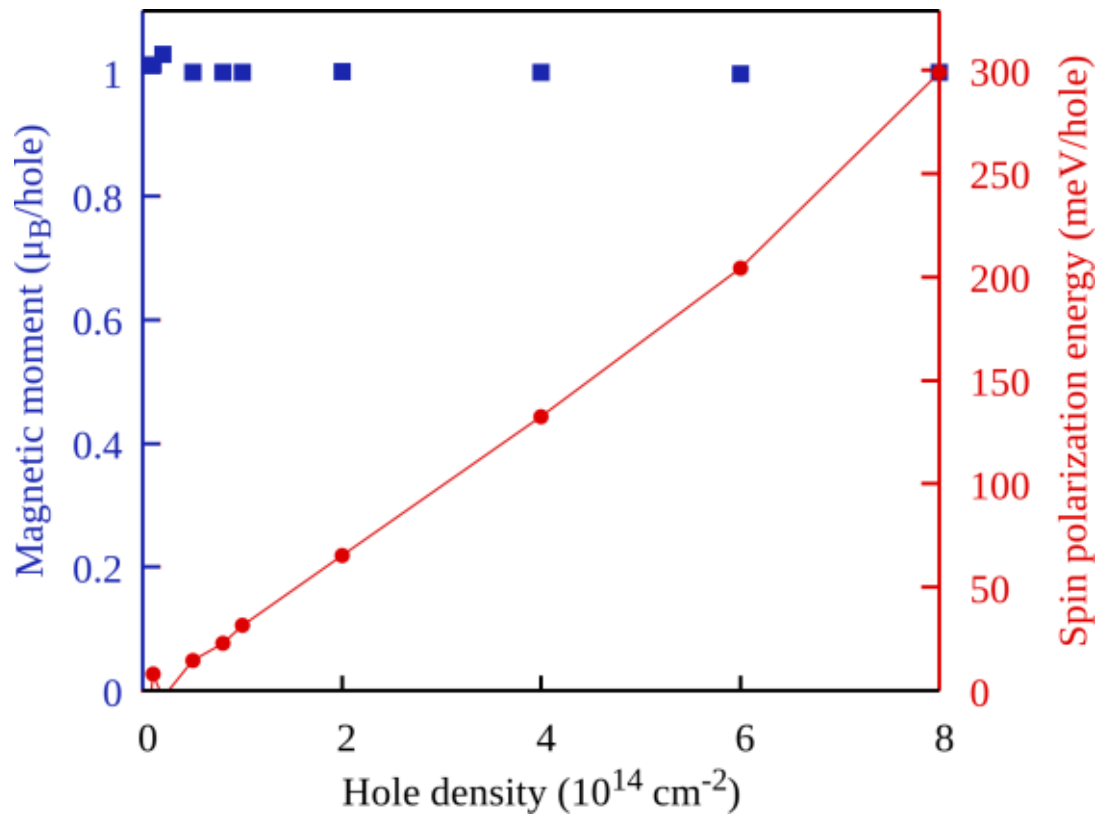

**Magnetic configurations and spin Hamiltonian**

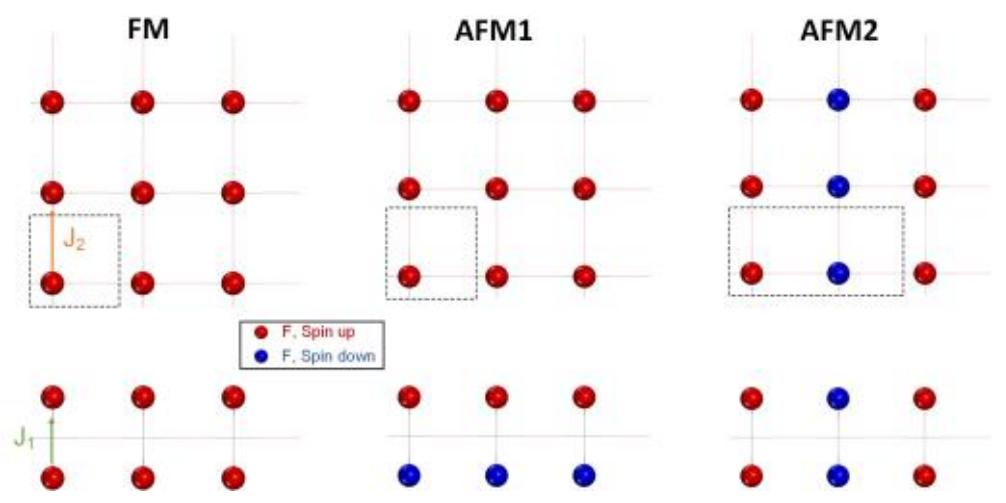

$$E_{FM} = E_0 - J_1S^2 - 4J_2S^2$$

$$E_{AFM1} = E_0 + J_1S^2 - 4J_2S^2$$

$$E_{AFM2} = E_0 - J_1S^2$$

**Magnetic exchange coupling parameters**

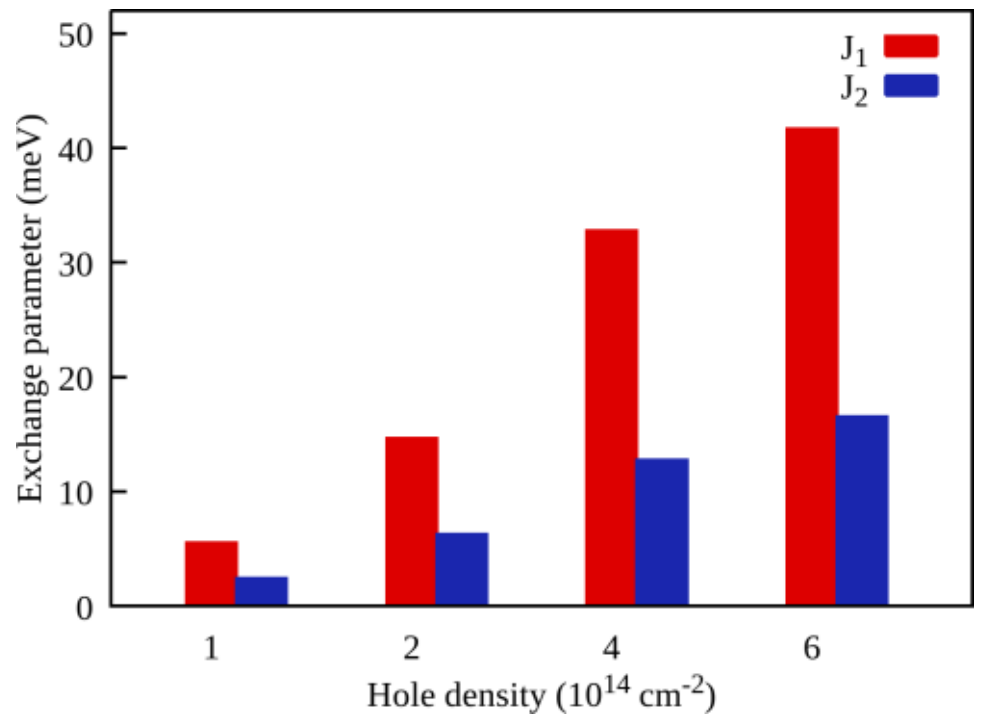

**Magnetic anisotropy energy (MAE, μeV) per magnetic atom**

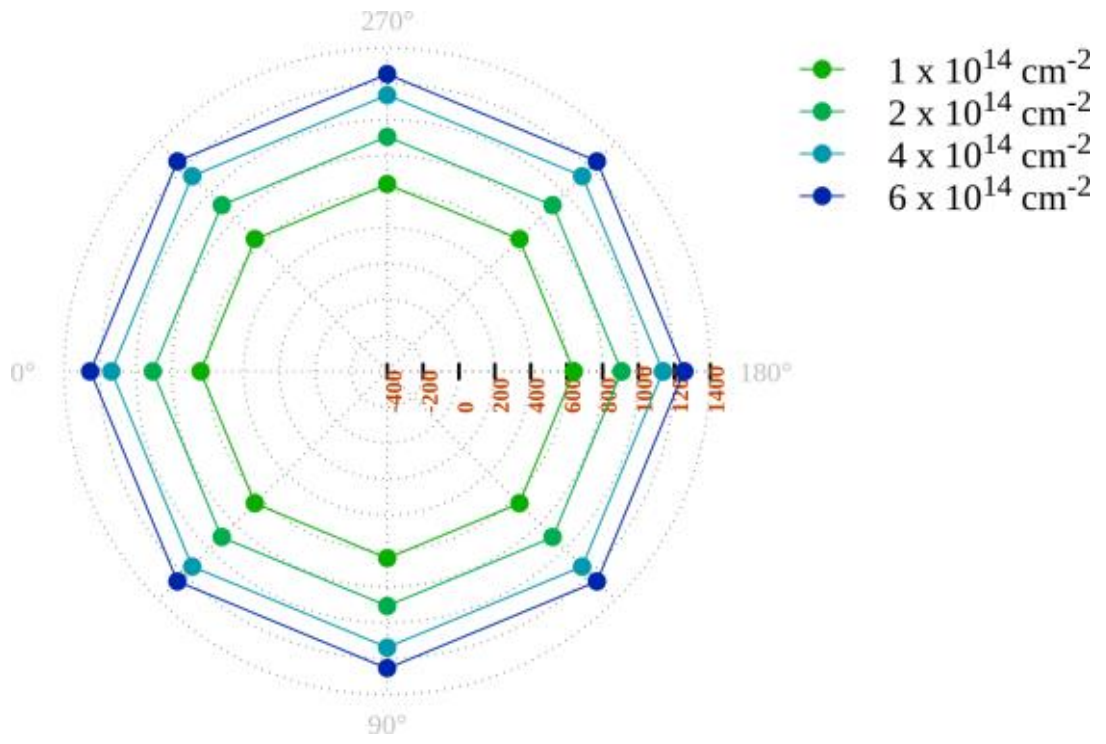

MAE = $E_\parallel - E_\perp$, a positive (negative) value of MAE indicates the off-plane (in-plane) easy axis.

**Monte Carlo simulations of the normalized magnetization of as a function of temperature**

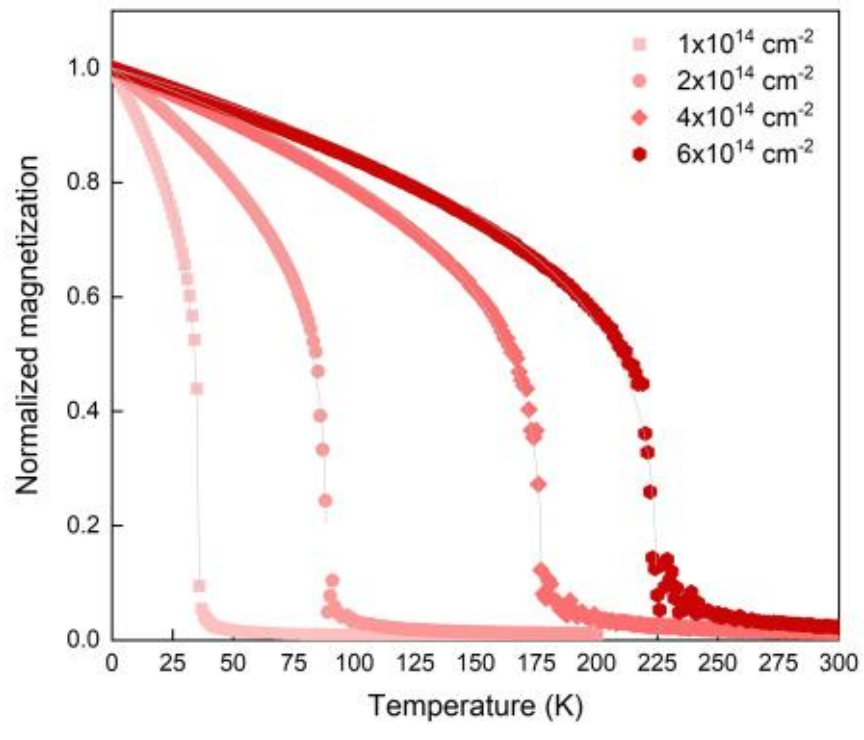

$T_c$: $1\times10^{14}$ cm$^{-2}$: 36 K $2\times10^{14}$ cm$^{-2}$: 88 K $4\times10^{14}$ cm$^{-2}$: 177 K $6\times10^{14}$ cm$^{-2}$: 224 K

# 96. $PbF_4$

| MC2D-ID | C2DB | 2dmat-ID | USPEX | Space group | Band gap (eV) |
|---|---|---|---|---|---|
| 150 | - | 2dm-4359 | - | P4/mmm | 2.48 |

**Convex hull** | **Atomic structure** | **Atomic coordinates** | **Phonon dispersion curve**

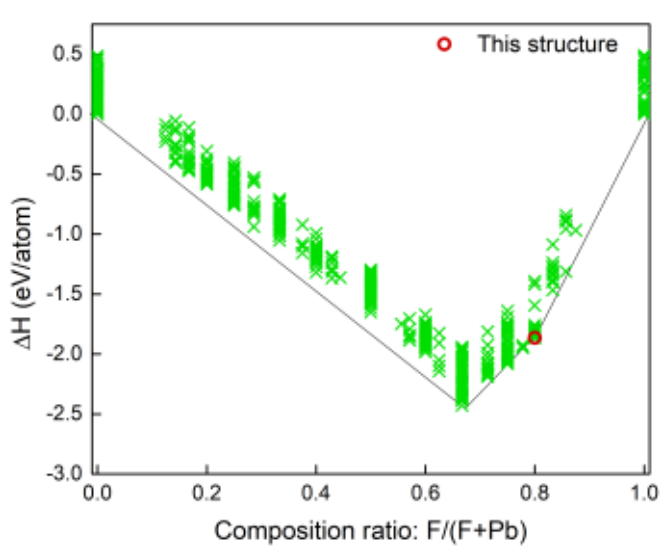

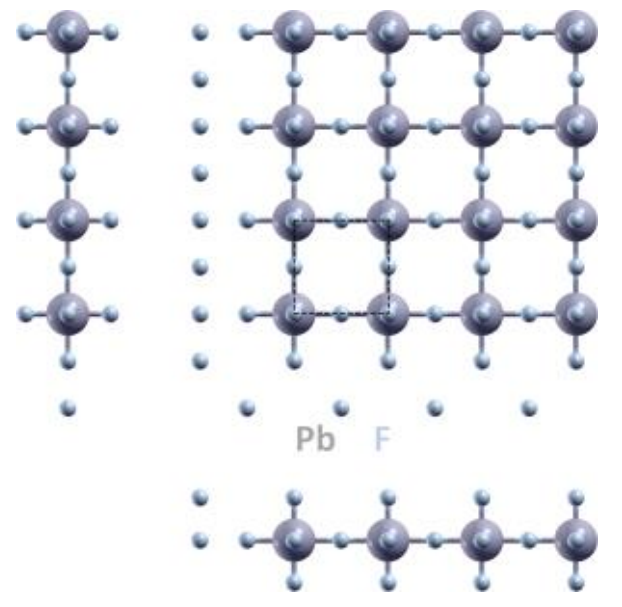

```
Pb1F4
   1.00000000000000
   4.3854459551064870    0.0000000000000000    0.0000000000000000
   0.0000000000000000    4.3854457720421047    0.0000000000000000
   0.0000000000000000    0.0000000000000000   23.9935650000000003
   Pb    F
   1     4
Direct
  0.9999999991334647  0.9999999973692013  0.5000000065320052
  0.0000000001832490  0.9999999998796909  0.5827573533406962
  0.0000000007667609  0.5000000032074752  0.5000000007589378
  0.5000000022286315  0.0000000008707701  0.5000000001661036
  0.9999999976879010  0.9999999986728767  0.4172426392022643
```

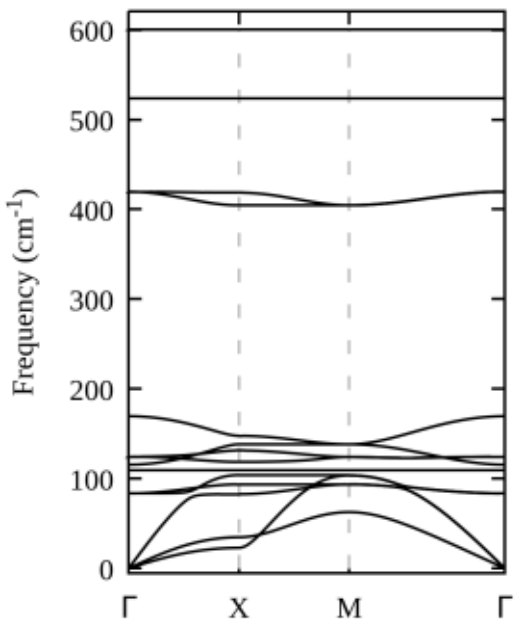

**Projected band structure and density of states**

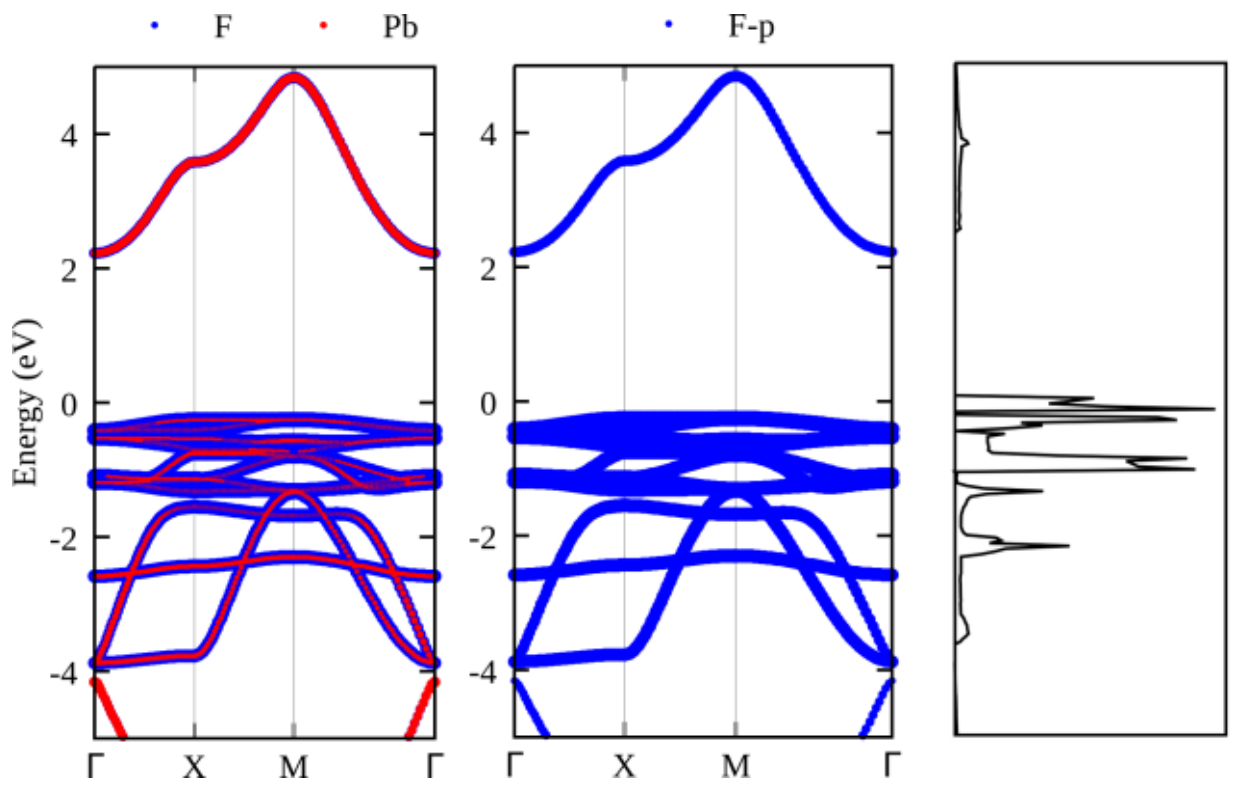

**Magnetic moment and spin polarization energy as a function of hole doping concentration**

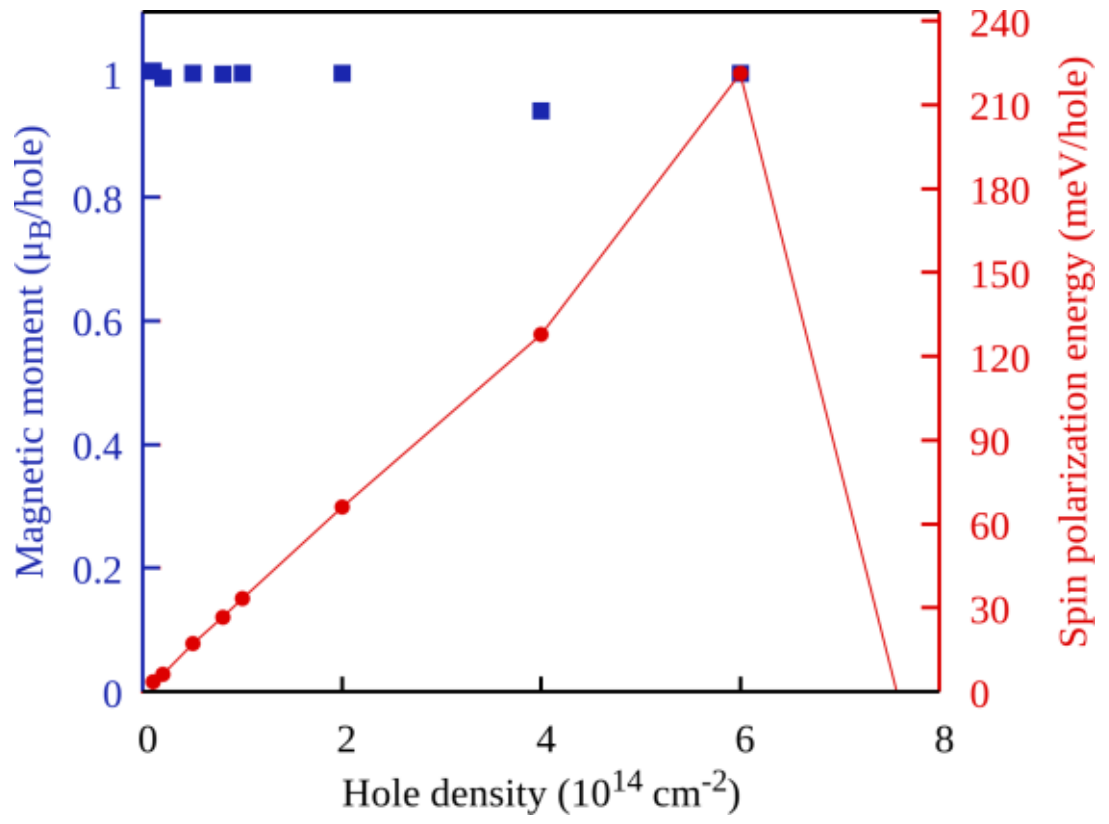

**Magnetic configurations and spin Hamiltonian**

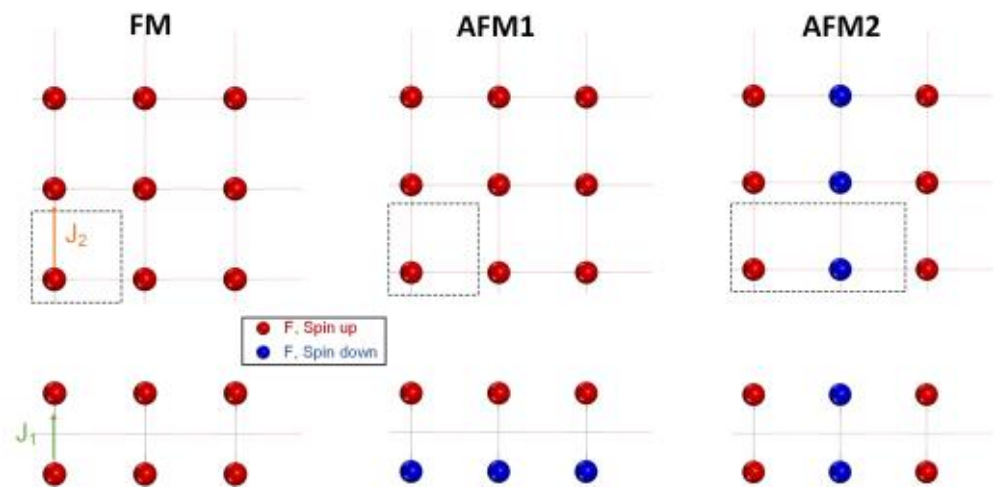

$$E_{FM} = E_0 - J_1S^2 - 4J_2S^2$$
$$E_{AFM1} = E_0 + J_1S^2 - 4J_2S^2$$
$$E_{AFM2} = E_0 - J_1S^2$$

**Magnetic exchange coupling parameters**

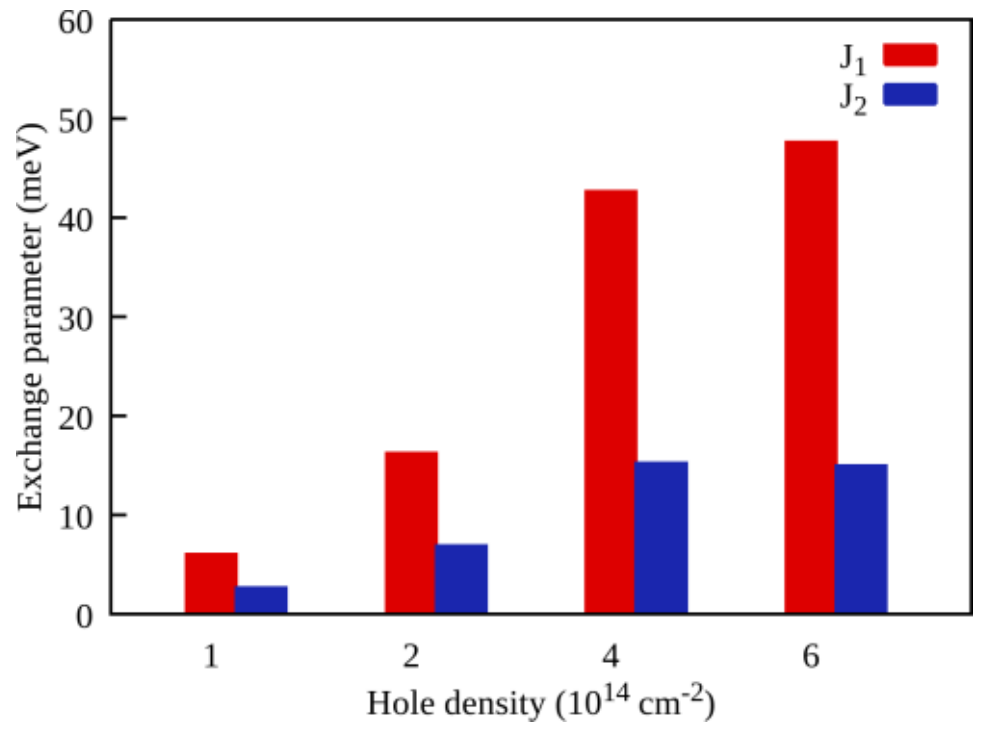

**Magnetic anisotropy energy (MAE, μeV) per magnetic atom**

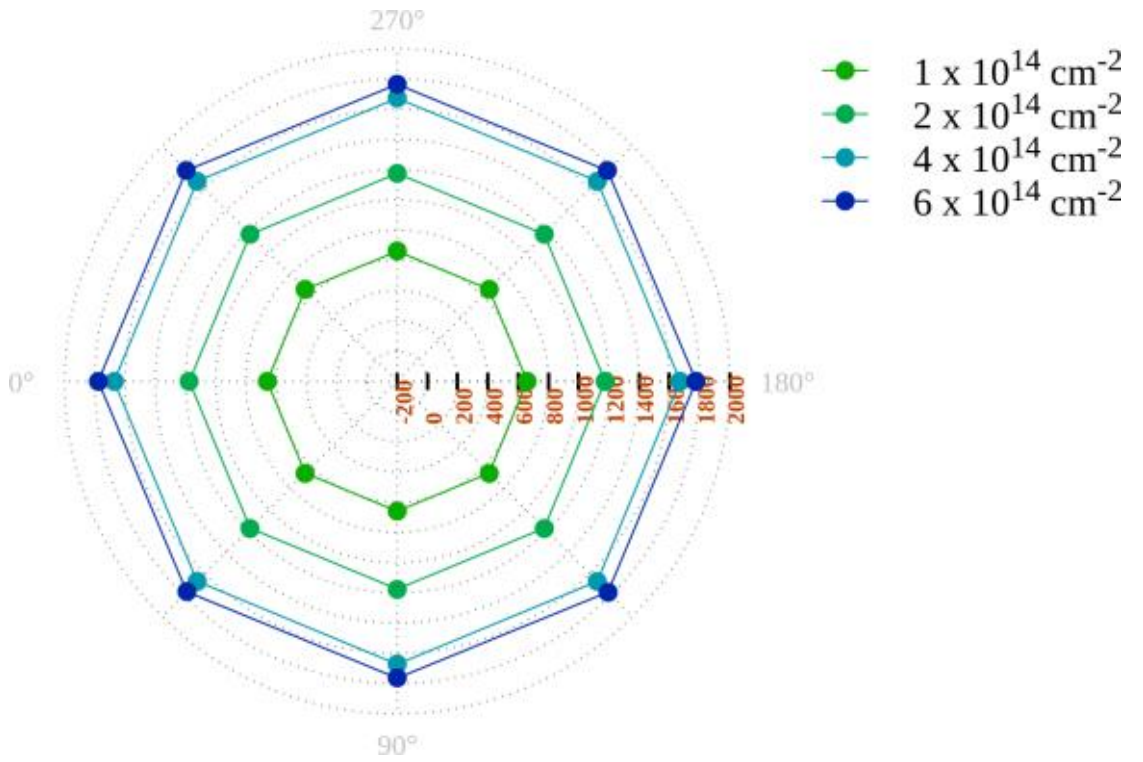

MAE = $E_{\parallel} - E_{\perp}$, a positive (negative) value of MAE indicates the off-plane (in-plane) easy axis.

**Monte Carlo simulations of the normalized magnetization of as a function of temperature**

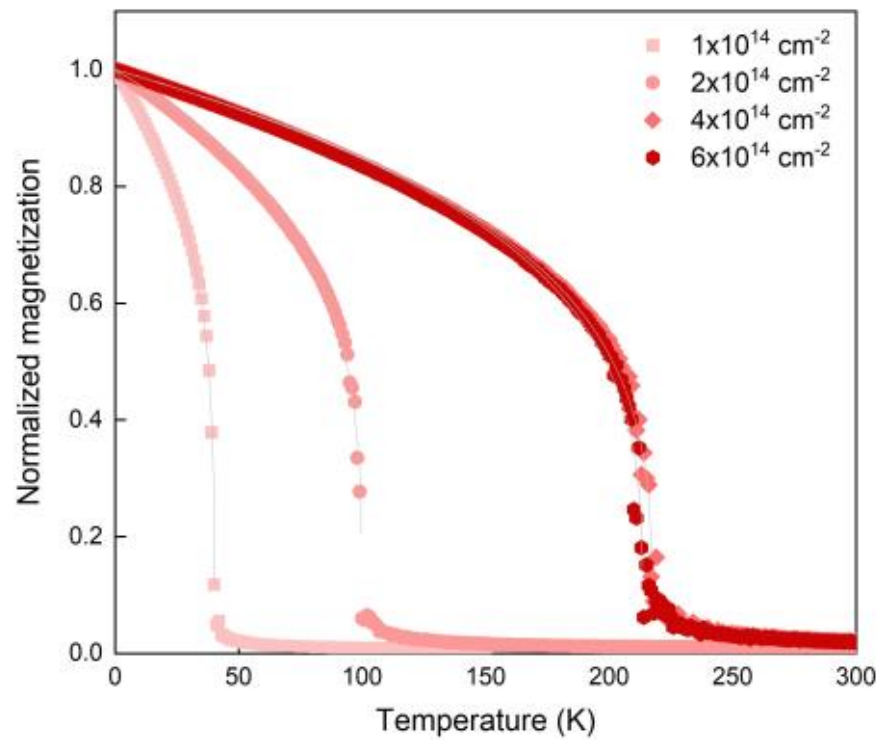

$T_c$: $1\times10^{14}$ cm$^{-2}$: 40 K  $2\times10^{14}$ cm$^{-2}$: 100 K  $4\times10^{14}$ cm$^{-2}$: 217 K  $6\times10^{14}$ cm$^{-2}$: 214 K

# 97.LiCl

| MC2D-ID | C2DB | 2dmat-ID | USPEX | Space group | Band gap (eV) |
|---|---|---|---|---|---|
| - | - | 2dm-6230 | - | P4/mmm | 5.35 |

**Convex hull**

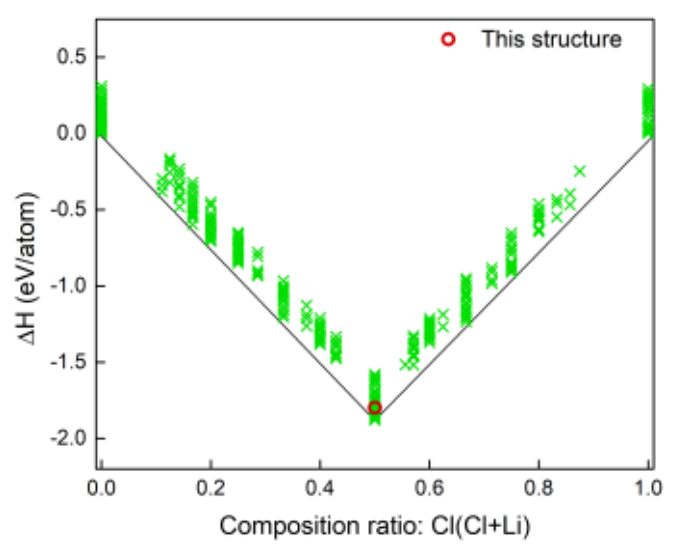

**Atomic structure**

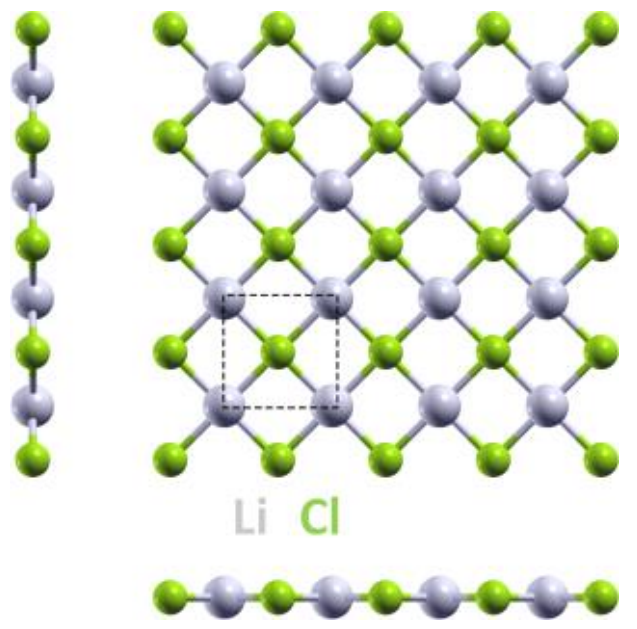

**Atomic coordinates**

```
Li1Cl1
   1.00000000000000
   3.4326989999999999   0.0000000000000000   0.0000000000000000
   0.0000000000000000   3.4326989999999999   0.0000000000000000
   0.0000000000000000   0.0000000000000000   20.0000000000000000
   Li   Cl
   1    1
Direct
  0.5000000000000000  0.5000000000000000  0.5000000000000000
  0.0000000000000000  0.0000000000000000  0.5000000000000000
```

**Phonon dispersion curve**

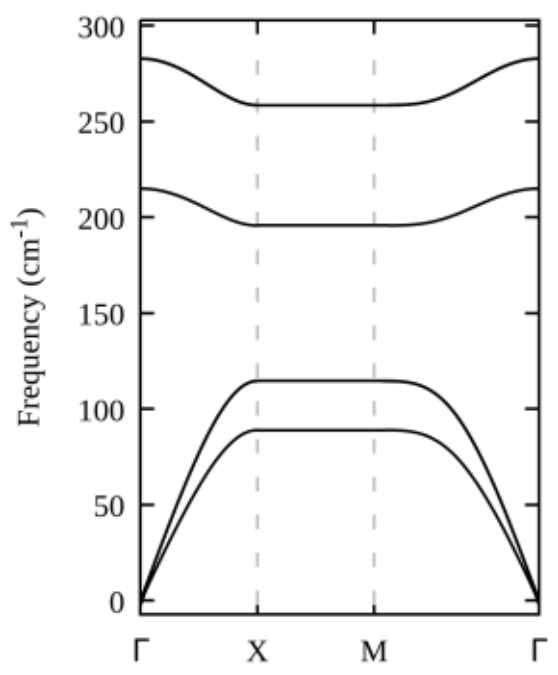

**Projected band structure and density of states**

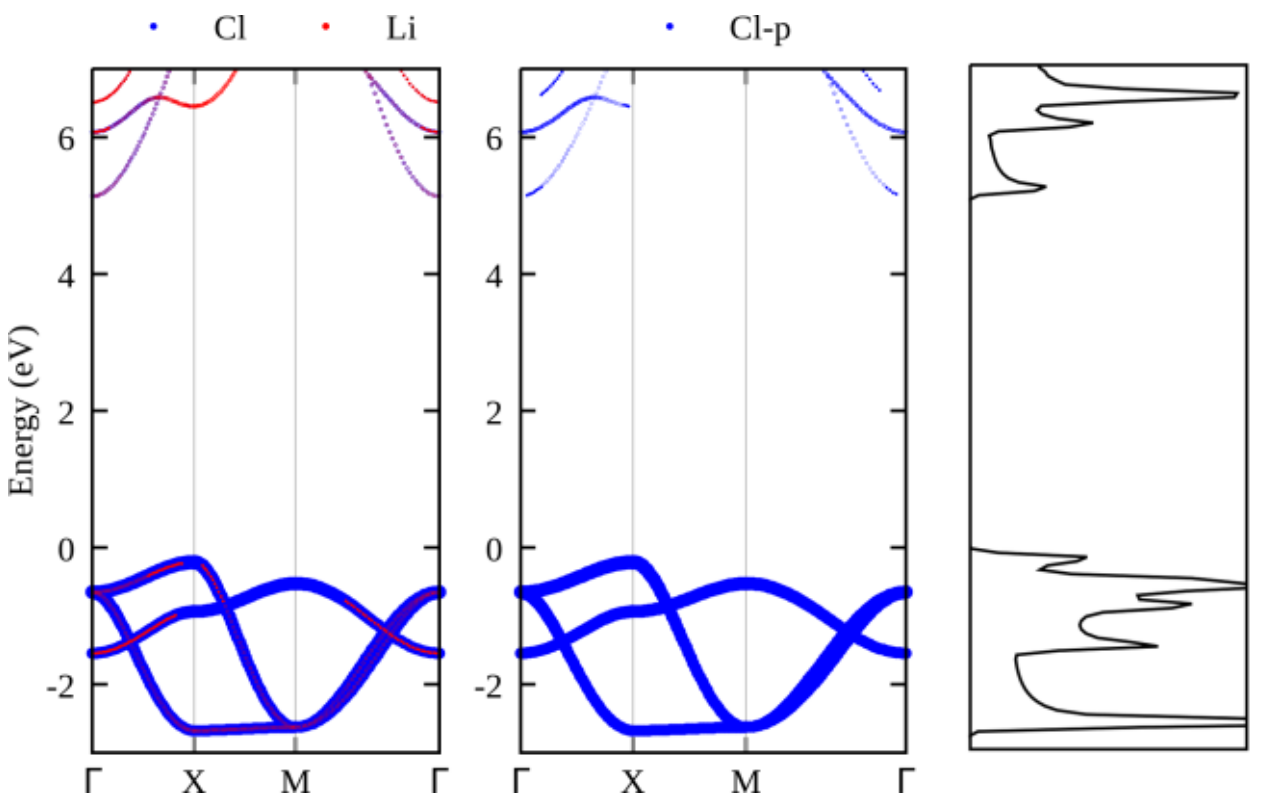

**Magnetic moment and spin polarization energy as a function of hole doping concentration**

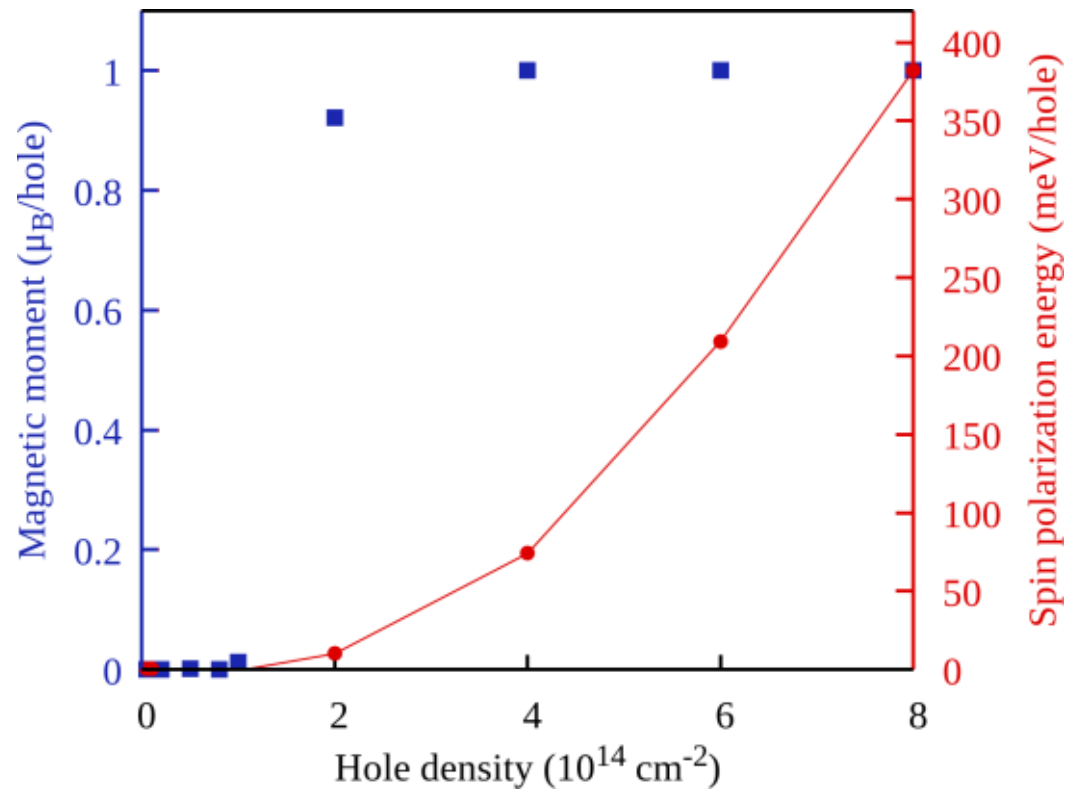

**Magnetic configurations and spin Hamiltonian**

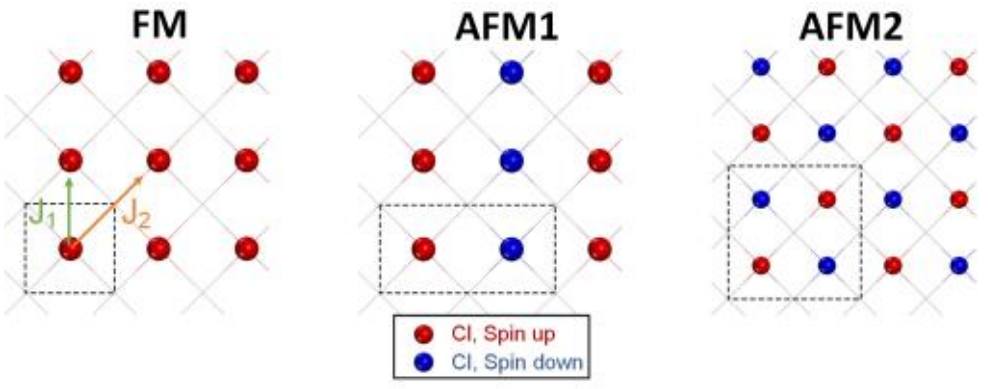

$$E_{FM} = E_0 - 2J_1S^2 - 2J_2S^2$$

$$E_{AFM1} = E_0 + 2J_2S^2$$

$$E_{AFM2} = E_0 + 2J_1S^2 - 2J_2S^2$$

**Magnetic exchange coupling parameters**

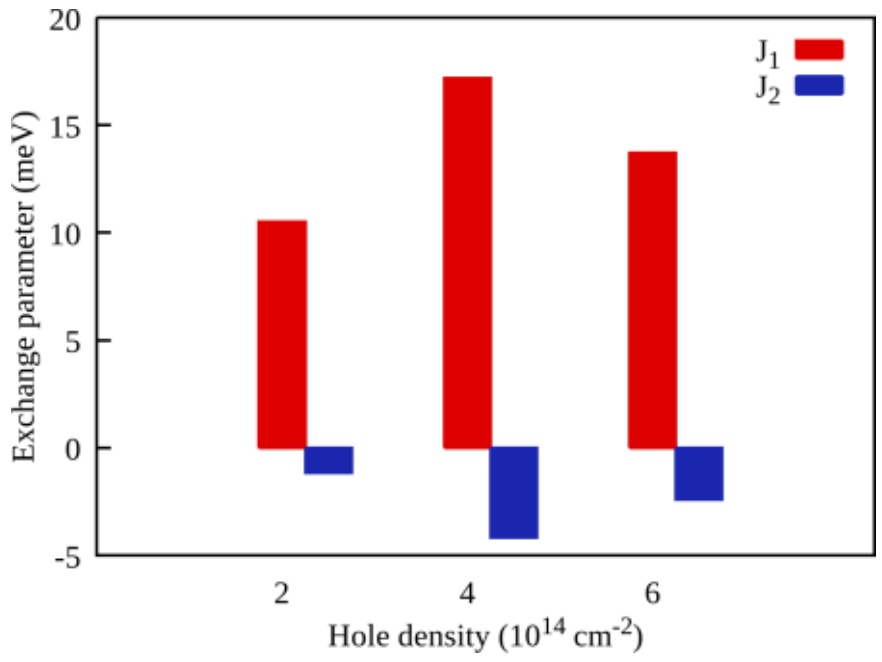

**Magnetic anisotropy energy (MAE, μeV) per magnetic atom**

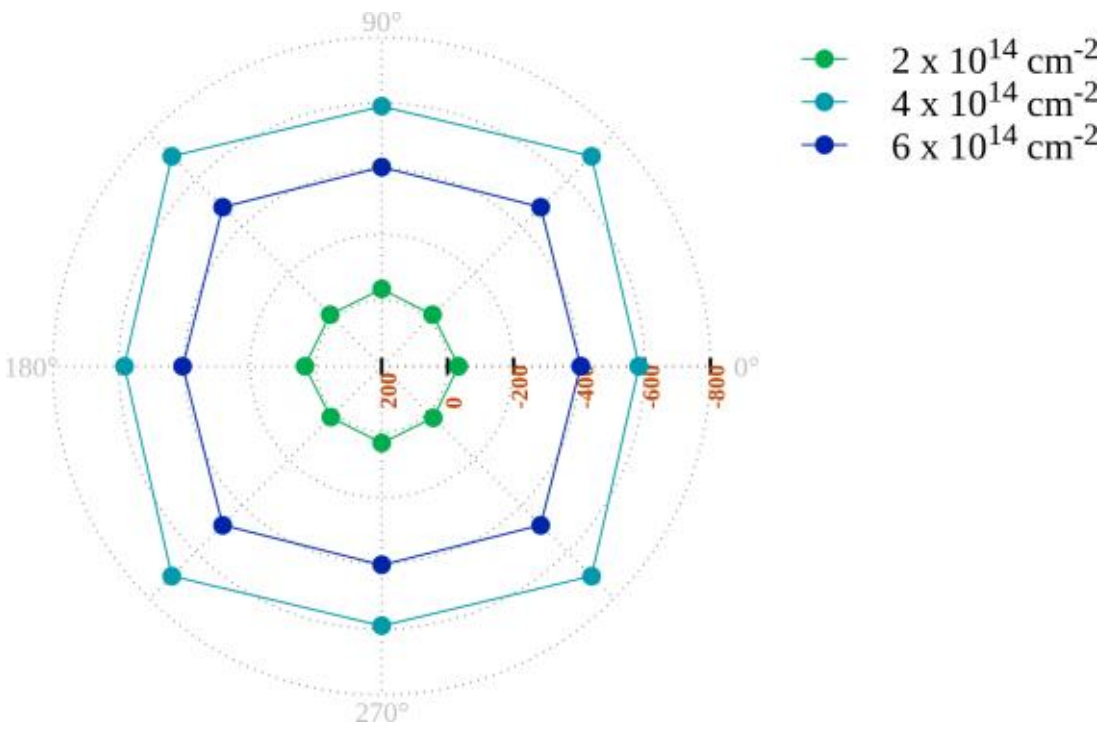

MAE = $E_{\parallel} - E_{\perp}$, a positive (negative) value of MAE indicates the off-plane (in-plane) easy axis.

**Monte Carlo simulations of the normalized magnetization of as a function of temperature**

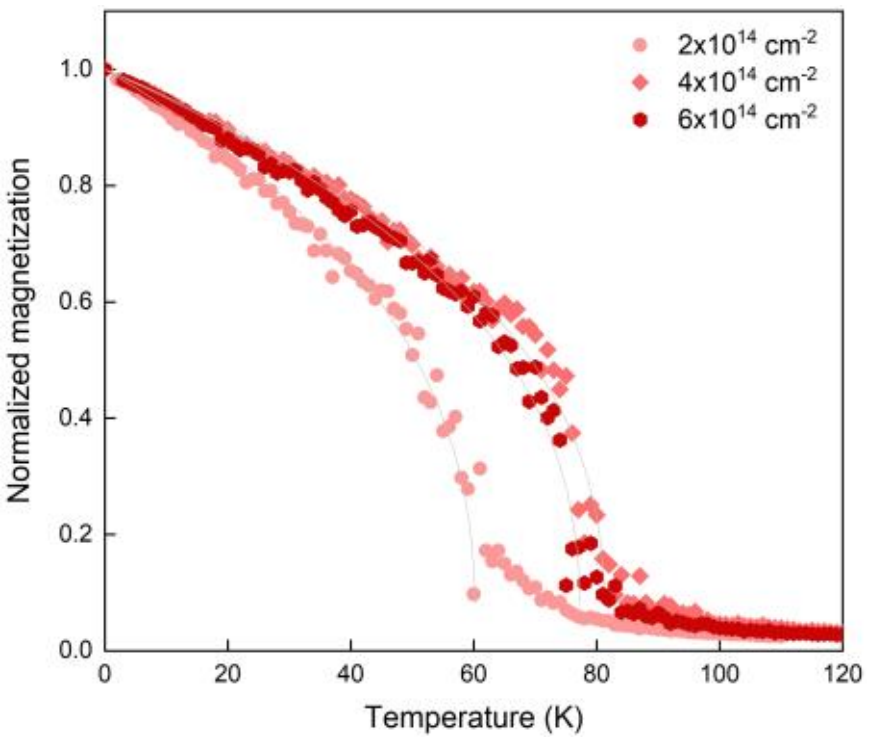

$T_c$: $1\times10^{14}$ cm$^{-2}$: - K  $2\times10^{14}$ cm$^{-2}$: 60 K

$4\times10^{14}$ cm$^{-2}$: 81 K  $6\times10^{14}$ cm$^{-2}$: 77 K

# 98.LiBr

| MC2D-ID | C2DB | 2dmat-ID | USPEX | Space group | Band gap (eV) |
|---|---|---|---|---|---|
| - | - | 2dm-6152 | - | P4/mmm | 4.64 |

**Convex hull** | **Atomic structure** | **Atomic coordinates** | **Phonon dispersion curve**

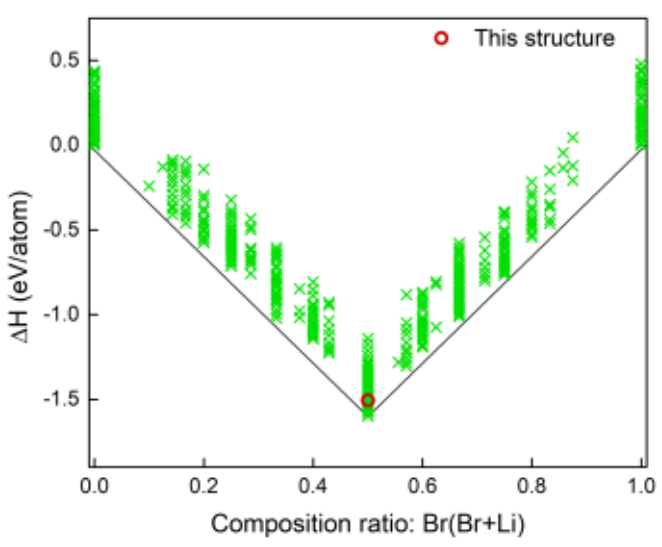

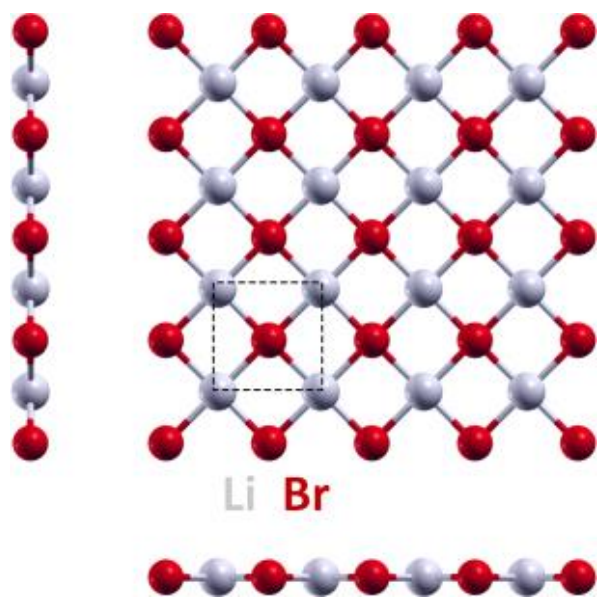

```
Li1Br1
  1.00000000000000
  3.6852198488649122   0.0000000000000000   0.0000000000000000
  0.0000000000000000   3.6852197352056466   0.0000000000000000
  0.0000000000000000   0.0000000000000000  20.0000000000000000
  Li   Br
  1    1
Direct
 0.4999999994756266  0.4999999998885158  0.4999999989455830
 0.0000000005243734  0.0000000001114842  0.5000000010544170
```

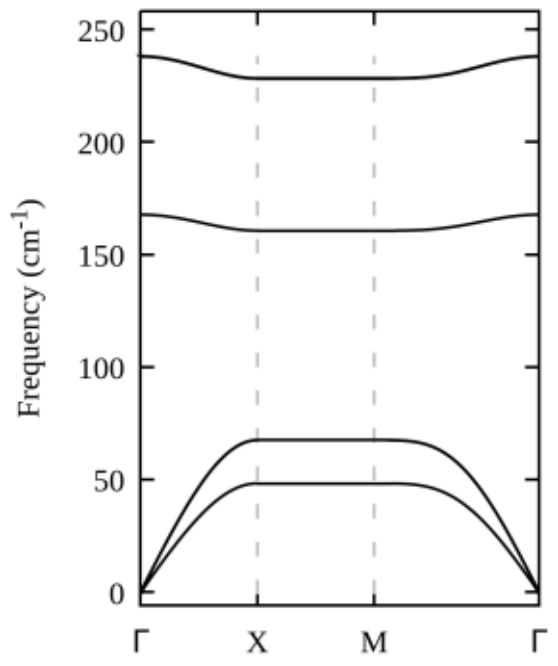

**Projected band structure and density of states**

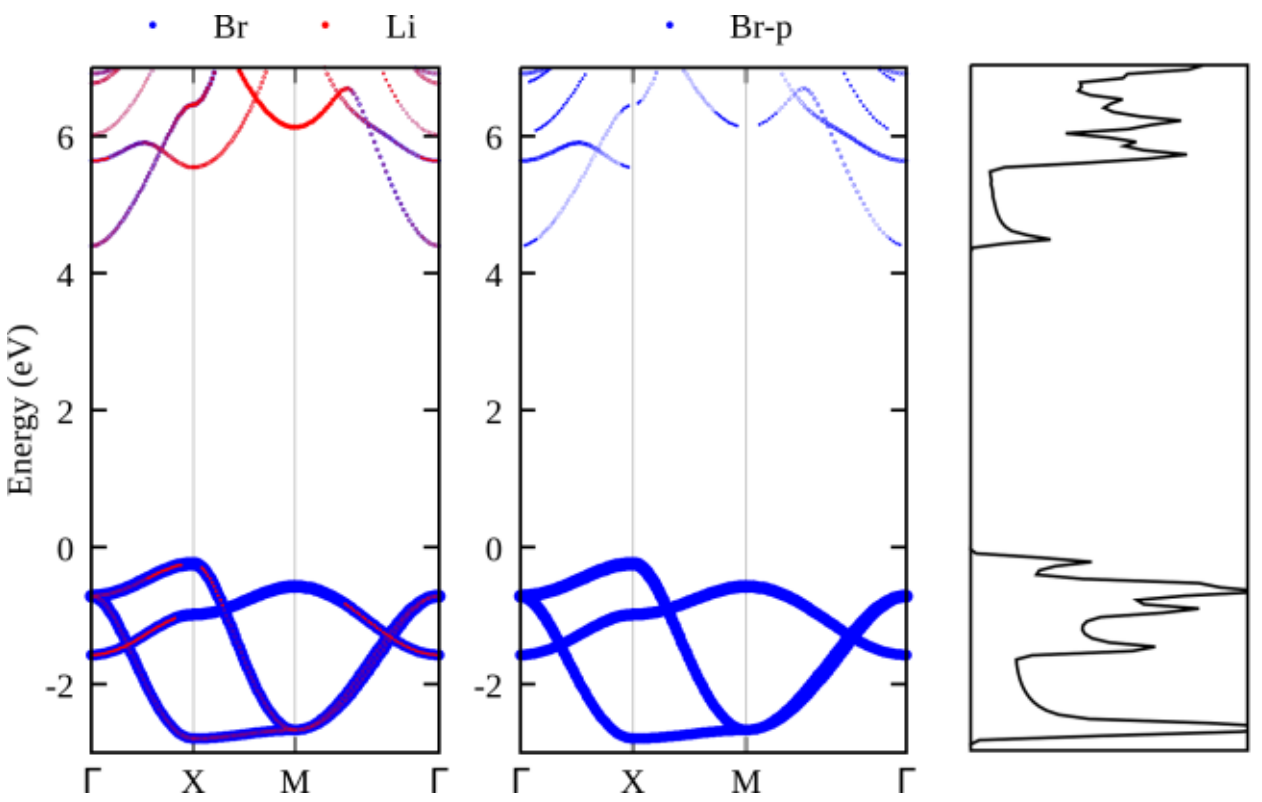

**Magnetic moment and spin polarization energy as a function of hole doping concentration**

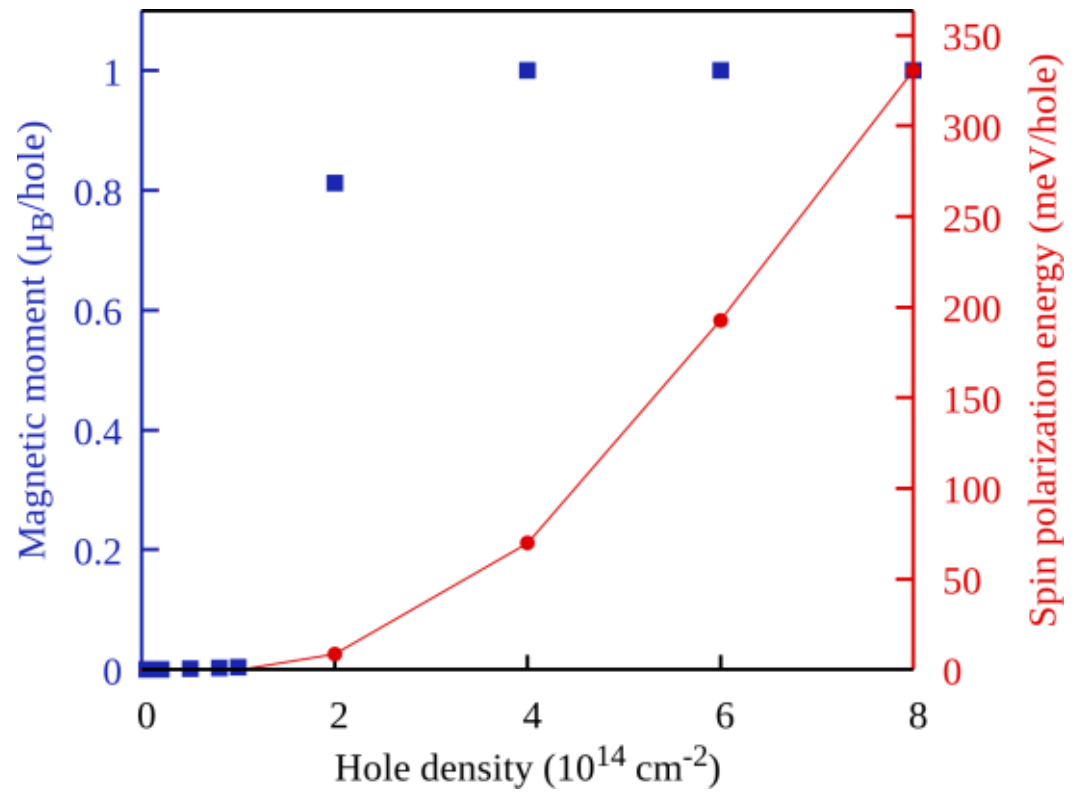

**Magnetic configurations and spin Hamiltonian**

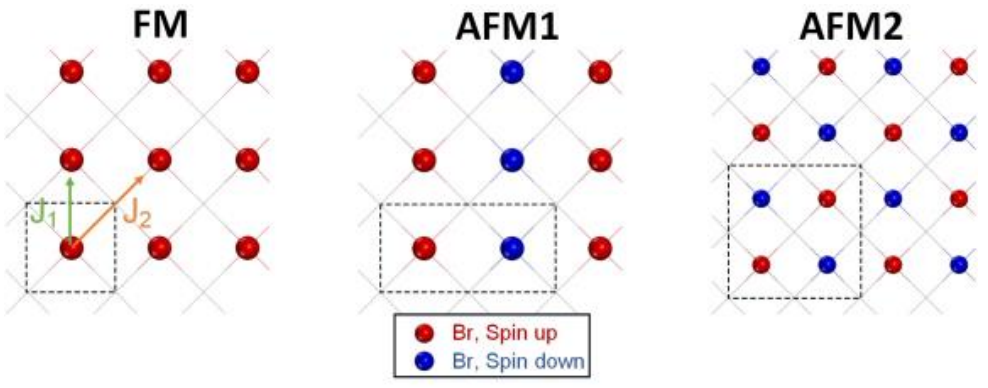

$$E_{FM} = E_0 - 2J_1S^2 - 2J_2S^2$$
$$E_{AFM1} = E_0 + 2J_2S^2$$
$$E_{AFM2} = E_0 + 2J_1S^2 - 2J_2S^2$$

**Magnetic exchange coupling parameters**

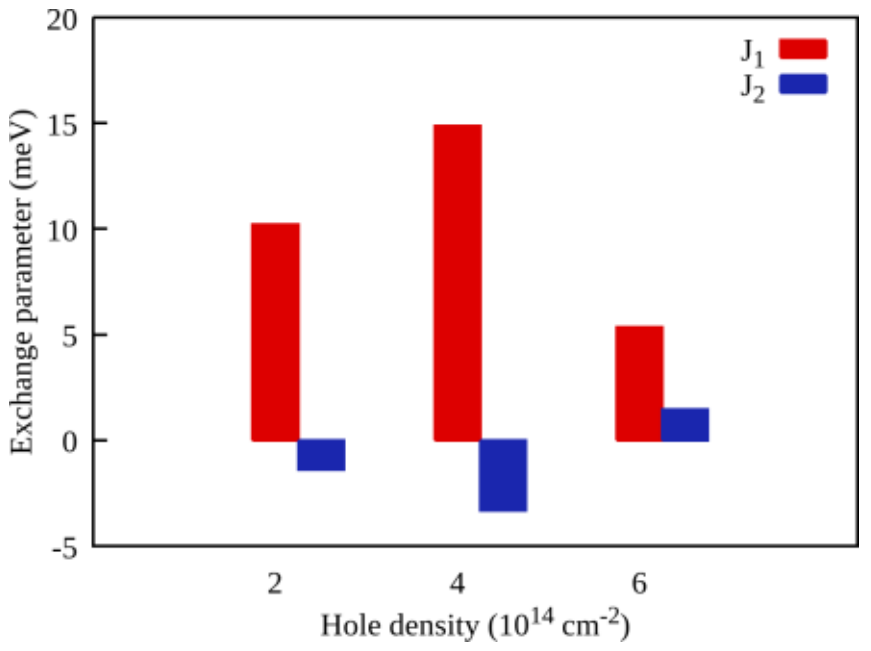

**Magnetic anisotropy energy (MAE, μeV) per magnetic atom**

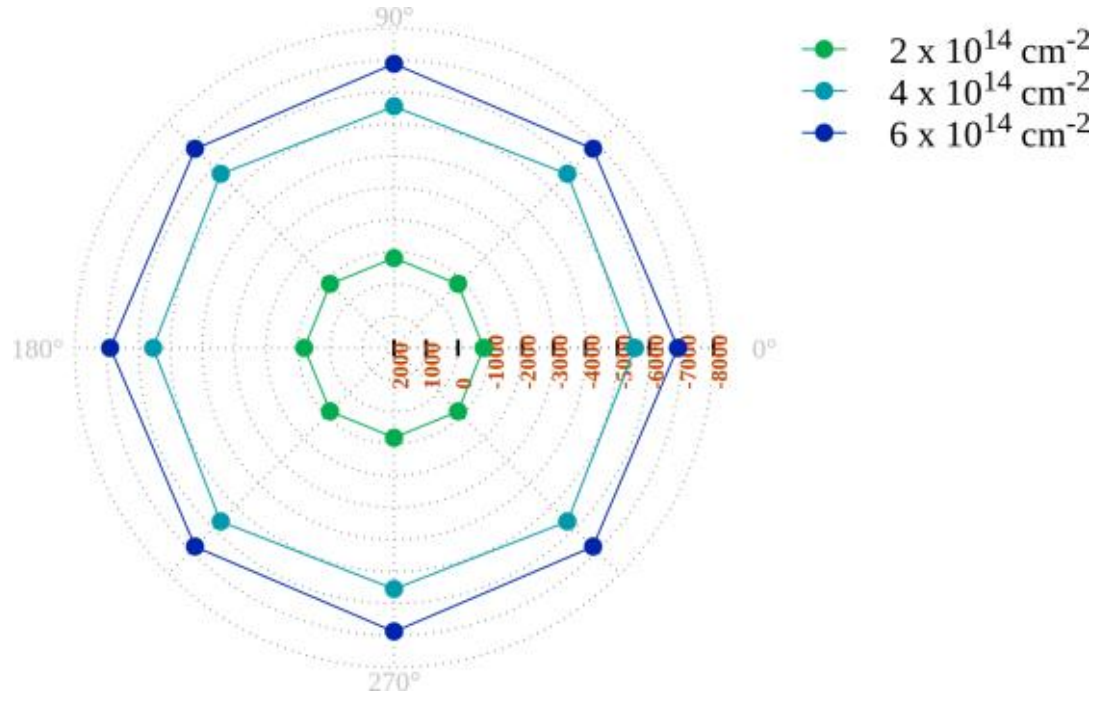

MAE = $E_{\parallel} - E_{\perp}$, a positive (negative) value of MAE indicates the off-plane (in-plane) easy axis.

**Monte Carlo simulations of the normalized magnetization of as a function of temperature**

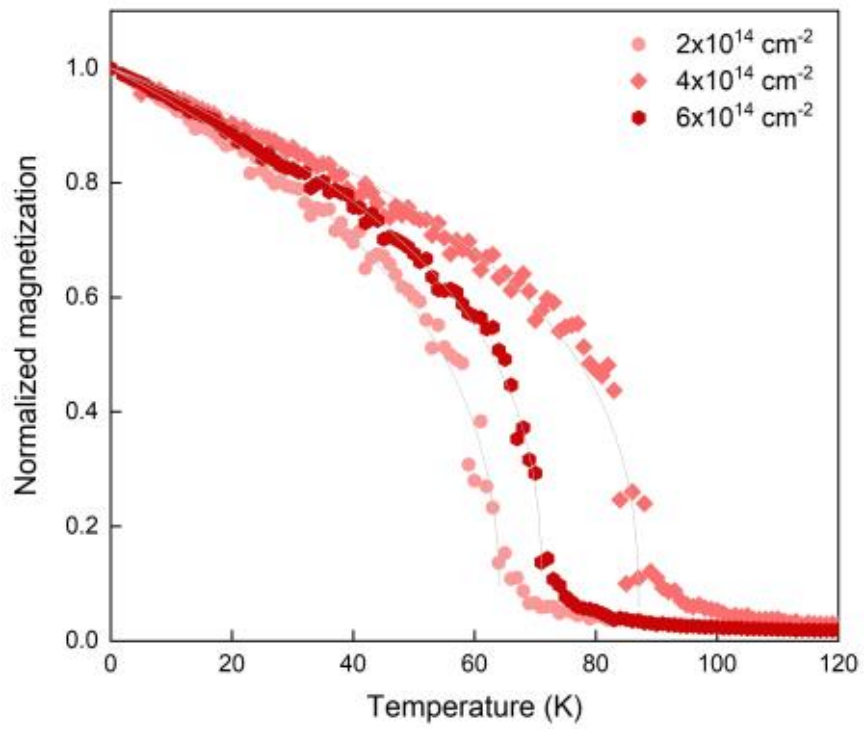

$T_c$: $1\times10^{14}$ cm$^{-2}$: - K  $2\times10^{14}$ cm$^{-2}$: 64 K

$4\times10^{14}$ cm$^{-2}$: 87 K  $6\times10^{14}$ cm$^{-2}$: 71 K

# 99.LiI

| MC2D-ID | C2DB | 2dmat-ID | USPEX | Space group | Band gap (eV) |
|---|---|---|---|---|---|
| - | - | 2dm-6194 | - | P4/mmm | 4.05 |

**Convex hull** | **Atomic structure** | **Atomic coordinates** | **Phonon dispersion curve**

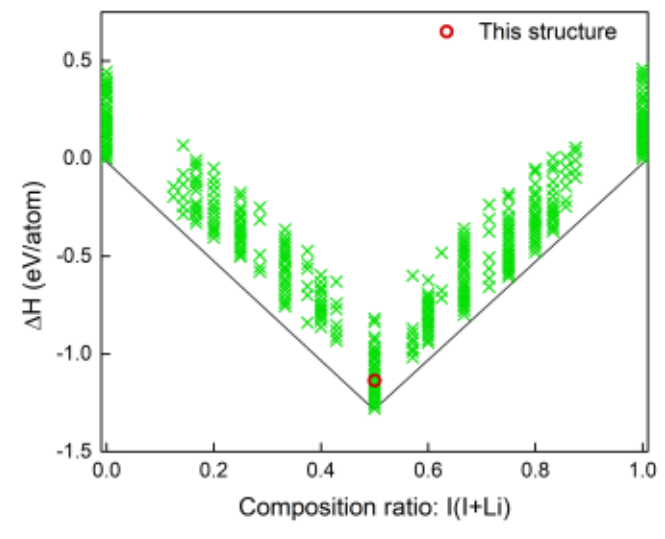

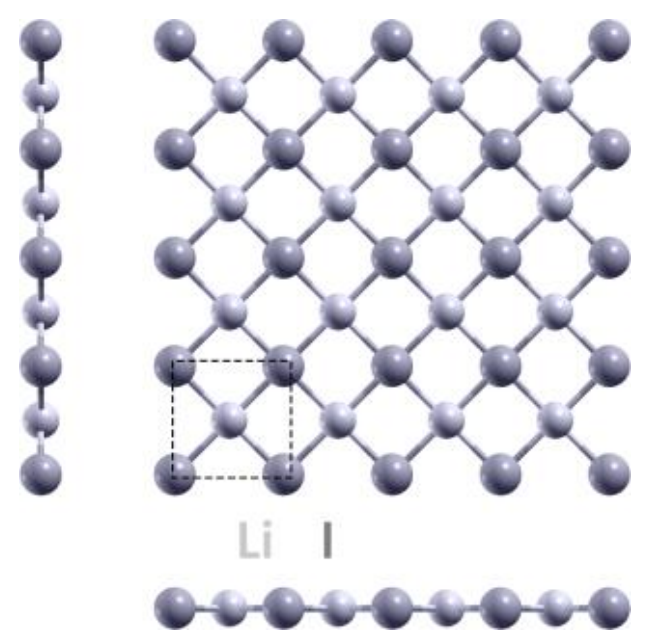

```
Li1I1
   1.00000000000000
   4.0410291989920726   0.0000000000000000   0.0000000000000000
   0.0000000000000000   4.0410296835011872   0.0000000000000000
   0.0000000000000000   0.0000000000000000  20.0000000000000000
   Li    I
   1     1
Direct
  0.5000000009804637  0.4999999989172252  0.5000000001712479
  0.9999999990195363  0.0000000010827748  0.4999999998287521
```

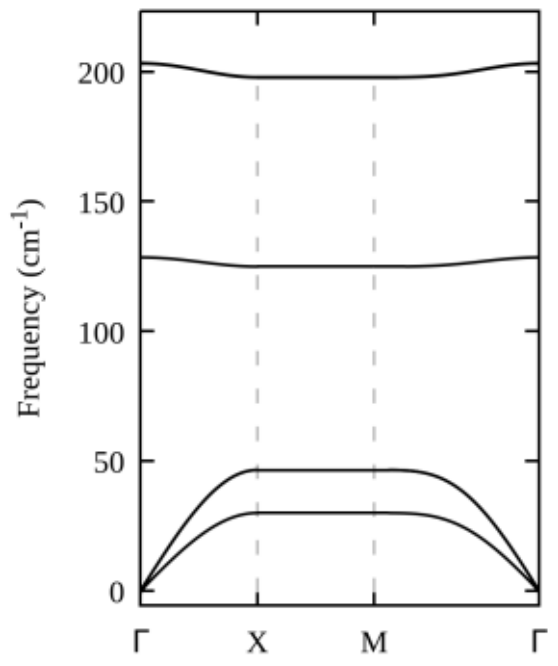

## Projected band structure and density of states

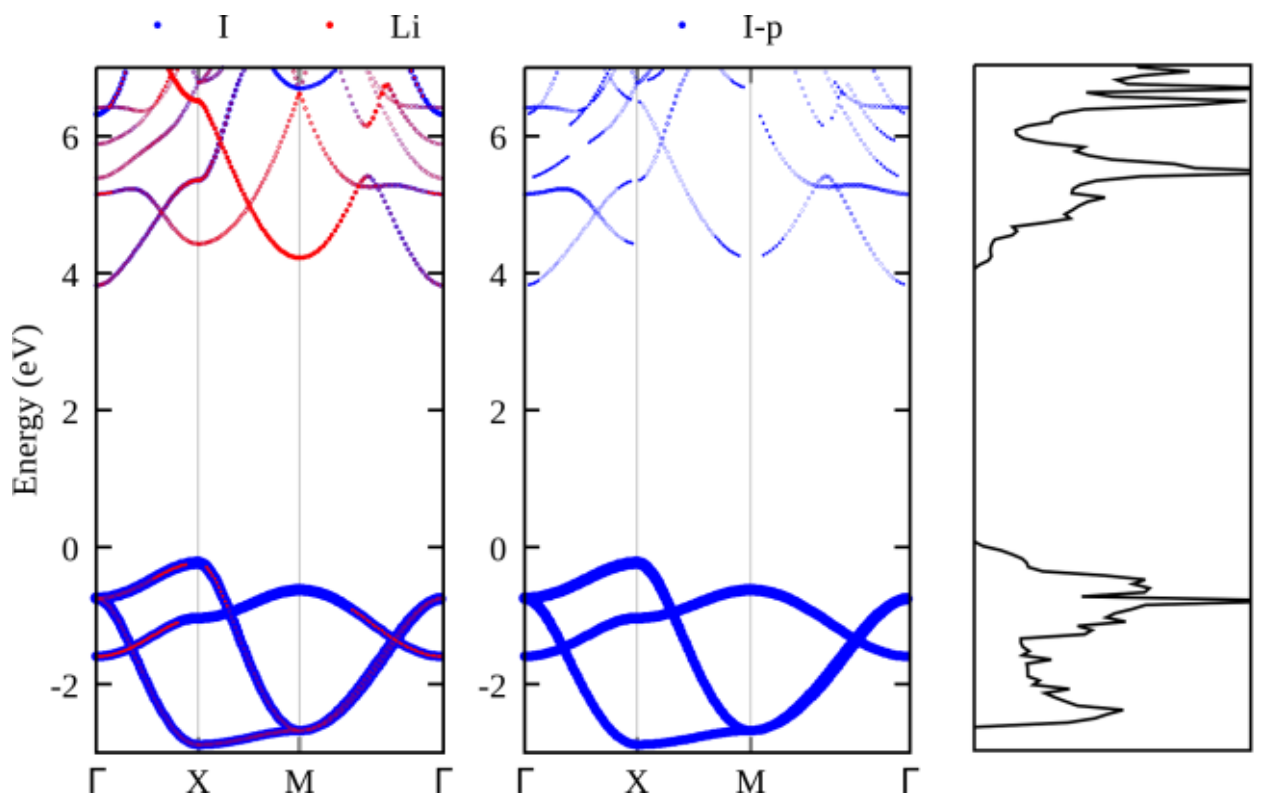

## Magnetic moment and spin polarization energy as a function of hole doping concentration

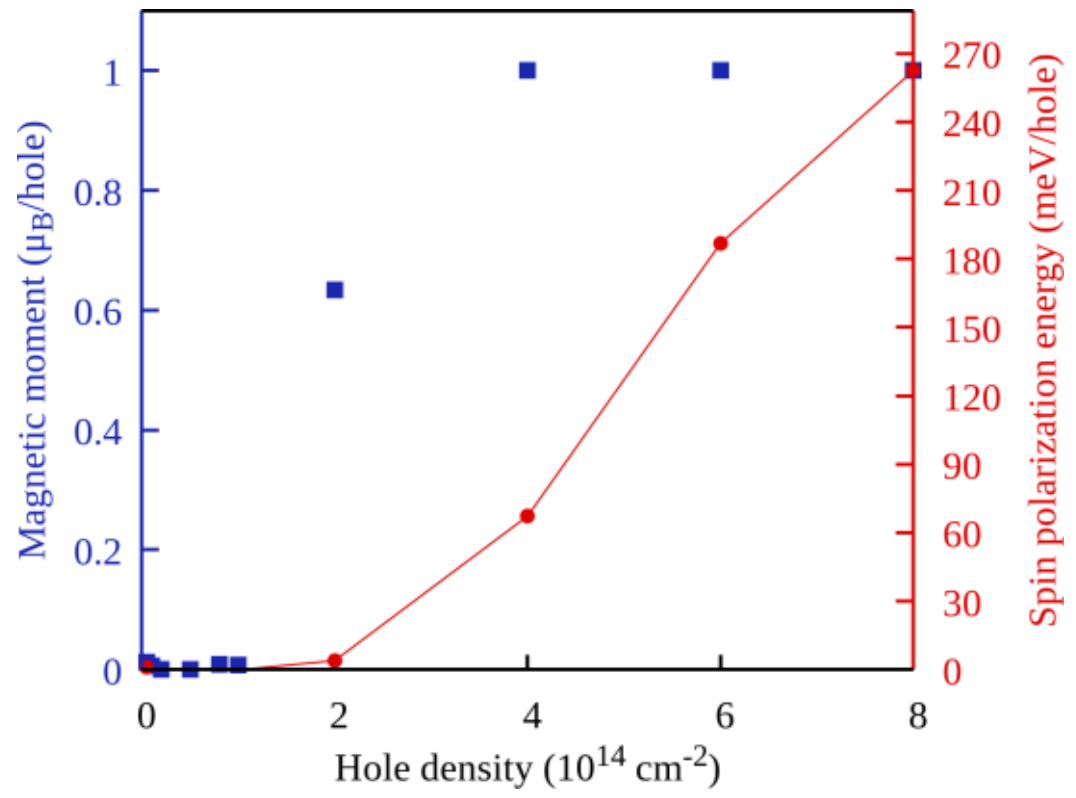

## Magnetic configurations and spin Hamiltonian

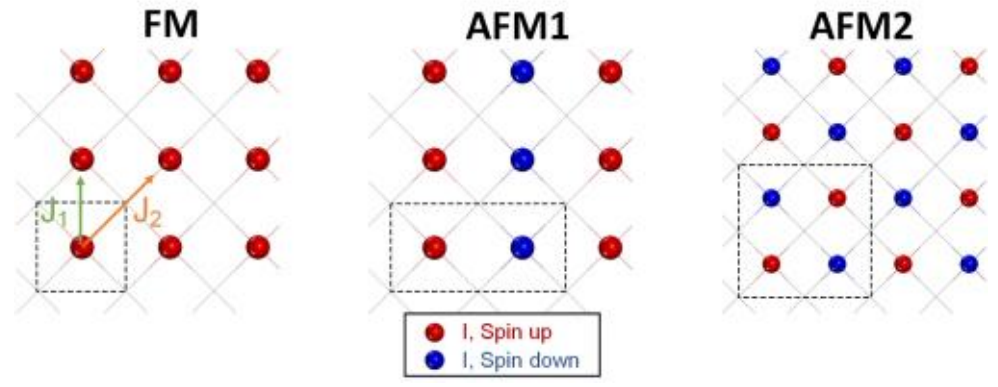

$$E_{FM} = E_0 - 2J_1S^2 - 2J_2S^2$$

$$E_{AFM1} = E_0 + 2J_2S^2$$

$$E_{AFM2} = E_0 + 2J_1S^2 - 2J_2S^2$$

## Magnetic exchange coupling parameters

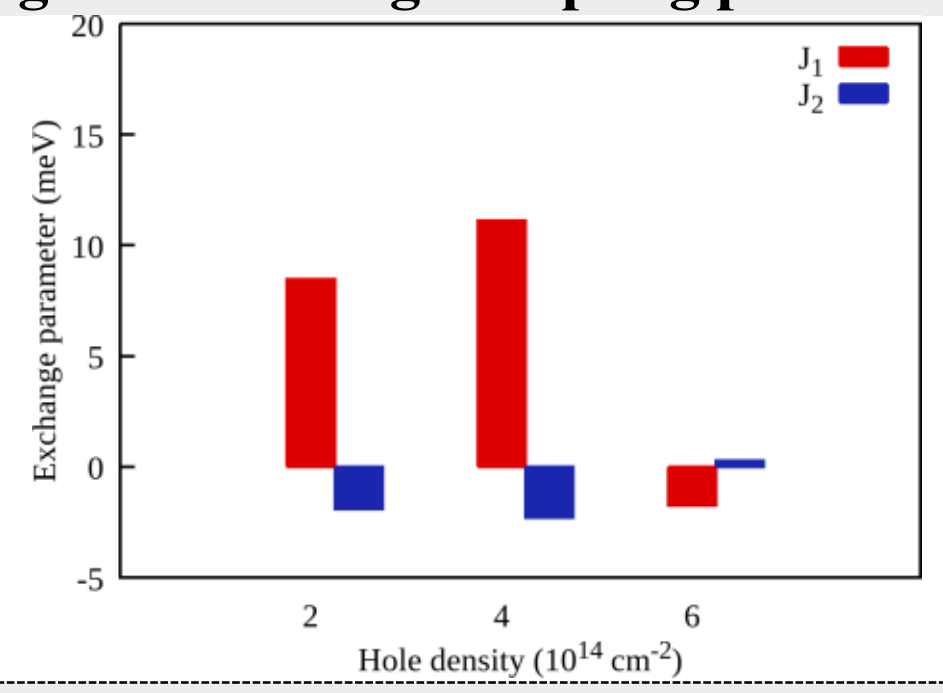

## Magnetic anisotropy energy (MAE, μeV) per magnetic atom

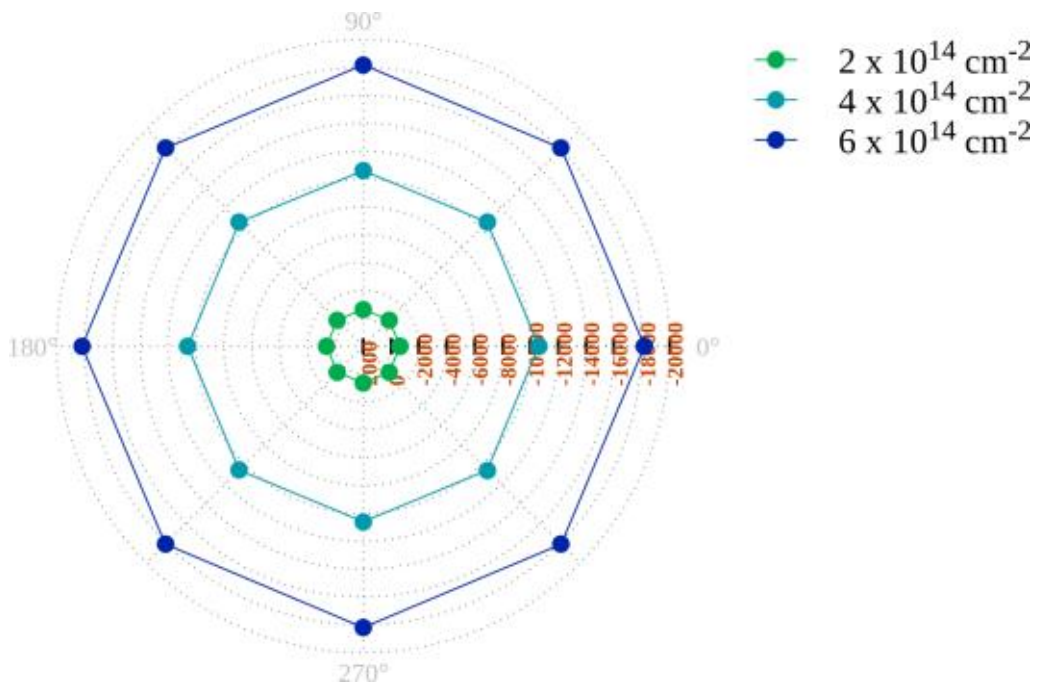

MAE = $E_{\parallel} - E_{\perp}$, a positive (negative) value of MAE indicates the off-plane (in-plane) easy axis.

## Monte Carlo simulations of the normalized magnetization of as a function of temperature

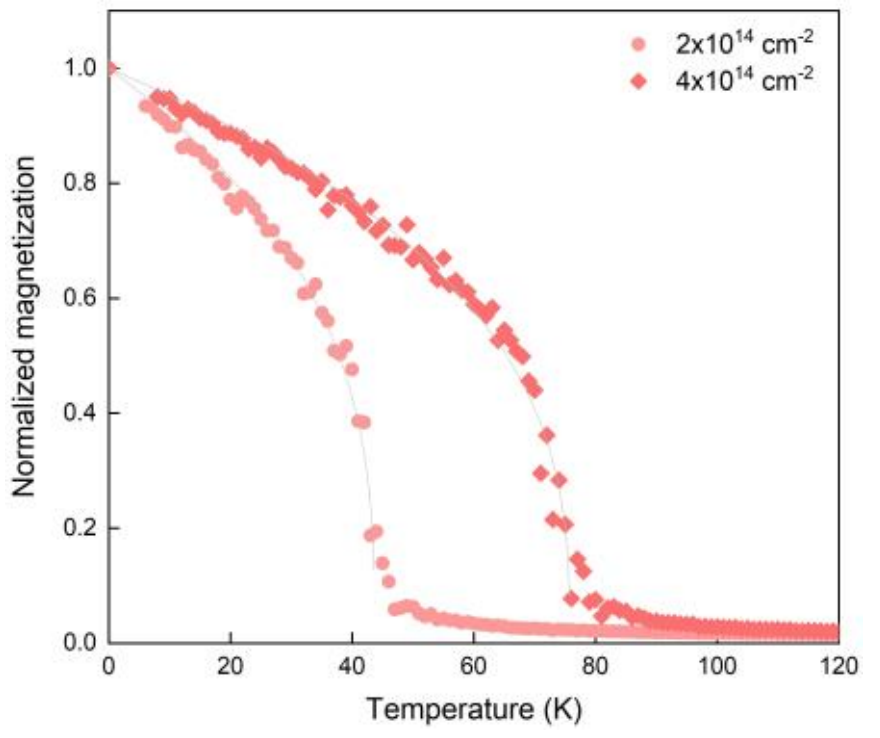

$T_c$: $1\times10^{14}$ cm$^{-2}$: - K   $2\times10^{14}$ cm$^{-2}$: 44 K

$4\times10^{14}$ cm$^{-2}$: 76 K   $6\times10^{14}$ cm$^{-2}$: - K

# 100. NaBr

| MC2D-ID | C2DB | 2dmat-ID | USPEX | Space group | Band gap (eV) |
|---|---|---|---|---|---|
| - | - | 2dm-6127 | - | P4/mmm | 4.36 |

**Convex hull**

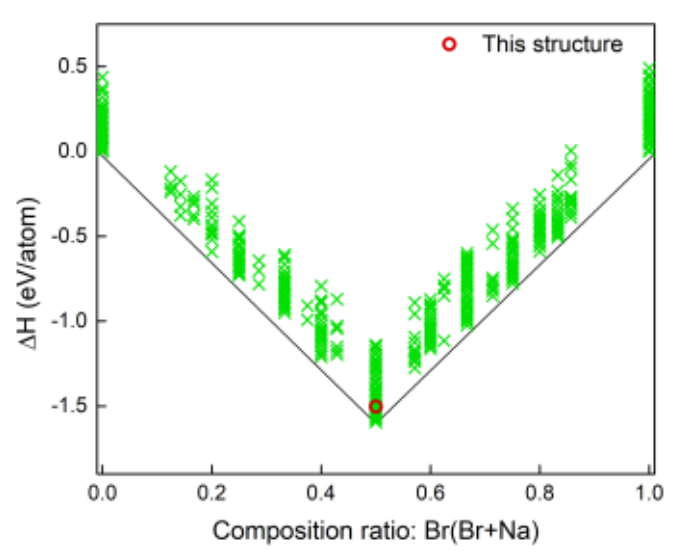

**Atomic structure**

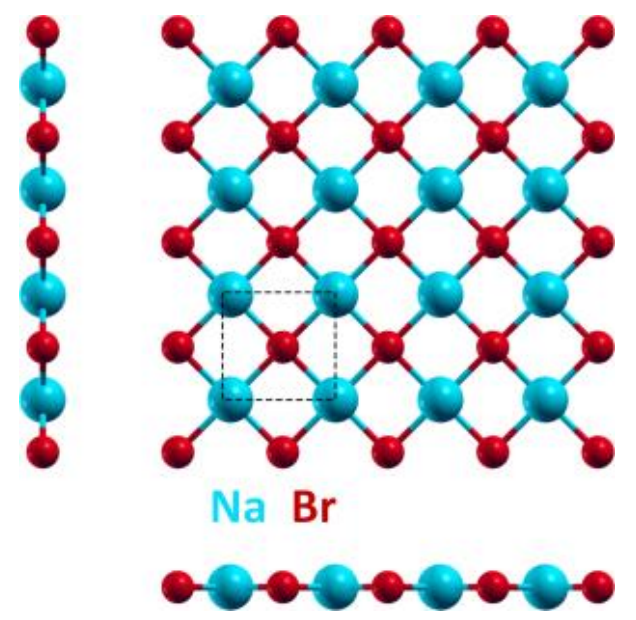

**Atomic coordinates**

```
Na1Br1
    1.00000000000000
    4.0826855104855753    0.0000000000000000    0.0000000000000000
    0.0000000000000000    4.0826855851440875    0.0000000000000000
    0.0000000000000000    0.0000000000000000   20.0000000000000000
    Na    Br
    1     1
Direct
   0.5000000000311715   0.4999999999283915   0.5000000000226166
   0.9999999999688285   0.0000000000716085   0.4999999999773834
```

**Phonon dispersion curve**

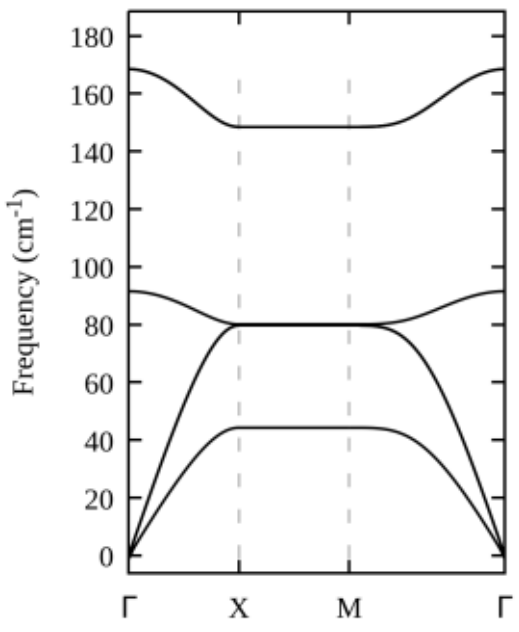

**Projected band structure and density of states**

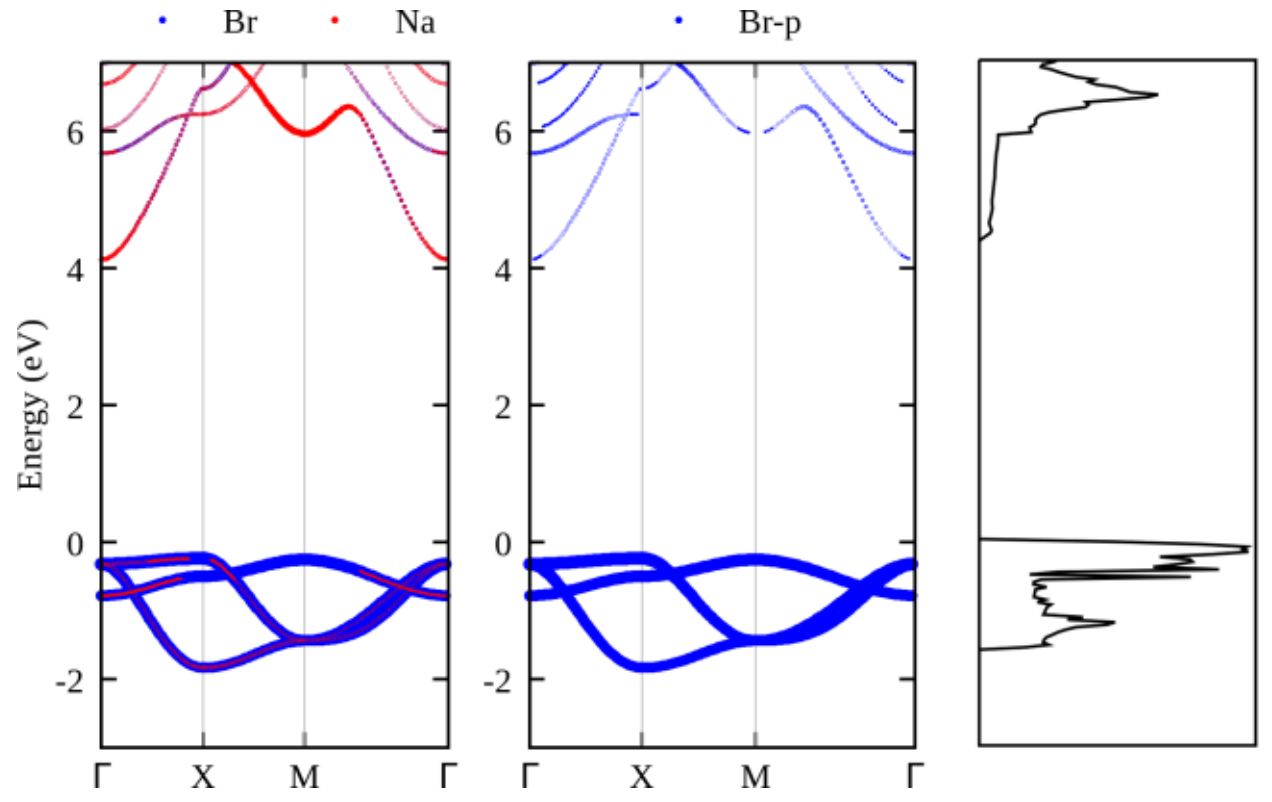

**Magnetic moment and spin polarization energy as a function of hole doping concentration**

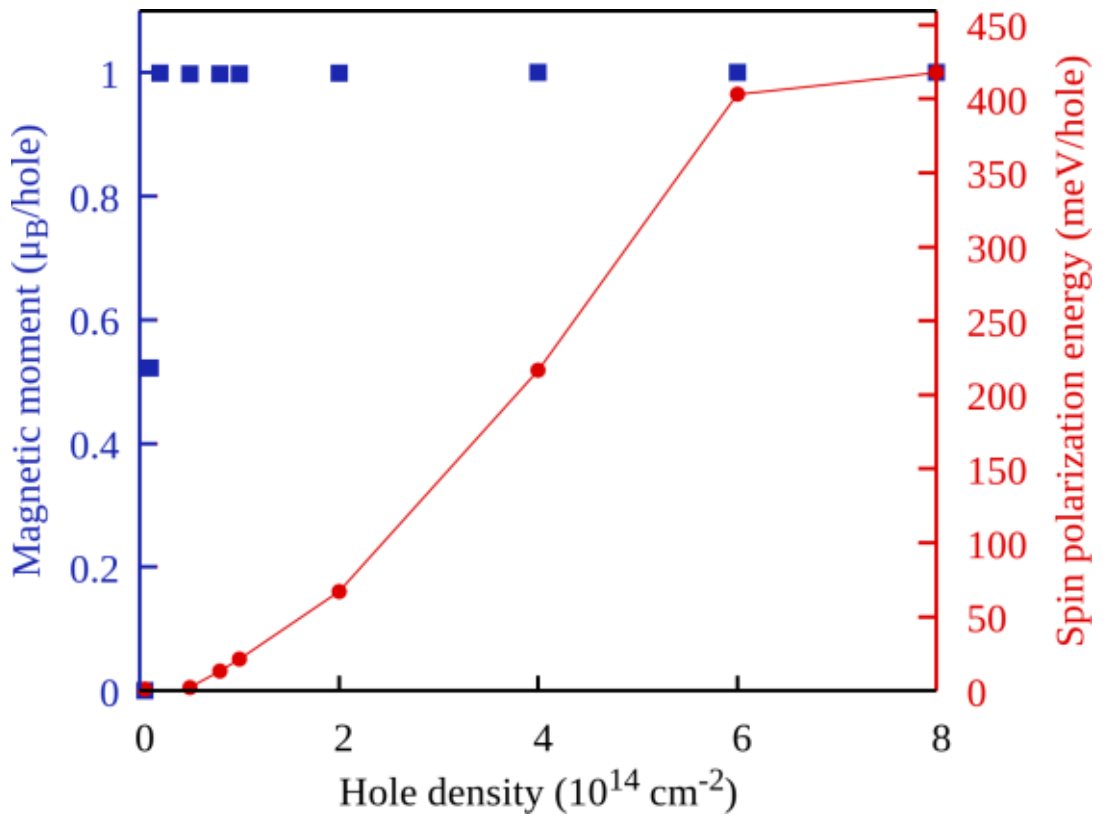

**Magnetic configurations and spin Hamiltonian**

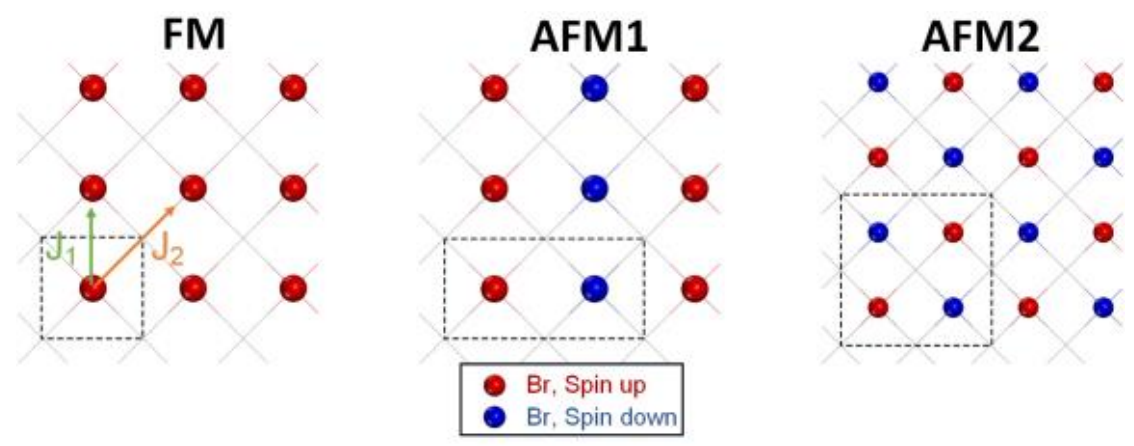

$$E_{FM} = E_0 - 2J_1S^2 - 2J_2S^2$$
$$E_{AFM1} = E_0 + 2J_2S^2$$
$$E_{AFM2} = E_0 + 2J_1S^2 - 2J_2S^2$$

**Magnetic exchange coupling parameters**

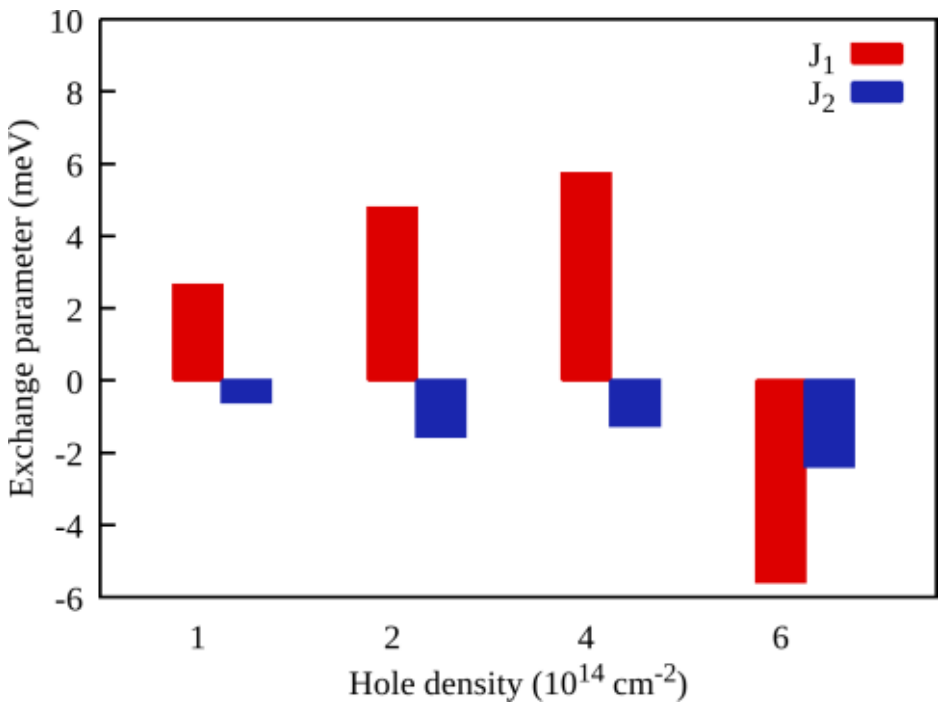

**Magnetic anisotropy energy (MAE, μeV) per magnetic atom**

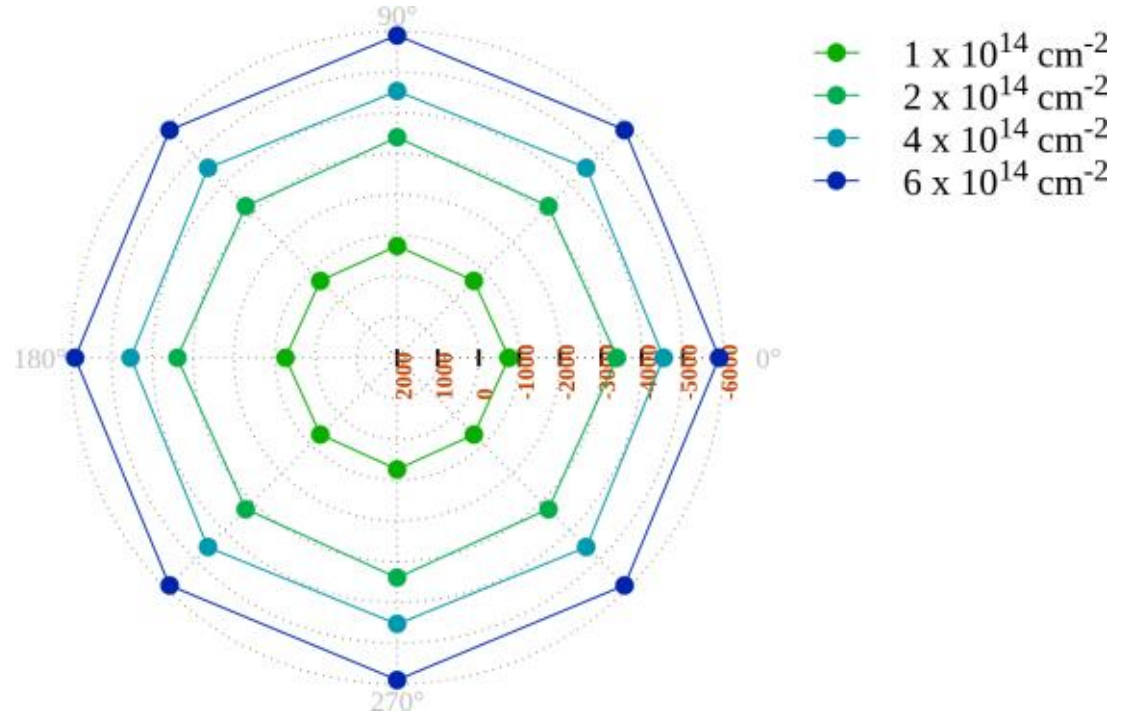

MAE = $E_{\parallel} - E_{\perp}$, a positive (negative) value of MAE indicates the off-plane (in-plane) easy axis.

**Monte Carlo simulations of the normalized magnetization of as a function of temperature**

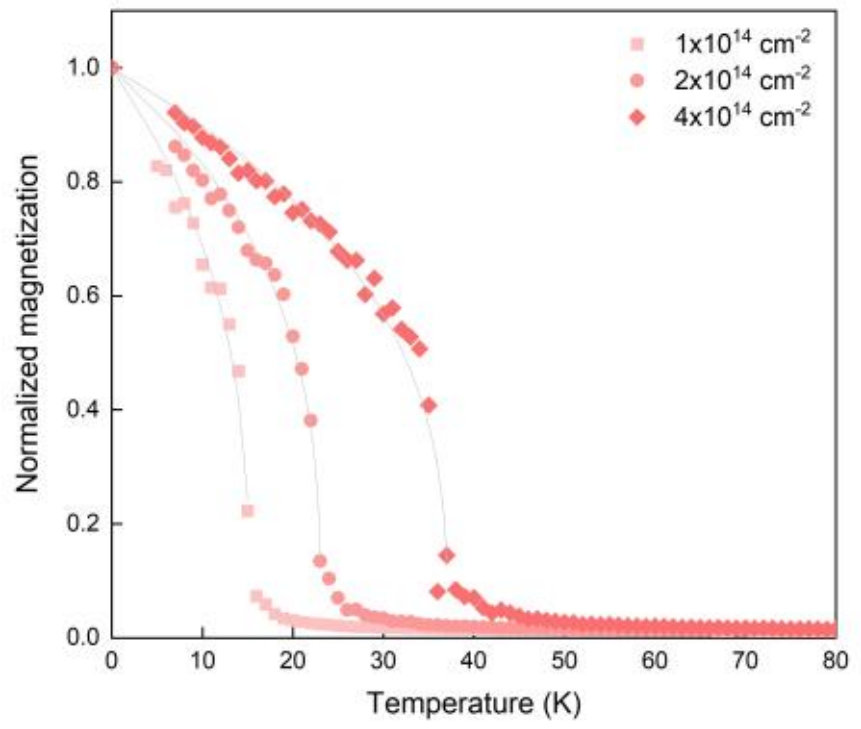

$T_c$: $1\times10^{14}$ cm$^{-2}$: 25 K  $2\times10^{14}$ cm$^{-2}$: 23 K

$4\times10^{14}$ cm$^{-2}$: 37 K  $6\times10^{14}$ cm$^{-2}$: - K

# 101. NaI

| MC2D-ID | C2DB | 2dmat-ID | USPEX | Space group | Band gap (eV) |
|---|---|---|---|---|---|
| - | - | 2dm-3215 | - | P4/mmm | 3.84 |

**Convex hull**

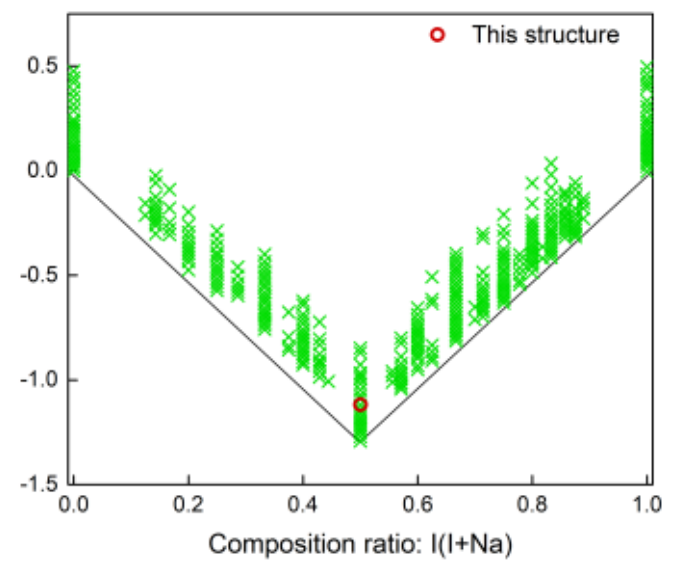

**Atomic structure**

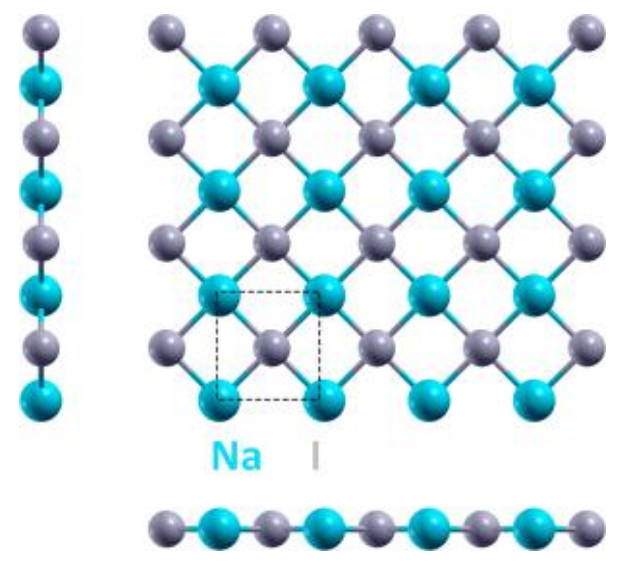

**Atomic coordinates**

```
Na1I1
   1.00000000000000
   4.4235681542498817   0.0000000000000000   0.0000000000000000
   0.0000000000000000   4.4235689204567565   0.0000000000000000
   0.0000000000000000   0.0000000000000000  20.0000000000000000
   Na   I
   1    1
Direct
  0.4999999996511733  0.5000000022245175  0.4999999994690754
  0.0000000003488267  0.9999999977754825  0.5000000005309246
```

**Phonon dispersion curve**

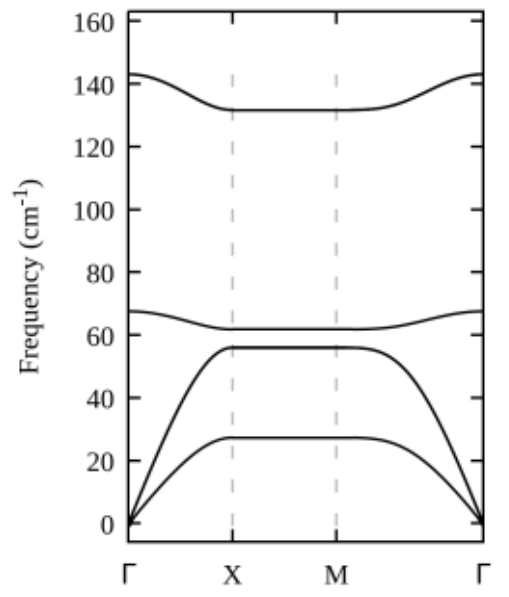

**Projected band structure and density of states**

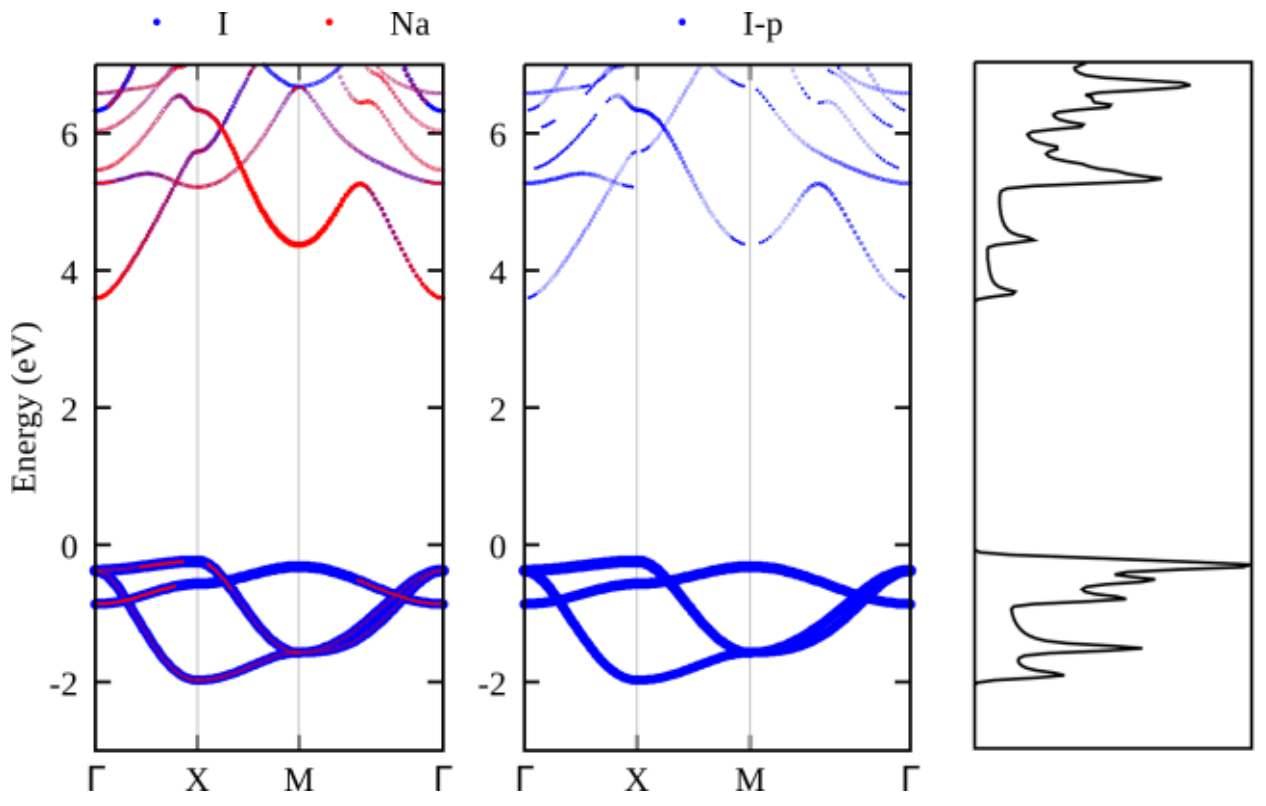

**Magnetic moment and spin polarization energy as a function of hole doping concentration**

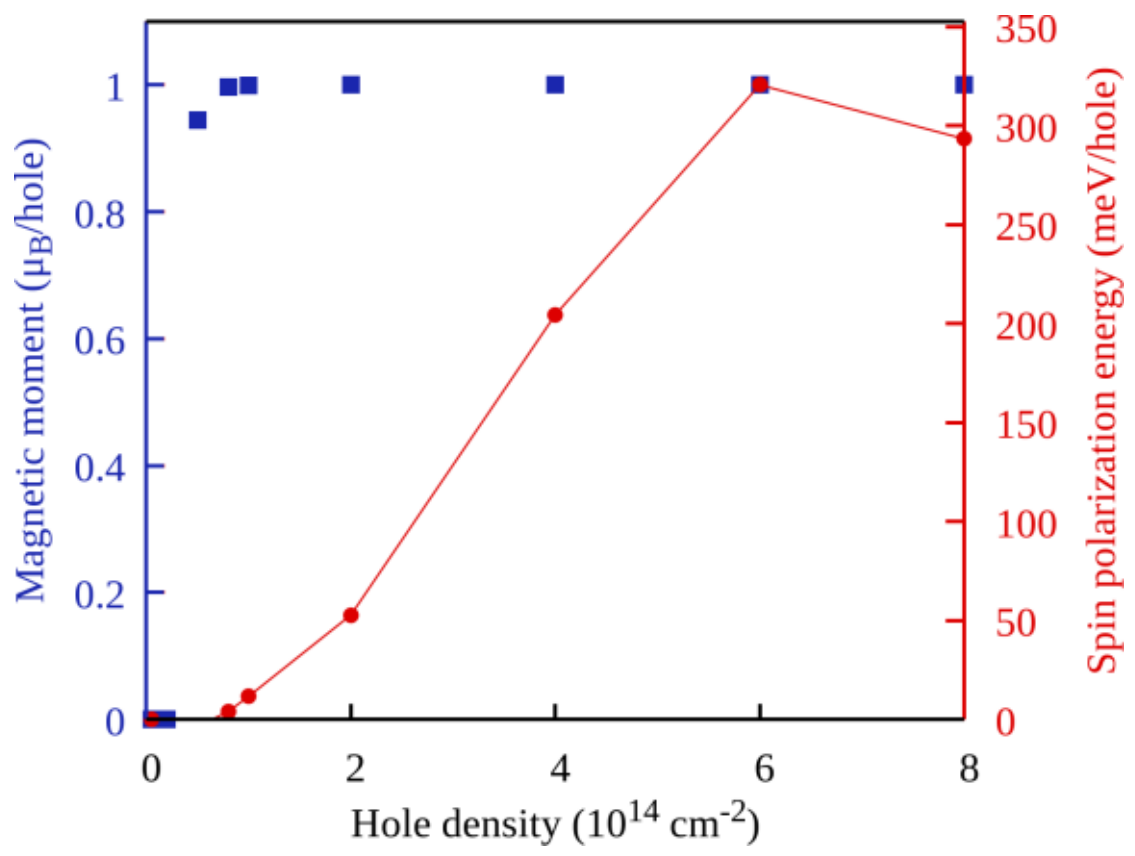

**Magnetic configurations and spin Hamiltonian**

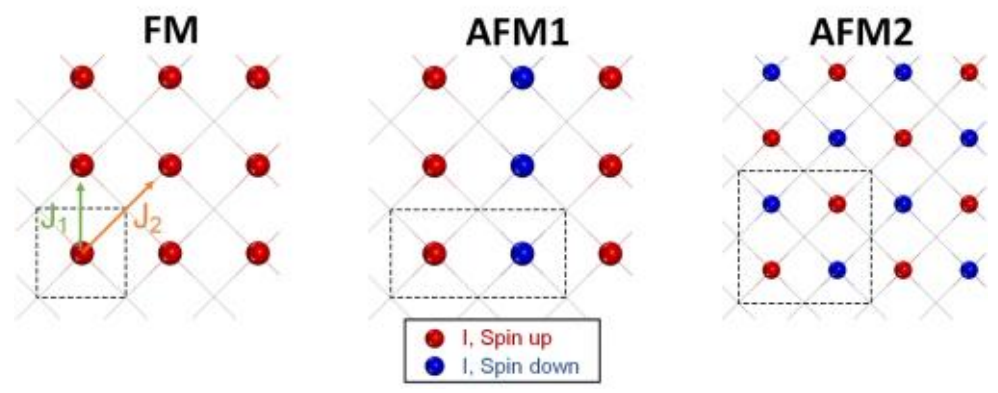

$$E_{FM} = E_0 - 2J_1S^2 - 2J_2S^2$$

$$E_{AFM1} = E_0 + 2J_2S^2$$

$$E_{AFM2} = E_0 + 2J_1S^2 - 2J_2S^2$$

**Magnetic exchange coupling parameters**

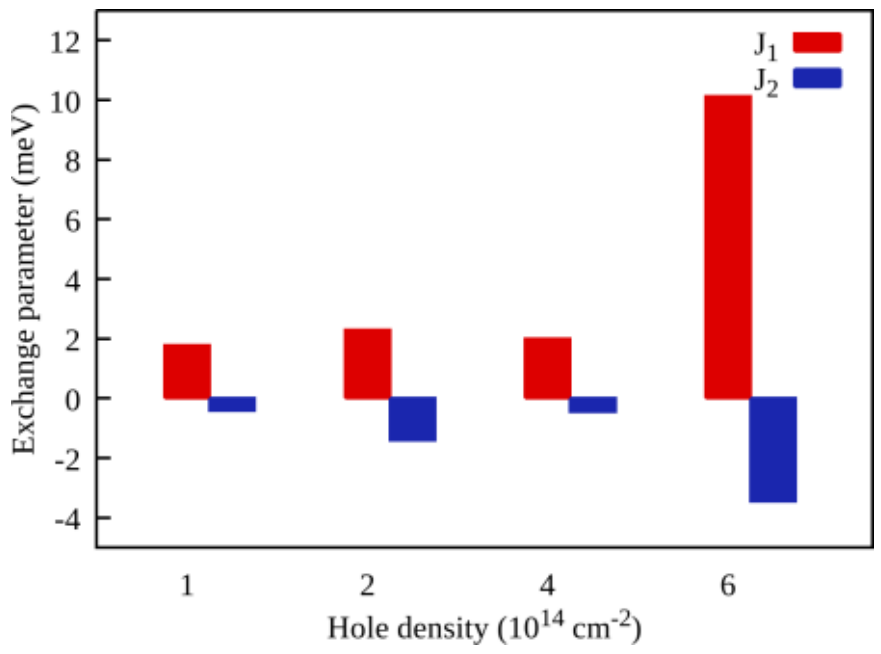

**Magnetic anisotropy energy (MAE, μeV) per magnetic atom**

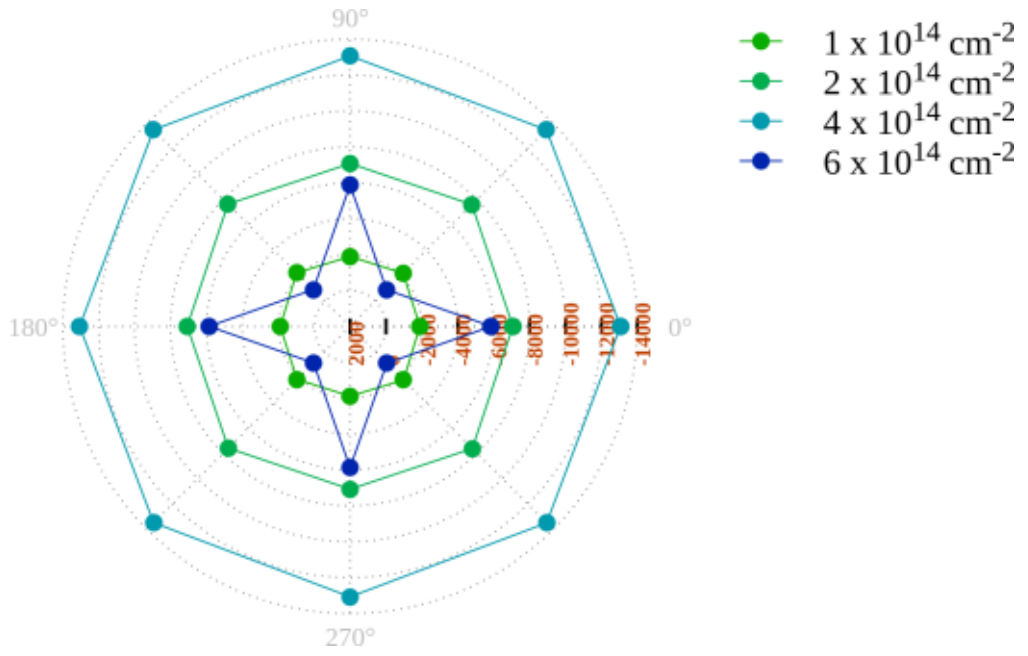

MAE = $E_{\parallel} - E_{\perp}$, a positive (negative) value of MAE indicates the off-plane (in-plane) easy axis.

**Monte Carlo simulations of the normalized magnetization of as a function of temperature**

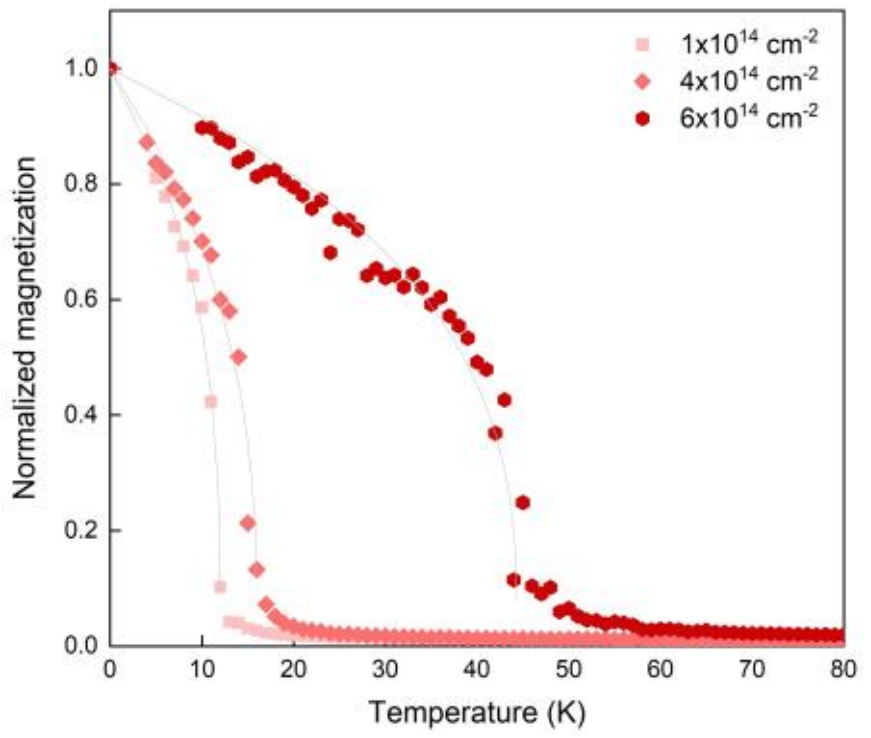

$T_c$: $1\times10^{14}$ cm$^{-2}$: 12 K    $2\times10^{14}$ cm$^{-2}$: - K

$4\times10^{14}$ cm$^{-2}$: 16 K    $6\times10^{14}$ cm$^{-2}$: 44 K

# 102. KBr

| MC2D-ID | C2DB | 2dmat-ID | USPEX | Space group | Band gap (eV) |
|---|---|---|---|---|---|
| - | - | 2dm-6264 | - | P4/mmm | 4.37 |

**Convex hull** | **Atomic structure** | **Atomic coordinates** | **Phonon dispersion curve**

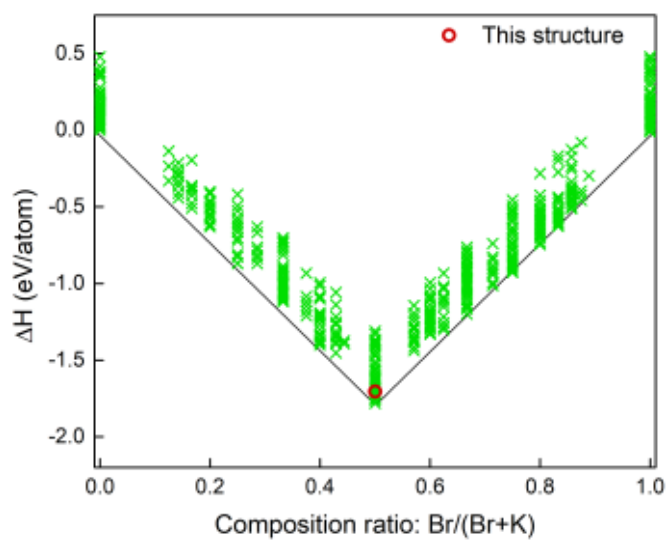

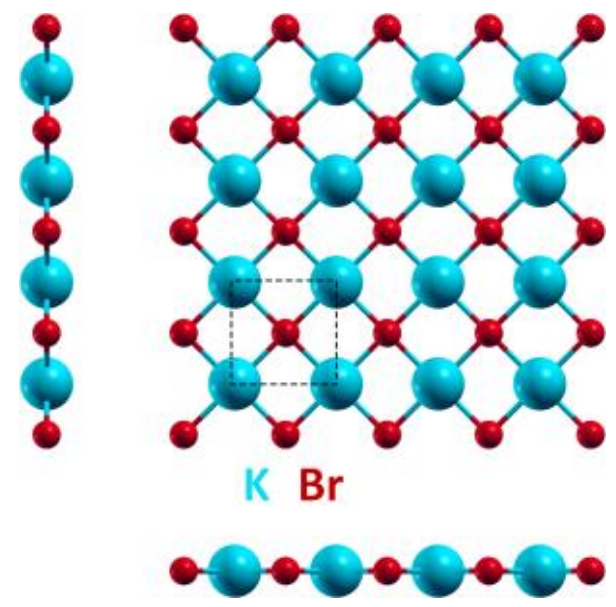

```
K1Br1
   1.00000000000000
   4.5696917733565918   0.0000000000000000   0.0000000000000000
   0.0000000000000000   4.5696918158573636   0.0000000000000000
   0.0000000000000000   0.0000000000000000  20.0000000000000000
   K    Br
   1    1
Direct
  0.4999999999747899  0.5000000000775557  0.4999999989178505
  0.0000000000252101  0.9999999999224443  0.5000000010821495
```

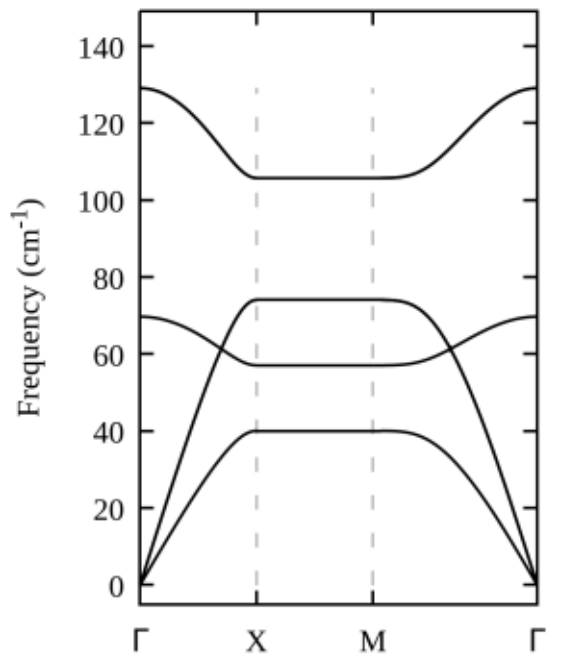

## Projected band structure and density of states

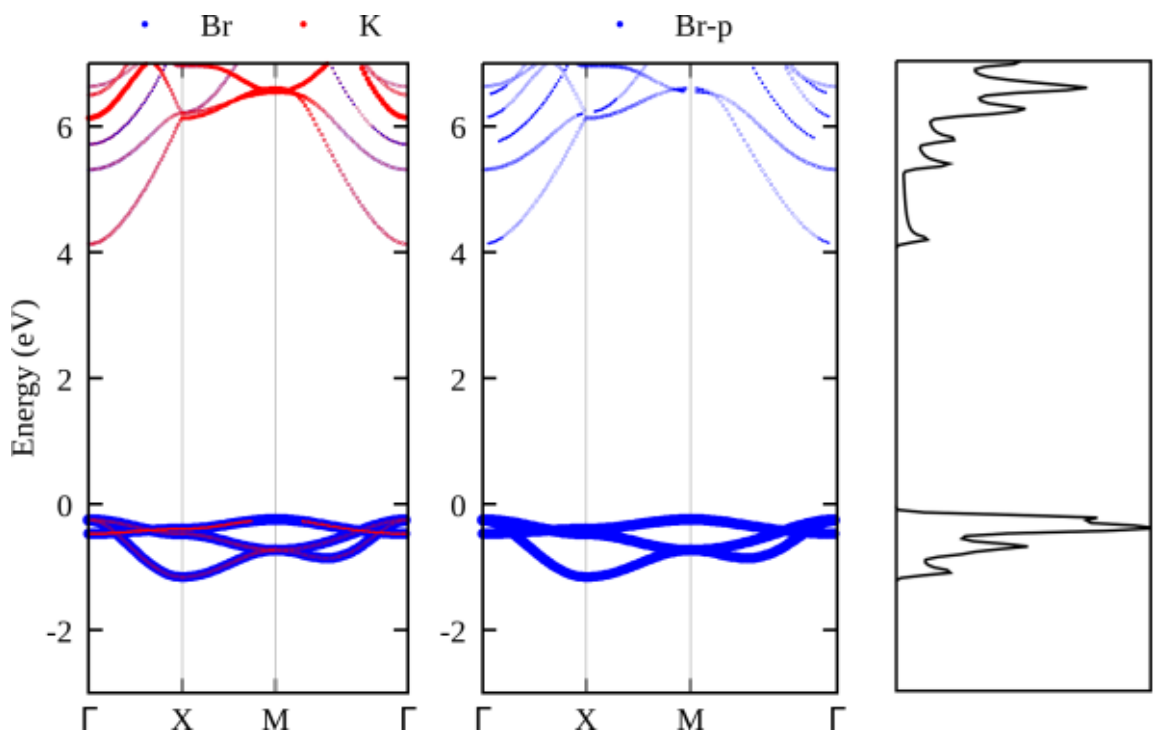

## Magnetic moment and spin polarization energy as a function of hole doping concentration

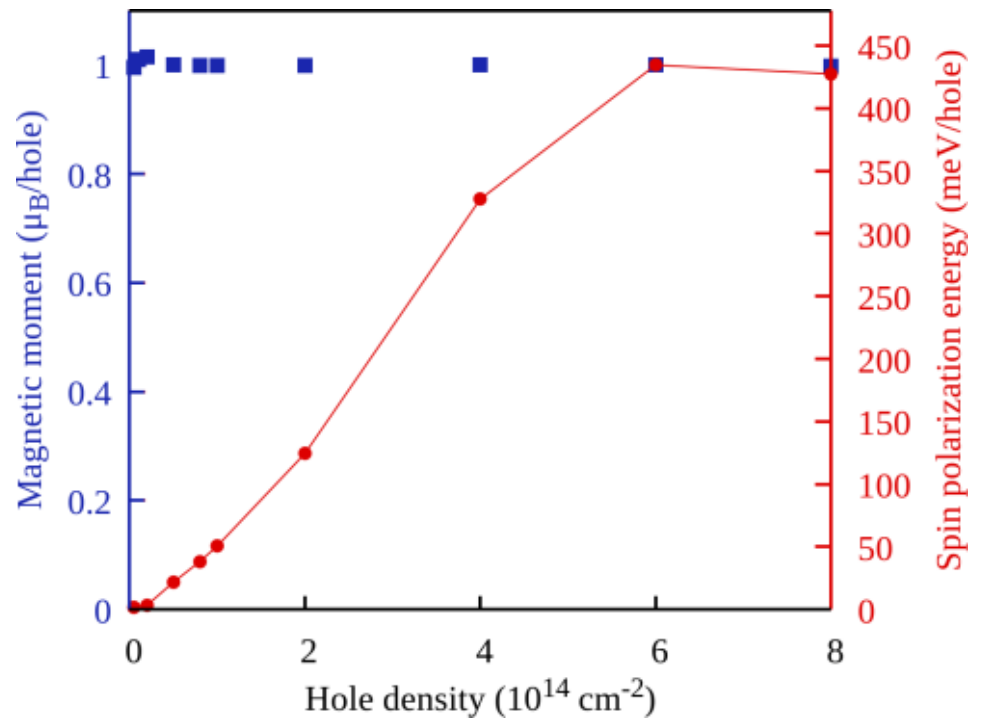

## Magnetic configurations and spin Hamiltonian

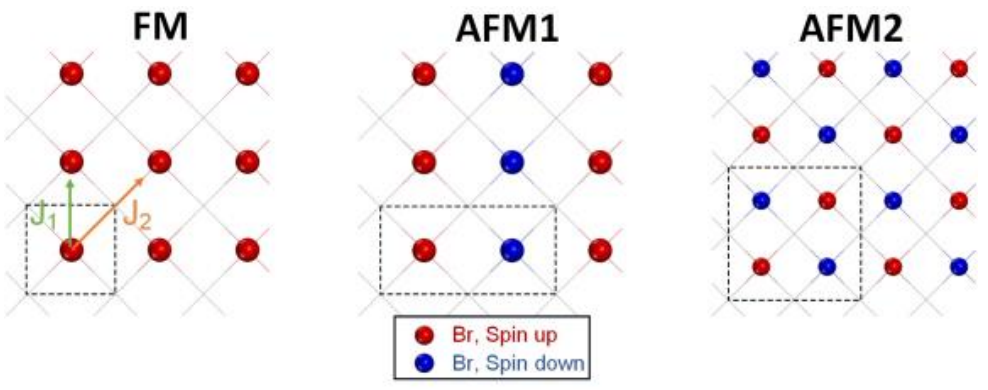

$$E_{FM} = E_0 - 2J_1S^2 - 2J_2S^2$$

$$E_{AFM1} = E_0 + 2J_2S^2$$

$$E_{AFM2} = E_0 + 2J_1S^2 - 2J_2S^2$$

## Magnetic exchange coupling parameters

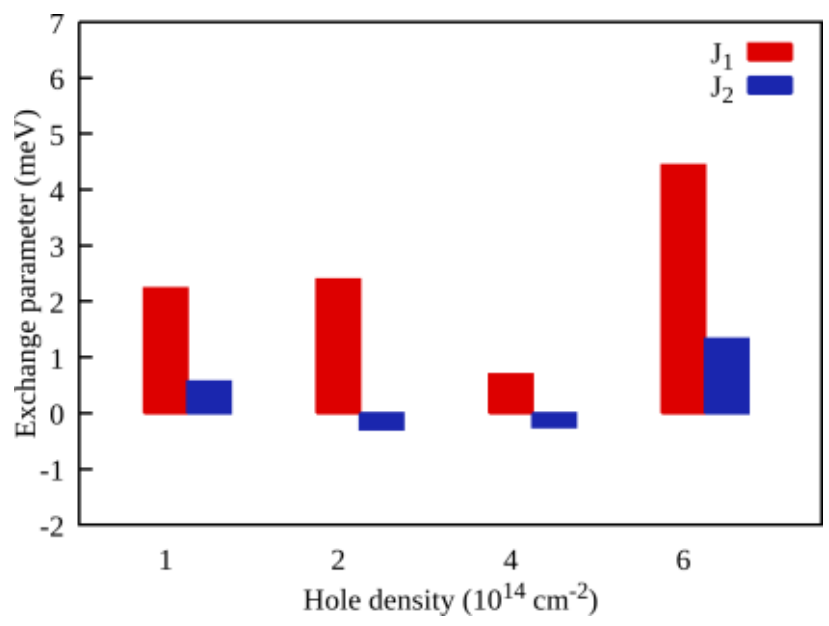

## Magnetic anisotropy energy (MAE, μeV) per magnetic atom

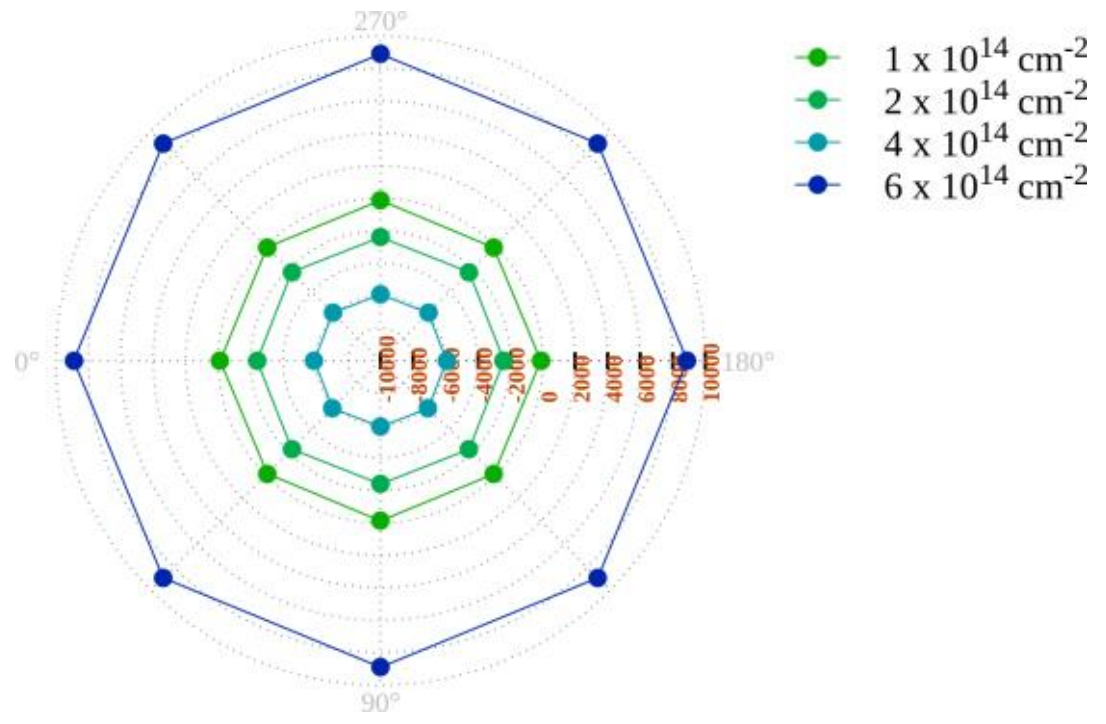

MAE = $E_{\parallel} - E_{\perp}$, a positive (negative) value of MAE indicates the off-plane (in-plane) easy axis.

## Monte Carlo simulations of the normalized magnetization of as a function of temperature

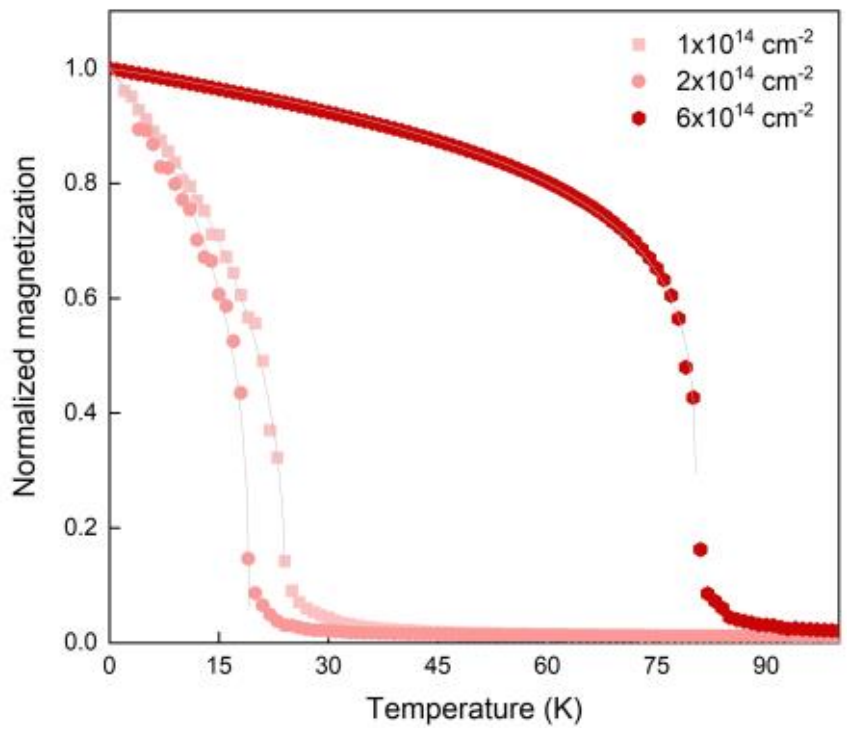

$T_c$: $1\times10^{14}$ cm$^{-2}$: 24 K  $2\times10^{14}$ cm$^{-2}$: 19 K  $4\times10^{14}$ cm$^{-2}$: - K  $6\times10^{14}$ cm$^{-2}$: 80 K

# 103. $SrO_2$

| MC2D-ID | C2DB | 2dmat-ID | USPEX | Space group | Band gap (eV) |
|---|---|---|---|---|---|
| - | - | 2dm-11 | - | P4/mmm | 2.94 |

**Convex hull** | **Atomic structure** | **Atomic coordinates** | **Phonon dispersion curve**

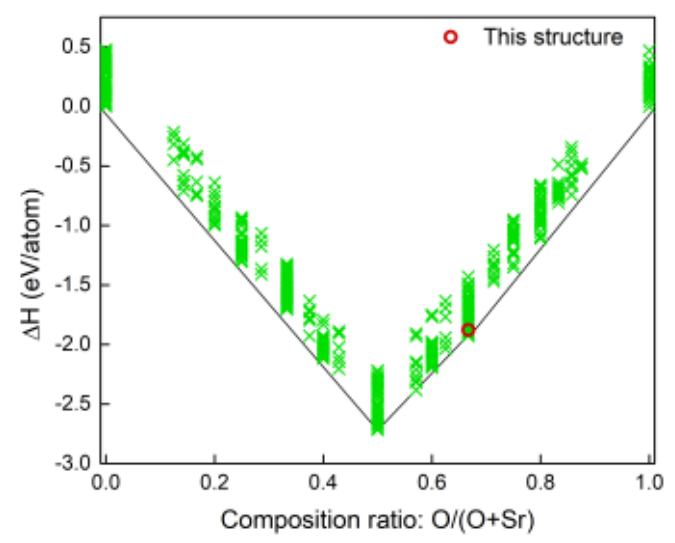

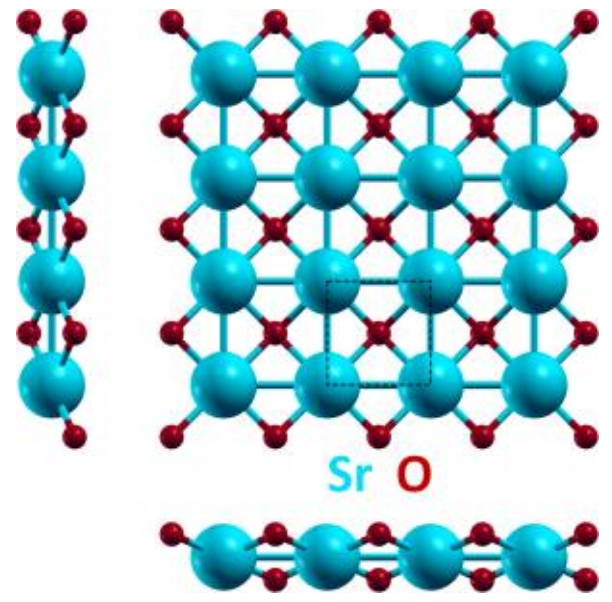

```
Sr1O2
   1.00000000000000
   3.4794399348957756    0.0000000000000000    0.0000000000000000
   0.0000000000000000    3.4794400036870723    0.0000000000000000
   0.0000000000000000    0.0000000000000000    21.5194009999999984
   Sr    O
   1     2
Direct
  0.4999999999049933  0.4999999998392468  0.0000000076882714
  0.9999999977551539  0.9999999955740932  0.9643033953589608
  0.0000000023398599  0.0000000045866528  0.0356965969527749
```

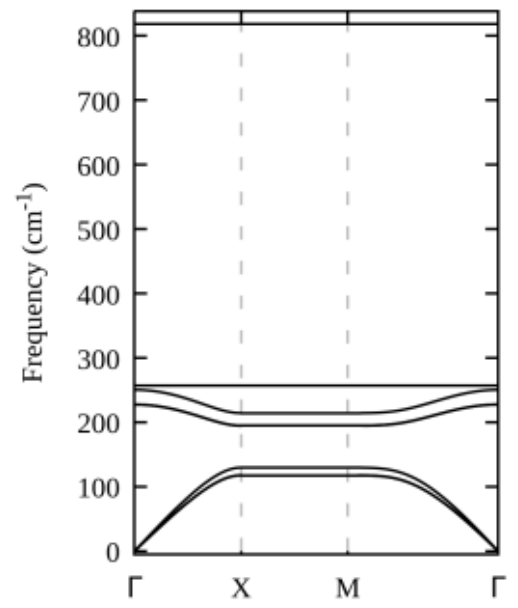

**Projected band structure and density of states** | **Magnetic moment and spin polarization energy as a function of hole doping concentration**

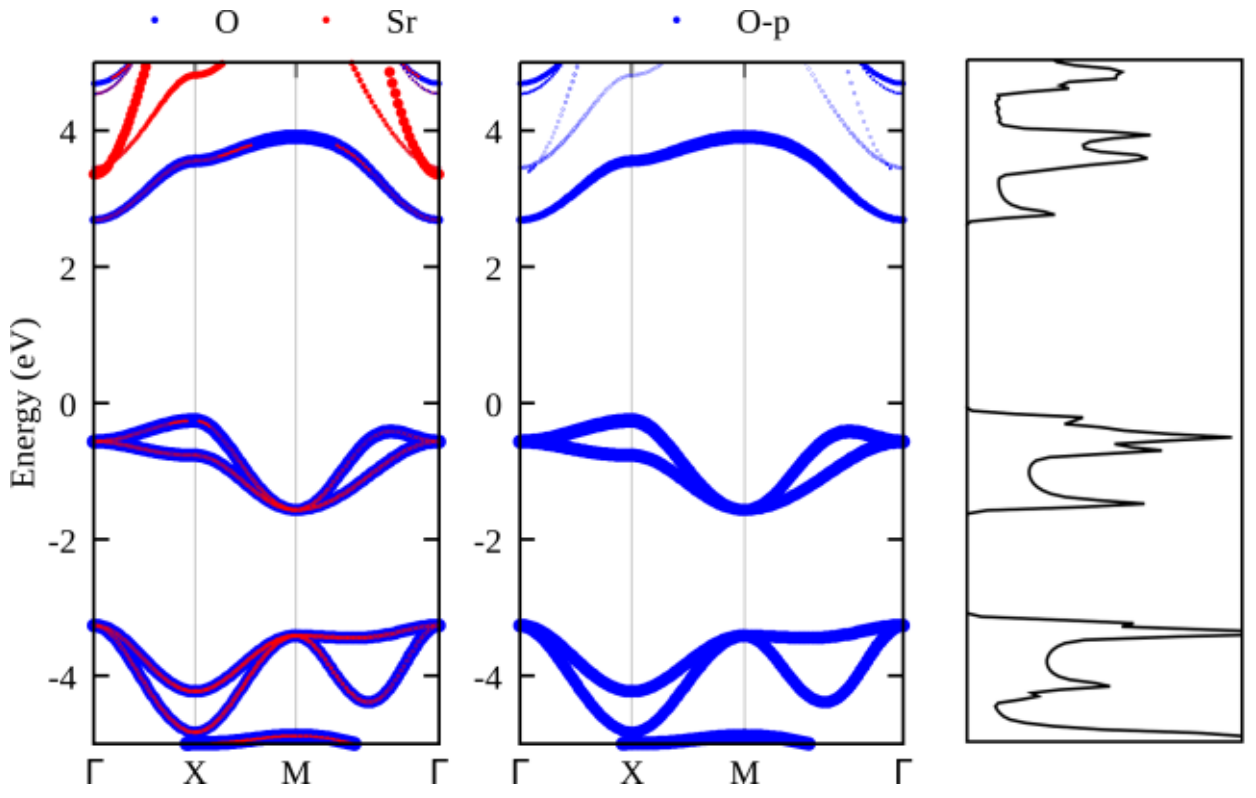

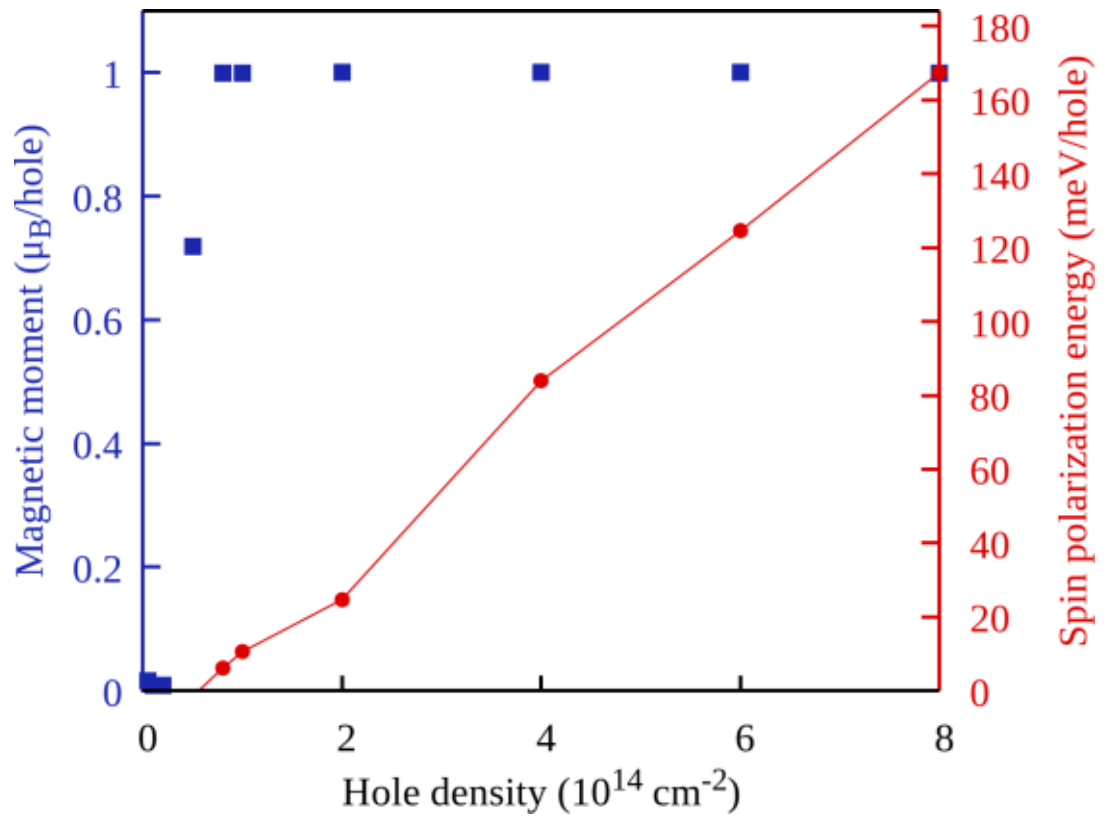

**Magnetic configurations and spin Hamiltonian** | **Magnetic exchange coupling parameters**

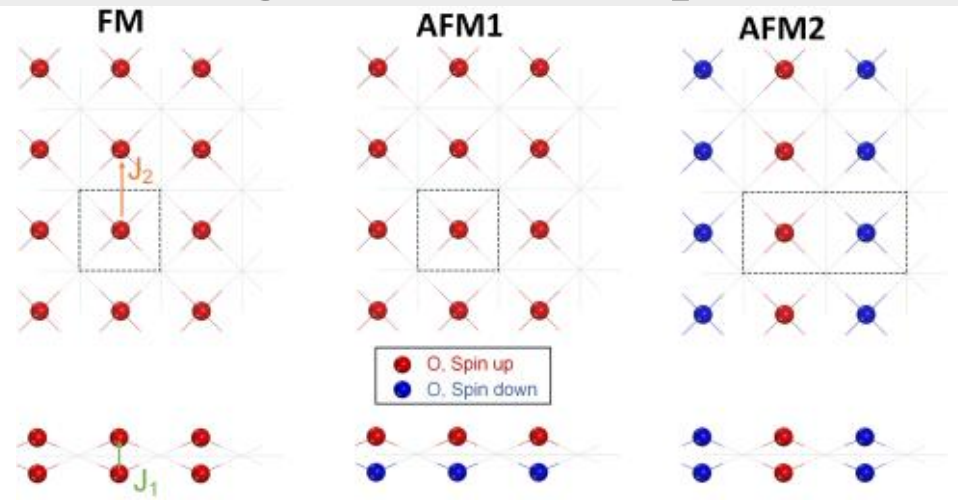

$$E_{FM} = E_0 - J_1S^2 - 4J_2S^2$$
$$E_{AFM1} = E_0 + J_1S^2 - 4J_2S^2$$
$$E_{AFM2} = E_0 - J_1S^2$$

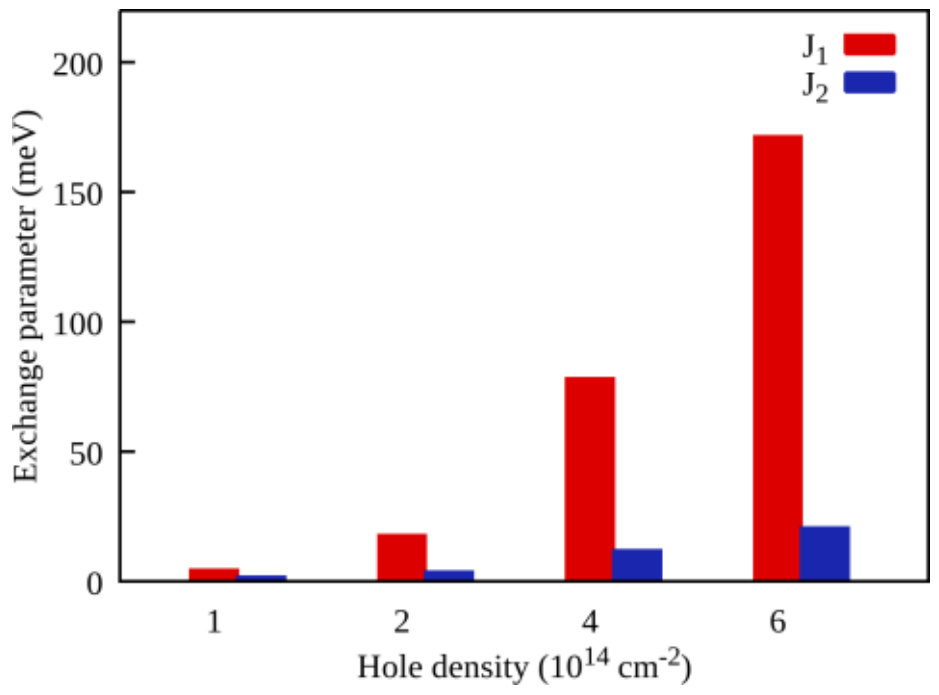

**Magnetic anisotropy energy (MAE, μeV) per magnetic atom** | **Monte Carlo simulations of the normalized magnetization of as a function of temperature**

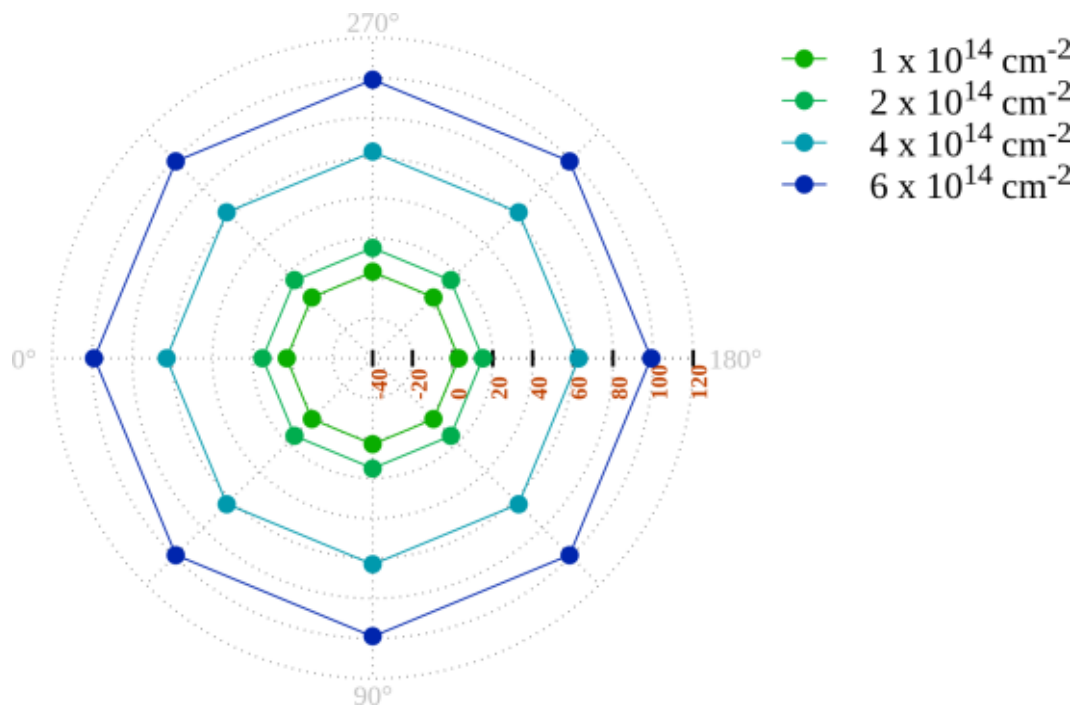

MAE = $E_{\parallel} - E_{\perp}$, a positive (negative) value of MAE indicates the off-plane (in-plane) easy axis.

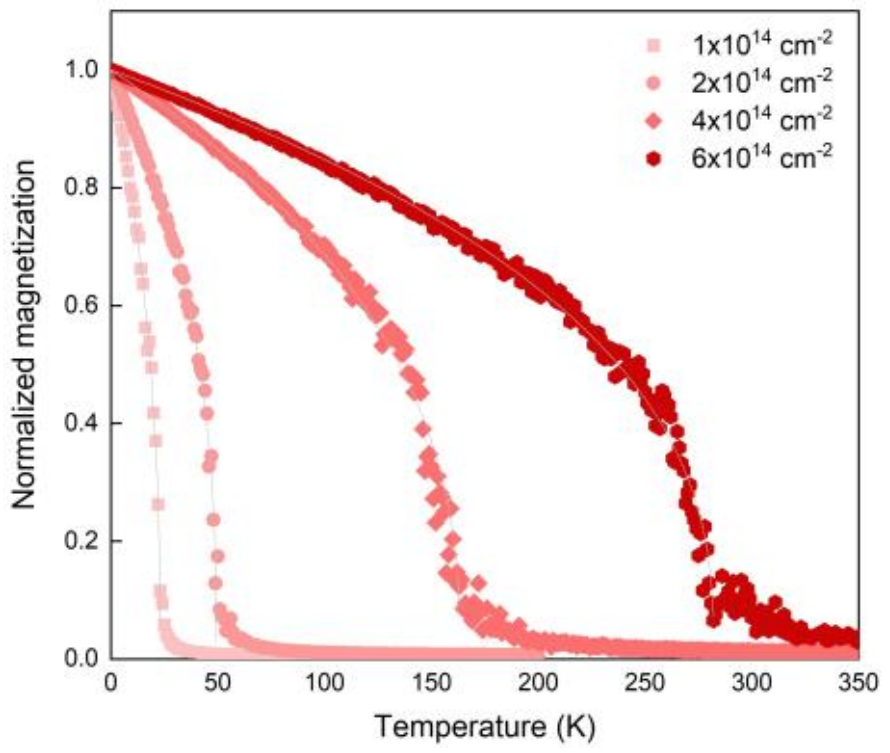

$T_c$: $1\times10^{14}$ cm$^{-2}$: 23 K   $2\times10^{14}$ cm$^{-2}$: 49 K

$4\times10^{14}$ cm$^{-2}$: 163 K   $6\times10^{14}$ cm$^{-2}$: 282 K

# 104. $Cd_2S_2$

| MC2D-ID | C2DB | 2dmat-ID | USPEX | Space group | Band gap (eV) |
|---|---|---|---|---|---|
| - | - | 2dm-2392 | - | P4/nmm | 2.26 |

**Convex hull**

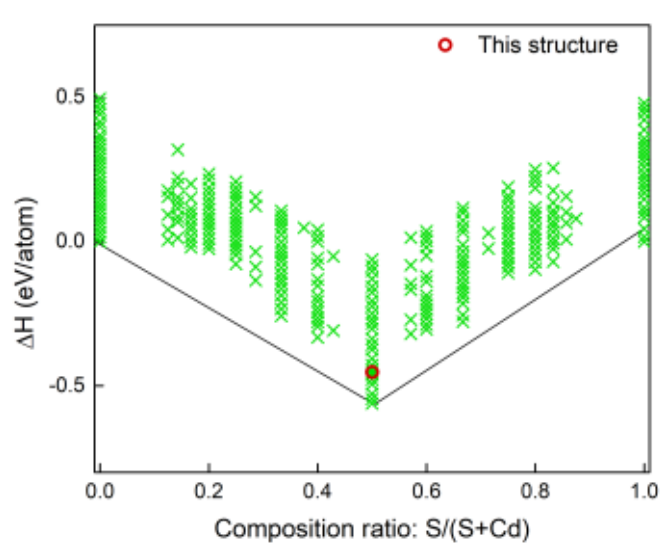

**Atomic structure**

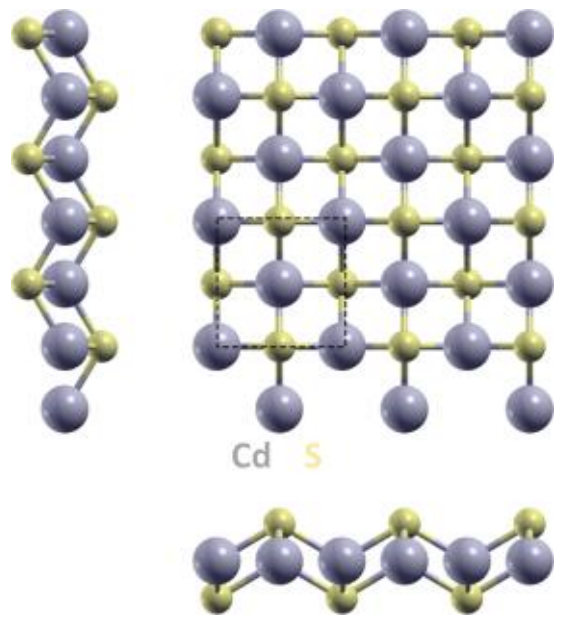

**Atomic coordinates**

Cd2S2
1.00000000000000
4.5311467153995970 0.0000000000000000 0.0000000000000000
0.0000000000000000 4.5311449053809385 0.0000000000000000
0.0000000000000000 0.0000000000000000 22.9900120000000001
Cd S
2 2
Direct
0.5000000090778087 0.4999999708515048 0.0000000259332467
0.9999999888466249 0.0000000341068613 0.9999999734893308
0.9999999556635117 0.5000000067003185 0.0583492951533486
0.5000000464120546 0.9999999883413224 0.9416507054240739

**Phonon dispersion curve**

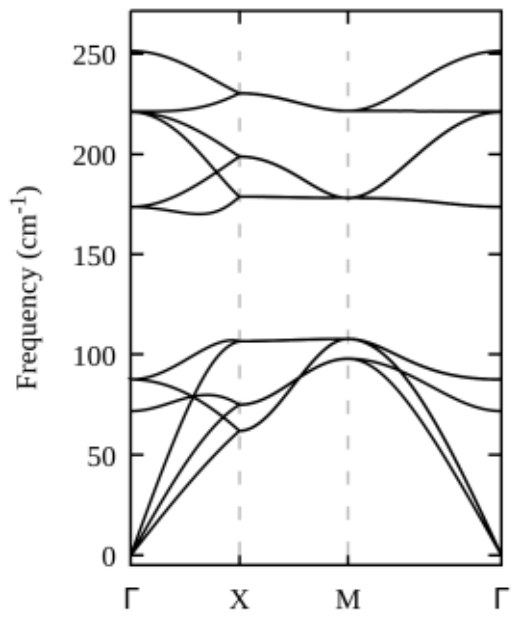

**Projected band structure and density of states**

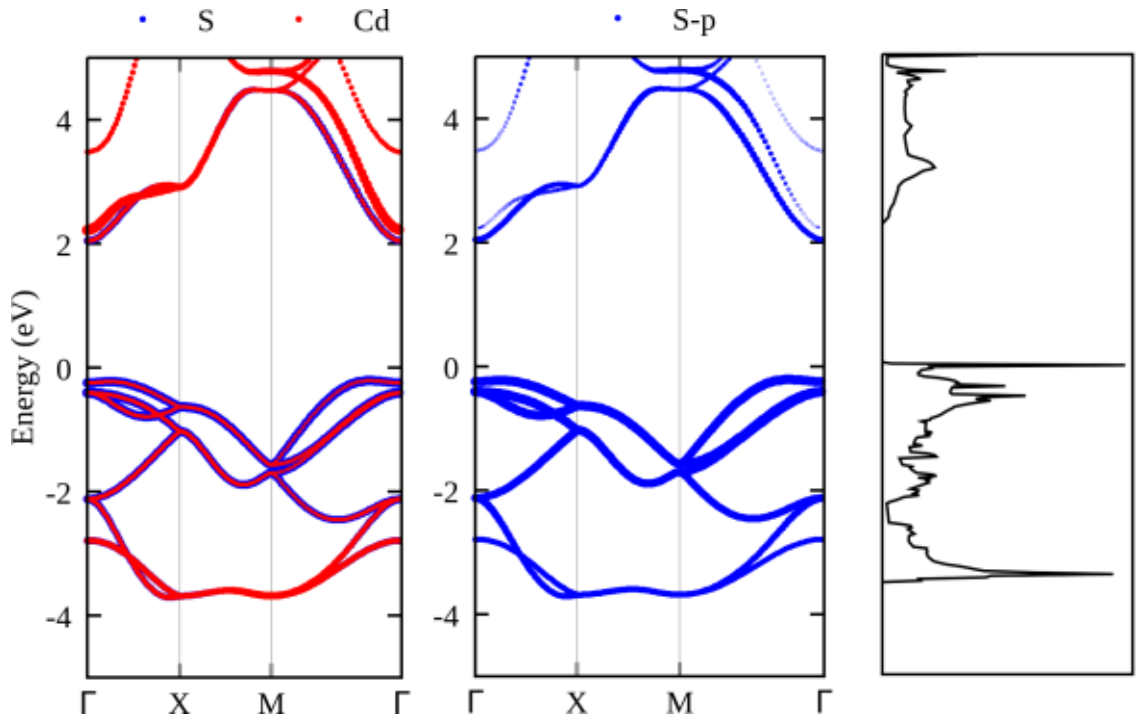

**Magnetic moment and spin polarization energy as a function of hole doping concentration**

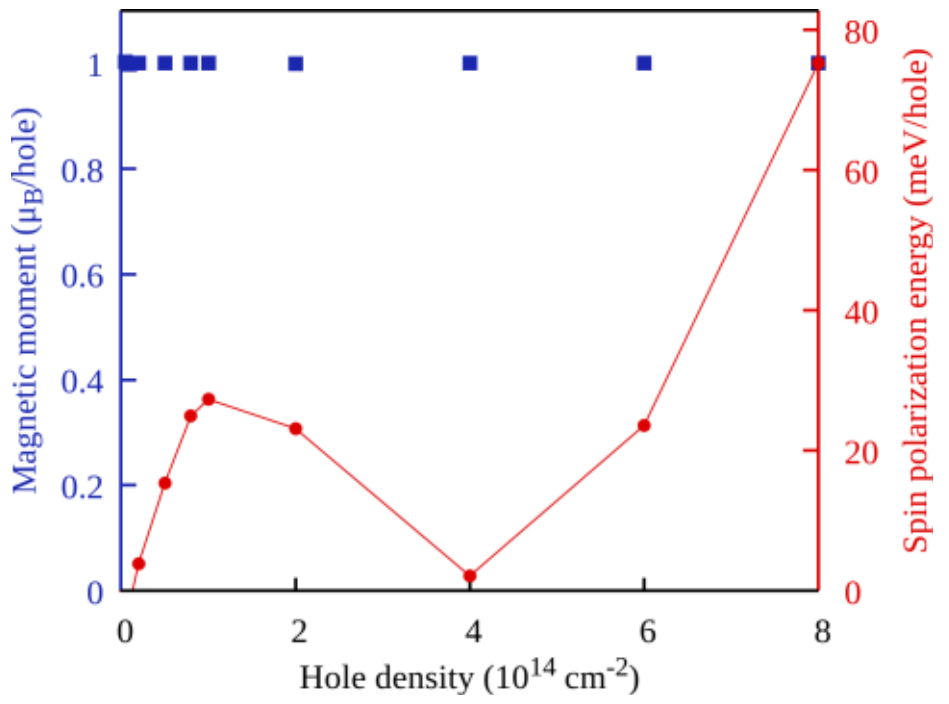

**Magnetic configurations and spin Hamiltonian**

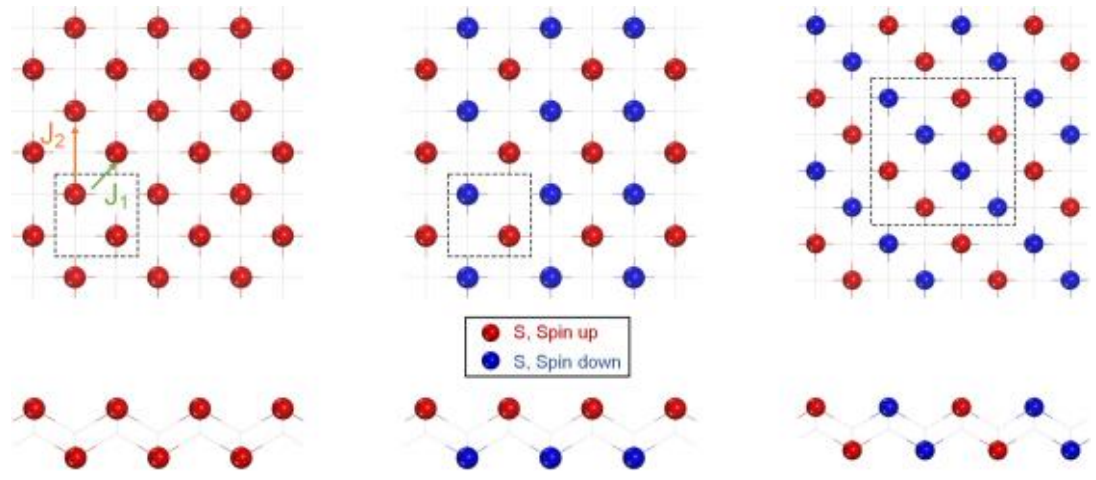

$$E_{FM} = E_0 - 4J_1S^2 - 4J_2S^2$$
$$E_{AFM1} = E_0 + 4J_1S^2 - 4J_2S^2$$
$$E_{AFM2} = E_0 + 4J_2S^2$$

**Magnetic exchange coupling parameters**

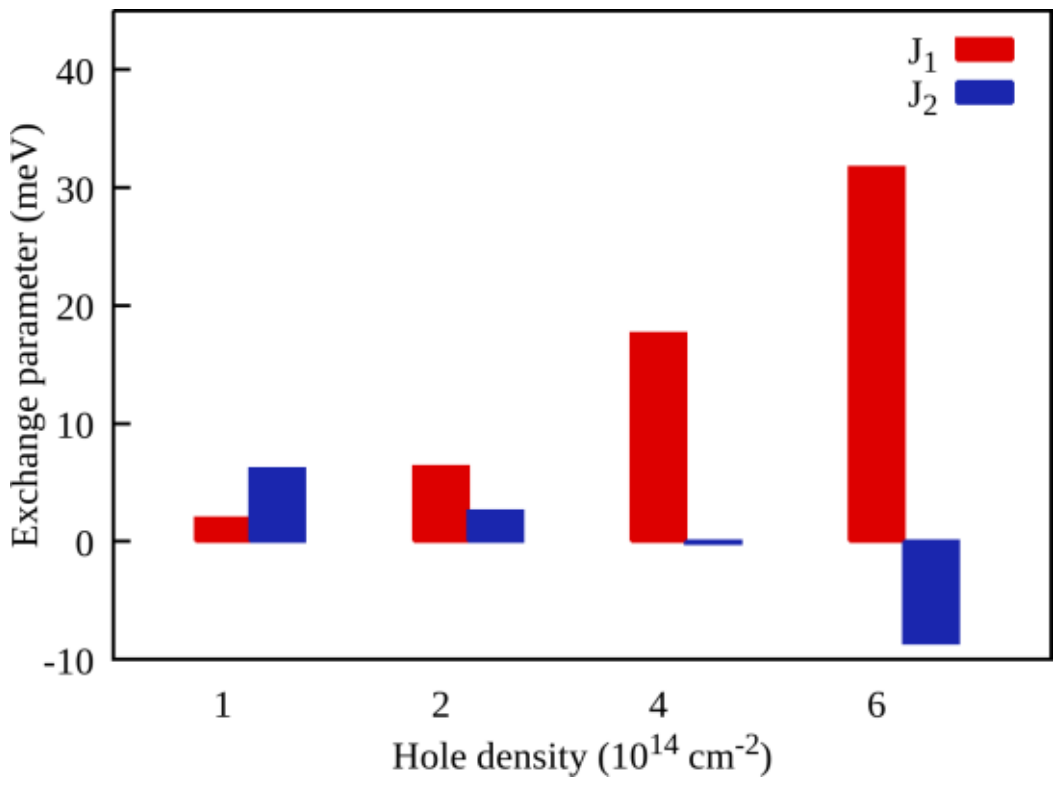

**Magnetic anisotropy energy (MAE, μeV) per magnetic atom**

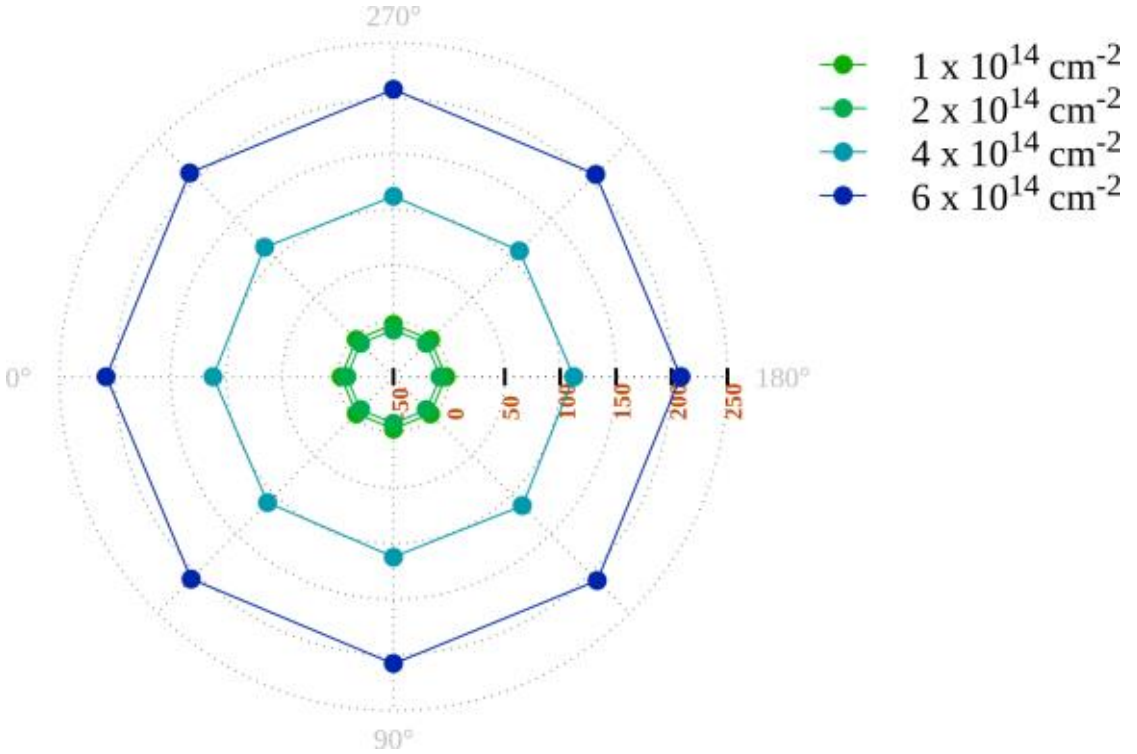

MAE = $E_{\parallel} - E_{\perp}$, a positive (negative) value of MAE indicates the off-plane (in-plane) easy axis.

**Monte Carlo simulations of the normalized magnetization of as a function of temperature**

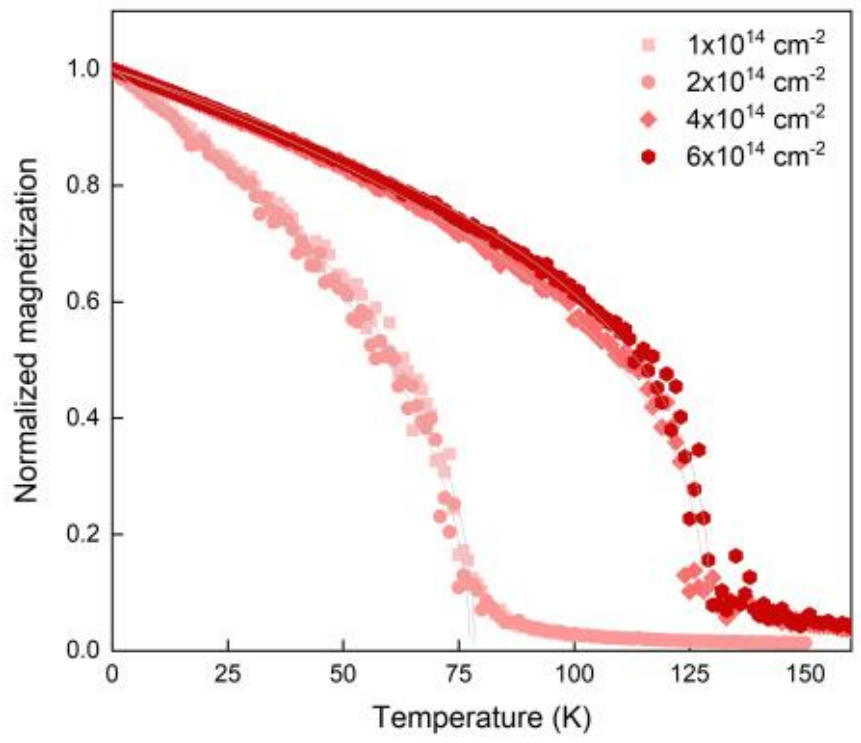

$T_c$: $1\times10^{14}$ cm$^{-2}$: 78 K  $2\times10^{14}$ cm$^{-2}$: 77 K

$4\times10^{14}$ cm$^{-2}$: 128 K  $6\times10^{14}$ cm$^{-2}$: 129 K

# 105. $Ge_2O_2$

| MC2D-ID | C2DB | 2dmat-ID | USPEX | Space group | Band gap (eV) |
|---|---|---|---|---|---|
| - | - | 2dm-831 | - | P4/nmm | 2.13 |

**Convex hull** **Atomic structure** **Atomic coordinates** **Phonon dispersion curve**

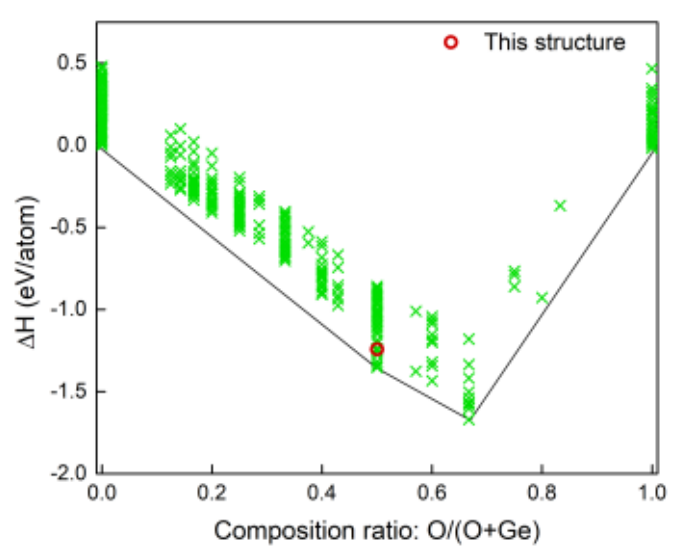

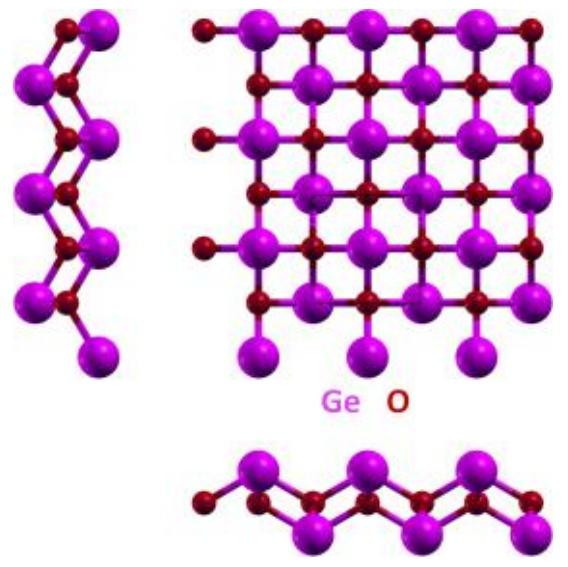

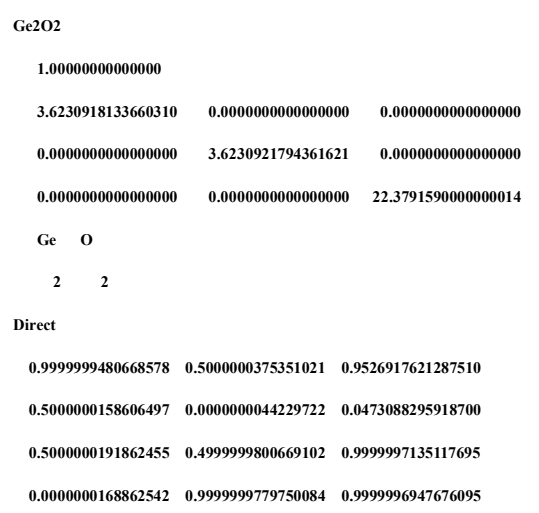

```
Ge2O2
   1.00000000000000
     3.6230918133660310    0.0000000000000000    0.0000000000000000
     0.0000000000000000    3.6230921794361621    0.0000000000000000
     0.0000000000000000    0.0000000000000000   22.3791590000000014
   Ge   O
     2     2
Direct
  0.9999999480668578  0.5000000375351021  0.9526917621287510
  0.5000000158606497  0.0000000044229722  0.0473088295918700
  0.5000000191862455  0.4999999800669102  0.9999997135117695
  0.0000000168862542  0.9999999779750084  0.9999996947676095
```

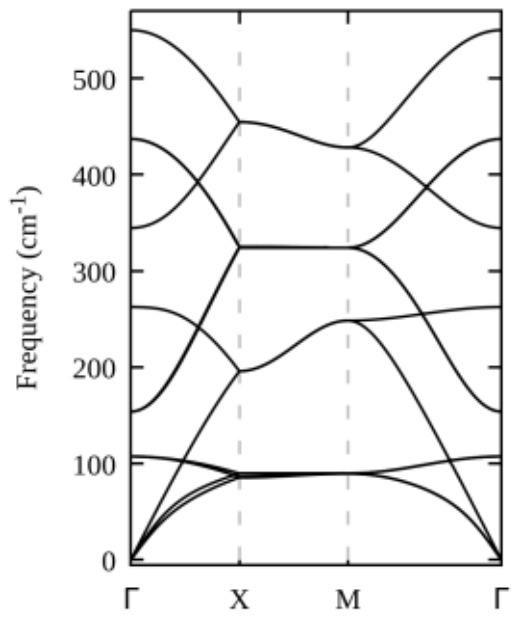

**Projected band structure and density of states** **Magnetic moment and spin polarization energy as a function of hole doping concentration**

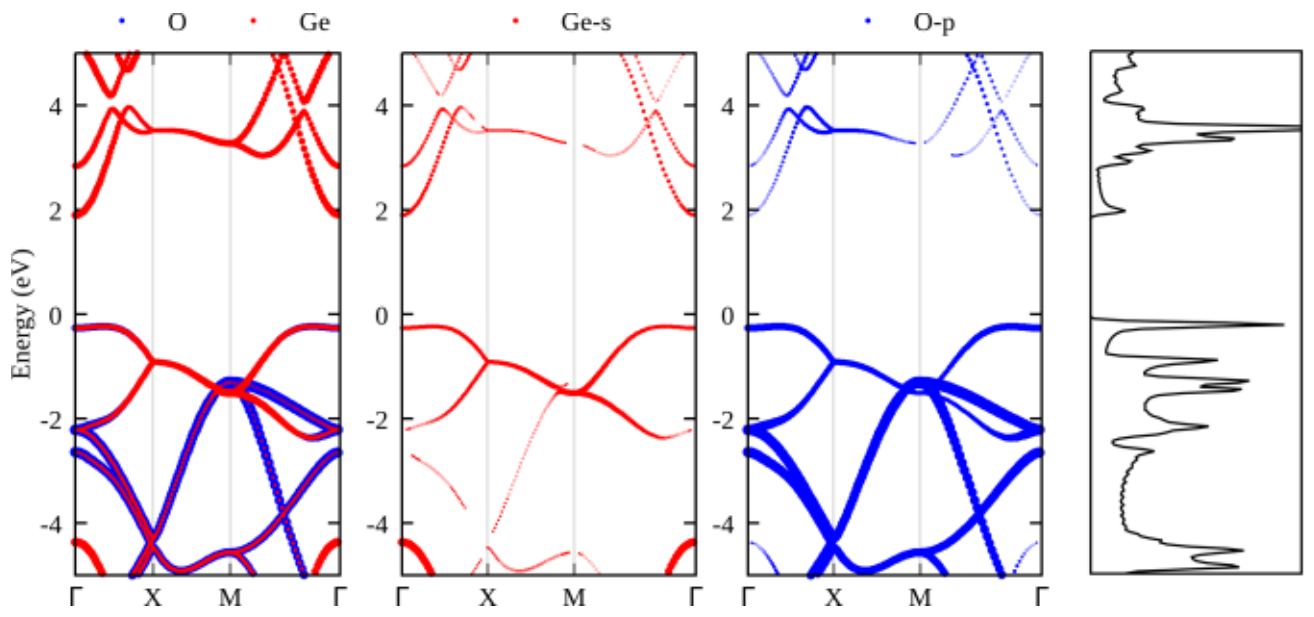

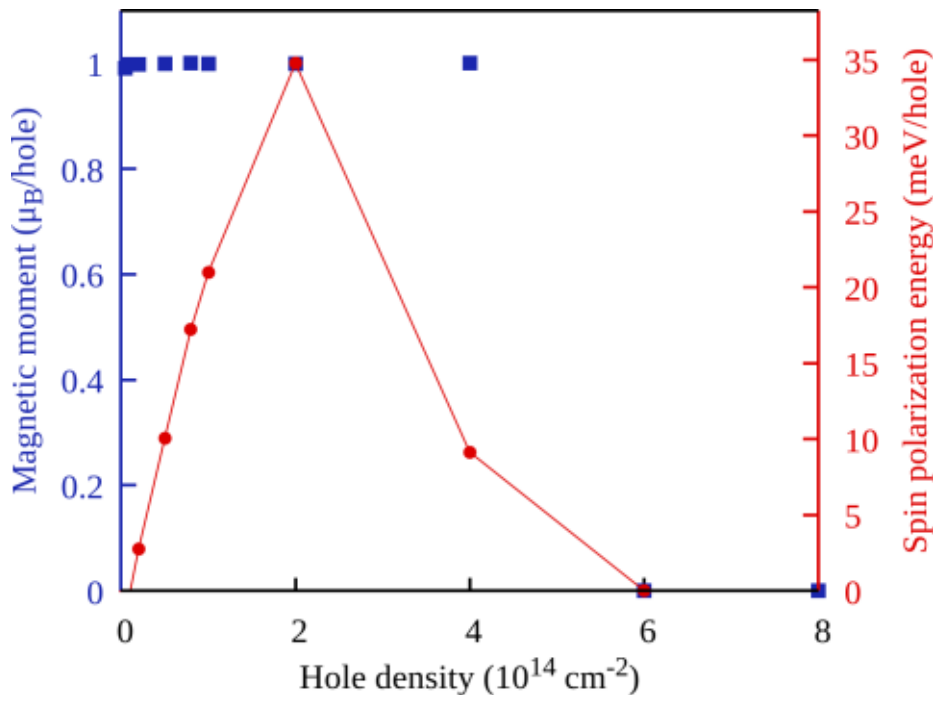

**Magnetic configurations and spin Hamiltonian** **Magnetic exchange coupling parameters**

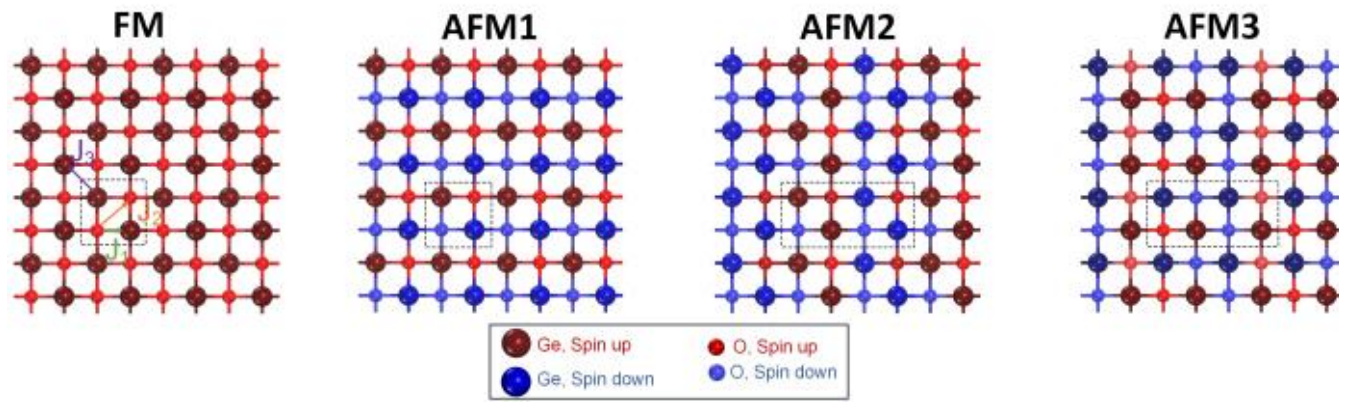

$$E_{FM} = E_0 - 8J_1S^2 - 4J_2S^2 - 4J_3S^2$$
$$E_{AFM1} = E_0 + 4J_2S^2 + 4J_3S^2$$
$$E_{AFM2} = E_0 + 4J_2S^2$$
$$E_{AFM3} = E_0 + 2J_1S^2 + 4J_3S^2$$

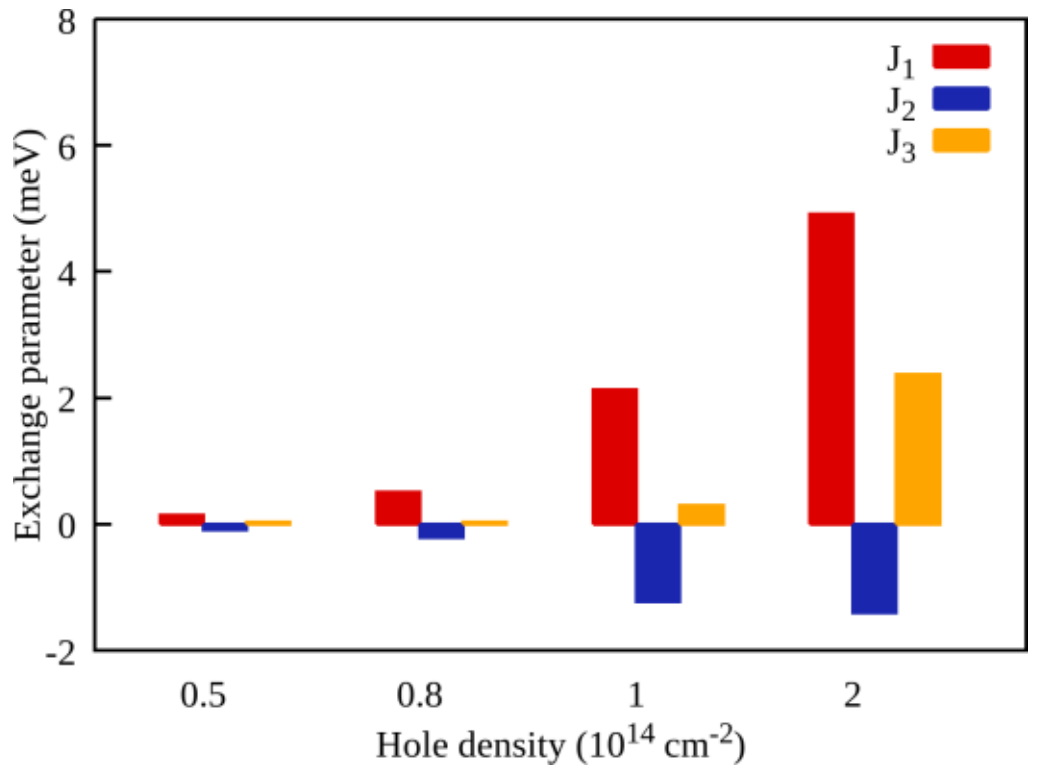

**Magnetic anisotropy energy (MAE, μeV) per magnetic atom** **Monte Carlo simulations of the normalized magnetization of as a function of temperature**

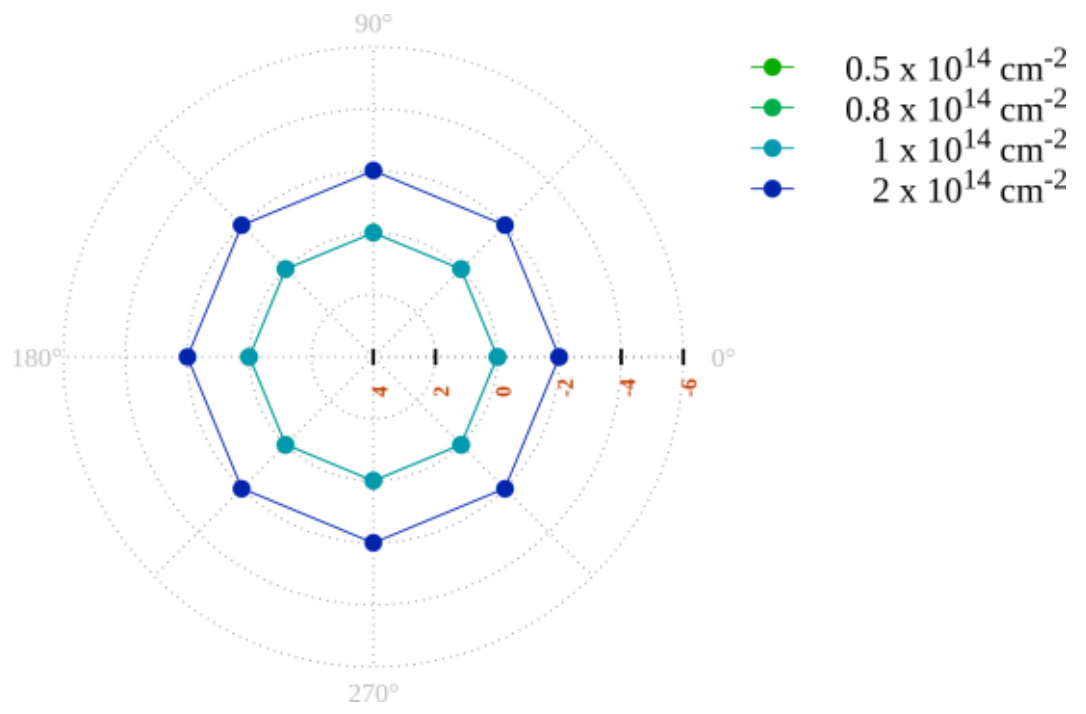

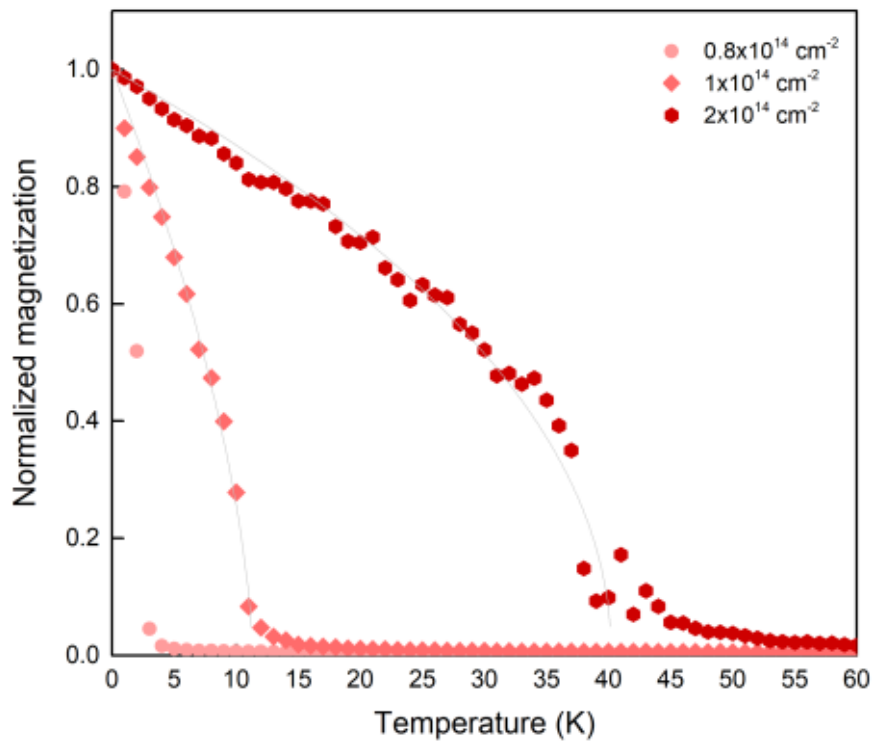

MAE = $E_\parallel - E_\perp$, a positive (negative) value of MAE indicates the off-plane (in-plane) easy axis.

$T_c$: $0.5\times10^{14}$ cm$^{-2}$: - K  $0.8\times10^{14}$ cm$^{-2}$: 3 K  $1\times10^{14}$ cm$^{-2}$: 11 K  $2\times10^{14}$ cm$^{-2}$: 40 K

# 106. $Sn_2O_2$

| MC2D-ID | C2DB | 2dmat-ID | USPEX | Space group | Band gap (eV) |
|---|---|---|---|---|---|
| 182 | - | 2dm-3629 | - | P4/nmm | 3.01 |

**Convex hull** | **Atomic structure** | **Atomic coordinates** | **Phonon dispersion curve**

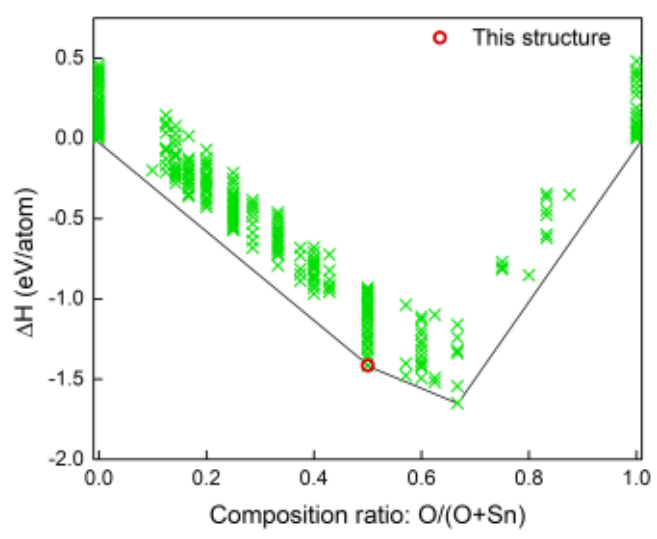

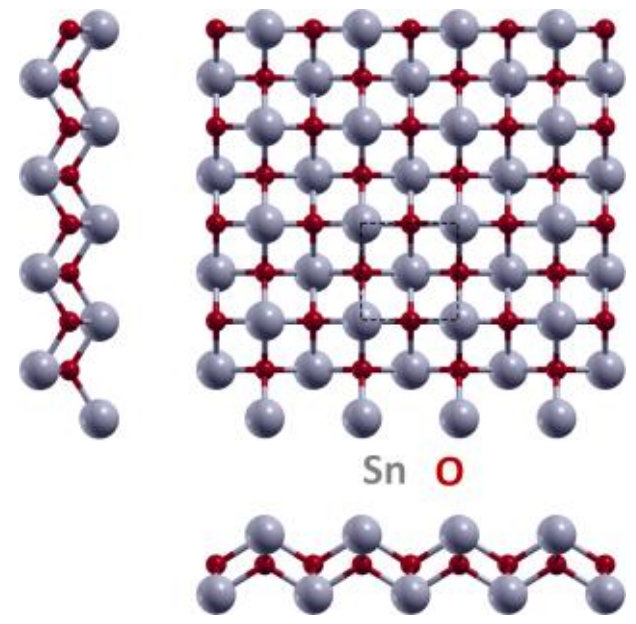

```
Sn2O2
   1.00000000000000
   3.8480553455705597   0.0000000000000000   0.0000000000000000
   0.0000000000000000   3.8480551394167248   0.0000000000000000
   0.0000000000000000   0.0000000000000000  22.3242170000000009
   Sn   O
   2    2
Direct
  0.0000000026401210  0.4999999951072667  0.4471324917123454
  0.4999999976294731  0.0000000023310065  0.5528675071911024
  0.0000000048212740  0.9999999976863023  0.4999999994032223
  0.4999999949091318  0.5000000048754174  0.5000000016933299
```

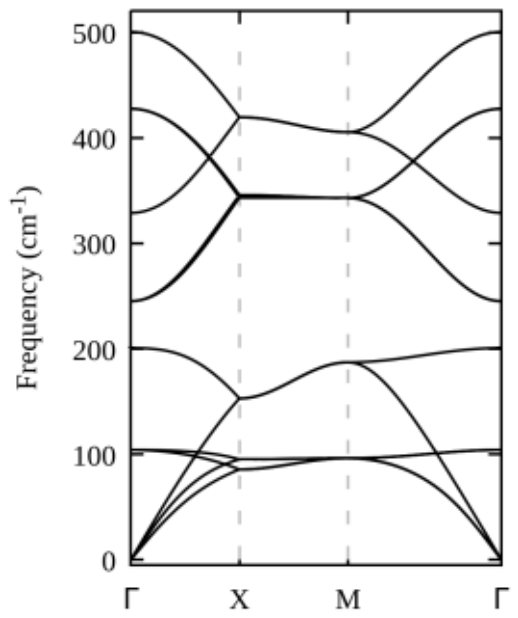

**Projected band structure and density of states** | **Magnetic moment and spin polarization energy as a function of hole doping concentration**

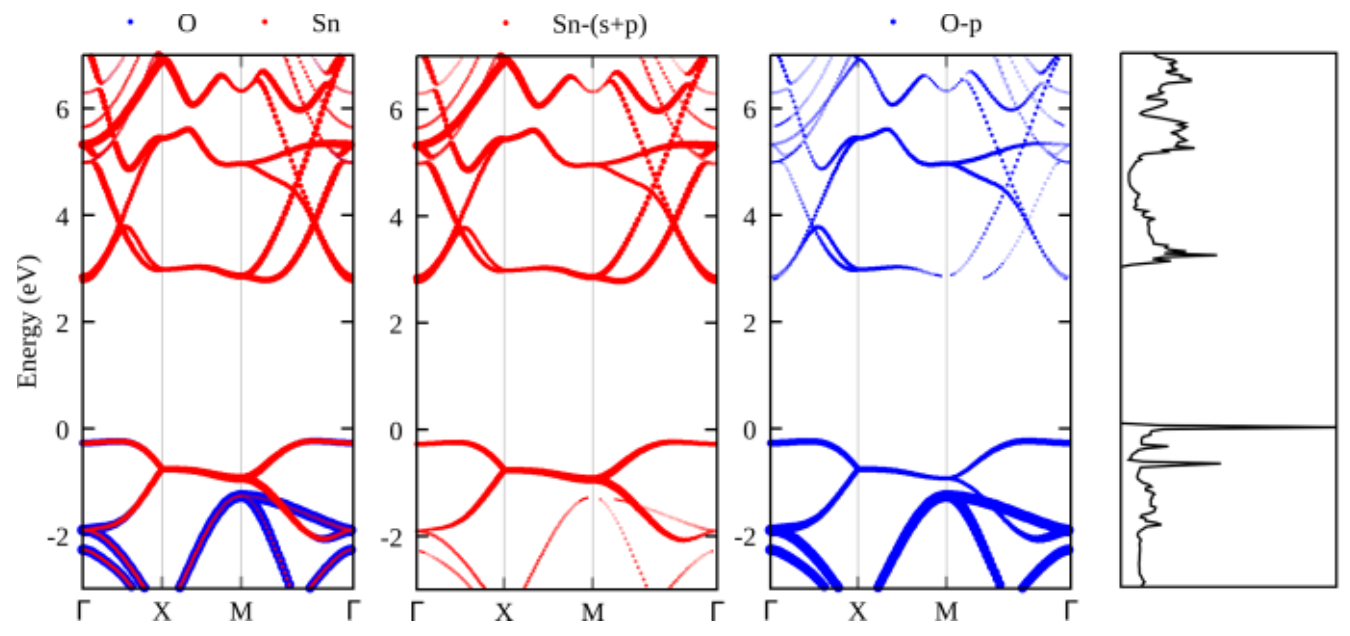

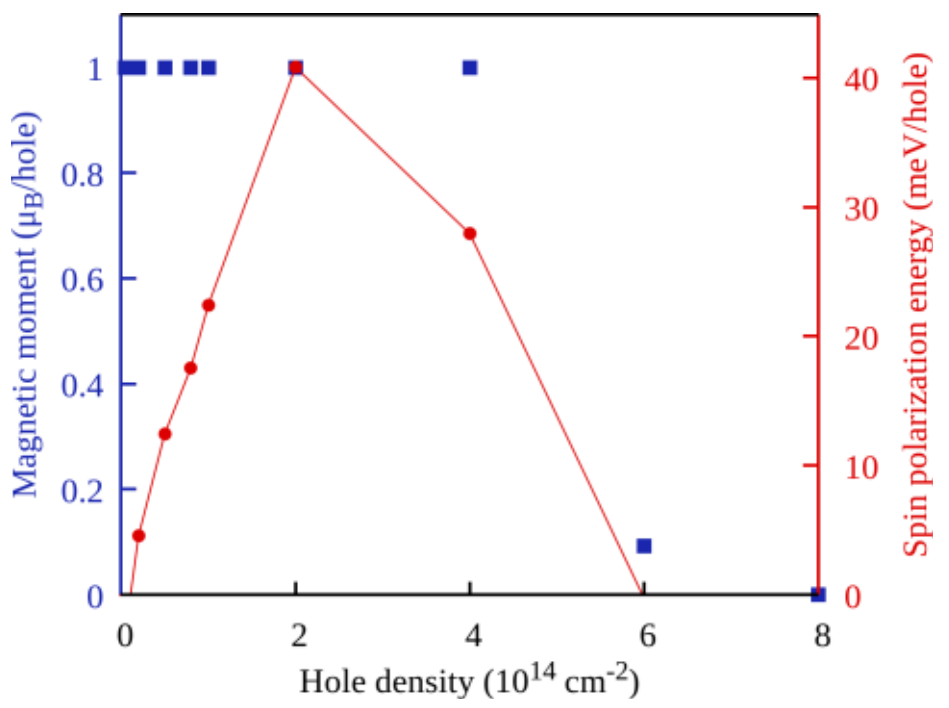

**Magnetic configurations and spin Hamiltonian** | **Magnetic exchange coupling parameters**

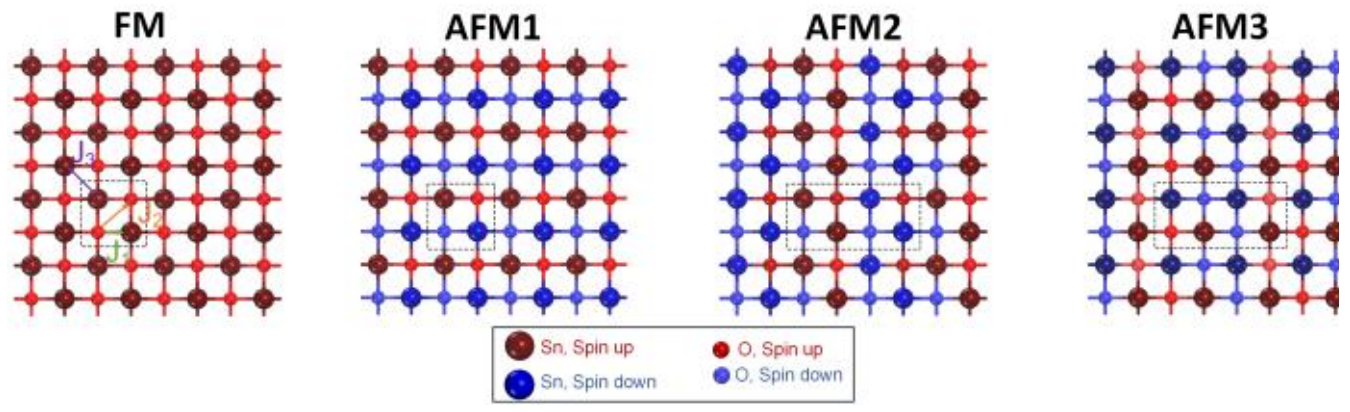

$$E_{FM} = E_0 - 8J_1S^2 - 4J_2S^2 - 4J_3S^2$$
$$E_{AFM1} = E_0 + 4J_2S^2 + 4J_3S^2$$
$$E_{AFM2} = E_0 + 4J_2S^2$$
$$E_{AFM3} = E_0 + 2J_1S^2 + 4J_3S^2$$

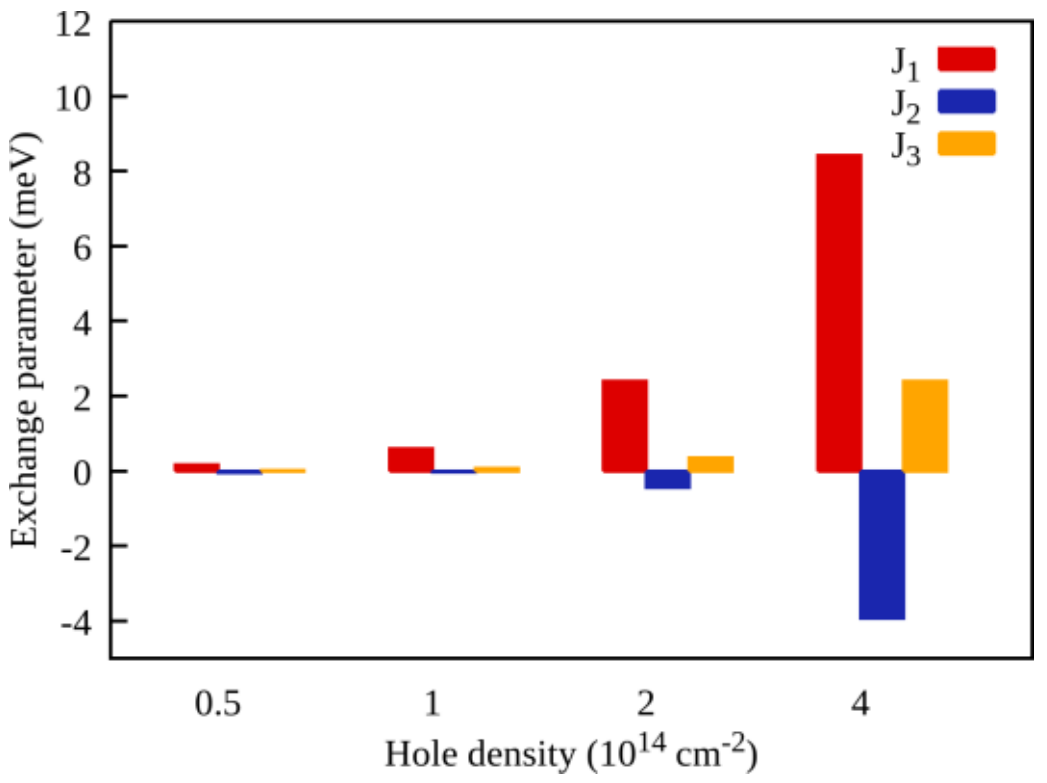

**Magnetic anisotropy energy (MAE, μeV) per magnetic atom** | **Monte Carlo simulations of the normalized magnetization of as a function of temperature**

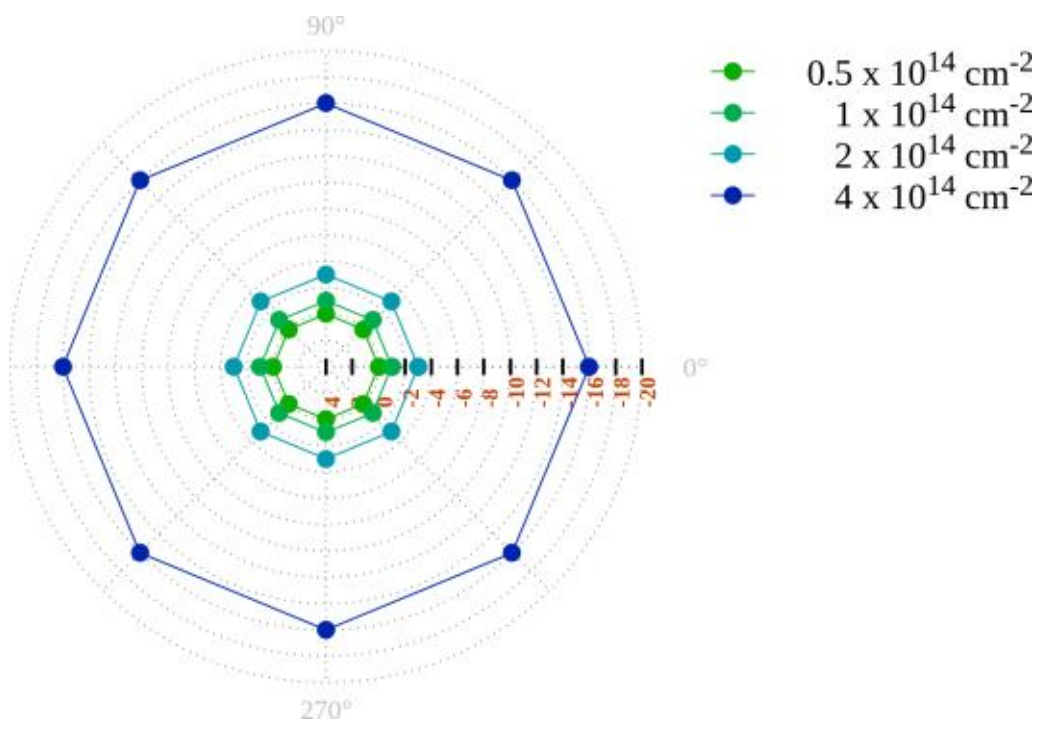

MAE = $E_{\parallel} - E_{\perp}$, a positive (negative) value of MAE indicates the off-plane (in-plane) easy axis.

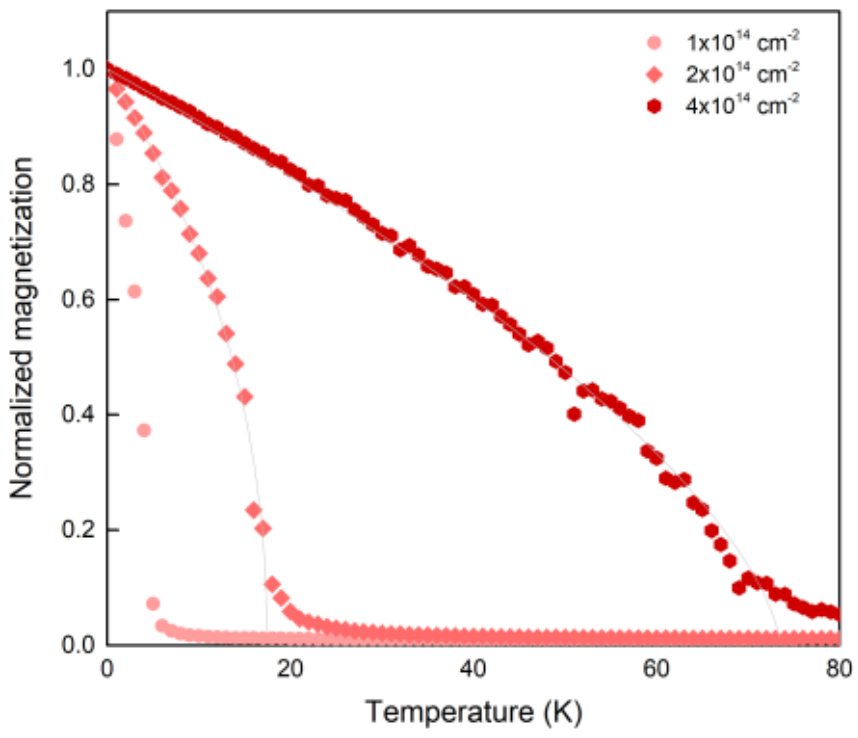

$T_c$: $0.5\times10^{14}$ cm$^{-2}$: 5 K  $1\times10^{14}$ cm$^{-2}$: 17 K  $2\times10^{14}$ cm$^{-2}$: 61 K  $4\times10^{14}$ cm$^{-2}$: - K

# 107. $Pb_2O_2$

| MC2D-ID | C2DB | 2dmat-ID | USPEX | Space group | Band gap (eV) |
|---|---|---|---|---|---|
| 153 | ✓ | 2dm-3561 | - | P4/nmm | 2.49 |

**Convex hull**

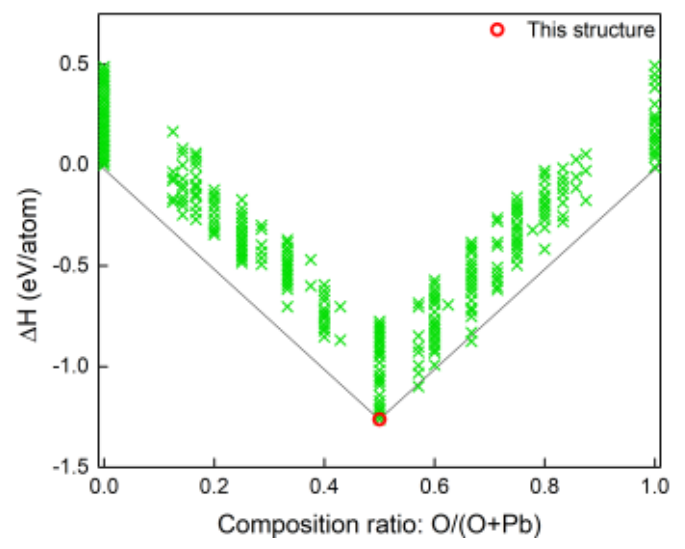

**Atomic structure**

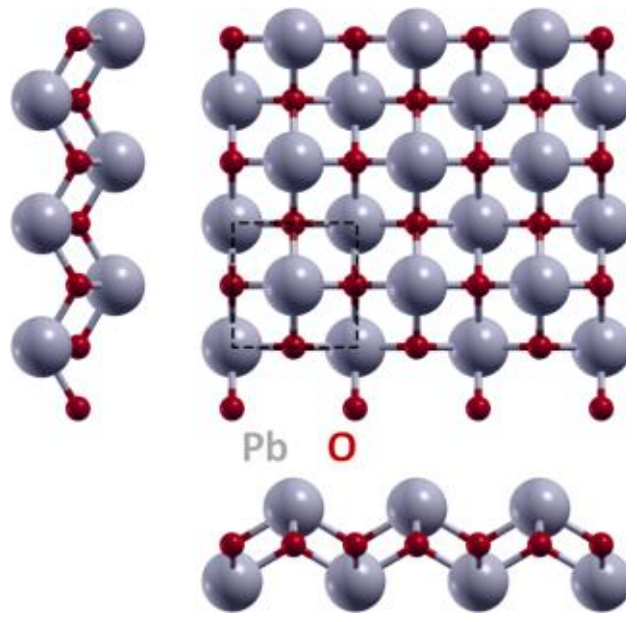

**Atomic coordinates**

```
Pb2O2
   1.00000000000000
   4.0492641769418585   0.0000000000000000   0.0000000000000000
   0.0000000000000000   4.0492642957581655   0.0000000000000000
   0.0000000000000000   0.0000000000000000  22.3791699999999985
   Pb   O
   2    2
Direct
  0.0000000109132117  0.4999999827697224  0.4464929309991490
  0.4999999860015976  0.9999999991165467  0.5535070723400963
  0.5000000029956198  0.5000000161566831  0.5000000014446400
  0.0000000000895710  0.0000000019570550  0.4999999952161147
```

**Phonon dispersion curve**

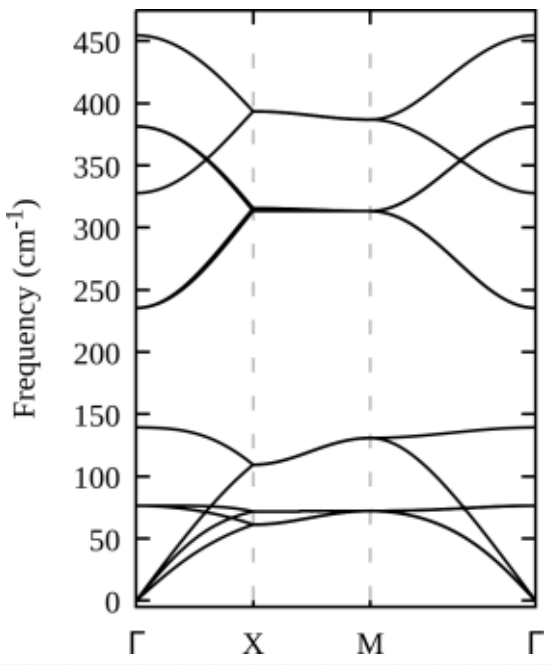

**Projected band structure and density of states**

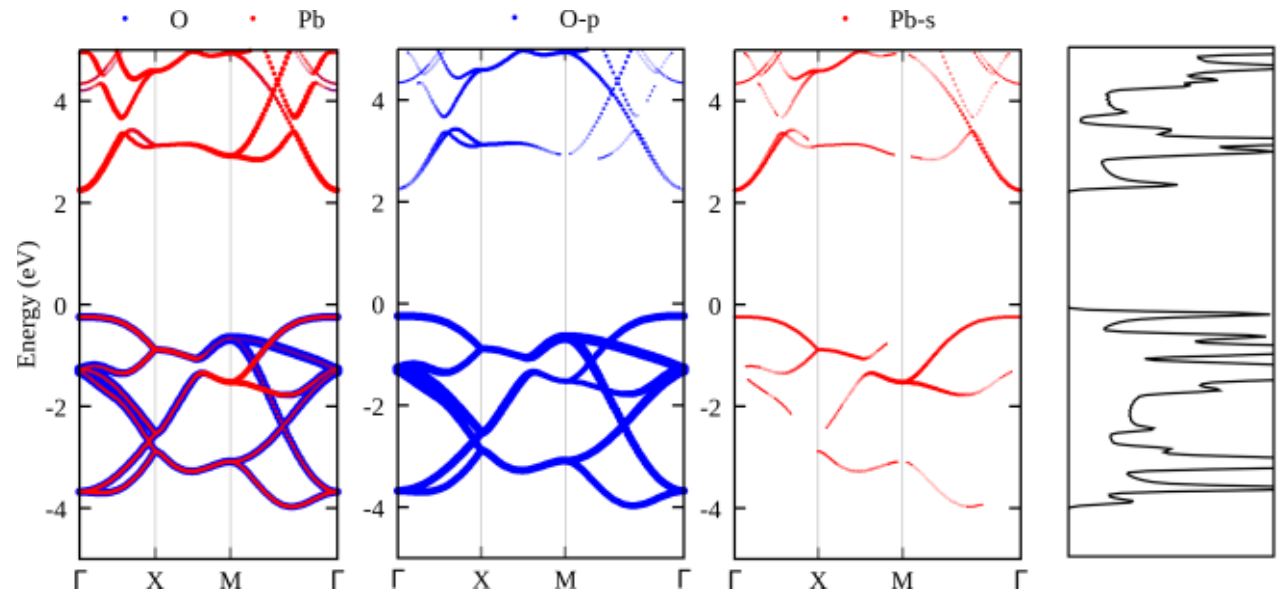

**Magnetic moment and spin polarization energy as a function of hole doping concentration**

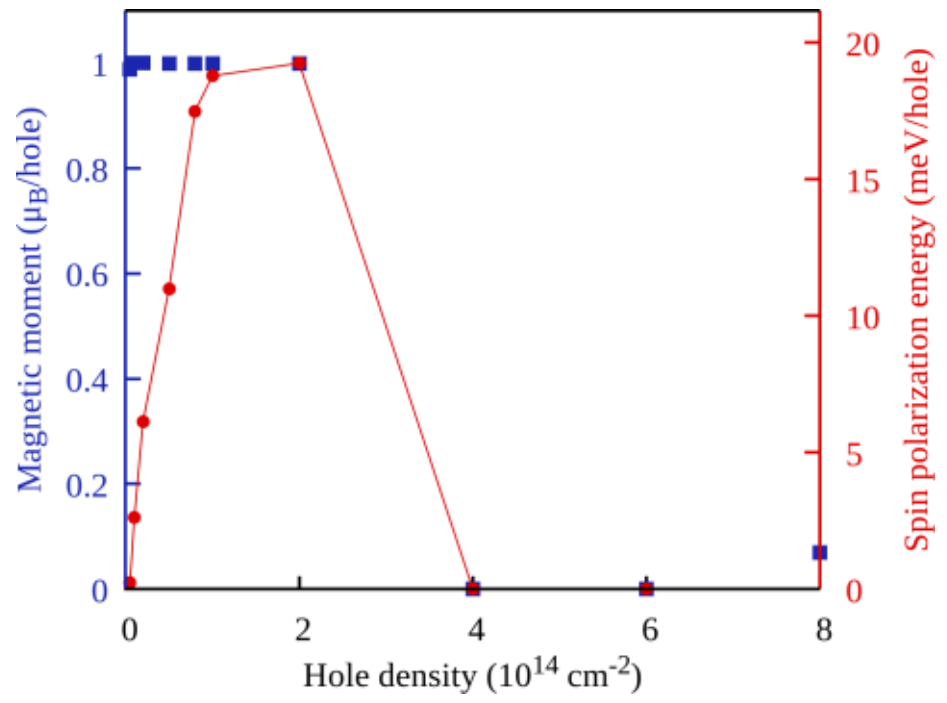

**Magnetic configurations and spin Hamiltonian**

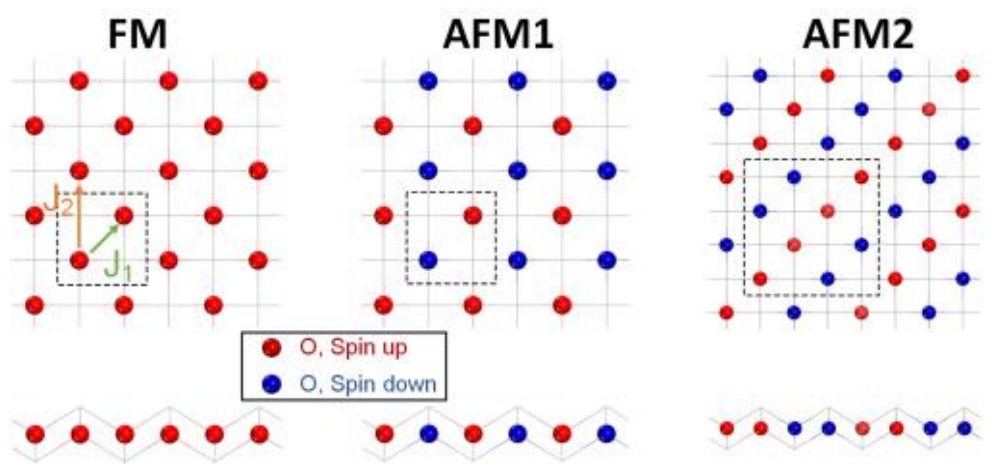

$$E_{FM} = E_0 - 4J_1S^2 - 4J_2S^2$$

$$E_{AFM1} = E_0 + 4J_1S^2 - 4J_2S^2$$

$$E_{AFM2} = E_0 + 4J_2S^2$$

**Magnetic exchange coupling parameters**

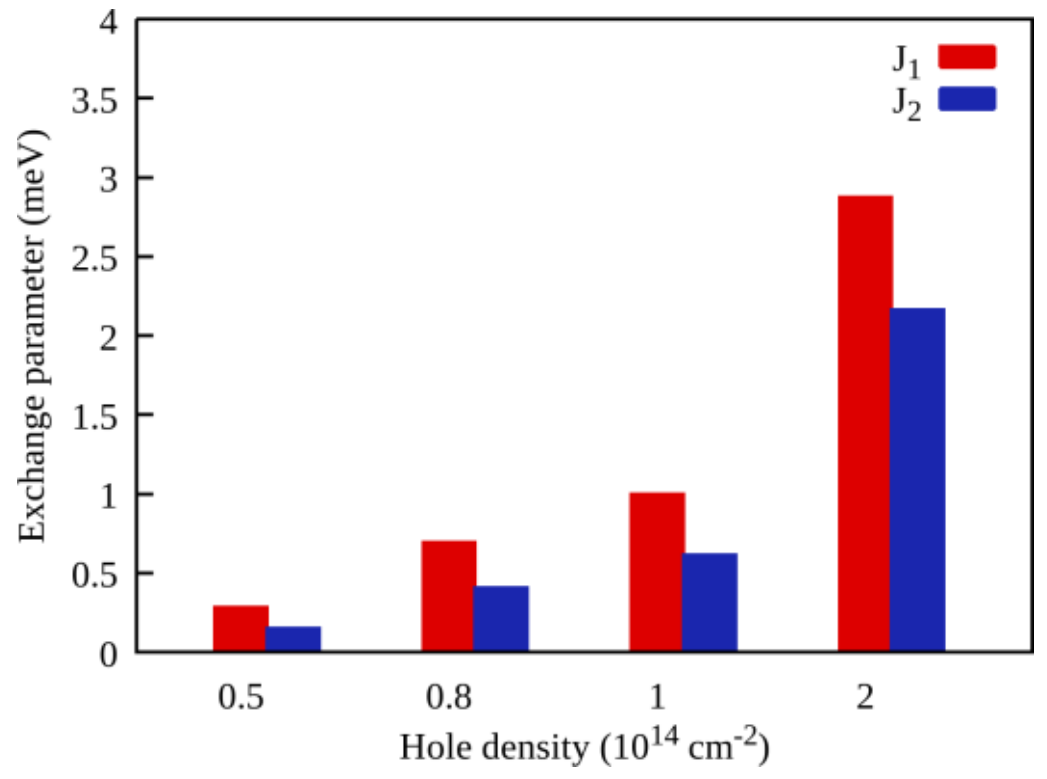

**Magnetic anisotropy energy (MAE, μeV) per magnetic atom**

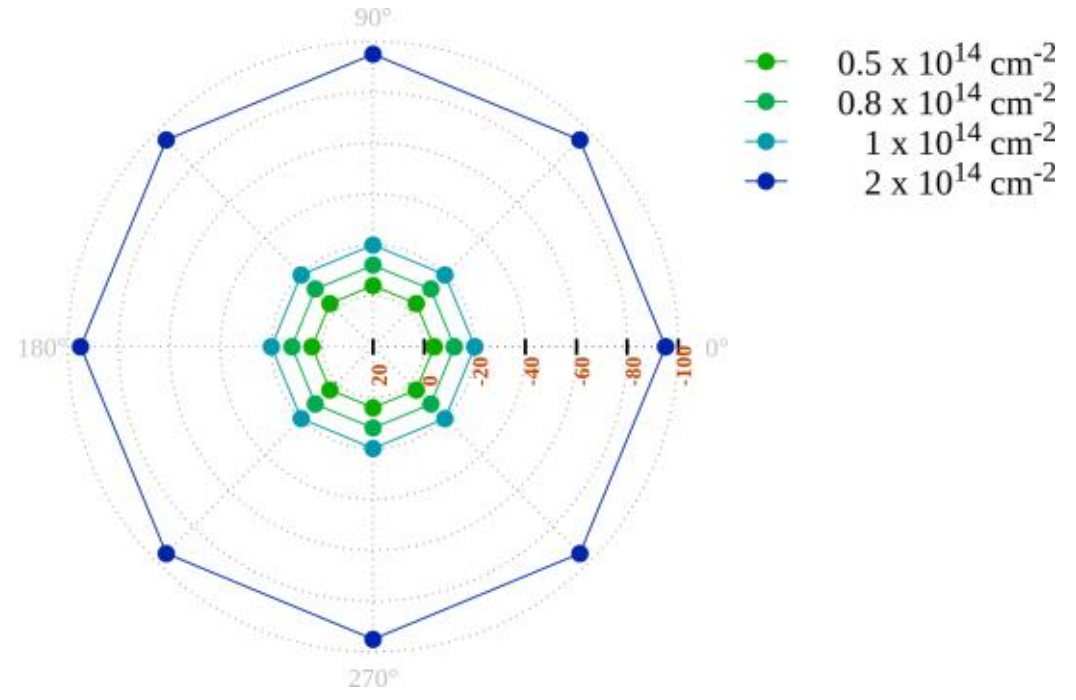

MAE = $E_{\parallel} - E_{\perp}$, a positive (negative) value of MAE indicates the off-plane (in-plane) easy axis.

**Monte Carlo simulations of the normalized magnetization of as a function of temperature**

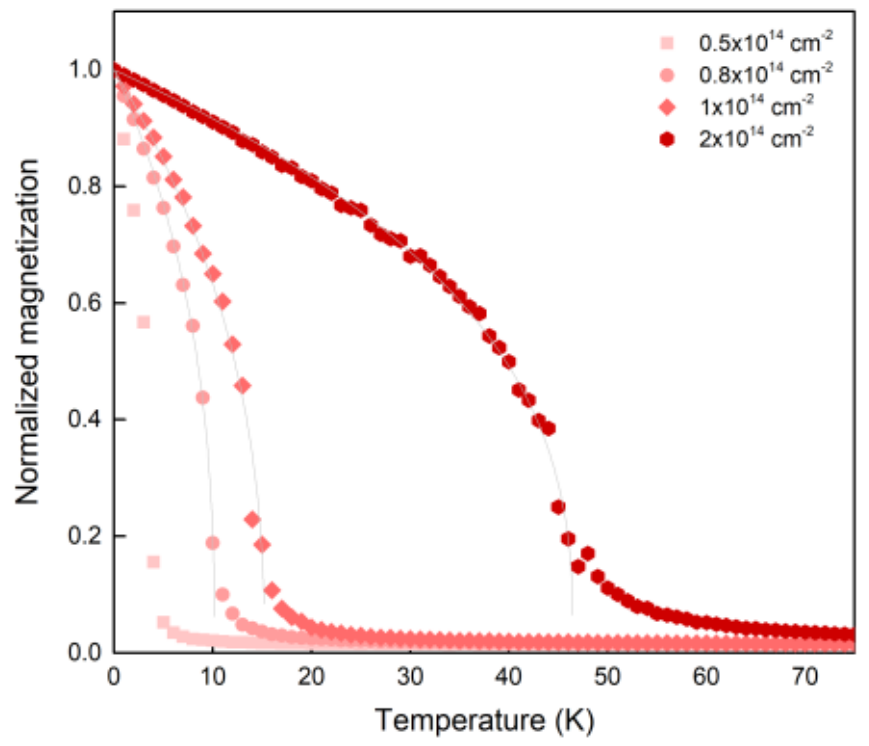

$T_c$: $0.5\times10^{14}$ cm$^{-2}$: 5 K  $0.8\times10^{14}$ cm$^{-2}$: 10 K

$1\times10^{14}$ cm$^{-2}$: 15 K  $2\times10^{14}$ cm$^{-2}$: 46 K

# 108. $Al_2N_2$

| MC2D-ID | C2DB | 2dmat-ID | USPEX | Space group | Band gap (eV) |
|---|---|---|---|---|---|
| - | - | 2dm-1029 | - | P4/nmm | 3.54 |

| Convex hull | Atomic structure | Atomic coordinates | Phonon dispersion curve |
|---|---|---|---|

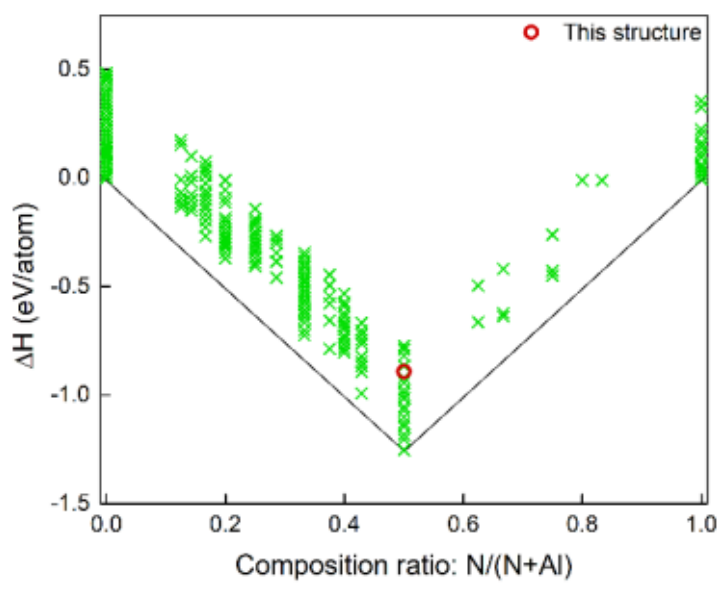

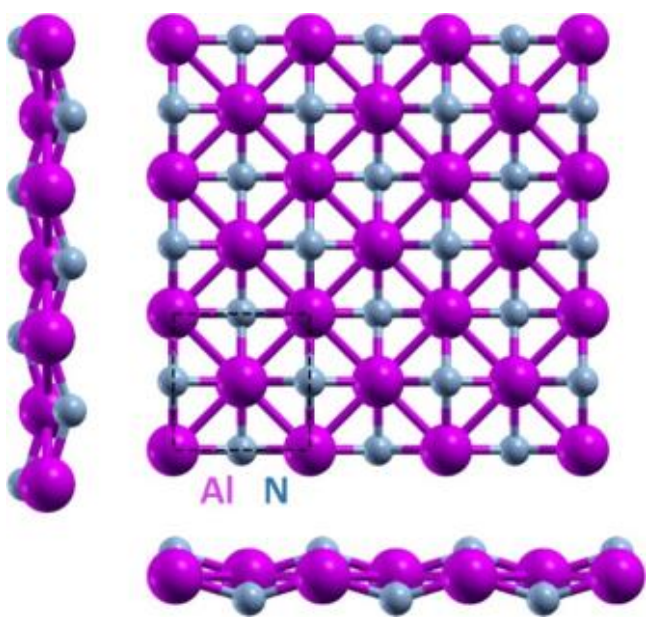

```
Al2N2
   1.00000000000000
   3.6327475818023207   0.0000000000000000   0.0000000000000000
   0.0000000000000000   3.6327475947125496   0.0000000000000000
   0.0000000000000000   0.0000000000000000 23.9423190000000012
   Al    N
   2      2
Direct
  0.0000000002115428   0.9999999991086526   0.9999999984051726
  0.4999999994827178   0.5000000007099743   0.9999999984286347
  0.5000000007493099   0.9999999975236946   0.9765657033252495
  0.9999999995564295   0.5000000026576856   0.0234342998409502
```

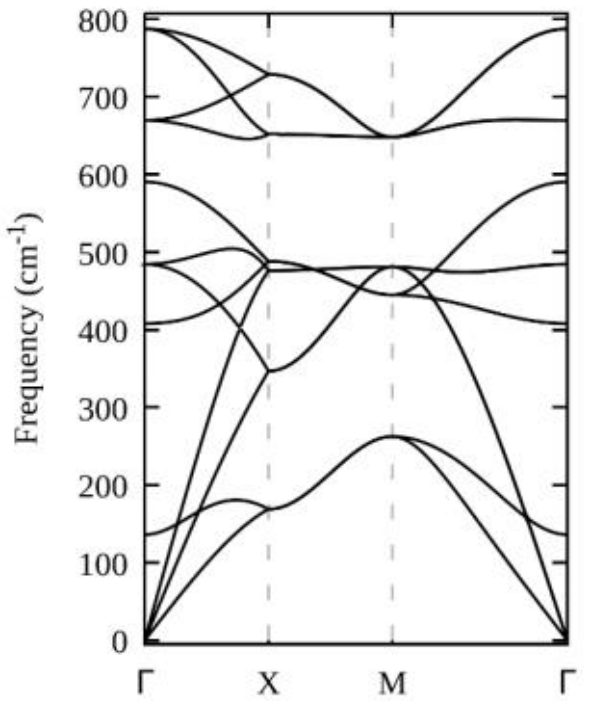

## Projected band structure and density of states

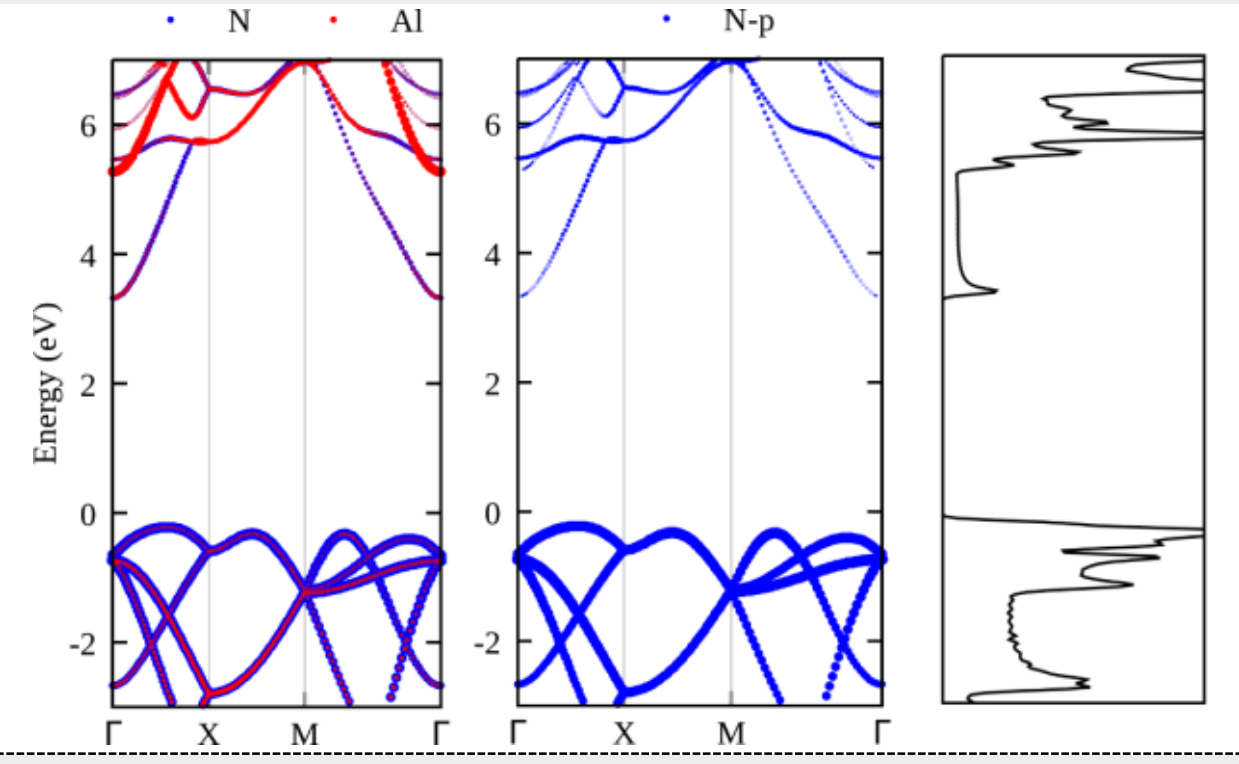

## Magnetic moment and spin polarization energy as a function of hole doping concentration

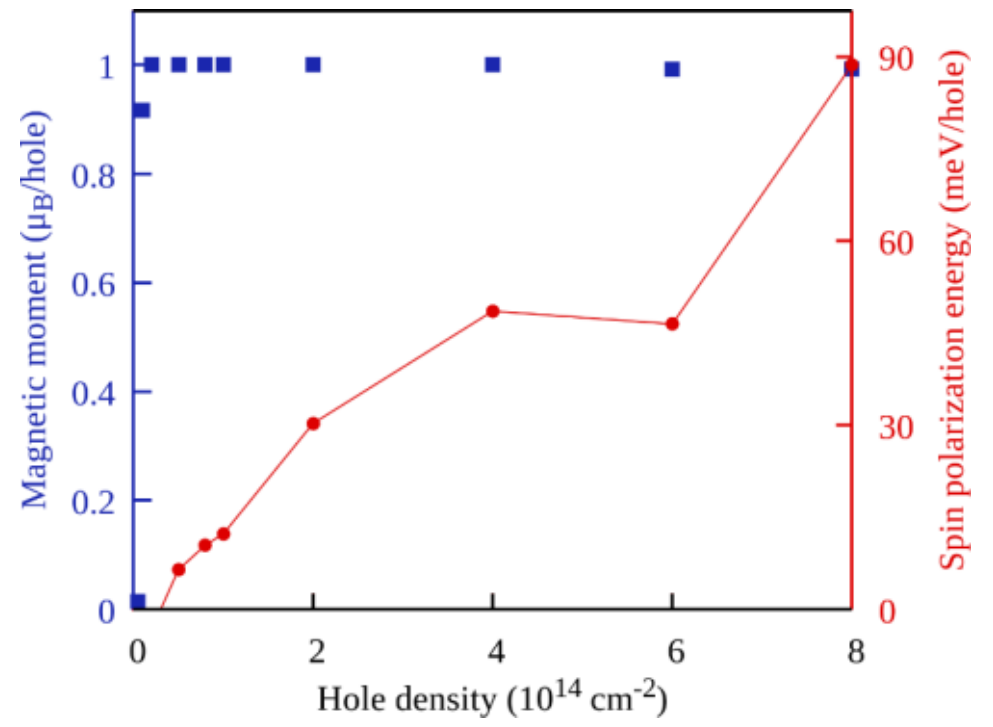

## Magnetic configurations and spin Hamiltonian

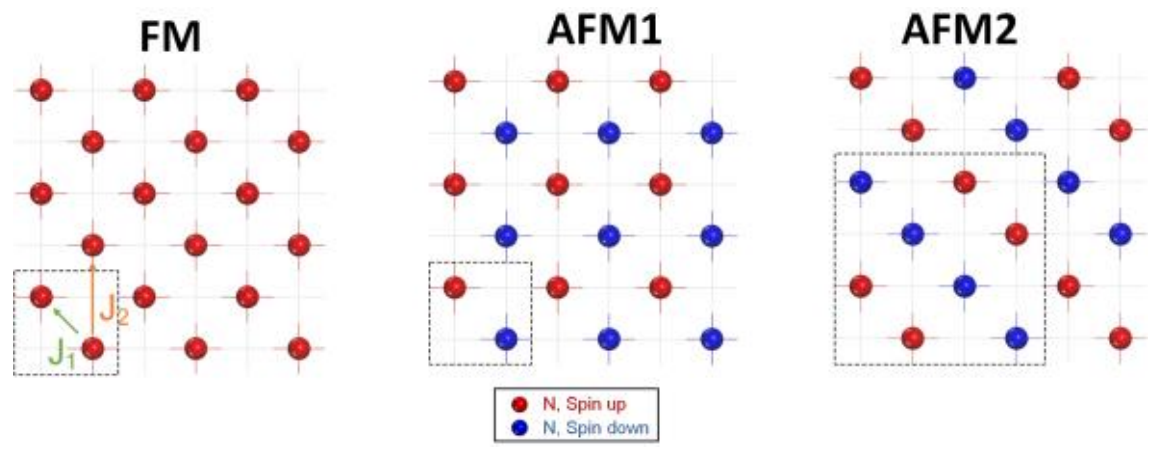

$$E_{FM} = E_0 - 4J_1S^2 - 4J_2S^2$$
$$E_{AFM1} = E_0 + 4J_1S^2 - 4J_2S^2$$
$$E_{AFM2} = E_0 + 4J_2S^2$$

## Magnetic exchange coupling parameters

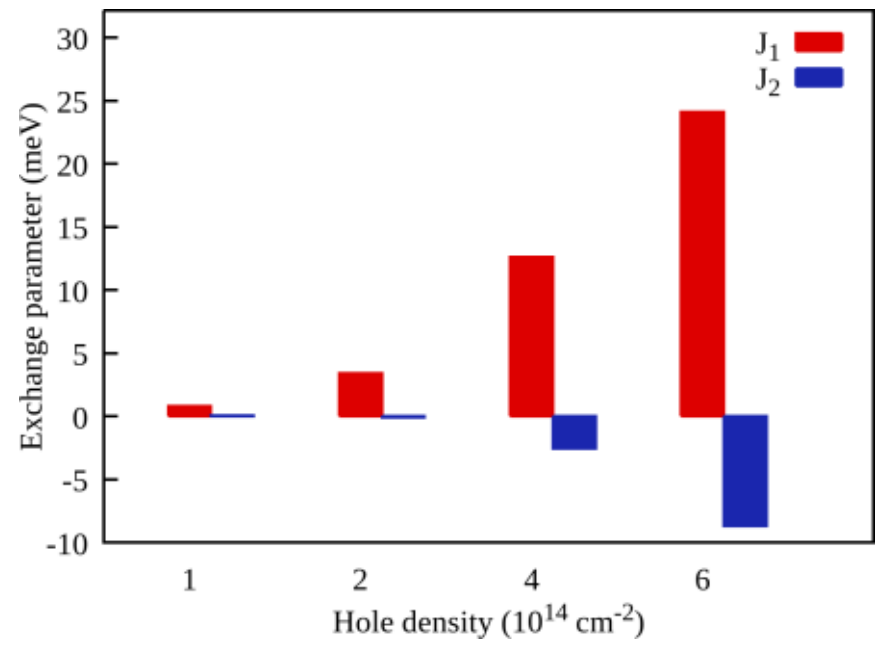

## Magnetic anisotropy energy (MAE, μeV) per magnetic atom

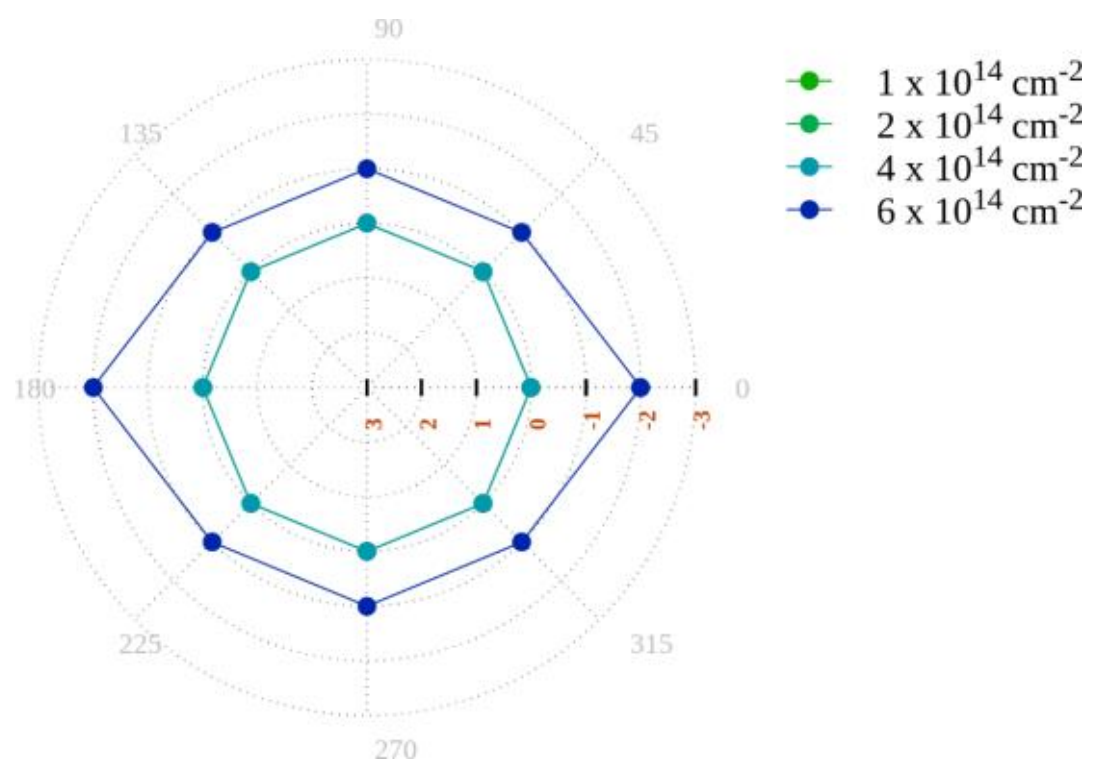

MAE = $E_{\parallel} - E_{\perp}$, a positive (negative) value of MAE indicates the off-plane (in-plane) easy axis.

## Monte Carlo simulations of the normalized magnetization of as a function of temperature

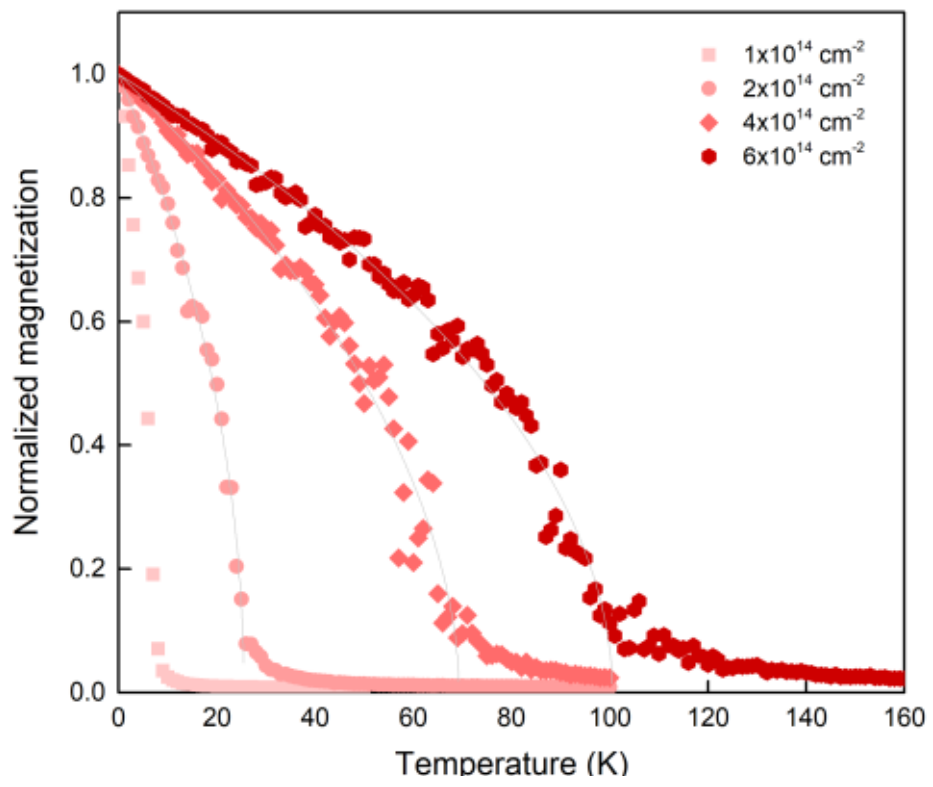

$T_C$: $1\times10^{14}$ cm$^{-2}$: 8 K  $2\times10^{14}$ cm$^{-2}$: 25 K

$4\times10^{14}$ cm$^{-2}$: 69 K  $6\times10^{14}$ cm$^{-2}$: 101 K

# 109. $Li_2(OH)_2$

| MC2D-ID | C2DB | 2dmat-ID | USPEX | Space group | Band gap (eV) |
|---|---|---|---|---|---|
| 111 | ✓ | 2dm-3650 | - | P4/nmm | 3.92 |

**Convex hull**

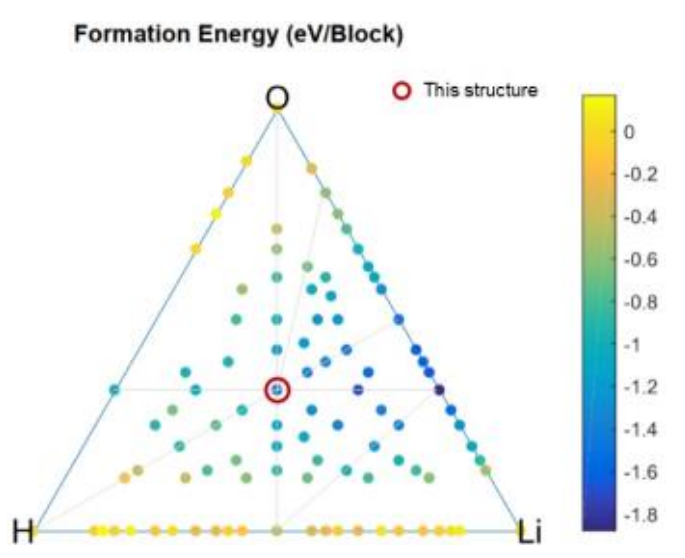

**Atomic structure**

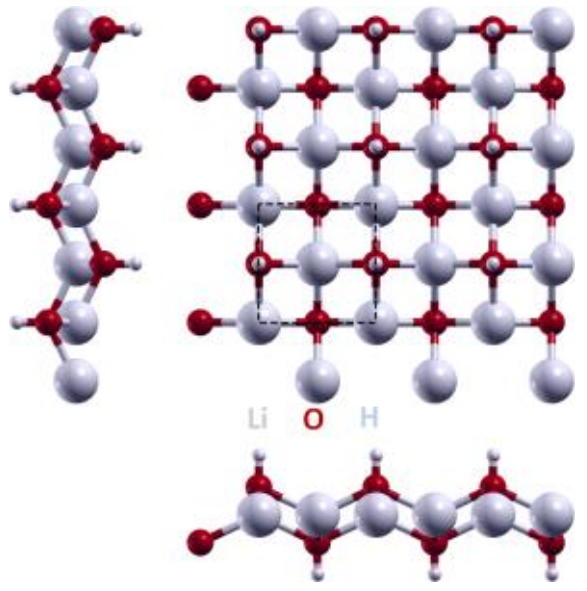

**Atomic coordinates**

```
Li2H2O2
  1.00000000000000
  3.5713938126153830   0.0000000000000000   0.0000000000000000
  0.0000000000000000   3.5713935372583849   0.0000000000000000
  0.0000000000000000   0.0000000000000000  23.6428140000000013
  Li   H   O
  2    2   2
Direct
 0.9999999940638915  0.0000000017331061  0.5000000001397638
 0.4999999962725070  0.5000000333215198  0.5000000065823755
 0.4999999832262390  0.9999999969004563  0.5765971880356275
 0.9999999969126421  0.5000000019982949  0.4234028051547583
 0.0000000847418349  0.5000000288072073  0.4643706448608924
 0.4999999447828856  0.9999999372394157  0.5356293552265896
```

**Phonon dispersion curve**

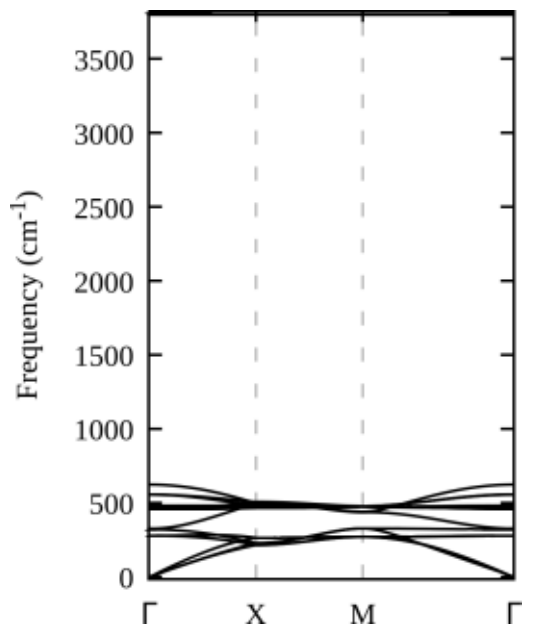

**Projected band structure and density of states**

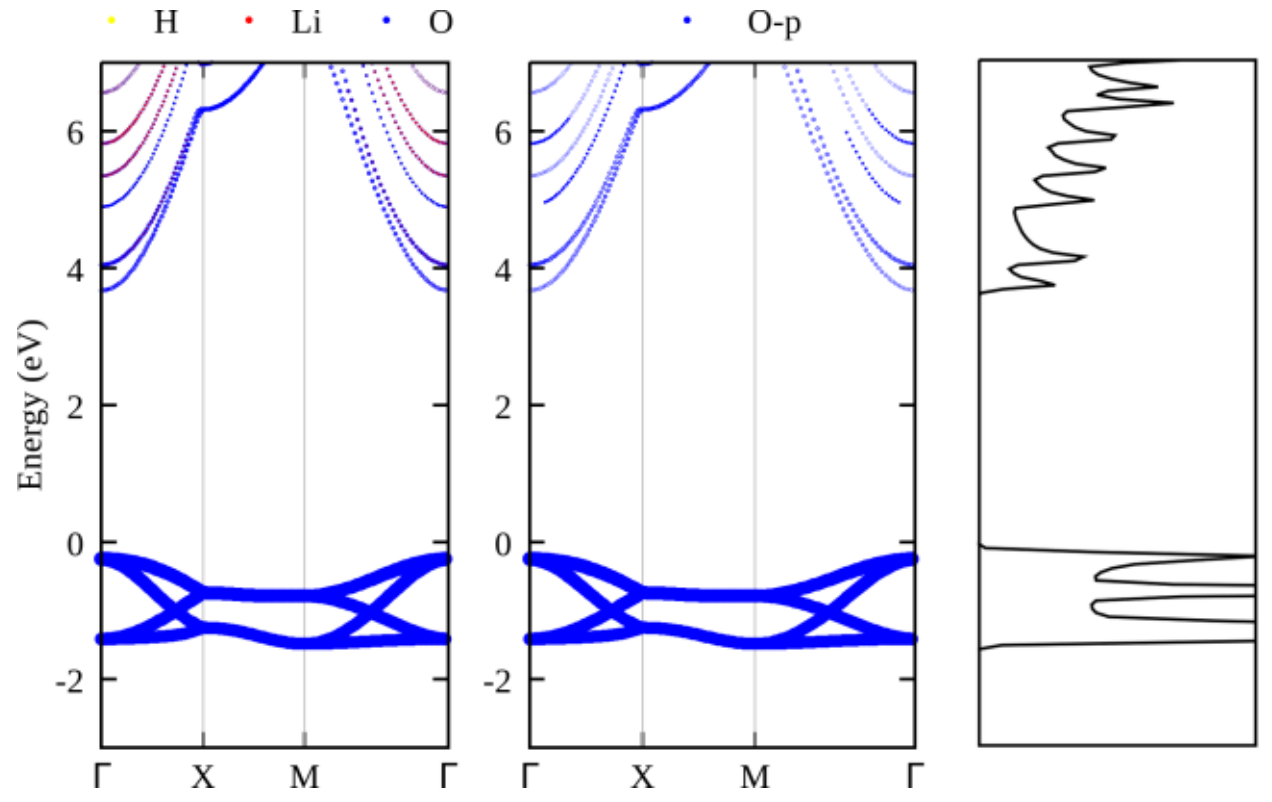

**Magnetic moment and spin polarization energy as a function of hole doping concentration**

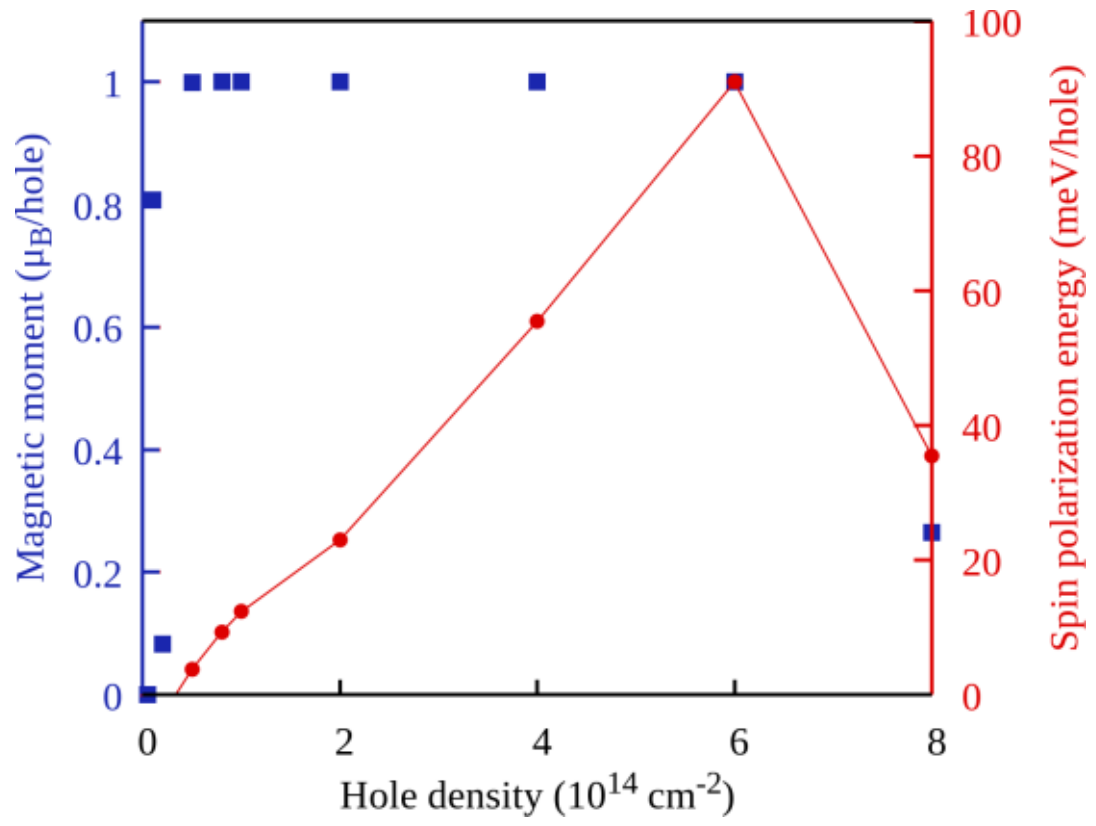

**Magnetic configurations and spin Hamiltonian**

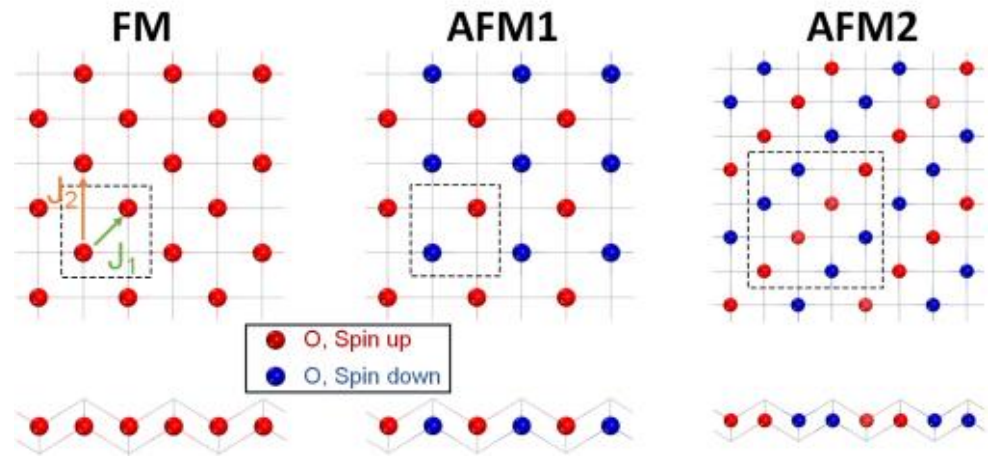

$$E_{FM} = E_0 - 4J_1S^2 - 4J_2S^2$$

$$E_{AFM1} = E_0 + 4J_1S^2 - 4J_2S^2$$

$$E_{AFM2} = E_0 + 4J_2S^2$$

**Magnetic exchange coupling parameters**

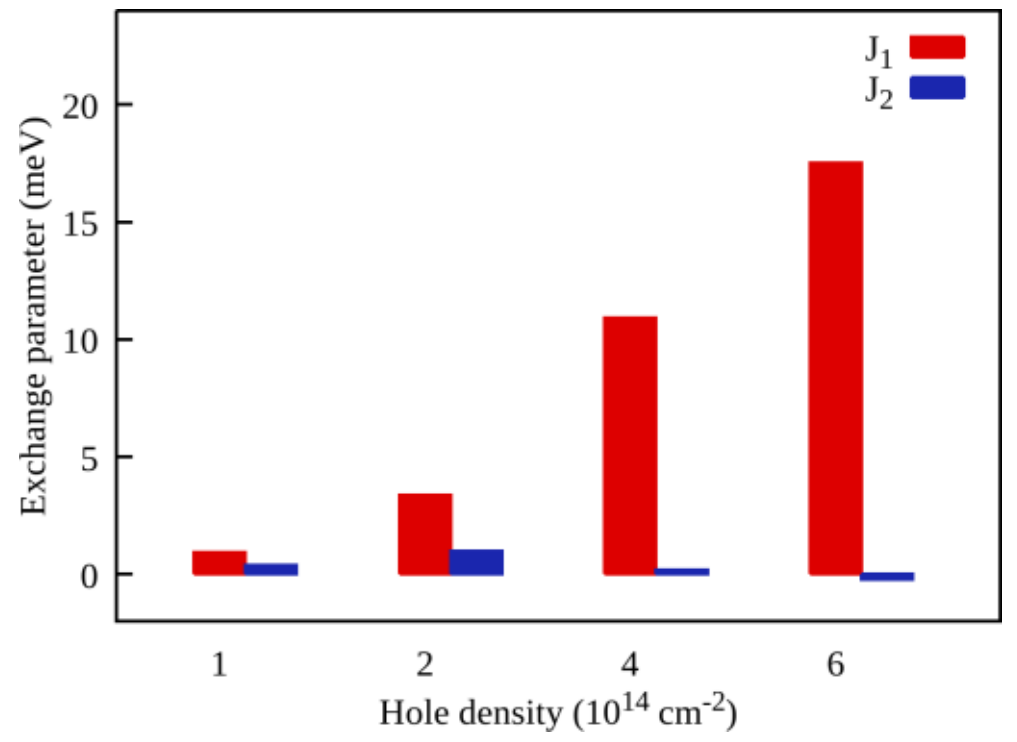

**Magnetic anisotropy energy (MAE, μeV) per magnetic atom**

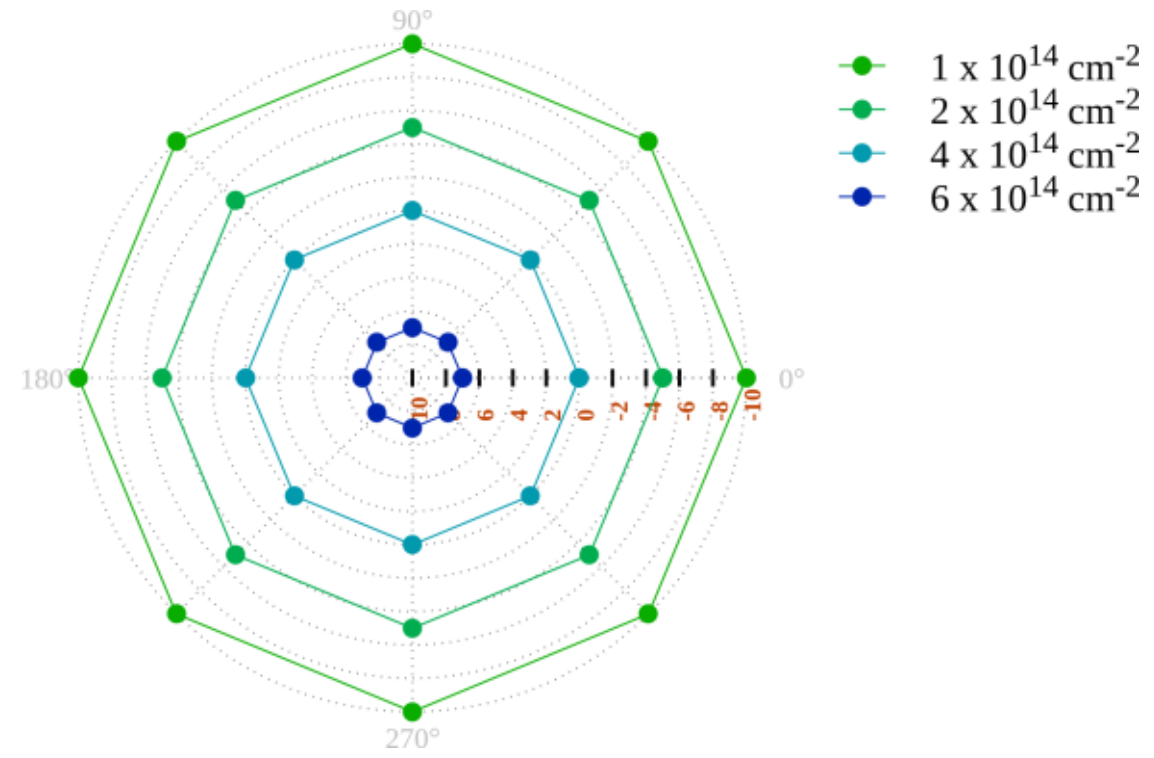

MAE = $E_{\parallel} - E_{\perp}$, a positive (negative) value of MAE indicates the off-plane (in-plane) easy axis.

**Monte Carlo simulations of the normalized magnetization of as a function of temperature**

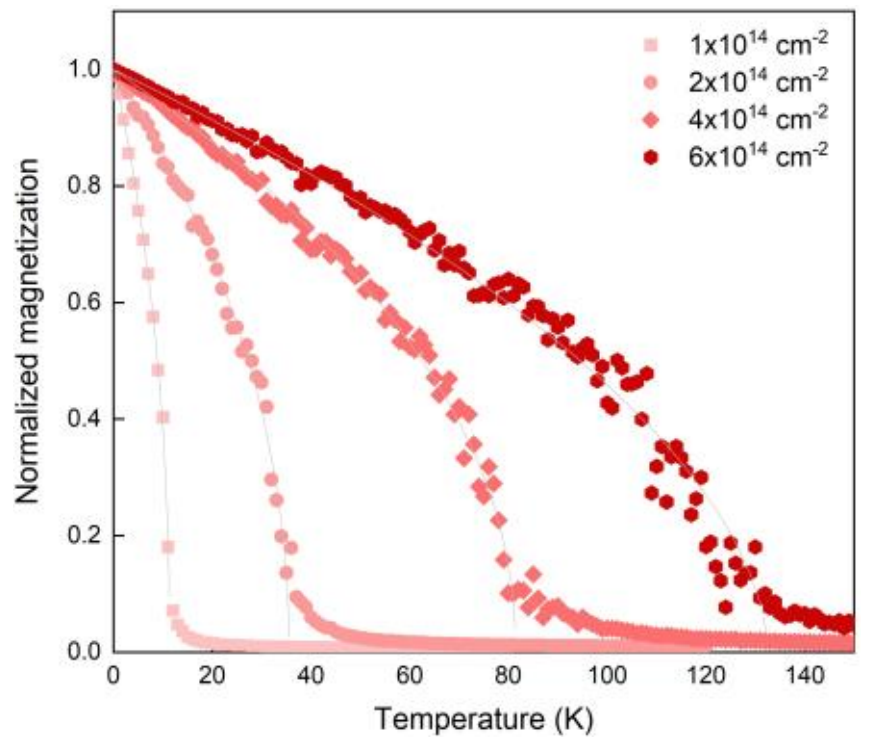

$T_c$: $1\times10^{14}$ cm$^{-2}$: 11 K  $2\times10^{14}$ cm$^{-2}$: 35 K

$4\times10^{14}$ cm$^{-2}$: 81 K  $6\times10^{14}$ cm$^{-2}$: 132 K

# 110. $Na_2(OH)_2$

| MC2D-ID | C2DB | 2dmat-ID | USPEX | Space group | Band gap (eV) |
|---|---|---|---|---|---|
| 131 | ✓ | 2dm-5304 | - | P4/nmm | 2.77 |

**Convex hull**

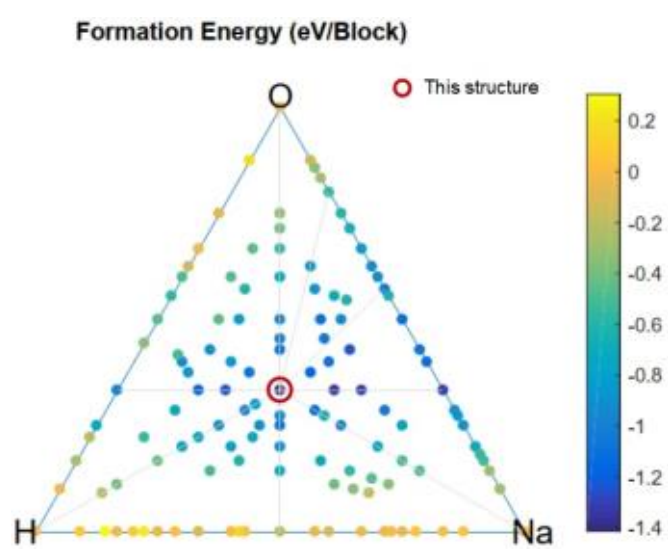

**Atomic structure**

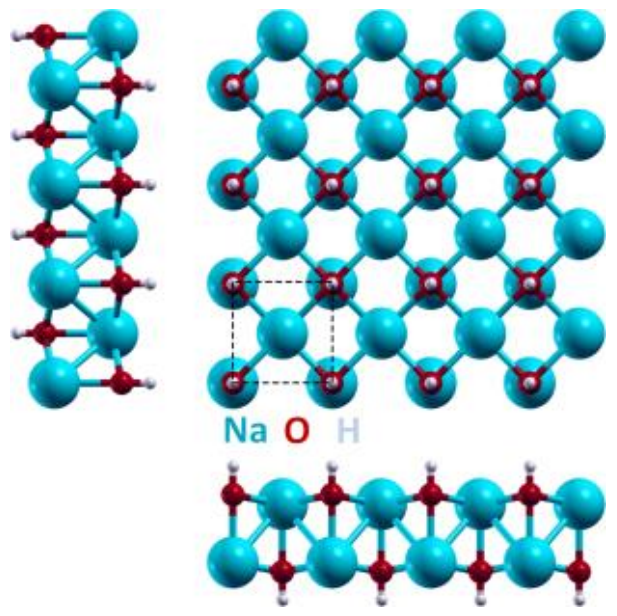

**Atomic coordinates**

```
Na2H2O2
1.00000000000000
3.4173975040401374    0.0000000000000000    0.0000000000000000
0.0000000000000000    3.4202648848351274    0.0000000000000000
0.0000000000000000    0.0000000000000000    24.4975050000000003
Na   H   O
2    2   2
Direct
0.8328904903480279   0.6813309661559188   0.5410498809907835
0.3328756579871879   0.1817173217661505   0.4589499776824937
0.8334356528785563   0.6819885831771160   0.4060455645891565
0.3334398469365425   0.1807410466727646   0.5939543914272249
0.8336403829060757   0.6818801626118756   0.4456891253749831
0.3336399689436220   0.1808349196161743   0.5543110599353582 8
```

**Phonon dispersion curve**

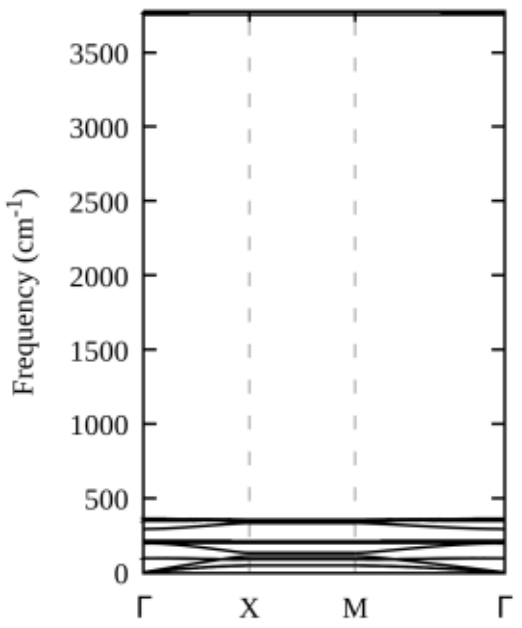

**Projected band structure and density of states**

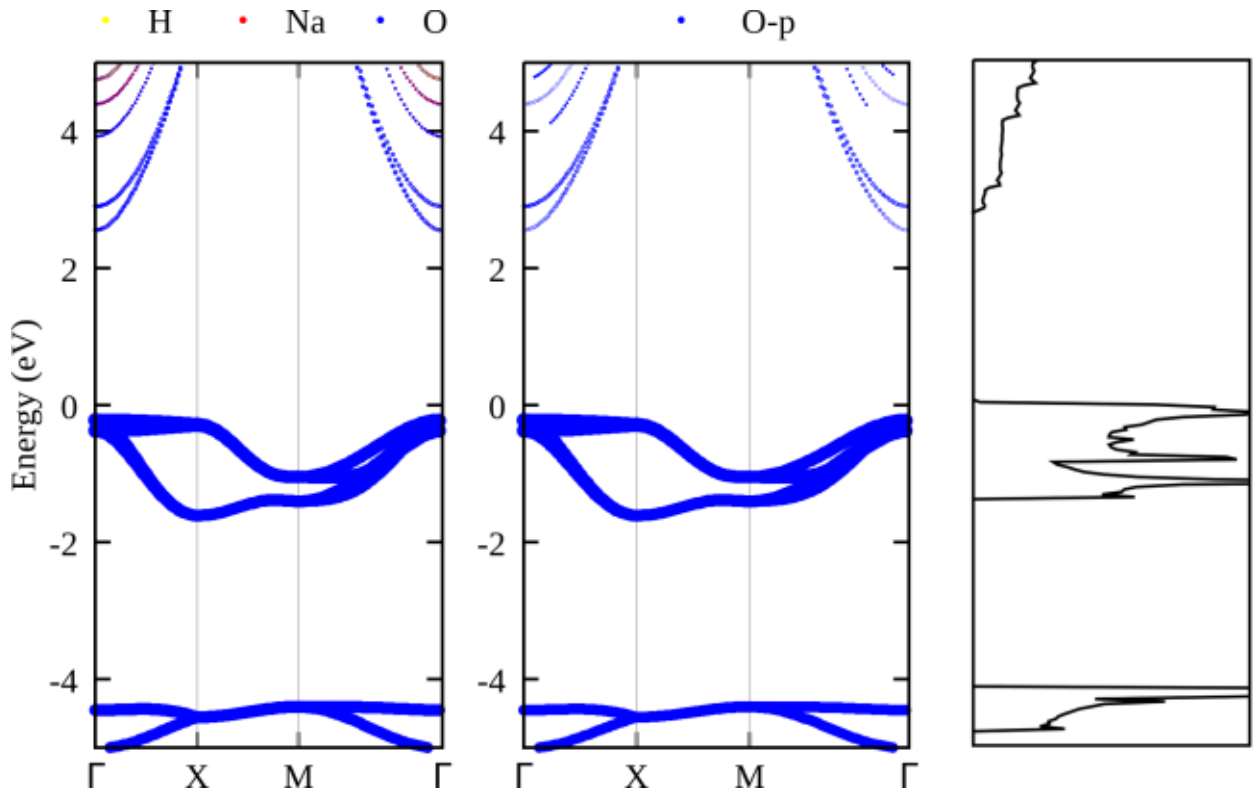

**Magnetic moment and spin polarization energy as a function of hole doping concentration**

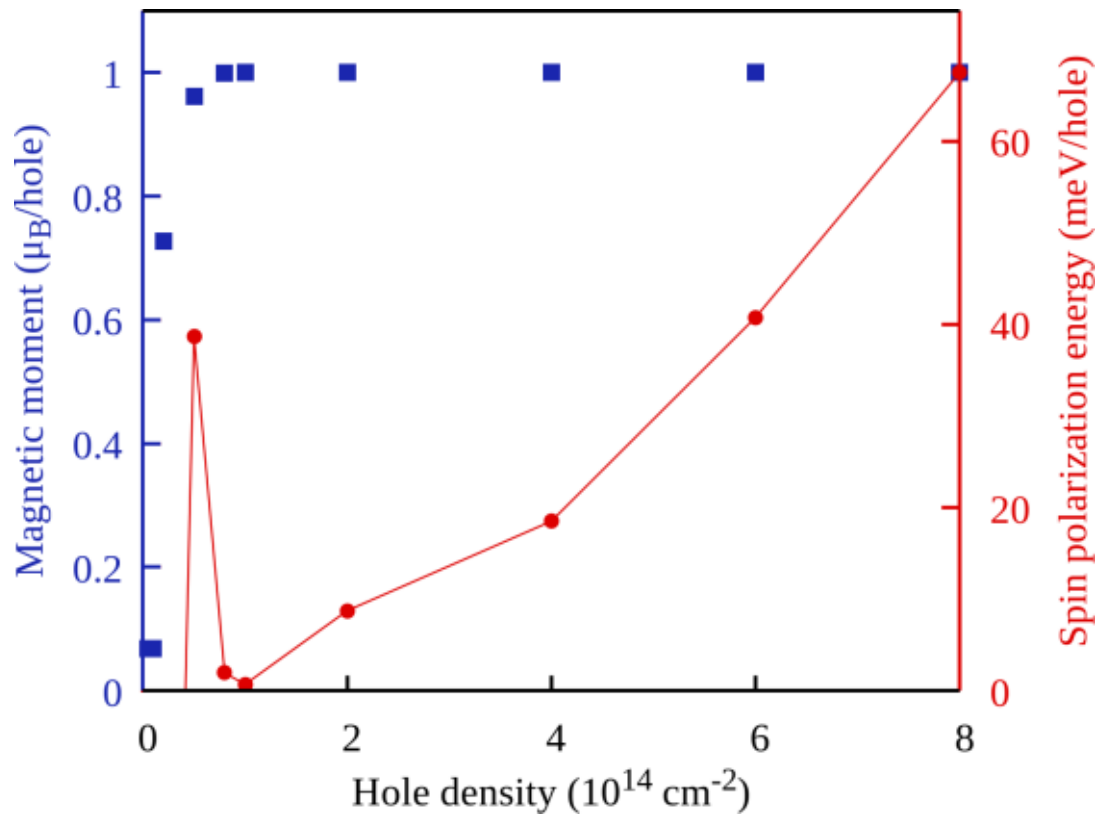

**Magnetic configurations and spin Hamiltonian**

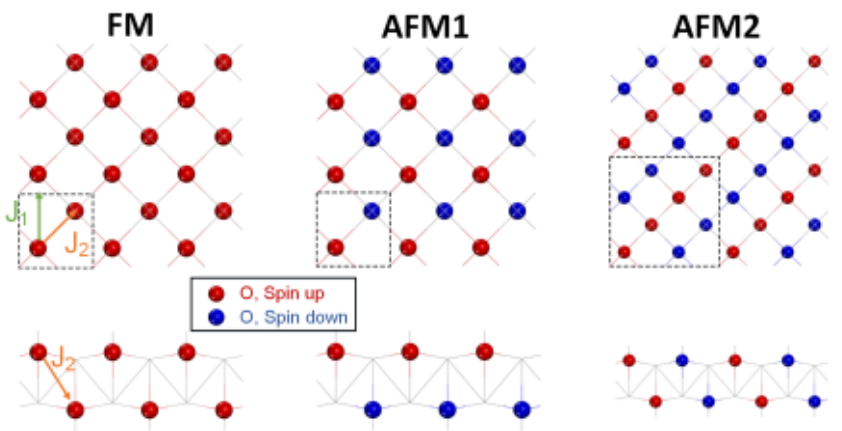

$$E_{FM} = E_0 - 4J_1S^2 - 4J_2S^2$$
$$E_{AFM1} = E_0 - 4J_1S^2 + 4J_2S^2$$
$$E_{AFM2} = E_0 + 4J_1S^2$$

**Magnetic exchange coupling parameters**

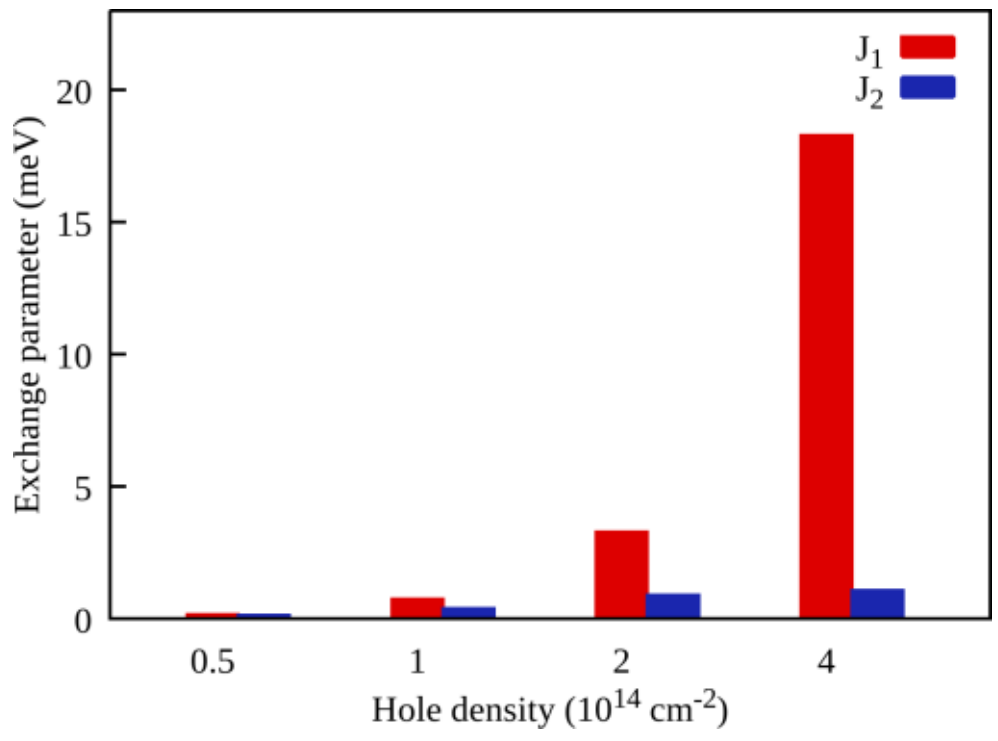

**Magnetic anisotropy energy (MAE, μeV) per magnetic atom**

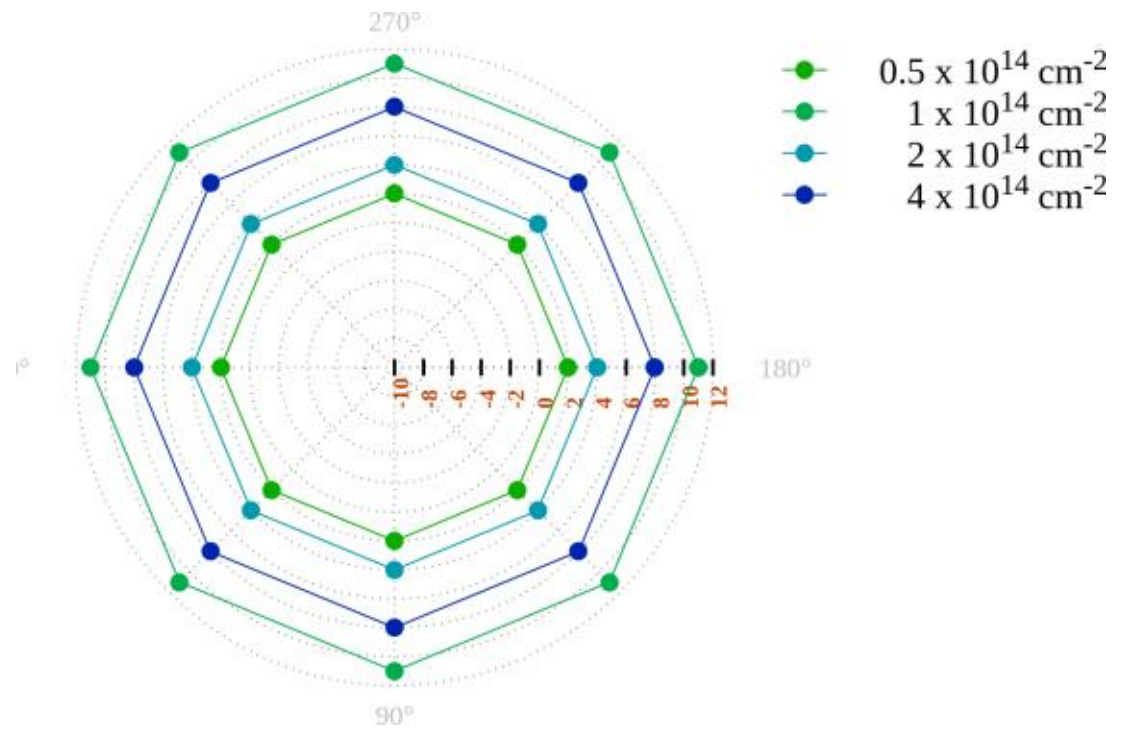

MAE = $E_{\parallel} - E_{\perp}$, a positive (negative) value of MAE indicates the off-plane (in-plane) easy axis.

**Monte Carlo simulations of the normalized magnetization of as a function of temperature**

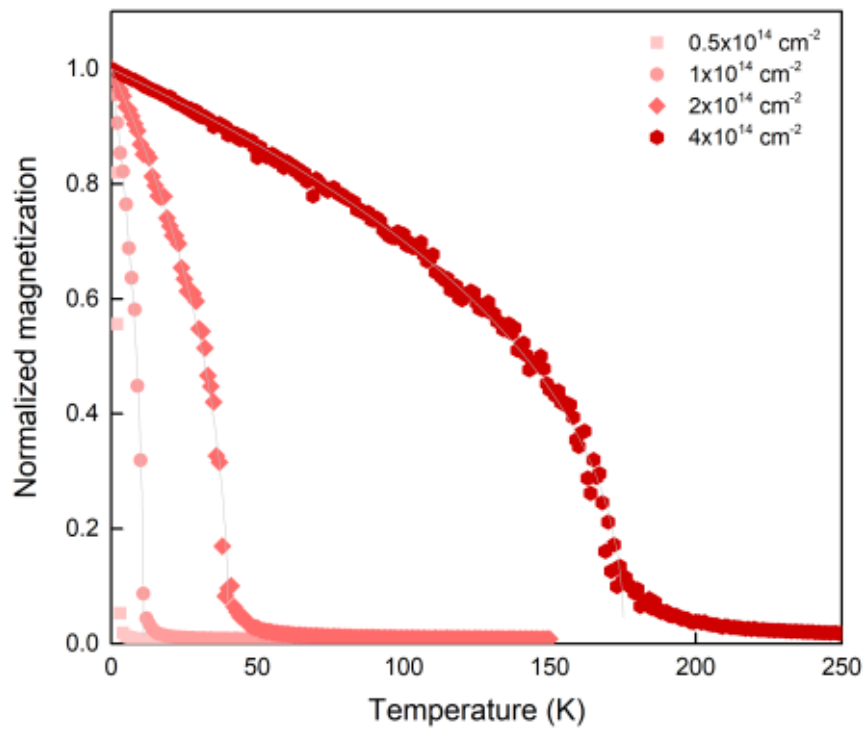

$T_c$: $0.5\times10^{14}$ cm$^{-2}$: 3 K  $1\times10^{14}$ cm$^{-2}$: 11 K

$2\times10^{14}$ cm$^{-2}$: 40 K  $4\times10^{14}$ cm$^{-2}$: 175 K

# 111. $Pb_2Br_2F_2$

| MC2D-ID | C2DB | 2dmat-ID | USPEX | Space group | Band gap (eV) |
|---|---|---|---|---|---|
| 148 | - | 2dm-3625 | - | P4/nmm | 3.04 |

**Convex hull**

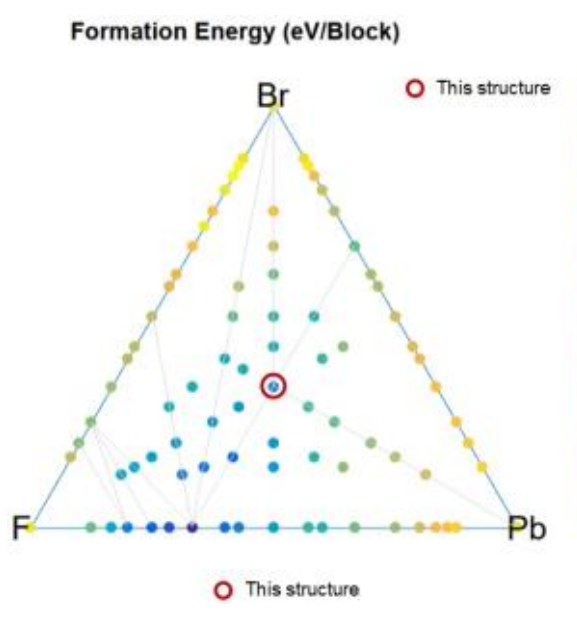

**Atomic structure**

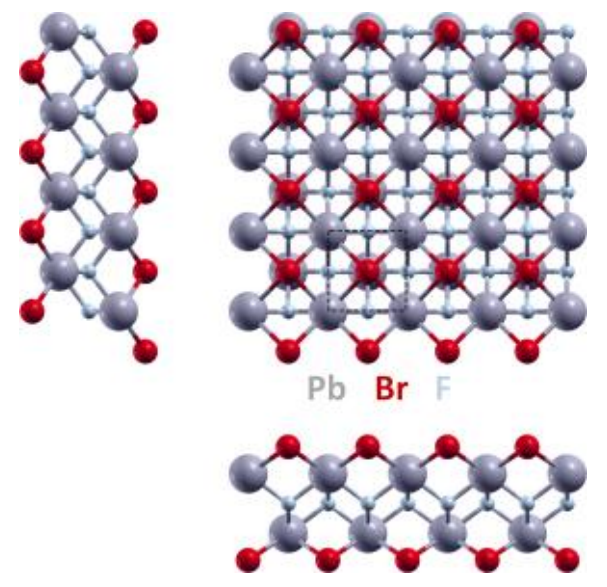

**Atomic coordinates**

```
Pb2Br2F2
1.00000000000000
4.1039841614003132 0.0000000000000000 0.0000000000000000
0.0000000000000000 4.1039841659102736 0.0000000000000000
0.0000000000000000 0.0000000000000000 25.4031884695999999
Pb Br F
2 2 2
Direct
0.0000000029635459 0.5000000040512447 0.5603851148380130
0.4999999961748287 0.9999999984268086 0.4396148853497124
0.4999999998419469 0.0000000001062332 0.6147554852933865
0.0000000002298322 0.5000000000353140 0.3852445149141914
0.9999999998432116 0.9999999975649132 0.4999999998302727
0.5000000009466348 0.4999999998154863 0.4999999997783604
```

**Phonon dispersion curve**

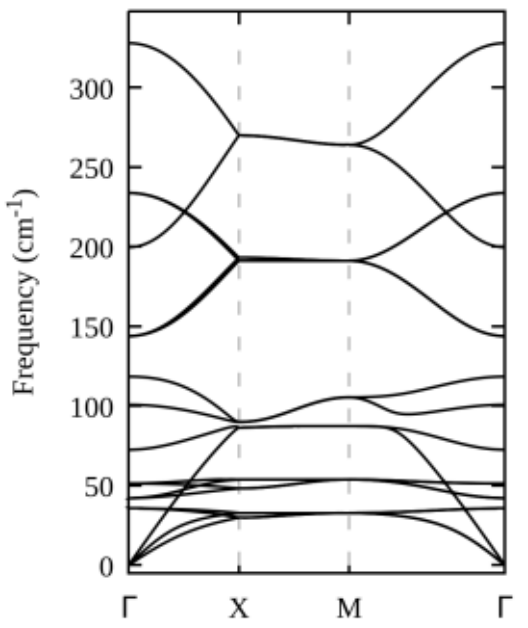

**Projected band structure and density of states**

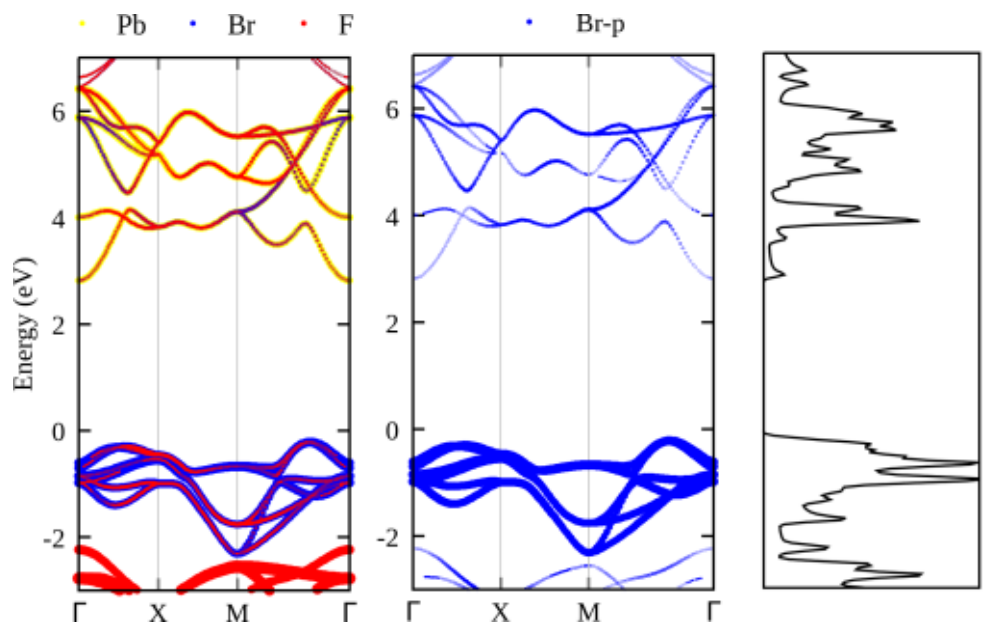

**Magnetic moment and spin polarization energy as a function of hole doping concentration**

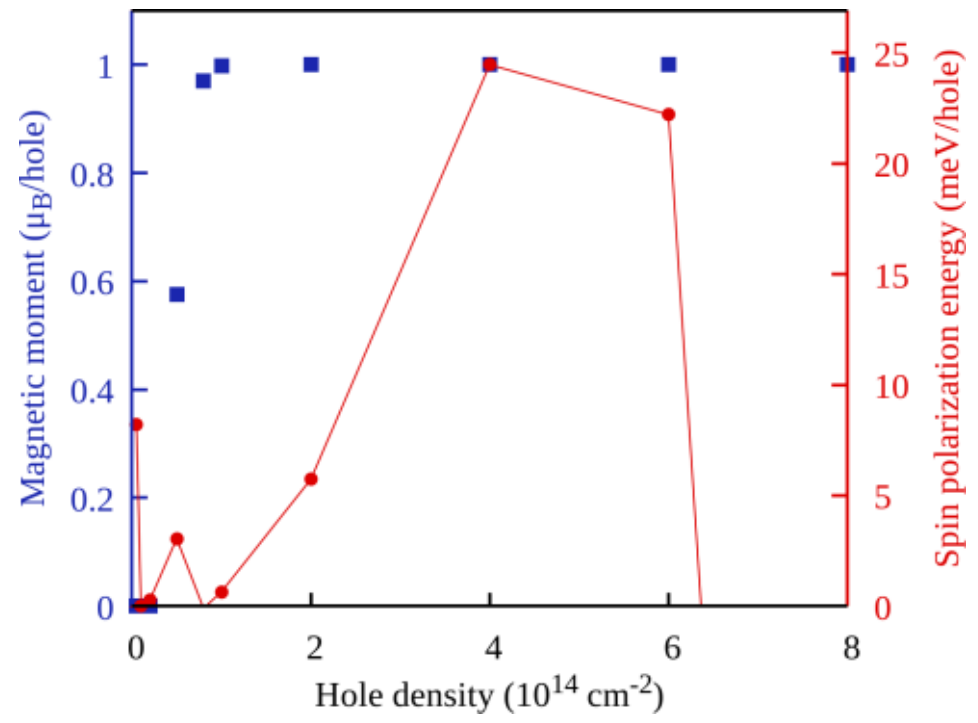

**Magnetic configurations and spin Hamiltonian**

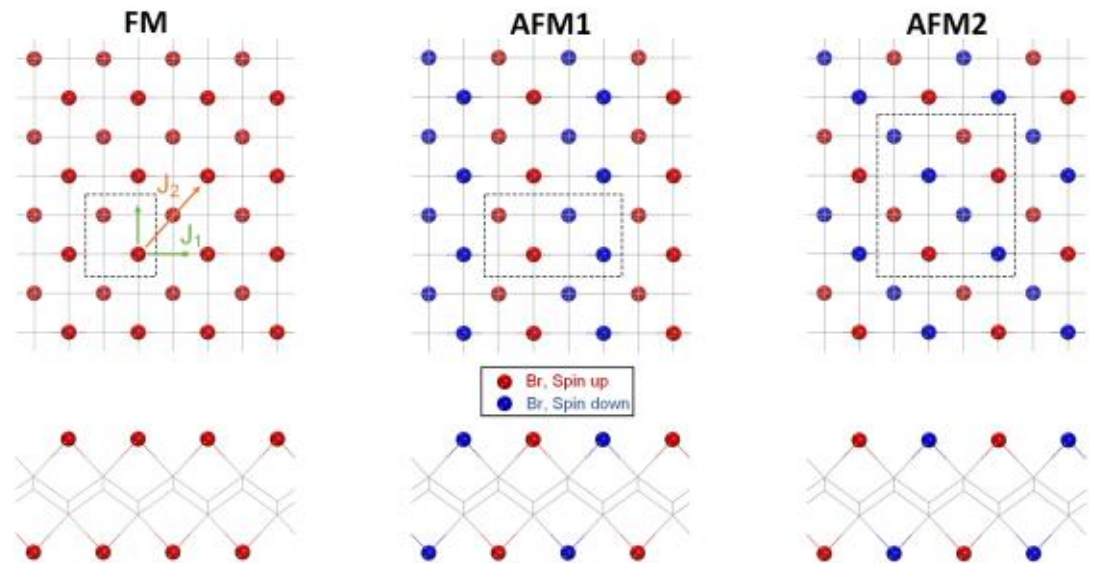

$$E_{FM} = E_0 - 4J_1S^2 - 4J_2S^2$$

$$E_{AFM1} = E_0 + 4J_2S^2$$

$$E_{AFM2} = E_0 + 4J_1S^2 - 4J_2S^2$$

**Magnetic exchange coupling parameters**

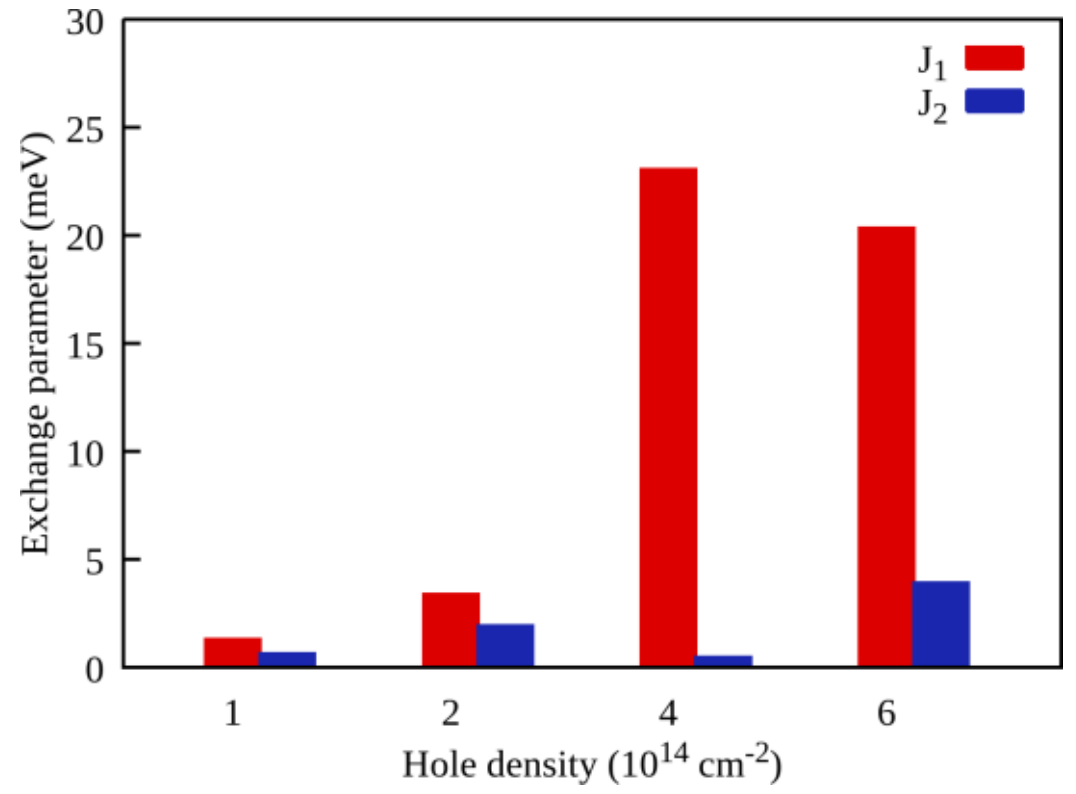

**Magnetic anisotropy energy (MAE, μeV) per magnetic atom**

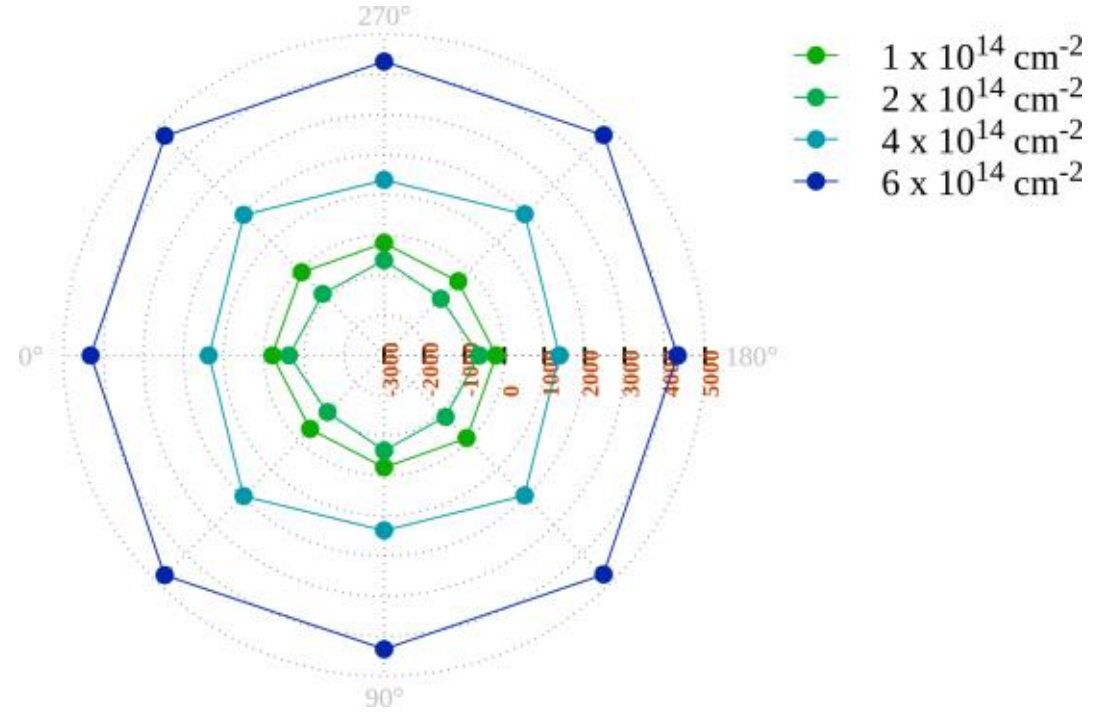

MAE = $E_{\parallel} - E_{\perp}$, a positive (negative) value of MAE indicates the off-plane (in-plane) easy axis.

**Monte Carlo simulations of the normalized magnetization of as a function of temperature**

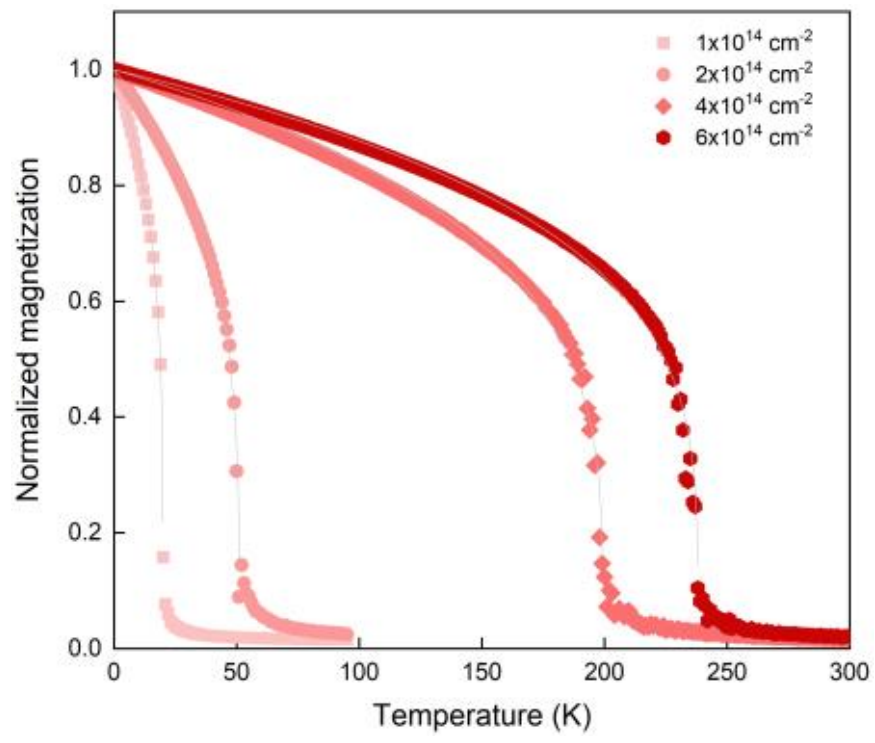

$T_c$: $1\times10^{14}$ cm$^{-2}$: 20 K  $2\times10^{14}$ cm$^{-2}$: 51 K

$4\times10^{14}$ cm$^{-2}$: 199 K  $6\times10^{14}$ cm$^{-2}$: 238 K

# 112. $Sc_2Br_2O_2$

| MC2D-ID | C2DB | 2dmat-ID | USPEX | Space group | Band gap (eV) |
|---|---|---|---|---|---|
| 176 | ✓ | 2dm-4564 | - | P4/nmm | 2.72 |

**Convex hull** | **Atomic structure** | **Atomic coordinates** | **Phonon dispersion curve**

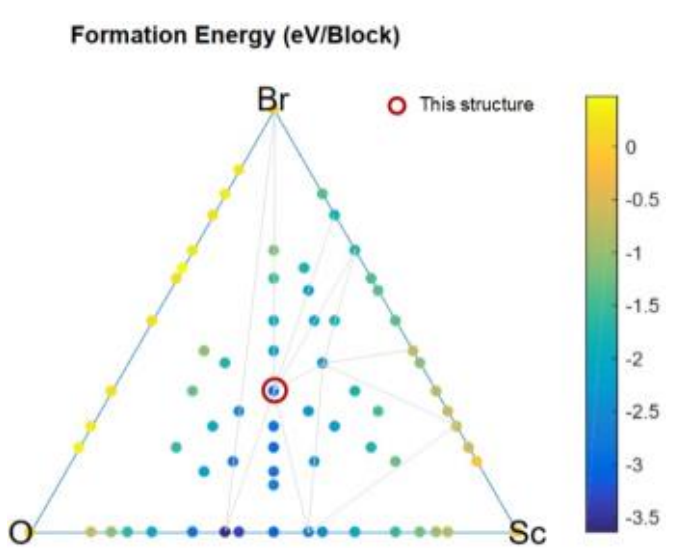

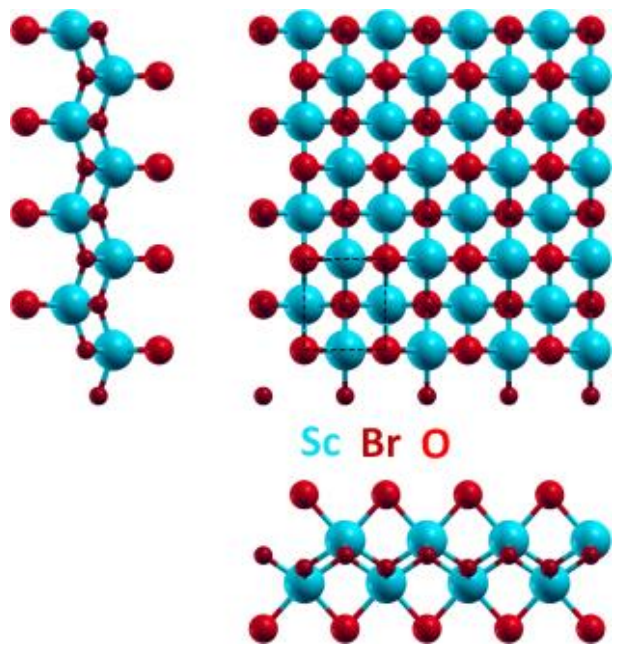

```
Sc2Br2O2
   1.00000000000000
   3.5745649225414975   0.0000000000000000   0.0000000000000000
   0.0000000000000000   3.9886947265690655   0.0000000000000000
   0.0000000000000000   0.0000000000000000  25.8493729999999999
   Sc   Br   O
   2    2    2
Direct
  0.9999998932254002  0.5000000070878201  0.5358870875154977
  0.5000001052406517  0.9999999966401631  0.4641129213449062
  0.4999999522509384  0.5000000653313492  0.6130380560660740
  0.0000000403175449  0.9999999274374076  0.3869618954813916
  0.4999999875516323  0.4999999459086979  0.4892888382271607
  0.0000000214138325  0.0000000575945620  0.5107112013649697
```

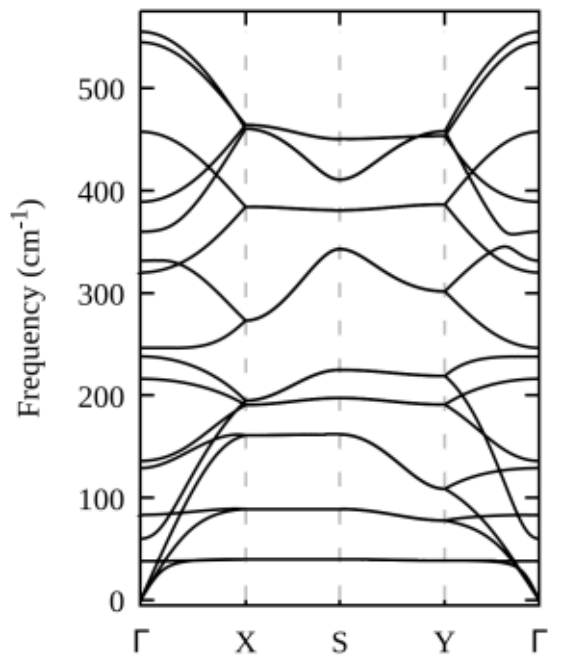

**Projected band structure and density of states** | **Magnetic moment and spin polarization energy as a function of hole doping concentration**

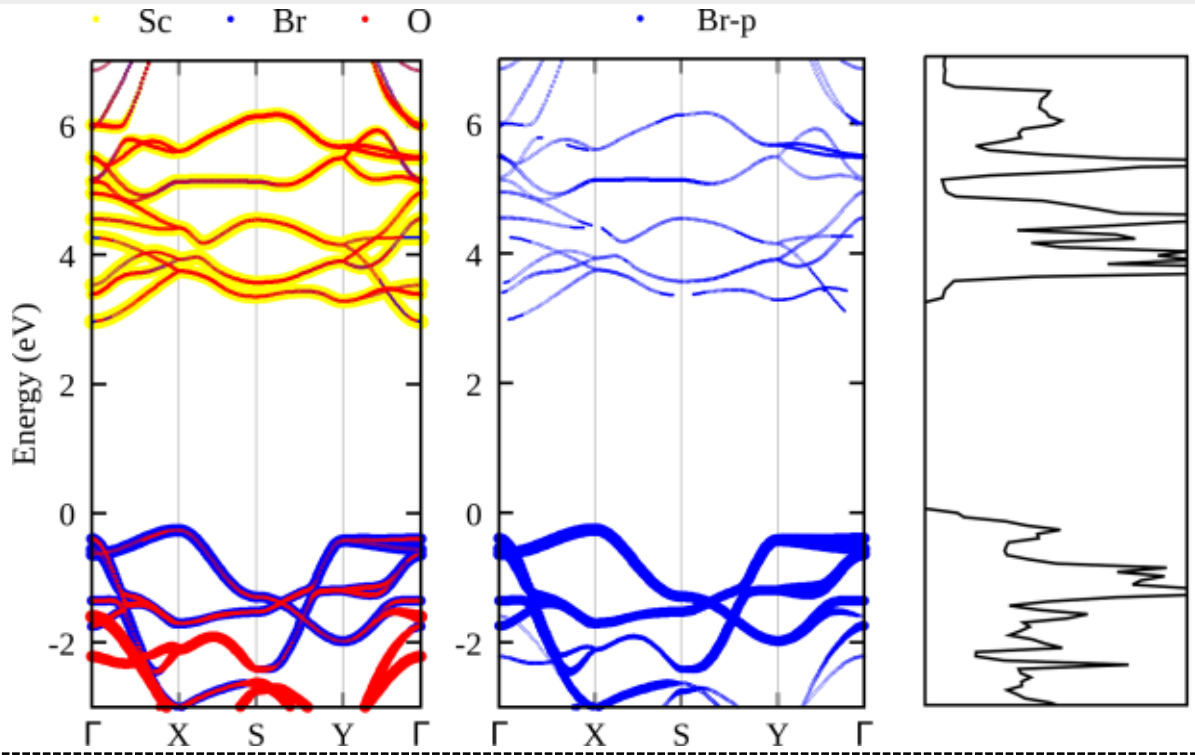

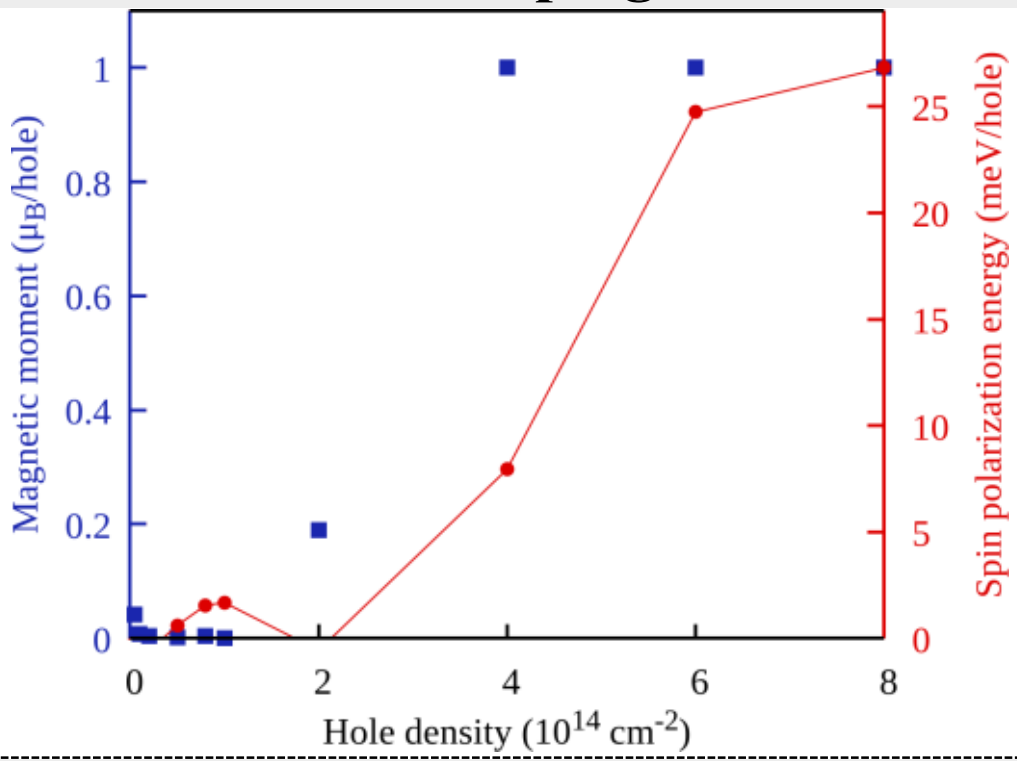

**Magnetic configurations and spin Hamiltonian** | **Magnetic exchange coupling parameters**

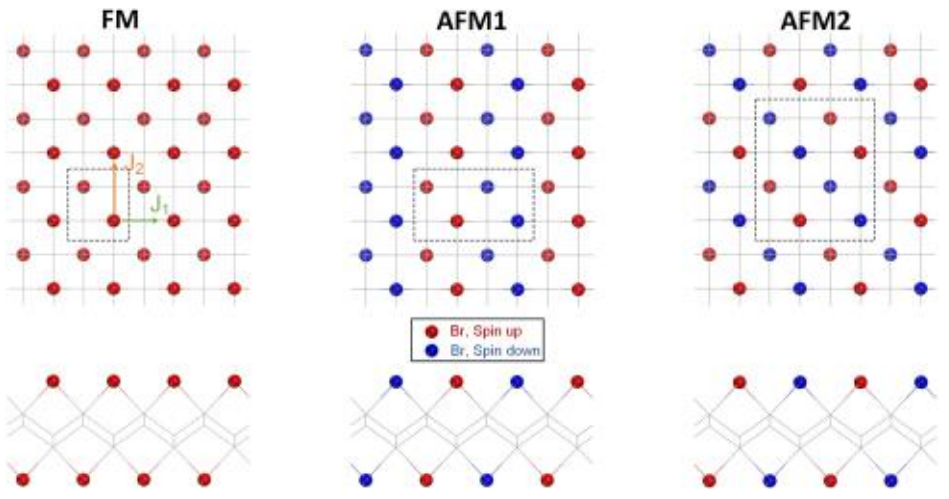

$$E_{FM} = E_0 - 2J_1S^2 - 2J_2S^2$$
$$E_{AFM1} = E_0 + 2J_1S^2 - 2J_2S^2$$
$$E_{AFM2} = E_0 + 2J_1S^2 + 2J_2S^2$$

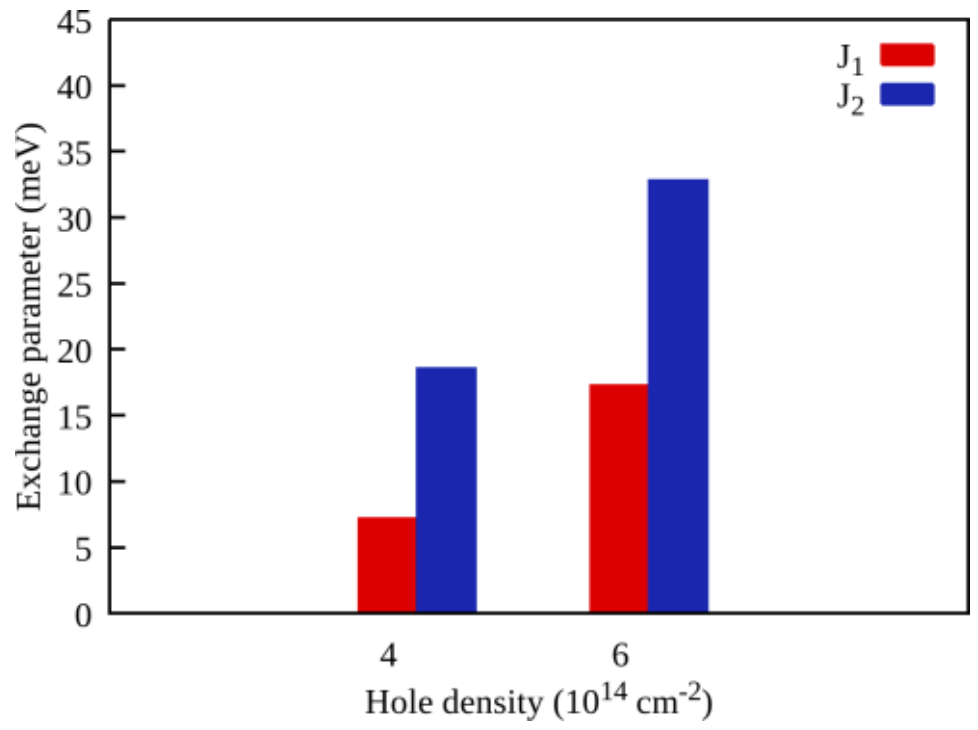

**Magnetic anisotropy energy (MAE, μeV) per magnetic atom** | **Monte Carlo simulations of the normalized magnetization of as a function of temperature**

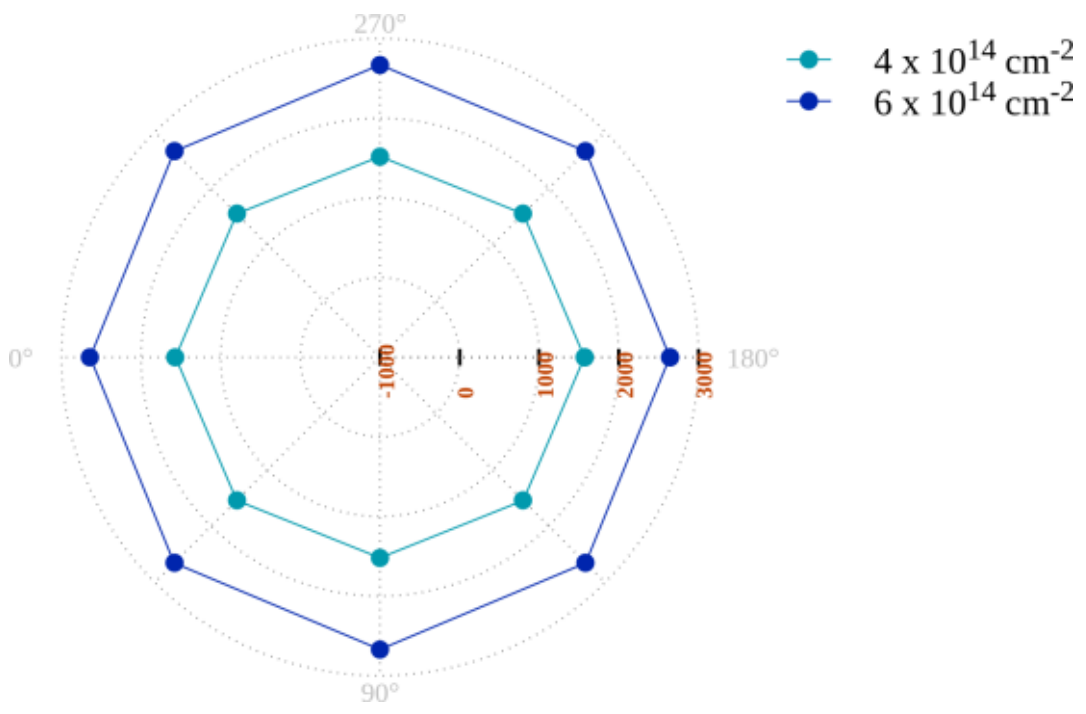

MAE = $E_{\parallel} - E_{\perp}$, a positive (negative) value of MAE indicates the off-plane (in-plane) easy axis.

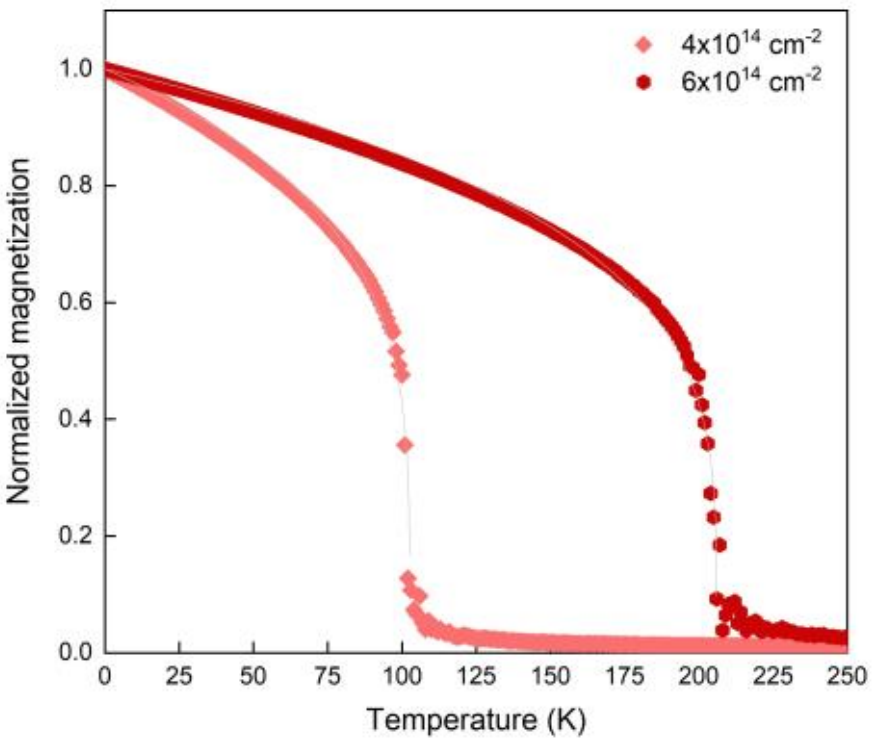

$T_c$: $1\times10^{14}$ cm$^{-2}$: - K  $2\times10^{14}$ cm$^{-2}$: - K

$4\times10^{14}$ cm$^{-2}$: 103 K  $6\times10^{14}$ cm$^{-2}$: 206 K

# 113. $Sr_2Br_2H_2$

| MC2D-ID | C2DB | 2dmat-ID | USPEX | Space group | Band gap (eV) |
|---|---|---|---|---|---|
| 188 | ✓ | - | - | P4/nmm | 4.34 |

**Convex hull** | **Atomic structure** | **Atomic coordinates** | **Phonon dispersion curve**

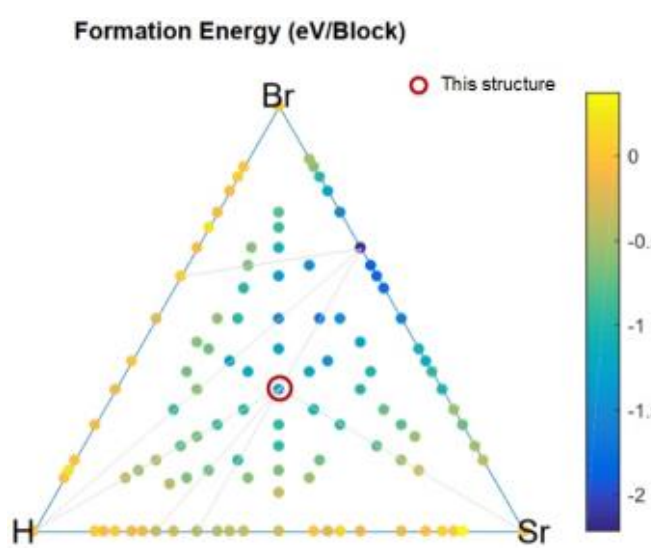

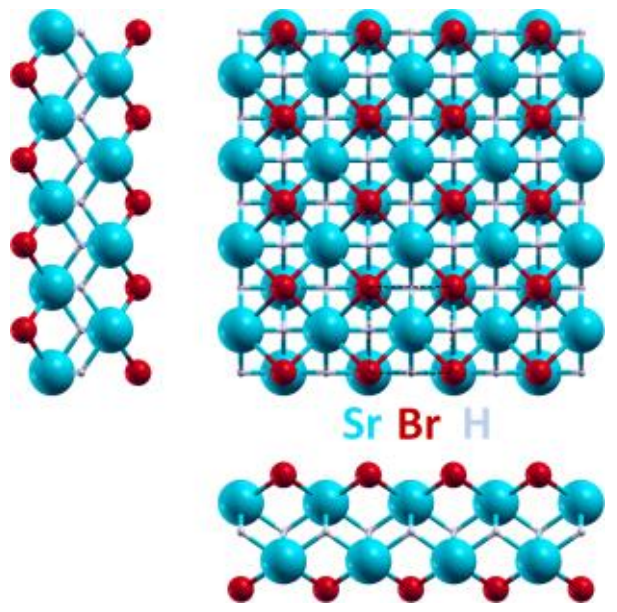

```
Sr2Br2H2
   1.00000000000000
   4.1041140556000002   0.0000000000000000   0.0000000000000000
   0.0000000000000000   4.1041140556000002   0.0000000000000000
   0.0000000000000000   0.0000000000000000   25.0283088683999999
   Sr   Br   H
   2    2    2
Direct
  0.0000000000000000  0.5000000000000000  0.5562024709999989
  0.5000000000000000  0.0000000000000000  0.4437975589999965
  0.5000000000000000  0.0000000000000000  0.6102313399999986
  0.0000000000000000  0.5000000000000000  0.3897686600000014
  0.0000000000000000  0.0000000000000000  0.5000000000000000
  0.5000000000000000  0.5000000000000000  0.5000000000000000
```

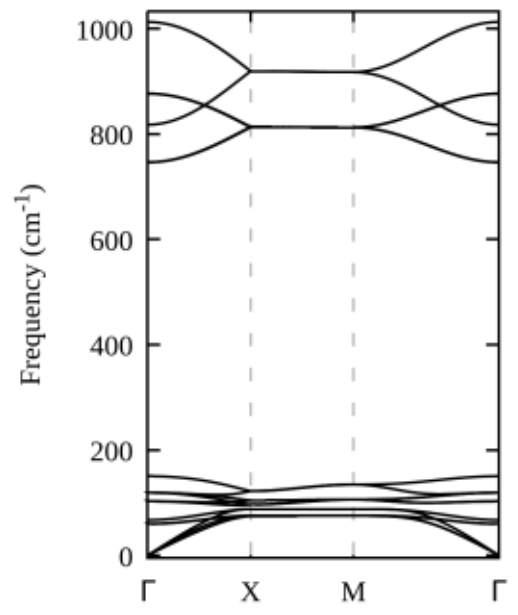

**Projected band structure and density of states** | **Magnetic moment and spin polarization energy as a function of hole doping concentration**

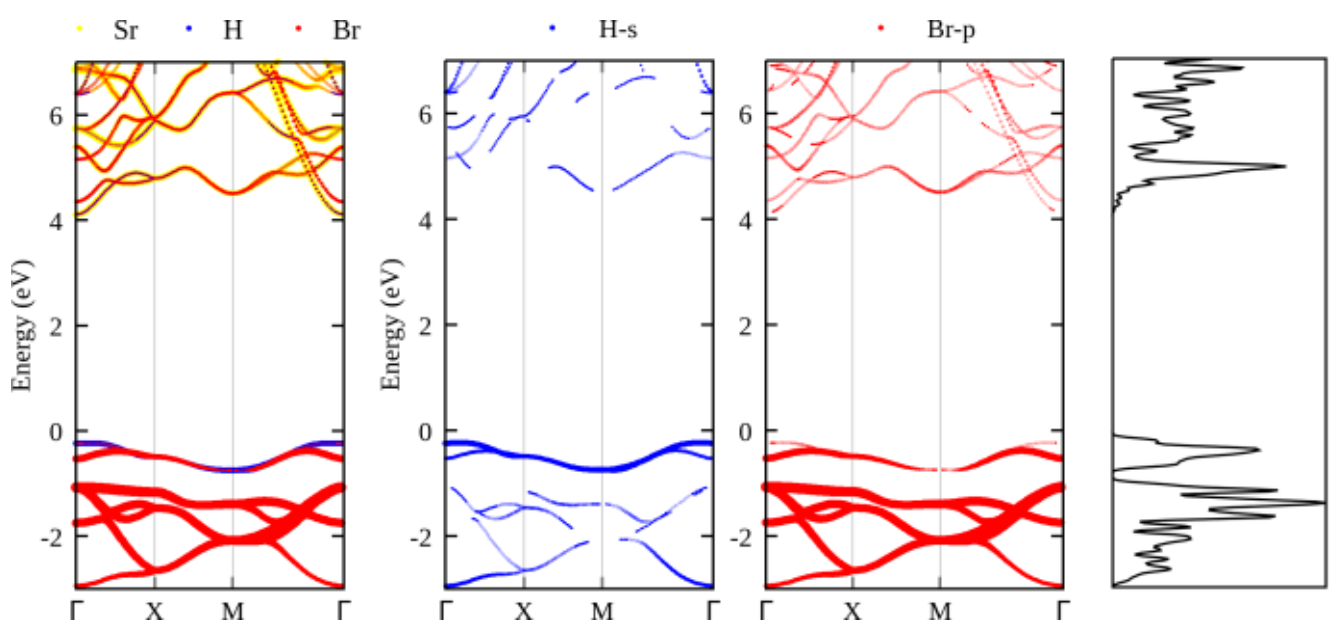

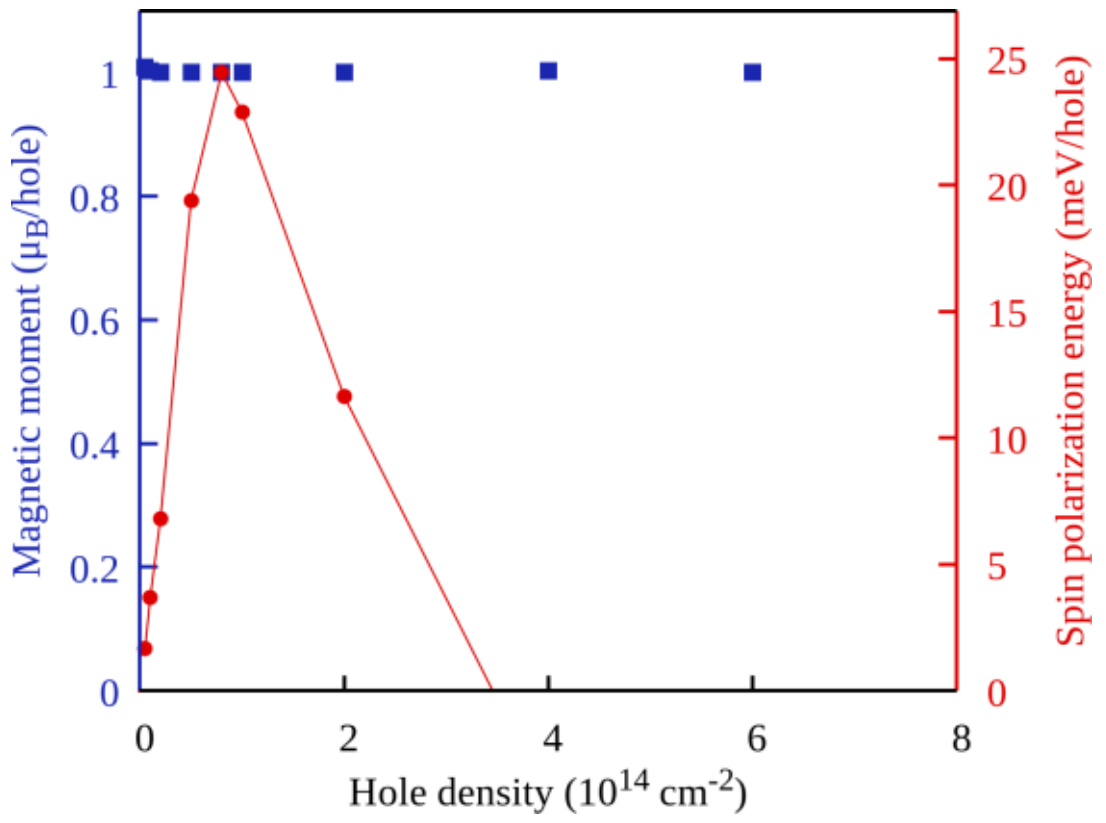

**Magnetic configurations and spin Hamiltonian** | **Magnetic exchange coupling parameters**

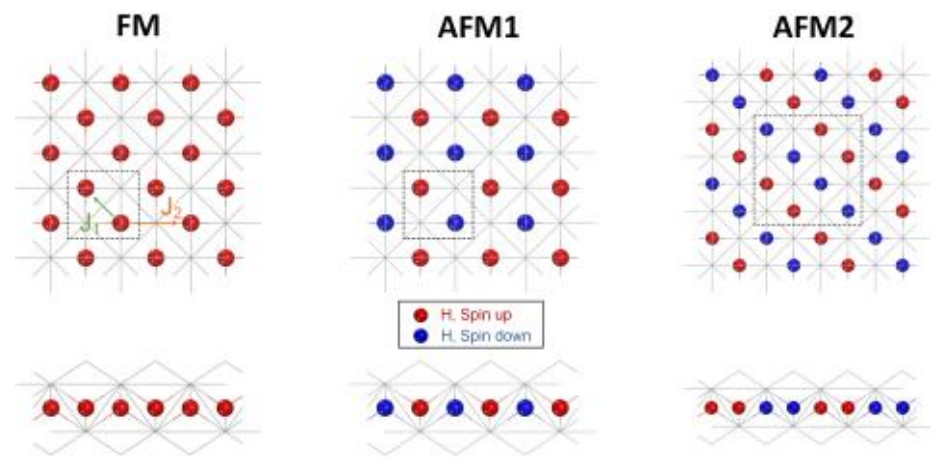

$$E_{FM} = E_0 - 4J_1S^2 - 4J_2S^2$$
$$E_{AFM1} = E_0 + 4J_1S^2 - 4J_2S^2$$
$$E_{AFM2} = E_0 + 4J_2S^2$$

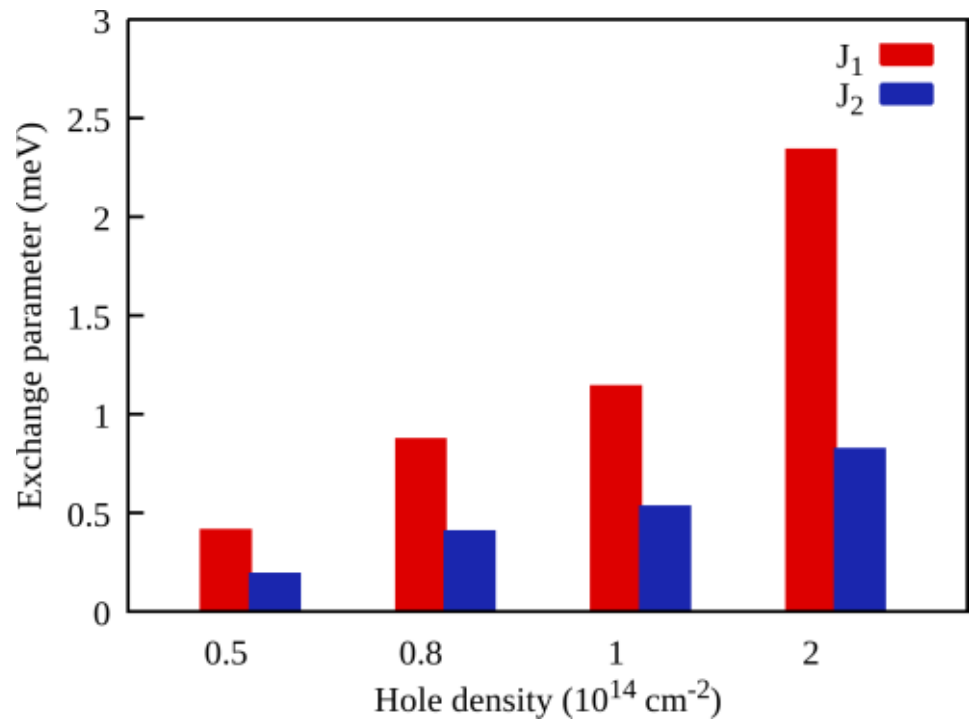

**Magnetic anisotropy energy (MAE, μeV) per magnetic atom** | **Monte Carlo simulations of the normalized magnetization of as a function of temperature**

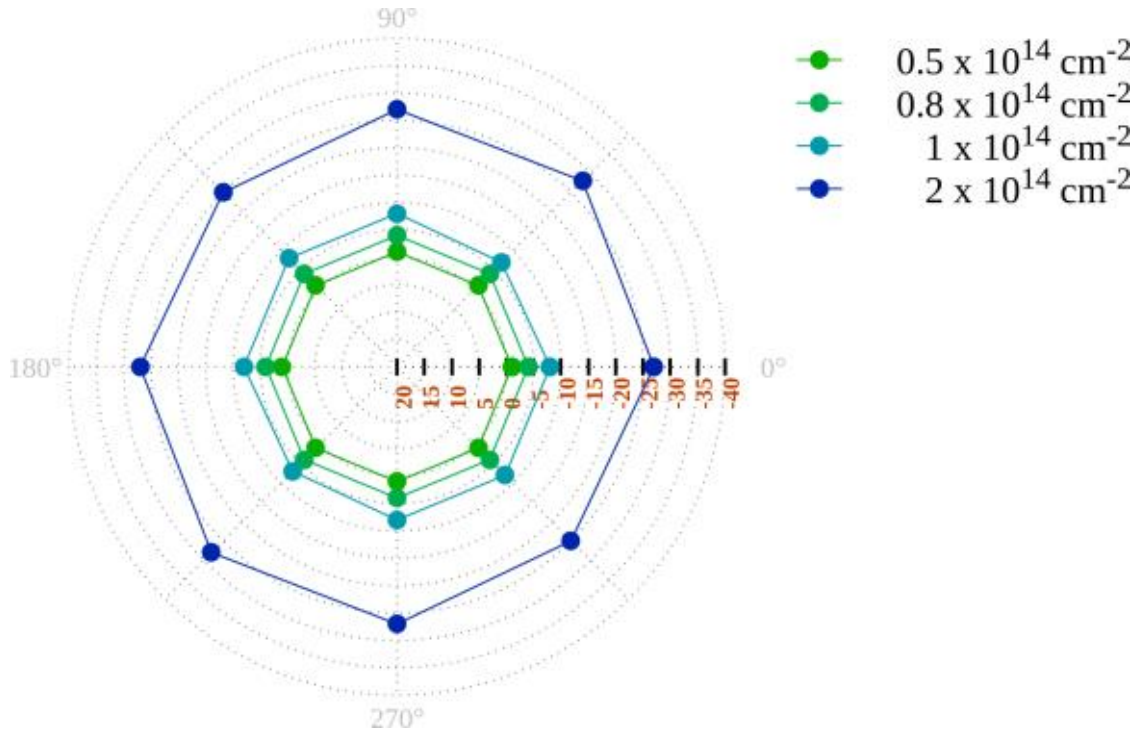

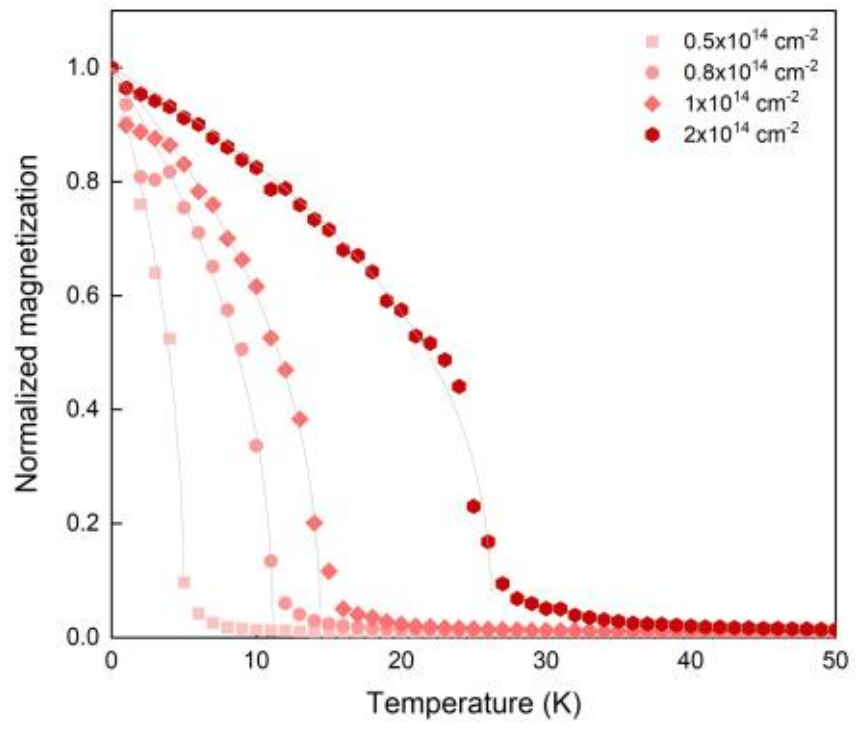

MAE = $E_{\parallel} - E_{\perp}$, a positive (negative) value of MAE indicates the off-plane (in-plane) easy axis.

$T_c$: $0.5\times10^{14}$ cm$^{-2}$: 5 K   $0.8\times10^{14}$ cm$^{-2}$: 11 K
$1\times10^{14}$ cm$^{-2}$: 14 K   $2\times10^{14}$ cm$^{-2}$: 26 K

# 114. $AlF_3$

| MC2D-ID | C2DB | 2dmat-ID | USPEX | Space group | Band gap (eV) |
|---|---|---|---|---|---|
| - | - | - | ✓ | Pmmm | 7.73 |

**Convex hull** | **Atomic structure** | **Atomic coordinates** | **Phonon dispersion curve**

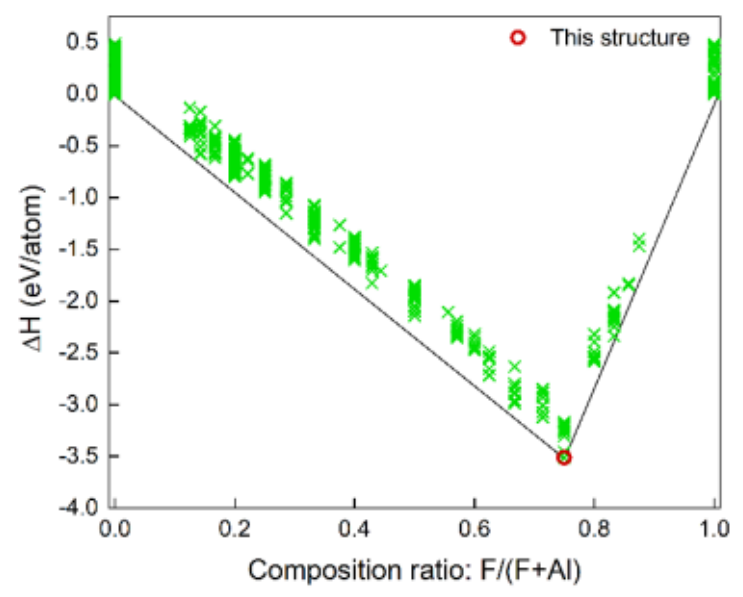

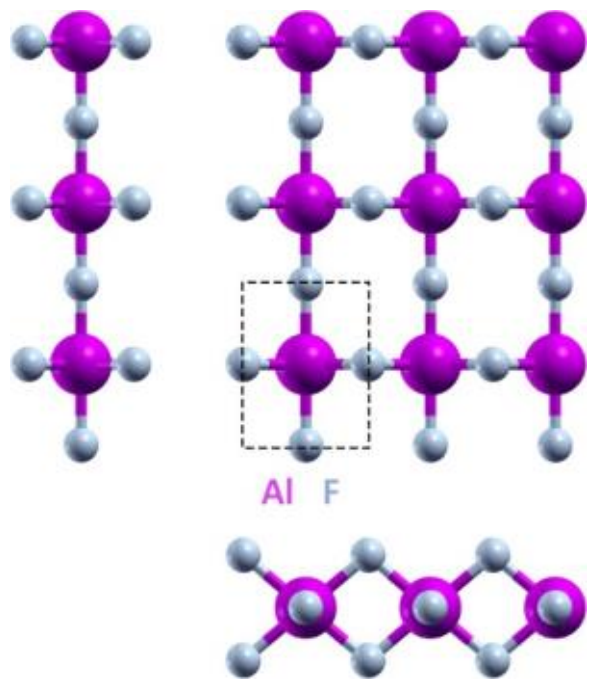

```
AlF3
    1.00000000000000
    2.8217698873637747    0.0000000000000000    0.0000000000000000
    0.0000000000000000    3.6404754518615086    0.0000000000000000
    0.0000000000000000    0.0000000000000000    20.0000000000000000
    Al    F
    1     3
Direct
   0.5011636113421850    0.5617777439934213    0.3749814757054186
   0.0011379468021243    0.5617020862032926    0.3155663691980434
   0.5011793557123678    0.0617761716906813    0.3750111511360803
   0.0011721077792330    0.5618489820967347    0.4344402663104455
```

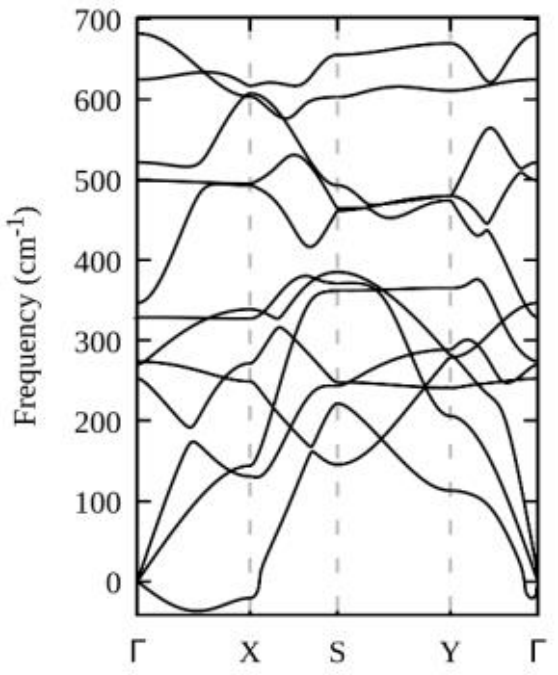

**Projected band structure and density of states**

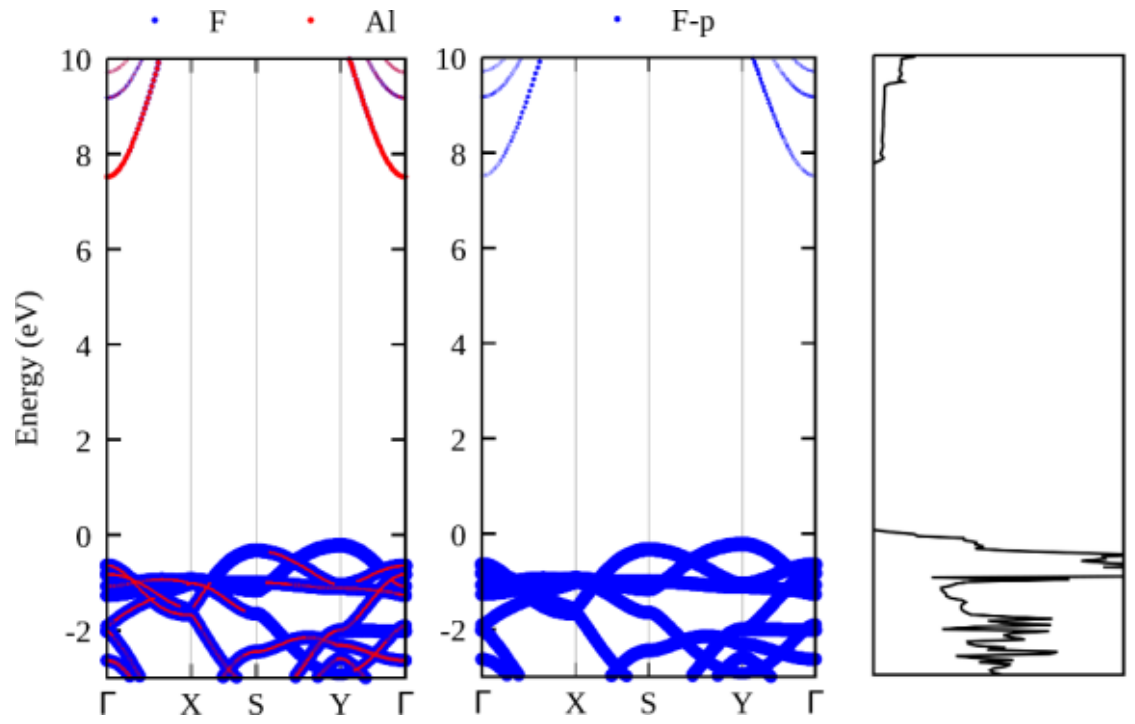

**Magnetic moment and spin polarization energy as a function of hole doping concentration**

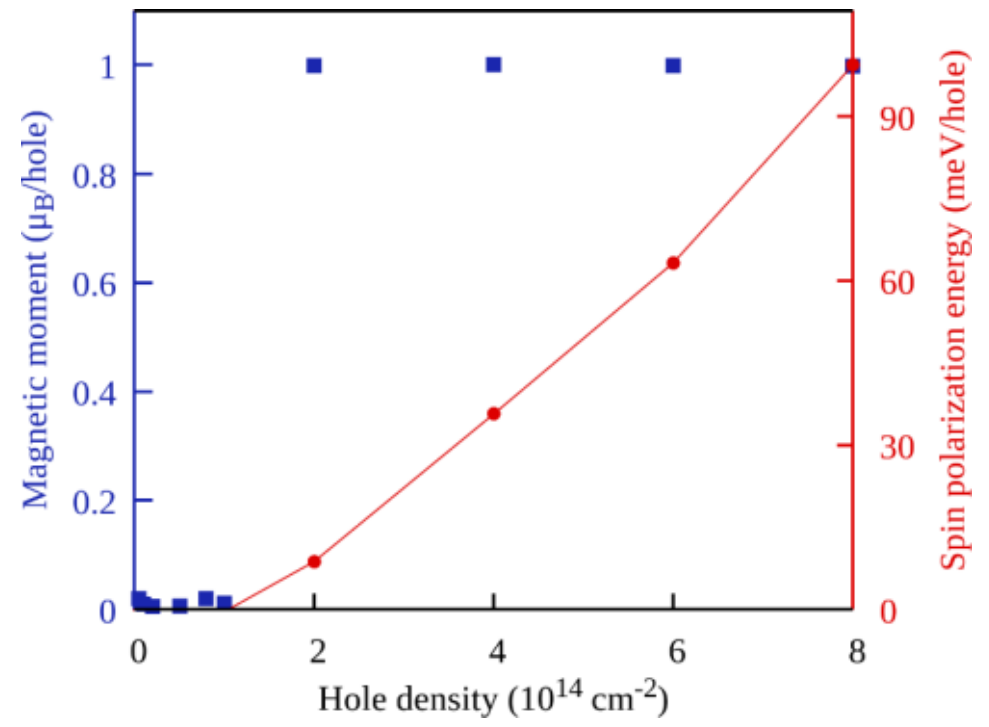

**Magnetic configurations and spin Hamiltonian**

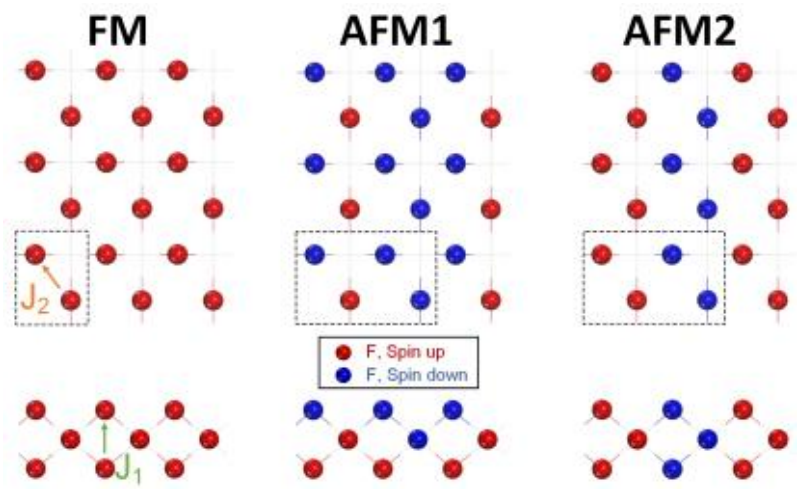

$$E_{FM} = E_0 - J_1S^2 - 8J_2S^2$$
$$E_{AFM1} = E_0 + J_1S^2$$
$$E_{AFM2} = E_0 - J_1S^2$$

**Magnetic exchange coupling parameters**

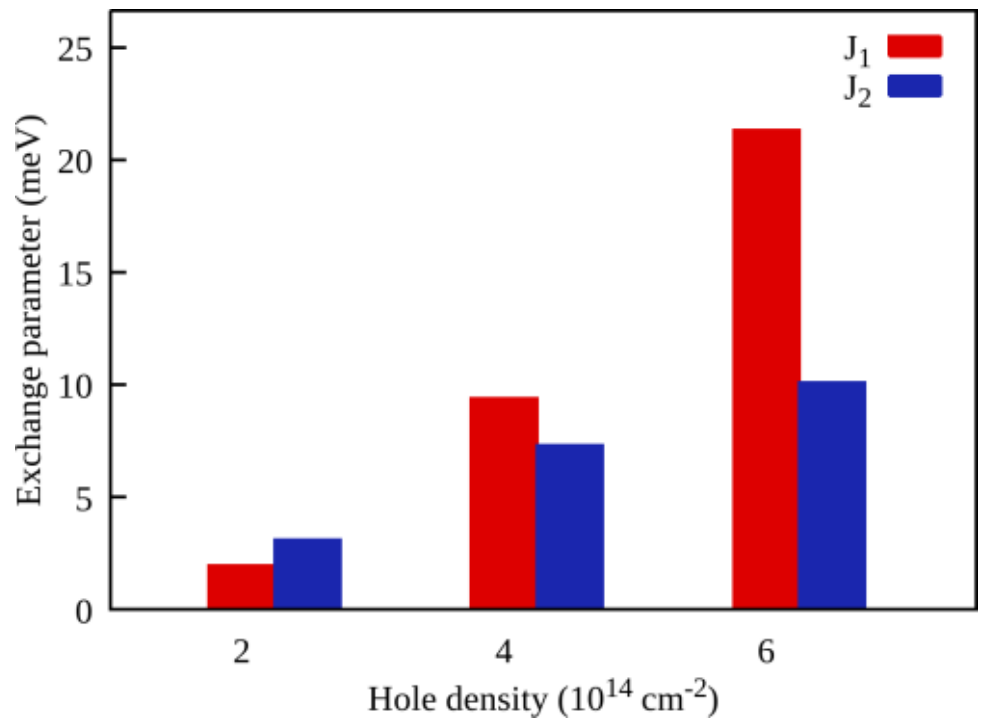

**Magnetic anisotropy energy (MAE, μeV) per magnetic atom**

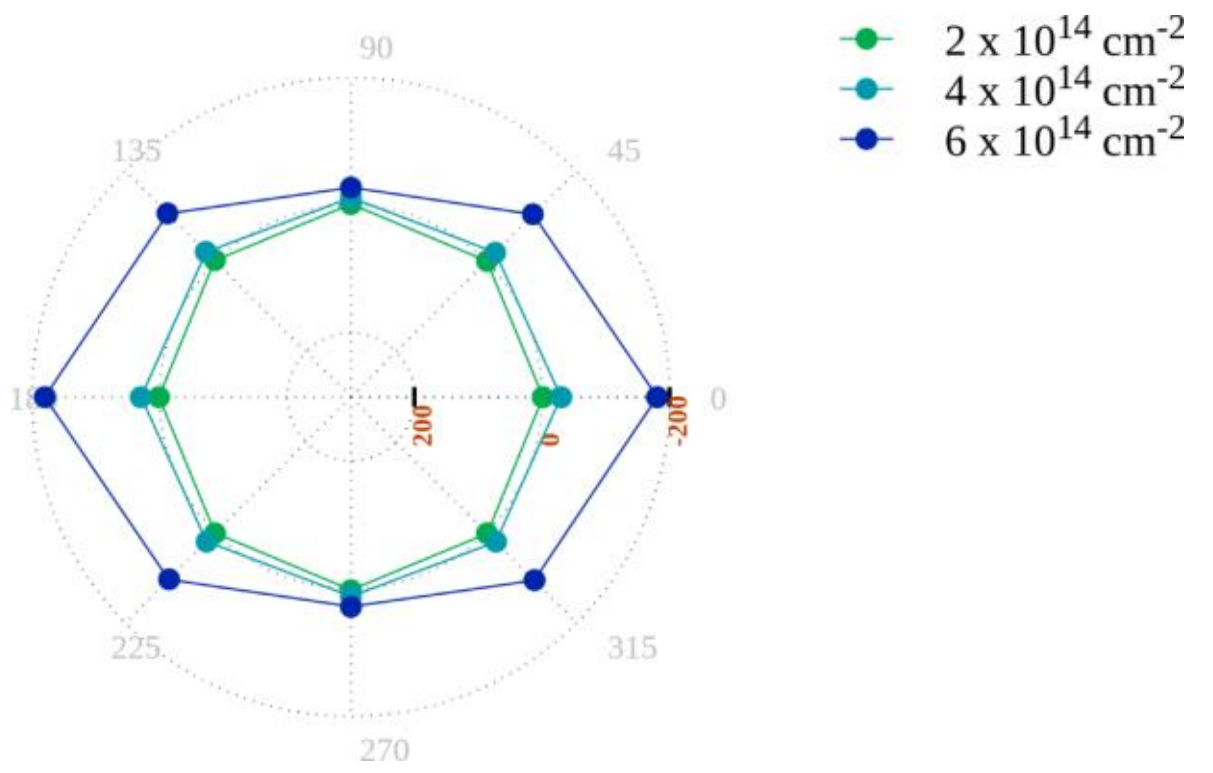

MAE = $E_{\parallel} - E_{\perp}$, a positive (negative) value of MAE indicates the off-plane (in-plane) easy axis.

**Monte Carlo simulations of the normalized magnetization of as a function of temperature**

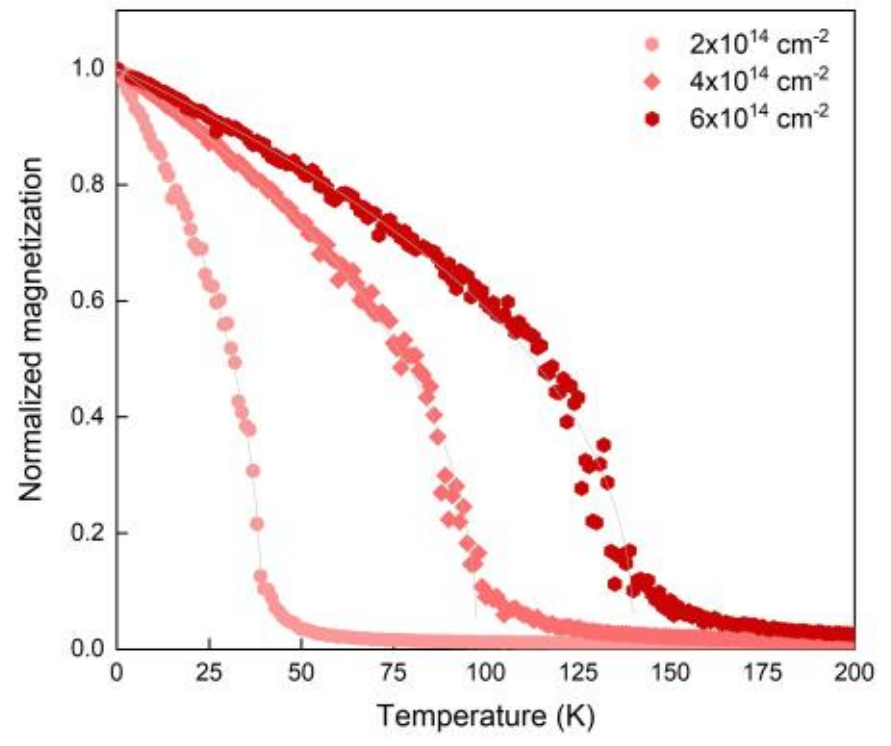

$T_c$: $1\times10^{14}$ cm$^{-2}$: - K    $2\times10^{14}$ cm$^{-2}$: 39 K

$4\times10^{14}$ cm$^{-2}$: 97 K    $6\times10^{14}$ cm$^{-2}$: 140 K

# 115. $Al_4N_4$

| MC2D-ID | C2DB | 2dmat-ID | USPEX | Space group | Band gap (eV) |
|---|---|---|---|---|---|
| - | - | 2dm-2590 | - | P4/mbm | 2.86 |

## Convex hull

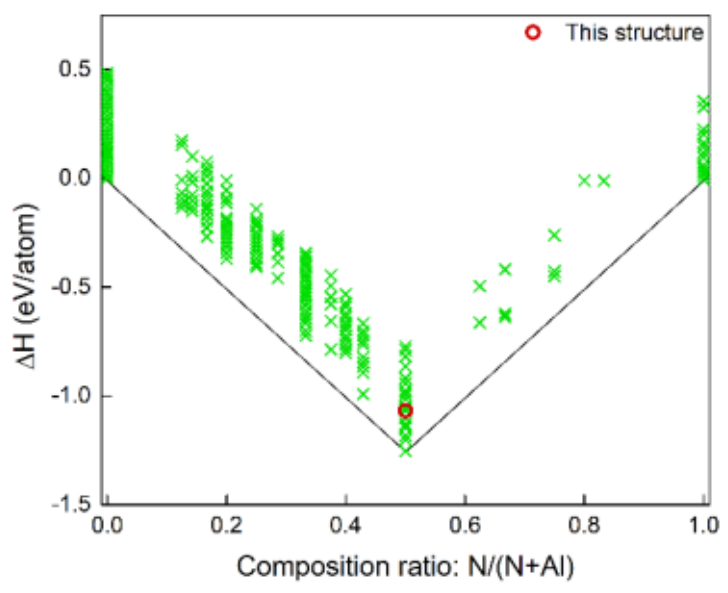

## Atomic structure

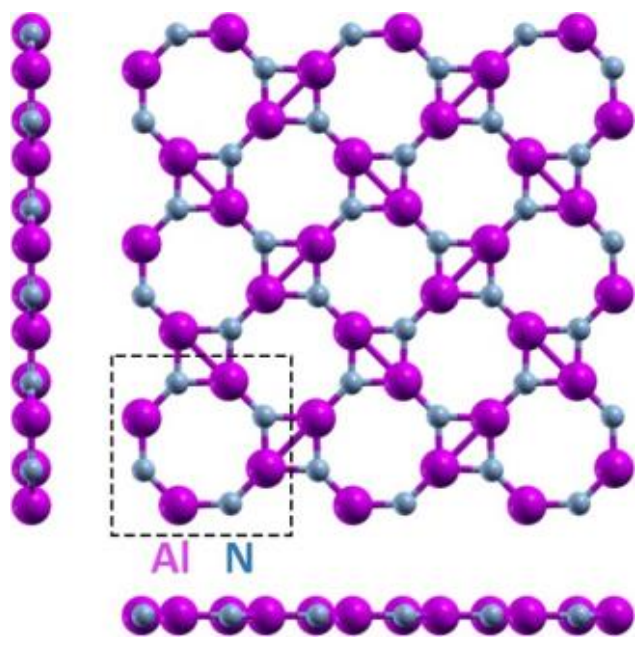

## Atomic coordinates

```
Al4 N4
   1.00000000000000
   6.156056481602916   0.000000000000000   0.000000000000000
   0.000000000000000   6.156056394760616 6 0.000000000000000
   0.000000000000000   0.000000000000000   20.000000000000000
   Al   N
   4    4
Direct
  0.3561749438853923  0.1438250552105913  0.6067160362650981
  0.1438250167987931  0.6438250168000081  0.6067160376887344
  0.8561749845307816  0.3561749838010471  0.6067160340132887
  0.6438250553214573  0.8561749445924391  0.6067160357455847
  0.1536257579985119  0.3463742399625076  0.6067159645347076
  0.8463742404299310  0.6536257593352417  0.6067159638015340
  0.6536257948346247  0.1536257898751501  0.6067159680978946
  0.3463742062005082  0.8463742104230079  0.6067159598531475
```

## Phonon dispersion curve

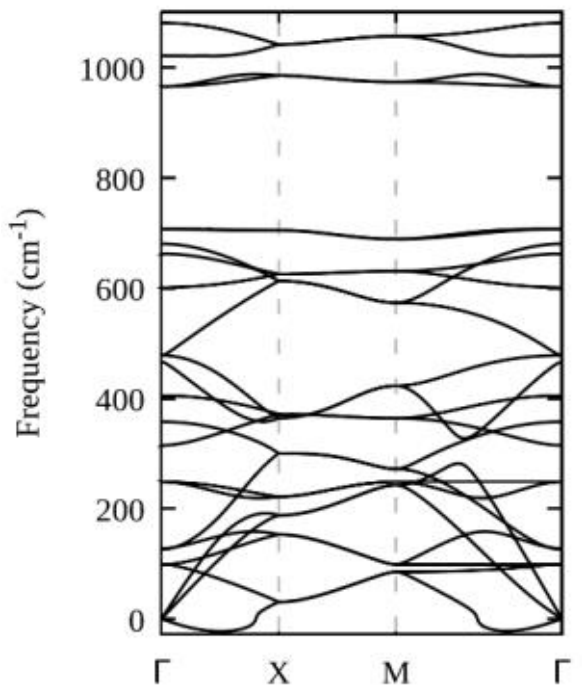

## Projected band structure and density of states

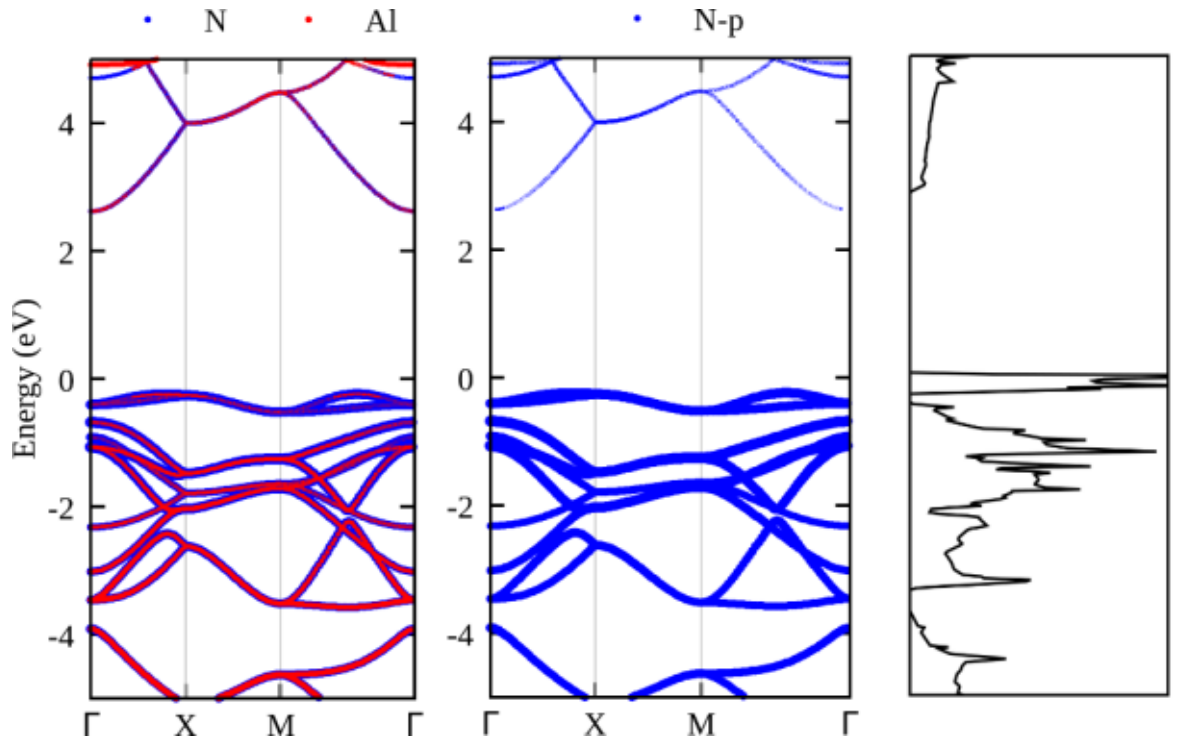

## Magnetic moment and spin polarization energy as a function of hole doping concentration

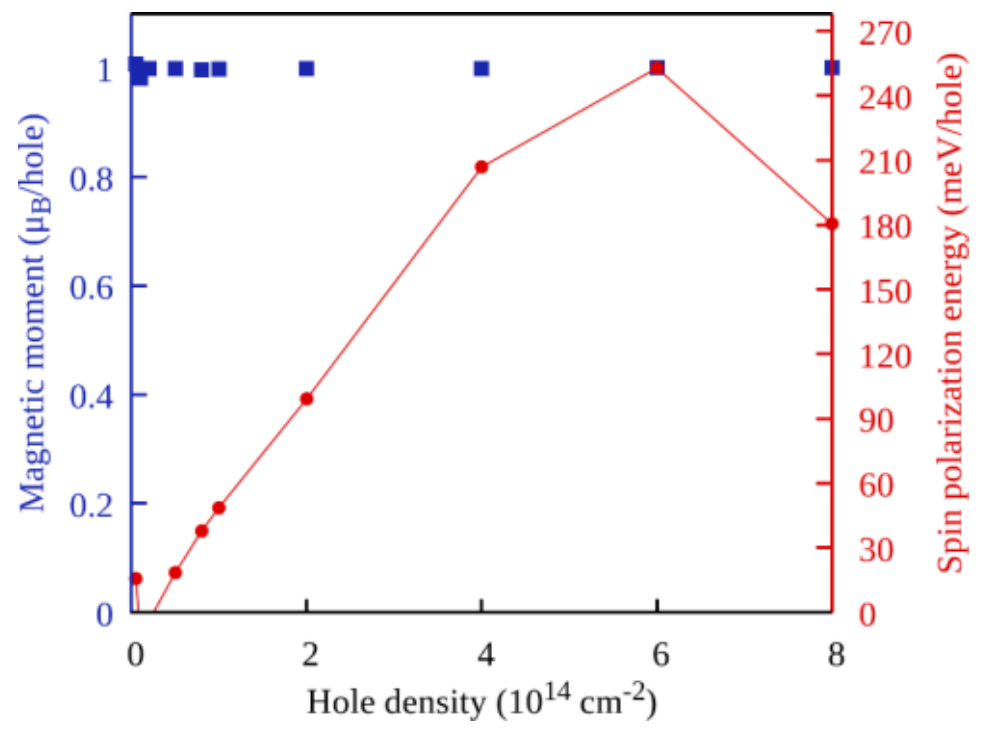

## Magnetic configurations and spin Hamiltonian

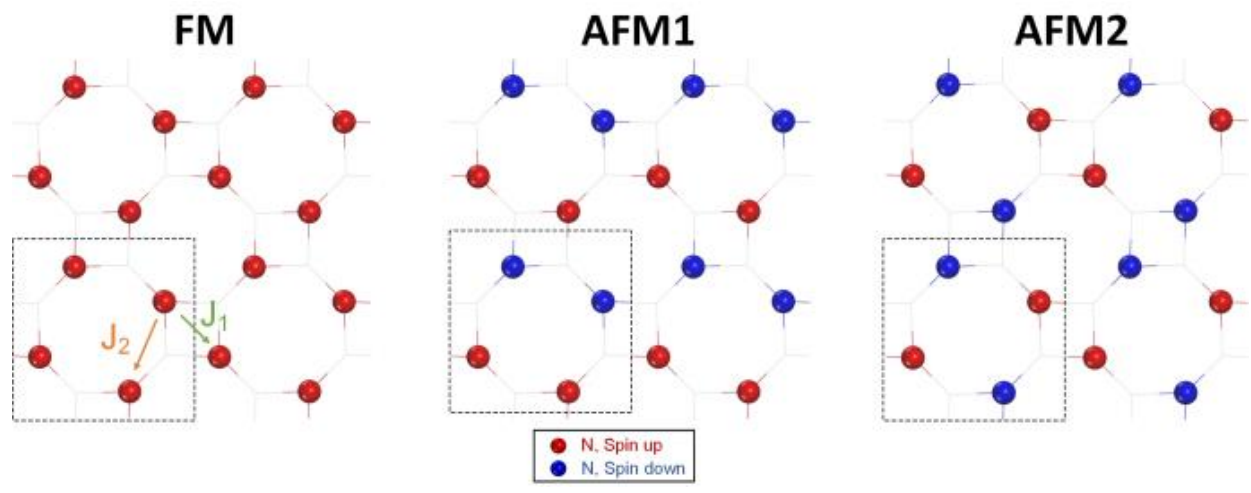

$$E_{FM} = E_0 - 2J_1S^2 - 8J_2S^2$$
$$E_{AFM1} = E_0 + 2J_1S^2$$
$$E_{AFM2} = E_0 - 2J_1S^2 + 8J_2S^2$$

## Magnetic exchange coupling parameters

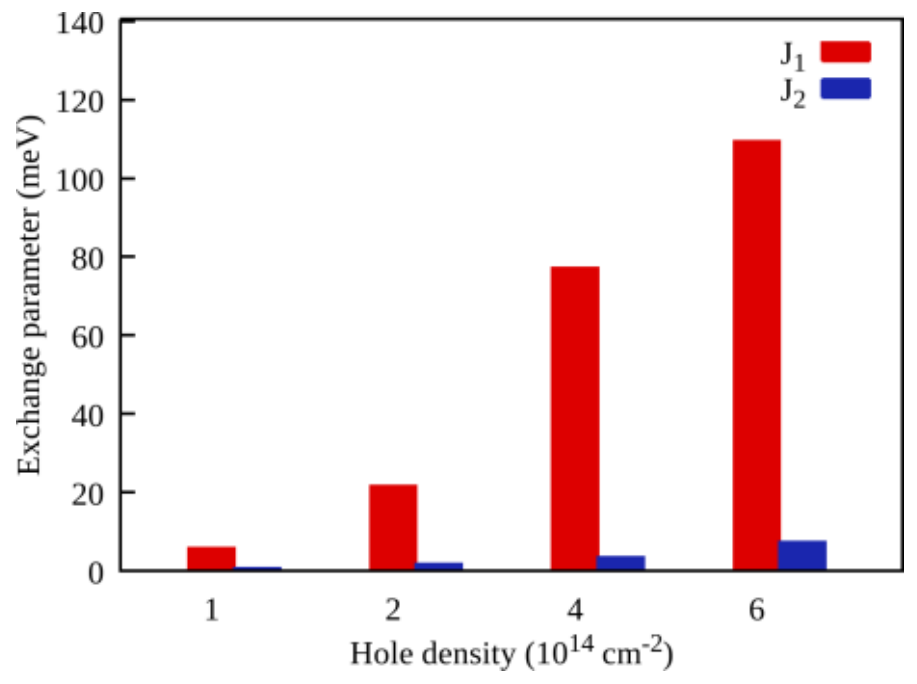

## Magnetic anisotropy energy (MAE, μeV) per magnetic atom

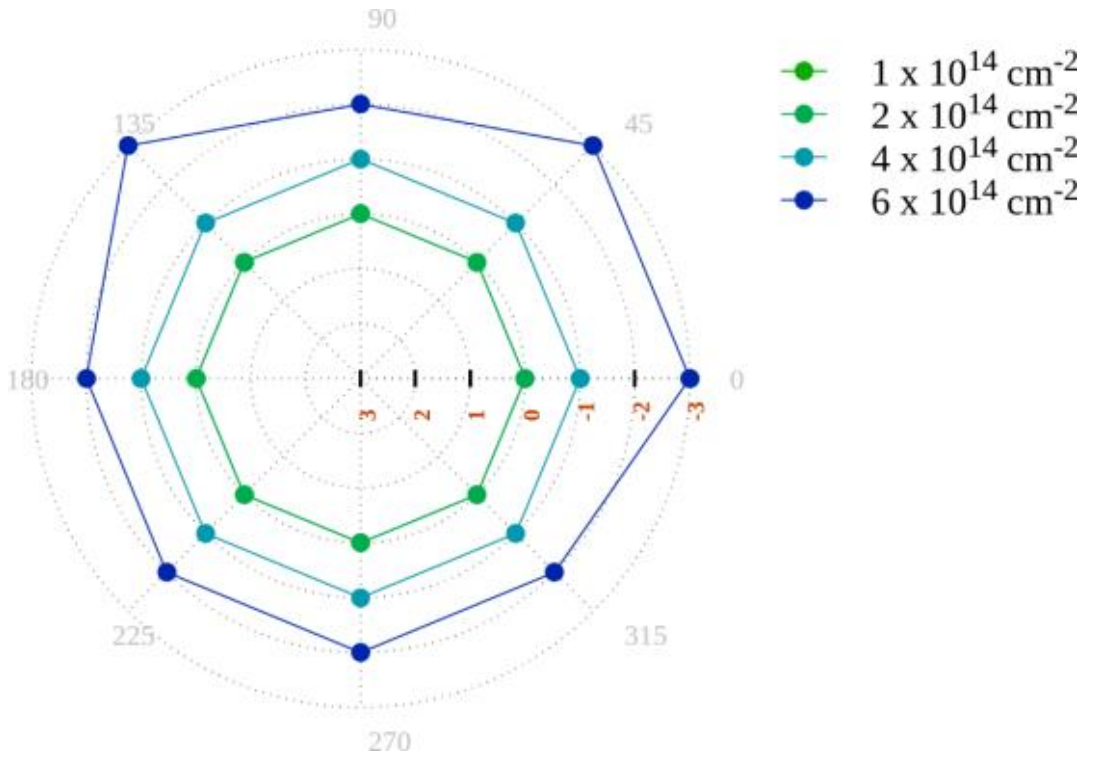

MAE = $E_{\parallel} - E_{\perp}$, a positive (negative) value of MAE indicates the off-plane (in-plane) easy axis.

## Monte Carlo simulations of the normalized magnetization of as a function of temperature

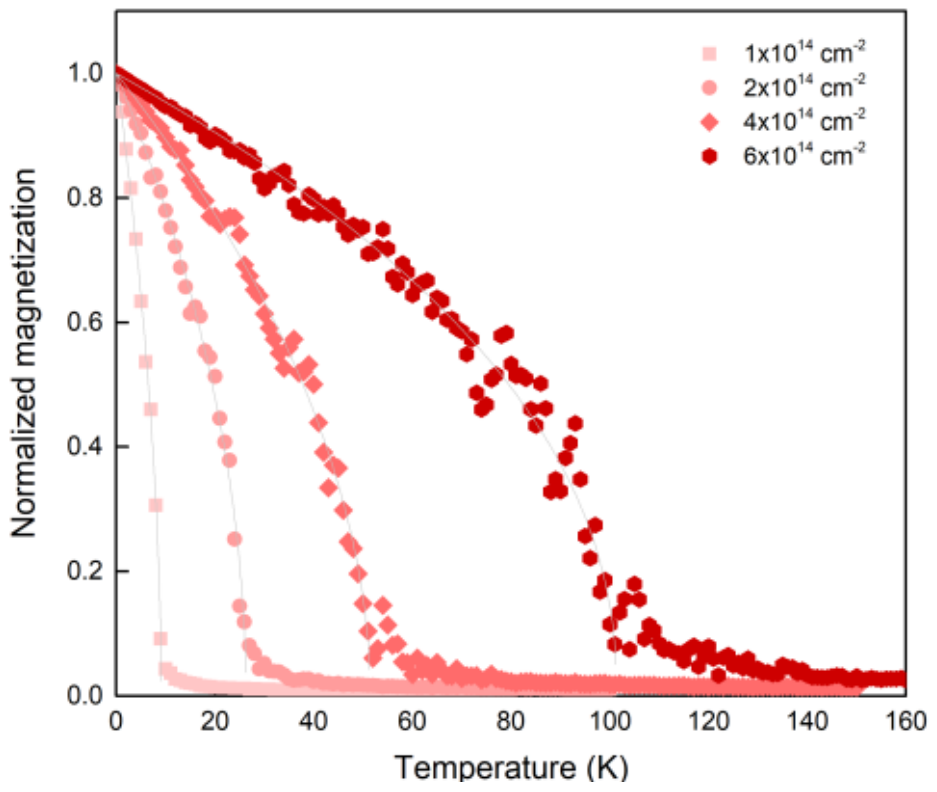

$T_c$: $1\times10^{14}$ cm$^{-2}$: 9 K　$2\times10^{14}$ cm$^{-2}$: 26 K　$4\times10^{14}$ cm$^{-2}$: 51 K　$6\times10^{14}$ cm$^{-2}$: 101 K

# 116. $Ga_4N_4$

| MC2D-ID | C2DB | 2dmat-ID | USPEX | Space group | Band gap (eV) |
|---|---|---|---|---|---|
| - | - | 2dm-2582 | - | P4/mbm | 2.05 |

**Convex hull**

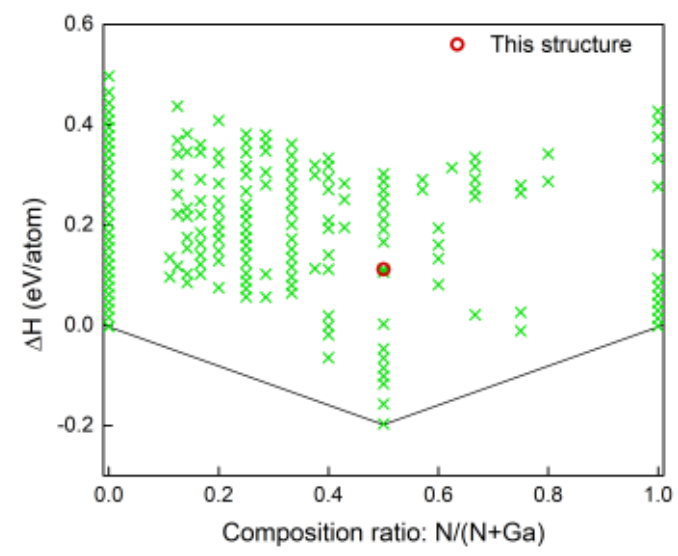

**Atomic structure**

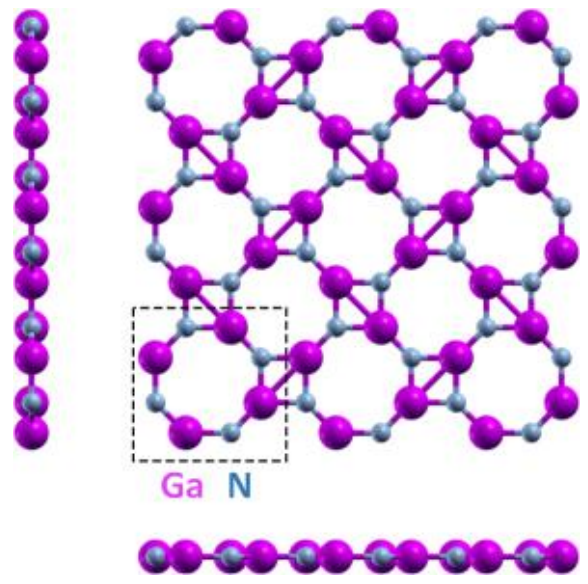

**Atomic coordinates**

```
Ga4N4
   1.00000000000000
   6.3251827589122840    0.0000000000000000    0.0000000000000000
   0.0000000000000000    6.3251971034337258    0.0000000000000000
   0.0000000000000000    0.0000000000000000   20.0000000000000000
   Ga    N
     4     4
Direct
  0.3503042731550039  0.1496967050798901  0.6067164713012545
  0.1496947757320584  0.6496939522736795  0.6067161534779828
  0.8503036384762837  0.3503025847410797  0.6067167725671752
  0.6496944167056071  0.8503064182194251  0.6067164587253018
  0.1490958482973710  0.3509043025197585  0.6067155413565217
  0.8509058038166657  0.6490947246571181  0.6067155360478154
  0.6490955845930699  0.1490957611569925  0.6067154072219907
  0.3509056592239332  0.8509055513520494  0.6067156593019476
```

**Phonon dispersion curve**

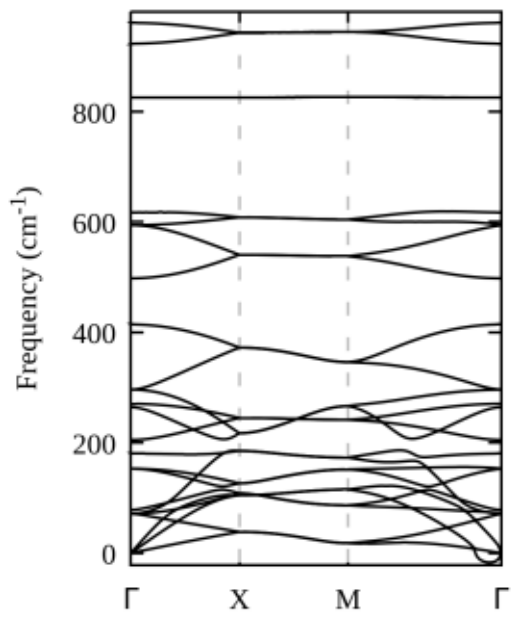

**Projected band structure and density of states**

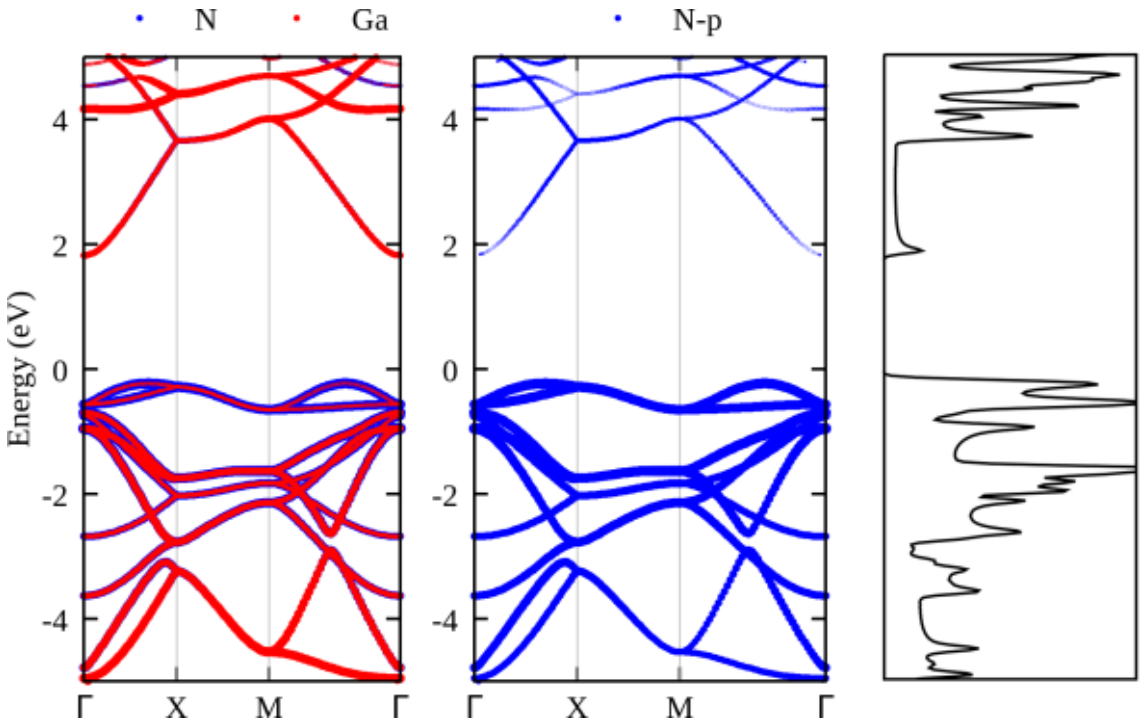

**Magnetic moment and spin polarization energy as a function of hole doping concentration**

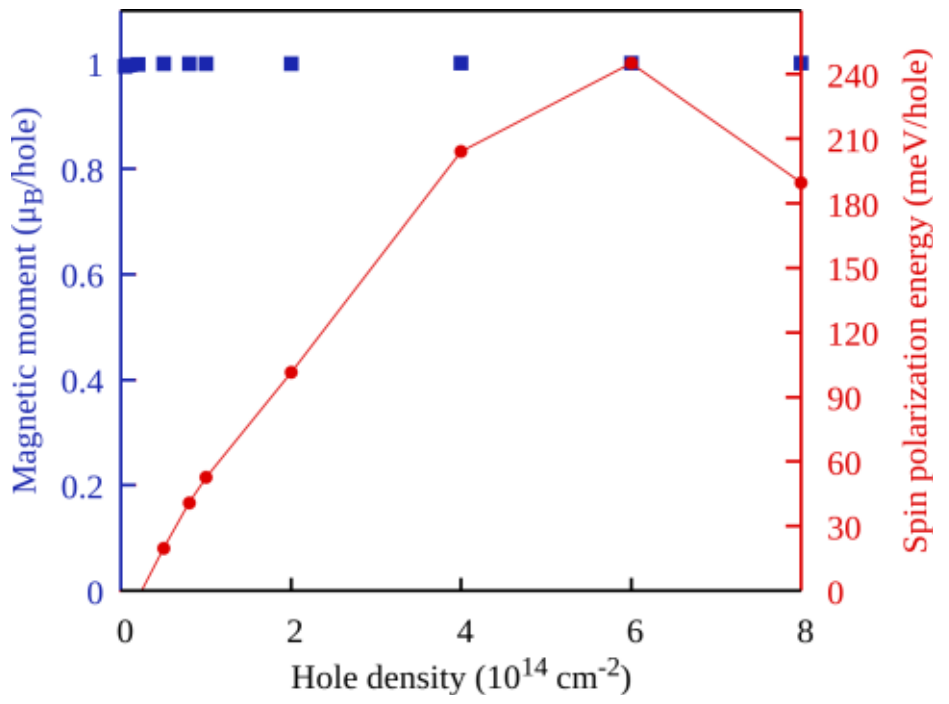

**Magnetic configurations and spin Hamiltonian**

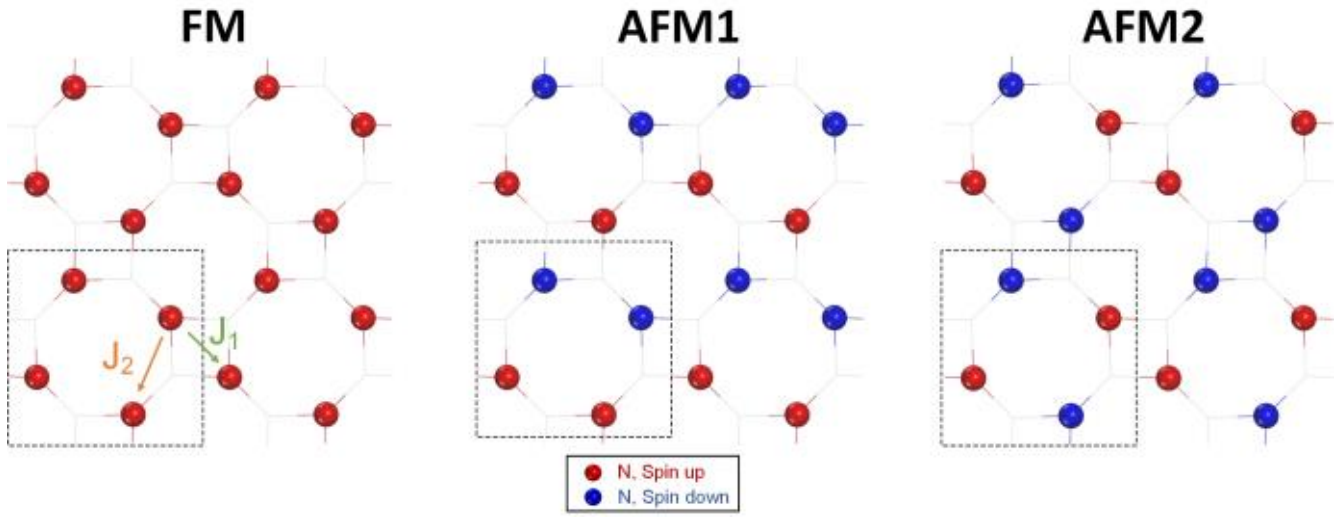

$$E_{FM} = E_0 - 2J_1S^2 - 8J_2S^2$$

$$E_{AFM1} = E_0 + 2J_1S^2$$

$$E_{AFM2} = E_0 - 2J_1S^2 + 8J_2S^2$$

**Magnetic exchange coupling parameters**

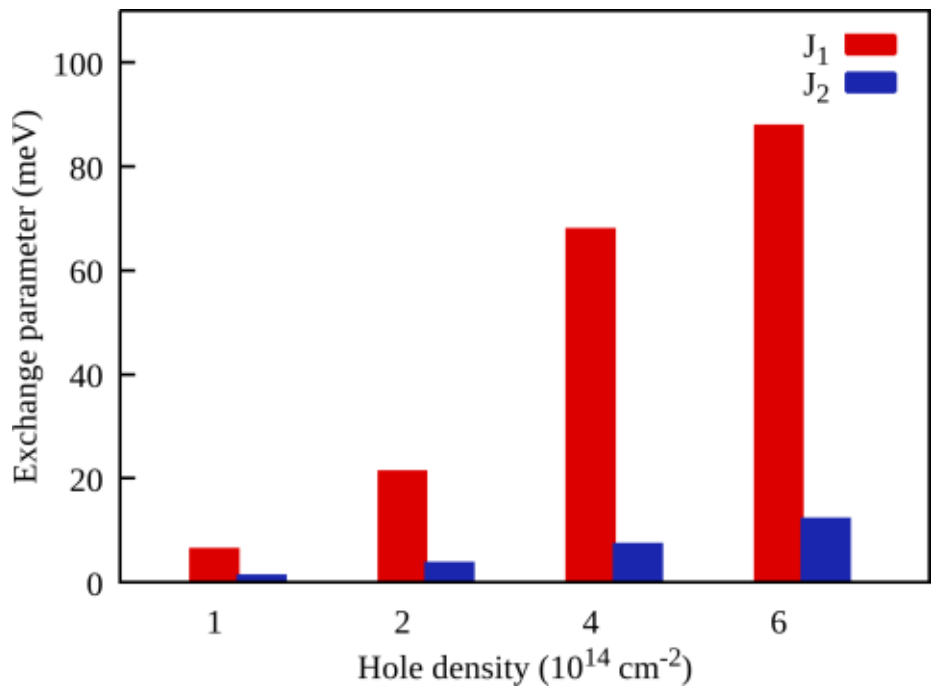

**Magnetic anisotropy energy (MAE, μeV) per magnetic atom**

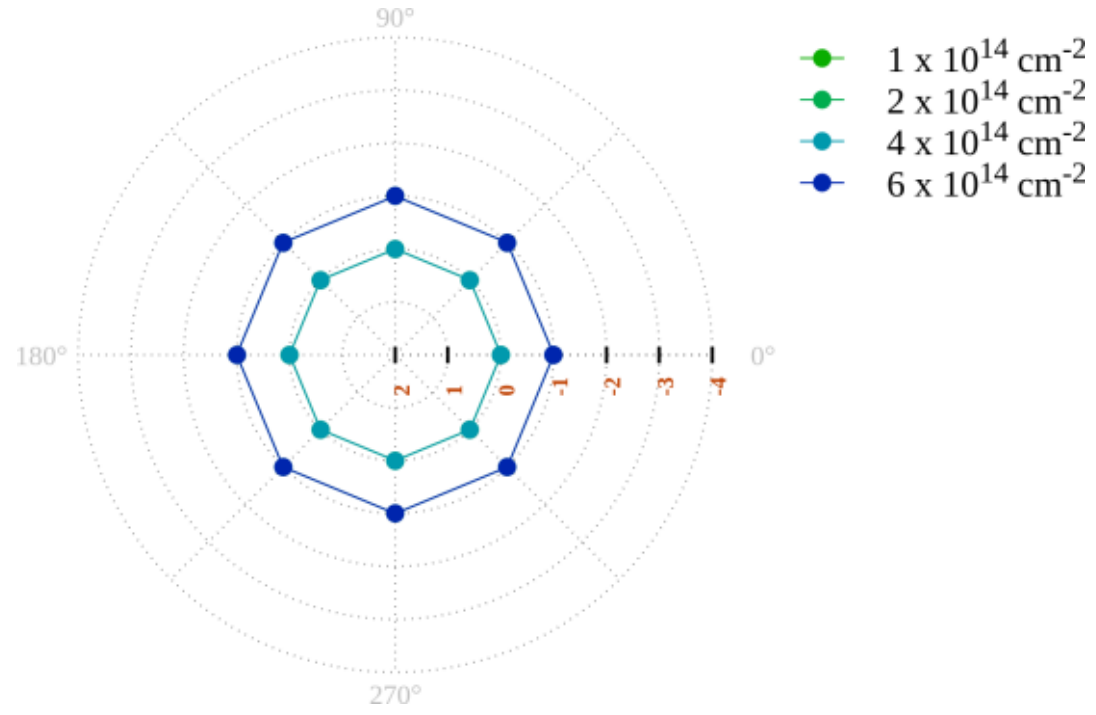

MAE = $E_{\parallel} - E_{\perp}$, a positive (negative) value of MAE indicates the off-plane (in-plane) easy axis.

**Monte Carlo simulations of the normalized magnetization of as a function of temperature**

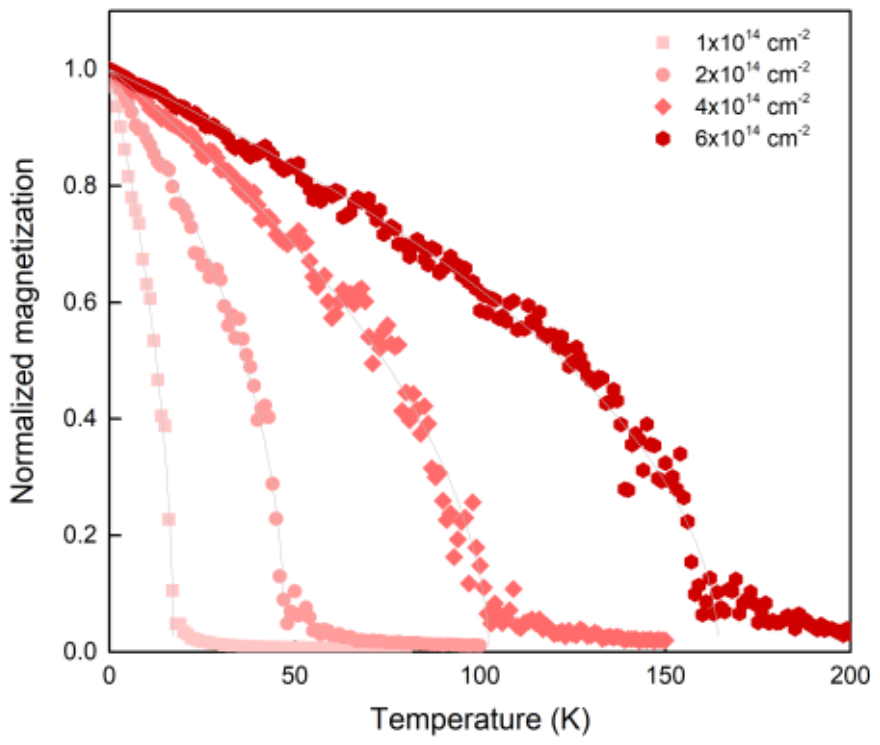

$T_c$: $1\times10^{14}$ cm$^{-2}$: 17 K  $2\times10^{14}$ cm$^{-2}$: 47 K

$4\times10^{14}$ cm$^{-2}$: 102 K  $6\times10^{14}$ cm$^{-2}$: 164 K

# 117. $B_2O_3$

| MC2D-ID | C2DB | 2dmat-ID | USPEX | Space group | Band gap (eV) |
|---|---|---|---|---|---|
| - | - | 2dm-33 | - | P62m | 5.22 |

**Convex hull** | **Atomic structure** | **Atomic coordinates** | **Phonon dispersion curve**

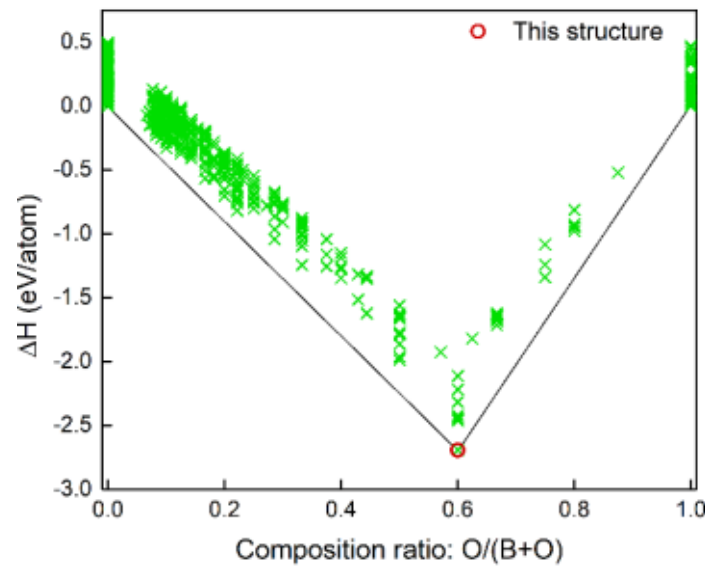

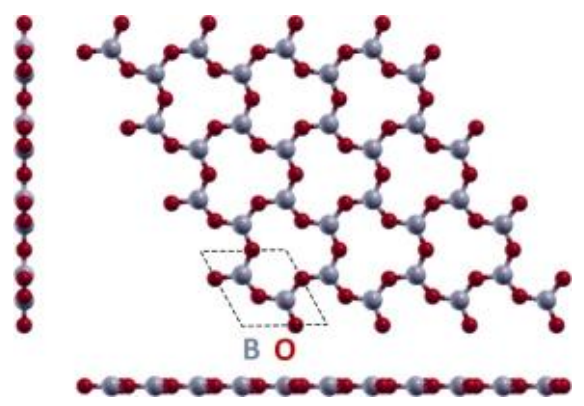

```
B2O3
  1.00000000000000
  4.4159446740658321   0.0000000000000000   0.0000000000000000
 -2.2079736202051823   3.8243201010166517   0.0000000000000000
  0.0000000000000000   0.0000000000000000  21.6555759999999999
    B    O
    2    3
Direct
   0.3333336116611321   0.6666666779422172   0.6291477653214699
   0.6666666758417463   0.3333335782488263   0.6291710961926356
   0.6166768581259490   0.0000003034604745   0.6291587479318892
   0.0000002949047229   0.6166768751406337   0.6291587490202772
   0.3833225594664498   0.3833225652078482   0.6291586415337420
```

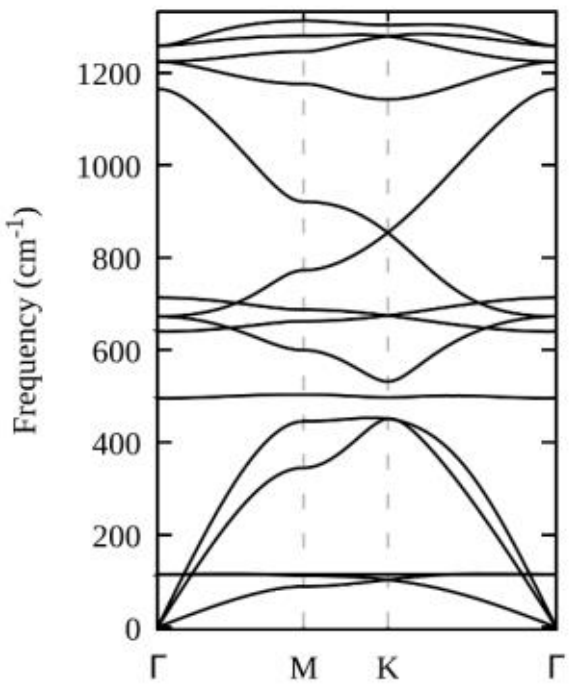

**Projected band structure and density of states** | **Magnetic moment and spin polarization energy as a function of hole doping concentration**

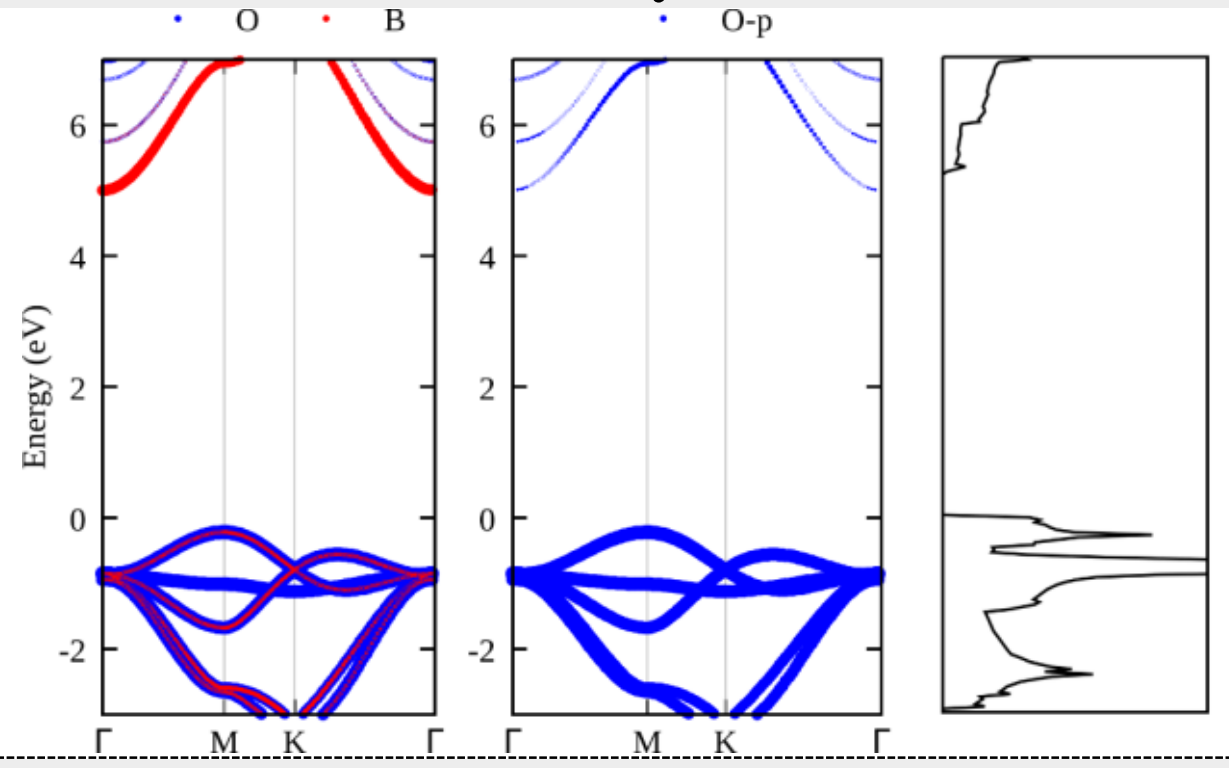

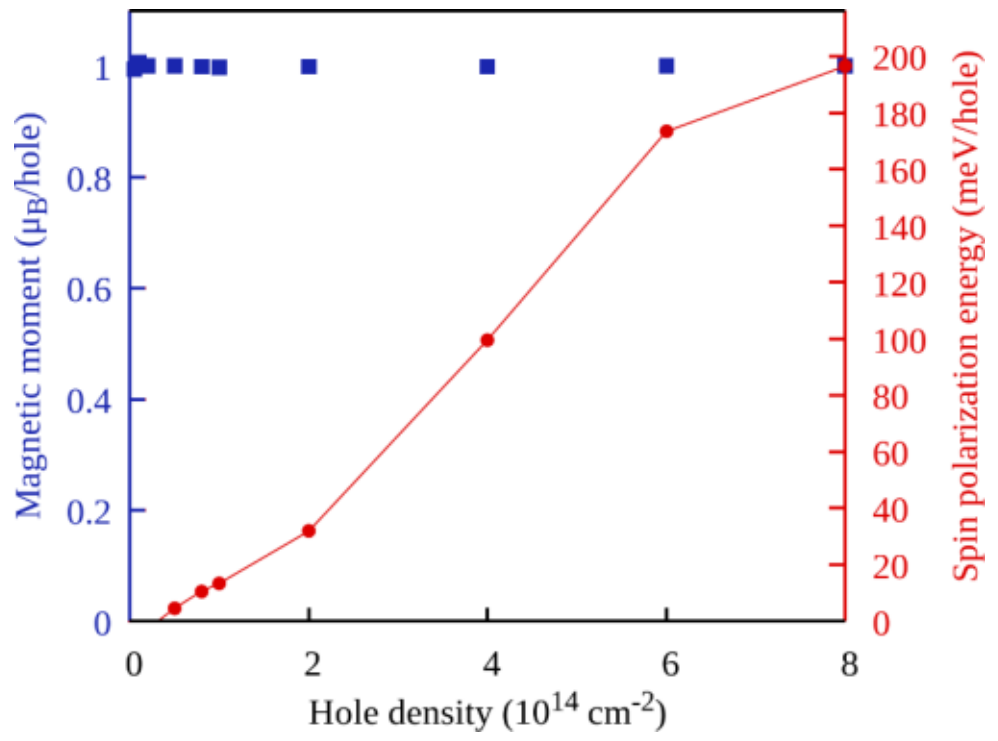

**Magnetic configurations and spin Hamiltonian** | **Magnetic exchange coupling parameters**

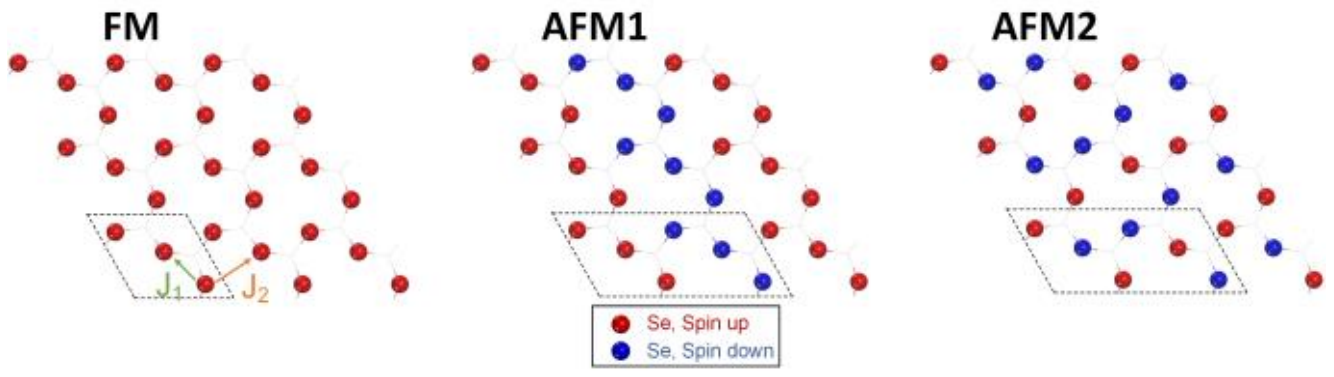

$$E_{FM} = E_0 - 6J_1S^2 - 3J_2S^2$$

$$E_{AFM1} = E_0 - 2J_1S^2 + J_2S^2$$

$$E_{AFM2} = E_0 + 2J_1S^2 + J_2S^2$$

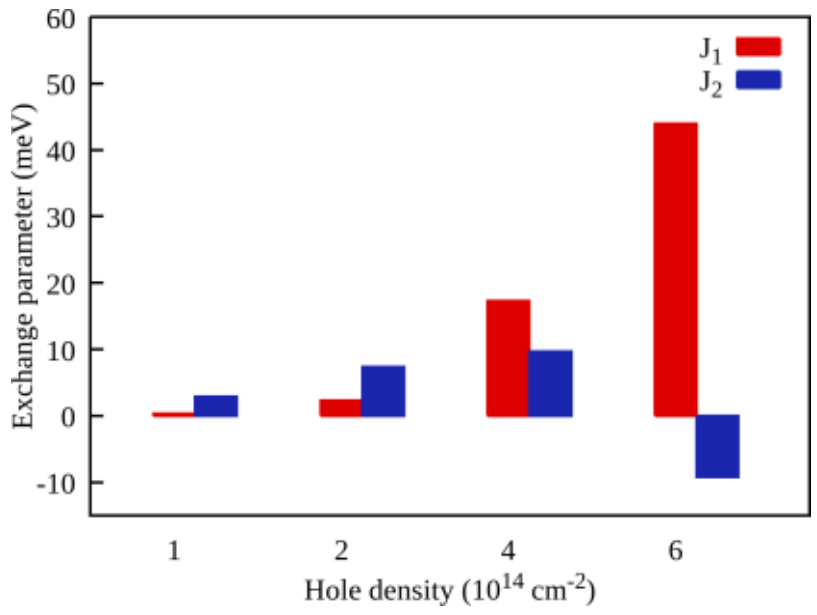

**Magnetic anisotropy energy (MAE, μeV) per magnetic atom** | **Monte Carlo simulations of the normalized magnetization of as a function of temperature**

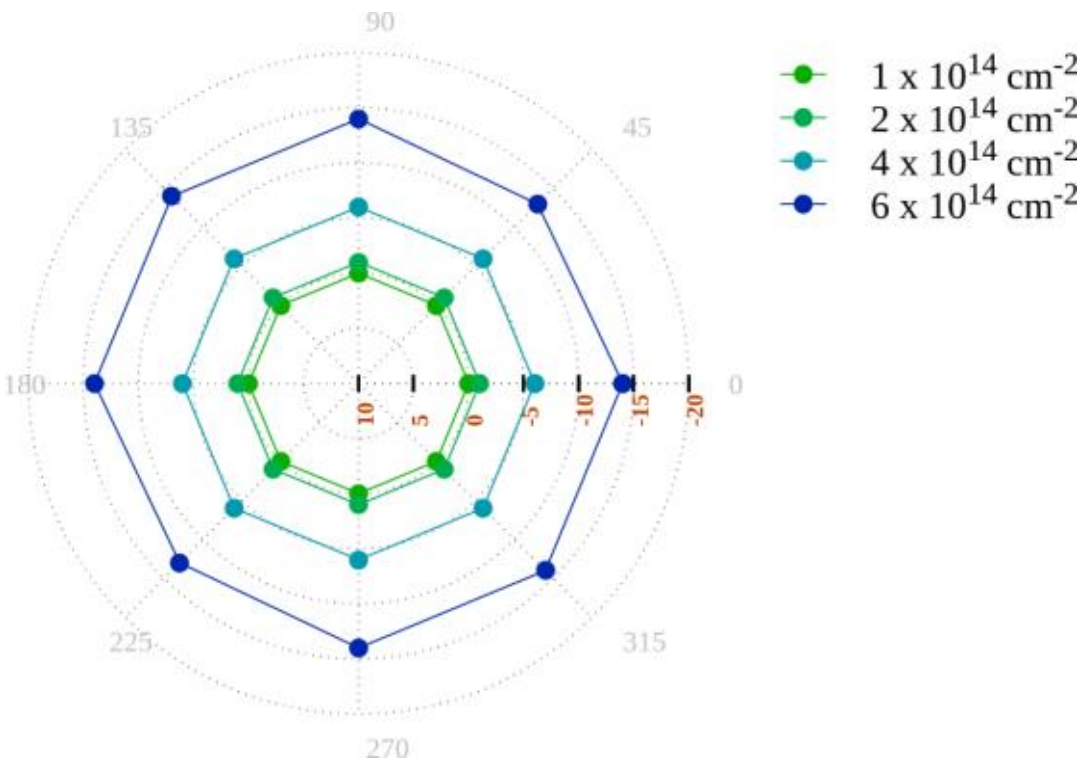

MAE = $E_{\parallel} - E_{\perp}$, a positive (negative) value of MAE indicates the off-plane (in-plane) easy axis.

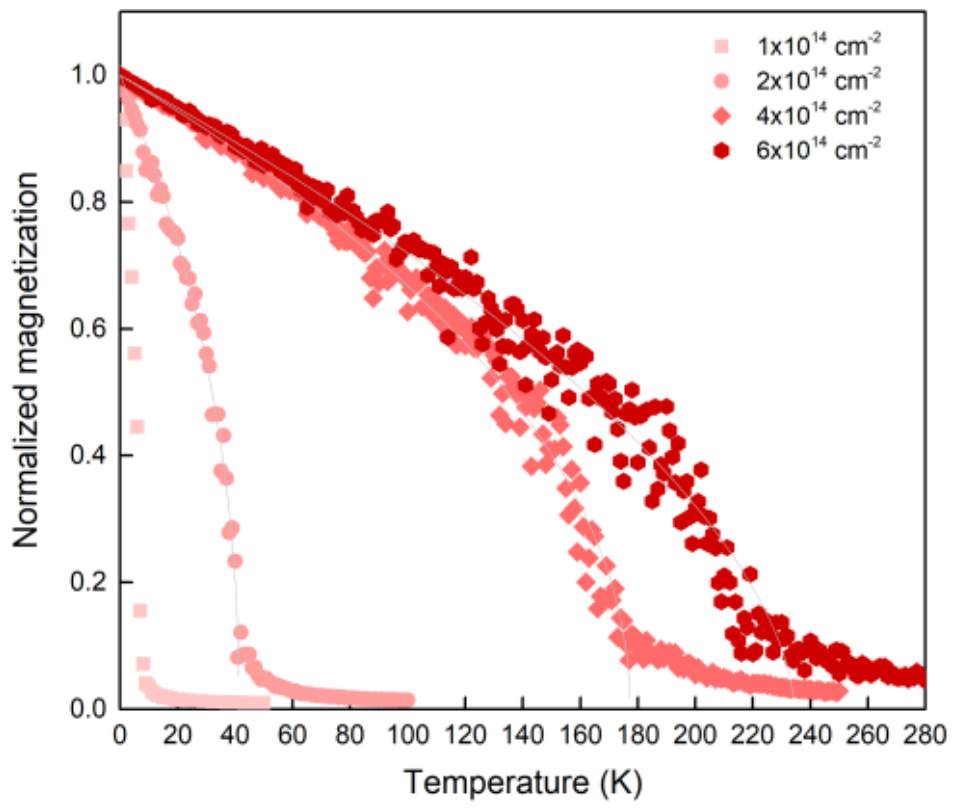

$T_c$: $1\times10^{14}$ cm$^{-2}$: 7K   $2\times10^{14}$ cm$^{-2}$: 41 K

$4\times10^{14}$ cm$^{-2}$: 177 K   $6\times10^{14}$ cm$^{-2}$: 234 K

# 118. $C_3N_4$

| MC2D-ID | C2DB | 2dmat-ID | USPEX | Space group | Band gap (eV) |
|---|---|---|---|---|---|
| - | - | 2dm-4847 | - | C2 | 2.25 |

## Convex hull

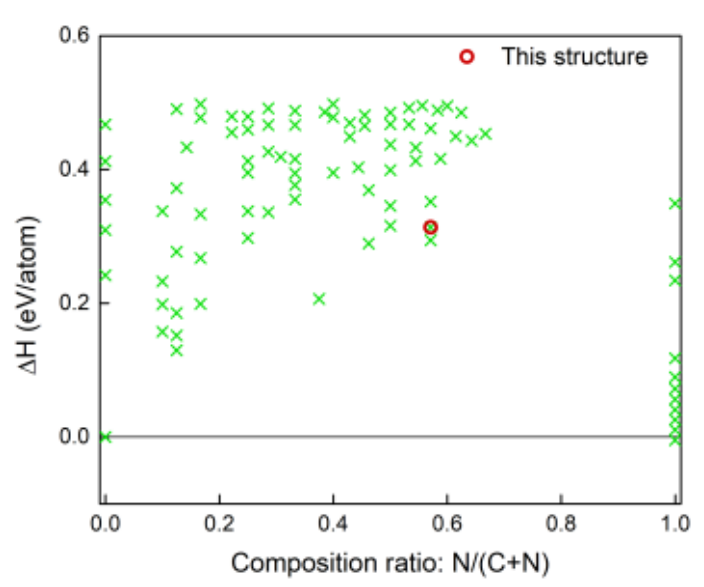

## Atomic structure

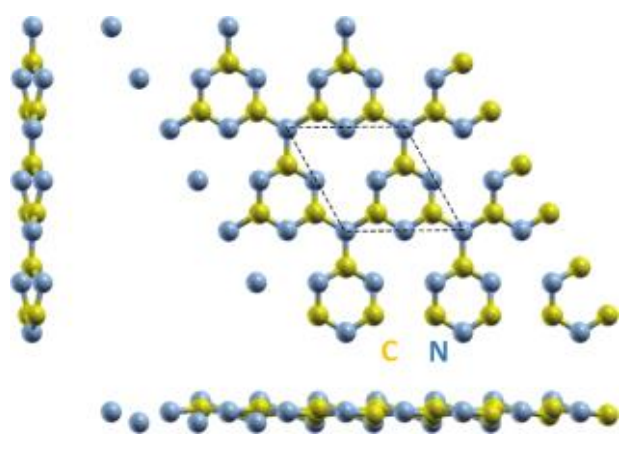

## Atomic coordinates

```
C3N4
  1.00000000000000
    4.7148016012078937   0.0000000000000000   0.0000000000000000
    2.3515188622351082   4.0710647115704353   0.0000000000000000
    0.0000000000000000   0.0000000000000000   20.7846499999999992
  C    N
  3    4
Direct
  0.8292154477102325  0.8656598237157468  0.4938500673793084
  0.3583144737507584  0.8640682525088920  0.4998956424374154
  0.8294419568887541  0.3911639484662840  0.5062309821806974
  0.5091577933182134  0.5520897669134094  0.5204099538574705
  0.9971507890925141  0.5444678953312492  0.5002984702008131
  0.5090099178357270  0.0251595474506033  0.4793819486384657
  0.0033326214037945  0.0414337656138267  0.4999349353058247
```

## Phonon dispersion curve

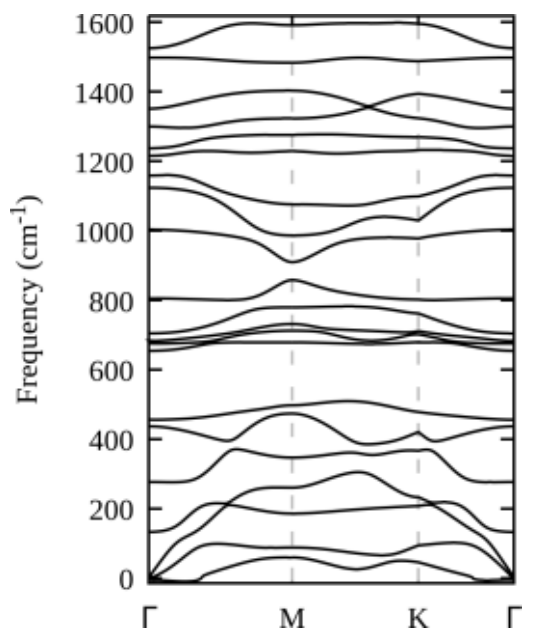

## Projected band structure and density of states

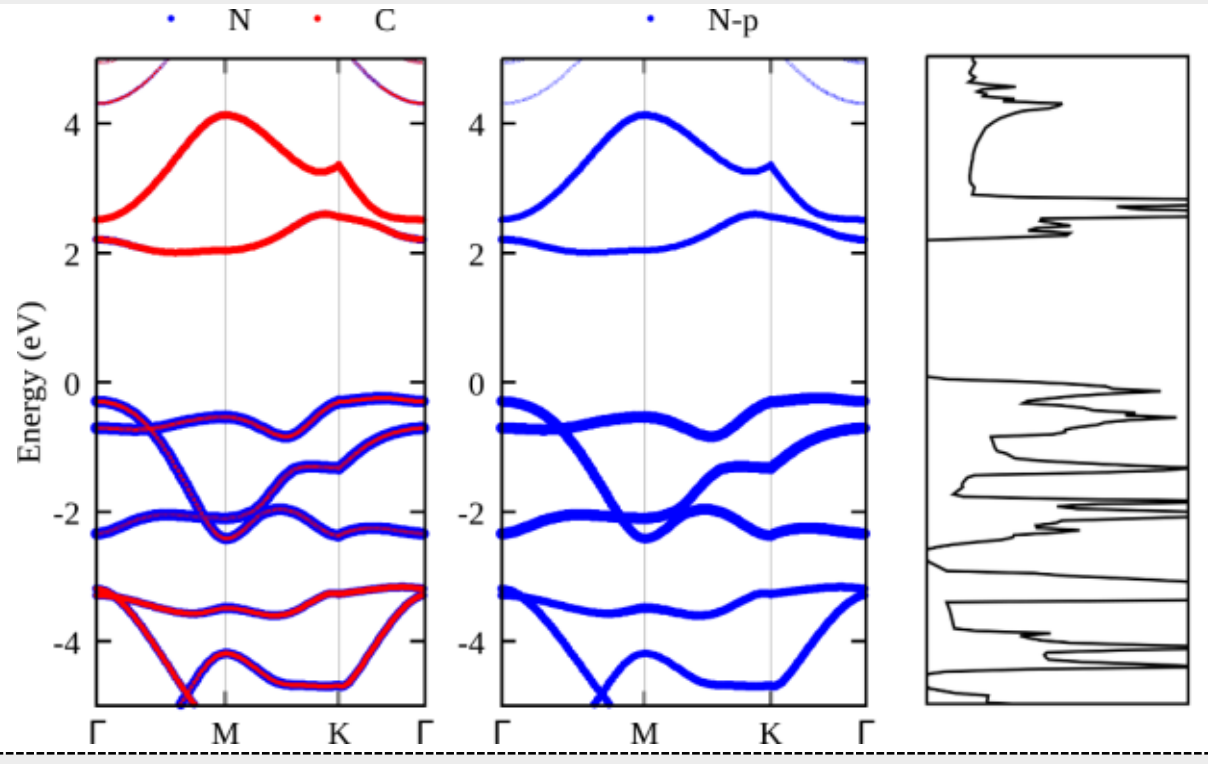

## Magnetic moment and spin polarization energy as a function of hole doping concentration

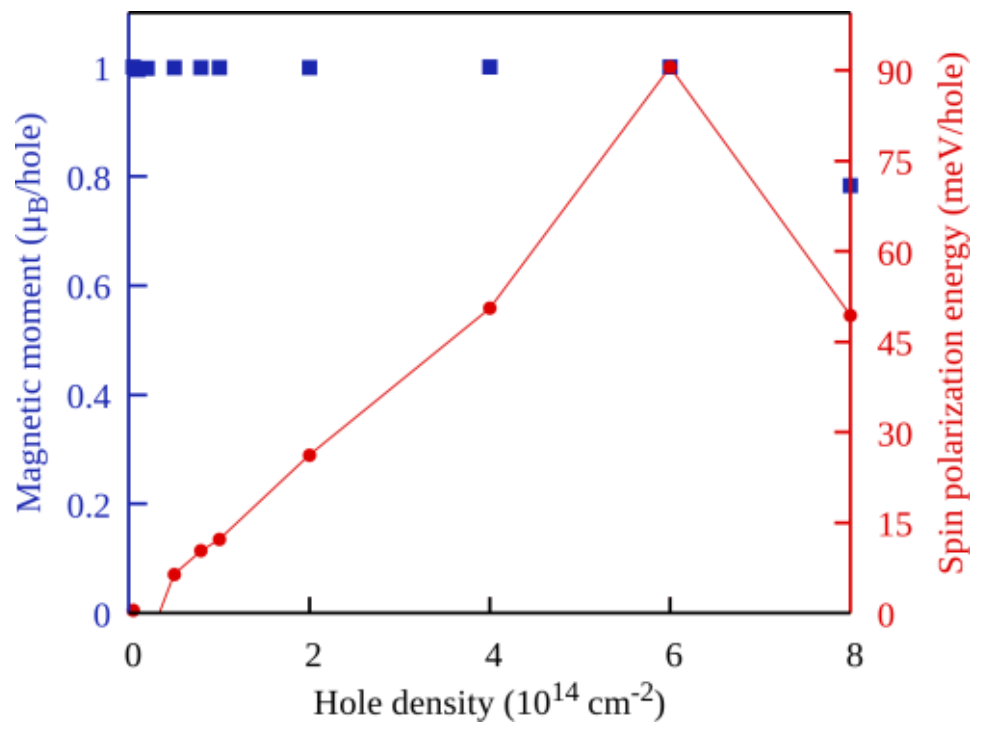

## Magnetic configurations and spin Hamiltonian

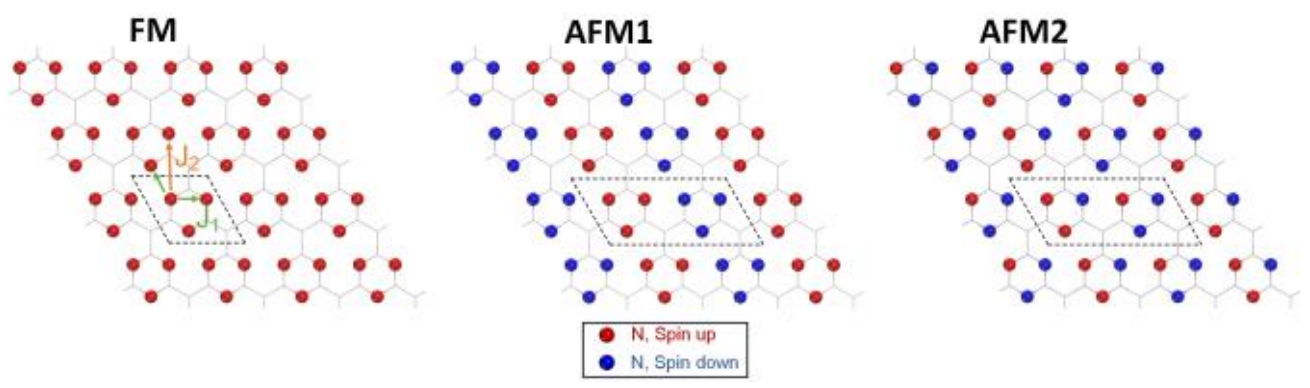

$$E_{FM} = E_0 - 6J_1S^2 - 6J_2S^2$$

$$E_{AFM1} = E_0 - 2J_1S^2 + 2J_2S^2$$

$$E_{AFM2} = E_0 + 2J_1S^2 + 2J_2S^2$$

## Magnetic exchange coupling parameters

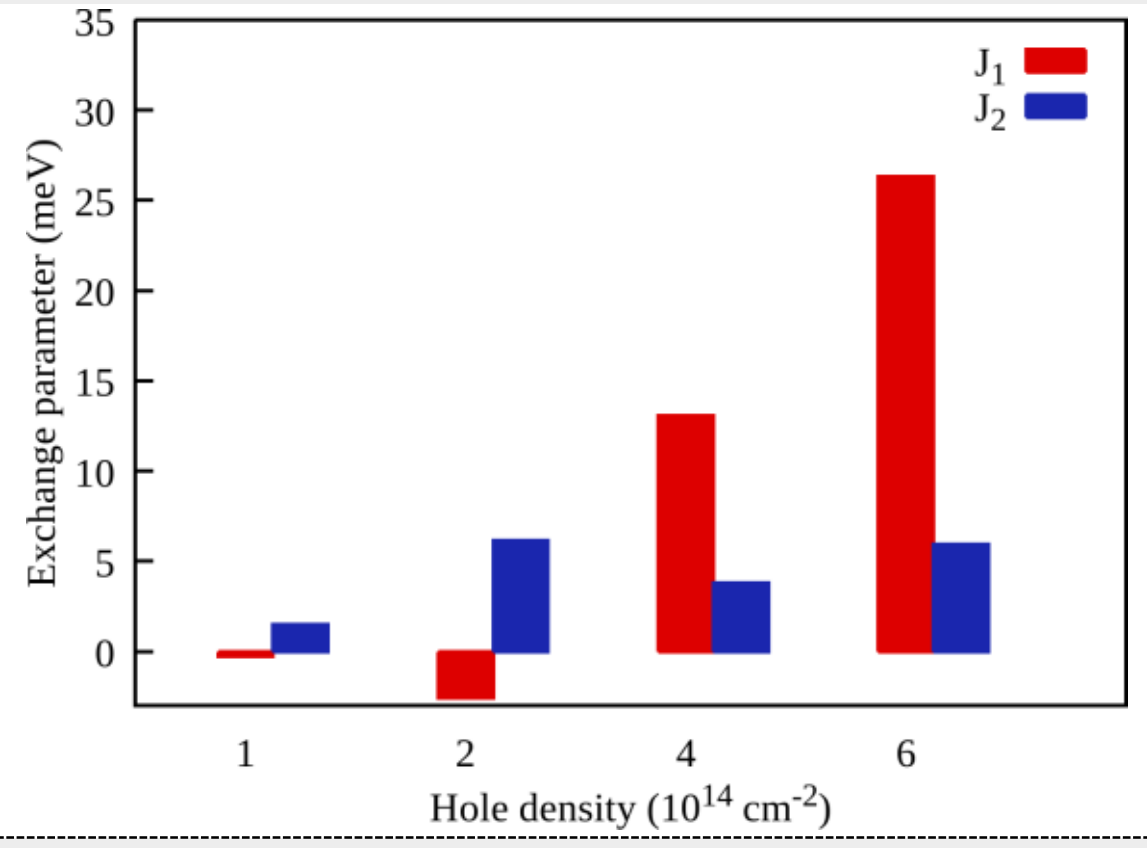

## Magnetic anisotropy energy (MAE, μeV) per magnetic atom

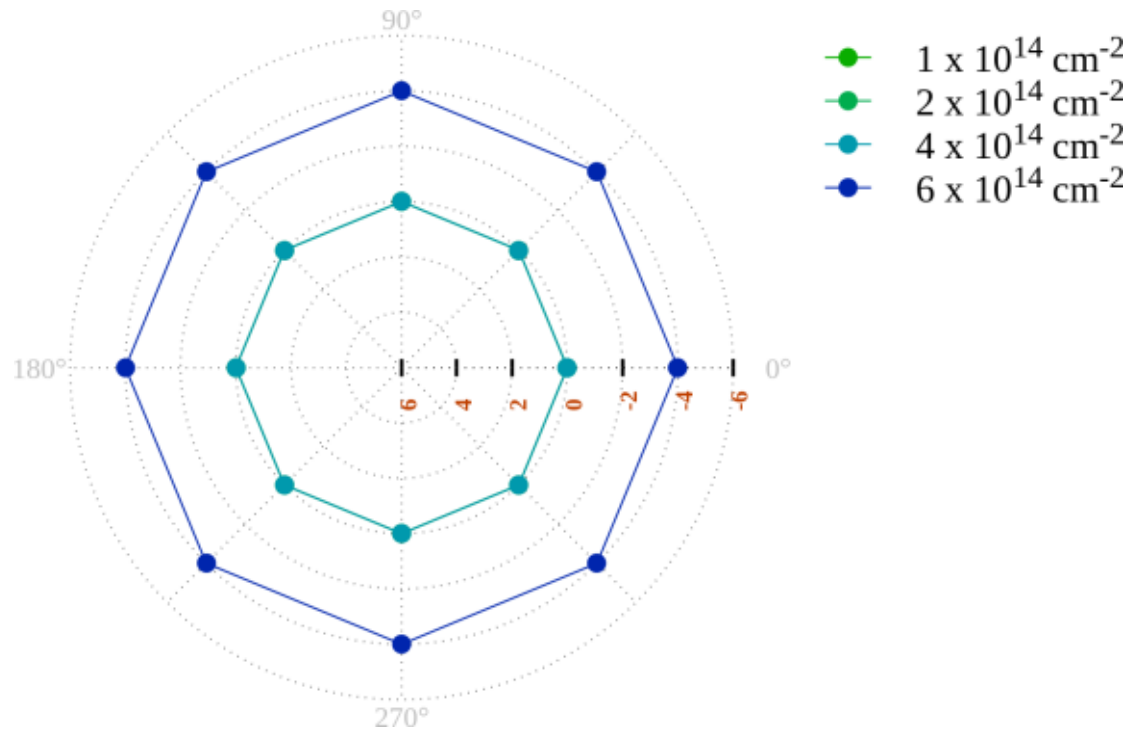

MAE = $E_{\parallel} - E_{\perp}$, a positive (negative) value of MAE indicates the off-plane (in-plane) easy axis.

## Monte Carlo simulations of the normalized magnetization of as a function of temperature

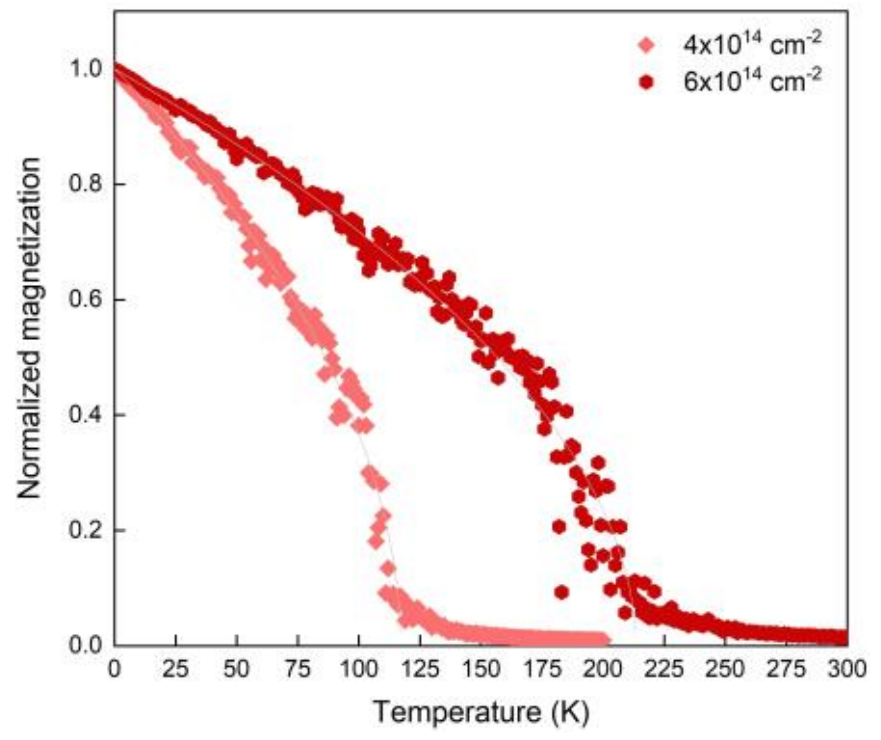

$T_c$: $1\times10^{14}$ $cm^{-2}$: - K  $2\times10^{14}$ $cm^{-2}$: - K

$4\times10^{14}$ $cm^{-2}$: 117 K  $6\times10^{14}$ $cm^{-2}$: 213 K

## 119. $In_2Cl_2O_2$

| MC2D-ID | C2DB | 2dmat-ID | USPEX | Space group | Band gap (eV) |
|---|---|---|---|---|---|
| - | ✓ | 2dm-3585 | - | Pmmn | 2.56 |

**Convex hull**

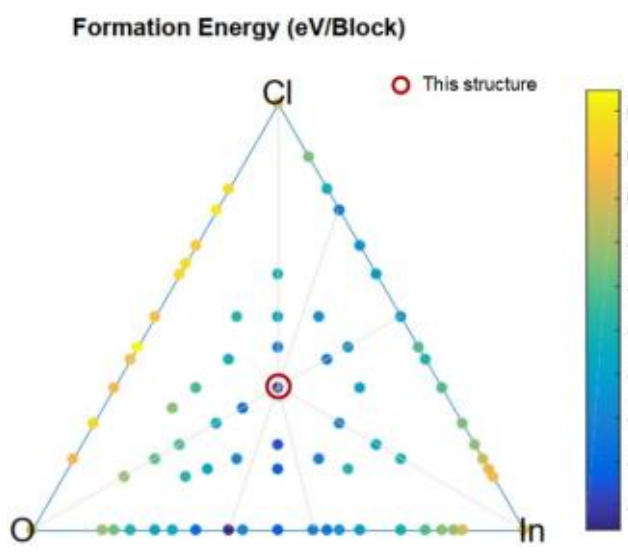

**Atomic structure**

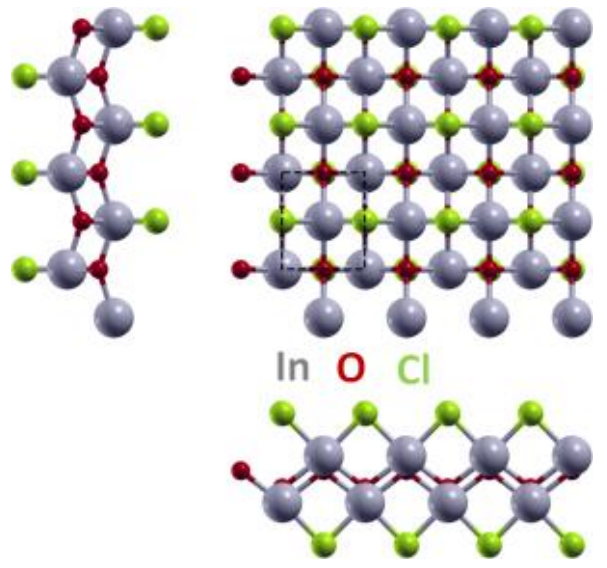

**Atomic coordinates**

```
In2Cl2O2
   1.00000000000000
   3.5731505241839181    0.0000000000000000    0.0000000000000000
   0.0000000000000000    4.1589447996631739    0.0000000000000000
   0.0000000000000000    0.0000000000000000   25.6634999999999991
   In    Cl    O
   2     2     2
Direct
  0.9999999987798702  0.0000000176380510  0.5393361404053039
  0.4999999997657127  0.5000000162422111  0.4606638625722042
  0.9999999998758398  0.4999999998113651  0.3895169892393682
  0.4999999980386818  0.9999999911125528  0.6104830086990489
  0.5000000012555006  0.9999999910346347  0.4855965158030031
  0.0000000022843949  0.4999999841611782  0.5144034832810718
```

**Phonon dispersion curve**

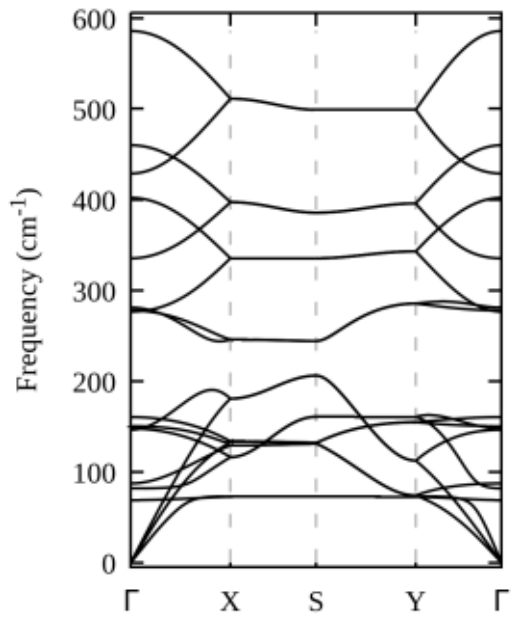

**Projected band structure and density of states**

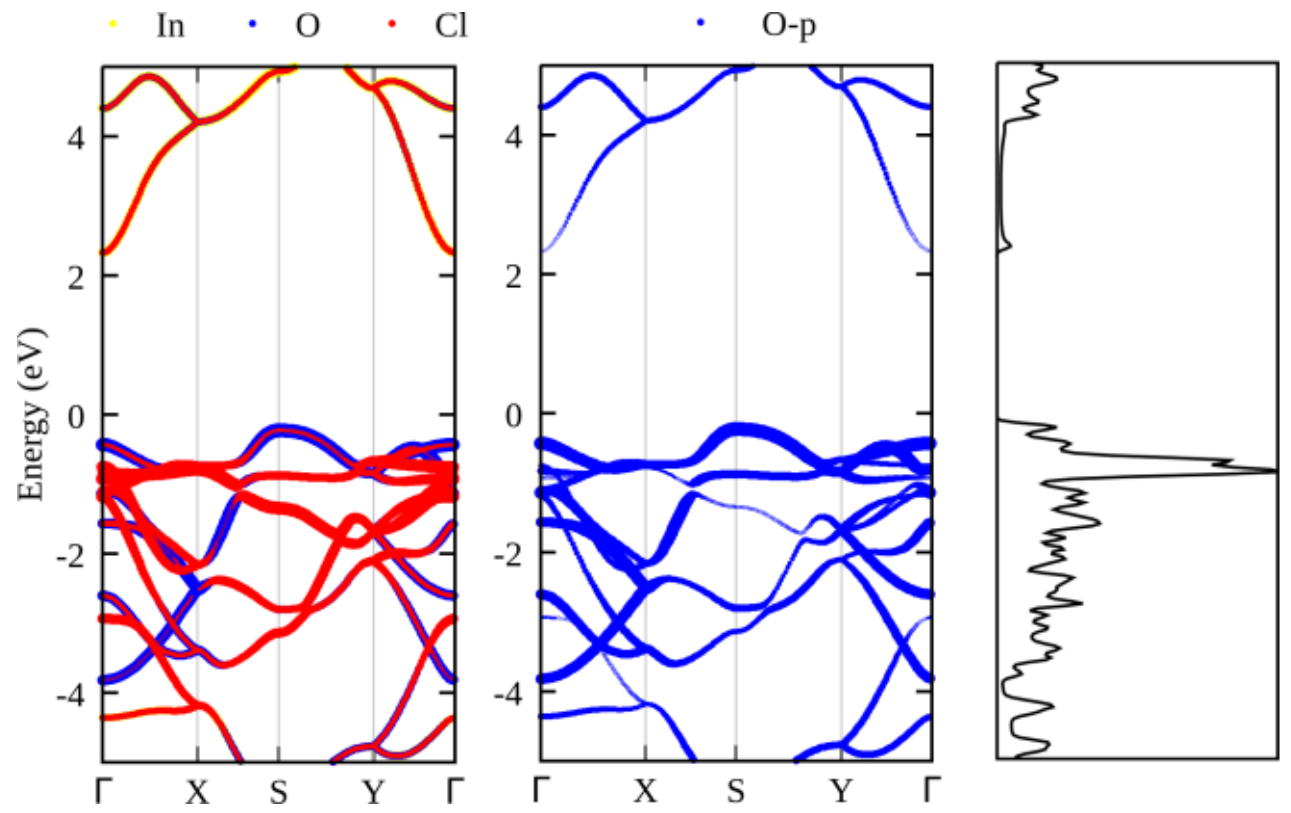

**Magnetic moment and spin polarization energy as a function of hole doping concentration**

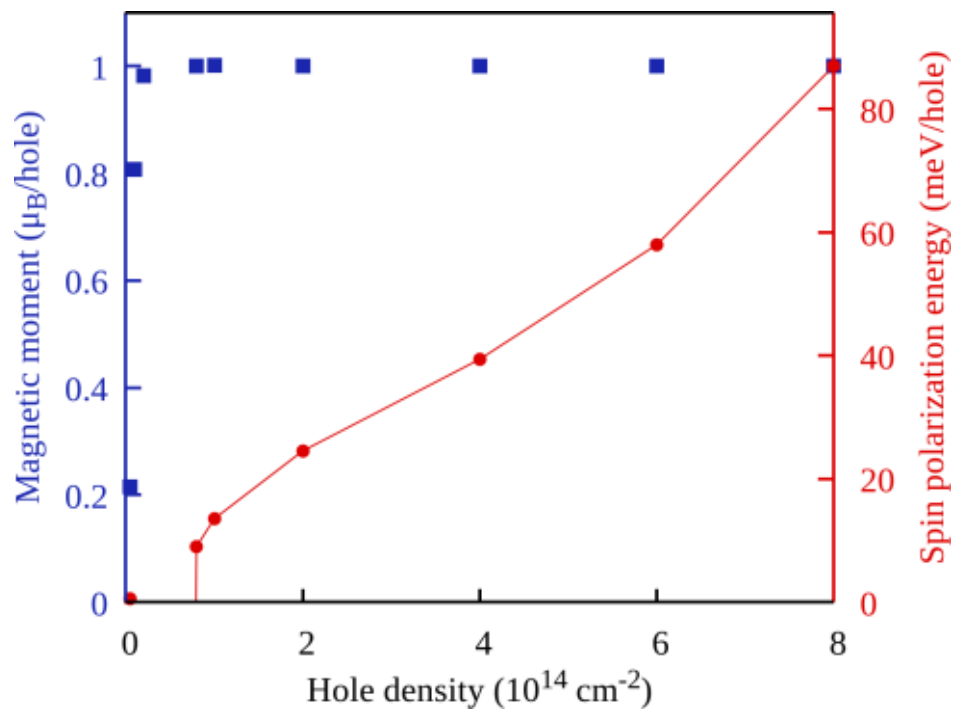

**Magnetic configurations and spin Hamiltonian**

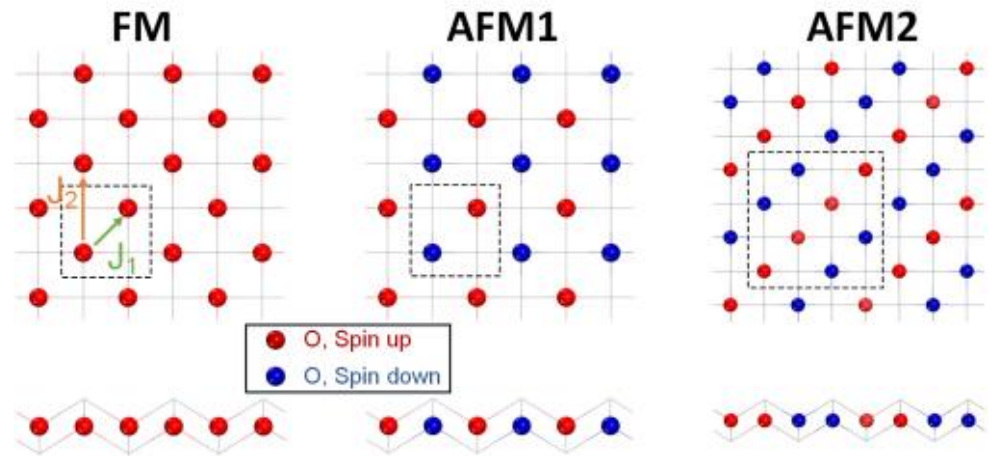

$$E_{FM} = E_0 - 4J_1S^2 - 4J_2S^2$$

$$E_{AFM1} = E_0 + 4J_1S^2 - 4J_2S^2$$

$$E_{AFM2} = E_0 + 4J_2S^2$$

**Magnetic exchange coupling parameters**

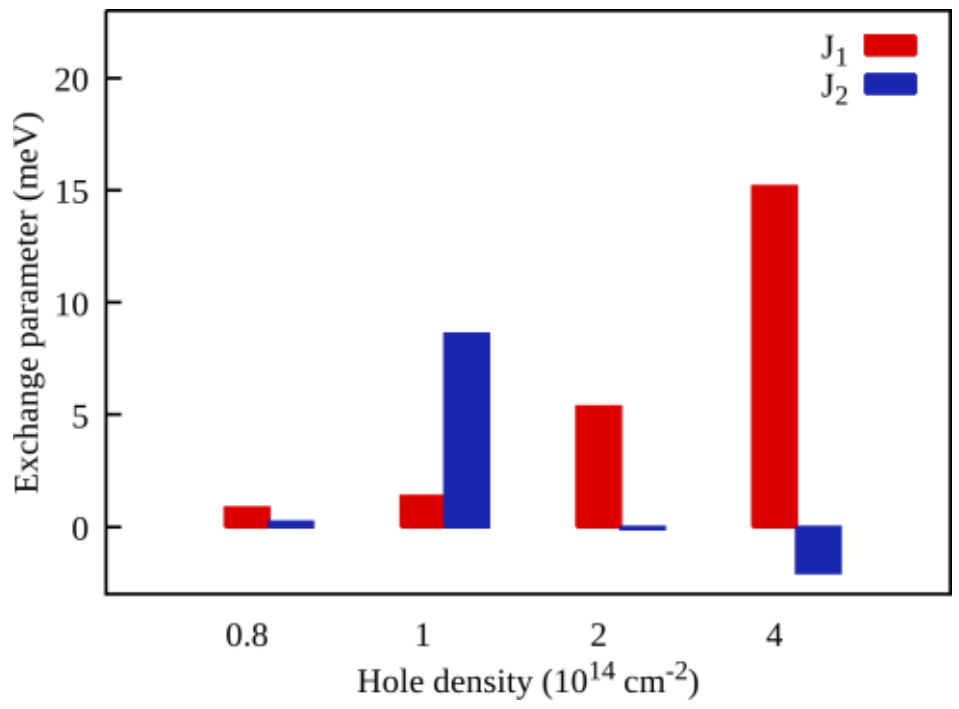

**Magnetic anisotropy energy (MAE, μeV) per magnetic atom**

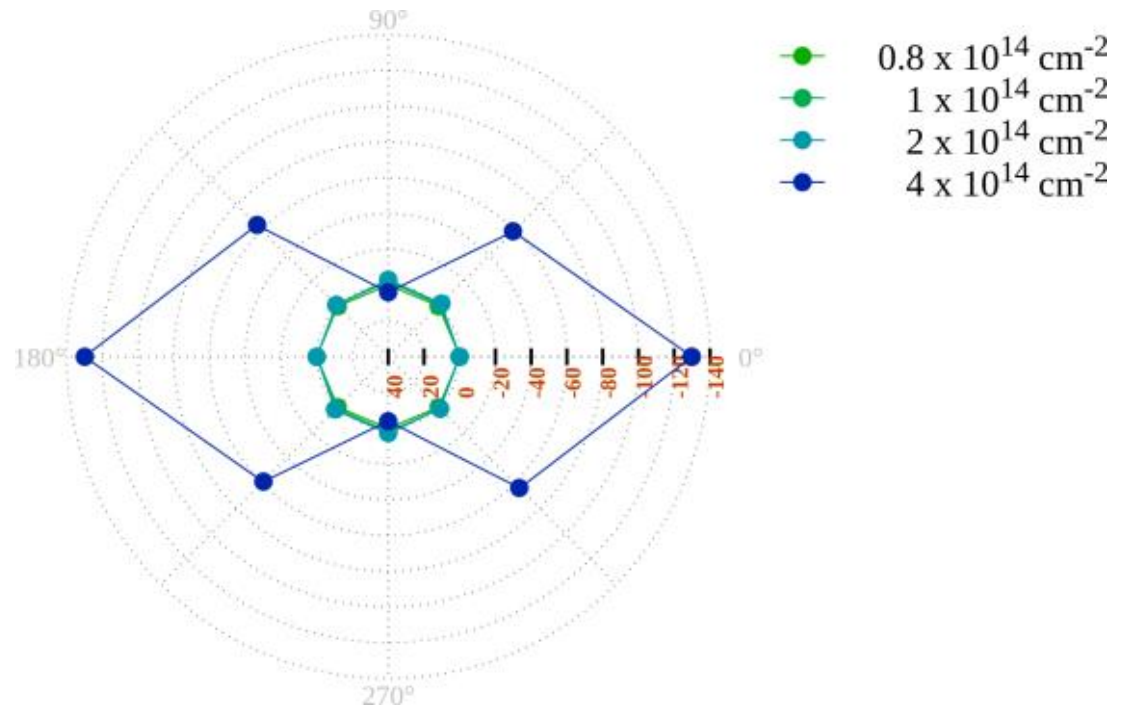

MAE = $E_{\parallel} - E_{\perp}$, a positive (negative) value of MAE indicates the off-plane (in-plane) easy axis.

**Monte Carlo simulations of the normalized magnetization of as a function of temperature**

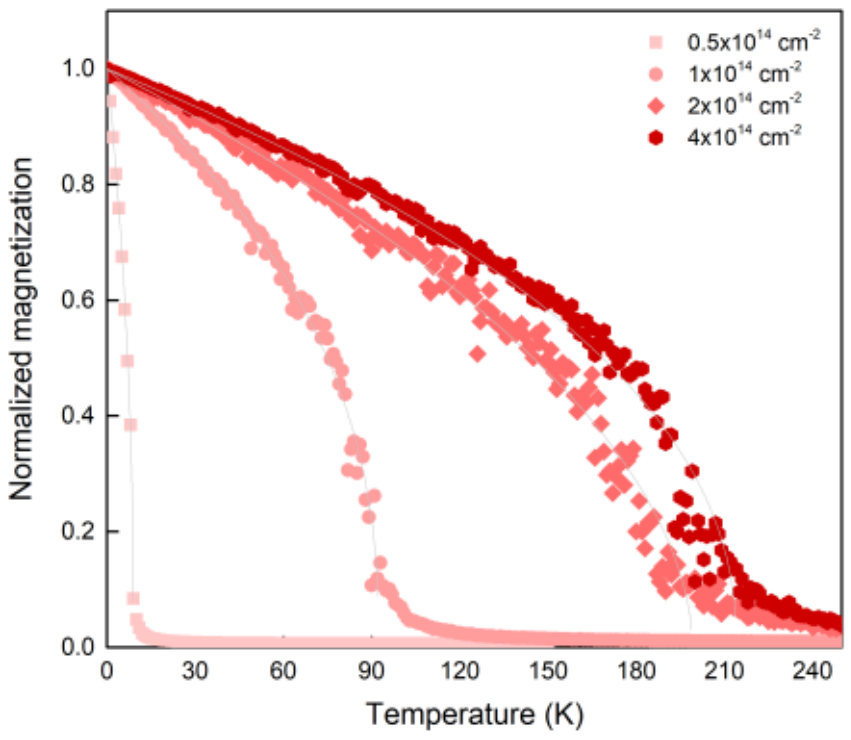

$T_c$: $0.8\times10^{14}$ cm$^{-2}$: 10 K  $1\times10^{14}$ cm$^{-2}$: 92 K

$2\times10^{14}$ cm$^{-2}$: 199 K  $4\times10^{14}$ cm$^{-2}$: 214 K

# 120. $In_2Br_2O_2$

| MC2D-ID | C2DB | 2dmat-ID | USPEX | Space group | Band gap (eV) |
|---|---|---|---|---|---|
| 91 | ✓ | 2dm-3667 | - | Pmmn | 2.29 |

**Convex hull**

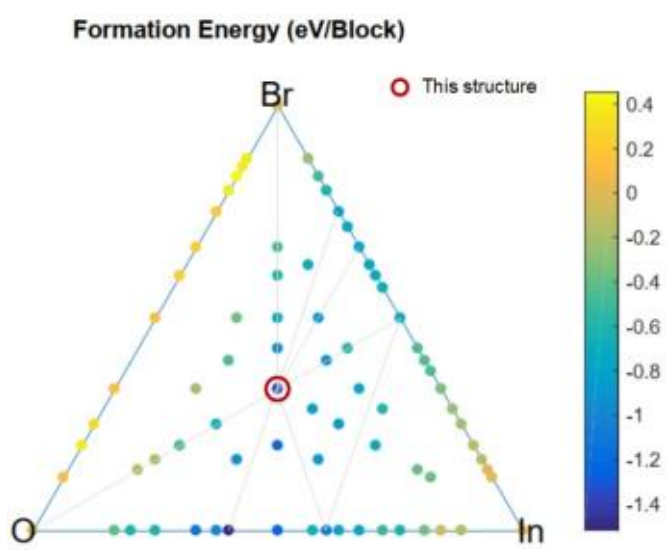

**Atomic structure**

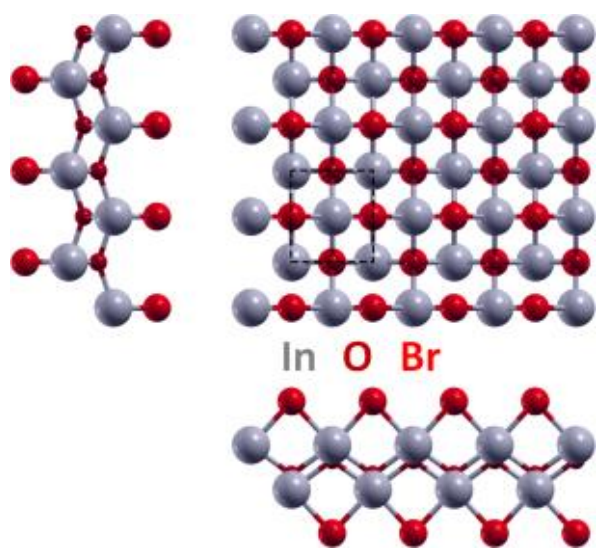

**Atomic coordinates**

```
I In2Br2O2
1.00000000000000
3.6777167334034697   0.0000000000000000   0.0000000000000000
0.0000000000000000   4.1352408658591528   0.0000000000000000
0.0000000000000000   0.0000000000000000   25.9786509999999993
In   Br   O
2    2    2
Direct
0.4999999980116812   0.4999999961174453   0.4612293562923924
0.0000000048975579   0.0000000030597818   0.5387706370112966
0.9999999991153103   0.4999999835151883   0.3850082613192214
0.5000000025816149   0.0000000170015824   0.6149917430419549
0.9999999987333652   0.5000000040808814   0.5125732056066994
0.4999999966604634   0.9999999962251280   0.4874267967284354
```

**Phonon dispersion curve**

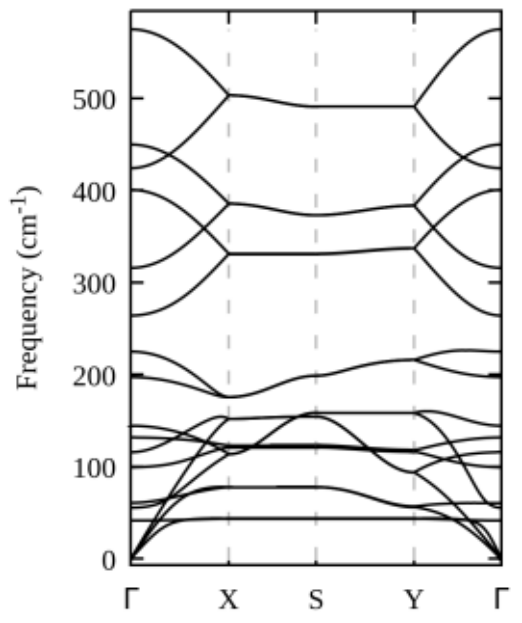

**Projected band structure and density of states**

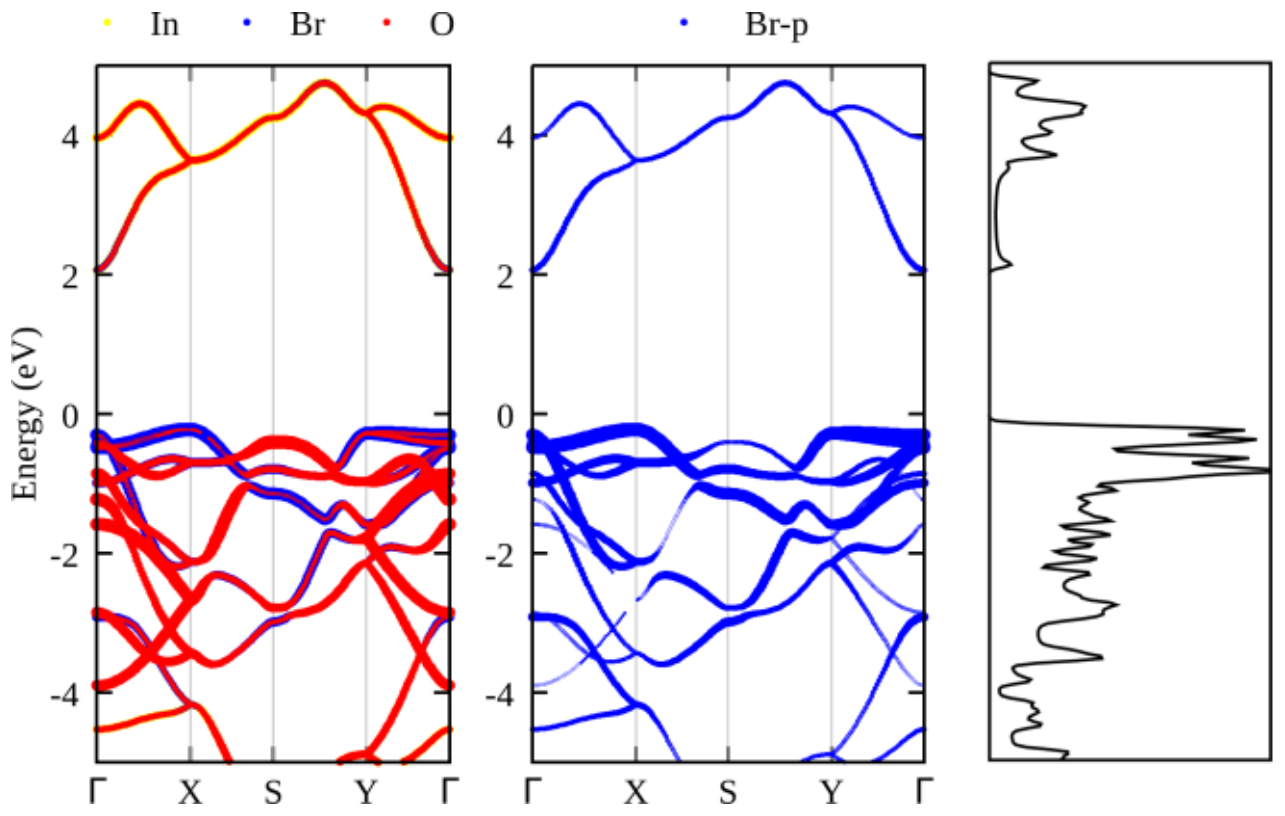

**Magnetic moment and spin polarization energy as a function of hole doping concentration**

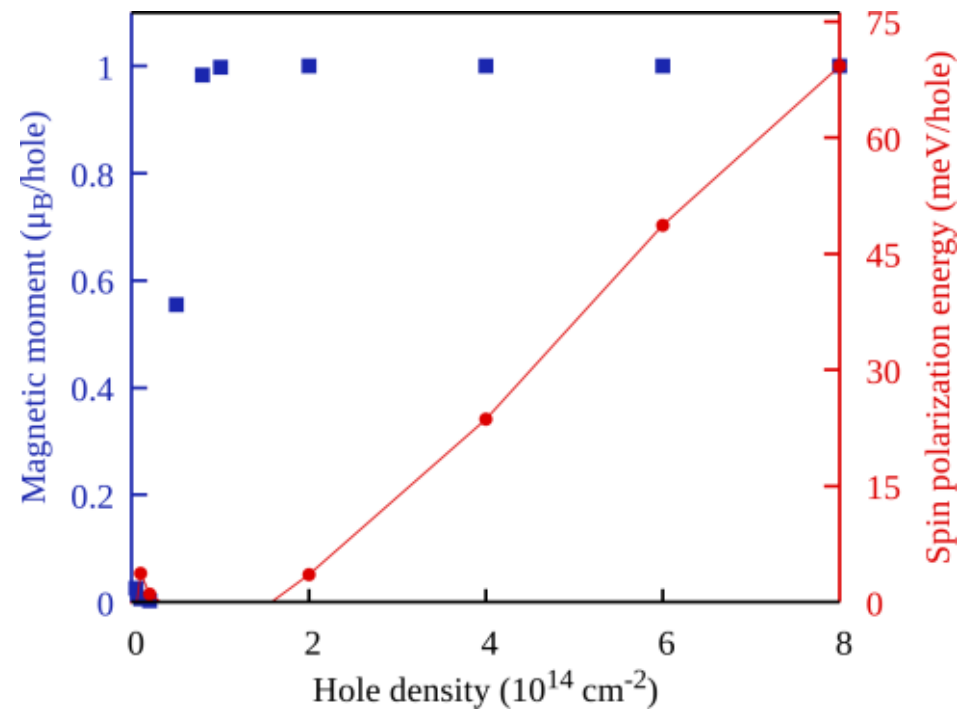

**Magnetic configurations and spin Hamiltonian**

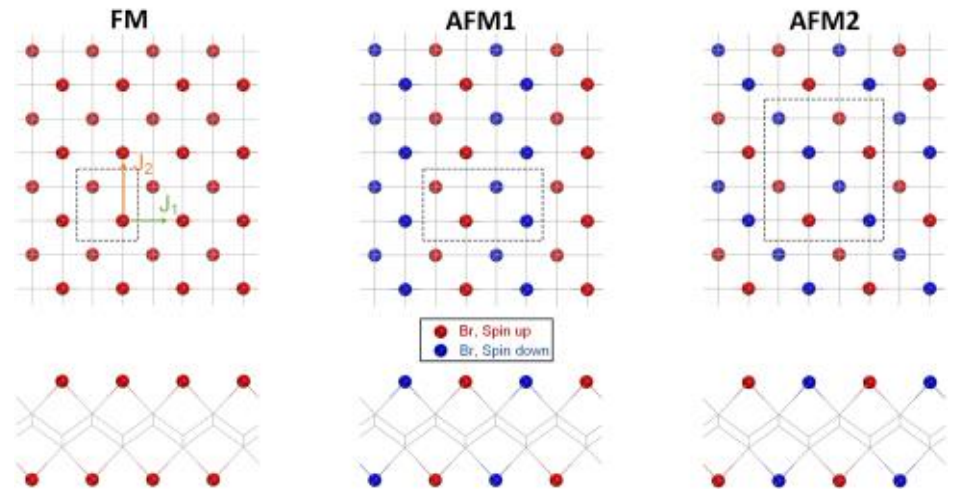

$$E_{FM} = E_0 - 2J_1S^2 - 2J_2S^2$$
$$E_{AFM1} = E_0 + 2J_1S^2 - 2J_2S^2$$
$$E_{AFM2} = E_0 + 2J_1S^2 + 2J_2S^2$$

**Magnetic exchange coupling parameters**

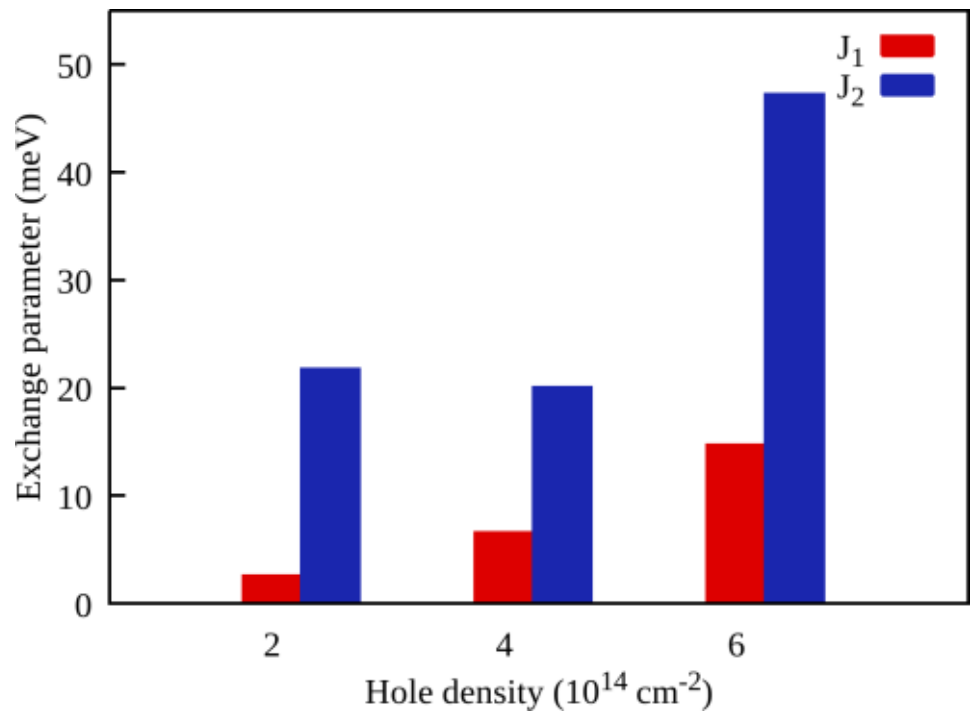

**Magnetic anisotropy energy (MAE, μeV) per magnetic atom**

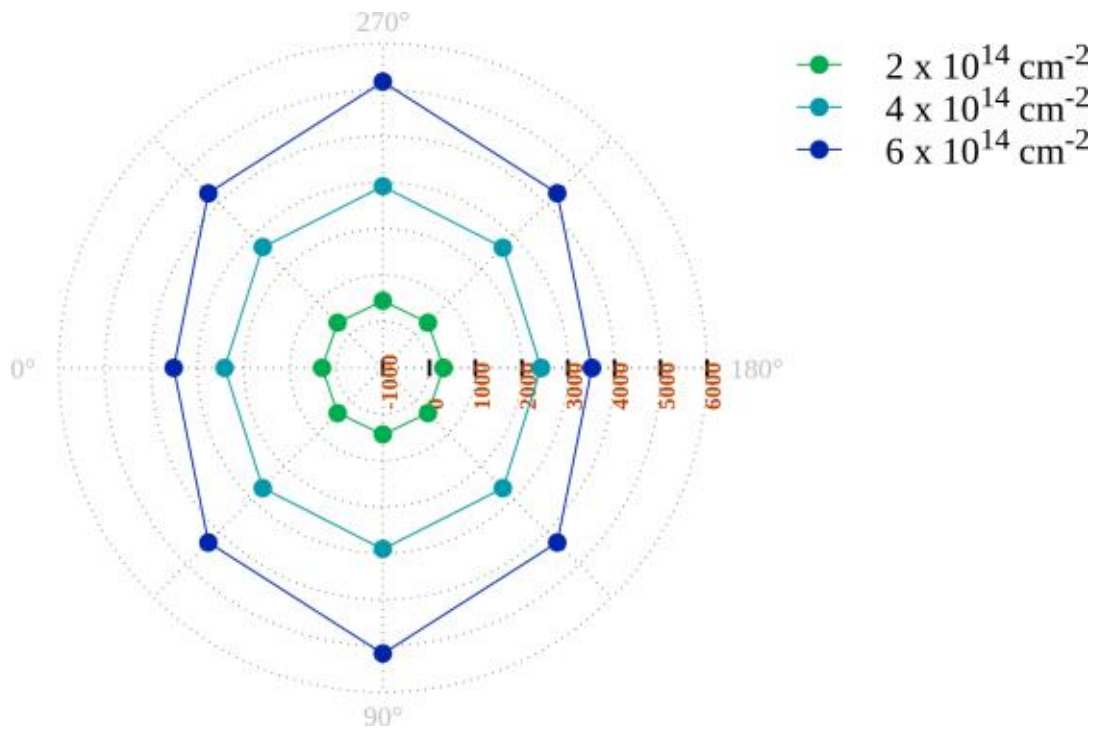

MAE = $E_{\parallel} - E_{\perp}$, a positive (negative) value of MAE indicates the off-plane (in-plane) easy axis.

**Monte Carlo simulations of the normalized magnetization of as a function of temperature**

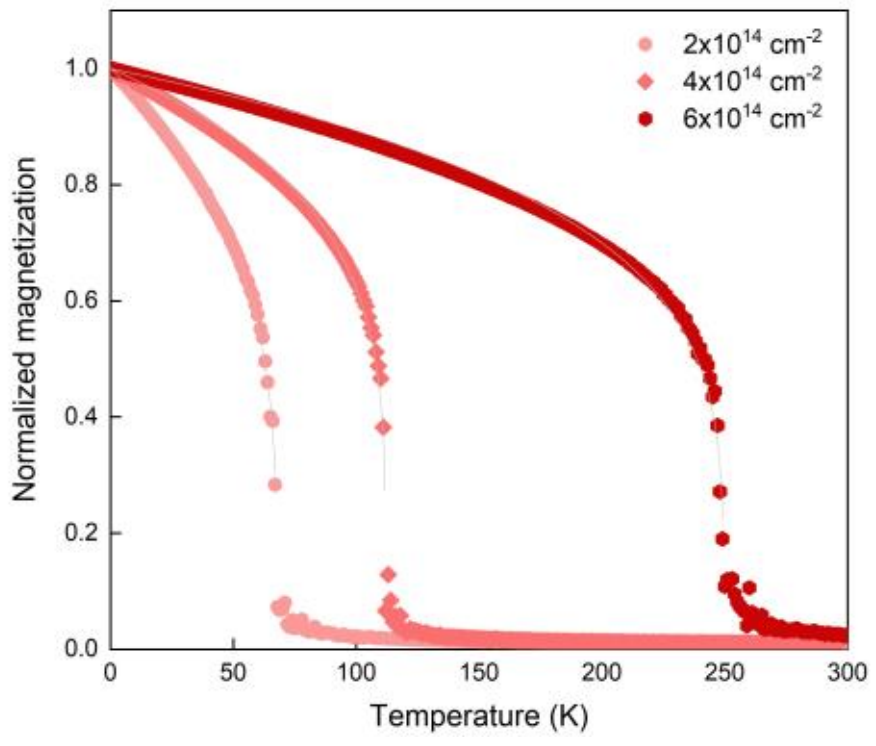

$T_c$: $1\times10^{14}$ cm$^{-2}$: - K　$2\times10^{14}$ cm$^{-2}$: 68 K

$4\times10^{14}$ cm$^{-2}$: 112 K　$6\times10^{14}$ cm$^{-2}$: 249 K

# 121. $Mg_2Ge_2O_6$

| MC2D-ID | C2DB | 2dmat-ID | USPEX | Space group | Band gap (eV) |
|---|---|---|---|---|---|
| - | - | 2dm-4548 | - | Pmma | 1.83 |

**Convex hull**

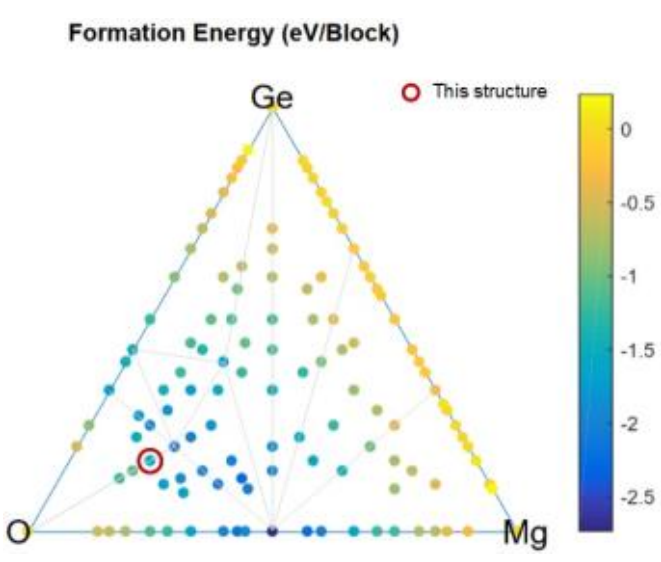

**Atomic structure**

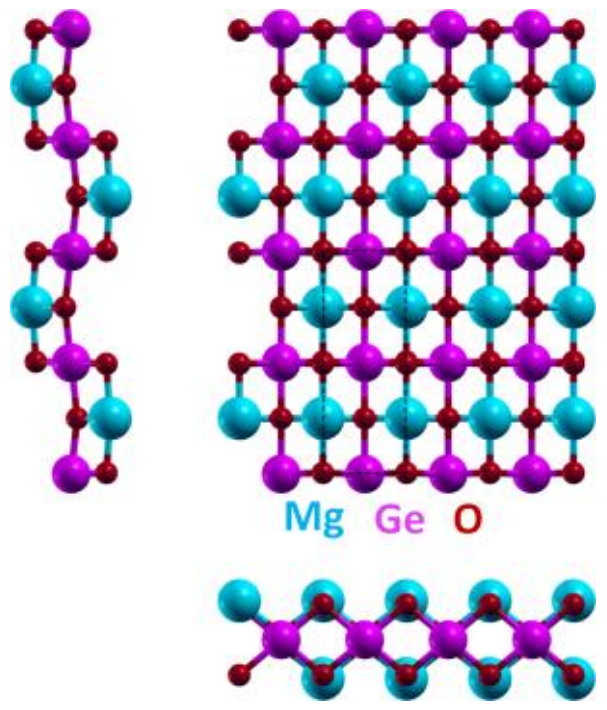

**Atomic coordinates**

```
Mg2Ge2O6
   1.00000000000000
   3.0059731683423987   0.0000000000000000   0.0000000000000000
   0.0000000000000000   8.0018786807623012   0.0000000000000000
   0.0000000000000000   0.0000000000000000  24.5345280000000017
   Mg   Ge   O
   2    2    6
Direct
  0.9999999854917405  0.7500000228972823  0.4428084404274983
  0.0000000032055354  0.2499999931306931  0.5571915758536861
  0.5000000212605755  0.5000000543986332  0.4999999655723784
  0.4999999934689470  0.9999999389468712  0.4999999701877158
  0.0000000583089417  0.9950844787043565  0.4481664650504484
  0.9999999378169875  0.0049156040384446  0.5518335574013022
  0.0000000420817869  0.4950843724664722  0.5518335522298301
  0.9999999689551089  0.5049155438453283  0.4481664516883015
  0.5000000016533050  0.2499999920652627  0.5089925584737145
  0.4999999877570787  0.7499999995066702  0.4910074631151176
```

**Phonon dispersion curve**

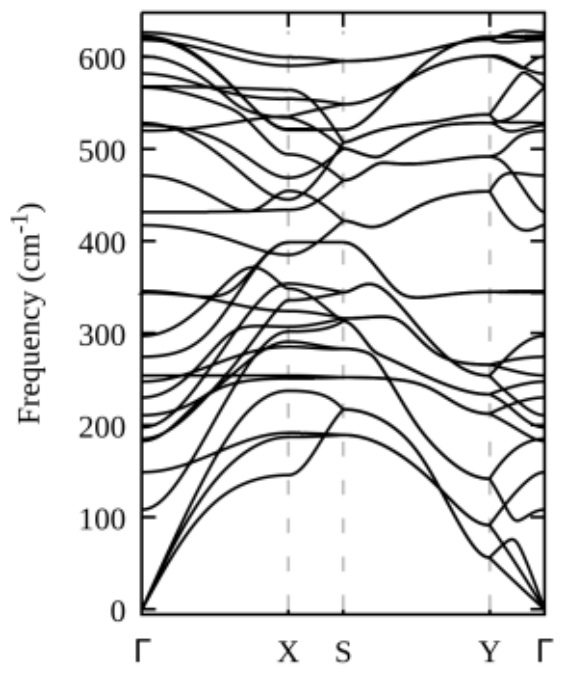

**Projected band structure and density of states**

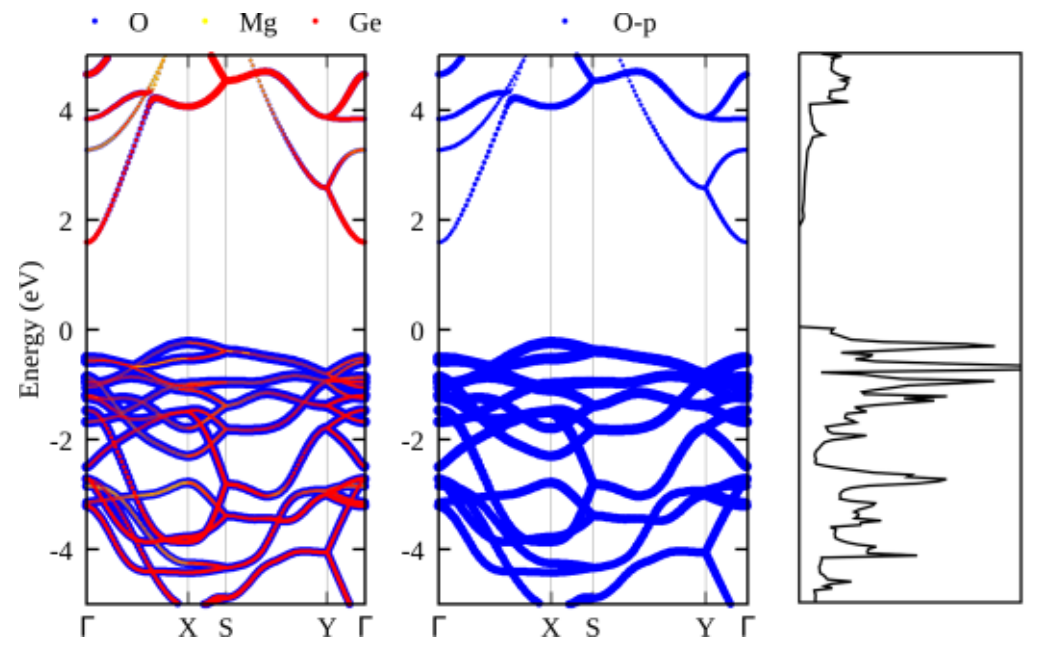

**Magnetic moment and spin polarization energy as a function of hole doping concentration**

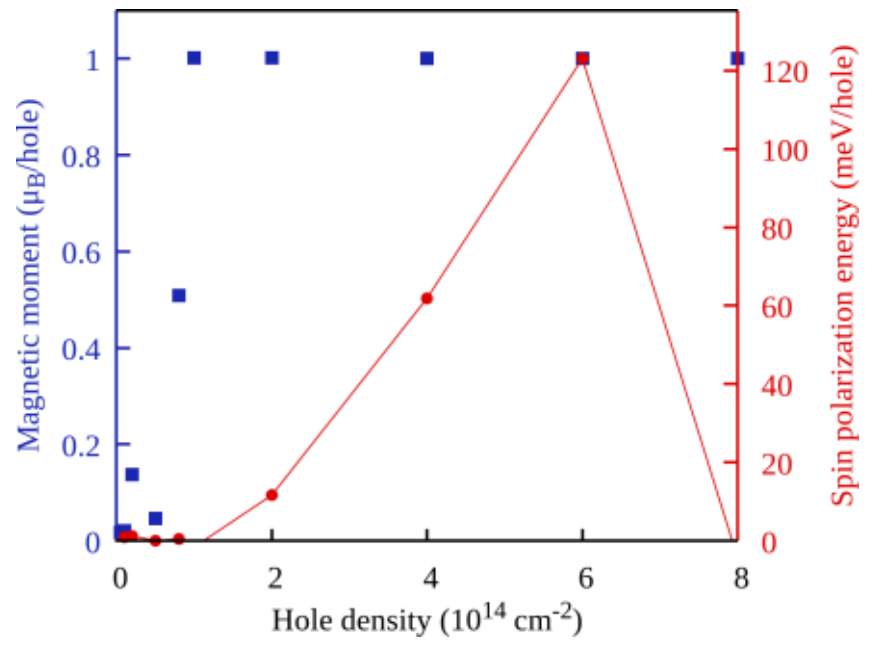

**Magnetic configurations and spin Hamiltonian**

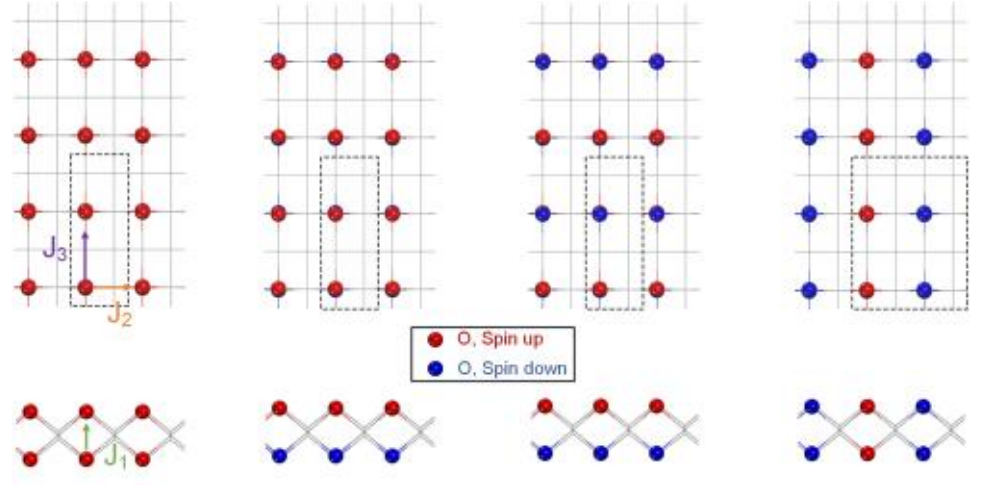

$$E_{FM} = E_0 - 2J_1S^2 - 4J_2S^2 - 4J_3S^2$$

$$E_{AFM1} = E_0 + 2J_1S^2 - 4J_2S^2 - 4J_3S^2$$

$$E_{AFM2} = E_0 + 2J_1S^2 - 4J_2S^2 + 4J_3S^2$$

$$E_{AFM3} = E_0 - 2J_1S^2 + 4J_2S^2 - 4J_3S^2$$

**Magnetic exchange coupling parameters**

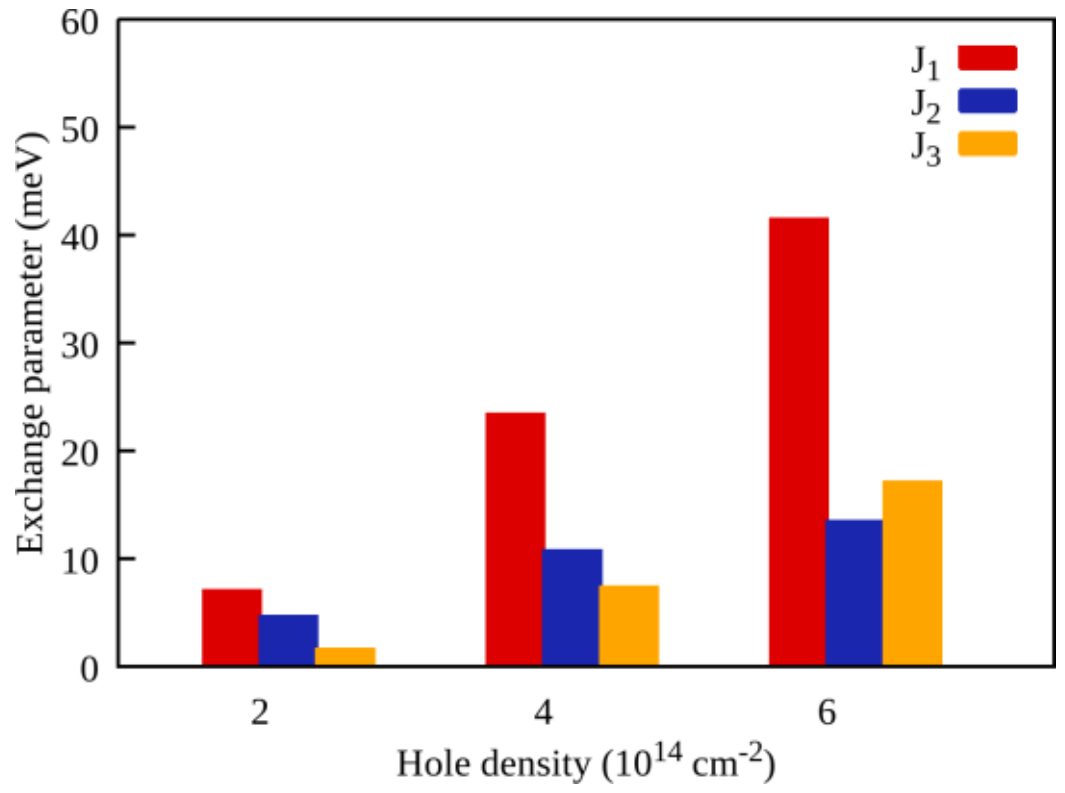

**Magnetic anisotropy energy (MAE, μeV) per magnetic atom**

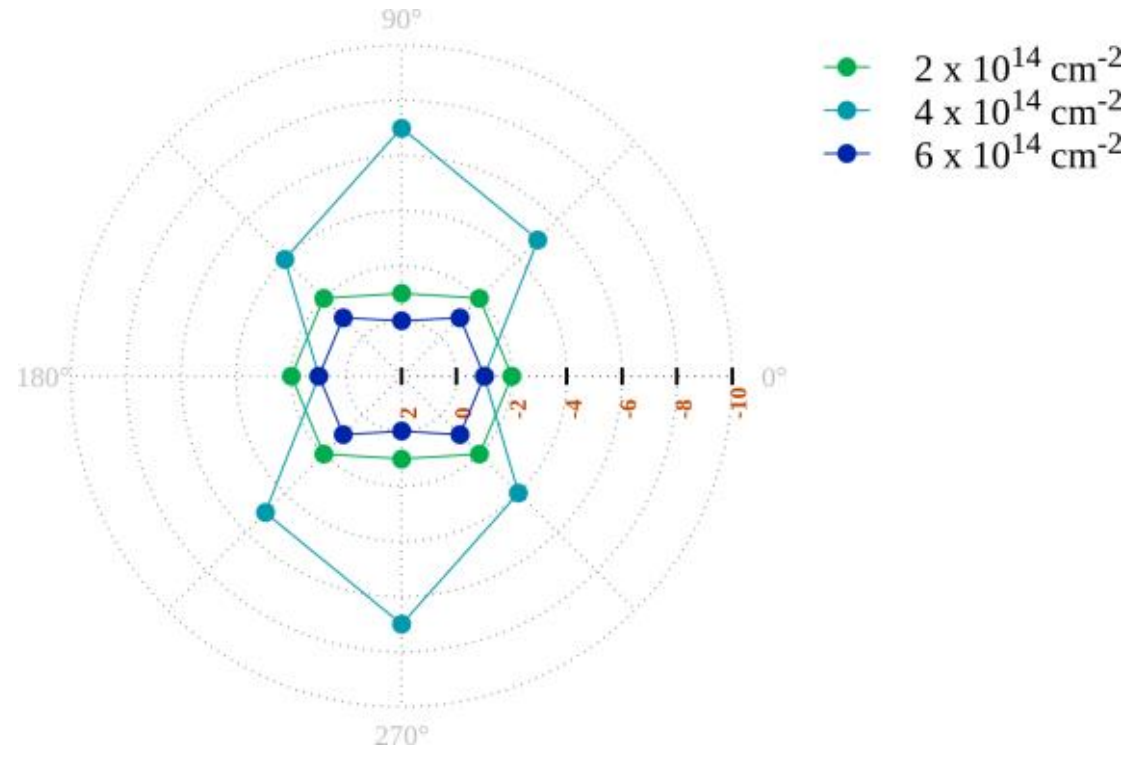

MAE = $E_{\parallel} - E_{\perp}$, a positive (negative) value of MAE indicates the off-plane (in-plane) easy axis.

**Monte Carlo simulations of the normalized magnetization of as a function of temperature**

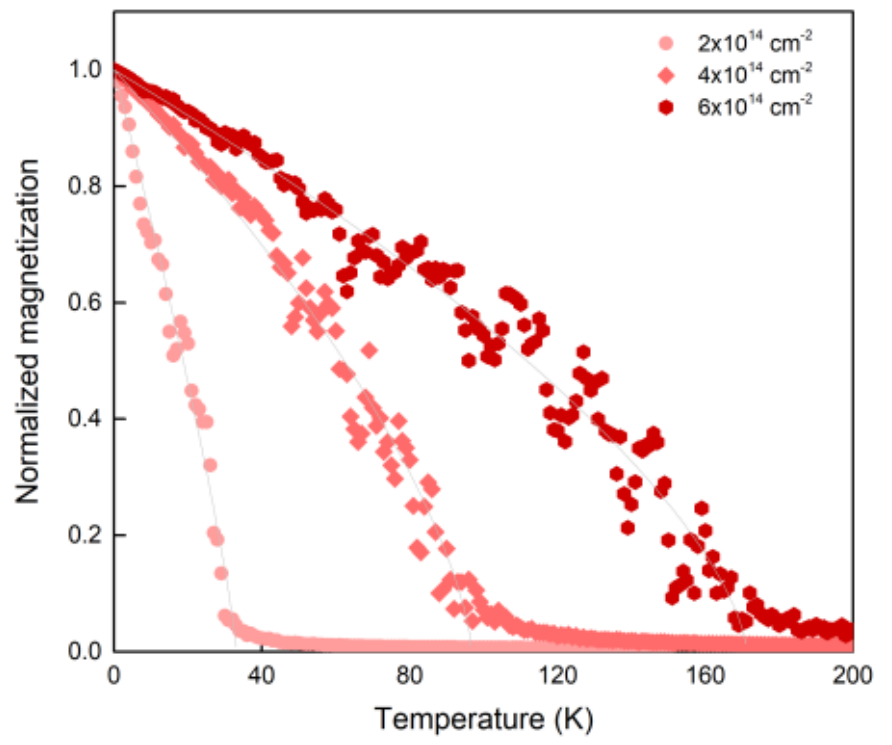

$T_c$: $1\times10^{14}$ cm$^{-2}$: - K  $2\times10^{14}$ cm$^{-2}$: 33 K

$4\times10^{14}$ cm$^{-2}$: 97 K  $6\times10^{14}$ cm$^{-2}$: 171 K

# 122. $TiPbO_3$

| MC2D-ID | C2DB | 2dmat-ID | USPEX | Space group | Band gap (eV) |
|---|---|---|---|---|---|
| - | | 2dm-5482 | - | P4mm | 2.65 |

**Convex hull** | **Atomic structure** | **Atomic coordinates** | **Phonon dispersion curve**

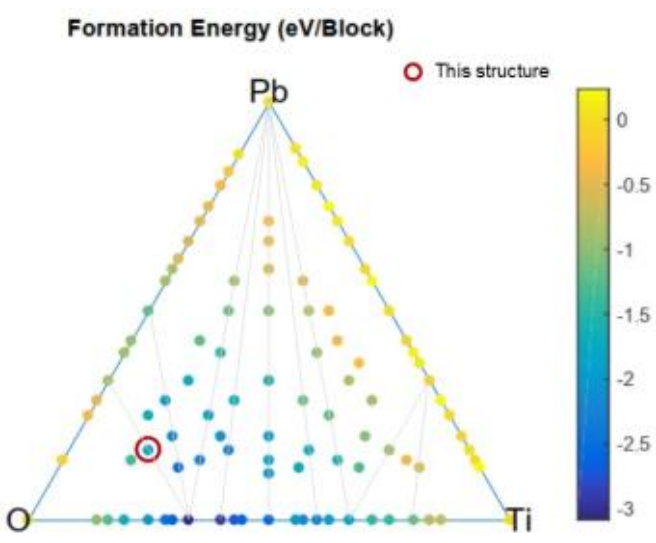

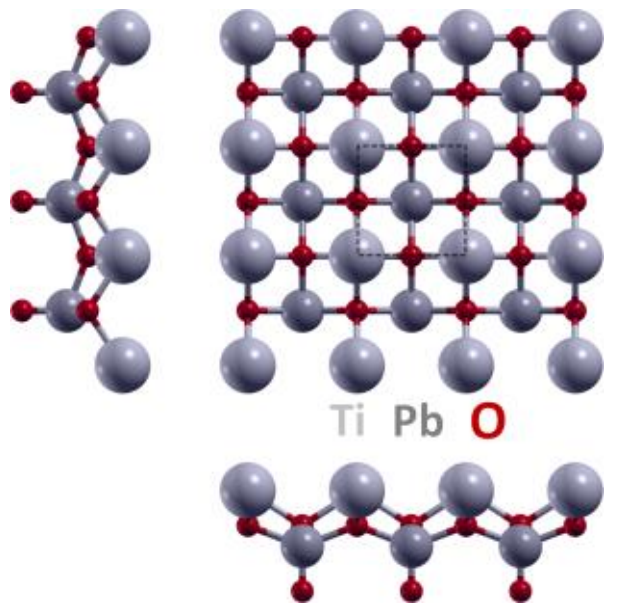

```
Ti1Pb1O3
  1.00000000000000
  3.9178046187640239   0.0000000000000000   0.0000000000000000
  0.0000000000000000   3.9178042947442169   0.0000000000000000
  0.0000000000000000   0.0000000000000000  23.8189570000000010
  Ti   Pb   O
  1    1    3
Direct
  0.4999999814381937  0.5000000060258714  0.4842658842307088
  0.0000000042042529  0.0000000083075875  0.5681751813756151
  0.5000000131192692  0.9999999961594170  0.5155089190967530
  0.0000000026095179  0.5000000029248710  0.5155089310102241
  0.4999999986287591  0.4999999865822531  0.4165420842866965
```

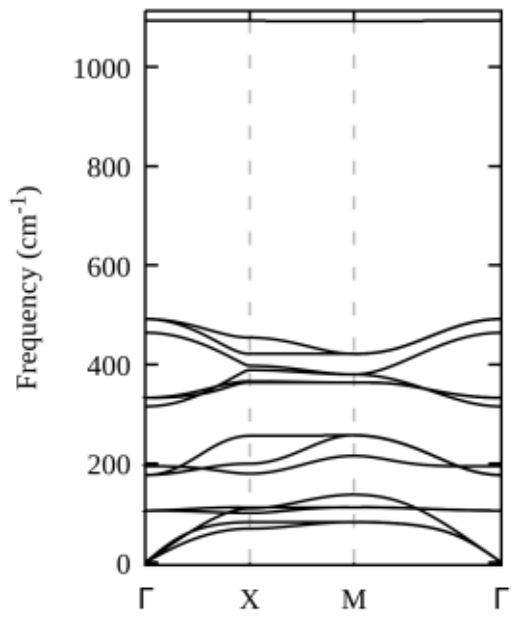

**Projected band structure and density of states** | **Magnetic moment and spin polarization energy as a function of hole doping concentration**

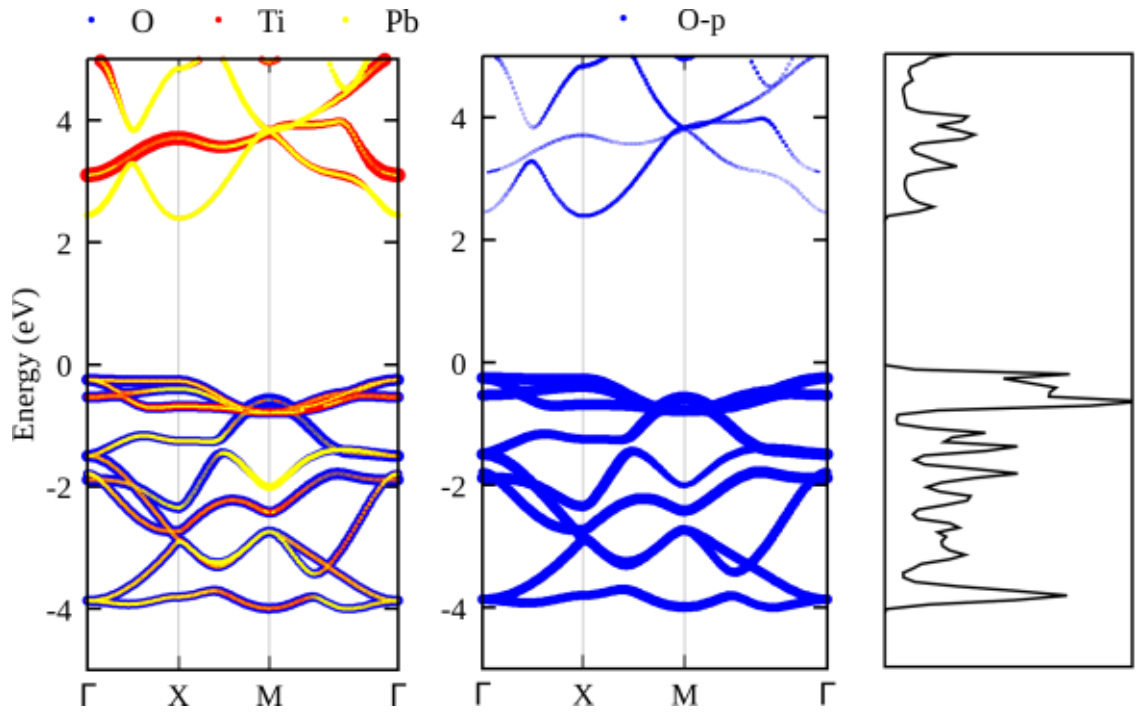

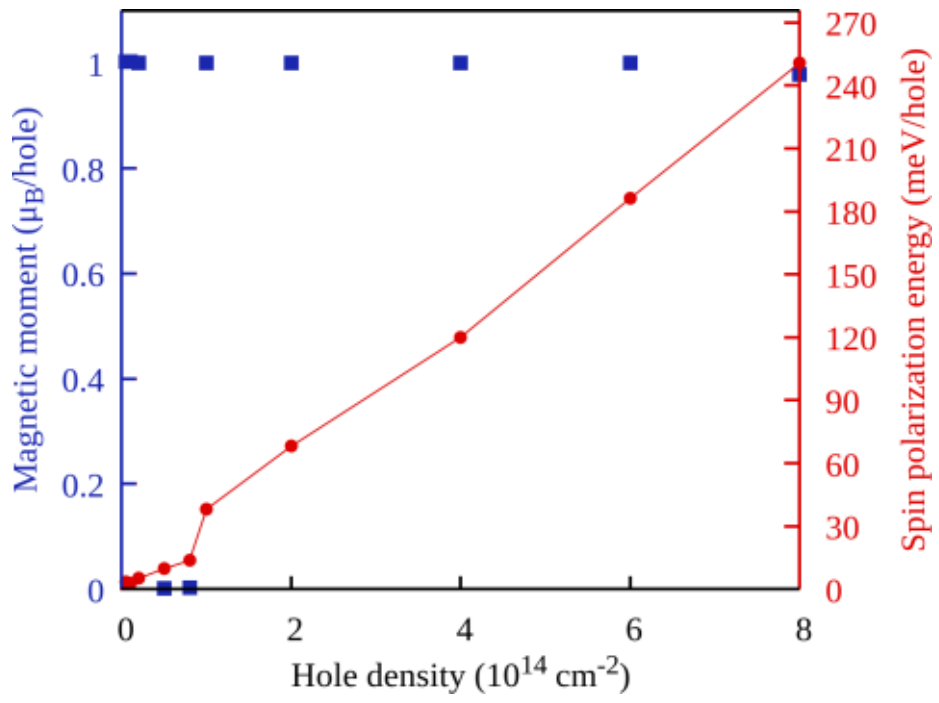

**Magnetic configurations and spin Hamiltonian** | **Magnetic exchange coupling parameters**

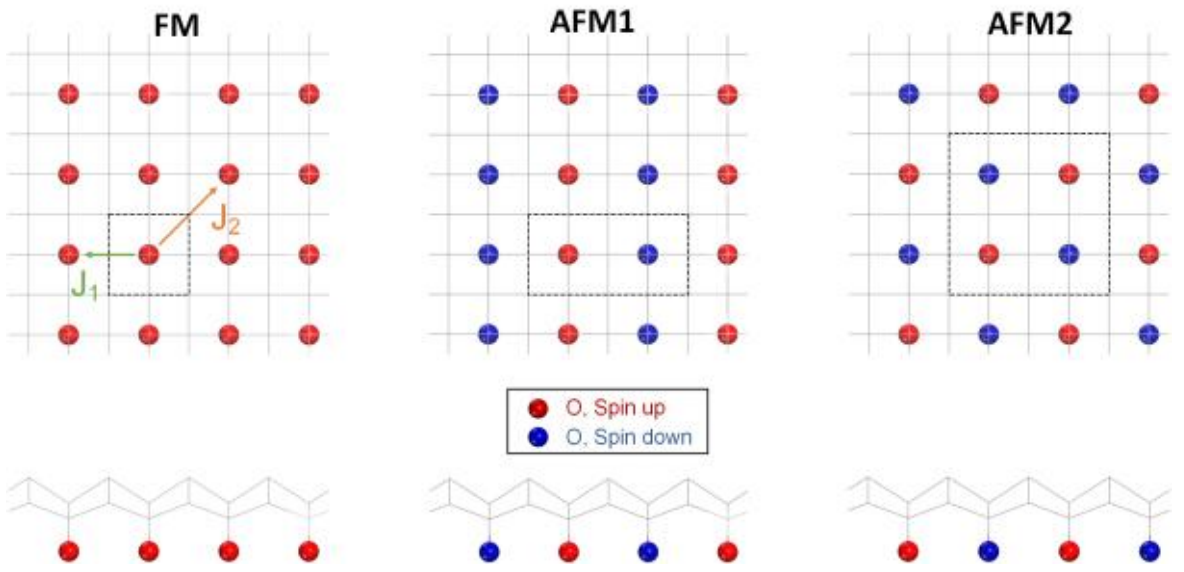

$$E_{FM} = E_0 - 2J_1S^2 - 2J_2S^2$$

$$E_{AFM1} = E_0 + 2J_2S^2$$

$$E_{AFM2} = E_0 + 2J_1S^2 - 2J_2S^2$$

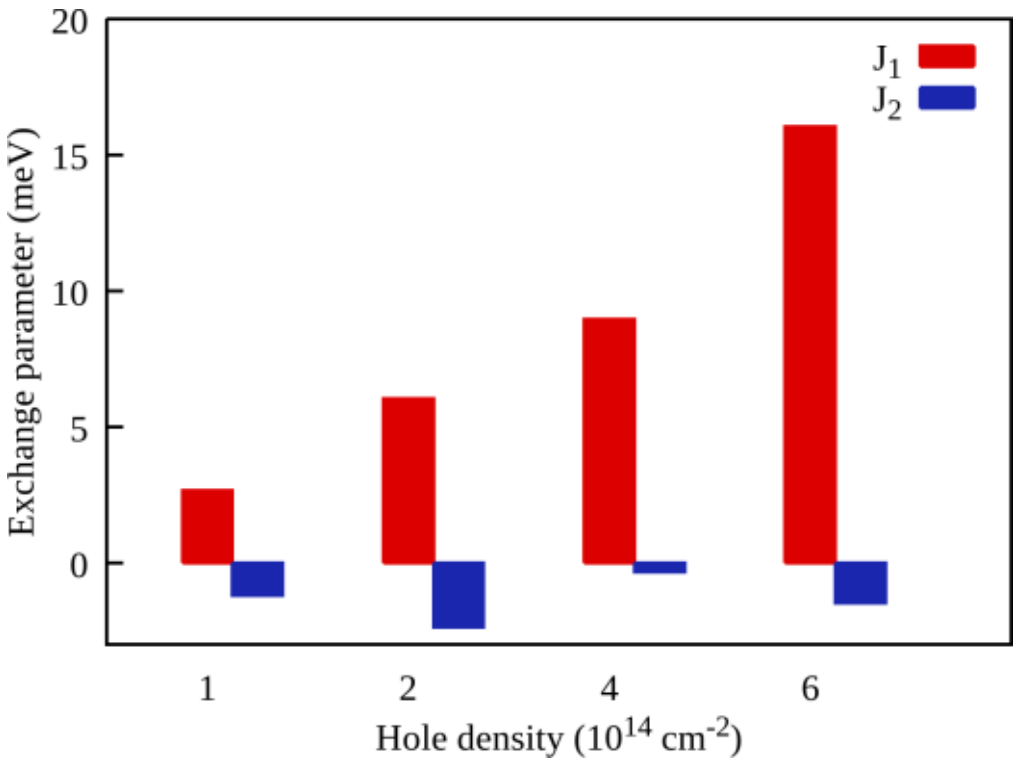

**Magnetic anisotropy energy (MAE, μeV) per magnetic atom** | **Monte Carlo simulations of the normalized magnetization of as a function of temperature**

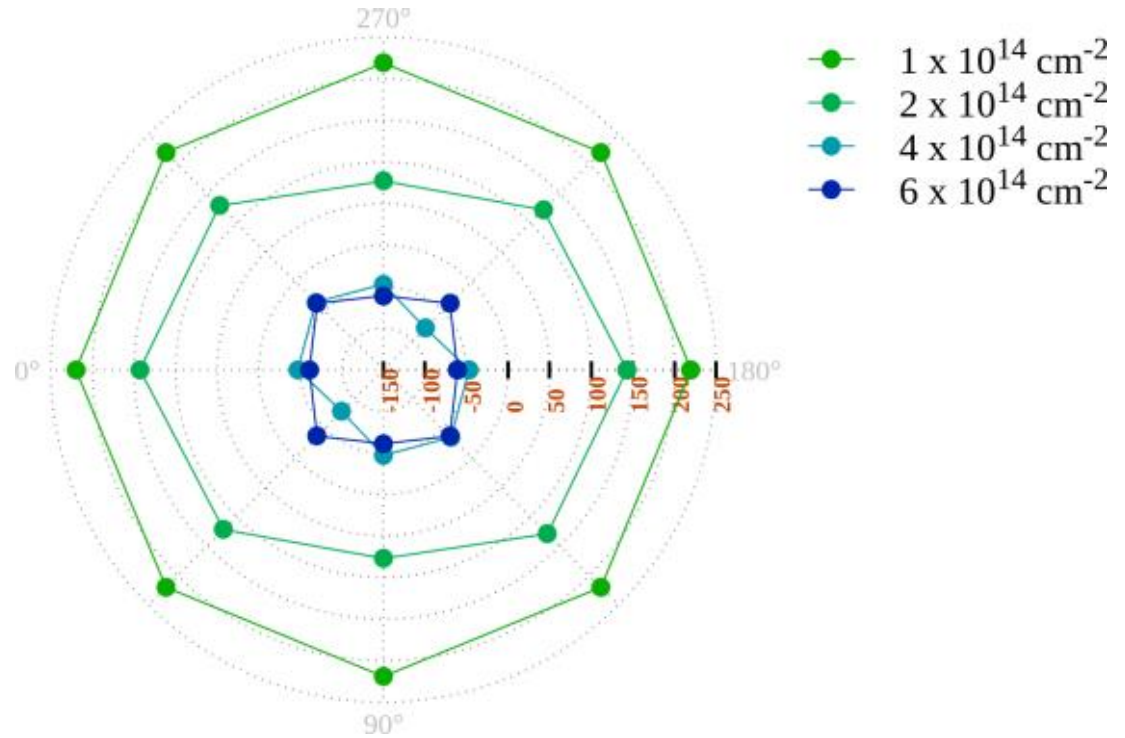

MAE = $E_{\parallel} - E_{\perp}$, a positive (negative) value of MAE indicates the off-plane (in-plane) easy axis.

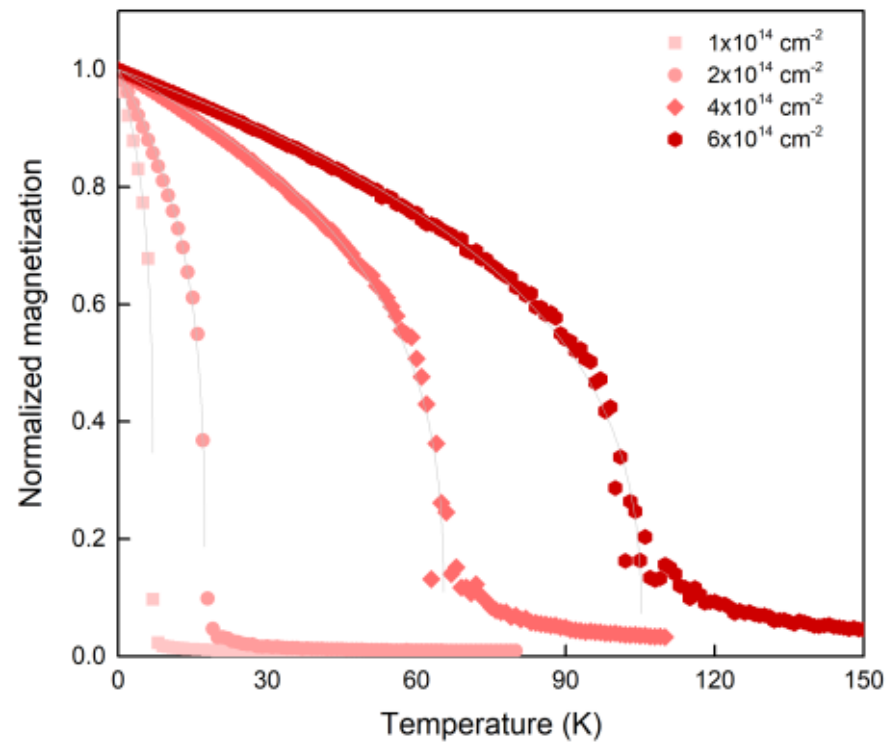

$T_c$: $1\times10^{14}$ cm$^{-2}$: 7 K  $2\times10^{14}$ cm$^{-2}$: 17 K

$4\times10^{14}$ cm$^{-2}$: 66 K  $6\times10^{14}$ cm$^{-2}$: 105 K

## Supplementary Reference